%% file: main.tex
\documentclass[12pt]{thesis}
\usepackage{epsfig,amssymb,amsmath,rotating,graphics}
\usepackage{indentfirst}
\usepackage{graphicx}
\usepackage{natbib}
\usepackage[colorlinks=true,linkcolor=black,citecolor=blue,urlcolor=blue]{hyperref}
\def\lsim{\lower.5ex\hbox{$\; \buildrel < \over \sim \;$}}
\def\gsim{\lower.5ex\hbox{$\; \buildrel > \over \sim \;$}}

\usepackage{makeidx}
\makeindex
\usepackage[explicit]{titlesec}
\usepackage{lipsum}
\usepackage{type1cm}
\usepackage{xcolor}
\usepackage{pdflscape}
\usepackage{marvosym}
\usepackage[T1]{fontenc}

\titleformat{\chapter}[display]
  {\normalfont\Large\rmfamily}
  {\sffamily\flushright\fontsize{60}{0}\textbf{\textcolor{blue!70}{{\Huge\chaptername}~\thechapter\vskip0pt\rule{\textwidth}{2pt}}}}{0pt}
  {\flushleft\fontsize{30}{0}{\Huge\textcolor{blue!80}{#1}}\vskip15pt}{\huge}
\titlespacing*{\chapter}
  {0pt}{-40pt}{0pt}

\begin{document}
\setcounter{section}{0}
\setcounter{figure}{0}
\setcounter{table}{0}
\vskip 3cm
\thispagestyle{empty}

\centerline{\Huge\bf Physics And Chemistry Of Star Forming}
\vskip 0.5cm
\centerline{\Huge\bf Region And Protoplanetary Disk}

\vskip 4cm
\begin{center}
{\large \bf Thesis submitted for the degree of\\
        {\Large Doctor of Philosophy} (Science)\\
            in Physics (Theoretical) \\}
\end{center}

\vskip 3.0cm

\centerline{\large by}
\vskip 0.5cm
\centerline{\Large\bf Mr. Milan Sil}

\vskip 5.0cm
\centerline{\large \bf Department of Physics}
\vskip 0.5cm

\centerline{\Large \bf University of Calcutta}

\vskip 0.5cm
\centerline{\large\bf 2021}

\newpage

\thispagestyle{empty}

\vspace*{8cm}

\begin{center}
{\Large \it Dedicated to my beloved family.}
\end{center}

\newpage
\pagestyle{newheadings}
\pagenumbering{roman}
\setcounter{page}{1}

\input{abstract.tex}
\input{ack.tex}

\input{publications.tex}

\newpage
\tableofcontents
\listoffigures
\listoftables

\newpage
\pagestyle{myheadings}
\pagenumbering{arabic}

\mainmatter
 \input{chap1.tex}
 \input{chap2.tex}
 \input{chap3.tex}
 \input{chap4.tex}

\input{chap5.tex}

\vfill\eject
 
\appendix
 \input{appA-chap3.tex}
 \input{glossary.tex}

\vfill\eject
\pagestyle{newheadings}

\backmatter
 \bibliographystyle{aasjournal}
 \scriptsize
 \bibliography{references}

\end{document}

%% file: abstract.tex
\chapter*{ABSTRACT}

My thesis work aims to study the inter-relation between various physical and chemical conditions in a wide range of astrophysical environments. Our studied regions range from the super-hot regions (i.e., nebular, photon-dominated, or photodissociation regions, diffuse area, through which the lights of the background stars can reach us) to the super-cold regions (i.e., dense molecular clouds, proto-planetary disks, etc. where interstellar dust particles absorb all background visible and ultraviolet lights).
The chemical complexity of the interstellar cloud gradually evolves due to the evolution in physical conditions.
The dense molecular clouds are the birth sites of star-formation, where a wide variety of complex organic molecules are observed.
Dust particles play an essential role in the formation of these complex organic molecules. During the warm-up stage of a star-forming region, the molecules formed during the cold phase start to return to the gas phase by various thermal and non-thermal evaporation processes.
These complex molecules again freeze out to the outer part of the proto-planetary disk during the further evolved stage according to their condensation temperature and form the so-called snow-lines. The binding energies of these molecules with the prevailing dust particles play a crucial role in determining the structural information of this disk. Thus the binding energy of the molecules is vital to understand several critical aspects of the star and planet formation processes. In this thesis, I will discuss the chemical
complexity obtained in a wide range of the star-forming region and whether this chemical complexity lead to biomolecules in space. \\

Chapter \ref{chap:intro} introduces the essential background of the research work (astrochemistry-related study) presented in this dissertation. The recent developments of this subject are also highlighted in this Chapter. \\

Features of the interstellar ice are investigated in Chapter \ref{chap:BE_ice}. Systematic
quantum chemical calculations have been performed to compute physisorption energies of several interstellar adsorbed species by considering different types of substrates mimicking interstellar dusts or grains or icy mantle layers. Assuming the binding energy values as appropriate for $\rm{H_2}$ substrate, we check the encounter desorption effect on the enrichment of surface species.
The presence of impurities in water ice affects the spectroscopic features of water itself. For this reason, a series of laboratory experiments and an extensive computational investigation are carried out to evaluate the effects of different amounts of representative impurities on the band strengths and absorption band profiles of interstellar ice. \\

Chapter \ref{chap:crab} deals with radiation-dominated environments such as the Crab nebula and diffuse cloud regions using a spectral synthesis code \textsc{Cloudy}. This Chapter attempts to model the chemistry of hydride and hydroxyl cations of the three most common noble gas atoms: argon, neon, and helium. Their various isotopologs are also considered to check the chemical evolution in space where the radiation is extreme. \\

Chapter \ref{chap:prebiotic} discusses the chemical treatment of some prebiotic molecules such as the aldimines, amines, three nitrogen-bearing species containing peptide-like bonds, and some phosphorous-bearing species in various star-forming regions of ISM, mainly where the radiation is highly attenuated and proposes the possibility to observe some of them in space. \\

Finally, the concluding remarks of the works presented in this thesis are made in Chapter \ref{chap:conclusions} along with my possible future research plan.

%% file: ack.tex
\chapter*{ACKNOWLEDGMENTS}

{\it Ph.D. is not about getting a certificate. It is about maturing to undertake Life and Science in unknown territories. It is becoming an adult scientist, still keeping the curiosity of a child within you. It is about keeping your vision and goals not fogged by day-to-day stupidies of politics in life. It is about dreaming beyond and working to leave knowledge behind for future generations. And finally, you are an adult in Knowledge Pursuit! \\

Sometimes the exploration is challenging in isolation. Voyage is beautiful when it is companied by wonderful people, and I am fortunate enough to have a bunch of such heads in this beautiful journey. There are numerous people I would like to personally acknowledge and thank for their assistance, encouragement, and support throughout the journey. \\

First of all, I would like to thank the whole research group ``Astrochemistry / Astrobiology'' at Indian Centre for Space Physics (ICSP), Kolkata, for an amicable, informal emancipated working atmosphere that
formed day-by-day an enjoyable period of residence. Minor issues like having stimulating lunch discussions or going out into a restaurant, friendly people could be found herein to balance scientific work. Moreover, I am thankful
for enabling travels to conferences and meetings. This level of expanding mindsets is priceless and crucial, both on a scientific and a personal level.
However, a couple of people deserve a mentioning by name to express my
deep gratitude. \\

I owe my sincere gratitude and appreciation to my Ph.D. supervisor, Dr. Ankan Das,
for all the support and encouragement he gave me throughout my Ph.D. work. To me,
it was an ideal degree of balancing inspiration, guidance, support, and ``letting me
do my things''. I am forever indebted for his appearances as an elder brother and advice like a guardian. He has always been there to guide me in the proper direction, whether academic or personal. This journey into the world of Astrochemistry was a multidisciplinary, collaborative science product, to which he introduced me so well! His belief humbles me in allowing me to be a part of his group, and I have tried my best to fulfill the expectations. Interactions have felt barrierless, more like a friendship than a strict professor-student relationship. This kind of freedom is the spore of scientific success! Thanks, Ankan da, for this
hardly describable atmosphere. I am looking forward to the future! \\

I am very grateful to my Ph.D. joint-supervisor, Prof. Sandip K. Chakrabarti, for his essential scientific and grammatical suggestions, beautiful notions, and cosmic concepts. In addition, creative ideas regarding visualization, graphs, etc., helped me a lot during various studies.
Without his guidance and constant feedback, the outcomes would not have been successful. \\

This research was performed in collaboration with other scientists all over the world. My thanks go to Professor Paola Caselli, Dr. Sergio Ioppolo, Dr. Jean-Christophe Loison, Professor Cristina Puzzarini, Professor Vicenzo Barone, Dr. Takashi Shimonishi, Dr. Naoki Nakatani, Dr. Kenji Furuya, Dr. Bhalamurugan Sivaraman, and Dr. Amit Pathak. They were always so helpful and provided me with their support and collaboration throughout my dissertation. \\

I would also like to thank my departmental group mates Dipen da, Prasanta da, Bratati, Suman Mondal da, Rana for their endless help, fruitful discussion, and contribution to this research as teamwork.
A special thanks indeed go to Dr. Gorai, from whom I learned so many things in Astrochemistry. Your presence in the lab was
something difficult to forget. We have shared many good times. I am highly grateful for your availability whenever I had problems or questions. I am happy to collaborate with my friend Satyam from Banaras Hindu University, Varanasi also contributed to my work. I would also like to thank Emmanuel E. Etim for his fruitful contribution to this research. \\

I am delighted to work at my institute, ICSP, where the atmosphere is precisely family-like, and thanks for that. All the academic and non-academic staff and faculty members, colleagues, and friends are like family members. I learn many things from Dr. Dipak Debnath, Dr. Ritabrata Sarkar, and Dr. Sudipta Sasmal, and thanks for all their support and cooperation. My knowledge in Astrophysics and Astronomy was initiated from the class taken by Ankan da, Dipak da, Rito da during my post-graduation days at Narendrapur. I heartily thank and express my love to all the past and present members of ICSP - Suman Ray da, Sourav da, Santanu da, Partha da, Sujay da, Arka da, Dusmanta da, Suman Chakraborty da, Tamal da, Shreeram da, Aslam da, Dipen da, Argha da, Prasanta da, Debjit da, Soujan da, Suman Mondal da, Bratati, Rana, Kaushik, Sujoy, Riya, Subrata, Swati, Abhijit, Abhrajit, Binayak, Rupnath, Sagar, Ashim, Shyam, and Pabitra. I feel melancholy thinking that my journey with you all is going to end here. I will miss our tea group at ICSP. I want to thank Rajkumar da, Ram da, Pavel da, Uttam da, Debashis da, Hriday da, Susanta da, Arnab da for every kind of help regarding administrative, official, and technical issues. On many occasions, we gathered, shared, and celebrated together. \\

I am lucky to get a chance to visit NASA's JPL and Caltech at Pasadena, California, and SOFIA Observatory, based at NASA's Armstrong Flight Research Center at Palmdale Regional Airport, California, during the 42nd COSPAR Scientific Assembly, July 2018, USA.
While staying abroad with the ICSP gang at Saga Motor Hotel, 
we enjoyed ourselves immensely, especially the breakfast and swimming pool.
It was my first visit abroad and a very precious moment of my life. \\

I want to thank all my friends during my school, graduate, and post-graduate studies, especially Apurba, Subhajit (Gopi), Saptarshi, Joy, Arnidam, Souvik, Debarghya, Abinash, Biswajit, Chandralina, Tania, Tamal, Monoj, Atanu, Satyajit, Tanumoy, Sumit, Sumana. Animesh, Snehashis, Surajit, Sujoy, Subhankar (Bappa), Hemanta, Soumen are my village locality friends since childhood. We played Cricket, Football, Carrom, Cards, etc., together. A special thanks to my Ph.D. time room/flat-mates, Ram da, Debjit da, Kaushik, and Sujoy.
We have shared many memorable moments like cooking, bike riding, restaurant hoping, outing, gossip, fun, birthday party, music, etc., to cherish forever. \\
     
I am fortunate to have many good teachers throughout my academic journey, starting from primary school to post-graduation studies. So my warm regards go to Koushik sir during my primary school days at Ichapur Prathamic Vidyalaya, Kaikala, Hooghly.
We were the first batch in class three when he joined our school. I can cherish the memory when I was ill and could not go to school; he came to see me in our house.
Haripal Gurudayal Institution is a place very close to me, and
still today, I enjoy remembering every possible moment that I spent there. The love that I
got from all the teachers there was something special. It is tough to
name all of them, but I must call a few. Rabin babu (Dr. Rabindranath Chattopadhyay), Swadesh babu, Uttam babu, Ujwal babu, Sougata babu, Pradip babu, Sourav babu, Biswajit Bag babu, Sanjit Kayal babu, Ritu di, Kakon di, Susanta sir (private teacher) are exceptional teachers during my high school days. My love for Chemistry starts building solely due to Dhruba sir and my Physics background today is for Rabin babu and Debdas sir. Swadesh babu is always been my favorite Mathematics teacher.
I delivered seminar lectures two times in my high school `Satabarshiki Hall' only for Rabin babu. He always inspires me for any scientific purpose.
AKM sir, TD sir, GSM sir, PS sir, SD sir, AB sir, DPG sir, KP sir, RB sir during my graduation days at Serampore College made my knowledge in Physics day-by-days.
Especially the theoretical and practical/experimental knowledge of TD sir inspires me a lot.
DJ sir (Prof. Debnarayan Jana) always encouraged and motivated me during the post-graduation days at Ramakrishna Mission Residential College, Narendrapur. All they are the builders of my academic root. \\

I am thankful to all my family members and relatives, especially my parents, for their unwavering support, no matter which path in life I chose to follow. I must express my sincere love and gratitude from the deepest of my heart to my maa, Mrs. Binapani Sil, and baba (bapi), Mr. Mridul Sil, for their endless unconditional love, prayers, care, sacrifices, and support. My maa taught me even during my high school days when I have no private teacher. Thank you for supporting me during all ups and downs of my career. Bapi worked overnight during our hard times. My didi, Mrs. Tithi Sil, and my brother-in-law Mr. Tapas Das helped our family and me immensely. Without their assistance, I could not be able to do my Ph.D. today. My dearest ``vagna'' \textsc{Shivam} was born (on 11th May 2016) during my Ph.D. days. He always keeps my family full of happiness and joy. I have shared so many beautiful, memorable, and adorable moments with \textsc{Shivam}. I will miss my loving uncle Late Sanat Sil whose sudden disappearance from the Earth (due to the deadly disease DLBCL cancer) during my Ph.D. shocked me severely. I always pray for your soul to rest in peace. Mentioning everyone in my family is difficult, but I thank you all for being there. \\

Finally, I would especially like to thank my adored wife, \textsc{Sudipta}, for all of her sacrifices, devotion, mental support, and love since the end of my graduate studies. Thank you also for supporting me when having had hard times in my life. After seven years of relationship, we have got married on 2nd February 2021. \\

I want to acknowledge the Department of Science and Technology (DST), the Government of India, for providing financial assistance through the DST-INSPIRE (Innovation in Science Pursuit for Inspired Research) Fellowship [IF160109] scheme to carry out my research during my Ph.D. I also thank DST-SERB (Science and Engineering Research Board) for the grant that I received under the International Travel Support (ITS) scheme to attend the overseas conference, 42nd COSPAR Scientific Assembly, July 2018, USA.
This research was possible in part due to a grant-in-aid from the Higher Education
Department of the Government of West Bengal. Acknowledges are also due to Indian Centre for Space Physics and the University of Calcutta for allowing me to do a Ph.D. I would also like to thank all the reviewers whose extensive comments helped improve the quality of the papers presented in this dissertation. \\

A special appreciation indeed goes to the complementary contributions of the internet, \textsc{Google}, and my desktop and laptop, which have accelerated my work. Last but by no means least, I would like to thank all those people I inadvertently forgot to mention, but whose help nevertheless was very much appreciated. Finally, it is worth noting the global pandemic situation due to Coronavirus COVID-19 affected the work severely. In the end, I wish to say that this roller-coaster journey could not be so decorative and gorgeous without all these supports.\\
\\
Cheers! \\

\begin{flushright}
Milan Sil \\
August 2021, Kolkata.
\end{flushright}
}

%% file: publications.tex
\chapter*{Scientific contributions}

\section*{Publications}

\subsection*{\bf List of Publications in Peer Reviewed Journals}
\begin{enumerate}

\item  \href{https://doi.org/10.1140/epjd/e2017-70610-4}{\bf Adsorption energies of H and H$_2$: A Quantum Chemical Study}, \underline{\bf Milan Sil}, Prasanta Gorai, Ankan Das, \& Sandip K. Chakrabarti, 
2017, \emph{The European Physical Journal D}, 71, 45. \textbf{(Journal Impact Factor 2020: 1.425)} \\

\item \href{https://doi.org/10.3847/1538-4357/aa984d}{\bf Chemical Modeling for Predicting the Abundances of Certain Aldimines and Amines in Hot Cores}, \underline{\bf Milan Sil}, Prasanta Gorai, Ankan Das, Bratati Bhat, \& Sandip, K, Chakrabarti, 2018, \emph{The Astrophysical Journal}, 853, 2. \textbf{(Journal Impact Factor 2020: 5.874)} \\

\item \href{https://doi.org/10.3847/1538-4365/aac886}{\bf An Approach to Estimate the Binding Energy of Interstellar Species}, Ankan Das, \underline{\bf Milan Sil}, Prasanta Gorai, Sandip K. Chakrabarti, \& J. C. Loison, 2018, \emph{The Astrophysical Journal Supplement Series}, 237, 9. \textbf{(Journal Impact Factor 2020: 8.136)} \\

\item \href{https://doi.org/10.3847/1538-4357/ab8871}{\bf Identification of Pre-biotic Molecules Containing Peptide-like Bond in a Hot Molecular Core, G10.47+0.03}, Prasanta Gorai, Bratati Bhat, \underline{\bf Milan Sil}, Suman K. Mondal, Rana Ghosh, Sandip K. Chakrabarti, \& Ankan Das, 2020, \emph{The Astrophysical Journal}, 895, 86. \textbf{(Journal Impact Factor 2020: 5.874)} \\

\item \href{https://doi.org/10.1021/acsearthspacechem.0c00098}{\bf Systematic Study on the Absorption Features of Interstellar Ices in the Presence of Impurities}, Prasanta Gorai, \underline{\bf Milan Sil}, Ankan Das, Bhalamurugan Sivaraman, Sandip K. Chakrabarti, Sergio Ioppolo, Cristina Puzzarini, Zuzana Kanuchova, Anita Dawes, Marco Mendolicchio, Giordano Mancini, Vicenzo Barone, Naoki Nakatani, Takashi Shimonishi, \& Nigel Mason, 2020, \emph{ACS Earth and Space Chemistry}, 4, 920. \textbf{(Journal Impact Factor 2020: 3.475)} \\

\item \href{https://doi.org/10.3847/1538-4357/abb5fe}{\bf Exploring the Possibility of Identifying Hydride and Hydroxyl Cations of Noble Gas Species in the Crab Nebula Filament}, Ankan Das, \underline{\bf Milan Sil}, Bratati Bhat, Prasanta Gorai, Sandip K. Chakrabarti, \& Paola Caselli, 2020, \emph{The Astrophysical Journal}, 902, 131. \textbf{(Journal Impact Factor 2020: 5.874)} \\

\item \href{https://doi.org/10.3389/fspas.2021.671622}{\bf Effect of binding energies on the encounter desorption}, Ankan Das, \underline{\bf Milan Sil}, Rana Ghosh, Prasanta Gorai, \& Sandip K. Chakrabarti, 2021, \emph{Frontiers in Astronomy and Space Sciences}, 8, 78. \\

\item \href{https://doi.org/10.3847/1538-3881/ac09f9}{\bf Chemical complexity of phosphorous bearing species in various regions of the Interstellar medium}, \underline{\bf Milan Sil}, Satyam Srivastav, Bratati Bhat, Suman Kumar Mondal, Prasanta Gorai, Rana Ghosh, Takashi Shimonishi, Sandip K. Chakrabarti, Bhalamurugan Sivaraman, Amit Pathak, Naoki Nakatani, Kenji Furuya, Ankan Das, 2021, \emph{The Astronomical Journal}, 162, 119. \textbf{(Journal Impact Factor 2020: 6.263)} \\

\item \href{https://ui.adsabs.harvard.edu/abs/2021arXiv210806240M/abstract}{\bf Is there any linkage between interstellar aldehyde and alcohol?} Suman Kumar Mondal, Prasanta Gorai, \underline{\bf Milan Sil}, Rana Ghosh, Ankan Das, Emmanuel E. Etim, Sandip K. Chakrabarti, Takashi Shimonishi, Naoki Nakatani, Kenji Furuya, Jonathan C. Tan, Ankan Das, 2021, \emph{The Astrophysical Journal}. \textbf{(Journal Impact Factor 2020: 5.874)}

\end{enumerate}

\subsection*{\bf Publication in Proceedings}
\begin{enumerate}
 \item \href{https://link.springer.com/chapter/10.1007/978-3-319-94607-8_38}{\bf Binding Energy and Isomerism: Two Important Aspects of Astrochemistry}, 
\underline{\bf Milan Sil}, 2018, Exploring the Universe: From Near Space to Extra-Galactic, 491-501, Astrophysics and Space Science Proceedings, vol 53. Springer, Cham., Online ISBN:978-3-319-94607-8.
\end{enumerate}

\clearpage

\section*{Oral presentations}

\begin{enumerate}

\item
Astrochemistry in the THz domain, October 2017,
Chennai, India

Title: ``Systematic study on the presence of impurities on interstellar ices''

\item
42nd COSPAR Scientific Assembly, July 2018, USA

 Title: ``\href{https://ui.adsabs.harvard.edu/abs/2018cosp...42E3124S/abstract}{Binding energy a key to defining interstellar volatile species}''
 
 Title: ``\href{https://ui.adsabs.harvard.edu/abs/2018cosp...42E3125S/abstract}{A Systematic Study of Pre-biotic Aldimines and Amines in Hot Cores}''
 
 Title: ``\href{https://ui.adsabs.harvard.edu/abs/2018cosp...42E3126S/abstract}{A Theoretical Prediction of Interstellar Bio-Molecule abundances}''

\item
\href{https://www.bose.res.in/Conferences/EXPUNIV2018/}{Exploring the Universe: Near Earth Space Science to
Extra-Galactic Astronomy}, November 2018, Kolkata, India

 Title: ``A New Set of Binding Energies for Astrochemical Modeling''
 
\item
43rd COSPAR Scientific Assembly (COSPAR-2021-Hybrid), 28 January - 4 February 2021, Sydney Australia.

 Title: ``\href{https://ui.adsabs.harvard.edu/abs/2021cosp...43E1920S/abstract}{Fate of identifying noble gas related species in the Crab nebula environment}''

 \end{enumerate}

\section*{Poster presentations}

\begin{enumerate}
\item
\href{https://sites.google.com/site/irastrondust2019/home}{International Conference on Infrared Astronomy and Astrophysical Dust}, October 2019, IUCAA Pune, India

 Title: ``Complex Organic Molecules in the Star Forming Region''
 
\item
43rd COSPAR Scientific Assembly (COSPAR-2021-Hybrid), 28 January - 4 February 2021, Sydney Australia.

 Title: ``\href{https://ui.adsabs.harvard.edu/abs/2021cosp...43E1996S/abstract}{Estimating realistic values of binding energy of species for astrochemical modeling}''

 \end{enumerate}

%% file: chap1.tex
\chapter{Introduction} \label{chap:intro}

Astrochemistry is a flourishing field of science intending to understand the
evolution of the universe toward molecular complexity.
It brings us closer to understanding our origins. Molecular astrophysics are often used as synonyms for astrochemistry to define an interdisciplinary field involving chemistry and astronomy, astrophysics, and a flavor of biology and geology. It is long anticipated that the seeds of life on the Earth might have originated from space through the infall of asteroids, comets, or meteorites \citep{chyb90,chyb92}.
We are far from confident if one or both exogenous delivery and
endogenous synthesis is correct.
Space is not empty. It is full of various simple and complex molecular species. Therefore, our observable universe may be called a molecular universe \citep{tiel13}. We can diagnose different aspects of this universe by observing various types of molecules. These molecules can be utilized as tracers to determine the physical and chemical conditions of different evolutionary stages of star formation.
The chemical model of a molecular cloud serves as a critical link in understanding the complexity of the star-forming region \citep{herb14}.
The ``interstellar medium'' (ISM) is a highly tenuous medium between the stars in a galaxy.
It is filled with ordinary matter, relativistic charged particles known as cosmic rays, and magnetic fields.
The ISM contributes to transport the elements synthesized in the interior of stars or outer layers of stars, in the form of atoms and molecules or of dust grains, to regions that are distant from the stellar birthplace and distribute them more uniformly through Supernovae explosions.
The ISM is enriched by diverse chemical processes that give rise to complex organic molecules (COMs). It stands for interstellar molecules with a different connotation, e.g., organic chemistry or biochemistry.
A carbon- (hydrogen, oxygen, and nitrogen are optional) bearing molecule containing six or more atoms is known as a COM.
Methanol (CH$_3$OH) is the prototypically most straightforward COM \citep{herb09}.
These COMs, or at least some segments of these molecules, are produced on dusts.
The hydrogenation in the cold stage and radical chemistry in the warm-up stage plays a crucial role in shaping this.
The icy covered grain mantles are further exposed to interstellar radiation field (ISRF)
and other dissociative radiation sources to form more extensive and more complex molecules.
ISRF has a substantial far-ultraviolet (far-UV) or vacuum-ultraviolet (VUV) component that readily dissociates most molecules.
Both far-UV and VUV emission are categorized as UV emission with shorter wavelengths than $200$ nm.
The resulting COMs take part in the star and planet formation process
and may eventually transfer the seeds of life on nascent planets. In addition, the study of prebiotic molecules is always fascinating as they are involved in forming amino acids, proteins, nucleobases,
and the basic building blocks of life \citep{chak00a,chak00b,chak15,maju12,maju13,maju15,garr13}.
However, the abundance of these complex prebiotic molecules is often below the observing limit of the present astronomical observation. Thus it is customary to observe their precursors to tentatively constrain the parameters that form the prebiotic species in space \citep{sil18}.

\begin{figure}[htbp]
\centering
\includegraphics[width=\textwidth]{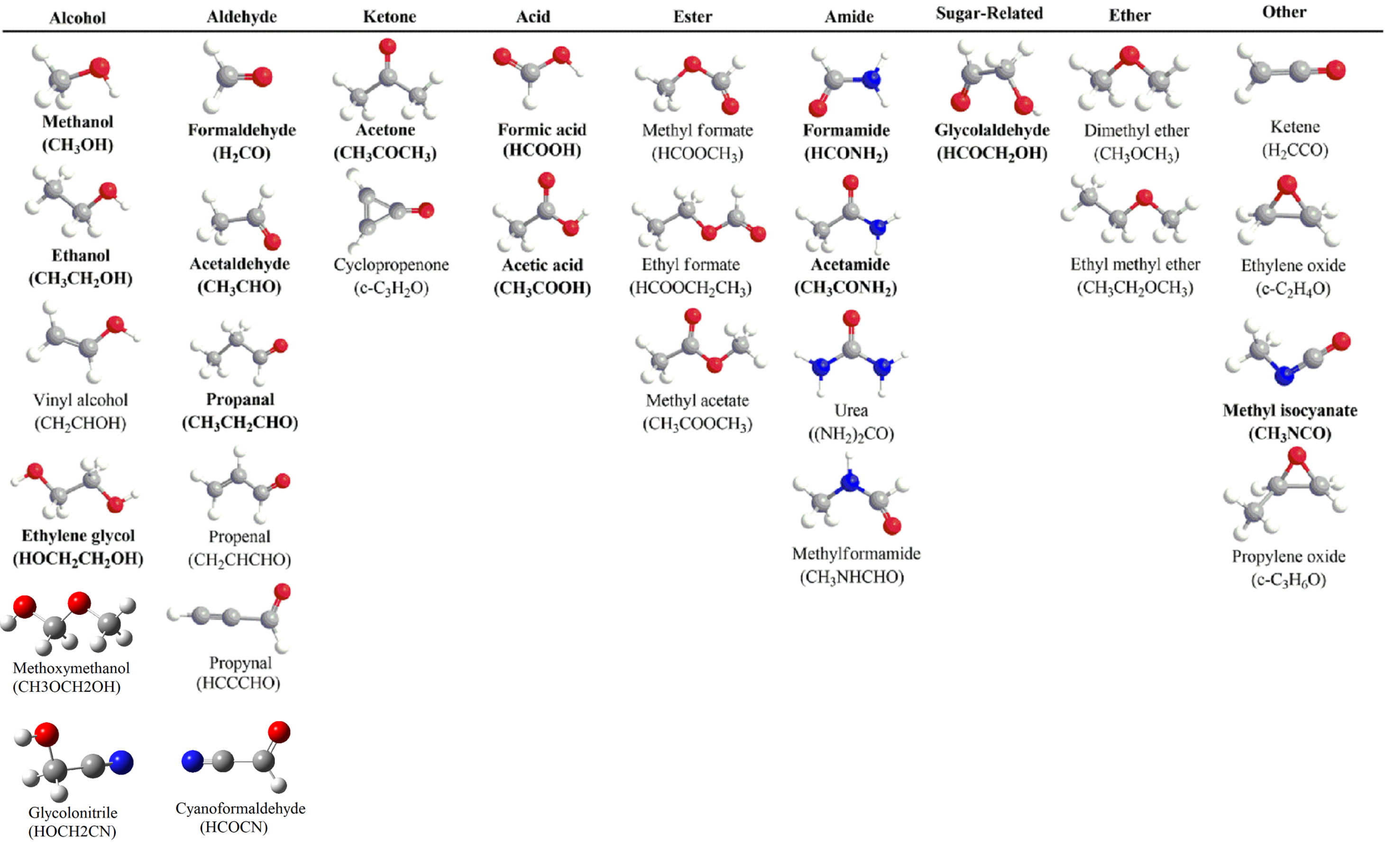}
\caption{Key classes of COMs detected in the ISM, which are primarily formed on ice-coated interstellar grains.}
\label{fig:COMs}
\end{figure}

Merely a hundred years ago, Sir Arthur Eddington was doubtful
about the existence of molecules in the vast interstellar space.
He pointed out that ``it is difficult to admit the existence of molecules in interstellar space because when once a molecule becomes dissociated there seems no chance of the atoms joining up again'' \citep[Bakerian lecture;][]{eddi26}.
As of July 2021, about 250 different gas-phase molecules are identified in the ISM (mainly in dense interstellar clouds) and circumstellar envelopes (CSEs, formed due to strong mass loss of stars in their later stages). These molecules are listed in Table \ref{tab:known_molecules}. These statistics are obtained by avoiding the isotopomers.
The rate of discovery continues at a rapid pace. Most of the molecules were observed in the gas phase by their rotational transitions in the centimeter, millimeter, and submillimeter portions of the electromagnetic spectrum.
Modern radio telescopes like Atacama Large Millimeter/submillimeter Array (ALMA) offer unique opportunities
to include more species in the catalog.
The dust particles and their icy mantles in colder regions can be
probed against absorption under background radiation by vibrational spectroscopy.
Among the observed complex species, about one-third of them are COMs, though we are far from having a complete inventory of the interstellar molecules. 
An up-to-date census by \cite{mcgu18}
reported a summary of the first detection of each molecular species, including the observational facility, wavelength range, transitions, enabling laboratory spectroscopic work, and listing tentative and disputed detections.
The ongoing web-based resources are of particular importance concerning the list of interstellar and circumstellar detections: the Astrochymist's {\it A Bibliography of Astromolecules}\footnote{\url{http://www.astrochymist.org/astrochymist_ism.html}} maintained by David Woon or the Cologne Database for Molecular Spectroscopy's {\it List of Molecules in Space}\footnote{\url{https://www.astro.uni-koeln.de/cdms/molecules}} \citep[CDMS;][]{mull01,mull05,endr16}, or community-supported Wikipedia's {\it List of interstellar and circumstellar molecules}\footnote{\url{https://en.wikipedia.org/wiki/List_of_interstellar_and_circumstellar_molecules}}.
Six key classes of COMs, namely, aldehydes (HCOR), ketones (RCOR$^\prime$), carboxylic acids (RCOOH), esters (RCOOR$^\prime$), amides (RCONH$_2$), nitriles (RCN), and isocyanates (RNCO), with R and R$^\prime$ being an alkyl group, detected in the ISM are shown in Figure \ref{fig:COMs}.
A substantial fraction ($\sim 33\%$) of known molecules have now been seen in external galaxies, while the numbers of molecules known in protoplanetary disks (23), interstellar ices (6), and
exoplanet atmospheres (5) are much smaller due to observational challenges.

\begin{landscape}
\begin{table}
\scriptsize
\caption{Known interstellar molecules ($\sim 251$ molecules; last updated: June 2021).}
\label{tab:known_molecules}
\vskip 0.2 cm
\hskip -2.5cm
\begin{tabular}{|ll|ll|ll|ll|l|l|l|ll|}
\hline
\multicolumn{2}{|c|}{\bf 2 atoms (45)} & \multicolumn{2}{|c|}{\bf 3 atoms (45)} & \multicolumn{2}{|c|}{\bf 4 atoms (31)} & \multicolumn{2}{|c|}{\bf 5 atoms (32)} & {\bf 6 atoms (24)} & {\bf 7 atoms (16)} & {\bf 8 atoms (16)} & \multicolumn{2}{c|}{\bf 9 atoms (15)} \\
\hline
$\rm{H_2}$ & SiS & $\rm{C_3}$ & $\rm{NH_2}$ & $\rm{c-C_3H}\ ^a$ & $\rm{c-SiC_3}\ ^a$ & $\rm{C_5}$ & $\rm{C_4H^-}$ & $\rm{C_5H}$ & $\rm{C_6H}$ & $\rm{CH_3C_3N}$ & $\rm{CH_3C_4H}$ & $\rm{C_3H_6}$ \\
AlF & CS & $\rm{C_2H}$ & $\rm{H_3^+}$ & $\rm{l-C_3H}\ ^b$ & $\rm{CH_3}$ & $\rm{C_4H}$ & HC(O)CN & $\rm{l-H_2C_4}\ ^b$ & $\rm{CH_2CHCN}$ & $\rm{HC(O)OCH_3}$ & $\rm{CH_3CH_2CN}$ & $\rm{CH_3CH_2SH}$ \\
AlCl & HF & $\rm{C_2O}$ & SiCN & $\rm{C_3N}$ & $\rm{C_3N^-}$ & $\rm{C_4Si}$ & HNCNH & $\rm{C_2H_4}$ & $\rm{CH_3C_2H}$ & $\rm{CH_3COOH}$ & $\rm{(CH_3)_2O}$ & $\rm{CH_3NHCHO}$ \\
$\rm{C_2}$ & HD & $\rm{C_2S}$ & AlNC & $\rm{C_3O}$ & $\rm{PH_3}$ & $\rm{l-C_3H_2}\ ^b$ & $\rm{CH_3O}$ & $\rm{CH_3CN}$ & $\rm{HC_5N}$ & $\rm{C_7H}$ & $\rm{CH_3CH_2OH}$ & $\rm{HC_7O}$ \\
CH & FeO & $\rm{CH_2}$ & SiNC & $\rm{C_3S}$& HCNO & $\rm{c-C_3H_2}\ ^a$ & $\rm{NH_4^+}$ & $\rm{CH_3NC}$ & $\rm{CH_3CHO}$ & $\rm{C_6H_2}$ & $\rm{HC_7N}$ & $\rm{H_2C_3HCCH}$ \\
$\rm{CH^+}$ & $\rm{O_2}$ & HCN & HCP & $\rm{C_2H_2}$ & HOCN & $\rm{H_2CCN}$ & $\rm{H_2NCO^+}$ & $\rm{CH_3OH}$ & $\rm{CH_3NH_2}$ & $\rm{CH_2OHCHO}$ & $\rm{C_8H}$ & $\rm{HC_3HCHCN}$ \\
CN & $\rm{CF^+}$ & HCO & CCP & $\rm{NH_3}$ & HSCN & $\rm{CH_4}$ & $\rm{NCCNH^+}$ & $\rm{CH_3SH}$ & $\rm{c-C_2H_4O}\ ^a$ & $\rm{l-HC_6H}\ ^b$ &$\rm{CH_3C(O)NH_2}$ &  $\rm{H_2CCHC_3N}$ \\
CO & SiH & $\rm{HCO^+}$ & AlOH & HCCN & $\rm{H_2O_2}$ & $\rm{HC_3N}$ & $\rm{CH_3Cl}$ & $\rm{HC_3NH^+}$ & $\rm{H_2CCHOH}$ & $\rm{CH_2CHCHO}$ & $\rm{C_8H^-}$ &  \\
\cline{12-13}
$\rm{CO^+}$& PO & $\rm{HCS^+}$ & $\rm{H_2O^+}$ & $\rm{HCNH^+}$ & $\rm{C_3H^+}$ & $\rm{HC_2NC}$ & $\rm{MgC_3N}$ & $\rm{HC_2CHO}$ & $\rm{C_6H^-}$ & $\rm{CH_2CCHCN}$ & {\bf 10 atoms (6)} & {\bf 11 atoms (7)} \\
\cline{12-13}
CP & AlO & $\rm{HOC^+}$ & $\rm{H_2Cl^+}$ & HNCO & HMgNC & HCOOH & $\rm{NH_2OH}$ & $\rm{NH_2CHO}$ & $\rm{CH_3NCO}$ & $\rm{H_2NCH_2CN}$ & $\rm{CH_3COCH_3}$ & $\rm{HC_9N}$ \\
SiC & $\rm{OH^+}$ & $\rm{H_2O}$ & KCN & HNCS & HCCO & $\rm{H_2CNH}$ & $\rm{HC_3O^+}$  & $\rm{C_5N}$ & $\rm{HC_5O}$ & $\rm{CH_3CHNH}$ & $\rm{HOCH_2CH_2OH}$ & $\rm{CH_3C_6H}$ \\
HCl & $\rm{CN^-}$ & $\rm{H_2S}$ & FeCN & $\rm{HOCO^+}$ & CNCN & $\rm{H_2C_2O}$ & $\rm{HC_3S^+}$ & $\rm{l-HC_4H}\ ^b$ & $\rm{HOCH_2CN}$ & $\rm{CH_3SiH_3}$ & $\rm{CH_3CH_2CHO}$ & $\rm{C_2H_5OCHO}$ \\
KCl & $\rm{SH^+}$ & HNC & $\rm{HO_2}$ & $\rm{H_2CO}$ & HONO & $\rm{H_2NCN}$ & $\rm{H_2CCS}$ & $\rm{l-HC_4N}\ ^b$ &$\rm{HC_4NC}$ & $\rm{(NH_2)_2CO}$ & $\rm{CH_3C_5N}$ & $\rm{CH_3OC(O)CH_3}$ \\
NH & SH & HNO & $\rm{TiO_2}$ & $\rm{H_2CN}$ & MgCCH & $\rm{HNC_3}$ & $\rm{C_4S}$ & $\rm{c-H_2C_3O}\ ^a$ & $\rm{HC_3HNH}$ & $\rm{HCCCH_2CN}$ & $\rm{CH_3CHCH_2O}$ & $\rm{CH_3COCH_2OH}$ \\
NO & $\rm{HCl^+}$ & MgCN & $\rm{C_2N}$& $\rm{H_2CS}$& HCCS & $\rm{SiH_4}$ & HC(O)SH  & $\rm{H_2CCNH}$ & $\rm{c-C_3HCCH}\ ^a$ & $\rm{HC_5NH^+}$ & $\rm{CH_3OCH_2OH}$ & $\rm{c-C_5H_6}\ ^a$ \\
NS & TiO & MgNC & $\rm{Si_2C}$ & $\rm{H_3O^+}$ &  & $\rm{H_2COH^+}$ & HC(S)CN  & $\rm{C_5N^-}$ & $\rm{l-H_2C_5}\ ^b$ & $\rm{CH_2CHCCH}$ & & $\rm{NH_2CH_2CH_2OH}$ \\
\cline{12-13}
NaCl & $\rm{ArH^+}$ & $\rm{N_2H^+}$ & $\rm{HS_2}$ &&&  &  & HNCHCN &  && {\bf 12 atoms (6)} & {\bf $>$12 atoms (8)}  \\
\cline{12-13}
OH & $\rm{N_2}$ & $\rm{N_2O}$ & NCO &&&&& $\rm{SiH_3CN}$ && & $\rm{c-C_6H_6}\ ^a$ & $\rm{HC_{11}N}$ \\
PN & $\rm{NO^+}$ & NaCN & HSC &&&&& $\rm{C_5S}$ && & $\rm{C_2H_5OCH_3}$ & $\rm{c-C_6H_5CN}\ ^a$  \\
SO & $\rm{NS^+}$ & OCS & HCS &&&&& $\rm{MgC_4H}$ && & $\rm{n-C_3H_7CN}$ & $\rm{1-C_{10}H_7CN}$  \\
$\rm{SO^+}$& VO & $\rm{SO_2}$ & CaNC &&&&& $\rm{CH_3CO^+}$ && & $\rm{i-C_3H_7CN}$ & $\rm{2-C_{10}H_7CN}$  \\
SiN & $\rm{HeH^+}$ & $\rm{c-SiC_2}\ ^a$ & NCS &&&&& $\rm{C_3H_3}$ && & $\rm{1-c-C_5H_5CN}\ ^a$ & $\rm{c-C_9H_8}\ ^a$ \\
SiO & & $\rm{CO_2}$ &&&&&& $\rm{H_2C_3S}$  && &$\rm{2-c-C_5H_5CN}\ ^a$ & $\rm{C_{60}}$ \\
&&&& &&&& $\rm{HCCCHS}$ &&&& $\rm{C_{60}^+}$ \\
&&&&&&&&&&&& $\rm{C_{70}}$ \\
\hline
\hline
\end{tabular} \\
\vskip 0.2cm
{\bf Notes:}
Total number of molecules for each category is provided in the parentheses.
These statistics are obtained by avoiding the isotopomers.
Deuterium isotopic species are given separately only if their method of detection is intrinsically different from that of pure hydrogen ones. \\
$^a$ The `c' refers to cyclic form. \\
$^b$ The `l' refers to cyclic form.
\end{table}
\end{landscape}

\section{Birth of stars and and the planetary systems}
\begin{figure}[htbp]
\begin{center}
\includegraphics[width=\textwidth]{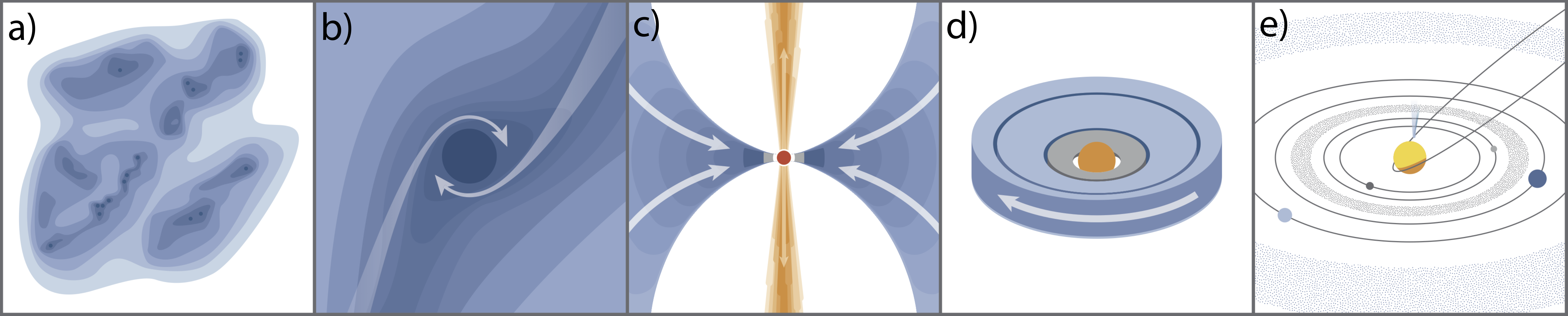}
\caption{Cartoon depiction of different stages characterizing low-mass (Solar-like) star and planet formation \citep[Courtesy:][]{ober21}.}
\label{fig:shu}
\end{center}
\end{figure}

\begin{figure}[htbp]
\begin{center}
\includegraphics[width=0.6\textwidth]{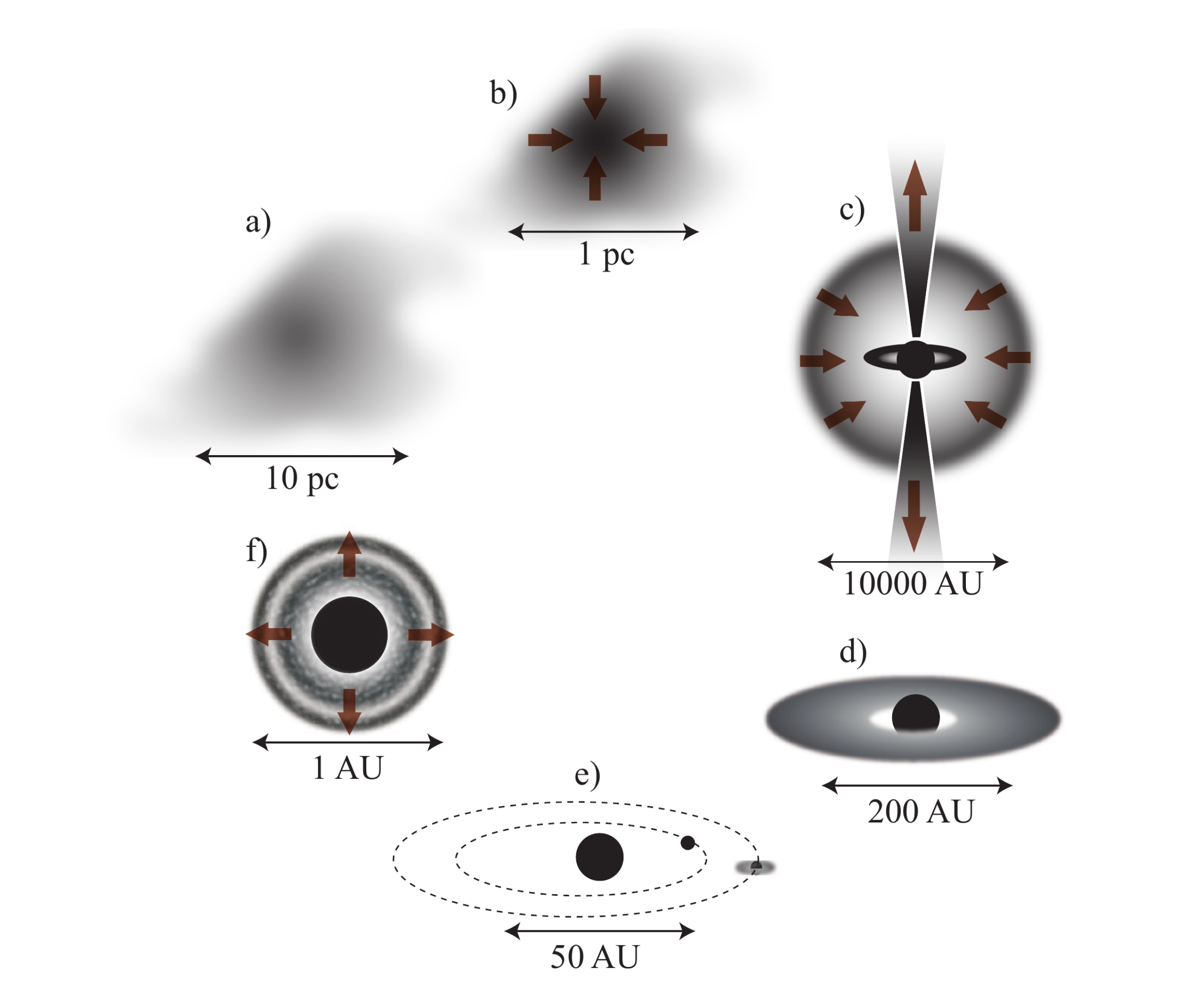}
\caption{The low-mass stellar life cycle from diffuse clouds and the beginning
of the star formation up to the death of the star.
The length scale of each stage is shown \citep[Courtesy:][]{ober09a}.}
\label{fig:star-formation-cyclic}
\end{center}
\end{figure}

Astrochemistry and its linkage to star formation have been the subject of several reviews \citep{vand98b,case12,cecc14,tiel13,vand14,ober16,jorg20,ober21}.
Stars span a large range of masses. The initial cloud mass is important beyond collapse
to finally form a star. Low-mass stars are the most relevant for determining planetary compositions. They constitute the vast majority of stars, and their long lifetimes are likely a requirement for the origin and sustainance of life. Several excellent reviews \citep{shu87,mcke07,luhm12}
discuss the low-mass star formation, illustrated in Figures \ref{fig:shu} and \ref{fig:star-formation-cyclic}.

Stars form in over-dense regions in inhomogeneous interstellar molecular clouds (Figure \ref{fig:shu}a), mainly associated with dense molecular clouds \citep{heye15}, where densities are $>10^5$ molecules cm$^{-3}$.
However, denser sub-structures having sizes of the order of a tenth of a parsec within the dense clouds are referred to as the dense cores.
They are characterized by orders of magnitude higher densities ($\sim 10^2-10^6$) and
lower temperatures ($\sim 10$ K) \citep{bens89,berg07}.

Some cores are dense and massive enough (fulfilling the Jeans mass criteria) that they can begin to collapse due to self-gravity \citep{shu77},
overcoming turbulence, thermal and magnetic pressure. Cores fated to collapse are referred to as the pre-stellar, as they will eventually form a star. This collapse is initially isothermal, keeping the core cool by molecules radiating away the heat produced by the collapse.
The collapse is not spherically symmetric due to the conservation of cloud angular momentum (Figure \ref{fig:shu}b).

As the collapse proceeds, at the center of the collapsing core, a dense condensation forms, and the collapse turns adiabatic due to its higher optical thickness. The condensation eventually heats up and begins to form a young star, a stage known as a protostar (Figure \ref{fig:shu}c).
During the protostellar stage, the central protostar consumes matter from the surrounding cloud core (i.e., protostellar envelope). Within the envelope, the temperature and density increase toward the center due to stellar heating.
When the collapsing gas reaches a temperature of $100-300$ K,
we refer to the material as a `hot corino', or,
if it is of high mass, the term `hot core' is used.
Some of the accreting material spreads out into a disk to conserve angular momentum, which simultaneously serves to funnel matter onto the star. Angular momentum is also removed from the system through the launch or ejection of a fraction of matter violently outward in the form of highly supersonic collimated jets and molecular outflows (Figure \ref{fig:shu}c). When the outflowing material encounters the quiescent gas of the envelope and the molecular cloud, it creates shocks, where the grain mantles and
refractory grains are (partially) sputtered and vaporized.
Subsequently, this stage is characterized by gas-phase chemistry,
as grains flow toward the protostar.
Once in the gas phase, molecules can be observed via their rotational lines.

As the protostellar system evolves, more and more mass is found in the star and disk than in the remnant envelope. The envelope is finally dispersed on time scales of $\sim$1 Myr, leaving a pre-main sequence star and a Keplerian disk (Figure \ref{fig:shu}d). The change of name from protostar to pre-main sequence star points to the fact that the central star became hot enough for fusion reaction to take place inside. The circumstellar disk is often referred to as a proto-planetary or planet-forming disk to indicate that these disks are the formation sites of planets \citep{will11}. However, recent observations suggest that planet formation may begin much earlier, already at the protostellar stage \citep{alma15,hars18}.

The protoplanetary disk stage lasts for $\sim 1-10$ Myrs depending on disk in stellar clusters. During this time, the disk material is accreted onto the star, onto planets, and dispersed through interactions with stellar photo-evaporative winds \citep{erco14}. Thus, what is left behind is a nascent planetary system (Figure \ref{fig:shu}e) that continues to evolve for hundreds of millions of years due to collisions between remaining the planets and planetesimals.

Understanding how the properties of protoplanetary disks link to planet formation is currently an active field of research. For example, it is theorized that ``snow lines'', which are regions far enough from the star to allow the freeze-out of gas-phase species \citep{kenn08}, are essential to the formation of planets around a new-born star.
As life exists in our Solar System, detailed knowledge of many different processes taking place at various locations in a protoplanetary disk is crucial to know.
A better understanding of the role of snow lines in the chemical composition of
a disk and in the planet formation process itself may ultimately show how
the building blocks of life arrived on Earth.
For example, through impacting
comets, which are small icy bodies and remnants of the protoplanetary material,
carry the chemical memory of the processes in diffuse and dense clouds \citep{chyb97,altw19}.

Over time, the star will grow out of the main sequence stage as it becomes depleted of hydrogen, essentially burning up its adjacent shells.
After several stages involving drastic temperature and pressure changes, for Sun-like stars, its outer layers are enhanced with heavy elements.
It will eventually be ejected into the diffuse ISM.
The ejected materials again collapse under the influence of gravity to form dense interstellar clouds, containing more overgrown and dark parts known as the cold cores.
The cycle starts over again, though more enriched in heavy elements than the previous cycle.

\section{The interstellar ingredients}

\begin{figure}[htbp]
\centering
\includegraphics[width=0.45\textwidth]{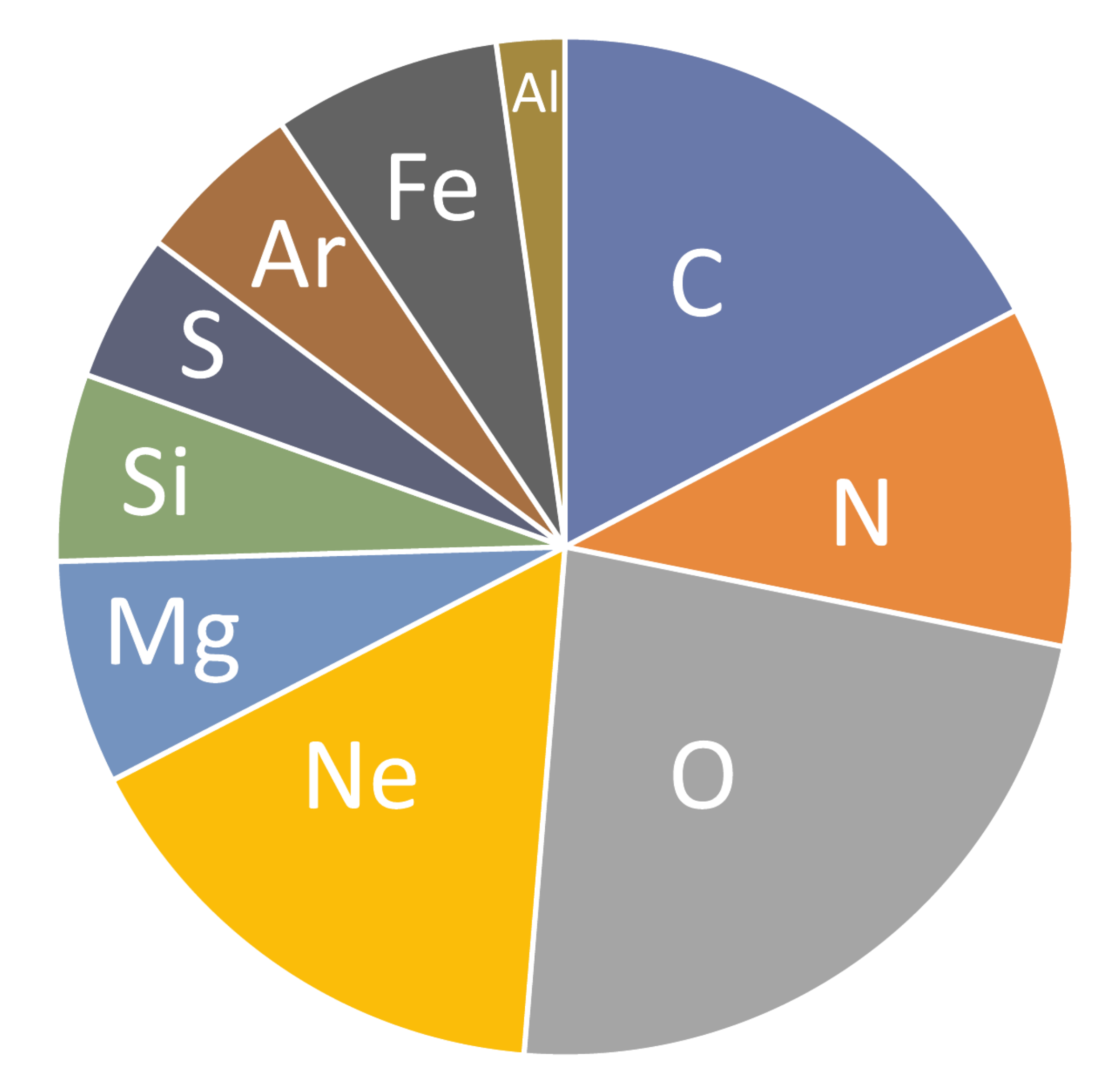}
\includegraphics[width=0.45\textwidth]{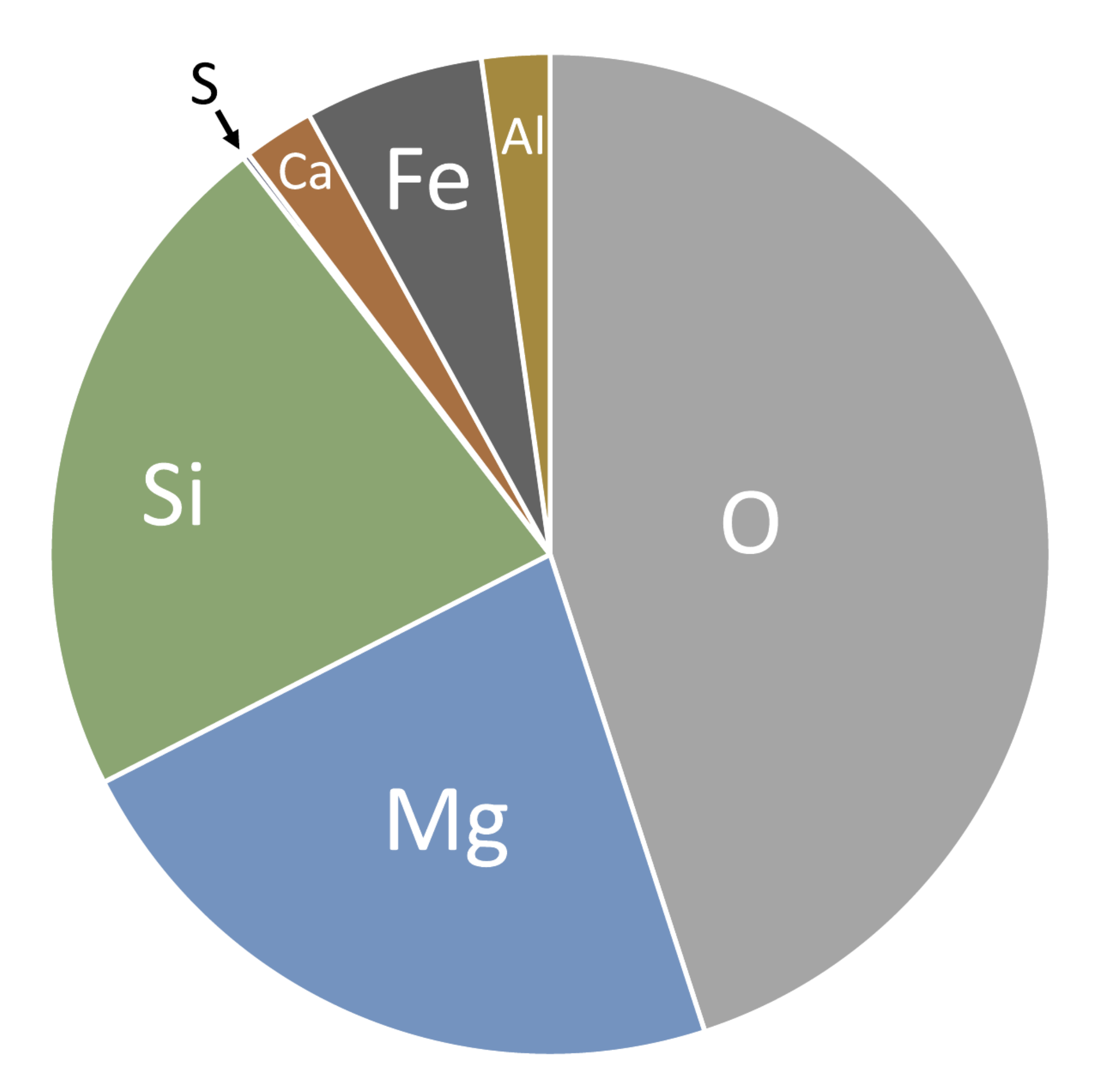}
\caption{Left: atomic abundances of the $10$ most common elements in the Universe besides H and He. Right: atomic abundances of the most common elements in the Earth’s mantle \citep[Courtesy:][]{fort20}.}
\label{fig:atomic_abundances}
\end{figure}

Astronomers have a somewhat restricted ``periodic
table'' than chemists. It is known as the ``astronomer's periodic table''. It only contains $11$ elements, which are the most abundant in the Universe \citep{mcca06}. The left panel of the pie graph (see Figure \ref{fig:atomic_abundances}) shows that after hydrogen ($90.8\%$ by number and $70.4\%$ by mass) and helium ($9.1\%$ by number and $28.1 \%$ by mass), the most abundant elements are oxygen, carbon, neon, nitrogen, etc. Except hydrogen and helium, all other elements are classified as ``metals'', even if they are not metal in a chemical sense. The abundances summarized in Table \ref{tab:solar} were derived for the solar photosphere by \cite{aspl09}.
These abundances are believed to be applicable to the ISM as well,
with a possibility of minor variations \citep{przy08}.
Before the formation of stars, the entire periodic table was based on hydrogen, little helium, trace amount of lithium (fittingly a metal).
In contrast, the elemental abundances of the Earth's mantle, shown on the right panel of Figure \ref{fig:atomic_abundances}, include roughly half of the oxygen, with the rest mostly made up of magnesium, silicon, iron, and some aluminum with other trace elements.

\begin{table}
\scriptsize
\centering
\caption{Solar elemental abundances \citep{aspl09}.}
\label{tab:solar}
\vskip 0.2cm
\begin{tabular}{cccc}
\hline
{\bf Element} & {\bf Abundance} & {\bf Element} & {\bf Abundance} \\
 \hline
 H  & 1.00 & Mg & $4.0\times10^{-5}$ \\
He & 0.085 & Al & $2.8\times10^{-6}$ \\
C & $2.7\times10^{-4}$ & Si & $3.2\times10^{-5}$ \\
N & $6.8\times10^{-5}$ & S & $1.3\times10^{-5}$ \\
O & $4.9\times10^{-4}$ & P & $2.6\times10^{-7}$ \\
Na & $1.7\times10^{-6}$ & Fe & $3.2\times10^{-5}$ \\
\hline
\end{tabular}
\end{table}

The physical conditions of ISM are different from occurring
naturally on the Earth, and they are challenging to recreate in laboratory experiments.
The ISM is highly heterogeneous with a broad range of temperatures, densities, and extinction parameters.
It is concentrated into regions of relatively dense gas and tiny dust particles known as the
interstellar clouds, with the dust-to-gas ratio of $0.01$ by mass
and $\approx 3\times10^{-12}$ by number.
Depending on several parameters, interstellar clouds are often classified as diffuse atomic, diffuse molecular, translucent, and dense \citep{snow06}.
An ISM contains gas with temperatures ranging from more than $10^6$ K
down to $10$ K and total hydrogen number density $n_H = 10^{-4}$ cm$^{-3}$ in
diffuse regions to $10^8$ cm$^{-3}$ in dense clouds.
The latter is more tenuous than a typical ultra-high vacuum laboratory experiment on Earth.
Considering an average $n_H = 10^2$ cm$^{-3}$, molecules get a scope to collide every 1.5 years in space, whereas in the Earth's atmosphere has a particle number density $\sim 10^{19}$ cm$^{-3}$, a collision occurs every $0.2\times10^{-9}$ seconds.
Thus, the interstellar space is a unique environment that can be studied under extreme conditions, which is totally different from the terrestrial conditions.

\subsubsection{Interstellar dust grains}
Interstellar grains or dust have size distributions from a few nanometers to a micrometer depending on the environment.
For example, there may be small grains in diffuse
clouds \citep{wein01}; more evolved, aggregated grains in dense clouds \citep{kohl12}; and more extensive, porous, and complex grains of $\mu$m to cm-size in protoplanetary disks \citep{blum18}.
Classical dust grains are characterized by a radius of $1000$ \AA, a density of 3 g cm$^{-3}$, and
$10^6$ surface sites for adsorption \citep{hase92}.
They include various components such as polycyclic aromatic hydrocarbons (PAH), particles of carbon (graphite, amorphous carbon, organic refractory material), silicate (olivine), or/and a mixture of both.
The olivine minerals are silicate structures having the general formula $\rm{M_2SiO_4}$, where M is magnesium (forsterite, $\rm{Mg_2SiO_4}$) or iron (fayalite, $\rm{Fe_2SiO_4}$).
The interstellar grain model, which has proven to be comparatively successful in conforming with the observed interstellar extinction, reflection, IR emission, and gas-phase abundances, consists of a combination of graphite and silicate grains. The $dn/da \propto a^{-­3.5}$ power­law, truncated at $a_{min} \approx 5$ nm and $a_{max} \approx 250$ nm,­ was first introduced in 1977 by \citeauthor*{math77}, and is often referred to as the ``MRN'' model.
Comprehensive discussions of the observations of interstellar dusts, properties, and models are provided in \cite{whit03,drai03,tiel10,drai11}.

Dust grains are not thought to have an active role as catalysts
in the chemical point of view, in which surface molecules accelerate the reaction. Instead, they provide a reservoir for the atoms and molecules, where they can react with others. Furthermore, due to the confined area, the reactions that are not possible in the gas phase can also process here.   
They thus enable reactions with activation barriers (minimum energy required to start a chemical reaction) that
are too slow in the gas, such as the hydrogenation of atomic O, C, and N.
Also, they act as a third body that absorbs the BE of a newly formed molecule, thereby stabilizing it before it can dissociate again.

Most astrochemical models \citep{garr06a,ruau16,cupp17,wake17}
mimicking the chemical evolution in the star-forming region have
now integrated the gas-grain processes.
Adsorption energy or BE ($E_d$) and diffusion energy ($E_b$) control the efficiency of grain surface reactions. $E_b$ can be estimated as a fraction of the $E_d$ of surfaces
species; the ratio of $E_d$ to $E_b$ often ranges from $0.3$ to $1.0$ \citep{garr07,cupp17}.
Species that are in the process of formation or trapped on
interstellar ice may transfer to the gas phase by various
desorption mechanisms \citep{cupp17} such as non-thermal
reactive desorption \citep[efficient desorption
mechanisms at low temperature;][]{garr07}, thermal desorption (efficient at high temperature), and energetic processes such as direct or indirect photoevaporation by photon or cosmic-ray particles. Interestingly, all the parameters causing  desorption are directly or indirectly related to the BE of the species.
The non-thermal chemical desorption is a mechanism in which part of the energy liberated by
the formation of the chemical bond is used to desorb the molecule.
It can also be impulsive spot heating by cosmic rays.
According to \cite{mini14}, the efficiency of non-thermal
chemical desorption mechanisms depends on four parameters: enthalpy of formation, degrees of freedom,
BE, and mass of newly formed molecules.
Quantities that govern the fraction of molecules returning to
the gas phase after a collision with an icy grain surface is sticking
coefficients ($\rm{S_T}$), residence times, and reaction probabilities.
The ratio of adsorbed to nonadsorbed incident atoms or molecules on a
surface during a given period after a collision is represented by the sticking coefficient.
The residence times represent the time that a particular
adsorbate (a substance that is adsorbed by adsorbent) stays on an adsorbent (a substance that adsorb other) surface after adsorption.
Both quantities are affected by the BE of adsorbate with an adsorbent surface.

\subsubsection{Interstellar ice composition}
Icy grain mantles are the building blocks of formation of COMs in the ISM and, therefore, worth providing some details. In clouds, protostellar
envelopes, and protoplanetary disks, such icy mantles constitute
the main reservoir of volatiles.
Although \cite{eddi37} proposed the existence of interstellar ice in 1937, there was a turning point more than 40 years later, when \cite{tiel82} introduced combined gas-grain chemistry for the chemical evolution of the ISM. Recently, it has been demonstrated that pre-biotic molecules can be produced in UV-irradiated astrophysical relevant ices \citep{woon02}. For instance, \citet{nuev14} experimentally showed that nucleobases could be formed by UV irradiation of pyrimidine in H$_2$O-rich ice mixtures containing NH$_3$, CH$_3$OH, and CH$_4$. Ices form by condensing
atoms and molecules from the gas phase and subsequent grain
surface chemistry \citep{tiel82}. O, C, N, and H accrete on grains to form H$_2$O, CH$_4$,
and NH$_3$ by hydrogenation.

\begin{figure}[htbp]
\begin{center}
\includegraphics[width=0.6\textwidth]{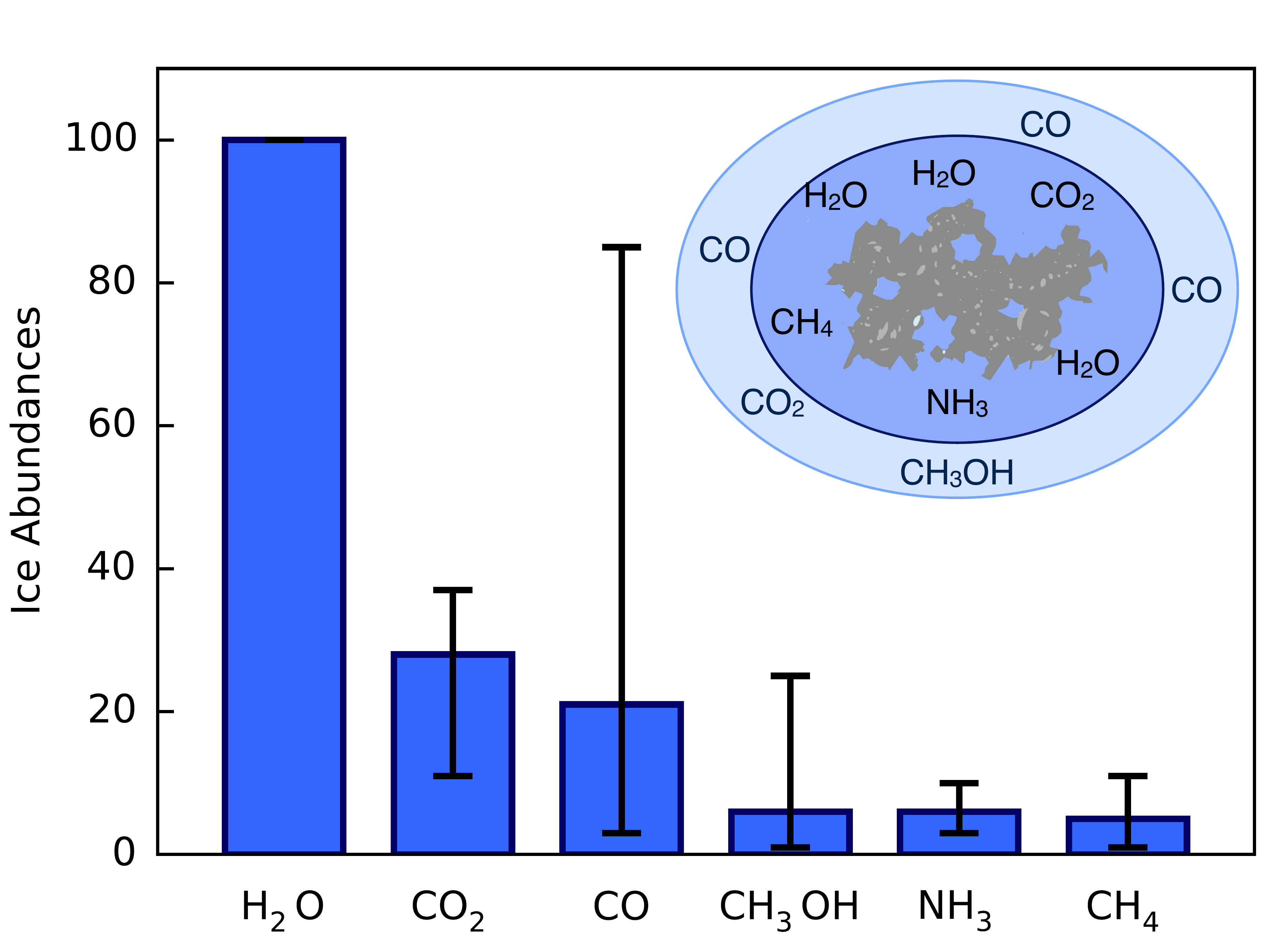}
\caption{The median composition of interstellar icy grain mantles
normalized to the most abundant ice species, water \citep{boog15} indicated
with the minimum and maximum abundance relative to water ice detected in that
line of sight \citep[Courtesy:][]{ober16}.}
\label{fig:ice-composition}
\end{center}
\end{figure}

Interstellar ice compositions are primarily studied through
IR absorption spectroscopy toward a background source
such as a protostar \citep{gibb04,ober11,boog15}.
Since the composition of grain mantles strongly depends on physical conditions \citep{das08a,das10,das11,das16}, the observed spectra vary along different lines of sight. However, there are some
general trends shown in Figure \ref{fig:ice-composition}.
The Figure shows the median ice abundances toward low-mass protostars,
as well as the minimum and maximum abundance concerning water ice, observed for
each of the other five species.
Water ice is the most abundant ice constituent in dense molecular clouds \citep{gibb04},
accounting for $\sim 60-70 \%$ of the icy mantels \citep{whit03}.
CO and CO2 follow it at $\sim 20-30\%$ each
relative to water. CH$_4$, NH$_3$, and CH$_3$OH are also detected
in many lines of sight at typical abundances of $\sim 5\%$ concerning water ice.

Water ice was first detected by comparing ground-based IR observations of its O-H stretching band at $3278.69$ cm$^{-1}$ ($3.05$ $\mu$m) toward Orion-KL \citep{gill73} and laboratory work by \citet{irvi68}. Since then, several ground-based observations were made to identify the signatures of water ice in different astrophysical environments, with further laboratory studies supporting such observations \citep{merr76,lege79,hage79}.

Only a few species are unambiguously detected in interstellar ices
due to difficulties in observing them in the mid-IR.
It is hard to detect interstellar atomic or molecular spectral signatures
in the IR regions using ground-based observation. Most of
the signals are attenuated by the Earth's atmosphere when they pass through it.
A background illuminating source such as a protostar or a field star is needed for absorption.
The energy required to excite a transition into
emission is likely to cause the molecule to desorb into the gas phase.
Furthermore, peak positions, intensities of molecular ice features, and line widths which are required to compare with laboratory spectra, further depend on ice temperature, the crystal structure of the ice, and mixing or layering with other species \citep{ehre97,schu99,cook16}.
However, CO is routinely observed, showing both polar and apolar band profiles in absorbance.
The CO ice-phase abundance may vary from $3\%$ to $20\%$ of the water-ice.
Based on laboratory work \citep{mant75}, the fundamental vibrational band of CO at 4.61 $\mu$m (2169.20 cm$^{-1}$) in absorption toward W33A was reported \citep{soif79}. The corresponding band profile includes a broad (polar) component and a narrow (non-polar) component \citep{chia95,chia98}.
$\rm{CO_2}$ absorption feature was observed toward several IRAS
sources based on laboratory work \citep{dhen89}. The ubiquitous nature of $\rm{CO_2}$ in ice mantles was found toward several astrophysical objects \citep{degr96,guer96,gera99} after the launch of ISO \citep{dart05}.
Ice-phase abundance of CH$_3$OH varies between $5\%$ and $30\%$
relative to $\rm{H_2O}$ \citep{baas88,grim91,alla92}. \cite{schu96} first attempted the observation of H$_2$CO observation toward the GL 2136 protostellar source and estimated the abundance to be $\sim 7\%$ relative to $\rm{H_2O}$. \cite{kess96,kean01,kess03} estimated the H$_2$CO abundance between $1\%$ and $3\%$. HCOOH was observed both in the solid and gas phase \citep{schu99,vand95,iked01}. Both gas and ice-phase CH$_4$ were simultaneously detected toward NGC 7538 IRS 9 \citep{lacy91}. \cite{ober08} observed and derived $\rm{CH_4}$ ice-phase abundance between $2\%$ and $8\%$, except for a few sources where it is $11-13\%$. \cite{knac82} claimed the first identification of $\rm{NH_3}$ in interstellar grains; a detection later proved to be wrong \citep{knac87}. Eventually, the detection of $\rm{NH_3}$ was reported toward NGC 7538 IRS 9 \citep{lacy98}. \cite{palu95,palu97} identified OCS mixed with $\rm{CH_3OH}$ toward several sources based on laboratory work.

An abundant $\rm{O_2}$ was discovered by the Rosetta Orbiter Spectrometer for Ion and Neutral Analysis 
(ROSINA) mass spectrometer onboard the European Space Agency’s (ESA) Rosetta spacecraft in the
comet 67P/Churyumov-Gerasimenko (hereafter 67P/C-G) coma.
The derived ratio was $\rm{O_2/H_2O}$ = $3.80\pm0.85\%$ \citep{biel15}.

Neutral mass spectrometer data obtained during the ESA's Giotto flyby are consistent with abundant amounts of $\rm{O_2}$ in the comet 1P/Halley coma, the $\rm{O_2/H_2O}$ ratio being evaluated to be $3.70\pm1.7\%$ \citep{rubi15}.
In the ISM, O$_2$ and N$_2$ are depleted on grains in the solid form and are nearly absent in the gas phase \citep{vand93}. Therefore, N$_2$ and O$_2$ cannot be detected using radio observations due to the absence of dipole moment. However, they might be seen by their weak IR active fundamental transition in the solid phase \citep{sand01,ehre97}.

Analysis of spectral profiles of the observed ice reveals that
interstellar ices are not perfectly mixed but rather, present in
at least two phases, water-rich, and water-poor.
The water-rich phase is observed to form first and
contains mostly water, CO$_2$, NH$_3$, and CH$_4$ ice.
A second CO-rich ice has the
remaining CO$_2$ ice and probably most of the CH$_3$OH (Figure \ref{fig:ice-composition}). Theoretically, these two separate phases form because ice formation is dominated by the hydrogenation of atoms at early times and by reactions involving CO at late times.

To date, IR observations suggest that the ice mantles in molecular clouds are positively composed of the few molecules mentioned earlier \citep{herb09}. However, more complex species, such as COMs, are also expected to be frozen on ice grains in dense cores. The low sensitivity or low resolution of available observations combined with spectral confusion in the IR region can cause weak features due to solid COMs hidden by more abundant ice species. The upcoming NASA's JWST\footnote{\url{https://jwst.stsci.edu}} space mission set to explore the molecular nature of the Universe and the habitability of planetary systems promises to be a giant leap forward in our quest to understand the origin of molecules in space. The high resolution of spectrometers onboard the JWST will enable the search for new COMs in interstellar ices and shed light on different ice morphologies, thermal histories, and mixing environments. JWST will map the sky and see right through and deep into massive clouds of gas and dust opaque in the visible window. However, the large amount of spectral data provided by JWST could be analyzed when extensive spectral laboratory and modeling data sets are available to interpret such data.

\subsubsection{Binding energy}
The binding energy (BE) or adsorption energy is an essential input parameter and thus crucial to improve the BE values in the models to better reproduce the astrophysical abundances.
A gas-phase species approaching a surface will feel at a large distance a weak
attraction due to van der Waals forces. These are due to mutually induced dipole
moments in the electron shells of the gas-phase species and the atoms on the
surface. At short ranges, forces associated with the overlap of the wave functions
of the approaching species and the surface atoms lead to much stronger binding.
BEs depend on the nature of the volatile
species, the chemical and morphological composition of the surface, and its coverage.
Several studies have reviewed known BE on various surfaces and more especially water ices \citep{fras01,coll04,nobl12,hama13,he16b,pent17,wake17,same17,shim18,sil17,das18,same18,molp20a,molp20b,dufl21,sil21}.

Most of the experimentally obtained BE values for stable molecules were
obtained from Temperature-Programmed Desorption (TPD) experiments \citep{fras01,ober05,boli05,acha07,burk10,hama13}.
In TPD experiments, the surface temperature is increased
rapidly, and the desorbing particles are collected.
Since the 1990s, the TPD technique has been used to determine the BE values experimentally.
The thermal desorption process of interstellar COMs is possible to simulate in the laboratory through TPD analysis, deriving essential parameters such as the thermal desorption temperatures ($T_d$) and desorption energies ($E_d$).
Although TPD measures the desorption energy, this energy essentially is the BE of the species
if there are no activated processes.
Molecules can interact on the grain surface through van der Waals-like forces and dipole-dipole interactions, as it has been routinely demonstrated through vibrational spectroscopic methods such as IR and Raman spectroscopy.
Moreover, molecules can diffuse inside the grains when the
submicron interstellar grains begin to accrete into hundreds of microns fluffy dust.
Up to now, TPD experiments are carried out mainly from
graphite and amorphous water ice surfaces \citep[e.g.,][]{coll04,hama11,shi15,chaa18}.
\cite{coll04} presented an extensive TPD study for a collection of $16$ astrophysically relevant molecular species.
The interaction and auto-ionization of HCl on low-temperature
($80-140$ K) water ice surfaces are studied by \cite{olan11} using low-energy ($5-250$ eV) electron-stimulated desorption (ESD) and TPD.
\cite{ward12} used TPD experiments coupled with time-of-flight mass spectrometry to determine the yield
of OCS and additionally yields a value for the computation of desorption energies of O atoms and OCS.
\cite{nobl12} presented an experimental study of CO, O$_2$, and CO$_2$ desorption from three different surfaces:
non-porous amorphous solid water (ASW), crystalline ice, and amorphous olivine-type silicates.
\cite{duli13} also performed TPD experiments to derive the BE of different species on the silicate substrate.
\cite{cora21}, for the first time, have performed TPD experiments of acetonitrile (CH$_3$CN) and acetaldehyde (CH$_3$COH).
They consider both pure and mixed with water from micrometric grains of silicate olivine $\rm{[(Mg,Fe)_2SiO_4}]$ used as dust analog on which the icy mixtures are condensed at $17$ K.

Experimentally determined BE values depend on the nature of the substrate from which the species desorb and on the properties of the deposited ices (i.e., pure, mixed, or layered), which impact the obtained BE values from TPD experiments.
Furthermore, TPD is not appropriate for reactive atoms or
radicals since they may spontaneously recombine to form stable
molecules at the surface before thermal desorption.
Consequently, the BEs of atoms or radicals are less documented.
Nevertheless, estimating the BE of these radicals is essential to map the chemical composition in the intermediate temperature ($40-80$ K).
To this effect, computational studies can provide faster
information as compared to experimental studies. Before the era of TPD experiments, some
estimated values of BE were used in gas-grain chemical models \citep{tiel82,hase93,char97}
based on the polarizability of molecules or atoms, determining the strength of van der Waals interaction with a bare grain surface.

For species with undefined BEs, as a rule of thumb, can be very crudely
estimated by an additive law, i.e., the BE of an unknown species being the sum of the BEs of the
reactants. However, this approximation of the BE may lead to very misleading results \citep{cupp17}.

\cite{pent17} have estimated the BEs based on the results
presented in \cite{coll04} for the deposition of each species on an H$_2$O substrate as,
\begin{equation}
 E_{bind,X}=\frac{T_{des,X}}{T_{des,H_2O}}\times E_{bind,H_2O},
\end{equation}
where $T_{des,X}$ is the desorption temperature of a species $X$
deposited on a H$_2$O film, T$_{des,H_2O}$ is the desorption temperature
of H$_2$O, and E$_{bind,H_2O}$ is the BE of H$_2$O.

The BEs of species are evaluated theoretically using high-performance computers.
A suitable computational model with reliable approximations is indispensable in practical simulations for interstellar chemistry.
For example, \cite{alha07} simulated the adsorption of H atoms to ASW using classical trajectory (CT) calculations, and the off-lattice kinetic MC approach was used by
\cite{kars14} to estimate the BE of CO and CO$_2$.
\cite{horn05} used CT calculations to simulate the adsorption
of H$_2$ with ASW and crystalline ice.
\cite{wake17} has proposed a proportional law
relating the interaction energy between the given species and one
water molecule and its BE on ASW.
Hybrid QM/MM (Quantum Mechanics/Molecular Mechanics)
approaches, such as the ONIOM method \citep{chun15},
recently used by \cite{same17,same18} on crystalline water ices and by \cite{dufl21} on both
hexagonal crystalline and amorphous ices, represented by a cluster of about 150
water molecules via classical MD and electronic structure methods.
\cite{molp20b} reported the MD simulations of the adsorption dynamics of the
nitrogen atoms on ASW.
They used extensive sampling on ab-initio accuracy.
In addition, \cite{shim18} has reported the BE of carbon, nitrogen, and oxygen atom on the low-temperature ASW based on quantum chemistry calculations and the experimental value found by \cite{mini16}.
\cite{molp20b} compared their findings with these studies.
\cite{molp21b} recently have studied the interaction of H$_2$ with crystalline and amorphous CO ice computationally, employing $ab\ initio$ MD simulations and providing BEs and sticking coefficients for this pair.
\cite{molp21a} have theoretically determined the BE distribution of the
reactants, reaction energies, and (when applicable) activation
energies of the chemical reactions involving P-hydrogenation sequence (P to PH$_3$) on top of a H$_2$O ice mantle.

In Chapter \ref{chap:BE_ice}, we \citep{sil17,das18,das21} present quantum chemical
approaches to determine the BE of several important species with astrophysical relevant surfaces.

\begin{figure}[htbp]
\centering
\includegraphics[width=0.5\textwidth]{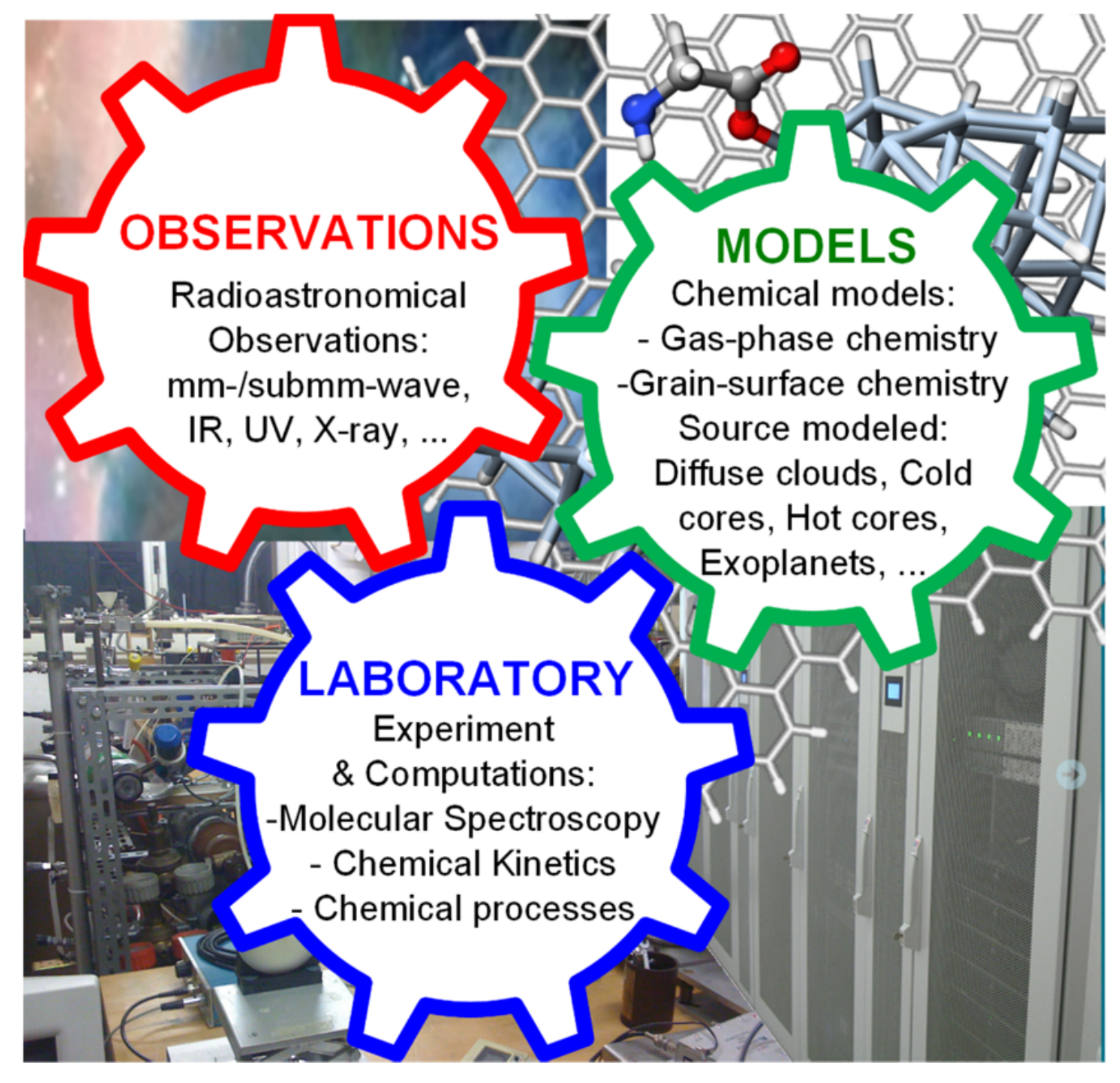}
\caption{Approaches in astrochemistry: observations, laboratory studies, and models are carried out to explain physico-chemical processes in astronomical environments \citep[Courtesy:][]{baro15a}.}
\label{fig:astrochemistry}
\end{figure}

\section{Current state-of-the-art of the astrochemistry}
The beginning of the astrochemistry community and its development history is well described by \cite{vand19,puzz20,fort20}. Presently, it has become a broad research field that can not be addressed in isolation.
It spans astronomical observations, modeling, and
theoretical or experimental laboratory-based investigations \citep{herb13,brow14} outlined in Figure \ref{fig:astrochemistry}.
It represents an example of how astrochemistry is developed nowadays, with such incredible research articles published in recent years, bringing new molecules, new ideas, and new paths for future research.
The removal of any limb of the astrochemical tripod would collapse the structure. 
The progress of observational studies revealed chemical diversity
in space, the source-to-source variation in chemical composition through a synergic interplay of
radio astronomical observations and laboratory spectroscopy \citep{mull05,mcgu18,vand19,puzz20,qasi20}.
The laboratory astrochemistry aims to simulate processes that occur over millions of years
in space within a few hours in the laboratory.
Abundances are extracted from the intensity of the observed/assigned molecular lines.
The interpretation of these abundances requires a comparison with model predictions \citep{garr08,das15a,das16,das19,das20,gora17a,gora20b,sil18,sil21}. Thus a strong interaction between different communities (such as, experimentalists, theoreticians, and observer) are involved in understanding the unknown universe.

\subsection{Astronomical observations}
\begin{figure}[htbp]
\centering
\includegraphics[width=\textwidth]{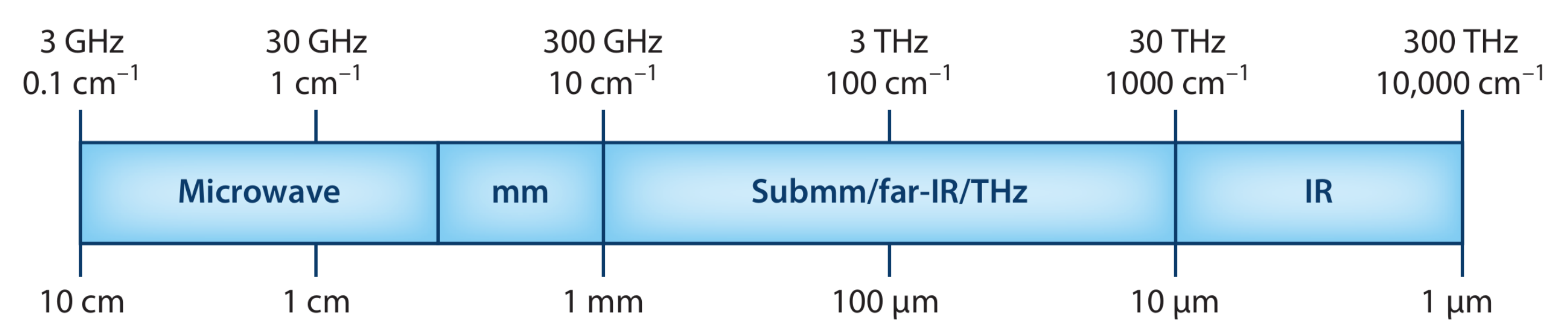}
\caption{Graphical representation of the microwave, millimeter (mm), submillimeter (Submm), and IR ranges of the electromagnetic spectrum \citep[Courtesy:][]{widi19}.}
\label{fig:EM_window}
\end{figure}

Detectable emission lines are arising from electronic, vibrational, torsional, and
rotational states of hundreds of species, which together can be used to probe the physical and chemical characteristics of specific regions of ISM.
In principle, astrochemical observations are possible at all wavelengths where molecules emit and absorb photons at discreet energies, i.e., from UV to radio wavelengths.
However, in practice, the vast majority of astrochemical observations are carried out at longer wavelengths, at infrared (IR), far-IR or terahertz (THz), submillimeter and millimeter, and
radio wavelengths (see Figure \ref{fig:EM_window}).
The electronic transitions (frequencies ranged from the visible to the UV region of the spectrum) are typically the most energetic and can stimulate photodissociation processes.
Vibrational transitions within the same electronic state are usually observed in the near- to mid-IR ($<20 \mu$m) wavelength range. Rotational transitions within a given
electronic and vibrational state, which are the weakest in energy, approximately range from
the submillimeter to centimeter range of the electromagnetic spectrum.

In the present situation, the field is driven by brand new significant astronomical facilities and technology \citep{jorg20}.
These include powerful ground-based telescopes and space-based telescopes.
Hubble Space Telescope (HST), X-ray Multi-Mirror Mission (XMM), Chandra, X-Ray Imaging and Spectroscopy Mission (XRISM, forthcoming)
are examples of UV and X-ray spectroscopy instruments perform from space. 
Far-IR region is studied with Spitzer Space Telescope, Herschel Space Observatory, and the Stratospheric Observatory for Infrared Astronomy (SOFIA).
The submillimeter telescopes in the Atacama desert include the Atacama Pathfinder
Experiment (APEX), the Atacama Submillimeter Telescope Experiment (ASTE), and
the Atacama Large Millimeter/submillimeter Array (ALMA).
All these together support observational astrochemistry.
With James Webb Space Telescope (JWST), Atmospheric Remote-sensing Infrared Exoplanet Large-survey (Ariel),
Extremely Large Telescope (ELT), and many other future missions on the horizon would provide increasingly sharp and sensitive data to the community.
Along with the advances in the detection of molecules, efficient codes like RADEX\footnote{\url{https://home.strw.leidenuniv.nl/~moldata/radex.html}}, which is a one-dimensional non-local thermodynamic equilibrium (non-LTE) radiative transfer code \citep{vand07}, have also been developed to infer the physical (excitation temperatures) and chemical (e.g., molecular abundances) conditions from observations.
Spectroscopic databases such as CDMS\footnote{\url{http://www.astro.uni-koeln.de/cdms/}} \citep{mull05,endr16}, Jet Propulsion Laboratory (JPL)\footnote{\url{http://spec.jpl.nasa.gov/home.html}} \citep{pick98}, SPLATALOGUE\footnote{\url{https://splatalogue.online//}}, HITRAN\footnote{\url{http://www.cfa.harvard.edu/hitran}} \citep{roth09,roth21}, EXOMOL\footnote{\url{https://www.exomol.com/}} line lists \citep{tenn16}, and Lovas/NIST\footnote{\url{http://physics.nist.gov/PhysRefData/Micro/Html/contents.html}} \citep{lova04} are invaluable resources for identifying lines.

\subsection{Theory and astrochemical models} \label{sec:theory_model}
Considerable progress in computational accuracy and pace continues in the theoretical understanding of astrochemical processes that rely on a range of modeling techniques \citep{chak00a,chak00b,chak06a,chak06b,cupp13,garr13b,das15a,das15b,das16,das19,das21,gora17a,gora17b,gora20b,sil18,sil21}.
The most computational expensive models are $ab\ initio$ quantum mechanical calculations that are employed
in astrochemistry to calculate collisional excitation coefficients, photodissociation cross-sections, reaction potentials,
and some molecule-grain surface interactions.
Collisional rate coefficients are needed to determine the excitation of interstellar molecules.
State-to-state collisional rate coefficients with $\rm{H_2}$ for pure rotational transitions of some speices like, $\rm{H_2O,\ H_2}$, CO, HCN, HNC, CN, CS, SO, $\rm{SO_2}$, CH, $\rm{CH_2}$, HF, HCl, $\rm{OH^+}$, $\rm{NH_2D,\ HC_3N,\ CH_3CN}$, and $\rm{CH_3OH}$ were computed.
There are also available rate coefficients with He for species like PN, SiH$^+$, ArH$^+$.
Collisions with electrons are also crucial for ions and molecules with large dipole moments such as HF, CH$^+$, and ArH$^+$.
The rate coefficients for vibration-rotation transitions also bring attention \citep[for example, new data are provided for CO with $\rm{H_2}$ and H;][]{song15} driven by IR spectra from ISO and Spitzer, and with JWST on the horizon.
The BASECOL\footnote{\url{https://basecol.vamdc.eu/}} \citep{dube13} and LAMDA\footnote{\url{https://home.strw.leidenuniv.nl/~moldata/}} \citep{scho05,vand07} databases are there to access the available collisional data.
A new revision of LAMDA, including its
current status, recent updates, and plans are described in \cite{vand20}.

Within astrochemistry, molecular dynamics (MD) simulations are used to uncover ice reaction mechanisms that can be parametrically
incorporated into higher-level modeling efforts.
More extended simulations require reaction probabilities parameterized, which is done through the so-called microscopic or kinetic Monte Carlo (KMC) models \citep{chan05,herb06,chak06a,chak06b,chan07,cupp07,das08a,das10,das11,mura08,cupp09,chan12,vasy13}.
Macroscopic stochastic or deterministic rate equation models are routinely employed
to carry out comprehensive gas-phase and solid-state interstellar chemistry models.
For example, in macroscopic Monte Carlo (MC)
models, the master equation for the combined gas-phase and surface chemistry is solved using the MC technique \citep{gill76,char98,char01,acha05,cupp07,vasy09}.
The theory and astrochemical applications of microscopic and macroscopic MC methods in astrochemistry were recently reviewed by \cite{cupp13}.
In terms of the number of species and reactions, the most comprehensive astrochemical models exclusively use the rate
equation approach or its modified version to consider the corrections obtained with the macroscopic MC methods
\citep{chak06a,chak06b,cupp07,das08a,garr08,das08a,das10,das11,das16,das21,sahu15,gora17a,gora17b,gora20b,sil18,sil21}.
The application of the rate equation approach to ice photochemistry under the astrophysical environments was recently reviewed by \cite{garr13b}.
The treatment of ice surfaces separate from ice layers
\citep{hase93} and so-called three-phase models (ice mantle,
ice surface, and gas-phase) are now standard \citep{vasy13,garr13}.
Besides, there is another class of chemical models labeled as thermochemical models \citep{ober21}
applied to the general ISM, dense clouds exposed to enhanced radiation fields, and protoplanetary disks.

The rate coefficients and activation energy barrier for gas-phase and ice-phase
chemical reactions are derived from experiments and theory.
Typically, chemical models of the ISM in use employ
databases of mainly 2-body (sometimes 3-body) chemical reactions;
these contain long lists of gas and ice-phase chemical reactions.
Reducing the degree of uncertainty for all the reactions would require an unfeasibly large number of laboratory experiments. Yet, chemical models entirely rely on the databases, making their accuracy a prime problem for astrochemistry.
Gas-phase reactions are the backbone of chemical networks as they are the primary routes for the formation and destructions of most molecules.
Gas-phase networks of COMs are incomplete \citep{chak00a,balu15,baro15c} and astrochemical models have been struggling to explain the observed abundances of COMs. Dust grain chemistry also plays a pivotal role in forming vital abundant species observed routinely in the ISM.
A couple of public databases are UMIST Database for Astrochemistry\footnote{\url{http://udfa.ajmarkwick.net/}} \citep[UDfA, which includes a total of $6173$ gas-phase reactions involving $467$ chemical species;][]{mcel13}
and KInetic Database for Astrochemistry\footnote{\url{http://kida.astrophy.u-bordeaux.fr/}} \citep[KIDA, consists of a total of
$7509$ reactions involving $489$ species;][]{wake15}, which continue to be updated with new results from
the chemical physics literature.
KIDA also provides astrochemical codes, such as the Nautilus gas-grain code\footnote{\url{http://kida.astrophy.u-bordeaux.fr/codes.html}}.
The latest version of Nautilus
allows simulating the chemical evolution of the ISM considering three phases, i.e., gas, grain surface, and grain mantles \citep{iqba18}. The gas chemistry uses the KIDA network \citep{wake15}, whereas the grain
chemical network is that presented in \cite{garr07,ruau15}. Such types of codes and
chemical models find essential applications in sources like cold cores, which in conjunction with submillimeter observations might explain the desorption and formation of COMs in icy mantles \citep{vasy17}.
\cite{heay17} have provided the wavelength-dependent photodissociation and photoionization cross sections for 102 astrophysically essential species. The cross-section and calculated rate of these species are available from the Leiden Observatory database of ``photodissociation and photoionization of astrophysically relevant molecules''\footnote{\url{http://www.strw.leidenuniv.nl/~ewine/photo}}.

\subsection{Laboratory experiments}
Laboratory experiments serve as the bridge between astrochemical models and observations.
For example, through the calculations of spectra of interstellar and circumstellar molecules, molecular and atomic excitations, etc.
Laboratory measurements of molecular spectra across the electromagnetic
spectrum are base for astrochemical purposes \citep{widi19}.
Laboratory facilities across the world include large synchrotron and advanced light sources and the most powerful computers.
There are innovative laboratory setups like
cavity ringdown and CHIRP spectroscopy, helium droplets, crossed beam
experiments, and UHV surface science techniques.
An overview of rotational spectroscopy and its relationship with observational
astronomy, as well as an overview of laboratory spectroscopic techniques
focusing on both historical approaches and new advancements,
are elaborated in \cite{widi19}.
\cite{smit11} summarizes different experimental methods developed to obtain collisional excitation data, gas-phase reaction rates, and product branching ratios.
Two significant developments of the past few decades are the so-called Selected Ion Flow Tube \citep[SIFT;][]{mart08}, CRESU \citep{sims93,chas01}, or crossed molecular beams \citep{mora11} method enables the measurements of gas-phase reaction rates.
Ice-phase reactions are characterized using surface-science apparatuses designed to mimic the high vacuum and range of temperatures (down to $4$ K) characteristic of interstellar environments.
The Reflection Absorption Infra-Red Spectroscopy (RAIRS)
is an IR technique that allows the identification of adsorbates from their vibrational frequencies and determines the orientation of the adsorbed species.
TPD experiments \citep{coll04} now routinely probe thermal desorption of both pure and mixed ices.
They provide binding energies (BEs) that can be used in models.
\cite{ober11} reviewed ice photochemistry experiments and discussed the qualitative and quantitative kinetic and mechanistic constraints.

\section{Some astrochemical tools for modelers}

\subsection{Quantum chemical calculations}
A technique called $ab\ initio$ quantum chemistry allows scientists to start from pure quantum mechanics. This theory describes behavior of the subatomic particles to calculate properties of a molecule based on the motions of the protons, neutrons, and electrons in the atoms that comprise it. On a super-computer or a server based computer, scientists can
run repeated simulations for a specific molecule, each time slightly adjusting its structure and the arrangement of its particles, and watch the results to find the optimal geometry of a compound.
But with quantum chemistry, we are limited by the size of the molecules. As a result, we need large amounts of computational power to do the calculations. In this thesis, the IR absorption spectra, rotational parameters, kinetic data, reaction data, etc. are mostly calculated quantum chemically using \textsc{Gaussian} 09 (G09) computational chemistry
program \citep{fris13} to study various features of species and include some of these physical parameters into our chemical model to study the evolution.

\subsubsection{The \textsc{Gaussian} 09 suite of programs}
\textsc{Gaussian} 09 is a Gaussian series of programs.
It provides state-of-the-art capabilities for a wide-ranging suite of the most advanced electronic structure modeling.
GaussView is a graphical user interface (GUI) designed to prepare input for submission to \textsc{Gaussian}
and to examine the output graphically that \textsc{Gaussian} produces.
The various types of applications of the \textsc{Gaussian} 09 software for our astrochemical
approach are briefly well documented in \cite{gora19}.
There are different $ab\ initio$ methods available such as Hartree-Fock (HF),
M\o ller-Plesset perturbation (MPn, n=2, 4), Coupled Cluster (CC), and
Configuration Interaction (CI). A most popular method of Density Functional Theory like B3LYP
and composite methods like Gaussian-4 (G4) are widely used in our calculations.
There are various basis sets available. For example,  minimal (STO-3G), split valence (people, can be modified with polarization and diffuse functions) and correlation consistent (cc-pVXZ, X = D for double, T for the triple, Q for quadruple, 5, 6, 7, can be modified with
diffuse functions using prefix aug-), which can be modified to obtain
better approximations to describe a system.
\textsc{Gaussian} 09 is a powerful software that can perform a multitude of calculations
for a given molecule, starting from the fundamental laws of quantum mechanics.
Geometry optimization, single-point energy, frequencies, and thermochemistry, potential energy surface (PES), solvation effect are the most valuable types of jobs to calculate molecular properties.
Atomic coordinates of the target molecule such as optimized atomic distances and angles,
Mulliken atomic charges, dipole moments, electron affinities, molecular orbitals, single-point energy, harmonic frequencies,
IR intensities, UV/visible spectra, temperature, pressure, isotopes used, molecular mass, thermal energy, free energy (sum of electronic and thermal free energies),
enthalpy (sum of electronic and thermal enthalpies), polarizability, susceptibility
are the most crucial information we get out of these calculations.

\subsection{Modeling radiation-dominated region}

\subsubsection{Diffuse and translucent clouds}
Diffuse atomic clouds represent the regime in the ISM that is fully exposed to the ISRF, and consequently, nearly all molecules are quickly destroyed by photodissociation.
It typically has a fairly low density of $\sim 10-100$ cm$^{-3}$ and temperatures of $30-100$ K.
There is an intermediate physical structure between
diffuse and dense clouds; this structure is called a translucent cloud and has a density of
$10^2 - 10^3$ cm$^{-3}$ and a visual extinction of $\approx 3$.
Several earlier models \citep{glas74,blac77,blac78b,vand86,jann88,vand88,vand89,vial86,vial88,heck93,lepe04,shaw06} have been performed to study the diffuse and translucent clouds.

\subsubsection{H\,{\sc ii} and photon-dominated regions}
\begin{figure}
\centering
\includegraphics[width=0.6\textwidth]{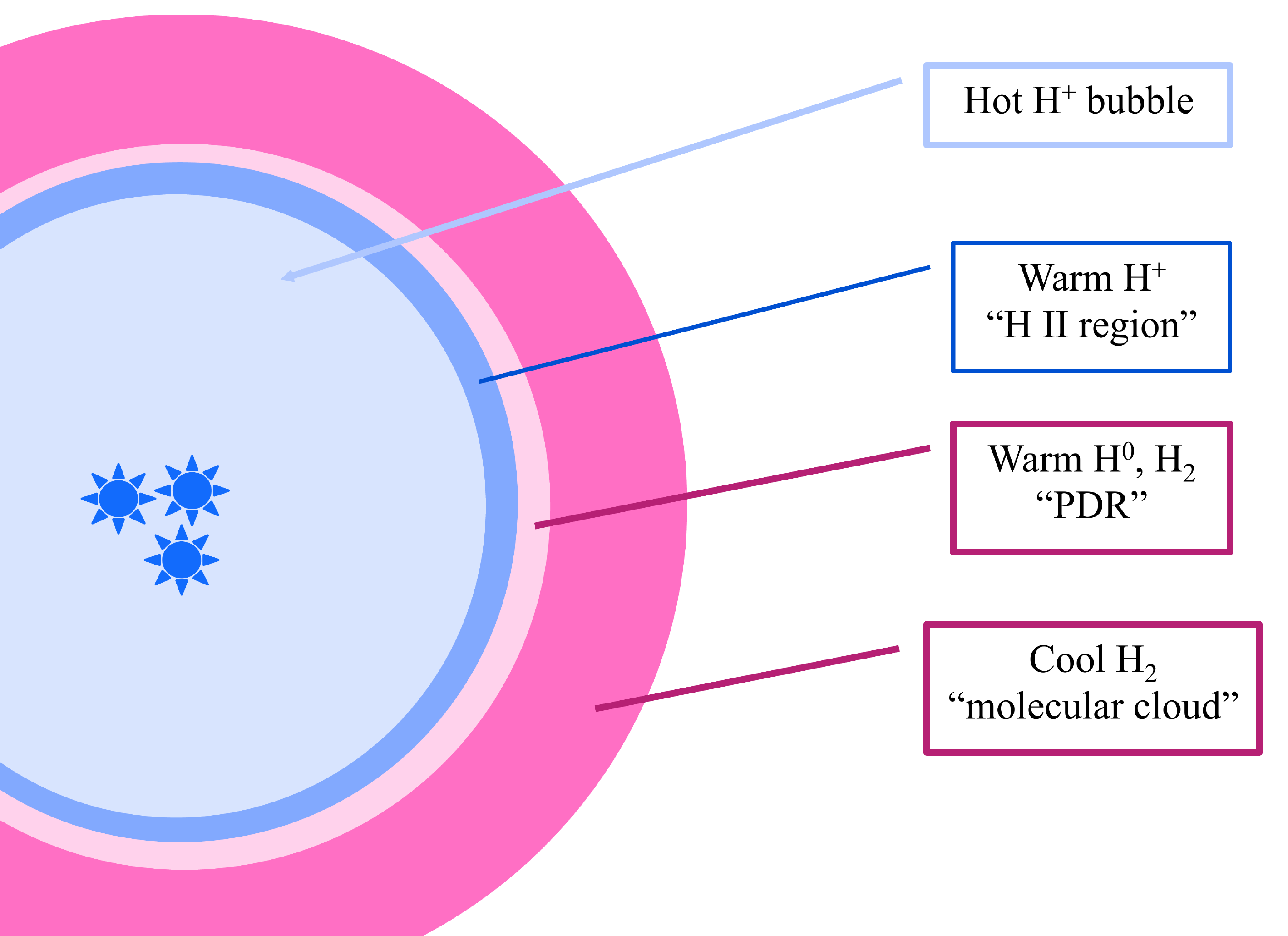}
\caption{Idealized structure of an H\,{\sc ii} region and PDR (Courtesy: \url{https://cloud9.pa.uky.edu/~gary/cloudy/CloudySummerSchool/2015_Pune/2_Thermal_equilibrium.pdf}).}
\label{fig:HII_PDR}
\end{figure}

H\,{\sc ii} regions are created when the extreme UV radiation from a star ionizes (H to H$^+$)
and heats the surrounding gas.
The gas in these regions is ionized and has a temperature of about $10^4$ K. Densities range from $10^3$ to $10^4$ cm$^{-3}$ for compact ($\sim0.5$ pc) H\,{\sc ii} regions such as the Orion Nebula to $\sim10$ cm$^{-3}$ for more diffuse and extended nebulae such as the North
America Nebula ($\sim10$ pc).
H\,{\sc ii} regions are formed by young
massive stars with spectral type earlier than about B1 ($T_{eff} > 25000$ K), which
emit copious amounts of photons beyond the Lyman limit ($h\nu > 13.6$ eV) and
ionize and heat their surrounding clouds.
All molecules are dissociated into atoms and ionized to form a high-temperature plasma.
An idealized structure of an H\,{\sc ii} region is indicated in Figure \ref{fig:HII_PDR}.

Photon-dominated or photodissociation regions (PDRs) are parts of the interstellar clouds that separate ionized H\,{\sc ii} region and cold molecular clouds (see Figure \ref{fig:HII_PDR}) near bright luminous O and B
stars \citep{holl99} and so are exposed to intense UV radiation enhanced by a factor of $10^5$ concerning average ISRF
and warmed by high levels of photons \citep{holl97}, which controls
both the physical and chemical state of the gas.
Here, the gas is heated by the
far-UV radiation ($6 < h\nu < 13.6$ eV, from the ambient UV field and hot stars) and cooled via the emission of spectral line radiation of atomic and molecular species and continuum emission by dust.

At the edge of the cloud,
the gas temperature becomes so high ($\sim 1000$ K) that endothermic processes and reactions with energy barriers can proceed. PDR’s are ubiquitous in the ISM.
The Orion Bar PDR and the Horsehead nebula in the same constellation Orion are prototype examples of well-studied dense PDRs.

\begin{figure}
\centering
\includegraphics[width=0.5\textwidth]{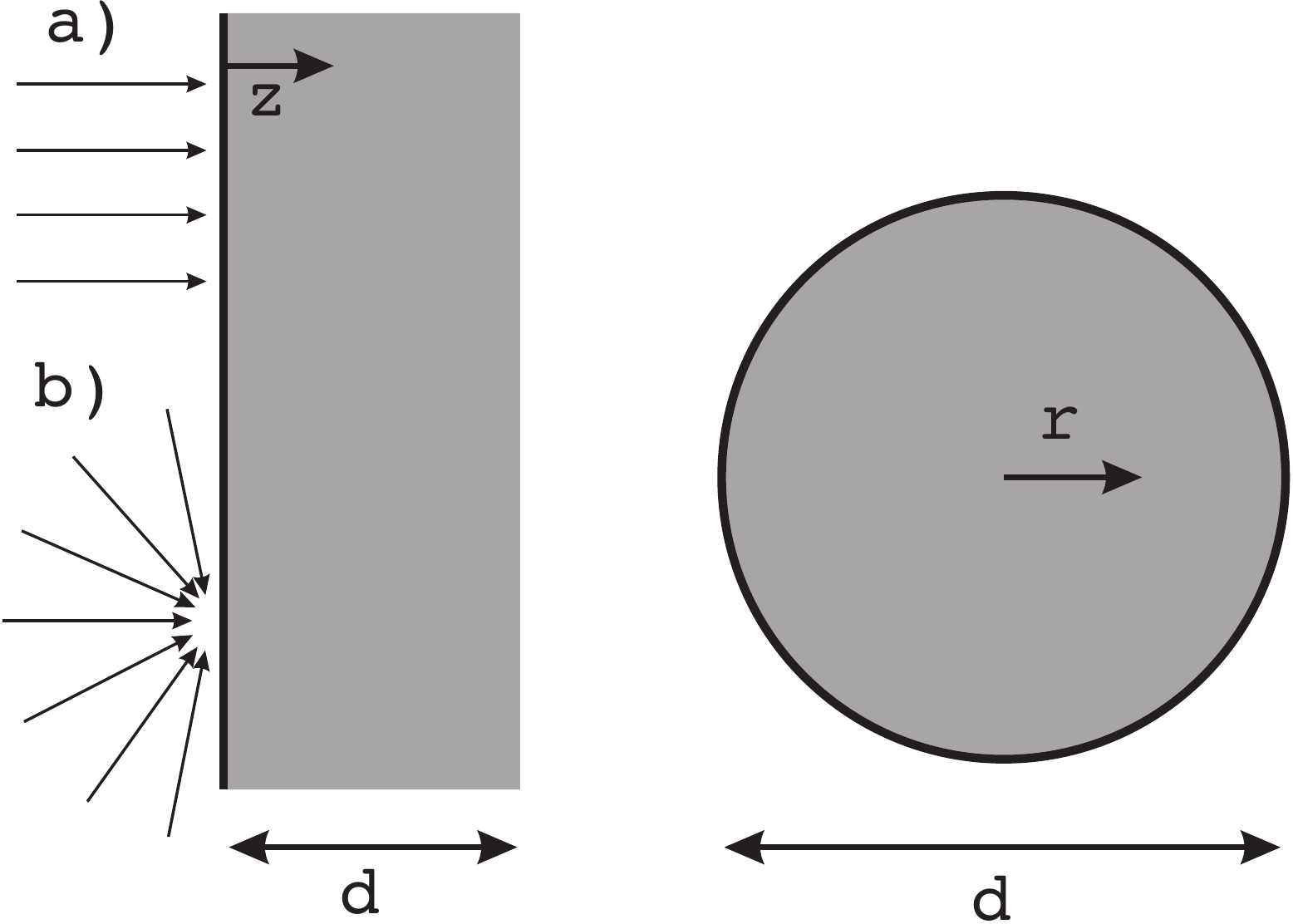}
\caption{Common geometry of the cloud considered for a PDR model calculations.
The surface of any plane-parallel or spherical cloud is illuminated either a) uni-directionally or b) isotropically. \citep[Courtesy:][]{roll07}.}
\label{fig:PDR_geometry}
\end{figure}

The standard PDR model is a one-dimensional system in which slabs of differing temperature
and density are considered, starting from the outside with low densities and high temperatures
and proceeding to the inside with lowering temperatures and increasing densities.
The excitation by electrons rather than $\rm{H_2}$ or H can become significant in PDRs.
Two typical geometrical setups of model PDRs are shown in Figure \ref{fig:PDR_geometry}.
Most PDR models consider a plane-parallel geometry, illuminated either from one side or both sides suggesting a directed illumination perpendicular to the cloud surface. Since most plane-parallel PDR models are infinite
perpendicular to the cloud depth $z$ it is also straightforward to
account for an isotropic far-UV irradiation within the pure 1-D formalism.
For a spherical geometry, one can exploit the model symmetry only for an far-UV
field isotropically striking onto the cloud.
The ambient far-UV is provided in units of Draine $\chi$ \citep{drai78} or Habing $G_0$ \citep{habi68} fields.

\begin{figure}
\centering
\includegraphics[width=0.8\textwidth]{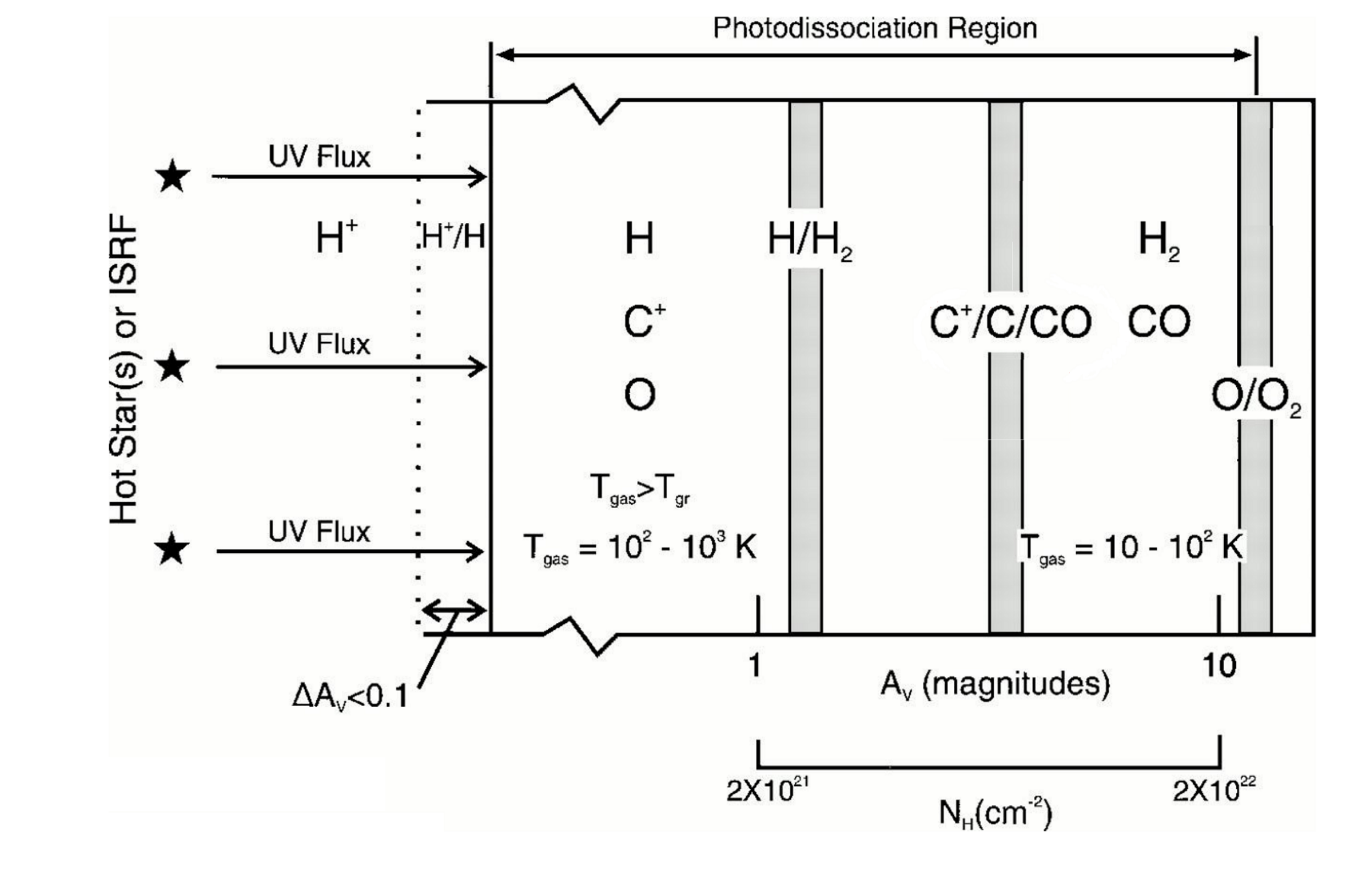}
\caption{A schematic view of a PDR structure. The PDR is illuminated from the left by a strong far-UV field \citep[Courtesy:][]{holl97}.}
\label{fig:PDR}
\end{figure}

PDRs play an essential role in interstellar chemistry.
They are responsible for the emission characteristics of ISM and dominate the IR and
submillimeter spectra of star-forming regions and galaxies.
Several earlier literature \citep{holl71,jura74,glas75,blac77,tiel85,vand88,holl91,fuen93,lebo93,ster95,jans95a,bert96,lee96,bake98,holl99,walm99,roll02,sava04,teys04,fuen05,meij05} have discussed numerous aspects of PDR chemistry in detail and
provided a comprehensive overview of the field.
PDR host the critical $\rm{H^+/H/H_2}$ and $\rm{C^+/C/CO}$ transition layers of the ISM.
As one traverses inward, the dominant species change once the chemistry is evolved (see, Figure \ref{fig:PDR}). For example, if we
consider the major forms of carbon: the outer layers are dominated by C$^+$, the innermost layers by CO, and some intermediate layers have significant amounts of neutral atomic carbon. Hydrogen
is atomic in the outer layers, but within a short distance inside an
efficient conversion to the molecular form is witnessed.
\cite{roll07} have benchmarked the physical and chemical structure of PDRs using several codes and compared among them.

\subsubsection{Supernova remnants}
Supernova remnants (SNRs) are formed when the material ejected in the explosion of an aged high-mass star surrounding ISM material
terminating the life of the star.
SNRs emit radiation in all electromagnetic wavelengths.
The Crab Nebula is a compact SNR.

\subsubsection{The spectral synthesis code}
\textsc{Cloudy} is a photo-ionization microphysics code, which simulates matter under a broad range of interstellar conditions. It is for general use under an open-source
license\footnote{\url{https://gitlab.nublado.org/cloudy/cloudy/-/wikis/home}}. \textsc{Cloudy} is designed (by numerical simulations) to understand complex physical environments starting from the first principles.
The equations of statistical equilibrium, charge conservation, and energy conservation are solved \citep{oste06}. \textsc{Cloudy} code deals with the range of temperatures that extends from the CMB temperature ($2.7$ K) up to $1.001\times10^{10}$ K, and the physical state ranges from fully molecular to bare nuclei.
It determines the physical conditions
within a non-equilibrium gas, possibly exposed to an external source of radiation, and determines the level of ionization, the particle density, the kinetic gas temperature, chemical state
of a cloud, populations of levels within atoms, and then predicts the resulting full-spectrum (which often includes hundreds of thousands of lines) and many observed quantities self-consistently by specifying only the minimum number of free parameters (i.e., the properties of the cloud and the radiation field exposed on it).
The level of ionization is calculated by balancing all ionization
(photo, Auger, and collisional ionization and charge transfer) and
recombination processes (low-temperature dielectronic,
high-temperature dielectronic, three-body recombination, charge transfer, and radiative). The free electrons are assumed to belong to a predominantly Maxwellian velocity distribution with a kinetic temperature, which is determined by the balance between heating (mechanical, photoelectric, cosmic-ray, etc.) and cooling (mainly inelastic collisions between the electrons and other particles) processes. The associated continuum and line radiative transfer is solved simultaneously.

The \textsc{Cloudy} code was first developed in August 1978 at the Institute of Astronomy, Cambridge.
Like all programs, \textsc{Cloudy} goes through versions as it is developed through  computing and compiling all the atomic data (line positions, oscillator
strengths, collisional and photoionization cross sections, recombination rate coefficients)
that enter the program \citep{ferl98,ferl13,ferl17}.
We use the latest version, 17.02, last reviewed by \cite{ferl17} for our models presented in this dissertation.

Here I describe some important aspects in the following subsections.

\subsubsection{What must be specified?}
\textsc{Cloudy} needs the following input parameters before it can predict conditions within a cloud: \\
1. The shape and intensity/luminosity of the external ISRF striking a cloud. \\
2. The total hydrogen density. \\
3. The chemical composition of the gas and whether grains are present. \\
4. The geometry of the cloud, including its radial extent (thickness).

In general, the radiation field striking the cloud can be specified either of the two ways: In the
luminosity case the total luminosity emitted by the central object and the inner radius of the cloud is needed,
whereas, in the intensity case, only the flux of radiation striking the cloud is specified.
Up to $100$ different radiation fields can be included in the total field striking the cloud.
However, there must be the same number of shapes and
intensity (luminosity) specifications.
The program considers the lightest 30 elements in detail. Abundances are
specified by number relative to hydrogen. Grains are not part of the default
composition but can be included. All stages of ionization are treated, and all published charge
exchange, radiative recombination, and dielectronic recombination processes are included as
recombination mechanisms. In addition, photoionization from the valence and inner shells is considered.
Many excited states and collisional ionization by both thermal and supra-thermal electrons, and charge
transfer, are included as ionization mechanisms.
The geometry is always considered to be one Dimensional and spherically symmetric but can be made effectively plane parallel by making the inner radius much
larger than the thickness of the cloud. The default is for the gas to have constant density and to fill its volume fully.

\subsubsection{Radiation fields}
\begin{figure}
\centering
\includegraphics[width=0.6\textwidth]{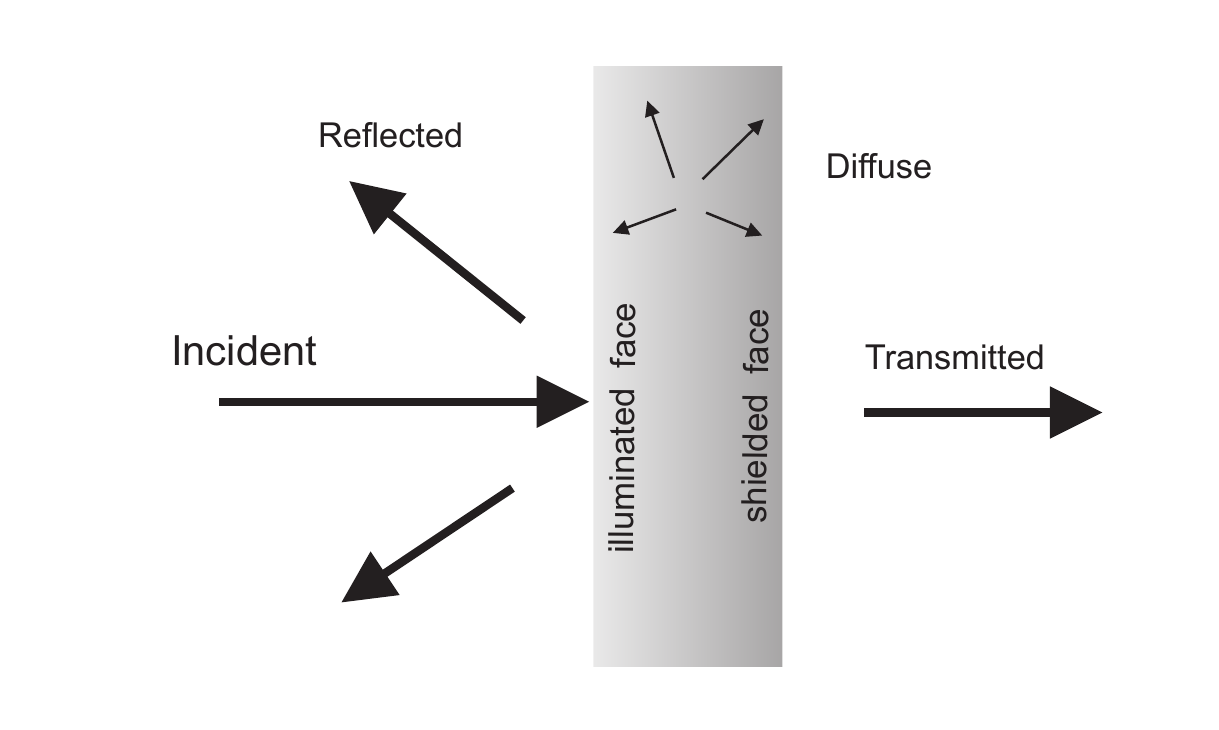}
\caption{Radiation fields that enter in the calculations \citep[Courtesy: Part 1 of \textsc{Hazy};][]{ferl17}.}
\label{fig:rad_field}
\end{figure}

Figure \ref{fig:rad_field} shows several of the radiation fields (incident, diffuse, transmitted, and reflected) computed in the calculation. The side of the cloud facing the source of the external radiation field is the illuminated face of the
cloud. The opposite side, in shadow, is the shielded face of the cloud. The luminous face is
generally hotter and more highly ionized than the covered face. The external radiation field emitted by the central object that strikes the illuminated face of the cloud is the incident radiation field, which is often the only energy source for the cloud. It is diminished by extinction as transmitted through the cloud. The spectral energy distribution (SED) of the incident radiation field is specified between
the energies of $3.040 \times 10^{-9}$ Ryd ($\lambda \approx 29.98$ m) and $100$ MeV $\approx 7.354 \times 10^6$ Ryd. The radiation field emitted by
gas and grains within the nebula is known as the diffuse radiation field (often referred to as the diffuse fields, e.g., the Lyman, Balmer, or two-photon continua emitted by hydrogen). These fields are very nearly isotropic and can be significant sources of ionizing radiation under some circumstances.
The transmitted radiation field is considered as the net emission emergent from the shielded face of the cloud.
Therefore, both the attenuated incident field and the diffuse radiation field are included.
On the other hand, the reflected radiation field is considered the emission from the illuminated face of the cloud into the
direction towards (i.e., within $2\pi$ sr of) the source of the external field. Thus, the reflected field
results from both backscattered incident radiation and diffuse emission emitted from the cloud
toward the source of ionizing radiation.

\begin{figure}
\centering
\includegraphics[width=0.6\textwidth]{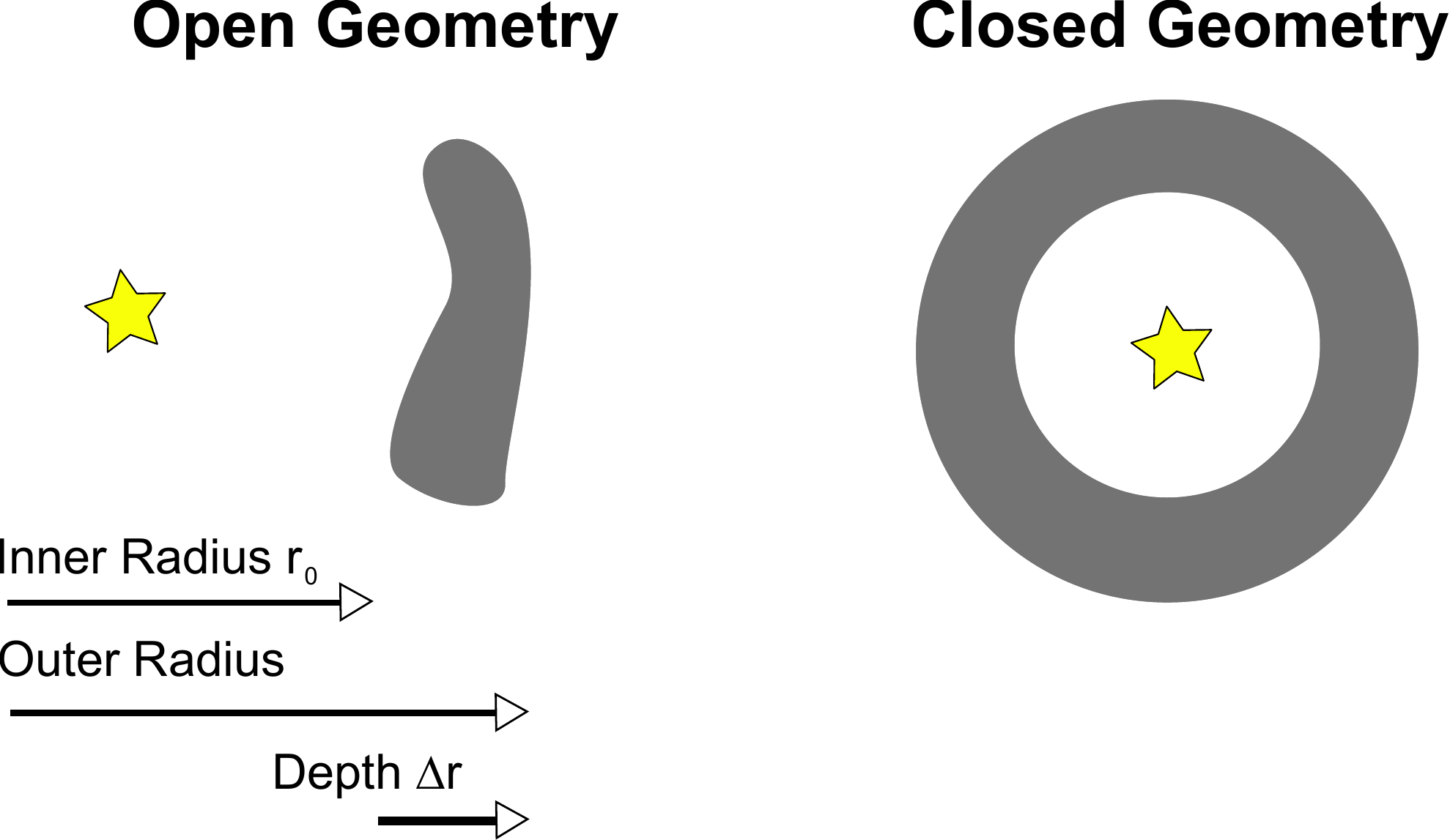}
\caption{Examples of types of geometry of the cloud which can be considered in the calculations \citep[Courtesy: Part 1 of \textsc{Hazy};][]{ferl17}.}
\label{fig:geometry}
\end{figure}

\subsubsection{Geometry}
Figure \ref{fig:geometry} shows two possible geometries and some terminologies used to describe them. This radius is
the distance from the center of symmetry, usually the center of the central object, to a given point.
The depth is the distance between the illuminated face of the cloud and a point within the cloud.
The inner radius is $r_0$, and the thickness or depth is $\Delta r$, and the current radius is $r$.
Figure \ref{fig:geometry} shows examples of the assumed two limiting cases, referred to as open and closed geometry, having a different influence upon the calculations. To know the terms better, we first need to understand what the covering factor is. The covering factor is the fraction of $4\pi$sr covered by the gas, as viewed from the central source of radiation. It is usually written as $\Omega/4\pi$, has the limits $0 \leq \Omega/4\pi \leq 1$, and is the fraction of the radiation field emitted by the central object strikes nebular gas. An open geometry is one in which the covering factor of the gas is negligible. Thus, all
radiation that escapes from the illuminated face towards the continuous radiation source escapes from the system without further interaction with gas.
Emission-line gas covers $\sim 4\pi$ sr as seen by the central object in a closed
geometry. The central object is small relative to the nebula, then all diffuse fields that
escape from the illuminated face of the cloud in the direction towards the central object will strike the far side of the nebula.
In Figure \ref{fig:geometry}, the source of
ionizing radiation is denoted by a star and the shaded area represents the cloud.
The inner radius ($r_0$) and thickness or depth ($\Delta r$) of the cloud determine whether the geometry is plane parallel ($\Delta r / r_0 <0.1$), a thick shell ($\Delta r / r_0 <3$), or spherical ($\Delta r / r_0 \geq 3$).

\subsubsection{Intensity and luminosity cases}
The external radiation field is usually specified with two different commands. One defines the
shape of the incident radiation field. The second command sets the brightness of the light.
The illumination of the radiation field striking the cloud can be specified as intensity or
luminosity.

In the case when intensity is considered as the input, the energy flux (erg cm$^{-2}$ s$^{-1}$) or photon flux (cm$^{-2}$ s$^{-1}$) striking a unit area of cloud is set. The inner radius does not need to be specified,
although it is possible to do so. If the starting radius is not specified, then an inner radius of $10^{30}$ cm is assumed. Usually it results in a plane-parallel geometry.
If the starting radius is specified, then the resulting geometry may be
spherical, plane-parallel, or a thick shell, depending on the thickness (depth) ratio to inner radii. The emission per unit area (erg cm$^{-2}$ s$^{-1}$) is predicted.

In the case when the luminosity is used as the input, the total luminosity of the central source of radiation is specified.
The inner or starting radius of the cloud must be set so that the flux of photons can be derived.
Finally, the luminosities of emission lines (erg s$^{-1}$) are predicted.

\begin{figure}
\centering
\includegraphics[width=0.4\textwidth]{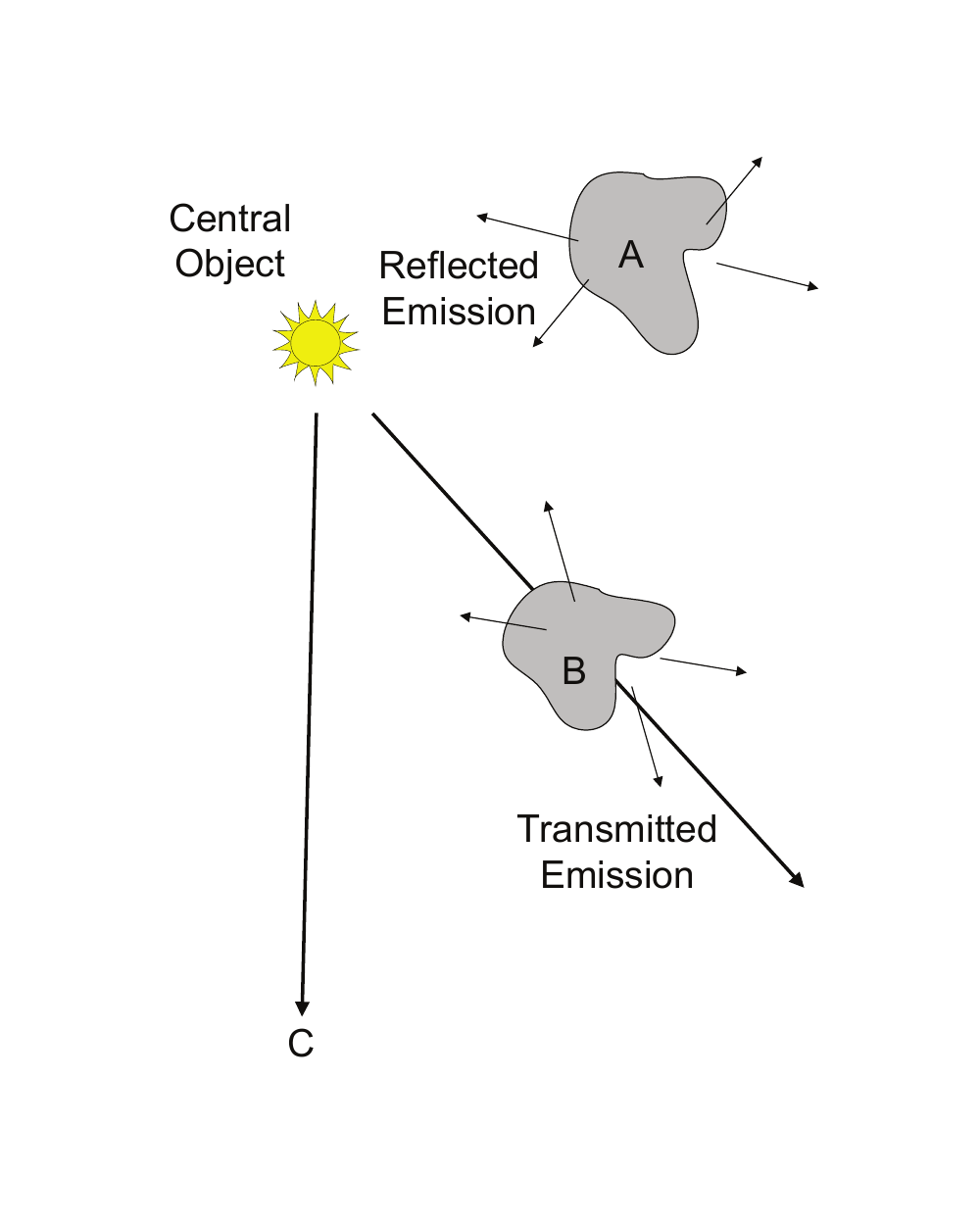}
\caption{Several components of the radiation field that enter in the calculations \citep[Courtesy: Part 1 of \textsc{Hazy};][]{ferl17}.}
\label{fig:geometry_radiation}
\end{figure}

\subsubsection{From the observation perspective}
Two lines of sight to the central object and two clouds are shown in Figure \ref{fig:geometry_radiation}. Each cloud produces both a reflected and transmitted emission component.
Three possible geometries, indicated by the letter on the Figure, occur depending on how the central source and the clouds are viewed:
(A) the central object cannot be directly observed, although
it can be seen by reflection from the illuminated face of a cloud,
(B) the transmitted continuum and the outward emission from the emitting cloud are observed, and
(C) the unattenuated continuum directly without absorption is observed.

\subsection{Modeling radiation-shielded region}

\subsubsection{The protostars and hot core/corino stage}
The dense and warmer places of high-mass star-forming regions are known as hot
molecular cores (HMCs), and their low mass analogs are known as the hot corinos.
The most recent studies contain up to three
stages to explain the chemical treatment of hot cores and corinos.
The initial (cold) stage occurs as the cold core ages and undergoes isothermal collapse.
At this time, an icy mantle is built up around the granular cores of silicates and carbon. The following
(warm-up) stage occurs during the passive warm-up of the inner envelope of the protostar to
temperatures of $100-300$ K and the subsequent sublimation of the ice mantles. Finally, the last (hot-core)
stage occurs at the temperature of the hot core or corino. During
the cold, warm-up, and hot-core stages, the individual chemistries produce zeroth-generation,
first-generation, and second-generation organic molecules, respectively \citep{herb09}.

Zeroth-generation species are large molecules formed in the gas and granular icy mantles during the cold core ages. An example is methanol (CH$_3$OH), produced by the surface hydrogenation of gas-phase carbon monoxide that accretes onto the cold mantle.
During the warm-up stage, a cold core becomes a young stellar object (YSO).
First-generation species are formed at least partially in and on granular mantles.
Here photodissociation of zeroth-generation species such as methanol and formaldehyde produces
radicals such as CH$_3$O and HCO.
The radicals can diffuse readily and associate to form larger molecules
such as methyl formate ($\rm{HCOOCH_3}$) at surface temperatures significantly above $10$ K.
Once the core become a hot core or corino, the second-generation species are formed.
In this stage, the temperature becomes $100-300$ K, which is high enough
to evaporate the icy mantles completely.
The gas-phase chemistry dominates via both $\rm{ion-molecule}$ and $\rm{neutral-neutral}$ reactions.
Because the ices on the grain mantles sublimate during the warm-up stage, the only chemistry
that can occur in the next, or hot-core, stage is the gas-phase chemistry. Because the temperature
is much larger than that of cold cores, many more reactions, including endothermic
processes and exothermic processes with barriers, can occur efficiently.
Eventually, complex molecules are destroyed by the gas-phase chemistry unless incorporated into disks.
Thus, this three-stage approach represents the most straightforward way of accounting for the history
of hot cores/corinos, an accounting in which physical conditions are time-dependent but not heterogeneous.

\subsubsection{The chemical model for molecular cloud code}
Assuming the gas and grains coupled through accretion, thermal, and non-
thermal desorption processes, we developed the chemical model for molecular cloud (hereafter, CMMC) code
\citep{sil18,sil21,das19,das21,gora20b,shim20}.
The gas-phase chemical network of the CMMC code mainly adopted from the UMIST
database \citep{mcel13}.
A cosmic-ray rate of $1.3 \times 10^{-17}$ s$^{-1}$ is considered in all our models.
For the grain surface reaction network, we mainly followed \cite{hase92,cupp07,ruau15,das15b,gora20b}.
The BE of the surface species is mostly considered from KIDA \citep{wake17}, and sometimes
from our own quantum chemical calculations \citep{das18,das21,sil21}.

\subsection{Radiative transfer modeling}
Radiative transfer calculation is a handy technique for extracting physical conditions
from observed line profiles or generating synthetic spectra.
From the radiative transfer calculation, we can get the molecular abundances from the telescopic data in IR and submillimeter wavelength.
The level of accuracy needed for radiative transfer calculation depends on the amount and nature of the available data like collisional rate coefficients discussed in subsection \ref{sec:theory_model}.
\cite{vand11} reviewed the radiative transfer calculations in detail.

\subsubsection{LTE model}
The most straightforward approach to continue the radiative transfer modeling calculation is to consider the Boltzmann distribution to estimate the level population taking the excitation
temperature ($T_{ex}$) equal to the kinetic temperature ($T_{kin}$). This approximation is called local thermodynamic equilibrium (LTE). But often this approximation does not work.
Nevertheless, this condition holds well around the high-density region where collision determines
the excitation. $T_{ex}$ is determined from a range of energy levels where several lines are observed for a molecule, the relative intensities of lines are plotted as a function of upper-level energy. From the slope of the graph, we can determine $T_{ex}$, and the diagram is called the ``rotation diagram'' \citep{gold99} for rotational energy levels.

\subsubsection{Non-LTE model}
In the low-density region, the number of molecules in each energy level is not
well-described by a Boltzmann distribution as the gas density lower than the density required to thermalize the population.
In that case, both the radiative (emission or absorption) processes and collisional (thermal) processes take place. As a result, $T_{ex}$ becomes less than $T_{kin}$ of the colliding gas.
This is known as the non-LTE formalism.
So a detailed calculation is required to take accounts of the radiative and collisional processes and balance between the excitation and de-excitation processes.
An overview of non-LTE radiative transfer methods is given by \cite{vanz02}.

\subsubsection{Radiative transfer codes}

RADEX \citep{vand07} is a one-dimensional non-LTE radiative transfer code that uses the escape probability formulation to assume
an isothermal and homogeneous medium without large-scale velocity fields.
It is publicly available as a part of LAMDA\footnote{\url{http://home.strw.
leidenuniv.nl/~moldata}}. In addition, this code provides an online version\footnote{\url{http://var.sron.nl/radex/radex.php}} for quick access.
It is a helpful way to analyze a large set of rapidly
observational data, providing constraints on some physical parameters density and kinetic
temperature. It provides an alternative to the widely used rotation diagram
method, which relies upon the availability of many optically thin emission
lines and is applicable only in roughly constraining the excitation temperature
in addition to the column density.

RATRAN \citep{hoge00} is another radiative transfer code to calculate the excitation of molecular lines.
The code can be implemented to all atoms or molecules for which collisional rate data is available and any axially symmetric source model.
Thus, it is a valuable tool in analyzing data from present and future IR and submillimeter telescopes. Furthermore, this code can treat both spherically symmetric (1D) and cylindrically symmetric (2D) source geometries.

LIME \citep{brin11} is a molecular excitation and non-LTE
spectral line radiative transfer code for 3D models for
arbitrary geometries. LIME can predict line strengths and profiles of molecular transitions and the intensity of the thermal continuum.
It can model disks around young stellar objects, proto-stellar envelopes, giant molecular clouds, outflows, and similar environments.
\cite{ober15,parf16,quen18b} have used this 3D non-LTE radiative transfer calculation
for modeling different astrophysical regions and spectral line emission properties from various species.

RADMC-3D\footnote{\url{http://www.ita.uni-heidelberg.de/dullemond/software/ radmc-3d}} is a radiative transfer code for dust and lines. It calculates for a geometrical distribution of gas and dust what its images and spectra would look like when viewed from a certain angle. Thus, it allows comparing their models with observed data. It can be applied to dusty molecular clouds, protoplanetary disks, CSEs, dusty tori near AGN (active galactic nuclei) and models of galaxies.                                                                                                                                                                                                                                                             

%% file: chap2.tex
\chapter{Interstellar Ice Features} \label{chap:BE_ice}

\section*{Overview}
The interstellar clouds belong to four different classes: diffuse atomic, diffuse molecular, translucent, and dense molecular \citep{snow06}. The size of these clouds vary from a few to
hundreds of light-years and are composed
of gas ($\sim 99\%$ by mass) and a tiny fraction of dust ($\sim 1\%$ by mass) particles. These clouds are
the birthplaces of stars. Our Sun is one of this kind, which was formed by this generic process and finally hosted the planets like our living home the Earth.
The colder regions of stellar outflows and supernova ejecta produce molecules and grains. 
Interstellar dust grains are composed of
crystalline and amorphous magnesium-rich silicates (e.g., olivine, pyroxene, etc.).
Carbonaceous compounds such as graphite and amorphous silicon carbide (SiC), and titanium carbide (TiC) clusters are also the constituents of dust grains.
At low dust temperatures ($\sim 10$ K), the atoms and molecules freeze out on dust grains, forming an icy layer or mantle. Freeze-out of the gas-phase elements onto cold grains builds up ice mantles ($\rm{H_2O,\ CO_2,\ CO,\ CH_3OH}$ are the major constituents).
Ices are the essential repository of the elements in dense regions of ISM.
In the broader astrophysical context, ices play crucial roles. It speeds up the grain coagulation process.
When the grain is heated up, these molecules sublime back into the gas.
This process is called thermal desorption process and it is controlled by the BEs or adsorption energies.

Thus, the determinations of
BEs of ices, either as pure or mixed ices, are essential for understanding the
observations of star- and planet-forming regions.
In this Chapter, we focus on two crucial directions of interstellar ices:
\begin{enumerate}
 \item  Binding energy: a key defining interstellar chemistry \citep{sil17,das18,das21}, and
 \item  Absorption features of interstellar ices in the presence of impurities \citep{gora20a}.
\end{enumerate}

\clearpage
\section{Binding energy: a key in defining interstellar chemistry}

One of the major obstacles to accurately modeling interstellar chemistry is an inadequate knowledge of the BE of interstellar species.
Abundances are governed by adsorption and desorption on grains. The chemical composition of the interstellar grain mantle is dependent mainly on the
BEs of the surface species. The BE of the interstellar species regulates the chemical complexity of the interstellar grain mantle. Since hydrogen is widespread either
in atomic or molecular form, we first review the variation
of the BEs of H and H$_2$ depending on the nature of the adsorbents.
Choice of adsorbents is based on the relative abundances of interstellar materials.
The abundance of interstellar ice constituents is generally expressed in terms of the water ice.
At the low temperature, the hydrogenation of oxygen is very active, and
$\rm{H_2O}$ molecules predominantly cover a significant part of the interstellar grain surface.
Therefore, the usage of BEs with bare grains is not useful in such scenarios.
So, the BE of surface species with the water is provided and is widely used in astrochemical modeling.
Thus the interaction of gas-phase species with water ice is essential to trace realistic physical and chemical processes.
We consider different clusters of water molecules to calculate the BE of several atoms, molecules, and radicals of astrochemical interest. Systematic studies are carried out to develop a relatively more accurate BE of astrophysically relevant species on water ice.
The calculated BEs are compared with the available experimental or theoretically obtained BE values.
$\rm{H_2}$ is also the dominant molecular species in the vast majority of interstellar environments.
Thus, its accretion rate on a grain is much higher in comparison to the other species.
In more dense regions, gas-phase H$_2$ easily accretes on the grain and mostly gets back to the gas phase due to their low sticking probability and lower BEs. But some H$_2$ could be trapped under some accreting species.
For example, one H$_2$ may accrete on another H$_2$ before it is desorbed. In this situation, the ``encounter desorption'' mechanism is proven to be an essential means of transportation of H$_2$ to the gas phase \citep{hinc15}. Usually, in the literature, only the encounter desorption of H$_2$ is considered due to its wide presence in the dense ISM. However, recently, \cite{chan21} considered the encounter desorption of the H atom as well.
So, we calculate the BE of several species considering $\rm{H_2}$ molecule as a substrate to examine the effect of encounter desorption on the availability of species on interstellar grain.
High-level quantum chemical calculations are performed to estimate the BEs of several relevant interstellar species. The obtained BEs are used in astrochemical models.

\subsection{Study of binding energies of H and H$_2$}
\label{sec:BE_H_H2_method}

Hydrogen is known as the most ubiquitous element in the Universe. The
spectral characteristics of molecular clouds in the far-UV and near-IR region are
strongly dominated by molecular hydrogen (H$_2$) due to its very high abundance \citep{tiel10}.
The H$_2$ formation process through gas-phase interactions between hydrogen atoms (H) or ions is too slow compared to the observed formation rate \citep{tiel10}. Therefore it is commonly believed that there must be other processes causing rapid H$_2$ formation. H$_2$ are formed by taking two H atoms and putting them together. Since the possibility of one single H atom finding a larger grain surface is
much higher than finding another H atom in the gas phase,
calculation of BE of H and H$_2$ with appropriate grain surfaces is essential
to understand their synthesis in the gas phase as well as interstellar icy grain mantle.
A larger grain surface may accrete many simple
species (atoms or molecules). There are two types of adsorption: (a) physisorption/molecular adsorption and (b) chemisorption/dissociative adsorption. Interstellar
grains play an essential role in molecule formation
at low-temperature regimes \citep{gree02}, and physisorption is always favored at low temperatures.
Incoming H atoms can stick with grain surfaces and form molecules via different
acceptable mechanisms like (a) Langmuir-Hinshelwood (LH), where two species can adsorb and react on neighboring sites \citep{lang22}, (ii) Eley-Rideal (ER), where formations can
happen by direct impingement of incoming species on an adsorbed species \citep{eley40,eley41} and
(iii) Hot-atom, where the incoming species with large impact parameters
may be trapped and react with other adsorbed species \citep{harr81}. Formation of H$_2$ on the grain surface has already been studied \citep{goul63,holl70,holl71,acha05,chak06a,chak06b,tiel10,sahu15}. Details of such formation
provide valued information about the physical and chemical evolution of the
space environments. H$_2$ is believed to be a precursor of many complex molecules.
One of its protonated forms, $\rm{H_3}^+$, acts as an essential coolant and saves itself along with other species from incoming radiation fields \citep{wake10}.

Here, we perform $ab\ initio$ calculations for the adsorbed
species (H and H$_2$) on benzene, silicate cluster, and water cluster systems to investigate the physisorption BE of H and H$_2$ and see their trend.

\subsubsection{Computational details and methodology}
The \textsc{Gaussian} 09 suite of programs \citep{fris13} is utilized
for all the BE quantum chemical calculations performed in our study.
In a periodic treatment of surface adsorption phenomena, the BE is related to the interaction energy ($\Delta$E), as:
\begin{equation} \label{eqn:BE_1}
    BE = -\Delta E
\end{equation}
For a bounded adsorbate, the BE is a positive quantity and is defined as:
\begin{equation} \label{eqn:BE_2}
    BE = (E_{surface}+E_{species})-E_{ss},
\end{equation}
where $\rm{E_{ss}}$ is the optimized energy for the complex system where a species is placed at a suitable distance from the grain surface through a weak bond so that a van der Waals interaction occurs during the optimization. $\rm{E_{surface}}$ and $\rm{E_{species}}$ are the optimized energies of the grain surface and species, respectively.

To find the optimized energy of all structures, we use a Second-order M\o ller-Plesset perturbation theory (MP2) level \citep{moll34} in conjunction with an aug-cc-pVDZ basis set.
`Aug' prefix is used to add diffusion function, and cc-pVDZ
is the Dunning's correlation consistent basis set \citep{dunn89} having double zeta function.
A fully optimized ground-state structure is verified
to be a stationary point (having non-imaginary frequency) by harmonic
vibrational frequency analysis.
Most of the calculations are performed without the inclusion of zero-point vibrational
energy (ZPVE) and basis set superposition error \citep[BSSE;][]{jans69,liu73} corrections.
Calculations are also performed by excluding BSSE using the counterpoise (CP) method \citep{boys70},
though the usage of CP corrections is highly debated.
CP correction is a method to limit an error that results when
studying an intermolecular reaction using an incomplete basis set.
In studies of weakly bound clusters, a BSSE can occur.
As monomer A approaches monomer B, the dimer (A \Cutline B) can be artificially stabilized as the monomer A utilizes the extra basis functions from the monomer B to describe its electron distribution and vice versa.
BSSEs are more pronounced for smaller basis sets.
One way to correct this is to use a larger basis set.
However, it is not always possible to use a larger basis set because
it is too computationally expensive.
Alternatively, one can calculate a CP correction, which approximates the bias
to the quality of the calculation that results in the intermediate range. 

We take the last iteration CP corrected energy of the dimer system $E_{CP-corrected}$ from
the \textsc{Gaussian} output file and calculate the BE,
including CP correction with the usual formula:
\begin{equation} \label{eqn:BE_3}
    BE_{CP} = (E_{surface}+E_{species})-E_{CP-corrected},
\end{equation}

\begin{figure}[htbp]
\centering
\includegraphics[width=0.5\textwidth]{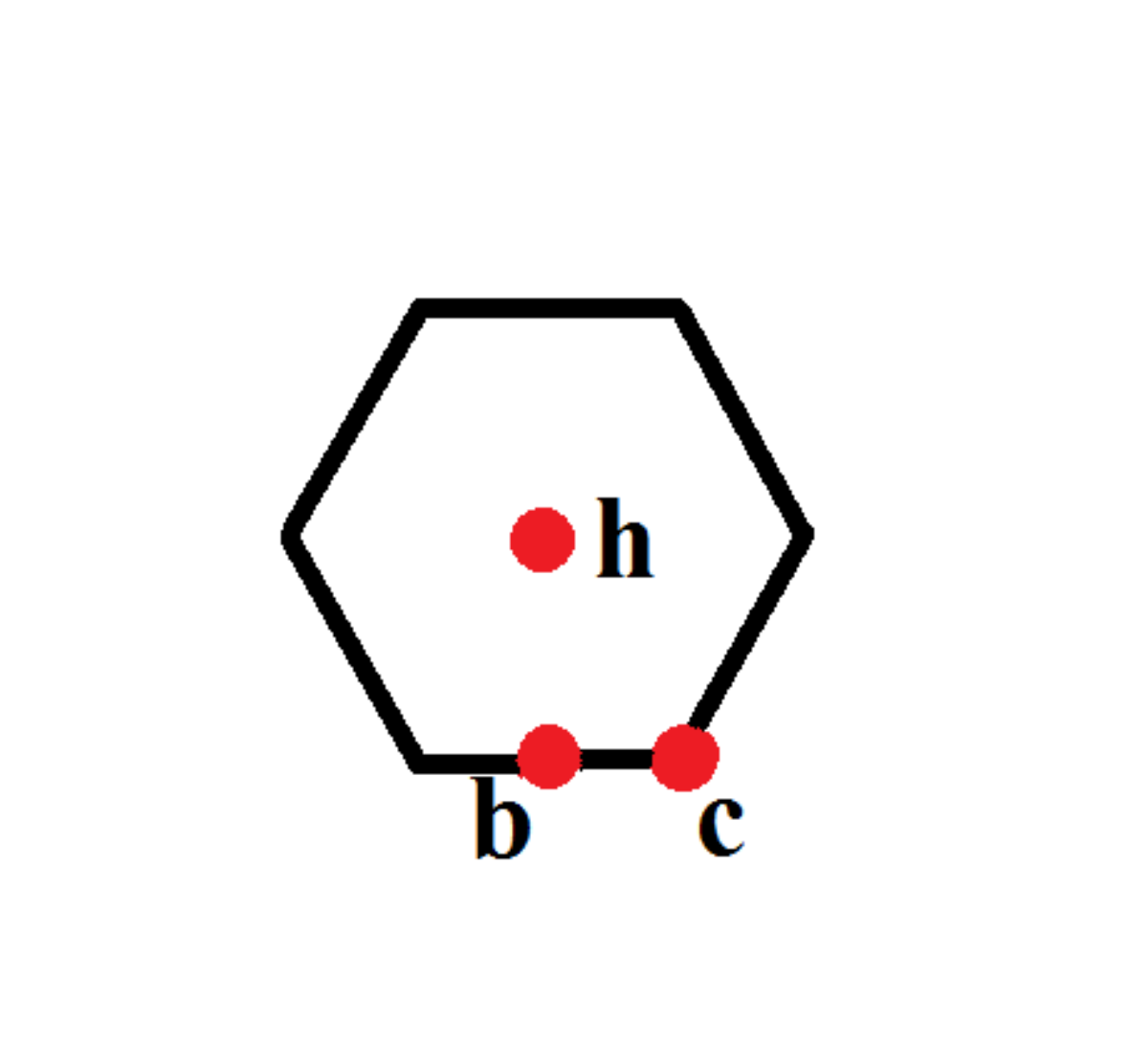}
\caption{Three favourable positions, c, h and b are shown \citep{sil17}. c position corresponds to the top of a carbon atom,
h (hollow) position is at the center of the hexagon, and b (bridge) position is at the mid-point
between two adjacent carbon atoms.}
\label{fig:BE_sites}
\end{figure}

\cite{avgu57,avgu60} discussed a general scheme for calculating the adsorption energy of non-polar molecules on a graphite surface. 
We also employ their relation for an educated estimation of the BEs of H and H$_2$.
As the input parameters, polarizability and diamagnetic susceptibility of H and H$_2$
are supplied by using quantum chemical calculations using the \textsc{Gaussian} 09 suite of programs with B3LYP/6-311g(d,p) level of theory.
For the placement of the adsorbed species on the lattice of graphite,
we use three different positions, as mentioned in \cite{barr37}.
First, over the carbon atom (position c); second, over the mid-point between two neighboring
carbon atoms (position b); and finally, over the center of the hexagon of carbon atoms (position h).
A schematic arrangement of H atoms on a simple unit of PAH molecule (benzene) for
these three different positions is shown in Figure \ref{fig:BE_sites}.

\cite{avgu57,avgu60} considered that H and H$_2$ are sticking with the
graphite as an adsorbent. They calculated the adsorption energy ($\Phi_i^{\prime\prime}$) of a segment of $i$ of adsorbate atom by,
\begin{equation}
 \Phi_i^{\prime\prime}(z)=-C_{i1}p_1z^{-q_1}-C_{i2}p_2z^{-q_2}-C_{i3}p_3z^{-q_3}+B_i^{\prime\prime}p_mz^{-q_m}
\end{equation}
where $B_{i}^{\prime\prime}$ is the constant associated with the repulsion, $z$ is the distance from the
$i^{th}$ segment, $p_1$, $q_1$, $p_2$, $q_2$, $p_3$, $q_3$, {\bf $p_m$, $q_m$}
are the constants available from the logarithmic graphs of \cite{avgu57,avgu60}
and $C_{i1}$, $C_{i2}$ and $C_{i3}$ are the constants.
For the computation of $C_{i1}$, $C_{i2}$, and $C_{i3}$, polarizability and
diamagnetic susceptibility of the adsorbate and adsorbent are required. Since we choose graphite as
the adsorbent, we use the values given in \cite{barr37} for the individual carbon atom
(polarizability $=0.937 \times 10^{-24}$ cm$^3$ and diamagnetic susceptibility $=-10.54 \times 10^{-30}$ cm$^3$).
For the adsorbates, the polarizability and diamagnetic susceptibility are calculated quantum chemically.
We obtain the values of mean polarizability $8.742887 \times 10^{-26}$ cm$^3$ for H atom
and $4.48999 \times 10^{-25}$ cm$^3$ for H$_2$ molecule and diamagnetic susceptibility of
$-4.07199302 \times 10^{-30}$ cm$^3$ for H atom and $-6.0225 \times 10^{-30}$ cm$^3$ for H$_2$ molecule.

We calculate the BEs of H and H$_2$ on graphite grain surfaces for various positions using the method mentioned above.
From Table \ref{tab:BE_values_sites}, it is clear that the most favorable adsorption
occurs at the hollow (h) site, which \cite{avgu60} had discussed earlier.
We perform additional calculations using the method presented by these authors to compare
with our calculated results.

\begin{table}
\scriptsize
\centering
\caption{BE of H and H$_2$ \citep{sil17} using the method used in \cite{avgu57}.}
\label{tab:BE_values_sites}
\begin{tabular}{cccc}
\hline
{\bf Species} & \multicolumn{3}{c}{\bf Binding energy (in Kelvin) for} \\
\cline{2-4}
& {\bf h position} & {\bf b position} & {\bf c position} \\
 \hline
 H  & 268 & 171 & 157 \\
H$_2$ & 944 & 594 & 545 \\
\hline
\end{tabular}
\end{table}

\subsubsection{Results and discussion}
Adsorption of species on the grain surface is the primary
process for any molecular formation mechanism. H and H$_2$ being the most abundant, it is essential to
know their BEs for various substrates.
Since BEs of the surface species mainly control surface chemistry,
an inaccurate estimation may cause wrong results.
This Section reports the results of our high-level quantum chemical calculations
for the BEs of H and H$_2$. Here, we use three types of substrates
as the grain materials: (1) benzene surface, (2) silica cluster, and (3) water cluster.

\begin{enumerate}

\item Benzene surface

\begin{table}
\scriptsize
\centering
\caption{BE of H and H$_2$ on the benzene surface \citep{sil17}.}
\label{tab:BE_benzene}
\vskip 0.2cm
\begin{tabular}{cccc}
\hline
{\bf Species} & \multicolumn{3}{c}{\bf Binding energy (in Kelvin) for} \\
\cline{2-4}
& {\bf h site}  & {\bf b site} & {\bf c site} \\
\hline
H  & 217 (336$^a$) & 117 & 105 \\
H$_2$  & 315 (1006$^a$) &   &  \\
\hline
\end{tabular} \\
\vskip 0.2cm
{\bf Note:} $^a$ without CP.
\end{table}

\begin{figure}[htbp]
\centering
\includegraphics[width=\textwidth]{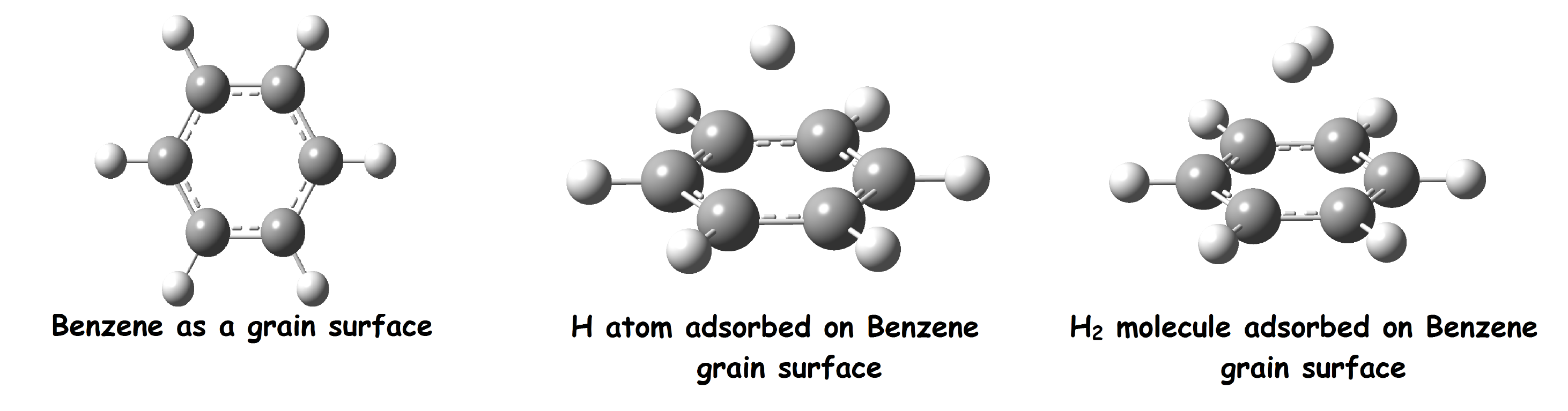}
\caption{Benzene as a representative grain surface \citep{sil17}.}
\label{fig:BE_benzene}
\end{figure}

The hydrogen-graphite and hydrogen-PAH systems are considered to investigate H$_2$ formation on realistic grain surfaces.
We consider benzene as a representative of graphite, graphene, or PAHs.
Although the coronene surface is more suited to mimic the interstellar carbonaceous grain, we use a benzene
surface to save computational time.
Benzene is a vital organic compound having the
chemical formula $\rm{C_6H_6}$.
\cite{bonf07} found that the CP corrected BE of H is $18.8$ meV with the
most favorable hollow site of benzene surface using MP2/aug-cc-pVDZ level of theory. They also calculated the CP corrected BE of H to be $18.7$ meV using the higher-level coupled-cluster single-double and perturbative triple excitation [CCSD(T)] method, taking the same basis set. The resulting differences between the MP2 and CCSD(T) methods were negligible in the broader astrophysical aspects. Thus to save computational
time, we also employ the MP2/aug-cc-pVDZ level of theory for our BE calculations for H and H$_2$.
The standard quantum chemical method is used for a $\rm{hydrogen-benzene}$ system to determine
physisorption BEs over the high-symmetry sites accurately.
Here, the BE of the H atom is calculated for three suitable
positions of the benzene surface.
For the calculations, CP correction is employed to minimize the BSSE.
Calculated BEs at three positions considered in Figure \ref{fig:BE_sites}
are presented in Table \ref{tab:BE_benzene}.
It is clear from Table \ref{tab:BE_benzene} that the hollow site
is the most favorable location because it has a higher BE.
We find the CP corrected BE of H is 217 K (18.7 meV) with hollow site of benzene surface.
The value is very similar to that obtained by \cite{bonf07}.
In the case of H$_2$, only the most favorable position (h site) is employed.
BE values with CP corrections and without CP corrections (shown in the parentheses)
are listed in Table \ref{tab:BE_benzene}. Figure \ref{fig:BE_benzene} shows
the structure of benzene, along with the adsorption of H and H$_2$ in
the hollow site. The BEs computed on benzene surfaces follow a similar
trend as theoretically calculated values presented in Table \ref{tab:BE_values_sites}.

\item Silica cluster

To calculate the BEs of H and H$_2$ on bare silicate grain (a compound containing
silicon), we consider
three silica molecules together. For better accuracy, it is required to increase the size of the cluster.
However, it would
take a substantial computational time to converge. Thus we restrict ourselves by considering
$(\rm{SiO_2})_3$ to mimic the silicate nature of
the grain surface. Figure \ref{fig:BE_silica} depicts the silica cluster and the position of adsorbed
species (H and H$_2$) with the silica cluster. In Table \ref{tab:BE_silica}, the calculated BE values are shown.
Here also, we find that BE of H$_2$ is higher than that of H.

\begin{figure}[htbp]
\centering
\includegraphics[width=\textwidth]{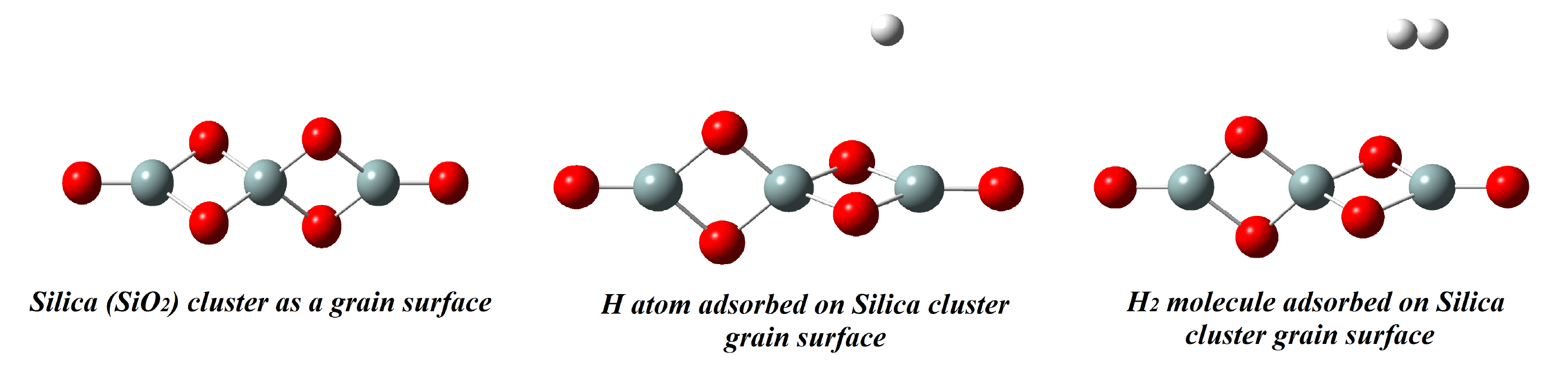}
\caption{Silica cluster as a representative grain surface \citep{sil17}.}
\label{fig:BE_silica}
\end{figure}

\begin{table}
\scriptsize
\centering
\caption{BE of H and H$_2$ on silica surface \citep{sil17}.}
\label{tab:BE_silica}
\vskip 0.2 cm
\begin{tabular}{cc}
\hline
{\bf Species} & {\bf Binding energy (in Kelvin)} \\
\hline
 H  & 580 \\
 H$_2$ & 1090 \\
\hline
\end{tabular}
\end{table}

\item Water cluster

To calculate the BE of H and H$_2$, here we use the
water cluster (six water molecules as a grain surface) to mimic the water ice grain mantle.
\cite{ohno05} calculated various
geometries of water clusters [$\rm{(H_2O)}_n$, for $n=1-8$].
They found that the most stable structure of the water hexamer is the chair configuration.
Thus, we consider the hexamer structure (chair) of the water molecule as a substrate here.
A significant portion ($\sim 70\%$ by mass) of the interstellar grain mantle is covered with
water molecules around the dense cloud region of ISM \citep{kean01,whit03,gibb04,das10,das11,das16}.
Thus the incoming gas species may be directly adsorbed onto the water ice. So, knowing the BEs
of the adsorbed species with water ice is essential to
estimate better the composition of interstellar grain mantle in dense cloud regions.
Figure \ref{fig:BE_water} shows the structure of the water cluster (c-hexamer chair configuration) along with the
adsorbed configuration of H and H$_2$. Table \ref{tab:BE_water} shows the calculated BEs (no CP correction is
considered here) with the water cluster.
As in other substrates, the water cluster also yields a similar trend between the BEs of H and H$_2$.
However, we find an exciting trend for the BEs obtained with the size of the water cluster ($n=1-6$).
As we increase the cluster size of the water molecules, we obtain higher values of BEs.
For example, the BEs of H and H$_2$ are $30$ K and $361$ K with the water monomer,
$125$ K and $528$ K with the water c-tetramer, and $181$ K and $545$ K with the water c-hexamer chair configuration. However, this trend is valid up to six water molecules clusters studied here.
Beyond this, a saturation with experimental values of BE is expected.

\begin{figure}[htbp]
\centering
\includegraphics[width=\textwidth]{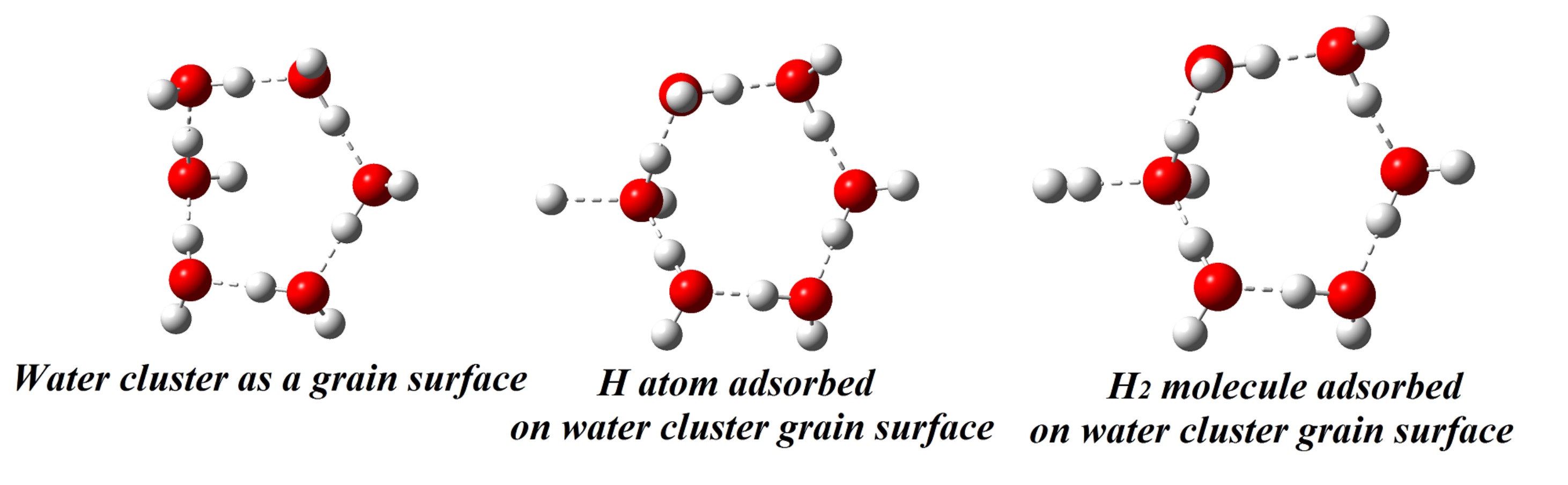}
\caption{Water cluster (c-hexamer chair configuration) as a grain surface \citep{sil17}.}
\label{fig:BE_water}
\end{figure}

\begin{table}
\scriptsize
\centering
\caption{BE of H and H$_2$ on water surface \citep{sil17}.}
\label{tab:BE_water}
\vskip 0.2 cm
\begin{tabular}{cc}
\hline
 {\bf Species} & {\bf Binding energy (in Kelvin)} \\
\hline
 H  & 181 \\
 H$_2$  & 545 \\
\hline
\end{tabular}
\end{table}

\end{enumerate}

\subsubsection{Binding energy of H and H$_2$ already available from the earlier studies}
Table \ref{tab:BE_sets} presents different sets of available theoretical and experimental values of BEs of H and H$_2$. Set 1 corresponds to the theoretical values obtained by \cite{alle77}.
For olivine grain (Set 2), they determined $373$ K and $314$ K BE values for H and H$_2$, respectively.
The desorption energies of H and H$_2$ were experimentally determined by \cite{katz99,pirr99} for amorphous carbon with values of $658$ K for H and $542$ K for H$_2$ (Set 3). \cite{ghio80} calculated the van der Waals
interaction between graphite and H with energy $503 \pm 6$ K. This value fully supports
the experimental data given by \cite{vida91}, where the best estimate was given as $499 \pm 3$ K.
\cite{han04} used the density functional theory (DFT) method to calculate
a value of $400$ K for H$_2$ interacting with carbonaceous material (carbon nanotube), while \cite{vida91}
gave the best estimate of $660 \pm 6$ K for $\rm{H_2-graphite}$ system. The estimated values of
\cite{vida91} are shown in Table \ref{tab:BE_sets} as Set 4.
\cite{duli05} found a large distribution of energies for H$_2$ using
TPD experiments. According to their study, the mean value of the BE of H$_2$ can be $520$ K for
the non-porous substrate, which is very close to the theoretical value of $550$ K \citep{holl70} and experimental value of $550 \pm 35$ K \citep{sand93}.
\cite{horn05} found the value for H$_2$ as $440$ K, which was also used by \cite{cupp08}.
For H atom on ice, there are three theoretical values for crystalline and amorphous ice:
\cite{holl70,buch91,alha02} determined energies
of $450$ K, $\approx500$ K, and $400\pm50$ K, respectively. \cite{cupp08} used $650$ K
calculated by \cite{alha07} for the BE of H on amorphous ice.
For sets 2 and 3, the BE of H$_2$ is less
than H, whereas the opposite is true for sets 1 and 4.
Interestingly, in all our calculations, we obtain the same trend between H and H$_2$.
BE of H$_2$ is found to be always greater than that of H for all types of
surfaces considered here.

\begin{table}
\scriptsize
\centering
\caption{Different sets of existing BE (in Kelvin) of H and H$_2$ \citep{sil17}.}
\label{tab:BE_sets}
\vskip 0.2 cm
\begin{tabular}{ccccc}
\hline
{\bf Species} & {\bf Set 1} & {\bf Set 2}  & {\bf Set 3} & {\bf Set 4} \\
& & {\bf for olivine} & {\bf for amorphous carbon} & \\
& \citep{alle77} & \citep{katz99} & \citep{katz99} & \citep{vida91} \\
\hline
 H  & 350 & 373 & 658 &  $499 \pm 3$\\
 H$_2$ & 450 & 314 & 542 & $660 \pm 6$ \\
\hline
\end{tabular}
\end{table}

\subsubsection{Chemical modeling}
\begin{figure}[htbp]
\centering
\includegraphics[width=0.8\textwidth]{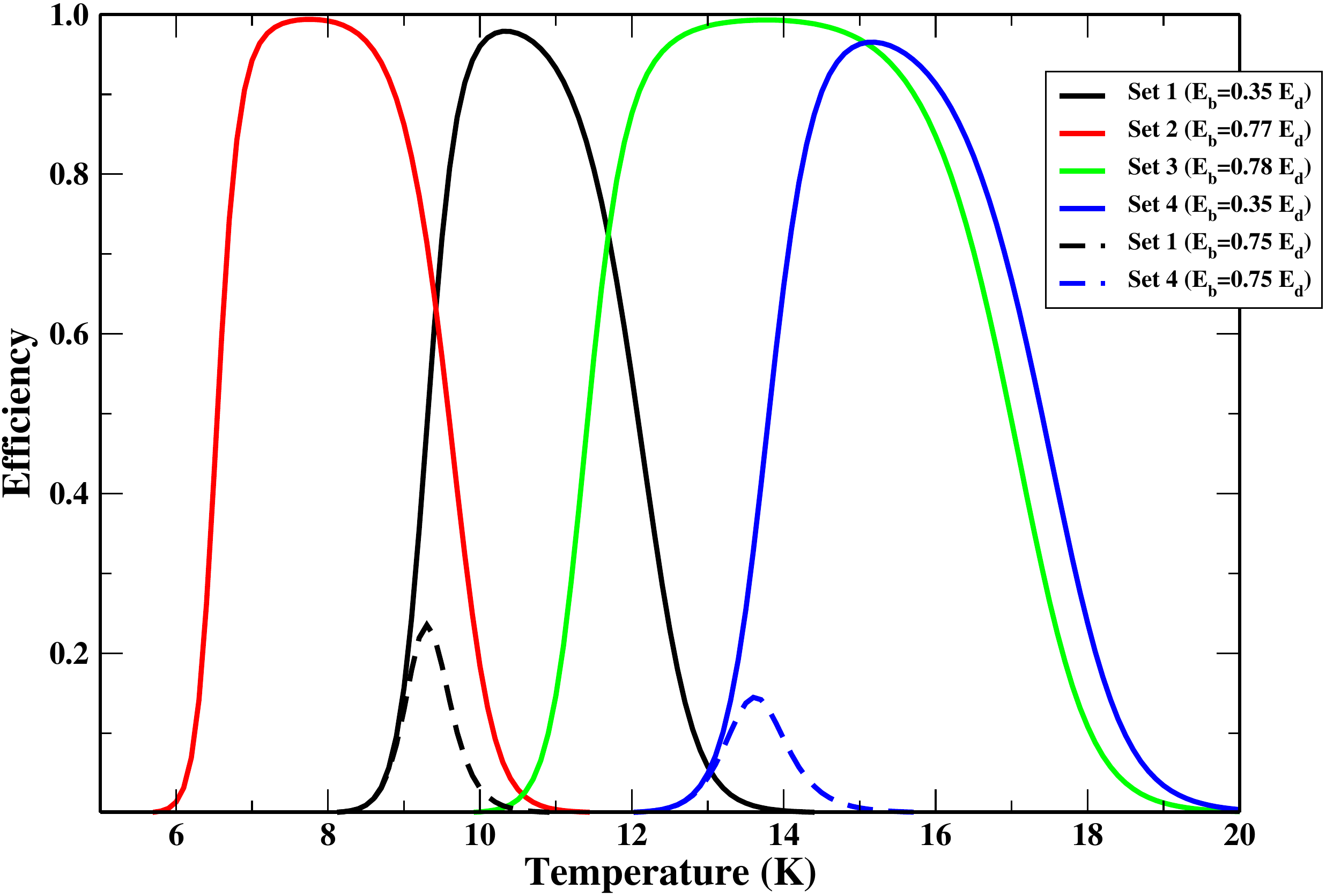}
\caption{Efficiency window for various sets of BE shown in Table \ref{tab:BE_sets} \citep{sil17}.}
\label{fig:BE_eff}
\end{figure}

H$_2$ formation is studied by various authors \citep{chak06a,chak06b,biha01,sahu15}.
\cite{biha01,katz99} discussed the steady-state conditions that may be reached by keeping the flux and
temperature fixed. Here, we use the steady-state relation obtained by the earlier authors
to calculate the H$_2$ formation efficiency window (see Figure \ref{fig:BE_eff}) for various
sets of BE values shown in Table \ref{tab:BE_sets}. For this case, we consider a flux of $0.18 \times 10^{-8}$
ML s$^{-1}$ falling on a grain of diameter $0.1$ $\mu$m. The density of the adsorption sites ($s_n$) is
assumed to be $2 \times 10^{14}$ cm$^{-2}$ for the olivine surface (Set 1 and Set 2 of Table \ref{tab:BE_sets}) and
$5 \times 10^{13}$ cm$^{-2}$ for the amorphous carbon surface (Set 3). In the case of Set 4 also, we assume
$s_n=5 \times 10^{13}$ cm$^{-2}$.

In Set 2 and Set 3, diffusion energies of the H atom were available.
In Set 1, obtained energy against diffusion barriers was $287$ K \citep{pirr97}. Thus the obtained ratio of the diffusion energy ($E_b$) and desorption energy ($E_d$) of H for Set 2 is $\sim 0.77$. For Set 3, this ratio is found to be  $\sim 0.78$.
This ratio is very poorly constrained in the literature, and various authors approximate their choices
in the range of $0.3-0.8$. Since in the case of Set 1 and Set 4, no  $E_b$ was suggested, we use
$E_b=0.75E_d$ (dashed curve of Set 1 and Set 4), close to the experimentally obtained ratio.
For another case of Set 1 and Set 4, we assume $E_b=0.35E_d$ (curves of Set 1 and Set 4).
As expected, it is found that as we are increasing the mobility of the H atom (by decreasing
the barrier against diffusion), the efficiency of formation increases (solid curves of Set 1 and Set 4), and
H$_2$ molecules are efficiently forming within the much wider zone. For the olivine grain (Set 2),
H$_2$ formation efficiency is maximum around $8-9$ K. In contrast, for the BE on
amorphous carbon grains (Set 3), it is around $13-14$ K. The efficiency window obtained for the amorphous
carbon grain (Set 3) is much broader than that of the olivine grain (Set 2). Considering the value
obtained by \cite{alle77} (Set 1), the H$_2$ formation efficiency peaks around $10$ K. For the Set 4
case, the peak efficiency is around $15$ K.

\subsubsection{Summary}
From our calculated physisorption BEs of H atom and H$_2$ molecule on
different grain surfaces, we find an exciting trend that
BE of H$_2$ always remains higher than H for all the cases. However, our calculated values of the
BEs of the H atom with the silica cluster and water cluster are very low compared to the experimentally
obtained values. In our calculations, only one benzene,
three SiO$_2$, and six H$_2$O molecules are considered to mimic the nature of
the carbonaceous, silicate, and water substrate, respectively,
which may not represent the actual astrophysical scenarios.
However, this is the starting point to
consider more complex and realistic systems. Assuming
a steady-state solution to the rate equation method,
we also present the H$_2$ formation efficiency window in various cases.

\clearpage

\subsection{Study of binding energies of several interstellar species}
Knowledge of the BE of
the interstellar species is crucial to understand the synthesis of molecules in the gas phase and interstellar icy grain mantle.
The BE depends on the volatile but also the
grain surface composition. The ISM mainly consists of silicate and carbonaceous type grains.
Numerous studies have been conducted in determining the composition of the interstellar grains and,
more significantly, icy grain mantles \citep{whit96,vand98a,ehre00,ehre03,pont08}.
These studies reveal that the denser regions
of molecular clouds are predominantly composed of H$_2$O in the amorphous phase, with the addition
of some other impurities, such as CO, CO$_2$, NH$_3$, CH$_3$OH, etc., in comparatively lower amounts.
Early experiments \citep{chak98} and astronomical observations \citep{malf98,mald03} suggest the presence of crystalline ice or vapor-deposited amorphous ice (ASW). It continues to attract more and more fundamental research \citep{duli13,hama13} due to its occurrences in astronomical environments such as icy satellites, comets, planetary rings, interstellar grains, etc.
Observations of the OH vibrational stretch in
the IR absorption show that the structure of ice in the ISM is
mostly amorphous \citep{hage81}.
Various surface processes such as adsorption, diffusion, tunneling reactions, and nuclear-spin conversion on interstellar ASW are summarized in
\cite{hama13}. ASW is by far the most significant component of the icy mantles, with
abundances of $\sim 10^{-4}$ concerning the total hydrogen
\citep{will02}, equivalent to coverages of up to $100$ mono-layers (MLs). The extinction
threshold ($A_V$) for H$_2$O mantle is $\sim 3.3$ mag \citep{whit88}.
In reality, the interstellar ices are thought to have low levels of
porosity, as they are continuously exposed to external radiation \citep{palu06}.
The ASW is known to have physical properties that are distinct from other
crystalline states, e.g., a lower thermal conductivity, a larger surface area, and
higher porosity \citep{dohn03}.
Here, we devise a computational approach to approximate the BE for some relevant interstellar species on ASW substrate.

\subsubsection{Computational Details and Methodology}
\begin{figure}
\centering
\includegraphics[width=\textwidth]{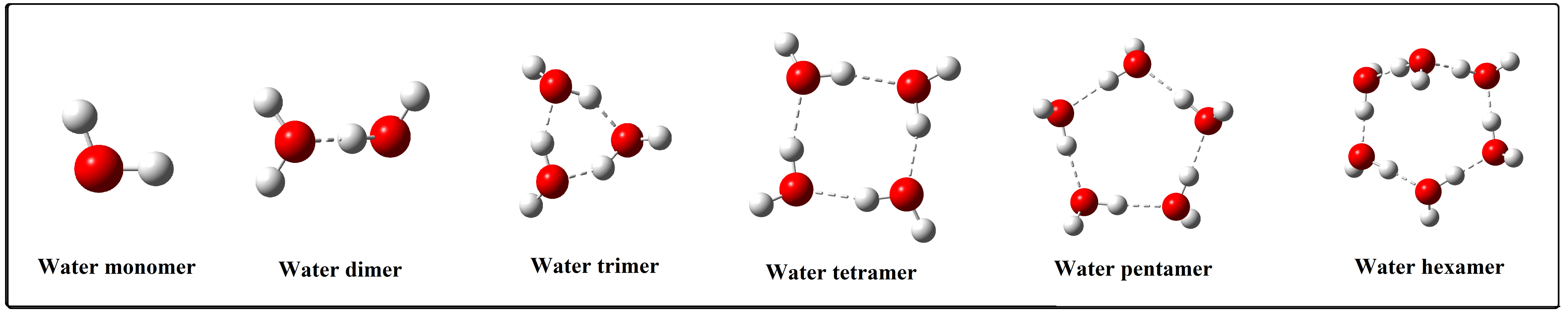}
\caption{Configurations of water molecule(s) used \citep{das18}.}
\label{fig:water_clusters}
\end{figure}

The BE is usually seen as a local property arising from the electronic
interaction between the grain surface and the species
deposited on its surface (adsorbate).
As mentioned earlier, a knowledge of the BE of the adsorbed species with water ice as adsorbent is essential
for building a realistic astrochemical model that studies the composition of the interstellar grain mantle.
Therefore, for investigation of the BEs, we use the most stable configurations of
water monomer, dimer, c-trimer, c-tetramer, c-pentamer, and c-hexamer (chair) \citep{ohno05} as the adsorbents (Figure \ref{fig:water_clusters}).

We calculate the BE of a species on the grain surface following the formulae \ref{eqn:BE_1} and \ref{eqn:BE_2} mentioned earlier in subsection \ref{sec:BE_H_H2_method}.
We carry out calculations using the \textsc{Gaussian} 09 suite of programs developed by \cite{fris13}.
The MP2/aug-cc-pVDZ level of theory is mainly used for computing the optimized
energy of all the species and complexes.
The CCSD(T)/aug-cc-pVTZ level of theory is also used to calculate the single-point energy by taking the optimized structure obtained with the MP2 method for tetramer and hexamer (water clusters).
The cc-pVDZ and cc-pVTZ are
the Dunnings-correlation-consistent basis set \citep{dunn89} with double and triple zeta functions, respectively.
The dependency of our computed BE values are reviewed on the implemented method and
basis set.
A fully optimized ground-state structure is verified as a stationary point (having non-imaginary frequency) by the harmonic vibrational frequency analysis.
Most of the calculations are performed without the corrections of ZPVE and BSSE.
Calculations are also performed by including ZPVE and excluding BSSE (using the CP method, see the formula \ref{eqn:BE_3}) to check their effect.

\subsubsection{Results and discussions on the BE values}
In this Section, the results of high-level quantum chemical calculations are presented and
discussed in detail.
\cite{wake17} selected $16$ stable species to compute the BEs by considering a water
monomer. Here, we employ a similar methodology and carry out calculations by increasing the
cluster size of water molecule to check its consequences.
Six different sets of adsorbents are used to see the effects of
cluster size on the computed BEs.
With the increase in cluster size, we find a trend of
increasing BE for most species considered here.
Most interestingly, the calculated BEs, by considering the water pentamer and hexamer configurations, seem to be closer to the experimentally obtained values than the calculated BEs with
the water monomer, dimer, trimer, or tetramer configurations.
The BEs of these $16$ stable species with various water cluster configurations are given in Table \ref{tab:BE_1}.
For the sake of comparison, in Table \ref{tab:BE_1}, we also show
the experimentally obtained values, estimated values from \cite{wake17},
and BEs from the UMIST Database\footnote{\url{http://udfa.ajmarkwick.net}} \citep{mcel13}.

\begin{figure}
\centering
\includegraphics[width=0.8\textwidth]{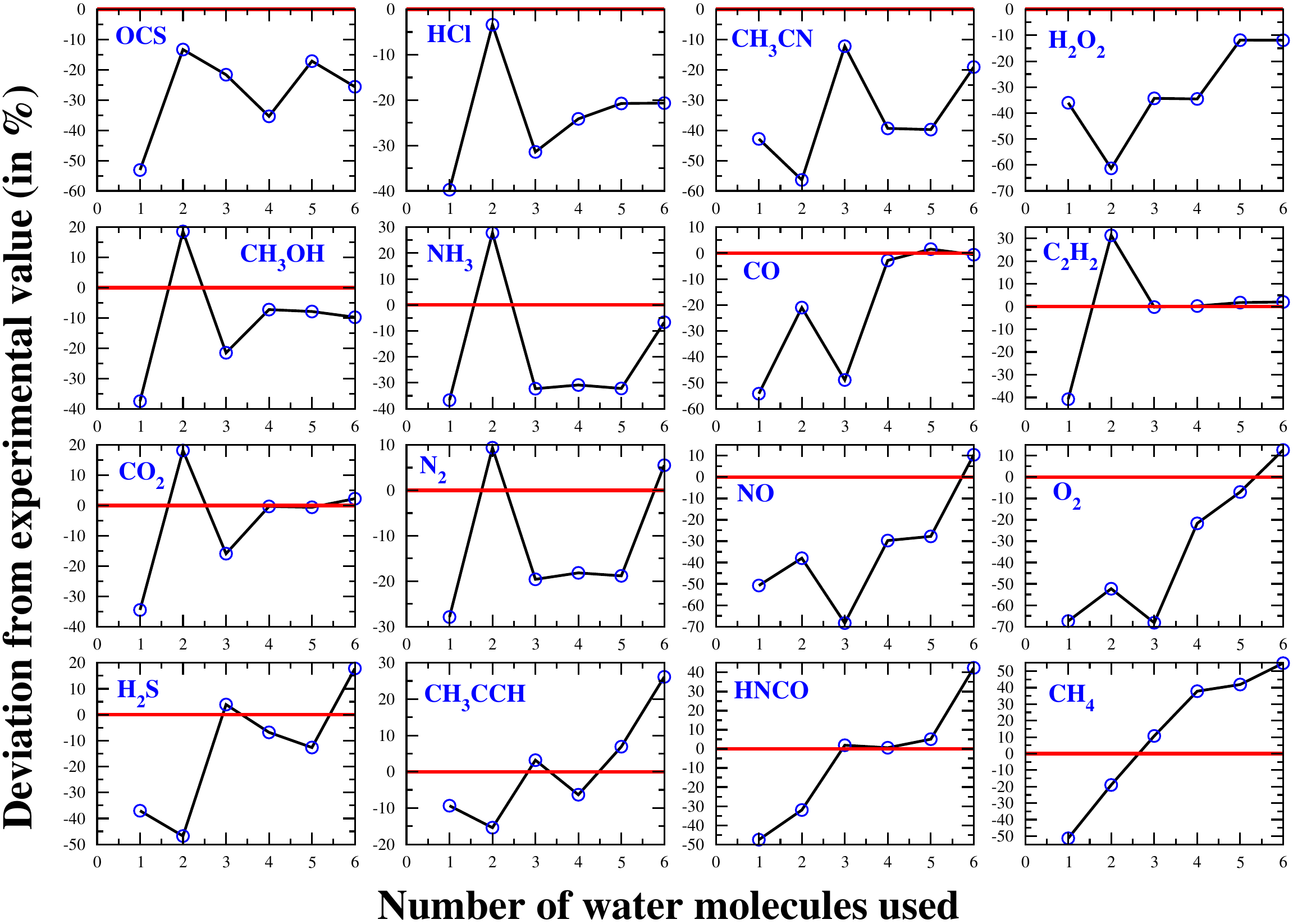}
\caption{Percentage deviations of BEs of $16$ stable species with increasing numbers of water clusters acting as the
grain surface \citep{das18}.}
\label{fig:BE_variation}
\end{figure}

\begin{figure}
\centering
\includegraphics[width=0.8\textwidth]{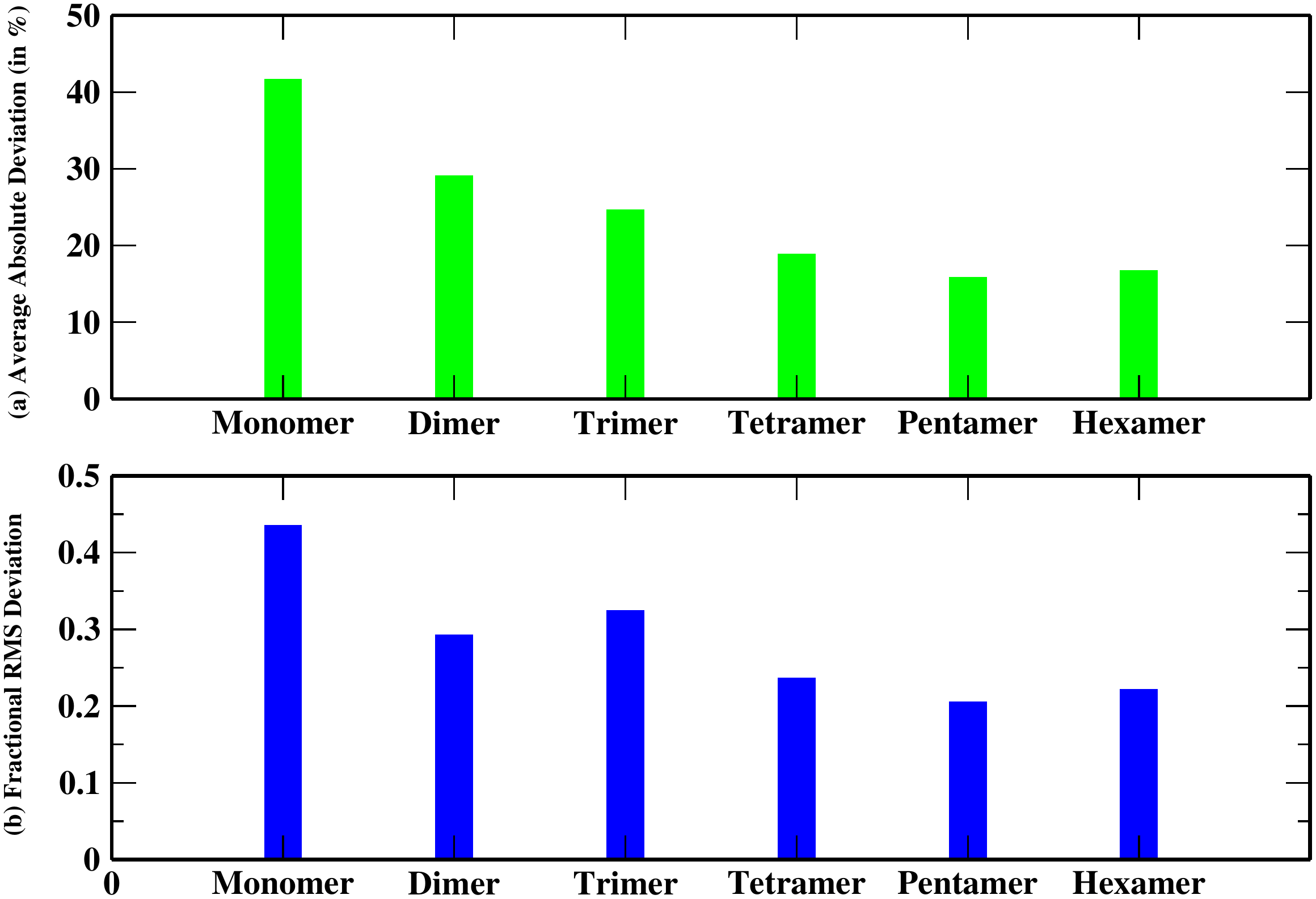}
\caption{(a) Average absolute percentage deviation and (b) fractional RMS deviation of our calculated values from experiments \citep{das18}.}
\label{fig:BE_deviation}
\end{figure}

\begin{figure}
\centering
\includegraphics[width=0.8\textwidth]{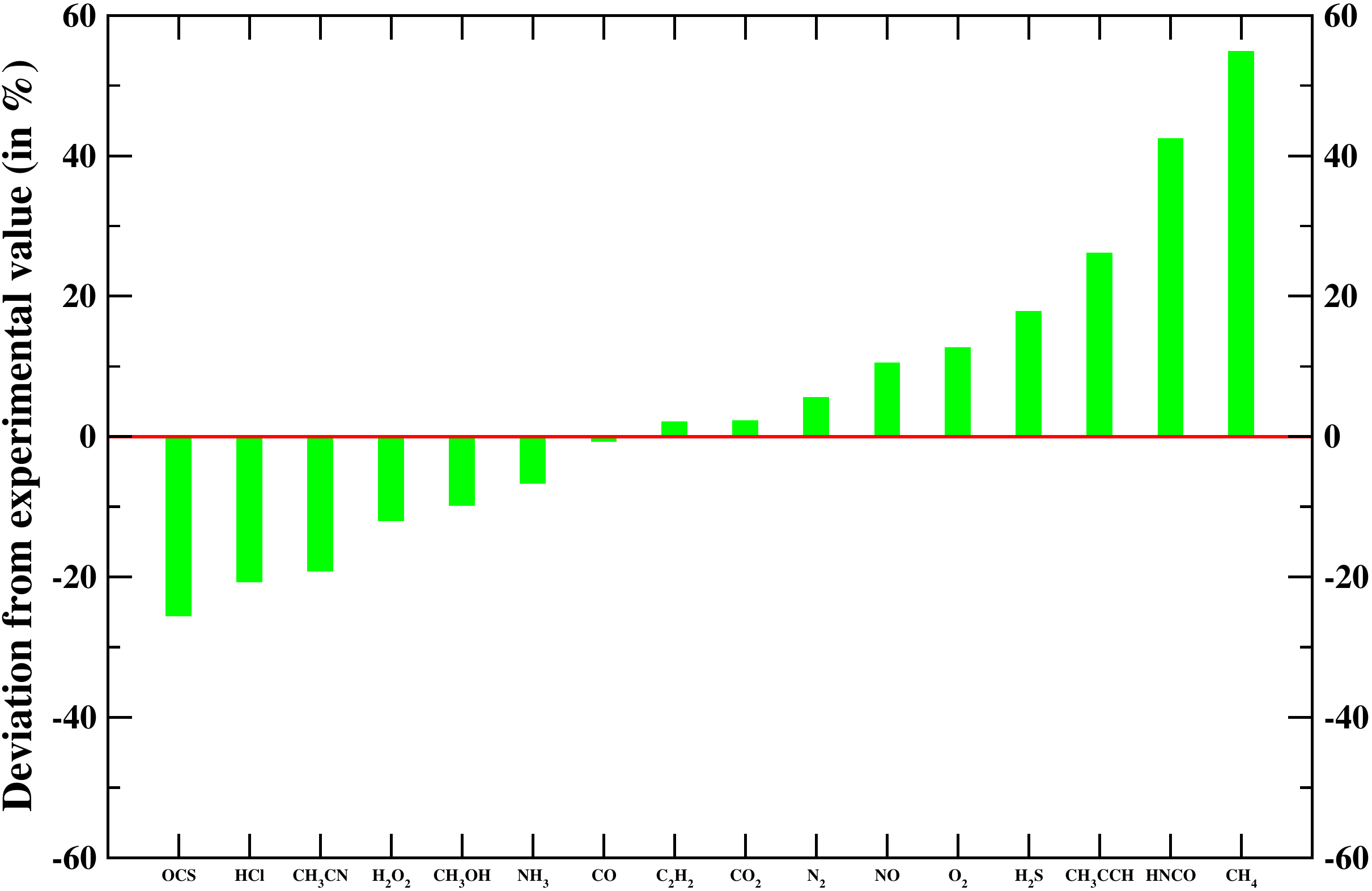}
\caption{Percentage deviation from experimental BE values of 16 stable species using the water c-hexamer (chair) cluster \citep{das18}.}
\label{fig:BE-FPD}
\end{figure}

Figure \ref{fig:BE_variation} is subdivided into $16$ blocks for $16$ different species.
We show the sizes of water clusters along the X-axis, and along the Y-axis, we show the percentage of deviation relative to experiments.
The red horizontal line in each block denotes the zero deviation.
The result clearly shows that mostly the percentage of deviation is comparatively higher (underestimated) when the water monomer is used and lower when the pentamer or hexamer
configuration of the water cluster is used.
However, we can see that the BEs with dimer configuration largely deviate from that obtained with the rest of the arrangements.
For example, in the case of OCS, HCl, CH$_3$OH, NH$_3$, C$_2$H$_2$, CO$_2$, and N$_2$, the BEs with the dimer configuration vastly increased compared to the BEs obtained with the other water structures.
On the contrary, BEs of CH$_3$CN, H$_2$O$_2$, H$_2$S, and CH$_3$CCH with the dimer configuration are minimal compared to the rest of the cases.
The percentage deviations are noted in parentheses in Tables \ref{tab:BE_1} and \ref{tab:BE_2} as well. Furthermore, we calculate the average absolute percentage deviation and root mean square (RMS) fractional deviation of these species presented in Tables \ref{tab:BE_1} and \ref{tab:BE_2}.
It is interesting to note (in Table \ref{tab:BE_1}) that while we gradually increase the cluster size, the average percentage of deviations are found to be $ \pm 41.6 \%$, $\pm 29.0 \%$, $ \pm 24.6 \%$, $ \pm 18.8 \%$, $ \pm 15.8 \%$, and  $ \pm 16.7 \%$, for monomer, dimer, trimer, tetramer, pentamer, and hexamer, respectively.
Similarly the fractional RMS deviations are found to be 0.435, 0.292, 0.324, 0.236, 0.205, and 0.221, respectively. These average absolute percentage deviations and fractional RMS deviations are shown in Figure \ref{fig:BE_deviation}(a) and (b), respectively.
The large deviations from experimental BE values in case of water dimer cluster
shown in Figure \ref{fig:BE_variation} are averaged out in Figure \ref{fig:BE_deviation}, which shows a similar trend. But in reality, it is not the case.
Figure \ref{fig:BE-FPD} depicts the percentage deviation from experimental BE values of each
species considering water hexamer.

\begin{landscape}
\begin{table}
\tiny
\caption{Calculated BEs of 16 stable species \citep{das18}.}
\label{tab:BE_1}
\vskip 0.2 cm
\hskip -3.0cm
\begin{tabular}{cccccccccccc}
\hline
{\bf Sl.} & {\bf Species} & \multicolumn{6}{c}{\bf Calculated BE (in Kelvin) on different water clusters using MP2/aug-cc-pVDZ} 
&{\bf Experimental} & \multicolumn{2}{c}{\bf \underline{BE from \cite{wake17}}}& {\bf BE (in Kelvin)} \\
\cline{3-8}
{\bf No.}&&{\bf Monomer}& {\bf Dimer} &{\bf Trimer}&{\bf Tetramer}&{\bf Pentamer}&{\bf Hexamer}&{\bf values of BE} 
&\underline{{\bf (in Kelvin)}} &\underline{{\bf (in Kelvin)}}&{\bf from}\\
& & \includegraphics[width=1.3cm, height=0.8cm]{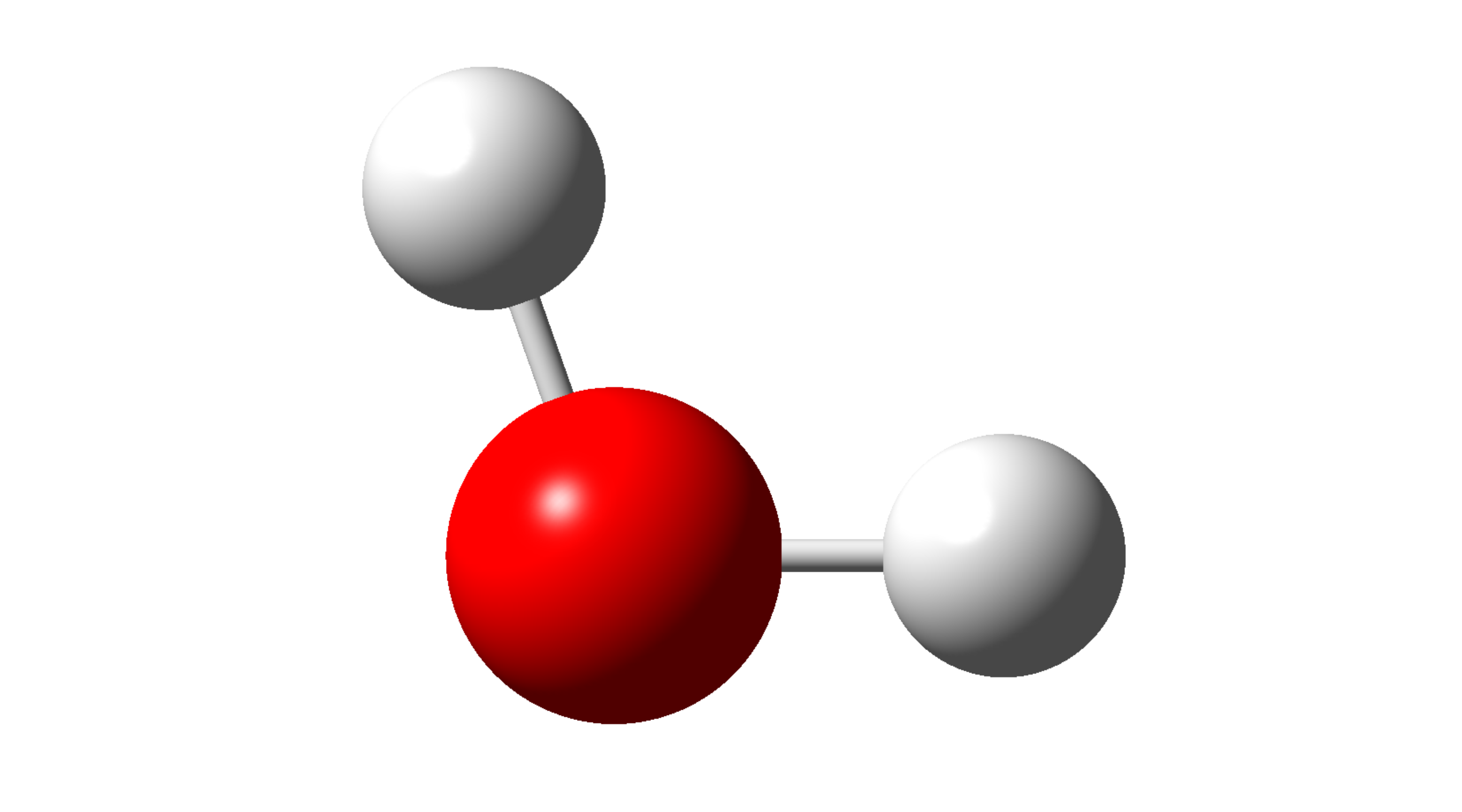} & \includegraphics[width=1.3cm, height=0.8cm]{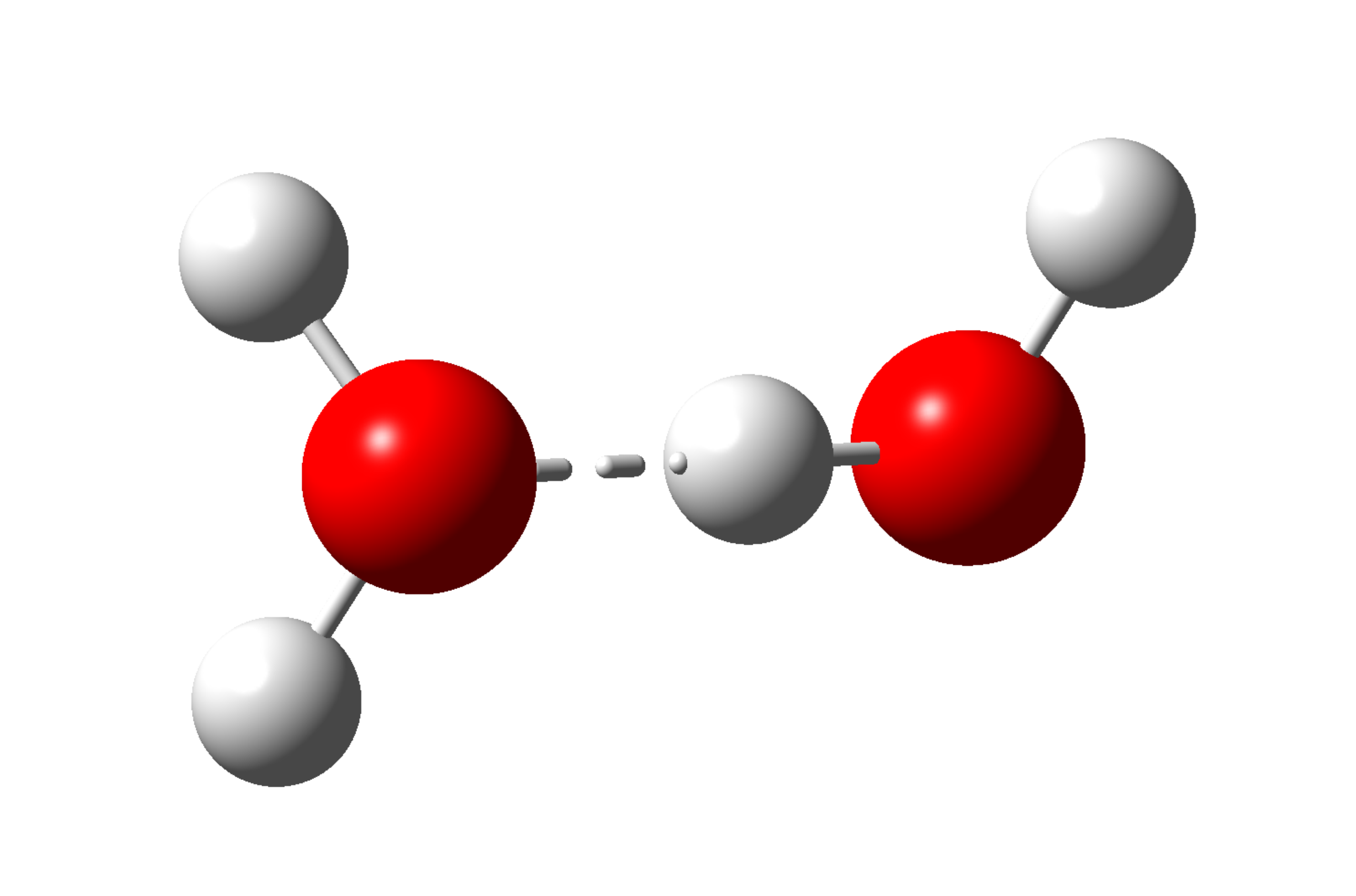} &
\includegraphics[width=1.5cm, height=1cm]{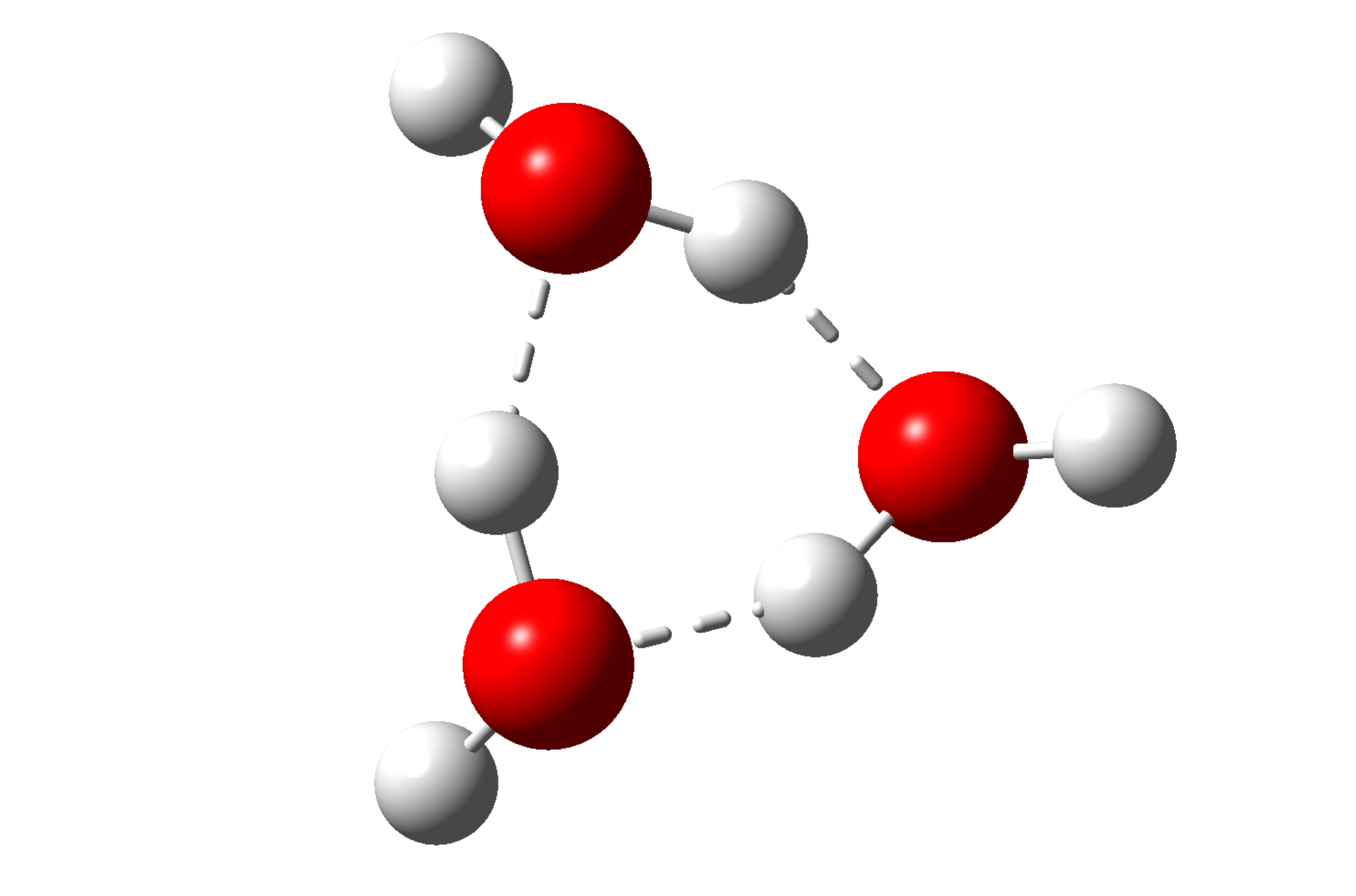} & \includegraphics[width=1.7cm, height=1.2cm]{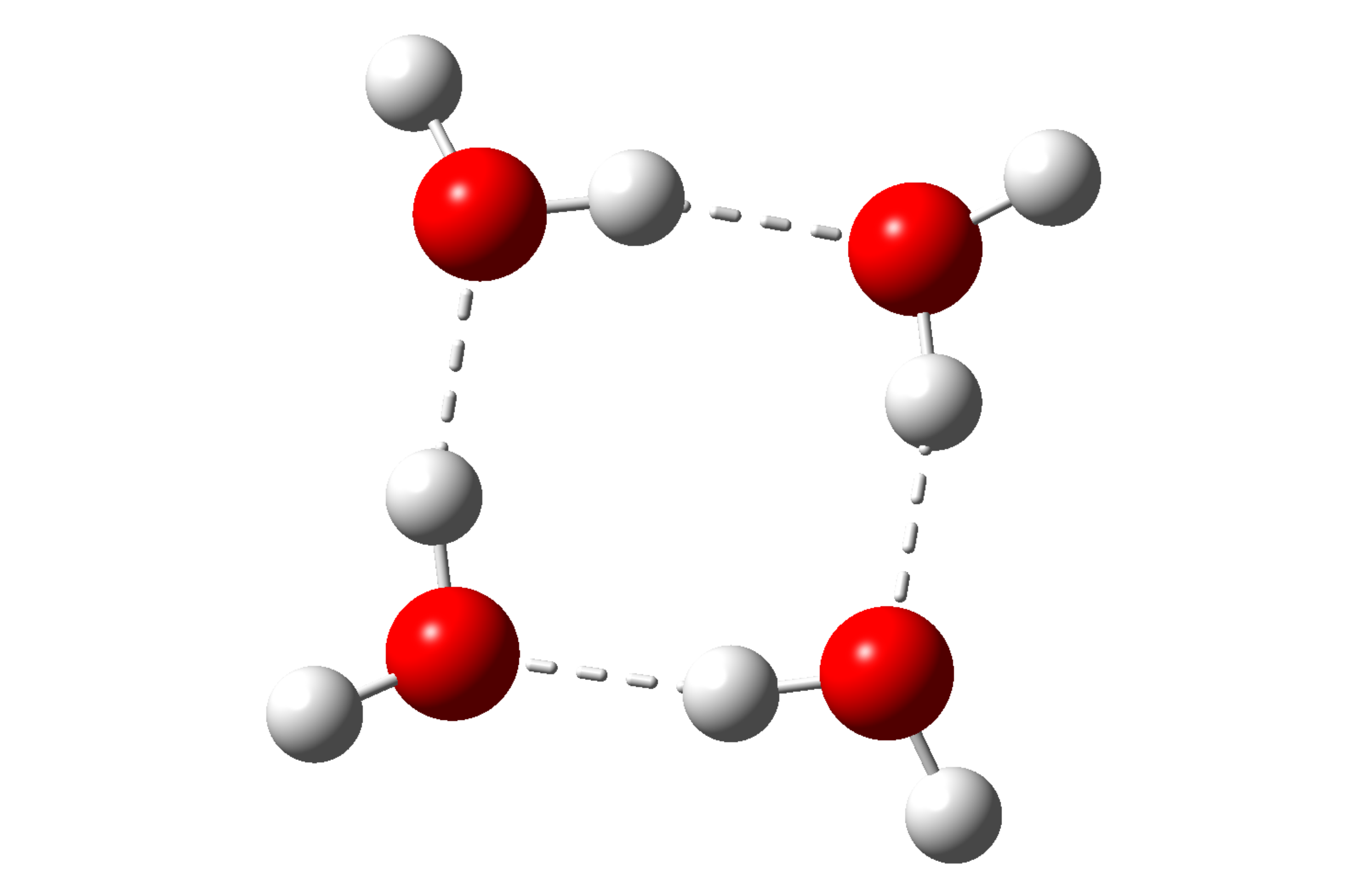}
& \includegraphics[width=1.7cm, height=1.2cm]{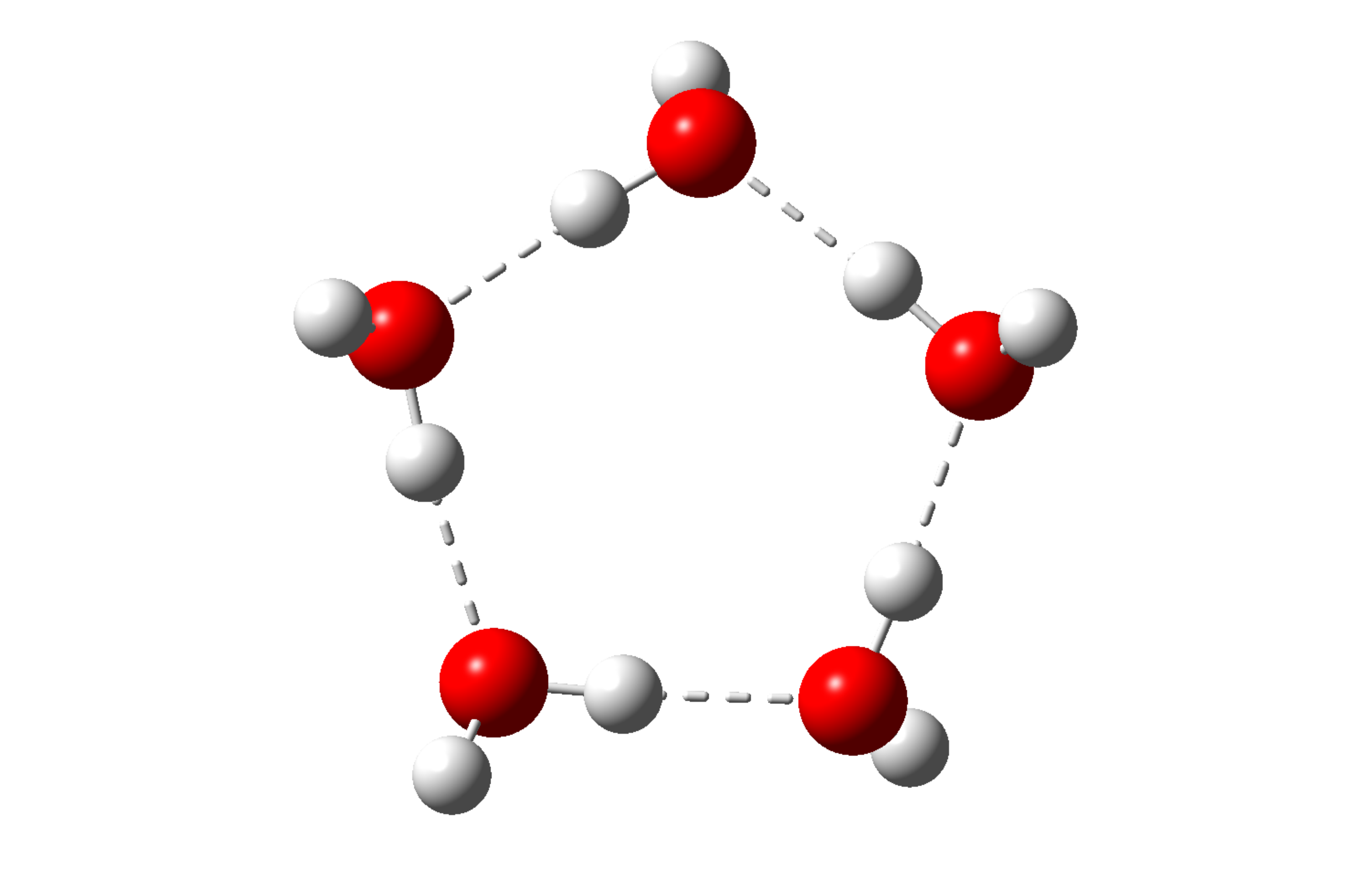} & \includegraphics[width=1.7cm, height=1.2cm]{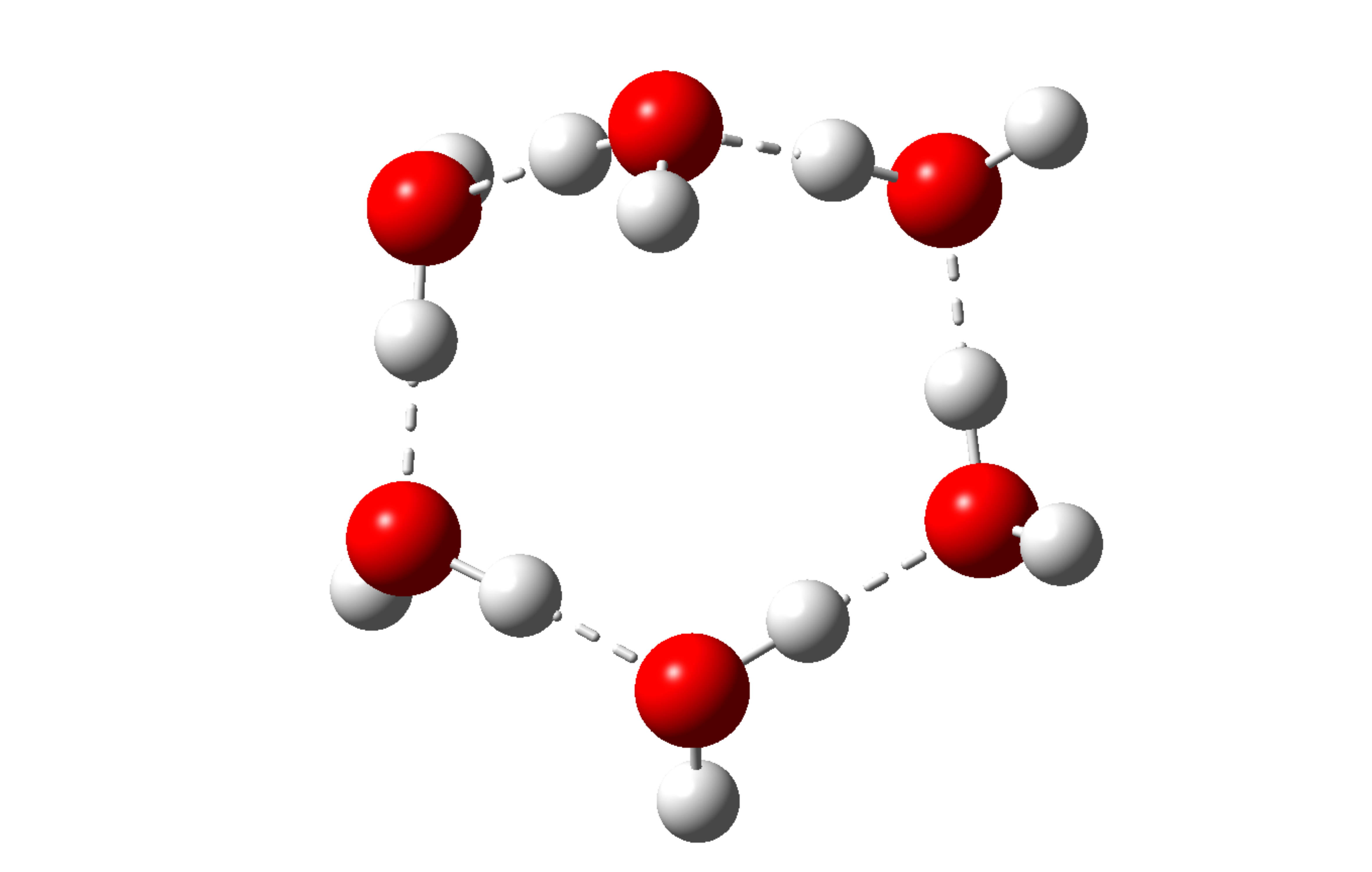} & 
{\bf (in Kelvin)} & {\bf using DFT} & {\bf using MP2}& {\bf UMIST database$^k$} \\
&&&&&&&& {\bf  M06-2X} & {\bf aug-cc-pVTZ}&  \\
\hline
1 & OCS & 1139 ($-53.1 \%$) & 2106 ($-13.3 \%$) &  1905 ($-21.6 \%$) & 1571 ($-35.3 \%$) & 2014 ($-17.1 \%$)& 1808 ($-25.5 \%$) & 2430 $\pm$ 24$^a$  & 2100 ($13.5 \%$) & --- & 2888 ($18.8 \%$) \\
2 & HCl & 3116 ($-39.7 \%$) & 4994 ($-3.4 \%$) & 3545 ($-31.4 \%$) & 3924 ($-24.1 \%$) & 4099 ($-20.7 \%$) & 4104 ($-20.6 \%$) & 5172$^b$  & 4800 ($7.1 \%$) & 5000 ($-3.3 \%$)  & 900 ($-82.5 \%$) \\
3 & CH$_3$CN & 2676 ($-42.8 \%$) & 2039 ($-56.4 \%$) & 4108 ($-12.2 \%$) & 2838 ($-39.3 \%$) & 2820 ($-39.7 \%$) & 3786 ($-19.1 \%$) & 4680$^c$ & 4300 ($8.1 \%$) & 4800 ($2.5 \%$) &  4680 ($0.00 \%$) \\
4 & H$_2$O$_2$ & 3838 ($-36.0 \%$) & 5632 ($-61.3\%$) &  3942 ($-34.3 \%$) & 3928 ($-34.5 \%$) & 5288 ($-11.8 \%$) & 5286 ($-11.9 \%$) & 6000$^d$  & 6800 ($-13.3 \%$) & 6100 ($1.6 \%$) & 5700 ($-5.0 \%$)\\
5 & CH$_3$OH & 3124 ($-37.5 \%$) & 5932 ($18.6 \%$) & 3924 ($-21.5 \%$)& 4368 ($-12.6 \%$) & 4607 ($-7.8 \%$) & 4511 ($-9.7 \%$) & 5000$^e$  & 4800 ($4.0 \%$) & 4850 ($-3.0 \%$)& 4930 ($-1.4 \%$) \\
6 & NH$_3$ & 3501 ($-36.6 \%$) & 5684 ($27.8 \%$) & 3745 ($-32.3 \%$) & 3825 ($-30.8 \%$) & 3751 ($-32.1 \%$) & 5163 ($-6.6 \%$) & 5530$^c$ & 5600 ($-1.2 \%$) & 5500 ($-0.5 \%$) &  5534 ($0.6 \%$) \\
7 & CO & 595 ($-54.2 \%$) & 1027 ($-21.0 \%$) &  664 ($-48.9 \%$) & 1263 ($-2.8 \%$) & 1320 ($1.5 \%$) & 1292 ($-0.6 \%$) & 1300$^{g}$ & 1300 ($0.0 \%$) & 1100 ($-15.3 \%$) & 1150 ($-11.5 \%$) \\
8 & C$_2$H$_2$ & 1532 ($-40.7 \%$) & 3399 ($31.4 \%$) & 2581 ($-0.2 \%$) & 2593 ($0.2 \%$) & 2633 ($1.7 \%$) & 2640 ($2.0 \%$) & 2587$^c$ & 2600 ($-0.5 \%$) & 2700 ($4.3 \%$) &  2587 ($0.0 \%$) \\
9 & CO$_2$ & 1506 ($-34.5 \%$) & 2719 ($18.2 \%$) & 1935 ($-15.9 \%$) & 2293 ($-0.3 \%$) & 2287 ($-0.5 \%$) & 2352 ($2.2 \%$) & 2300$^g$ & 3100 ($34.7 \%$) & 2600 ($13.0 \%$) &  2990 ($30.0 \%$) \\
10 & N$_2$ & 793 ($-27.9 \%$) & 1203 ($9.4 \%$) & 884 ($-19.6 \%$)& 900 ($-18.1 \%$) & 893 ($-18.8 \%$) & 1161 ($5.5 \%$) & 1100$^e$ & 1100 ($0.0 \%$) & 1400 ($27.2 \%$) & 790 ($-28.1 \%$) \\
11 & NO & 886 ($-50.7 \%$) & 1115 ($-38.0 \%$) & 568 ($-68.4 \%$)& 1265 ($-29.7 \%$) & 1300 ($-27.7 \%$) & 1988 ($10.4 \%$) & 1800$^e$ & 1600 ($11.1 \%$) & 1700 ($-5.5 \%$) &  1600 ($-11.1 \%$) \\
12 & O$_2$ & 391 ($-67.4 \%$) & 572 ($-52.3 \%$) & 382 ($-68.2 \%$) & 940 ($-21.6 \%$) & 1116 ($-7.0 \%$) & 1352 ($12.6 \%$) & 1200$^g$ & 1000 ($16.6 \%$) & 900 ($-25.0 \%$) & 1000 ($-16.6 \%$) \\
13 & H$_2$S & 1727 ($-37.0 \%$) & 1460 ($-46.8 \%$) & 2849 ($3.9 \%$) & 2556 ($-6.8 \%$) & 2396 ($-12.6 \%$) & 3232 ($17.8 \%$) & 2743$^{c,k}$ & 2700 ($1.5 \%$) & 2550 ($-7.0 \%$) &  2743 ($0.0 \%$) \\
14 & CH$_3$CCH & 2266 ($-9.3 \%$) & 2115 ($-15.4 \%$) &  2580 ($3.2 \%$)& 2342 ($-6.3 \%$) & 2673 ($6.9 \%$) & 3153 ($26.1 \%$) & 2500 $\pm$ 40$^h$ & 3800 ($-52.0 \%$) & 3800 ($52.0 \%$) & 2470 ($-1.2 \%$) \\
15 & HNCO & 2046 ($-47.5 \%$) & 2653 ($-32.0 \%$) &  3973 ($1.9 \%$) & 3922 ($0.5 \%$) & 4097 ($5.0 \%$) & 5554 ($42.4 \%$) & 3900$^i$  & 4850 ($-24.3 \%$) & ... &  2850 ($-26.9 \%$) \\
16 & CH$_4$ & 469 ($-51.2 \%$) & 780 ($-19.0 \%$) & 1066 ($10.7 \%$) & 1327 ($37.7 \%$) & 1366 ($41.8 \%$) & 1491 ($54.8 \%$) & 963$^j$ & 800 ($16.9 \%$) & 1000 ($3.8 \%$) & 1090 ($13.1 \%$) \\
\hline
\multicolumn{2}{c}{\bf  Average absolute deviation}& $\pm 41.6 \%$ & $\pm 29.0 \%$ & $\pm 24.6 \%$
&  $\pm 18.8 \%$ & $\pm 15.8 \%$ & $\pm 16.7 \%$ &  & $\pm 12.8 \%$
& $\pm 11.7 \%$ &  $\pm 15.4 \%$ \\
\hline
\multicolumn{2}{c}{\bf Fractional RMS deviation} &  0.435 & 0.292 & 0.324 &  0.236
&  0.205 &0.221 &  &  0.189 &  0.182 & 0.254\\
\hline
\end{tabular} \\
\\
{\bf Note:} \\
Percentage deviations from experimental BE values (Column 8) are shown in parentheses for Columns 3, 4, 5, 6, 7, 8, 10, 11, and 12. \\
$^a$ \cite{ward12}, $^b$ \cite{olan11}, $^c$ \cite{coll04}, $^d$ \cite{duli13}, $^e$ \cite{wake17},
$^g$ \cite{mini16}, $^h$ \cite{kimb14}, $^i$ \cite{nobl15}, $^j$ \cite{raut07},
$^k$ UMIST database (\url{http://udfa.ajmarkwick.net}).
\end{table}

\begin{table}
\scriptsize
\caption{Comparison of calculated BEs using water monomer (adsorbent) with experimentally used BEs \citep{das18}.}
\label{tab:BE_2}
\vskip 0.2 cm
\hskip -2.5cm
\begin{tabular}{cccccccc}
\hline
{\bf Sl.} & {\bf Species} & \multicolumn{5}{c}{\bf BE (in Kelvin) using different methods and basis sets 
including ($+$) or excluding ($-$) ZPVE and BSSE} & {\bf Experimental} \\
\cline{3-7}
{\bf No.}&&{\bf MP2/aug-cc-pVDZ}&{\bf MP2/aug-cc-pVDZ}&{\bf MP2/aug-cc-pVDZ}&{\bf MP2/aug-cc-pVTZ} & {\bf CCSD(T)/aug-cc-pVTZ} & {\bf values of BE} \\
&& {\bf $-$ ZPVE and $-$ BSSE} & {\bf $+$ ZPVE but $-$ BSSE} & {\bf $-$ ZPVE but $+$ BSSE} & {\bf $-$ ZPVE and $-$ BSSE} & {\bf $-$ ZPVE and $-$ BSSE} & {\bf (in Kelvin)} \\
\hline
1 & OCS & 1139 ($-53.1 \%$) & 683 ($-71.9 \%$) & 803 ($-66.9 \%$) &  1074 ($-55.8 \%$) & 1086 ($-55.3 \%$) & 2430 $\pm$ 24$^a$ \\
2 & HCl & 3116 ($-39.7 \%$) & 2113 ($-59.1 \%$) & 2627 ($-49.2 \%$) &  2975 ($-42.5 \%$) & 2777 ($-46.3 \%$) & 5172$^b$ \\
3 & CH$_3$CN & 2676 ($-42.8 \%$) & 1970 ($-57.9 \%$) & 2242 ($-52.1 \%$) & 2676 ($-42.8 \%$) & 2635 ($-43.7 \%$) & 4680$^c$ \\
4 & H$_2$O$_2$ & 3838 ($-36.0 \%$) & 2647 ($-55.9 \%$) & 3204 ($-46.6 \%$) & 3775 ($-37.1 \%$) & 3802 ($-36.6 \%$) & 6000$^d$  \\
5 & CH$_3$OH & 3124 ($-37.5 \%$) & 2149 ($-57.0 \%$) & 2586 ($-48.3 \%$) & 3021 ($-39.6 \%$) & 2988 ($-40.2 \%$) & 5000$^e$ \\
6 & NH$_3$ & 3501 ($-36.6 \%$) & 2368 ($-57.2 \%$) & 2941 ($-46.8 \%$) & 3375 ($-39.0 \%$) & 3332 ($-39.7 \%$) & 5530$^c$  \\
7 & CO & 595 ($-54.2 \%$) & 236 ($-81.8 \%$) & 349 ($-73.2 \%$) & 565 ($-56.5 \%$) & 662 ($-49.1 \%$) & 1300$^g$  \\
8 & C$_2$H$_2$ & 1532 ($-40.7 \%$) & 950 ($-63.3 \%$) & 1111 ($-57.1 \%$) & 1519 ($-41.3 \%$) & 1444 ($-44.2 \%$) & 2587$^c$   \\
9 & CO$_2$ & 1506 ($-34.5 \%$) & 1109 ($-51.8 \%$) & 1159 ($-49.6 \%$) & 1417 ($-38.4 \%$) & 1511 ($-34.3 \%$) & 2300$^g$  \\
10 & N$_2$ & 793 ($-27.9 \%$) & 340 ($-69.1 \%$) & 534 ($-51.4 \%$) & 791 ($-28.1 \%$) & 759 ($-31.0 \%$) & 1100$^e$  \\
11 & NO & 886 ($-50.7 \%$) & 353 ($-80.4 \%$) & 249 ($-86.2 \%$) &  876 ($-51.3 \%$) & 780 ($-56.7 \%$) & 1800$^e$   \\
12 & O$_2$ & 391 ($-67.4 \%$) & 258 ($-78.5 \%$) & 191 ($-84.1 \%$) & 385 ($-67.9 \%$) & 419 ($-65.1 \%$) & 1200$^g$   \\
13 & H$_2$S & 1727 ($-37.0 \%$) & 971 ($-64.6 \%$) & 1305 ($-52.4 \%$) & 1662 ($-39.4 \%$) & 1598 ($-41.7 \%$) & 2743$^{c,k}$  \\
14 & CH$_3$CCH & 2266 ($-9.3 \%$) & 1548 ($-38.1 \%$) & 1675 ($-33.0 \%$) & 2175 ($-13.0 \%$) & 2083 ($-16.7 \%$) & 2500 $\pm$ 40$^h$ \\
15 & HNCO & 2046 ($-47.5 \%$) & 1376 ($-64.7 \%$) & 1644 ($-57.8 \%$) & 3260 ($-16.4 \%$) & 2058 ($-47.2 \%$) & 3900$^i$  \\
16 & CH$_4$ & 469 ($-51.2 \%$) & 145 ($-84.9 \%$) & 265 ($-72.5 \%$) & 374 ($-61.2 \%$) & 401 ($-58.4 \%$) & 963$^j$ \\
\hline
\multicolumn{2}{c}{\bf Average abs deviation}  & $\pm$ 41.6 \% & $\pm$ 64.7 \% & $\pm$ 58 \% & $\pm$ 41.9 \% & $\pm$ 44.1 \% & --- \\
\hline
\multicolumn{2}{c}{\bf Frac. RMS deviation}  & 0.435 & 0.659 & 0.597 & 0.443 & 0.456 & --- \\
\hline
\end{tabular} \\
\\
{\bf Note:}\\
Percentage deviations from experimental BE values (Column 8) are shown in parentheses for Columns 3, 4, 5, 6, and 7. \\
$^a$ \cite{ward12}, $^b$ \cite{olan11}, $^c$ \cite{coll04}, $^d$ \cite{duli13}, $^e$ \cite{wake17},
$^g$ \cite{mini16}, $^h$ \cite{kimb14}, $^i$ \cite{nobl15}, $^j$ \cite{raut07},
$^k$ UMIST database (\url{http://udfa.ajmarkwick.net}).
\end{table}
\end{landscape}

\begin{table}
\scriptsize
\centering
\caption{Calculated BEs using water hexamer (adsorbent) to check the effect of basis set superposition error (BSSE) using the
counterpoise (CP) method \citep{das18}.}
\label{tab:BE_3}
\vskip 0.2cm
\begin{tabular}{cccc}
\hline
{\bf Sl.} & {\bf Species} & \multicolumn{2}{c}{\bf BE (in Kelvin) using MP2/aug-cc-pVDZ level of theory} \\
\cline{3-4}
{\bf No.}&&{\bf Values without BSSE correction} & {\bf BSSE corrected values} \\
\hline
1 & OCS & 1808 ($-25.5 \%$) & 1294 ($-46.7 \%$) \\
2 & HCl & 4104 ($-20.6 \%$) & 3777 ($-26.9 \%$) \\
3 & $\rm{CH_3CN}$ & 3786 ($-19.1 \%$) & 2194 ($-53.1 \%$) \\
4 & $\rm{H_2O_2}$ & 5286 ($-11.9 \%$) & 4161 ($-30.6 \%$) \\
5 & $\rm{CH_3OH}$ & 4511 ($-9.7 \%$) & 3550 ($-29.0 \%$) \\
6 & $\rm{NH_3}$ & 5163 ($-6.6 \%$) & 3082 ($-44.2 \%$) \\
7 & CO & 1292 ($-0.6 \%$) & 840 ($-35.3 \%$) \\
8 & $\rm{C_2H_2}$ & 2640 ($2.0 \%$) & 1890 ($-26.9 \%$) \\
9 & $\rm{CO_2}$ & 2352 ($2.2 \%$) & 1624 ($-29.3 \%$) \\
10 & N$_2$ & 1161 ($5.5 \%$) & 568 ($-48.3 \%$) \\
11 & NO & 1988 ($10.4 \%$) & 911 ($-49.3 \%$) \\
12 & O$_2$ & 1352 ($12.6 \%$) & 519 ($-56.7 \%$) \\
13 & $\rm{H_2S}$ & 3232 ($17.8 \%$) & 1954 ($-28.8 \%$) \\
14 & $\rm{CH_3CCH}$ & 3153 ($26.1 \%$) & 1382 ($-44.7 \%$) \\
15 & HNCO & 5554 ($42.4 \%$) & 5017 ($28.6 \%$) \\
16 & $\rm{CH_4}$ & 1491 ($54.8 \%$) & 653 ($-32.1 \%$) \\
\hline
\multicolumn{2}{c}{\bf Average absolute deviation} & $\pm 16.7 \%$ & $\pm 34.6 \%$ \\
\hline
\multicolumn{2}{c}{\bf Fractional RMS deviation} & 0.221 & 0.395 \\
\hline
\end{tabular} \\
\vskip 0.2 cm
{\bf Note:} Percentage deviation from experimental BE values are shown in parentheses for Columns 3 and 4.
\end{table}

Table \ref{tab:BE_1} presents the average absolute percentage of deviation
of calculated  BE values from \cite{wake17}.
We see that, on an average, the predicted/scaled values of \cite{wake17} deviate from the experimental
values by $ \pm 12.8 \%$ and $ \pm 11.7 \%$, when DFT/M06-2X and MP2/aug-cc-pVTZ level of theories were used, respectively.
The BE values in the UMIST database deviate from the experimental values by $ \pm 15.4 \%$.
Our calculations with pentamer and hexamer configurations produce
average deviations of $\pm 15.8 \%$ and $\pm 16.7 \%$, respectively, from the experimental values.
No proportional law fitting \citep[as in][]{wake17} is required for our calculations. 
Since the calculations with the pentamer and hexamer configurations
can roughly estimate the experimental values, it is suggested to use
these configurations to evaluate the BE of species without any existing experimental values.

Table \ref{tab:BE_2} shows a comparison between the results obtained by
considering BSSE corrections using the CP method (Column 5)
and without BSSE corrections (Column 3)
considering water monomer as adsorbent. By considering the water hexamer structure,
the same comparison is performed noted in Table \ref{tab:BE_3}.
BSSE-corrected BE values are lower than those without BSSE corrections, implying that
the basis set leads to significant BSSE.
Table \ref{tab:BE_2} also shows a comparison between the results obtained by
including ZPVE (Column 4) and without including ZPVE (Column 3), considering water monomer.
Notably, results obtained without ZPVE and BSSE corrections are closer to the experimental values.
To check the dependency on the level of theories,
BE calculations are performed using a higher-level theory (CCSD(T)/aug-cc-pVTZ)
by single-point energy calculations considering water monomer.
The results are shown in Column 7 of Table \ref{tab:BE_2}.
The same higher-level theory is employed to evaluate BEs of some selected species
(N, O, O$_2$, H$_2$O, CO, and N$_2$) considering water tetramer noted in Table \ref{tab:BE_4}.
We notice a minor change of BE values considering BSSE, ZPVE corrections, and higher-level theory.
\cite{wake17} noticed a slightly better fitting of the proportional law
without including ZPVE because the ZPVE is roughly proportional to the BE.
After the inclusion of the ZPVE, the parameters for the fits are not the same,
but the results of the fits show similar deviations.
However, we notice that the MP2 method in conjunction with the aug-cc-pVDZ basis set
(without considering ZPVE and BSSE corrections) is comparatively best among all
the methods and basis sets used in our computation.
A lower-level theory (without considering ZPVE and BSSE corrections) results in better
agreement with the experimental result than those from a higher-level theory.
This is likely because effects various approximations at lower levels cancel each other.
It is possibly a coincidence but is helpful as it makes it easier to process many systems.

\begin{table}
\scriptsize
\centering
\caption{Calculated BEs using water tetramer (adsorbent) to check the influence of the higher-order quantum chemical level of theory \citep{das18}.}
\label{tab:BE_4}
\vskip 0.2 cm
\begin{tabular}{cccc}
\hline
{\bf Sl.} & {\bf Species} & \multicolumn{2}{c}{\bf BE (in Kelvin) using water tetramer} \\
\cline{3-4}
{\bf No.}&&{\bf MP2/aug-cc-pVDZ}&{\bf CCSD(T)/aug-cc-pVTZ} \\
\hline
1 & N & 269 & 273 \\
2 & O & 1002 & 1024 \\
3 & O$_2$ & 940 & 853 \\
4 & H$_2$O & 2670 & 2632 \\
5 & CO & 1263 & 1196 \\
6 & N$_2$ & 900 & 854 \\
\hline
\end{tabular}
\end{table}

BE calculations considering water pentamer and hexamer cluster configuration (as a substrate)
are computationally expensive and sometimes take a long time to converge;
thus, it is not easy to apply for a large set of species.
Table \ref{tab:BE_A1} provides our calculated BE values
for 21 species with the hexamer configuration.
Since BE values taking the water hexamer cluster configuration 
deviate by $\sim \pm 16.7 \%$, they can be scaled by $1 \pm 0.167$.
We also provide a large set of BE values for $100$ crucial interstellar
or circumstellar species considering the water tetramer configuration as a substrate.
In this case, the deviation from the experimental values is $\sim \pm 18.8 \%$ and
can be scaled by $1 \pm 0.188$.
Our calculations taking tetramer configuration of water cluster are an exciting
alternative to complete calculations or the fitting model used by \cite{wake17}.
Some strange values of BE of some species
are due to induced deviation from the global minimum.
The mobility of atoms is significant for grain chemistry in dense molecular clouds
and depends mainly on their BE with the grain surface.
The BEs of some relevant atoms (such as H, He, C, N, O, Na, Mg, Si, P, and S) 
are also included in Table \ref{tab:BE_A1}.
Our calculations with the tetramer configuration likely deviate from the experimental BE values
due to the long-range interactions (interactions with water molecules not close to the species, as visible from Figures \ref{fig:BE_T1}$-$\ref{fig:BE_T5}), which seems crucial for cases with low BEs. 
Thus, using only a scale factor without offset from the tetramer configuration may 
likely underestimate the real BE, particularly for low BE cases.

\begin{table}
\tiny
\caption{Calculated and available list of BEs of various species \citep{das18}.}
\label{tab:BE_A1}
\vskip 0.2 cm
\hskip -0.8 cm
\begin{tabular}{cccccccc}
\hline
{\bf Sl.} & {\bf Species} & {\bf Ground } & {\bf BE (in Kelvin) } & {\bf BE (in Kelvin)} & \multicolumn{2}{c}{\bf \underline{BE from KIDA}} & {\bf BE (in Kelvin)} \\
{\bf No.} &&{\bf state used} & {\bf on water tetramer}&{\bf on water hexamer}  & {\bf Old values} & {\bf New values} & {\bf from other} \\
& &  & & & {\bf (in Kelvin)} & {\bf (in Kelvin)} & {\bf literature sources} \\
\hline
1 & H & Doublet & 125 & 181 & 450$^g$ & 650 $\pm$ 195 & 650 $\pm$ 100$^c$ \\
2 & H$_2$ & Singlet & 528 & 545 & 450 & 440 $\pm$ 132 & 500 $\pm$ 100$^c$ \\
3 & He & Singlet & 113 &  & 100 & & 100 $\pm$ 50$^c$  \\
4 & C & Triplet & 660 &  & 800 & 10000 $\pm$ 3000 & 715 $\pm$ 360$^c$, 14100 $\pm$ 420$^k$ \\
5 & N & Quartet & 269 & 619 & 800 & 720 $\pm$ 216 & 715 $\pm$ 358$^c$, 400 $\pm$ 30$^k$ \\
6 & O & Triplet & 1002 & 660 & 1660 $\pm$ 60$^a$ & 1600 $\pm$ 480 & 1660 $\pm$ 60$^c$, \\
&&&&&&& 1504 $\pm$ 12$^j$, 1440 $\pm$ 160$^k$ \\
7 & Na & Doublet & 2214  && 11800 & & 10600 $\pm$ 500$^c$ \\
8 & Mg & Singlet & 654 && 5300 & & 4750 $\pm$ 500$^c$ \\
9 & Si & Triplet & 6956 && 2700 & 11600 $\pm$ 3480 & 2400 $\pm$ 500$^c$ \\
10 & P & Quartet & 616 && 1100 & & 750 $\pm$ 375 \\
11 & S & Triplet & 1428 && 1100 & 2600 $\pm$ 780 & 985 $\pm$ 495$^c$  \\
12 & NH & Triplet & 1947 && 2378 & 2600 $\pm$ 780 & 542 $\pm$ 270$^c$ \\
13 & OH & Doublet & 3183 && 2850$^g$ & 4600 $\pm$ 1380 & 3210 $\pm$ 1550$^c$ \\
14 & PH & Triplet & 944 && && 800 $\pm$ 400 \\
15 & C$_2$ & Triplet & 9248 && 1600 & & 1085 $\pm$ 500$^c$ \\
16 & HF & Singlet & 5540 &&&& 500 $\pm$ 250 \\
17 & HCl & Singlet & 3924 & 4104 & 5174 $\pm$ 1$^b$ & 5172 $\pm$ 1551.6 & 900 $\pm$ 450 \\
18 & CN & Doublet & 1736 & & 1600 & & 1355 $\pm$ 500$^c$ \\
19 & N$_2$ & Singlet & 900 & 1161 & 1000$^g$ & 1100 $\pm$ 330 & 990 $\pm$ 100$^c$, 1000$^f$ \\
20 & CO & Singlet & 1263 & 1292 & 1150$^g$ & 1300 $\pm$ 390 & 1100 $\pm$ 250$^c$, 1300$^e$ \\
21 & SiH & Doublet & 8988 & & 3150 & 13000 $\pm$ 3900 & 2620 $\pm$ 500$^c$  \\
22 & NO & Doublet & 1265 & 1988 & 1600 & 1600 $\pm$ 480 & 1085 $\pm$ 500$^c$ \\
23 & O$_2$ & Triplet & 940 & 1352 & 1000 & 1200 $\pm$ 360 & 898 $\pm$ 30$^c$, 1200$^e$, 1000$^f$ \\
24 & HS & Doublet & 2221 & & 1450 & 2700 $\pm$ 810 & 1350 $\pm$ 500$^c$ \\
25 & SiC & Triplet & 5850 & & 3500 & & 3150 $\pm$ 500$^c$ \\
26 & CP & Doublet & 1699  & & 1900 & & 1050 $\pm$ 500 \\
27 & CS & Singlet & 2217  && 1900 & 3200 $\pm$ 960 & 1800 $\pm$ 500$^c$ \\
28 & NS & Doublet & 2774 && 1900 & & 1800 $\pm$ 500$^c$ \\
29 & SO & Triplet & 2128  && 2600 & 2800 $\pm$ 840 & 1800 $\pm$ 500$^c$ \\
30 & S$_2$ & Triplet & 1644 && 2200 & & 2000 $\pm$ 500$^c$ \\
31 & CH$_2$ & Triplet & 1473 && 1050 & 1400 $\pm$ 420 & 860 $\pm$ 430$^c$ \\
32 & NH$_2$ & Doublet & 3240  && 3956 & 3200 $\pm$ 960 & 770 $\pm$ 385$^c$ \\
33 & H$_2$O & Singlet & 2670  & 4166 & 5700$^g$ & 5600 $\pm$ 1680 & 4800 $\pm$ 100$^c$ \\
34 & PH$_2$ & Doublet & 1226  && 2000 && 850 $\pm$ 425 \\
35 & C$_2$H & Doublet & 2791 && 2137 & 3000 $\pm$ 900 & 1330 $\pm$ 500$^c$ \\
36 & N$_2$H & Doublet & 3697 && 1450 & & \\
37 & O$_2$H & Doublet & 5778 && 3650 & 5000 $\pm$ 1500 & 1510 $\pm$ 500$^c$ \\
38 & HS$_2$ & Doublet & 4014 && 2650 & & 2300 $\pm$ 500$^c$ \\
39 & HCN & Singlet & 2352 && 2050 & 3700 $\pm$ 1110 & 1580 $\pm$ 500$^c$ \\
40 & HNC & Singlet & 5225 && 2050 & 3800 $\pm$ 1140 & 1510 $\pm$ 500$^c$ \\
41 & HCO & Doublet & 1857 && 1600$^g$ & 2400 $\pm$ 720 & 1355 $\pm$ 500$^c$ \\
42 & HOC & Doublet & 5692 && & & \\
43 & HCS & Doublet & 2713 && 2350 & 2900 $\pm$ 870 & 2000 $\pm$ 500$^c$ \\
44 & HNO & Singlet & 2988 && 2050 & 3000 $\pm$ 900 & 1510 $\pm$ 500$^c$ \\
45 & H$_2$S & Singlet & 2556 & 3232 & 2743$^f$ & 2700 $\pm$ 810 & 2290 $\pm$ 90$^c$ \\
46 & C$_3$ & Singlet & 2863 && 2400 & 2500 $\pm$ 750 & 2010 $\pm$ 500$^c$ \\
47 & O$_3$ & Singlet & 2545 && 1800 & 2100 $\pm$ 630 & 2100 $\pm$ 100$^c$ \\
48 & C$_2$N  & Doublet & 1281 && 2400 & & 2010 $\pm$ 500$^c$ \\
49 & C$_2$S  & Triplet & 2477 && 2700 & & 2500 $\pm$ 500$^c$ \\
50 & OCN & Doublet & 3085 && 2400 & & 1805 $\pm$ 500$^c$ \\
51 & CO$_2$ & Singlet & 2293 & 2352 & 2575 & 2600 $\pm$ 780 & 2267 $\pm$ 70$^c$, 2300$^e$ \\
52 & OCS & Singlet & 1571 & 1808 & 2888 & 2400 $\pm$ 720 & 2325 $\pm$ 95$^c$ \\
53 & SO$_2$ & Singlet & 3745 && 3405 & 3400 $\pm$ 1020 & 3010 $\pm$ 110$^c$ \\
54 & CH$_3$ & Doublet & 1322  && 1175 & 1600 $\pm$ 480 & 1040 $\pm$ 500$^c$ \\
55 & NH$_3$ & Singlet & 3825  & 5163 & 5534 & 5500 $\pm$ 1650 & 2715 $\pm$ 105$^c$, 5530$^f$ \\
\hline
\end{tabular}
\end{table}

\begin{table}
\tiny
\hskip -0.8 cm
\begin{tabular}{cccccccc}
\hline
{\bf Sl.} & {\bf Species} & {\bf Ground} & {\bf BE (in Kelvin)} & {\bf BE (in Kelvin)} & \multicolumn{2}{c}{\bf \underline{BE from KIDA}} & {\bf BE (in Kelvin)} \\
{\bf No.} & & {\bf state used} & {\bf on water tetramer} & {\bf on water hexamer} & {\bf Old values} & {\bf New values} & {\bf from other} \\
& & & & & {\bf (in Kelvin)} & {\bf (in Kelvin)} & {\bf literature sources} \\
\hline
\hline
56 & SiH$_3$ & Doublet & 1269 && 4050 & & 3440 $\pm$ 500$^c$ \\
57 & C$_2$H$_2$ & Singlet & 2593 & 2640 & 2587$^f$ & 2587 $\pm$ 776.1 & 2090 $\pm$ 85$^c$ \\
58 & N$_2$H$_2$ & Singlet & 3183 &&& & \\
59 & H$_2$O$_2$ & Singlet & 3928 & 4248 & 5700 & 6000 $\pm$ 1800 & 6000 $\pm$ 100$^c$, 5410$^l$ \\
60 & H$_2$S$_2$ & Singlet & 4368 && 3100 & & 2600 $\pm$ 500$^c$ \\
61 & H$_2$CN & Doublet & 2984 && 2400 & & 2400 $\pm$ 500$^c$ \\
62 & CHNH & Doublet & 3742 &&& & \\
63 & H$_2$CO & Singlet & 3242 && 2050$^g$ & 4500 $\pm$ 1350 & 3260 $\pm$ 60$^c$  \\
64 & CHOH & Triplet & 4800 &&& & \\
65 & HC$_2$N & Triplet & 3289 &&& & 2270 $\pm$ 500$^c$ \\
66 & HC$_2$O & Doublet & 2914 && 2400 & & 2010 $\pm$ 500$^c$ \\
67 & HNCO & Singlet & 3922 & 5554 & 2850 & 4400 $\pm$ 1320 & 2270 $\pm$ 500$^c$, 3900$^h$ \\
68 & H$_2$CS & Singlet & 3110 && 2700 & 4400 $\pm$ 1320 & 2025 $\pm$ 500$^c$ \\
69 & C$_3$O & Singlet & 3542 && 2750 & & 2520 $\pm$ 500$^c$ \\
70 & CH$_4$ & Singlet & 1327 & 2321 & 1300$^g$ & 960 $\pm$ 288 & 1250 $\pm$ 120$^c$ \\
71 & SiH$_4$ & Singlet & 1527 && 4500 & & 3690 $\pm$ 500$^c$ \\
72 & C$_2$H$_3$ & Doublet & 2600 && 3037 & 2800 $\pm$ 840 & 1760 $\pm$ 500$^c$  \\
73 & CHNH$_2$ & Singlet & 7069 &&& & \\
74 & CH$_2$NH & Singlet & 4352 & & 5534 & & 1560 $\pm$ 500$^c$ \\
75 & CH$_3$N & Triplet & 2194 &&& & \\
76 & c-C$_3$H$_2$ & Singlet & 3892 && 3387 & 5900 $\pm$ 1770 & 2110 $\pm$ 500$^c$ \\
77 & H$_2$CCN & Doublet & 3730  && 4230 &  & 2470 $\pm$ 500$^c$ \\
78 & H$_2$CCO & Singlet & 2847 && 2200 & 2800 $\pm$ 840 & 2520 $\pm$ 500$^c$ \\
79 & HCOOH & Singlet & 3483 && 5570$^g$ & & 4532 $\pm$ 150$^c$ \\
80 & CH$_2$OH & Doublet & 4772 && 5084 & 4400 $\pm$ 1320 & 2170 $\pm$ 500$^c$ \\
81 & NH$_2$OH & Singlet & 4799 && & & 2770 $\pm$ 500$^c$ \\
82 & C$_4$H & Doublet & 2946 && 3737 & & 2670 $\pm$ 500$^c$ \\
83 & HC$_3$N & Singlet & 2925  && 4580 & & 2685 $\pm$ 500$^c$ \\
84 & HC$_3$O & Doublet & 2619 &&& & \\
85 & C$_5$ & Singlet & 2403 && 4000 & & 3220 $\pm$ 500$^c$ \\
86 & C$_2$H$_4$ & Singlet & 2052 && 3487 & 2500 $\pm$ 750 & 2010 $\pm$ 500$^c$ \\
87 & CH$_2$NH$_2$ & Doublet & 3831 && 5534$^d$ & & \\
88 & CH$_3$NH & Doublet & 3414 &&& & 1760 $\pm$ 500$^c$ \\
89 & CH$_3$OH & Singlet & 4368 & 4511 & 5534 & 5000 $\pm$ 1500 & 3820 $\pm$ 135$^c$, 5530$^f$ \\
90 & CH$_2$CCH & Doublet & 2726 & & 3837 & 3300 $\pm$ 990 & 3840 $\pm$ 500$^c$ \\
91 & CH$_3$CN & Singlet & 2838 & 3786 & 4680$^f$ & 4680 $\pm$ 1404 & 3790 $\pm$ 130$^c$ \\
92 & H$_2$C$_3$N & Doublet & 2637 &&& & \\
93 & H$_2$C$_3$O & Singlet & 3006 &&& & \\
94 & C$_6$ & Quintet & 3226 && 4800 & & 3620 $\pm$ 500$^c$ \\
95 & CH$_3$NH$_2$ & Singlet & 4434 && 6584 & & 5130 $\pm$ 500$^c$, 4269$^m$ \\
96 & C$_2$H$_5$ & Doublet & 1752 && 3937 & 3100 $\pm$ 930 & 2110 $\pm$ 500$^c$ \\
97 & CH$_3$CCH & Singlet & 2342 & 3153 & 4287 & 3800 $\pm$ 1140 & 4290 $\pm$ 500$^c$, 2500 $\pm$ 40$^i$ \\
98 & CH$_2$CCH$_2$ & Singlet & 2775 && & 3000 $\pm$ 900 & 4290 $\pm$ 500$^c$ \\
99 & CH$_3$CHO & Singlet & 3849 && 2450 & 5400 $\pm$ 1620 & 2870 $\pm$ 500$^c$ \\
100 & C$_7$ & Singlet & 4178 && 5600 & & 4430 $\pm$ 500$^c$ \\
\hline
\hline
\end{tabular} \\
\vskip 0.2cm
{\bf Note:} \\
The average deviations from experiments for the cases of tetramer (Column 4) and hexamer (Column 5) are 
$\sim \pm 18.8 \%$ and $\sim \pm 16.7 \%$ respectively. \\
$^a$\cite{he15},
$^b$\cite{olan11},
$^c$\cite{pent17},
$^d$\cite{ruau15},
$^e$\cite{mini16},
$^f$\cite{coll04},
$^g$\cite{garr06b},
$^h$\cite{nobl15},
$^i$\cite{kimb14},
$^j$\cite{ward12},
$^k$\cite{shim18},
$^l$\cite{lamb17}.
$^m$\cite{chaa18}.
\end{table}

\begin{figure}
\centering
\includegraphics[width=\textwidth]{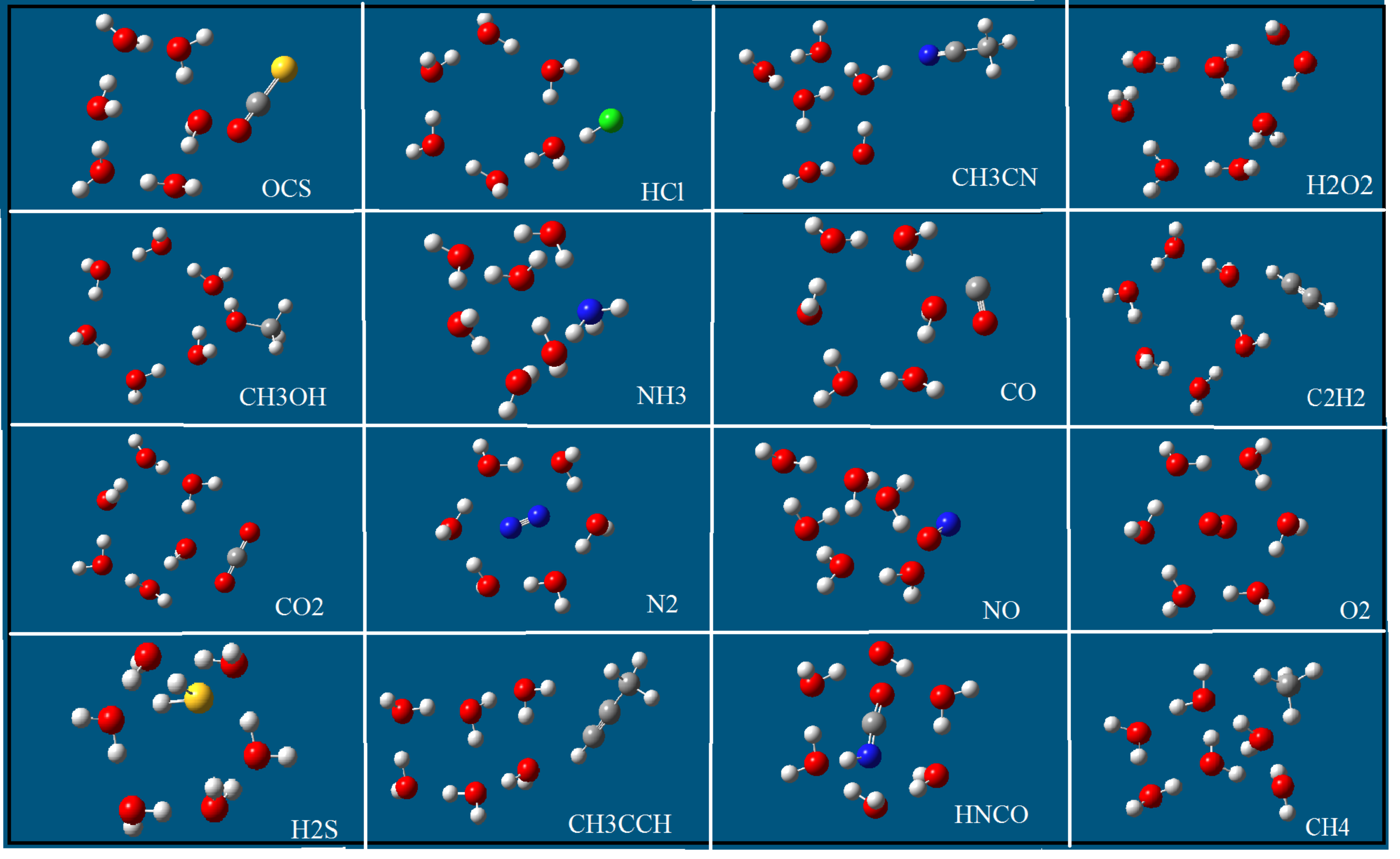}
\caption{Optimized geometries with the c-hexamer (chair) configuration \citep{das18}.}
\label{fig:BE_H1}
\end{figure}

\begin{figure}
\centering
\includegraphics[width=\textwidth]{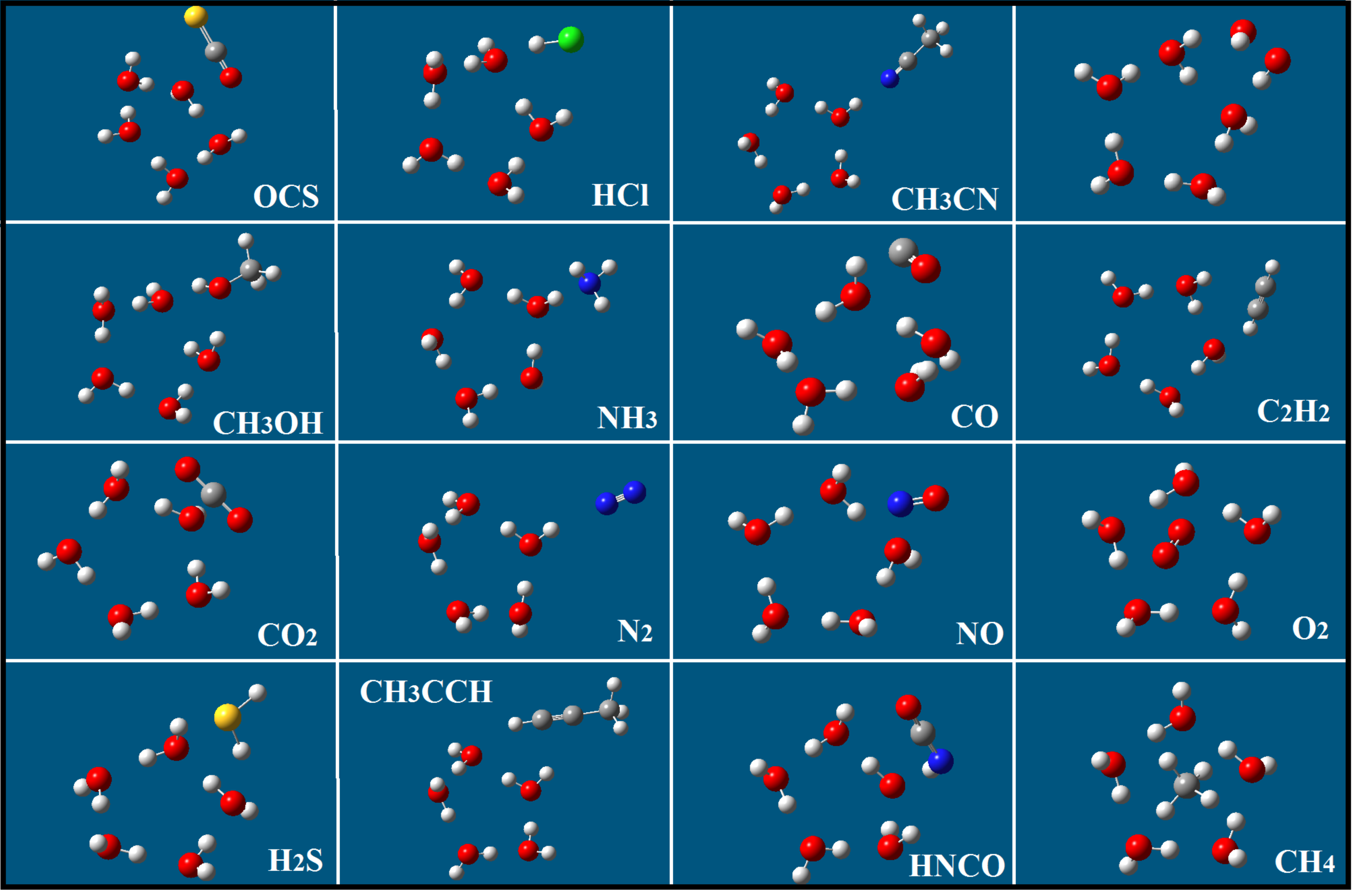}
\caption{Optimized geometries with the c-pentamer configuration \citep{das18}.}
\label{fig:BE_P1}
\end{figure}

\begin{figure}
\centering
\includegraphics[width=\textwidth]{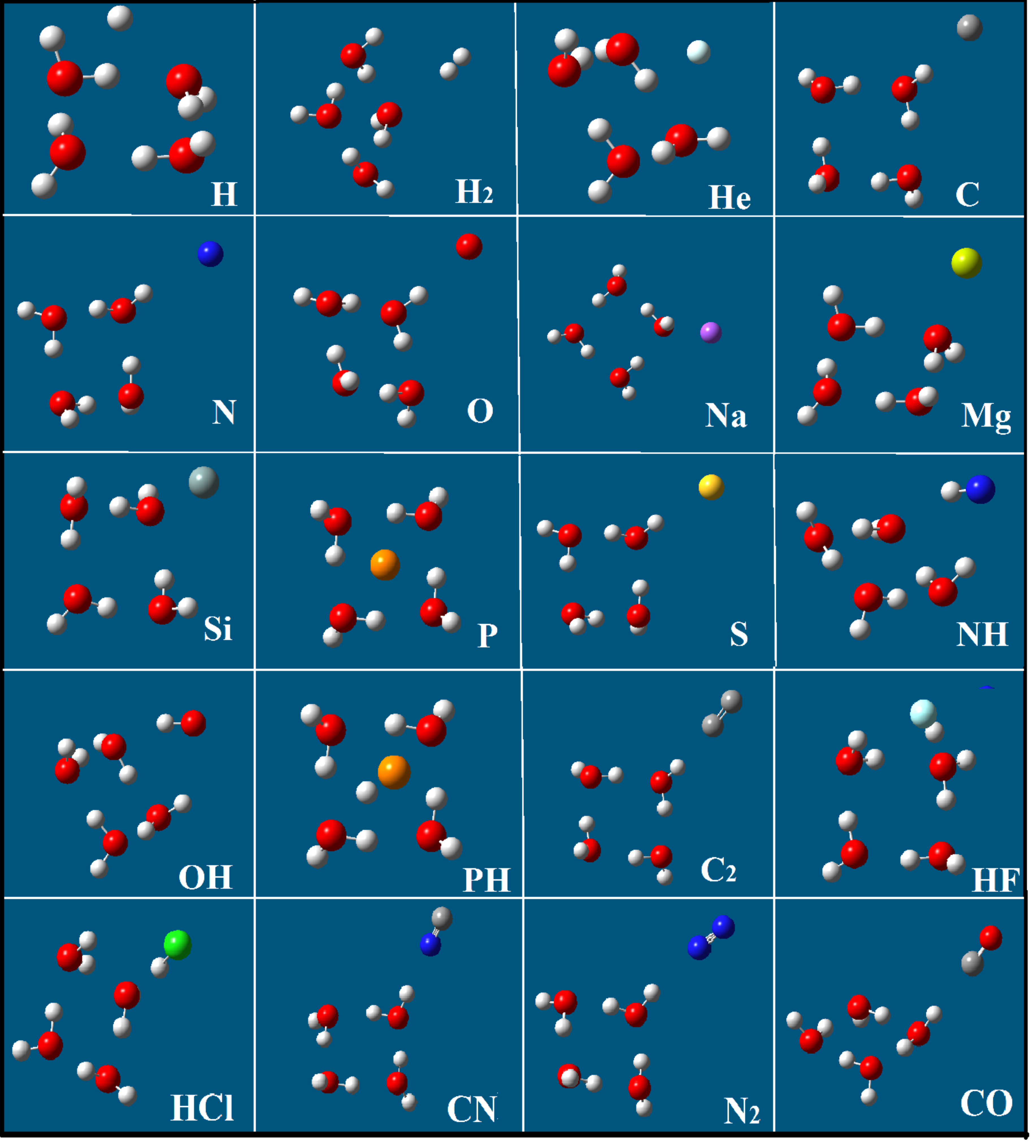}
\caption{Optimized geometries with the c-tetramer configuration \citep{das18}.}
\label{fig:BE_T1}
\end{figure}

\begin{figure}
\centering
\includegraphics[width=\textwidth]{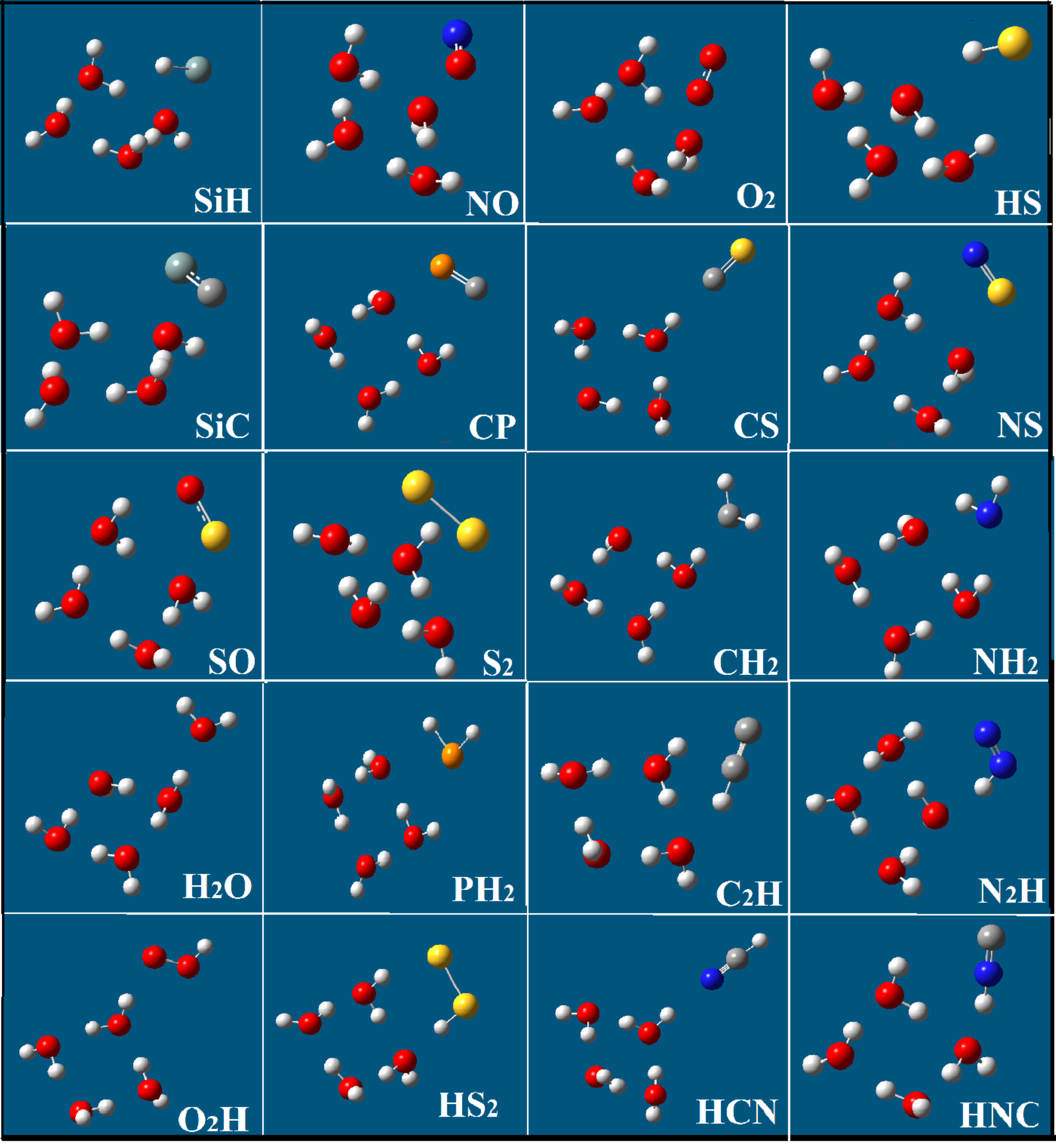}
\caption{Optimized geometries with the c-tetramer configuration \citep{das18}.}
\label{fig:BE_T2}
\end{figure}

\begin{figure}
\centering
\includegraphics[width=\textwidth]{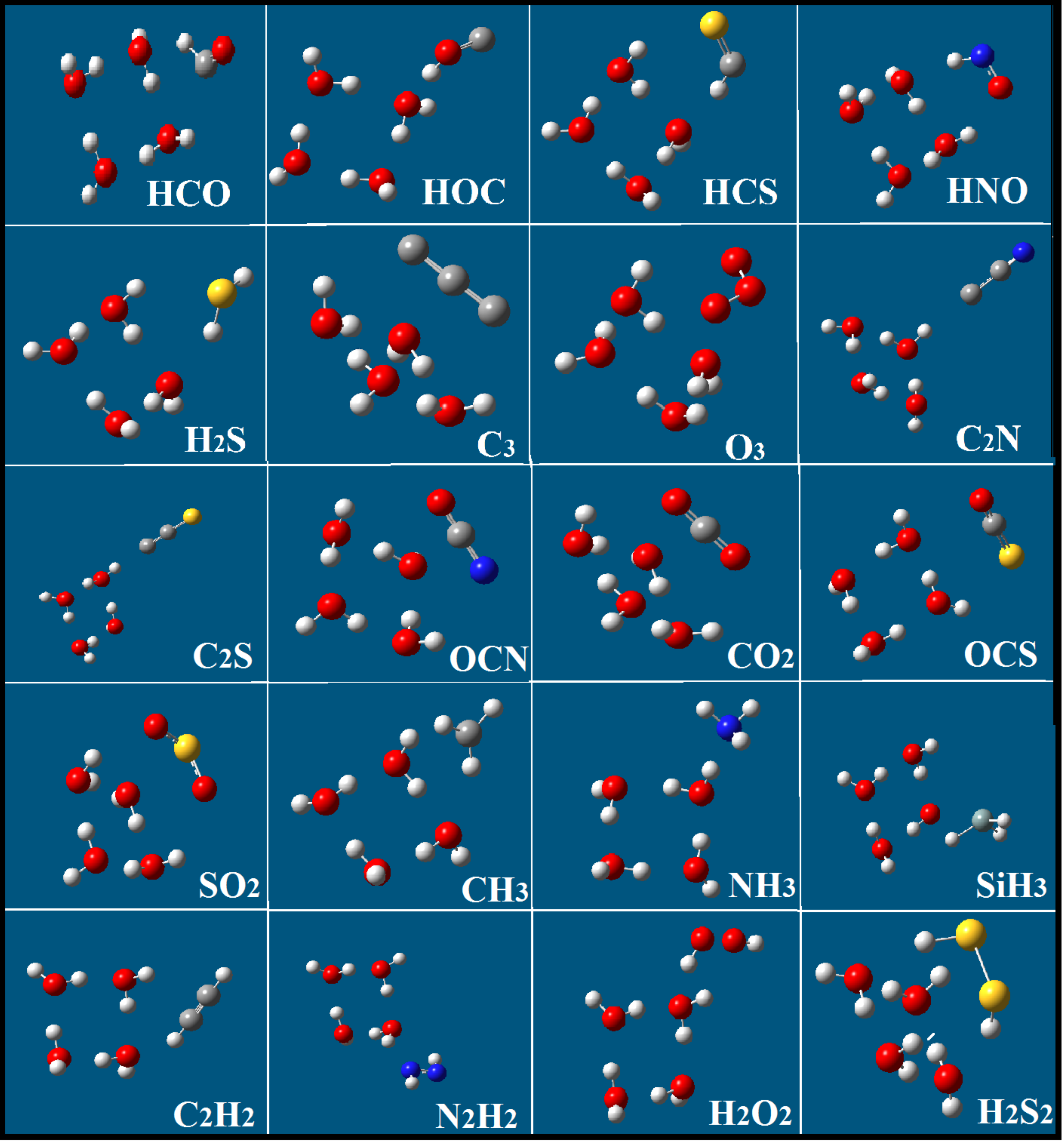}
\caption{Optimized geometries with the c-tetramer configuration \citep{das18}.}
\label{fig:BE_T3}
\end{figure}

\begin{figure}
\centering
\includegraphics[width=\textwidth]{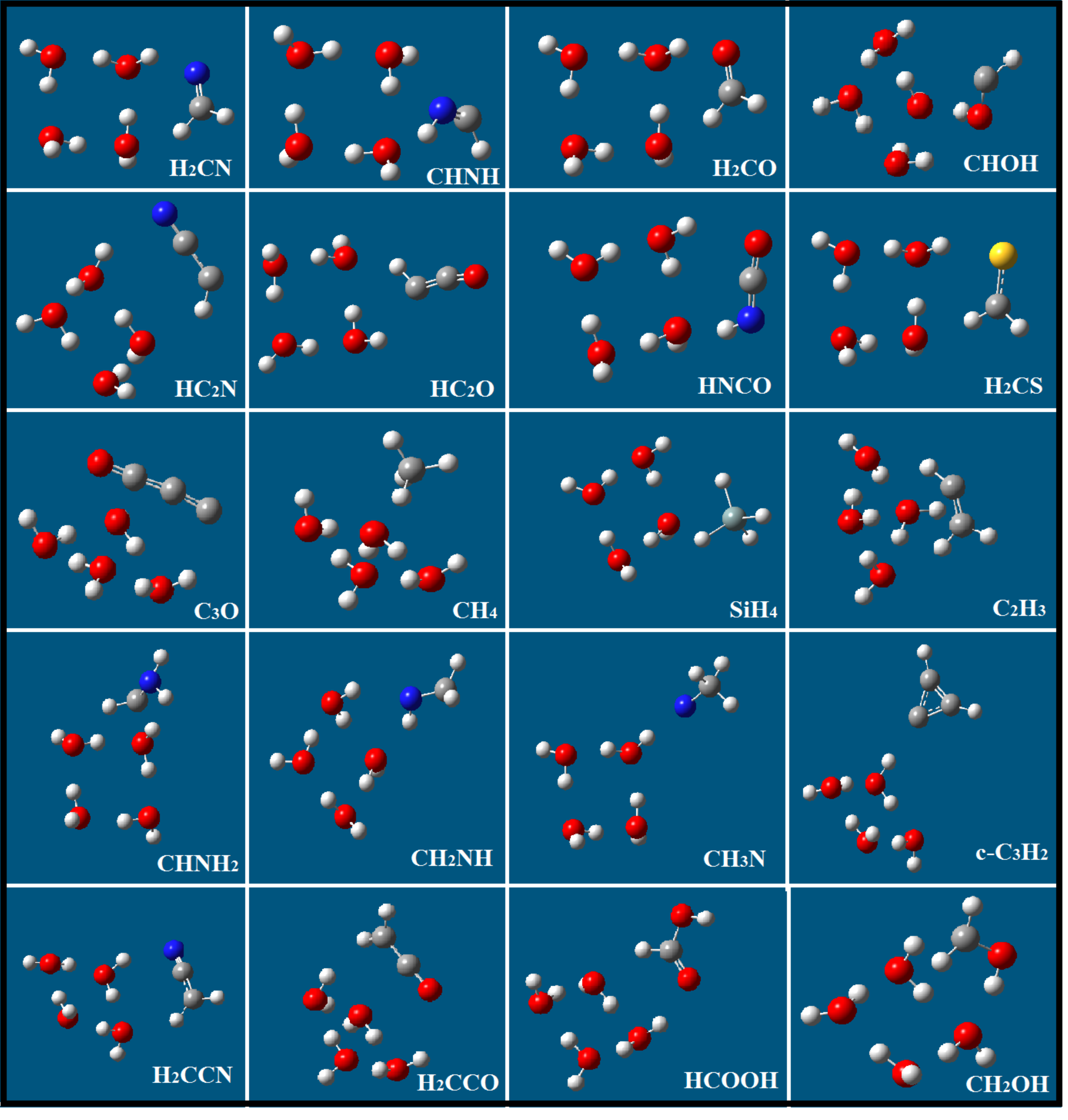}
\caption{Optimized geometries with the c-tetramer configuration \citep{das18}.}
\label{fig:BE_T4}
\end{figure}

\begin{figure}
\centering
\includegraphics[width=\textwidth]{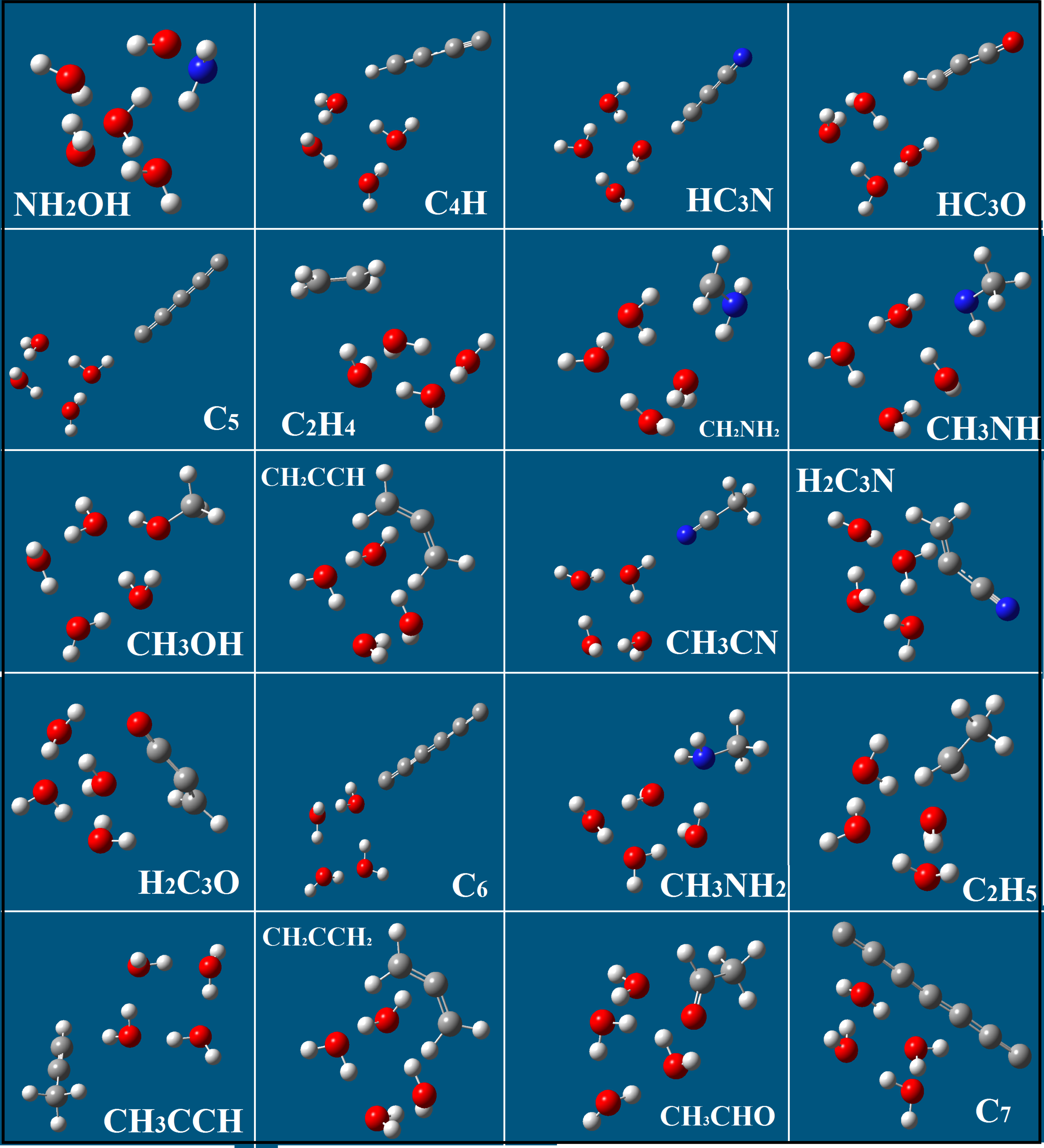}
\caption{Optimized geometries with the c-tetramer configuration \citep{das18}.}
\label{fig:BE_T5}
\end{figure}

Previously used and recently proposed BE values by the KIDA\footnote{\url{http://kida.obs.u-bordeaux1.fr}} \citep{wake17} and other sources are provided in Columns 6, 7, and 8, respectively, of Table \ref{tab:BE_A1}.
Among the sixteen stable species shown in Tables \ref{tab:BE_1} and \ref{tab:BE_2},
there are both polar (OCS, CH$_3$OH, NH$_3$, H$_2$S, CH$_3$CCH, CH$_3$CN, H$_2$O$_2$, HCl, HNCO, NO) and non-polar molecules (CO, CO$_2$, N$_2$, O$_2$, CH$_4$, C$_2$H$_2$).
We notice that for the non-polar species, BE calculation sometimes overestimates the attractive interaction if we choose the position of the species at the center of a water cluster.
The species (like HCl, CH$_3$OH, etc.) leading to hydrogen bonds with H$_2$O
are localized on one water molecule with H-bonding.
In most such cases, the interaction is well reproduced by our calculations.
The case with the oxygen atom is found as an intermediate one.
The BEs of the oxygen atom are found to be $1002$ K and $660$ K (given in Table \ref{tab:BE_A1})
for the tetramer and hexamer configurations, respectively.
In contrast, the current value is $1660 \pm 60$ K in KIDA \citep{wake17} desorption energy on an ASW surface. This is because the complex containing water hexamer with an oxygen atom is significantly modified,
so the interaction with water is weaker.
It seems that the interaction with the oxygen atom does not compensate for this.
However, the values are more accurate in the complex containing water tetramer with an oxygen atom.
The likely cause could be that water interaction is not so perturbed in that case.
The BE of hydrogen atom is found to be $125$ K with tetramer and $181$ K with hexamer (see Table \ref{tab:BE_A1}).
Strangely, it leads to a much smaller
interaction compared to the average value ($\sim 650$ K) computed by \cite{alha07}.
Thus, though a very good approximation (on an average) is found by using the pentamer
and hexamer configurations, we should keep in mind that using only one adsorption site
geometry may induce significant errors in BE values.
For a targeted species, it is recommended to consider different
adsorption sites and corresponding optimized geometries and then take an average to come up with
a more accurate approximation. 
In our case, when various binding sites are found, we try to choose the one with the
closest available experimental values.
We provide the complex geometries used to determine the BEs as
\href{https://cfn-live-content-bucket-iop-org.s3.amazonaws.com/journals/0067-0049/237/1/9/1/apjsaac886.tar.gz?AWSAccessKeyId=AKIAYDKQL6LTV7YY2HIK&Expires=1623772201&Signature=S\%2BDGvwW3CEdHh\%2FkVDB6fWplAxbs\%3D}{supplementary materials} \citep{das18}.
Figures \ref{fig:BE_H1}$-$\ref{fig:BE_T5} provide the optimized geometries for the species considered here with hexamer, pentamer, and tetramer configurations,
respectively. The BEs are very much sensitive to the chosen ground state spin multiplicity.
Therefore, the ground states used for each species are also provided in Column 3 of Table \ref{tab:BE_A1}.
The ground state spin multiplicities of each species are calculated
using the \textsc{Gaussian} 09 suite of programs.
For this purpose, separate calculations (job type ``opt+freq") are performed,
each with different spin multiplicities,
and then compare. The lowest energy electronic state solution of the chosen spin multiplicity is the ground state noted for the species in Table \ref{tab:BE_A1}.

\clearpage
\subsubsection{Summary}
A series of quantum chemical calculations are performed systematically to compute
physisorption BE of various interstellar adsorbed species by considering the monomer, dimer,
c-trimer, c-tetramer, c-pentamer, and c-hexamer (chair) configurations of water clusters as substrates. The benefit of using two or more water molecules
is that one can consider the H-bonding donor and acceptor behavior of some species.
We note that our computed BEs approach gradually to the experimental values when the size of the cluster is increased. More specifically, when the c-pentamer and
c-hexamer (chair) configurations are considered, our computed BE values deviate from experimental values on average of $\sim \pm 15.8 \%$ and $\sim \pm 16.7 \%$, respectively.
It is worth mentioning that the computed BE would be saturated to the experimental values upon
considering a sufficient number of binding sites.
For a broad set of interstellar or circumstellar species, we provide the BE values that deviate, on an average, $\sim \pm 18.8 \%$ with the c-tetramer configuration. The comparisons show that we can safely use our procedure to compute the BEs of any such molecules with a reasonable accuracy.

\clearpage

\subsection{Effect of binding energies on the encounter desorption}

The encounter desorption mechanism was initially introduced by \cite{hinc15} to eliminate the overestimation of the abundance of $\rm{H_2}$ on the grain.
The desorption process occurs during surface diffusion and is induced by the presence of repulsive inter-H$_2$ forces, effectively reducing the BE of $\rm{H_2}$.
They considered $\rm{gH_2+gH_2 \rightarrow H_2 + gH_2}$, where ``g'' denotes the surface species. They obtained a striking match between the microscopic MC and rate equation approach when this unique approach was implemented.
The MC method is best suited for monitoring the chemical composition of a grain mantle.
However, it is a time-consuming \citep{chak06a,chak06b,das08a,das10,das11,das16,cupp07}.
Recently, \cite{chan21} considered a similar process and included desorption of H by a similar mechanism. They considered $\rm{gH+gH_2 \rightarrow H + gH_2}$, which means that the surface H desorbs with a certain probability whenever the surface H meets one surface H$_2$. They reported a significant deviation between the formation of some key surface species with the inclusion of this treatment.

The encounter desorption effect is important mainly during the cold prestellar phase. Classifying the various evolutionary phases of star formation is one of the essential intricacies of astronomy and astrophysics. A thorough understanding of the star formation process is yet to be fully established. Briefly,  stars are formed by a long condensation process \citep{paga12}. In the beginning, warm diffuse material ($\sim 8000$ K) converts into a cold neutral atomic gas ($\sim 100$ K and $\sim 10 - 100$ cm$^{-3}$). Later, it transforms into a more dense region ($10^2-10^4$ cm$^{-3}$ and $\sim 10-20$ K). In the absence of the other heating sources, a dense core ($> 10^4$ cm$^{-3}$) develops in some places, which further evolves into prestellar cores ($> 10^5$ cm$^{-3}$) \citep{berg07,keto08}. It further evolves to form a protostar. The gas-phase abundance in the colder and denser regions decreases due to their depletion. On the contrary, ice mantles develop. In the low-temperature regime, the hydrogenation reaction mainly controls the chemical complexity. The chemical composition of the bulk ice varies with the evolutionary stages of the star formation. Thus, it is expected that the ice composition would be very different in various places. However, from the IR observations, it appears that significant repositories of interstellar hydrogen, oxygen, carbon, and nitrogen are water (H$_2$O), methanol (CH$_3$OH), carbon monoxide (CO), carbon dioxide (CO$_2$), formaldehyde (H$_2$CO), methane (CH$_4$), and ammonia (NH$_3$) \citep{boog15}.

BE values are mainly obtained from the TPD studies on various model substrates like graphite, diamond-like carbon, amorphous or crystalline silica, silicates, water, and other ice surfaces \citep{coll04,acha07,ward12,nobl12,duli13}. But, adequate information regarding BE of the species with $\rm{H_2}$ substrate is lacking. \cite{vida91} estimated the BE of H atom on $\rm{H_2}$ substrate $\sim 45$ K. \cite{cupp07} estimated the BEs of several species (e.g., O, OH, $\rm{H_2,\ O_2,\ H_2O,\ O_3,\ O_2H,\ and \ H_2O_2}$) on $\rm{H_2}$ substrate by scaling its obtained BE on H$_2$O substrate with the ratio of BE between H on $\rm{H_2}$ and on H$_2$O substrates.
Thus, a vital impediment in examining the encounter desorption with other species is the shortage of information about the BE of these species with $\rm{H_2}$ molecule. Here, we employ the quantum chemical calculations to determine the BE of $95$ interstellar species considering $\rm{H_2}$ substrate. Obtained BEs are used in our chemical model \citep{das21}.

\subsubsection{Computational details and methodology}

\begin{table}
\scriptsize
\centering
\caption{Calculated BE (with MP2/aug-cc-pVDZ level of theory) of various species with H$_2$ monomer surface \citep{das21}.
\label{tab:BE}}
\vskip 0.2cm
\begin{tabular}{|c|c|c|c|c|c|}
\hline
{\bf Sl.}& {\bf Species} & {\bf Optimized} & {\bf Ground} & \multicolumn{2}{c|}{\bf Binding Energy} \\
\cline{5-6}
 {\bf No.} & & {\bf Structures} & {\bf State} & {\bf in K} & {\bf in kJ/mol} \\
\hline
1 & H & \includegraphics[width=0.2\textwidth]{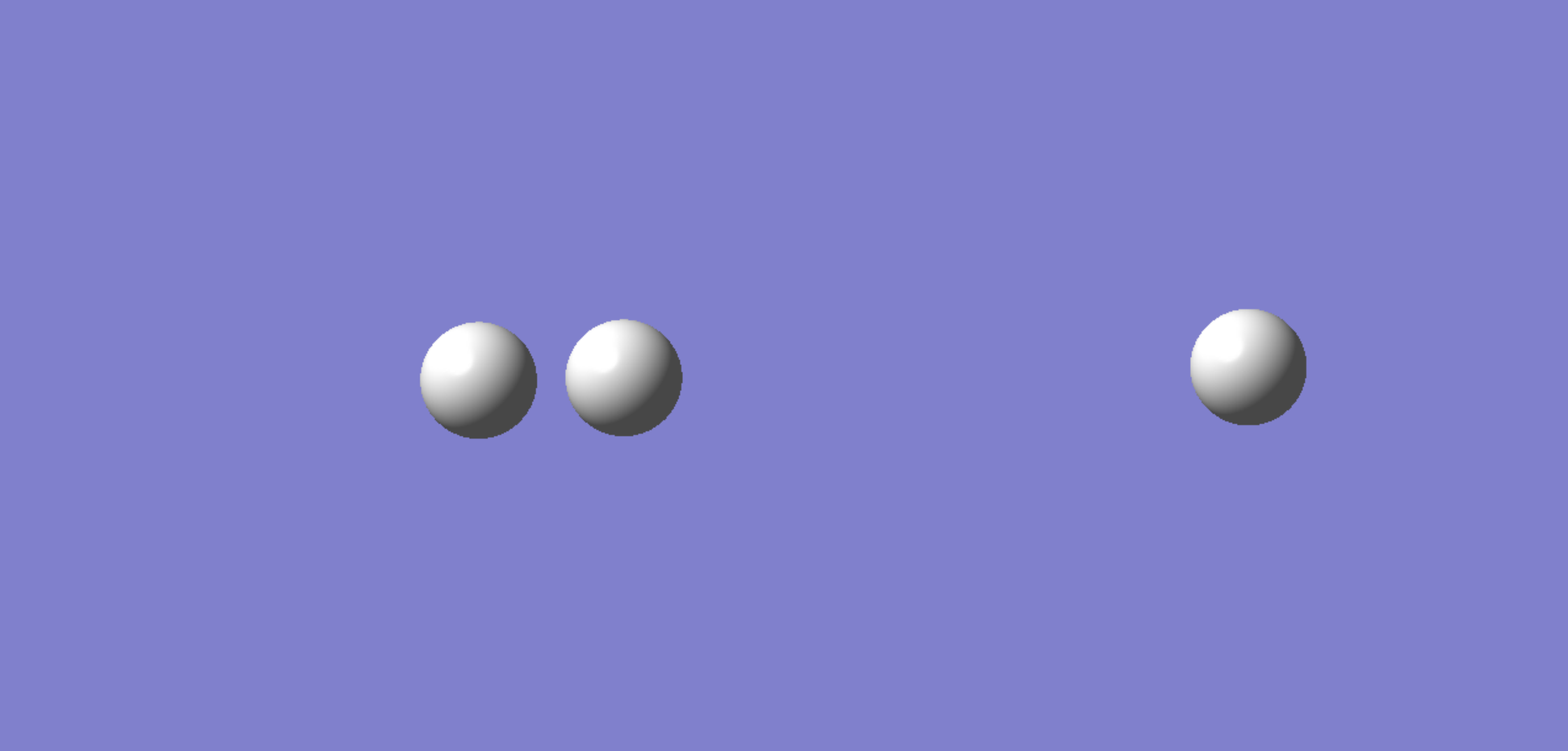} & Doublet & 23 (25$^a$), 45$^c$ & 0.189 (0.210$^a$) \\
2 & H$_2$ & \includegraphics[width=0.2\textwidth]{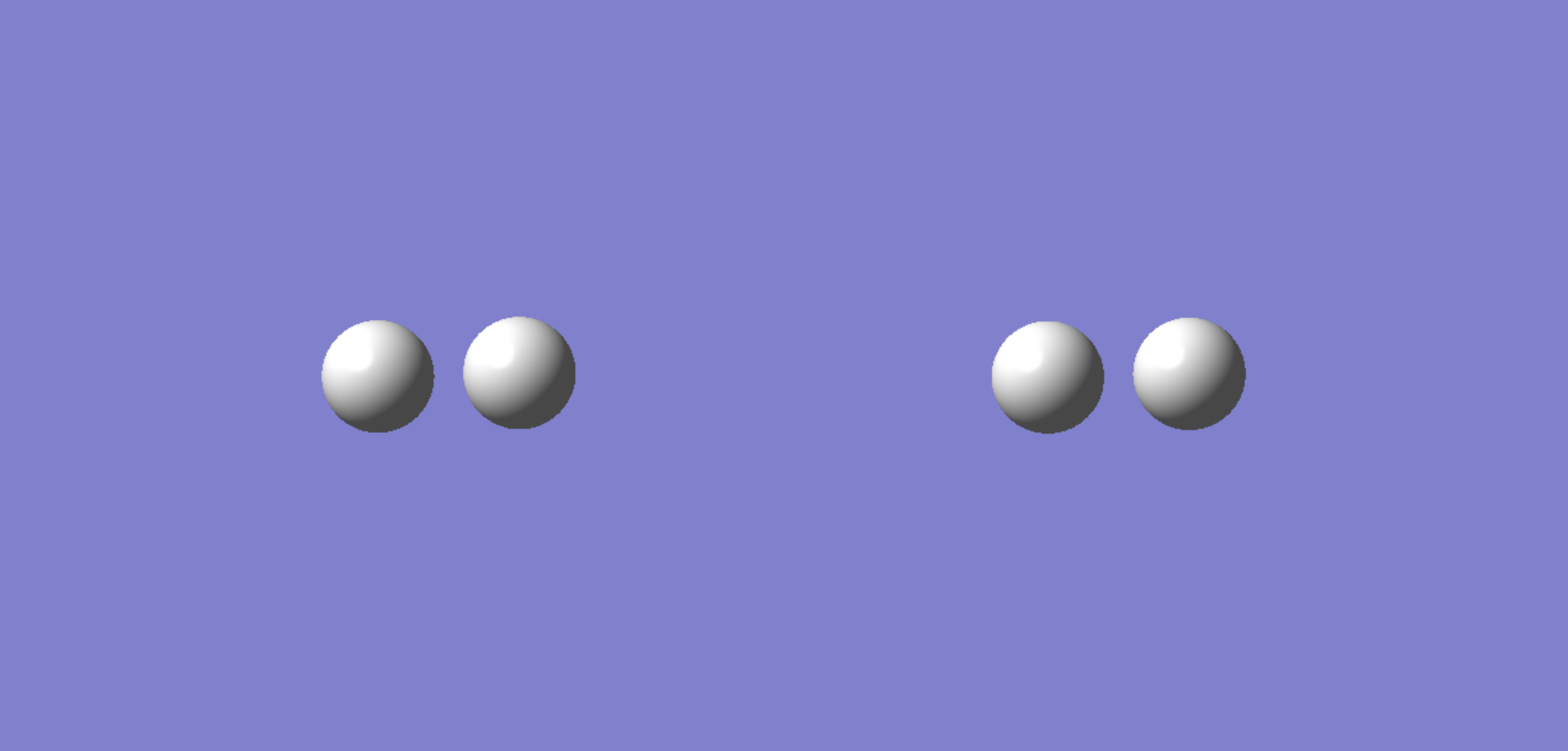} & Singlet & 67 (79$^a$), 23$^c$, 100$^d$ & 0.549 (0.659$^a$) \\
3&He & \includegraphics[width=0.2\textwidth]{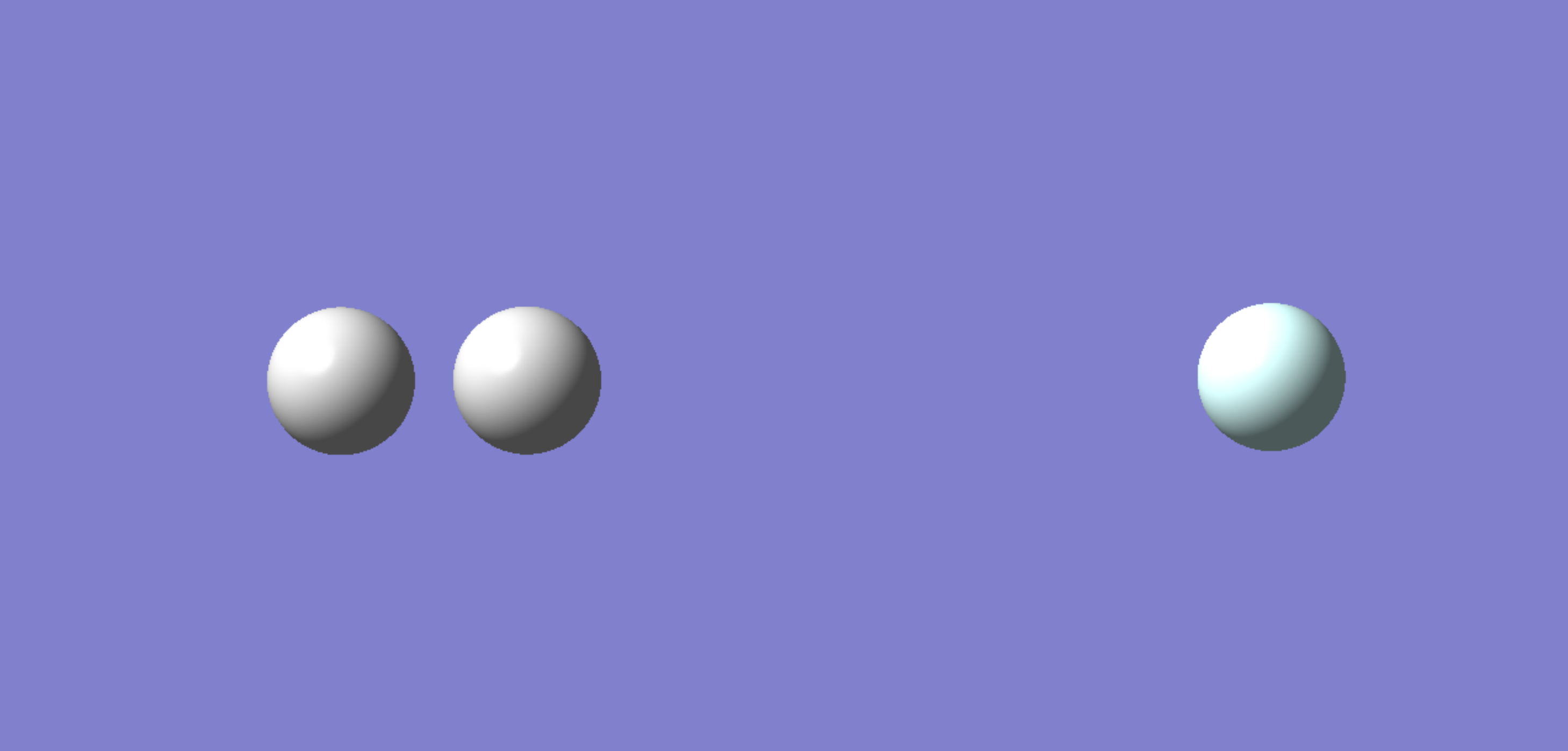} & Singlet &27& 0.226 \\
4&C & \includegraphics[width=0.2\textwidth]{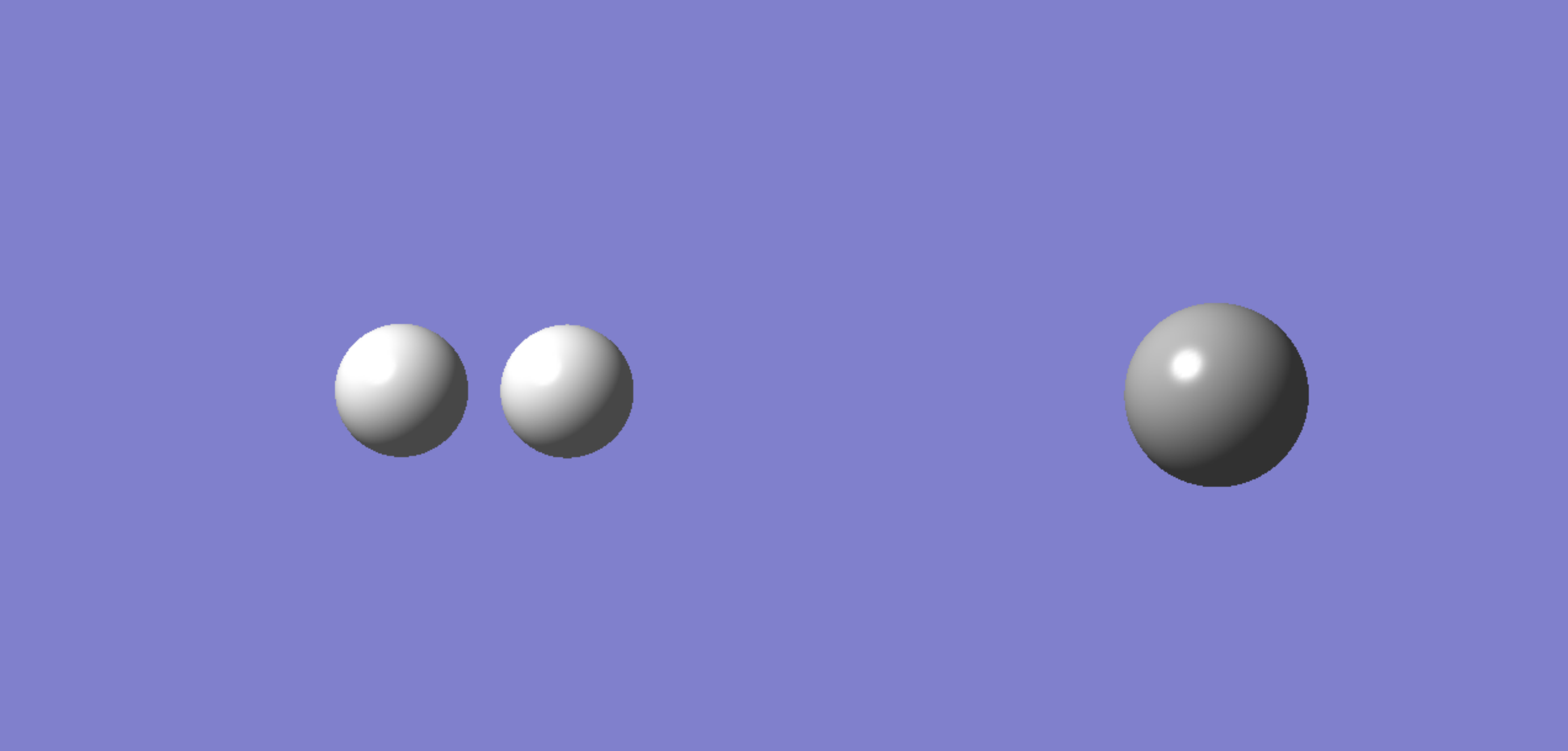} & Triplet &50 & 0.417  \\
5&N & \includegraphics[width=0.2\textwidth]{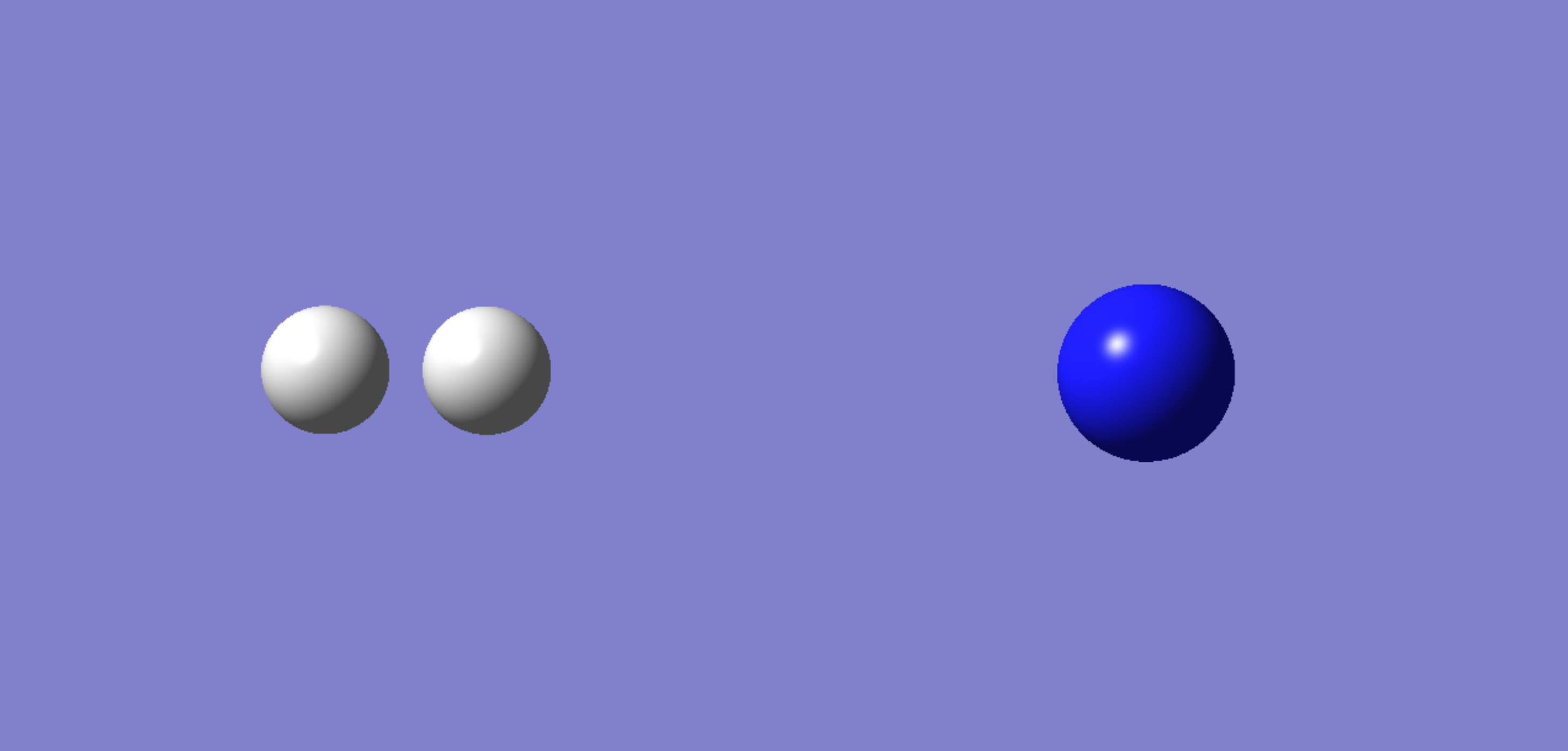} & Quartet &83 (78$^a$)& 0.690 (0.651$^a$)  \\
6&O & \includegraphics[width=0.2\textwidth]{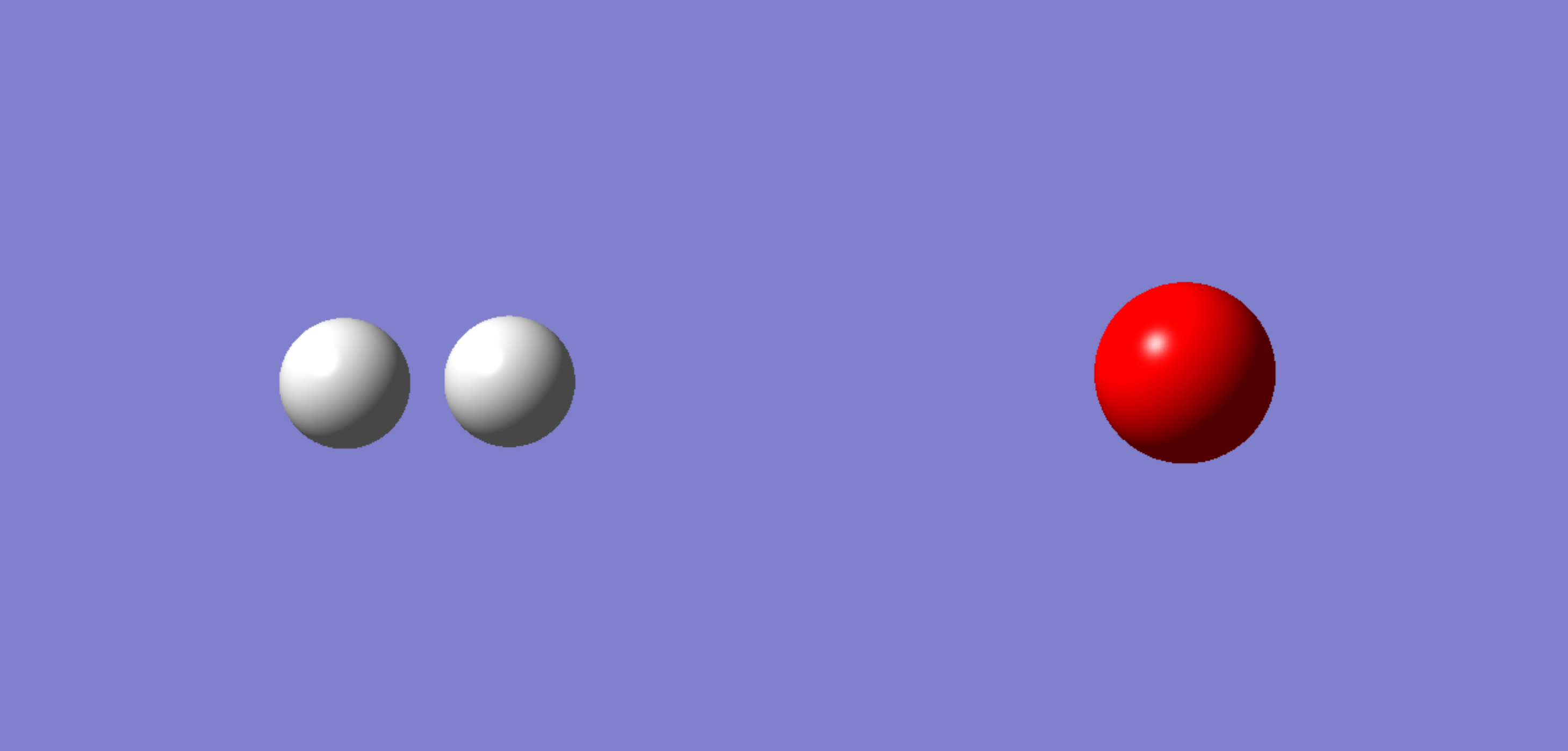} & Triplet &46, 55$^c$ & 0.386 \\
7&Na& \includegraphics[width=0.2\textwidth]{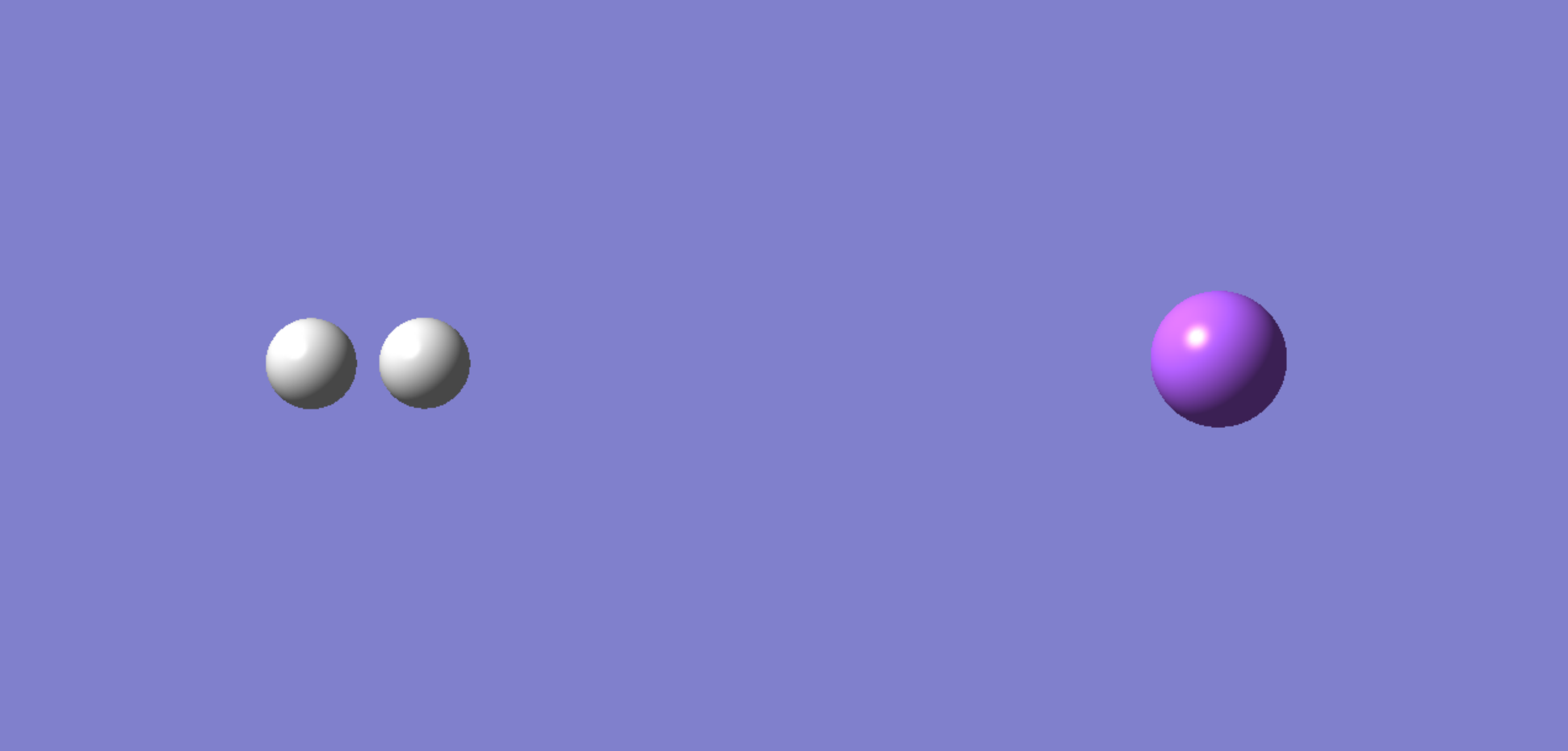} & Doublet &22& 0.184  \\
8&Mg& \includegraphics[width=0.2\textwidth]{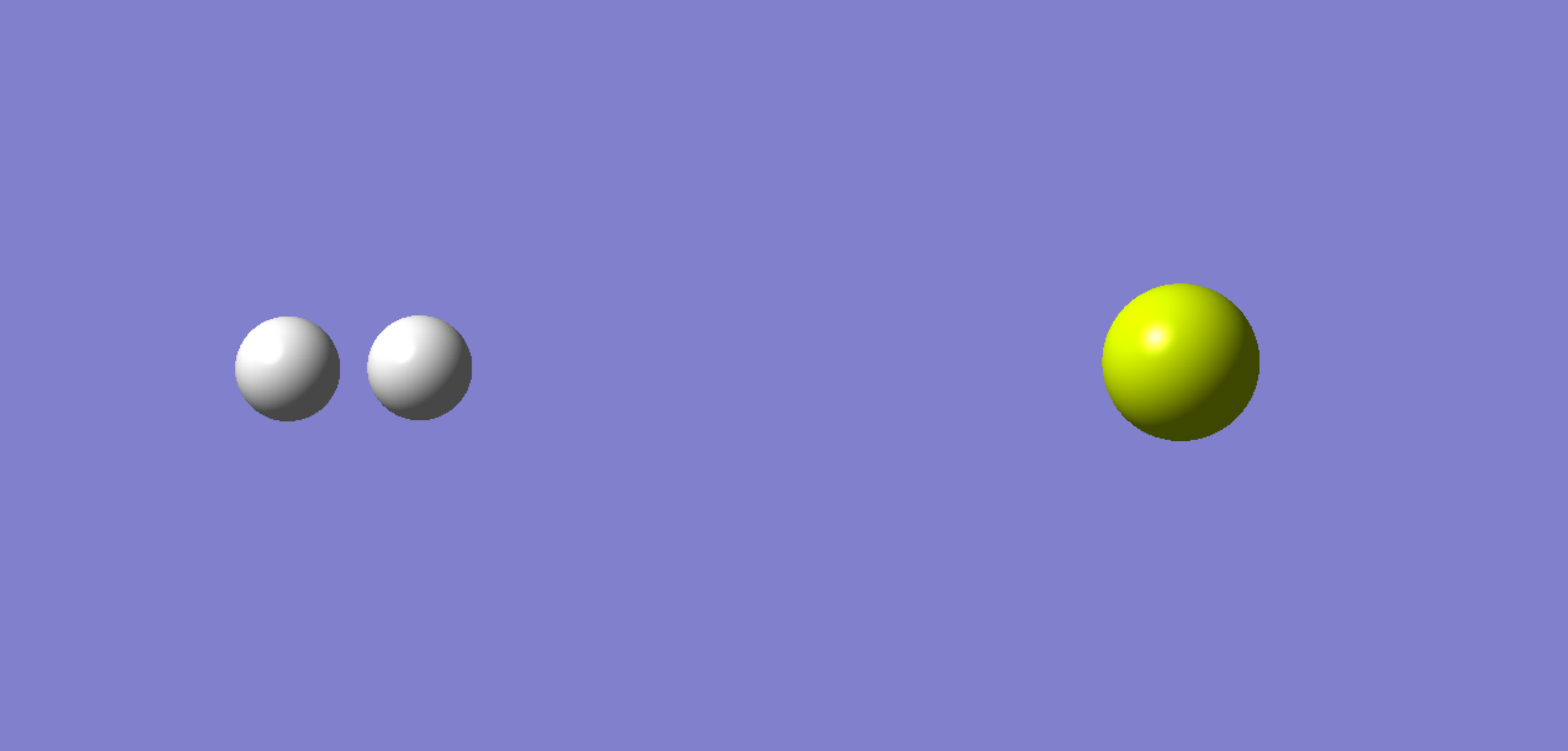} & Singlet &62& 0.514  \\
9&Si& \includegraphics[width=0.2\textwidth]{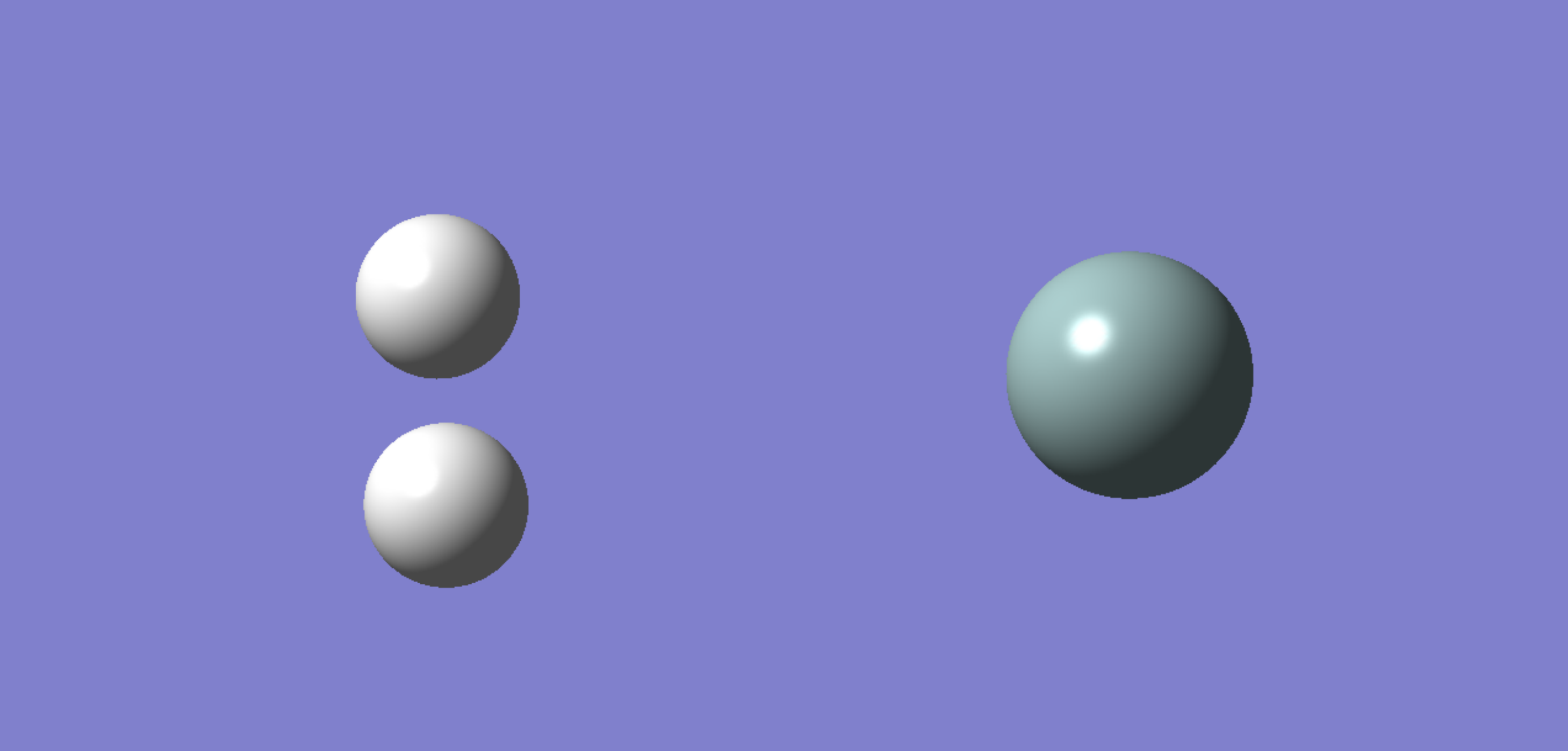} & Triplet &642& 5.343  \\
10&P& \includegraphics[width=0.2\textwidth]{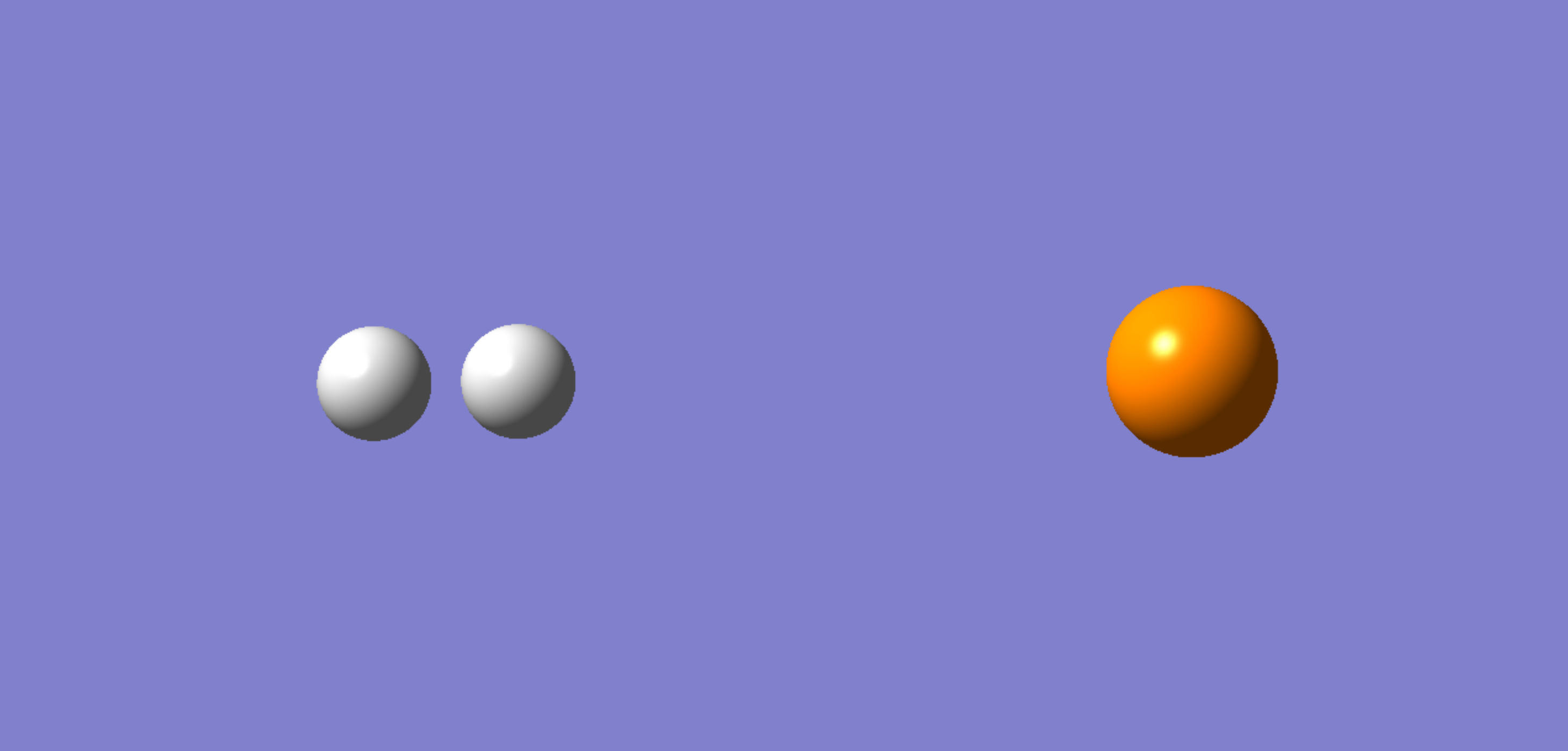} & Quartet &107 & 0.887 \\
11&S& \includegraphics[width=0.2\textwidth]{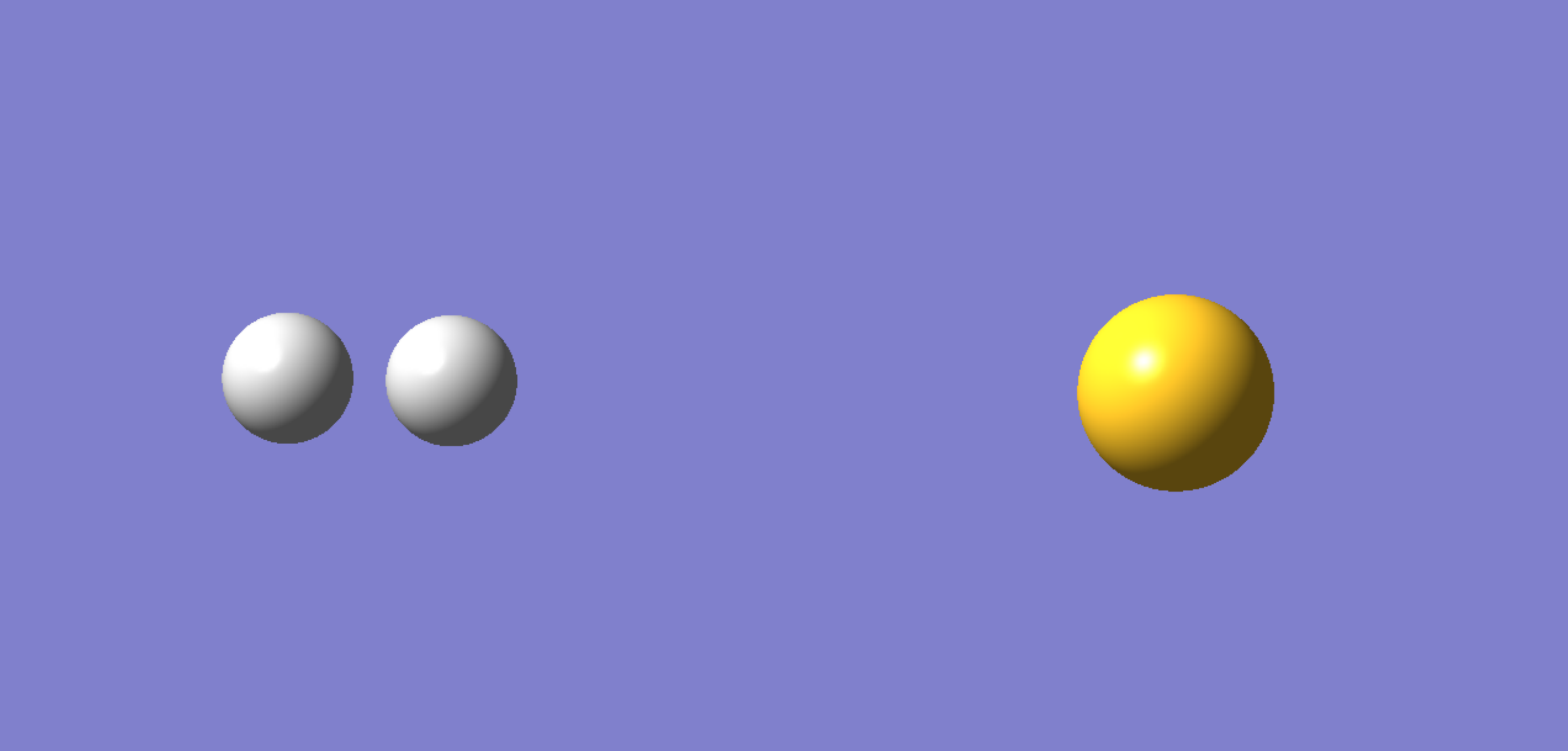} & Triplet & 88 & 0.732  \\
12&NH& \includegraphics[width=0.2\textwidth]{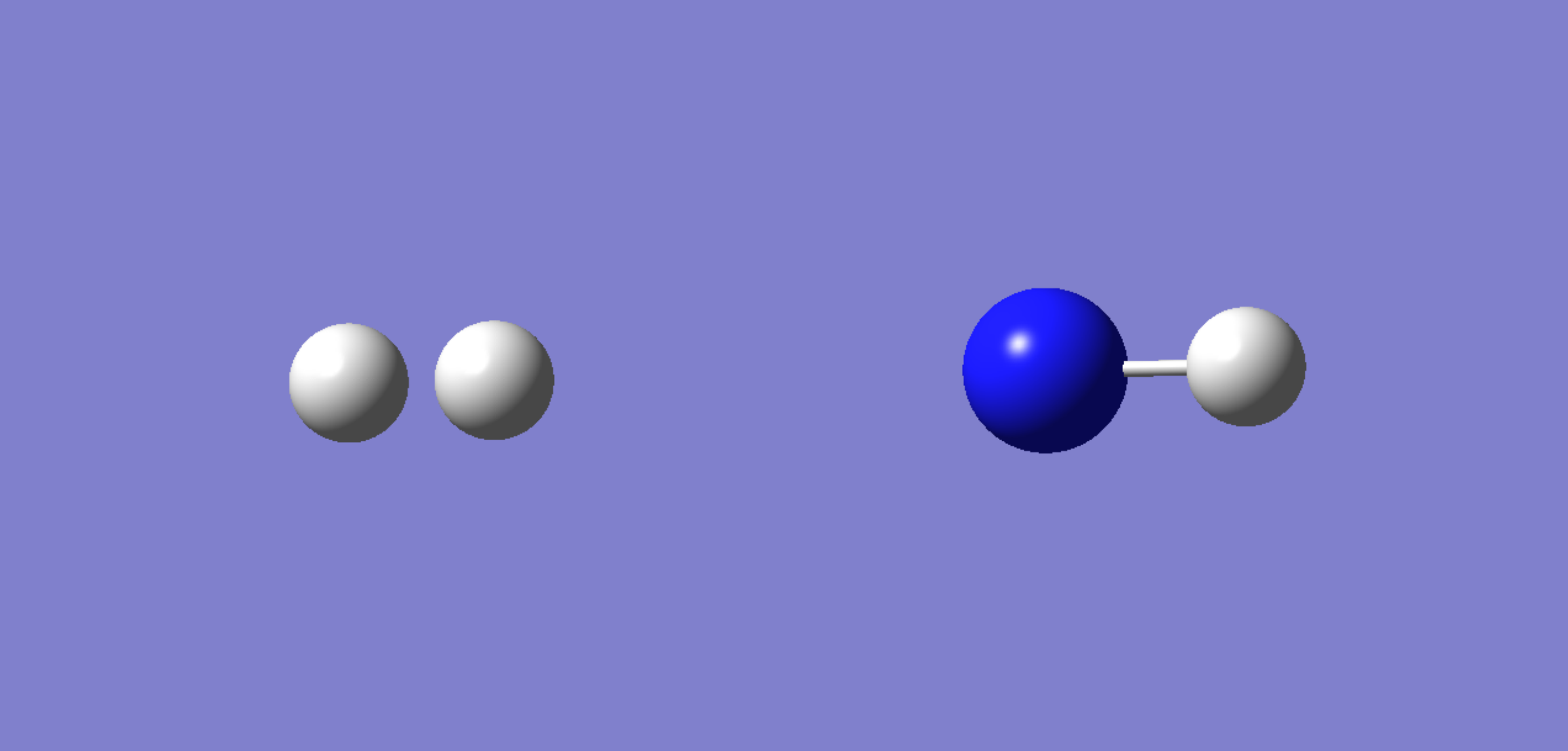} & Triplet &286& 2.381  \\
\hline
\end{tabular}
\end{table}

\begin{table}
\scriptsize
\centering
\begin{tabular}{|c|c|c|c|c|c|}
\hline
{\bf Sl.}& {\bf Species} & {\bf Optimized} & {\bf Ground} & \multicolumn{2}{c|}{\bf Binding Energy} \\
\cline{5-6}
 {\bf No.} & & {\bf Structures} & {\bf State} & {\bf in K} & {\bf in kJ/mol} \\
\hline
13&OH& \includegraphics[width=0.2\textwidth]{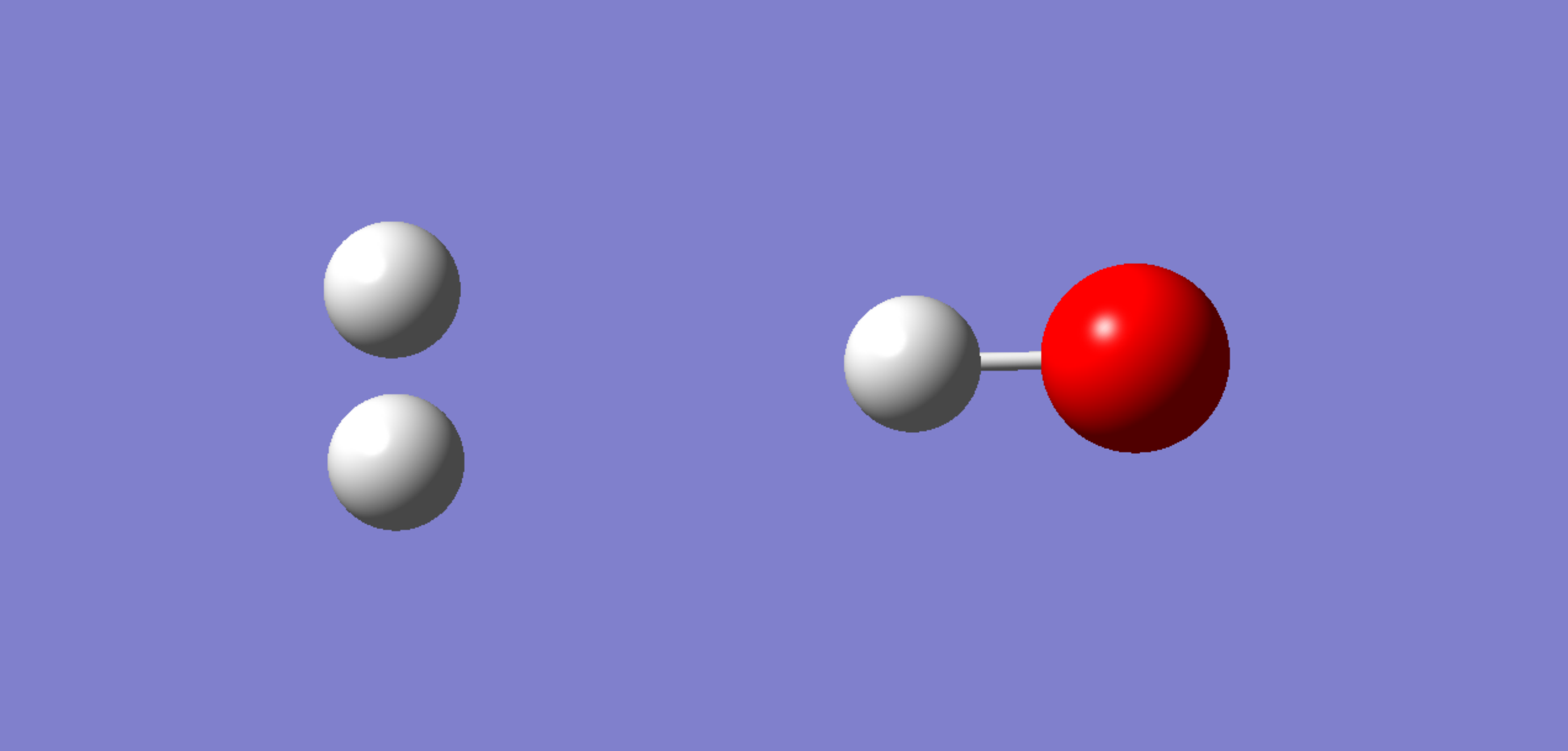} & Doublet &380, 240$^c$ & 3.158  \\
&& \includegraphics[width=0.2\textwidth]{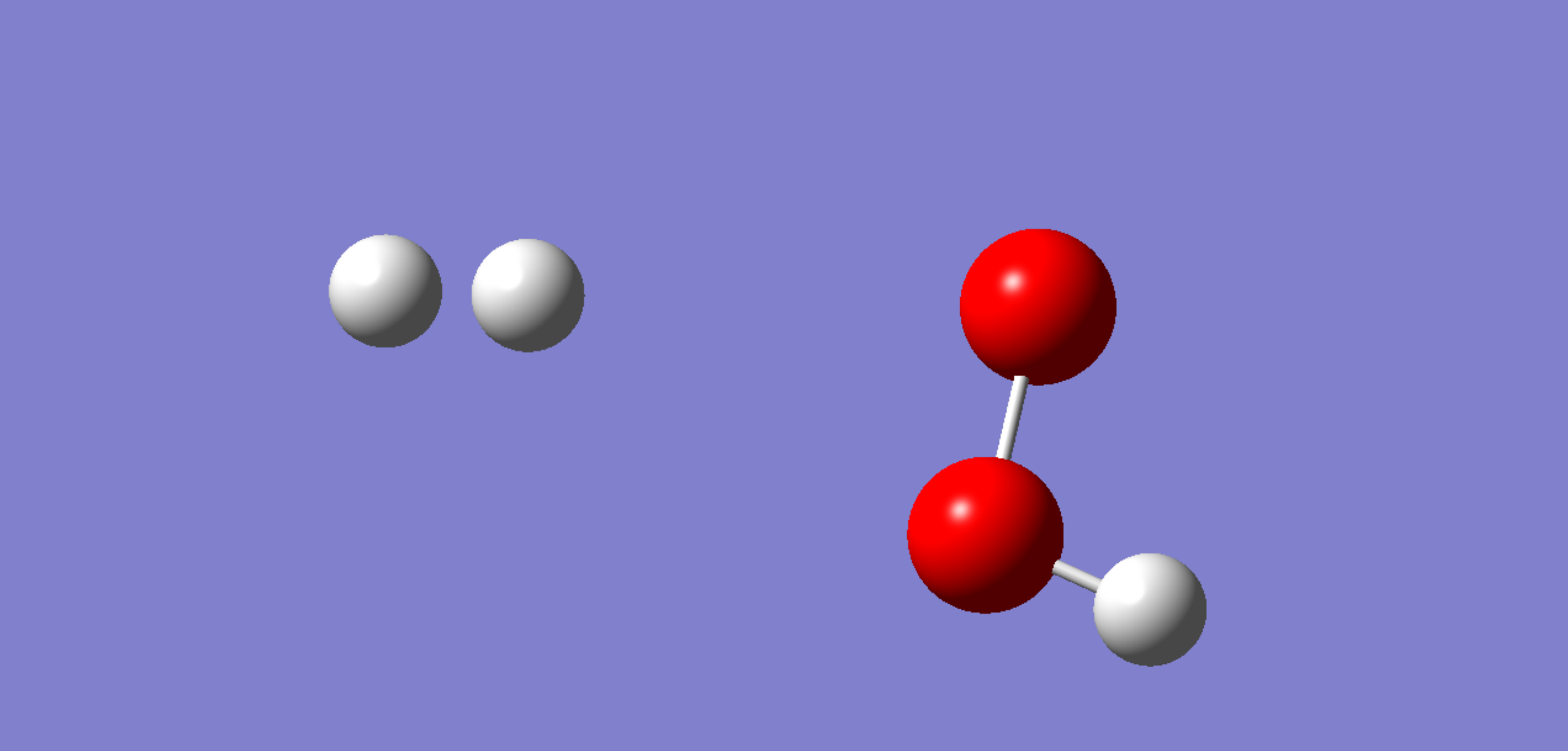} & & 380$^b$ & 3.158$^b$ \\
14&PH& \includegraphics[width=0.2\textwidth]{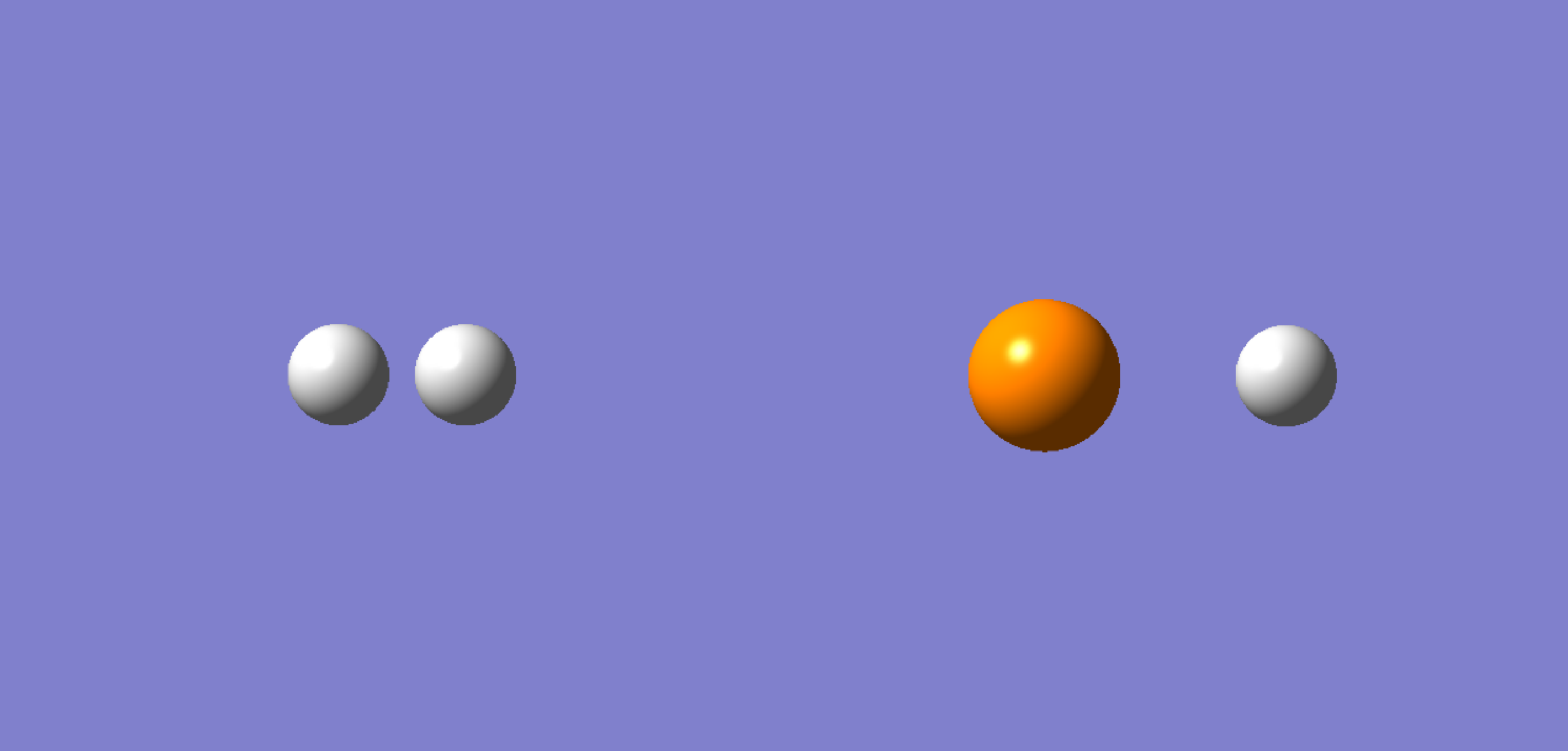} & Triplet &151& 1.258  \\
15&C$_2$& \includegraphics[width=0.2\textwidth]{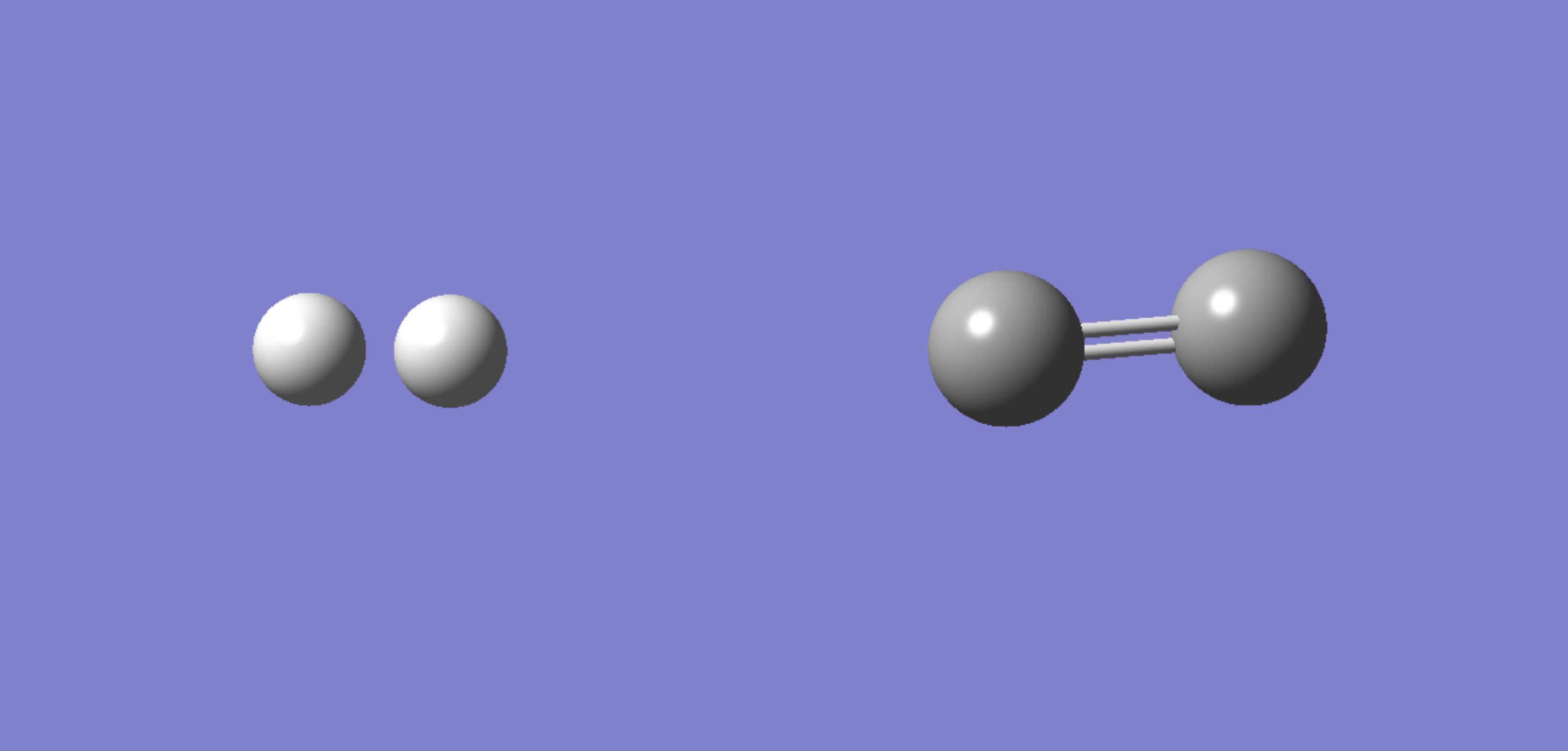} & Triplet &204& 1.696  \\
16&HF& \includegraphics[width=0.2\textwidth]{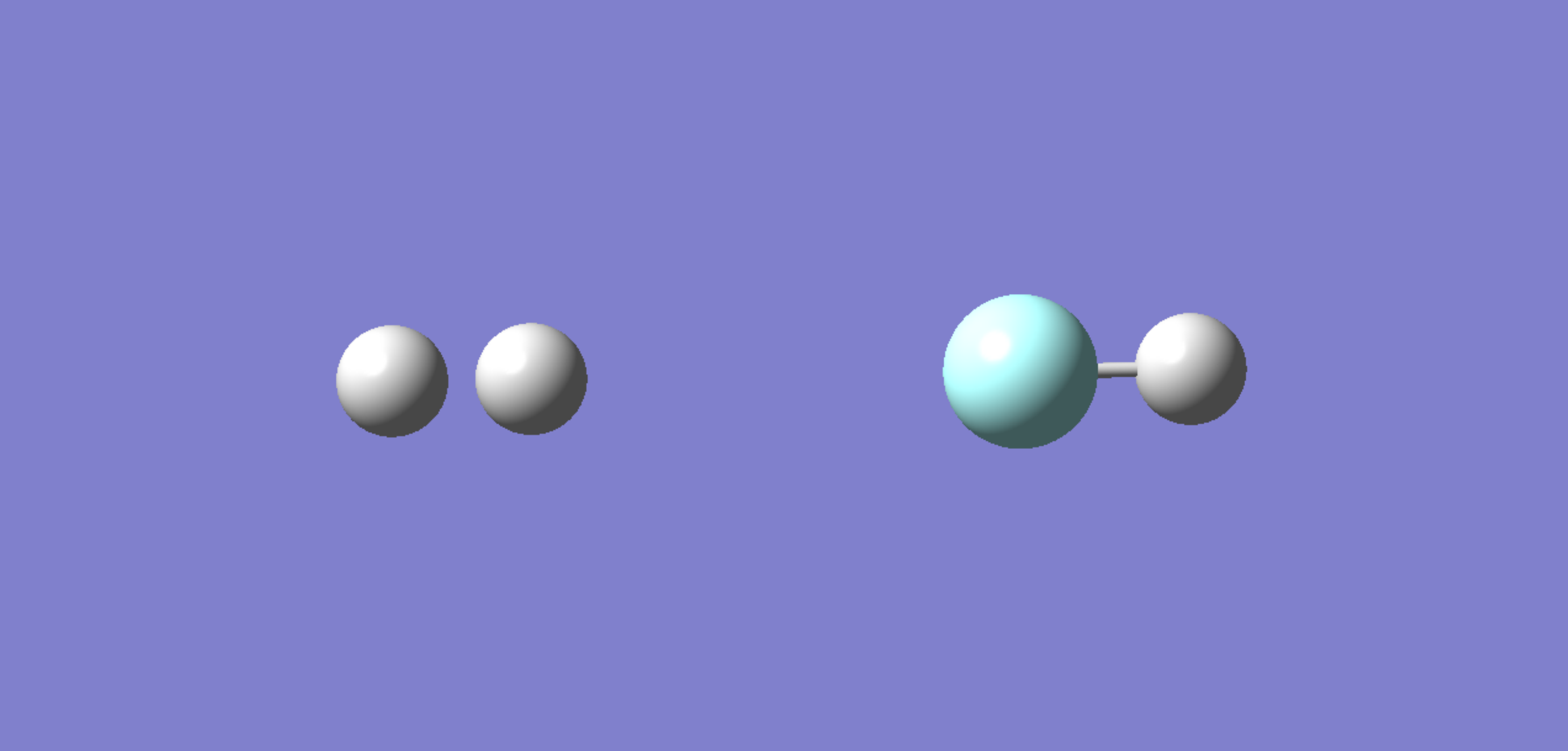} & Singlet &287& 2.386  \\
17&HCl& \includegraphics[width=0.2\textwidth]{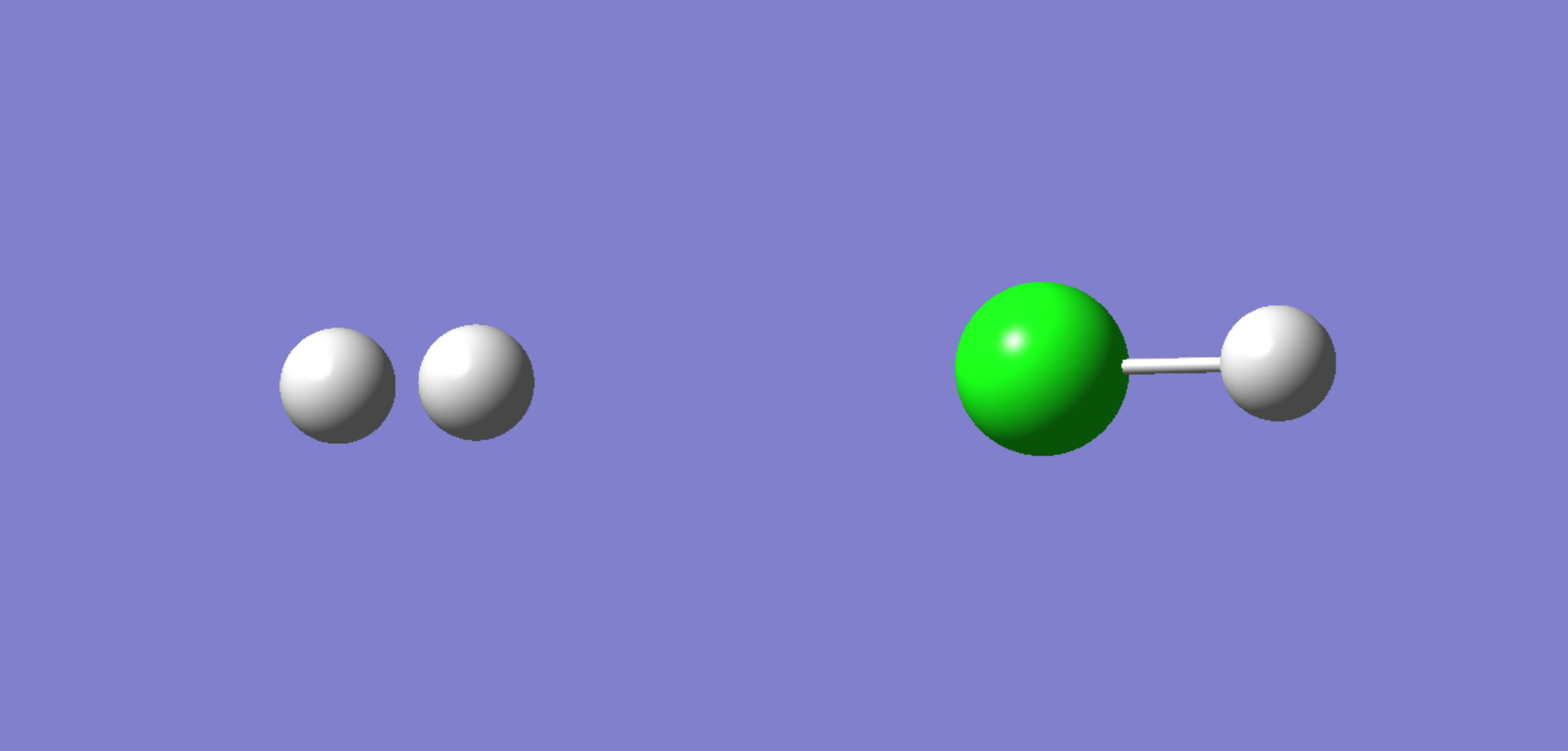} & Singlet &162& 1.350  \\
18&CN& \includegraphics[width=0.2\textwidth]{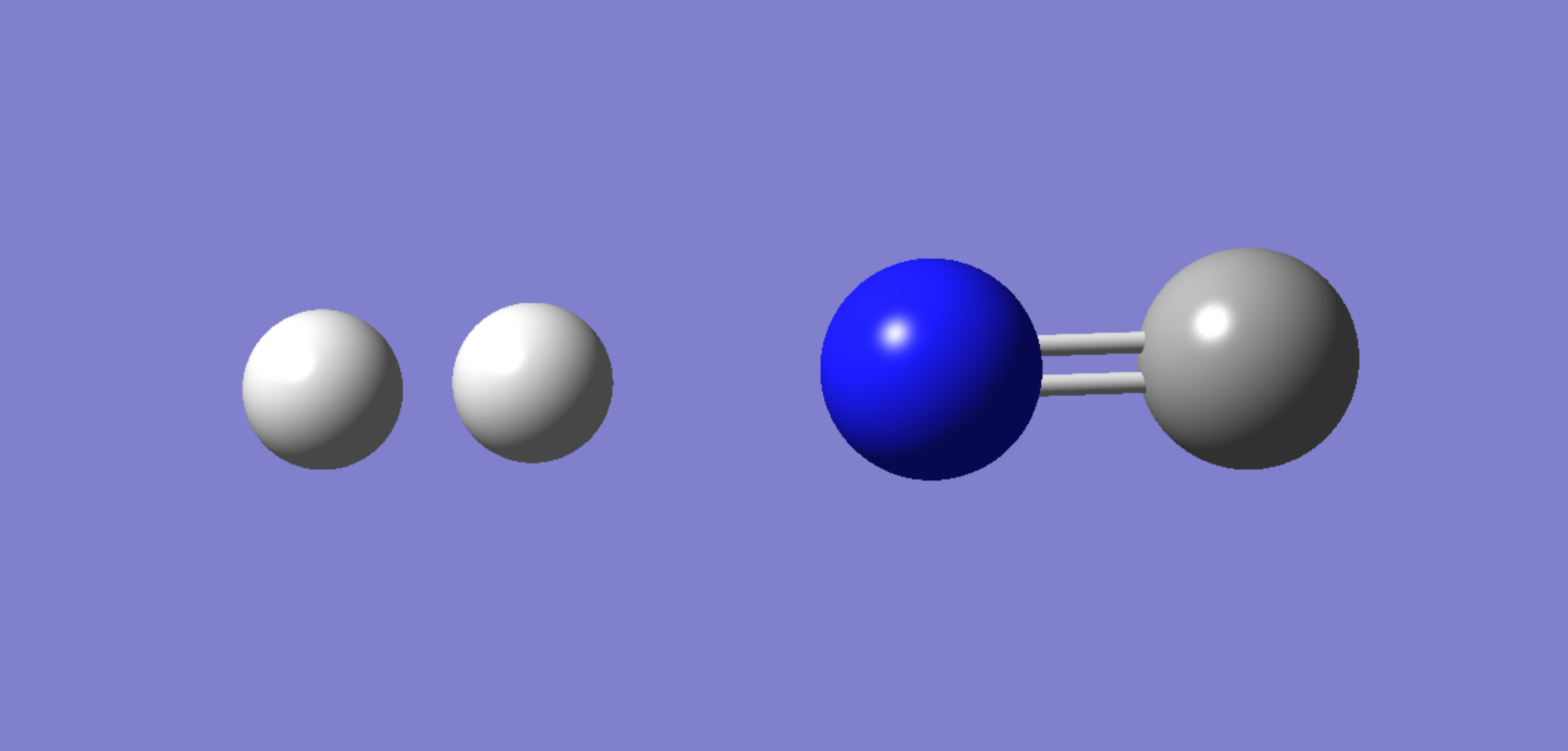} & Doublet &4695& 39.041  \\
19&N$_2$& \includegraphics[width=0.2\textwidth]{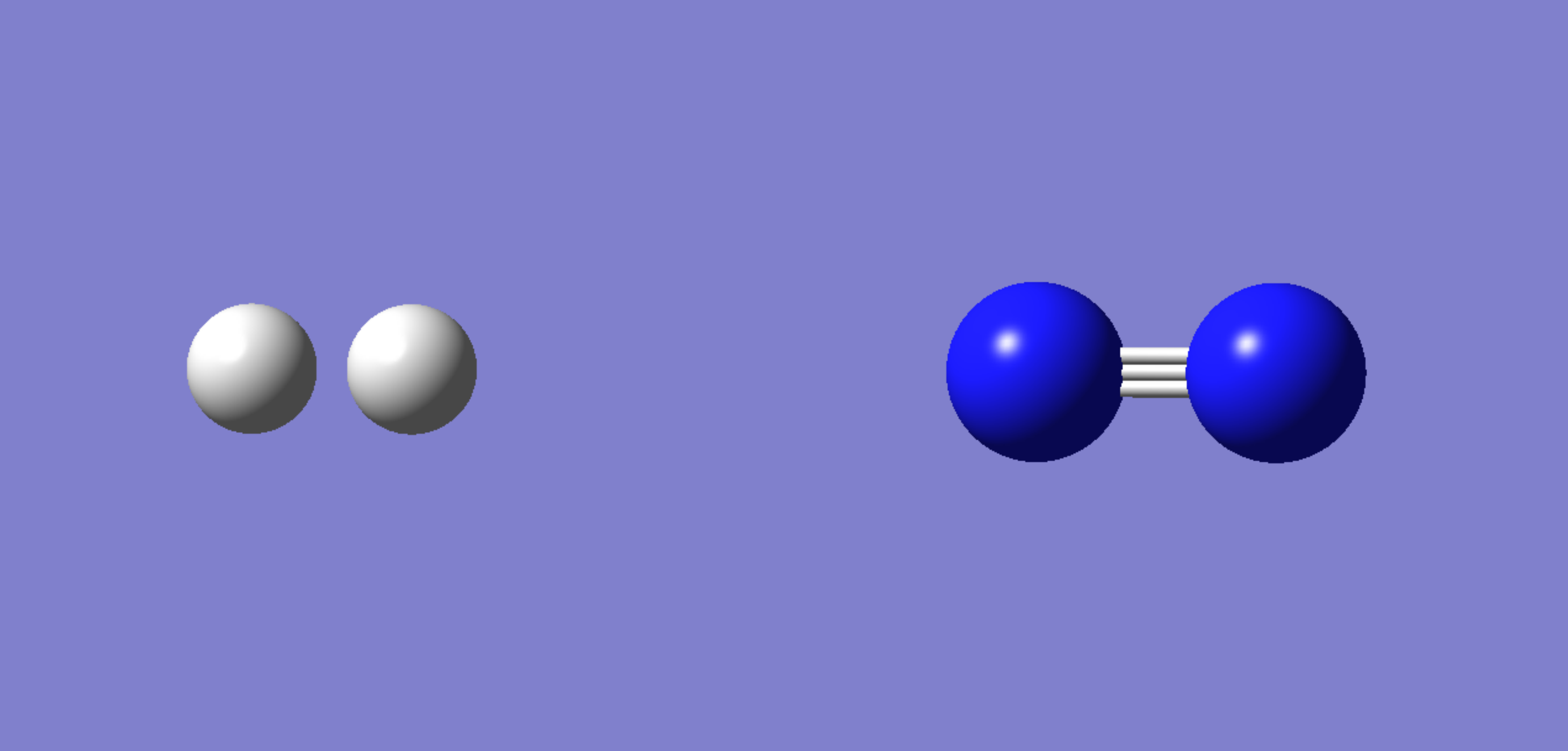} & Singlet &198 & 1.649 \\
20 & CO & \includegraphics[width=0.2\textwidth]{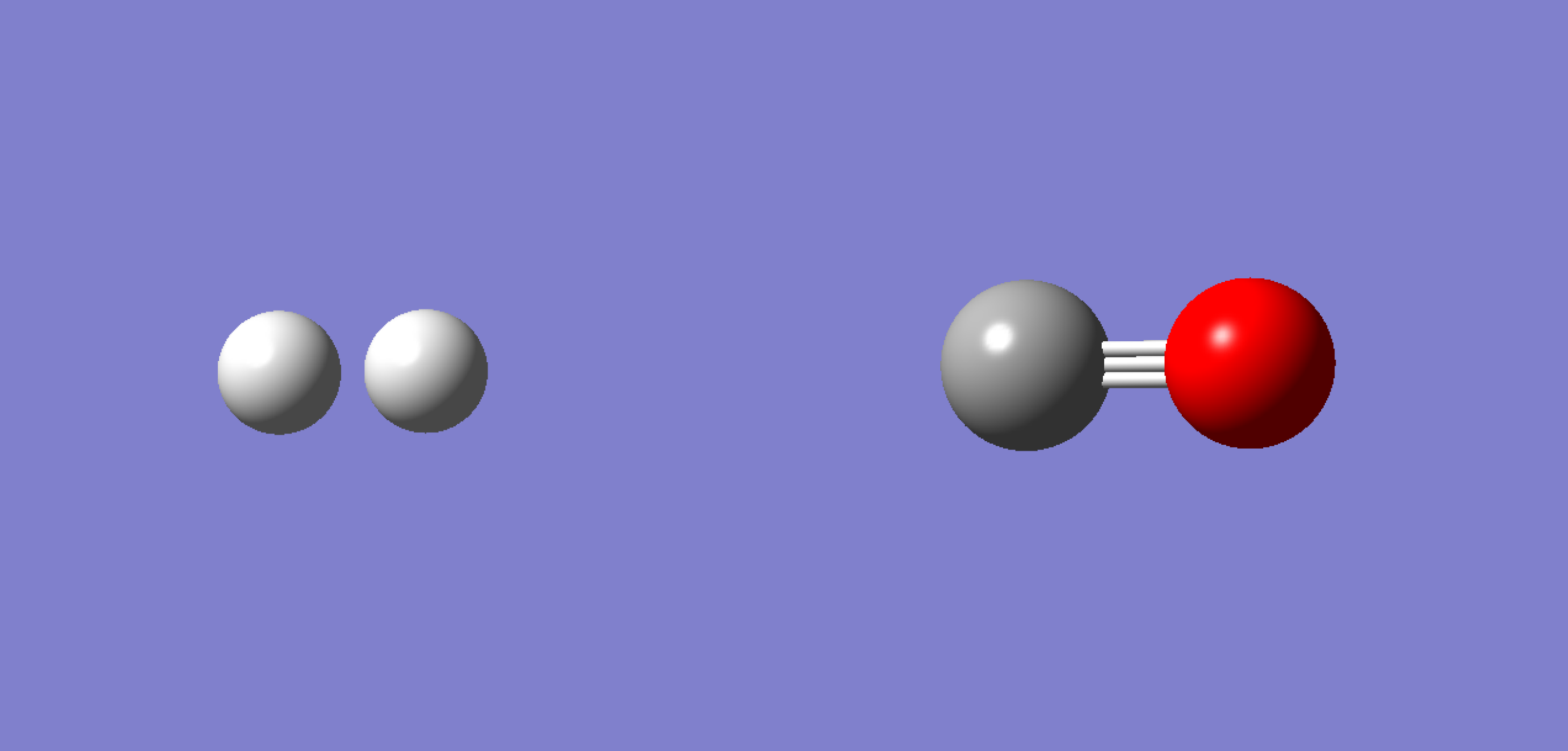} & Singlet &215& 1.788  \\
21 & SiH & \includegraphics[width=0.2\textwidth]{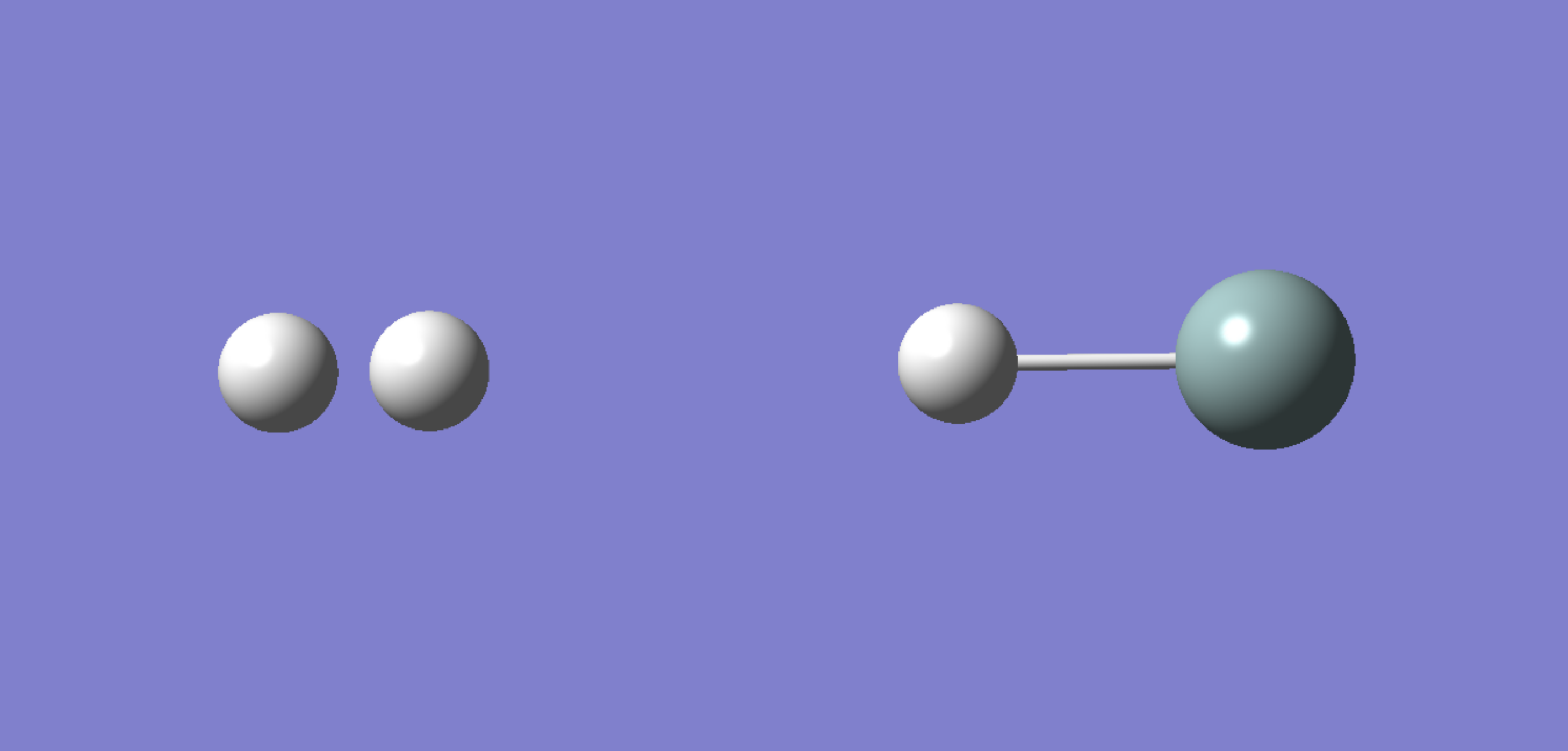} & Doublet & 188 & 1.562  \\
22 & NO & \includegraphics[width=0.2\textwidth]{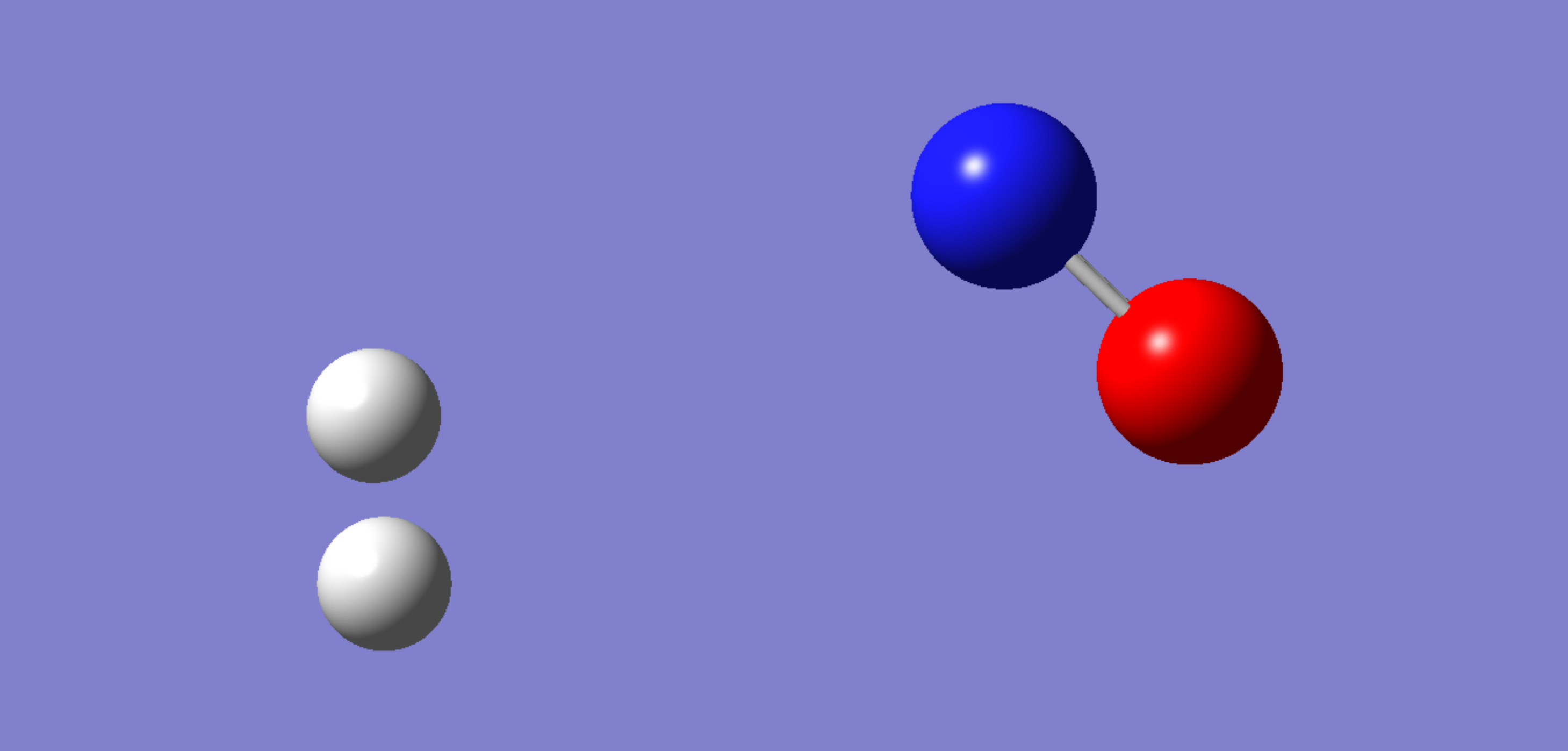} & Doublet & 159 & 1.321  \\
23 &O$_2$& \includegraphics[width=0.2\textwidth]{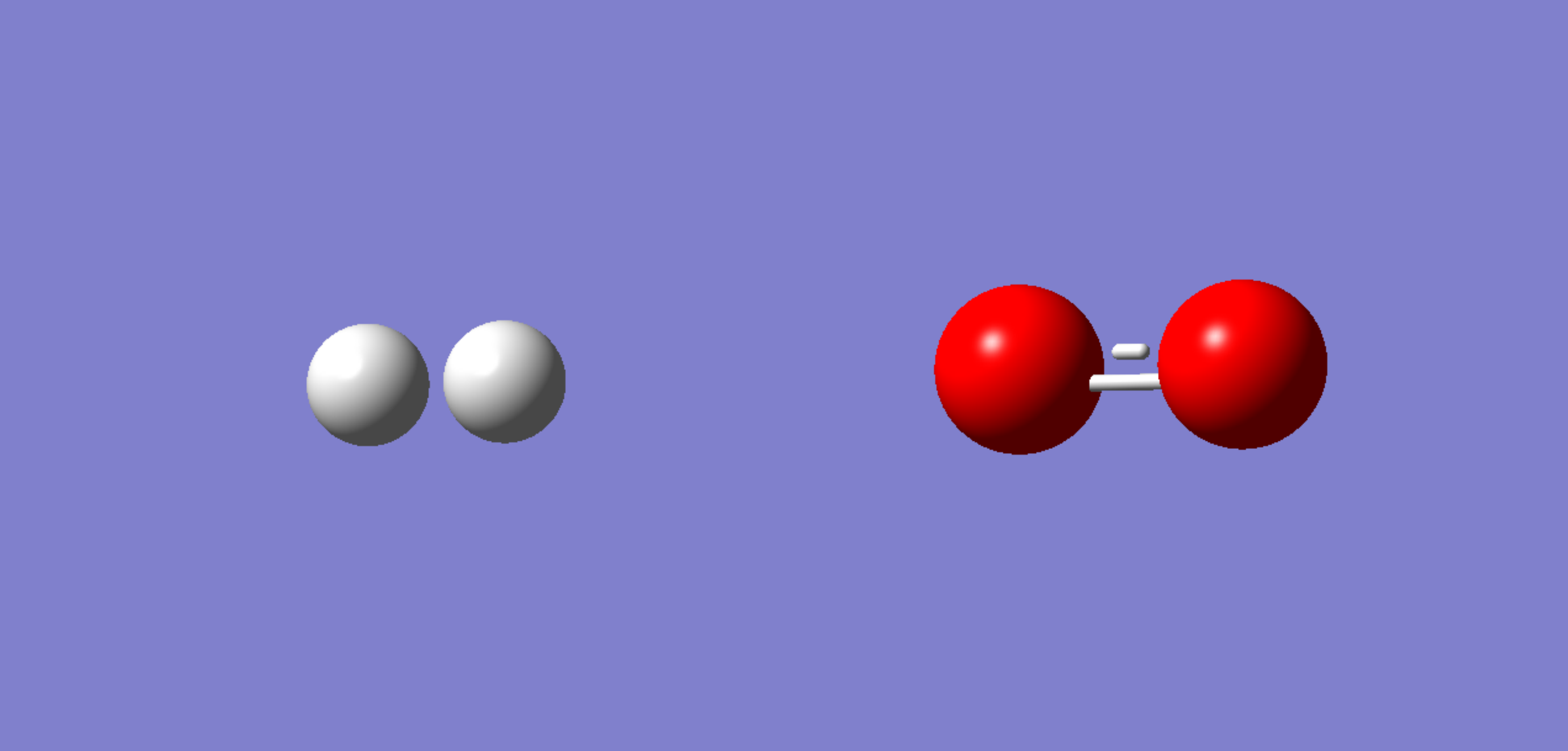} & Triplet &159, 69$^c$ & 1.321  \\
24&HS& \includegraphics[width=0.2\textwidth]{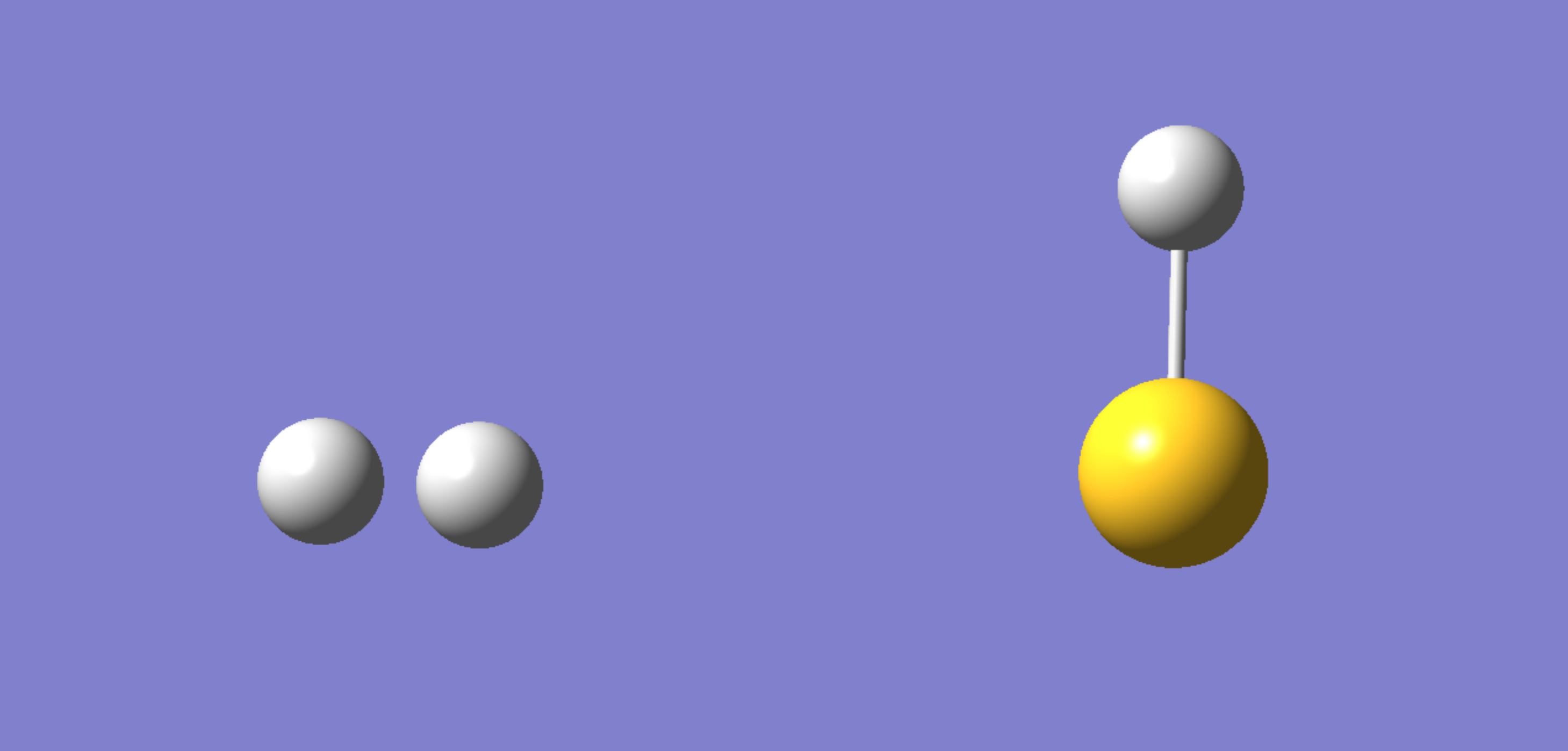} & Doublet &222&1.848  \\
25&SiC& \includegraphics[width=0.2\textwidth]{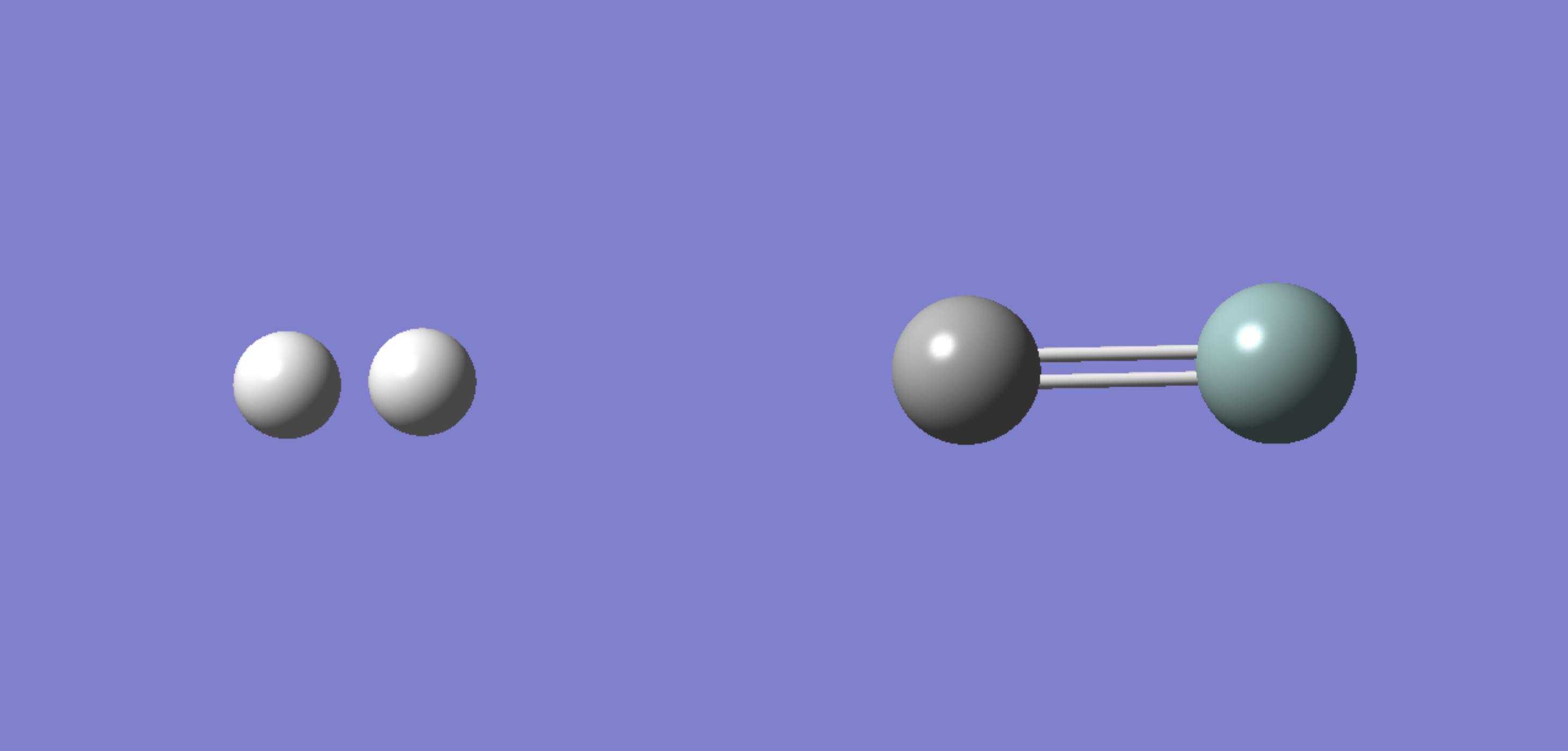} & Triplet &212& 1.759 \\
\hline
\end{tabular}
\end{table}

\begin{table}
\scriptsize
\centering
\begin{tabular}{|c|c|c|c|c|c|}
\hline
{\bf Sl.}& {\bf Species} & {\bf Optimized} & {\bf Ground} & \multicolumn{2}{c|}{\bf Binding Energy} \\
\cline{5-6}
 {\bf No.} & & {\bf Structures} & {\bf State} & {\bf in K} & {\bf in kJ/mol} \\
\hline
26&CP& \includegraphics[width=0.2\textwidth]{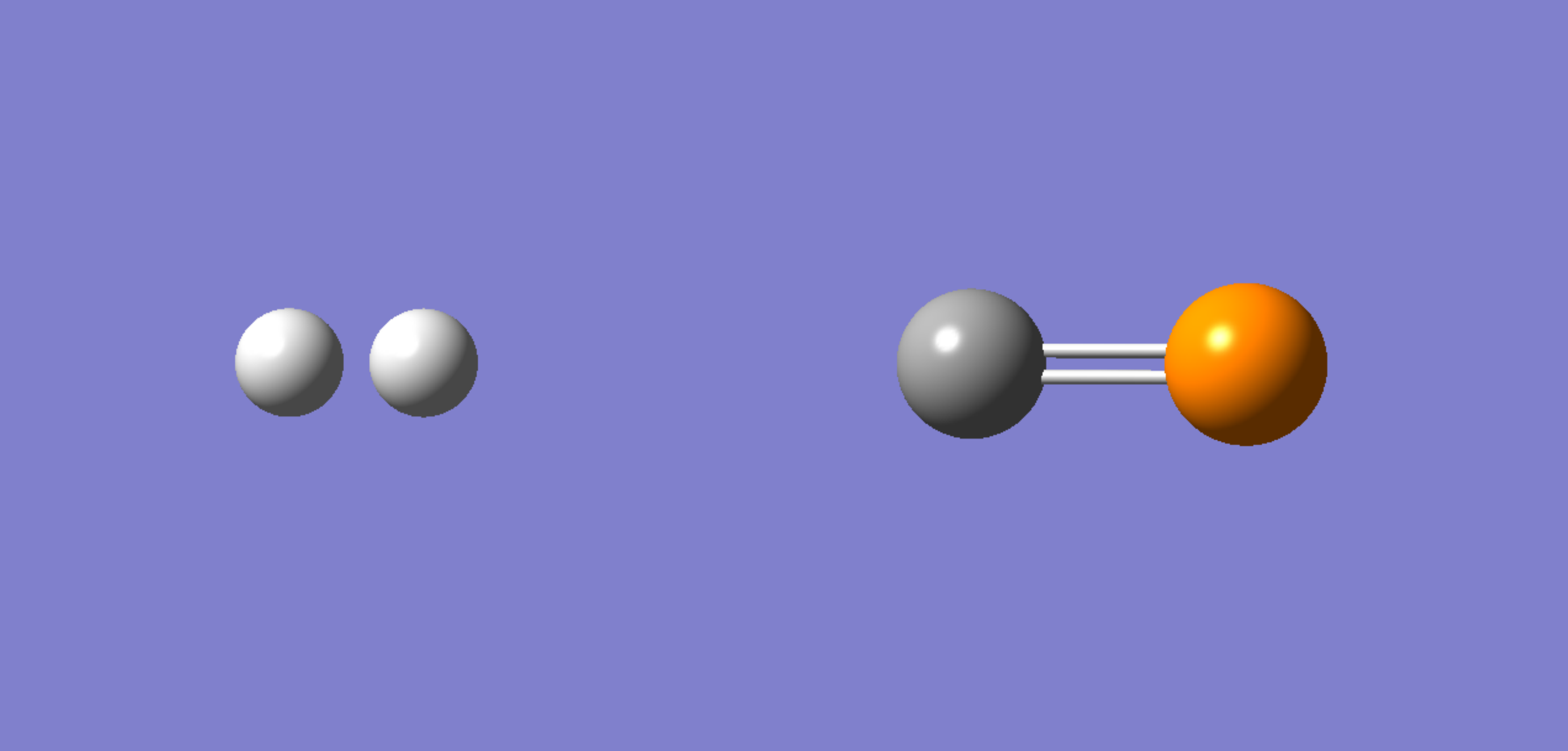} & Doublet &165&1.373 \\
27&CS& \includegraphics[width=0.2\textwidth]{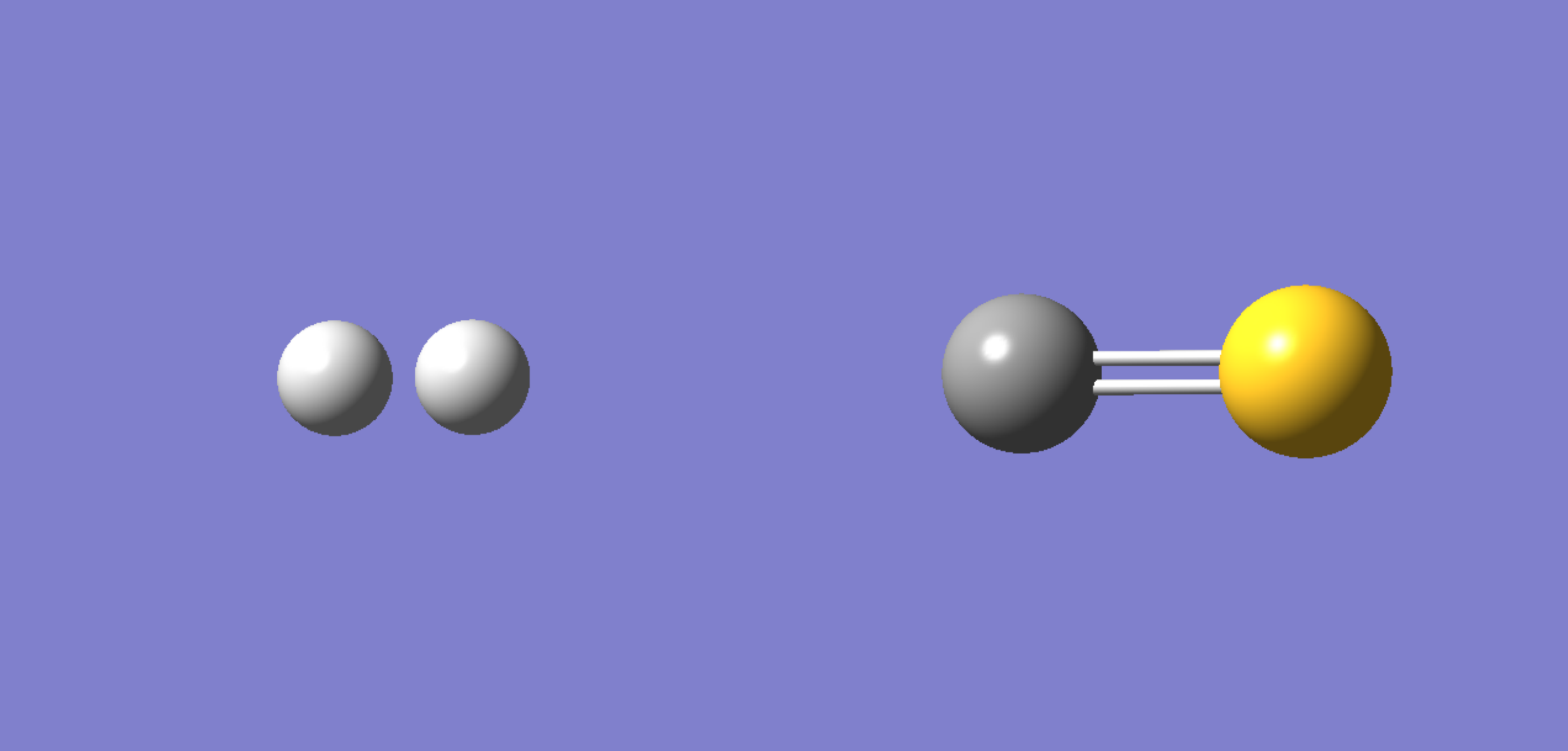} & Singlet &337&2.804  \\
28&NS& \includegraphics[width=0.2\textwidth]{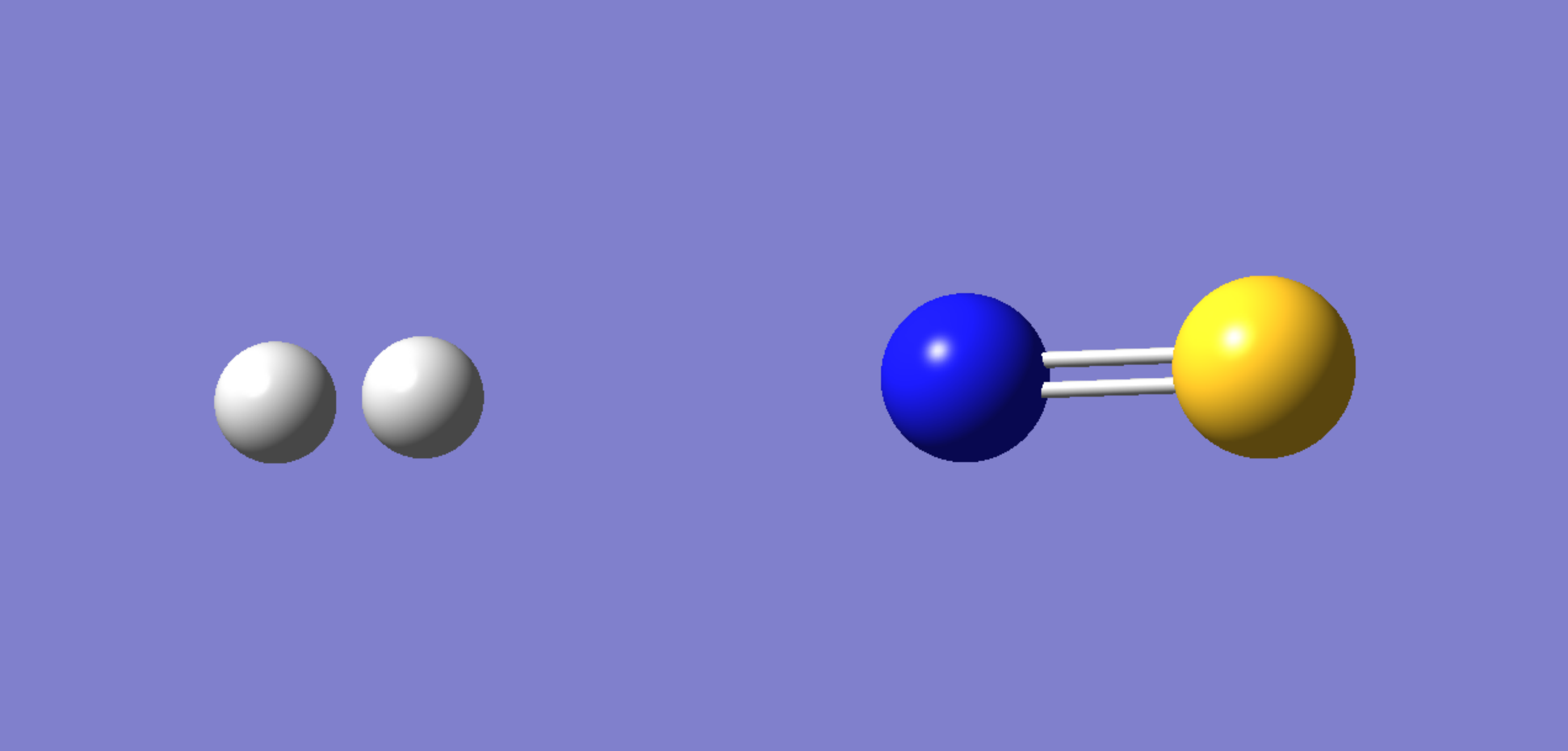} & Doublet & 353 & 2.938  \\
&& \includegraphics[width=0.2\textwidth]{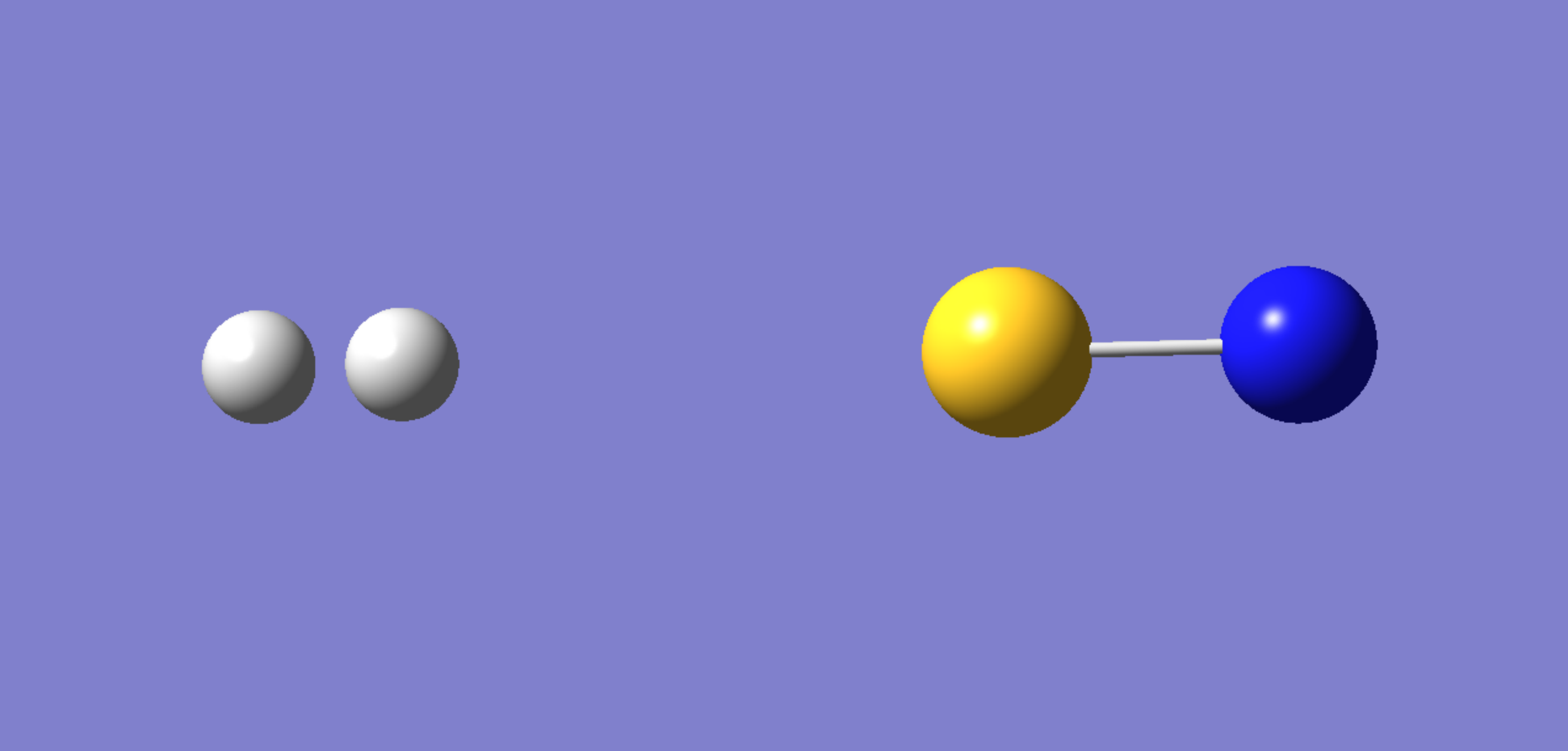} & & 171$^b$ & 1.423$^b$ \\
29&SO& \includegraphics[width=0.2\textwidth]{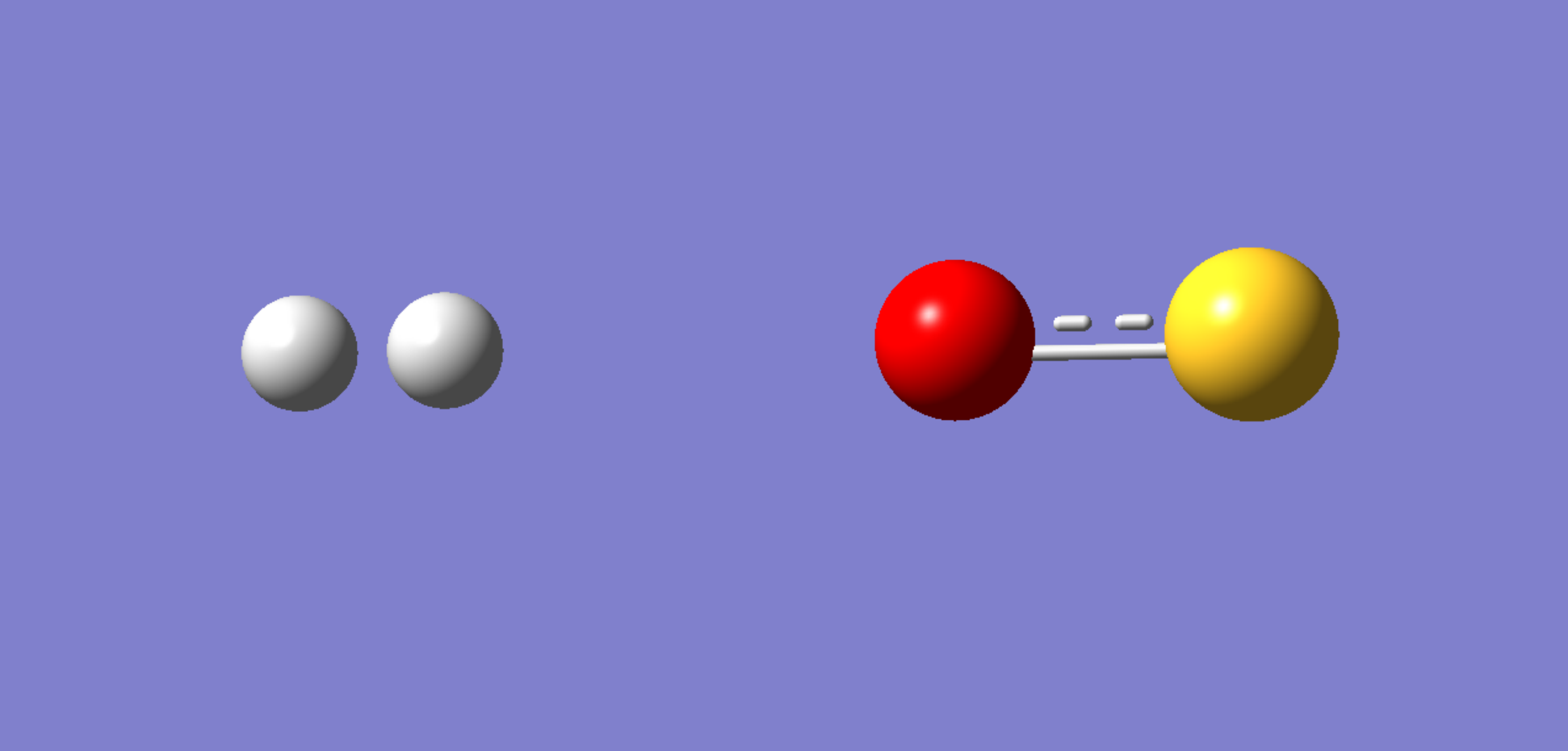} & Triplet &337&2.801  \\
30&S$_2$& \includegraphics[width=0.2\textwidth]{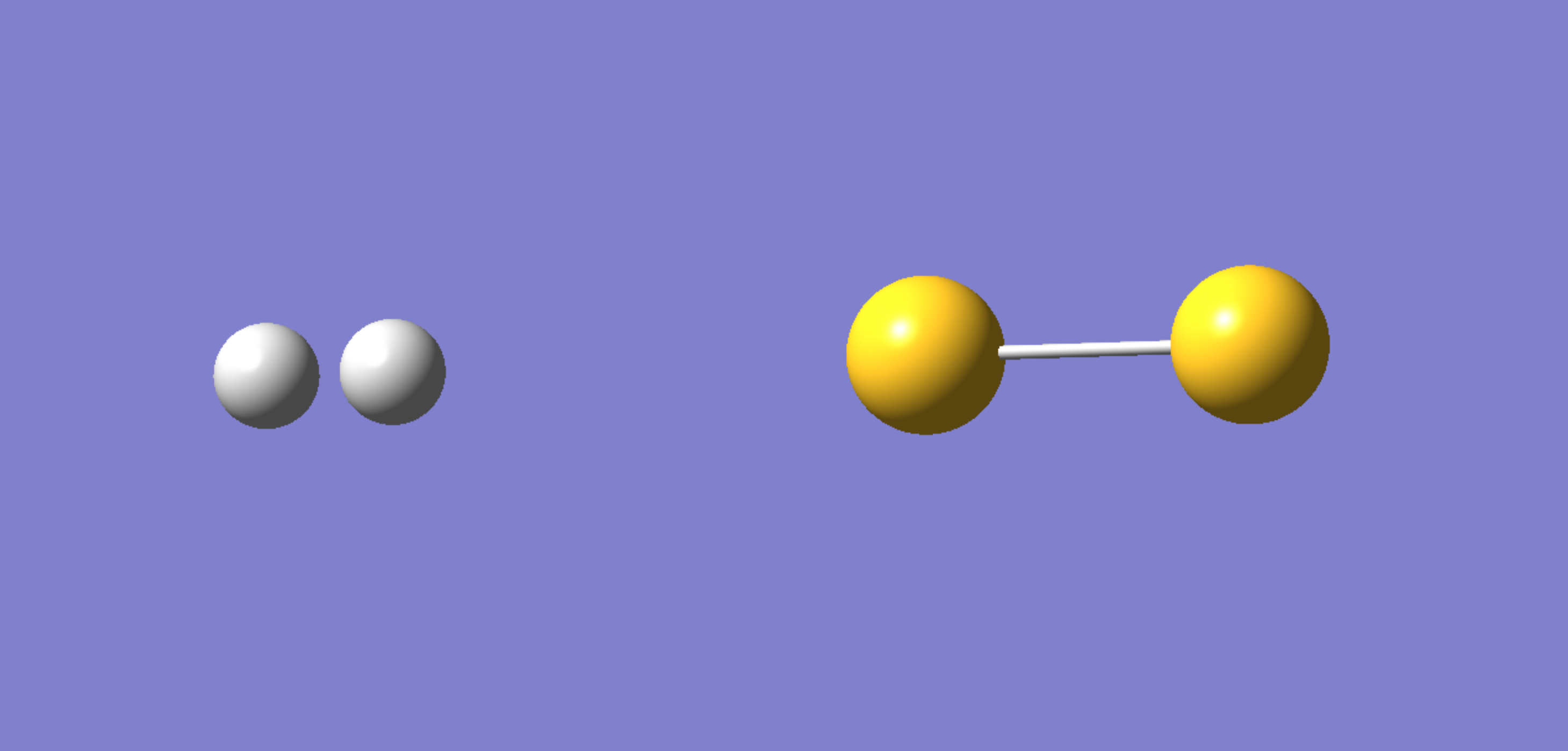} & Triplet &187&1.552  \\
31&CH$_2$& \includegraphics[width=0.2\textwidth]{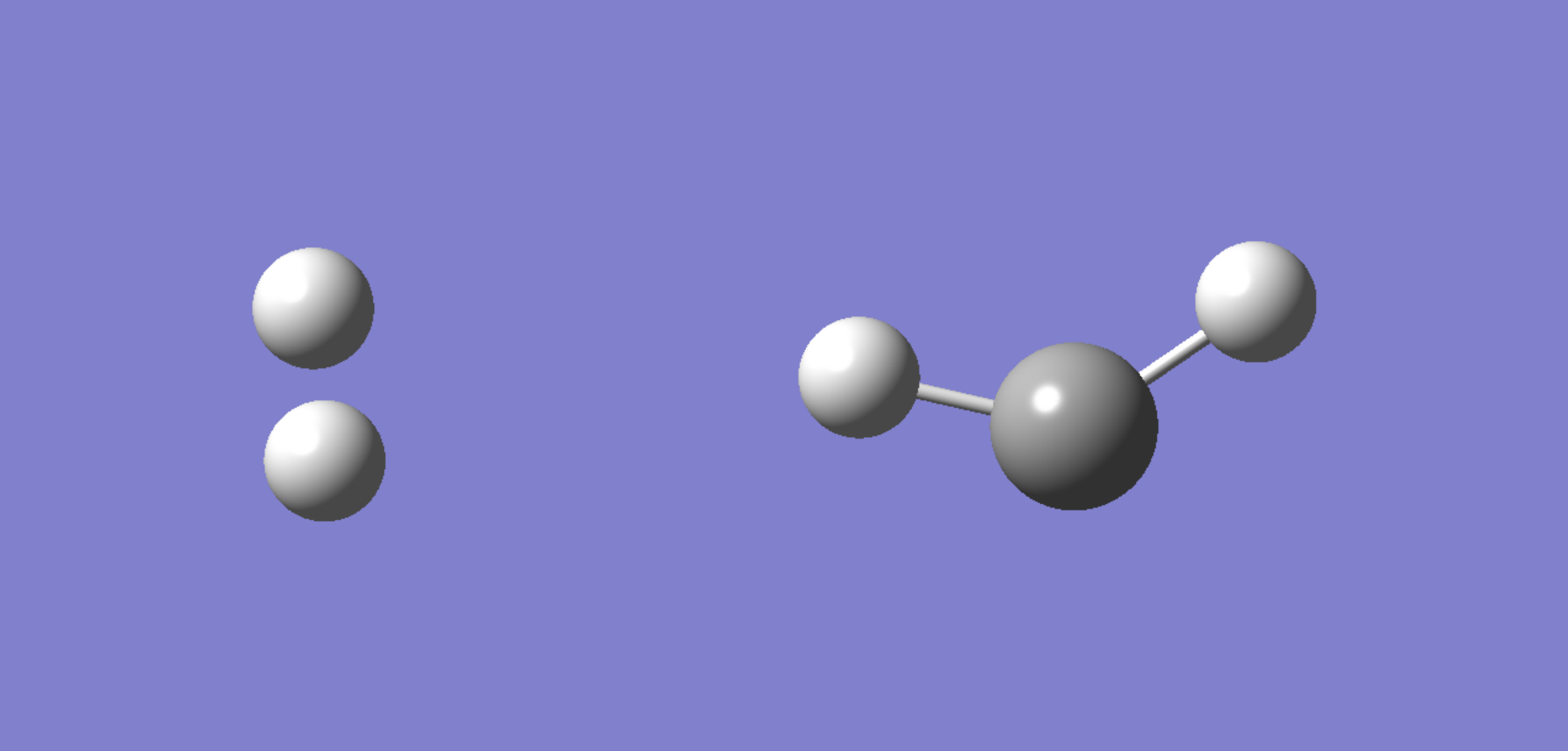} & Triplet &165&1.376  \\
32&NH$_2$& \includegraphics[width=0.2\textwidth]{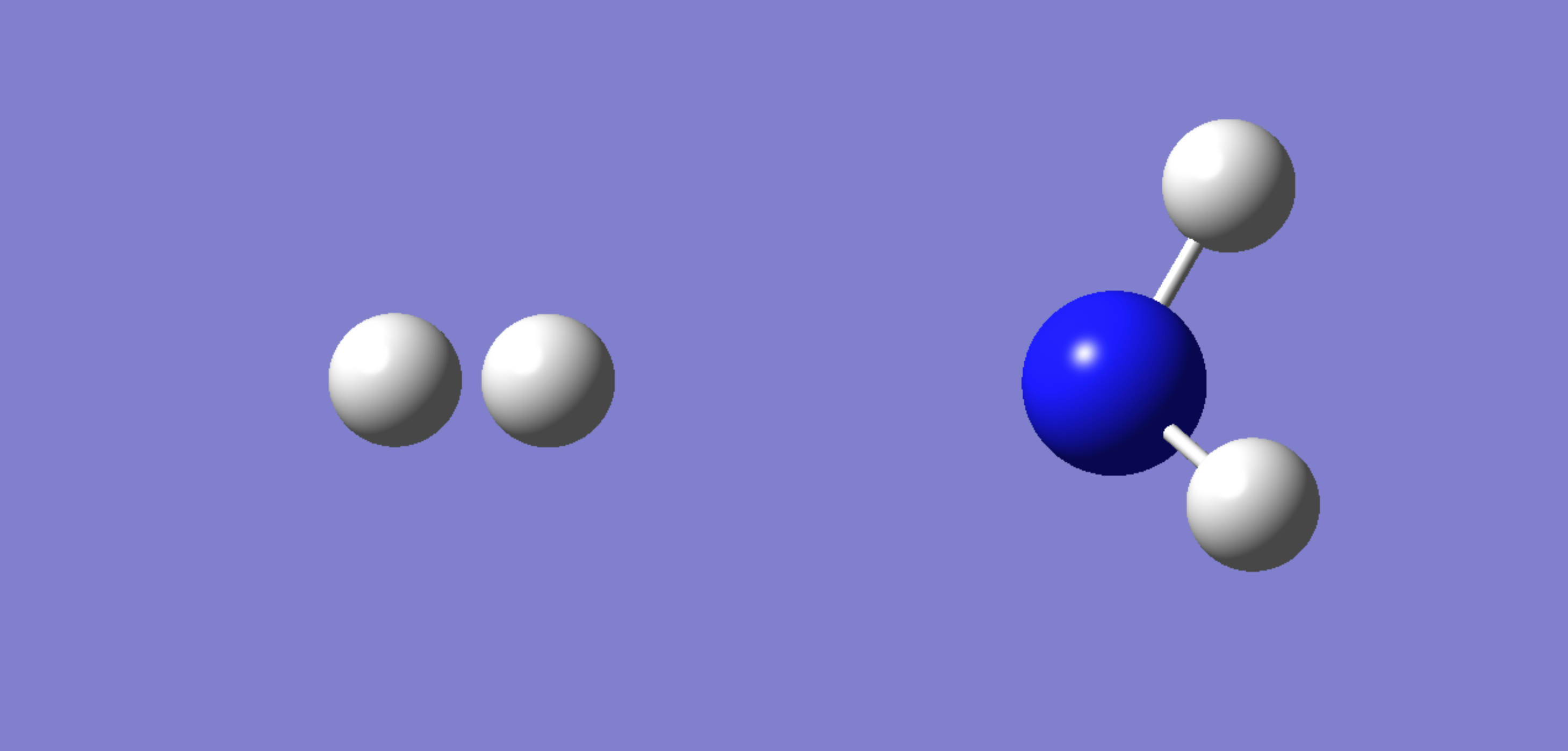} & Doublet &347& 2.888  \\
33&H$_2$O& \includegraphics[width=0.2\textwidth]{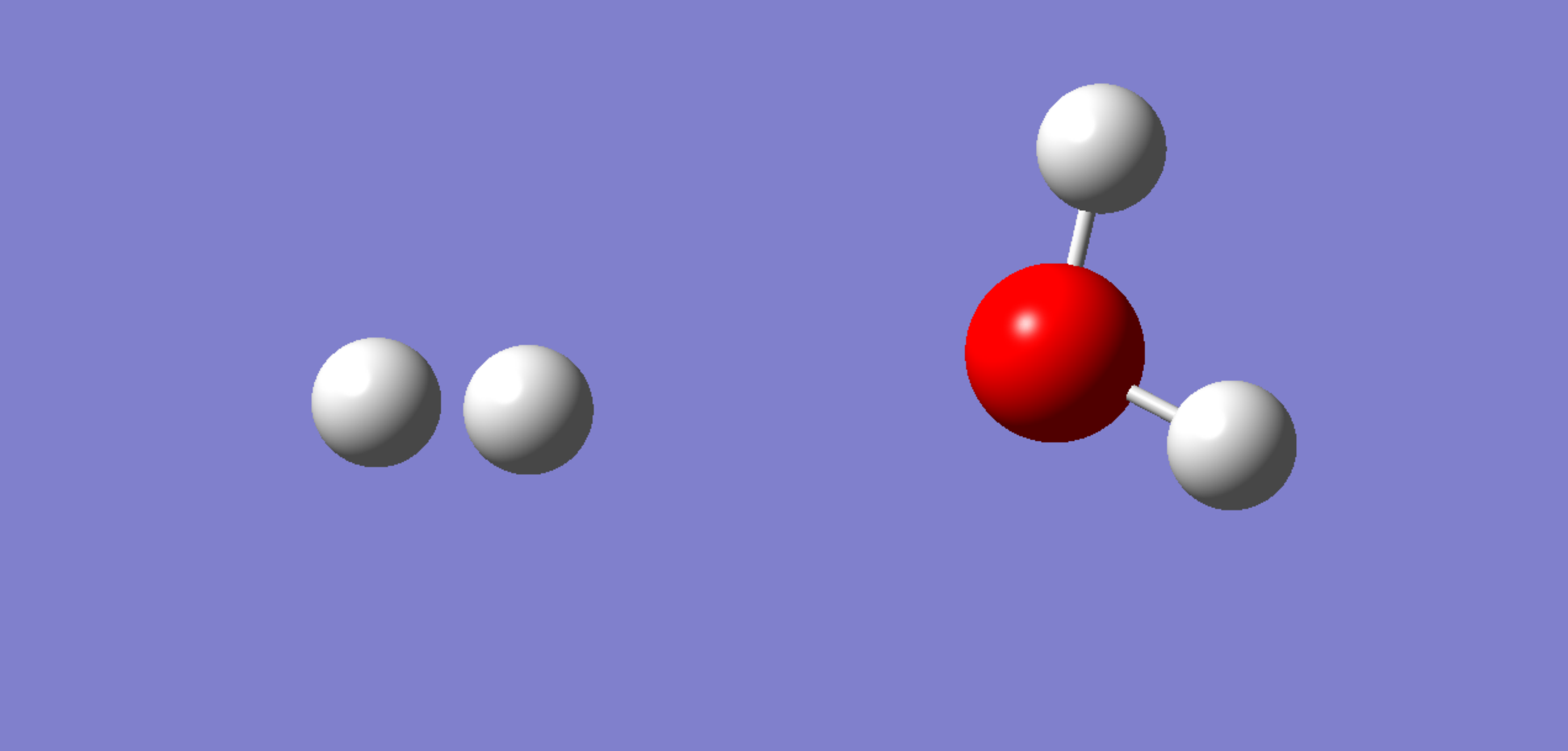} & Singlet &360, 390$^c$& 2.993 \\
&& \includegraphics[width=0.2\textwidth]{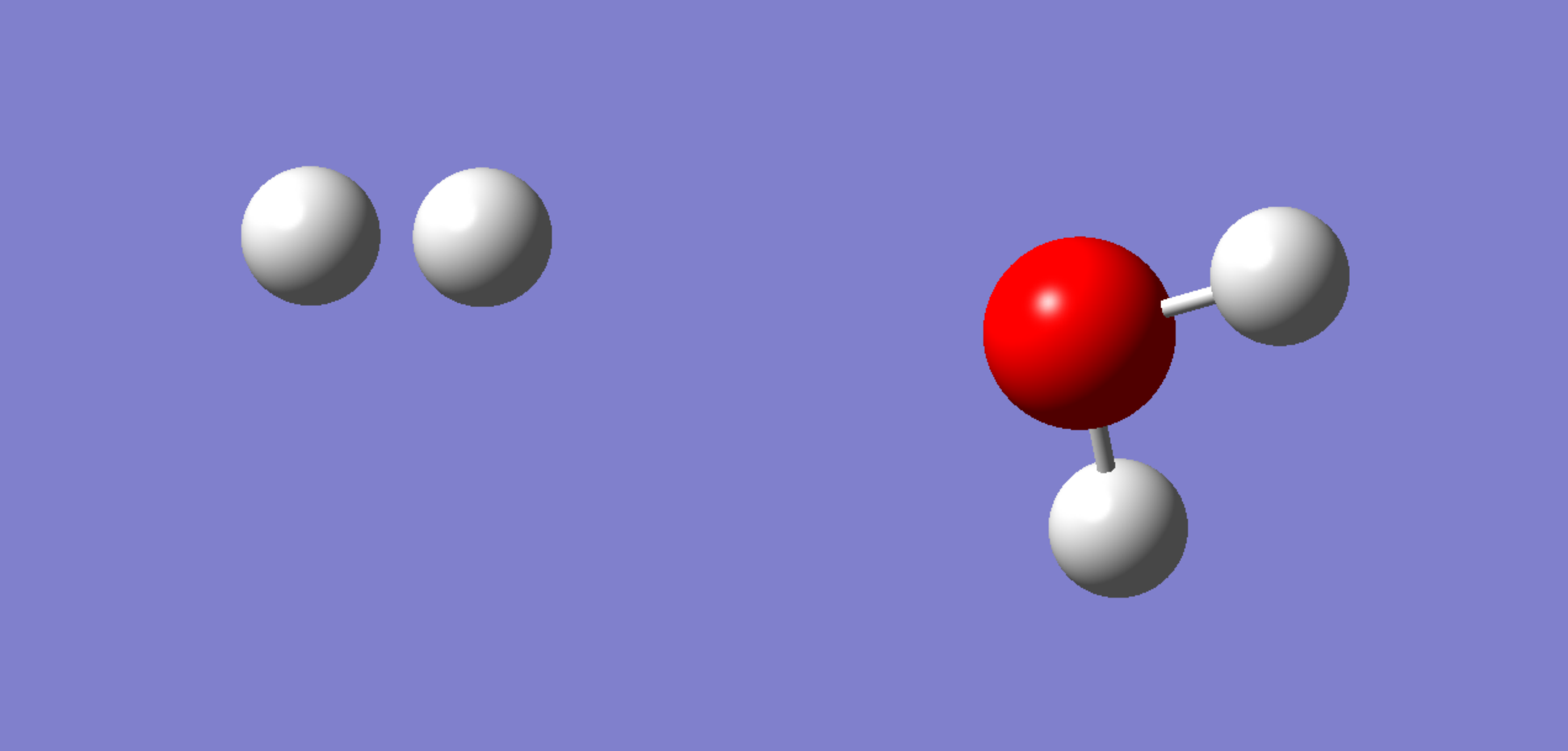} & & 360$^b$ & 2.993$^b$ \\
34&PH$_2$& \includegraphics[width=0.2\textwidth]{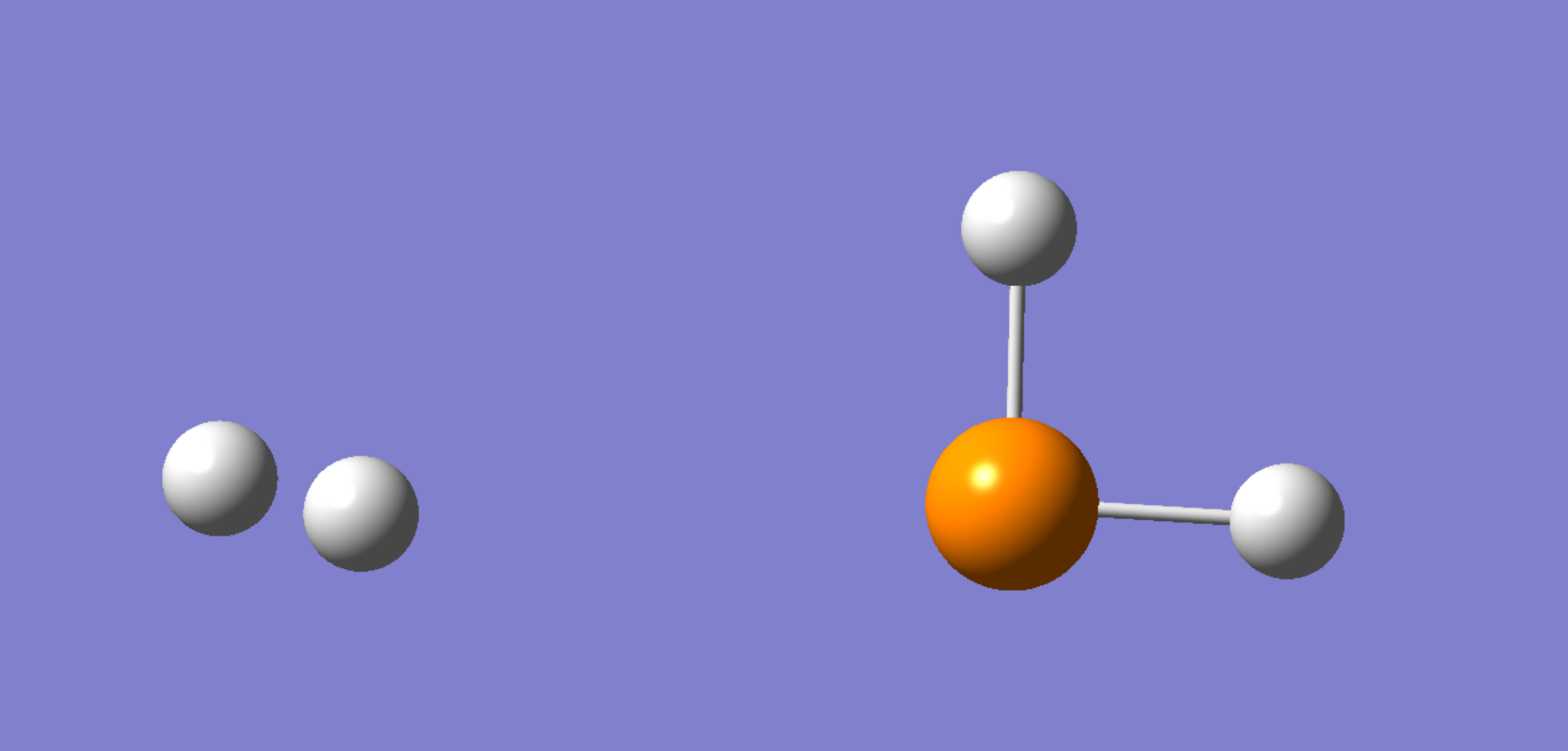} & Doublet &178& 1.483  \\
35&C$_2$H& \includegraphics[width=0.2\textwidth]{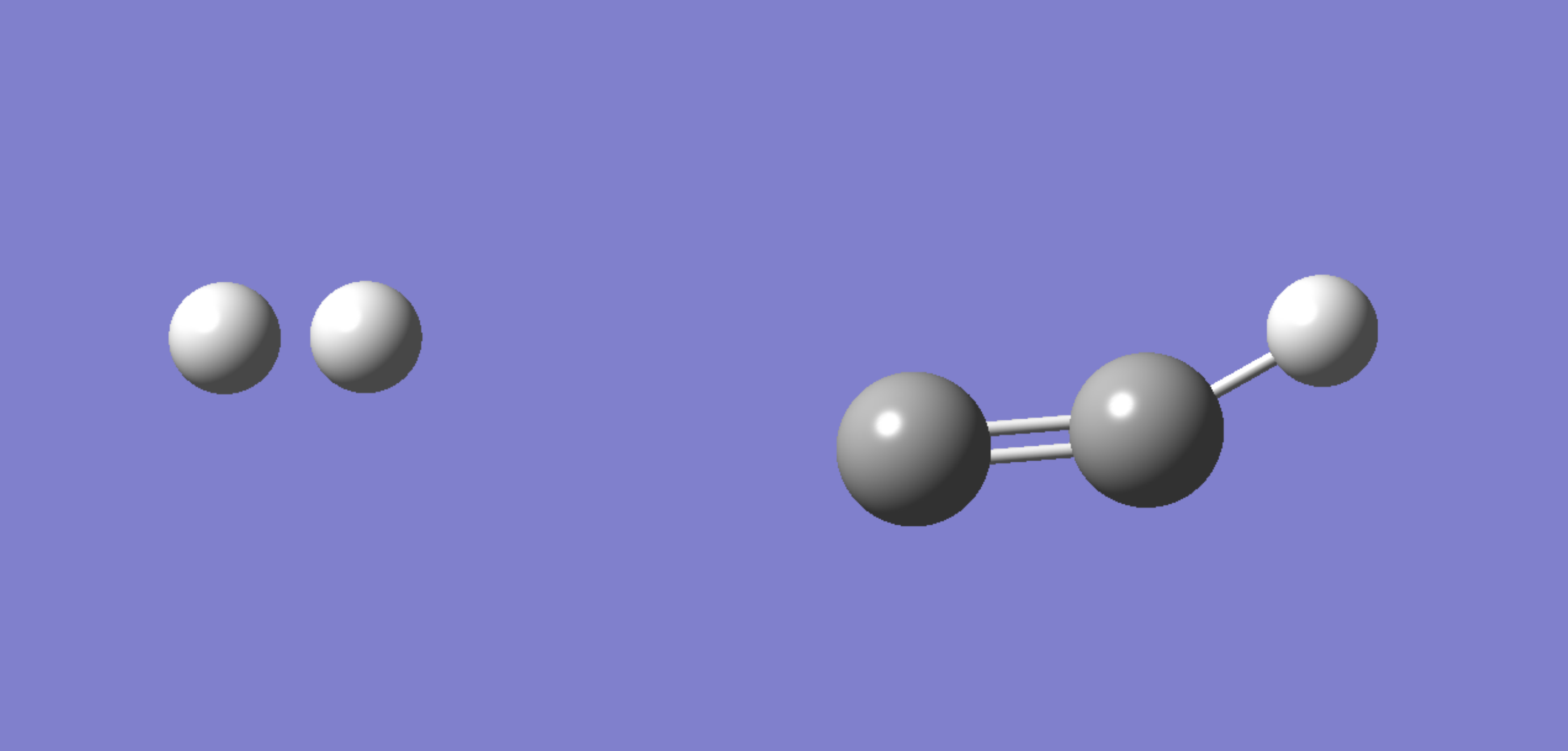} & Doublet & 242 & 2.014  \\
36&N$_2$H& \includegraphics[width=0.2\textwidth]{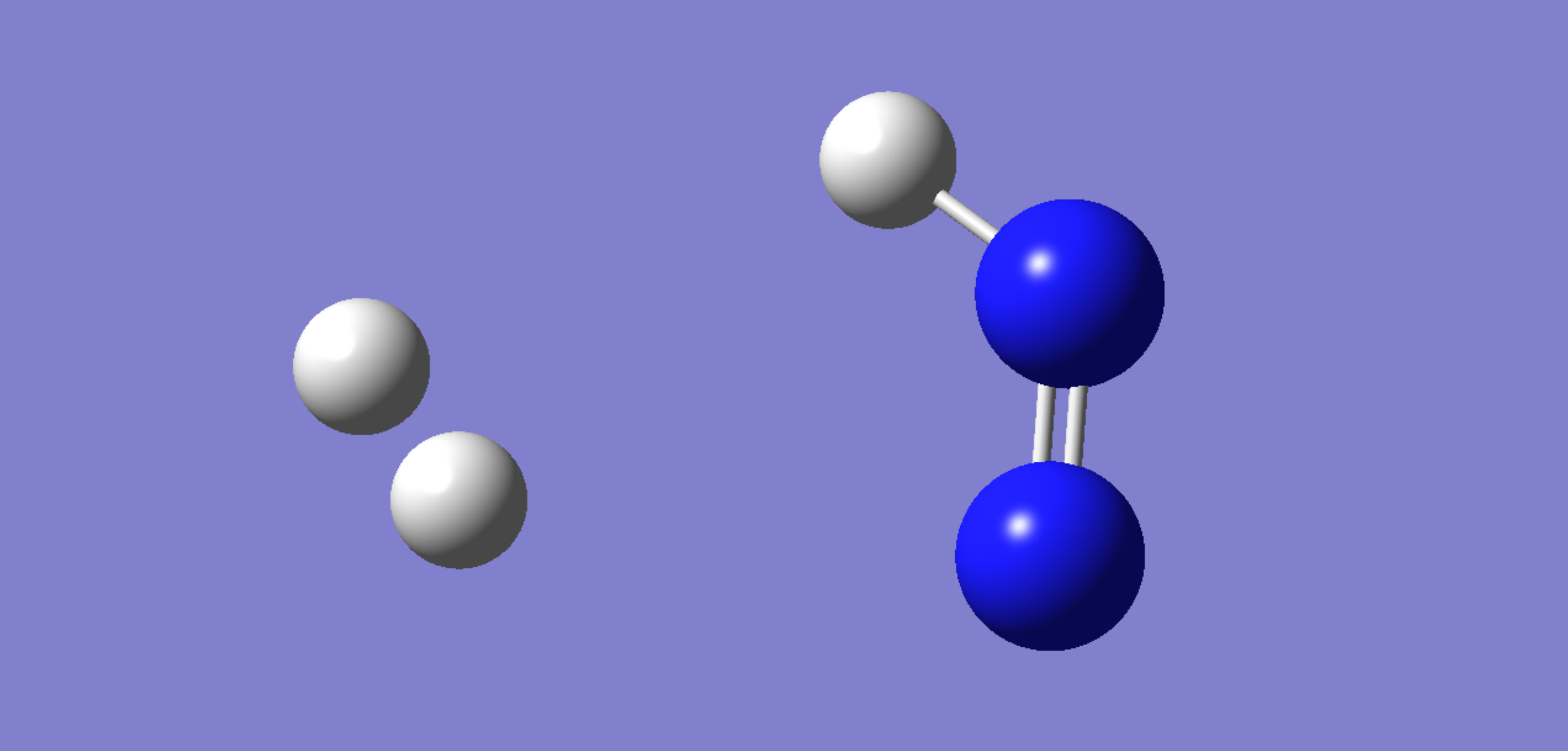} & Doublet &432 & 3.589  \\
\hline
\end{tabular}
\end{table}

\begin{table}
\scriptsize
\centering
\begin{tabular}{|c|c|c|c|c|c|}
\hline
{\bf Sl.}& {\bf Species} & {\bf Optimized} & {\bf Ground} & \multicolumn{2}{c|}{\bf Binding Energy} \\
\cline{5-6}
 {\bf No.} & & {\bf Structures} & {\bf State} & {\bf in K} & {\bf in kJ/mol} \\
\hline
37&O$_2$H& \includegraphics[width=0.2\textwidth]{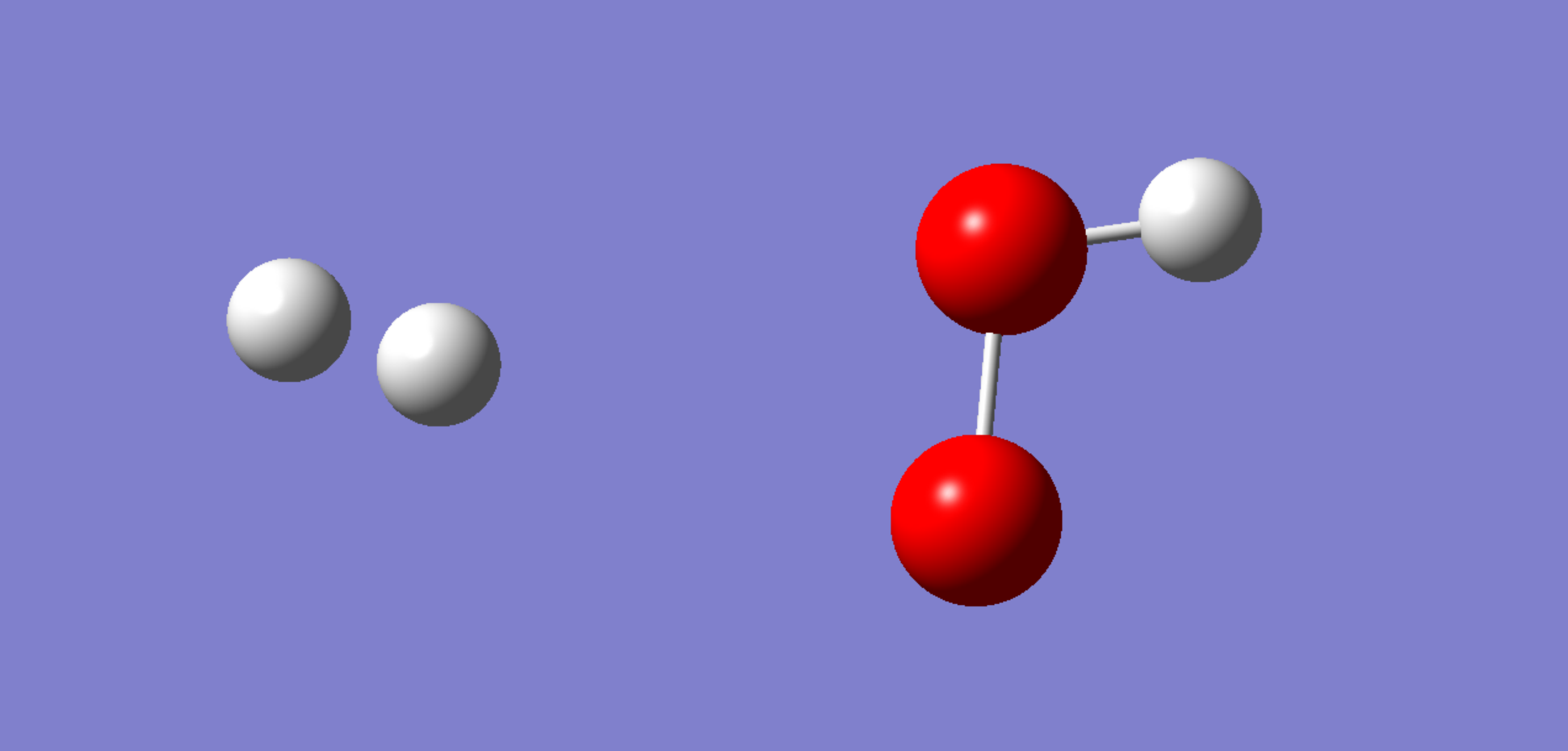} & Doublet &339, 300$^c$ & 2.819  \\
&& \includegraphics[width=0.2\textwidth]{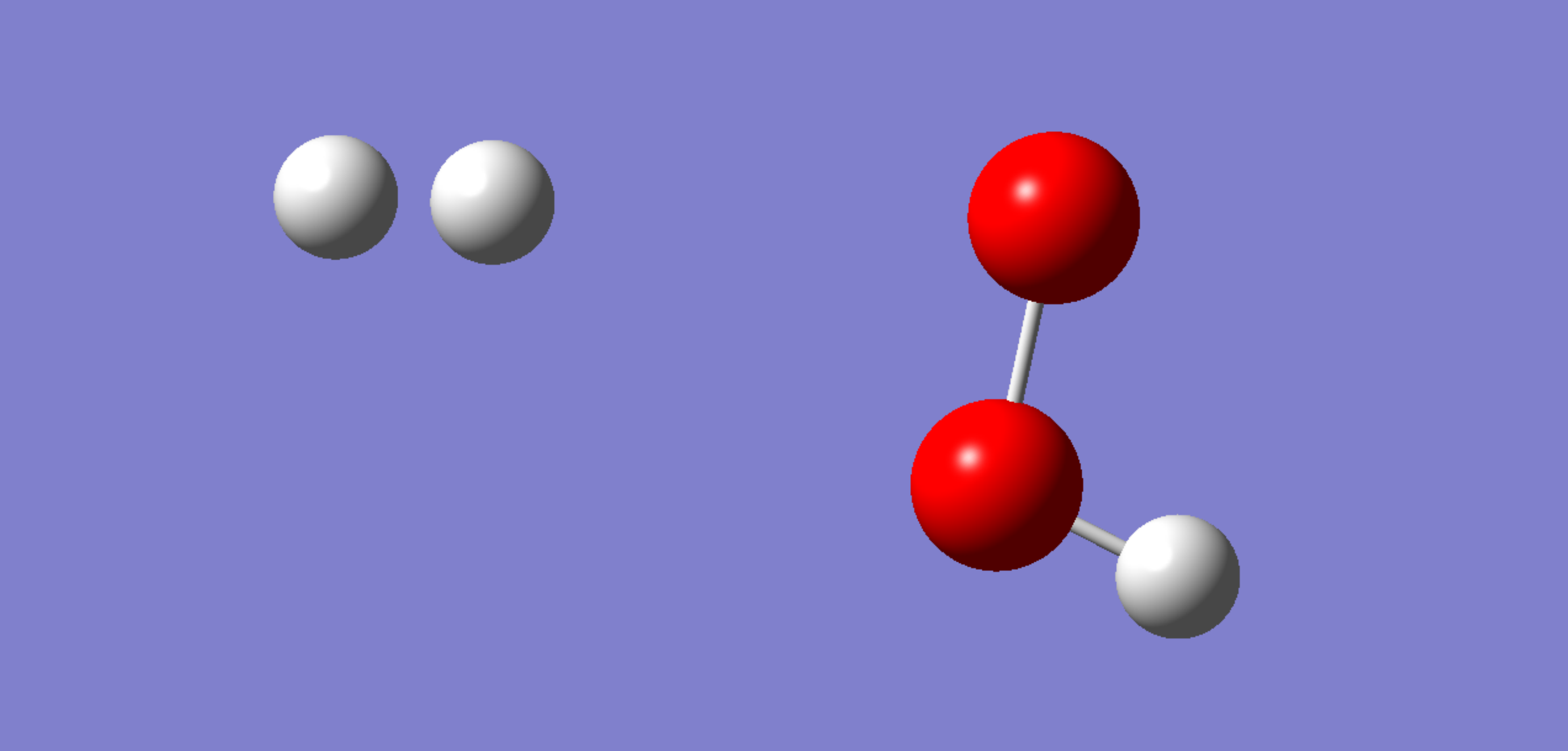} & & 339$^b$ & 2.819$^b$ \\
38&HS$_2$& \includegraphics[width=0.2\textwidth]{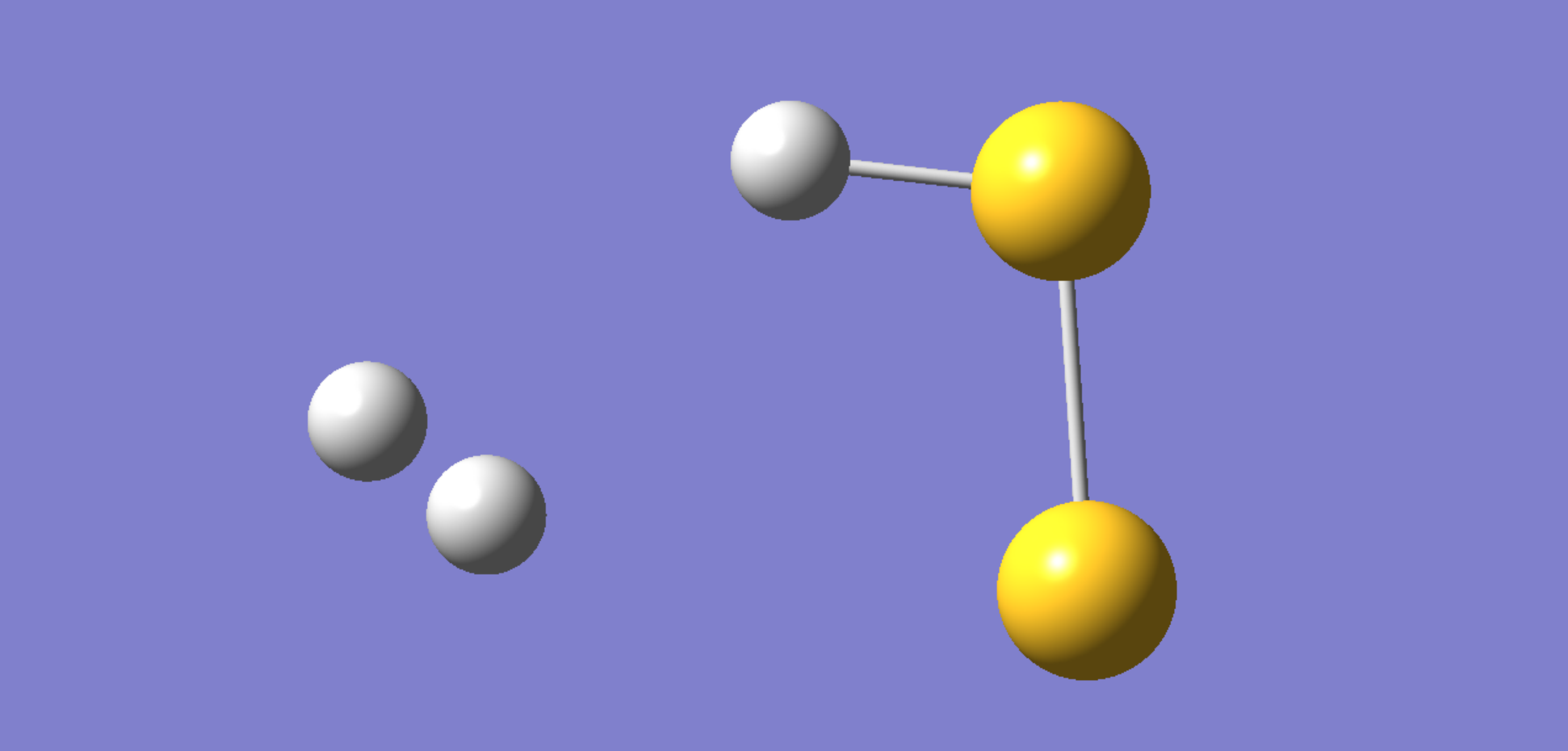} & Doublet &660 &5.487  \\
39&HCN & \includegraphics[width=0.2\textwidth]{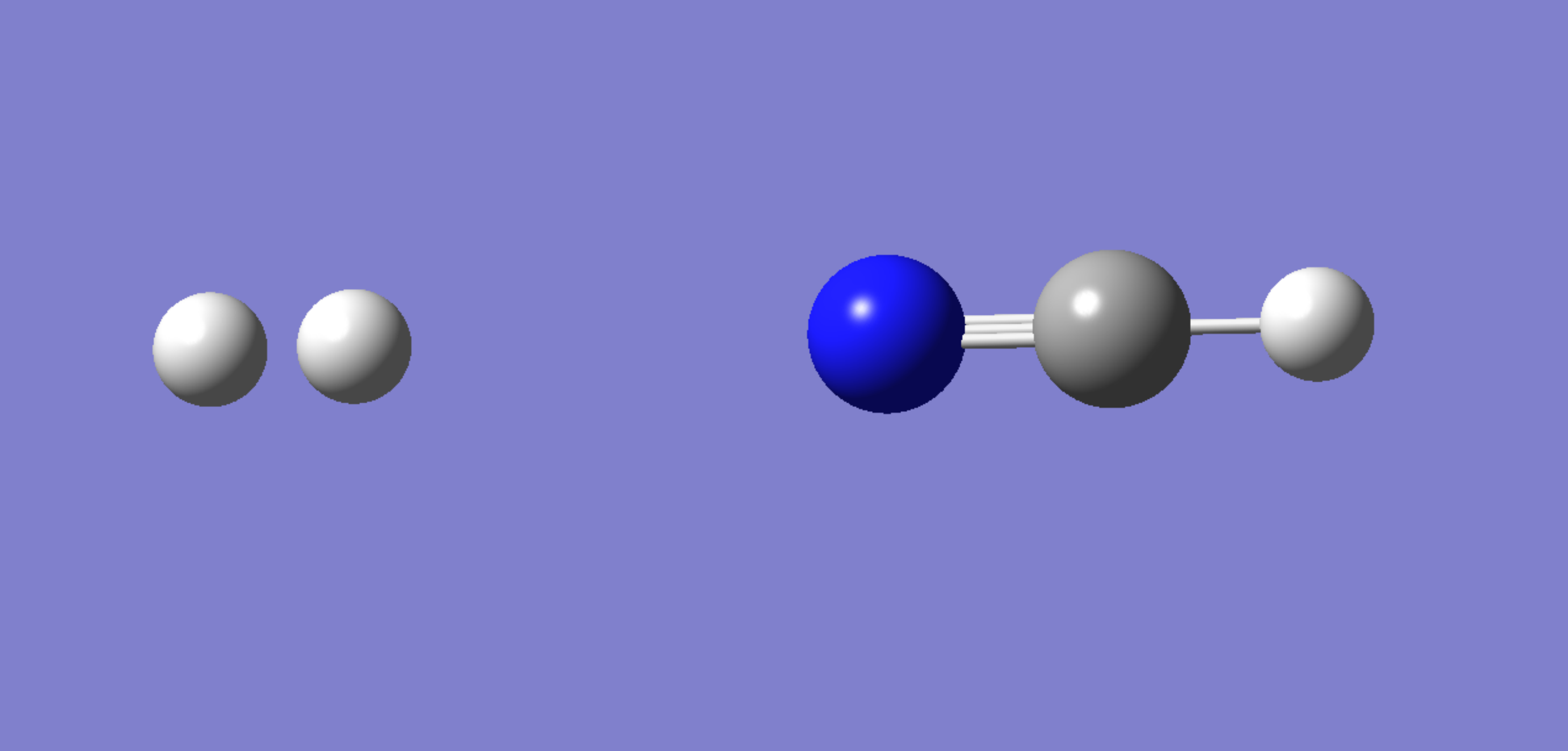} & Singlet &395 &3.282  \\
40&HNC & \includegraphics[width=0.2\textwidth]{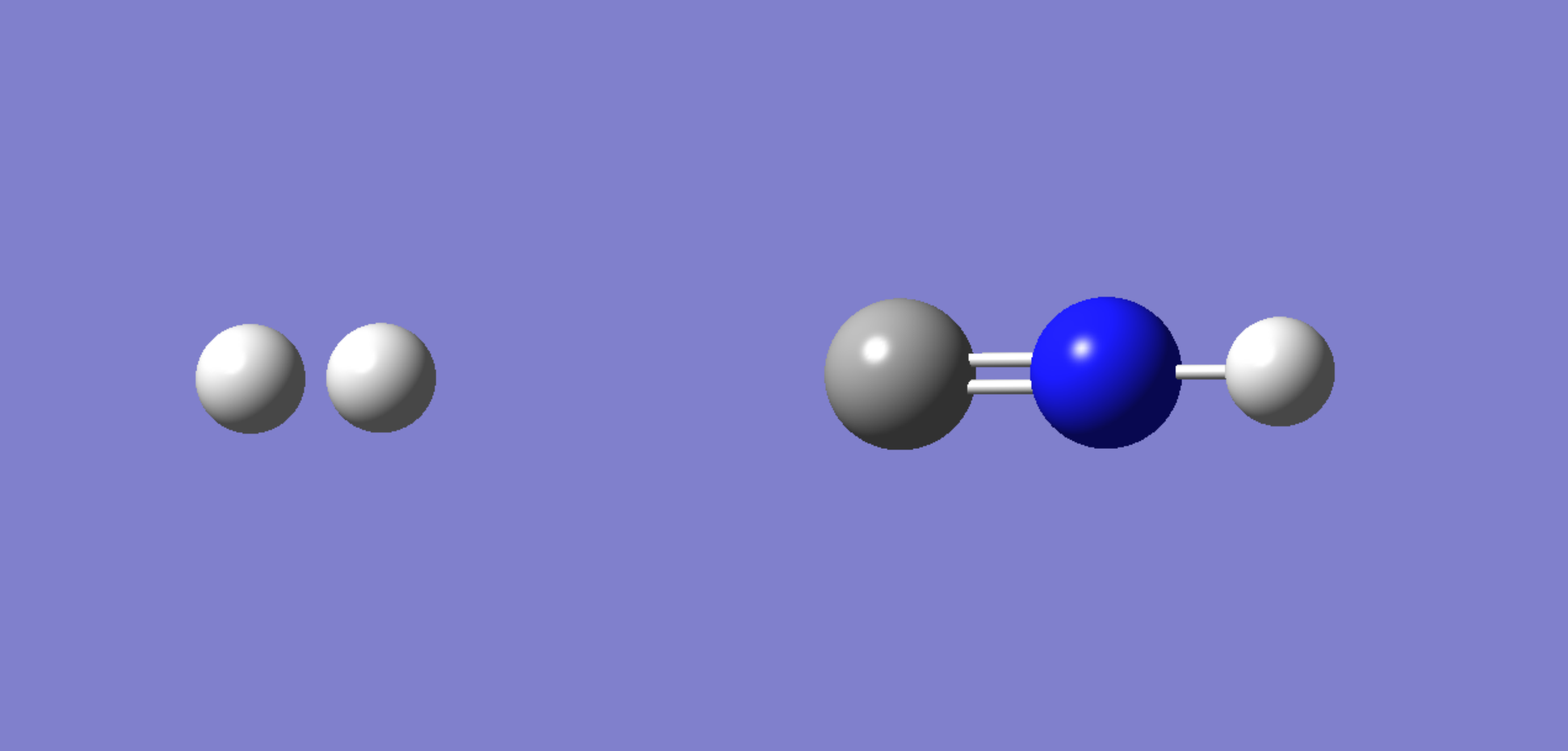} & Singlet &338 & 2.814  \\
41&HCO & \includegraphics[width=0.2\textwidth]{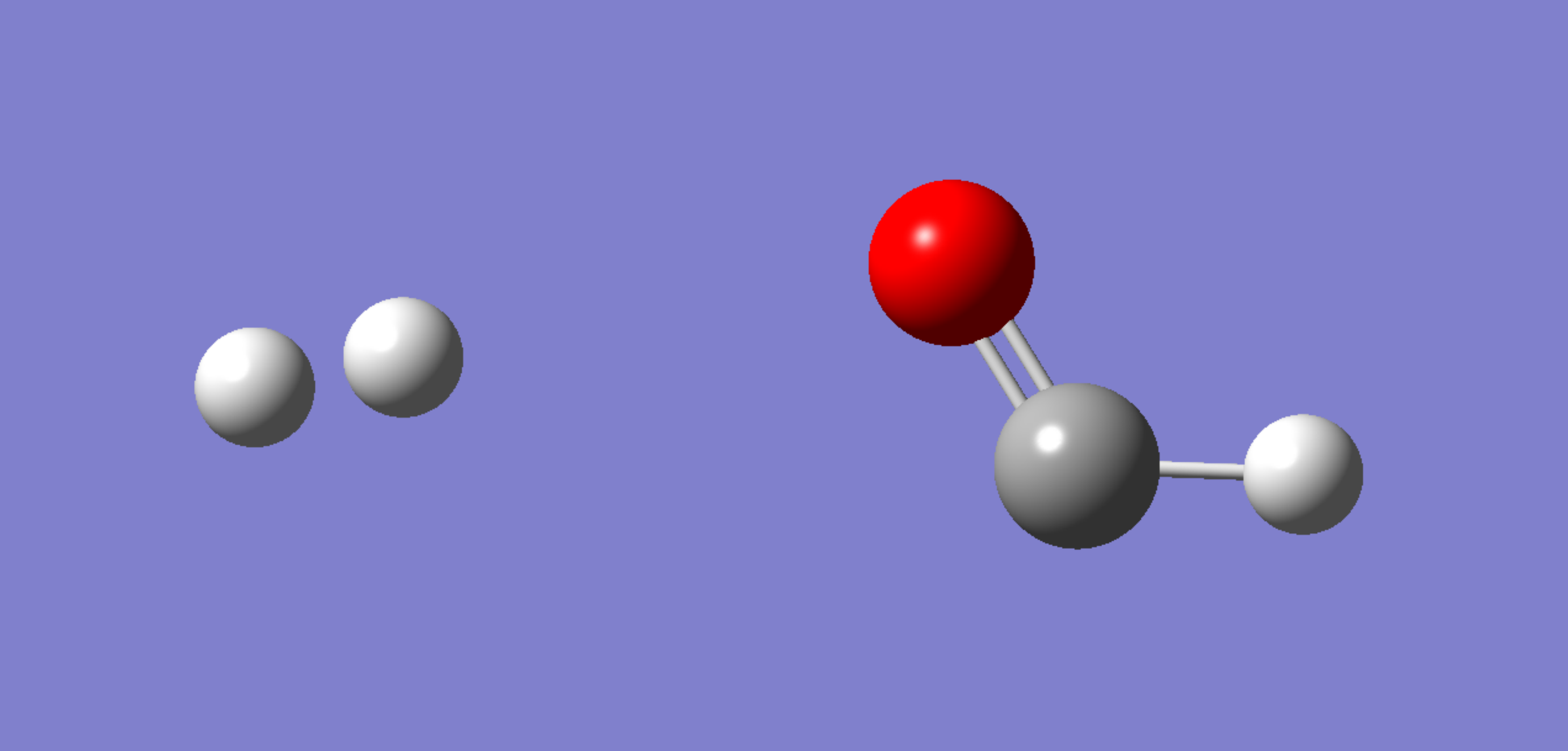} & Doublet &243 & 2.019  \\
42&HOC & \includegraphics[width=0.2\textwidth]{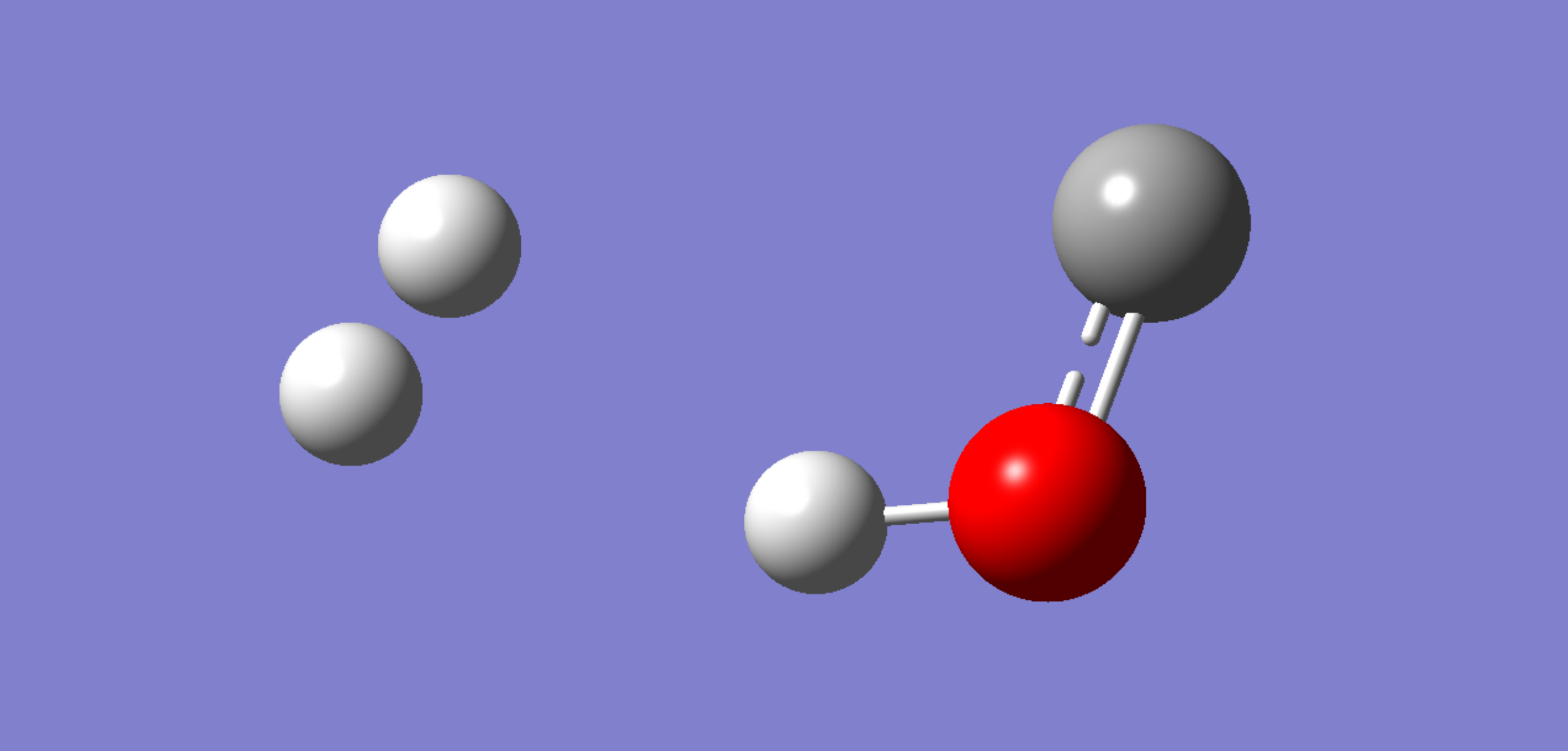} & Doublet & 769 & 6.396   \\
43&HCS & \includegraphics[width=0.2\textwidth]{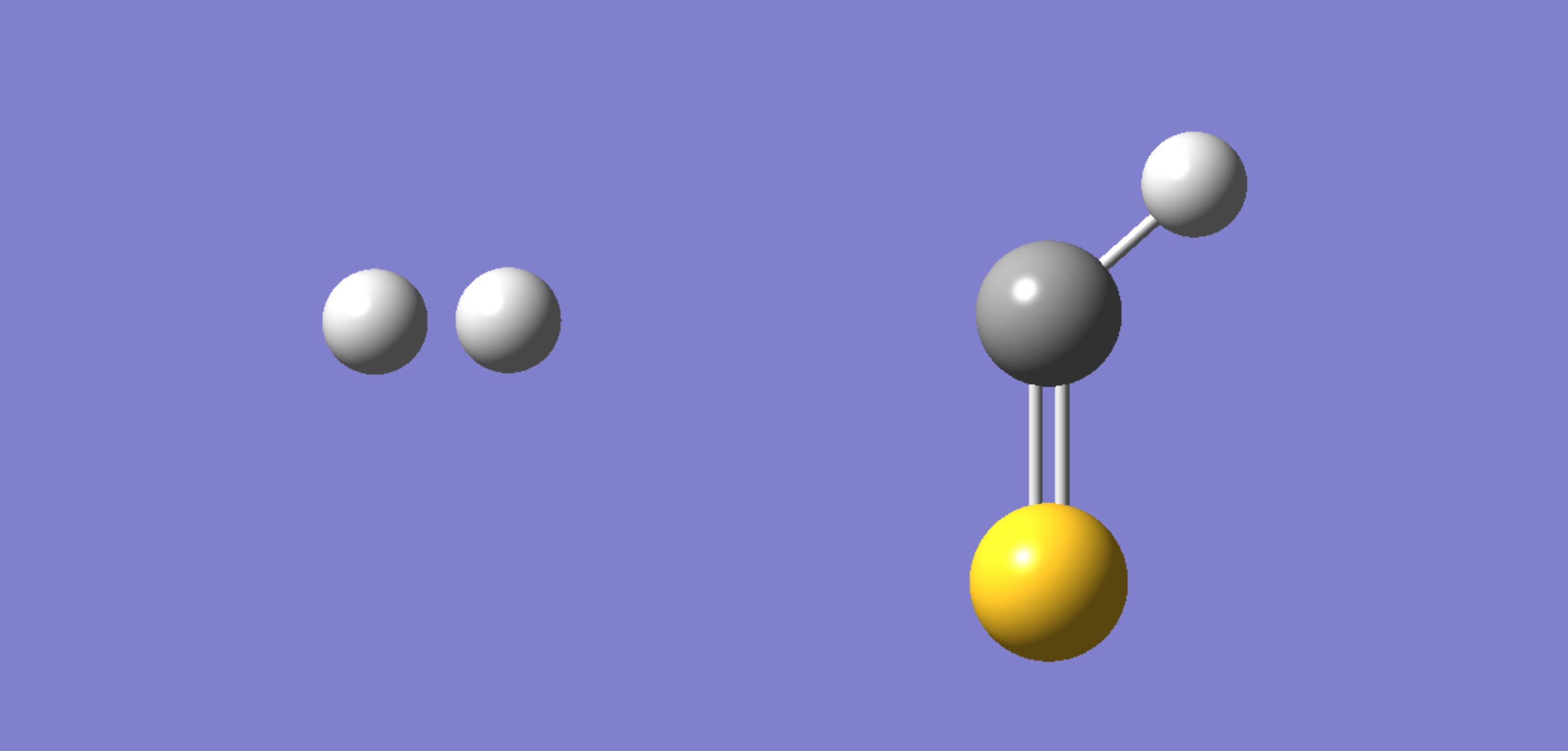} & Doublet & 334 & 2.780  \\
&& \includegraphics[width=0.2\textwidth]{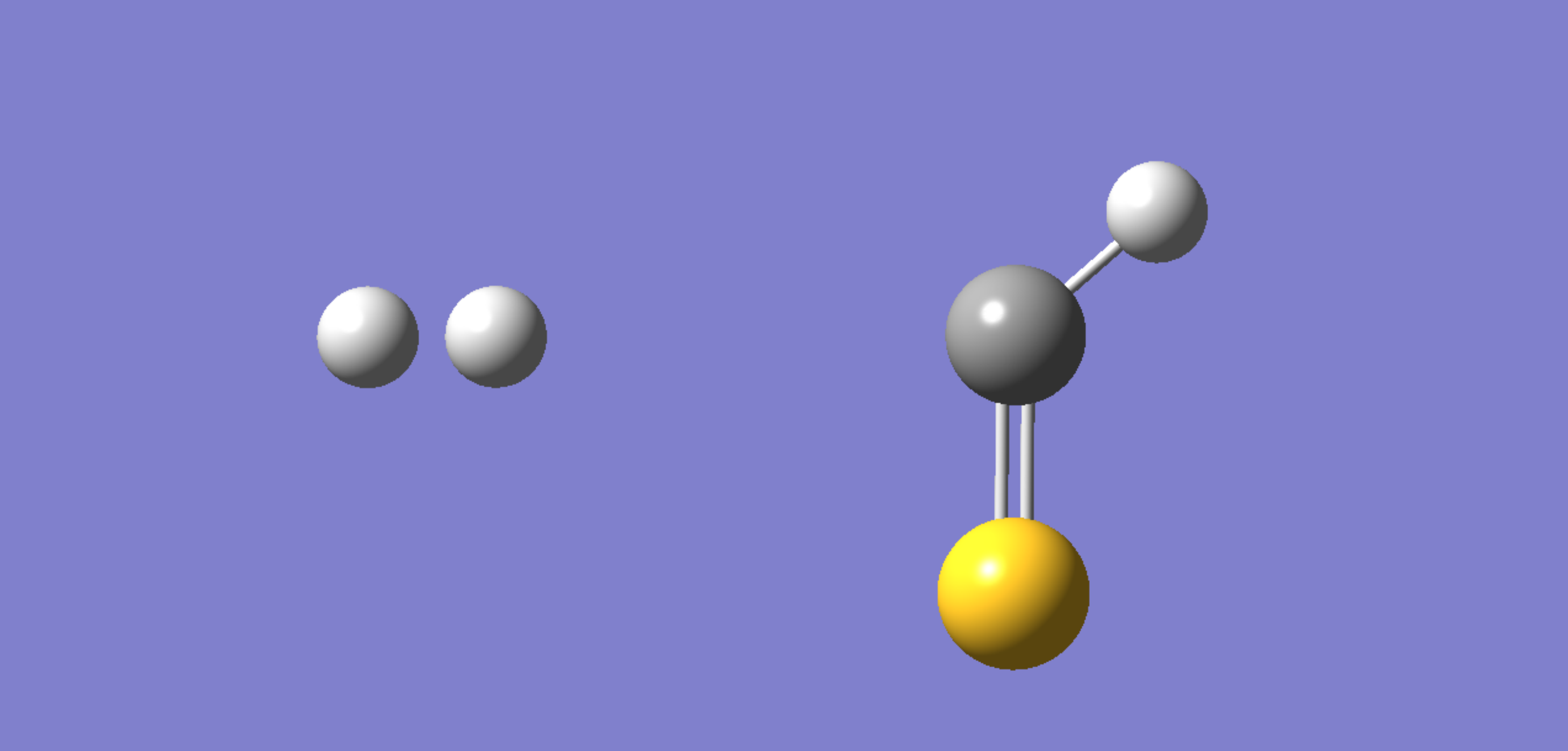} & & 334$^b$ & 2.780$^b$ \\
44&HNO & \includegraphics[width=0.2\textwidth]{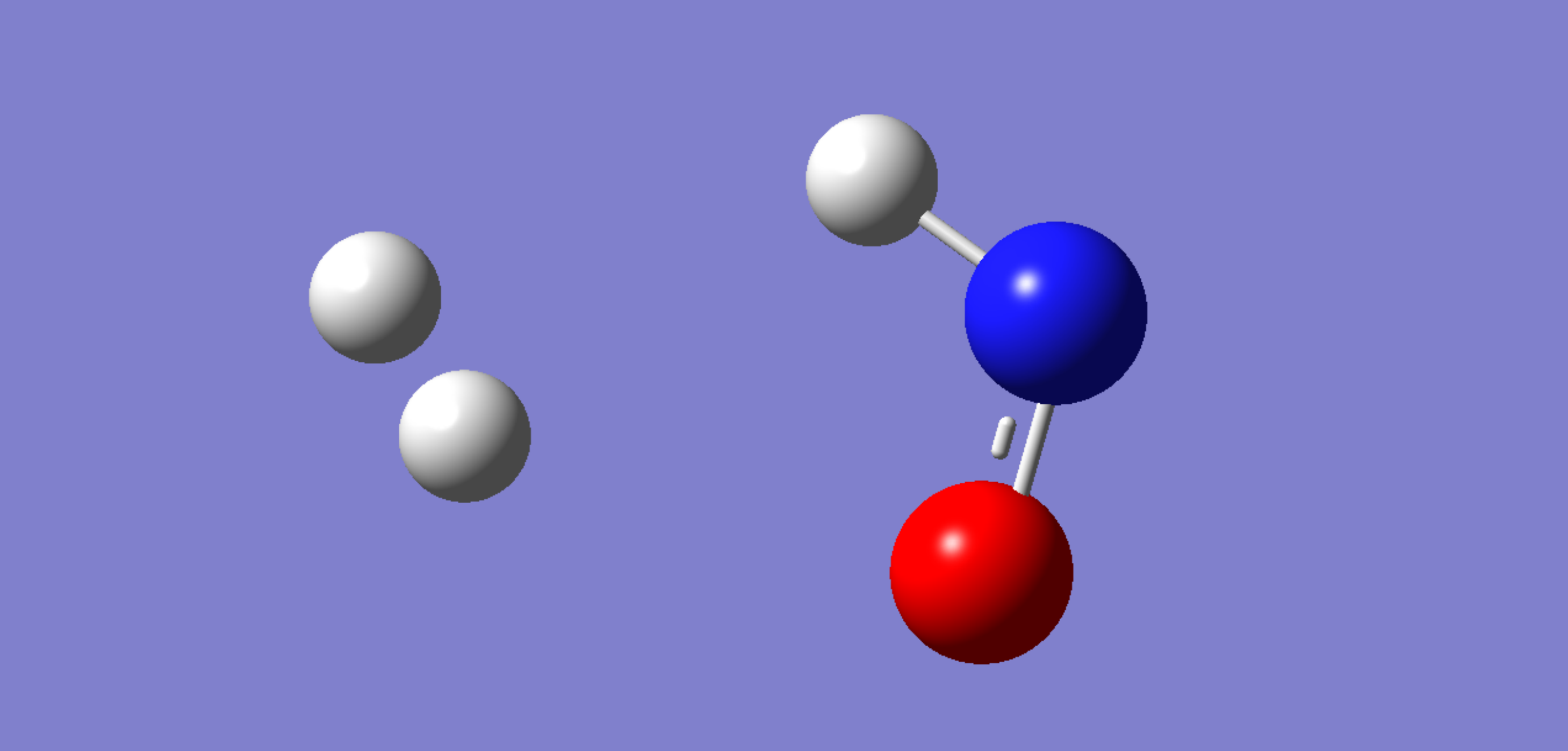} & Singlet &574 & 4.773  \\
45&H$_2$S & \includegraphics[width=0.2\textwidth]{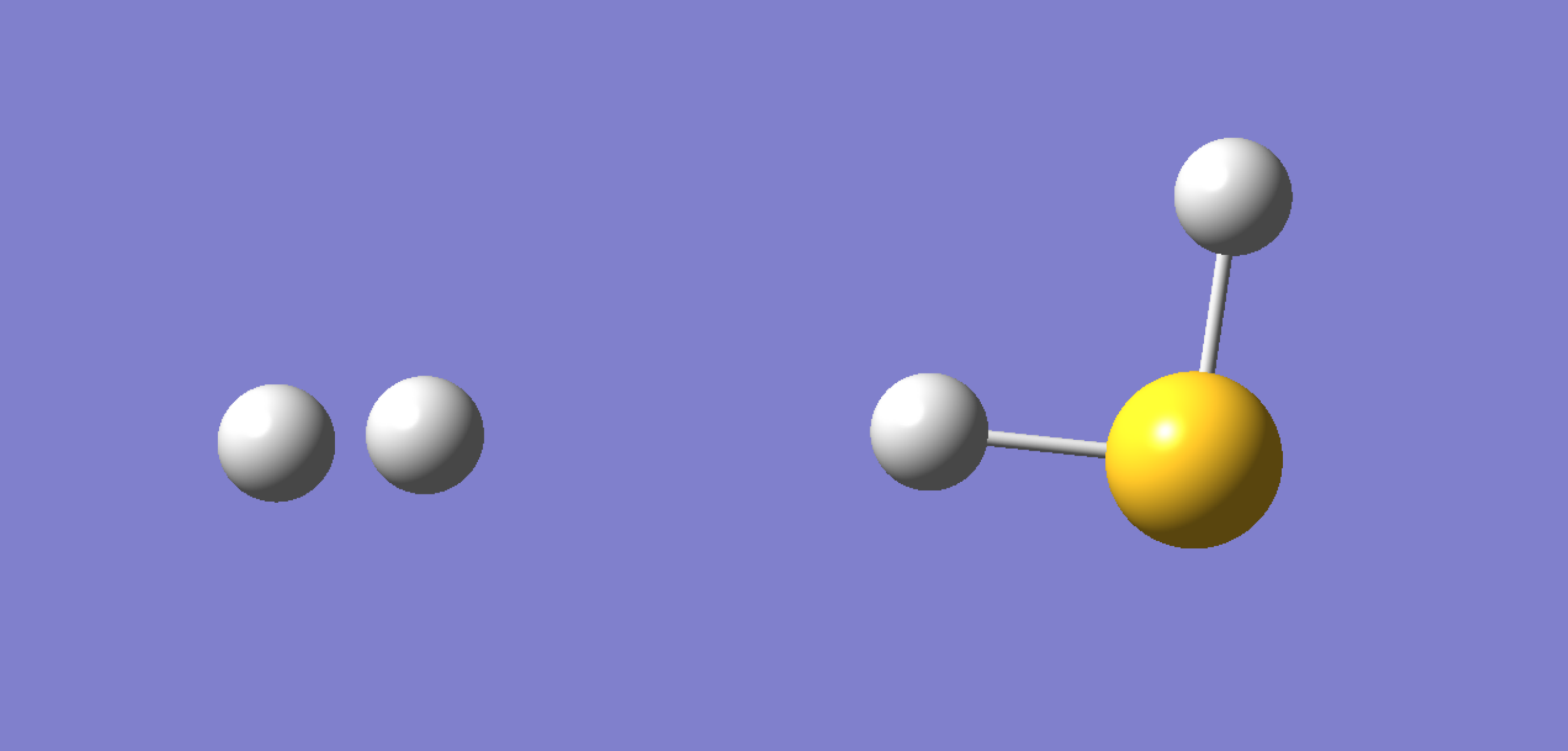} & Singlet &99 & 0.824  \\
46&C$_3$& \includegraphics[width=0.2\textwidth]{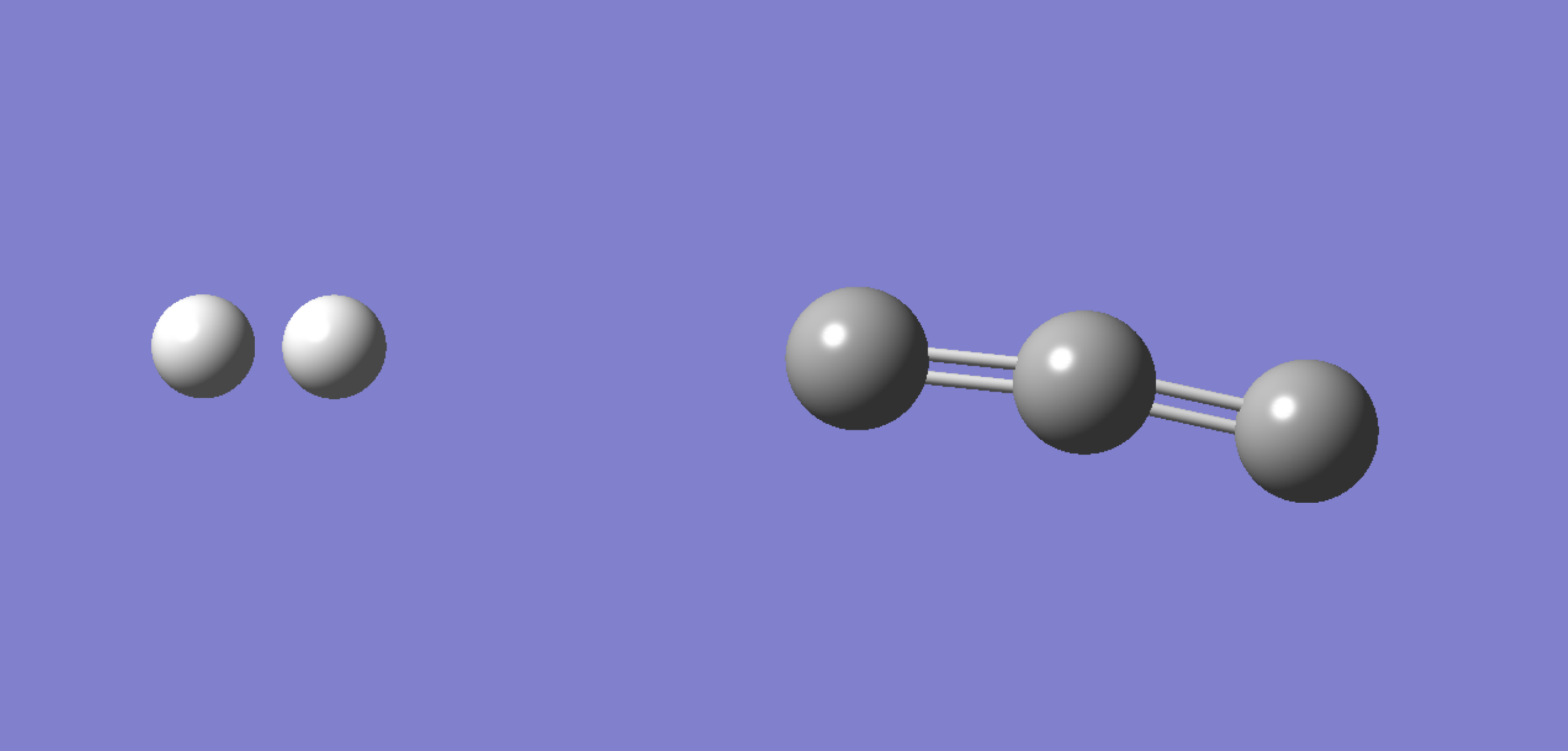} & Singlet &295 & 2.455 \\
47&O$_3$& \includegraphics[width=0.2\textwidth]{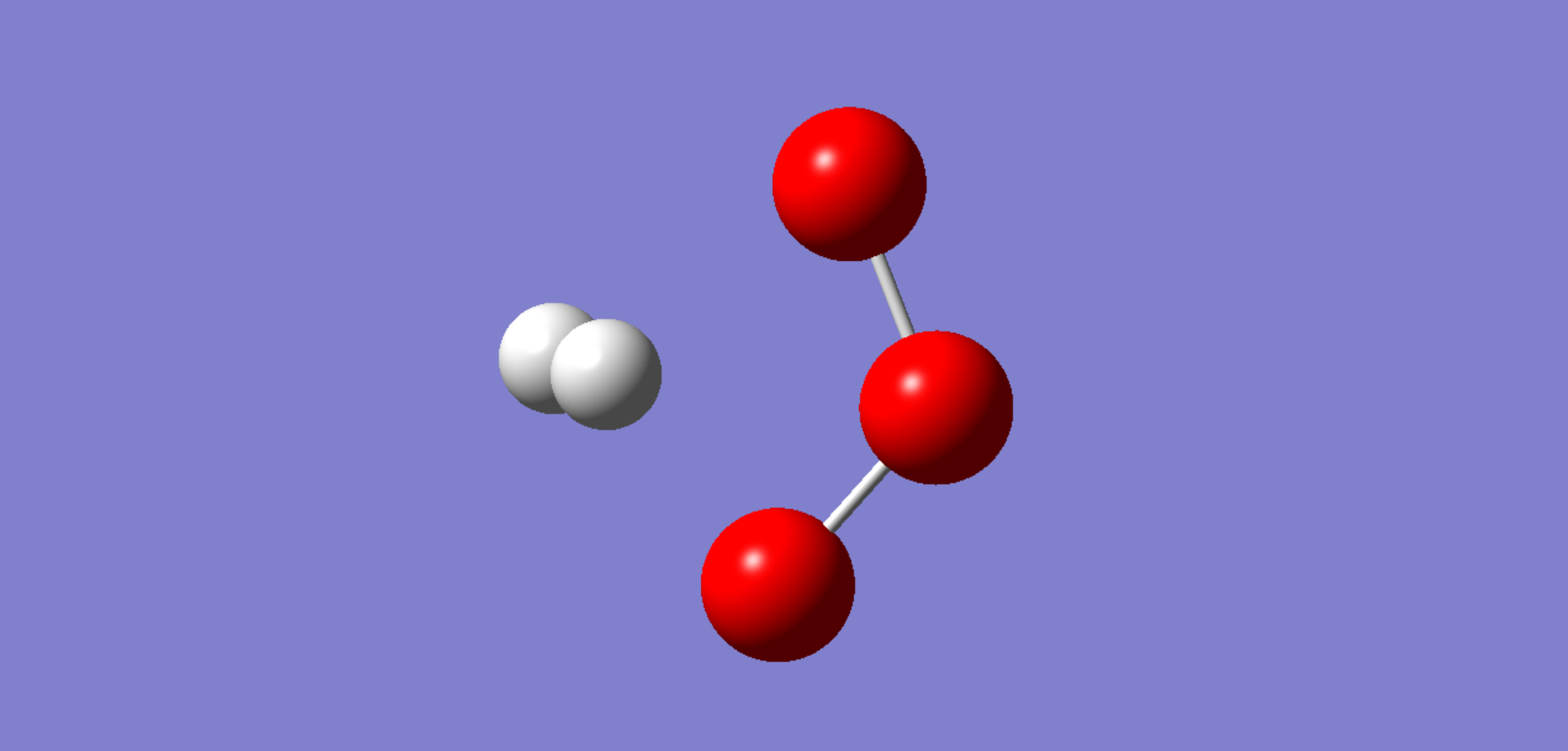} & Singlet & 381, 120$^c$ & 3.169\\
&& \includegraphics[width=0.2\textwidth]{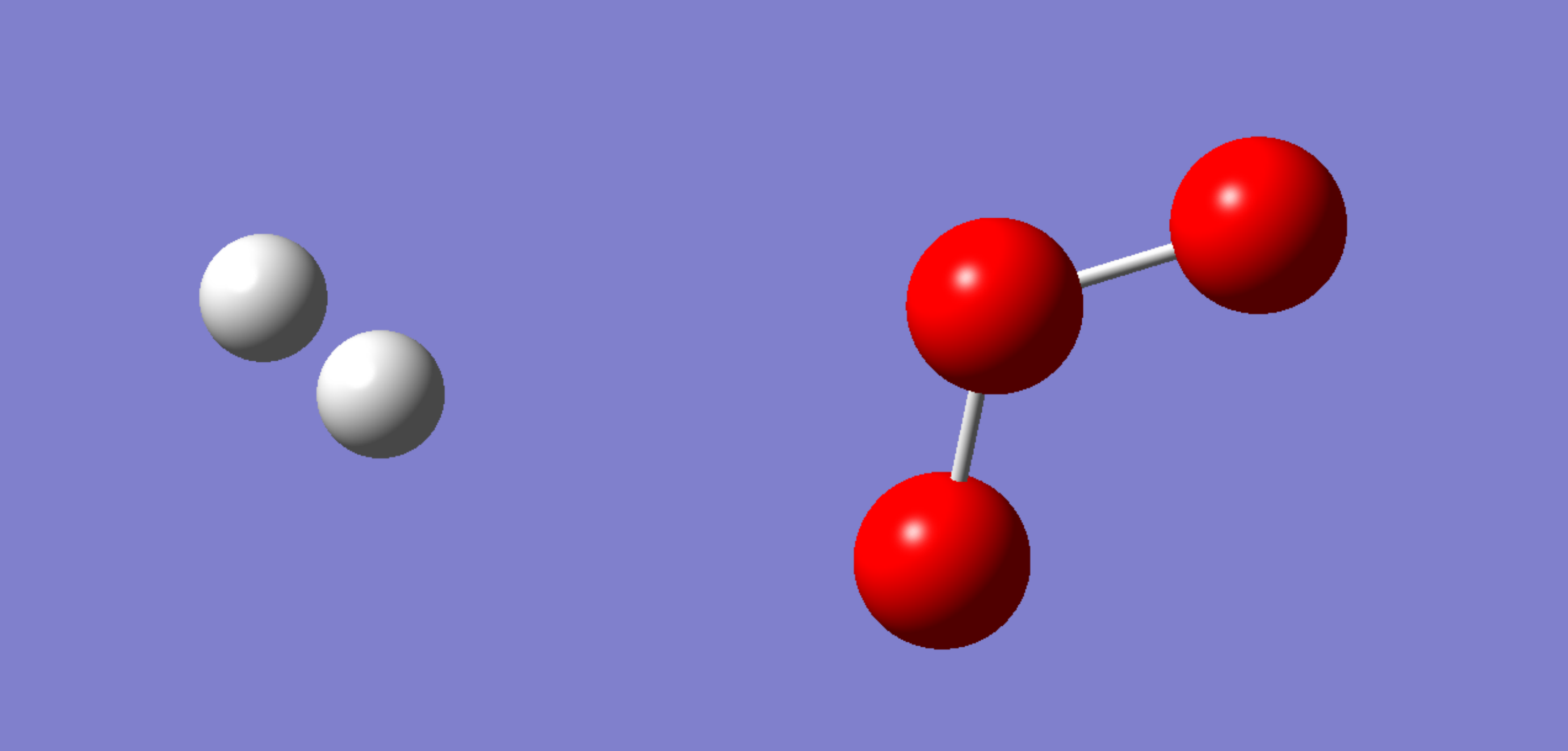} & & & \\
\hline
\end{tabular}
\end{table}

\begin{table}
\scriptsize
\centering
\begin{tabular}{|c|c|c|c|c|c|}
\hline
{\bf Sl.}& {\bf Species} & {\bf Optimized} & {\bf Ground} & \multicolumn{2}{c|}{\bf Binding Energy} \\
\cline{5-6}
 {\bf No.} & & {\bf Structures} & {\bf State} & {\bf in K} & {\bf in kJ/mol} \\
\hline
48&C$_2$N& \includegraphics[width=0.2\textwidth]{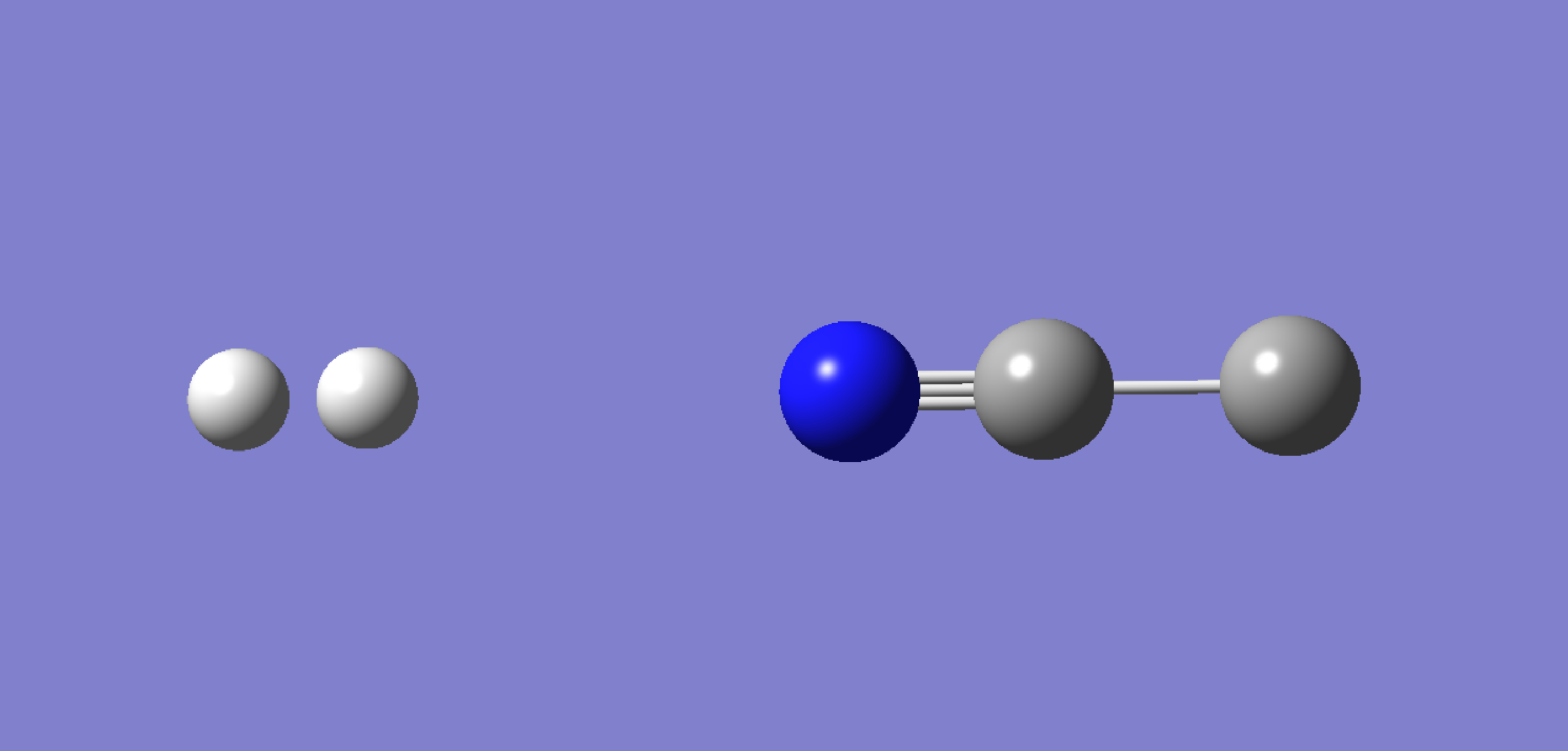} & Doublet &339 & 2.817 \\
49&C$_2$S& \includegraphics[width=0.2\textwidth]{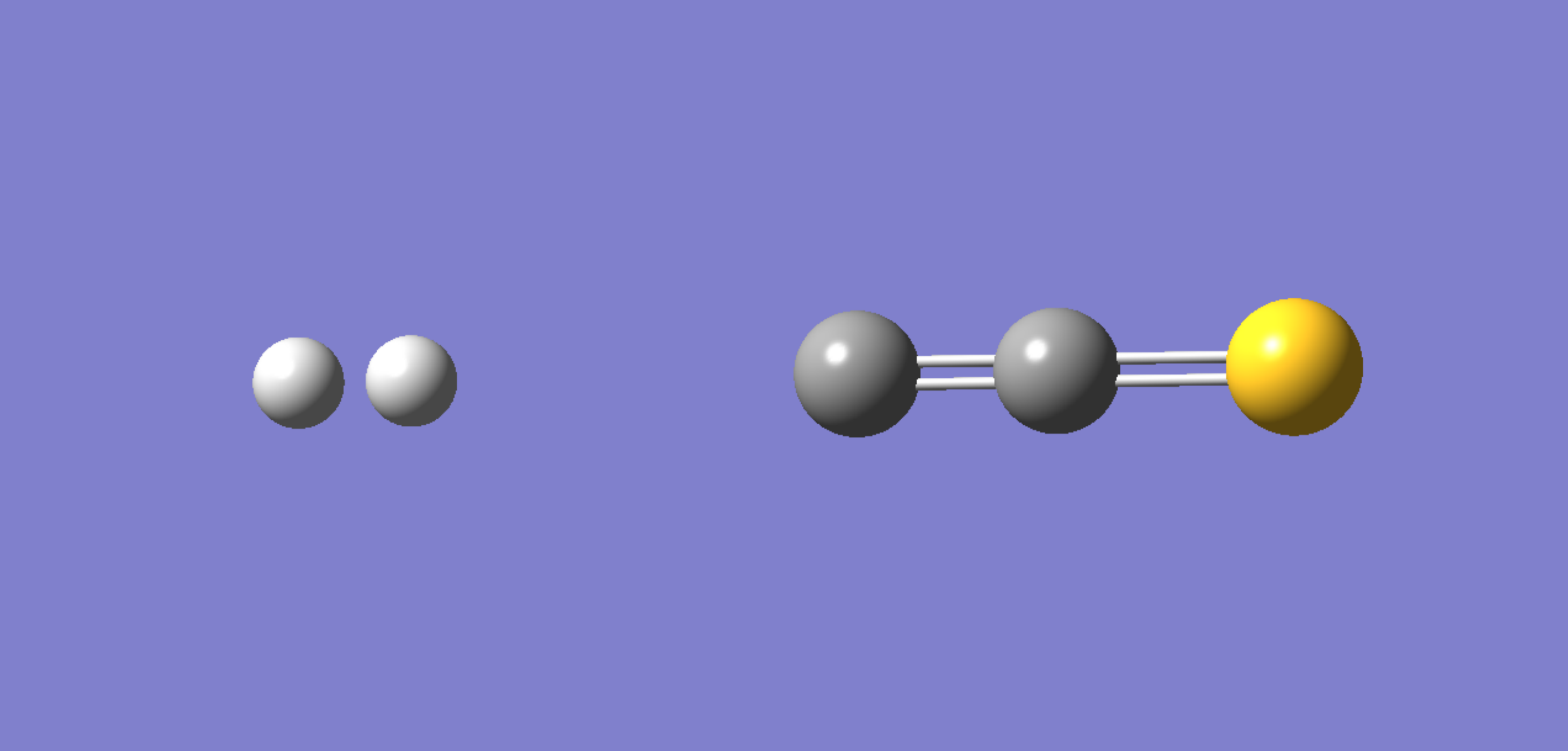} & Triplet & 355 & 2.951 \\
50&OCN& \includegraphics[width=0.2\textwidth]{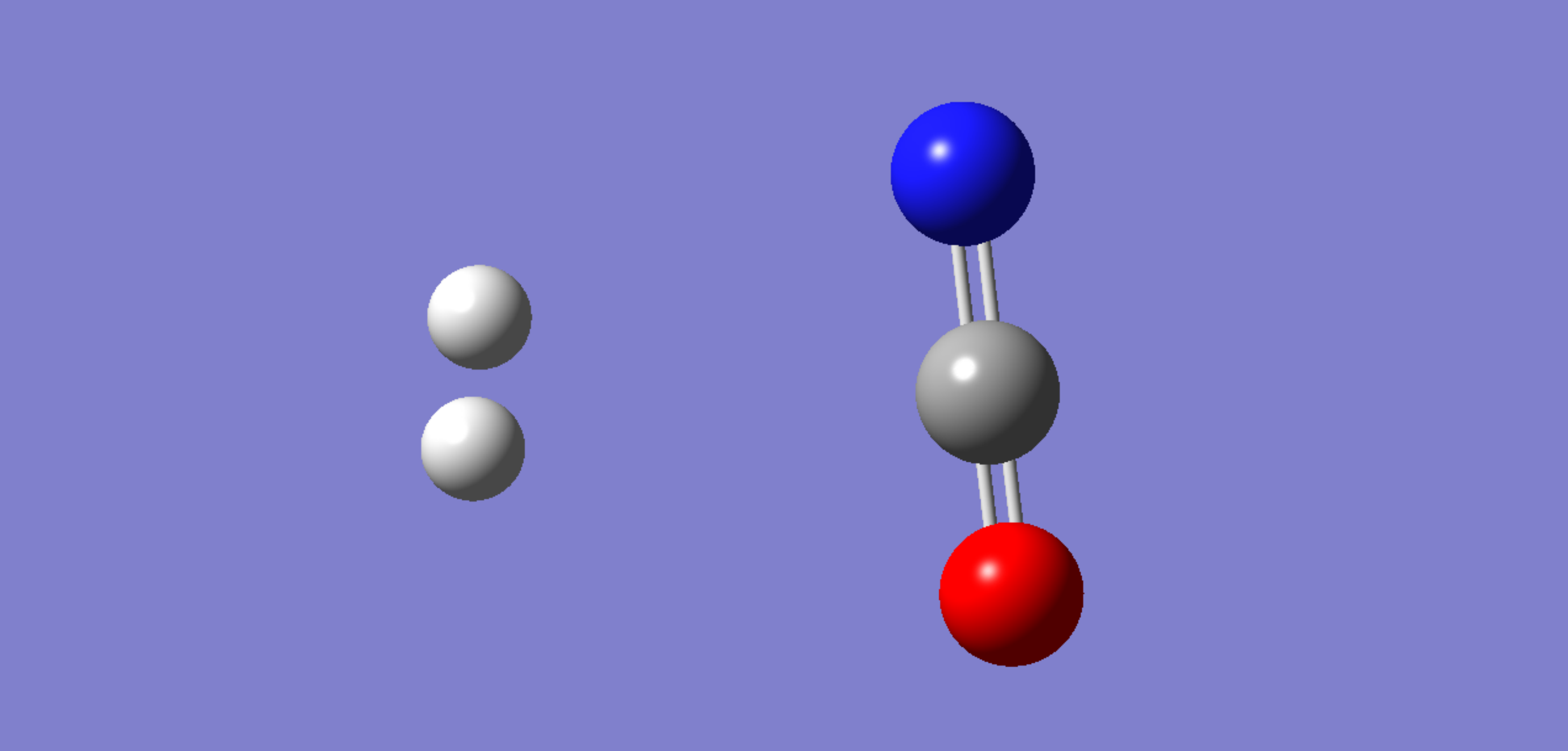} & Doublet & 422 & 3.510 \\
51&CO$_2$&  \includegraphics[width=0.2\textwidth]{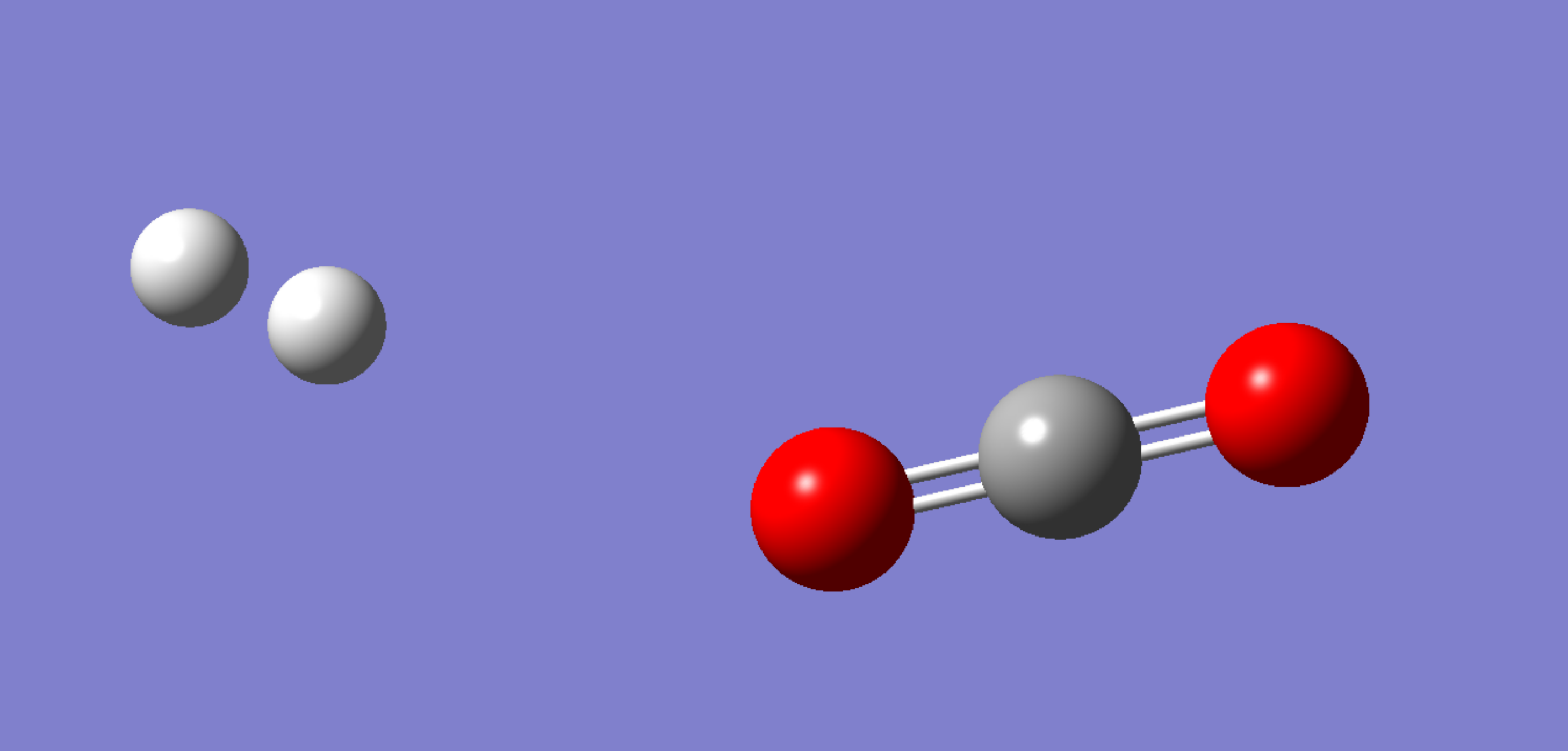} & Singlet &241 & 2.003 \\
 52&OCS&  \includegraphics[width=0.2\textwidth]{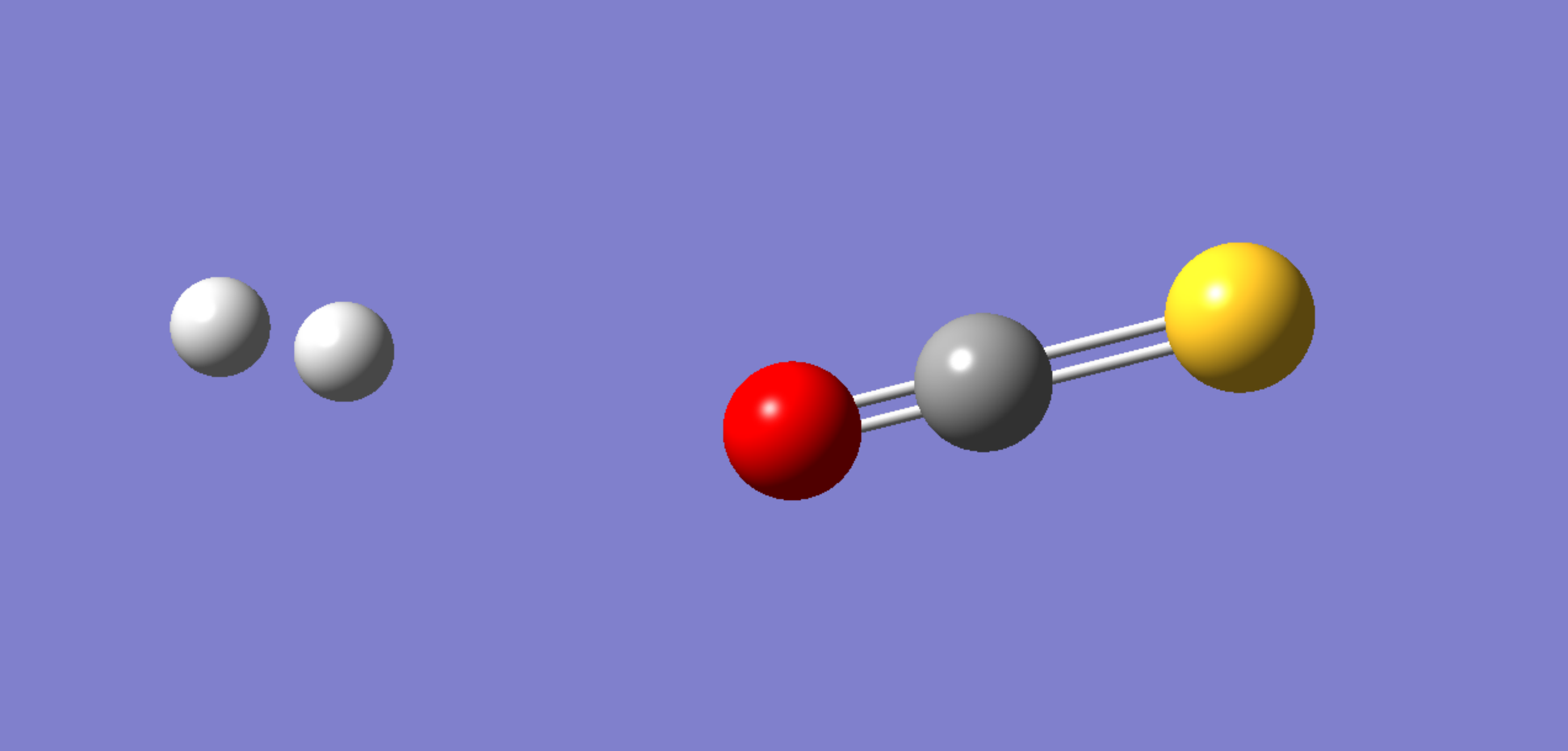} & Singlet &257& 2.137 \\
53&SO$_2$&  \includegraphics[width=0.2\textwidth]{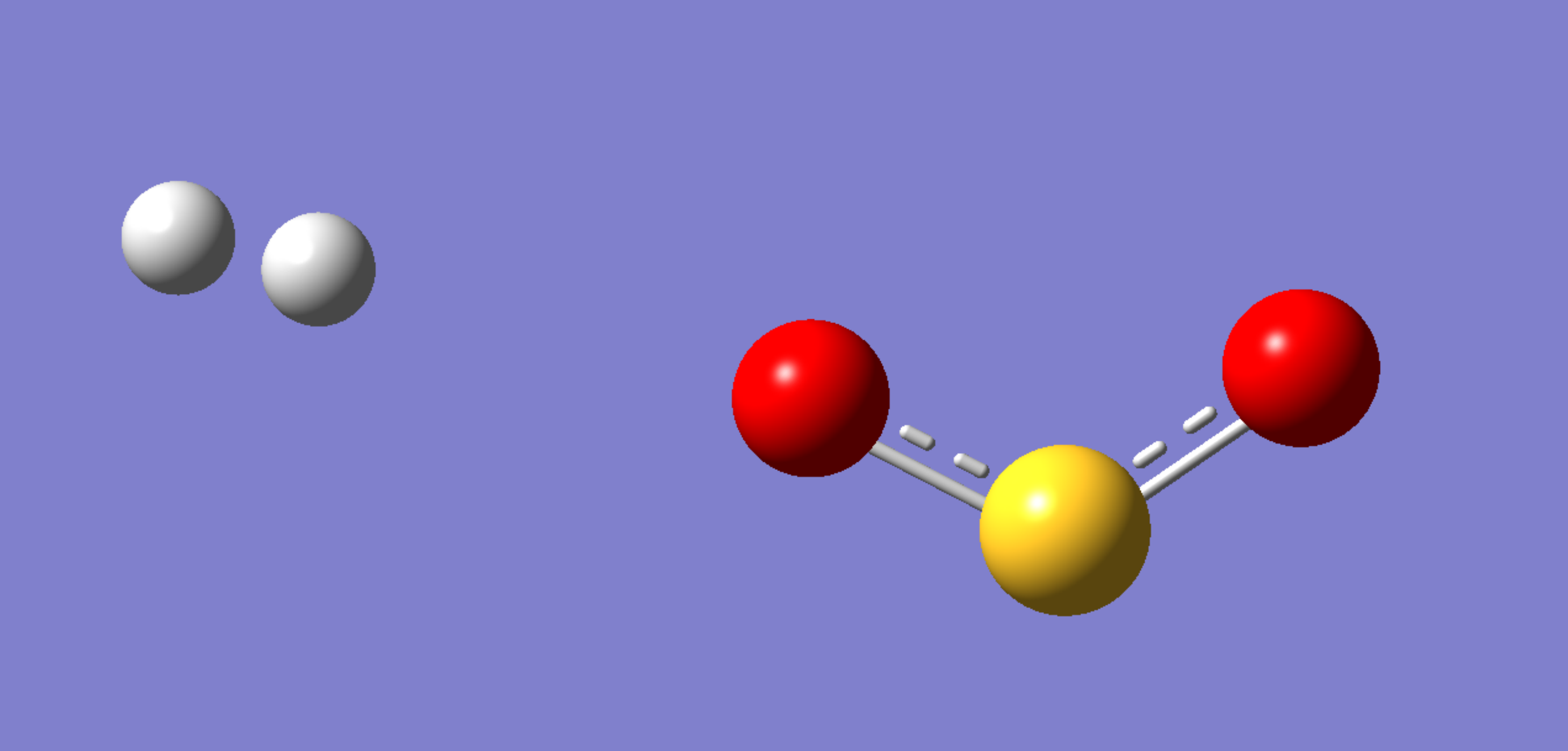} & Singlet &324& 2.691 \\
54&CH$_3$&  \includegraphics[width=0.2\textwidth]{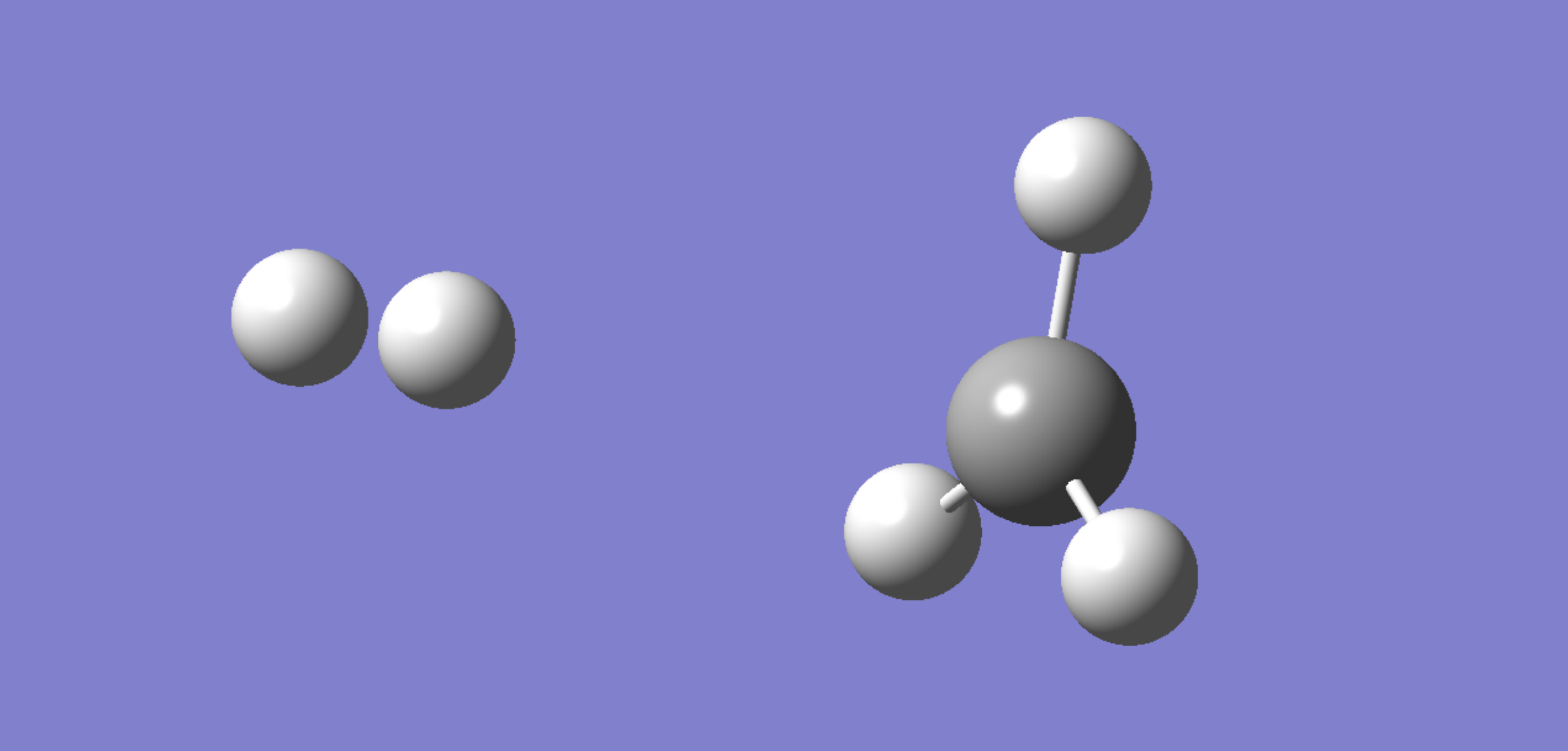} & Doublet &198& 1.644 \\
 55&NH$_3$&  \includegraphics[width=0.2\textwidth]{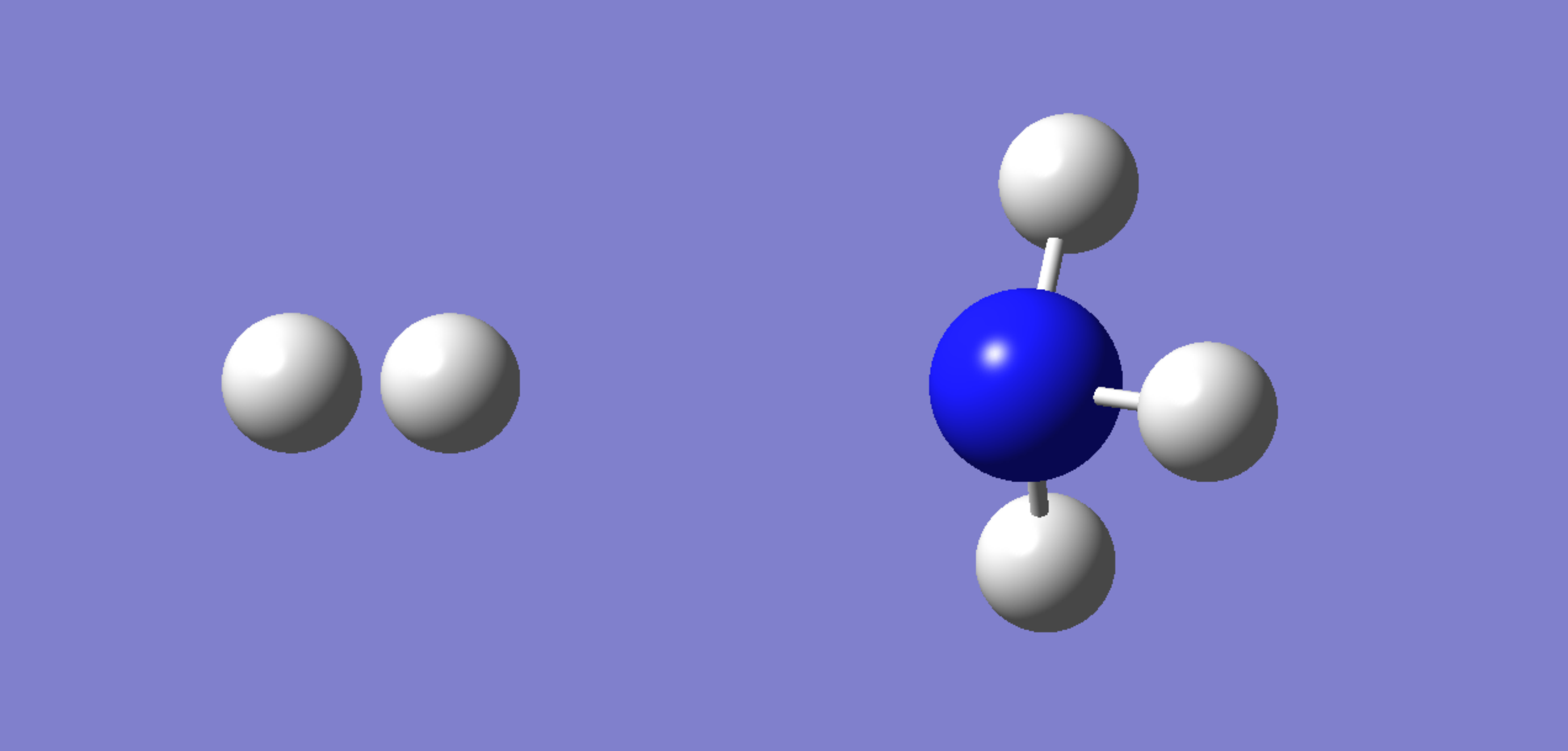} & Singlet &455 & 3.781 \\
 56 & SiH$_3$ &  \includegraphics[width=0.2\textwidth]{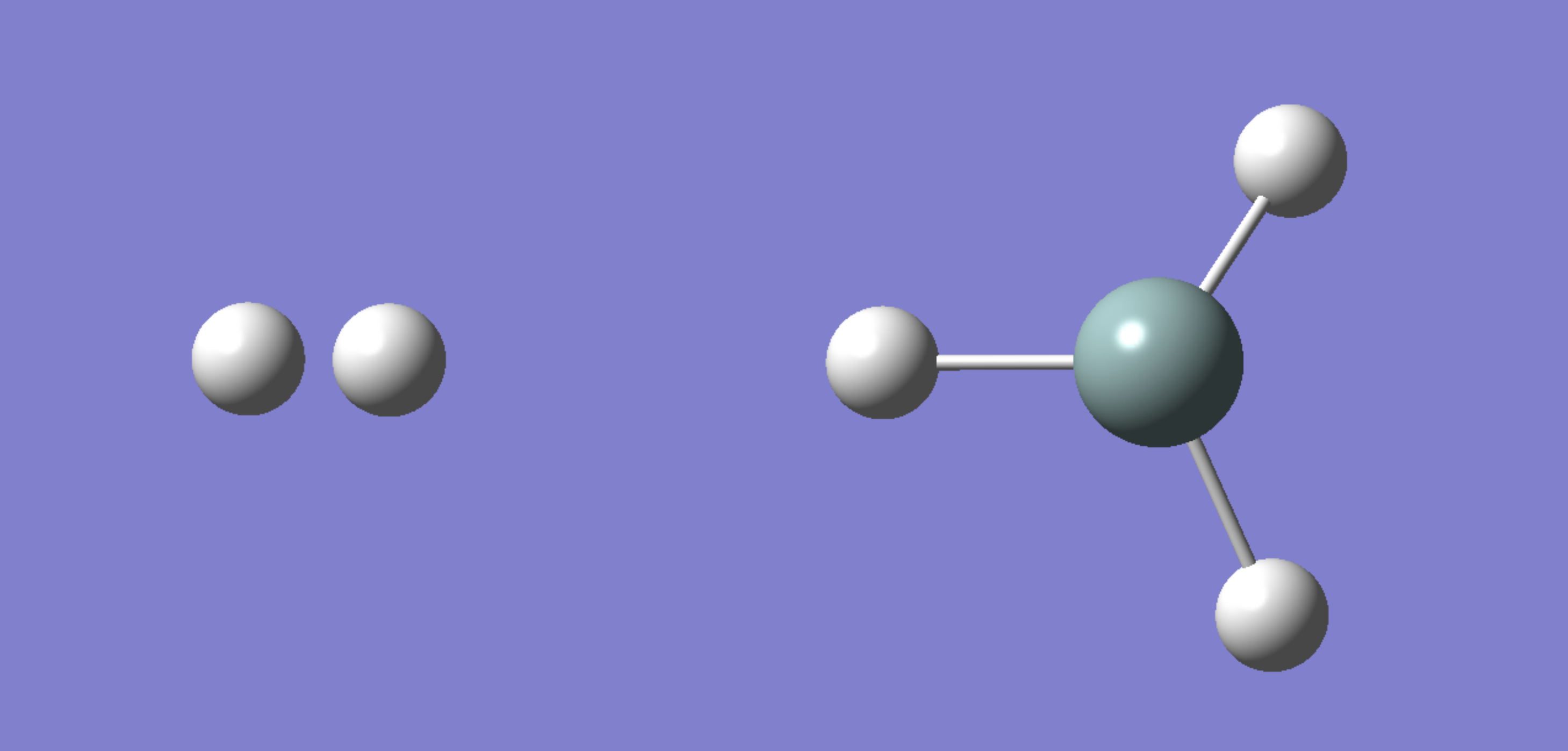} & Doublet & 159 & 1.321 \\
 57 & C$_2$H$_2$ &  \includegraphics[width=0.2\textwidth]{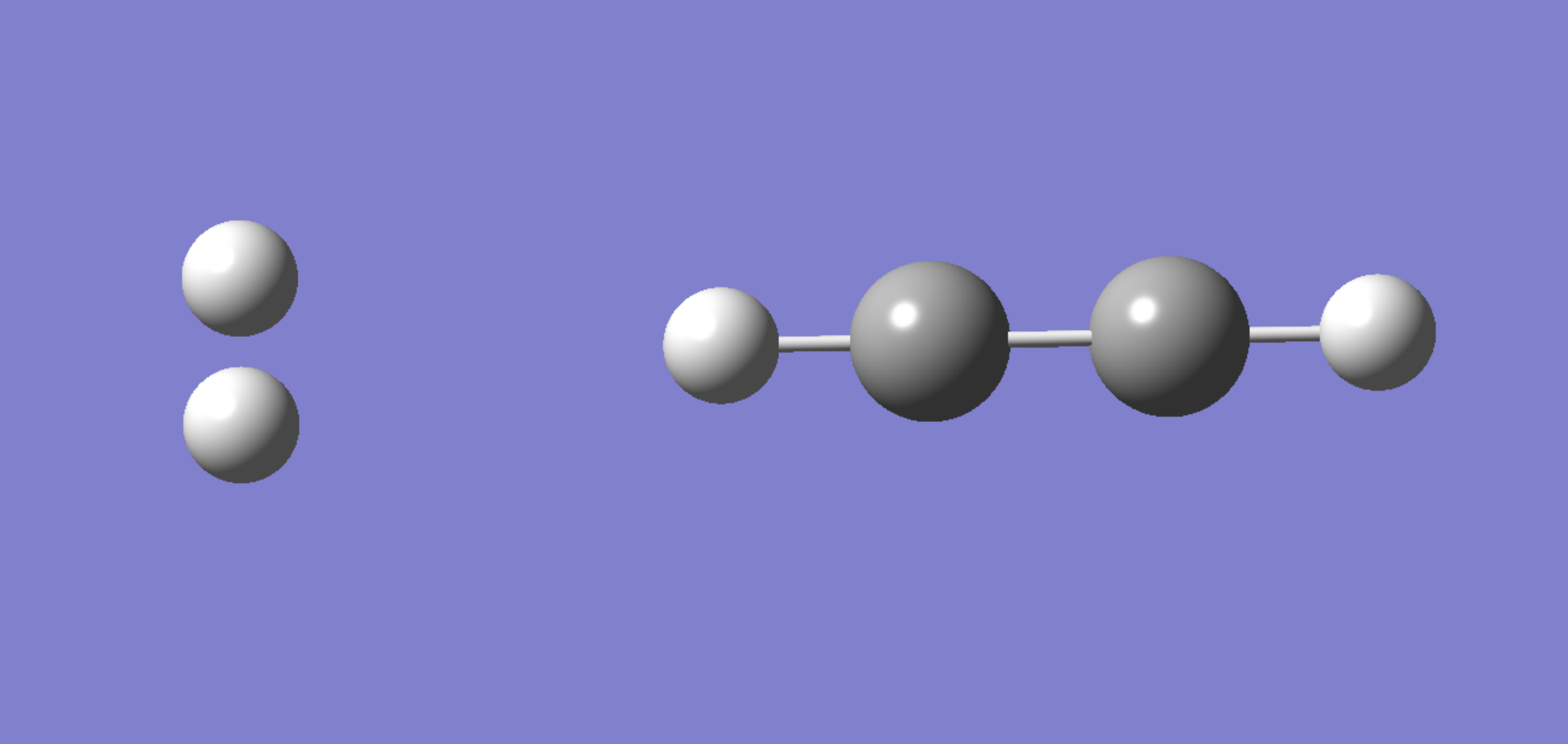} & Singlet & 337 & 2.799 \\
 58 & N$_2$H$_2$ &  \includegraphics[width=0.2\textwidth]{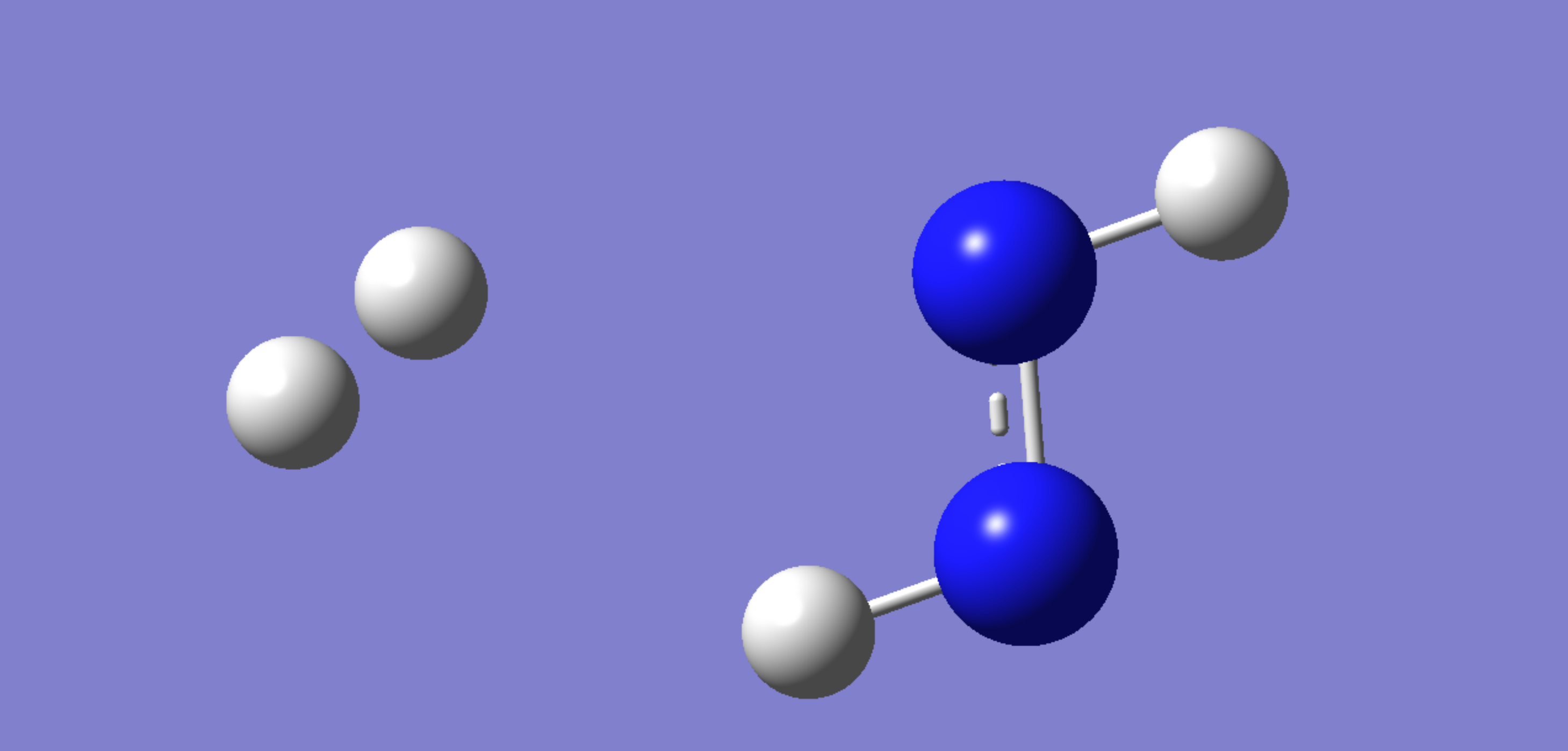} & Singlet & 608 & 5.059 \\
 59 & H$_2$O$_2$ &  \includegraphics[width=0.2\textwidth]{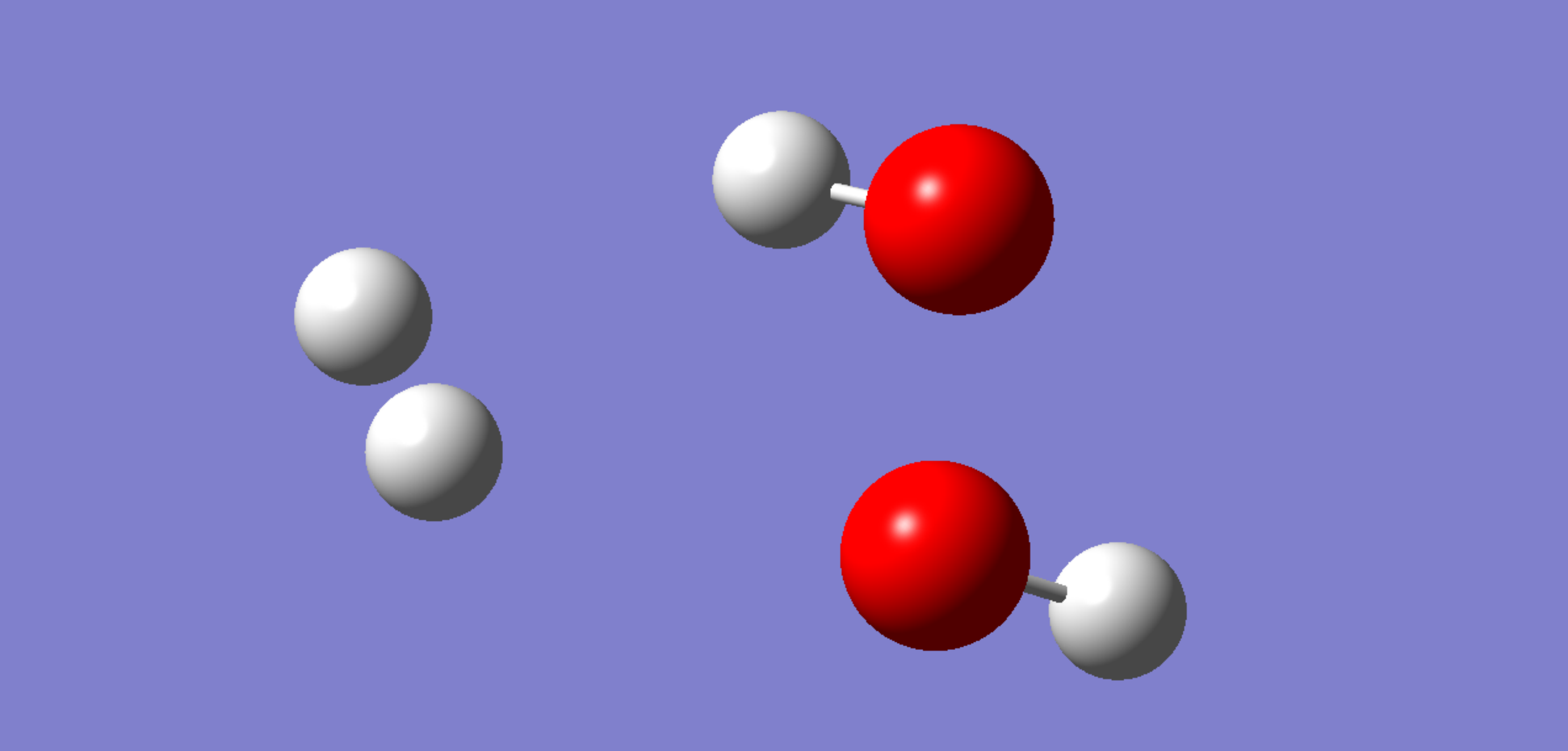} & Singlet & 628, 340$^c$ & 5.222 \\
60 & H$_2$S$_2$ &  \includegraphics[width=0.2\textwidth]{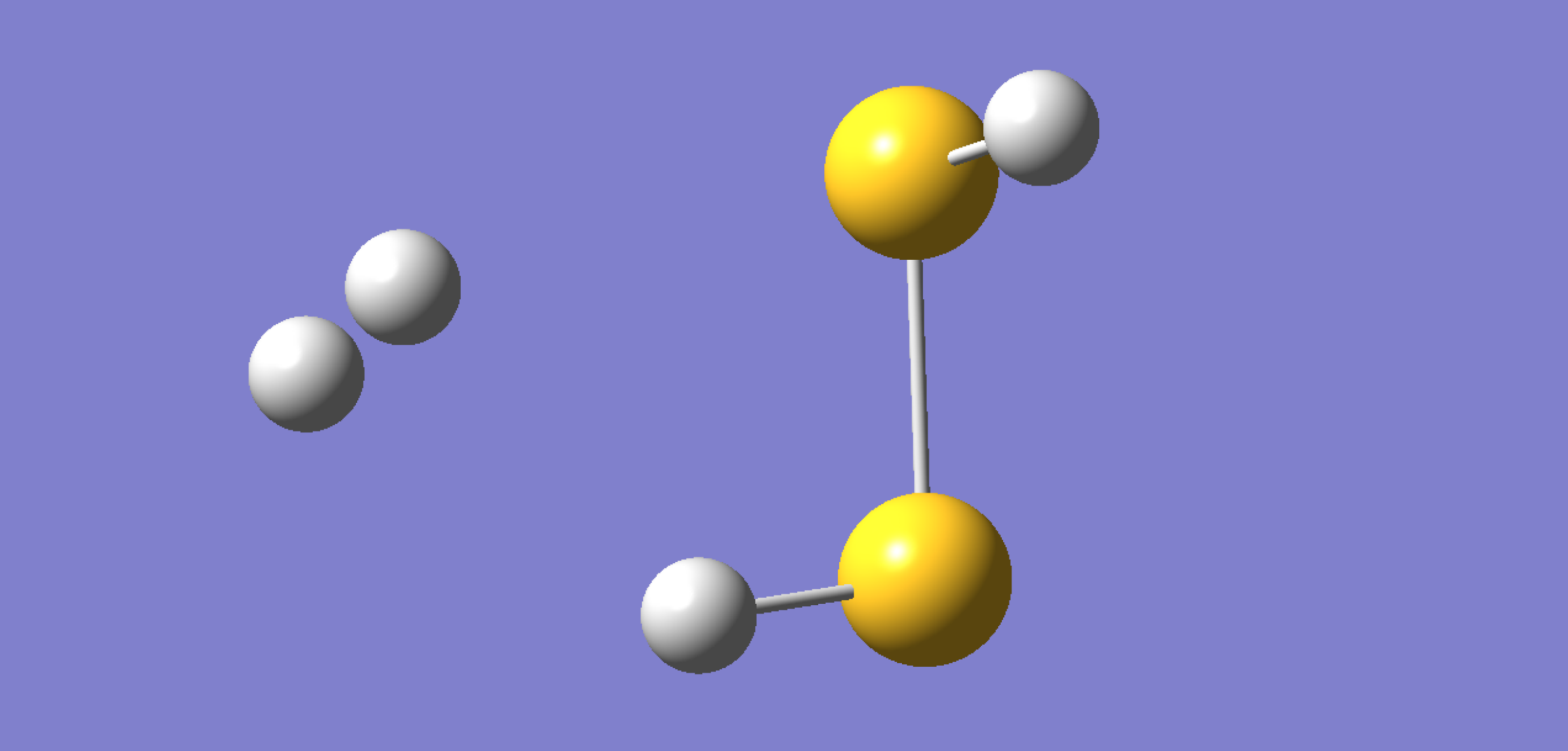} & Singlet & 573 & 4.763 \\
61 & H$_2$CN &  \includegraphics[width=0.2\textwidth]{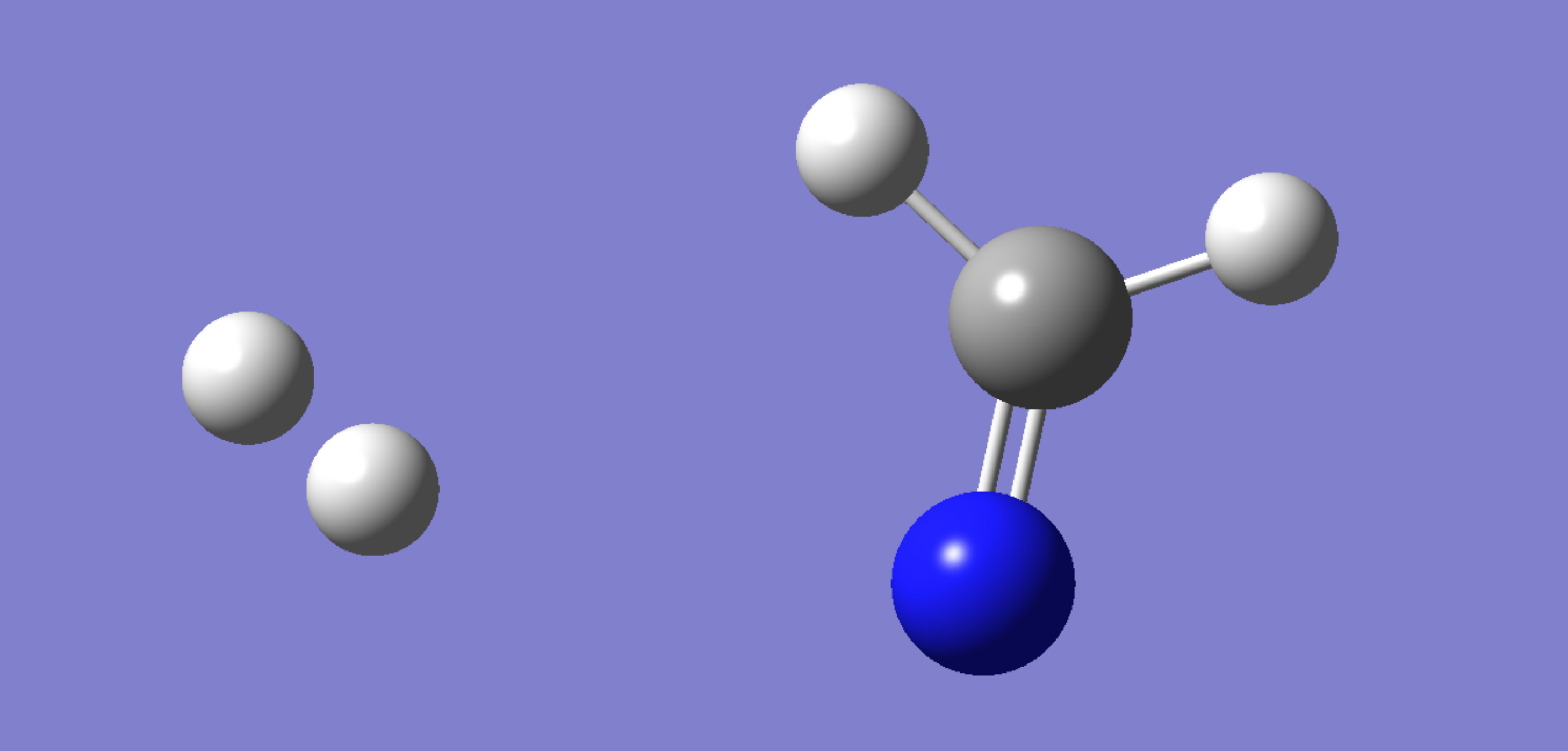} & Doublet & 376 & 3.130 \\
\hline
\end{tabular}
\end{table}

\begin{table}
\scriptsize
\centering
\begin{tabular}{|c|c|c|c|c|c|}
\hline
{\bf Sl.}& {\bf Species} & {\bf Optimized} & {\bf Ground} & \multicolumn{2}{c|}{\bf Binding Energy} \\
\cline{5-6}
 {\bf No.} & & {\bf Structures} & {\bf State} & {\bf in K} & {\bf in kJ/mol} \\
\hline
62 & H$_2$CO &  \includegraphics[width=0.2\textwidth]{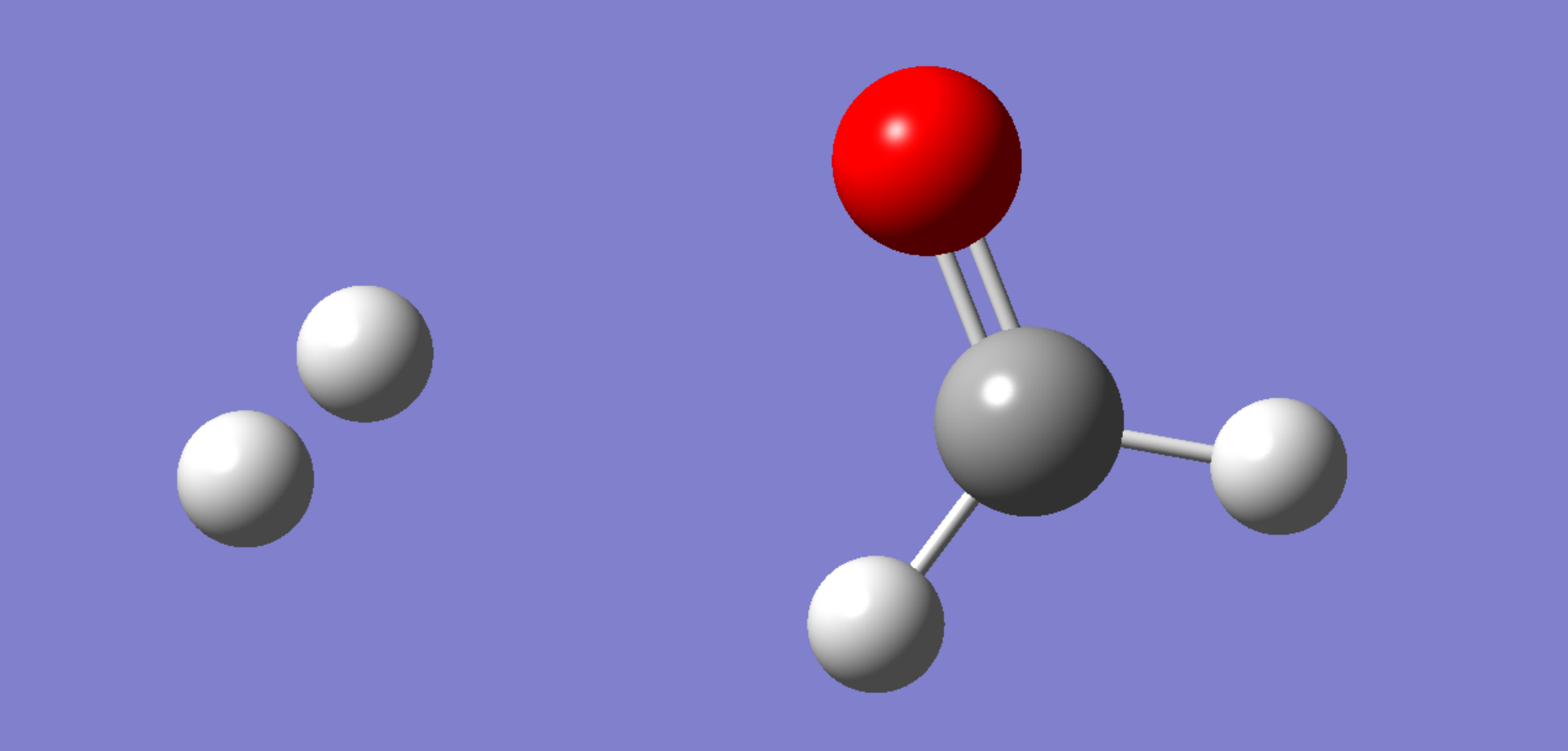} & Singlet& 507 & 4.219 \\
 63 & HC$_2$N &  \includegraphics[width=0.2\textwidth]{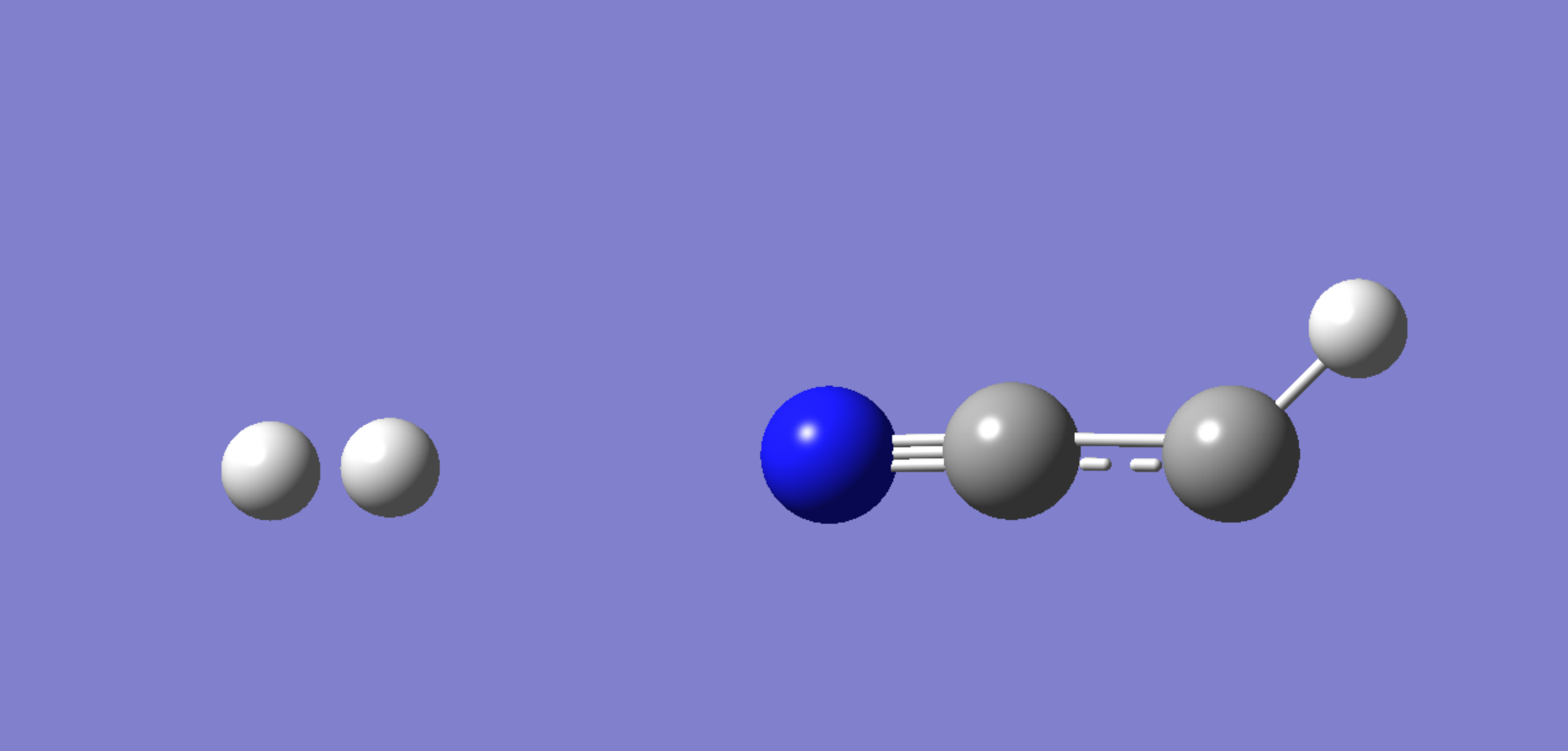} & Triplet &413 & 3.434 \\
 64&HC$_2$O &  \includegraphics[width=0.2\textwidth]{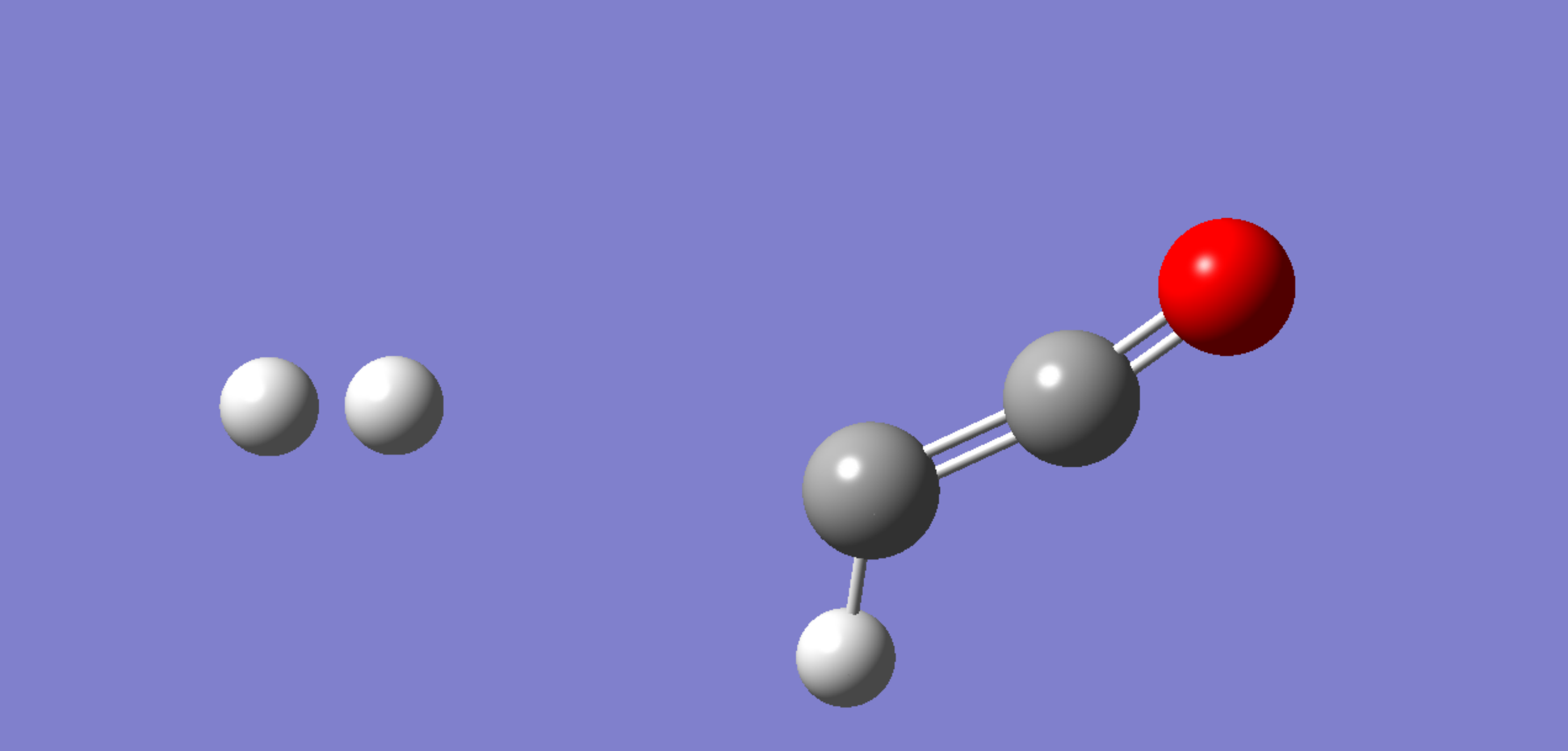} & Doublet&326 & 2.712 \\
 65 & HNCO &  \includegraphics[width=0.2\textwidth]{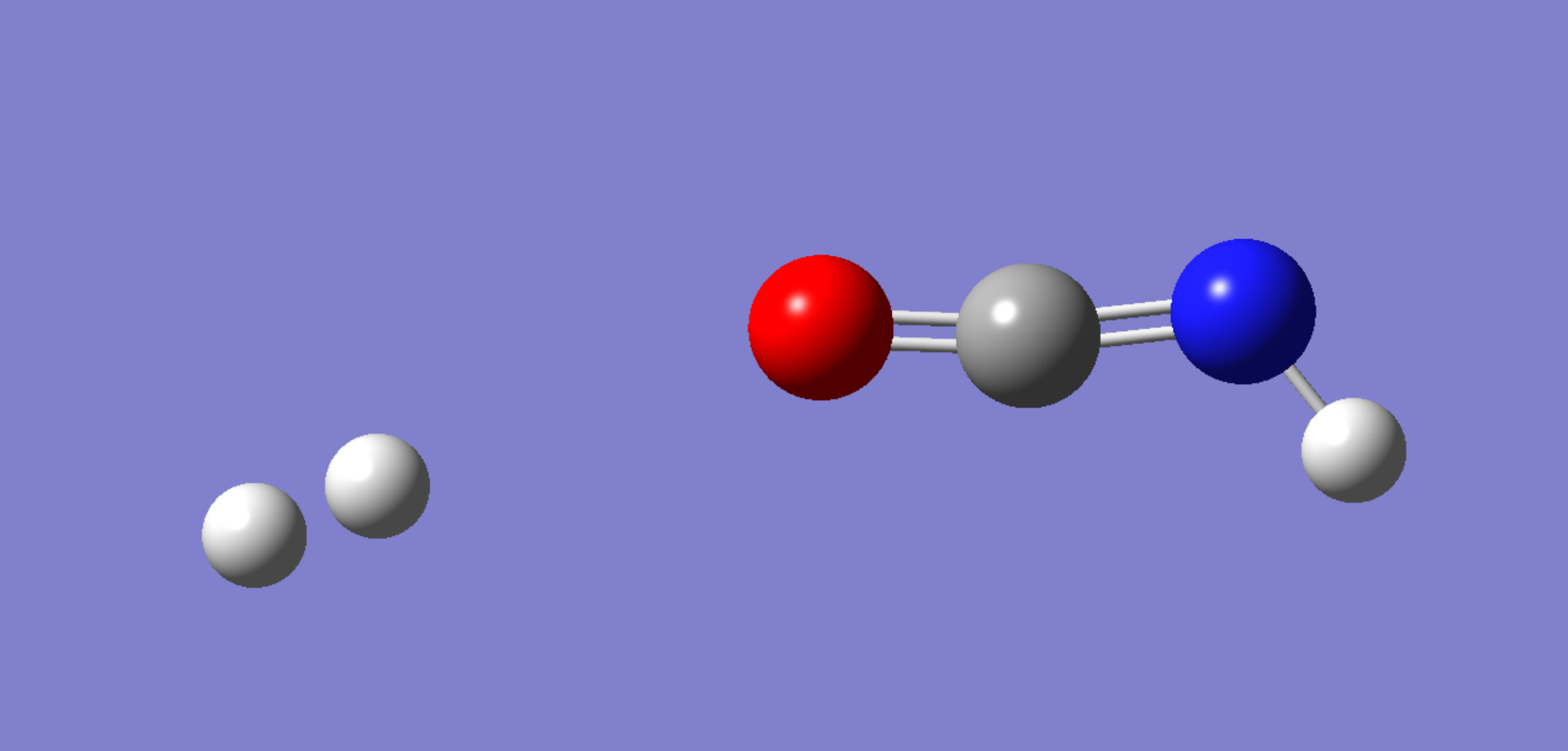} & Singlet& 289 & 2.405 \\
66 & H$_2$CS &  \includegraphics[width=0.2\textwidth]{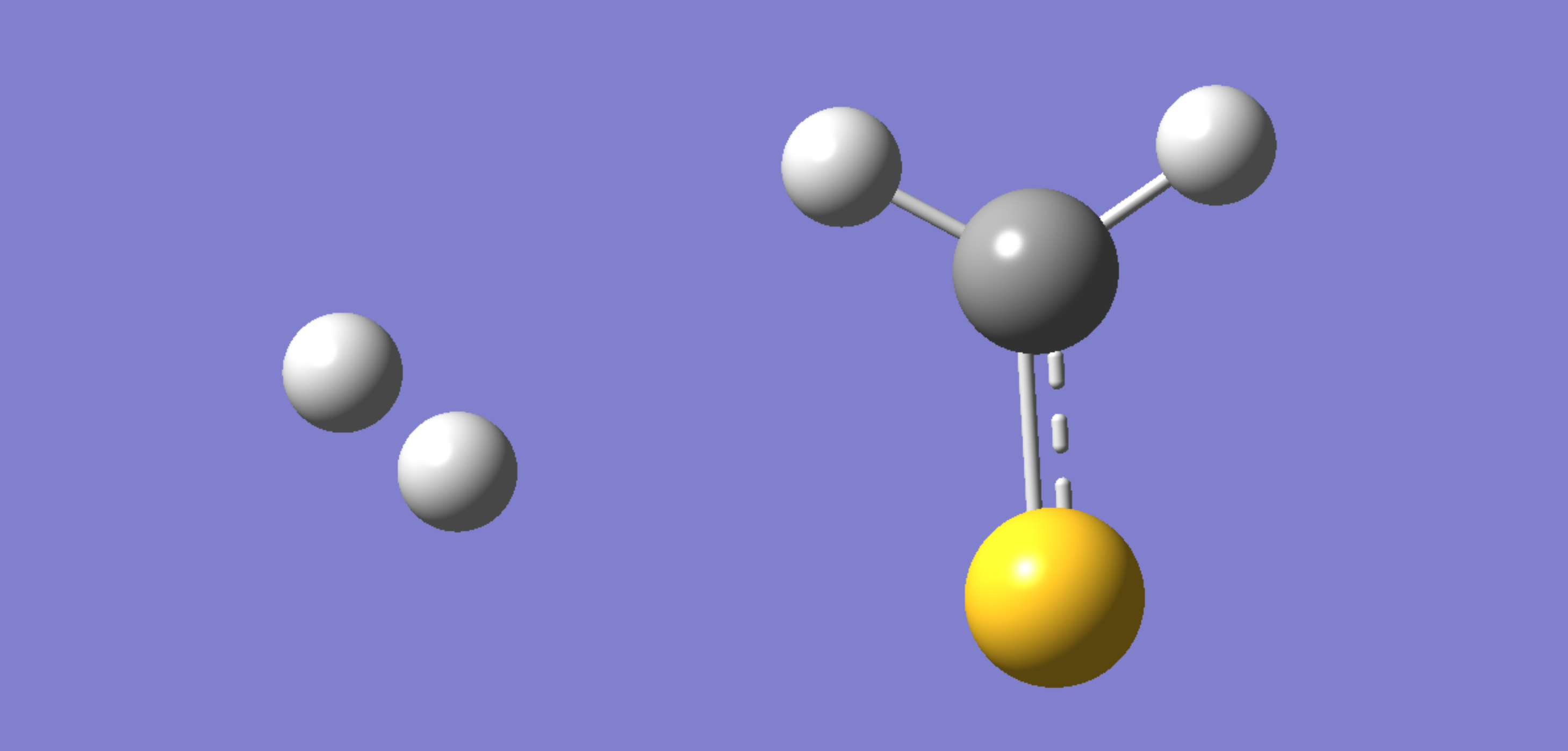} & Singlet&545 & 4.532 \\
 67&C$_3$O & \includegraphics[width=0.2\textwidth]{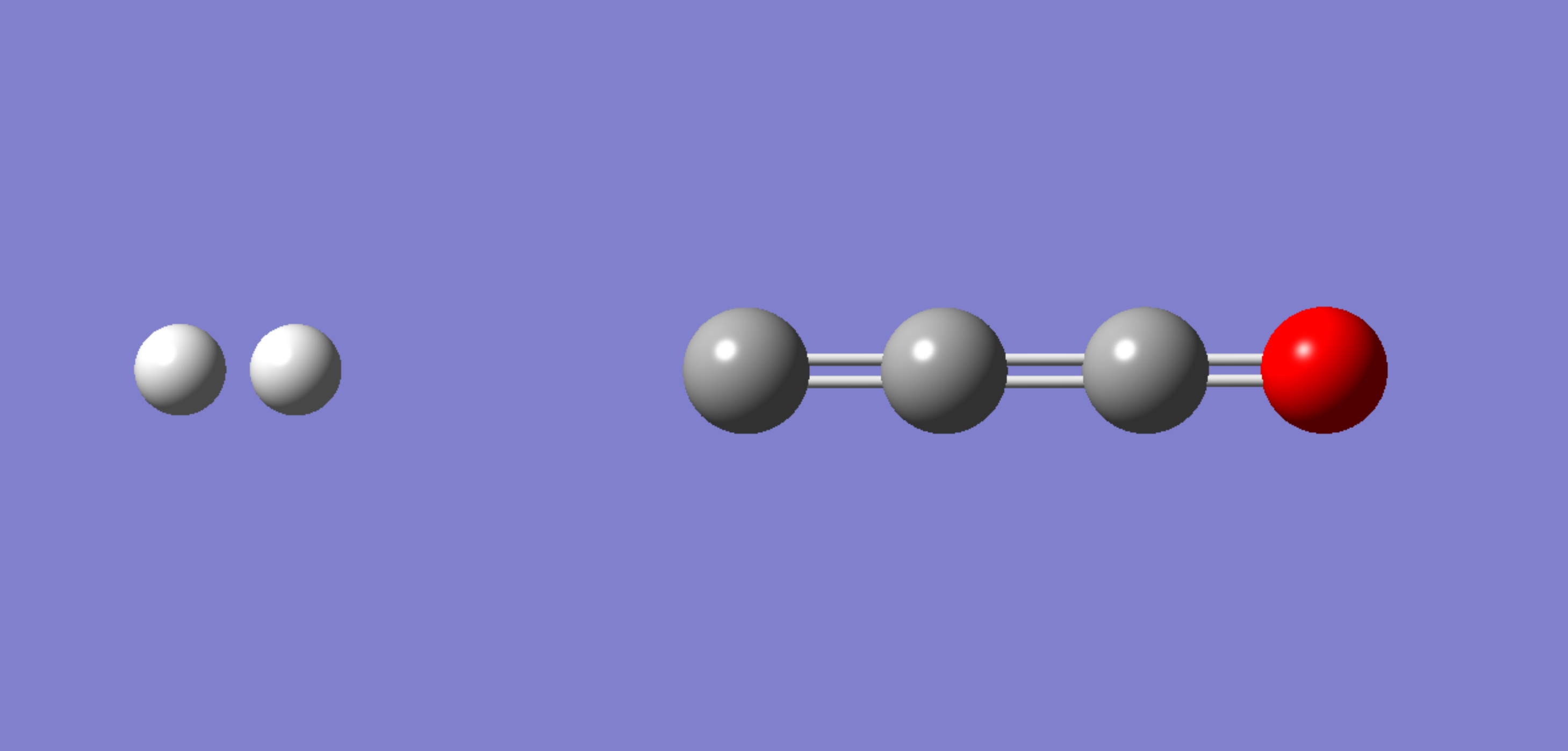} & Singlet & 414 & 3.442 \\
 68&CH$_4$ & \includegraphics[width=0.2\textwidth]{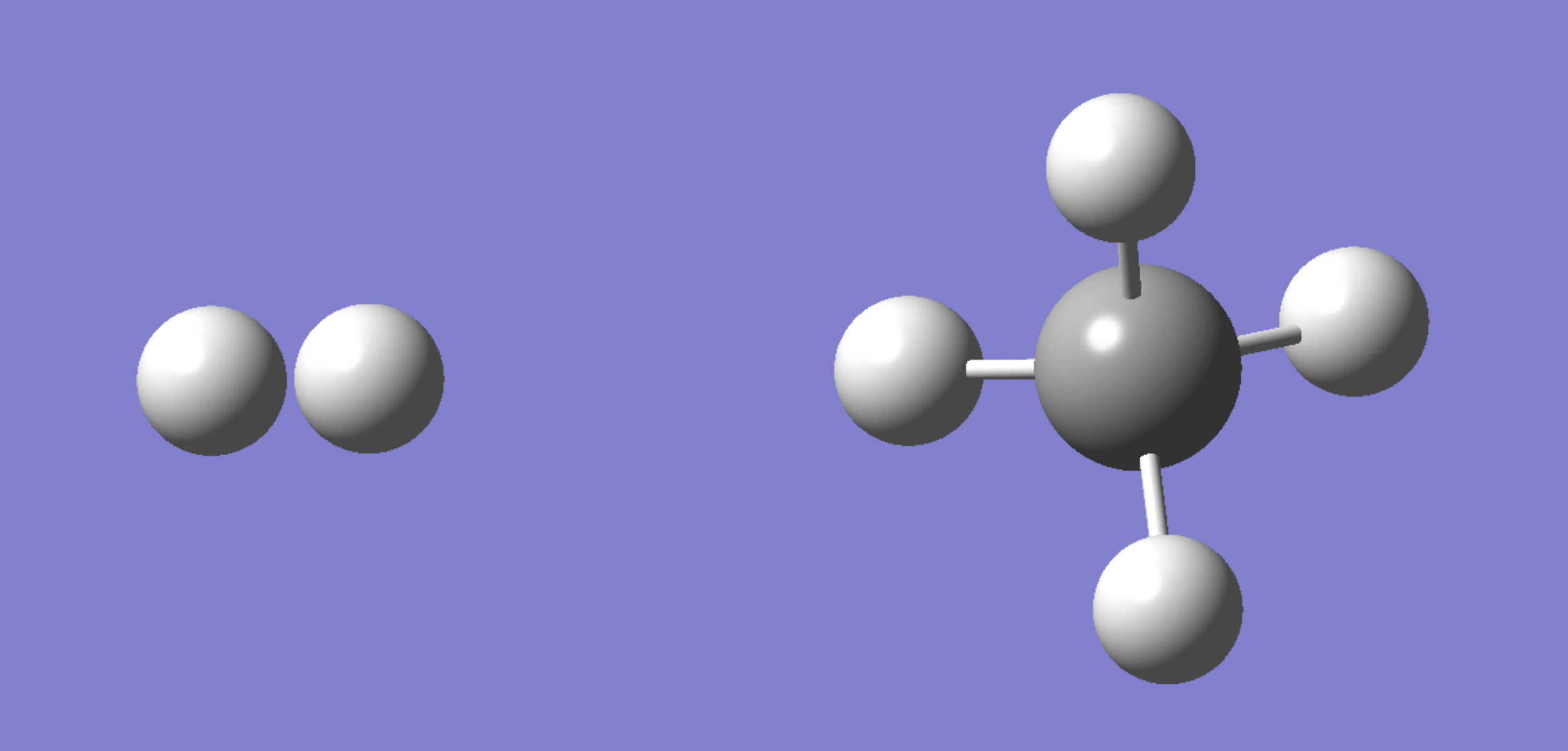} & Singlet&138 & 1.150 \\
69&SiH$_4$ & \includegraphics[width=0.2\textwidth]{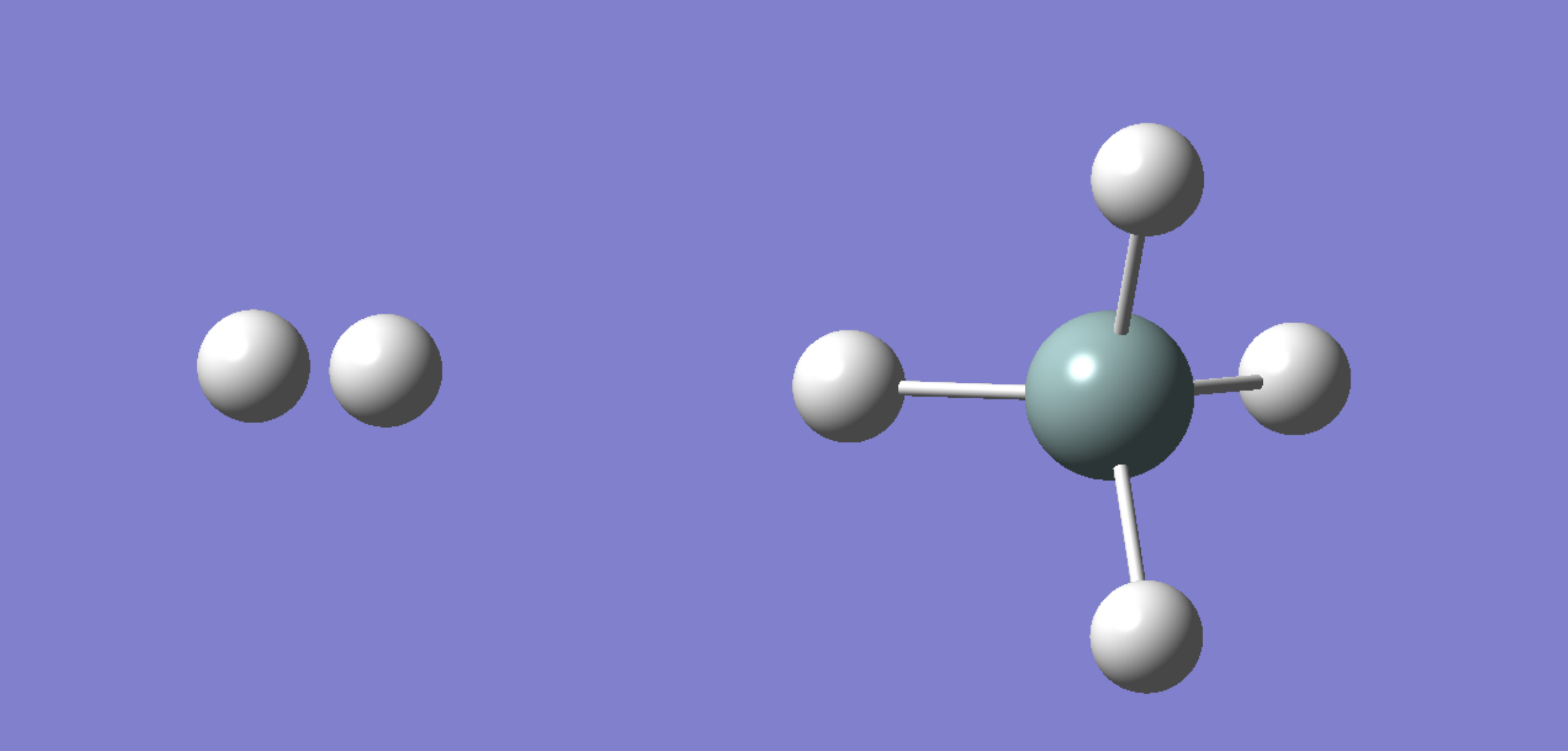} & Singlet&165 & 1.370 \\
 70&C$_2$H$_3$ & \includegraphics[width=0.2\textwidth]{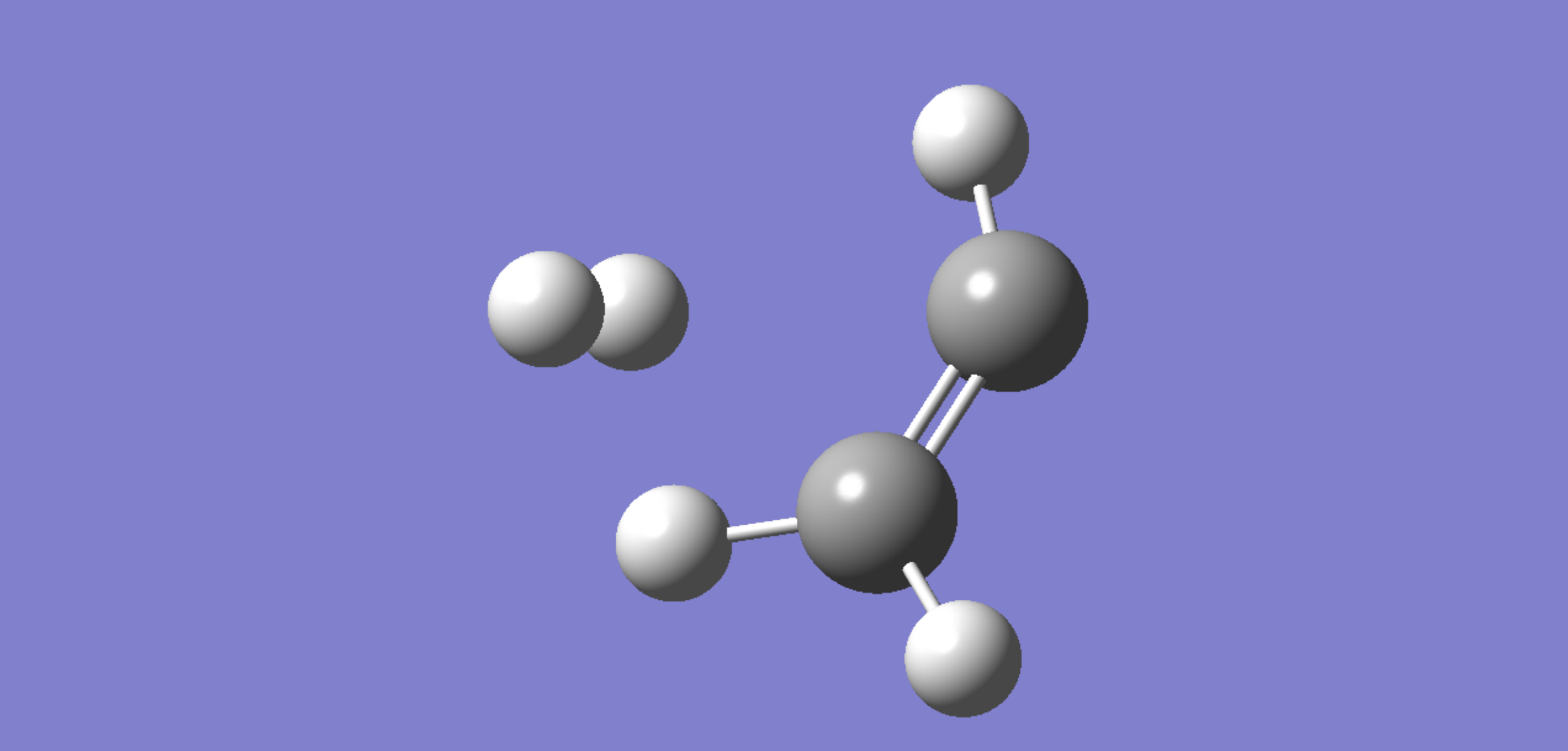} & Doublet &265 & 2.200 \\
 71 & CHNH$_2$ & \includegraphics[width=0.2\textwidth]{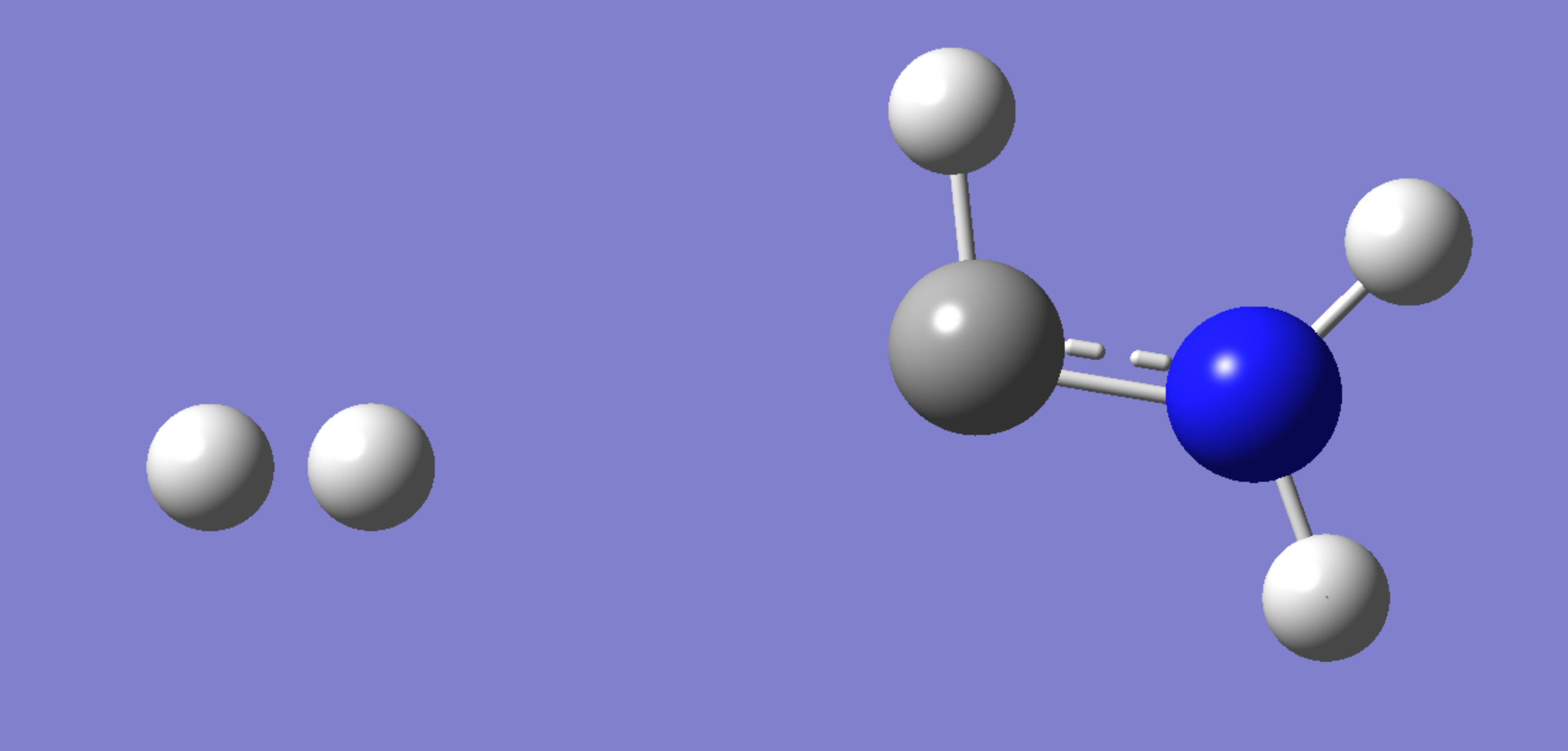} & Singlet & 858 & 7.133 \\
  && \includegraphics[width=0.2\textwidth]{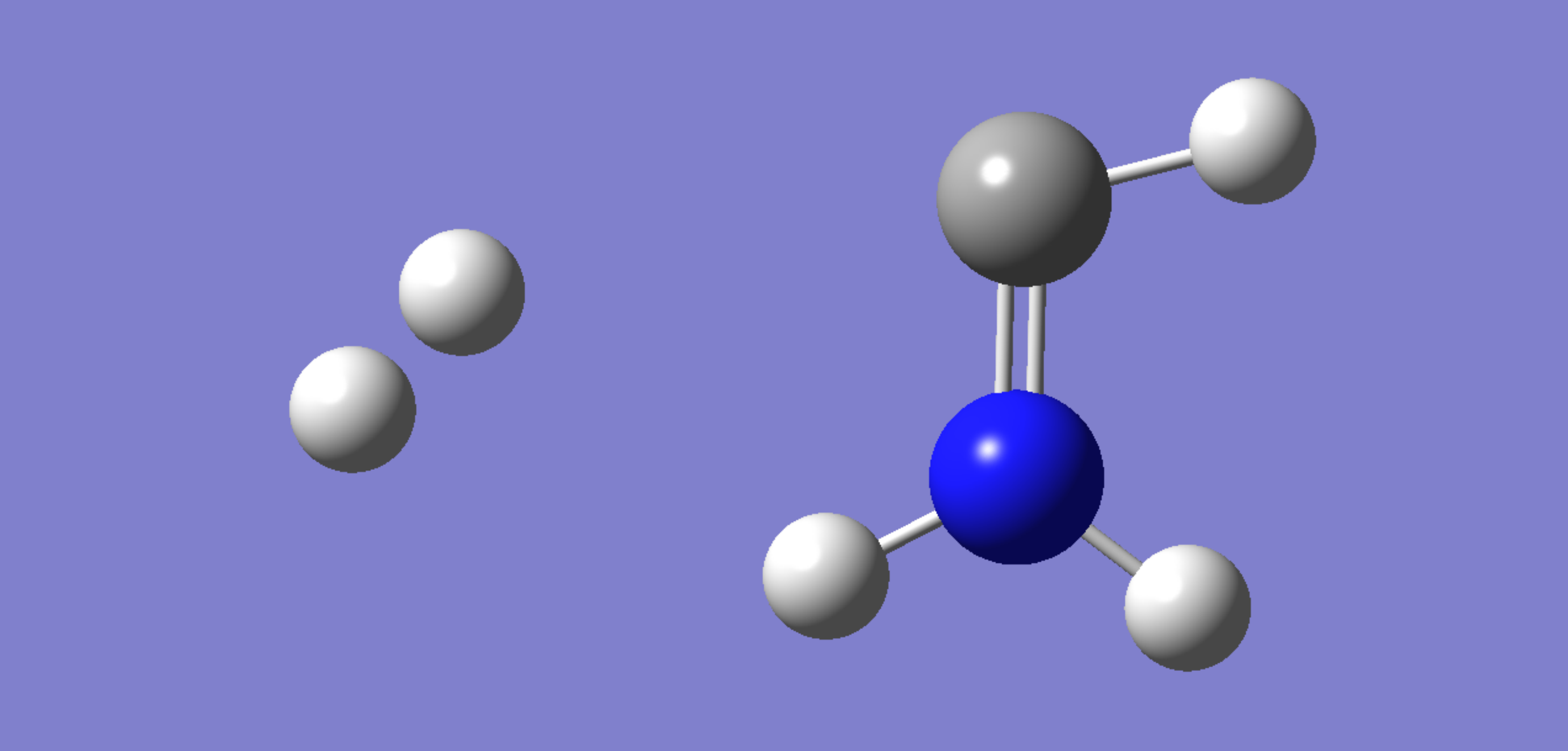} & & 463$^b$ & 3.846$^b$ \\
72 & CH$_2$NH & \includegraphics[width=0.2\textwidth]{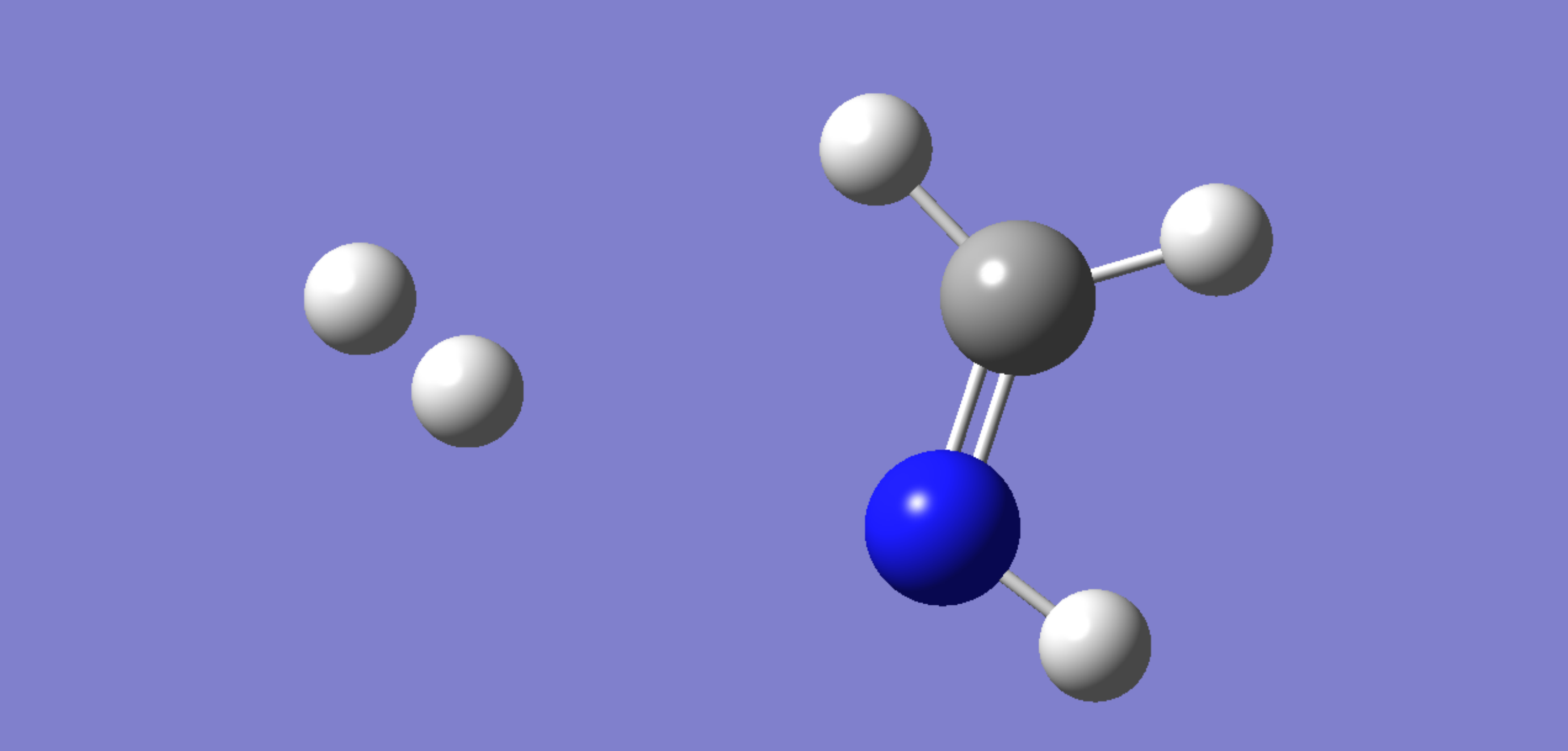} & Singlet& 602 & 5.007 \\
 73 &c-C$_3$H$_2$ & \includegraphics[width=0.2\textwidth]{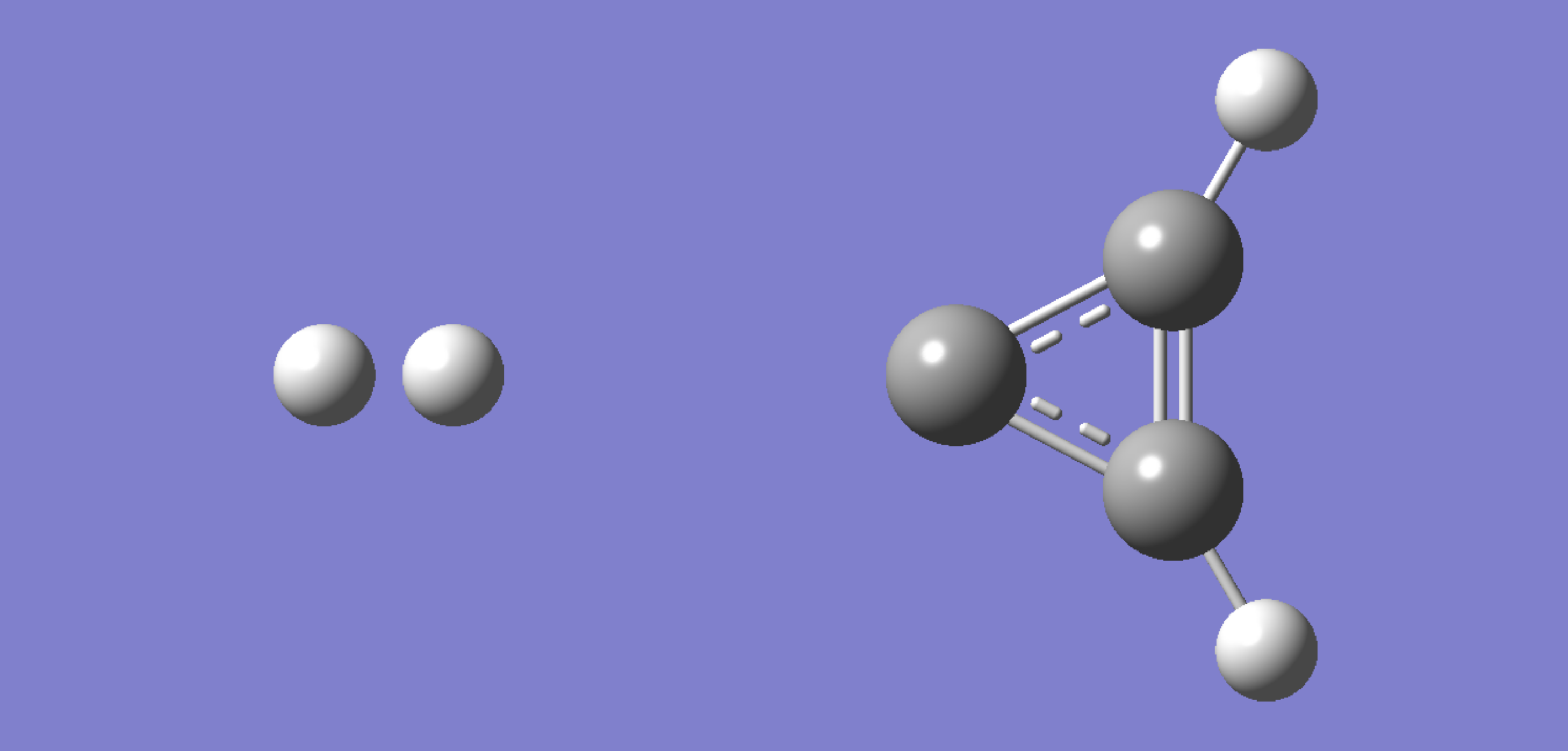} & Singlet & 472 & 3.925 \\
 \hline
\end{tabular}
\end{table}

\begin{table}
\scriptsize
\centering
\begin{tabular}{|c|c|c|c|c|c|}
\hline
{\bf Sl.}& {\bf Species} & {\bf Optimized} & {\bf Ground} & \multicolumn{2}{c|}{\bf Binding Energy} \\
\cline{5-6}
 {\bf No.} & & {\bf Structures} & {\bf State} & {\bf in K} & {\bf in kJ/mol} \\
\hline
 74&CH$_2$CN & \includegraphics[width=0.2\textwidth]{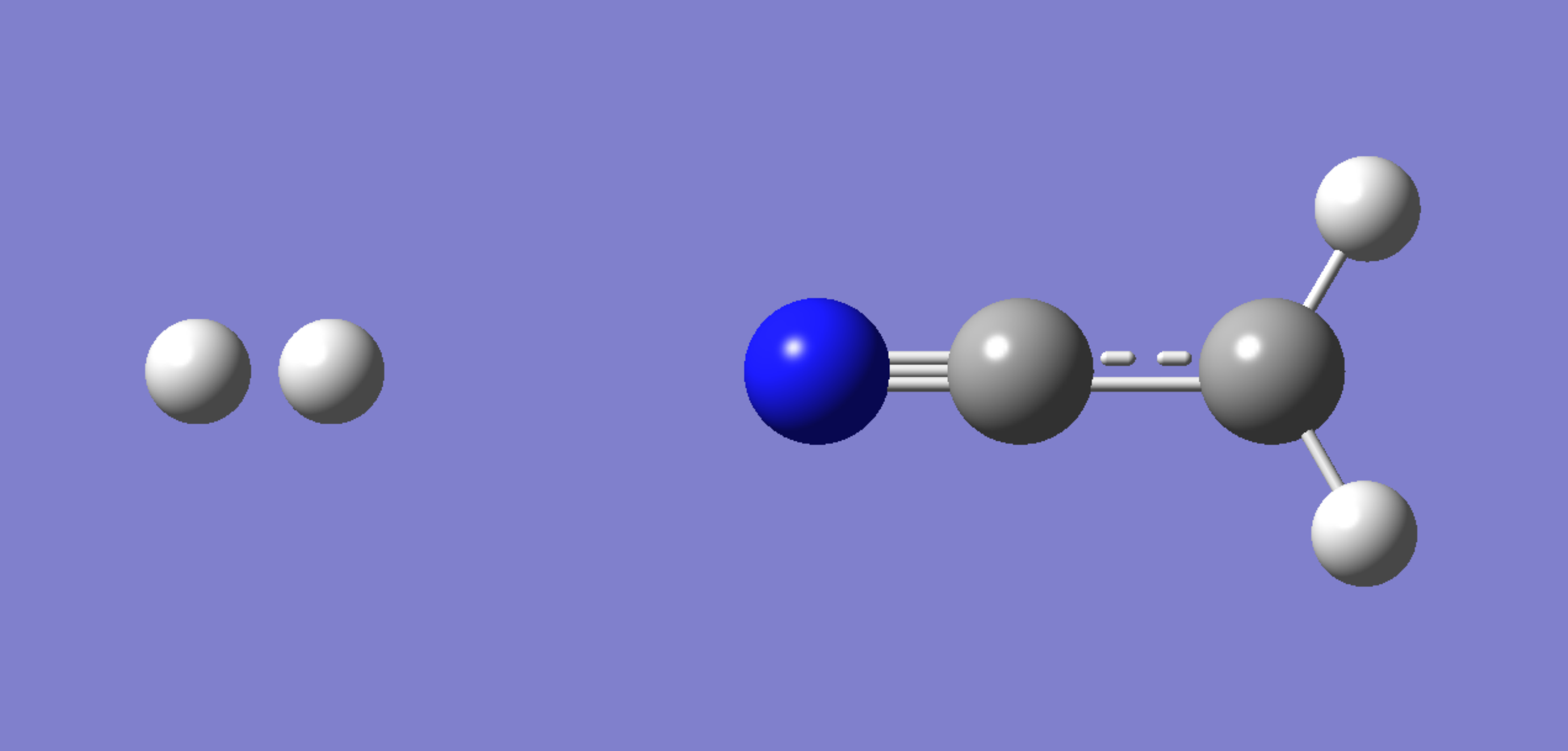} & Doublet & 440 & 3.662 \\
75&CH$_2$CO & \includegraphics[width=0.2\textwidth]{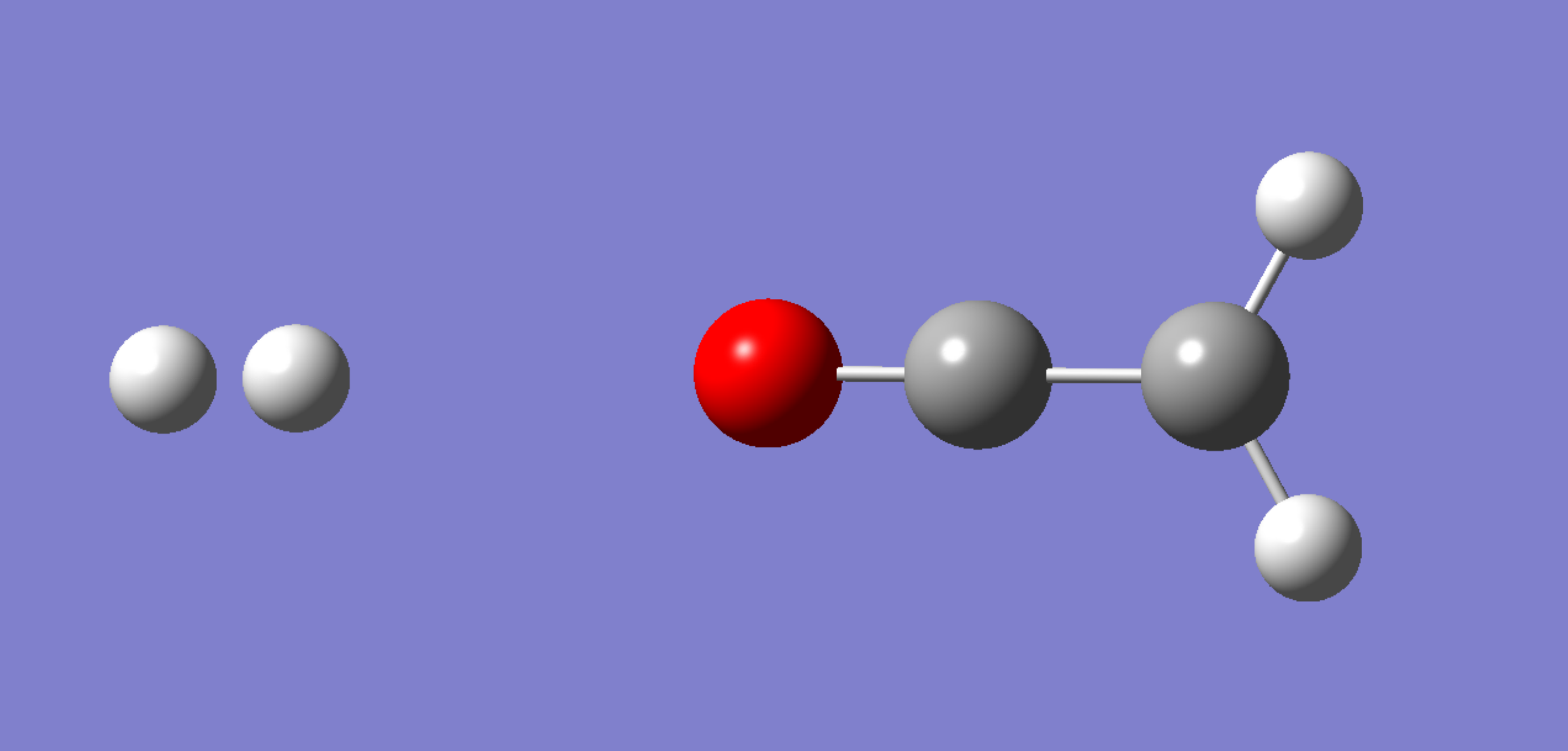} & Singlet&276 & 2.297 \\
76&HCOOH & \includegraphics[width=0.2\textwidth]{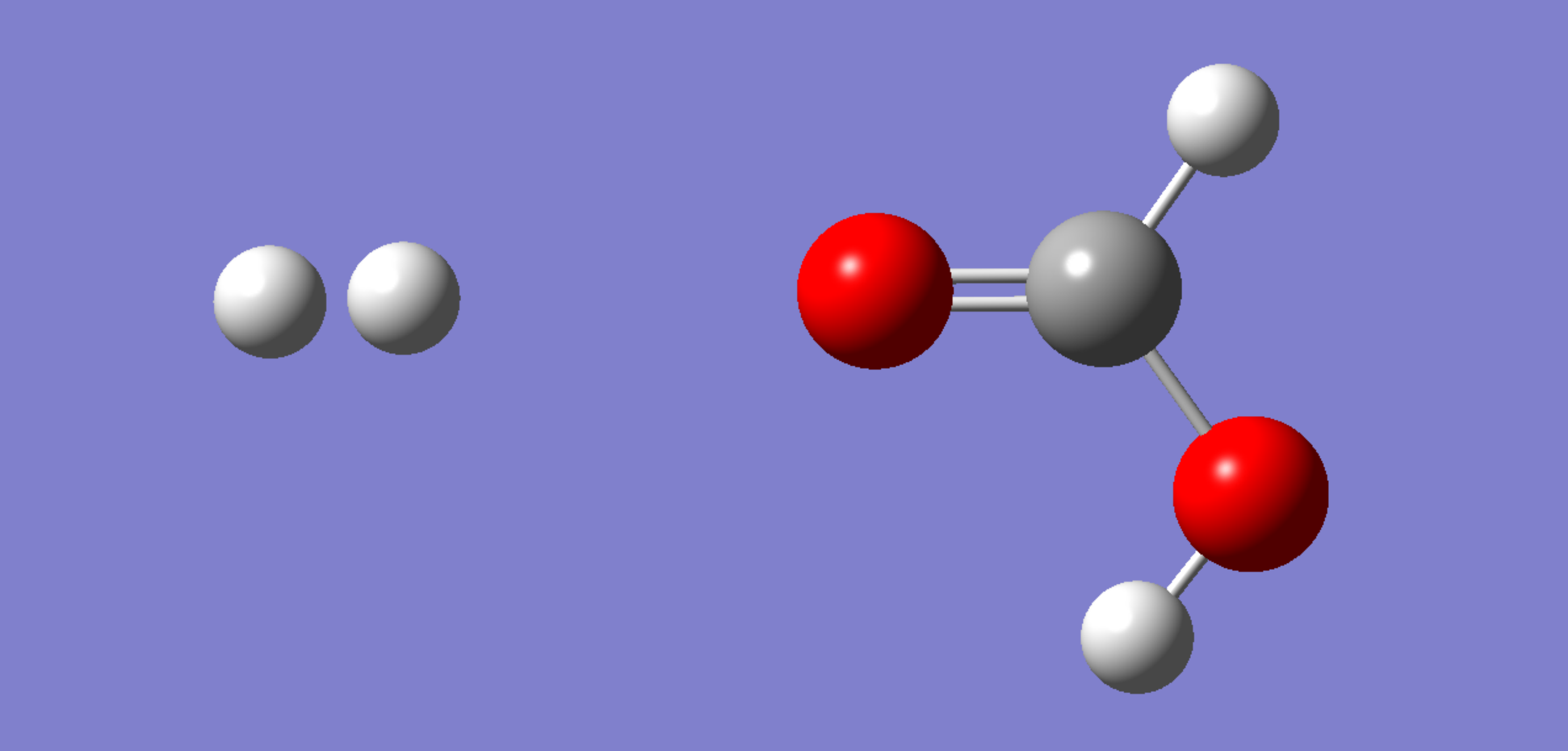} & Singlet&369 & 3.066 \\
 77 & CH$_2$OH & \includegraphics[width=0.2\textwidth]{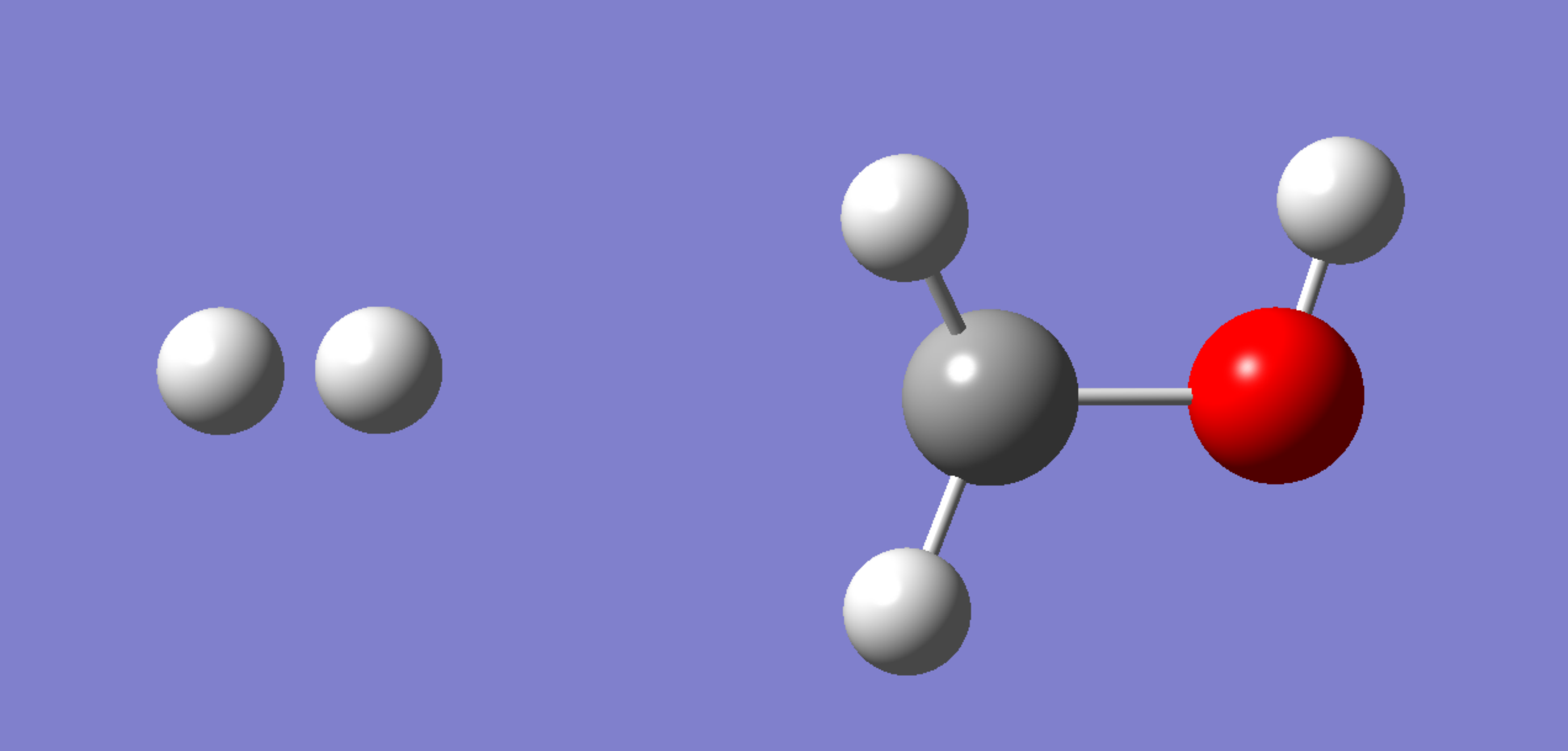} & Doublet & 272 & 2.263 \\
78 & NH$_2$OH & \includegraphics[width=0.2\textwidth]{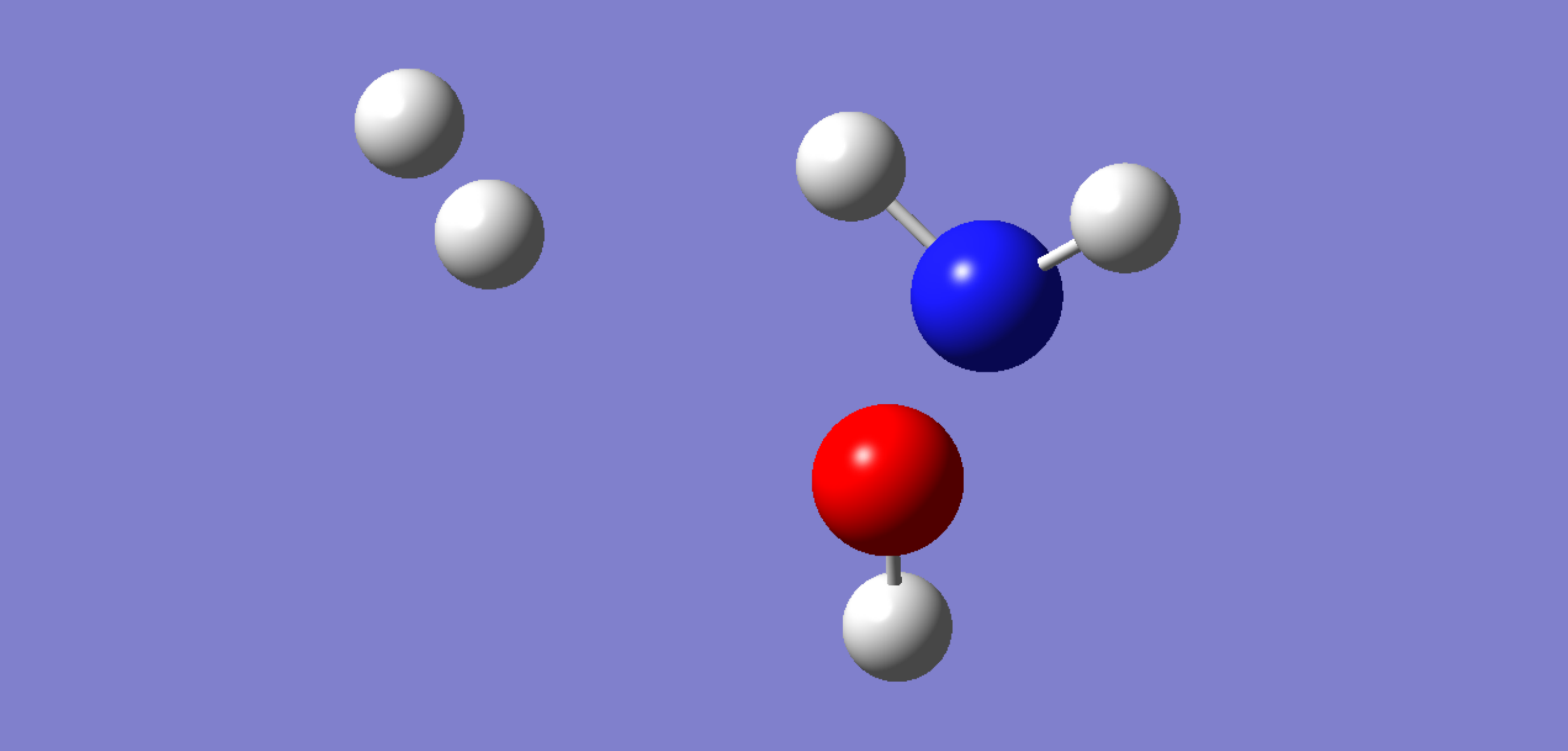} & Singlet & 2770 & 23.028 \\
 79 & HC$_3$N & \includegraphics[width=0.2\textwidth]{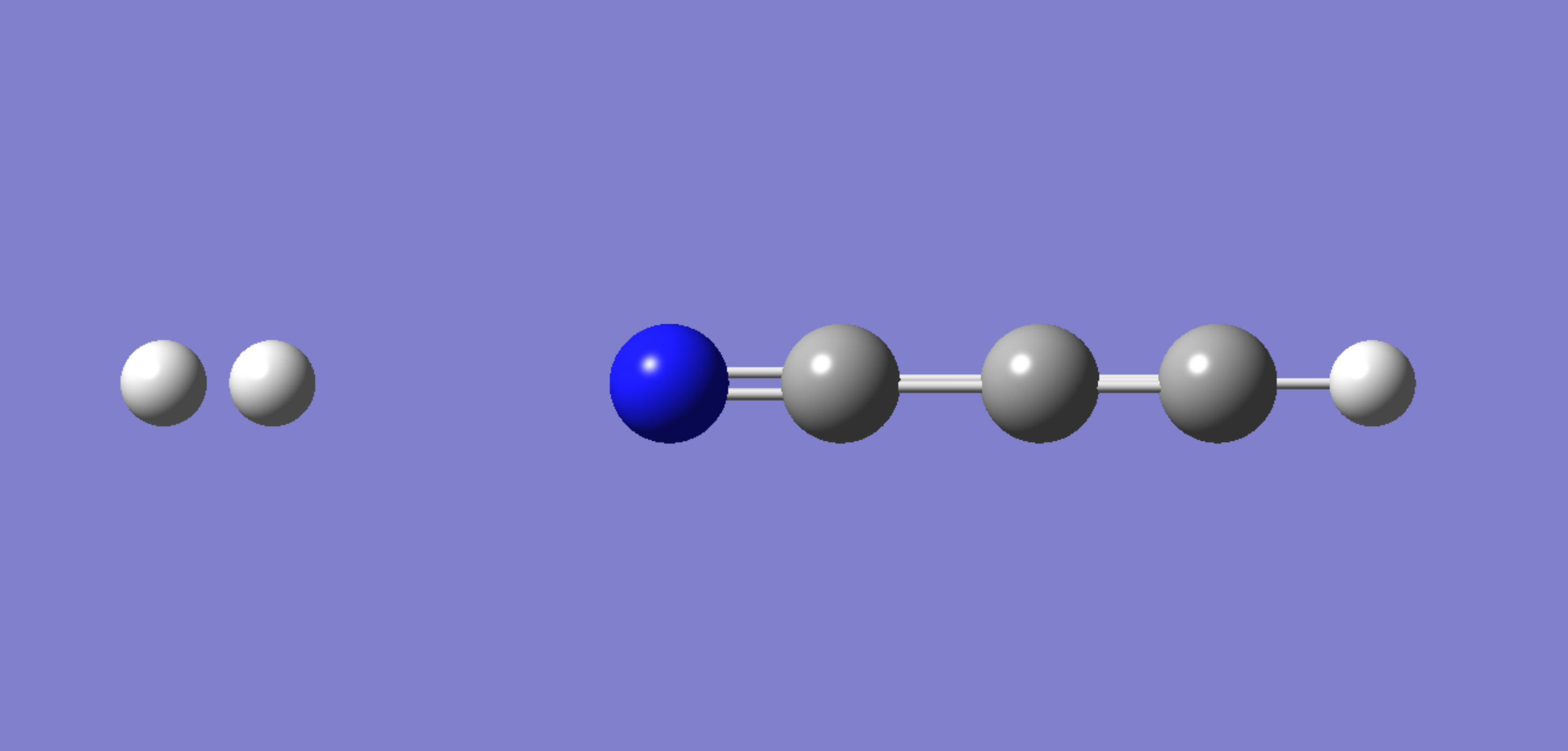} & Singlet & 427 & 3.555 \\
 80&C$_5$ & \includegraphics[width=0.2\textwidth]{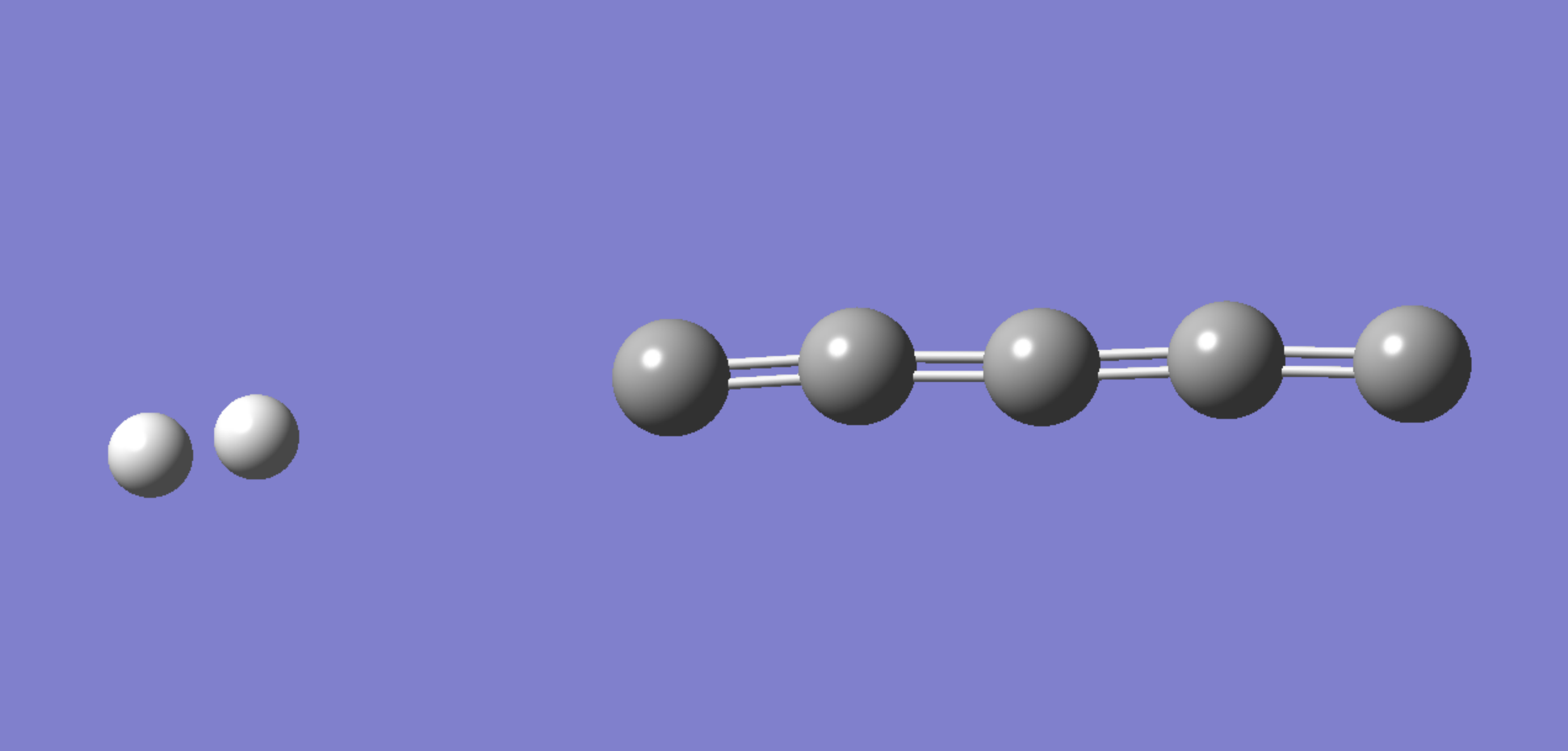} & Singlet &379 & 3.156 \\
 81&C$_2$H$_4$ & \includegraphics[width=0.2\textwidth]{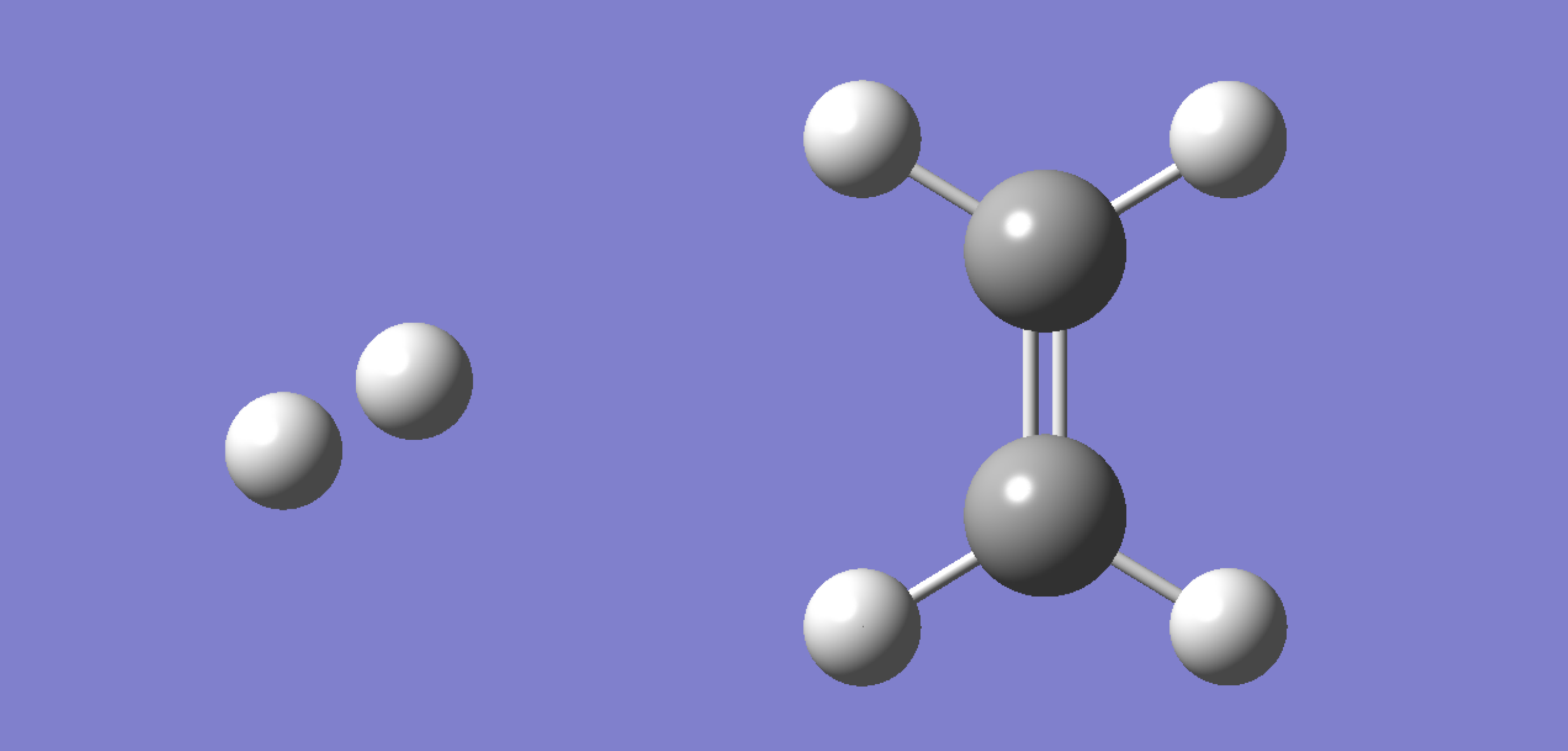} & Singlet&250 & 2.079 \\
82&CH$_2$NH$_2$ & \includegraphics[width=0.2\textwidth]{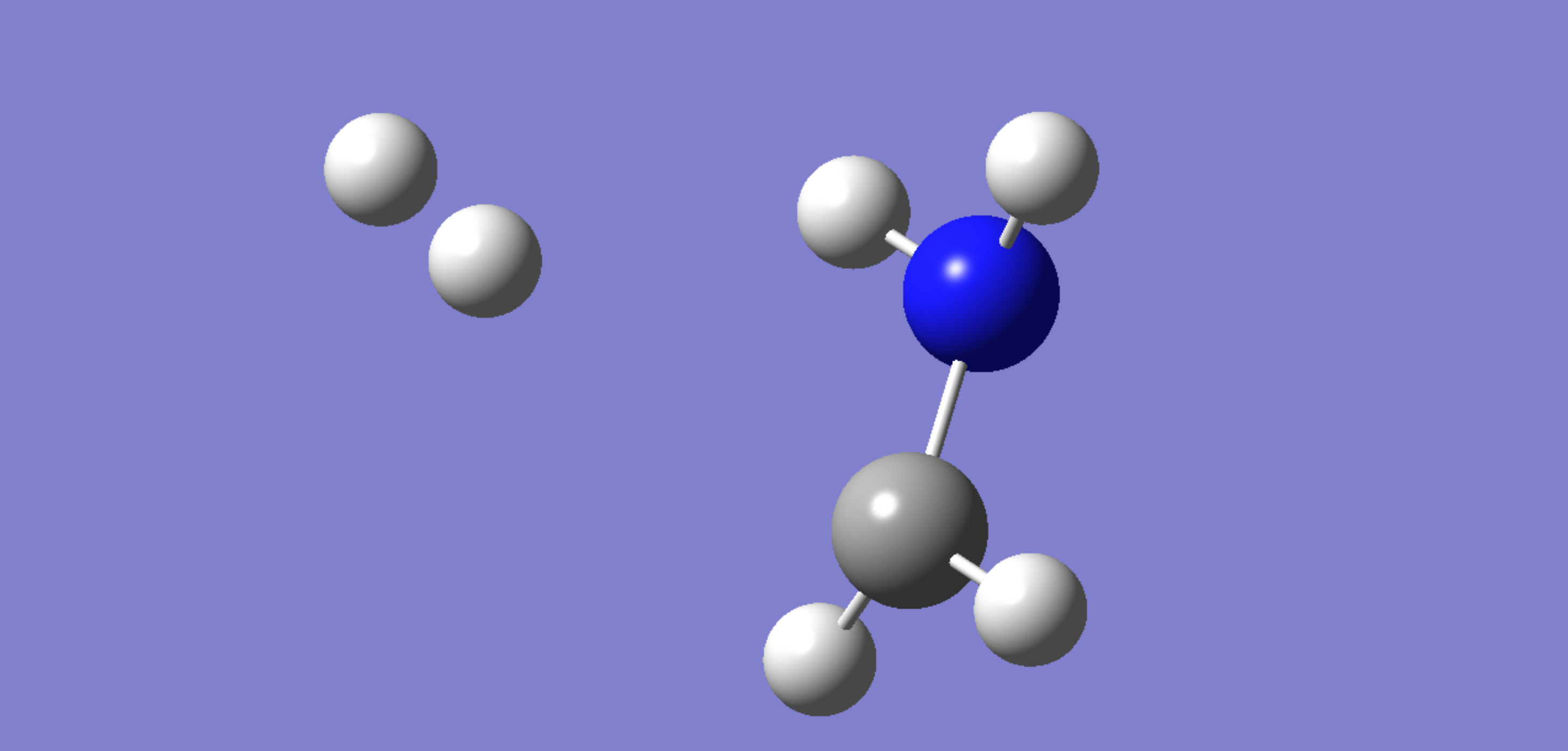} & Doublet &428 & 3.560 \\
 83&CH$_3$OH & \includegraphics[width=0.2\textwidth]{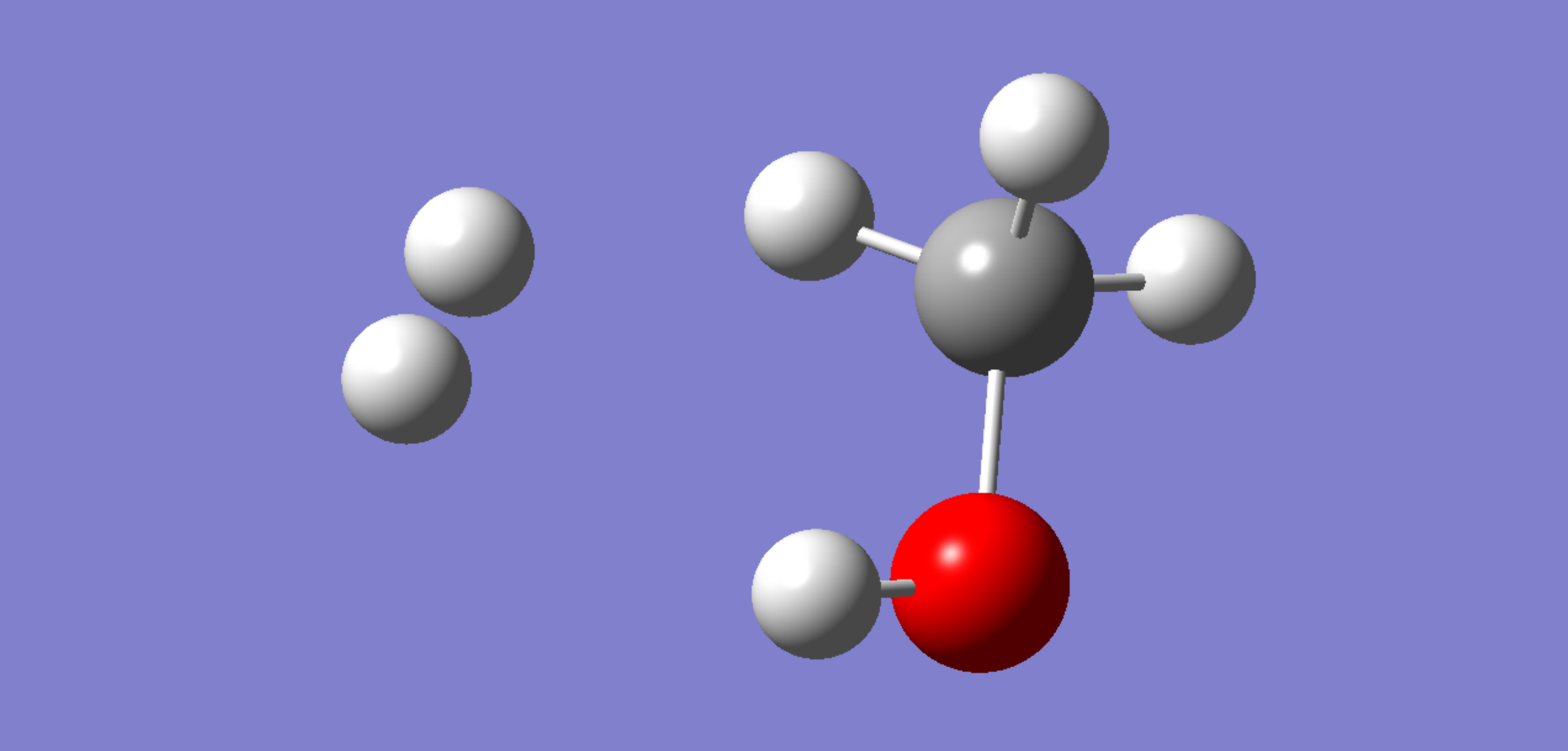} & Singlet & 414 & 3.445 \\
 && \includegraphics[width=0.2\textwidth]{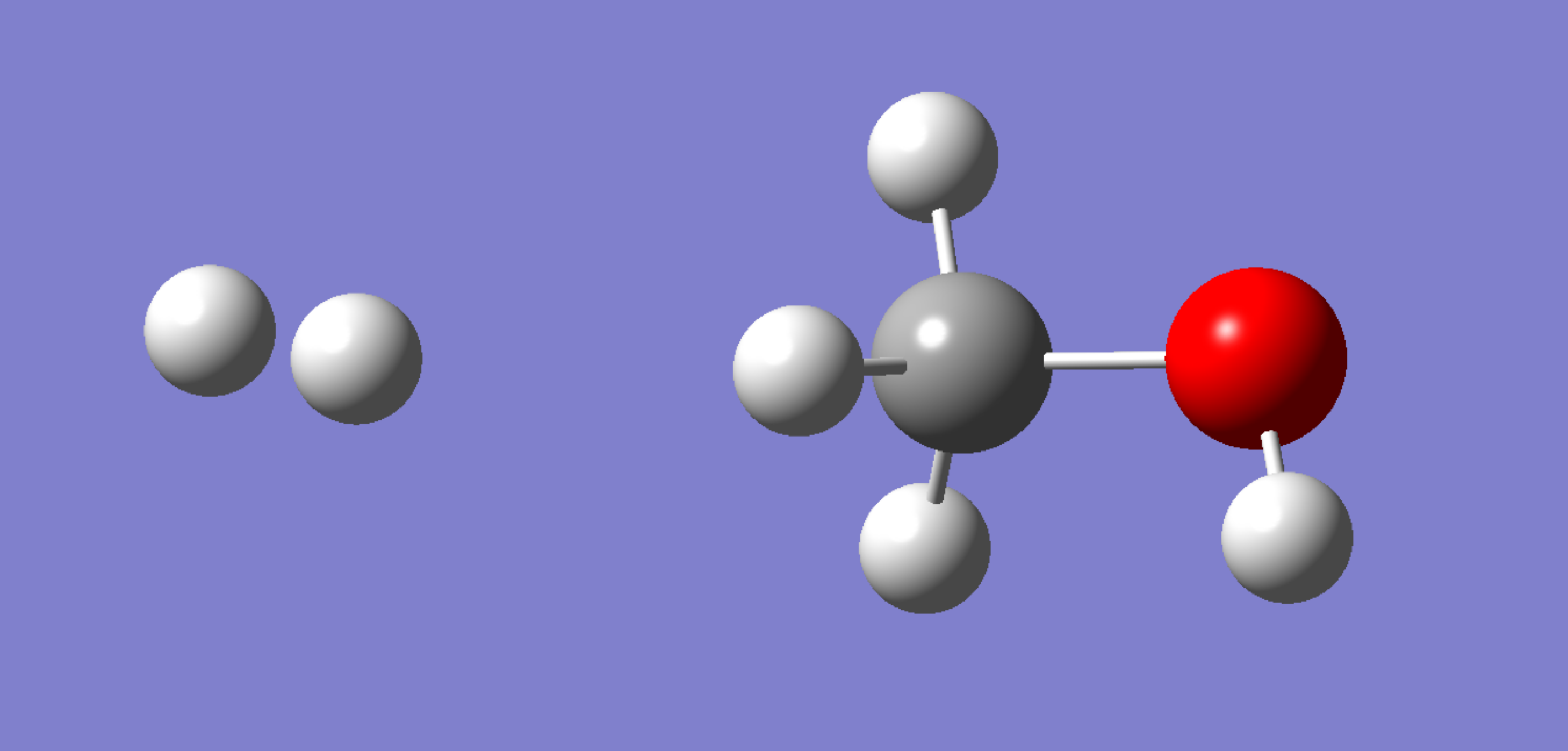} & & 258$^b$ & 2.145$^b$ \\
 84 & CH$_2$CCH & \includegraphics[width=0.2\textwidth]{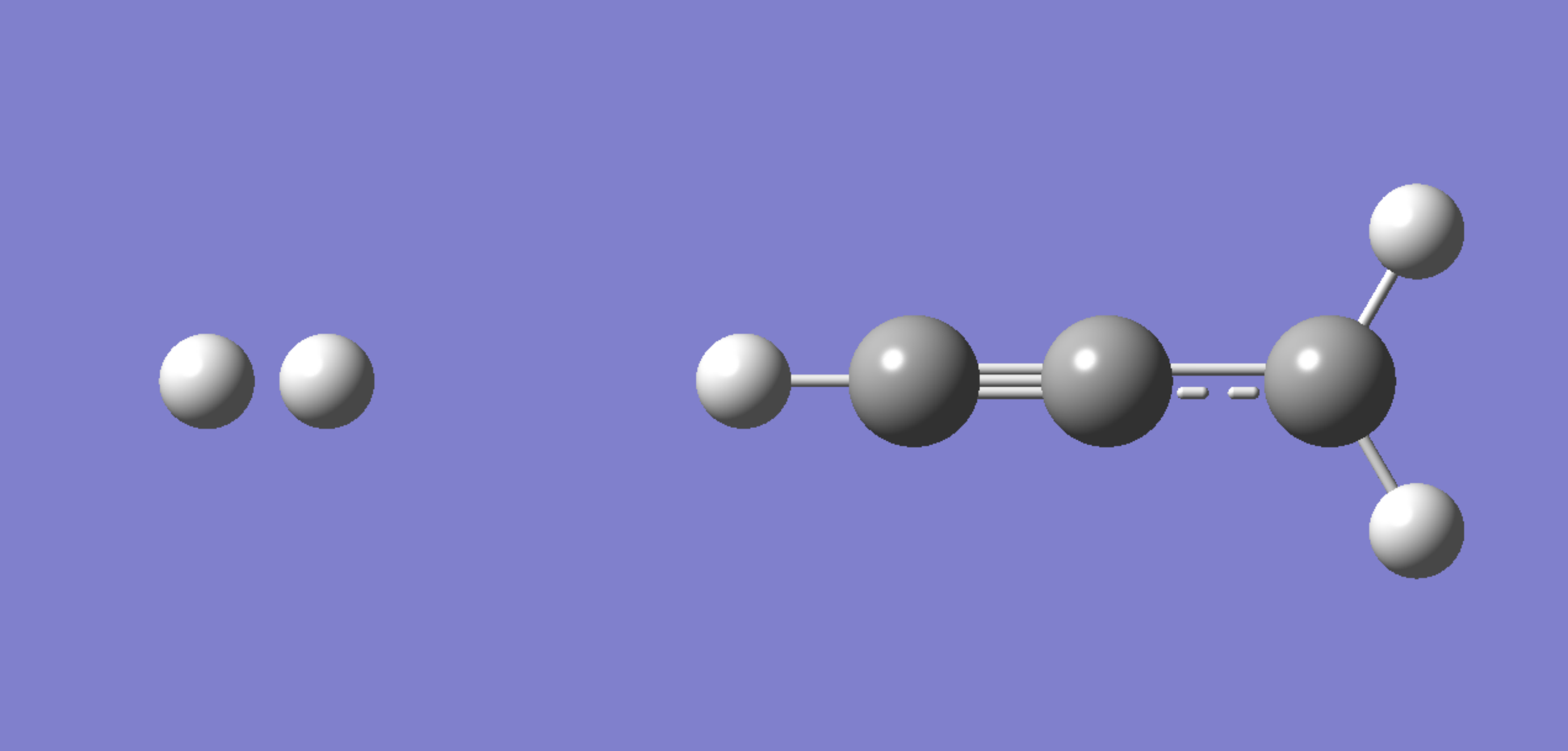} & Doublet & 105 & 0.872 \\
 85&CH$_3$CN & \includegraphics[width=0.2\textwidth]{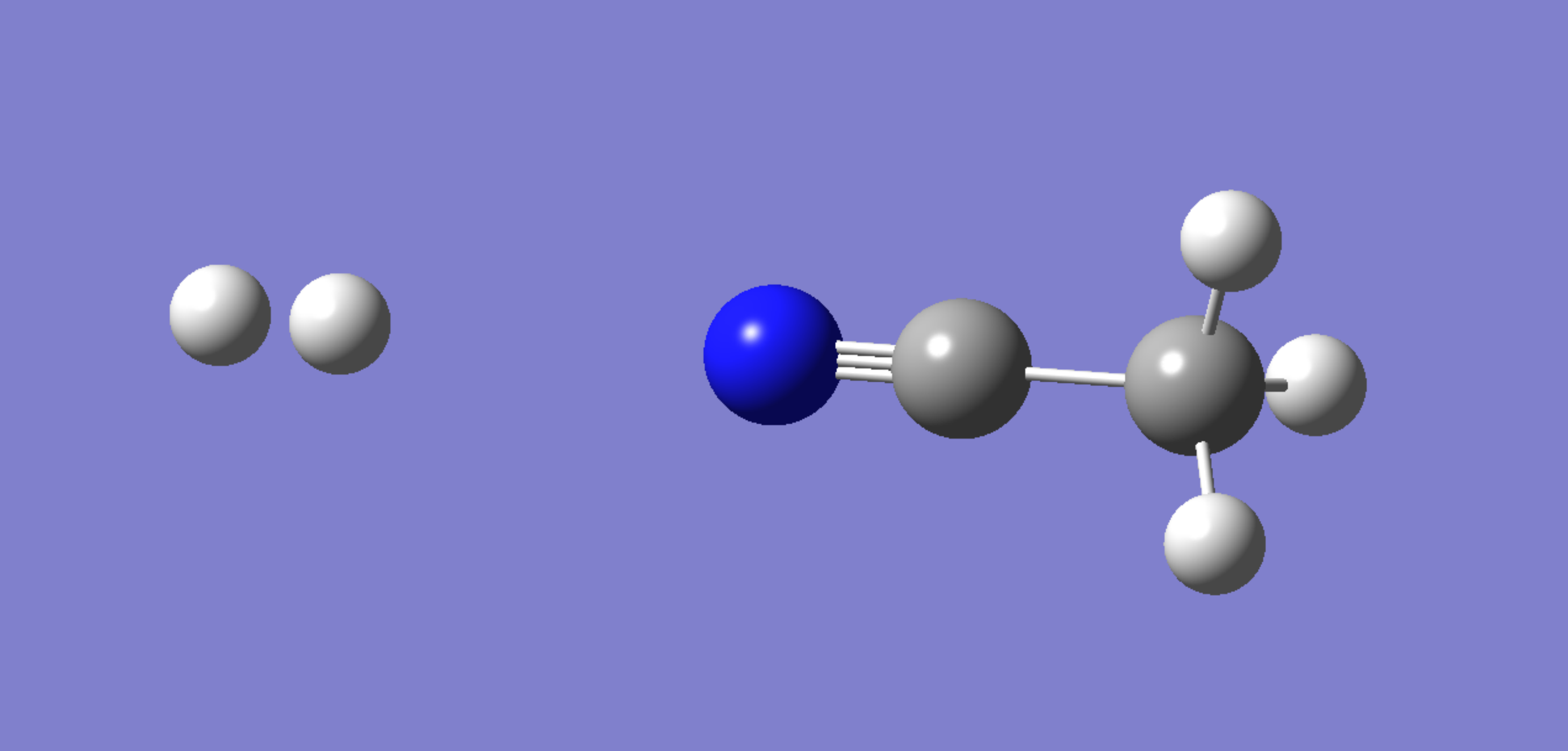} & Singlet &453 & 3.765 \\
 \hline
 \end{tabular}
\end{table}

\begin{table}
\scriptsize
{\centering
\begin{tabular}{|c|c|c|c|c|c|}
\hline
{\bf Sl.}& {\bf Species} & {\bf Optimized} & {\bf Ground} & \multicolumn{2}{c|}{\bf Binding Energy} \\
\cline{5-6}
 {\bf No.} & & {\bf Structures} & {\bf State} & {\bf in K} & {\bf in kJ/mol} \\
\hline
 86&CH$_3$NH$_2$ & \includegraphics[width=0.2\textwidth]{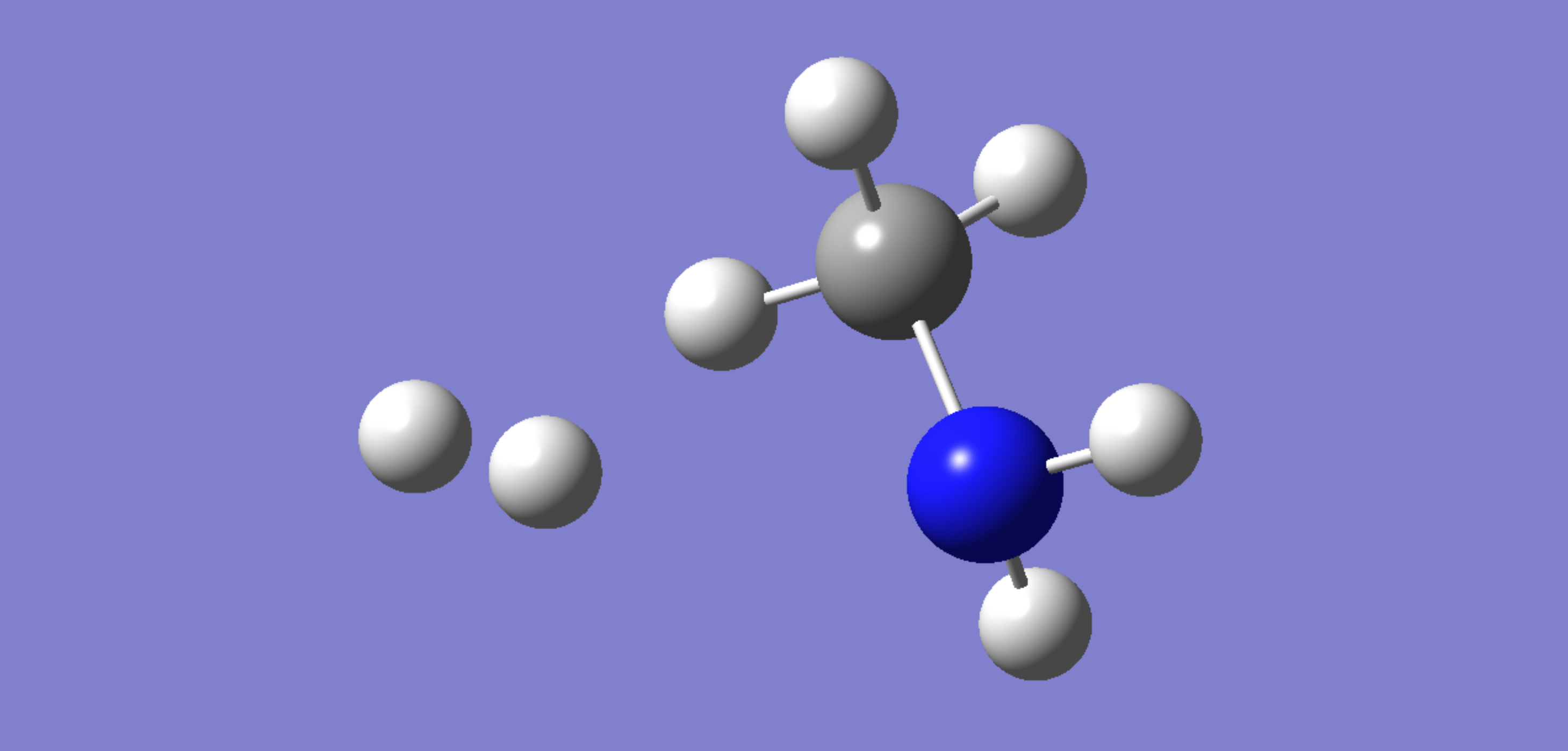} & Singlet & 610 & 5.072 \\
 87&C$_2$H$_5$ & \includegraphics[width=0.2\textwidth]{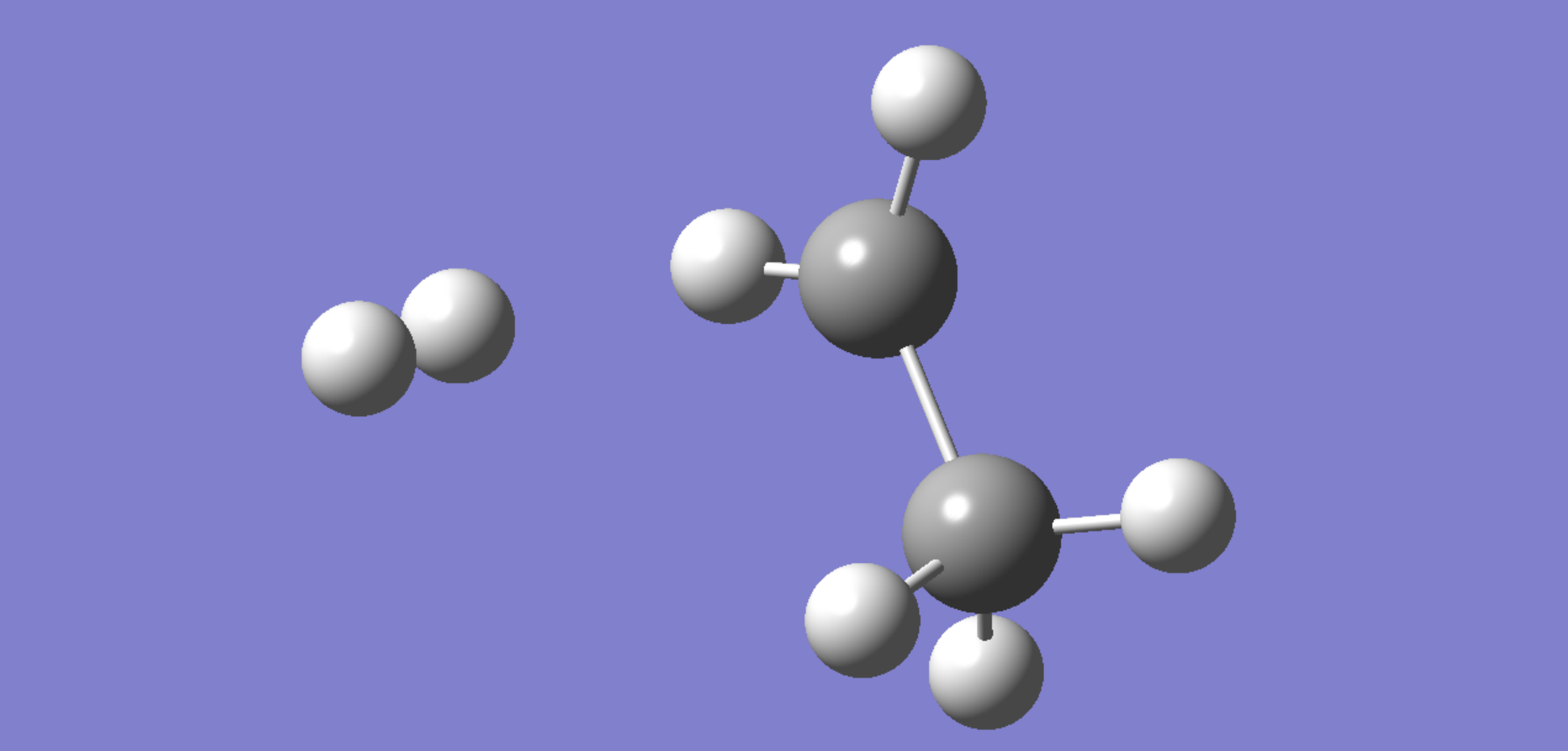} & Doublet &327 & 2.720 \\
 88&CH$_3$CCH & \includegraphics[width=0.2\textwidth]{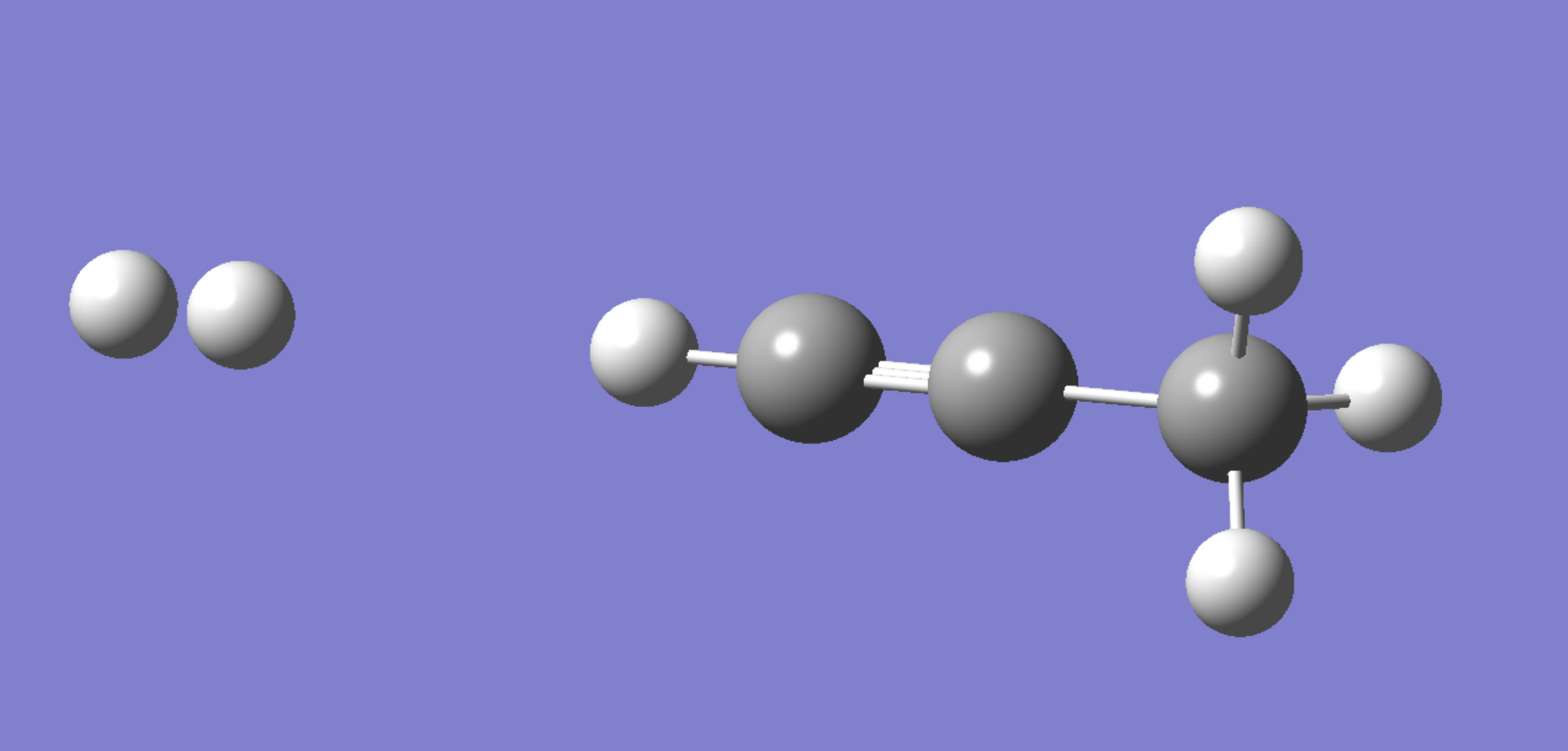} & Singlet &125 & 1.040 \\
89&CH$_2$CCH$_2$ & \includegraphics[width=0.2\textwidth]{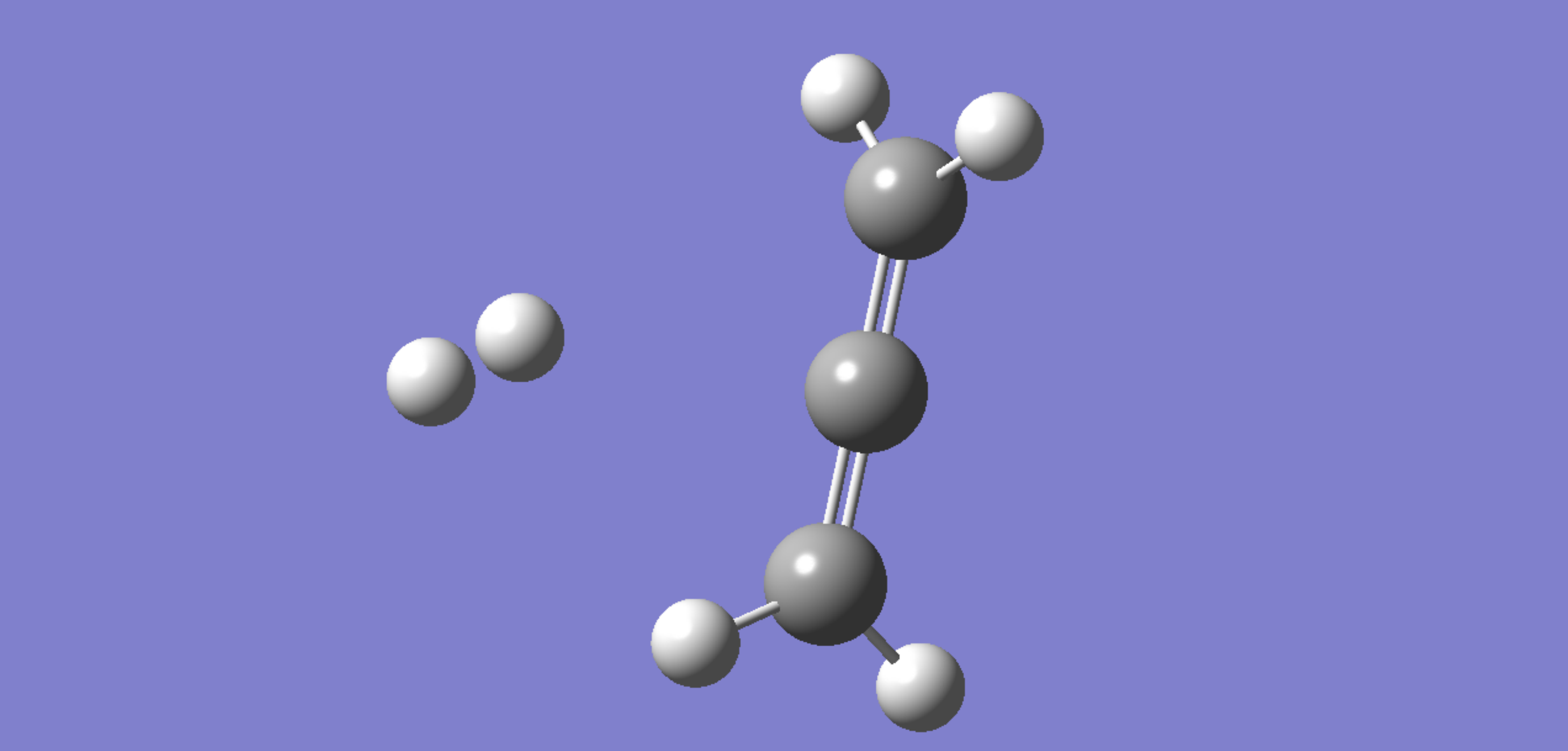} & Singlet &489 & 4.070 \\
90&CH$_3$CHO & \includegraphics[width=0.2\textwidth]{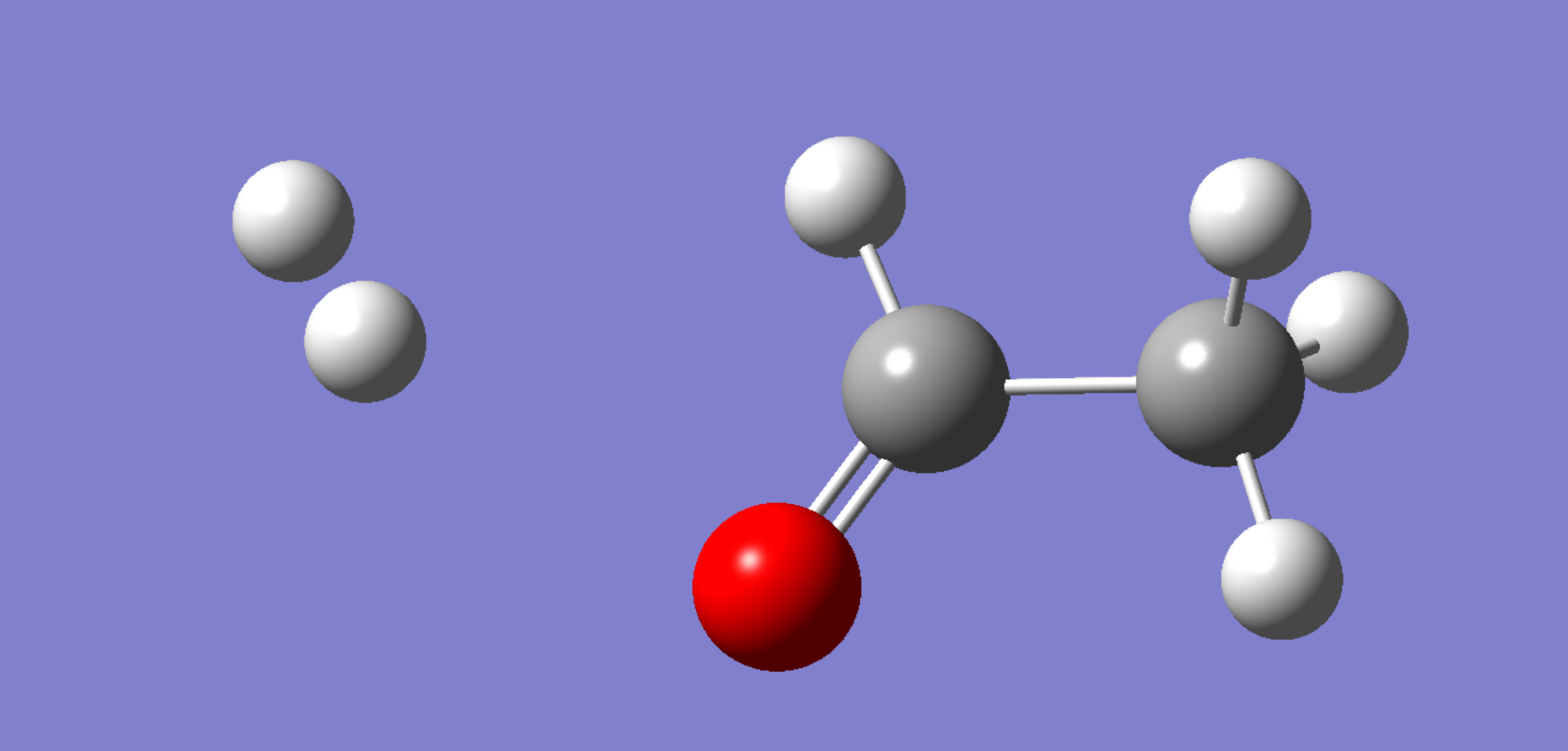} & Singlet &573 & 4.765 \\
91&PN & \includegraphics[width=0.2\textwidth]{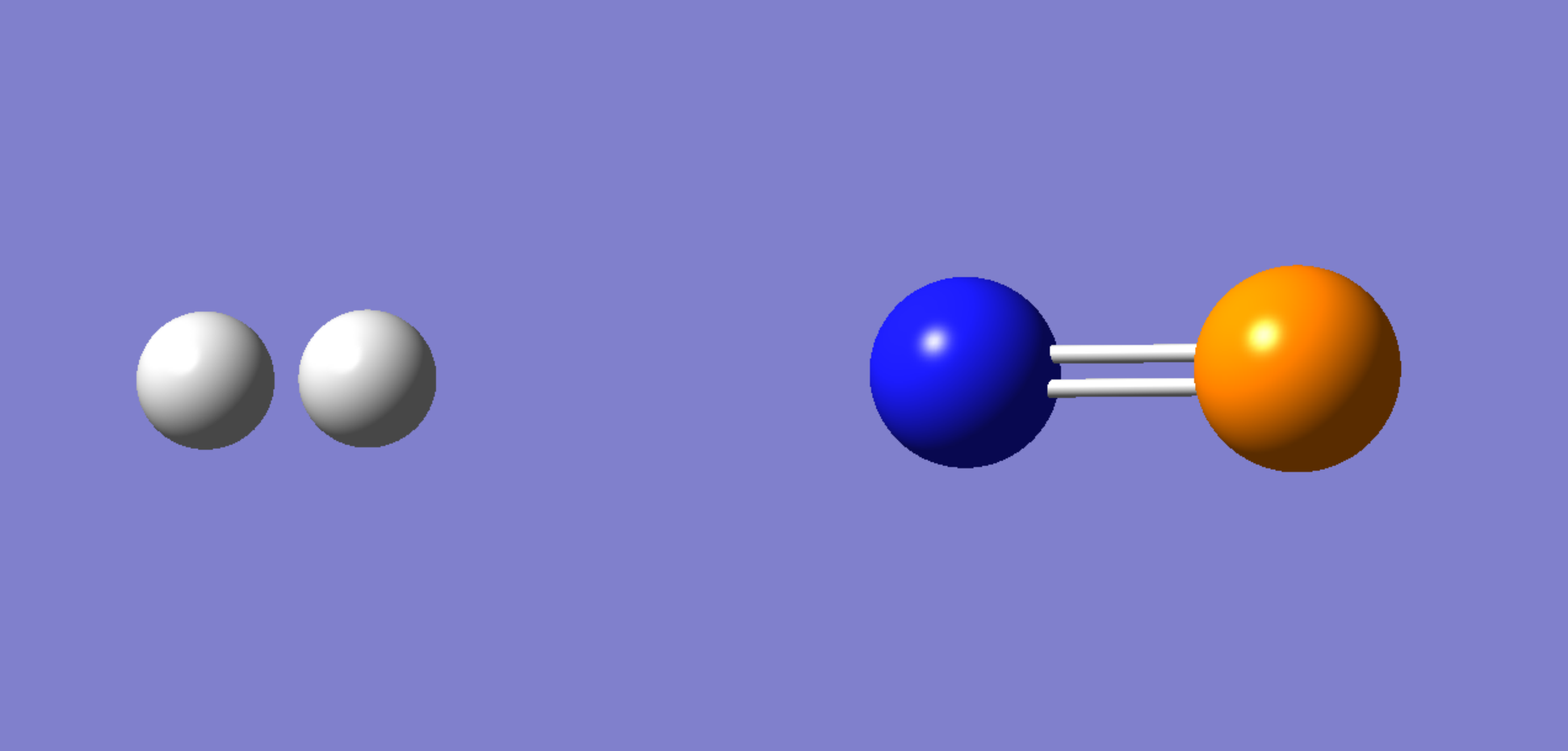} & Singlet &399 & 3.324 \\
 92 &PO & \includegraphics[width=0.2\textwidth]{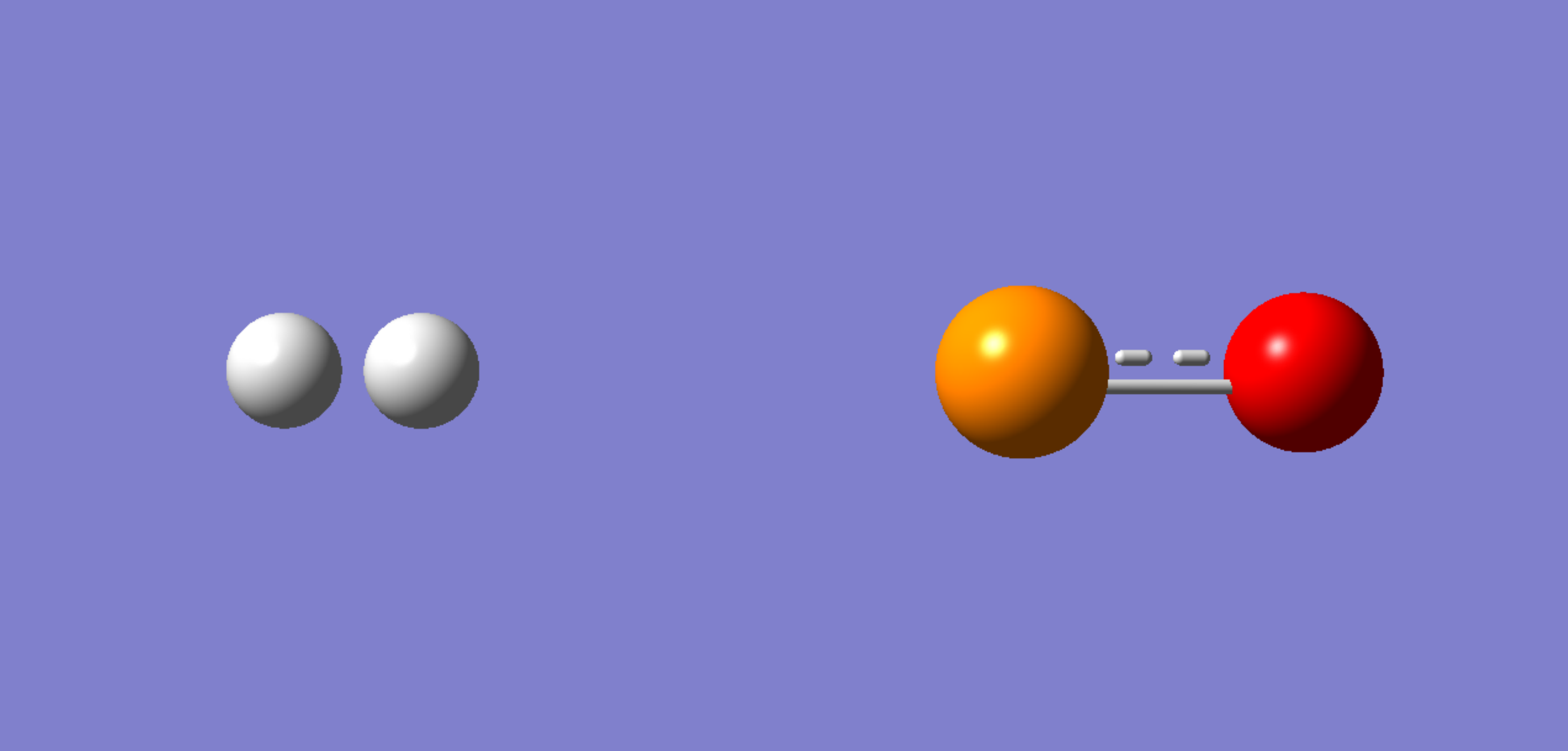} & Doublet &509 & 4.230  \\
 93 & SiN & \includegraphics[width=0.2\textwidth]{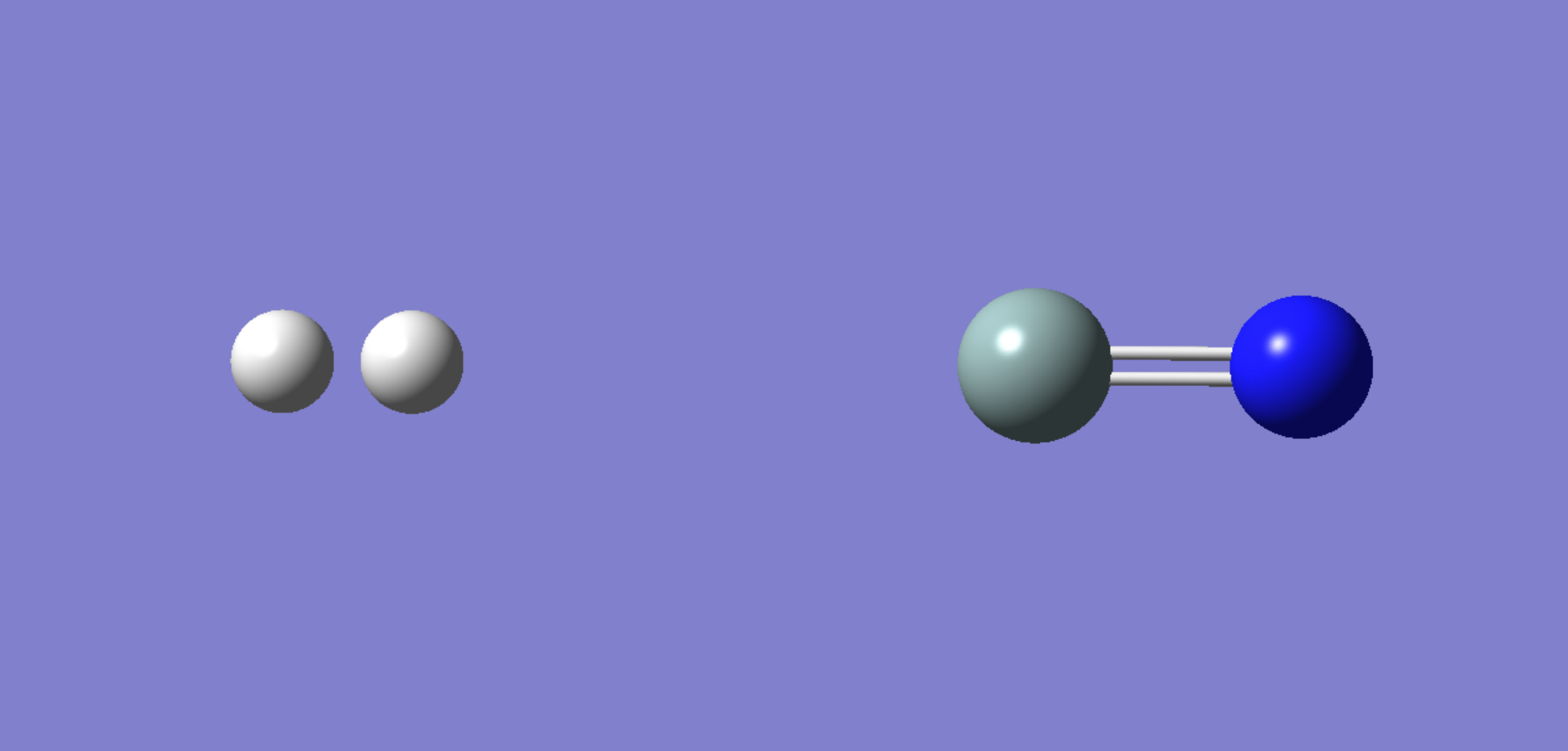} & Doublet & 154 & 1.281 \\
94&F & \includegraphics[width=0.2\textwidth]{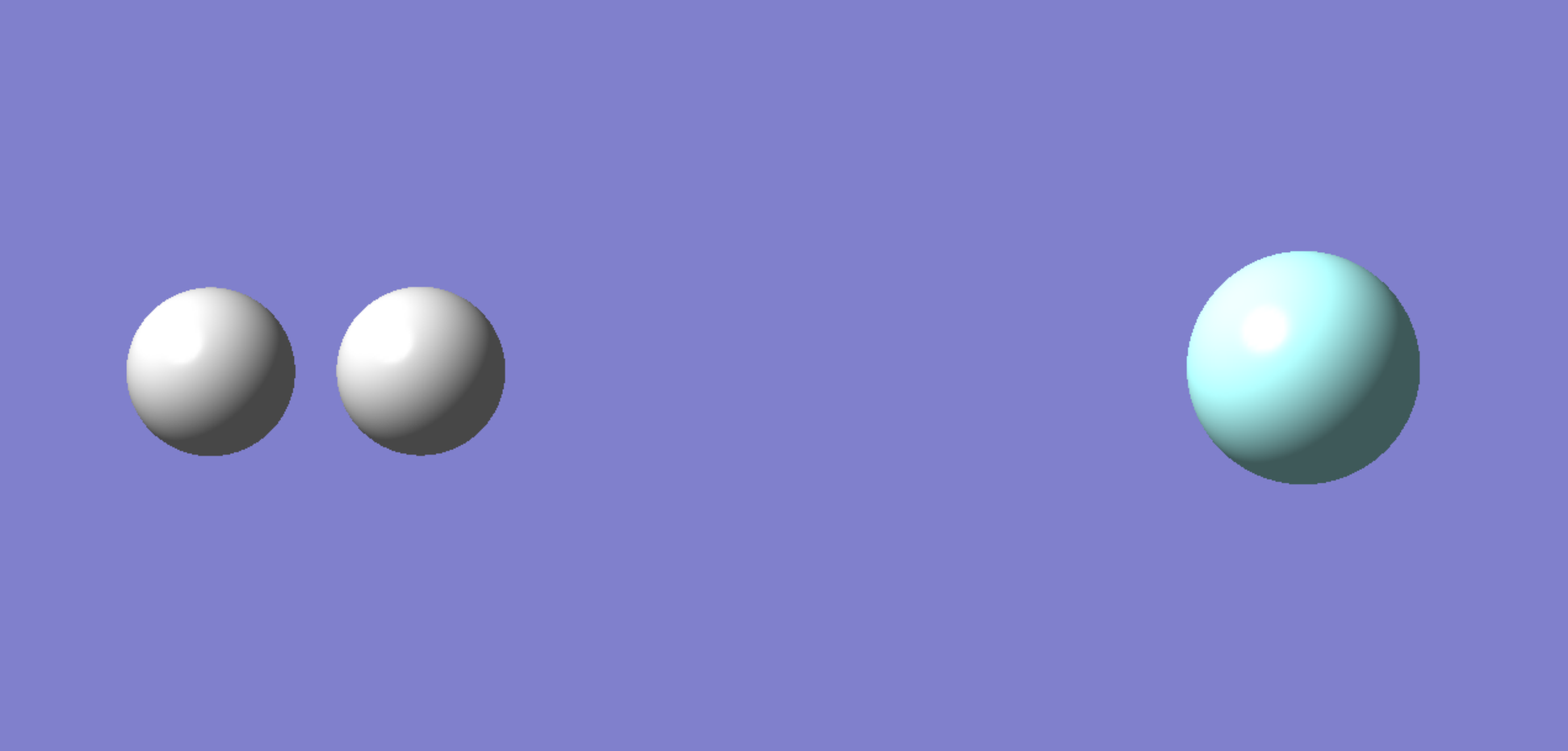} & Doublet &24 & 0.202 \\
95& C$_2$H$_5$OH & \includegraphics[width=0.2\textwidth]{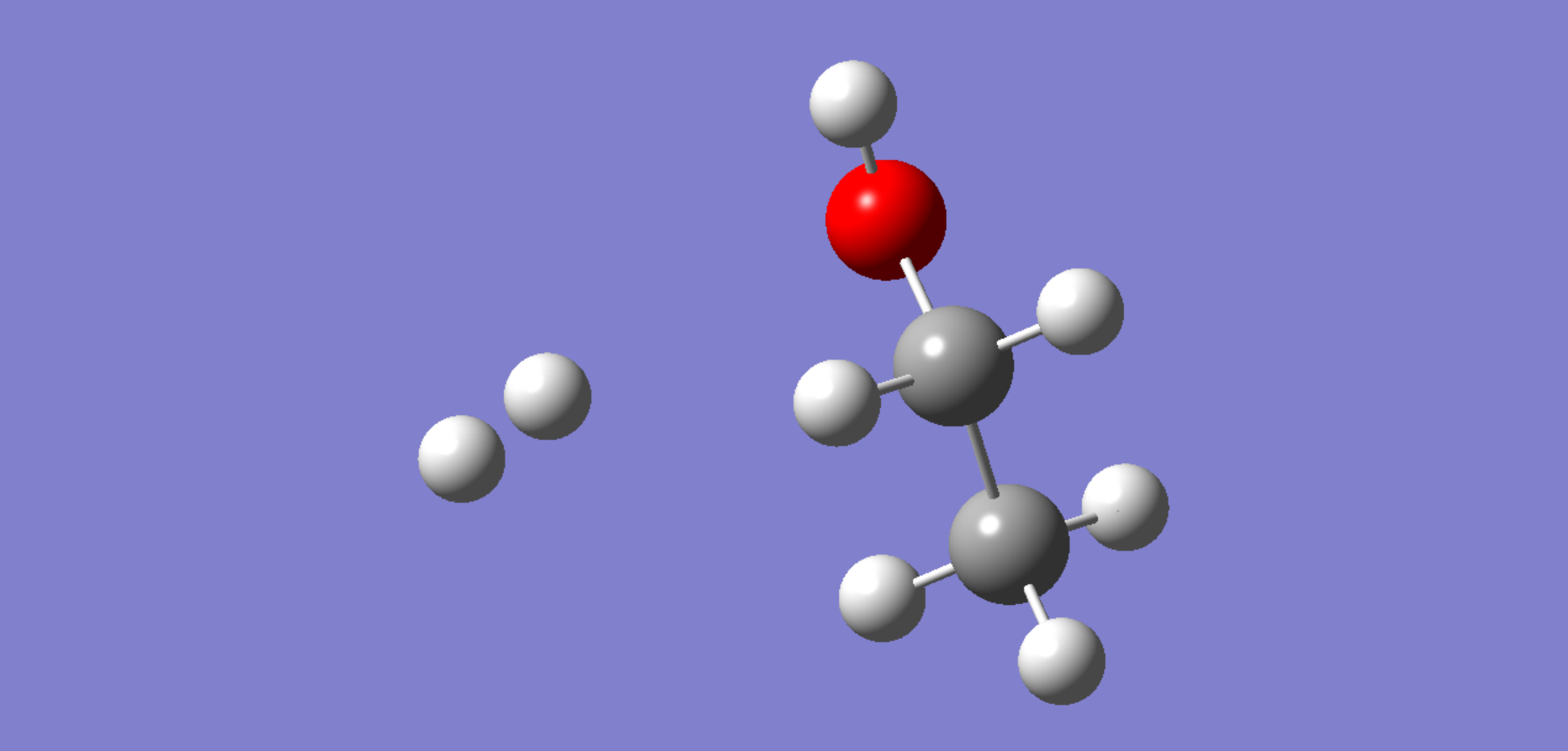} & Singlet & 590 & 4.906 \\
\hline
\end{tabular} \\
}
\vskip 0.2cm
$^a$ The BE values for the adsorbates H, H$_2$, and N with the adsorbent as H$_2$ considering IEFPCM model are noted in parentheses. \\
$^b$ Alternative BE values are for different binding sites.\\
$^c$ \cite{cupp07}. \\
$^d$ \cite{sand93}.
\end{table}

\subsubsection{Quantum chemical calculations}
We calculate the BE
of a species on the grain surface following Equations \ref{eqn:BE_1} and \ref{eqn:BE_2} mentioned earlier in subsection \ref{sec:BE_H_H2_method}.
The optimized energy of all structures is obtained
using the MP2/aug-cc-pVDZ level of theory \citep{dunn89}.
For this computation, a monomer configuration of the H$_2$ molecule is considered an adsorbent. Each adsorbate noted in Table \ref{tab:BE} is placed with a weak bond to sustain a van der Waals interaction during the optimization process. The optimized geometries are furnished in Table \ref{tab:BE} and supplementary information\footnote{The supplementary material of this article is available online at:
\url{https://www.frontiersin.org/articles/10.3389/fspas.2021.671622/full\#supplementary-material}} as well. It is to be noted that the interstellar species considered in this study are often larger than the H$_2$. The estimated BEs depend on the adsorption sites. This may be misleading. Thus, it is necessary to consider the average of the BEs obtained with various binding sites. Following the BE calculations carried out by \cite{das18} (see Tables \ref{tab:BE_2} and \ref{tab:BE_3}), here,
we do not consider the ZPVE and BSSE corrections for our BE calculations.
All the obtained BE values considering $\rm{H_2}$ substrate are presented in Table \ref{tab:BE}. Interestingly, except for phosphorous, the calculated BEs of most of the abundant atoms (H, C, N, O, and S) are $<100$ K.

The BE values reported here are obtained by considering a free-standing H$_2$ interacting with a species. But in reality, this H$_2$ would be pre-adsorbed and can fill the surface. Thus, it would yield a different BE than the previous case.
To check the influence of condensed H$_2$O, we carry out our computation by considering the species embedded in the continuum solvation field. The local impacts and the Integral Equation Formalism variant of the Polarizable Continuum Model (IEFPCM) are examined \citep{canc97,toma05} with water as a solvent. The obtained values for H, H$_2$, N with the IEFPCM model are noted in Table \ref{tab:BE} (in parentheses). However, the two calculations significantly differ. For example, the BE of H, H$_2$, and N are obtained $\sim$ 23 K, 67 K, and 83 K, respectively, with the free-standing H$_2$, whereas with the IEFPCM model, we obtain $\sim$ 25 K, 79 K, and 78 K, respectively. So, the free-standing H$_2$ underestimates the BE of H and H$_2$ by 2 K and 12 K, respectively. In contrast, it overestimates BE of N by $\sim 5$ K. We also provide the BEs from literature \citep{cupp07,sand93} in Table \ref{tab:BE} (if available) for the comparison.

The BEs of the $16$ stable interstellar species considering the various configurations of water molecules are provided in Table \ref{tab:BE_1} \citep{das18}.
Table \ref{tab:BE} shows their BEs with the H$_2$ monomer.
The corresponding ground state spin multiplicities are also noted.
We notice that the computed BEs with H$_2$ monomer substrate are much smaller than the BE with the water configurations.

  \begin{table}
  \scriptsize
    \caption{Initial elemental abundances considered in this study \citep{das21}.}
    \centering
    \vskip 0.5 cm
    \begin{tabular}{cc}
    \hline
    Species& Abundances\\
    \hline
        H$_2$ & $5.00 \times 10^{-1}$ \\
        He & $9.00 \times 10^{-2}$\\
        N& $7.60 \times 10^{-5}$\\
        O& $2.56 \times 10^{-4}$\\
        C$^+$&$1.20 \times 10^{-4}$ \\
        S$^+$& $8.00 \times 10^{-8}$\\
        Si$^+$ &$8.00 \times 10^{-9}$\\
        Fe$^+$& $3.00 \times 10^{-9}$\\
        Na$^+$& $2.00 \times 10^{-9}$\\
        Mg$^+$&$7.00 \times 10^{-9}$\\
        Cl$^+$&$2.00 \times 10^{-10}$\\
        HD&$1.60 \times 10^{-5}$\\
        \hline
    \end{tabular}
    \label{tab:init}
\end{table}

\subsubsection{Astrochemical model}
We include the encounter desorption phenomenon in our CMMC code \citep{das21} to study its effect.
The surface chemistry network of our model is mainly adopted from \cite{ruau15,das15b,gora20b}.
The gas-phase network of the CMMC code is mainly adopted from the UMIST database \citep{mcel13}. Additionally, we also include the deuterated gas-phase chemical network from the UMIST.
A cosmic-ray rate of $1.3 \times 10^{-17}$ s$^{-1}$ is considered in all our models.
In addition, cosmic ray-induced desorption and non-thermal desorption rate with a fiducial parameter
of $0.01$ are considered. We adopt a photodesorption rate of
$1 \times 10^{-4}$ per incident UV photon \citep{ruau15} for all the surface species.
Except for H and H$_2$, a sticking coefficient of $1.0$ is considered for all the neutral species.
The sticking coefficients of H and H$_2$ are considered by following the relation proposed by \cite{chaa12}.
Following \cite{garr11}, here, we implement the competition between diffusion, desorption, and reaction.
For the diffusion energy ($E_b$), we consider $R\ \times$ desorption energy ($E_d$).
Here, the scaling factor, $R$, can vary between 0.35 and 0.8 \citep{garr07}.
The BE of the surface species is mainly considered from \cite{wake17}, and a few from \cite{das18}. Table \ref{tab:init} refers to the adopted initial abundances concerning the total hydrogen nuclei in all forms. Except for the value of HD in Table \ref{tab:init}, elemental abundances are taken from \cite{seme10}. We consider the initial abundances of HD from \cite{robe00}.

The effect of encounter desorption was first introduced by \cite{hinc15}. The rate of encounter desorption of H$_2$ on the surface of H$_2$ is defined as:
\begin{equation}
    En_{H_2}= \frac{1}{2}\ k_{H_2,H_2} \ gH_2\  gH_2\ P(H_2,H_2),
    \label{eqn:enH2}
\end{equation}
where $gH_2$ is the surface concentration of H$_2$ molecules in cm$^{-3}$, $P(H_2,H_2)$ represents the probability of desorption over the diffusion, and
$k_{H_2,H_2}$ is the diffusion rate coefficient over the H$_2$O substrate.
$k_{H_2,H_2}$ is defined as follows \citep{hase92}:
\begin{equation}
k_{H_2,H_2}= \kappa (R_{diffH_2} + R_{diffH_2})/n_d \ \ cm^3s^{-1}.
\end{equation}
In the above equation, $n_d$ is the dust-grain number density, $\kappa$ is the probability of the occurrence of the reaction (unity for the exothermic reaction without activation energy), and $R_{diff}$ is the diffusion of the species.  $P(H_2,H_2)$ in Equation \ref{eqn:enH2} is defined as:
\begin{equation}
\scriptsize
     P(H_2,H_2)=\frac{Desorption \ rate \  of \ H_2 \ on \ H_2 \ substrate}{Desorption \ rate \ of \ H_2 \ on \ H_2 \ substrate \ + \ Diffusion \ of \ H_2 \ on \ H_2 \ substrate}.
    \label{eqn:prob}
\end{equation}
There would be various desorption factors (thermal, reactive, cosmic-ray etc.).
The thermal desorption is defined as: $\nu\ exp(-E_d(H_2,H_2)/T)$ s$^{-1}$, where $T$ is the temperature of the dust. Similarly, there would be various diffusion processes, but the thermal diffusion would be the dominating. It is defined as: $\nu\ exp(-E_b(H_2,H_2)/T)/S$ (s$^{-1}$) $= thermal\ hopping\ rate\ /\ number\ of\ sites$ (s$^{-1}$).
Recently, \cite{chan21} extended this work considering the encounter desorption of H atom. They defined the encounter desorption of H$_2$, (in Equation \ref{eqn:prob}) by the hopping rate of H$_2$ on H$_2$ substrate instead of the diffusion rate of H$_2$ on H$_2$ substrate. Following the prescription defined in \cite{chan21}, the encounter desorption of species X is defined as:
\begin{equation}
 En_{X,H_2} \ = \ \frac{h_{X,H_2}}{S} gX \ gH_2 \ P(X,H_2) \ P_X,
\end{equation}
where $h_{X,H_2}$ is the hopping rate over H$_2$O surface, $P(X,H_2)$ is the desorption probability of $gX$ while encountering $gH_2$, and $P_X$ denotes the probability of $gX$ for migrating at the location of $gH_2$ over the H$_2$O substrate. $P(X,H_2)$ and $P_X$ are defined as,
\begin{equation}
\scriptsize
 P(X,H_2)=\frac{Desorption \ rate \  of \ X \ on \ H_2 \ substrate}{Desorption \ rate \ of \ X \ on \ H_2 \ substrate \ + \ Hopping\ rate \ of \ X \ on \ H_2 \ substrate},
        \label{eqn:chan}
        \end{equation}
\begin{equation}
\scriptsize
P_X =\frac{Hopping \ rate \ of \ X \ on \ H_2O \ substrate}{Hopping \ rate\  of \ X \ on \ H_2O \ substrate \ + \ Hopping \ rate \ of \ H_2 \ on \ H_2O \ substrate}.
  \end{equation}

\subsubsection{Results and discussions}

\subsubsection{Encounter desorption of H$_2$} \label{sec:H2-EN}
\begin{figure}
    \centering
    \includegraphics[width=0.6\textwidth,angle=-90]{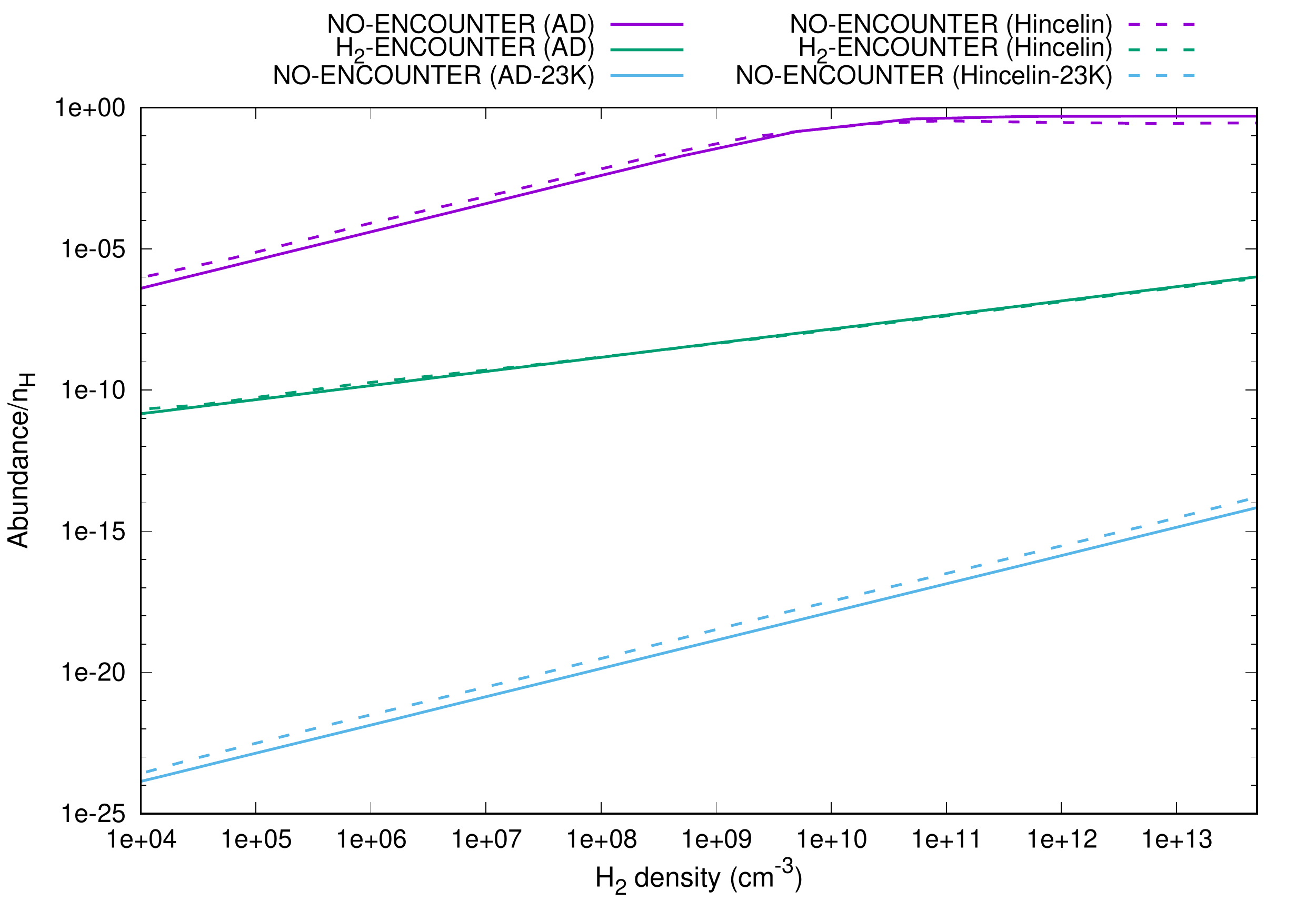}
    \caption{A comparison between the cases obtained in \cite{hinc15} and \cite{das21}. Figure 2 of \cite{hinc15} is extracted using the online tool of \cite{roha20}. Three distinct cases are shown: (A) no encounter desorption is considered with $E_d(H_2,H_2O)=440$ K, (B) no encounter desorption is considered with $E_d(H_2,H_2)=23$ K, and (C) encounter desorption of H$_2$ was considered with $E_d(H_2,H_2O)=440$ K and $E_d(H_2,H_2)=23$ K. An excellent match between the our calculated (solid curves) steady-state abundance of H$_2$ on grain surface and that obtained in \cite{hinc15} (dashed curves) is noticed.}
    \label{fig:comp}
\end{figure}

We benchmark our model with \cite{hinc15}.
In Figure \ref{fig:comp}, we compare our results with those obtained in \cite{hinc15}.
For this comparison, following \cite{hinc15}, we use $T=10$ K, $E_d(H_2,H_2O)= 440$ K, $E_d(H,H_2O)=450$ K, $E_d(H_2,H_2)=23$ K, and $R=0.5$.
Solid curves in Figure \ref{fig:comp} represent our \citep{das21} obtained cases, and the rest are extracted from \cite{hinc15} using the online tool of \cite{roha20}.
Our results considering and avoiding encounter desorption show an excellent match with \cite{hinc15}. Presently in the KIDA, more updated BE values are listed \citep{wake17}.
It suggests that  $E_d(H,H_2O)=650$ K. The results obtained from our quantum chemical calculations shown in Table \ref{tab:BE} represent the estimated BE values with the H$_2$ substrate.
We use these updated BE values, and the effects of their changes are discussed.

  \begin{table}
\tiny
    \caption{The obtained abundance of gH$_2$, gH, gH$_2$O and gCH$_3$OH for the effect of encounter desorption of H$_2$ under various situation with $R=0.35$, $n_H=10^7$ cm$^{-3}$, and $T=10$ K \citep{das21}.}
   \vskip 0.2 cm
   \hskip -0.5 cm
   \begin{tabular}{|c|c|c|c|c|c|}
    \hline
   Case & Case specification & \multicolumn{4}{|c|}{Abundance at $10^6$ years with ${\rm n_H}=10^{7}$ cm$^{-3}$}\\
   \cline{3-6}
    No. && gH$_2$ & gH & gH$_2$O & gCH$_3$OH \\
   &&&& (\% increase) &  (\% increase) \\
    \hline
    \multicolumn{6}{|c|}{$\rm{E_d(H,H_2O})=450$ K}\\
    \hline
        1& No encounter desorption &$  2.0011749 \times 10^{-4}$&$2.311976  \times 10^{-24}$&$  8.8375039 \times 10^{-5}$&$ 8.8416179 \times 10^{-6}$\\
        &&&& (0.00) & (0.00) \\
        2& ${\rm E_d(H_2,H_2)=23}$ K \citep{hinc15}& $1.183836 \times 10^{-11}$& $ 3.2799079 \times 10^{-24}$&$ 8.927643 \times 10^{-5}$ &$  6.3629209 \times 10^{-6}$\\
        &&&& ($1.02$) & ($-28.03$) \\
         3& ${\rm E_d(H_2,H_2)=23}$ K \citep{chan21}&$2.7660509 \times 10^{-11}$&$3.253898 \times 10^{-24}$&$8.9327989 \times 10^{-5}$ & $ 6.4358829 \times 10^{-6}$\\
         &&&& (1.08) & ($-27.21$) \\
         4& ${\rm E_d(H_2,H_2)=67}$ K \citep{chan21}& $1.051303 \times 10^{-10}$&$2.25004 \times 10^{-24}$&$1.02707 \times 10^{-4}$ & $6.119676 \times 10^{-6}$\\
         &&&& (16.22) & ($-30.79$) \\
         5& ${\rm E_d(H_2,H_2)=79}$ K \citep{chan21}& $  1.5474109 \times 10^{-10}$&$  2.2527589 \times 10^{-24}$&$ 1.028505 \times 10^{-4}$ & $6.2066129 \times 10^{-6}$\\
         &&&& (16.38) & ($-29.8$) \\
         \hline
        \multicolumn{6}{|c|}{$\rm{E_d(H,H_2O)=650}$ K}\\
        \hline
        6& No encounter desorption & $2.00117 \times 10^{-4}$&$ 1.467684 \times 10^{-21}$&$9.434889 \times 10^{-5}$&$ 5.799469 \times 10^{-6}$\\
        &&&& (0.00) & (0.00) \\
         7& ${\rm E_d(H_2,H_2)=67}$ K \citep{chan21}&$1.051293 \times 10^{-10}$&$2.0551489 \times 10^{-21}$&$ 9.361905 \times 10^{-5}$ &$4.559358 \times 10^{-6}$\\
         &&&& ($-0.77$) & ($-29.80$) \\
         \hline
    \end{tabular}
    \label{tab:H2abunnew}
\end{table}

  \begin{figure}
    \centering
    \includegraphics[width=0.5\textwidth,angle=-90]{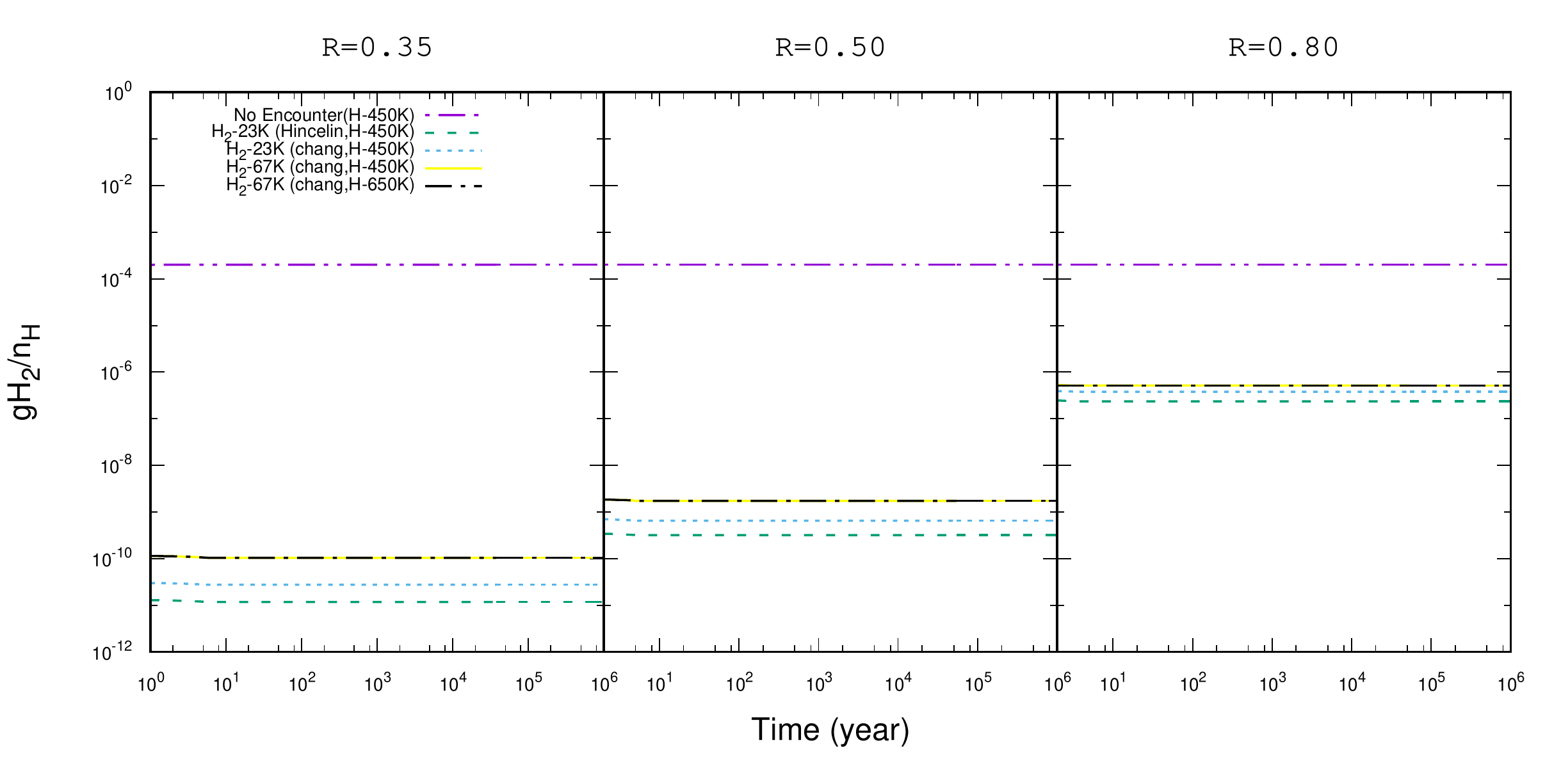}
    \caption{Time evolution of the abundances of gH$_2$ with $\rm{n_H=10^7}$ cm$^{-3}$ and $T=10$ K are shown for $R=0.35$, $0.5$, and $0.8$ \citep{das21}. The dash-dotted purple curve shows time evolution of $gH_2$ abundance without encounter desorption [with $E_d(H,H_2O) = 450$ K]. gH$_2$ abundance remains roughly constant with the changes in R. However when encounter desorption is introduced, gH$_2$ abundance increases with the $R$. The time evolution of the gH$_2$ abundance with $E_d(H_2,H_2) =23$ K and $E_d(H,H_20) = 450$ K is shown with the green dashed line when the method of \cite{hinc15} is used and blue dotted line when the method of \cite{chan21} is used. $gH_2$ abundances obtained with our estimated BE value [i.e., $E_d(H_2,H_2)= 67$ K] are shown with a solid yellow line. For this case, we have used $E_d(H,H_2O)=450$ K and the method used in \cite{chan21}. The black dash-dotted line shows the time evolution of $gH_2$ abundance with $E_d(H,H_2O)=650$ K and using the method of \cite{chan21}. A significant difference is obtained when we use different energy barriers and different methods \citep{hinc15,chan21}. Obtained values of $gH_2$ are further noted in Table \ref{tab:H2abunnew} for better understanding.}
    \label{fig:H2-abun}
\end{figure}

\subsubsection{$gH_2$}
\begin{figure}
    \centering
    \includegraphics[width=0.2\textwidth,angle=-90]{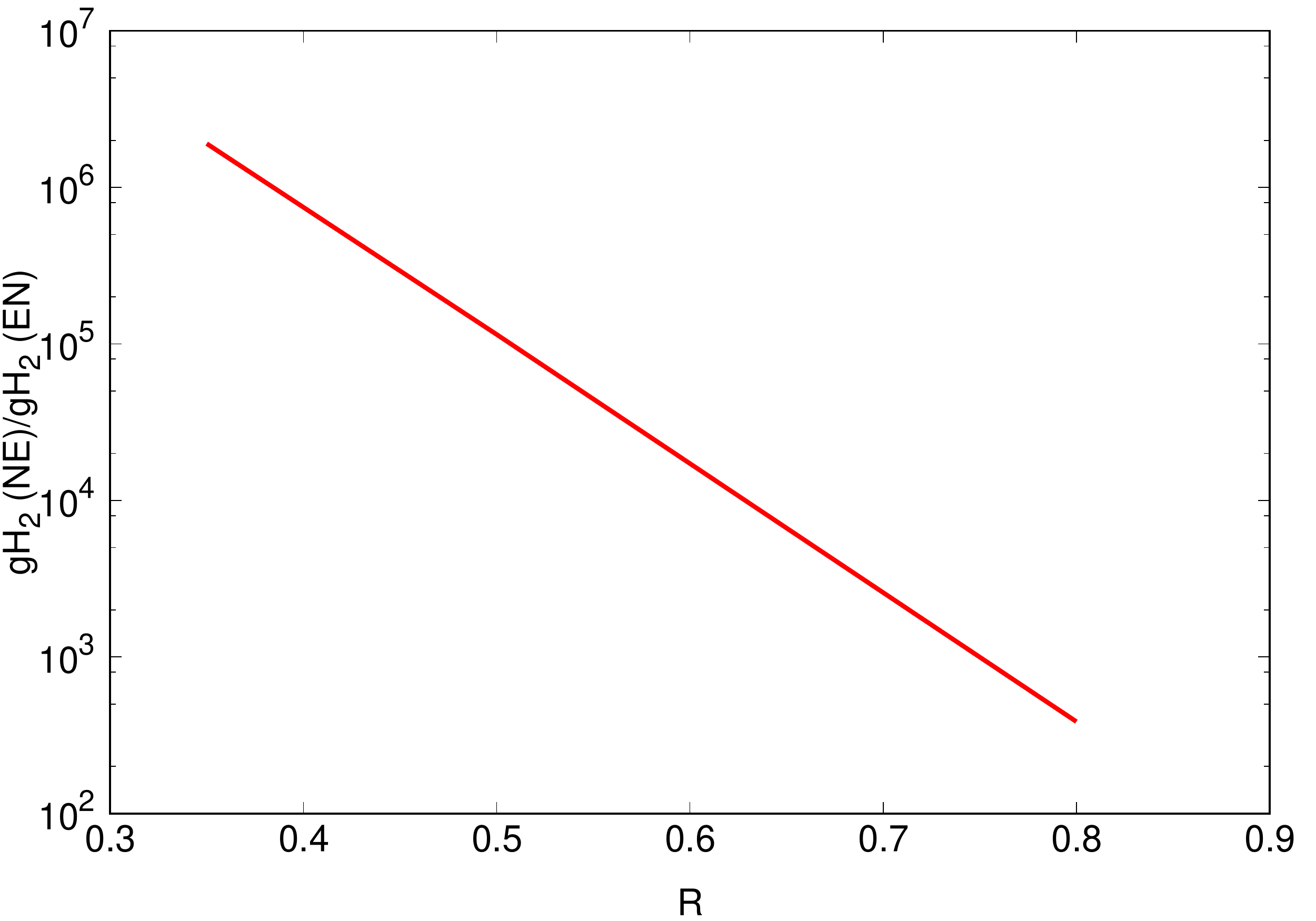}
     \includegraphics[width=0.2\textwidth,angle=-90]{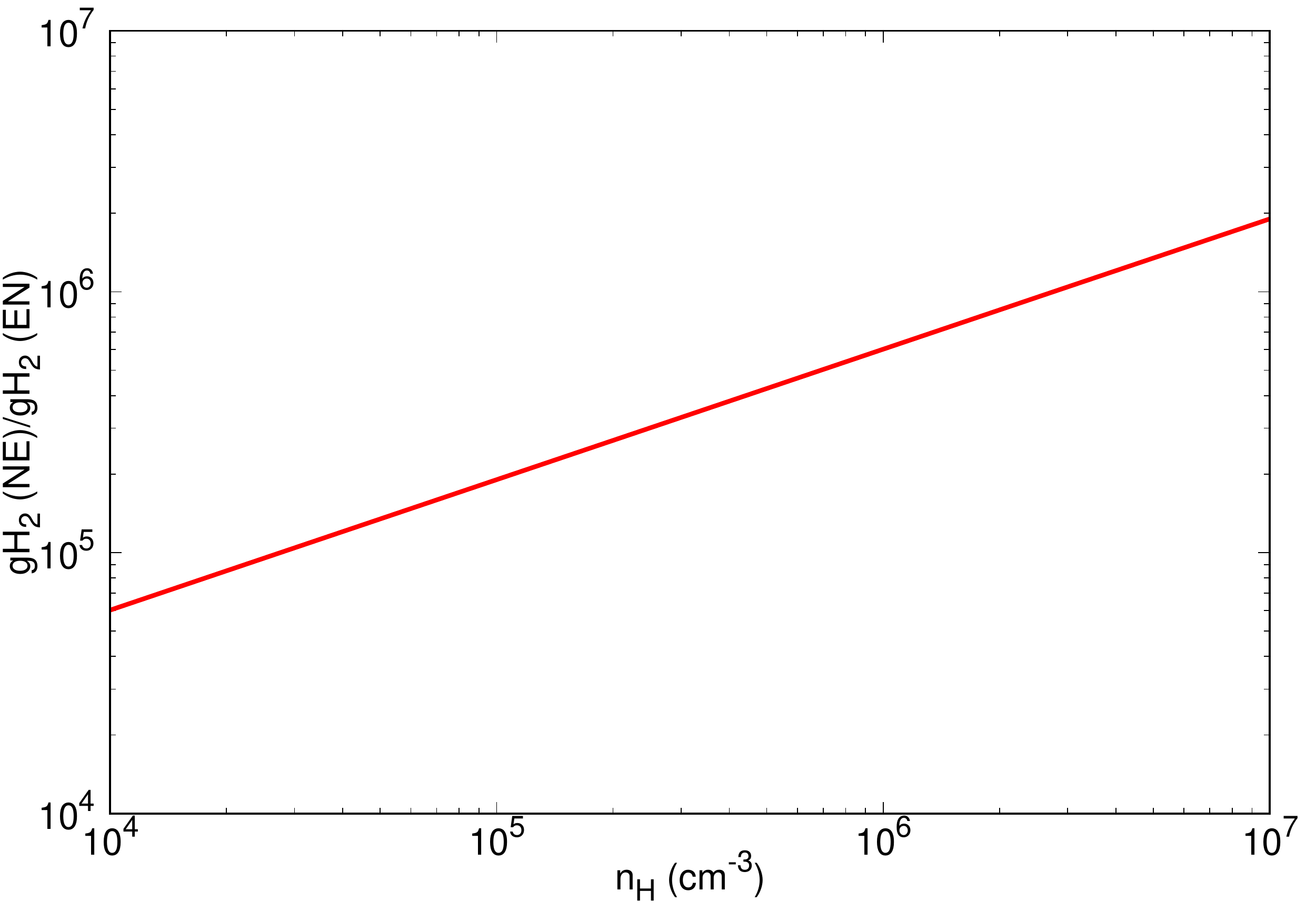}
      \includegraphics[width=0.2\textwidth,angle=-90]{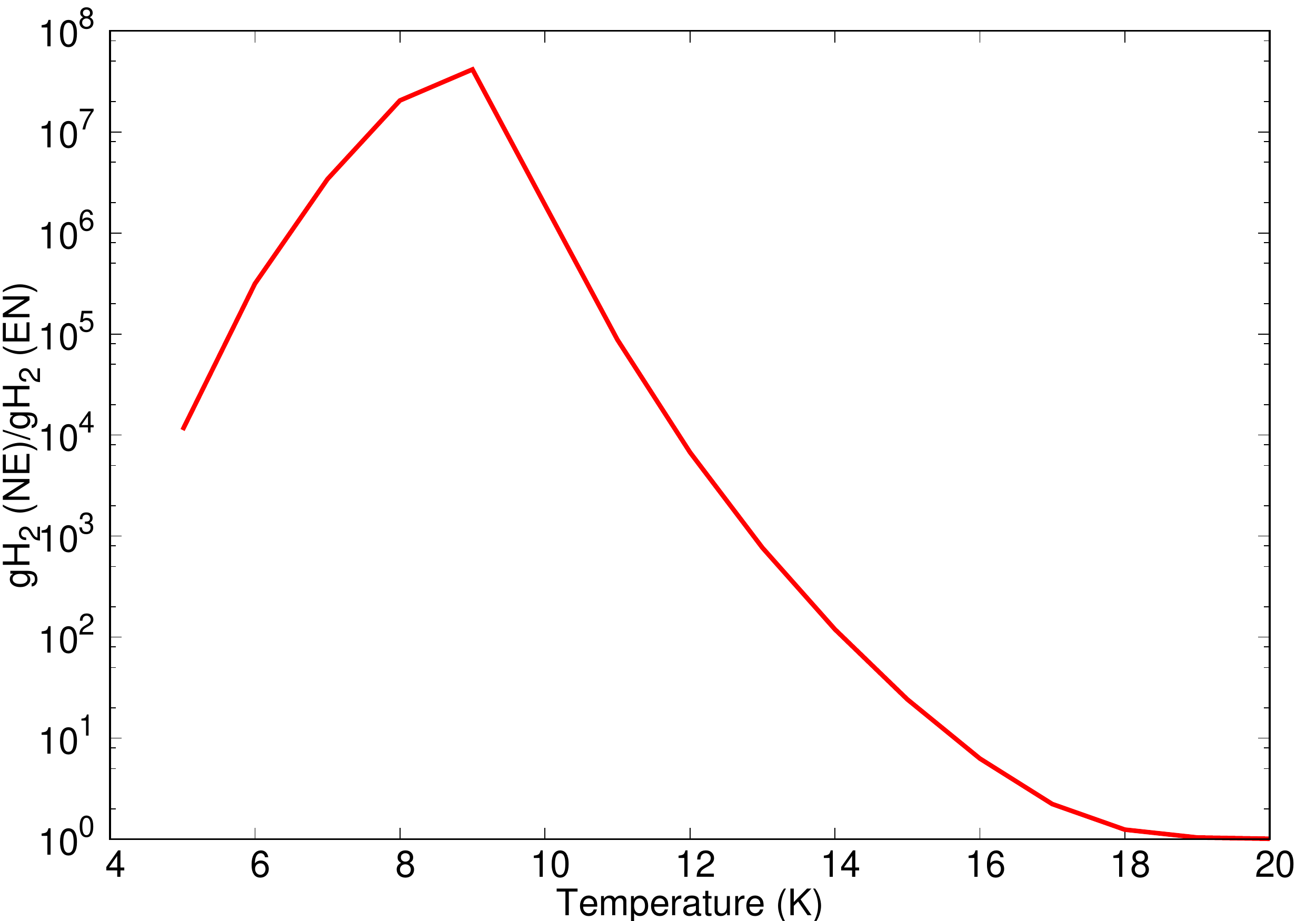}
    \caption{The ratio between the final abundances of $gH_2$ obtained with the no encounter (NE) desorption and encounter desorption (EN) is shown \citep{das21}. From left to right, it shows the variation of this ratio with $R$, $n_H$, and temperature, respectively.}
    \label{fig:H2-ratio}
\end{figure}

\begin{figure}
    \centering
    \includegraphics[width=0.5\textwidth,angle=-90]{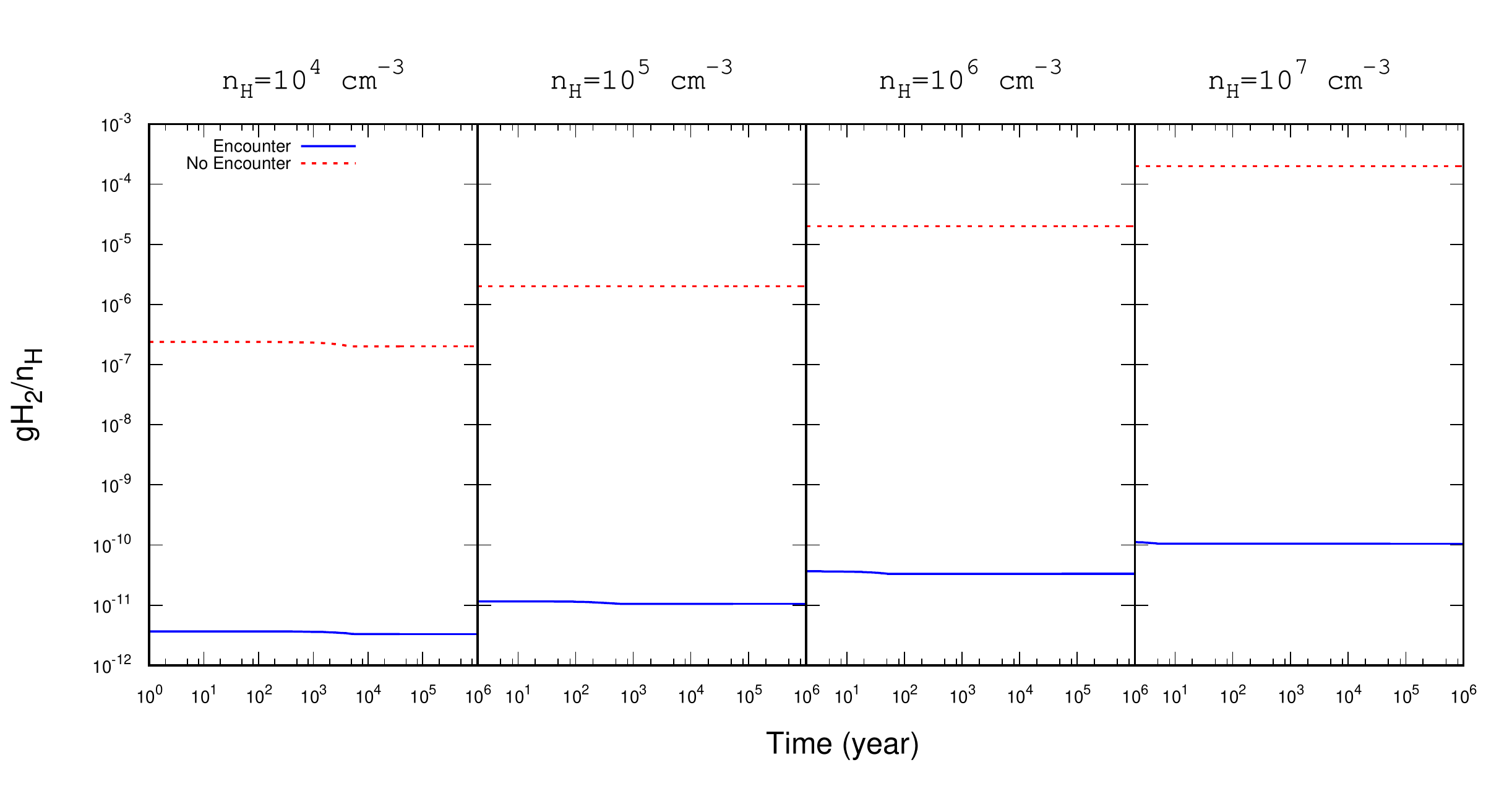}
    \caption{Time evolution of $gH_2$ with $R=0.35$ and various $n_H$ ($10^4, \ 10^5,$ $10^6$, and $10^7$ cm$^{-3}$) are shown \citep{das21}. The effect of encounter desorption increases with the increase in density.}
    \label{fig:H2-abun-den}
\end{figure}

\begin{figure}
    \centering
    \includegraphics[width=0.5\textwidth,angle=-90]{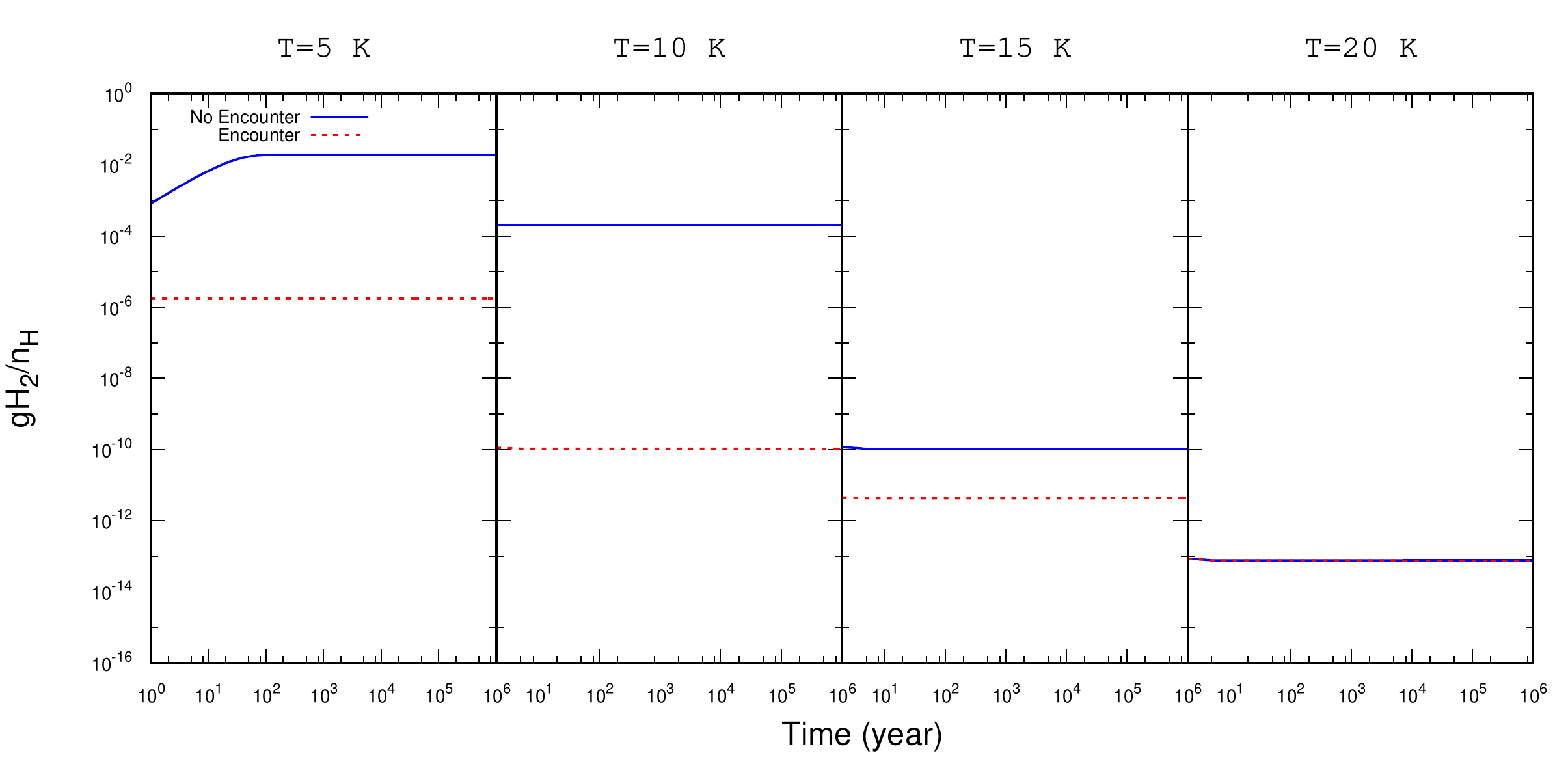}
    \caption{Time evolution of $gH_2$ with $R=0.35$, $n_H=10^7$ cm$^{-3}$ and various temperatures ($5,\ 10,\ 15$, and $20$ K) are shown \citep{das21}. The effect of encounter desorption decreases with the increase in temperature.}
    \label{fig:H2-abun-temp}
\end{figure}

Figure \ref{fig:H2-abun} shows the time evolution of $gH_2$ by considering $n_H=10^7$ cm$^{-3}$, $T=10$ K and $R = 0.35-0.80$.
Interestingly the abundance of $gH_2$ seems to be constant with the changes of $R$, whereas it strongly depends on $R$ in encounter desorption. This is because lower value of $R$ means a quicker hopping, whereas a  higher value represents a delayed hopping rate.
With the increase in $R$, $gH_2$ abundance raises for the encounter desorption case. However, it means that as we raise the value of $R$, the encounter desorption effect depreciates.
The left panel of Figure \ref{fig:H2-ratio} exposes that with the increase in $R$, a steady decrease in the ratio between  $gH_2$ abundance with no encounter desorption case (NE) and with the encounter desorption case (EN) is noticed. The probability of the encounter desorption is found to be inversely proportional to the rate of diffusion (Equation \ref{eqn:prob}) or hopping (Equation \ref{eqn:chan}). Since the increase in  $R$ induces faster diffusion and hopping, it lowers the encounter desorption probability of H$_2$ as expected.
Figure \ref{fig:H2-abun-den} shows the time evolution of $gH_2$ with the NE and EN for $R=0.35$, $T=10$ K and $n_H = 10^4 - 10^7$ cm$^{-3}$.
In both cases, the abundances of $gH_2$ increase with the density.
The middle panel of Figure \ref{fig:H2-ratio} shows that the gH$_2$ abundance ratio between NE and EN with density. It shows that the effect of encounter desorption is more pronounced for the higher density.
Figure \ref{fig:H2-abun-temp} shows that the $gH_2$ abundances when we use $n_H=10^7$, $R=0.35$, and $T=5-20$ K. In the right panel of Figure \ref{fig:H2-ratio}, we show the $gH_2$ abundance ratio obtained between NE and EN with the temperature changes.
The figures show that the effect of encounter desorption is maximum toward the lower temperature ($\sim 10$ K). It ceases around $20$ K. This is similar to the H$_2$ formation efficiency discussed in \cite{chak06a,chak06b} for olivine grain.
As the temperature decrease, mobility of H atoms decreases. Thus, the formation rate decreases. Conversely, with the increase in temperature, the hopping rate increases, increasing the formation efficiency, but simultaneously, the residence time of H atoms decreases, affecting the H$_2$ formation efficiency. As a result, the H$_2$ formation efficiency is maximum at around $\sim 10$ K, and the effect of the encounter desorption is pronounced at the peak hydrogen formation efficiency.

For a better understanding, the obtained abundances with $R=0.35$, $T=10$ K, and $\rm{n_H=10^7}$ cm$^{-3}$ are noted in Table \ref{tab:H2abunnew} at the end of the total simulation time ($\sim 10^6$ years).
\cite{chan21} considered the competition between hopping rate and desorption rate of H$_2$ (Equation \ref{eqn:chan}), whereas \cite{hinc15} considered the battle between the diffusion and desorption rate of H$_2$ (Equation \ref{eqn:prob}). This difference in consideration resulting $\sim$ two times higher abundance of $gH_2$ with the consideration of \cite{chan21} compared to \cite{hinc15} (see cases 2 and 3 of Table \ref{tab:H2abunnew} and Figure \ref{fig:H2-abun}).
Our quantum chemical calculation yields
$E_d(H_2,H_2)=67$ K, which is higher than it was used in the earlier literature value of $23$ K \citep{cupp07,hinc15,chan21}. The computed BE is further increased to $79$ K when we consider the IEFPCM model. Table \ref{tab:H2abunnew} shows that an increase in the BE ($E_d(H_2,H_2)=67$ K, and 79 K, for cases 4 and 5 of Table \ref{tab:H2abunnew}) results in sequentially higher surface coverage of $gH_2$ than it was with $E_d(H_2,H_2)=23$ K (case 3 of Table \ref{tab:H2abunnew}).
In case 5 of Table \ref{tab:H2abunnew}, we note the abundance of $gH_2$ in the absence of the encounter desorption effect, but a higher BE of H atom is used ($E_d(H,H_2O)=650$ K). Case 6 of Table \ref{tab:H2abunnew} also considers this BE of H atom along with $E_d(H_2,H_2)=67$ K, and the method of \cite{chan21} is used. A comparison between the abundance of $gH_2$ obtained with cases 4 and 6 (BEs of $gH$ are different) shows a marginal decrease in the abundance of $gH_2$ when a higher BE of $gH$ is used.

\subsubsection{$gH$}
The abundance of $gH$ is noted in Table \ref{tab:H2abunnew}. Its abundance is marginally decreased in \cite{chan21} compared to \cite{hinc15}. The use of higher $E_d(H_2,H_2)$ ($\sim 67$ K and $79$ K) lowers the abundance of $gH$ (compared to case 2). However,  $E_d(H)=650$ K can increase the $gH$ abundance by a couple of orders of magnitude (see case 7 of Table \ref{tab:H2abunnew}).

\subsubsection{$gH_2O,\ gCH_3OH$}
The encounter desorption effect on the other principal surface species ($gH_2O$ and $gCH_3OH$) is also shown in Table \ref{tab:H2abunnew}. The increase (percentage) in their abundances from the case where no encounter desorption is considered [for $E_d(H,H_2O)=450$ K and $650$ K, respectively] are noted in the bracketed term.
Table \ref{tab:H2abunnew} shows that the consideration of encounter desorption of H$_2$ can significantly change (decrease by $\sim 27-30\%$) the methanol abundance (cases 3 and 7) from cases 1 and 6 (i.e., no encounter desorption). However, the percentage changes in the surface abundance of water are minimal ($\sim \pm 1\%$) for the addition of the encounter desorption of H$_2$. These (increase or decrease) are highly dependent on the BE of H, temperature, density, and the value of $R$ ($\sim 0.35$ noted in Table \ref{tab:H2abunnew}).
The $E_d(H_2,H_2)$ from $23$ K to $67$ K can alter the methanol and water abundance on the grain. For example, Table \ref{tab:H2abunnew} shows that there is a significant increase ($\sim 15\%$) in the abundance of $gH_2O$ when higher BE ($E_d(H_2,H_2)=67$ K) is used (case 3 and case 4). However, this higher BE can marginally under-produce the methanol on the grain.
In brief, from Table \ref{tab:H2abunnew}, the encounter desorption may influence the chemical composition of the grain surfaces. These changes are highly dependent on the adopted BE with the water and H$_2$ ice and adopted physical parameters ($n_H,\ R,\ T$).

\begin{figure}
    \centering
    \includegraphics[width=0.4\textwidth,angle=-90]{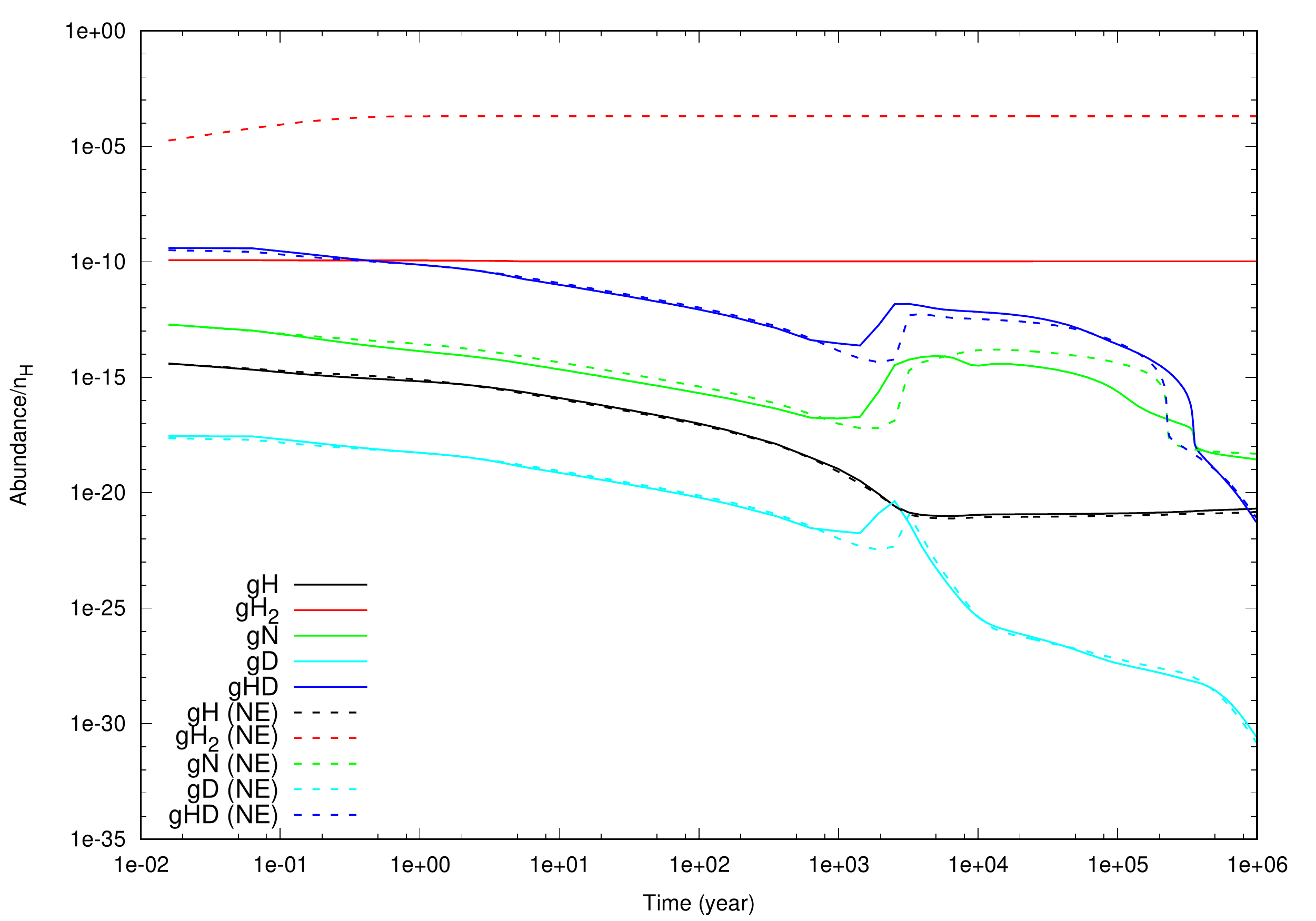}
    \caption{Time evolution of the abundances of H, H$_2$, D, HD, and N obtained from our simulation \citep{das21} is shown. Solid curves depict the cases by considering the encounter desorption [with $E_d(H_2,H_2)=67$ K] of H$_2$ and no encounter desorption (dashed curves) with $E_d(H,H_2O)=650$ K, $n_H=10^7$ cm$^{-3}$, $T=10$ K, and $R=0.35$.}
    \label{fig:abun}
\end{figure}

\begin{figure}
    \centering
    \includegraphics[height=7cm,width=6cm,angle=-90]{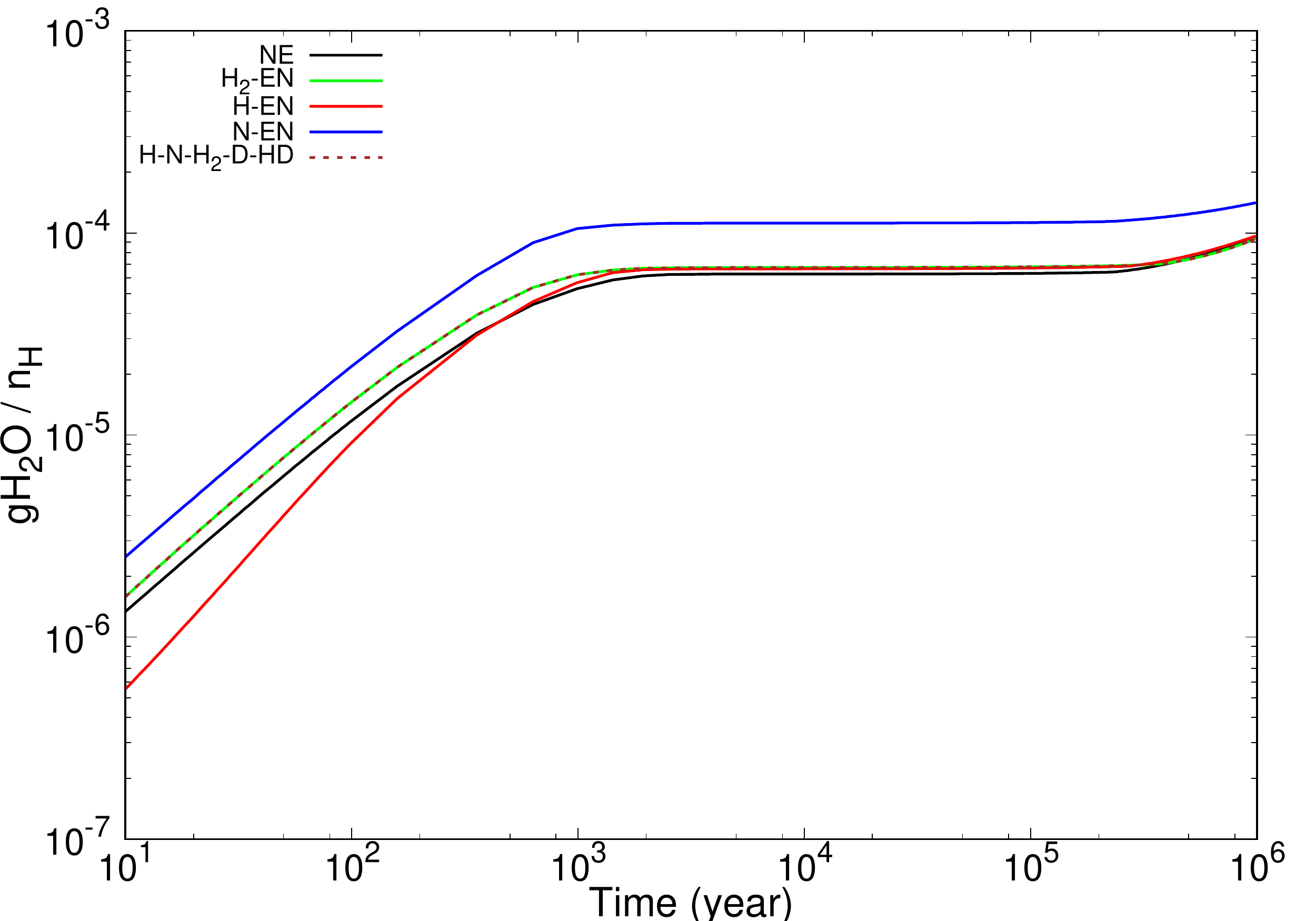}
    \includegraphics[height=7cm,width=6cm,angle=-90]{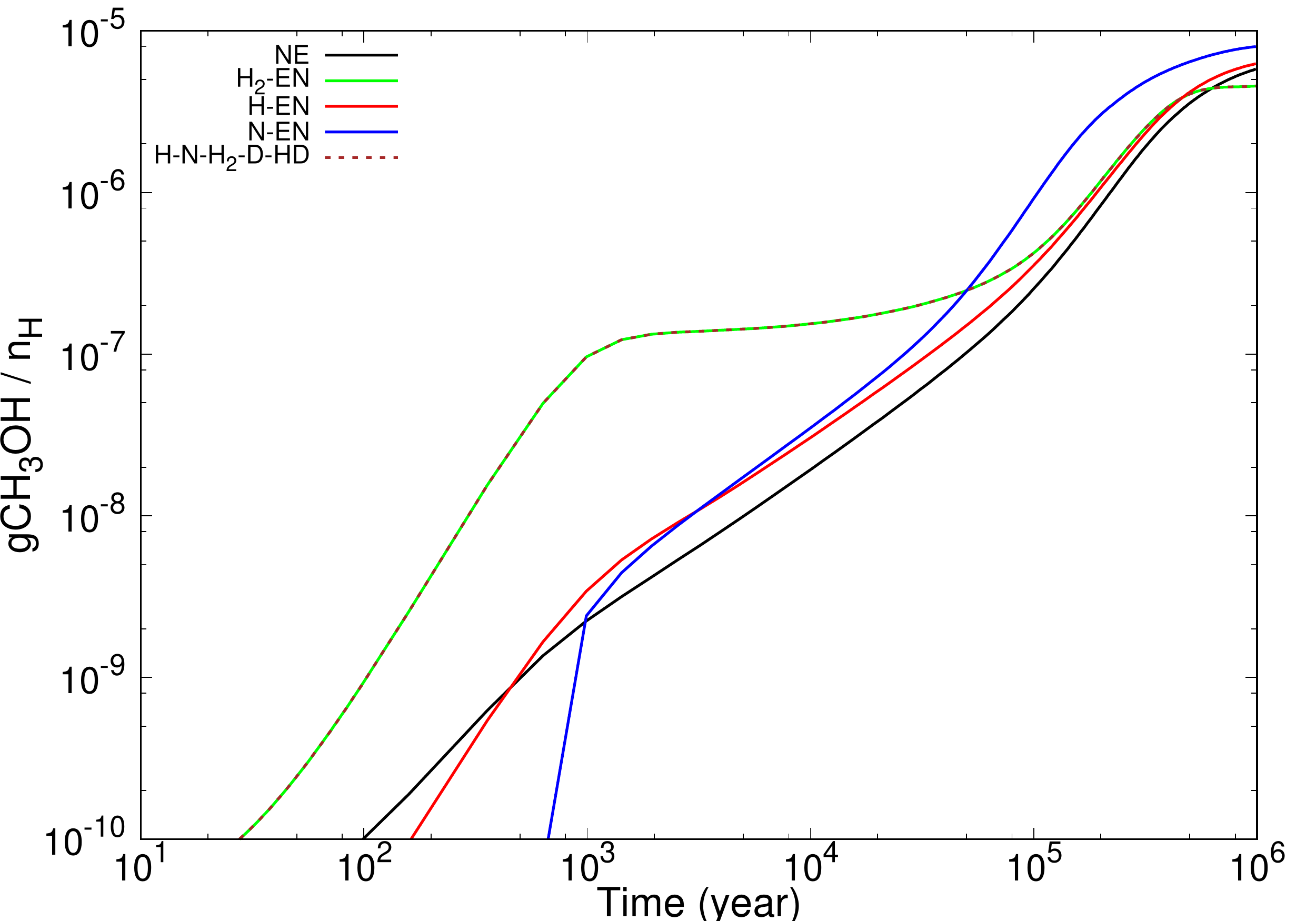}
    \includegraphics[height=7cm,width=6cm,angle=-90]{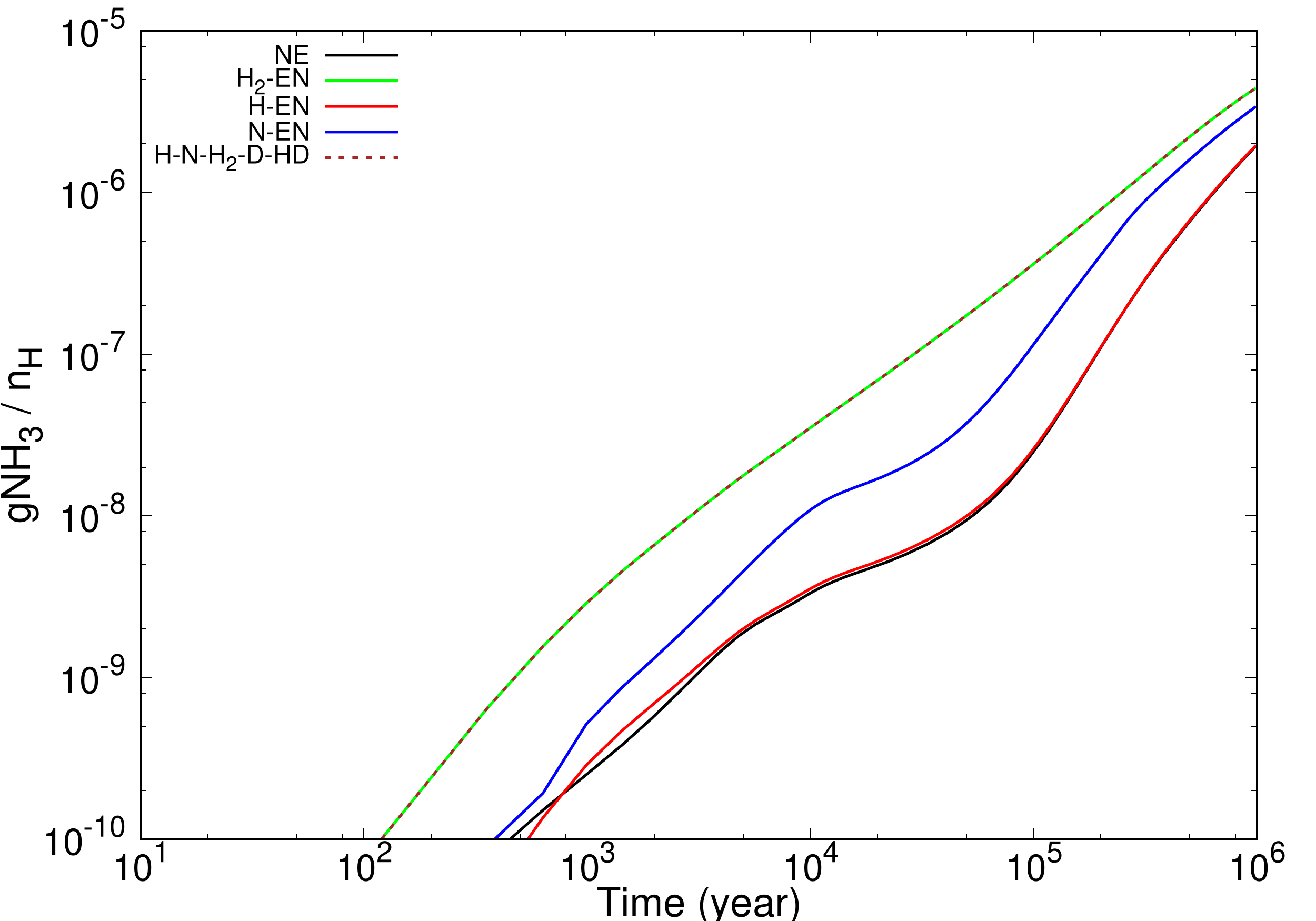}
    \caption{Time evolution of the abundances of ice-phase water (first panel), methanol (second panel) and ammonia (third panel) is shown for n$_H=10^7$ cm$^{-3}$, $T=10$ K, and $R=0.35$ \citep{das21}. A considerabale difference between the consideration of encounter desorption (solid green line for H$_2$, solid red line for H, and solid blue line for N) and without encounter desorption (black line) is shown. The encounter desorption of H, N, H$_2$, D, and HD are considered (brown dotted line) and depict that it marginally deviates from the encounter desorption of H$_2$.
    \label{fig:others}}
\end{figure}

\begin{figure}
    \centering
    \includegraphics[height=7cm,width=6cm,angle=-90]{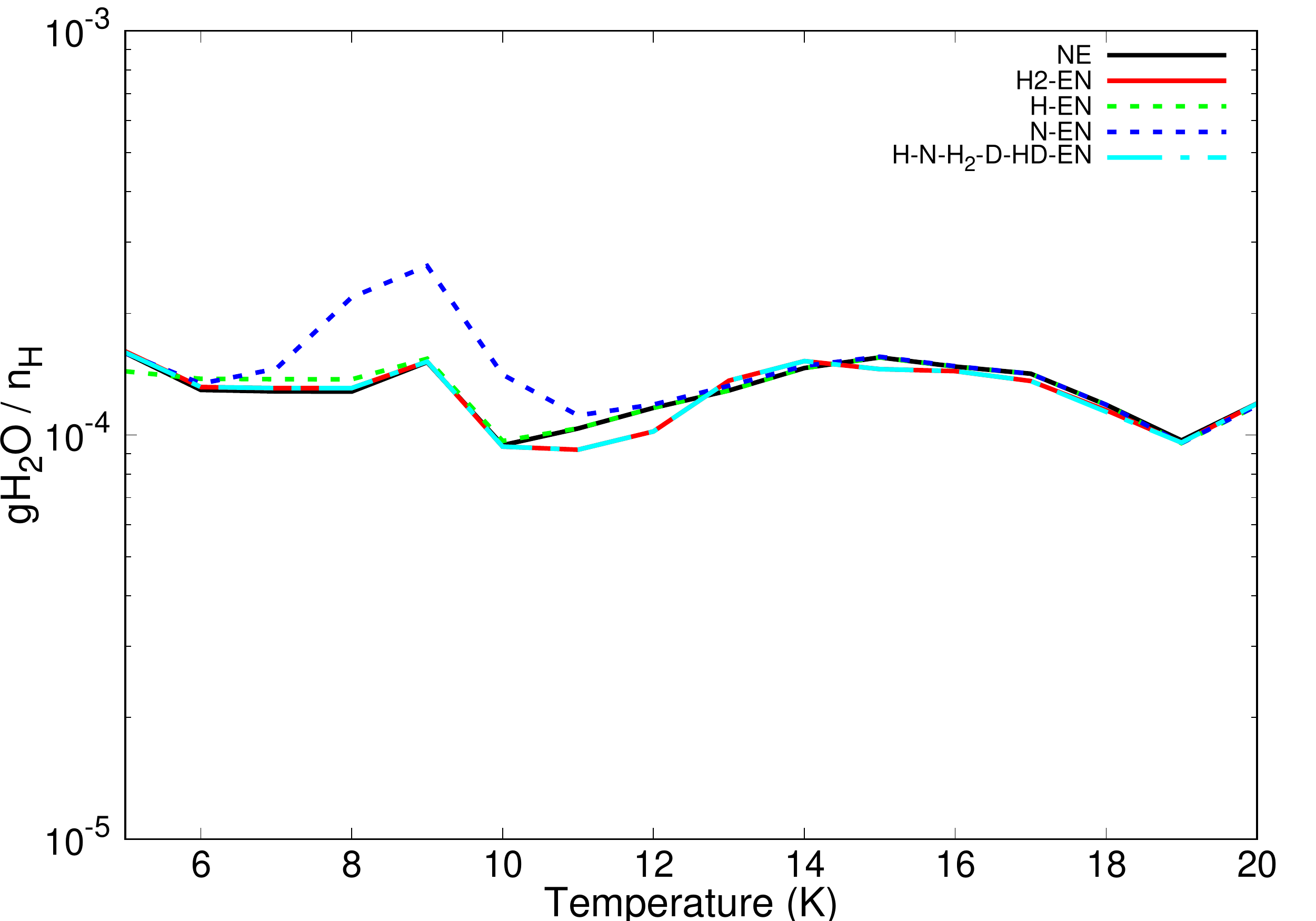}
    \includegraphics[height=7cm,width=6cm,angle=-90]{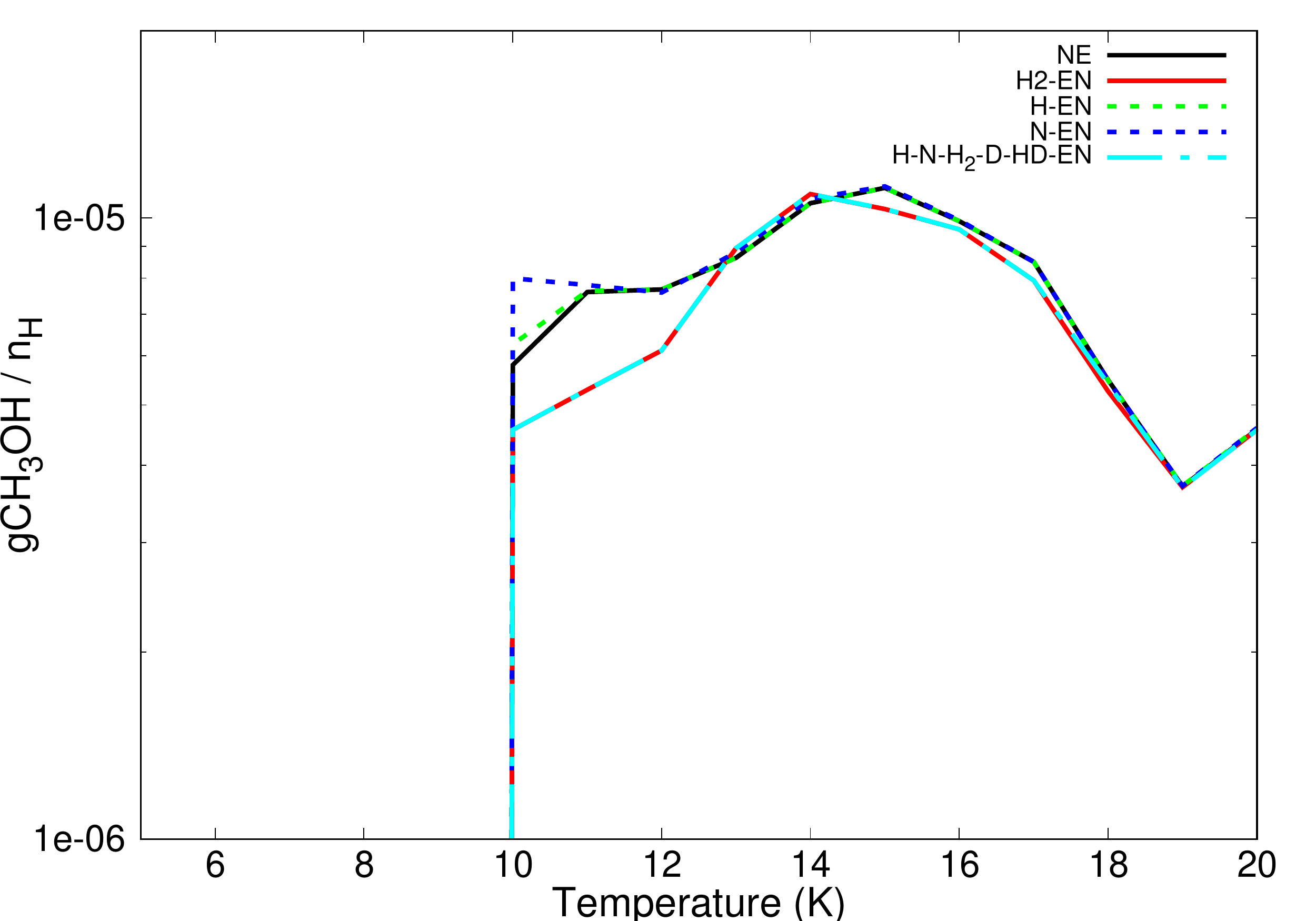}
    \includegraphics[height=7cm,width=6cm,angle=-90]{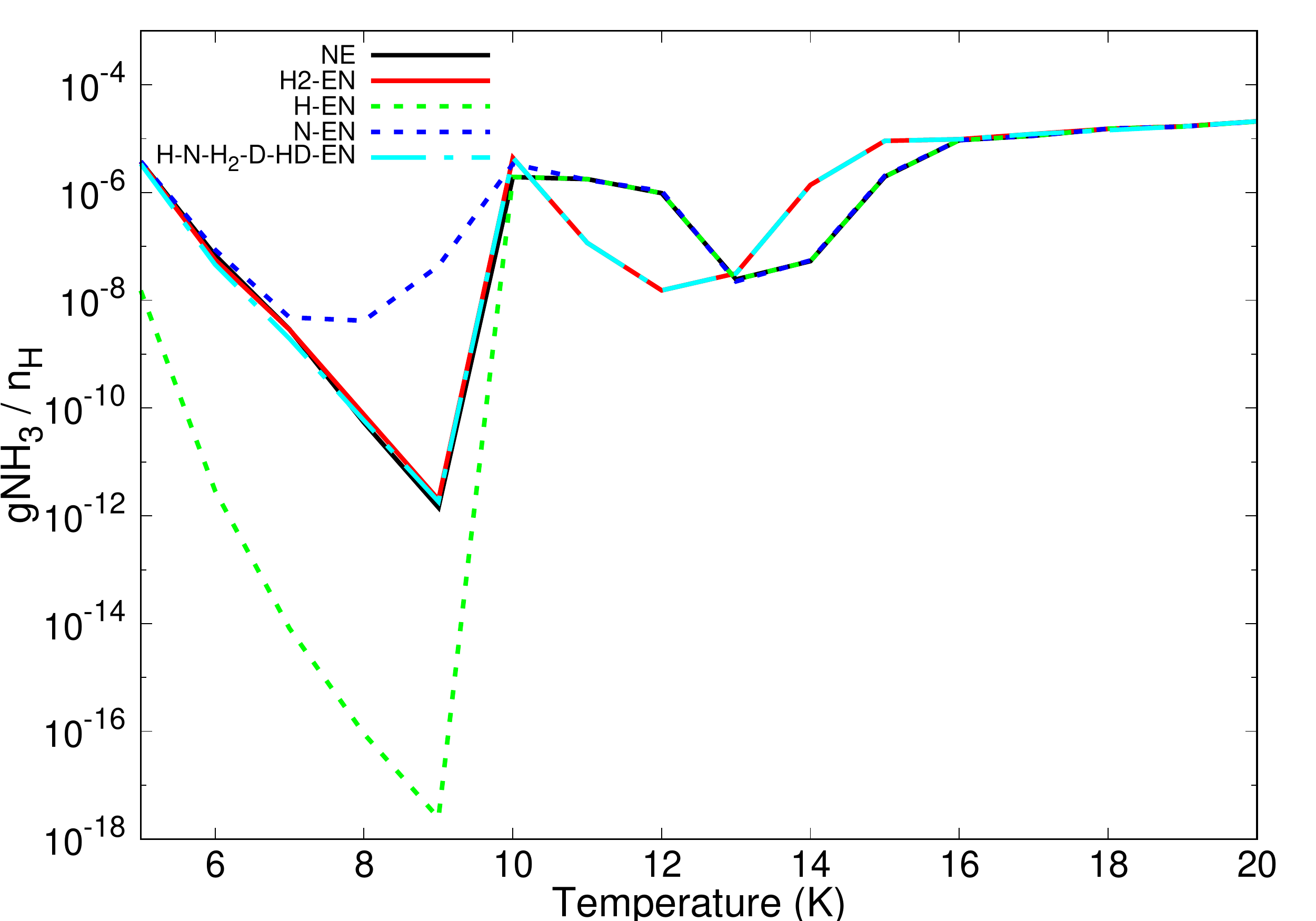}
    \caption{Temperature variation of the abundances of ice-phase water (first panel), methanol (second panel), and ammonia (third panel) is shown for $n_H=10^7$ cm$^{-3}$ and $R=0.35$ \citep{das21}. A significant variation between the consideration of encounter desorption and without encounter desorption (black line) is shown. The encounter desorption of H, N, H$_2$, D, and HD are collectively considered, and as like Figure \ref{fig:others}, it marginally varies from the encounter desorption of H$_2$.
    \label{fig:others-T}}
\end{figure}

\subsubsection{Encounter desorption of other species}
The notion of encounter desorption \citep{hinc15} primarily arose to eliminate the enhanced surface coverage of H$_2$ in the relatively denser and colder medium. Because H$_2$ has lower BE with the water surface ($\sim 440$ K), it could move fast on the surface and occupy a position on the top of another H$_2$ molecule. Comparatively, the chances of this occurrence enhance in the colder and denser region. Furthermore, since the H$_2$ molecule on H$_2$ has negligible BE  \citep[23 K used in][]{cupp07,hinc15}, it could easily desorb to the gas phase. All the other surface species can, of course, meet with H$_2$. Still, implementing this encounter desorption is essential when the species can occupy a position on the top of the H$_2$ molecule. For example, a carbon (C) atom has a BE of $10000$ K \citep{wake17}. H$_2$ could quickly meet one C atom on the grain surface, but due to the lower mobility of the C atom at a low temperature, H$_2$ will be on the top of the C atom every time. Since the whole $\rm{C-H_2}$ system is linked to the water substrate, this will not meet the encounter desorption probability.
Among the other key elements considered in this study, $gH,\ gN$, and $gF$ have the BE of 650 K \citep{wake17}, 720 K \citep{wake17}, and 800 K (listed in the original gas-grain code of Eric Herbst group in 2006 of OSU), respectively, with the water ice. Thus, it yields a reasonable timescale for hopping at a low grain temperature ($\sim 10$ K). However, since the initial elemental abundance of F is very small, we can ignore its contribution. 
The time scale of hopping is heavily dependent on the considered $R$.
For example, with $R=0.35$, at $10$ K, the hopping timescale for $gH$ and $gN$ is $1.12 \times 10^4$ years [with $E_d(H,H_2O)=650$ K] and $4.61 \times 10^{-3}$ years [with $E_d(N,H_2O)=720$ K], respectively. It changes to $1.9$ and $226$ years for H and N atoms, respectively, with $R=0.5$.
Since the usual lifetime of a dark cloud is $\sim 10^6$ years, the encounter desorption criterion is often satisfied. Among the di-atomic species, H$_2$ only has a faster-swapping rate (having BE $440$ K, which corresponds to a hopping time scale of $\sim 1.24 \times 10^{-7}$ and $9 \times 10^{-5}$ years, respectively, with $R=0.35$ and $R=0.5$). Considering the faster hopping rate and their abundances on the grain surface, we consider the encounter desorption of these species.
We consider $\rm{gX+gH_2 \rightarrow X + gH_2}$, where X refers to gas-phase H$_2$, H, and N.

In Figure \ref{fig:abun}, we show the time evolution of the abundances of $gH$, $gH_2$, $gN$, $gD$, and $gHD$ with $n_H=10^7$ cm$^{-3}$, $T=10$ K, and $R=0.35$. The encounter desorption of H$_2$ and without the encounter desorption effect are shown to show the differences.
Figure \ref{fig:abun} depicts that the abundances of $gN$, $gH$, and $gH_2$, have a reasonably high surface coverage. Since these species have a considerable hopping rate at a low temperature, desorption of these species needs to be considered. Here, we include the encounter desorption of these species sequentially to check their influence on some key surface species (gH$_2$O, gCH$_3$OH, and gNH$_3$). To check the effect of the other species, we gradually include the encounter desorption of H$_2$, H, and N. Figure \ref{fig:others} shows the time evolution of the encounter desorption of $gH_2O,\ gCH_3OH$, and $gNH_3$. We have already noted the encounter desorption of $gH_2$ in Section \ref{sec:H2-EN}. Figure \ref{fig:others} shows that when we include the encounter desorption of the H and N atoms, the time evolution of the abundances shows considerable changes in the abundance.
It depicts that the encounter desorption of N atom can substantially increase the abundances of $gH_2O,\ gCH_3OH$, and $gNH_3$ for the physical condition considered here ($n_H=10^7$ cm$^{-3}$, $T=10$ K, and $R=0.35$).
We further consider the encounter desorption of D and HD by considering the same BE as it obtained for H and H$_2$ with the H$_2$ substrate. The cumulative effect (considering the encounter desorption of H, H$_2$, N, D, and HD together) is shown with the dotted curve. We notice that the abundance profile with the cumulative effect shows a notable difference from the absence of the encounter desorption. But the cumulative effect minutely differs from the encounter desorption effect of H$_2$. 
Figure \ref{fig:others-T} shows the temperature variation of the final abundances of water, methanol, and ammonia to total hydrogen nuclei in all forms. It shows that the ice-phase abundances of methanol, water, and ammonia can strongly deviate from the no encounter desorption case. As in Figure \ref{fig:others}, we also see that the cumulative effect of the encounter desorption deviates from the encounter desorption of H$_2$. Around $20$ K, a great match between the cumulative encounter desorption case (dash-dotted cyan curve), H$_2$ encounter desorption case (solid red line), and no encounter desorption case (solid black line) is obtained. The right panel of Figure \ref{fig:H2-ratio} shows that for a  temperature beyond $10$ K, the effect of the encounter desorption of H$_2$ starts to decline. Around $20$ K, it roughly vanishes. Since the cumulative effect reflects the nature of H$_2$ encounter desorption, it also matches the no encounter desorption case at $\sim 20$ K.

\clearpage

\subsubsection{Summary}
We provide realistic BEs of several interstellar species considering H$_2$ monomer as a substrate.
Furthermore, our quantum chemical calculation finds a lower BE value considering H$_2$ monomer
as a substrate than it obtained earlier with the monomer (on average $\sim 7$ times),
tetramer (on average $\sim 9$ times), and hexamer (on average $\sim 11$ times)
water cluster as a substrate \citep{das18}.
$E_d(H_2,H_2)=23$ K \citep{cupp07,hinc15,chan21} and $E_d(H,H_2)=45$ K \citep{cupp07,chan21} are
used earlier in the literature. Our quantum chemical computations find an opposite trend with $E_d(H_2,H_2)=67$ K and $E_d(H,H_2)=23$ K, i.e., the BE of the H$_2$ molecule always remains higher than that of the H atom, supporting our earlier work \citep{sil17}. Supported with these reported  BE values, we further implement our CMMC code to check the encounter desorption effect of H$_2$, H, and N on the interstellar ices. The consideration of these updated BEs shows a significant deviation in the abundances of the grain surface species.
It suggests that the inclusion of the encounter desorption of the H, H$_2$, and N may influence the abundances of the principal surface constituents such as water, methanol, and ammonia. The cumulative effect almost resembles a similar abundance with that obtained considering the encounter desorption of H$_2$ only. The encounter desorption effect of H$_2$ diminishes for a bit higher temperature ($\sim 20$ K). Then, the encounter desorption of the cumulative cases matches precisely with the no encounter desorption case.

\clearpage

\section{Absorption features of interstellar ices in the presence of impurities}
Spectroscopic studies play a crucial role in identifying and analyzing interstellar ices and their structure. 
IR spectroscopy is used for identifying interstellar species, particularly in condensed phases. However, it requires that vibrations are IR-active.
The criterion is fulfilled when the dipole moment changes during vibration.
The IR feature of a water cluster is one of the primary tools to analyze the features of the aggregation processes in a water matrix \citep{ohno05,bouw07,ober07}.
The four vibrational modes of water (libration, bending, bulk stretching, and free OH stretching) are crucial to extract relevant information about the water cluster itself in various astrophysical environments \citep{gera95,ohno05,bouw07,ober07}.
Many molecules are identified within the interstellar ices either as pure, mixed, or layered structures.
Since water is the principal component of the interstellar ice matrix, the latter is considered composed of water molecules, with the other species being impurities or pollutants.
Absorption band features of water ice can significantly change with the presence of different types of impurities (e.g., CO, $\rm{CO_2,\ CH_3OH,\ H_2CO}$, etc.).
Many studies have been devoted to vibrational spectra of dilute aqueous solutions corresponding to the polar impurities. However, the intention is usually focused on the vibrational properties of the solute and not on the solvent \citep{choi11,capp11,blas13}. In a broader context, some IR studies are available for the whole solubility range of some species \citep{max03,max04}.

\subsection{Methodology}

There are three established structures of water ice (with a local density of $0.94$ g cm$^{-3}$) formed by vapor deposition at low pressure. Two of them are crystalline (hexagonal and cubic), and one is a low-density amorphous form \citep{jenn94}. High-density amorphous water ice (with a local density $1.07$ g cm$^{-3}$) also exists and can be formed by the vapor deposition at low temperatures \citep{jenn94}. \cite{prad12} experimentally analyzed the number of water molecules needed to generate the smallest ice crystal. According to that study, the appearance of crystallization is first observed for $275 \pm 25$ and $475 \pm 25$ water molecules, these aggregates showing the well-known band of crystalline ice around $3200$ cm$^{-1}$ (in the OH-stretching region). \cite{jenn94} found experimentally that the onset of crystallization occurs at $148$ K. Since we aim to validate our calculations for the low-temperature and low-pressure regime, we focus on the amorphous ices showing a peak in the IR spectrum at around $3400$ cm$^{-1}$. However, since it is not clear how many water molecules are necessary to mimic the amorphous nature of water ice, we consider pure water clusters of different sizes and study their absorption spectra. For this purpose, we optimize water clusters of increasing size at different levels of theory. The water clusters considered are 2H$_2$O (dimer), 4H$_2$O (tetramer), 6H$_2$O (hexamer), 8H$_2$O (octamer), and 20H$_2$O, with their structures being optimized with three different methods (B3LYP, B2PLYP, and QM/MM) as explained in the next computational details Section. The specific choice of these cluster models is based on experimental outcomes. Experimentally, it has been demonstrated that the water dimer has a nearly linear hydrogen-bonded structure \citep{odut80}. The water clusters with 4H$_2$O molecules are cyclic in the gas phase \citep{vian97}.  For the 6H$_2$O cluster, different structures are available: a three-dimensional cage in the gas phase \citep{liu97} and cyclic (chair) in the liquid helium droplet \citep{naut00}. For 8H$_2$O, an octamer cube is found in gaseous states \citep{ohno05}. Finally, the 20H$_2$O cluster
is considered to check the direct effect of the environment, with more details being
provided in the computational detail Section.

The harmonic frequencies are computed by using the optimized structures of the series of clusters above. The band strengths of the four fundamental modes are calculated by assuming the integration bounds, as shown in Table \ref{tab:integration_bounds}. Similar integration bounds (except for free OH stretching mode) were considered in \citet{bouw07} and \citet{ober07}. Similar absorption profiles of four fundamental modes of pure water are obtained from our calculations.
Their intensity, band positions, and strengths vary for the different cluster sizes and levels of theory used. It is thus essential to find the best compromise between accuracy and computational cost. It means understanding which is the smallest cluster and the cheapest level of theory
to provide a reliable description of water ice. To this aim, we compare the band positions and the corresponding band strengths of the four vibrational fundamental modes of water obtained with different cluster sizes and different methodologies to experimental work.

\begin{table}
\centering
\scriptsize
\caption{Integration bounds for the four fundamental modes of vibration \citep{gora20a}.}
\label{tab:integration_bounds}
\vskip 0.2 cm
\begin{tabular}{cccc}
 \hline
&&\multicolumn{2}{c}{\bf Integration bounds}\\
\cline{3-4}
{\bf Species} & {\bf Assignment} & {\bf Lower (cm$^{-1}$)} & {\bf Upper (cm$^{-1}$)}\\
 \hline
 & $\mathrm{\nu_{libration}}$&500 &1100\\
 & $\mathrm{\nu_{bending}}$&1100  &1900\\
H$_2$O& $\mathrm{\nu_{bulk-stretching}}$&3000&3600  \\
 &$\mathrm{\nu_{free OH-stretching}}$ &3600&4000\\
 \hline
\end{tabular}
\end{table}

While the outcome of this comparison will be discussed later in the text, here, we anticipate that the 4H$_2$O cluster in the c-tetramer configuration will be chosen as a water ice unit. To investigate the effect of impurities, we consider several impurity molecules that are added to obtain the desired ratio, as shown in Table \ref{tab:ice_mixture_composition}. For example, to get a $2:1$ ratio of water: impurity(x), we consider four water molecules hooked up with two ``$\rm{x}$'' molecules. However, for some systems, we needed to consider more water molecules to have more realistic features of the water cluster. Since it is known that the water ice clusters containing six $\rm{H_2O}$ molecules are the form of all-natural snow and ice on Earth \citep{abas05}, we also present a case with six H$_2$O molecules together as a unit. The cyclic hexamer (chair) configuration of the water cluster containing six $\rm{H_2O}$ is the most stable \citep{ohno05} and considered in our calculations.

\begin{table}
\centering
\scriptsize
\caption{Ice mixture composition details \citep{gora20a}.}
\label{tab:ice_mixture_composition}
\vskip 0.2 cm
\begin{tabular}{cccc}
 \hline
{\bf H$_2$O:X} & {\bf Total no.} & {\bf No. of} & {\bf No. of}\\
& {\bf of molecules} & {\bf water molecules} & {\bf pollutant molecules} \\
\hline
1:0.25 & 5 & 4 (80.0\%) & 1 (20.0\%) \\
1:0.50 & 6 & 4 (66.7\%) & 2 (33.3\%) \\
1:0.75 & 7 & 4 (57.1\%) & 3 (42.9\%) \\
1:1.00 & 8 & 4 (50.0\%) & 4 (50.0\%) \\
\hline
\end{tabular}\\
\vskip 0.2 cm
{\bf Note:} Contributions in percentage are provided in the parentheses.
\end{table}

\begin{figure}
\centering
\includegraphics[width=\textwidth]{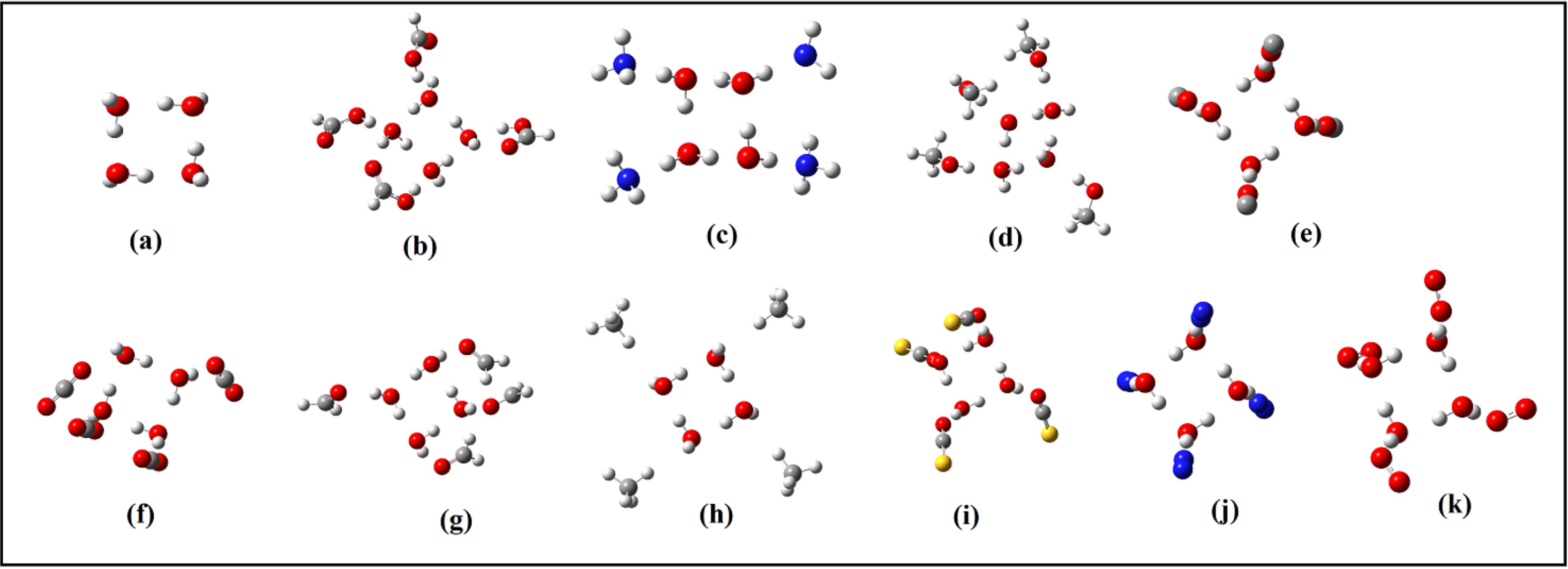}
\caption{Optimized structures for (a) pure water and for the $4:4$ concentration ratio: (b) $\rm{H_2O-HCOOH}$, (c)
$\rm{H_2O-NH_3}$, (d) $\rm{H_2O-CH_3OH}$,  (e) $\rm{H_2O-CO}$, (f) $\rm{H_2O-CO_2}$, (g) $\rm{H_2O-H_2CO}$, (h) $\rm{H_2O-CH_4}$,
(i) $\rm{H_2O-OCS}$, (j) $\rm{H_2O-N_2}$, and (k) $\rm{H_2O-O_2}$ clusters \citep{gora20a}.}
\label{fig:optimized_structure}
\end{figure}

In Figure \ref{fig:optimized_structure}a, we present the optimized water clusters for the c-tetramer configuration. The same structure was considered by \cite{ohno05} and others \citep{sil17,das18,nguy19}. Since four H atoms are available for interacting with the impurities through hydrogen bonds, in our
calculations, we can reach a $1:1$ ratio between the water and the impurity (i.e., we can reach up to $50\%$ concentration of the impurity in the ice mixture).

To understand the effect of impurities on the band strengths of the four fundamental bands considered, we calculate the area under the curve for each band for different mixtures of pure water and pollutants. The band strength is derived using the following relation \citep[introduced by][]{bouw07,ober07}
\begin{equation}
A_{H_2O:x=1:y}^{band} = \int_{band} I_{H_2O:x=1:y} \times \frac{A_{band}^{H_2O}}{ \int_{band} I_{H_2O}},
\end{equation}
where $A_{H_2O:x}^{band}$ is the calculated band strength of the vibrational water mode in the $1:y$ mixture, $I_{H_2O:x=1:y}$ is its integrated area, $A_{band}^{H_2O}$ is the band strength of the water modes available from the literature, and $\int_{band} I_{H_2O}$ is the integrated area under the vibrational mode for pure water ice. The experimental absorption band strengths of the three modes of pure water ice are taken from \citet{gera95}, who carried out measurements with amorphous water at $14$ K. The adopted values are $2\times10^{-16}$, $1.2\times10^{-17}$, and $3.1\times10^{-17}$ $\mathrm{cm \ molecule^{-1}}$, for the bulk stretching ($3280$ cm$^{-1}$), bending ($1660$ cm$^{-1}$), and libration mode ($760$ cm$^{-1}$), respectively. Our ab initio calculations refer to the temperature at $0$ K. To calculate the band strengths, we consider the strongest feature of that band. Since for the free OH stretching
mode, no experimental values exist, we consider the result $A_{free OH}^{H_2O}= 2.09 \times 10^{-17}$ and $2.52 \times 10^{-17}$ $\mathrm{cm \ molecule^{-1}}$ for the c-tetramer and hexamer water clusters, respectively.

\subsubsection{Computational details}

As already mentioned, quantum chemical calculations are performed to evaluate the changes of the absorption features of four different fundamental modes, namely, (i) libration, (ii) bending, (iii) bulk stretching, and (iv) free OH stretching of water in the presence of impurities (CO, $\rm{CO_2}$, $\rm{CH_3OH}$, $\rm{H_2CO}$, HCOOH, $\rm{CH_4}$, NH$_3$, OCS, $\rm{N_2}$, and $\rm{O_2}$). High-level quantum chemical calculations (such as CCSD(T) method and hybrid force field method) are proven to be the best suited for reproducing the experimental data \citep{puzz14a,baro15a}. However, due to the dimensions of our targeted species, these levels of theory are hardly applicable.

As already anticipated, different DFT functionals are tested. Most computations
are carried out using the B3LYP hybrid functional \citep{beck88,lee88} in conjunction with the 6-31G(d) basis set \citep[\textsc{Gaussian} 09 package;][]{fris13}. Some test computations are also performed by using the B2PLYP double-hybrid functional \citep{grim06} in conjunction with the m-aug-cc-pVTZ basis set \citep{papa09} in which the $d$ functions are removed on hydrogen atoms (m-aug-cc-pVTZ-$d$H). In this case, harmonic force fields are obtained employing analytic first and second derivatives \citep{bicz10} available in the \textsc{Gaussian} 16 suite of programs \citep{fris16}. The reliability and effectiveness of this computational model in the evaluation of vibrational frequencies and intensities are documented in several studies \citep[see, for example,][]{baro15b}. We also perform the anharmonic calculations (with the B3LYP/6-31G(d) level) for the $\rm{H_2O-CO}$ and $\rm{H_2O-NH_3}$ systems to check the impact of anharmonicity on the band strength profiles of the four water fundamental modes.

The spectral features of the astrophysical ices can alter active (direct) and passive (bulk) ways. Following a consolidated practice \citep{sand20}, to include the passive contribution of the bulk ice on the spectral properties of ice mixtures, we embed our explicit cluster in a continuum solvation field to represent local effects on the ice mixture. To this end, we resort to the IEF variant of the PCM \citep{toma05}. The solute cavity has been built by using a set of interlocking spheres centered on the atoms with the following radii (in \AA): $1.443$ for hydrogen, $1.925$ for carbon, $1.830$ for nitrogen, and $1.750$ for oxygen, each of them scaled by a factor of $1.1$, which is the default value in \textsc{Gaussian}. For the ice dielectric constant, bulk water ($\varepsilon=78.355$) is used, although any dielectric constant larger than $\sim 30$ would lead to very similar results.

\begin{figure}
\centering
\begin{minipage}{0.23\textwidth}
\includegraphics[width=\textwidth]{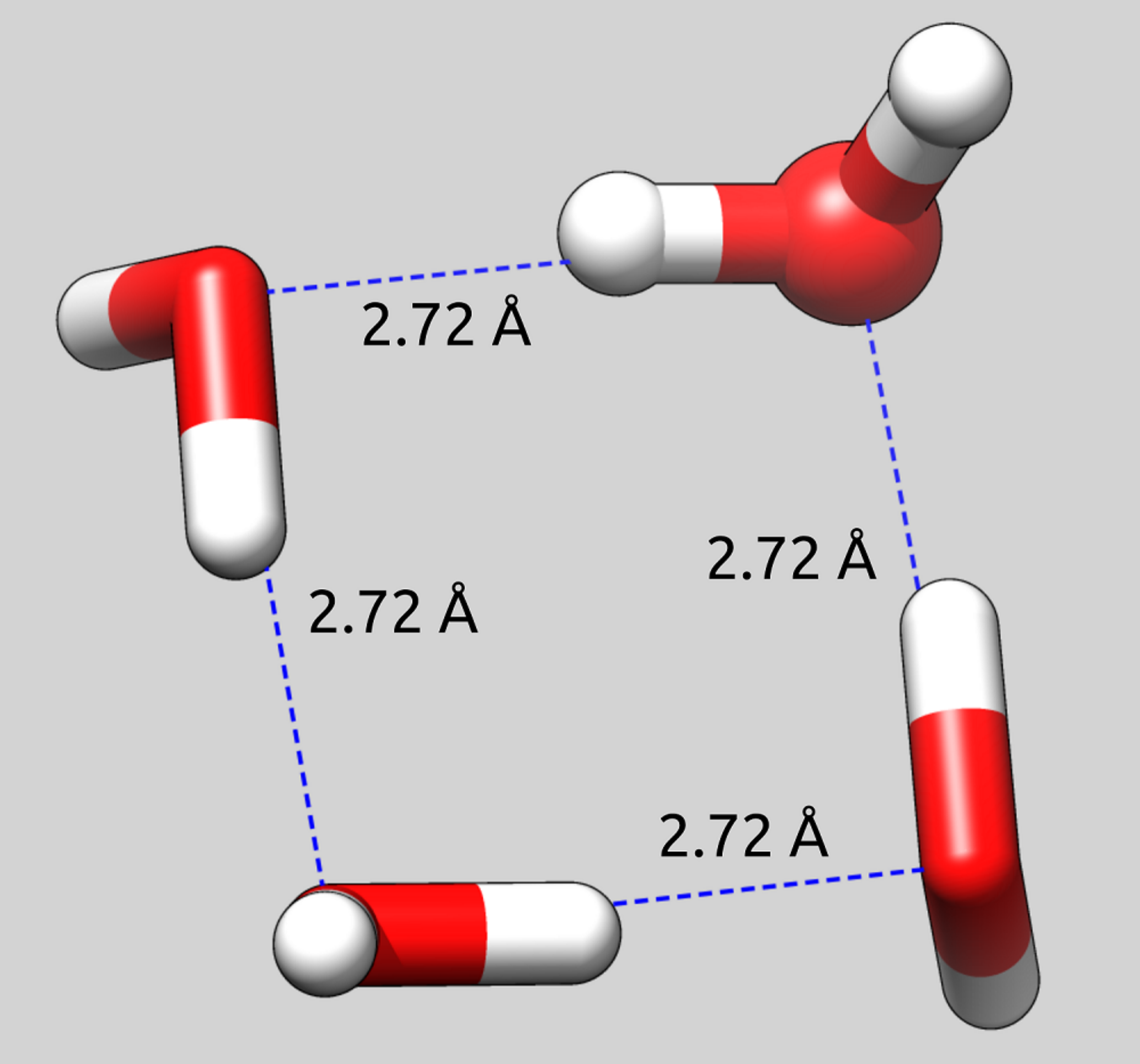}
\end{minipage}
\hskip 2cm
\begin{minipage}{0.50\textwidth}
\includegraphics[width=\textwidth]{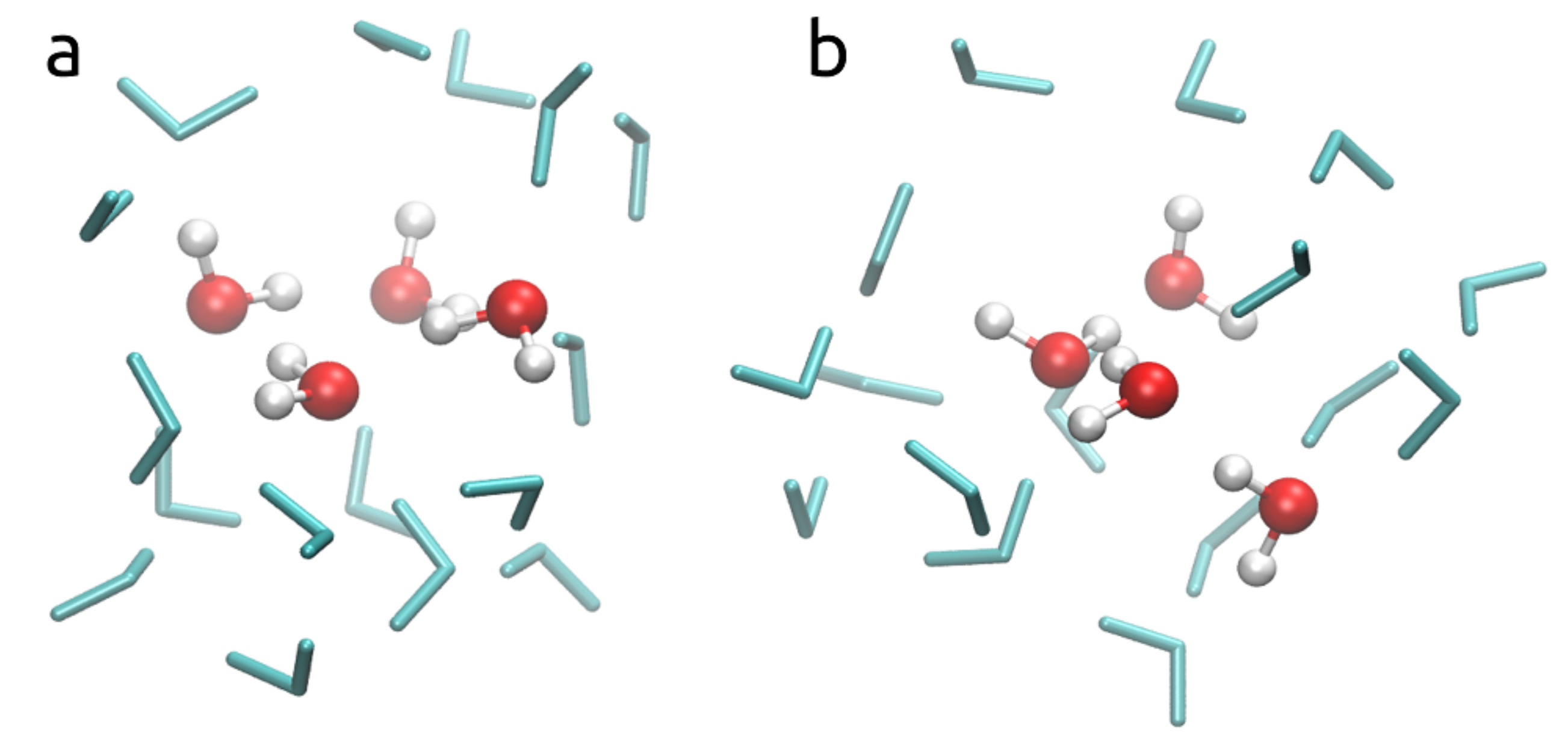}
\end{minipage}
\caption{ {\bf Left:} four-water system; the single QM water molecule is depicted in a ball- and-stick representation
    and the 3 MM molecules in a licorice representation; $\rm{O-H}$ distances are indicated too. It is to be noted that
    the four H$_2$O can be considered equivalent. {\bf Right:} innermost water molecules described at the QM level
    (ball and stick) and surrounding molecules described at the MM level (lines) for the 20H$_2$O system. (a) Configuration 1.
    (b) Configuration 2 \citep{gora20a}.}
\label{fig:qm-mm}
\end{figure}

In addition, we also perform QM/MM geometry optimizations of a pure water cluster containing 4H$_2$O molecules in which all but one molecule at the square vertexes are placed in the MM layer (see Figure \ref{fig:qm-mm}, left panel). A pure water cluster system containing 20H$_2$O molecules has also been considered. For this, we start from the coordinates of the full quantum mechanical (QM) optimization and select two alternative sets of four innermost molecules at the center of the cluster with a complete hydrogen bond network \citep[determined with a geometric criterion;][]{pagl17} with the first neighbor water molecules.
The remaining 16 molecules are described at the MM level (see Figure \ref{fig:qm-mm}, right panel).

It is difficult and highly time-consuming and disk/memory-consuming to accurately calculate
the structure and properties of large and complex molecular systems using pure quantum mechanical
(QM) methods. Among many efforts made to make such calculations feasible, one is the hybrid method,
otherwise known as a multiscale method. In this method, a large molecule is divided into multiple
fragments. Various theoretical methods, ranging from very expensive and accurate methods to less
expensive and accurate methods, are then applied to the different fragments. Thus, combining expensive QM and inexpensive classical molecular mechanics (MM) force fields was initiated \citep{honi71} and formalized in a ``generic'' QM/MM scheme \citep{wars76}, often called an ``additive'' scheme \citep{senn09}.
In this formalism, the entire molecular system is divided into two parts:
the model and environment systems. The small ``model'' part is the key chemically important layer.
It is treated with the accurate and expensive QM method.
The remaining environment (``env'') part is treated with the less accurate but more
efficient MM method.
In this scheme, as shown in Equation \ref{eqn:qm-mm}, the total energy of the whole system ($E_{QM/MM}$) is a sum of the energy of the model system by the QM method ($E_{QM}$), the energy of the environment system by the MM method ($E_{MM}$), and the interactions ($E_{QM-MM}$) between
the QM model system and the MM environment system.
\begin{equation} \label{eqn:qm-mm}
 E_{QM/MM}=E_{QM} + E_{MM} + E_{QM-MM}
\end{equation}
All QM/MM \citep{chun15} calculations are carried out with the \textsc{Gaussian} 16 \citep{fris16} code (rev. C01) using the hybrid B3LYP functional in conjunction with the 6-31 G(d) basis set. Atom types and force field parameters for water molecules in the MM layer are assigned according to the SPC-Fw flexible water model \citep{wu06}.
The choice is driven by (i) the necessity for a flexible, three-body classical water model and (ii) the accuracy with which the selected model reproduces ice I$_h$ properties. The solvent effects are considered by using PCM \citep{canc97}.
The vibrational analysis results from QM/MM calculations are provided in Tables \ref{tab:4H2O_MM-QM}, \ref{tab:20H2O_MM-QM_Conf_1}, and
\ref{tab:20H2O_MM-QM_Conf_2}.

\begin{table}
\scriptsize
\centering
\caption{Harmonic IR frequencies and intensities of the H$_2$O$^{QM}$ + 3 H$_2$O$^{MM}$ complex (Figure 2, left panel) \citep{gora20a}.}
\label{tab:4H2O_MM-QM}
\vskip 0.2 cm
\begin{tabular}{ccc}
    \hline 
    {\bf NM} &$\omega$ {[\bf cm$^{-1}$]} & {\bf IR Int. [km/mol]} \\
    \hline
    1 &      39.0097 &       3.5702 \\
    2 &      70.0478 &       0.9941 \\
    3 &     107.4593 &     248.8304 \\
    4 &     114.6669 &      81.5288 \\
    5 &     134.1712 &     121.9418 \\
    6 &     251.1220 &      19.7407 \\
    7 &     253.1088 &      18.8069 \\
    8 &     270.4875 &       7.7664 \\
    9 &     286.0773 &      10.8931 \\
    10 &     331.3394 &     156.5948 \\
    11 &     417.2017 &      11.4416 \\
    12 &     483.9590 &     107.2607 \\
    13$^t$ &     502.7496 &     211.3432 \\
    14$^t$ &     539.1224 &      59.4593 \\
    15$^t$ &     620.5928 &       2.9662 \\
    16$^t$ &     670.5374 &     209.1029 \\
    17$^t$ &     734.0061 &     476.6369 \\
    18$^t$ &     870.1429 &     201.9038 \\
    19$^b$ &    1407.2265 &     191.4554 \\
    20$^b$ &    1422.8362 &     445.8394 \\
    21$^b$ &    1456.6860 &      64.4541 \\
    22$^b$ &    1764.1323 &     143.3391 \\
    23$^s$ &    3399.8219 &     161.5281 \\
    24$^s$ &    3533.5836 &     119.6470 \\
    25$^s$ &    3561.1437 &     374.8984 \\
    26$^s$ &    3579.9423 &     110.7922 \\
    27$^f$ &    3643.1130 &      84.3280 \\
    28$^f$ &    3647.6874 &     450.2363 \\
    29$^f$ &    3651.3875 &      90.6885 \\
    30$^f$ &    3797.0095 &      37.5965 \\
\hline
\end{tabular} \\
\vskip 0.2 cm
{\bf Note:} $^t$OH torsion; $^b$OH scissoring; $^s$OH stretching; $^f$free OH
\end{table}

\begin{table}
\scriptsize
\caption{Harmonic IR frequencies and intensities of the 
first 4 H$_2$O$^{QM}$+16 H$_2$O$^{MM}$ complex (see Figure \ref{fig:qm-mm}(a), right panel,  Configuration 1) 
evaluated at the B3LYP/6-31g(d) level \citep{gora20a}.}
\label{tab:20H2O_MM-QM_Conf_1}
\vskip 0.2 cm
\hskip -1.0 cm
\begin{tabular}{ccccccccc}
\hline
{\bf NM} &$\omega$ {[\bf cm$^{-1}$]} & {\bf IR Int. [km/mol]} & {\bf NM} &$\omega$ {[\bf cm$^{-1}$]} & {\bf IR Int. [km/mol]} & {\bf NM} &$\omega$ {[\bf cm$^{-1}$]} & {\bf IR Int. [km/mol]} \\
\hline  
            1 &    28.5053 &     1.6212 & 59 &   334.6995 &     4.4240 & 117$^{b}$ &  1406.8655 &   454.2350 \\
            2 &    32.4112 &     0.3452 & 60 &   347.3954 &     4.9640 & 118$^{b}$ &  1412.4834 &   218.9410 \\
            3 &    40.5682 &     1.9422 & 61 &   387.1876 &     8.8358 & 119$^{b}$ &  1415.0050 &   217.1485 \\
            4 &    42.6937 &     0.9658 & 62 &   426.8869 &    78.1914 & 120$^{b}$ &  1416.7673 &   553.3807 \\
            5 &    45.5125 &     2.8291 & 63 &   432.7664 &    47.4011 & 121$^{b}$ &  1423.8136 &   320.7421 \\
            6 &    47.0457 &     2.4796 & 64 &   449.0251 &    78.8810 & 122$^{b}$ &  1427.0557 &   239.2493 \\
            7 &    51.1545 &     1.5206 & 65 &   455.8202 &     9.0749 & 123$^{b}$ &  1431.3527 &   280.4839 \\
            8 &    53.0040 &     0.4852 & 66 &   458.4089 &    52.9014 & 124$^{b}$ &  1433.6066 &   107.3301 \\
            9 &    57.4113 &     4.9283 & 67 &   461.3395 &     7.1542 & 125$^{b}$ &  1436.1459 &   143.6834 \\
            10 &    58.9090 &     1.5004 & 68 &   468.0311 &    44.1926 & 126$^{b}$ &  1442.3811 &    59.3969 \\
            11 &    62.8440 &     3.3091 & 69 &   476.0688 &     5.2943 & 127$^{b}$ &  1456.0330 &    69.7821 \\
            12 &    64.6577 &     0.5053 & 70 &   483.8486 &    73.9300 & 128$^{b}$ &  1465.8850 &   261.4679 \\
            13 &    67.1132 &     0.4600 & 71 &   500.1090 &    30.3953 & 129$^{b}$ &  1481.4184 &   129.4706 \\
            14 &    71.0670 &     1.0092 & 72 &   500.9839 &    36.5412 & 130$^{b}$ &  1490.6689 &    54.4350 \\
            15 &    72.1009 &     5.5430 & 73 &   508.1366 &    20.0694 & 131$^{b}$ &  1759.3916 &   113.0751 \\ 
            16 &    78.5377 &     1.2279 & 74 &   512.3840 &    13.9454 & 132$^{b}$ &  1781.9431 &   109.2325 \\
            17 &    79.5809 &     2.8592 & 75 &   532.3995 &    46.0624 & 133$^{b}$ &  1805.1440 &   185.2649 \\
            18 &    84.0856 &     0.3164 & 76 &   535.4481 &     4.4789 & 134$^{b}$ &  1821.5031 &    81.3961 \\
            19 &    88.1866 &     0.5336 & 77 &   543.6866 &    60.5730 & 135$^{s}$ &  3277.1620 &  1061.3522 \\
            20 &    91.6244 &     1.1620 & 78 &   545.4049 &    57.7011 & 136$^{s}$ &  3379.4017 &  1450.1515 \\
            21 &    95.9146 &     1.9115 & 79 &   557.0745 &    41.0189 & 137$^{s}$ &  3415.3773 &   260.6407 \\
            22 &   131.1634 &   139.0258 & 80 &   560.6706 &    18.2338 & 138$^{s}$ &  3441.1055 &    54.0971 \\
            23 &   147.4509 &     6.2284 & 81 &   581.7986 &    25.4989 & 139$^{s}$ &  3501.0930 &   137.3189 \\
            24 &   155.1556 &     1.5936 & 82 &   585.3805 &   214.1458 & 140$^{s}$ &  3510.2084 &   280.1448 \\
            25 &   157.0881 &    12.4196 & 83 &   588.1325 &    72.1128 & 141$^{s}$ &  3515.4119 &   164.7163 \\
            26 &   163.8254 &     4.8492 & 84$^{t}$ &   593.3498 &     6.1937 & 142$^{s}$ &  3527.1563 &    98.4213 \\
            27 &   169.5773 &     3.8618 & 85$^{t}$ &   598.2759 &   204.0196 & 143$^{s}$ &  3532.0223 &   374.1860 \\
            28 &   173.1802 &    11.6779 & 86$^{t}$ &   620.7530 &   378.5822 & 144$^{s}$ &  3535.9755 &   331.7897 \\
            29 &   177.8697 &    22.1882 & 87$^{t}$ &   623.3143 &    58.9759 & 145$^{s}$ &  3539.5696 &   667.4359 \\
            30 &   179.4412 &     3.0508 & 88$^{t}$ &   629.3486 &   219.1103 & 146$^{s}$ &  3541.9090 &    42.2047 \\
            31 &   189.5523 &    12.2882 & 89$^{t}$ &   639.6399 &    56.7824 & 147$^{s}$ &  3546.9082 &   185.0747 \\
            32 &   193.8849 &    18.7990 & 90$^{t}$ &   642.7590 &   240.6043 & 148$^{s}$ &  3549.9374 &   209.7596 \\
            33 &   199.5633 &    30.2086 & 91$^{t}$ &   648.3193 &    65.9335 & 149$^{s}$ &  3555.5498 &   189.4891 \\
            34 &   203.7815 &    44.7590 & 92$^{t}$ &   658.9772 &    36.8759 & 150$^{s}$ &  3558.8959 &    76.1871 \\
            35 &   205.2272 &    29.8834 & 93$^{t}$ &   676.8776 &   176.4931 & 151$^{s}$ &  3563.2840 &   189.8014 \\
            36 &   210.7469 &    42.9633 & 94$^{t}$ &   679.7026 &   594.2500 & 152$^{s}$ &  3568.4433 &   217.3753 \\
            37 &   212.8473 &    18.5861 & 95$^{t}$ &   685.0436 &    47.0063 & 153$^{s}$ &  3575.4761 &    97.1563 \\
            38 &   215.6034 &   104.3096 & 96$^{t}$ &   690.6345 &   554.1850 & 154$^{s}$ &  3576.5391 &    86.9267 \\
            39 &   219.2947 &     9.1183 & 97$^{t}$ &   695.3495 &   353.6124 & 155$^{s}$ &  3579.7822 &    19.6952 \\
            40 &   221.2358 &   136.0590 & 98$^{t}$ &   699.7148 &   156.7689 & 156$^{s}$ &  3589.8658 &    38.6004 \\
            41 &   223.7398 &    14.6743 & 99$^{t}$ &   704.2320 &    32.3584 & 157$^{s}$ &  3590.2134 &    57.5165 \\
            42 &   226.9991 &    92.7095 & 100$^{t}$ &   721.7785 &   334.5258 & 158$^{s}$ &  3603.0421 &   149.2000 \\
            43 &   229.4667 &    11.8754 & 101$^{t}$ &   725.6449 &   111.7286 & 159$^{s}$ &  3618.1897 &   266.3486 \\
            44 &   231.1955 &   180.5173 & 102$^{t}$ &   728.6757 &   454.3782 & 160$^{s}$ &  3620.8200 &   224.9253 \\
            45 &   235.9253 &     6.3462 & 103$^{t}$ &   747.1205 &   168.5362 & 161$^{s}$ &  3622.0658 &   233.3283 \\
            46 &   239.8566 &   146.4113 & 104$^{t}$ &   753.5614 &   192.6962 & 162$^{s}$ &  3623.0520 &   397.9328 \\
            47 &   249.2394 &     2.9212 & 105$^{t}$ &   765.5413 &   275.9737 & 163$^{s}$ &  3624.0959 &    35.3011 \\
            48 &   254.8166 &    52.0837 & 106$^{t}$ &   766.1653 &   425.8765 & 164$^{s}$ &  3626.6346 &   276.1415 \\
            49 &   260.4354 &    12.2059 & 107$^{t}$ &   772.0067 &   356.3800 & 165$^{s}$ &  3630.3037 &   955.2942 \\
            50 &   270.0093 &    13.8498 & 108$^{t}$ &   803.0436 &   327.2798 & 166$^{s}$ &  3631.8424 &    16.4537 \\
            51 &   280.9175 &    13.8528 & 109$^{t}$ &   818.6290 &   277.6746 & 167$^{s}$ &  3633.8572 &   248.7481 \\
            52 &   284.5581 &     1.3055 & 110$^{t}$ &   839.8062 &   313.0187 & 168$^{f}$ &  3646.1348 &   108.4691 \\
            53 &   286.3104 &    31.7468 & 111$^{t}$ &   850.6360 &   671.1597 & 169$^{f}$ &  3650.9996 &   175.2420 \\
            54 &   307.5132 &     4.0551 & 112$^{t}$ &   929.4103 &   415.2180 & 170$^{f}$ &  3652.3329 &   128.0414 \\
            55 &   308.3710 &     4.8097 & 113$^{t}$ &   955.5738 &   180.7041 & 171$^{f}$ &  3652.8838 &   205.0487 \\
            56 &   314.4016 &    25.2868 & 114$^{t}$ &   966.2260 &    55.4464 & 172$^{f}$ &  3653.3844 &    33.2033 \\
            57 &   319.3857 &     8.4817 & 115$^{b}$ &  1401.1608 &   328.9992 & 173$^{f}$ &  3653.8800 &   205.2072 \\
            58 &   330.7679 &     1.3826 & 116$^{b}$ &  1404.9892 &   279.6521 & 174$^{f}$ &  3654.0193 &    86.2440 \\
\hline
\end{tabular} \\
\vskip 0.2 cm
{\bf Note:} $^t$OH torsion; $^b$OH scissoring; $^s$OH stretching; $^f$free OH.
\end{table}

\begin{table}
\scriptsize
\caption{Harmonic IR frequencies and intensities of the 
first 4 H$_2$O$^{QM}$+16 H$_2$O$^{MM}$ complex (Figure \ref{fig:qm-mm}(b), right panel,  Configuration 2) evaluated at the B3LYP/6-31g(d) level \citep{gora20a}.}
\label{tab:20H2O_MM-QM_Conf_2}
\vskip 0.2 cm
\hskip -1.0 cm
\begin{tabular}{ccccccccc}
\hline
{\bf NM} &$\omega$ {[\bf cm$^{-1}$]} & {\bf IR Int. [km/mol]} & {\bf NM} &$\omega$ {[\bf cm$^{-1}$]} & {\bf IR Int. [km/mol]} & {\bf NM} &$\omega$ {[\bf cm$^{-1}$]} & {\bf IR Int. [km/mol]} \\
\hline
         1 &    29.4016 &     1.2924 & 59 &   338.1423 &     9.0936 & 117$^{b}$ &  1404.8169 &   265.9060 \\
         2 &    34.4375 &     0.6290 & 60 &   366.0320 &     9.0877 & 118$^{b}$ &  1411.9292 &    63.9287 \\
         3 &    40.5656 &     1.6279 & 61 &   381.8881 &     2.8750 & 119$^{b}$ &  1414.1855 &   237.2331 \\
         4 &    44.5388 &     1.2326 & 62 &   425.0003 &    72.9104 & 120$^{b}$ &  1414.6760 &   480.1803 \\
         5 &    47.5478 &     0.5182 & 63 &   433.1753 &    46.5032 & 121$^{b}$ &  1424.0139 &   304.7906 \\
         6 &    49.2499 &     4.4372 & 64 &   449.7627 &    54.6515 & 122$^{b}$ &  1426.0145 &   322.1644 \\
         7 &    52.7161 &     1.1206 & 65 &   453.1267 &    32.8363 & 123$^{b}$ &  1430.3473 &   297.6835 \\
         8 &    53.7670 &     0.7604 & 66 &   454.2652 &    31.3116 & 124$^{b}$ &  1434.2284 &    77.0402 \\
         9 &    58.1400 &     3.8023 & 67 &   457.6722 &    30.6715 & 125$^{b}$ &  1442.9373 &   116.7008 \\
        10 &    62.8023 &     2.1220 & 68 &   461.6648 &    46.8946 & 126$^{b}$ &  1444.1623 &   105.5626 \\
        11 &    65.0382 &     1.4900 & 69 &   472.4824 &     9.8892 & 127$^{b}$ &  1459.2322 &    55.3003 \\
        12 &    65.4195 &     1.0334 & 70 &   485.9271 &    79.1355 & 128$^{b}$ &  1465.8496 &   252.0918 \\
        13 &    68.9622 &     1.6328 & 71 &   489.0285 &    21.2511 & 129$^{b}$ &  1482.2478 &   140.3077 \\
        14 &    71.1636 &     2.8146 & 72 &   504.9767 &    27.4262 & 130$^{b}$ &  1505.1507 &    42.7361 \\
        15 &    75.3263 &     4.0105 & 73 &   512.8253 &     6.2202 & 131$^{b}$ &  1775.8533 &    82.3784 \\
        16 &    77.6513 &     1.4318 & 74 &   520.2208 &    42.0235 & 132$^{b}$ &  1780.7273 &    68.7098 \\
        17 &    79.5630 &     1.9564 & 75 &   531.5264 &    37.2553 & 133$^{b}$ &  1801.6979 &   332.8605 \\
        18 &    86.9075 &     2.3428 & 76 &   534.9918 &     6.2757 & 134$^{b}$ &  1807.9136 &    66.8586 \\
        19 &    88.7674 &     0.0132 & 77 &   548.5225 &    54.6906 & 135$^{s}$ &  3312.7170 &   920.6313 \\
        20 &    92.6829 &     0.4316 & 78 &   551.0483 &    47.1921 & 136$^{s}$ &  3377.7305 &   741.7304 \\
        21 &    97.2538 &     2.3252 & 79 &   557.6667 &    22.4631 & 137$^{s}$ &  3421.1305 &   761.8361 \\
        22 &   132.9698 &   136.0213 & 80 &   567.0486 &    44.8651 & 138$^{s}$ &  3445.3287 &   168.4186 \\
        23 &   144.3282 &     7.0450 & 81 &   577.5475 &   247.0326 & 139$^{s}$ &  3485.4299 &   175.0443 \\
        24 &   150.9225 &     8.8323 & 82 &   582.7791 &   105.7169 & 140$^{s}$ &  3509.6340 &   116.0370 \\
        25 &   160.4446 &    11.5643 & 83 &   589.4278 &    39.6372 & 141$^{s}$ &  3525.2414 &    27.1764 \\
        26 &   163.9774 &    15.3603 & 84$^{t}$ &   593.7719 &    63.8280 & 142$^{s}$ &  3527.9472 &    11.8392 \\
        27 &   166.9931 &    18.7482 & 85$^{t}$ &   602.0192 &   207.8189 & 143$^{s}$ &  3529.5426 &   406.7703 \\
        28 &   174.8622 &     3.9430 & 86$^{t}$ &   607.5690 &   235.3652 & 144$^{s}$ &  3536.2221 &   608.4735 \\
        29 &   178.6797 &     9.5964 & 87$^{t}$ &   620.9062 &   128.9013 & 145$^{s}$ &  3539.1759 &   168.8773 \\
        30 &   186.2172 &     5.2561 & 88$^{t}$ &   630.0995 &   226.7652 & 146$^{s}$ &  3542.1608 &   517.6770 \\
        31 &   189.8802 &    22.8351 & 89$^{t}$ &   642.3136 &    46.5019 & 147$^{s}$ &  3544.0015 &   105.2415 \\
        32 &   196.0450 &    18.6789 & 90$^{t}$ &   646.4580 &   155.7202 & 148$^{s}$ &  3551.6796 &   422.3966 \\
        33 &   200.1828 &    70.0050 & 91$^{t}$ &   656.2139 &   245.2327 & 149$^{s}$ &  3556.2138 &   104.4402 \\
        34 &   201.5062 &     8.6800 & 92$^{t}$ &   659.7475 &    92.3092 & 150$^{s}$ &  3559.4252 &   140.9715 \\
        35 &   202.0414 &     8.0519 & 93$^{t}$ &   668.0591 &    88.0836 & 151$^{s}$ &  3562.5905 &   344.5012 \\
        36 &   208.0234 &    21.2606 & 94$^{t}$ &   678.1031 &   752.6557 & 152$^{s}$ &  3566.5079 &    71.2504 \\
        37 &   211.9547 &    44.9818 & 95$^{t}$ &   687.1452 &    86.4196 & 153$^{s}$ &  3568.8607 &    31.6257 \\
        38 &   216.4440 &   154.2248 & 96$^{t}$ &   702.2325 &   266.3036 & 154$^{s}$ &  3577.3367 &   111.0585 \\
        39 &   217.8326 &   186.4806 & 97$^{t}$ &   704.8812 &    97.3595 & 155$^{s}$ &  3577.8583 &   220.6935 \\
        40 &   218.1099 &    26.3216 & 98$^{t}$ &   711.5869 &   144.5530 & 156$^{s}$ &  3583.4466 &   183.5092 \\
        41 &   221.6619 &    28.3099 & 99$^{t}$ &   715.1892 &    66.1333 & 157$^{s}$ &  3590.2972 &    17.7387 \\
        42 &   224.3879 &    13.6861 & 100$^{t}$ &   725.8863 &   598.8691 & 158$^{s}$ &  3591.3690 &    14.5938 \\
        43 &   227.4102 &     2.3297 & 101$^{t}$ &   726.0798 &    18.1883 & 159$^{s}$ &  3611.7321 &     8.8719 \\
        44 &   229.1919 &    70.6417 & 102$^{t}$ &   730.7761 &   246.5783 & 160$^{s}$ &  3618.8579 &   235.6198 \\
        45 &   230.9956 &    94.8355 & 103$^{t}$ &   736.1656 &   277.1486 & 161$^{s}$ &  3623.1189 &   386.5676 \\
        46 &   243.1731 &   117.8833 & 104$^{t}$ &   744.0336 &   554.7832 & 162$^{s}$ &  3624.3597 &    98.4237 \\
        47 &   249.4187 &     0.5086 & 105$^{t}$ &   761.4750 &   271.4461 & 163$^{s}$ &  3627.3709 &   210.1150 \\
        48 &   253.8914 &    22.9344 & 106$^{t}$ &   775.7715 &   563.7792 & 164$^{s}$ &  3628.9811 &    88.7231 \\
        49 &   257.2356 &    35.1484 & 107$^{t}$ &   786.5512 &   479.5887 & 165$^{s}$ &  3630.5375 &   639.2414 \\
        50 &   267.5344 &    24.9850 & 108$^{t}$ &   805.5061 &   212.9173 & 166$^{s}$ &  3631.3807 &   548.9593 \\
        51 &   280.0676 &     6.8173 & 109$^{t}$ &   826.6848 &   229.7506 & 167$^{s}$ &  3632.6677 &   305.9009 \\
        52 &   285.7365 &    23.7669 & 110$^{t}$ &   845.7629 &   191.2681 & 168$^{f}$ &  3646.2015 &   105.3221\\
        53 &   296.6659 &    17.1441 & 111$^{t}$ &   887.6214 &   524.4212 & 169$^{f}$ &  3651.1285 &   183.1653 \\
        54 &   304.8541 &     3.0164 & 112$^{t}$ &   923.2407 &   210.8728 & 170$^{f}$ &  3652.3160 &   143.6466 \\
        55 &   308.4159 &     0.7582 & 113$^{t}$ &   950.0147 &   288.1648 & 171$^{f}$ &  3652.6720 &   169.0597 \\
        56 &   314.4300 &     7.5522 & 114$^{t}$ &  1014.9949 &   128.1520 & 172$^{f}$ &  3653.3819 &    24.7682 \\
        57 &   317.0437 &     9.2531 & 115$^{b}$ &  1401.2013 &   312.9877 & 173$^{f}$ &  3653.5148 &    97.5307 \\ 
        58 &   327.0967 &     1.0243 & 116$^{b}$ &  1403.0314 &   627.4395 & 174$^{f}$ &  3654.0199 &   236.1947\\
\hline
\end{tabular} \\
\vskip 0.2 cm
{\bf Note:} $^t$OH torsion; $^b$OH scissoring; $^s$OH stretching; $^f$free OH.
\end{table}

\clearpage
\subsection{Experimental methods} \label{sec:experiment}
Laboratory data are used from the literature whenever possible to constrain simulations \citep{bouw07,ober07}. In the cases of formic acid, ammonia, and methanol in water ice, new experiments are performed using the high-vacuum (HV) portable astrochemistry chamber (PAC) at the Open University (OU) in the United Kingdom. A detailed description of the system is described elsewhere \citep{dawe16}. Briefly, the main chamber is a commercial conflat flange cube (Kimball Physics Inc.) connected to a molecular turbopump (300 L/s), a custom-made stainless steel dosing line through an all-metal leak valve, a cold finger of a closed-cycle He cryostat (Sumitomo Cryogenics), and two ZnSe windows suitable for IR spectroscopy. During operation,  the base pressure in the chamber is in the $10^{-9}$ mbar range, and the base temperature of the cold finger is $20$ K. In thermal contact with the cryostat, the substrate is a ZnSe window (20 mm $\times$ 2 mm). A DT-670 silicon diode temperature sensor (LakeShore Cryotronics) is connected to the substrate to measure its temperature, while a Kapton flexible heater (Omegalux) is used to change its temperature. The diode and heater are both connected to an external temperature controller (Oxford Instruments).

Gaseous samples are prepared and mixed in a pre-chamber (dosing line) before being dosed into the main chamber through an all-metal leak valve. A mass-independent pressure transducer is used to control the amount of gas components mixed in the pre-chamber. Chemicals are purchased at Sigma-Aldrich with the highest purity available [HCOOH ($>95\%$), NH$_3$ ($99.95\%$), and CH$_3$OH ($99.8\%$)]. Ices are grown in situ by direct vapor deposition onto the substrate at normal incidence via a $3$ mm nozzle $20$ mm away from the sample. IR spectroscopy is performed in transmission mode using a Fourier transform infrared (FTIR; Nicolet Nexus 670) spectrometer with an external mercury cadmium telluride (MCT) detector. A background spectrum comprising $512$ co-added scans are acquired before deposition at $20$ K and used as a reference spectrum for all the spectra collected after deposition to remove all the IR signatures along the beam pathway that did not originate from the ice sample. Each IR spectrum is a collection of 256 co-added scans. The IR path is purged with dry compressed air to remove water vapor.

\subsection{Results and Discussion}

In this Section, firstly, the pure water ice will be addressed to establish the best compromise between accuracy and computational cost to describe the water ice unit cell. To this aim, we will resort to the comparison with the experiment. Then, we will move to the ice containing impurities. To further proceed with the validation of our protocol, water ice containing HCOOH, NH$_3$, CH$_3$OH, CO, and CO$_2$ as impurities will be investigated, thus exploiting the comparison between experiment and computations. This will also involve, as mentioned above, new measurements. Finally, in the last part, our protocol will be extended to the study of water ices with H$_2$CO, CH$_4$, N$_2$, and O$_2$ as
impurities.

\begin{table}
\tiny
{\centering
\caption{Absorption band strengths and band positions (within parentheses) of pure water ice$^a$ \citep{gora20a}.}
\label{tab:band_strength_pure_water}
\vskip 0.2 cm
\begin{tabular}{cccccc}
\hline
{\bf Vibration}&{\bf Experiment}& {\bf Dimer} & {\bf c-Tetramer} & {\bf c-Hexamar (chair)} & {\bf Octamer (cube)}\\
{\bf mode} & {\bf \cite{gera95}} & {\bf B3LYP-631G(d)} & {\bf B3LYP-631G(d)} & {\bf B3LYP-631G(d)} & {\bf B3LYP-631G(d)} \\
\hline
Libration &$3.1\times 10^{-17}$ (760)&$2.70 \times 10^{-17} (670)$&$2.40\times 10^{-17} (733)$&$1.17 \times 10^{-16}(870)$&$1.27 \times 10^{-17} (848)$\\
Bending   &$1.2\times 10^{-17}$ (1660)&$2.23 \times 10^{-17} (1710)$&$4.18\times 10^{-17} (1714)$&$4.62 \times 10^{-17} (1730)$&$6.85 \times 10^{-17} (1717)$\\
Stretching &$2.0\times 10^{-16}$ (3280)&$8.27 \times 10^{-17}(3540)$&$3.11\times 10^{-16} (3298)$&$5.53 \times 10^{-16} (3220)$&$3.89 \times 10^{-16} (3320)$\\
Free-OH & $-$                  & $1.35 \times 10^{-17} (3810)$&$2.10\times 10^{-17} (3775)$&$2.52 \times 10^{-17} (3780)$&$1.52 \times 10^{-17} (3788)$\\
\end{tabular}
\begin{tabular}{ccccc}
\hline
{\bf Vibration}&{Experiment}&  {\bf c-Tetramer} & {\bf 1H$_2$O(QM)+3H$_2$O(MM)} & {\bf 4H$_2$O(QM)+16H$_2$O(MM)}  \\
{\bf mode} & {\bf \cite{gera95}} & {\bf B2PLYP/m-aug-cc-pVTZ} & {\bf B3LYP-631G(d)} & {\bf B3LYP-631G(d)} \\
\hline
Libration &$3.1\times 10^{-17}$ (760) &$3.94 \times 10^{-17} (714)$&$7.91\times 10^{-17} (669)$&$4.9 \times 10^{-17} (720)$ \\
Bending   &$1.2\times 10^{-17}$ (1660)&$2.93 \times 10^{-17} (1635)$&$2.37\times 10^{-17} (1431)$&$1.88 \times 10^{-17} (1426)$ \\
Stretching &$2.0\times 10^{-16}$ (3280) &$2.97 \times 10^{-16}(3477)$&$2.68\times 10^{-17} (3593)$&$1.76\times 10^{-16} (3565)$ \\
Free-OH & ...                  & $2.40 \times 10^{-17} (3865)$&$6.24\times 10^{-18} (3797)$&$2.52 \times 10^{-17} (3636)$ \\
\hline
\end{tabular} \\
}
\vskip 0.2 cm
{\bf Note:} $^a$ Computed values in cm molecule$^{-1}$ and band position in cm$^{-1}$.
\end{table}

\begin{table}
\scriptsize
\caption{Comparison of band positions and band intensities as obtained by different quantum-chemical level of theories \citep{gora20a}.}
\label{table:comparison-different-water-cluster}
\vskip 0.2 cm
\hskip -1.0 cm
\begin{tabular}{ccccccccc}
\hline
&\multicolumn{2}{c}{4H$_2$O+PCM,}&\multicolumn{2}{c}{4H$_2$O+PCM,}&\multicolumn{2}{c}{1H$_2$O(QM)+3H$_2$O(MM)}
&\multicolumn{2}{c}{ 4H$_2$O(QM)+16H$_2$O(MM)} \\
&\multicolumn{2}{c}{B3PLYP/6-31G(d)}&\multicolumn{2}{c}{B2PLYP/m-aug-cc-pVTZ}&\multicolumn{2}{c}{+PCM, B3PLYP/6-31G(d)}&\multicolumn{2}{c}{+PCM, B3PLYP/6-31G(d)}\\
\cline{2-9}
{Assignment}&{Wavenumber} & {Intensity } &{Wavenumber} & {Intensity }&{ Wavenumber} & {Intensity } &{Wavenumber} & {Intensity}\\
&(cm$^{-1}$)&(km/mol)&(cm$^{-1}$) &(km/mol)&(cm$^{-1}$) &(km/mol)&(cm$^{-1}$)&(km/mol)\\
\hline
1&&&&&&&28.5053&1.6212\\
2&&&&&&&32.4112& 0.3452\\
3&&&39.3404 &0.0000&39.0097&3.5702&40.5682&1.9422\\
4&42.31&0.000&&&&&42.6937&0.9658\\
5&&&&&&&45.5125&2.8291\\
6&&&&&&&47.0457&2.4796\\
7&&&&&&&51.1545&1.5206\\
8&&&&&&&53.0040&0.4852\\
9&&&&&&&57.4113&4.9283\\
10&&&&&&&58.9090&1.5004\\
11&&&&&&&62.8440&3.3091\\
12&&&&&&&64.6577&0.5053\\
13&&&&&&&67.1132& 0.4600\\
14&&&&&70.0478&0.9941&71.0670&1.0092\\
15&&&&&&&72.1009&5.5430\\
16&&&&&&&    78.5377 &     1.2279\\
17&&&&&&  &79.5809 &     2.8592\\
18&&&85.5169 & 4.1241&& &84.0856 &     0.3164\\
19 &&&&&&&88.1866 &     0.5336\\
20 &&&&&&&91.6244 &     1.1620\\
21 &101.53&0.5953&&&107.4593&248.8304&95.9146 &     1.9115\\
22 &&&&&114.669&81.5288&131.1634 &   139.0258\\
23 &&&&&134.1712&121.9418&147.4509 &     6.2284\\
24 &&&&&&&155.1556 &     1.5936\\
25 &&&&&&            &   157.0881 &    12.4196\\
26 &&&&&&            &   163.8254 &     4.8492\\
27 &&&&&&            &   169.5773 &     3.8618\\
28 &&&&&&            &   173.1802 &    11.6779\\
29 &&&&&&            &   177.8697 &    22.1882\\
30 &&&&&&            &   179.4412 &     3.0508\\
31 &&&&&&             &   189.5523 &    12.2882\\
32 &&&&&&            &   193.8849 &    18.7990\\
33 &&&&&&       &199.5633 &    30.2086\\
34 &&&202.1393 & 0.0000&&  &   203.7815 &    44.7590\\
35 &&&&&& &   205.2272 &    29.8834\\
36 &&&&&&     &   210.7469 &    42.9633\\
37 &&&&&& &   212.8473 &    18.5861\\
38 &&&&&&  &   215.6034 &   104.3096\\
39 &&&&&& &   219.2947 &     9.1183\\
40 &&&222.9069&104.7500&&    &   221.2358 &   136.0590\\
41 &&&22.29069&104.7500&&  &   223.7398 &    14.6743\\
42 &&&&&&  &   226.9991 &    92.7095\\
43 &&&&&&  &   229.4667 &    11.8754\\
44 &&&231.0105&94.8938&& &   231.1955 &   180.5173\\
45 &&&&&& &   235.9253 &     6.3462\\
46 &243.67&0.000&240.0832&0.6734&&  &   239.8566 &   146.4113\\
47 &&&246.4688&287.8694&251.1220&19.7407&249.2394 &2.9212\\
48 &&&246.4688&287.8694&253.1088&18.8069&254.8166 &52.0837\\
49 &&&&&&&260.4354 &12.2059\\
50 &277.10&206.9799&274.1815&0.000&270.4875&7.7664&270.0093&13.8498\\
51 &277.10&206.9799&&&286.0773&10.8931&280.9175&13.8528\\
52 &281.08&0.3532&&&&&284.5581 &1.3055\\
53 &&&&&&&286.3104 &31.7468\\
54 &301.85&0.1654&&&&&307.5132 &4.0551\\
55 &307.74&298.2083&&&&&308.3710 &4.8097\\
56 &307.74&298.2083&&&&&314.4016 &25.2868\\
57 &&&&&&&319.3857 &8.4817\\
58 &327.52&0.0000&&&331.3394&156.5948& 330.7679 &     1.3826\\
59 &&&&&&&334.6995&4.4240\\
60 &&&&&&&347.3954&4.9640\\
\hline
\end{tabular}
\end{table}

\begin{table}
\scriptsize
\hskip -1.0 cm
\begin{tabular}{ccccccccc}
\hline
&\multicolumn{2}{c}{4H$_2$O+PCM,}&\multicolumn{2}{c}{4H$_2$O+PCM,}&\multicolumn{2}{c}{1H$_2$O(QM)+3H$_2$O(MM)}
&\multicolumn{2}{c}{ 4H$_2$O(QM)+16H$_2$O(MM)} \\
&\multicolumn{2}{c}{B3PLYP/6-31G(d)}&\multicolumn{2}{c}{B2PLYP/m-aug-cc-pVTZ}&\multicolumn{2}{c}{+PCM, B3PLYP/6-31G(d)}&\multicolumn{2}{c}{+PCM, B3PLYP/6-31G(d)}\\
\cline{2-9}
{Assignment}&{Wavenumber} & {Intensity } &{Wavenumber} & {Intensity }&{ Wavenumber} & {Intensity } &{Wavenumber} & {Intensity}\\
&(cm$^{-1}$)&(km/mol)&(cm$^{-1}$) &(km/mol)&(cm$^{-1}$) &(km/mol)&(cm$^{-1}$)&(km/mol)\\
\hline
61&&&398.6011&0.000&&&387.1876 & 8.8358\\
62&&&427.5608&44.2017&417.2017&11.4416&426.8869 &78.1914\\
63&&&&&&&432.7664&47.4011\\
64 &&&443.1204&79.6160&&&449.0251&78.8810\\
65 &&&443.1204&79.6160&&&455.8202 &9.0749\\
66 &&&&&&&458.4089 &    52.9014\\
67 &&&&&&&   461.3395 &     7.1542\\
68 &&&&&&&   468.0311 &    44.1926\\
69 &473.37&0.000&&&&&   476.0688 &     5.2943\\
70 &&&&&483.9590&107.2607&483.8486 &    73.9300\\
71 &&&&&&        &   500.1090 &    30.3953\\
72 &&&&&502.7496$^{t}$&211.3432&  500.9839 &    36.5412\\
73 &&&&&&            &   508.1366 &    20.0694\\
74 &&&&&&&512.3840&13.9454\\
75 &523.50&58.1147&&&&&532.3995&46.0624\\
76 &523.76&0.9980&&&539.1224$^{t}$&59.4593&535.4481&4.4789\\
77 &523.76&0.9980&&&&&543.6866&60.5730\\
78 &&&&&&          &   545.4049 &    57.7011\\
79 &&&&&&            &   557.0745 &    41.0189\\
80 &&&&&&           &   560.6706 &    18.2338\\
81&&&&&&             &   581.7986 &    25.4989\\
82 &&&&&&            &   585.3805 &   214.1458\\
83 &&&&&&           &   588.1325 &    72.1128\\
84$^{t}$ &&&&&&&593.3498 & 6.1937\\
85$^{t}$ &&&&&&&598.2759 &204.0196\\
86$^{t}$ &&&&&620.5928$^{t}$&2.9662&620.7530 &   378.5822\\
87$^{t}$ &&&&&&&623.3143 &58.9759\\
88$^{t}$ &&&&&&&629.3486 &219.1103\\
89$^{t}$ &&&&&&&639.6399 &56.7824\\
90$^{t}$ &&&&&&&642.7590 &240.6043\\
91$^{t}$ &&&&&&&648.3193 &65.9335\\
92$^{t}$ &&&&&&&658.9772 &36.8759\\
93$^{t}$ &&&&&670.5374$^{t}$&209.1029&676.8776 &   176.4931\\
94$^{t}$ &&&&&&&679.7026&594.2500\\
95$^{t}$ &&&683.4168&312.1539&&&685.0436 &47.0063\\
96$^{t}$  &&&&&&&690.6345&554.1850\\
97$^{t}$  &&&&&&&695.3495&353.6124\\
98$^{t}$  &&&&&&&699.7148&156.7689\\
99$^{t}$  &&&&&&&704.2320&32.3584\\
100$^{t}$ &&&&&&&721.7785&334.5258\\
101$^{t}$ &737.11$^{t}$&144.3682&&&734.0061$^{t}$&476.6369&725.6449 &   111.7286\\
102$^{t}$ &&&&&&&728.6757&454.3782\\
103$^{t}$ &&&&&&&747.1205&168.5362\\
104$^{t}$ &&&&&&&753.5614&192.6962\\
105$^{t}$ &&&&&&&765.5413&275.9737\\
106$^{t}$ &&&778.3957$^{t}$ &237.5051&&    &   766.1653 &   425.8765\\
107$^{t}$ &&&778.3957$^{t}$ &237.5051&&   &   772.0067 &   356.3800\\
108$^{t}$ &&&&&&     &   803.0436 &   327.2798\\
109$^{t}$ &&&&&&     &   818.6290 &   277.6746\\
110$^{t}$ &&&&&&   &   839.8062 &   313.0187\\
111$^{t}$ &894.95$^{t}$&342.8199&&&870.1429$^{t}$&201.9038&   850.6360 &   671.1597\\
112$^{t}$ &894.95$^{t}$&342.8199&934.3601$^{t}$ &0.000&&   &   929.4103 &   415.2180\\
113$^{t}$ &&&&&&  &   955.5738 &   180.7041\\
114$^{t}$ &1127.57$^{t}$&0.000&&&&    &   966.2260 &    55.4464\\
115$^{b}$ &&&&&&  &  1401.1608 &   328.9992\\
116$^{b}$ &&&&&&    &  1404.9892 &   279.6521\\
117$^{b}$ &&&&&1407.2265$^{b}$&191.4554&  1406.8655 &   454.2350\\
118$^{b}$ &&&&&&     &  1412.4834 &   218.9410\\
119$^{b}$ &&&&&&   &  1415.0050 &   217.1485\\
120$^{b}$ &&&&&&    &  1416.7673 &   553.3807\\
\hline
\end{tabular}
\end{table}

\begin{table}
\scriptsize
\hskip -1.0 cm
\begin{tabular}{ccccccccc}
\hline
&\multicolumn{2}{c}{4H$_2$O+PCM,}&\multicolumn{2}{c}{4H$_2$O+PCM,}&\multicolumn{2}{c}{1H$_2$O(QM)+3H$_2$O(MM)}
&\multicolumn{2}{c}{ 4H$_2$O(QM)+16H$_2$O(MM)} \\
&\multicolumn{2}{c}{B3PLYP/6-31G(d)}&\multicolumn{2}{c}{B2PLYP/m-aug-cc-pVTZ}&\multicolumn{2}{c}{+PCM, B3PLYP/6-31G(d)}&\multicolumn{2}{c}{+PCM, B3PLYP/6-31G(d)}\\
\cline{2-9}
{Assignment}&{Wavenumber} & {Intensity } &{Wavenumber} & {Intensity }&{ Wavenumber} & {Intensity } &{Wavenumber} & {Intensity}\\
&(cm$^{-1}$)&(km/mol)&(cm$^{-1}$) &(km/mol)&(cm$^{-1}$) &(km/mol)&(cm$^{-1}$)&(km/mol)\\
\hline
 121$^{b}$&&&&&1422.8362$^{b}$&445.8394&  1423.8136 &   320.7421\\
 122$^{b}$&&&&&&    &  1427.0557 &   239.2493\\
 123$^{b}$&&&&&&    &  1431.3527 &   280.4839\\
 124$^{b}$&&&&&&    &  1433.6066 &   107.3301\\
 125$^{b}$&&&&&&    &  1436.1459 &   143.6834\\
 126$^{b}$&&&&&&    &  1442.3811 &    59.3969\\
 127$^{b}$&&&&&1456.6860$^{b}$&64.4541 &  1456.0330 &    69.7821\\
 128$^{b}$&&&&&&    &  1465.8850 &   261.4679\\
 129$^{b}$&&&&&&    &  1481.4184 &   129.4706\\
 130$^{b}$&&&&&&    &  1490.6689 &    54.4350\\
 131$^{b}$&1711.28$^{b}$&251.8157&1630.3695$^{b}$&176.7723&1764.1323$^{b}$&143.3391&  1759.3916 &   113.0751\\
 132$^{b}$&1723.10$^{b}$&82.9912&1647.3535$^{b}$&74.0296&&&  1781.9431 &   109.2325\\
 133$^{b}$&1723.10$^{b}$&82.9911&1647.3536$^{b}$&74.0296&&&  1805.1440 &   185.2649\\
 134$^{b}$&1741.00$^{b}$&0.0000&1672.6698$^{b}$&0.000&&&1821.5031 &    81.3961\\
 135$^{s}$&3182.21$^{s}$&0.0000&&&& &3277.1620 &  1061.3522\\
 136$^{s}$&3297.07$^{s}$&1871.6907&&&&    &  3379.4017 &  1450.1515\\
 137$^{s}$&3297.07$^{s}$&1871.6907&3411.5900$^{s}$&0.000&3399.8219$^{s}$&161.5281&  3415.3773 &   260.6407\\
 138$^{s}$&3366.39$^{s}$&50.2255&3477.3996$^{s}$&1793.2478&&    &  3441.1055 &    54.0971\\
 139$^{s}$&&&3477.3996$^{s}$&1793.2478&&    &  3501.0930 &   137.3189\\
 140$^{s}$&&&&&&    &  3510.2084 &   280.1448\\
 141$^{s}$&&&&&&    &  3515.4119 &   164.7163\\
 142$^{s}$&&&3522.3300$^{s}$&2.0149&&  &  3527.1563 &    98.4213\\
 143$^{s}$&&&&&3533.5836$^{s}$&119.6470&  3532.0223 &   374.1860\\
 144$^{s}$&&&&&&&  3535.9755 &   331.7897\\
 145$^{s}$&&&&&&    &  3539.5696 &   667.4359\\
 146$^{s}$&&&&&&    &  3541.9090 &    42.2047\\
 147$^{s}$&&&&&&    &  3546.9082 &   185.0747\\
 148$^{s}$&&&&&&    &  3549.9374 &   209.7596\\
 149$^{s}$&&&&&&    &  3555.5498 &   189.4891\\
 150$^{s}$&&&&&&    &  3558.8959 &    76.1871\\
 151$^{s}$&&&&&3561.1437$^{s}$&374.8984    &  3563.2840 &   189.8014\\
 152$^{s}$&&&&&&    &  3568.4433 &   217.3753\\
 153$^{s}$&&&&&&    &  3575.4761 &    97.1563\\
 154$^{s}$&&&&&&    &  3576.5391 &    86.9267\\
 155$^{s}$&&&&&3579.9423$^{s}$&110.7922&  3579.7822 &    19.6952\\
 156$^{s}$&&&&&&    &  3589.8658 &    38.6004\\
 157$^{s}$&&&&&&    &  3590.2134 &    57.5165\\
 158$^{s}$&&&&&&    &  3603.0421 &   149.2000\\
 159$^{s}$&&&&&&    &  3618.1897 &   266.3486\\
 160$^{s}$&&&&&&    &  3620.8200 &   224.9253\\
 161$^{s}$&&&&&&    &  3622.0658 &   233.3283\\
 162$^{s}$&&&&&&    &  3623.0520 &   397.9328\\
 163$^{s}$&&&&&&    &  3624.0959 &    35.3011\\
 164$^{s}$&&&&&&    &  3626.6346 &   276.1415\\
 165$^{s}$&&&&&&    &  3630.3037 &   955.2942\\
 166$^{s}$&&&&&&    &  3631.8424 &    16.4537\\
 167$^{s}$&&&&&&    &  3633.8572 &   248.7481\\
 168$^{f}$&&&&&3643.1130$^{f}$&84.3280 &  3646.1348 &108.4691\\
 169$^{f}$&&&&&3647.6874$^{f}$&450.2363& 3650.9996 &175.2420\\
 170$^{f}$&&&&&3651.3875$^{f}$&90.6885&3652.3329 &128.0414\\
 171$^{f}$&3778.38$^{f}$&126.4545&3864.6871$^{f}$&144.7620 &&&3652.8838&205.0487\\
 172$^{f}$&3779.79$^{f}$&52.8022&3865.9912$^{f}$&138.6232&&&3653.3844&33.2033\\
 173$^{f}$&3779.79$^{f}$&52.8022&3865.9912$^{f}$&138.6232&&&3653.8800&205.2072\\
 174$^{f}$&3781.36$^{f}$	&0.0000&3867.1847$^{f}$&0.0000&3797.0095$^{f}$&37.5965&3654.0193&86.2440\\
\hline
\end{tabular}
\vskip 0.2cm
{\bf Note:}\\
$^t$OH torsion; $^b$OH scissoring; $^s$OH stretching; $^f$free OH. \\
For the conversion of km/mol to cm molecule$^{-1}$, intensity values need to be multiplied by a factor $\rm{1.6603\times10^{-19}}$.
\end{table}

\subsubsection{Part 1. Validation}

\subsubsection{Band strength of pure water}
\label{band_strength_pure_water}

\begin{figure}
\centering
\includegraphics[width=0.45\textwidth]{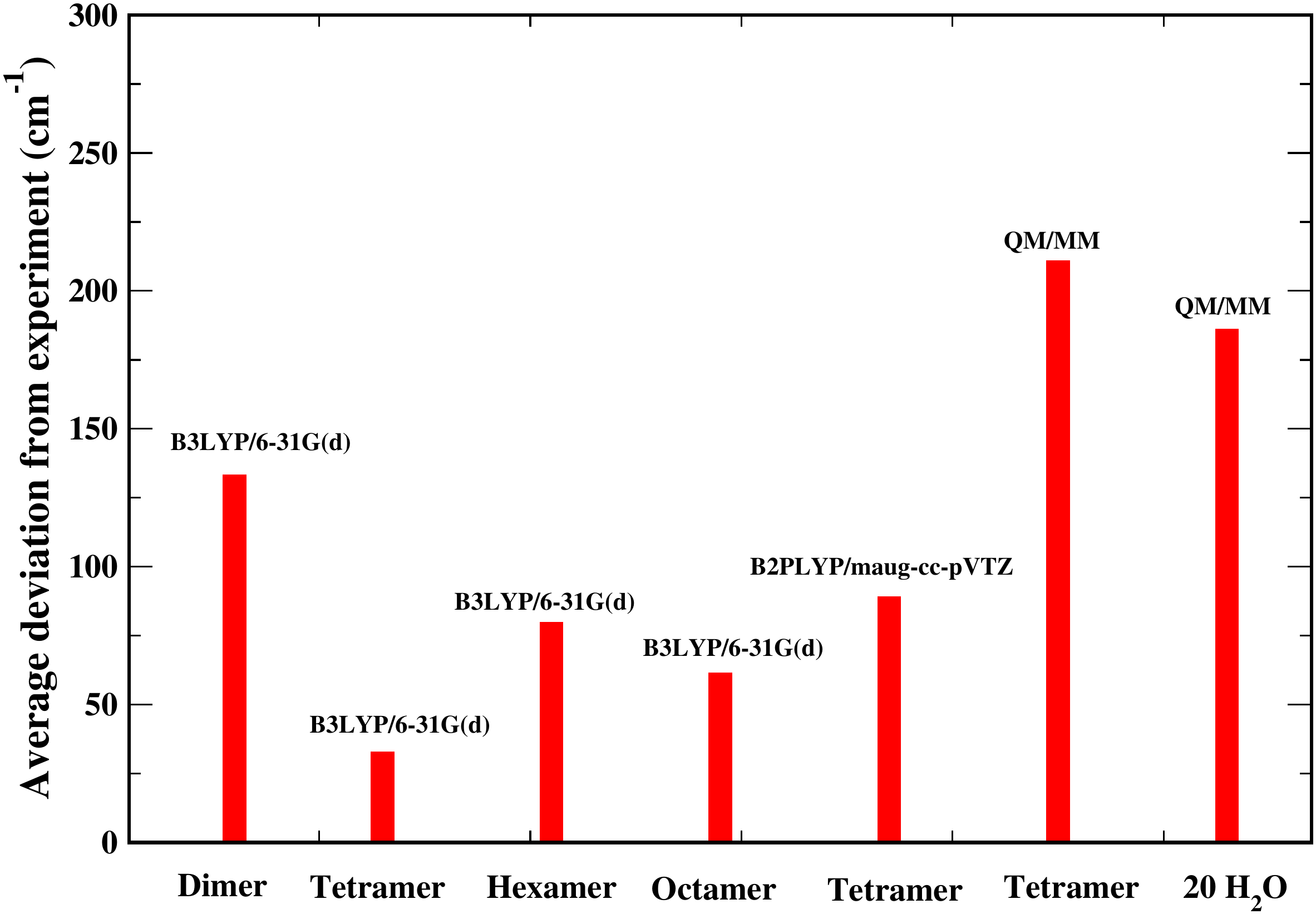}
\includegraphics[width=0.45\textwidth]{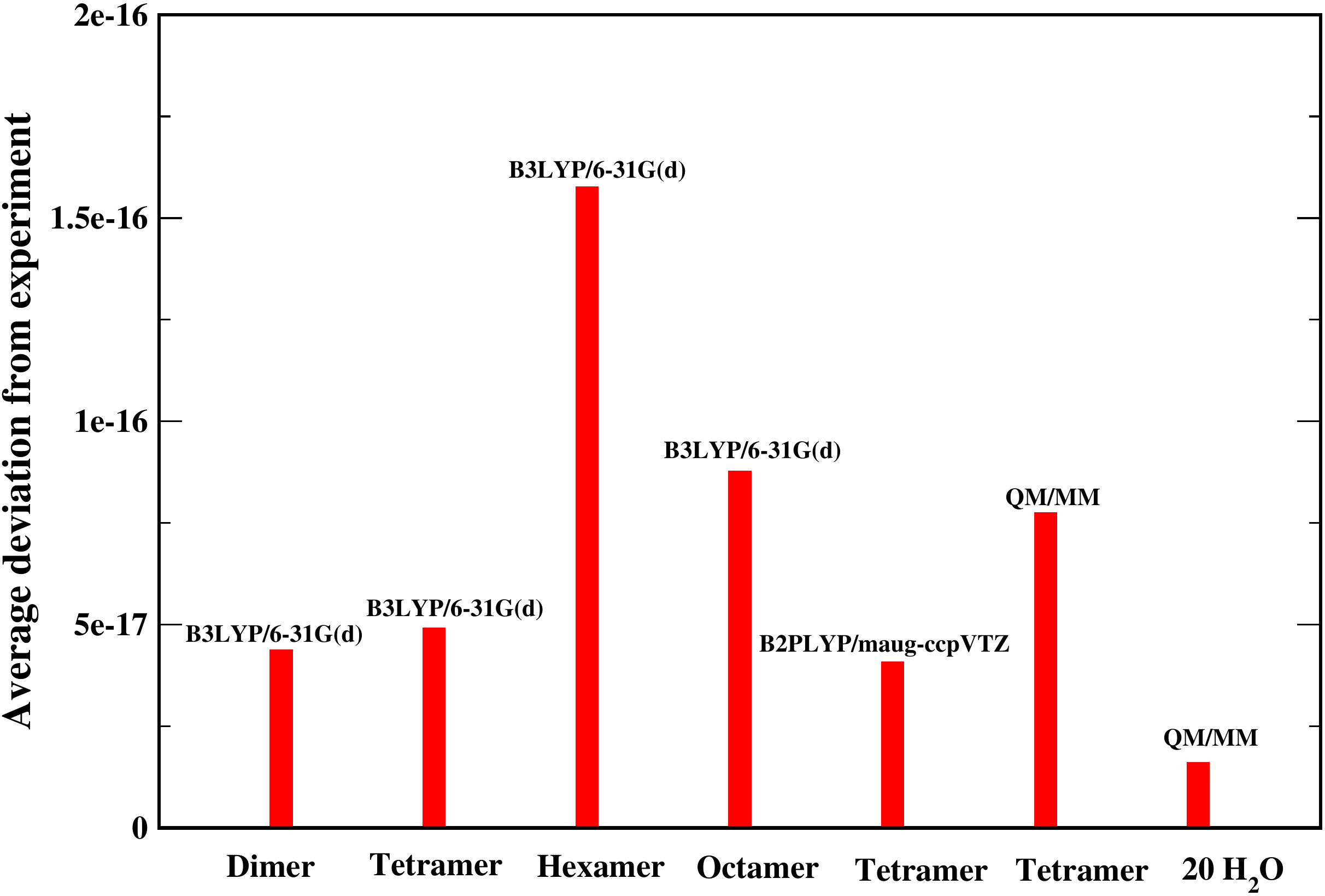}
\caption{Deviation of computed band positions (left panel) and band strengths (right panel) from experiments \citep{gora20a}.}
\label{fig:histo-h2o-cluster-compare}
\end{figure}

In Table \ref{tab:band_strength_pure_water}, the water band positions obtained with different methods and different sizes of the water cluster are compared with experimental data. Our computations provide several frequencies corresponding to a single mode of vibration. For the comparison, we report the computed frequencies of the four fundamental modes after convolving them with a Gaussian function with an adequate width \citep{lica15} (all transition frequencies are collected in Table \ref{table:comparison-different-water-cluster}). The comparison of Table \ref{tab:band_strength_pure_water} is graphically summarized in Figure \ref{fig:histo-h2o-cluster-compare}.
The left panel shows the average deviation of the band position of three fundamental modes of water (libration, bending, and stretching) from the experimental counterpart \citep{gera95}. Interestingly, the band positions obtained using the tetramer configuration and the B3LYP/6-31G(d) level of theory provide the best agreement. The right panel shows the average deviation of the band strengths
from experiments. QM/MM calculations for the 20 water molecule clusters (described in the computational details Section above) show the minimum deviation from experimental data. The results obtained for the tetramer configuration, both at the B3LYP and B2PLYP levels, also provide minor variations. Based on the comparison results, the B3LYP/6-31G(d) level of theory and the tetramer configuration are suitable combinations to describe the water cluster with a little computational cost.

\begin{table}
\tiny
\centering
\caption{Frequencies, integral absorbance coefficients, and normal modes of vibration of $\rm{H_2O-X}$ ($X =$ HCOOH, $\rm{NH_3}$, $\rm{CH_3OH}$, CO, $\rm {CO_2}$, $\rm{H_2CO}$, $\rm{CH_4}$, OCS, $\rm{N_2}$, and $\rm{O_2}$) \citep{gora20a}.}
\label{tab:H2O_X}
\vskip 0.2cm
\begin{tabular}{cccc}
 \hline
 \hline
{\bf  $\rm{H_2O:X}$} & {\bf Frequency at peak}& {\bf Integral absorbance} & {\bf Modes of} \\
&{\bf in cm$^{-1}$ ($\mu$m)} & {\bf coefficient in} & {\bf vibration} \\
&& {\bf cm molecule$^{-1}$} & \\
\hline
\hline
\multicolumn{4}{c} {\bf HCOOH} \\
\hline
\hline
1:0.25&875.47 (11.42)&$5.98\times10^{-17}$& OH torsion \\
&1698.42 (5.89)&$5.24\times10^{-17}$& H$_2$O scissoring \\
&3326.84 (3.00)&$1.88\times10^{-16}$&OH stretching\\
&3635.02 (2.75)&$4.11\times10^{-17}$&free OH\\
\hline
1:0.50&859.72 (11.63)&$6.18\times10^{-17}$& OH torsion \\
&1696.3 (5.90)&$5.37\times10^{-17}$& H$_2$O scissoring \\
&3341.67 (2.99)&$2.08\times10^{-16}$&OH stretching\\
&3632.99 (2.75)&$4.19\times10^{-17}$&free OH\\
\hline
1:0.75&878.28 (11.38)&$4.91\times10^{-17}$& OH torsion \\
&1688.62 (5.92)&$6.27\times10^{-17}$& H$_2$O scissoring \\
&3260.11 (3.07)&$2.46\times10^{-16}$&OH stretching\\
&3611.81 (2.77)&$6.46\times10^{-17}$&free OH\\
\hline
1:1.00&747.49 (13.38)&$9.21\times10^{-17}$& OH torsion \\
&1685.24 (5.93)&$7.56\times10^{-17}$& H$_2$O scissoring \\
&3343.53 (2.99)&$2.08\times10^{-16}$&OH stretching\\
&3608.22 (2.77)&$7.15\times10^{-17}$&free OH\\
\hline
\hline
\multicolumn{4}{c} {\bf NH$_3$} \\
\hline
\hline
1:0.25&886.69 (11.28)&$7.31\times10^{-17}$& OH torsion \\
&1717.36 (5.82)&$3.58\times10^{-17}$& H$_2$O scissoring \\
&3055.07 (3.27)&$2.20\times10^{-16}$&OH stretching\\
&3778.71 (2.65)&$1.16\times10^{-17}$&free OH\\
\hline
1:0.50&877.63 (11.39)&$8.92\times10^{-17}$& OH torsion \\
&1725.39 (5.80)&$3.28\times10^{-17}$&H$_2$O scissoring \\
&3127.37 (3.20)&$4.38\times10^{-16}$&OH stretching\\
&3737.41 (2.65)&$7.02\times10^{-18}$&free OH\\
\hline
1:0.75&859.21 (11.64)&$4.99\times10^{-17}$& OH torsion \\
&1730.52 (5.78)&$2.47\times10^{-17}$& H$_2$O scissoring \\
&3264.44 (3.06)&$4.27\times10^{-16}$&OH stretching\\
&3777.53 (2.65)&$5.89\times10^{-18}$&free OH\\
\hline
1:1.00&938.38 (10.66)&$9.35\times10^{-17}$& OH torsion \\
&1751.53 (5.71)&$4.43\times10^{-17}$& H$_2$O scissoring \\
&3270.01 (3.06)&$4.90\times10^{-16}$&OH stretching\\
&--&--&free OH\\
\hline
\hline
\multicolumn{4}{c} {\bf CH$_3$OH} \\
\hline
\hline
1:0.25&751.68 (13.30)&$5.60\times10^{-17}$& OH torsion \\
&1703.98 (5.87)&$3.85\times10^{-17}$& H$_2$O scissoring \\
&3252.5 (3.07)&$2.33\times10^{-16}$ &OH stretching\\
&3774.67 (2.65)&$1.73\times10^{-17}$&free OH\\
\hline
1:0.50&719.11 (13.91)&$7.67\times10^{-17}$& OH torsion \\
&1654.59 (6.04)&$3.62\times10^{-17}$& H$_2$O scissoring \\
&3207.15 (3.12)&$4.57\times10^{-16}$ &OH stretching\\
&3843.18 (2.60)&$2.26\times10^{-17}$&free OH\\
\hline
1:0.75&702.71 (14.23)&$5.49\times10^{-17}$& OH torsion \\
&1690.03 (5.92)&$2.92\times10^{-17}$& H$_2$O scissoring \\
&3351.12 (2.96)&$2.25\times10^{-16}$ &OH stretching\\
&3783.4 (2.64)&$1.72\times10^{-17}$&free OH\\
\hline
1:1.00&906.29 (11.03)&$6.61\times10^{-17}$& OH torsion \\
&1693.39 (5.90)&$2.96\times10^{-17}$& H$_2$O scissoring \\
&3295.67 (3.03)&$3.97\times10^{-16}$ &OH stretching\\
&3774.54 (2.65)&$2.03\times10^{-17}$&free OH\\
\hline
\hline
\multicolumn{4}{c} {\bf CO} \\
\hline
\hline
1:0.25 & 895.46 (11.17) & $5.70\times10^{-17}$ & OH torsion \\
& 1711.46 (5.84) & $4.14\times10^{-17}$ & H$_2$O scissoring \\
& 3290.66 (3.04) & $3.14\times10^{-16}$ & OH stretching\\
& 3791.25 (2.64) & $2.39\times10^{-17}$ & free OH\\
\hline
1:0.50 & 898.71 (11.13) & $5.64\times10^{-17}$ & OH torsion \\
&1711.73 (5.84) &$4.19\times10^{-17}$ & H$_2$O scissoring \\
&3285.68 (3.04) &$3.22\times10^{-16}$ & OH stretching\\
&3791.07 (2.64) &$2.42\times10^{-17}$ & free OH\\
\hline
1:0.75&898.22 (11.13)&$5.71\times10^{-17}$& OH torsion \\
&1713.47 (5.84)&$4.18\times10^{-17}$& H$_2$O scissoring \\
&3290.55 (3.04)&$3.16\times10^{-16}$ &OH stretching\\
&3789.19 (2.64)&$4.93\times10^{-17}$&free OH\\
\hline
1:1.00&899.44 (11.12)&$5.89\times10^{-17}$& OH torsion \\
&1715.94 (5.83)&$4.22\times10^{-17}$& H$_2$O scissoring \\
&3293.09 (3.04)&$3.13\times10^{-16}$ &OH stretching\\
&3789.31 (2.64)&$6.92\times10^{-17}$&free OH\\
\hline
\hline
\end{tabular}
\end{table}

\begin{table}
\tiny
\centering
\begin{tabular}{cccc}
 \hline
 \hline
{\bf  $\rm{H_2O:X}$} & {\bf Frequency at peak}& {\bf Integral absorbance} & {\bf Modes of} \\
&{\bf in cm$^{-1}$ ($\mu$m)} & {\bf coefficient in} & {\bf vibration} \\
&& {\bf cm molecule$^{-1}$} & \\
\hline
\hline
\multicolumn{4}{c} {\bf CO$_2$} \\
\hline
\hline
1:0.25&893.79 (11.19)&$5.63\times10^{-17}$& OH torsion \\
&1710.62 (5.84)&$4.23\times10^{-17}$& H$_2$O scissoring \\
&3288.08 (3.04)&$3.10\times10^{-16}$ &OH stretching\\
&3766.88 (2.65)&$3.85\times10^{-17}$&free OH\\
\hline
1:0.50&890.45 (11.23)&$5.72\times10^{-17}$& OH torsion \\
&1713.11 (5.84)&$4.22\times10^{-17}$& H$_2$O scissoring \\
&3278.84 (3.05)&$3.26\times10^{-16}$ &OH stretching\\
&3767.04 (2.65)&$4.21\times10^{-17}$&free OH\\
\hline
1:0.75&891.90 (11.21)&$5.98\times10^{-17}$& OH torsion \\
&1709.11 (5.85)&$4.20\times10^{-17}$& H$_2$O scissoring \\
&3294.04 (3.04)&$3.17\times10^{-16}$ &OH stretching\\
&3766.76 (2.65)&$5.31\times10^{-17}$&free OH\\
\hline
1:1.00& 896.53 (11.15)&$5.61\times10^{-17}$& OH torsion \\
&1709.45 (5.85)&$4.41\times10^{-17}$& H$_2$O scissoring \\
&3296.79 (3.03)&$3.05\times10^{-16}$ &OH stretching\\
&3765.50 (2.66)&$7.28\times10^{-17}$&free OH\\
\hline
\hline
\multicolumn{4}{c} {\bf H$_2$CO} \\
\hline
\hline
1:0.25&890.85 (11.22)&$6.00\times10^{-17}$& OH torsion \\
&1712.86 (5.84)&$3.97\times10^{-17}$& H$_2$O scissoring \\
&3274.97 (3.05)&$3.06\times10^{-16}$&OH stretching\\
&3607.83 (2.77)&$8.46\times10^{-17}$&free OH\\
\hline
1:0.50&886.56 (11.28)&$5.44\times10^{-17}$& OH torsion \\
&1710.12 (5.85)&$4.17\times10^{-17}$& H$_2$O scissoring \\
&3246.10 (3.08)&$3.27\times10^{-16}$&OH stretching\\
&3601.67 (2.78)&$9.12\times10^{-17}$&free OH\\
\hline
1:0.75&937.07 (10.67)&$5.41\times10^{-17}$& OH torsion \\
&1767.86 (5.66)&$2.81\times10^{-17}$& H$_2$O scissoring \\
&3279.19 (3.05)&$3.23\times10^{-16}$&OH stretching\\
&3607.45 (2.77)&$1.35\times10^{-17}$&free OH\\
\hline
1:1.00&904.32 (11.06)&$5.51\times10^{-17}$& OH torsion \\
&1713.74 (5.84)&$3.51\times10^{-17}$& H$_2$O scissoring \\
&3338.99 (2.99)&$2.96\times10^{-16}$&OH stretching\\
&3612.36 (2.77)&$9.67\times10^{-17}$&free OH\\
\hline
\hline
\multicolumn{4}{c} {\bf CH$_4$} \\
\hline
\hline
1:0.25&903.11 (11.07)&$5.96\times10^{-17}$& OH torsion \\
&1710.29 (5.85)&$4.14\times10^{-17}$& H$_2$O scissoring \\
&3287.91 (3.04)&$3.06\times10^{-16}$&OH stretching\\
&3777.98 (2.65)&$2.03\times10^{-17}$&free OH\\
\hline
1:0.50&910.32 (10.98)&$5.97\times10^{-17}$& OH torsion \\
&1706.49 (5.86)&$4.14\times10^{-17}$& H$_2$O scissoring \\
&3280.67 (3.05)&$3.30\times10^{-16}$ &OH stretching\\
&3778.16 (2.65)&$2.16\times10^{-17}$&free OH\\
\hline
1:0.75&905.03 (11.05)&$6.05\times10^{-17}$& OH torsion\\
&1706.58 (5.86)&$4.10\times10^{-17}$& H$_2$O scissoring \\
&3296.35 (3.03)&$3.22\times10^{-16}$ &OH stretching\\
&3778.86 (2.65)&$2.18\times10^{-17}$&free  OH\\
\hline
1:1.00&893.41 (11.19)&$6.23\times10^{-17}$& OH torsion \\
&1705.91 (5.86)&$4.11\times10^{-17}$& H$_2$O scissoring \\
&3312.78 (3.02)&$3.41\times10^{-16}$ &OH stretching\\
&3777.04 (2.65)&$2.40\times10^{-17}$&free OH\\
\hline
\hline
\multicolumn{4}{c} {\bf OCS} \\
\hline
\hline
1:0.25&897.72 (11.14)&$5.74\times10^{-17}$& OH torsion \\
&1711.96 (5.84)&$4.11\times10^{-17}$&H$_2$O scissoring \\
&3288.8 (3.04)&$3.10\times10^{-16}$&OH stretching\\
&3772.38 (2.65)&$4.82\times10^{-17}$&free OH\\
\hline
1:0.50&889.91 (11.24)&$5.80\times10^{-17}$&OH torsion \\
&1712.15 (5.84)&$4.30\times10^{-17}$&H$_2$O scissoring \\
&3281.28 (3.05)&$3.27\times10^{-16}$&OH stretching\\
&3772.57 (2.65)&$6.60\times10^{-17}$&free OH\\
\hline
1:0.75&901.72 (11.09)&$5.77\times10^{-17}$&OH torsion \\
&1713.88 (5.83)&$4.40\times10^{-17}$&H$_2$O scissoring \\
&3290.61 (3.04)&$3.26\times10^{-16}$&OH stretching\\
&3771.53 (2.65)&$9.02\times10^{-17}$&free OH\\
\hline
1:1.00&897.75 (11.14)&$5.75\times10^{-17}$& OH torsion \\
&1713.97 (5.83)&$4.47\times10^{-17}$&H$_2$O scissoring \\
&3299.75 (3.03)&$3.29\times10^{-16}$&OH stretching\\
&3770.63 (2.65)&$1.11\times10^{-16}$&free OH\\
\hline
\hline
\end{tabular}
\end{table}

\begin{table}
\tiny
\centering
\begin{tabular}{cccc}
 \hline
 \hline
{\bf  $\rm{H_2O:X}$} & {\bf Frequency at peak}& {\bf Integral absorbance} & {\bf Modes of} \\
&{\bf in cm$^{-1}$ ($\mu$m)} & {\bf coefficient in} & {\bf vibration} \\
&& {\bf cm molecule$^{-1}$} & \\
\hline
\hline
\multicolumn{4}{c} {\bf N$_2$} \\
\hline
\hline
1:0.25& 894.31 (11.18) &$5.63\times10^{-17}$& OH torsion \\
&1709.81 (5.85)&$4.16\times10^{-17}$& H$_2$O scissoring \\
&3292.23 (3.04)&$3.16\times10^{-16}$ &OH stretching\\
&3778.08 (2.65)&$2.34\times10^{-17}$&free OH\\
\hline
1:0.50&889.34 (11.24)&$5.98\times10^{-17}$& OH torsion \\
&1716.33 (5.83)&$4.54\times10^{-17}$& H$_2$O scissoring \\
&3290.41 (3.04)&$3.10\times10^{-16}$ &OH stretching\\
&3778.76 (2.65)&$2.36\times10^{-17}$&free OH\\
\hline
1:0.75&892.07 (11.21)&$5.98\times10^{-17}$& OH torsion \\
&1716.67 (5.82)&$4.69\times10^{-17}$& H$_2$O scissoring \\
&3295.86 (3.03)&$3.05\times10^{-16}$ &OH stretching\\
&3784.28 (2.64)&$5.14\times10^{-17}$&free OH\\
\hline
1:1.00&899.90 (11.11)&$5.73\times10^{-17}$& OH torsion \\
&1714.47 (5.83) &$4.52\times10^{-17}$& H$_2$O scissoring \\
&3297.32 (3.03)&$3.15\times10^{-16}$ &OH stretching\\
&3780.99 (2.64)&$7.59\times10^{-17}$&free OH\\
\hline
\hline
\multicolumn{4}{c} {\bf O$_2$} \\
\hline
\hline
1:0.25&893.78 (11.19)&$5.65\times10^{-17}$& OH torsion \\
&1711.86 (5.84)&$4.11\times10^{-17}$& H$_2$O scissoring \\
&3286.24 (3.04)&$3.10\times10^{-16}$ &OH stretching\\
&3729.93 (2.68)&$4.41\times10^{-17}$&free OH\\
\hline
1:0.50& 934.71 (10.70)&$6.49\times10^{-17}$& OH torsion \\
&1700.17 (5.88)&$3.78\times10^{-17}$& H$_2$O scissoring \\
&3225.04 (3.10)&$4.51\times10^{-16}$ &OH stretching\\
&3774.84 (2.65)&$2.77\times10^{-17}$&free OH\\
\hline
1:0.75&892.96 (11.20)&$5.75\times10^{-17}$& OH torsion \\
&1717.35 (5.82)&$4.04\times10^{-17}$& H$_2$O scissoring \\
&3286.59 (3.04)&$3.12\times10^{-16}$ &OH stretching\\
&3729.19 (2.68)&$8.72\times10^{-17}$&free OH\\
\hline
1:1.00&900.52 (11.10)&$5.67\times10^{-17}$& OH torsion\\
&1718.90 (5.82)&$4.18\times10^{-17}$& H$_2$O scissoring \\
&3299.52 (3.03)&$3.19\times10^{-16}$ &OH stretching\\
&3727.75 (2.68)&$1.08\times10^{-16}$&free OH\\
\hline
\hline
\end{tabular}
\end{table}

In the following sections, the results for water ice with HCOOH and NH$_3$ as impurities are first reported and discussed, thereby exploiting the outcomes of new experiments. Then, we move to the $\rm{CH_3OH-H_2O}$ ice, for which new experimental results are obtained. Finally, for the last two cases addressed, namely $\rm{CO-H_2O}$ and $\rm{CO_2-H_2O}$, the experimental data for the comparison are taken from the literature. Unless otherwise stated, we use the c-tetramer configuration for the rest of our calculations.

\subsubsection{HCOOH ice}
\label{HCOOH_ice}
IR spectra are measured for various mixtures of H$_2$O and HCOOH ice deposited at $20$ K, as explained in the experimental details Section (see Section \ref{sec:experiment}). These, normalized to the $\rm{O-H}$ stretch, are shown in Figure \ref{fig:experiment}a. Minor contamination due to CO$_2$ is detected in some experiments. However, in all experiments, the amount of CO$_2$ deposited in the ice is 1000 and more than 100 times less abundant than H$_2$O and HCOOH, respectively. Therefore, we do not expect that the CO$_2$ contamination affects the recorded IR spectra profiles.

The mixture ratios are determined from the fit of the spectrum of a selected mixture, measurement of the area of the water band at $3333.33$ cm$^{-1}$ (3.00 $\mu$m), and comparing with the pure water counterparts. For HCOOH, the absorption area is measured at 1700 cm$^{-1}$. In fact, HCOOH has the strongest mode at $1694.92$ cm$^{-1}$ ($5.90$ $\mu$m), which corresponds to its $\rm{C=O}$ stretching mode. However, the feature overlaps with the OH bending mode of solid water at $1666.67$ cm$^{-1}$ ($6.00$ $\mu$m). Therefore, the water bending mode contribution at $\sim$1700 cm$^{-1}$ has been subtracted from the total area before the band strength mentioned above being used to calculate the amount of HCOOH in the ice mixture. The band strengths used here are $2.0\times10^{-16}$ for H$_2$O \citep{gera95} and $6.7\times10^{-17}$ for HCOOH \citep{mare87,schu99}. Another relatively weaker mode of HCOOH at $1388.89$ cm$^{-1}$ ($7.20$ $\mu$m) is also considered because the corresponding region is free from interfering transitions \citep{schu99}. As seen in Figure \ref{fig:experiment}a, the HCOOH:H$_2$O ratios cover the 0.05 to 3.46 range. It is worth noting that abundances of solid-phase HCOOH in the interstellar ices vary between $1\%$ and $5\%$ relative to the $\rm{H_2O}$ ice \citep{biss07}.

Figure \ref{fig:optimized_structure}b shows how the HCOOH molecules are bonded to the water molecules to form the $4:4$ $\rm{H_2O-HCOOH}$ mixture used in our calculations. The absorption band profiles of the $\rm{H_2O-HCOOH}$ clusters with different impurity concentrations are shown in Figure \ref{fig:H2O-HCOOH}.
The transition frequencies and the corresponding most vital intensity values,
obtained at the B3LYP/6-31G(d) level, are given in Table \ref{tab:H2O_X}. Calculations are also carried out using the B2PLYP functional.
The results are summarized in Tables \ref{tab:4H2O_B2PLYP}$-$\ref{tab:4H2O_4HCOOH_B2PLYP}.

\begin{table}
\scriptsize
\centering
\caption{Harmonic IR frequencies and intensities of the complex 4H$_2$O evaluated at the B2PLYP/maug-cc-pVTZ level \citep{gora20a}.}
\label{tab:4H2O_B2PLYP}
\vskip 0.2 cm
\begin{tabular}{ccc}
\hline
{\bf NM} &$\omega$ {[\bf cm$^{-1}$]} & {\bf IR Int. [km/mol]} \\
\hline
       1 &      39.3404 &       0.0000 \\
       2 &      85.5169 &       4.1241 \\
       3 &     202.1393 &       0.0000 \\
       4 &     222.9069 &     104.7500 \\
       5 &     222.9069 &     104.7500 \\
       6 &     231.0105 &      94.8938 \\
       7 &     240.0832 &       0.6734 \\
       8 &     246.4688 &     287.8694 \\
       9 &     246.4688 &     287.8694 \\
      10 &     274.1815 &       0.0000 \\
      11 &     398.6011 &       0.0000 \\
      12 &     427.5608 &      44.2017 \\
      13 &     443.1204 &      79.6160 \\
      14 &     443.1204 &      79.6160 \\
      15 &     683.4168 &     312.1539 \\
      16$^t$ &     778.3957 &     237.5051 \\
      17$^t$ &     778.3957 &     237.5051 \\
      18$^t$ &     934.3601 &       0.0000 \\
      19$^b$ &    1630.3695 &     176.7723 \\
      20$^b$ &    1647.3535 &      74.0296 \\
      21$^b$ &    1647.3536 &      74.0296 \\
      22$^b$ &    1672.6698 &       0.0000 \\
      23$^s$ &    3411.5900 &       0.0000 \\
      24$^s$ &    3477.3996 &    1793.2478 \\
      25$^s$ &    3477.3996 &    1793.2478 \\
      26$^s$ &    3522.3300 &       2.0149 \\
      27$^f$ &    3864.6871 &     144.7620 \\
      28$^f$ &    3865.9912 &     138.6232 \\
      29$^f$ &    3865.9912 &     138.6232 \\
      30$^f$ &    3867.1847 &       0.0000 \\
\hline
\end{tabular}
\vskip 0.2 cm
{\bf Note:} $^t$OH torsion; $^b$OH scissoring; $^s$OH stretching; $^f$free OH.
\end{table}

\begin{table}
\scriptsize
\centering
\caption{Harmonic IR frequencies and intensities of the complex 4H$_2$O/HCOOH evaluated at the B2PLYP/maug-cc-pVTZ level \citep{gora20a}.}
\label{tab:4H2O_1HCOOH_B2PLYP}
\vskip 0.2 cm
\begin{tabular}{ccc}
\hline
{\bf NM} &$\omega$ {[\bf cm$^{-1}$]} & {\bf IR Int. [km/mol]} \\
\hline
       1 &       7.4652 &       2.8797 \\
       2 &      21.6554 &       0.4836 \\
       3 &      31.8296 &       0.8007 \\
       4 &      45.1056 &       0.3993 \\
       5 &      61.0883 &       3.2003 \\
       6 &      99.9662 &       7.7802 \\
       7 &     176.7899 &      26.3202 \\
       8 &     190.2330 &      26.8829 \\
       9 &     212.6031 &      80.9785 \\
      10 &     214.8008 &      65.5508 \\
      11 &     229.1782 &     167.5034 \\
      12 &     238.5224 &      82.0767 \\
      13 &     255.2508 &     249.2937 \\
      14 &     283.6779 &      26.8213 \\
      15 &     291.4533 &      55.7068 \\
      16 &     356.4031 &      83.8068 \\
      17 &     417.6885 &      49.4695 \\
      18 &     428.1868 &      57.7373 \\
      19 &     460.8606 &      57.1370 \\
      20 &     508.1693 &      55.1486 \\
      21 &     660.2176 &     284.6155 \\
      22 &     690.3556 &      83.8851 \\
      23$^t$ &     754.6078 &     231.9718 \\
      24$^t$ &     824.4685 &     178.7024 \\
      25$^{t*}$ &     939.4045 &      72.3782 \\
      26$^{t*}$ &     992.4637 &     143.3107 \\
      27 &    1085.0300 &      25.7493 \\
      28 &    1202.3494 &     398.5316 \\
      29 &    1404.3401 &       4.1326 \\
      30 &    1450.6121 &      11.3947 \\
      31$^b$ &    1630.0596 &     190.6522 \\
      32$^b$ &    1643.5475 &      74.8871 \\
      33$^b$ &    1650.3969 &      67.4629 \\
      34$^b$ &    1671.1011 &       2.8373 \\
      35 &    1727.4428 &     643.7374 \\
      36 &    3082.2813 &     261.4632 \\
      37$^{s*}$ &    3180.6017 &    2093.2108 \\
      38$^{s*}$ &    3305.3198 &     591.1162 \\
      39$^{s}$ &    3439.5143 &    1097.5194 \\
      40$^{s}$ &    3501.1192 &     960.3923 \\
      41$^{s}$ &    3594.2178 &     538.8467 \\
      42$^{f}$ &    3844.3939 &     126.6809 \\
      43$^{f}$ &    3864.5499 &     125.4762 \\
      44$^{f}$ &    3865.8634 &     156.4890 \\
      45$^{f}$ &    3866.9802 &      55.7713 \\
\hline
\end{tabular} \\
\vskip 0.2 cm
{\bf Note:} $^t$OH torsion; $^b$OH scissoring; $^s$OH stretching; $^f$free OH; *vibrations contaminated by HCOOH modes.
\end{table}

\begin{table}
\scriptsize
\centering
\caption{Harmonic IR frequencies and intensities of the complex 4H$_2$O/2HCOOH evaluated at the B2PLYP/maug-cc-pVTZ level \citep{gora20a}.}
\label{tab:4H2O_2HCOOH_B2PLYP}
\vskip 0.2 cm
\begin{tabular}{ccc}
\hline
{\bf NM} &$\omega$ {[\bf cm$^{-1}$]} & {\bf IR Int. [km/mol]} \\
\hline
       1 &       4.6054 &       0.4281 \\
       2 &      13.0063 &       2.9178 \\
       3 &      16.5189 &       0.8051 \\
       4 &      19.0314 &       2.3774 \\
       5 &      20.9593 &       2.6393 \\
       6 &      33.4976 &       0.4467 \\
       7 &      36.7713 &       1.8500 \\
       8 &      42.5055 &       1.8790 \\
       9 &      57.0477 &       3.6453 \\
      10 &      87.6438 &       1.4306 \\
      11 &      98.3938 &       9.9521 \\
      12 &     109.3757 &       0.0813 \\
      13 &     179.0412 &      24.1400 \\
      14 &     190.5090 &      28.2740 \\
      15 &     214.0975 &      84.9229 \\
      16 &     218.8524 &      55.4812 \\
      17 &     230.2931 &     171.1476 \\
      18 &     238.6575 &      89.3090 \\
      19 &     267.3732 &     231.1914 \\
      20 &     283.3954 &      21.6224 \\
      21 &     299.6074 &      70.4602 \\
      22 &     358.3291 &      97.5174 \\
      23 &     418.1841 &      53.6749 \\
      24 &     433.0143 &      50.5761 \\
      25 &     453.3466 &      44.2133 \\
      26 &     511.7339 &      57.2963 \\
      27 &     626.1257 &      81.2731 \\
      28 &     663.5082 &     202.0801 \\
      29 &     664.5570 &     266.9661 \\
      30 &     690.0285 &      82.0272 \\
      31$^{t}$ &     746.7845 &     238.4794 \\
      32$^{t}$ &     832.5630 &     181.3816 \\
      33$^{t*}$ &     936.5175 &   ` 69.8256 \\
      34$^{t*}$ &     992.1307 &     142.3809 \\
      35 &    1086.4569 &      25.8243 \\
      36 &    1100.7355 &       4.5531 \\
      37 &    1117.2066 &     421.1594 \\
      38 &    1202.1232 &     398.7897 \\
      39 &    1301.5992 &      19.3272 \\
      40 &    1403.7379 &       3.7351 \\
      41 &    1441.6206 &       5.0048 \\
      42 &    1449.0432 &      11.3813 \\
      43$^{b}$ &    1630.1114 &     189.4041 \\
      44$^{b}$ &    1643.0592 &      75.7767 \\
      45$^{b}$ &    1650.4443 &      71.8600 \\
      46$^{b}$ &    1671.6065 &       3.7550 \\
      47 &    1727.3571 &     640.6790 \\
      48 &    1740.9364 &     573.8795 \\ 
      49 &    3082.1544 &     256.8737 \\
      50$^{s*}$ &    3108.5982 &       9.0680 \\
      51$^{s*}$ &    3182.8047 &    2058.2321 \\
      52$^{s*}$ &    3311.8175 &     593.5776 \\
      53$^{s}$ &    3443.5223 &    1123.8585 \\
      54$^{s}$ &    3495.5861 &     975.1779 \\
      55$^{s}$ &    3593.0205 &     538.1736 \\
      56 &    3722.9218 &     111.6916 \\
      57$^{f}$ &    3843.8168 &     126.4406 \\
      58$^{f}$ &    3860.4740 &     132.4083 \\
      59$^{f}$ &    3863.8070 &     104.3051 \\
      60$^{f}$ &    3865.7512 &     106.7733 \\
\hline
\end{tabular} \\
\vskip 0.2 cm
{\bf Note:} $^t$OH torsion; $^b$OH scissoring; $^s$OH stretching; $^f$free OH; *vibrations contaminated by HCOOH modes.
\end{table}

\begin{table}
\scriptsize
\centering
\caption{Harmonic IR frequencies and intensities of the complex 4H$_2$O/3HCOOH evaluated at the B2PLYP/maug-cc-pVTZ level \citep{gora20a}.}
\label{tab:4H2O_3HCOOH_B2PLYP}
\vskip 0.2 cm
\begin{tabular}{cccccc}
\hline
{\bf NM} &$\omega$ {[\bf cm$^{-1}$]} & {\bf IR Int. [km/mol]} & {\bf NM} &$\omega$ {[\bf cm$^{-1}$]} & {\bf IR Int. [km/mol]} \\
\hline
       1 &       4.8748 &       0.5689 & 51 &    1400.7389 &       2.9534 \\
       2 &       8.5054 &       0.7327 & 52 &    1440.0730 &       5.2611 \\
       3 &      10.9091 &       1.7844 & 53 &    1445.8193 &      12.3688 \\
       4 &      14.1204 &       2.2282 & 54 &    1445.9512 &       8.9357 \\
       5 &      16.8372 &       1.6310 & 55$^{b}$ &    1630.5379 &     192.0093 \\
       6 &      19.1432 &       3.8603 & 56$^{b}$ &    1645.2809 &      74.7331 \\
       7 &      22.9673 &       0.5549 & 57$^{b}$ &    1649.3090 &      90.0689 \\
       8 &      28.0702 &       2.8070 & 58$^{b}$ &    1670.3929 &       1.3964 \\
       9 &      34.3674 &       0.6429 & 59 &    1729.0788 &     813.5933 \\
      10 &      40.7401 &       1.8661 & 60 &    1729.3363 &     445.0961 \\
      11 &      45.9916 &       0.3401 & 61 &    1741.9079 &     575.8421 \\
      12 &      54.8514 &       5.3280 & 62 &    3085.0941 &     261.1337 \\
      13 &      66.6228 &       5.5016 & 63 &    3085.5154 &      98.0625 \\
      14 &      89.0988 &       1.1463 & 64 &    3110.4187 &       7.6777 \\
      15 &      99.3575 &       2.6618 & 65$^{s*}$ &    3215.3096 &    2288.1686 \\
      16 &     110.9935 &       9.9385 & 66$^{s*}$ &    3232.6395 &    1833.2038 \\
      17 &     174.0921 &      44.5063 & 67$^{s*}$ &    3306.6741 &     311.8714 \\
      18 &     177.9118 &      30.6049 & 68$^{s}$ &    3451.6305 &     978.9670 \\
      19 &     188.6541 &      24.0948 & 69$^{s}$ &    3473.6093 &     894.3564 \\
      20 &     209.4056 &      83.4103 & 70$^{s}$ &    3609.3777 &     516.6451 \\
      21 &     212.1508 &      55.0788 & 71 &    3723.4615 &     111.6932 \\
      22 &     234.2838 &      28.4231 & 72$^{f}$ &    3838.2715 &     172.2543 \\
      23 &     245.2791 &     210.0525 & 73$^{f}$ &    3841.3048 &      97.0635 \\
      24 &     261.2989 &      35.4759 & 74$^{f}$ &    3857.9521 &     119.5655 \\
      25 &     284.3732 &      27.8615 & 75$^{f}$ &    3865.5487 &     121.3331 \\
      26 &     314.1459 &     145.9808 & && \\
      27 &     352.7412 &      77.0879 & && \\
      28 &     410.7784 &      41.6288 & && \\
      29 &     418.4381 &     142.3769 & && \\
      30 &     451.8321 &      62.9268 & && \\
      31 &     486.2526 &      26.6238 & && \\
      32 &     515.8604 &      43.5007 & && \\
      33 &     626.3452 &      78.8764 & && \\
      34 &     658.3748 &     265.4010 & && \\
      35 &     664.0178 &     196.3924 & && \\
      36 &     686.9044 &      88.1977 & && \\
      37 &     687.6686 &      73.9457 & && \\
      38$^{t*}$ &     798.2680 &     156.7957 & && \\
      39$^{t}$ &     803.7700 &     211.1567 & && \\
      40$^{t*}$ &     926.7114 &      95.5456 & && \\
      41 &     949.5463 &     177.0601 &&& \\
      42 &    1005.8276 &     134.0566 &&& \\
      43 &    1083.1613 &      20.4516 &&& \\
      44 &    1083.9241 &      23.7748 &&& \\
      45 &    1098.4056 &       4.7221 &&& \\
      46 &    1118.0234 &     421.3278 &&& \\
      47 &    1198.2543 &     380.9330 &&& \\
      48 &    1198.3514 &     418.6655 &&& \\
      49 &    1303.1561 &      18.8667 &&& \\ 
      50 &    1400.6093 &       4.0266 &&& \\
\hline
\end{tabular} \\
\vskip 0.2 cm
{\bf Note:} $^t$OH torsion; $^b$OH scissoring; $^s$OH stretching; $^f$free OH; *vibrations contaminated by HCOOH modes.
\end{table}   

\begin{table}
\scriptsize
\centering
\caption{Harmonic IR frequencies and intensities of the complex 4H$_2$O/4HCOOH evaluated at the B2PLYP/maug-cc-pVTZ level \citep{gora20a}.}
\label{tab:4H2O_4HCOOH_B2PLYP}
\vskip 0.2 cm
\begin{tabular}{cccccc}
\hline
{\bf NM} &$\omega$ {[\bf cm$^{-1}$]} & {\bf IR Int. [km/mol]}  & {\bf NM} &$\omega$ {[\bf cm$^{-1}$]} & {\bf IR Int. [km/mol]} \\
\hline
       1 &       5.2521 &       0.5902 & 51 &    1082.2723 &      15.9752 \\
       2 &       6.6193 &       1.2706 & 52 &    1082.8090 &      29.1817 \\
       3 &       8.8369 &       0.8598 & 53 &    1083.2383 &       8.1740 \\
       4 &       9.6756 &       1.0009 & 54 &    1096.9310 &       4.7996 \\
       5 &      14.2363 &       3.0419 & 55 &    1118.5677 &     421.7245 \\
       6 &      15.7986 &       1.1225 & 56 &    1194.7613 &     402.1943 \\
       7 &      18.3559 &       3.0357 & 57 &    1196.6655 &     678.8383 \\
       8 &      19.2707 &       2.7650 & 58 &    1196.9390 &     124.3700 \\
       9 &      22.1555 &       0.9922 & 59 &    1303.5269 &      19.2431 \\
      10 &      27.4779 &       3.1848 & 60 &    1398.1664 &       2.8858 \\
      11 &      30.7377 &       0.4104 & 61 &    1399.8325 &       2.3693 \\
      12 &      36.6940 &       0.3940 & 62 &    1400.1243 &       3.6218 \\
      13 &      42.5989 &       1.4605 & 63 &    1438.1358 &       3.4116 \\
      14 &      50.4443 &       2.9039 & 64 &    1443.1291 &      10.6650 \\
      15 &      56.1885 &       2.0333 & 65 &    1444.6815 &       8.9685 \\
      16 &      67.8597 &       3.6648 & 66 &    1445.0393 &      12.6099 \\
      17 &      79.8514 &      11.6336 & 67$^{b}$ &    1633.0557 &     204.1450 \\
      18 &      88.6848 &       0.8582 & 68$^{b}$ &    1645.9171 &      94.0994 \\
      19 &     101.4052 &       1.0586 & 69$^{b}$ &    1651.0001 &      68.5795 \\
      20 &     120.4063 &       2.7295 & 70$^{b}$ &    1672.8271 &       2.1359 \\
      21 &     173.0431 &      53.0876 & 71 &    1729.8409 &     773.0317 \\
      22 &     174.6165 &      79.9014 & 72 &    1729.9568 &     580.0062 \\
      23 &     181.6174 &      16.9912 & 73 &    1730.9646 &     527.5802 \\
      24 &     193.2247 &       8.7228 & 74 &    1742.3118 &     572.3800 \\
      25 &     207.4059 &      98.2327 & 75 &    3086.2791 &     171.0952 \\
      26 &     209.4436 &      56.3457 & 76 &    3087.0204 &     182.6391 \\
      27 &     214.9922 &      15.9518 & 77 &    3087.9488 &     105.7546 \\
      28 &     251.9141 &      48.3664 & 78 &    3114.3577 &       6.5557 \\
      29 &     260.0519 &      27.0848 & 79$^{s*}$ &    3235.8092 &    2226.0999 \\
      30 &     277.2414 &      21.6313 & 80 &    3251.6769 &    2145.0628 \\
      31 &     294.7764 &     138.3103 & 81$^{s*}$ &    3269.8018 &    1271.6200 \\
      32 &     370.6813 &      53.2423 & 82$^{s*}$ &    3338.7804 &     329.7573 \\
      33 &     411.5523 &     157.3717 & 83$^{s}$ &    3451.7939 &     830.3998 \\
      34 &     426.8010 &     160.0370 & 84$^{s}$ &    3506.2315 &     849.9293 \\
      35 &     432.4841 &      46.3694 & 85$^{s}$ &    3580.2425 &     496.8387 \\
      36 &     474.7455 &       4.1043 & 86 &    3723.2687 &     112.7727 \\
      37 &     500.7327 &      25.1590 & 87$^{f}$ &    3837.0764 &     199.2805 \\
      38 &     506.2341 &      58.6370 & 88$^{f}$ &    3839.9016 &     135.3088 \\
      39 &     626.8822 &      79.1196 & 89$^{f}$ &    3841.2858 &      85.4332 \\
      40 &     662.8116 &     197.7640 & 90$^{f}$ &    3859.2614 &     129.2712 \\
      41$^{t*}$ &     680.2702 &     165.0245 &&& \\
      42$^{t*}$ &     685.5338 &      74.2193 &&& \\
      43$^{t*}$ &     685.8711 &     129.0114 &&& \\
      44$^{t}$ &     703.1086 &     134.5453 &&& \\
      45$^{t}$ &     777.4948 &     157.2018 &&& \\
      46$^{t*}$ &     836.4350 &     145.9547 &&& \\
      47 &     918.5370 &      51.5171 &&& \\
      48 &     934.3780 &     130.8642 &&& \\
      49 &     940.6167 &     399.9756 &&& \\
      50 &    1005.3540 &      55.4131 &&& \\
\hline
\end{tabular} \\
\vskip 0.2 cm
{\bf Note:} $^t$OH torsion; $^b$OH scissoring; $^s$OH stretching; $^f$free OH; *vibrations contaminated by HCOOH modes.
\end{table}

\begin{table}
\scriptsize
{\centering
\caption{Linear fit coefficients for the $\rm{H_2O-X}$ (X =  HCOOH, $\rm{NH_3}$, $\rm{CH_3OH}$, CO, $\rm {CO_2}$, $\rm{H_2CO}$, $\rm{CH_4}$, OCS,
$\rm{N_2}$, and $\rm{O_2}$) mixtures$^a$ \citep{gora20a}.}
\label{tab:linear_coeff}
\vskip 0.2cm
\begin{tabular}{cccc}
 \hline
{\bf Mixture} & {\bf Vibrational} & \multicolumn{2}{c}{\bf Linear coefficients} \\
\cline{3-4}
& {\bf mode} & {\bf Constant} & {\bf Slope} \\
 & & [$\rm 10^{-16}$ cm $\rm molecule^{-1}$] & [$\rm 10^{-19}$ cm $\rm molecule^{-1}$]\\
\hline
$\rm{H_2O-HCOOH}$&$\mathrm{\nu_{libration}}$&2.45 { (0.26)$^b$}&132.73 { (0.90)$^b$}\\
&$\mathrm{\nu_{bending}}$&0.58 { (0.05)$^b$}&184.25 { (14.40)$^b$}\\
&$\mathrm{\nu_{stretching}}$&1.80 { (1.90)$^b$} &48.60 { (-7.30)$^b$}\\
&$\mathrm{\nu_{free OH}}$&0.20 { (0.16)$^b$} &9.80 { (-0.40)$^b$} \\
\hline
$\rm{H_2O-NH_3}$&$\mathrm{\nu_{libration}}$&0.27 { (0.34)$^b$}&6.11 { (5.00)$^b$}\\
&$\mathrm{\nu_{bending}}$&0.09 { (0.12)$^b$}&5.48 { (2.20)$^b$}\\
&$\mathrm{\nu_{stretching}}$&1.90 {(2.38)$^b$}&0.41 { (-14.4)$^b$}\\
&$\mathrm{\nu_{free OH}}$&0.21 { (0.12)$^b$}&-4.21 { (-2.1)$^b$}\\
\hline
$\rm{H_2O-CH_3OH}$ &$\mathrm{\nu_{libration}}$&0.25&10.0\\
&$\mathrm{\nu_{bending}}$&0.12&2.00\\
&$\mathrm{\nu_{stretching}}$&1.92&32.00\\
&$\mathrm{\nu_{free OH}}$&0.26&2.65\\
\hline
$\rm{H_2O-CO}$ & $\mathrm{\nu_{libration}}$ & 0.30 (0.30$\pm$0.02)$^c$ & -0.32 (-2.1$\pm$0.4)$^c$ \\
&$\mathrm{\nu_{bending}}$&0.12 (0.13$\pm$0.02)$^c$ &-0.016 (-1.0 $\pm$0.3)$^c$\\
&$\mathrm{\nu_{stretching}}$&1.98 (2.0$\pm$0.1)$^c$ &-3.2 (-16$\pm$3)$^c$ \\
&$\mathrm{\nu_{free OH}}$&0.18 (0.0)$^c$ &5.69 (1.2 $\pm$0.1)$^c$ \\
\hline
$\rm{H_2O-CO_2}$ &$\mathrm{\nu_{libration}}$ & 0.3 (0.32$\pm$0.02)$^d$ &2.07 (-3.2$\pm$0.4)$^d$ \\
&$\mathrm{\nu_{bending}}$&0.11 (0.14$\pm$0.01)$^d$ & 0.12 (-0.5$\pm$0.2)$^d$ \\
&$\mathrm{\nu_{stretching}}$&2.02 (2.1$\pm$0.1)$^d$ & -0.22 (-22$\pm$2)$^d$ \\
&$\mathrm{\nu_{free OH}}$&0.19 (0.0)$^d$ & 10.02 (1.62$\pm$0.07)$^d$ \\
\hline
$\rm{H_2O-H_2CO}$&$\mathrm{\nu_{libration}}$&0.26&5.73\\
&$\mathrm{\nu_{bending}}$&0.10&4.59\\
&$\mathrm{\nu_{stretching}}$&1.92&0.10\\
&$\mathrm{\nu_{free OH}}$&0.13&16.53\\
\hline
$\rm{H_2O-CH_4}$ &$\mathrm{\nu_{libration}}$&0.31&0.53\\
&$\mathrm{\nu_{bending}}$&0.11&1.18\\
&$\mathrm{\nu_{stretching}}$&2.01&3.39\\
&$\mathrm{\nu_{free OH}}$&0.20&0.52\\
\hline
$\rm{H_2O-OCS}$&$\mathrm{\nu_{libration}}$&0.30&0.42\\
&$\mathrm{\nu_{bending}}$&0.11&0.23\\
&$\mathrm{\nu_{stretching}}$&1.96&2.18\\
&$\mathrm{\nu_{free OH}}$&0.17&0.13\\
\hline
$\rm{H_2O-N_2}$&$\mathrm{\nu_{libration}}$&0.31&-0.30\\
&$\mathrm{\nu_{bending}}$&0.12&0.17\\
&$\mathrm{\nu_{stretching}}$&0.12&0.11\\
&$\mathrm{\nu_{free OH}}$&0.17&7.75\\
\hline
$\rm{H_2O-O_2}$&$\mathrm{\nu_{libration}}$&0.31&-0.23\\
&$\mathrm{\nu_{bending}}$&0.12&-0.13\\
&$\mathrm{\nu_{stretching}}$&2.02&4.71\\
&$\mathrm{\nu_{free OH}}$&0.13&13.80\\
\hline
\end{tabular} \\
}
\vskip 0.2cm
{\bf Note:} \\
$^a$ Experimental values are provided in the parentheses. \\
$^b$ This work. \\
$^c$ \cite{bouw07}. \\
$^d$ \cite{ober07}.
\end{table}

The data are fitted with a linear function  $A_{eff} = a[X] + b$
(where $X =$ HCOOH, NH$_3$, $\rm{CH_3OH}$, CO, $\rm{CO_2}$, $\rm{H_2CO}$, CH$_4$, OCS, N$_2$, and O$_2$) to investigate the variation of the band strength with impurity concentrations. The coefficient $a$ provides the information
whether the band strength increases or decreases by increasing the concentration of X, $[X]$. The coefficient $b$ indicates the band strength of
the vibration mode in the absence of impurities. The fitting coefficients for all impurities considered are provided in Table \ref{tab:linear_coeff}.
In Figure \ref{fig:band_strength}a, the band strength profile as a function of the concentration of HCOOH is shown.

\begin{table}
\scriptsize
\centering
\caption{Harmonic IR frequencies and intensities of the complex 4H$_2$O/NH$_3$ evaluated at the B2PLYP/maug-cc-pVTZ level \citep{gora20a}.}
\label{tab:4H2O_1NH3_B2PLYP}
\vskip 0.2 cm
\begin{tabular}{ccc}
\hline
{\bf NM} &$\omega$ {[\bf cm$^{-1}$]} & {\bf IR Int. [km/mol]} \\
\hline
       1     &      17.4723 &       1.6404 \\
       2     &      24.2573 &       2.7053 \\
       3     &      42.9718 &       9.2433 \\
       4     &      56.2828 &      14.2890 \\
       5     &      77.4675 &       5.3857 \\
       6     &     176.7386 &      14.5514 \\
       7     &     205.9991 &      61.5386 \\
       8     &     210.2974 &      51.1840 \\
       9     &     220.3754 &     138.5385 \\
      10     &     234.6966 &      41.4586 \\
      11     &     240.2432 &     175.3537 \\
      12     &     253.5984 &      22.5694 \\
      13     &     262.9908 &     199.8180 \\
      14     &     269.0541 &      36.5738 \\
      15     &     288.0112 &      25.8883 \\
      16     &     411.0477 &      10.8123 \\
      17     &     438.1929 &      57.2835 \\
      18     &     462.1627 &     115.4840 \\
      19     &     649.2280 &      90.3859 \\
      20$^t$ &     673.7378 &     245.4033 \\
      21$^t$ &     761.0067 &     286.3846 \\
      22$^t$ &     785.5912 &     213.8984 \\
      23$^t$ &     855.6009 &     110.1452 \\
      24 &         973.8257 &      54.6453 \\
      25 &        1133.1043 &     207.9712 \\
      26$^b$ &    1635.5747 &     161.1828 \\
      27$^b$ &    1648.9602 &      81.2871 \\
      28$^{b*}$ &    1654.1532 &      68.4123 \\
      29        &    1659.8070 &      23.9527 \\
      30$^{b*}$ &    1664.0270 &      13.9746 \\
      31$^{b*}$ &    1687.1189 &       4.9894 \\
      32$^s$    &    3278.5997 &    1998.0299 \\
      33$^s$    &    3358.0759 &     412.2786 \\
      34$^s$    &    3442.1676 &    1170.0672 \\
      35        &    3483.9230 &       0.6553 \\
      36$^s$    &    3499.4250 &     959.9493 \\
      37$^s$    &    3600.3604 &     714.2057 \\
      38        &    3600.7264 &      13.1052 \\
      39        &    3601.5192 &      18.8438 \\
      40$^f$    &    3865.2764 &     131.5539 \\
      41$^f$    &    3866.4890 &      96.2050 \\
      42$^f$    &    3867.7812 &      62.0566 \\
\hline
\end{tabular} \\
\vskip 0.2 cm
{\bf Note:} $^t$OH torsion; $^b$OH scissoring; $^s$OH stretching; $^f$free OH; *vibrations contaminated by NH$_3$ modes.
\end{table}

\begin{table}
\scriptsize
\centering
\caption{Harmonic IR frequencies and intensities of the complex 4H$_2$O/2NH$_3$ evaluated at the B2PLYP/maug-cc-pVTZ level \citep{gora20a}.}
\label{tab:4H2O_2NH3_B2PLYP}
\vskip 0.2 cm
\begin{tabular}{ccc}
\hline
{\bf NM} &$\omega$ {[\bf cm$^{-1}$]} & {\bf IR Int. [km/mol]} \\
\hline
       1 &      13.2137 &       3.5154 \\
       2 &      35.8031 &       4.5193 \\
       3 &      37.7152 &      15.3999 \\
       4 &      42.6399 &       0.6394 \\
       5 &      44.0908 &       1.0833 \\
       6 &      50.2282 &      25.2398 \\
       7 &      62.5104 &       1.7351 \\
       8 &      74.8511 &       7.5826 \\
       9 &     174.7854 &       0.3931 \\
      10 &     184.1028 &      49.5918 \\
      11 &     202.3520 &      59.3483 \\
      12 &     215.2501 &       3.8987 \\
      13 &     233.8400 &     170.4567 \\
      14 &     236.8583 &      13.8898 \\
      15 &     250.5424 &      28.2234 \\
      16 &     252.5110 &      43.5646 \\
      17 &     263.8678 &      24.6139 \\
      18 &     264.6765 &     198.4026 \\
      19 &     280.7045 &      30.8723 \\
      20 &     288.8026 &      34.1641 \\
      21 &     434.2362 &       2.5448 \\
      22 &     465.8322 &     178.0575 \\
      23 &     636.4757 &       5.8561 \\
      24 &     665.1100 &      92.9446 \\
      25$^t$    &     688.2585 &     220.0675 \\
      26$^t$    &     740.4480 &     369.3099 \\
      27$^t$    &     786.8028 &     188.6747 \\
      28$^t$    &     823.9109 &     101.1914 \\
      29$^t$    &     894.7322 &     182.6960 \\
      30$^t$    &     973.9338 &      41.1551 \\
      31        &    1131.1976 &     200.1314 \\
      32        &    1131.3535 &     216.3485 \\
      33$^b$    &    1643.5894 &     150.8608 \\
      34$^b$    &    1651.7575 &     100.8326 \\
      35$^{b*}$ &    1656.3567 &      54.5510 \\
      36$^{b*}$ &    1659.4699 &       6.3334 \\
      37        &    1660.0271 &      21.6415 \\
      38        &    1660.6689 &      23.4365 \\
      39$^{b*}$ &    1680.0902 &      25.5517 \\
      40$^{b*}$ &    1690.1602 &       1.1013 \\
      41$^s$    &    3300.1629 &     879.2415 \\
      42$^s$    &    3307.6453 &    2808.4413 \\
      43$^s$    &    3377.2024 &     646.7542 \\
      44$^s$    &    3391.5768 &     729.5638 \\
      45        &    3483.7965 &       0.9990 \\
      46        &    3483.9555 &       1.0499 \\
      47$^s$    &    3578.8775 &    1603.7262 \\
      48$^s$    &    3582.2181 &      63.0838 \\
      49        &    3600.8734 &      18.3990 \\
      50        &    3600.9540 &      18.6472 \\     
      51        &    3601.2568 &      18.6290 \\
      52        &    3601.5982 &      18.6555 \\
      53$^f$    &    3866.9165 &      77.0140 \\
      54$^f$    &    3867.0372 &      90.8323 \\
\hline
\end{tabular} \\
\vskip 0.2 cm
{\bf Note:} $^t$OH torsion; $^b$OH scissoring; $^s$OH stretching; $^f$free OH; *vibrations contaminated by NH$_3$ modes.
\end{table}

\begin{table}
\scriptsize
\centering
\caption{Harmonic IR frequencies and intensities of the complex 4H$_2$O/3NH$_3$ evaluated at the B2PLYP/maug-cc-pVTZ level \citep{gora20a}.}
\label{tab:4H2O_3NH3_B2PLYP}
\vskip 0.2 cm
\begin{tabular}{cccccc}
\hline
{\bf NM} &$\omega$ {[\bf cm$^{-1}$]} & {\bf IR Int. [km/mol]} & {\bf NM} &$\omega$ {[\bf cm$^{-1}$]} & {\bf IR Int. [km/mol]} \\
\hline
       1 &      12.5470 &       2.6010 & 51$^s$    &    3360.1503 &    1320.9055 \\
       2 &      23.2847 &      18.0759 & 52$^s$    &    3374.8469 &    2291.7299 \\
       3 &      28.7571 &      23.4249 & 53$^s$    &    3393.0086 &     199.4039 \\
       4 &      35.7710 &       0.3934 & 54$^{s*}$ &    3478.9512 &     630.0948 \\
       5 &      38.9667 &       0.0081 & 55        &    3483.7372 &       1.4326 \\
       6 &      40.5045 &       3.8999 & 56        &    3483.7543 &       6.5600 \\
       7 &      44.8244 &       0.2126 & 57$^{s*}$ &    3484.7724 &     142.9342 \\
       8 &      46.5267 &      10.5312 & 58$^s$    &    3520.0887 &    1009.6190 \\
       9 &      50.6992 &       8.8616 & 59$^s$    &    3596.2284 &     723.7876 \\
      10 &      69.7890 &       5.0744 & 60 &    3600.8160 &      17.3429 \\
      11 &      72.6524 &       9.5767 & 61 &    3600.9676 &      17.3826 \\
      12 &     166.7922 &       5.9217 & 62 &    3601.0701 &      18.0350 \\
      13 &     180.8039 &      37.7349 & 63 &    3601.4302 &      17.6940 \\
      14 &     191.1116 &      39.0754 & 64 &    3601.4607 &      18.0960 \\
      15 &     207.3981 &      10.9113 & 65 &    3601.5560 &      18.1230 \\
      16 &     226.1385 &      96.0046 & 66$^f$ &    3867.3234 &      81.2889 \\
      17 &     239.4714 &      55.1417 &&& \\
      18 &     245.0773 &       9.5193 &&& \\
      19 &     245.6175 &      62.3353 &&& \\
      20 &     248.1708 &      26.4263 &&& \\
      21 &     249.2122 &      68.1865 &&& \\
      22 &     258.4416 &      35.6050 &&& \\
      23 &     269.5283 &      80.6776 &&& \\
      24 &     272.4683 &      30.5828 &&& \\
      25 &     285.7225 &      20.2189 &&& \\
      26 &     453.0415 &      74.3332 &&& \\
      27 &     625.3188 &      29.7964 &&& \\
      28 &     647.1033 &     149.3482 &&& \\
      29 &     670.2563 &     144.8666 &&& \\
      30 &     705.1954 &     132.9061 &&& \\
      31$^t$    &     747.2466 &     332.8596 &&& \\
      32$^t$    &     768.5970 &     229.7125 &&& \\
      33$^t$    &     809.9923 &      25.6758 &&& \\
      34$^t$    &     842.7130 &     230.7932 &&& \\
      35$^t$    &     871.0453 &     187.0537 &&& \\
      36$^t$    &     977.2613 &      31.3718 &&& \\
      37        &    1126.3891 &     215.3444 &&& \\
      38        &    1128.4031 &     174.0675 &&& \\
      39        &    1128.4565 &     237.3456 &&& \\ 
      40$^b$    &    1645.9709 &     133.6692 &&& \\
      41$^{b*}$ &    1654.7182 &      86.0037 &&& \\
      42$^{b*}$ &    1655.6708 &      67.1056 &&& \\
      43$^{b*}$ &    1658.8667 &       1.5603 &&& \\
      44 &    1659.6806 &      25.9239 &&& \\
      45 &    1660.1304 &      26.0074 &&& \\
      46 &    1660.3043 &      24.5183 &&& \\ 
      47$^{b*}$ &    1672.4596 &      26.6593 &&& \\
      48$^{b*}$ &    1678.8129 &      23.6462 &&& \\
      49$^{b}$  &    1693.1435 &       0.5825 &&& \\
      50$^s$    &    3302.1390 &    1539.8513 &&& \\
\hline
\end{tabular} \\
\vskip 0.2 cm
{\bf Note:} $^t$OH torsion; $^b$OH scissoring; $^s$OH stretching; $^f$free OH; *vibrations contaminated by NH$_3$ modes. 
\end{table}

\begin{table}
\scriptsize
\centering
\caption{Harmonic IR frequencies and intensities of the complex 4H$_2$O/4NH$_3$ evaluated at the B2PLYP/maug-cc-pVTZ level \citep{gora20a}.}
\label{tab:4H2O_4NH3_B2PLYP}
\vskip 0.2 cm
\begin{tabular}{cccccc}
\hline
{\bf NM} &$\omega$ {[\bf cm$^{-1}$]} & {\bf IR Int. [km/mol]} & {\bf NM} &$\omega$ {[\bf cm$^{-1}$]} & {\bf IR Int. [km/mol]} \\
\hline
       1 &       4.9193 & 4.9769 & 51 &    1659.5005 & 30.4815 \\
       2 &       9.0775 & 2.0942 & 52 &    1659.5303 & 21.8453 \\
       3 &      17.7847 & 11.5017 & 53 &    1660.0190 & 25.0784 \\
       4 &      19.1211 & 1.0599 & 54 &    1660.1436 & 28.1216 \\
       5 &      22.4392 & 0.6988 & 55$^{b*}$ &    1673.3466 & 67.4843 \\
       6 &      25.9838 & 0.4032 & 56$^{b*}$ &    1674.2718 & 0.3138 \\
       7 &      27.6254 & 26.7443 & 57$^{b}$ &    1688.6077 & 33.8790 \\
       8 &      35.4482 & 23.6215 & 58$^{b}$ &    1691.9499 & 0.0133 \\
       9 &      39.7909 & 2.7928 & 59$^s$ &    3369.6234 & 0.2819 \\
      10 &      42.1429 & 0.1113 & 60$^s$ &    3385.6965 & 2574.9240 \\
      11 &      46.9852 & 0.0030 & 61$^s$ &    3387.6855 & 2801.0441 \\
      12 &      62.8166 & 0.0005 & 62$^s$ &    3394.0812 & 0.7615 \\
      13 &      72.1392 & 0.0251 & 63$^{s*}$ &    3473.3538 & 0.1108 \\
      14 &      74.9466 & 18.9444 & 64 &    3483.3149 & 13.0568 \\
      15 &     158.5797 & 0.0024 & 65 &    3483.4935 & 11.1778 \\
      16 &     174.6287 & 36.1323 & 66 &    3483.6384 & 2.6130 \\
      17 &     178.2147 & 35.4550 & 67$^{s*}$ &    3484.6015 & 0.0846 \\
      18 &     201.4197 & 0.0105 & 68$^{s}$ &    3507.3380 & 1567.4274 \\
      19 &     233.9639 & 3.4824 & 69$^{s}$ &    3510.2809 & 1820.7051 \\
      20 &     235.9429 & 83.2509 & 70$^{s}$ &    3536.4083 & 0.1714 \\ 
      21 &     237.2040 & 33.3809 & 71 &    3600.7487 & 17.4948 \\
      22 &     239.2784 & 72.0949 & 72 &    3600.7879 & 17.3727 \\
      23 &     239.6302 & 3.0360 & 73 &    3600.9757 & 17.5063 \\
      24 &     242.2423 & 80.4390 & 74 &    3601.0932 & 17.3475 \\
      25 &     243.0833 & 36.6777 & 75 &    3601.1747 & 17.5682 \\
      26 &     246.3526 & 2.0432 & 76 &    3601.2866 & 17.5694 \\
      27 &     261.6961 & 0.8343 & 77 &    3601.4175 & 17.2943 \\
      28 &     266.9875 & 0.6015 & 78 &    3601.5786 & 17.6734 \\
      29 &     269.0885 & 42.6136 &&& \\
      30 &     273.9278 & 71.7557 &&& \\ 
      31$^t$ & 613.8454 & 0.0116 &&& \\
      32 &     630.4192 & 212.2155 &&& \\
      33 &     666.2987 & 0.0041 &&& \\
      34 &     670.4647 & 0.0071 &&& \\
      35 &     700.3570 & 199.6747 &&& \\
      36 &     744.5970 & 379.8388 &&& \\
      37$^t$ & 755.9618 & 361.3020 &&& \\
      38$^t$ & 799.7584 & 0.0050 &&& \\
      39$^t$ & 815.2370 & 0.3721 &&& \\
      40$^t$ & 825.5015 & 342.7098 &&& \\
      41$^t$ & 846.8651 & 297.2880 &&& \\
      42$^t$ & 946.2925 & 0.0017 &&& \\
      43 &    1125.1644 & 153.1328 &&& \\
      44 &    1125.2155 & 143.7533 &&& \\
      45 &    1125.4081 & 503.2784 &&& \\
      46 &    1125.4766 & 35.6108 &&& \\
      47$^{b*}$ &    1654.5148 & 120.1771 &&& \\
      48$^{b*}$ &    1654.8637 & 41.6854 &&& \\
      49$^{b*}$ &    1657.2551 & 58.5445 &&& \\
      50$^{b*}$ &    1657.8279 & 12.4900 &&& \\
\hline
\end{tabular} \\
\vskip 0.2 cm
{\bf Note:} $^t$OH torsion; $^b$OH scissoring; $^s$OH stretching; $^f$free OH; *vibrations contaminated by NH$_3$ modes.
\end{table}

\subsubsection{NH$_3$ ice}
\label{NH3_ice}
Most of the intense modes of ammonia overlap with the dominant features due to water and silicates. However, when ammonia is mixed with $\rm{H_2O}$ ice, it forms hydrates showing an intense mode at $2881.84$ cm$^{-1}$ ($3.47$ $\mu$m) \citep{dart05}.
Another characteristic feature of ammonia is the umbrella
mode at $1111.11$ cm$^{-1}$ (9.00 $\mu$m), which is relatively intense.
Still, it often overlaps with the $\rm{CH_3}$ rocking mode of methanol,
thus leading to an overestimation of the abundance of ammonia.

\begin{figure}
\centering
\includegraphics[width=0.49\textwidth]{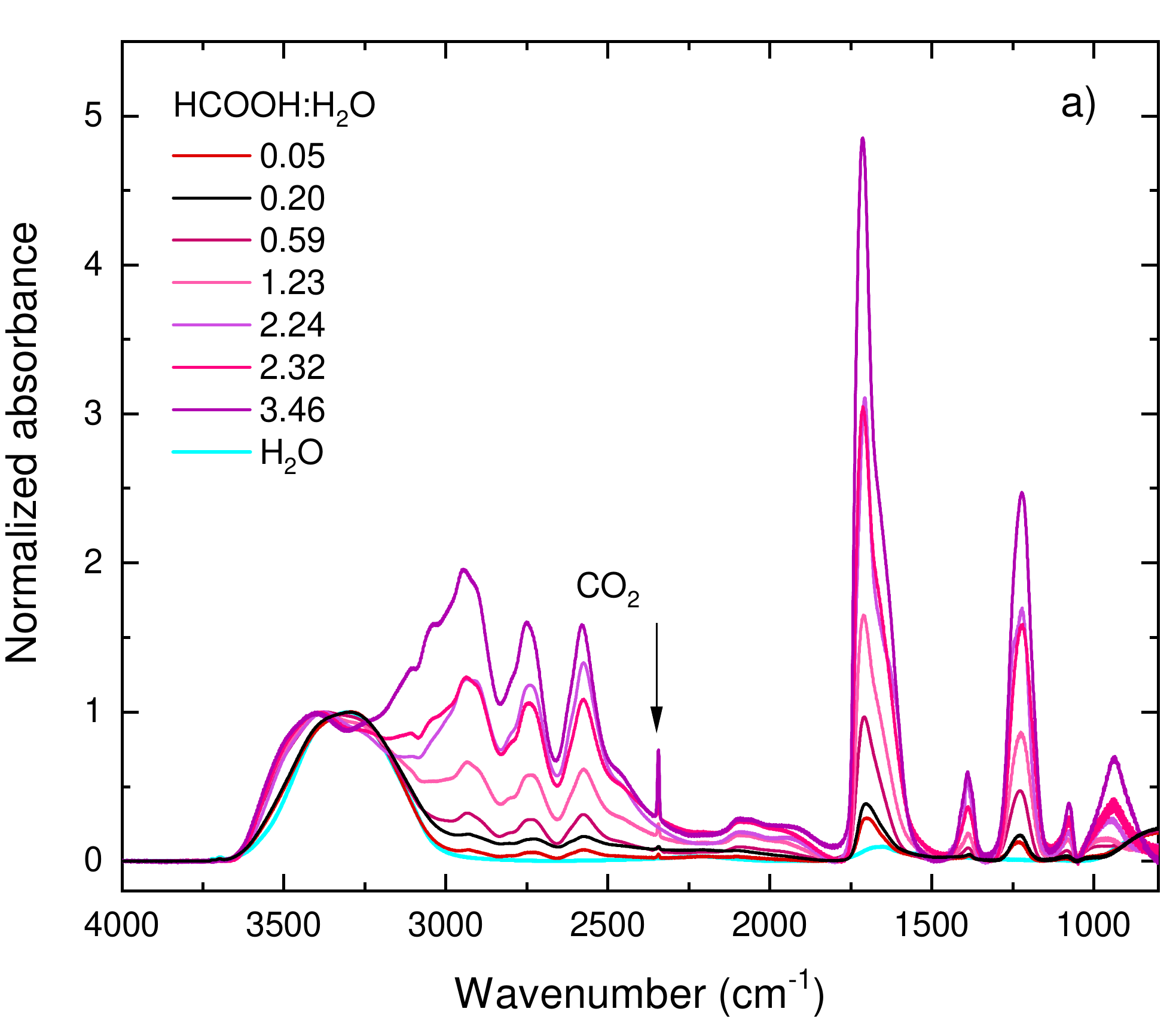}
\includegraphics[width=0.49\textwidth]{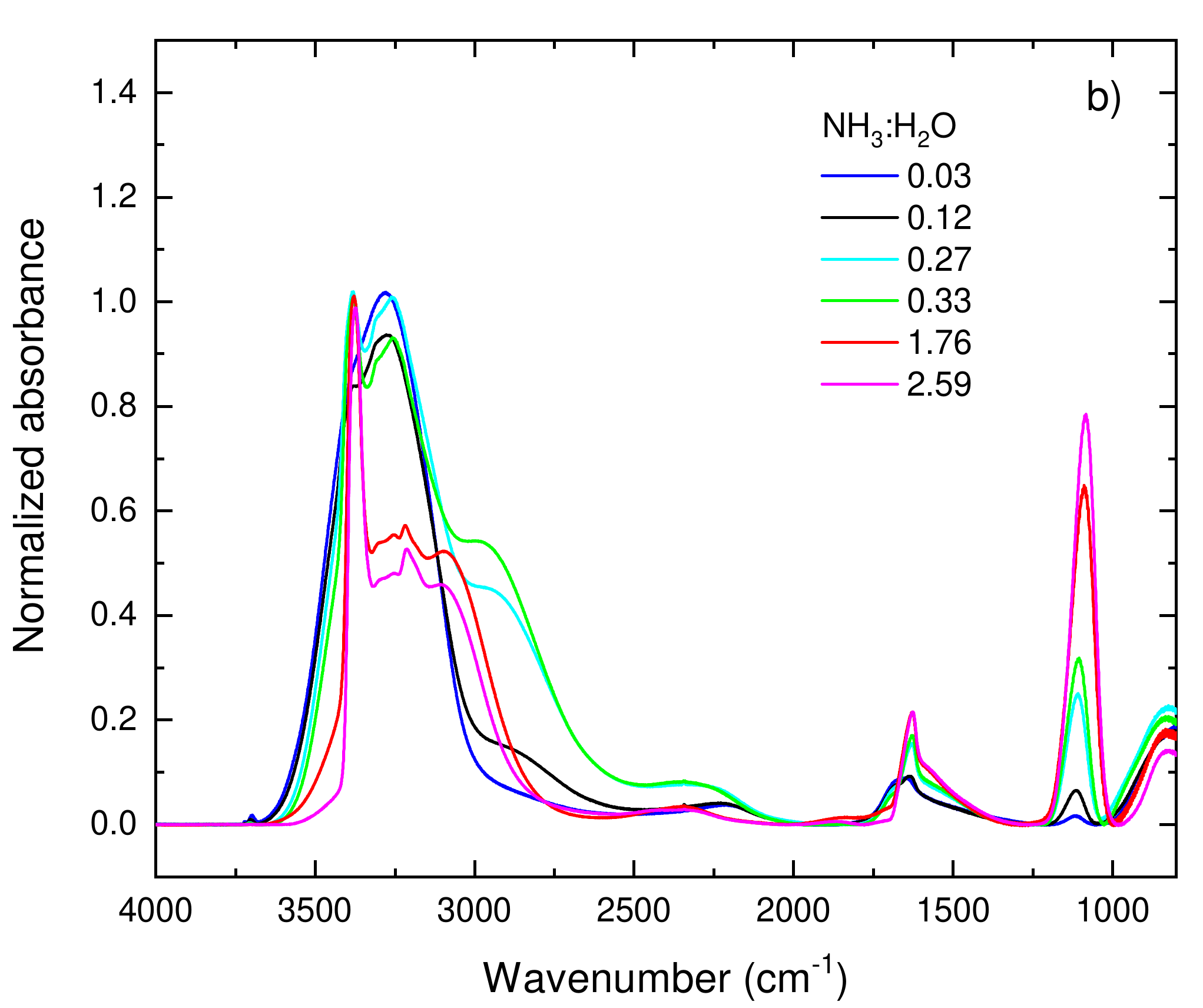}
\vskip 0.5cm
\includegraphics[width=0.49\textwidth]{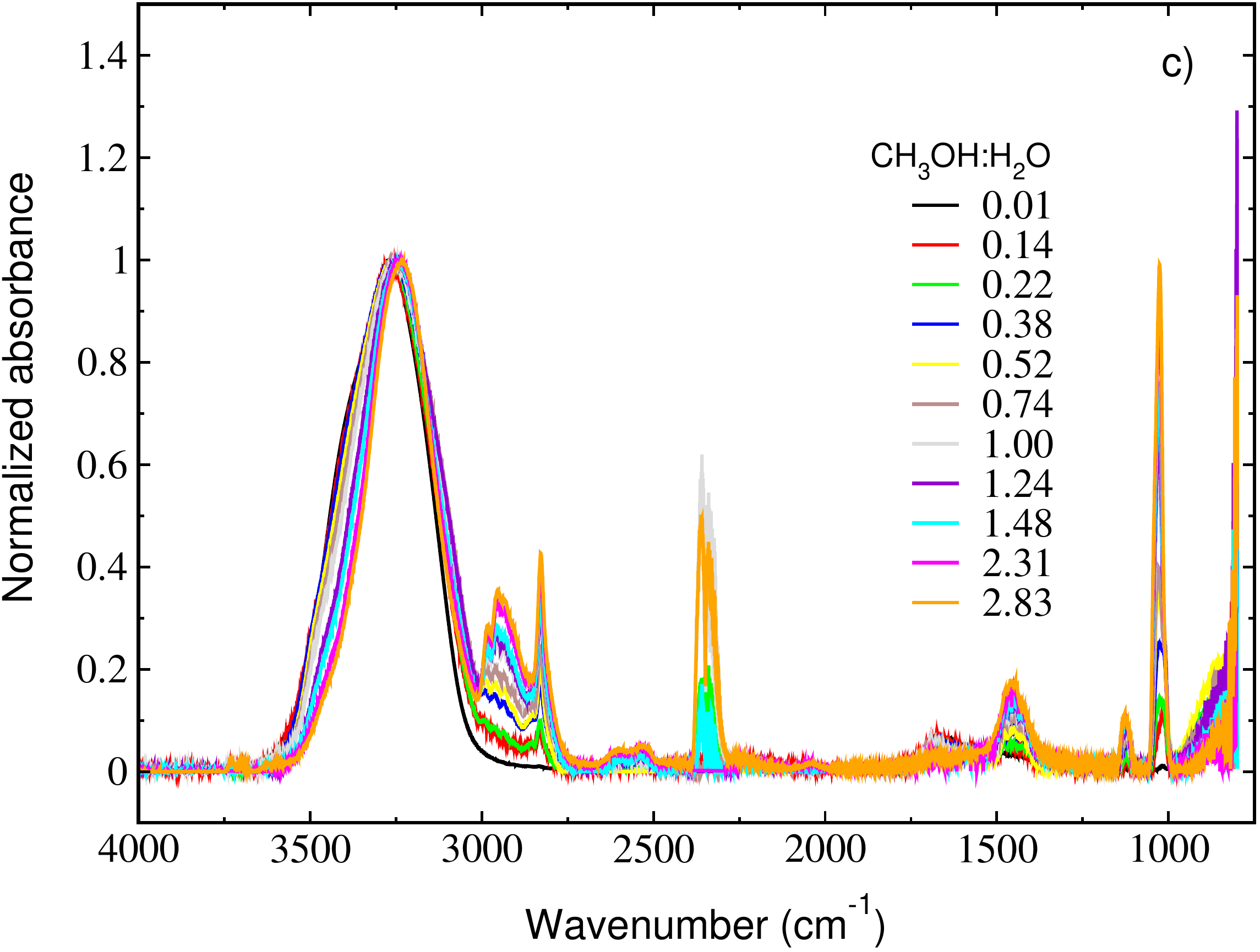}
\caption{(a) IR spectra for different $\rm{HCOOH:H_2O}$ ice mixtures deposited at $T=20$ K. (b) IR spectra for different $\rm{NH_3:H_2O}$ ice mixtures deposited at $T=20$ K. (c) IR spectra for different $\rm{CH_3OH:H_2O}$ ice mixtures deposited at $T=30$ K. The color legend is explained in the insets. All IR spectra are normalized with respect to the $\rm{O-H}$ stretching band \citep{gora20a}.}
\label{fig:experiment}
\end{figure}

\begin{figure}
\centering
\includegraphics[width=0.68\textwidth]{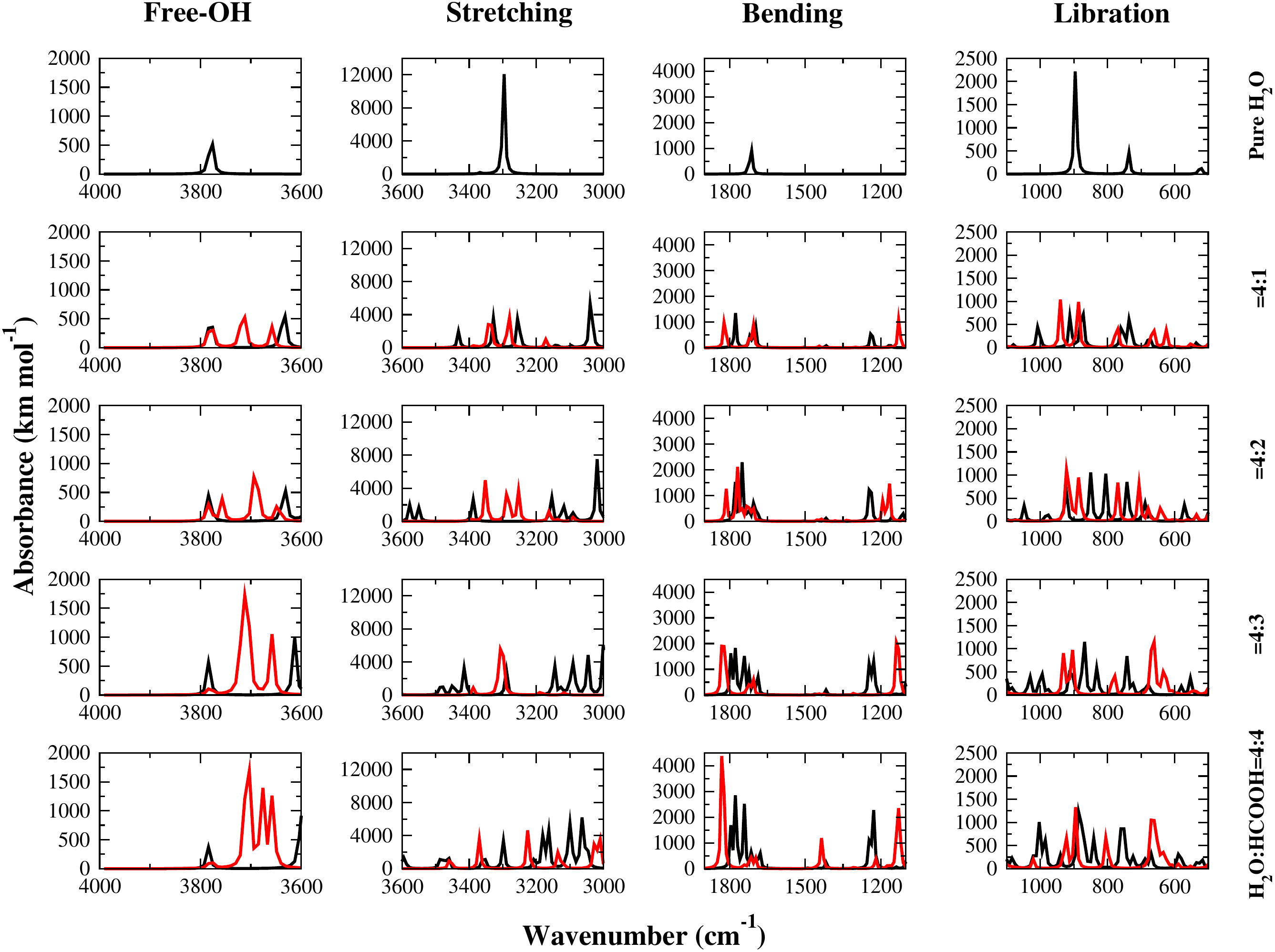}
\caption{Absorption spectra of the four modes for water ice for the five measured compositions, ranging
from pure water ice (top) to $4:4$ $\rm{H_2O-HCOOH}$ mixture (bottom). The black line represents the absorbance spectra of various
concentration of $\rm{H_2O-HCOOH}$, where HCOOH is used as a hydrogen bond donor, and for the red line, HCOOH is used as a
hydrogen bond  acceptor \citep{gora20a}.}
\label{fig:H2O-HCOOH}
\end{figure}

In this work, IR spectra are recorded for various mixing ratios of $\rm{H_2O-NH_3}$ ice deposited at $20$ K. IR spectra, normalized to the most intense band (i.e., the $\rm{O-H}$ stretching mode) are shown in Figure \ref{fig:experiment}b. Mixing ratios are derived by measuring the areas of the selected bands for the H$_2$O band (at 2220 cm$^{-1}$) \citep{mast09} and for NH$_3$ (umbrella mode band at 1070 cm$^{-1}$) \citep{dhen86}, with a procedure analogous to that introduced in the previous Section for HCOOH.

\begin{figure}
\centering
\begin{minipage}{0.4\textwidth}
\includegraphics[width=\textwidth]{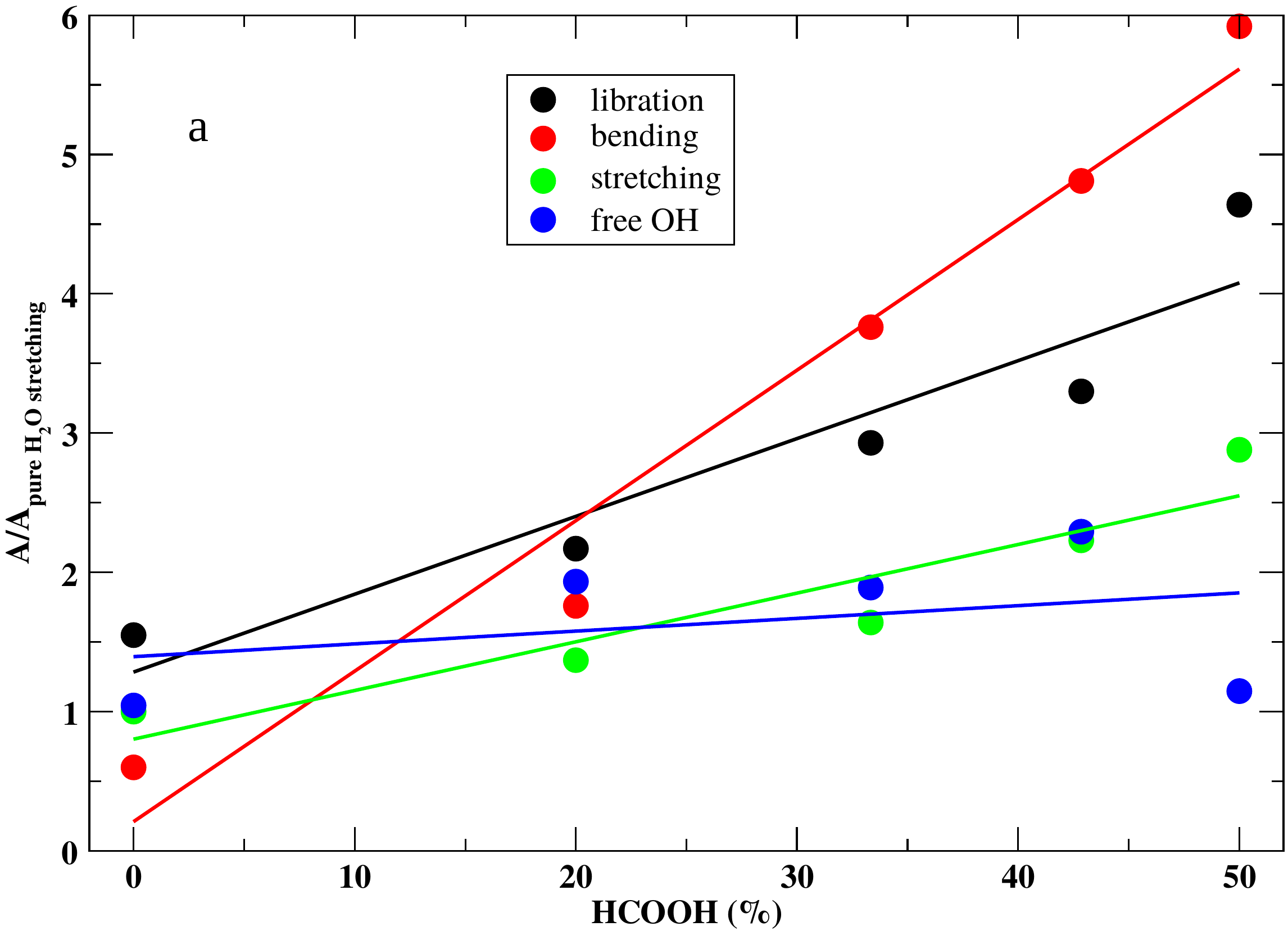}
\end{minipage}
\begin{minipage}{0.4\textwidth}
\includegraphics[width=\textwidth]{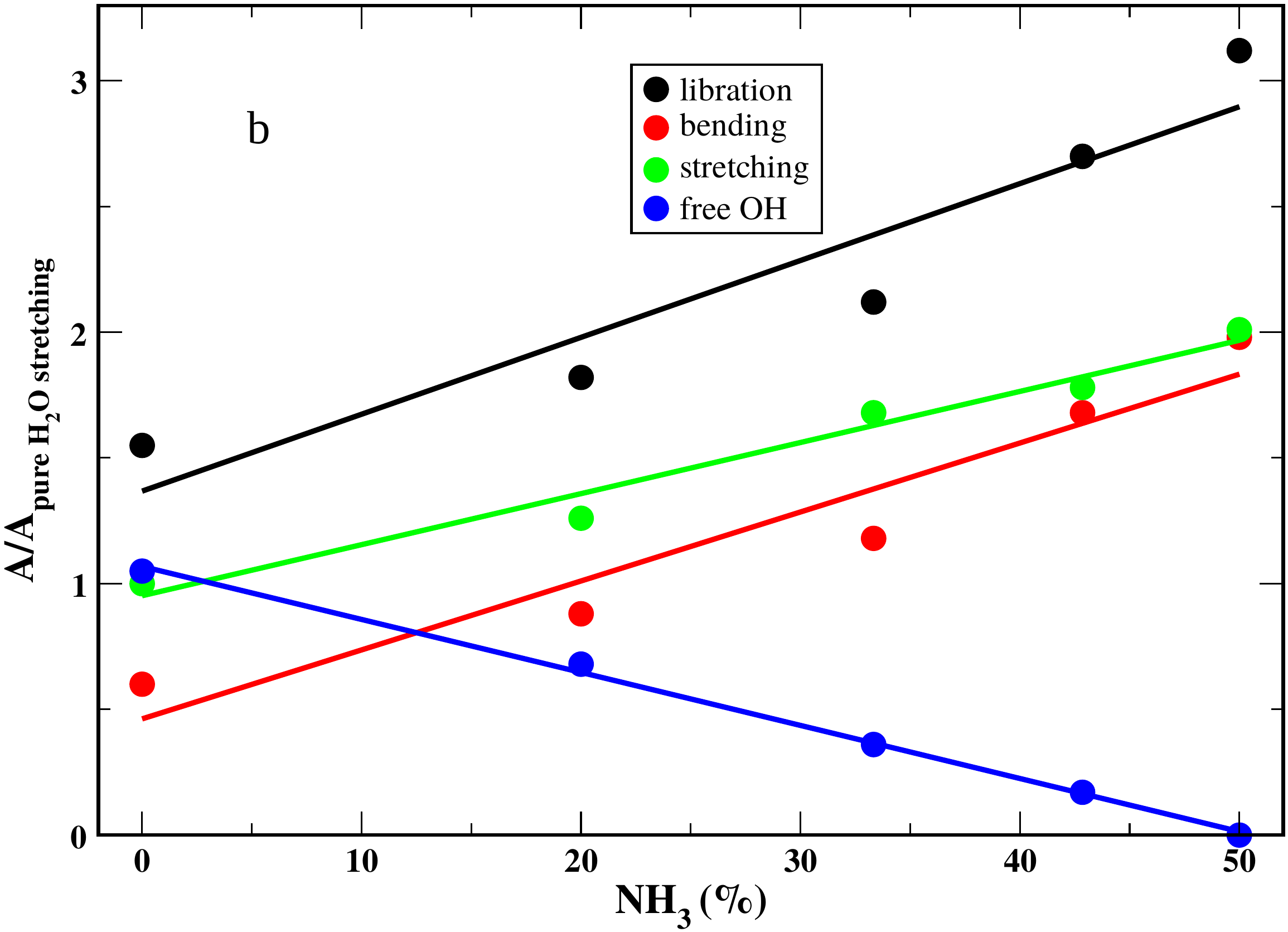}
\end{minipage}
\begin{minipage}{0.4\textwidth}
\includegraphics[width=\textwidth]{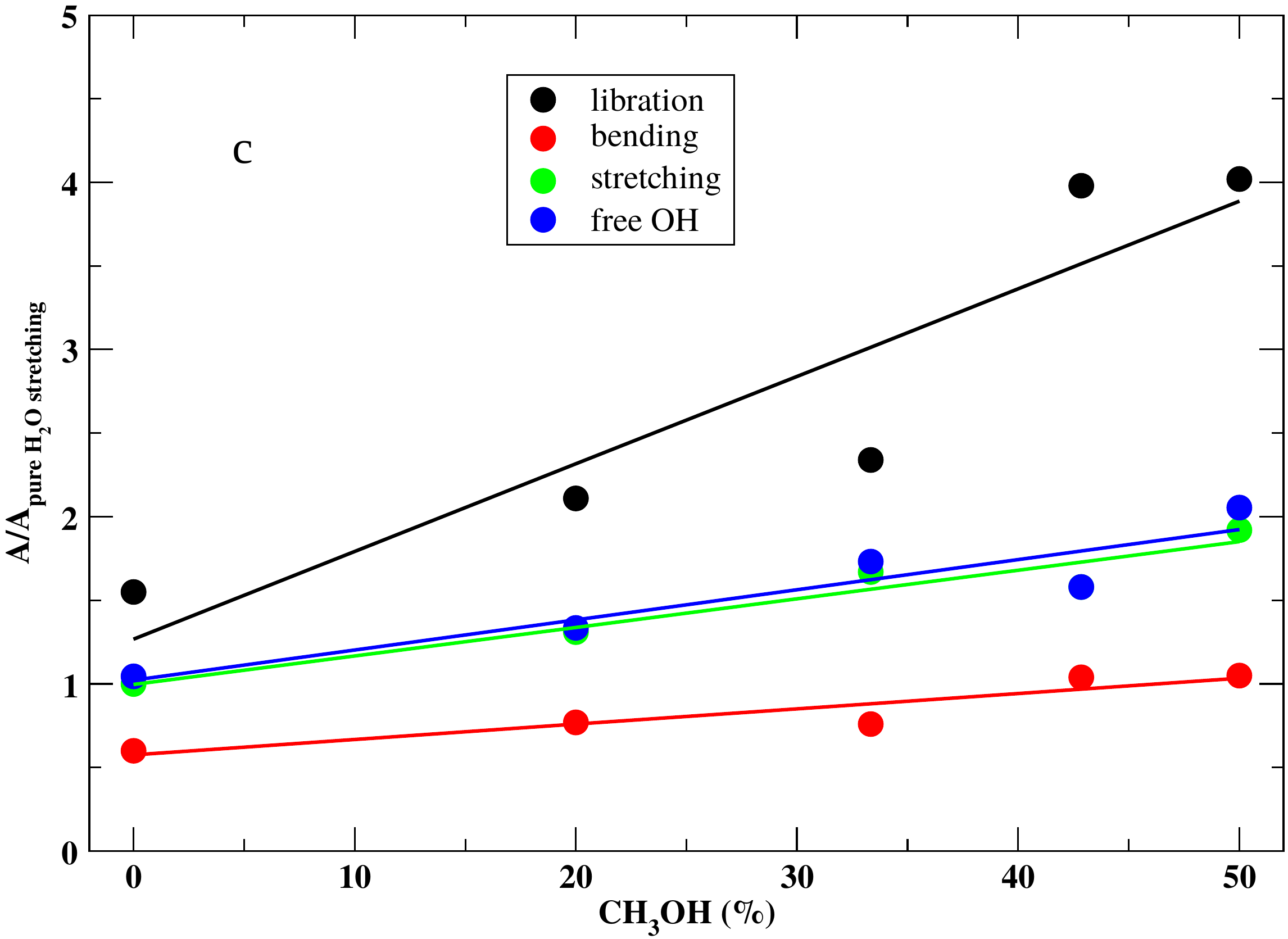}
\end{minipage}
\begin{minipage}{0.4\textwidth}
\includegraphics[width=\textwidth]{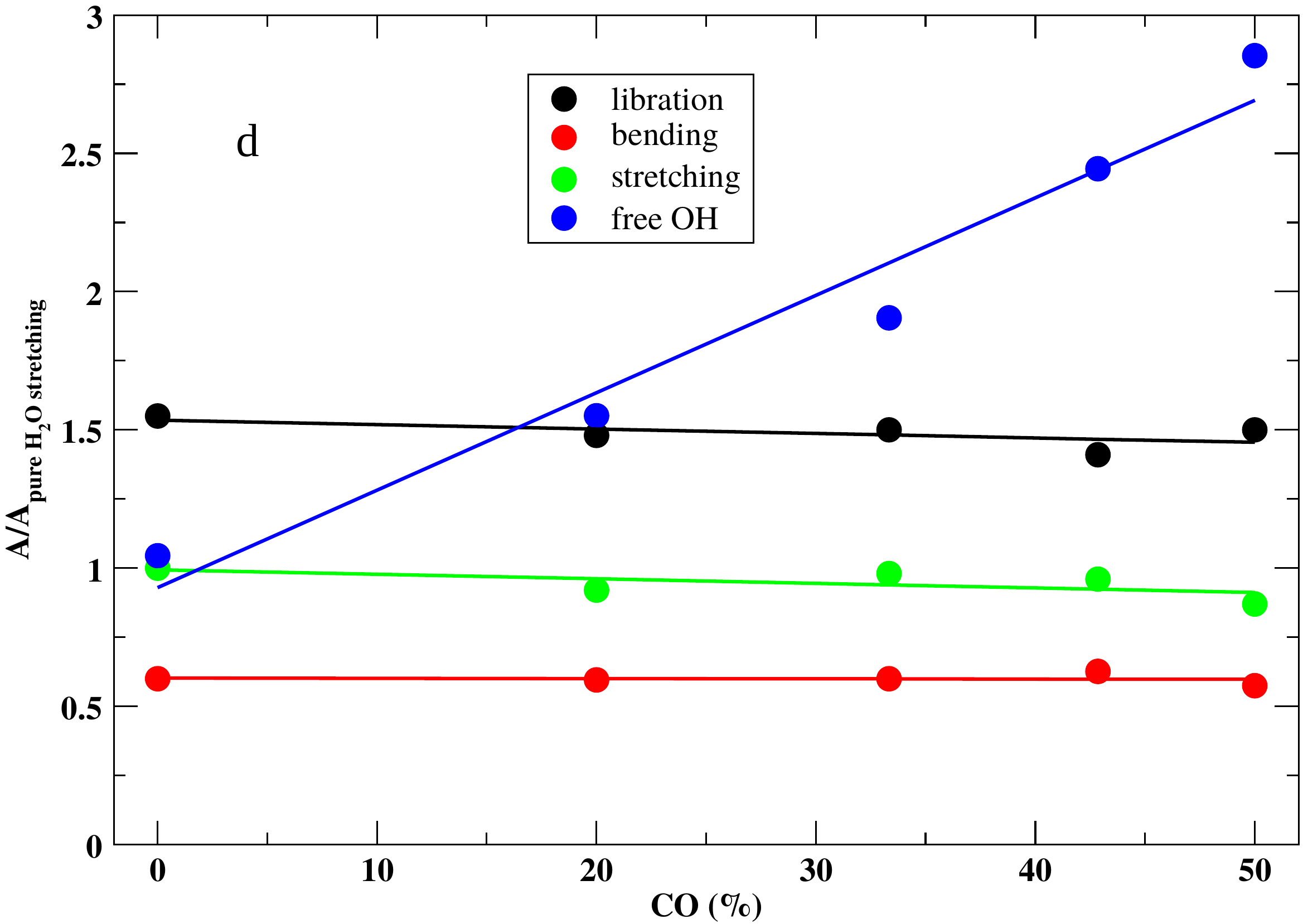}
\end{minipage}
\begin{minipage}{0.4\textwidth}
\includegraphics[width=\textwidth]{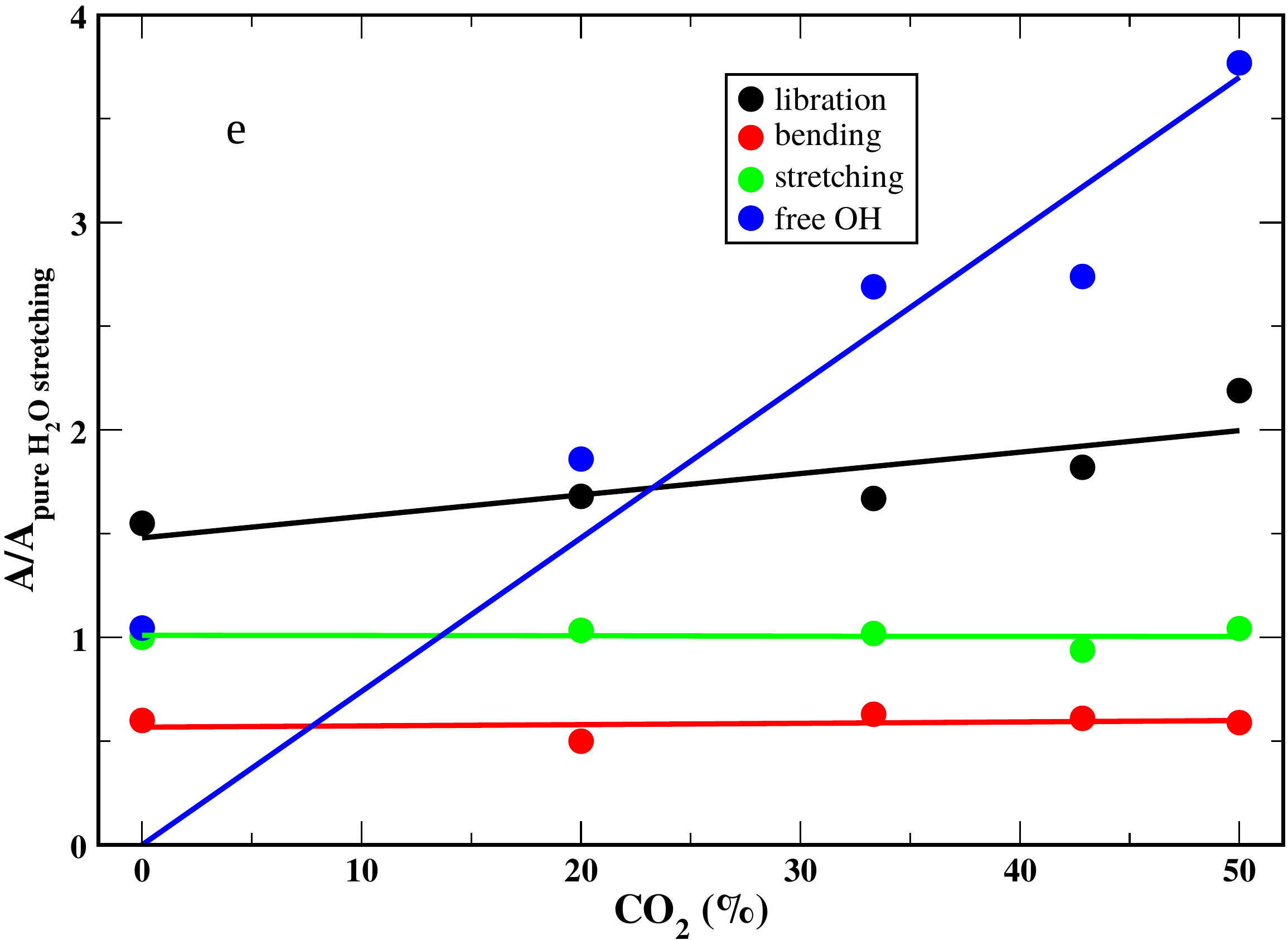}
\end{minipage}
\begin{minipage}{0.4\textwidth}
\includegraphics[width=\textwidth]{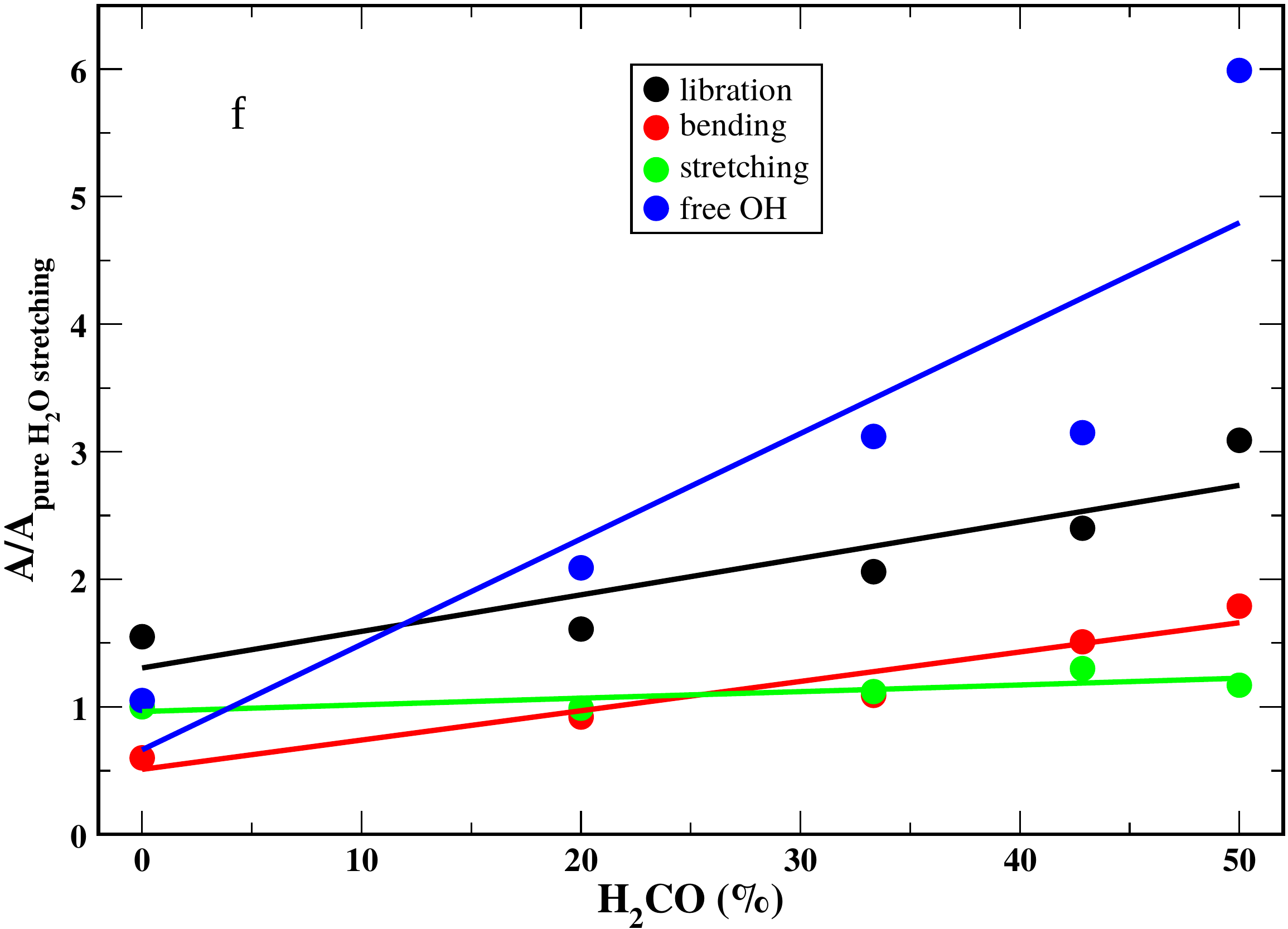}
\end{minipage}
\begin{minipage}{0.4\textwidth}
\includegraphics[width=\textwidth]{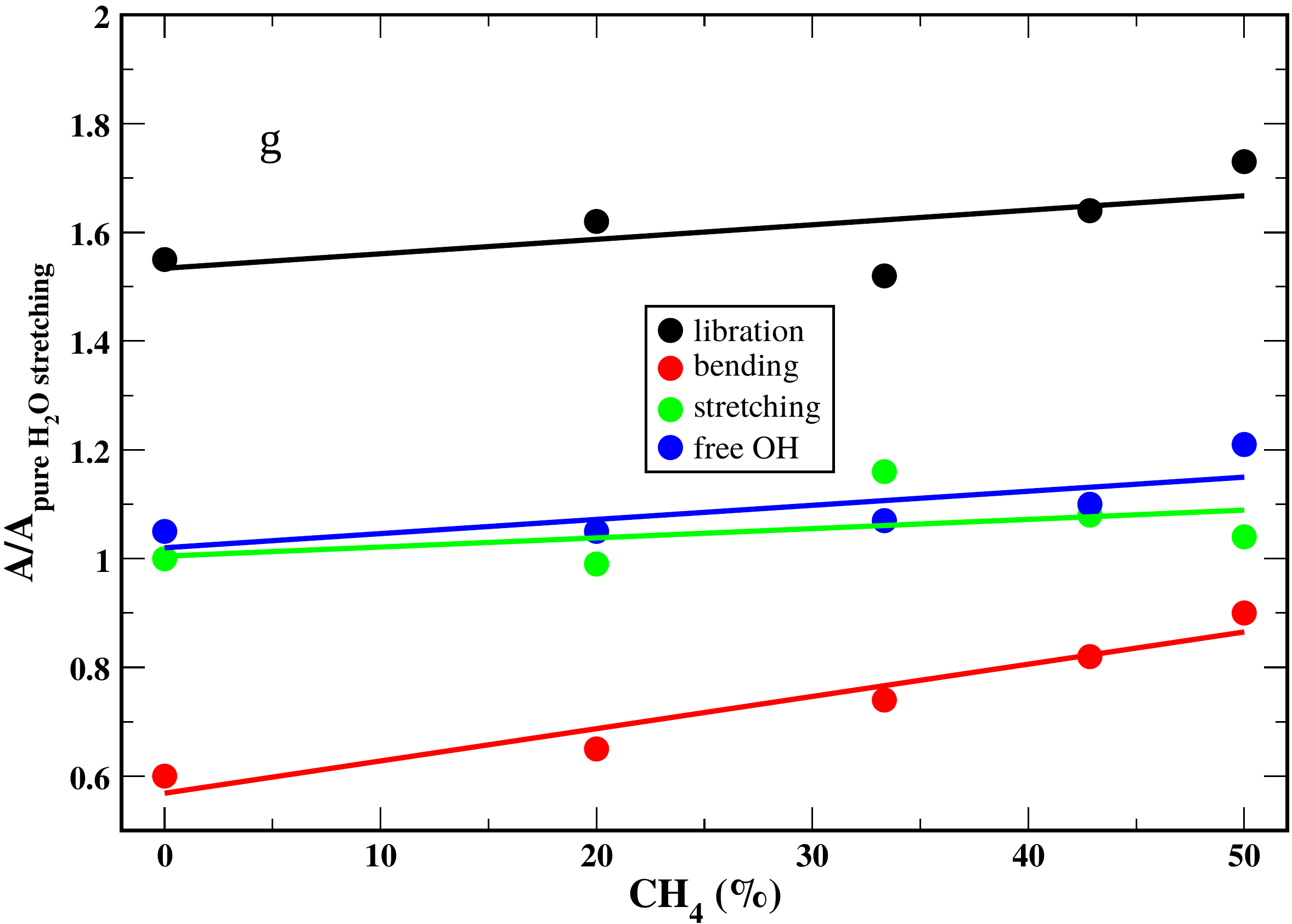}
\end{minipage}
\begin{minipage}{0.4\textwidth}
\includegraphics[width=\textwidth]{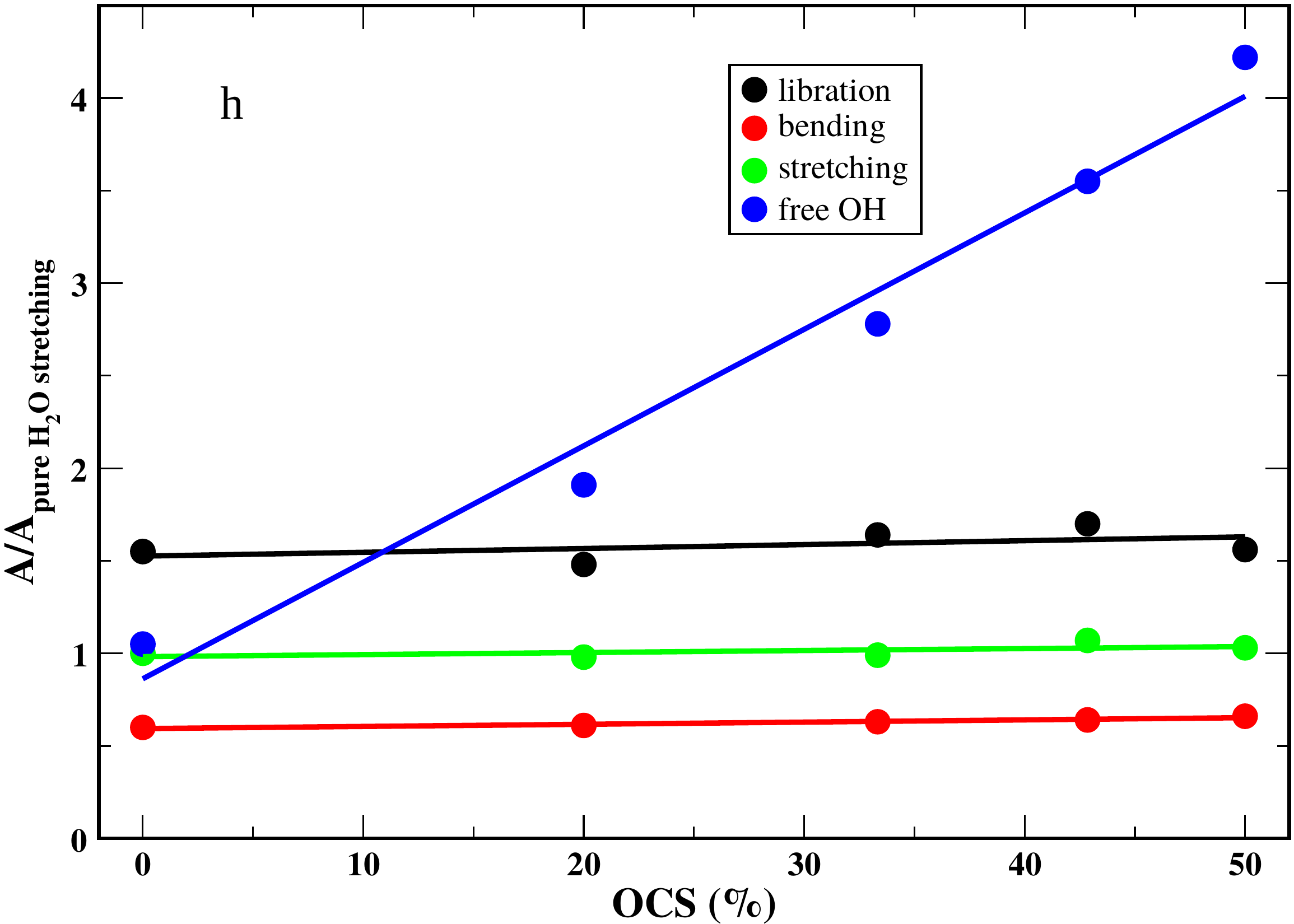}
\end{minipage}
\begin{minipage}{0.4\textwidth}
\includegraphics[width=\textwidth]{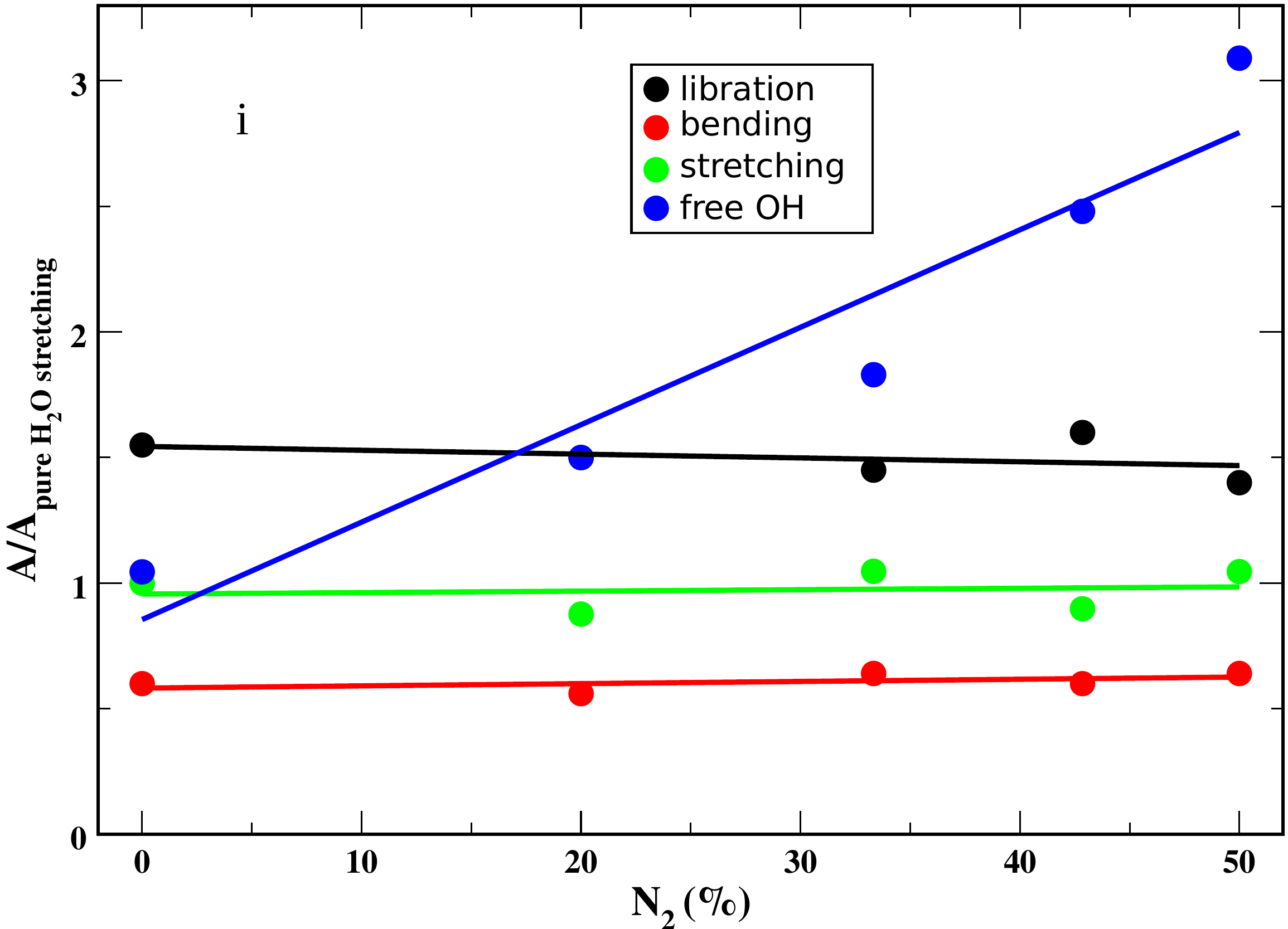}
\end{minipage}
\begin{minipage}{0.4\textwidth}
\includegraphics[width=\textwidth]{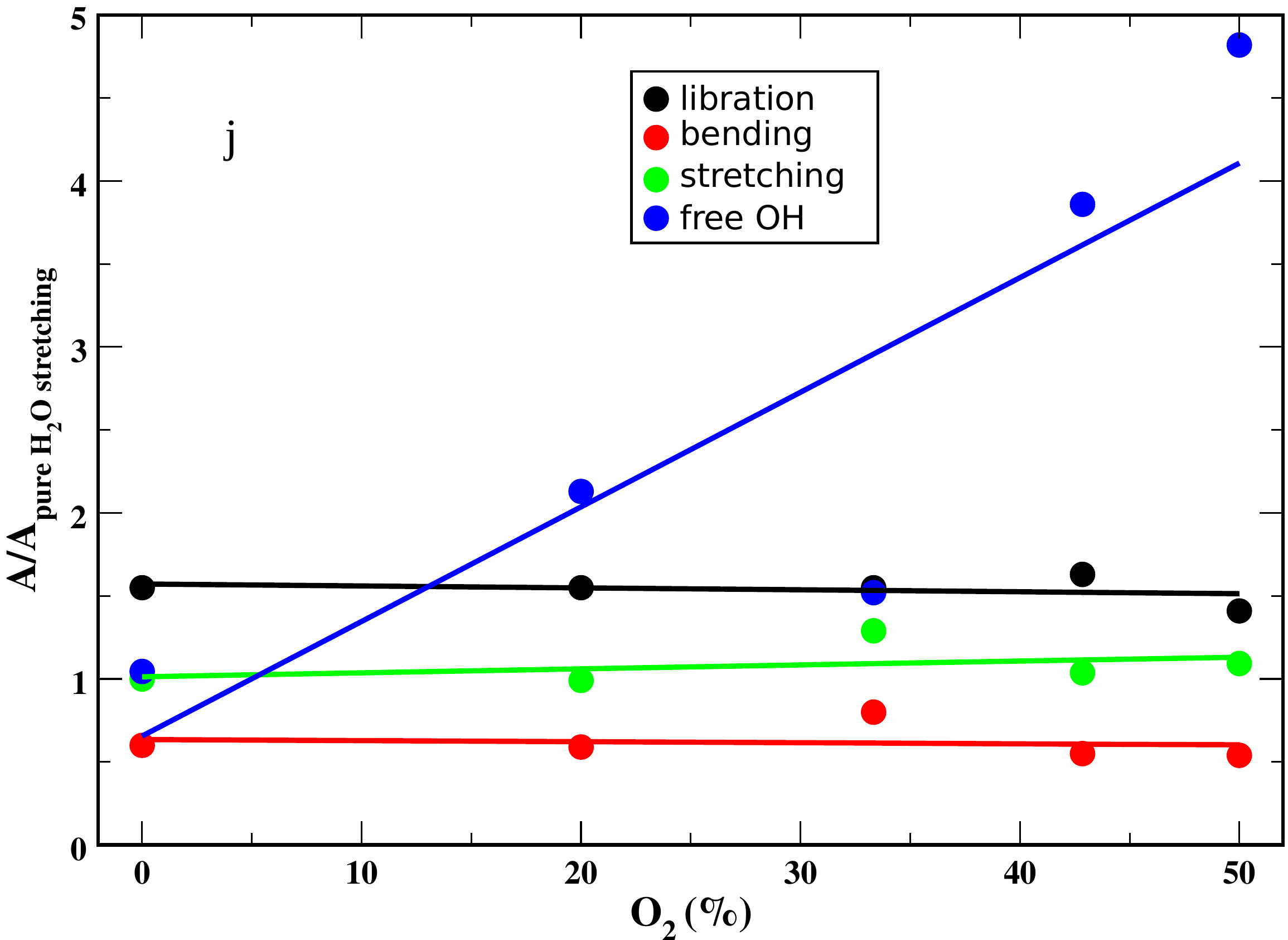}
\end{minipage}
\caption{Band strengths of the four fundamental vibration modes of water for
(a) $\rm{H_2O-HCOOH}$, (b) $\rm{H_2O-NH_3}$, (c) $\rm{H_2O-CH_3OH}$, (d) $\rm{H_2O-CO}$, (e) $\rm{H_2O-CO_2}$, (f) $\rm{H_2O-H_2CO}$, (g) $\rm{H_2O-CH_4}$, (h) $\rm{H_2O-OCS}$, (i) $\rm{H_2O-N_2}$, and (j) $\rm{H_2O-O_2}$ clusters with various concentrations. The water c-tetramer configuration was used for pure water \citep{gora20a}.}
\label{fig:band_strength}
\end{figure}

\begin{figure}
\centering
\includegraphics[width=0.68\textwidth]{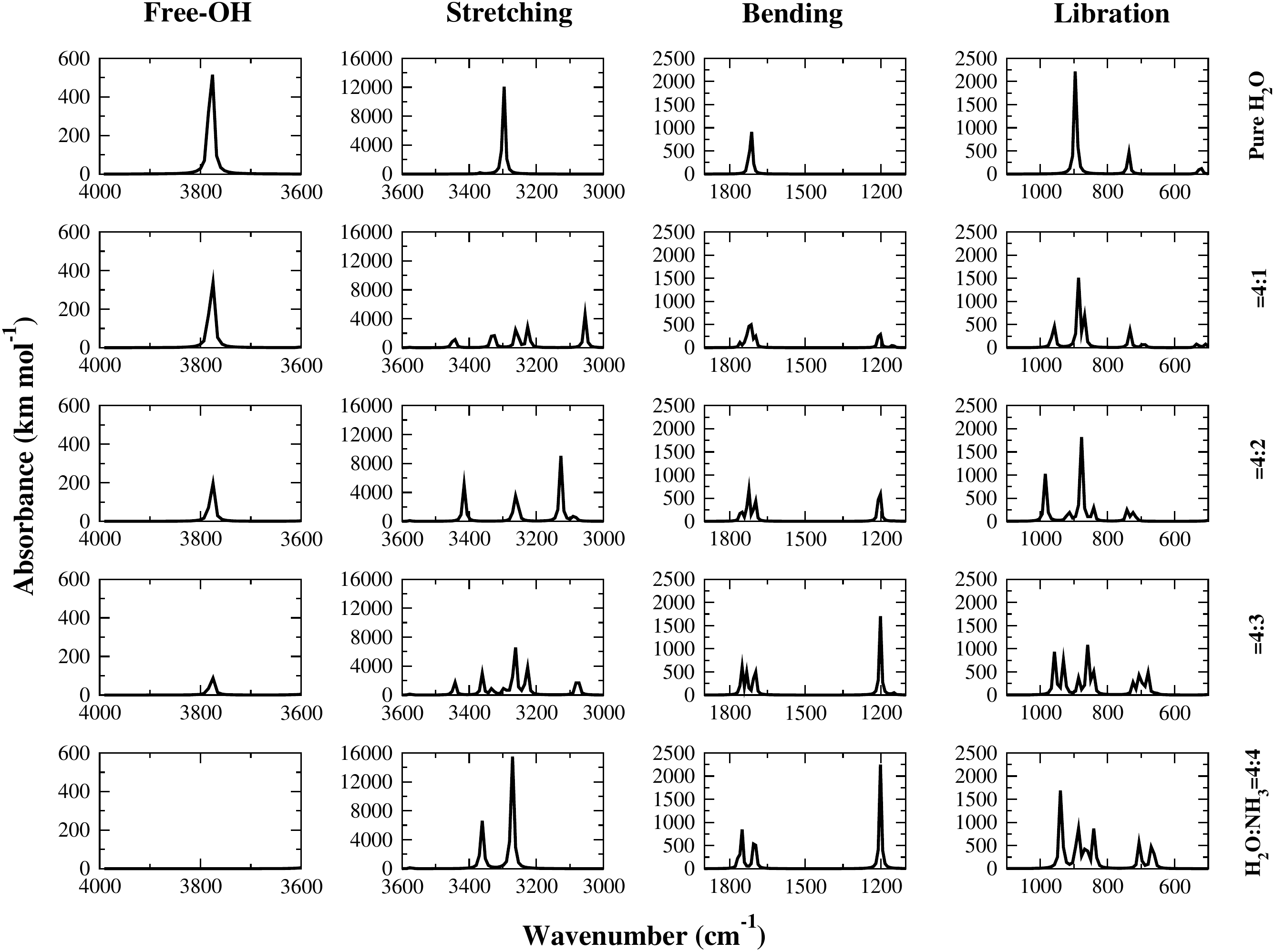}
\caption{Absorption spectra of the four modes for water ice for the five measured compositions, ranging
from pure water ice (top) to $4:4$ $\rm{H_2O-NH_3}$ mixture (bottom) \citep{gora20a}.}
\label{fig:H2O-NH3}
\end{figure}

Figure \ref{fig:optimized_structure}c shows the optimized geometry of the $\rm{H_2O-NH_3}$ system with a $4:4$ ratio as obtained from our quantum chemical calculations.
Figure \ref{fig:H2O-NH3} depicts the absorption band profiles of $\rm{H_2O-NH_3}$ mixtures
with various concentrations. The transition frequencies and the corresponding intensity values are provided in Table \ref{tab:H2O_X} as well. The vibrational analysis is also carried out at a higher-level theory, thereby using the B2PLYP functional.
The results are reported in Tables \ref{tab:4H2O_1NH3_B2PLYP}$-$\ref{tab:4H2O_4NH3_B2PLYP}. Figure \ref{fig:band_strength}b shows the band strengths as a function of the impurity
concentration under consideration, i.e., NH$_3$.

\subsubsection{Comparison between experiment and simulations}
\label{comparison}

\begin{figure}
\centering
\includegraphics[width=0.6\textwidth]{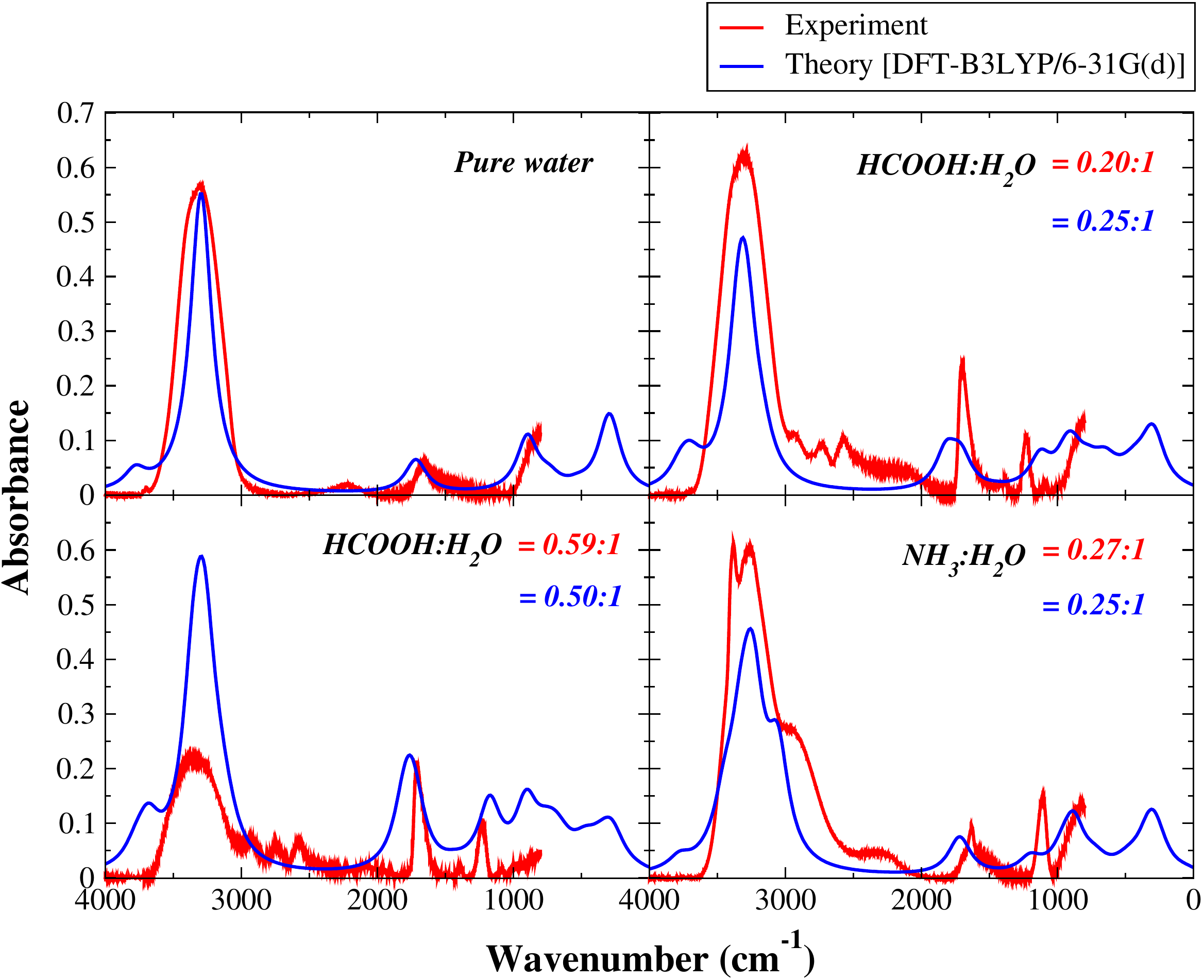}
\caption{Comparison of computed and experimental IR spectra ($0 - 4000$ cm$^{-1}$) for pure water as well as water with HCOOH and NH$_3$ as impurities. We have used harmonic frequencies for the computed spectra, and the intensity is scaled with a factor 1000 to have the best match with the experimental one \citep{gora20a}.}
\label{fig:comparison_IR}
\end{figure}
\begin{figure}
\centering
\begin{minipage}{0.49\textwidth}
\vskip 0.8cm
\includegraphics[width=\textwidth]{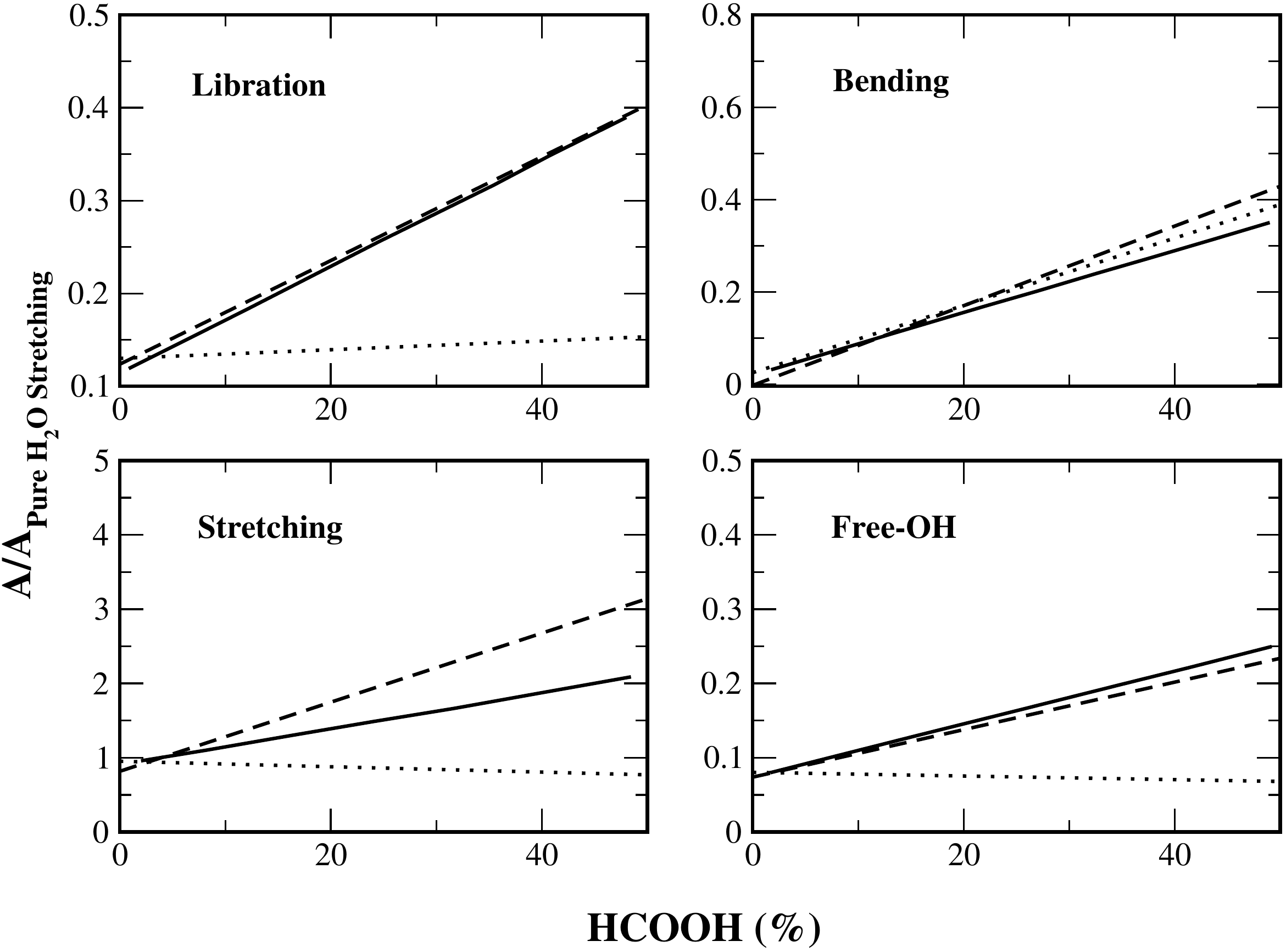}
\end{minipage}
\begin{minipage}{0.49\textwidth}
\includegraphics[width=\textwidth]{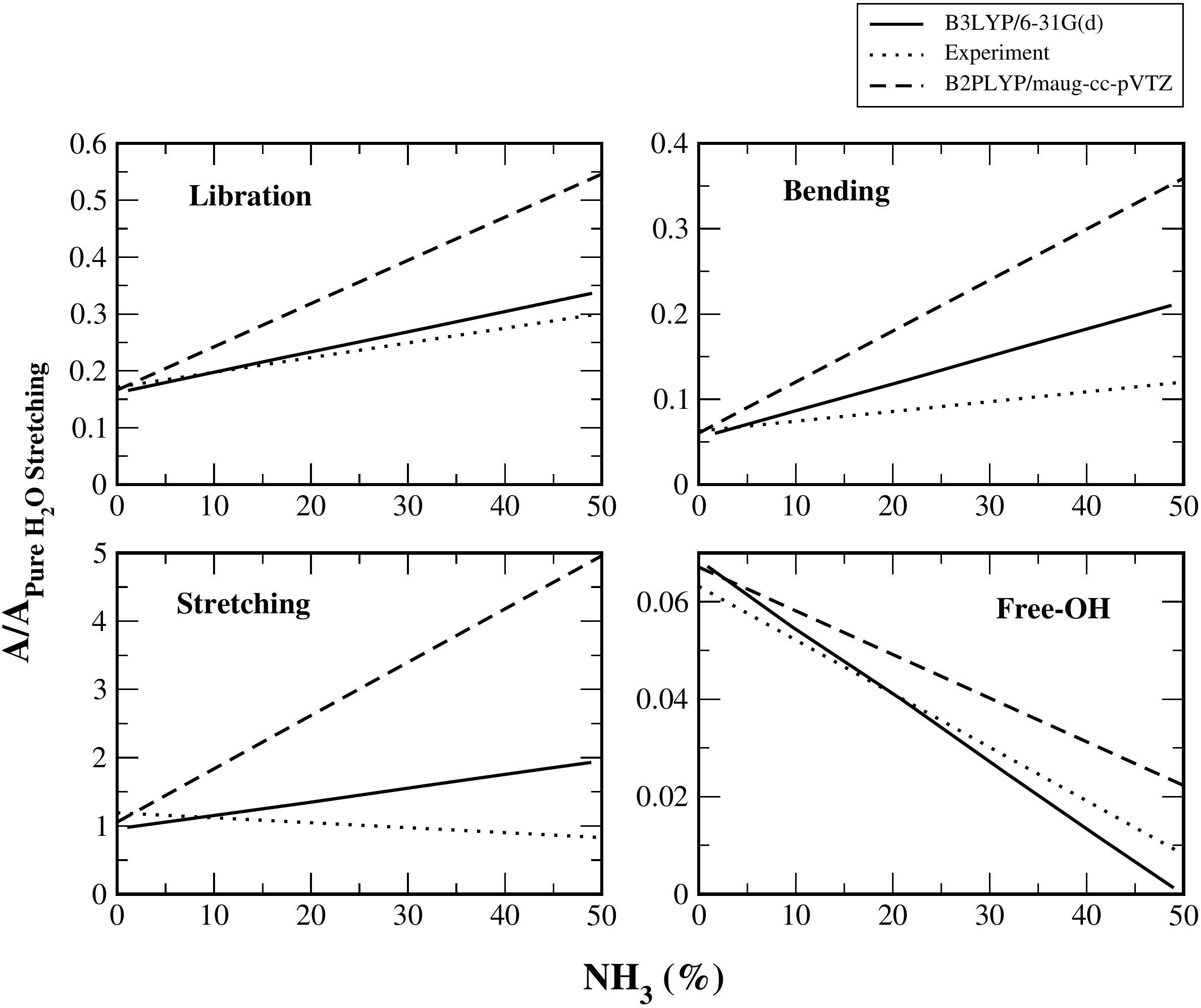}
\end{minipage}
\caption{Comparison between the calculated and experimental band strength profiles with various concentration of HCOOH and NH$_3$ \citep{gora20a}.}
\label{fig:HCOOH-NH3_band_strength}
\end{figure}
\begin{figure}
\centering
\includegraphics[width=0.4\textwidth]{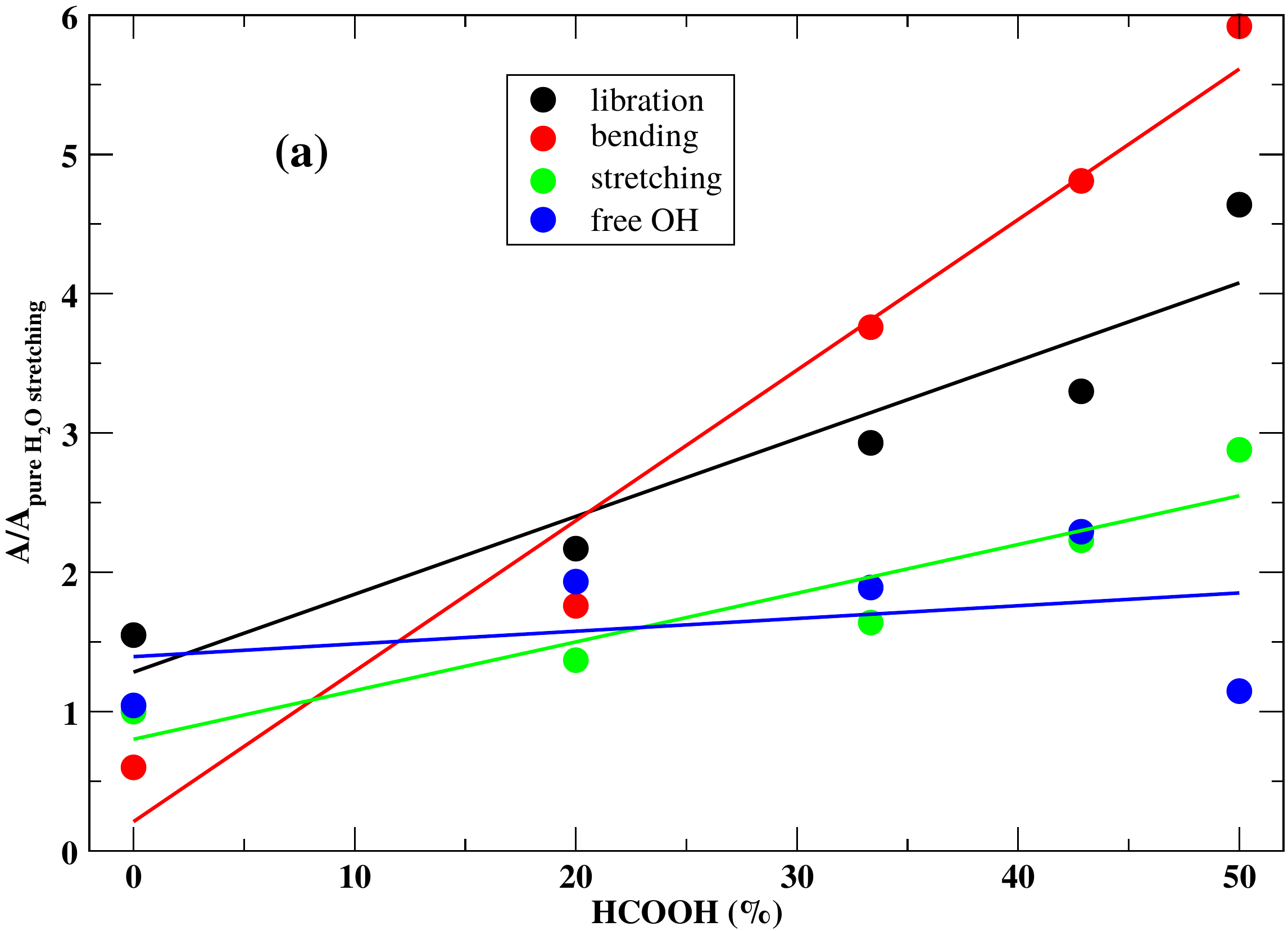}
\includegraphics[width=0.4\textwidth]{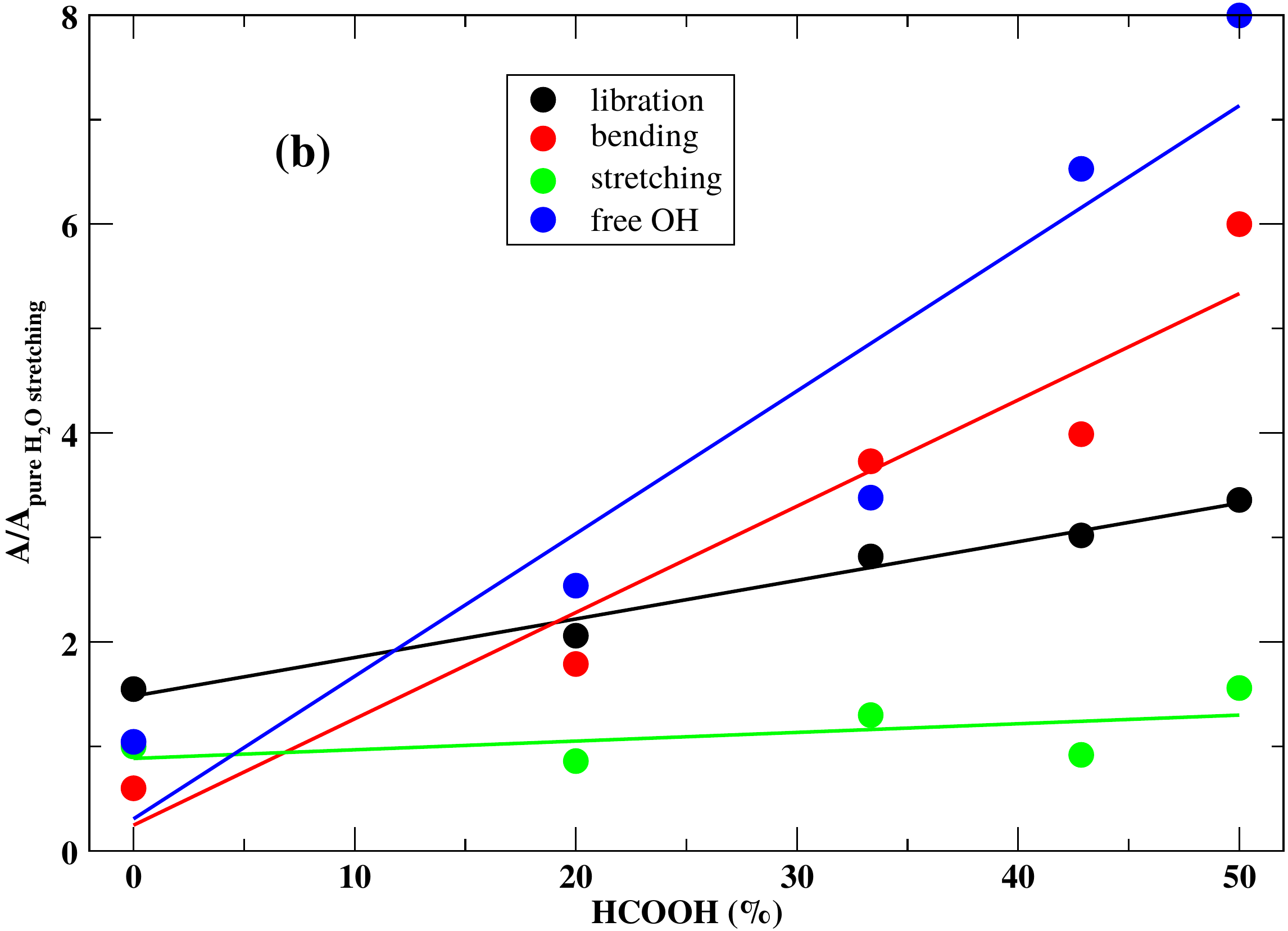}
\caption{Band strength for $\rm{H_2O-HCOOH}$ mixtures: (a) HCOOH as an H-bond donor and (b) HCOOH as an H-bond acceptor \citep{gora20a}.}
\label{fig:hcooh-donor-acceptor}
\end{figure}

In Figure \ref{fig:comparison_IR}, the comparison between experimentally obtained spectra and our computed spectra for pure water, the $\rm{H_2O-HCOOH}$ mixture, and $\rm{H_2CO-NH_3}$ mixture is shown. We note a good agreement between experimental and theoretical absorption spectra. Figure \ref{fig:HCOOH-NH3_band_strength} shows the comparison between the experimental (dotted lines) and theoretical (solid and dashed lines) band strengths of the four water bands as a function of the concentration of HCOOH and NH$_3$. From Figure \ref{fig:HCOOH-NH3_band_strength}, it is evident that the experimental strength of the libration and bending modes increases by increasing the concentration of HCOOH. On the contrary, the strength of the stretching and free OH modes shows a decreasing trend. These behaviors should be compared with the B2PLYP/mug-cc-pVTZ (dashed) and B3PLYP/6-31G(d) (solid) trends. For the libration and bending band strength profiles, there is a qualitative agreement with experiments. However, in the stretching and free OH modes, theoretical band strength profiles deviate from experimental work. The lack of experimental data in the $3600-4000$ cm$^{-1}$ range (see Figure \ref{fig:experiment}a) may contribute to this disagreement. By comparing the two levels of theory, it is noted that there is a rather good agreement. In the $\rm{H_2O-HCOOH}$ mixture, HCOOH can act as both an H-bond donor and an H-bond acceptor. We consider both interactions. We note that if we consider HCOOH as an H-bond acceptor, the band strengths of the three modes (libration, bending, and stretching) are lower relative to the case where HCOOH is treated as the H-bond donor. However, in the free OH mode, the band strength slope increases (See Figure \ref{fig:hcooh-donor-acceptor}).

\begin{figure}
\centering
\includegraphics[width=0.4\textwidth]{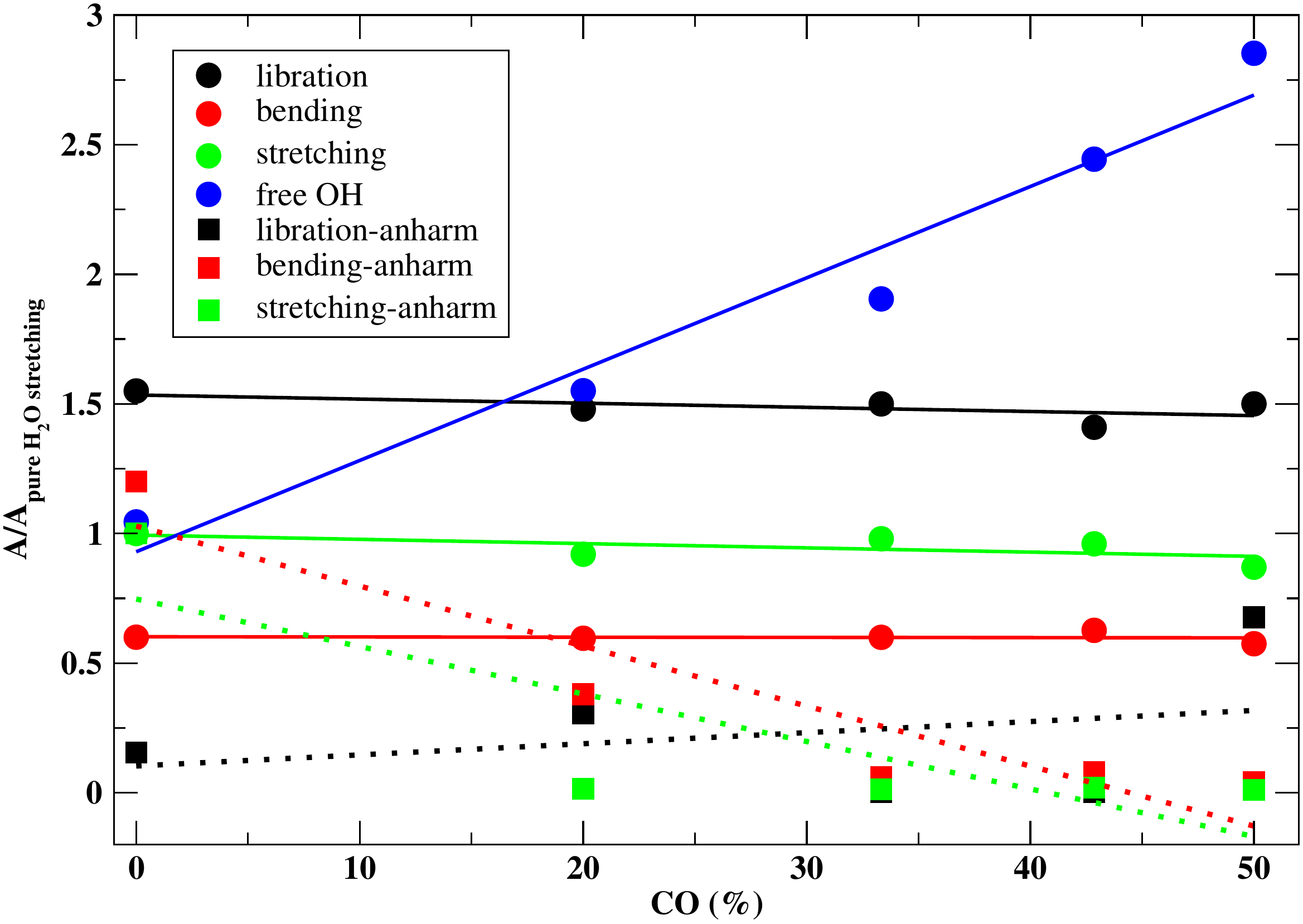}
\includegraphics[width=0.4\textwidth]{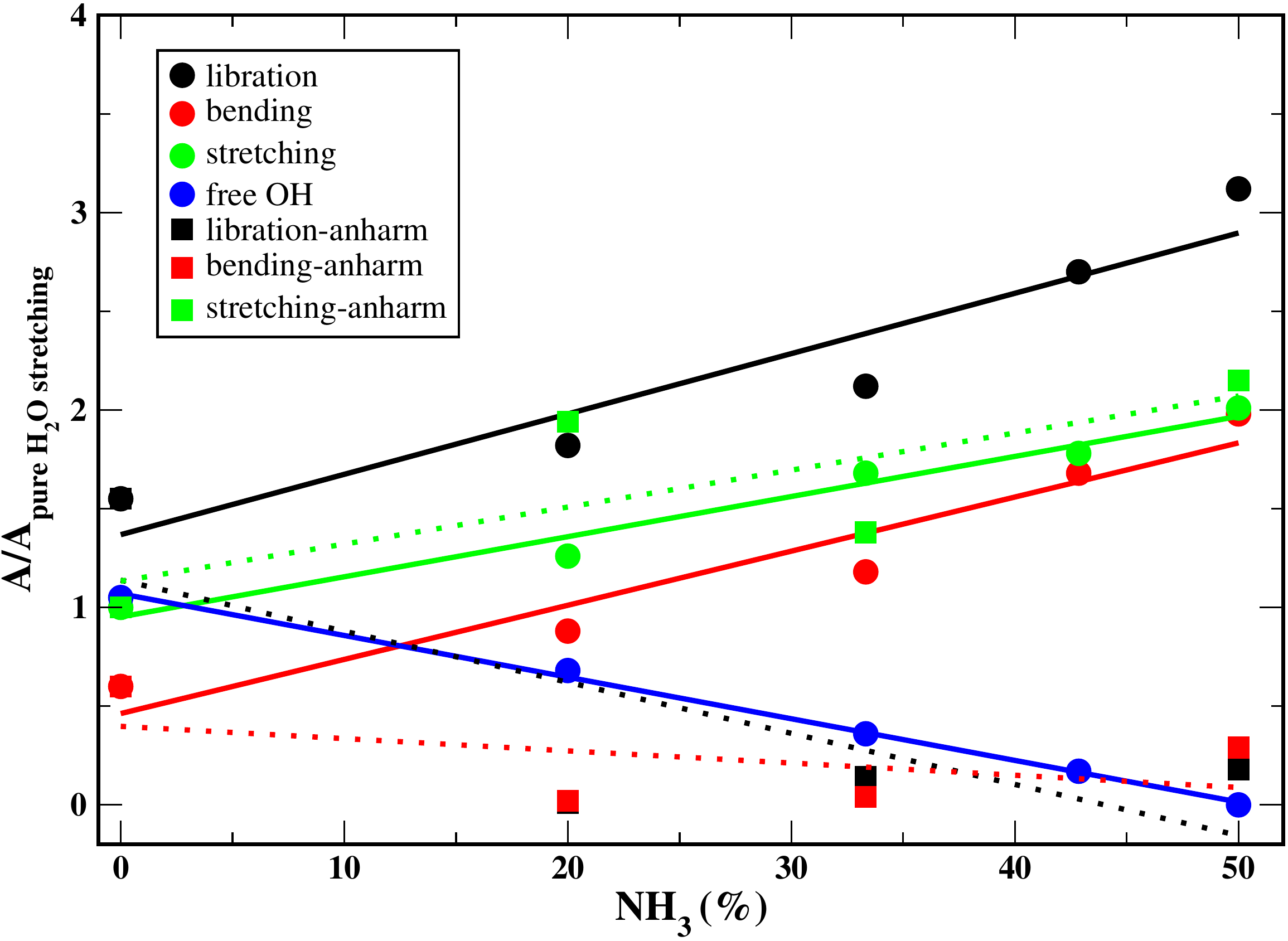}
\caption{Filled circles are the data points where we considered harmonic frequencies and the corresponding fitted profiles are the solid lines.
Solid filled squares represent the data sets where we considered anharmonic frequencies and the corresponding fitted results are the dotted
lines \citep{gora20a}.}
\label{fig:harm-anharm-compare}
\end{figure}

Moving to ammonia, the experimental data of Figure \ref{fig:HCOOH-NH3_band_strength} show that the band strength of the free OH stretching mode nearly vanishes when a $50\%$ concentration of the impurity (NH$_3$) is reached. This feature interestingly supports our calculated spectra shown in the last panel of Figure \ref{fig:H2O-NH3}. Instead, libration and bending modes have an opposite trend, with the band strength increasing by increasing the concentration of NH$_3$. The band strength of the stretching mode shows a slightly decreasing trend with the concentration of NH$_3$. From the inspection of Figure \ref{fig:HCOOH-NH3_band_strength}, it is evident that both sets of theoretical results (B3LYP and B2PLYP) are in reasonably good agreement with experimental data for the libration, bending, and free OH modes. Interestingly, the results obtained using the lower level of theory are in better agreement with experiments. In Figure \ref{fig:harm-anharm-compare}, the comparison of band strengths evaluated using (a) harmonic and (b) anharmonic calculations is shown. We only consider fundamental bands in the 0 to 3600 cm$^{-1}$ frequency range to investigate the effect of anharmonicity on the band strengths. From our experimental study on the $\rm{H_2O-NH_3}$ system, as already mentioned, we obtain an increasing trend of the band strength for the libration, bending, and stretching modes with the increase in the concentration of NH$_3$.
In contrast, the band strength decreases for the free OH mode and becomes zero with a $50\%$ concentration of NH$_3$. We find similar trends to those obtained from the experiment using harmonic calculations for all four fundamental modes. However, if we consider anharmonic estimates, only the behavior of the stretching mode is well reproduced. All other modes deviate from the experimental results. While we are not claiming that harmonic calculations are better than the anharmonic ones, this comparison suggests that the former show a better error compensation. A similar outcome is obtained for the $\rm{H_2O-CO}$ system and will be briefly addressed later in the text.

Based on the comparisons discussed above, the B3LYP/6-31G(d) level of theory provides reliable results. Therefore, it is employed in the following investigations. First of all, we compare the computed band strengths with the experiments for the $\rm{H_2O-CH_3OH,\ CO-H_2O}$, and $\rm{CO_2-H_2O}$ mixtures to support their suitability.

\begin{figure}
\centering
\includegraphics[width=0.68\textwidth]{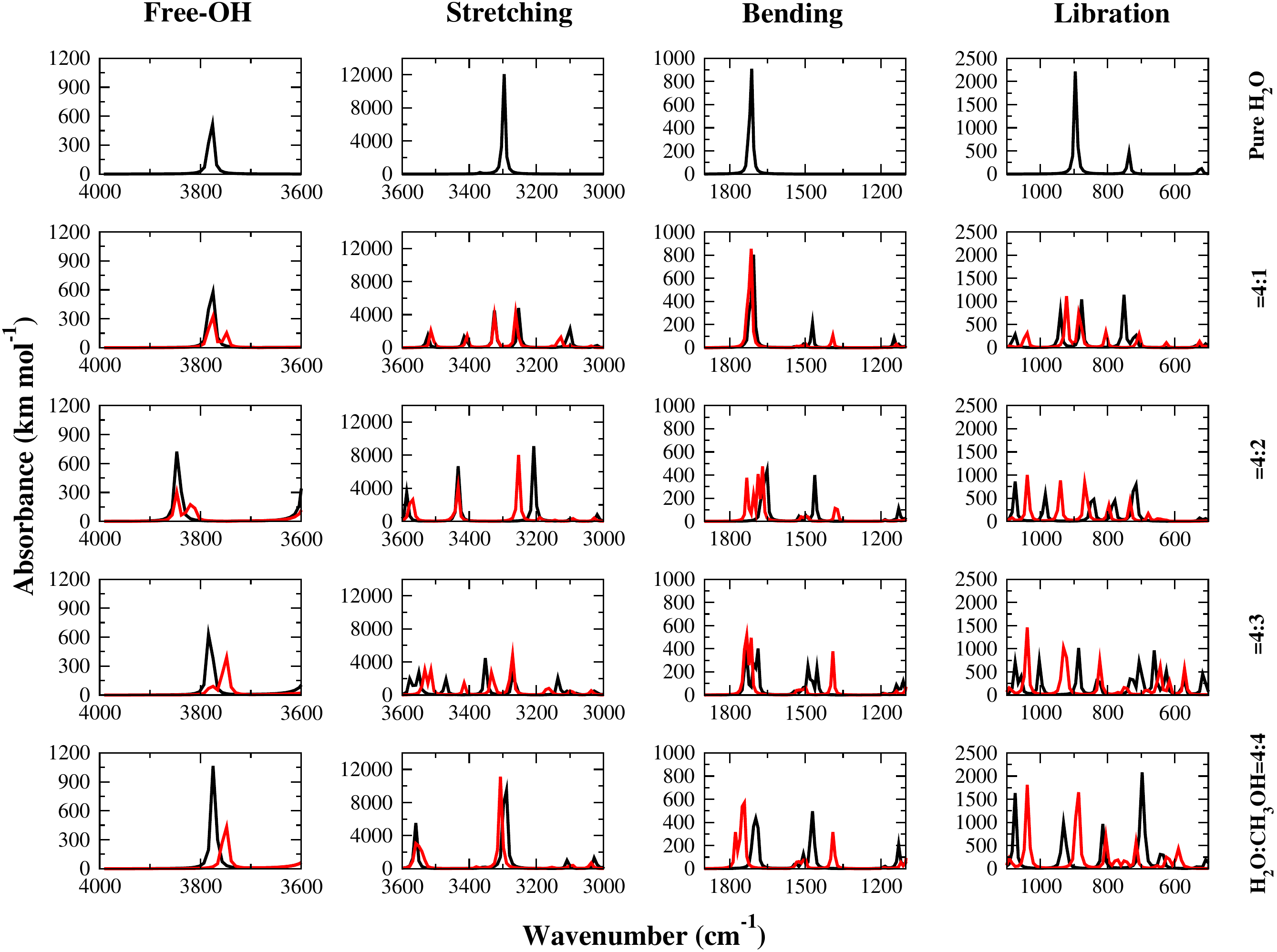}
\caption{Absorption spectra of the four modes for water ice for the five measured compositions, ranging from pure water ice (top) to $4:4$
$\rm{H_2O-CH_3OH}$ mixture (bottom). The black line represents the absorbance spectra of various
concentrations of H$_2$O-CH$_3$OH, where CH$_3$OH is used as a hydrogen bond donor, and for the red line, CH$_3$OH is used as a
hydrogen bond  acceptor \citep{gora20a}.}
\label{fig:H2O-CH3OH}
\end{figure}

 \begin{figure}
 \centering
 \includegraphics[width=0.7\textwidth]{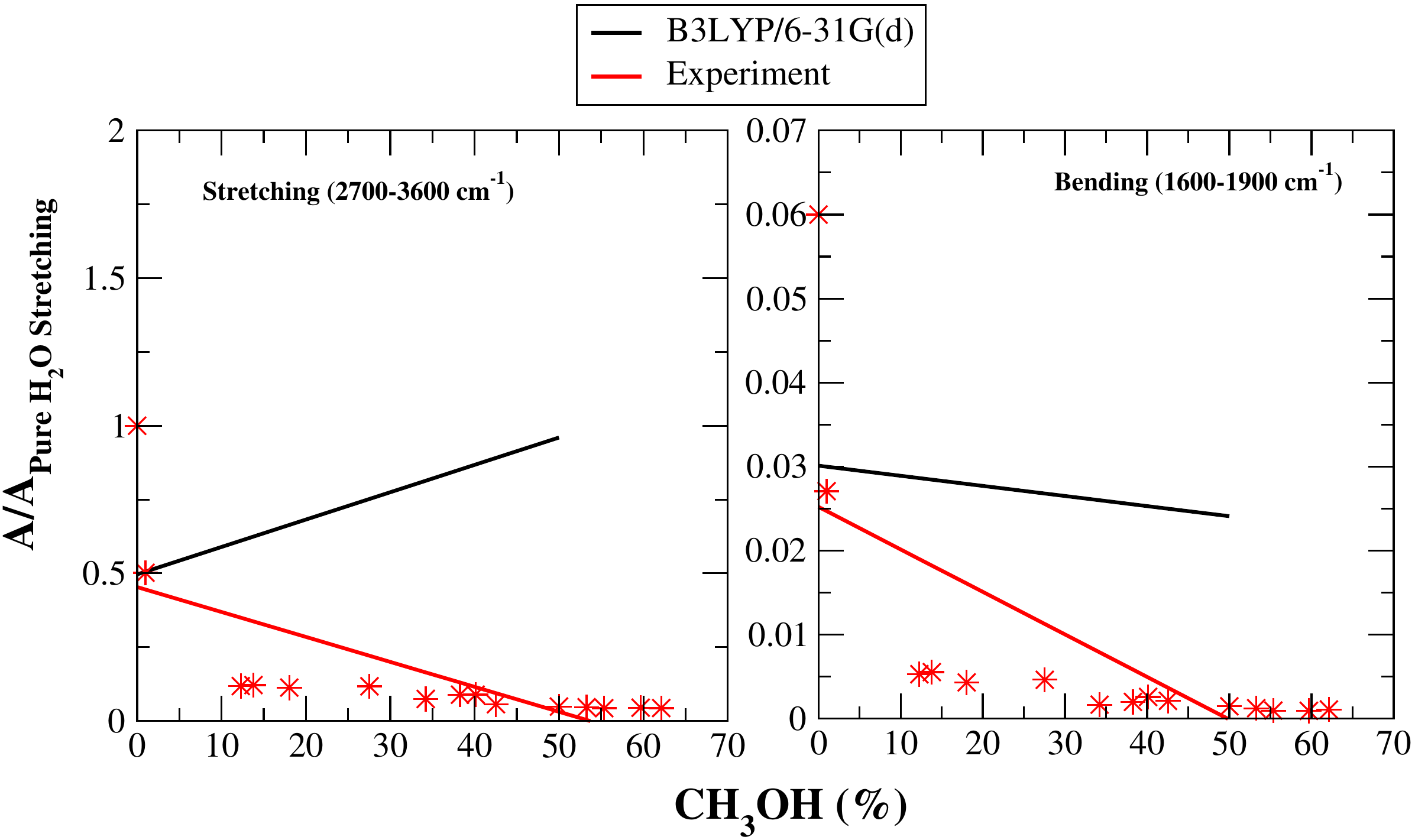}
 \caption {Comparison between calculated and experimentally fitted band strength profiles as a function of CH$_3$OH concentration. Stars represent the experimental data points \citep{gora20a}.}
 \label{fig:CH3OH-BS-COMP}
 \end{figure}

\subsubsection{CH$_3$OH ice}
\label{CH3OH_ice}
The effect of the CH$_3$OH concentration on the band profiles of water ice has been experimentally investigated. In the case of methanol, CO$_2$ gas is still present in the system (i.e., outside the vacuum chamber) in quantities that vary in time, causing negative and/or positive contributions to CO$_2$ gas-phase absorption features concerning the background spectrum, as is evident in Figure \ref{fig:experiment}c at $\sim$2340 cm$^{-1}$. Such contamination is most likely due to the dosing line, but its negligible amount should not affect the final results. Figure \ref{fig:experiment}c shows the experimental absorption spectra for various $\rm{CH_3OH-H_2O}$ ice mixtures deposited at $T=30$ K. The spectra are normalized to $1$ with respect to the maximum of the $\rm{O-H}$ stretch band.

 \begin{figure}
\centering
\includegraphics[width=0.4\textwidth]{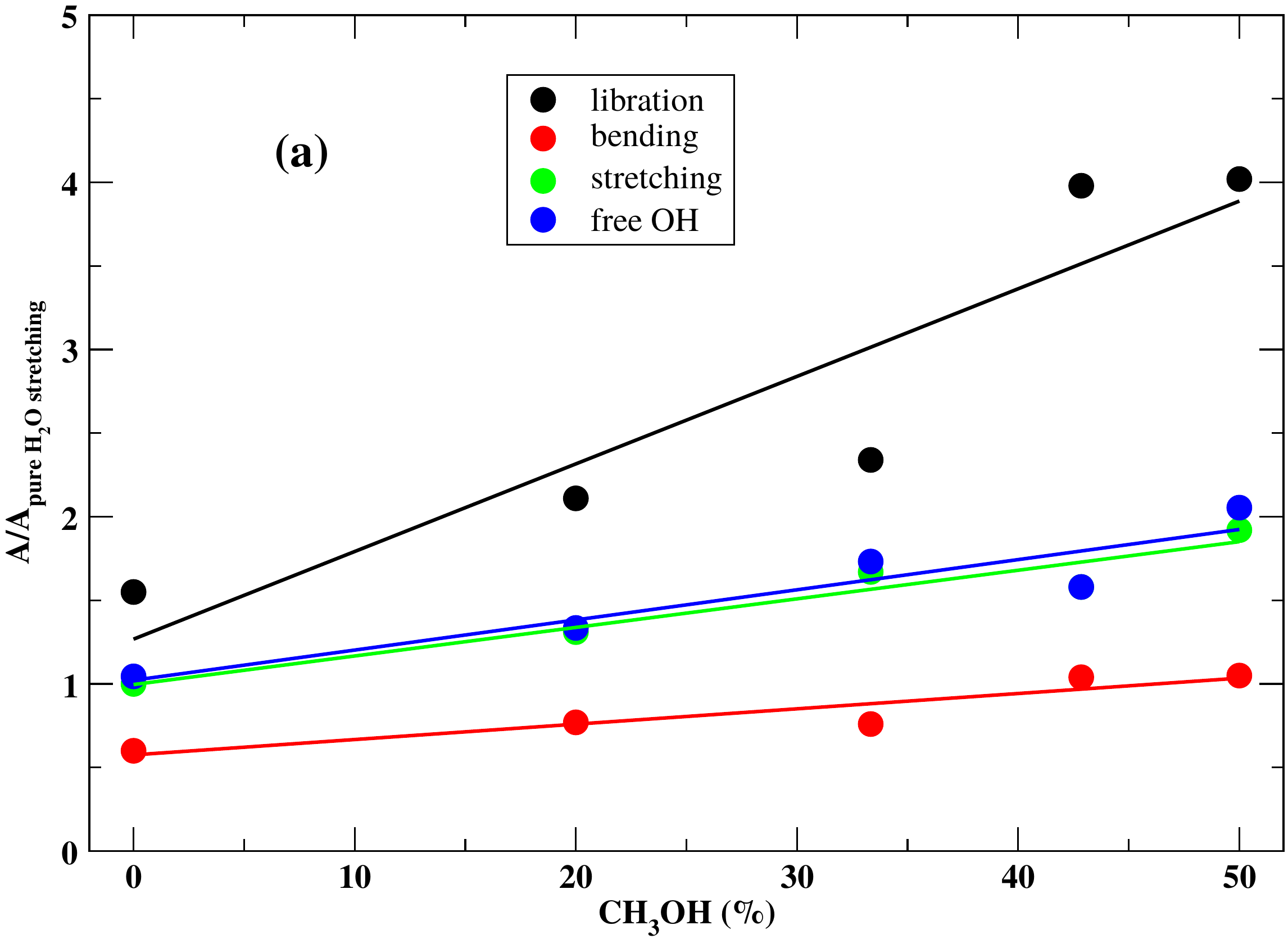}
\includegraphics[width=0.4\textwidth]{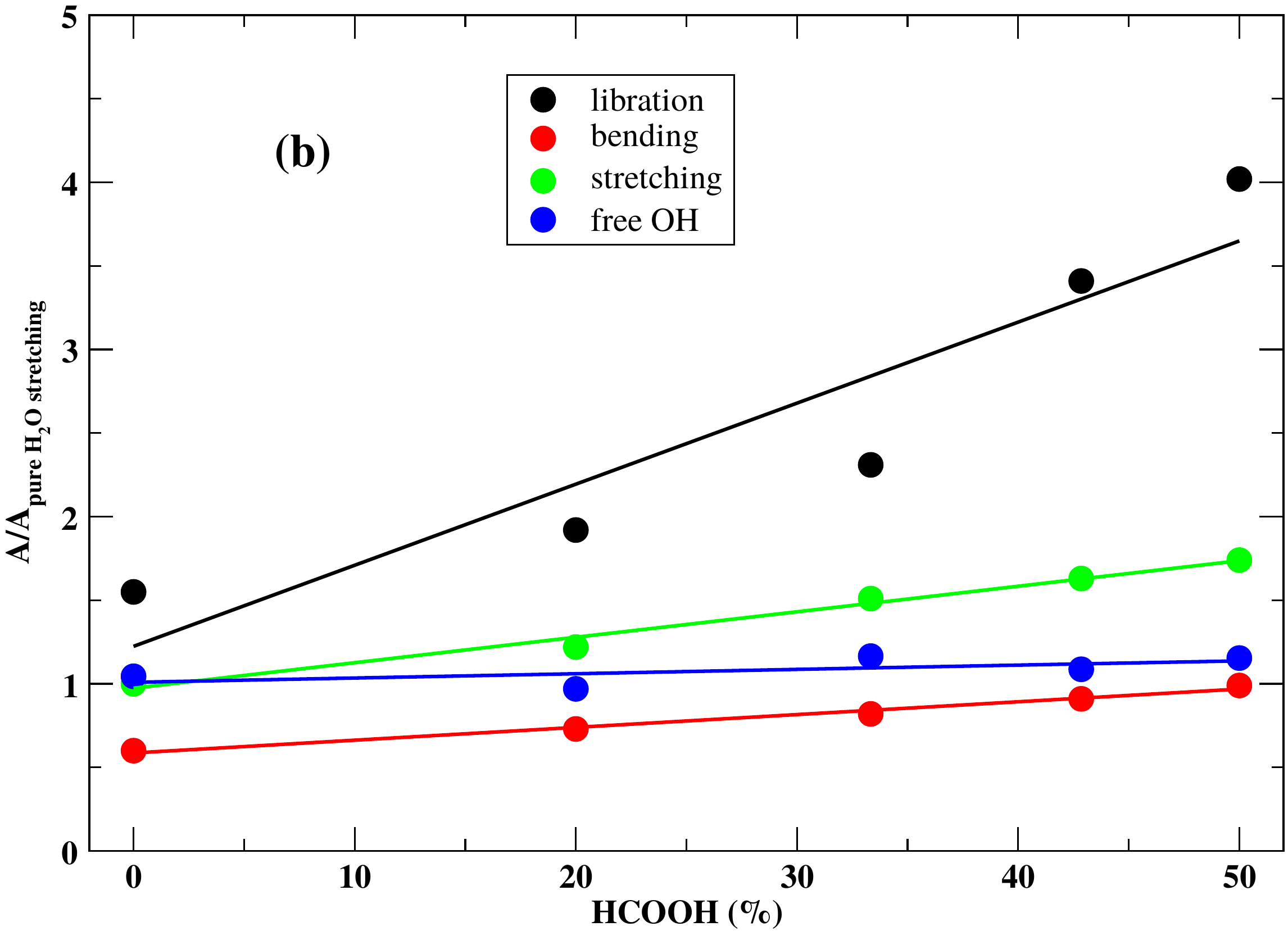}
\caption{Band strength for $\rm{H_2O-CH_3OH}$ mixtures: (a) CH$_3$OH as a hydrogen bond donor and (b) CH$_3$OH as a hydrogen bond acceptor \citep{gora20a}.}
\label{fig:ch3oh-donor-acceptor}
\end{figure}
 
Figure \ref{fig:optimized_structure}d shows the optimized structure of the $\rm{H_2O-CH_3OH}$ mixture with a $4:4$ concentration ratio. It is noted that a weak hydrogen bond is expected to be formed. The simulated IR spectra for different concentrations are shown in Figure \ref{fig:H2O-CH3OH}. Peak positions, integral absorption coefficients, and band assignments for various H$_2$O-CH$_3$OH mixtures are collected in Table \ref{tab:H2O_X}. The computed band strengths as a function of different concentrations are shown in Figure \ref{fig:band_strength}c.
The calculated strength of the bending mode gradually increases with $\rm{CH_3OH}$ concentration (see Figure \ref{fig:CH3OH-BS-COMP}; right panel), which is in qualitative agreement with the experimental results \cite{dawe16}. In the stretching mode, computationally, a slightly increasing trend of the band strength is noted, whereas experimental results show an opposite trend (see Figure \ref{fig:CH3OH-BS-COMP}; left panel). Because of the lack of experimental spectra, we cannot compare the band strength of the libration and free OH modes. In $\rm{H_2O-CH_3OH}$ mixtures, methanol can act as an H-bond donor and an H-bond acceptor. We consider both possibilities and find that if we consider methanol as an H-bond donor, the band strength of all four modes shows an increasing trend. On the other hand, if we consider methanol as an H-bond acceptor, the band strengths of three modes, namely, libration, bending, and stretching, present trends similar to the previous case (where methanol acts as an H-bond donor), while the free OH band shows a less pronounced behavior (see Figure \ref{fig:ch3oh-donor-acceptor}).

\begin{figure}
\centering
\includegraphics[width=0.68\textwidth]{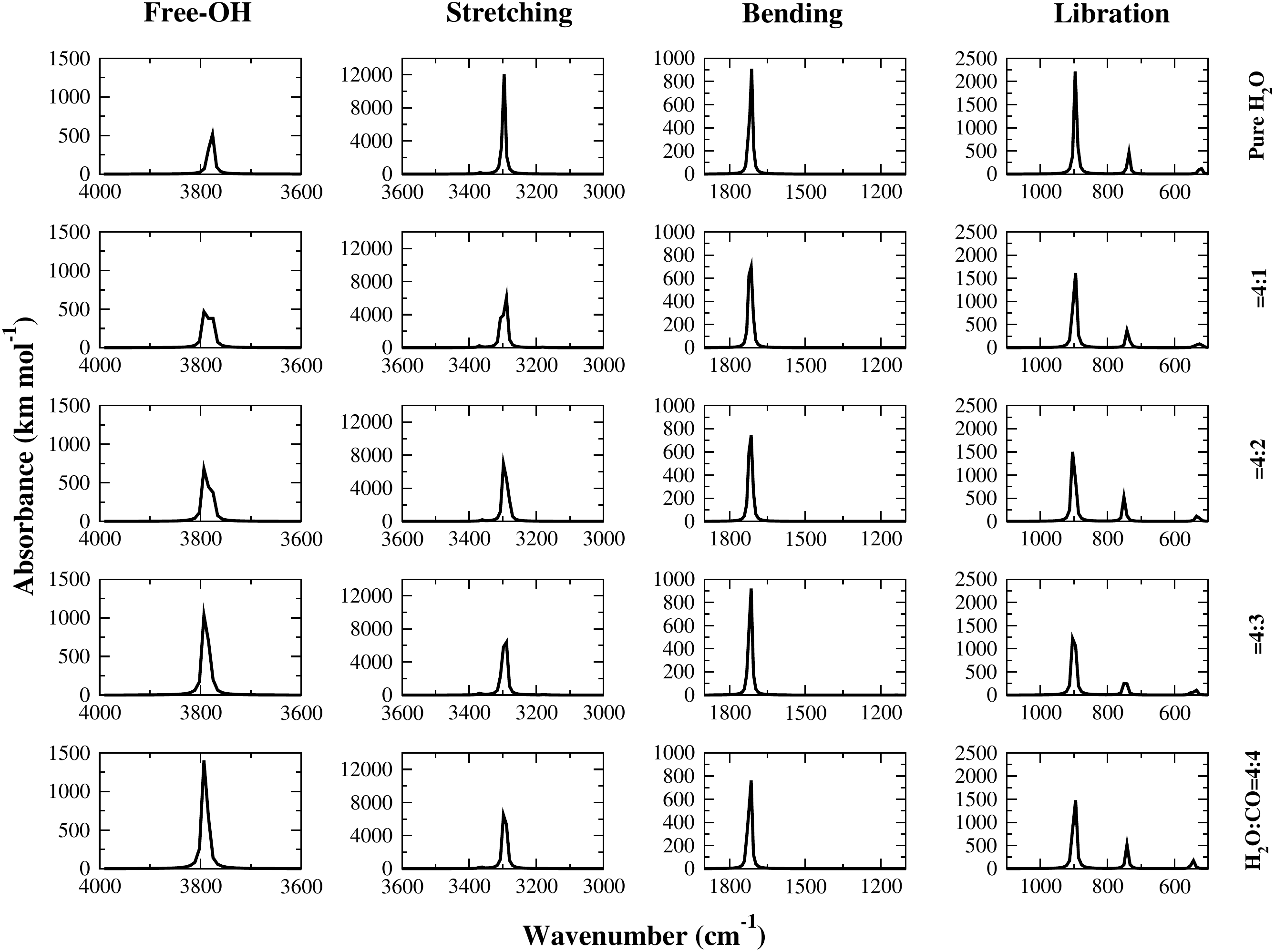}
\caption{Absorption spectra of the four modes for water ice for the five measured compositions, ranging from pure water ice (top) to $4:4$ $\rm{H_2O-CO}$ mixture (bottom) \citep{gora20a}.}
\label{fig:H2O-CO}
\end{figure}

\subsubsection{CO ice}
\label{CO_ice}
Figure \ref{fig:optimized_structure}e depicts the $\rm{H_2O-CO}$ optimized structure with a $4:4$ concentration ratio: the four CO molecules interact with the H atoms of the $\rm{H_2O}$ molecules not involved in the H-bond (interaction of the O atom of CO with the H atom of $\rm{H_2O}$). However, for the $\rm{H_2O-CO}$ system, the interaction can occur through both O and C atoms of CO with the H atom of H$_2$O \citep{zami18}. Therefore, we consider both types of interaction and evaluated their effects on the band strengths. However, we do not find any considerable difference. Thus, we only discuss the band strength of the $\rm{H_2O-CO}$ mixture with the interaction on the O side of CO. For completeness, it should be mentioned that there is also another type of interaction, which occurs between the $\pi$ bond of CO and one water$-$hydrogen, and it gives rise to a T-shaped complex \citep{coll14}. However, according to a computational study by \cite{coll14}, this has a negligible effect on IR vibrational bands. As a consequence, we do not investigate in detail this kind of complex. The simulated IR absorption spectra of the four fundamental vibrational modes for various compositions are shown in Figure \ref{fig:H2O-CO}. The four fundamental frequencies of water ice change significantly by increasing the concentration of CO. The most intense peak positions and the corresponding integral abundance coefficients for different $\rm{H_2O-CO}$ mixtures are provided in Table \ref{tab:H2O_X}. In Figure \ref{fig:band_strength}d, the integrated intensities of water vibrational modes are plotted as a function of the CO concentration. It is noted that the strength of the libration, bending, and stretching modes decreases with the concentration of CO.
However, the free OH mode shows a sharp increase in the band strength with increasing CO concentration. In Table \ref{tab:linear_coeff}, the resulting linear fit coefficients are collected together with the available experimental values for $\rm{H_2O-CO}$ mixtures deposited at $15$ K \citep{bouw07}. It is noted that theoretical band strength slopes are in relatively good agreement with experimental results \citep{bouw07}. For the $\rm{H_2O-CO}$ system, anharmonic calculations are also carried out. While the band strengths of the bending and stretching modes have a similar trend to experimental data, a deviation is noted for the libration mode (see Figure \ref{fig:harm-anharm-compare}).

\begin{figure}
 \centering
\begin{minipage}{0.49\textwidth}
\includegraphics[width=\textwidth]{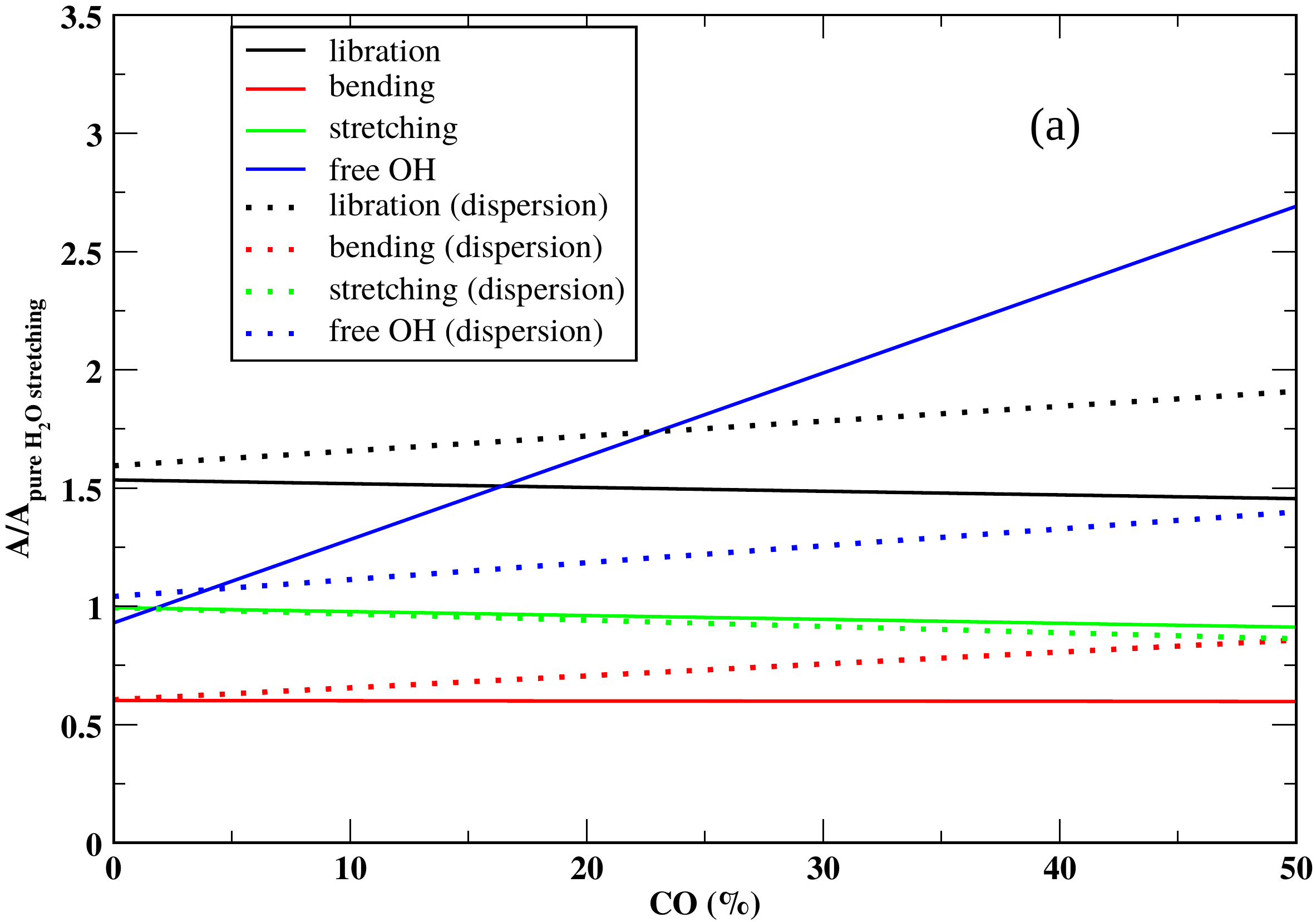}
\end{minipage}
\begin{minipage}{0.49\textwidth}
\includegraphics[width=\textwidth]{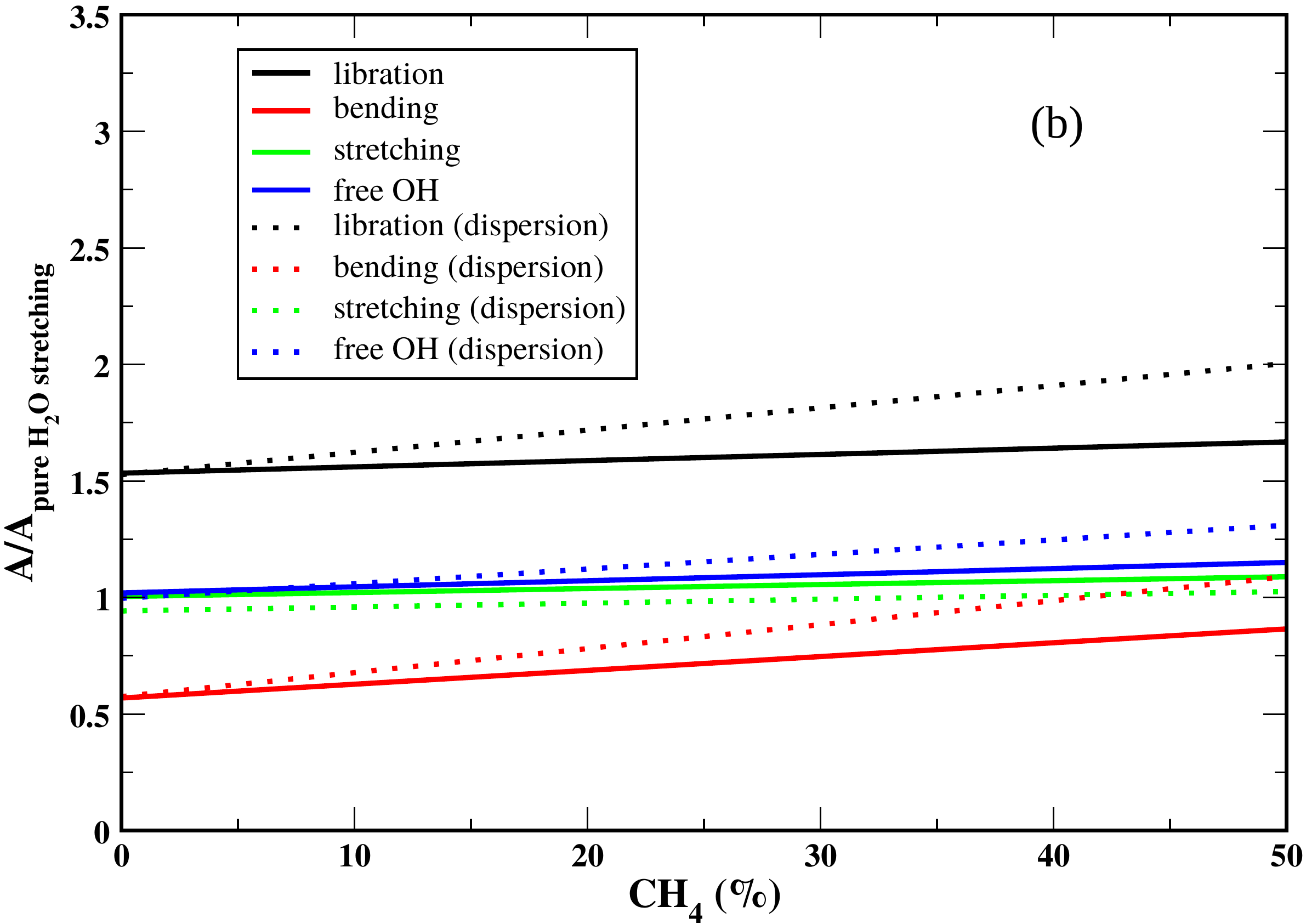}
\end{minipage}
\begin{minipage}{0.49\textwidth}
\includegraphics[width=\textwidth]{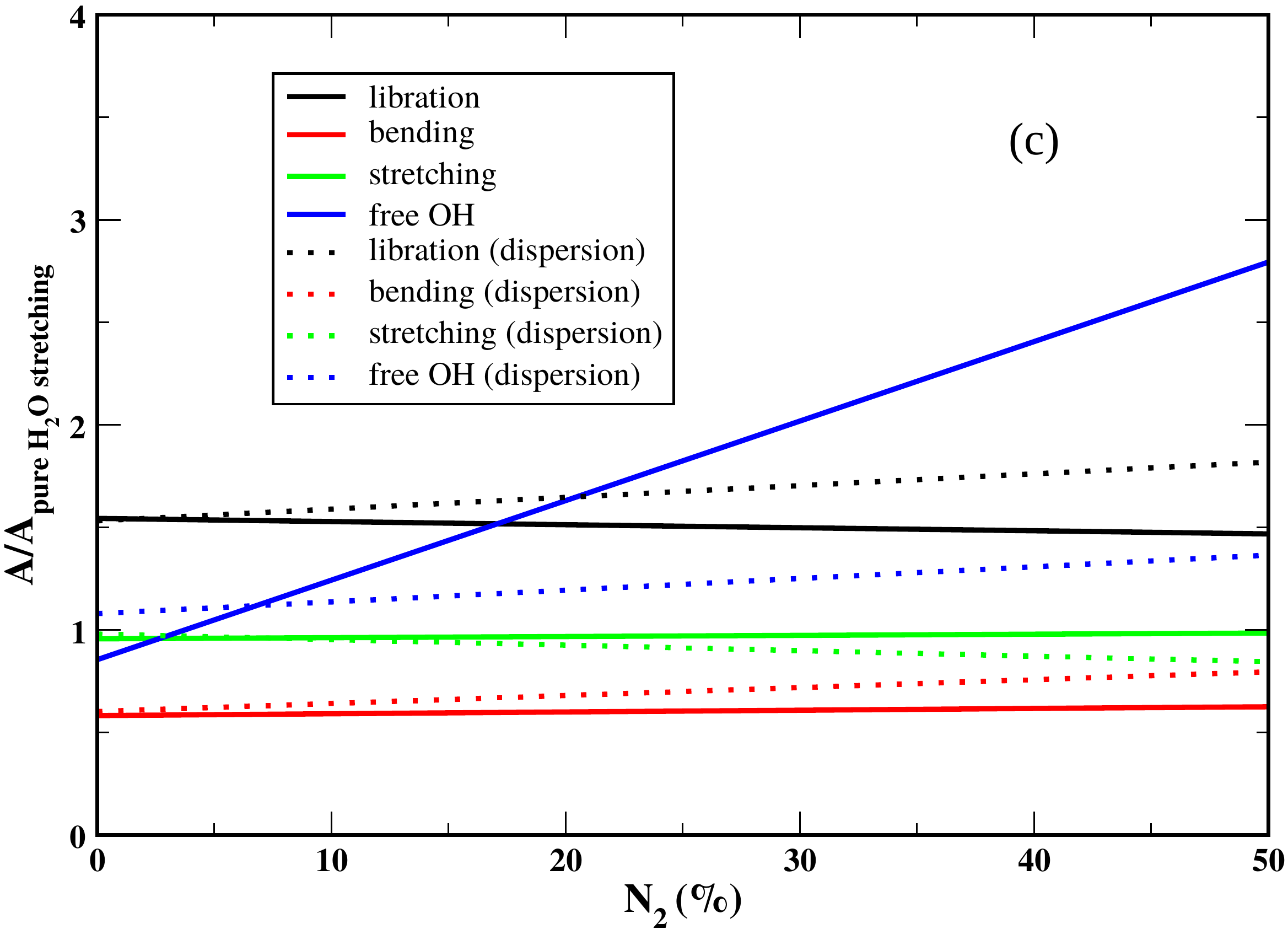}
\end{minipage}
\begin{minipage}{0.49\textwidth}
\includegraphics[width=\textwidth]{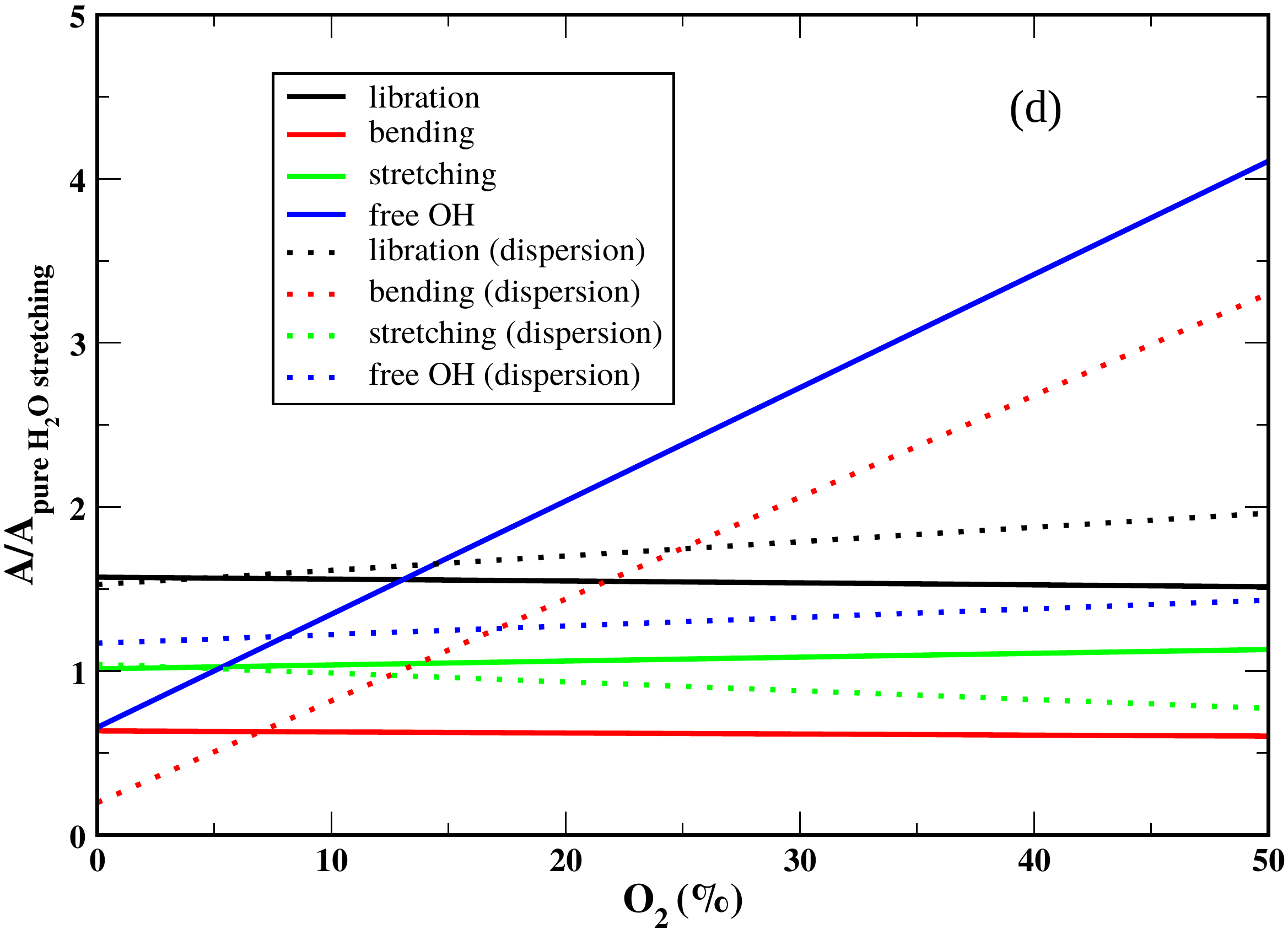}
\end{minipage}
\caption{Comparison of the band strengths of the four fundamental modes of water for various mixtures of (a) $\rm{H_2O-CO}$, (b) $\rm{H_2O-CH_4}$, (c) $\rm{H_2O-N_2}$, and (d) $\rm{H_2O-O_2}$ by considering or not the dispersion effect \citep{gora20a}.}
\label{fig:comp-norm-dis}
\end{figure}

To check the influence of dispersion, B3LYP-D3/6-31G(d) calculations are performed, with D3 denoting the correction for dispersion effects \citep{grim10}. B3LYP-D3 calculations are carried out for $\rm{H_2O-CO}$, $\rm{H_2O-CH_4}$, $\rm{H_2O-N_2}$, and $\rm{H_2O-O_2}$ systems. Figure \ref{fig:comp-norm-dis}a shows the comparison of the band strengths of different vibrational modes of water with and without the dispersion correction for the $\rm{H_2O-CO}$ system. The overall conclusion is that there is a good agreement with the experimental band strengths when the dispersion effect is not considered. On the contrary, our computed band strength profile shows a different trend when the dispersion correction is included. The libration and bending modes present a positive slope with the increase in impurity concentration, whereas
experimental results show a negative slope. For the free OH mode, a slightly increasing trend of the band strength is obtained, whereas the experimental band strength presents a sharp increase with CO concentration. The band strength of the stretching mode has similar behavior to dispersion and without dispersion and agrees with the experimental result \citep[][see Figure 3]{bouw07}. Thus, in summary, while we are not claiming that the dispersion effects are not important for the systems investigated, we note that by neglecting them, we obtain a consistent description of the experimental behavior (probably due to a fortuitous error compensation).

\begin{figure}
\centering
\includegraphics[width=0.68\textwidth]{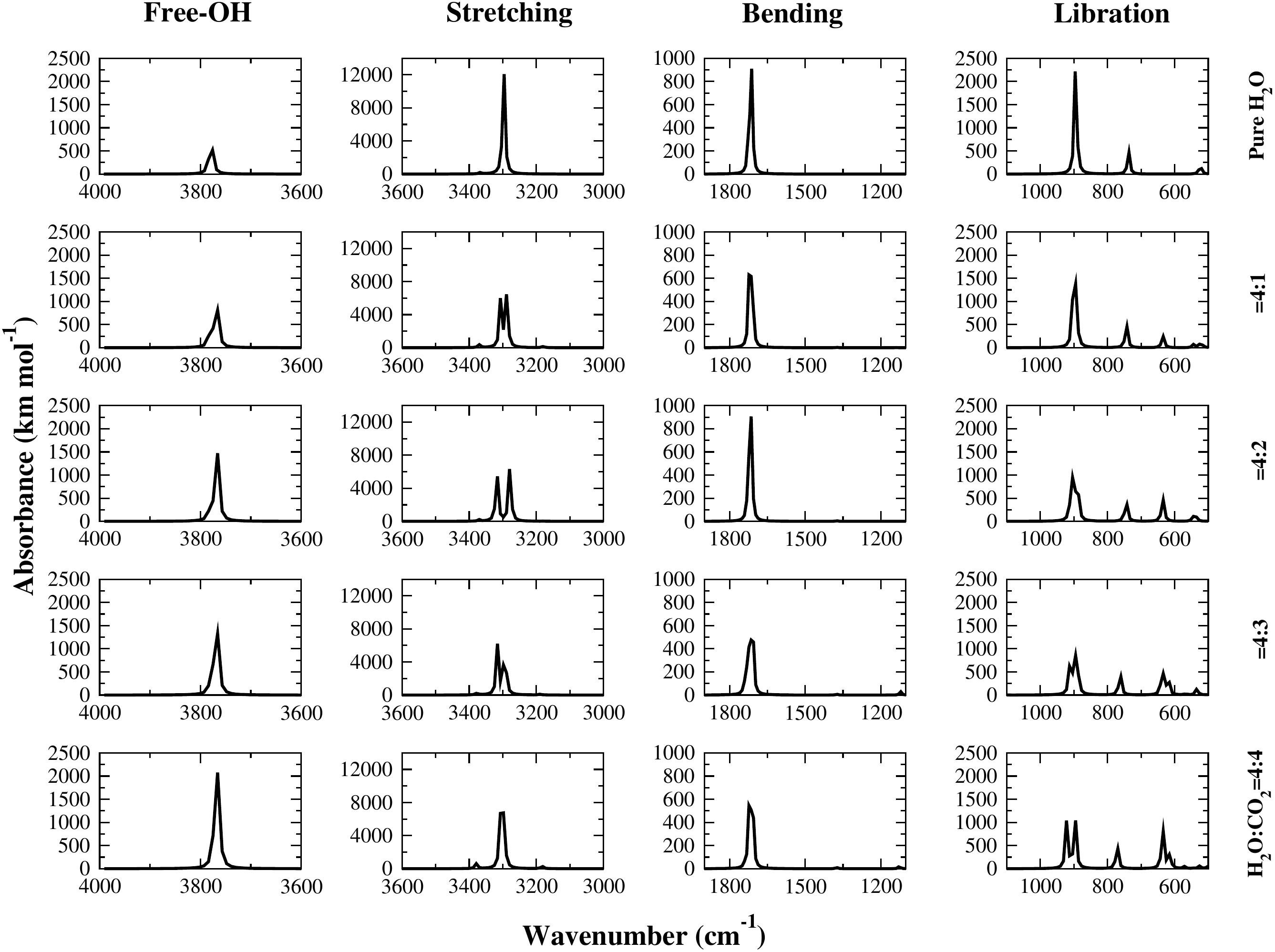}
\caption{Absorption spectra of the four modes for water ice for the five measured compositions, ranging from pure water ice (top) to $4:4$ $\rm{H_2O-CO_2}$ mixture (bottom) \citep{gora20a}.}
\label{fig:H2O-CO2}
\end{figure}

\subsubsection{CO$_2$ ice}
\label{CO2_ice}
Figure \ref{fig:optimized_structure}f shows the optimized geometry of the $4:4$ mixture of $\rm{H_2O-CO_2}$ system. The absorption features of water ice for different $\rm{CO_2}$ concentrations are shown in Figure \ref{fig:H2O-CO2}. The most intense frequencies for the various $\rm{H_2O-CO_2}$ mixtures are also summarized in Table \ref{tab:H2O_X}. The trend of the band strength as a function of CO$_2$ concentrations is shown in Figure \ref{fig:band_strength}e. For the free OH mode, a rapid increase with CO$_2$ concentration is noted, which agrees with the experimental results by \cite{ober07}. Computed band strengths of the libration and bending modes also increase by increasing the $\rm{CO_2}$ concentration, which is, however, in contrast with the available experimental data \citep{ober07}. The band strength of the bulk stretching mode decreases instead with $\rm{CO_2}$ concentration, in reasonably good agreement with the available experiments \citep{ober07}. FTIR spectroscopy of the matrix-isolated molecular complex $\rm{H_2O-CO_2}$ shows that CO$_2$ does not form a weak hydrogen bond with H$_2$O \citep{tso85}, but instead, CO$_2$ destroys the bulk hydrogen bond network. This may cause a significant decrease in the band strength of the bulk stretching mode, while the intermolecular $\rm{O-H}$ bond strength increases with the CO$_2$ concentration. Therefore, the disagreement between calculated and experimental band strengths could be due to the
cluster size of water molecules. In Table \ref{tab:linear_coeff}, the resulting linear fit coefficients are reported, together with the available experimental values for the $\rm{H_2O-CO_2}$ mixture deposited at $15$ K \citep{ober07}.

\subsubsection{Part 2. Applications}

The results discussed in previous Sections suggest that the water c-tetramer structure and harmonic B3LYP/6-31G(d) calculations can predict the experimental results presented here and the literature data. Thus, to study the effect of other impurities ($\rm{H_2CO}$, CH$_4$, OCS, N$_2$, and O$_2$) on pure water ice, we further exploit this methodology. Additionally, the effect of impurities on the band strengths of the four fundamental bands is also studied by considering the c-hexamer (chair) structure.

\begin{figure}
\centering
\includegraphics[width=0.68\textwidth]{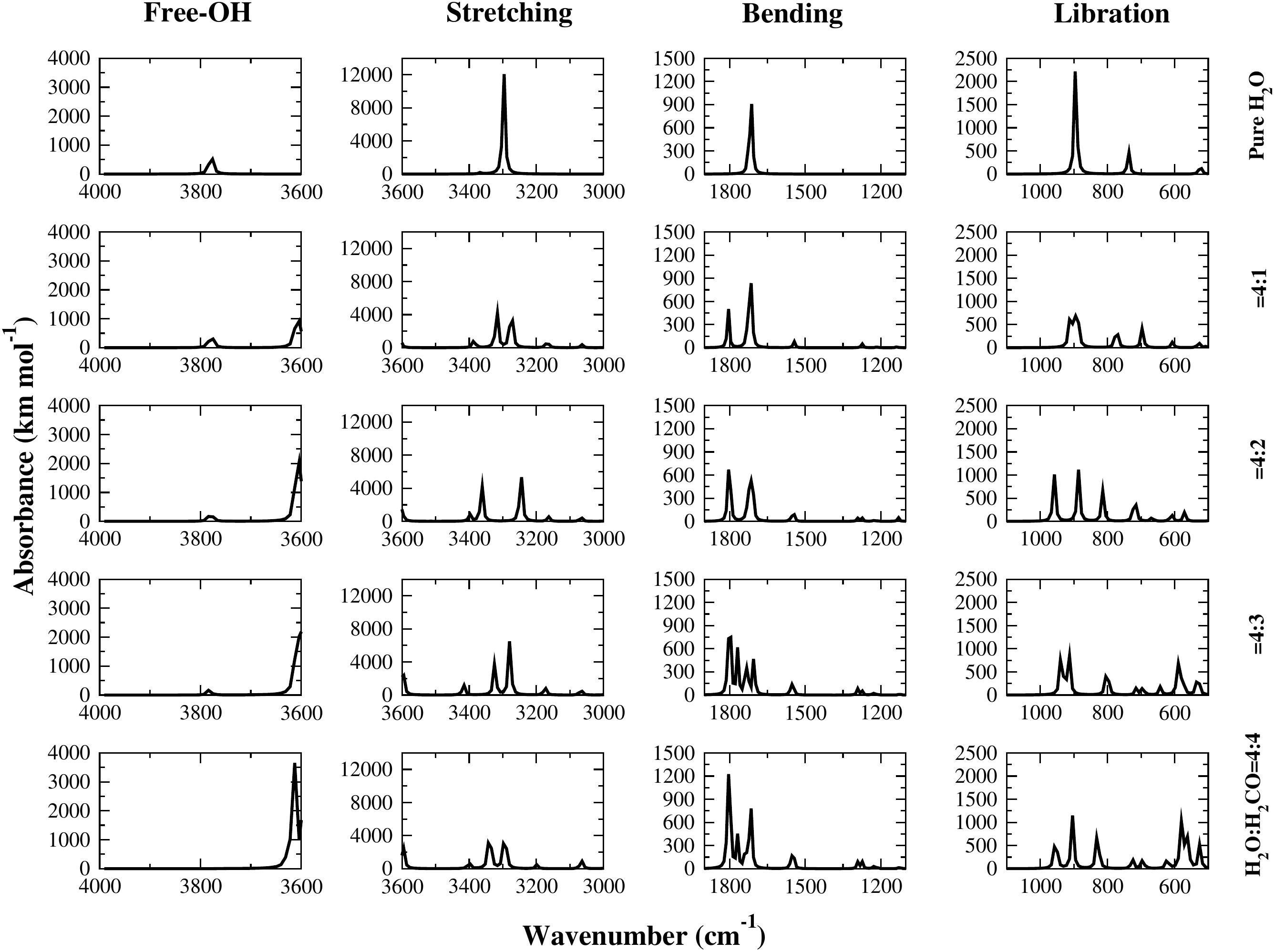}
\caption{Absorption spectra of the four modes for water ice for the five measured compositions, ranging from pure water ice (top) to $4:4$ $\rm{H_2O-H_2CO}$ mixture (bottom) \citep{gora20a}.}
\label{fig:H2O-H2CO}
\end{figure}

\subsubsection{H$_2$CO ice}
\label{H2CO_ice}
The strongest modes of formaldehyde ($\rm{H_2CO}$) lie at $1724.14$ cm$^{-1}$ ($5.80$ $\mu$m) and $1497.01$ cm$^{-1}$ ($6.68$ $\mu$m). Figure \ref{fig:optimized_structure}g depicts the optimized structure of the $4:4$ $\rm{H_2O-H_2CO}$ mixture. The desired ratio is attained upon forming the H-bond between the O atom of $\rm{H_2CO}$ and the dangling H atoms of $\rm{H_2O}$. The effect of formaldehyde on the water IR spectrum is shown in Figure \ref{fig:H2O-H2CO}. Frequencies, integral absorption coefficients, and mode assignments are also reported in Table \ref{tab:H2O_X}. The band strength profiles as a function of the concentration of H$_2$O are shown in Figure \ref{fig:band_strength}f. Like the $\rm{methanol-water}$ mixture, all band strengths increase with the concentration of formaldehyde, the free OH stretching mode being the most affected.

\begin{figure}
\centering
\includegraphics[width=0.68\textwidth]{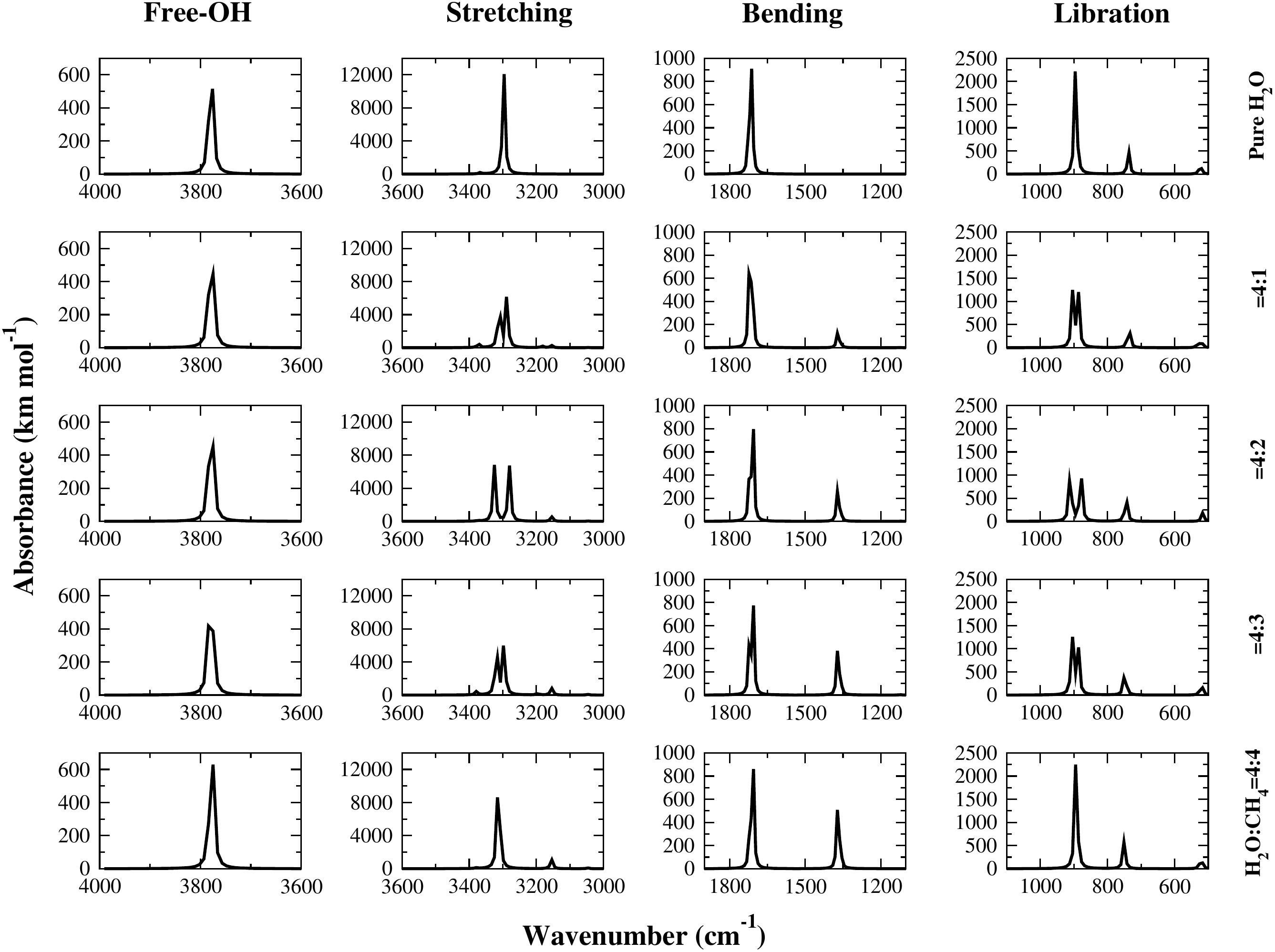}
\caption{Absorption spectra of the four modes for water ice for the five measured compositions, ranging from pure water ice (top) to $4:4$ $\rm{H_2O-CH_4}$ mixture (bottom) \citep{gora20a}.}
\label{fig:H2O-CH4}
\end{figure}

\subsubsection{CH$_4$ ice}
\label{CH4_ice}
$\rm{CH_4}$ cannot be observed through rotational spectroscopy since it has no permanent dipole moment. The optimized structure of the $\rm{H_2O-CH_4}$ system with a $4:4$ ratio is shown in Figure \ref{fig:optimized_structure}h. The absorption IR spectra for different $\rm{H_2O-CH_4}$ mixtures are depicted in Figure \ref{fig:H2O-CH4}. Peak positions, integral absorption coefficients, and band assignments are also provided in Table \ref{tab:H2O_X}. Figure \ref{fig:band_strength}g shows the band strength variations with the concentration of $\rm{CH_4}$. All band strengths marginally increase with the $\rm{CH_4}$ concentration. Figure \ref{fig:comp-norm-dis}b shows the comparison of the band strengths with and without incorporating corrections for accounting for dispersion effects. For all the four fundamental modes, differences are minor.

\begin{figure}
\centering
\includegraphics[width=0.68\textwidth]{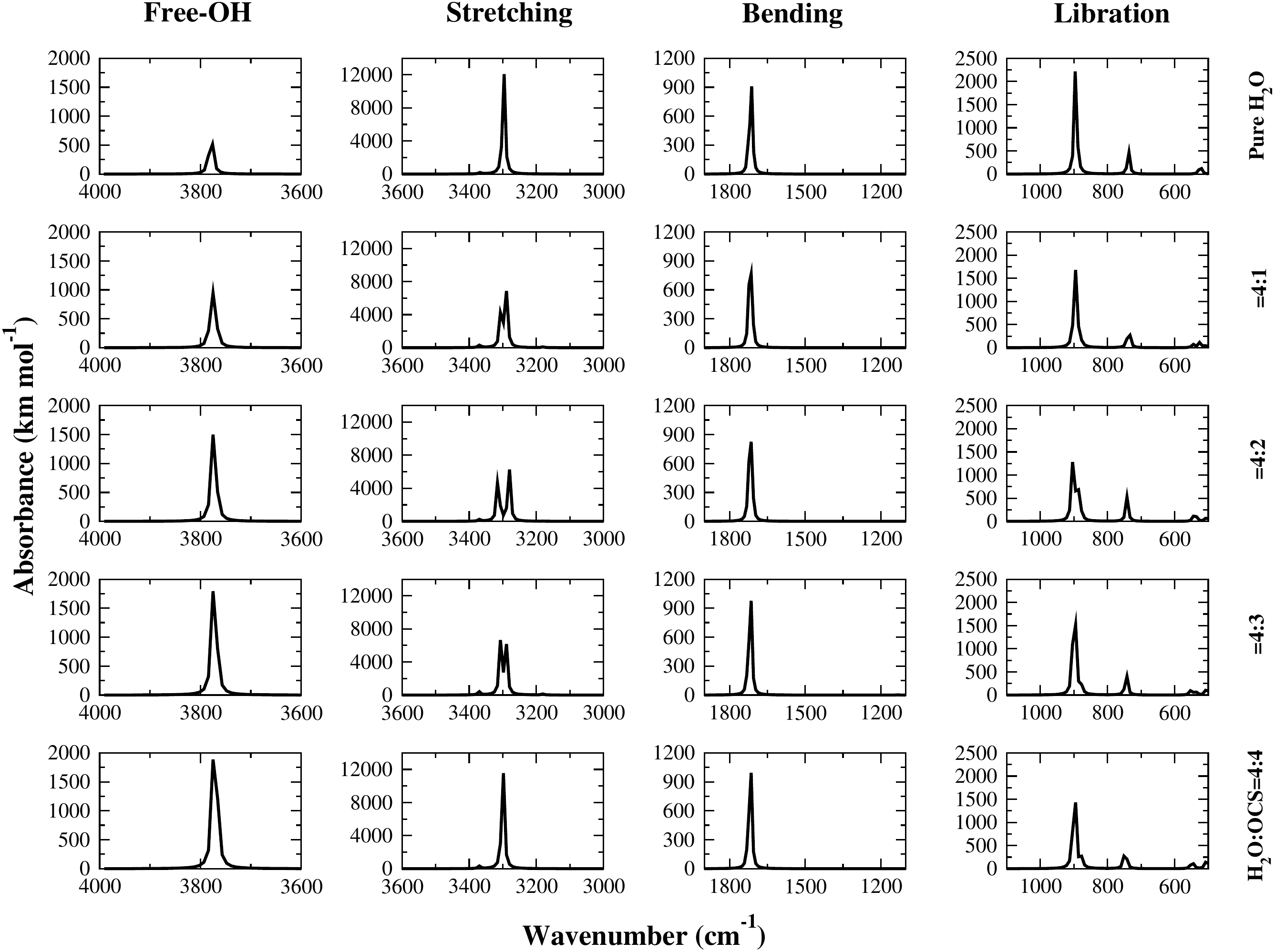}
\caption{Absorption spectra of the four modes for water ice for the five measured compositions, ranging from pure water ice (top) to $4:4$ $\rm{H_2O-OCS}$ mixture (bottom) \citep{gora20a}.}
\label{fig:H2O-OCS}
\end{figure}

\subsubsection{OCS ice}
\label{OCS_ice}
\citet{garo10} proposed that carbonyl sulfide (OCS) is a key ingredient of the grain surface. Its abundance in the ice phase may vary between $0.05 \%$ and $0.15 \%$ \citep{dart05}. Figure \ref{fig:optimized_structure}i shows the optimized structure of the $4:4$ $\rm{H_2O-OCS}$. Since oxygen is more electronegative than sulfur, the O atom of the OCS molecule is H-bonded to the water-free hydrogens. Figure \ref{fig:H2O-OCS} shows the absorption IR band spectra for $\rm{H_2O-OCS}$ clusters with various concentrations. Figure \ref{fig:band_strength}h depicts the band strengths as a function of the concentration of OCS. Here, the free OH mode is the most affected, and its band strength increases with the concentration of OCS. All other modes roughly remain unaffected by varying the amount of impurity.

\begin{figure}
\centering
\includegraphics[width=0.68\textwidth]{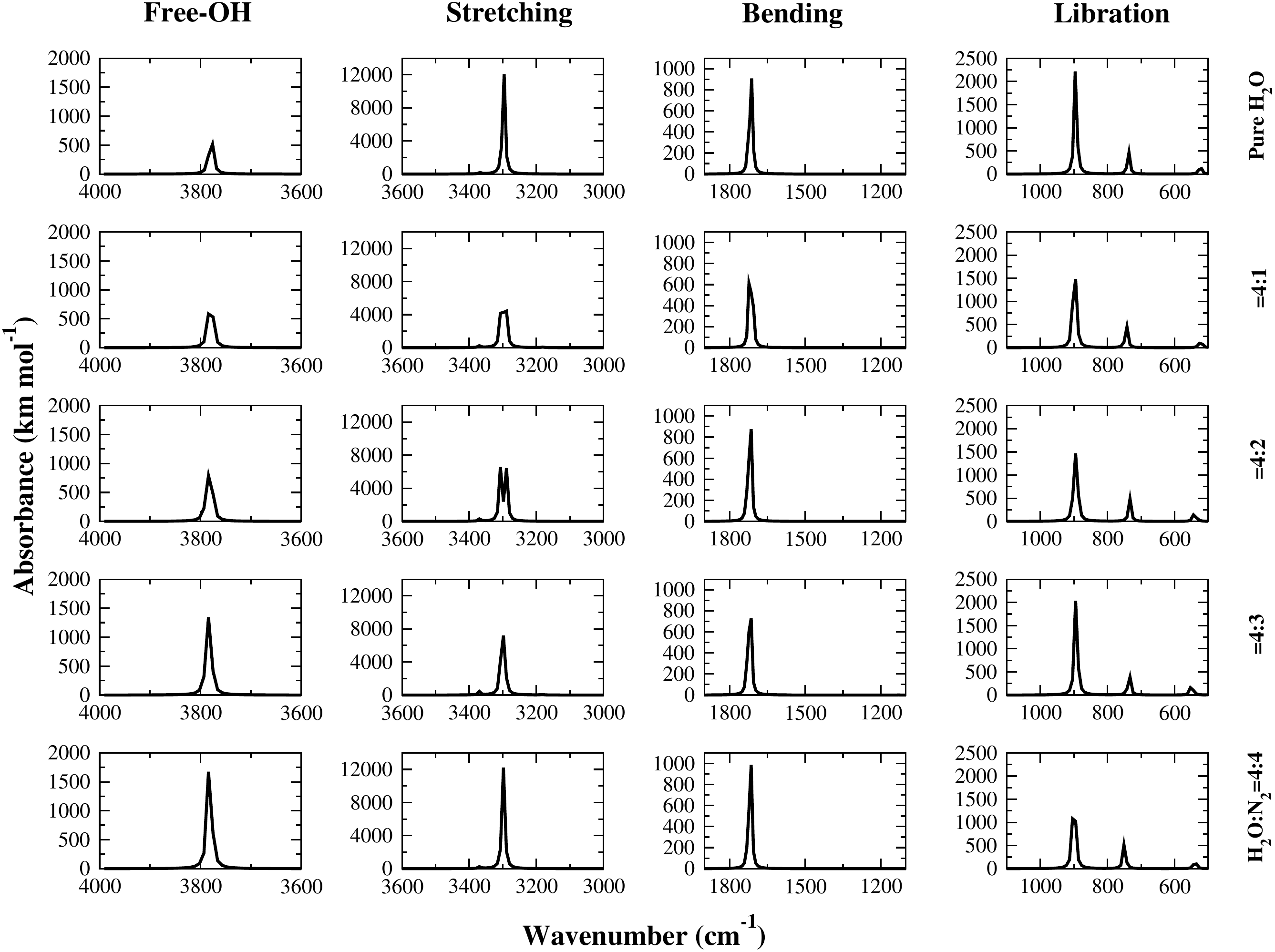}
\caption{Absorption spectra of the four modes for water ice for the five measured compositions, ranging from pure water ice (top) to $4:4$ $\rm{H_2O-N_2}$ mixture (bottom) \citep{gora20a}.}
\label{fig:H2O-N2}
\end{figure}

\subsubsection{N$_2$ ice}
\label{N2_ice}
N$_2$ is a stable homonuclear molecule, and due to its symmetry, it is IR-inactive. However, when
embedded in an ice matrix, the crystal field breaks the symmetry, and an IR transition is activated
around $2325.58$ cm$^{-1}$ ($4.30$ $\mu$m). Figure \ref{fig:optimized_structure}j shows the optimized geometry of the $\rm{H_2O-N_2}$ system with a $4:4$ ratio. The IR absorption spectra of water ice containing different amounts of $\rm{N_2}$ are shown in Figure \ref{fig:H2O-N2}.
The corresponding peak frequencies and intensities are also provided in Table \ref{tab:H2O_X}. The dependence of the band strengths on the N$_2$ concentration is depicted in Figure \ref{fig:band_strength}i. It is noticed that the slope of the band strength of the libration mode decreases, whereas the bending, stretching, and free OH modes show an increasing trend with the concentration of N$_2$. The linear fitting coefficients are provided in Table \ref{tab:linear_coeff}.
Figure \ref {fig:comp-norm-dis}c shows the comparison of band strengths with and without considering
the dispersion effects. It is noted that the inclusion of the dispersion effect leads to minor changes.

\begin{figure}
\centering
\includegraphics[width=0.68\textwidth]{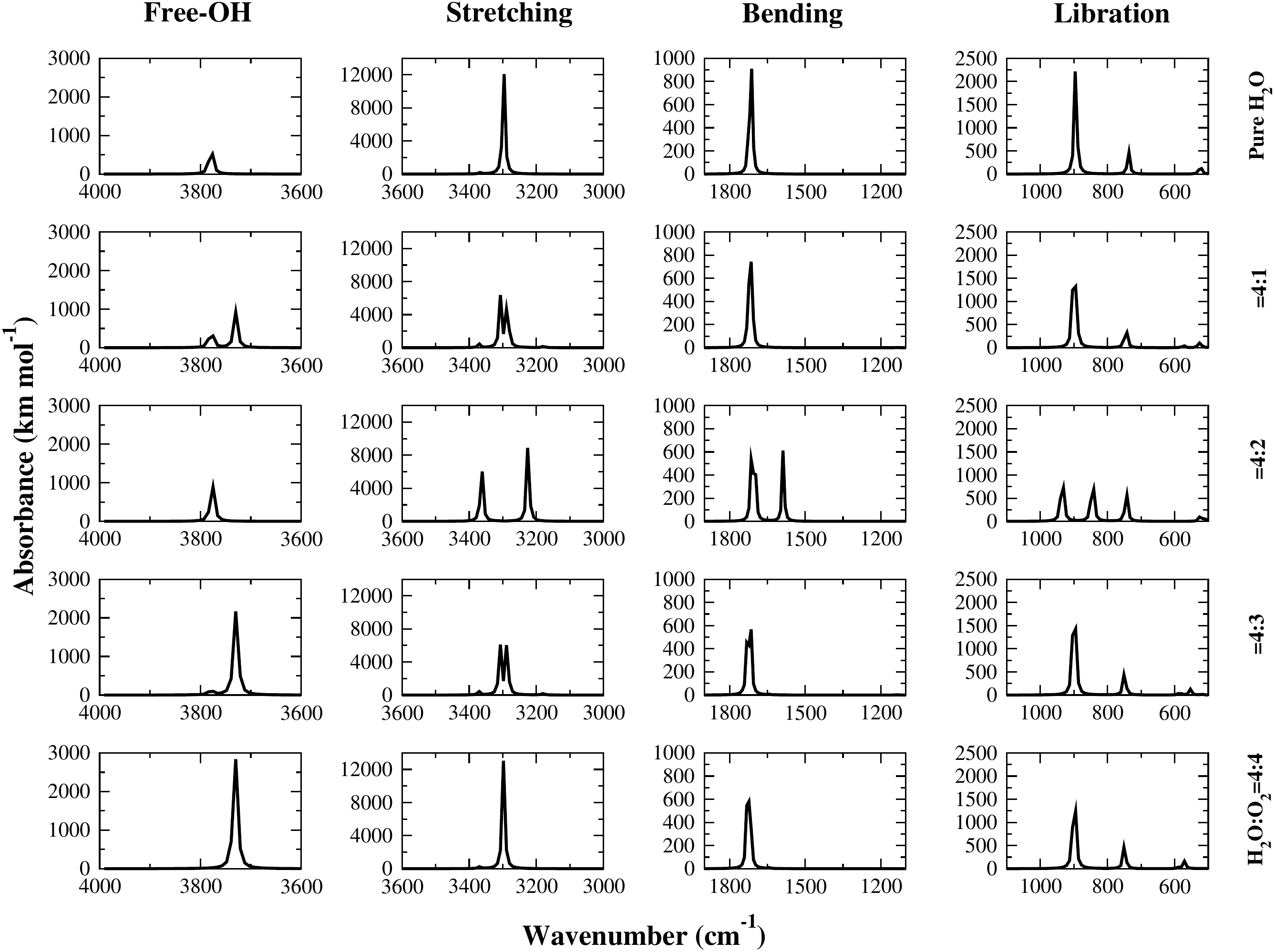}
\caption{Absorption spectra of the four modes for water ice for the five measured compositions, ranging from pure water ice (top) to $4:4$ $\rm{H_2O-O_2}$ mixture (bottom) \citep{gora20a}.}
\label{fig:H2O-O2}
\end{figure}

\subsubsection{O$_2$ ice}
\label{O2_ice}
Analogous to $\rm{N_2}$, $\rm{O_2}$ is a homonuclear molecule, which is IR-inactive except when it is embedded in an ice matrix \citep{ehre92,ehre98}, thus giving rise to an absorption band around $1550.39$ cm$^{-1}$ ($6.45$ $\mu$m). $\rm{O_2}$ ice is not very abundant because the largest part of the oxygen budget in the dense molecular clouds is locked in the form of $\rm{CO_2}$, CO, water ice, and silicates.
The optimized geometry of the $4:4$ $\rm{H_2O-O_2}$ ratio is shown in Figure \ref{fig:optimized_structure}k. IR spectra for different concentrations and the corresponding peak frequencies and intensities are provided in Figure \ref{fig:H2O-O2} and Table \ref{tab:H2O_X}, respectively. The dependence of band strengths upon O$_2$ concentration is shown in Figure \ref{fig:band_strength}j.
Similar to the $\rm{N_2-water}$ case, the free OH mode is the most affected. The slope of the band strength of the libration and bending modes decreases. In contrast, the stretching and free OH modes show an increasing trend with the concentration of O$_2$. The fitting coefficients for different $\rm{H_2O-N_2}$ mixtures are provided in Table \ref{tab:linear_coeff}.

Figure \ref{fig:comp-norm-dis}d depicts the comparison of the band strengths with and without the inclusion of dispersion effects for the $\rm{H_2O-O_2}$ system. It is evident that the trend of the band strength with the impurity concentration slightly increases for the libration mode. In contrast, it slightly decreases for the stretching mode when corrections for dispersion effects are present. The band strength rapidly increases in the bending mode, whereas it rapidly decreases for the free OH mode.

\subsubsection{Comparison between various mixtures}
\label{comparison_between_various_mixtures}

\begin{figure}
\centering
\begin{minipage}{\textwidth}
\includegraphics[width=\textwidth]{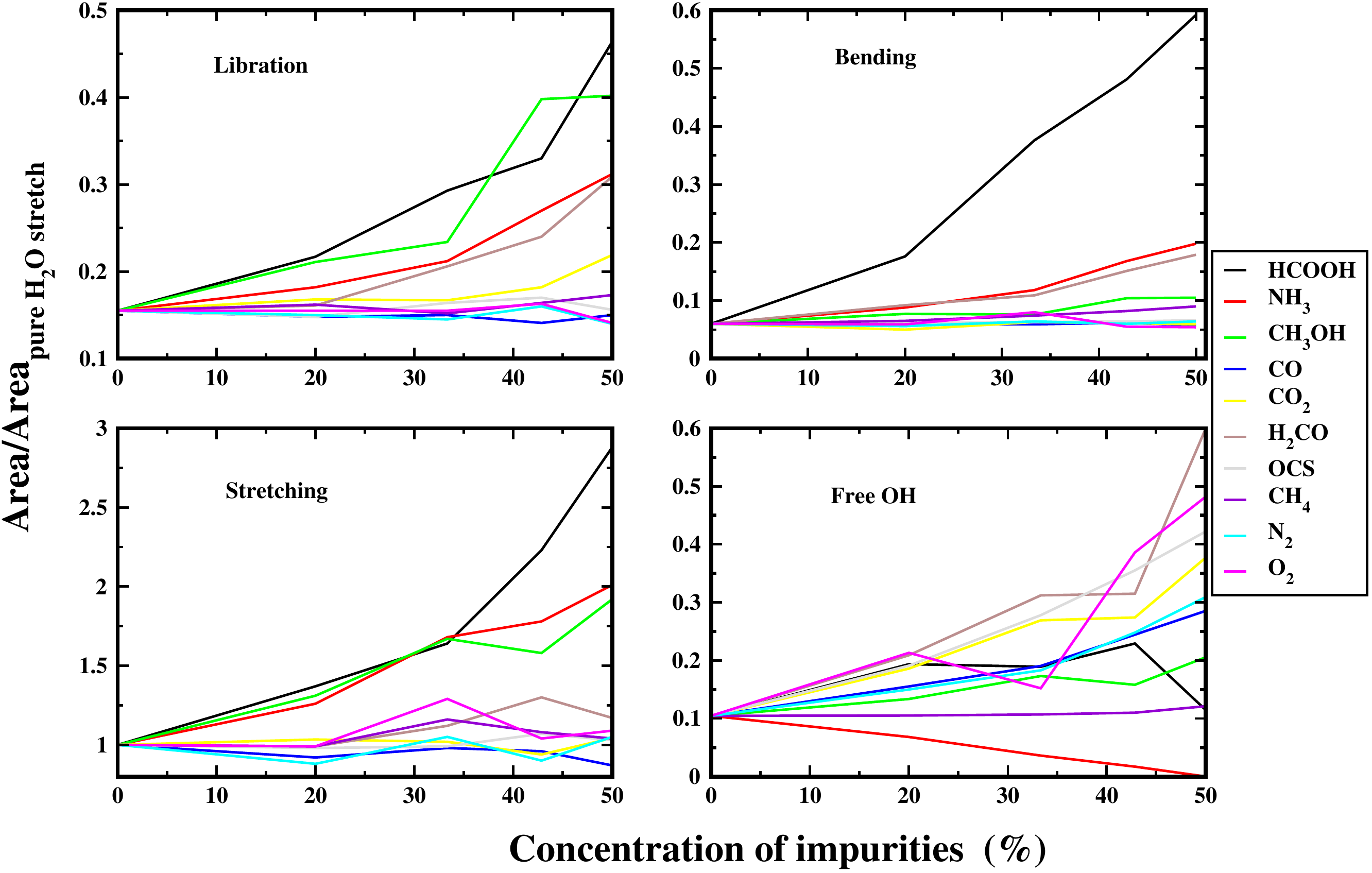}
\end{minipage}
\begin{minipage}{\textwidth}
\vskip 0.8cm
\includegraphics[width=\textwidth]{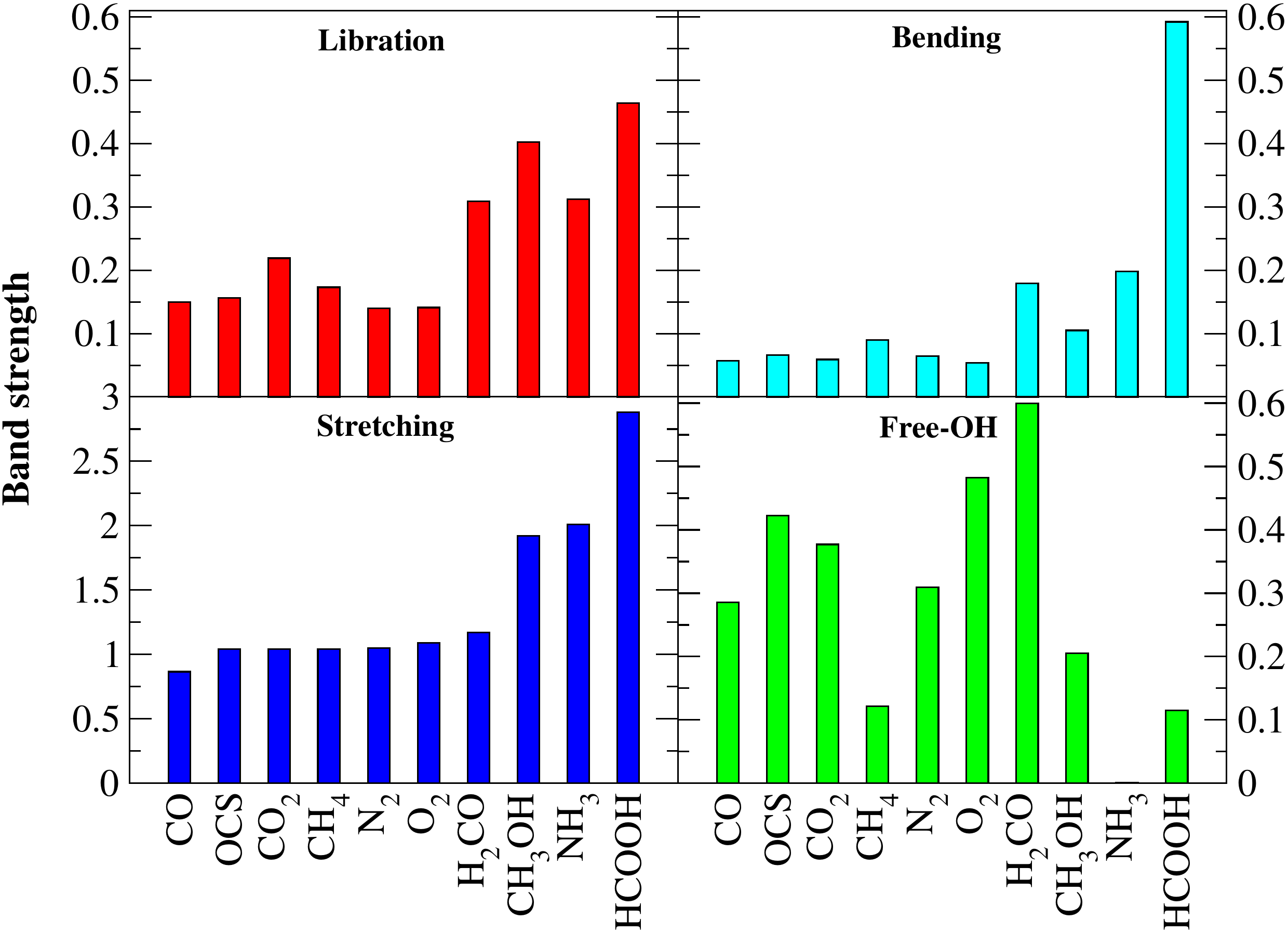}
\end{minipage}
\caption{Top panel: effect of impurities on the four fundamental vibrational modes of water. Bottom panel: comparison of the band strengths for the four fundamental vibrational modes as affected by impurities \citep{gora20a}.}
\label{fig:comparison_between_four_modes}
\end{figure}

To compare the effect of all impurities considered in this study on the band strength, we plot the band profiles of the four fundamental modes of water ice as a function of the concentration of impurities, the results being shown in Figure \ref{fig:comparison_between_four_modes}, top panel. For all basic modes, band strengths increase with the concentration of $\rm{CH_3OH}$, $\rm{H_2CO}$, HCOOH, $\rm{CH_4}$.
In Figure \ref{fig:comparison_between_four_modes}, bottom panel, we report the relative band strengths for the $4:4$ ratio mixtures to better understand their effect. From this, it is clear that the libration, bending, and stretching modes are affected mainly by formic acid, while the free OH mode is mainly affected by formaldehyde. An interesting feature is found for the free OH mode for the $\rm{NH_3-H_2O}$ system. By increasing the $\rm{NH_3}$ concentration to that of pure water, the band strength of the free OH mode decreases and disappears when the $4:4$ concentration ratio is reached.

\begin{figure}
\centering
\includegraphics[width=\textwidth]{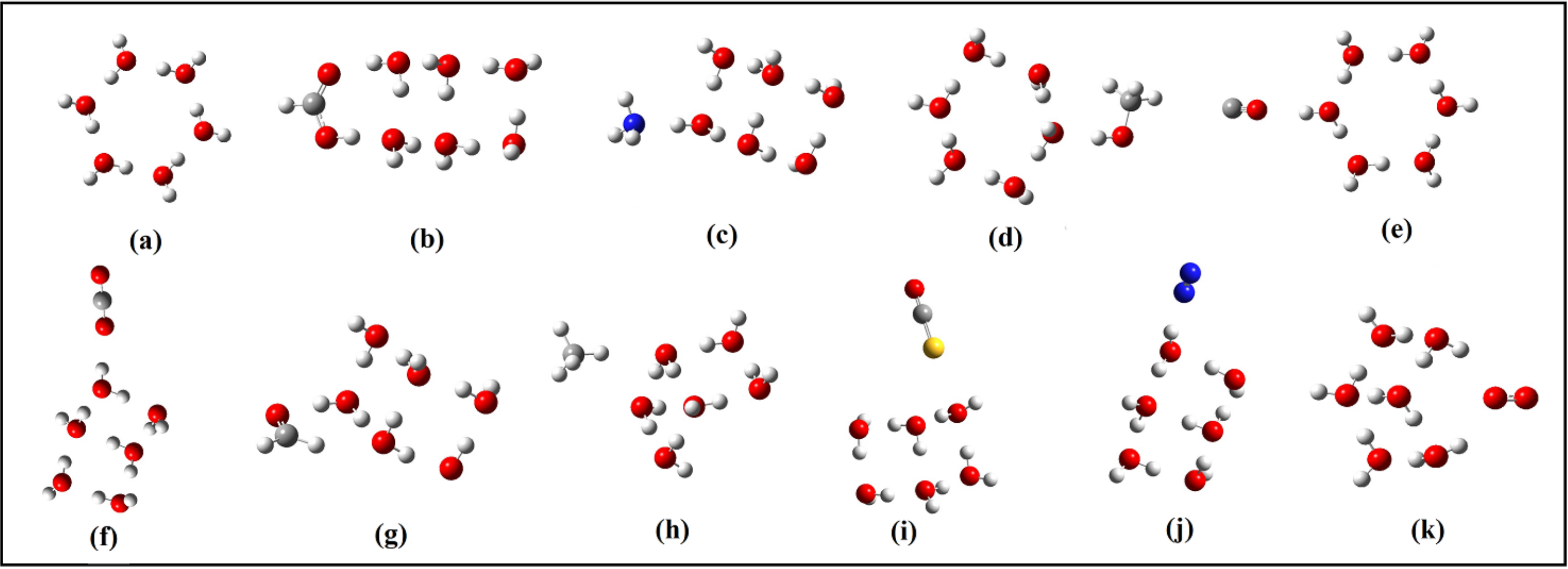}
\caption{Optimized structures of (a) pure water, (b) $\rm{H_2O-HCOOH}$, (c) $\rm{H_2O-NH_3}$, (d) $\rm{H_2O-CH_3OH}$, (e) $\rm{H_2O-CO}$, 
(f) $\rm{H_2O-CO_2}$, (g) $\rm{H_2O-H_2CO}$, (h) $\rm{H_2O-CH_4}$, (i) $\rm{H_2O-OCS}$, (j) $\rm{H_2O-N_2}$, and (k) $\rm{H_2O-O_2}$ clusters 
with a $6:1$ concentration ratio \citep{gora20a}.}
\label{fig:optimized_structure_6H2O}
\end{figure}

\begin{figure}
\centering
\begin{minipage}{0.4\textwidth}
\includegraphics[width=\textwidth]{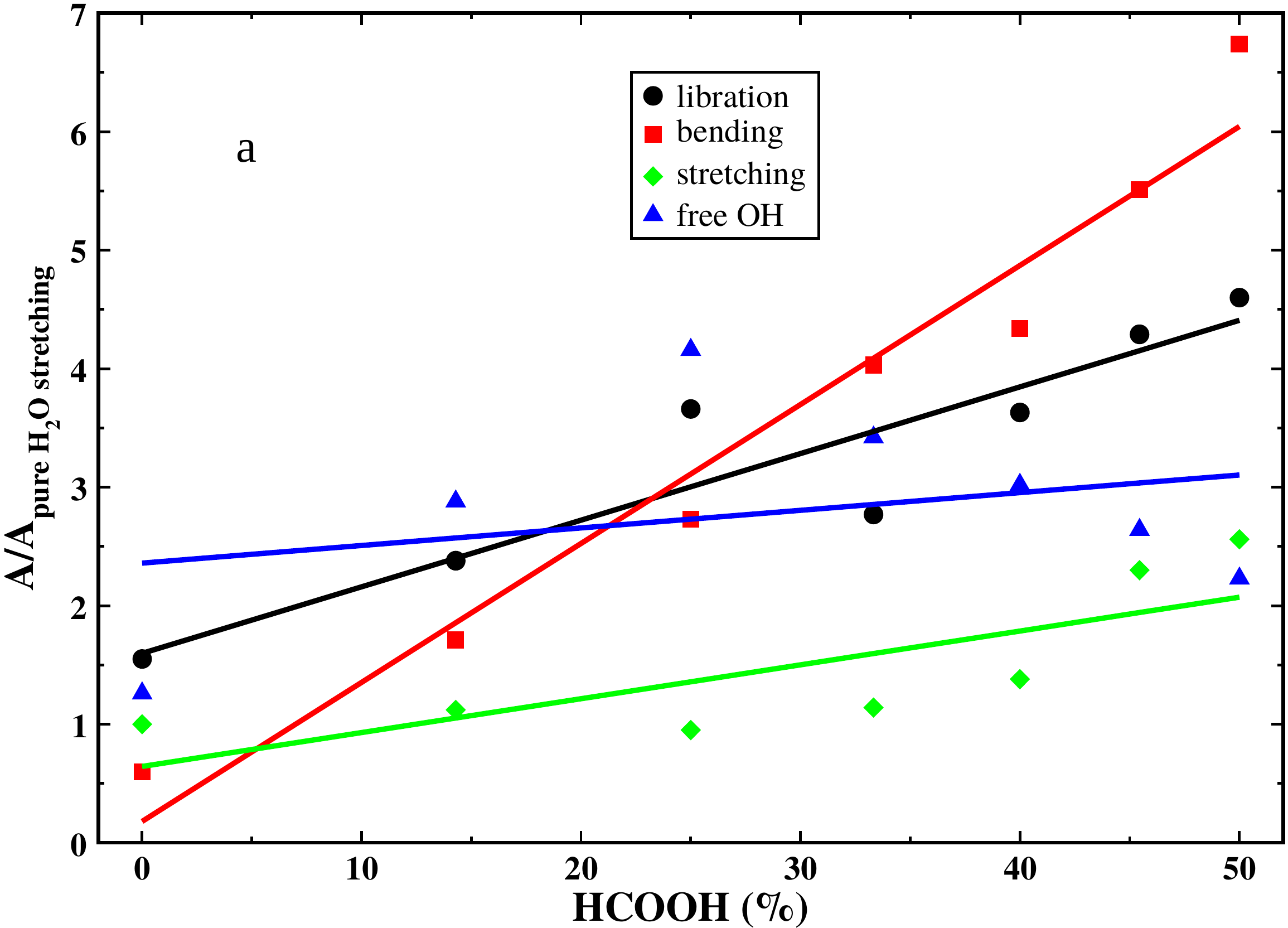}
\end{minipage}
\begin{minipage}{0.4\textwidth}
\includegraphics[width=\textwidth]{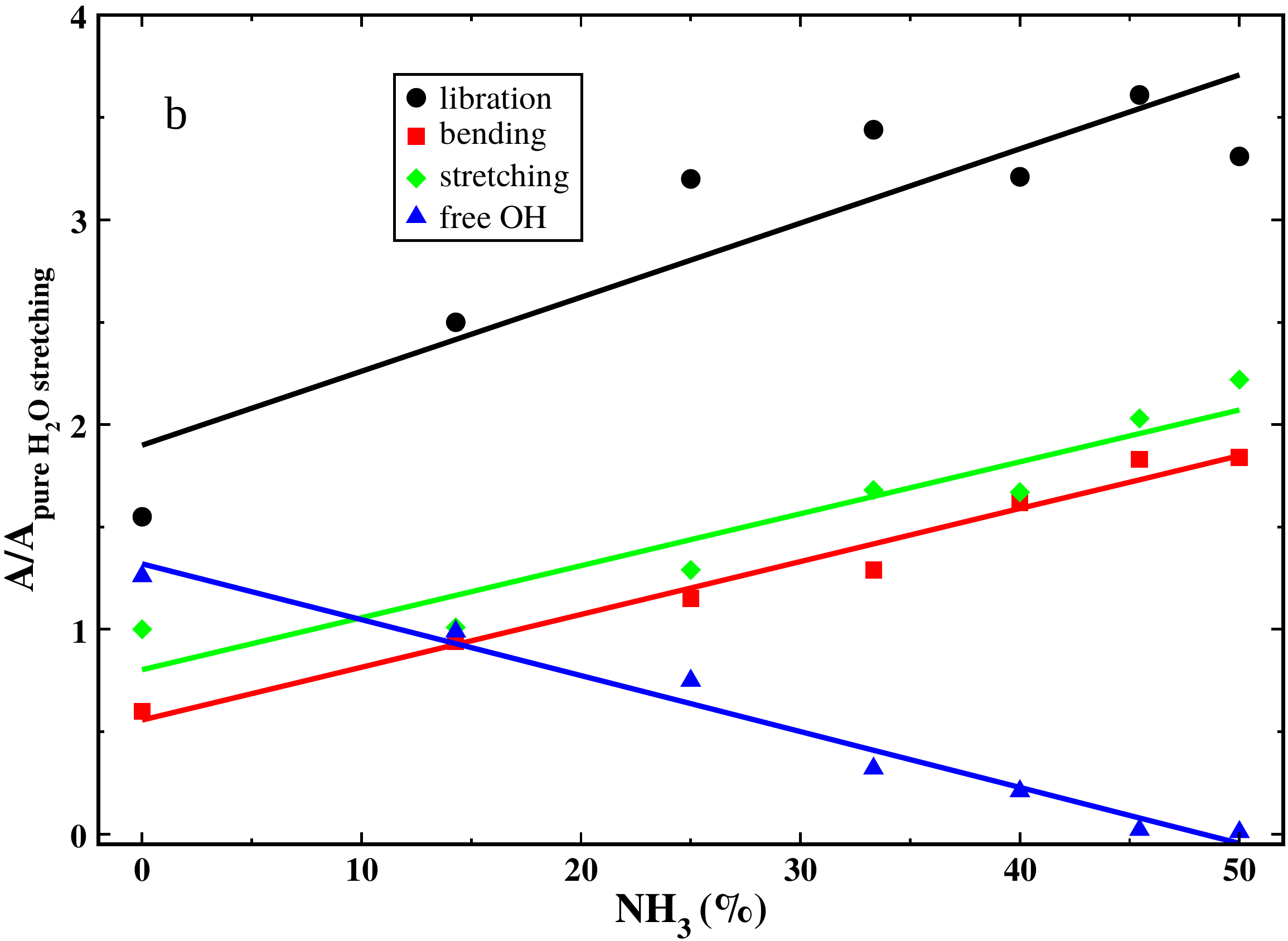}
\end{minipage}
\begin{minipage}{0.4\textwidth}
\includegraphics[width=\textwidth]{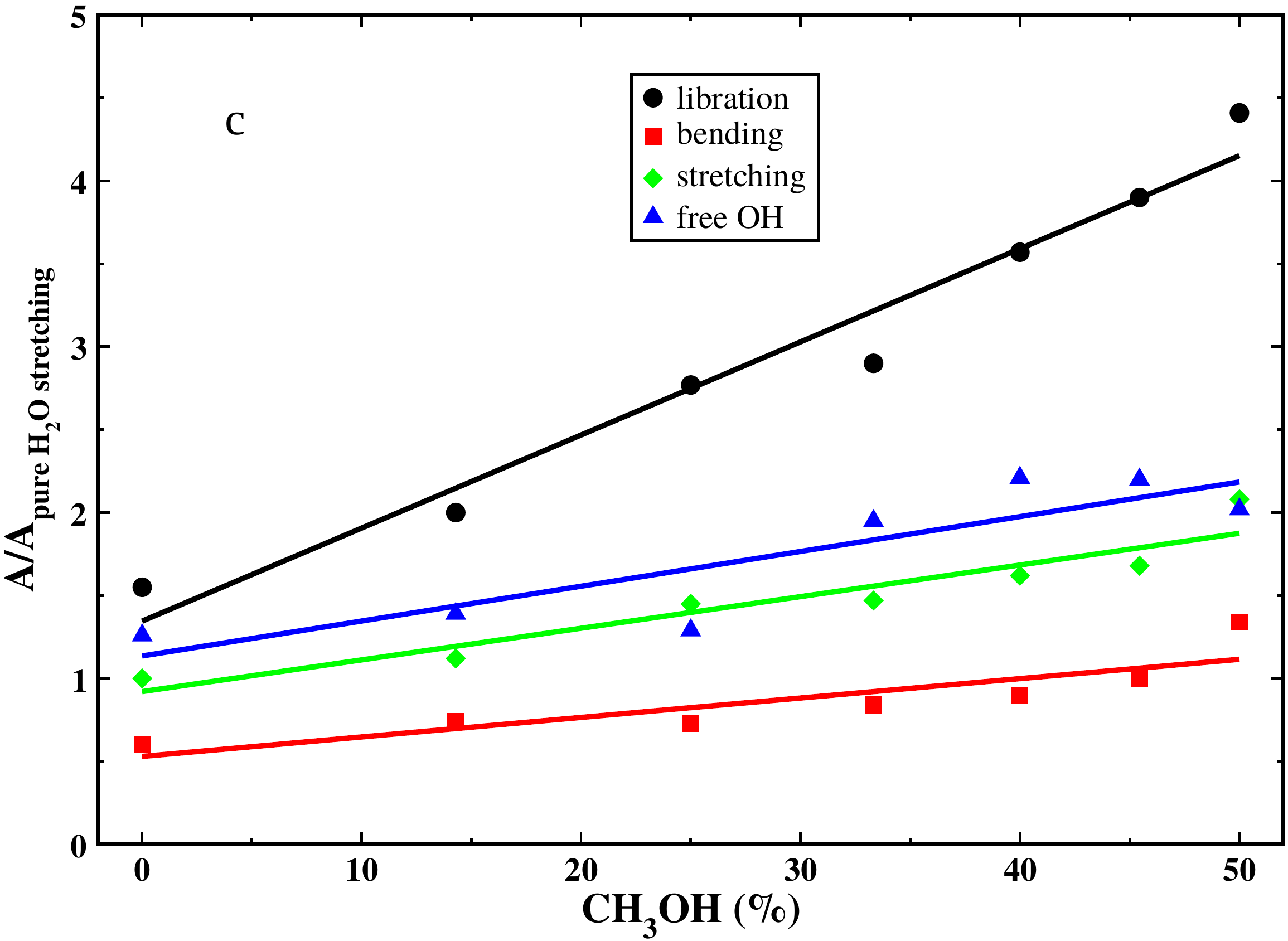}
\end{minipage}
\begin{minipage}{0.4\textwidth}
\includegraphics[width=\textwidth]{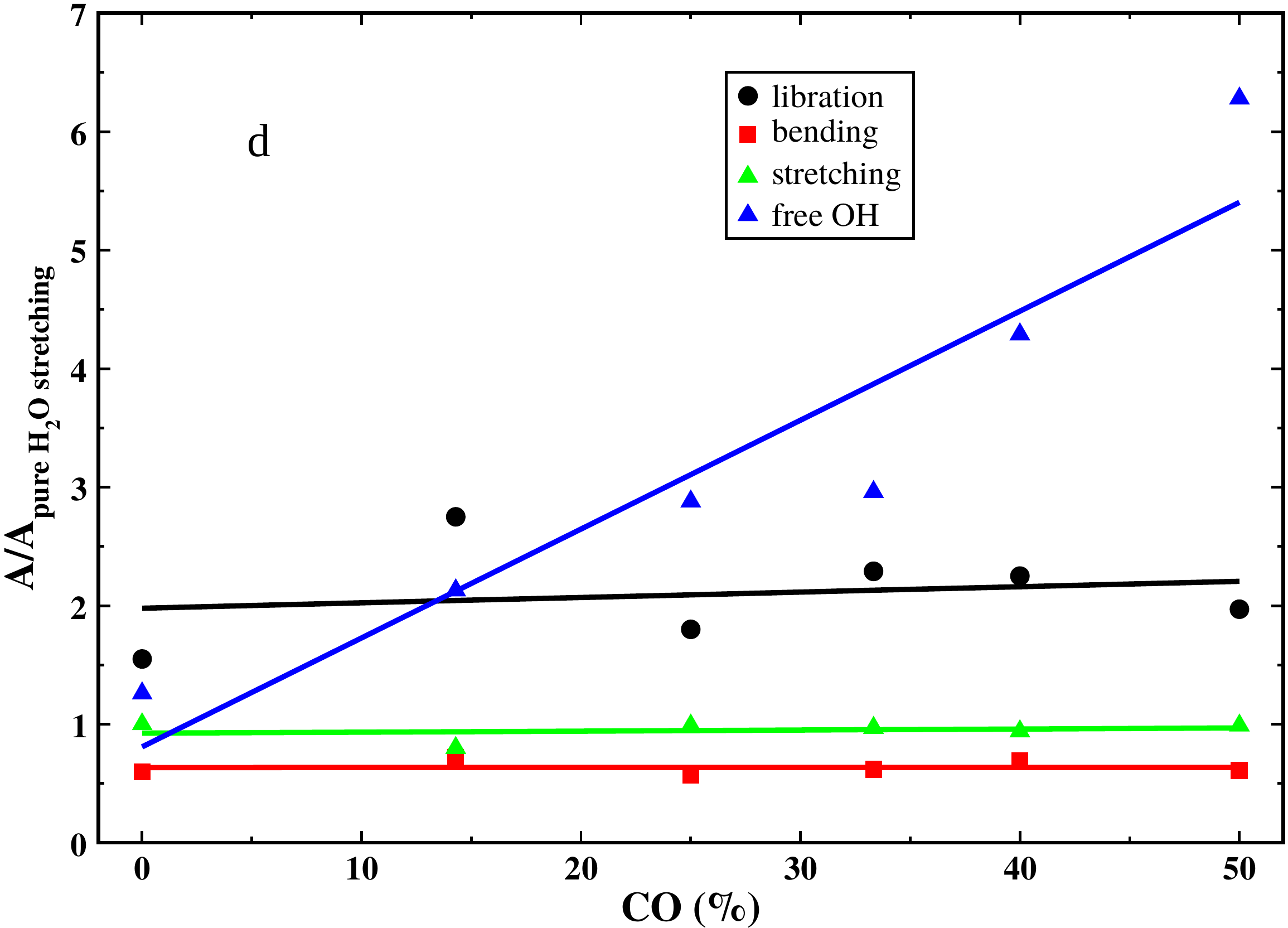}
\end{minipage}
\begin{minipage}{0.4\textwidth}
\includegraphics[width=\textwidth]{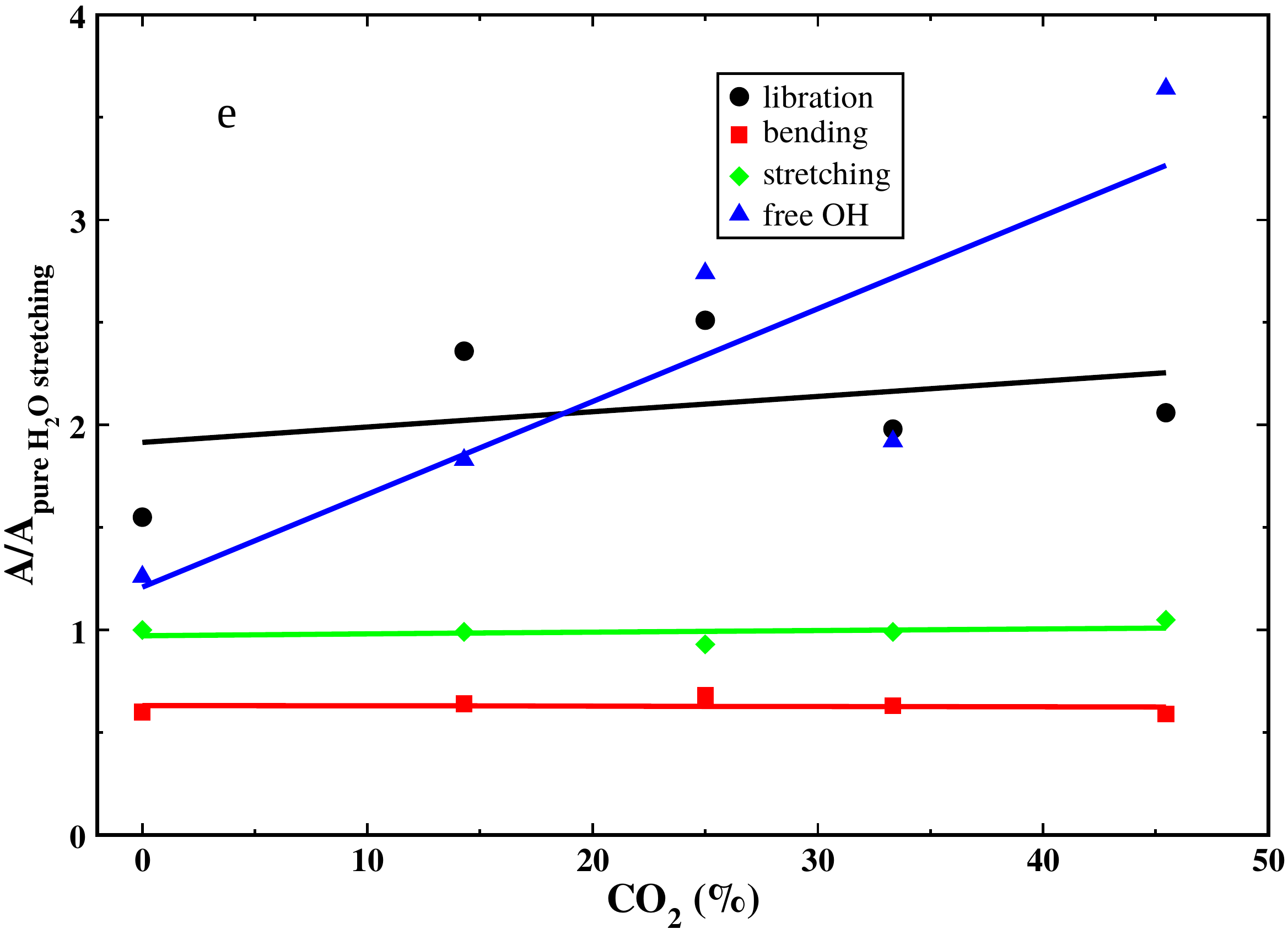}
\end{minipage}
\begin{minipage}{0.4\textwidth}
\includegraphics[width=\textwidth]{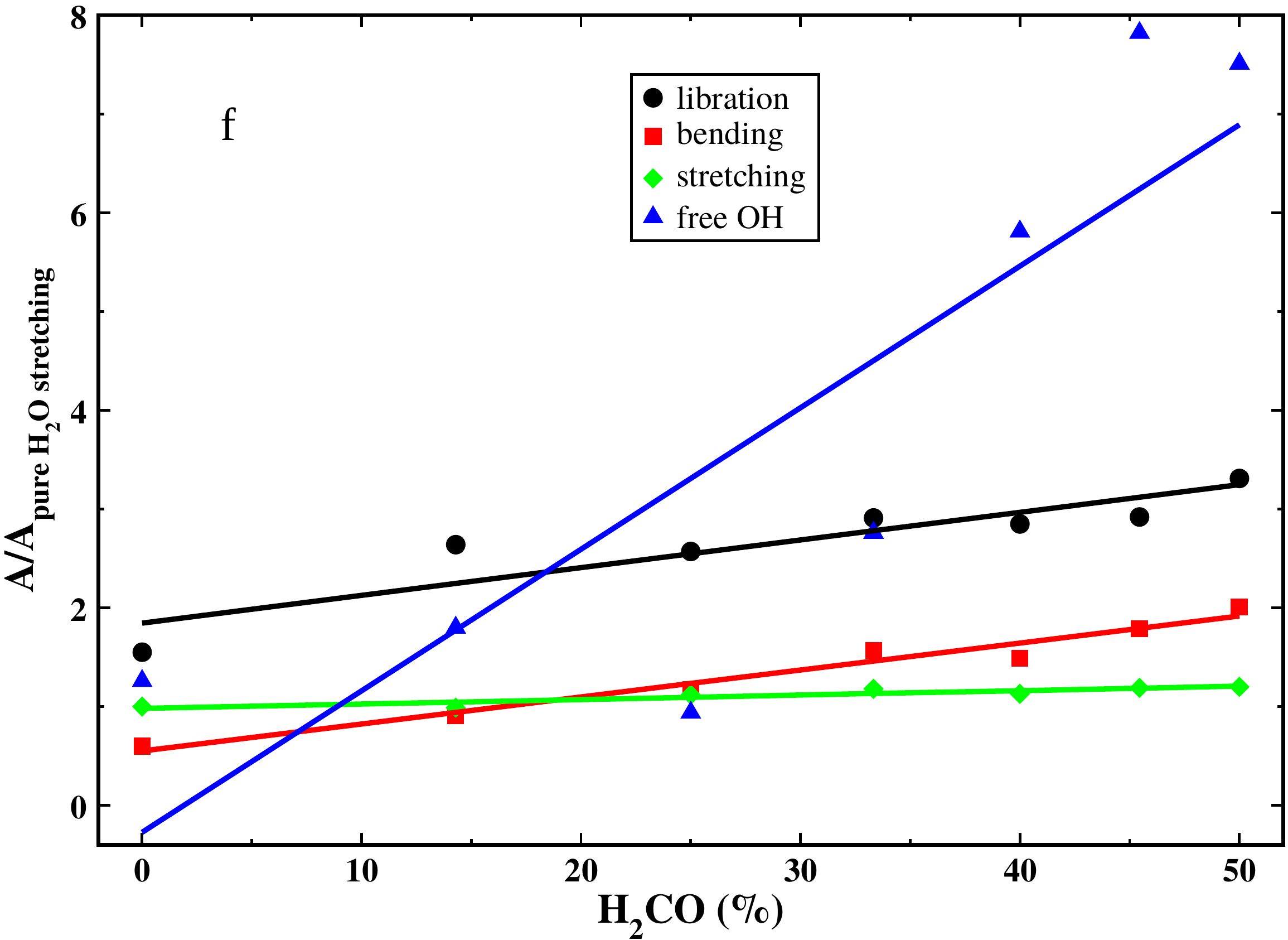}
\end{minipage}
\begin{minipage}{0.4\textwidth}
\includegraphics[width=\textwidth]{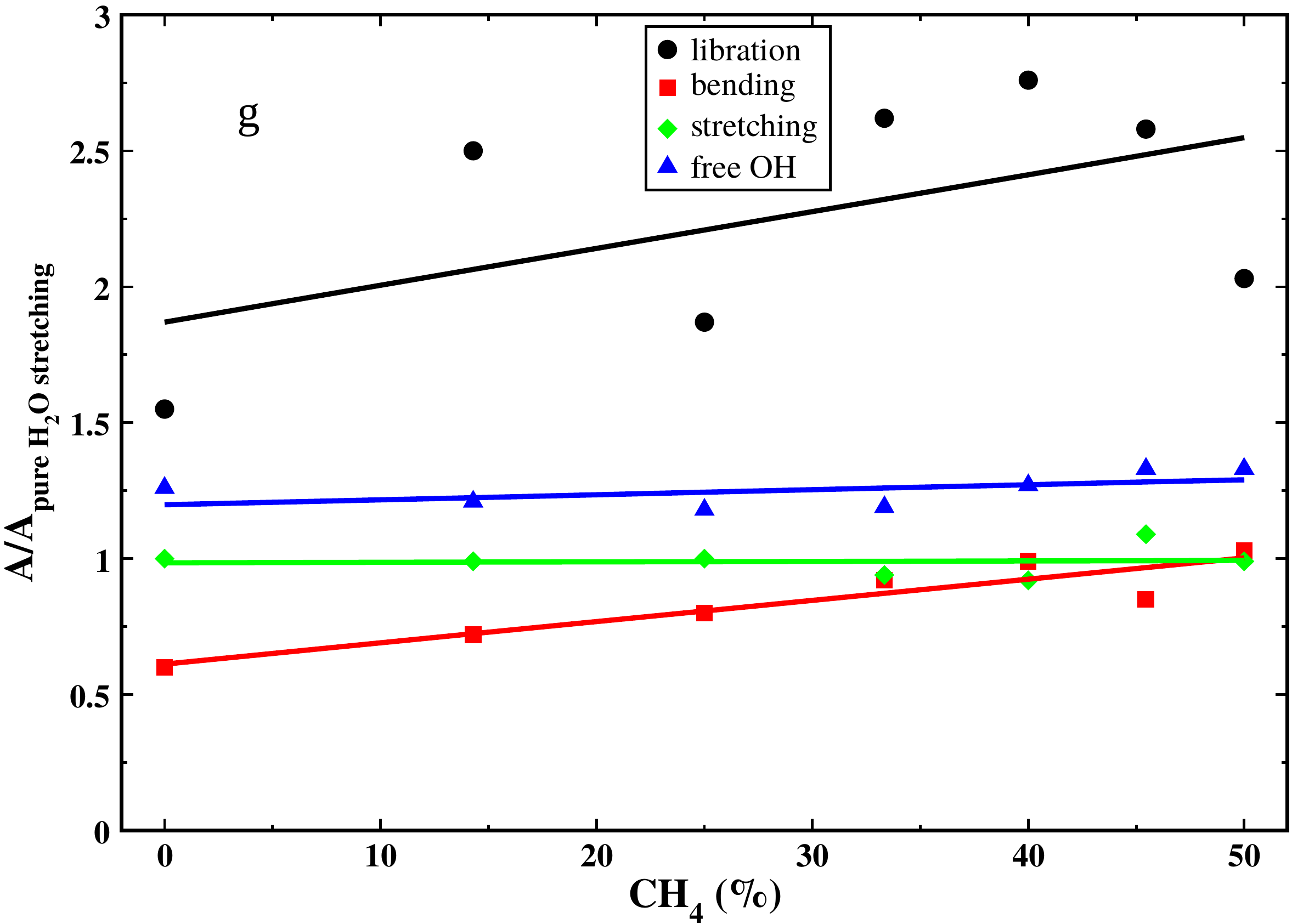}
\end{minipage}
\begin{minipage}{0.4\textwidth}
\includegraphics[width=\textwidth]{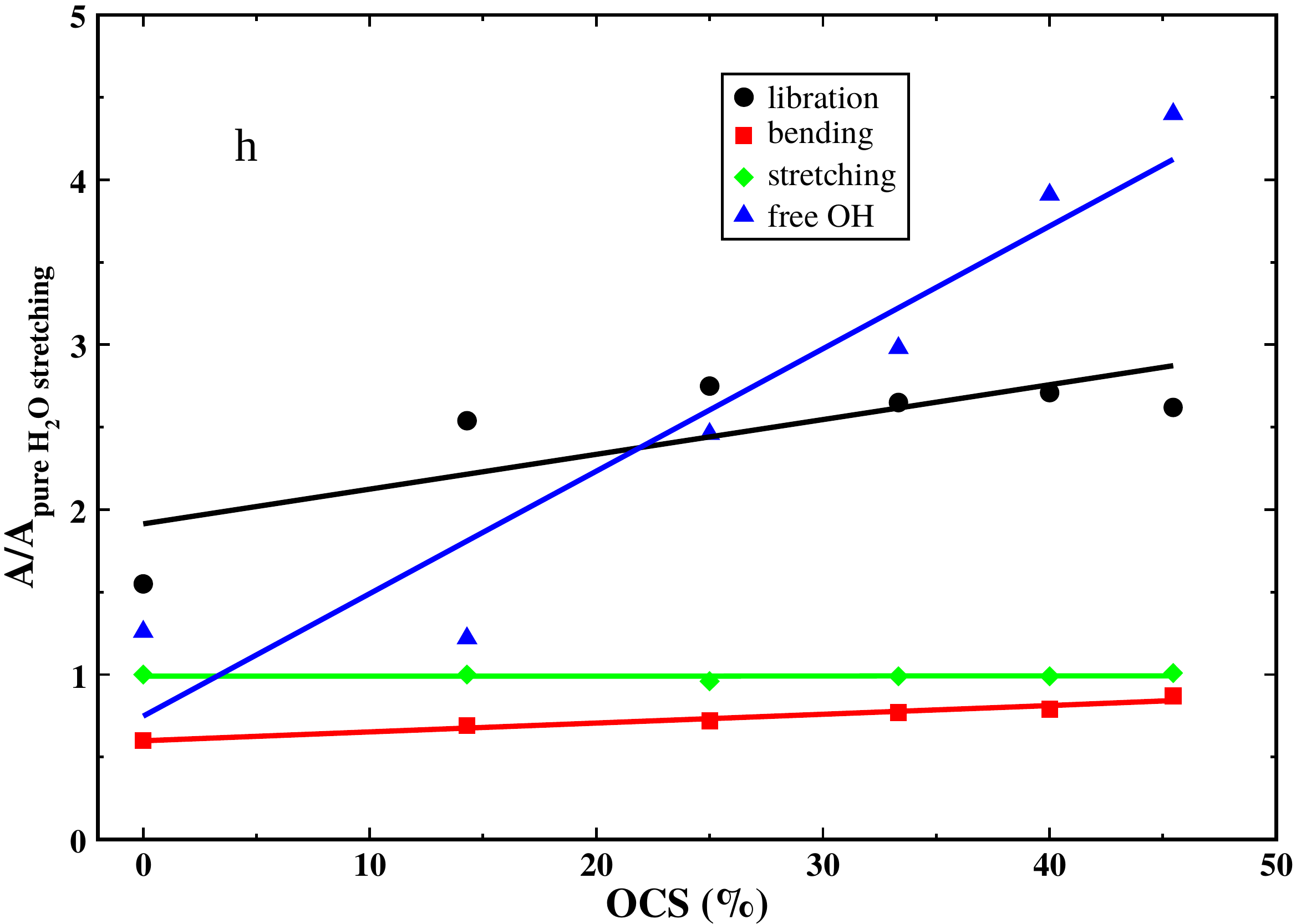}
\end{minipage}
\begin{minipage}{0.4\textwidth}
\includegraphics[width=\textwidth]{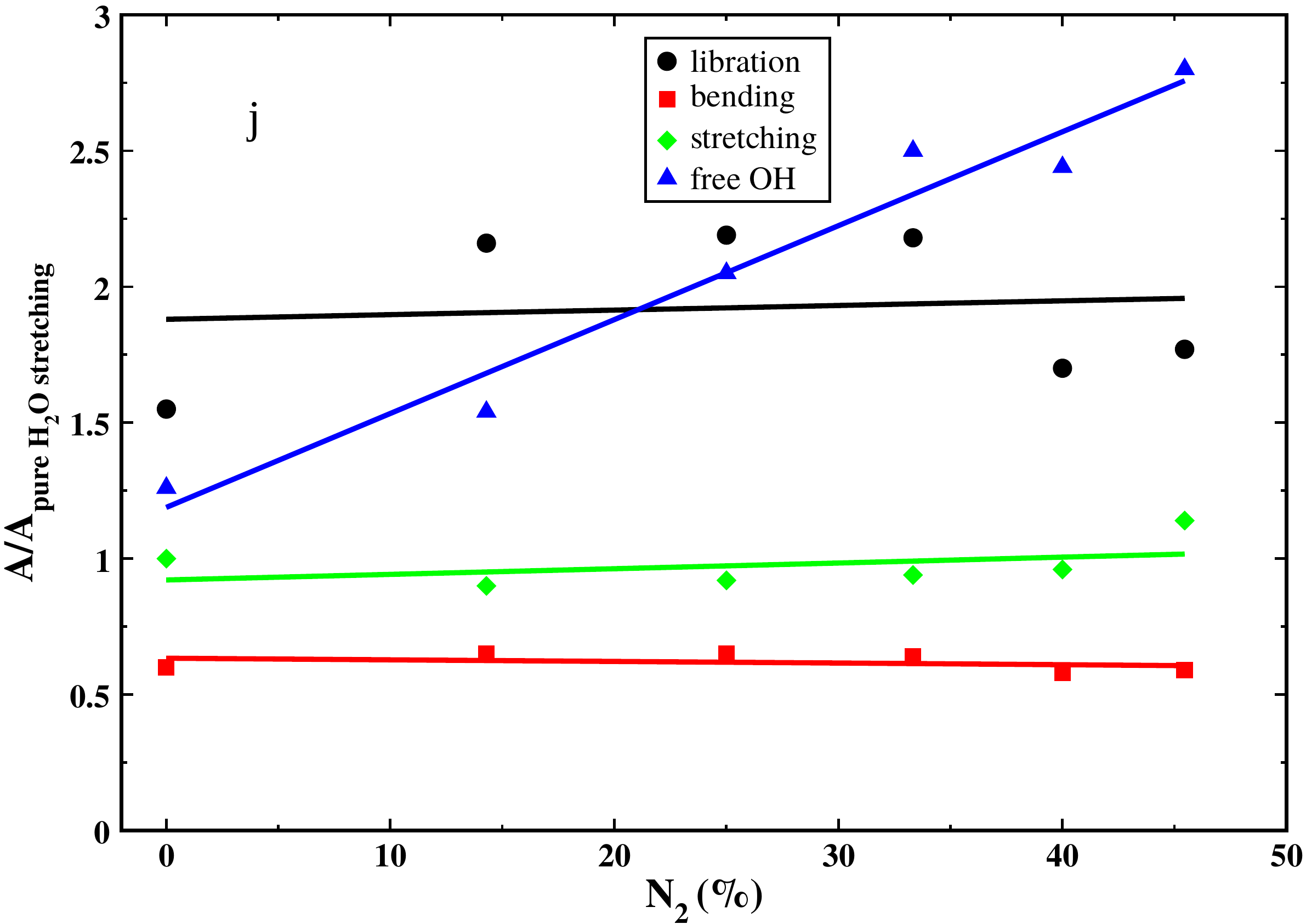}
\end{minipage}
\begin{minipage}{0.4\textwidth}
\includegraphics[width=\textwidth]{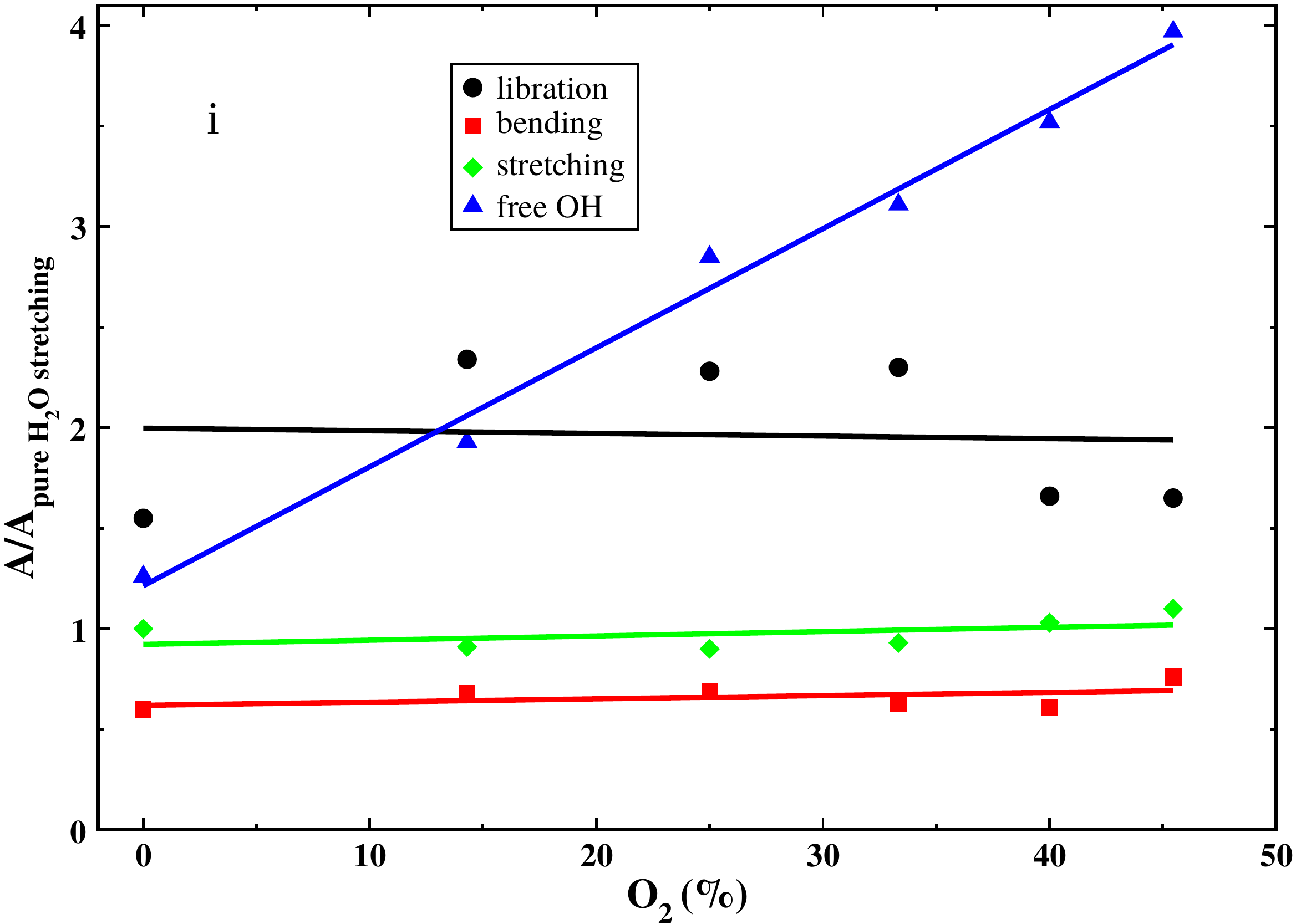}
\end{minipage}
\caption{Band strengths of the four fundamental vibration modes of water for 
(a) $\rm{H_2O-HCOOH}$, (b) $\rm{H_2O-NH_3}$, (c) $\rm{H_2O-CH_3OH}$, (d) $\rm{H_2O-CO}$, (e) $\rm{H_2O-CO_2}$,
(f) $\rm{H_2O-H_2CO}$, (g) $\rm{H_2O-CH_4}$, (h) $\rm{H_2O-OCS}$,
(i) $\rm{H_2O-N_2}$, and (j) $\rm{H_2O-O_2}$ clusters with various concentrations. The water c-hexamer (chair) configuration gas been used for pure water \citep{gora20a}.}
\label{fig:6h2o_x_band_strength}
\end{figure}

\begin{figure}
\centering
\includegraphics[width=0.7\textwidth]{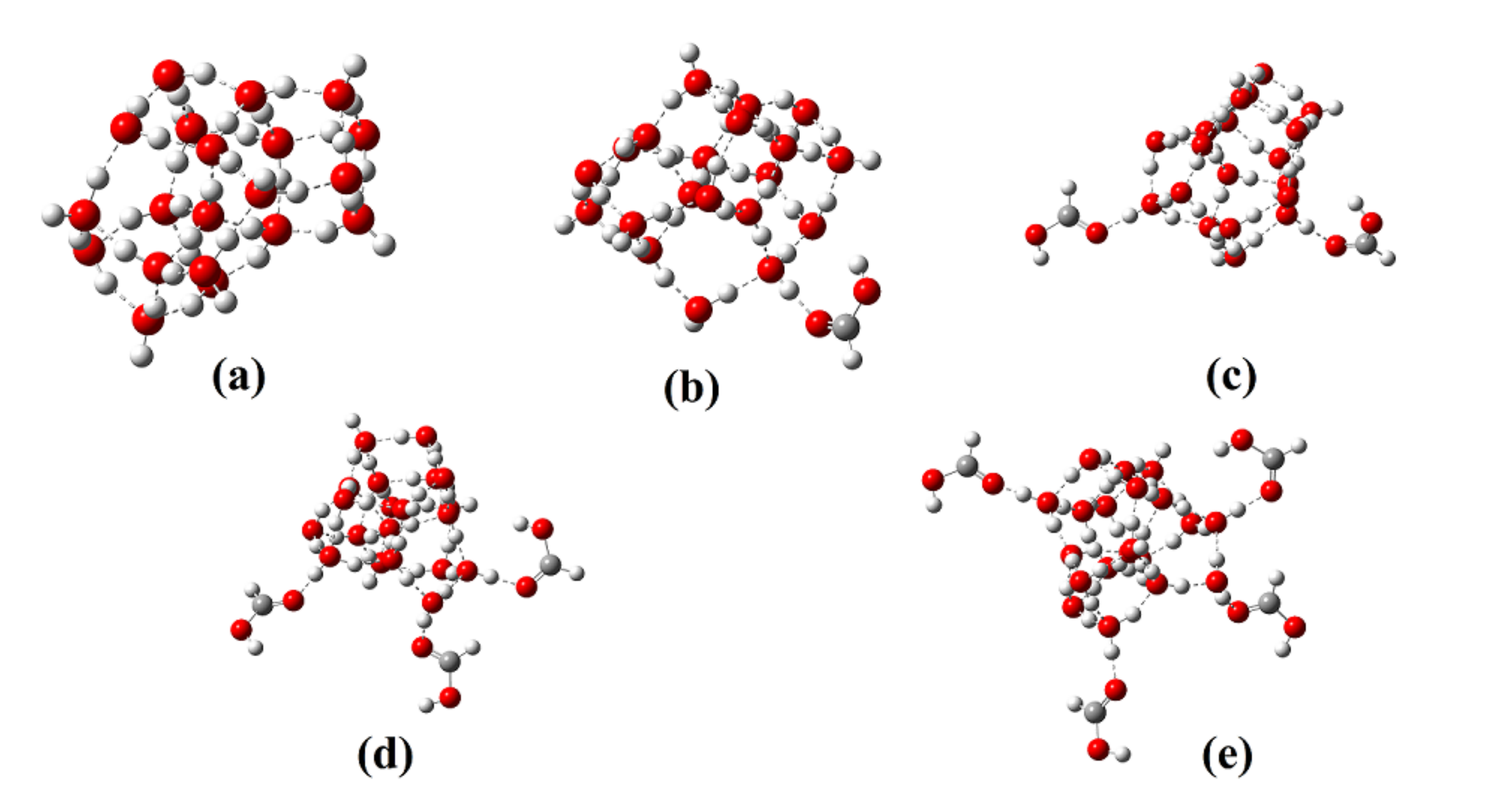}
\caption{Structure of water clusters containing $\rm{20H_2O}$ molecules with HCOOH as impurity in different concentration ratio:
(a) pure water, (b) $\rm{H_2O:HCOOH=20:1}$, (c) $\rm{H_2O:HCOOH=10:1}$, (d) $\rm{H_2O:HCOOH=6.67:1}$,
(e) $\rm{H_2O:HCOOH=5:1}$ \citep{gora20a}.}
\label{fig:20H2O-HCOOH}
\end{figure}

\begin{figure}
\centering
\includegraphics[width=0.7\textwidth]{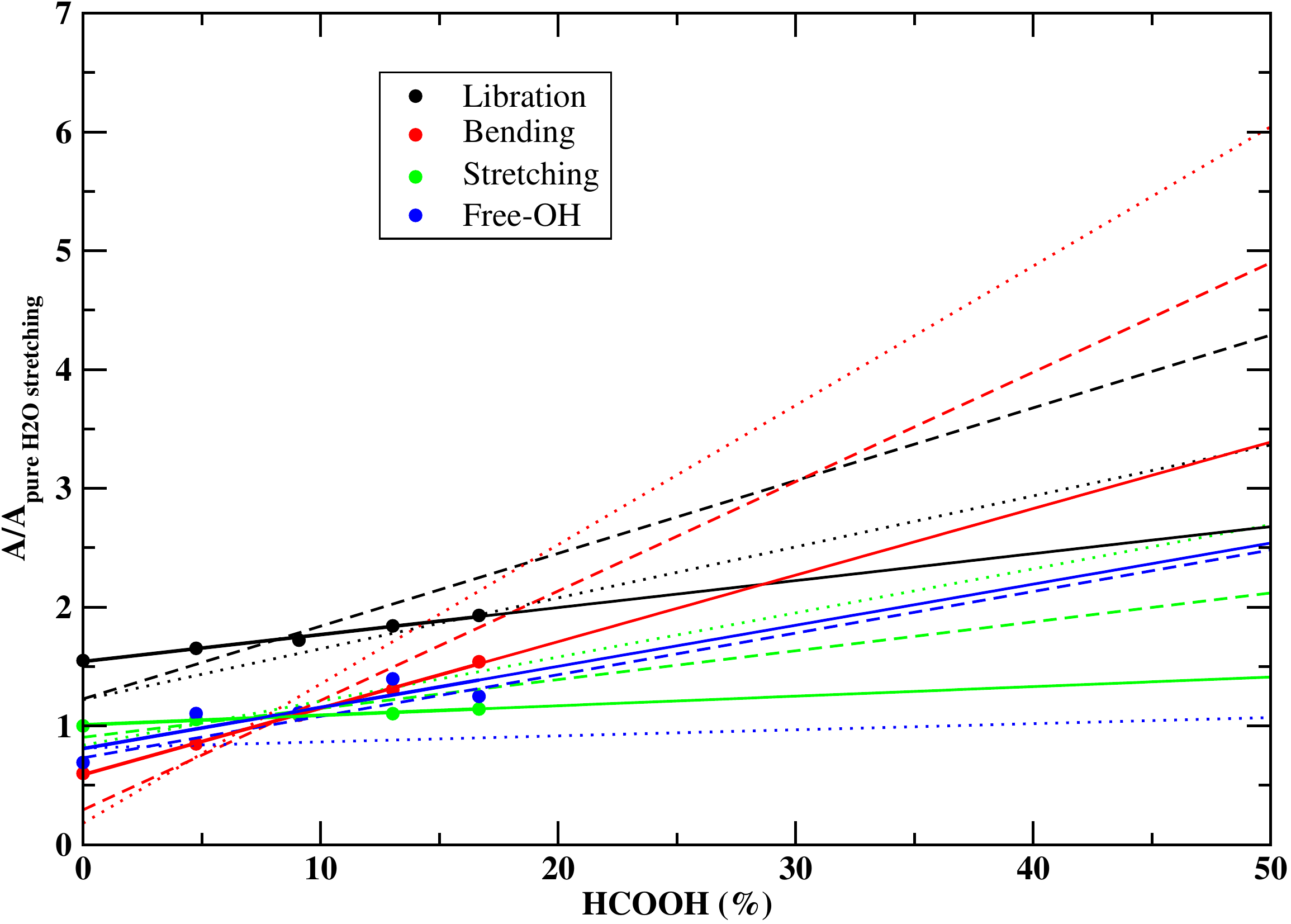}
\caption{Comparison of the band strength of the four fundamental vibrational modes for water clusters containing 20$\rm{H_2O}$, 6H$_2$O, and 4H$_2$O molecules with HCOOH as an impurity in different concentrations. Solid lines represent the band strength profiles for 20H$_2$O cluster, dotted 
lines for the water c-hexamer (chair) (6H$_2$O), and dashed lines for water c-tetramer (4H$_2$O) \citep{gora20a}.}
\label{fig:water_clusters-HCOOH}
\end{figure}

Figure \ref{fig:optimized_structure_6H2O} depicts the optimized structures of the pure water c-hexamer (chair) configuration along with those obtained for a $6:1$ concentration ratio. Figure \ref{fig:6h2o_x_band_strength}, analogous to Figure \ref{fig:band_strength}, collects the results for the band strength variations where c-hexamer (chair) water cluster configuration is considered.
The geometries of water clusters containing $20$ water molecules with HCOOH as an impurity in various concentrations are shown in Figure \ref{fig:20H2O-HCOOH}.
The corresponding variations of the band strengths with increasing concentration of HCOOH are depicted in
Figure \ref{fig:water_clusters-HCOOH}.
This figure also reports the comparison of band strength profiles for different water clusters. The structures of the 20-water-molecule clusters are taken from \cite{shim18}.
The structures as a model of the ASW surface were obtained by MD-annealing calculations using classical force fields.
The comparison shown in Figure \ref{fig:water_clusters-HCOOH} demonstrates
that the 4H$_2$O model provides results similar to those obtained with 6 and 20 water molecules.
This furthermore confirms the validity of our approach.

\begin{figure}
\centering
\includegraphics[width=0.7\textwidth]{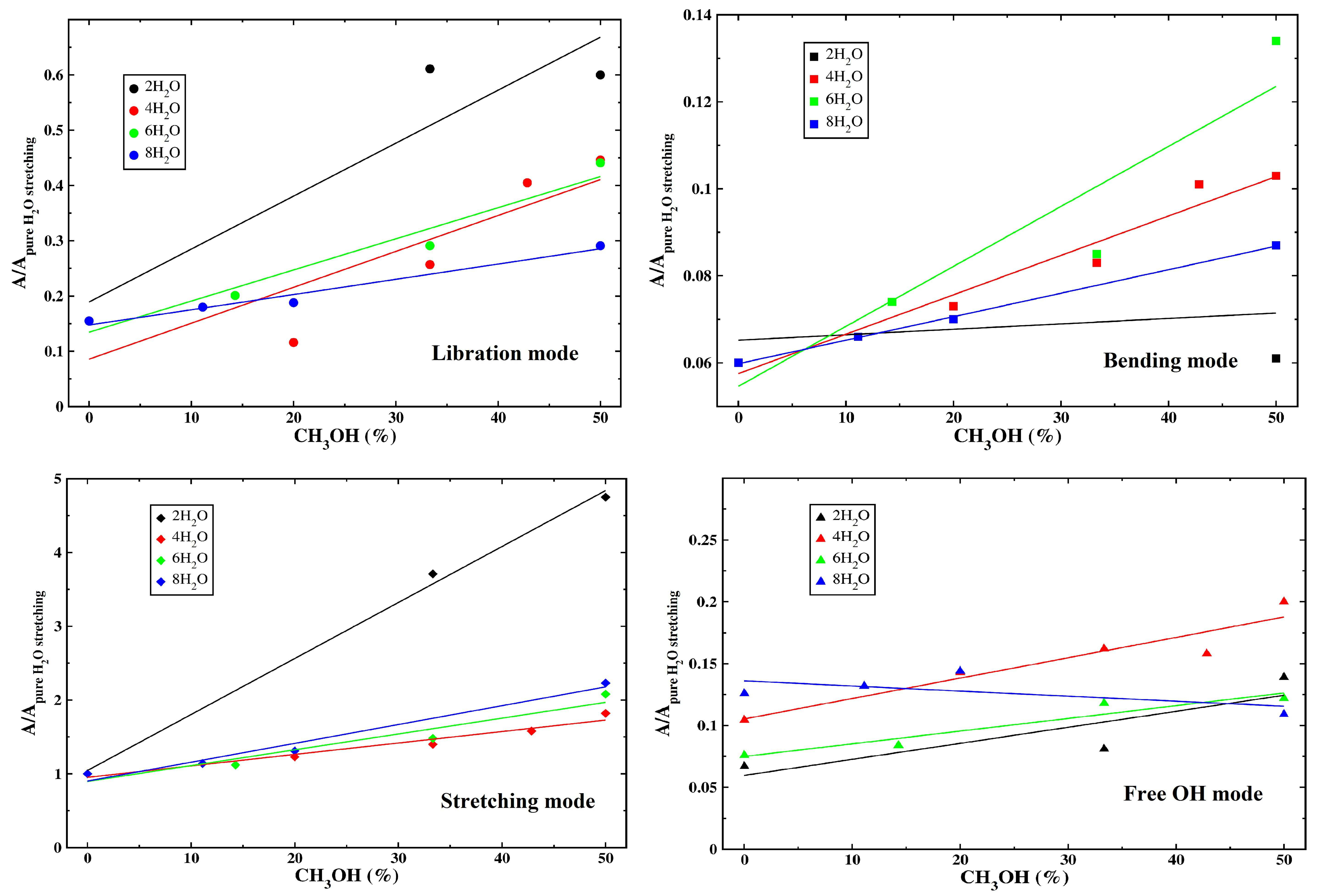}
\caption {Effect of the cluster size on the band strength profile \citep{gora20a}.}
\label{fig:CH3OH_varied_cluster_size}
\end{figure}

\subsection{Summary}

Water ice is known to be the major constituent of interstellar icy grain mantles. Interestingly, there are several astronomical observations \citep{boog00,kean01} of the OH stretching and HOH bending modes at $3278.69$ cm$^{-1}$ ($3.05$ $\mu$m) and $1666.67$ cm$^{-1}$ ($6.00$ $\mu$m), respectively.
However, it is noteworthy that the intensity ratio of these two bands is very different from that obtained in laboratory experiments for pure water ice. This suggests that the presence of impurities in water ice affects the spectroscopic features of water itself. For this reason, a series of laboratory experiments are carried out to explain the discrepancy between observations and experiments.
Furthermore, these observations have prompted us to perform an extensive computational investigation to evaluate the effect of different amounts of representative impurities on the band strengths and absorption band profiles of interstellar ice.

A systematic study of the four fundamental vibrational modes of water in various molecular species with different concentration ratios is carried out.
We select the most abundant interstellar ice impurities (HCOOH, $\rm{NH_3,\ CH_3OH}$, CO,
$\rm{CO_2,\ H_2CO,\ CH_4,\ OCS,\ N_2}$, and $\rm{O_2}$) and study their impact on four fundamental vibrational bands of pure water ice by employing different cluster models.
Specifically, we examine the effect on the libration, bending, bulk stretching, and free OH stretching modes. The theoretical calculations are supported and complemented by some IR spectroscopy experiments
to verify the effect of HCOOH, $\rm{NH_3}$, and $\rm{CH_3OH}$ on the band profiles of pure $\rm{H_2O}$ ice. Indeed, both the experimental and theoretical peak positions might differ from the astronomical observations. This is because the grain shape, size, constituents, surrounding physical conditions, and impurities play a crucial role in tuning the ice spectroscopic features.

Although most of the computations are performed for a cluster containing only four water molecules as
a model system (to find a trend in the absorption band strength), we demonstrate that increasing the size of the cluster would change the band strength profile only marginally.
From the band strength profiles shown in Figure \ref{fig:CH3OH_varied_cluster_size}, it is apparent
that the stretching mode is the most affected, and the bending mode is the least affected
by the presence of impurities. Libration, bending, and bulk stretching modes are the most affected by HCOOH impurity, followed by $\rm{CH_3OH}$ and $\rm{H_2CO}$.
Another interesting point to be noted is that the band strength of the free OH stretching mode decreases with the increasing concentration of $\rm{NH_3}$ and completely vanishes when the concentration of NH$_3$ becomes $50 \%$. Interestingly, the experimental free OH band profile shows a decreasing trend when water is mixed with $\rm{NH_3}$ (Figure \ref{fig:HCOOH-NH3_band_strength}, right panel), similar to
that obtained computationally.

Finally, the work presented here aims at gaining information on the effect of intermolecular interactions in interstellar relevant ices, thus providing some valuable new laboratory and computed absorption spectra of water-rich ices. These will be useful for interpreting future detailed space-borne observations using the upcoming JWST mission \citep{gibb04} in the mid-IR spectral region.

%% file: chap3.tex
\chapter{Radiation Dominated Region} \label{chap:crab}

\section*{Overview}

Two noble gas molecular cations, argonium ($\rm{ArH^+}$) and hydro-helium or helenium ($\rm{HeH^+}$), are discovered toward the two radiation-dominated environments in space:
the Crab Nebula SNR and the planetary nebula (NGC 7027), respectively. Though the elemental abundance of neon is lower than helium, it is higher than that of the Argon. However, neonium cation ($\rm{NeH^+}$) remains undetected in space to date.
Hydroxyl radicals ($-\rm{OH}$) are very abundant in neutral and cationic forms. Consequently, they are found predominantly in radiation-dominated regions. Still, hydroxyl cations of such noble gases (i.e., ArOH$^+$, NeOH$^+$, and HeOH$^+$) are yet to be detected in space.
This Chapter attempts to model the noble gas chemistry containing hydride and hydroxyl cations (ArH$^+$, NeH$^+$, HeH$^+$, ArOH$^+$, NeOH$^+$, and HeOH$^+$). We also consider various isotopologs of these hydride and hydroxide cations ($^{36}$Ar, $^{38}$Ar, $^{40}$Ar, $^{20}$Ne, and $^{22}$Ne). The chemical evolution of these species under the diffuse and exotic environment (the Crab Nebula filamentary region) is studied \citep{das20}. The intrinsic line surface brightness (SB) is calculated to find a favorable parameter space that can explain the observational features for the condition suitable in the Crab filamentary region.

\clearpage
The Crab Nebula (M1 = NGC 1952) is the freely expanding remnant of the historical core-collapse supernova of AD 1054 (SN 1054). It is too young to be polluted by interstellar or circumstellar material. The Crab pulsar is situated at a distance of $3.37$ kpc \citep{fras19} from the Sun with RA and DEC 05$^h$ 34$^m$ 31.935$^s$ and $+22^\circ 0' 52''.19$ respectively \citep{kapl08}. The Crab, laying about $200$ pc away from the Galactic plane in a low-density region, contains both atomic and molecular hydrogen, electrons, and an area of enhanced ionized argon emission.

\cite{loh10,loh11} identified numerous strong
H$_2$-emitting (2.12 $\mu$m) knots in the Crab, and \cite{rich13} modeled emission features of $\rm{H_2}$-emitting gas in the Crab knot 51 filamentary region.
The kinetic gas and dust temperature around the knots of the Crab is around $\sim 2000-3000$ K \citep{loh12,rich13} and $\sim 28-63$ K \citep{gome12}, respectively. The strong radiation in the Crab enhances the abundance of the electrons, which can readily convert H atoms into H$^-$ by radiative attachment reaction ($\rm{H + e^- \rightarrow H^- + \gamma}$), and H$^-$ eventually react with H atoms again to form the
H$_2$ molecules by associative detachment reaction ($\rm{H^- + H \rightarrow H_2 + e^-}$).
There can be some physisorption and
chemisorption \citep{caza04} pathways that may lead to the formation of $\rm{H_2}$.
The majority of the H$_2$ molecules are formed inside the cleanest knot (knot 51) of the Crab
by associative detachment reaction rather than by the usual grain catalysis ($\rm{2H + grain \rightarrow H_2 + grain}$) method \citep{rich13}.

In the atmosphere of the Earth, Argon is the third most abundant element, having an isotopic ratio of $^{40}$Ar/$^{38}$Ar/$^{36}$Ar is $1584/1.00/5.30$ \citep{lee06}. $^{40}$Ar isotope is mainly produced from the decay of potassium-$40$ in the Earth's crust. Strikingly, the ratio obtained in the Jupiter family comet, 67P/C-G, by ROSINA
mass spectrometer is similar of about $^{36}$Ar/$^{38}$Ar $\sim 5.4\pm1.4$ \citep{altw16}.
In the solar wind, the ratio of $^{40}$Ar/$^{38}$Ar/$^{36}$Ar has been
measured to 0.00/1.00/5.50 \citep{mesh07}, whereas, in the ISM, the $^{36}$Ar isotope is  the most abundant ($\sim 84.6$\%), followed by $^{38}$Ar ($\sim 15.4$\%) and traces of $^{40}$Ar \citep[$\sim 0.025$\%;][]{wiel02}.
The $^{36}$Ar isotope is mainly produced during the core collapse of supernovae events by the explosive nucleosynthesis reactions in massive stars.
Using the observed data from the Spectral and Photometric Image REceiver (SPIRE)
of the Herschel satellite, \cite{barl13} reported
$J = 1 \rightarrow 0$ ($617.5$ GHz) and $J = 2 \rightarrow 1$ ($1234.6$ GHz) emission of $^{36}$ArH$^+$ along with the strongest fine-structure component of the OH$^+$ ion ($971.8$ GHz) toward the Crab.
They predicted the limits of the abundance ratios to be $^{36}$ArH$^+$/$^{38}$ArH$^+>2$ and $^{36}$ArH$^+$/$^{40}$ArH$^+>4-5$.
Molecular excitations mainly occur
due to the collision with electrons in the Crab region with a density of about $\sim 10^2$ cm$^{-3}$.
ArH$^+$ can be used as a unique tracer of $\rm{H_2}$ (by anticorrelation) as well as atomic gas (by correlation) in specific environments \citep{barl13,schi14,neuf16}.
\cite{schi14} assigned $J = 1 \rightarrow 0$ transition of both the isotopologs of ArH$^+$ ($^{36}$ArH$^+$ and $^{38}$ArH$^+$) in absorption. These transitions were observed with the Heterodyne Instrument for the Far Infrared (HIFI) on board the Herschel satellite toward Sagittarius (Sgr) B2(M) and only the primary isotopolog ($^{36}$ArH$^+$) toward numerous prominent continuum sources, e.g., Sgr B2(N), W51e, W49N, W31C, and G34.26+0.15. Their Herschel survey also covered the $J = 1 \rightarrow 0$ NeH$^+$ transition at $1039.3$ GHz but failed to report any detection despite neon being much more abundant than Argon.
\cite{mull15} also detected $^{36}$ArH$^+$ and $^{38}$ArH$^+$ in the absorption of a foreground galaxy at $z = 0.89$ along with two different lines of sight toward PKS $1830-211$ with band 7 of the ALMA interferometer. \cite{hami16} described the excitation of ArH$^+$ in the Crab by collisions with electrons
through radiative transfer calculations. They found that the $2\rightarrow1$ and $1\rightarrow0$
emissions ratio is consistent with an ArH$^+$ column density of $1.7 \times 10^{12}$ cm$^{-2}$. \cite{prie17} performed a combined photoionization and photodissociation study for the emission of ArH$^+$, HeH$^+$, and OH$^+$ in the Crab filament. This filamentary region is subjected to synchrotron radiation and a high flux of charged particles. Their model successfully reproduced the observation of \cite{barl13} when they considered total hydrogen densities between $1900$ and $2 \times 10^4$ cm$^{-3}$.
Though they predicted HeH$^+$ emission above detection thresholds, the formation time-scale for this molecule is much longer than the age of the Crab.

Helium is the second most abundant species (after hydrogen) in the universe, having an abundance of $1/10$ relative to hydrogen nuclei. Because Argon, neon, and helium are noble gases, they do not usually form stable molecules, but they can form stable ions. After a few hundred thousand years of the Big Bang, when the universe cools sufficiently under $4000$ K, helium
was the first neutral element produced in the universe due to its highest ionization potential.
Shortly after the first helium
atom formed, the first chemical bond in the universe formed through the radiative association reaction between the neutral helium atom and a proton during the Age of Recombination. It developed HeH$^+$ by photon emission, which is believed to be a major conduit of photons for the observable cosmic microwave background (CMB) radiation.
The radiative association between He and He$^+$ was also possible to form He$^{2+}$.
But the constant presence of H$^+$ when most of the helium was neutralized keeps the population of HeH$^+$ relatively higher than He$^{2+}$. Due to this fact, HeH$^+$ is considered to be the first molecular ion formed in the universe, and its bond is regarded as the first chemical bond of the universe \citep{lepp02,gall13,fort19}.
The helium hydride ion, HeH$^+$, was first observed in the laboratory nearly a century ago in 1925 \citep{hogn25}, and parts of its vibrational spectrum were first speculated in the mid-1970s \citep{blac78}. Despite these early measurements and predictions, recently, \cite{gust19} reported the first astrophysical identification of HeH$^+$ based on the advances in terahertz spectroscopy and high-altitude observation. They used the German REceiver for
Astronomy at Terahertz Frequencies (GREAT) on board the SOFIA. They were successful in identifying HeH$^+$ by its rotational ground-state transition at a
wavelength of $149.137$ $\mu$m ($2010.184$ GHz) in a young and dense planetary nebula (NGC 7027) located in the Cygnus constellation. Very recently, \cite{neuf20} identified the $v = 1 - 0$ P(1) at 3.51629 $\mu$m and $v = 1-0$ P(2) at 3.60776 $\mu$m) of HeH$^+$ in emission.
They detected these transitions toward the planetary nebula
NGC 7027 using the iSHELL spectrograph on NASA's Infrared Telescope Facility (IRTF) situated at Maunakea. They further confirmed the early discovery of this species reported in \cite{gust19}.

\section{Physical conditions} \label{sec:physical_cond}

Because the physical and chemical processes are inter-linked, it is customary to use appropriate physical parameters to constrain the abundances of the noble gas species.
Here, we construct two models for the Crab Nebula filamentary region: Model A and Model B, to explain the various aspects of the Crab using the
\textsc{Cloudy} code\footnote{\url{https://gitlab.nublado.org/cloudy/cloudy/-/wikis/home}} \citep[version 17.02, last described by][]{ferl17}.

Earlier, \cite{owen15} modeled the properties of dust and gas densities by fitting the predicted SED to the multi-wavelength observations.
Based on their results, amorphous carbon grain is used to mimic the dust
pertaining inside the Crab. For this purpose, the optical constants are taken from \cite{zubk96}, and a mass density of $1.85$ g cm$^{-3}$ is adopted. We modify the default grain size distribution of \textsc{Cloudy} and assume that it would maintain a power-law distribution $n(a) \propto a^{-\alpha}$ with $\alpha$ =  2.7,
$a_{min}$ = 0.005 $\mu$m, and $a_{max}$ = 0.5 $\mu$m following
the clumpy model VI of \cite{owen15}.
We use a higher dust-to-gas mass ratio \citep[$\frac{M_d}{M_g} = 0.027$;][]{owen15} suitable for the Crab. The $A_V/N(H)$ ratio is self-consistently
calculated based on the dust-to-gas mass ratio in the \textsc{Cloudy} code. We obtain $A_V/N(H)$ $\sim 2.094 \times 10^{-20}$ mag cm$^2$.
\cite{prie17} used a similar dust-to-gas mass ratio in their model, but they kept their extinction-to-gas ratio $A_V/N(H)$ at the
standard interstellar value ($6.289 \times 10^{-22}$ mag cm$^2$), which is about two orders of magnitude lower than the (more realistic) value used here.
We assume a shell of matter having a thickness ($dr$) of $3.5 \times 10^{16}$ cm located at $2.5$ pc from the central point source \citep[i.e., inner radius, $r_{in}=2.5$ pc;][]{prie17}.
Because we consider $r_{in}>> dr$, in principle, a plane-parallel geometry is assumed.
Following \cite{shaw05}, the extensive model for the H$_2$ is considered. We include the physics of PAHs in our model. Also, the photoelectric heating and collisional processes are included.

\begin{figure}
\begin{center}
\includegraphics[width=0.65\textwidth]{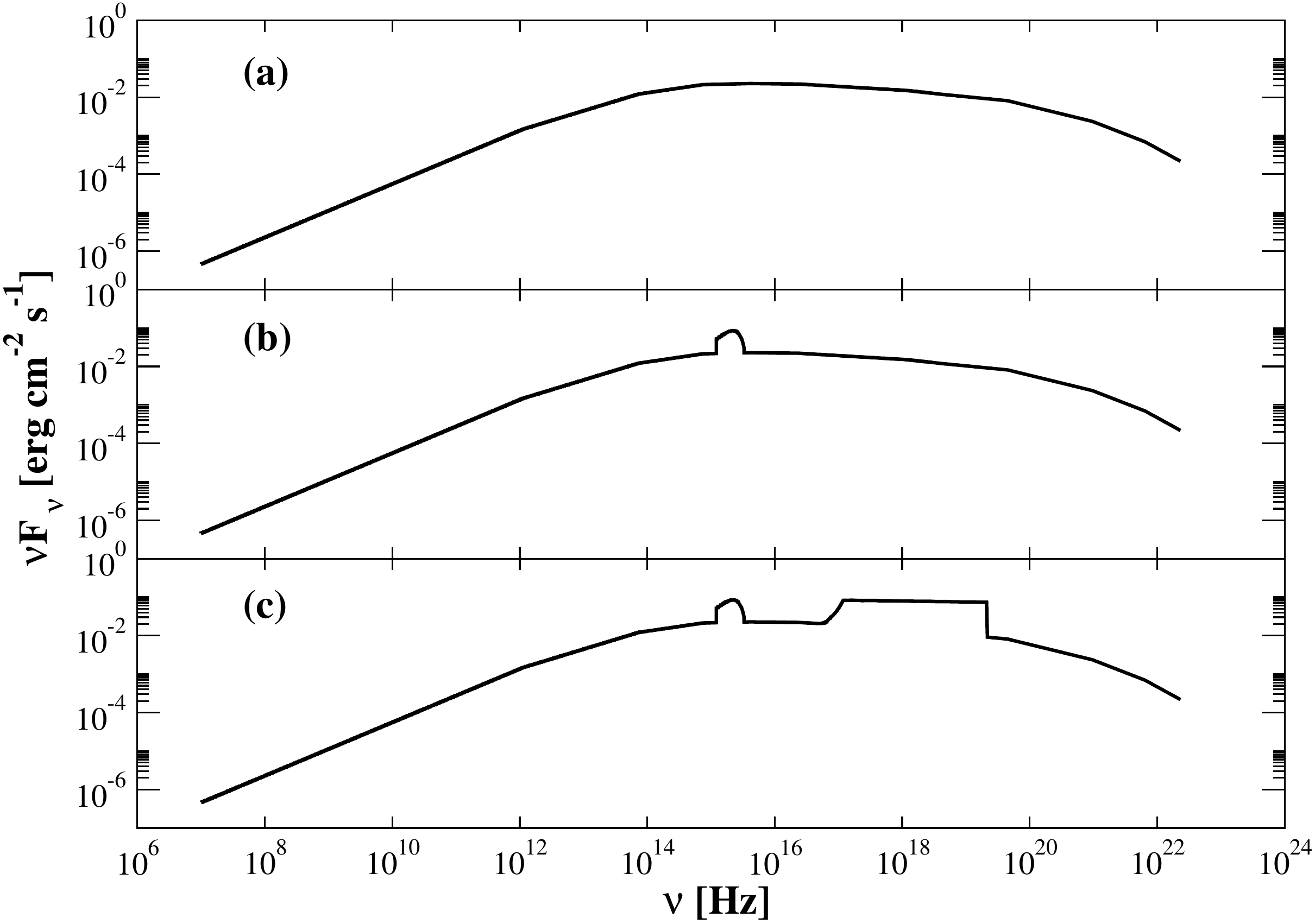}
\caption{Shape and intensity of the resulting incident SED \citep{das20}. The three panels of this figure show the modifications of the SED sequentially. The SED obtained from \cite{hest08} is shown in panel (a), panel (b) shows the SED after the inclusion of the Galactic background radiation field of $31$ Draine units, and finally, panel (c) shows the resulting complete SED after the inclusion of the X-ray spectrum from Figure 1 of \cite{prie17}.}
\label{fig:sed}
\end{center}
\end{figure}

\begin{figure}
\begin{center}
\includegraphics[width=0.65\textwidth]{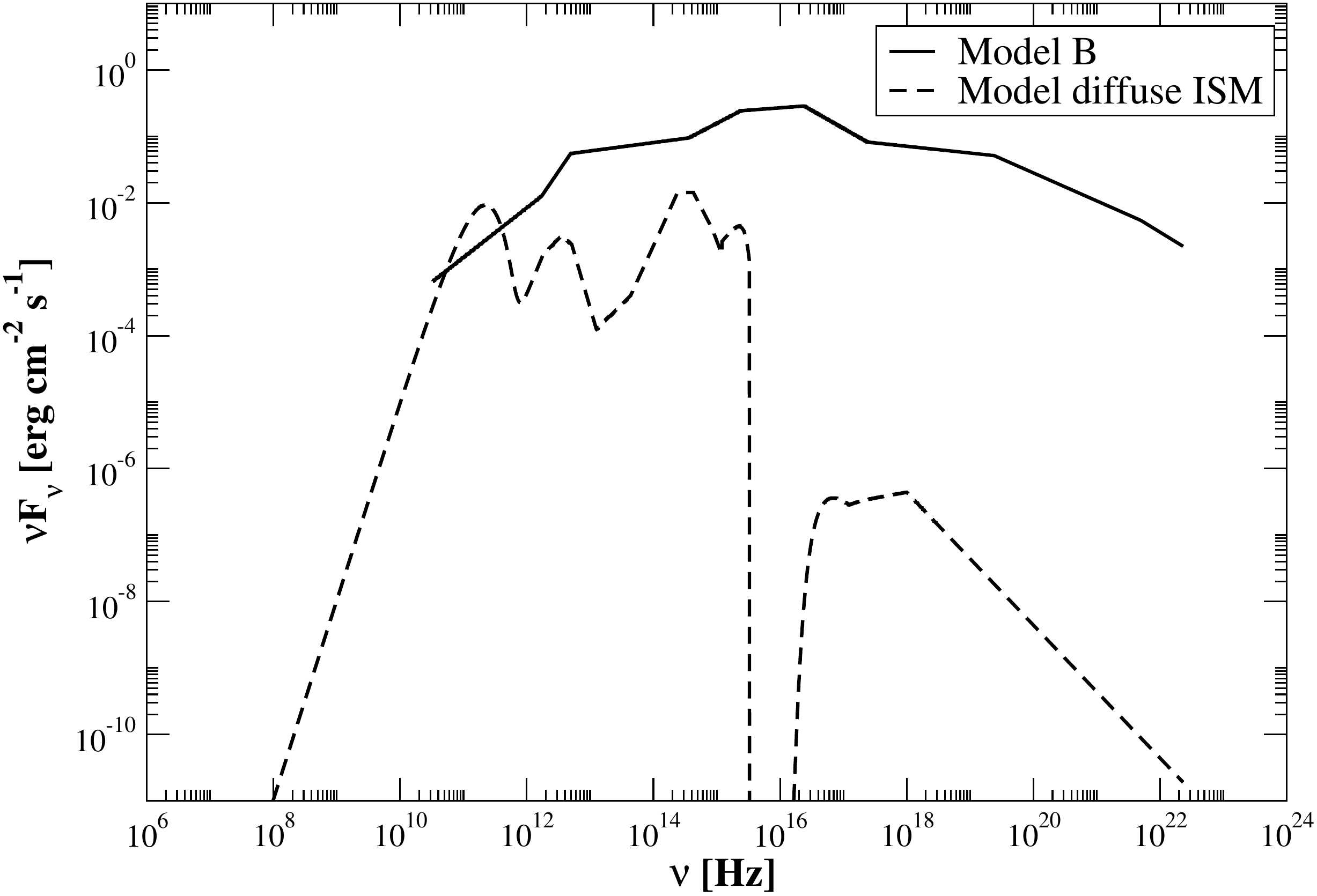}
\caption{Shape and intensity of the incident SED \citep{davi85} considered for Model B are shown with the solid line.
The incident SED considered for the diffuse ISM case is shown with the dashed line \citep{das20}.}
\label{fig:sed_richardson}
\end{center}
\end{figure}

We adopt the shape of the SED mentioned in \cite{hest08}. We consider the luminosity ($L$) $\sim 1.3 \times 10^{38}$ erg s$^{-1}$ as the synchrotron emission from the pulsar at the center of the Crab has a spin-down luminosity $\sim 10^5$ times that of the Sun \citep{hest08}. Because our object is located $2.5$ pc away from the central source,
the intensity of the external radiation field striking a unit surface area of the cloud
($\frac{L}{4 \pi r_{in}^2}$) is $\sim 0.174$ erg cm$^{-2}$ s$^{-1}$. The obtained shape and intensity
of the SED are shown in Figure \ref{fig:sed}a. The Galactic background radiation field proposed
by \cite{bert96} is also included to modify our SED. This radiation field is only defined over a
narrow wavelength range. The strength of this radiation field is $31$ Draine units \citep[i.e., $31 \times$ the ISRF in Draine's units $\approx 31 \times 2.7 \times 10^{-3}$ erg s$^{-1}$ cm$^{-2}$,][]{drai78}.
The resulting SED with the Galactic background radiation field included is
shown in Figure \ref{fig:sed}b. We digitally extract \citep[using the online tool of][]{roha20} the output X-ray spectrum (i.e., Figure 1) of \cite{prie17} and include an X-ray flux of 0.35 erg cm$^{-2}$ s$^{-1}$ from 0.1 to 100 $\mathrm{\AA}$ in our SED (Figure \ref{fig:sed}c).
The shape and intensity of the final SED used in the case of the Crab are shown in Figure \ref{fig:sed}c. All of the parameters discussed here are considered as the physical input parameters of Model A.

\cite{rich13} studied the nature of the H$_2$-emitting gas in knot 51 of the Crab. They mentioned that Davidson's SED \citep{davi85} is an excellent fit to reproduce observations. In Figure \ref{fig:sed_richardson}, the SED of \cite{davi85} is shown with a solid curve for the ionizing particle model (Model B) following \cite{rich13}. Additionally, for diffuse ISM case, a SED is considered shown in Figure \ref{fig:sed_richardson} with a dashed curve. Details about this SED and modeling results are discussed in Section \ref{diff_ISM}.

\begin{table}
\scriptsize
\centering
\caption{Adopted physical parameters for the Crab filament \citep{das20}. \label{table:model}}
\vskip 0.2 cm
\begin{tabular}{l c}
\hline
{\bf Physical parameters} & {\bf Adopted values} \\
\hline
\multicolumn{2}{c}{\bf  Model A \citep[adopted from][]{prie17}} \\
\hline
Inner radius ($r_{in}$) & $2.5$ pc = $7.715 \times 10^{18}$ cm\\
Shell thickness ($dr$) & $3.5 \times 10^{16}$ cm\\
Luminosity ($L$) & $1.3 \times 10^{38}$ erg s$^{-1}$\\
ISRF & 31 Draine units \\
SED & \cite{hest08} + X-ray from \\
& Figure 1 of \cite{prie17} \\
Type of grain & Amorphous carbon \\
Dust-to-gas mass ratio & $0.027$ \citep{owen15}\\
\hline
\multicolumn{2}{c}{\bf  Model B \citep[adopted from][]{rich13}} \\
\hline
Incident ionizing photon & $10^{10.06}$ cm$^{-2}$s$^{-1}$ \\
 flux on the slab ($\Phi$(H)) & \\
Thickness & $10^{16.5}$ cm \\
Additional heating & $\zeta_H/\zeta_0 = 10^{5.3}$ \\
$\rm{n_{H(min)}}$ & $10^{3}$ cm$^{-3}$ \\
$\rm{n_{H(core)}}$ & $10^{5.25}$ cm$^{-3}$ \\
SED & \cite{davi85} \\
Type of grain & Mix of graphite and silicate \\
Dust-to-gas mass ratio & $0.003$ \\
\hline
\end{tabular}
\end{table}

\begin{table}
\scriptsize
{\centering
\caption{Initial gas-phase elemental abundances with respect
to total hydrogen nuclei in all forms for the Crab filament \citep{das20}. \label{table:abun}}
\vskip 0.2 cm
\begin{tabular}{cccc}
\hline
\hline
{\bf  Element} & {\bf  Abundance} & {\bf  Element} & {\bf  Abundance} \\
\hline
\multicolumn{4}{c}{\bf  Model A \citep[adopted from][]{owen15}} \\
\hline
H & 1.00 & $^{36}$Ar & $1.00 \times 10^{-5}$ \\
He & 1.85 & $^{38}$Ar & $1.82 \times 10^{-6}$ \\
C & $1.02 \times 10^{-2}$ & $^{40}$Ar & $2.90 \times 10^{-9}$ \\
N & $2.50 \times 10^{-4}$ & $^{20}$Ne & $4.90 \times 10^{-3}$ \\
O & $6.20 \times 10^{-3}$ & $^{22}$Ne & $3.60 \times 10^{-4}$  \\
\hline
\multicolumn{4}{c}{\bf Model B \citep[adopted from][]{rich13}} \\
\hline
H & 1.00 & Si & $8.91 \times 10^{-6}$ \\
He & $2.95 \times 10^{-1}$ & S & $1.95 \times 10^{-5}$ \\
C & $3.98 \times 10^{-4}$ & Cl & $4.68 \times 10^{-8}$ \\
N & $5.62 \times 10^{-5}$ & $^{36}$Ar & $4.79 \times 10^{-6}$ \\
O & $5.25 \times 10^{-4}$ & $^{38}$Ar & $8.70 \times 10^{-7}$ \\
$^{20}$Ne & $1.82 \times 10^{-4}$ & $^{40}$Ar & $1.39 \times 10^{-9}$  \\
$^{22}$Ne & $1.34 \times 10^{-5}$ & Fe & $2.45 \times 10^{-5}$ \\
Mg & $2.00 \times 10^{-5}$ & & \\
\hline
\hline
\end{tabular} \\
}
\vskip 0.2 cm
{\bf Note:} For the initial isotopic ratio of argon and neon, we have used $^{36}$Ar/$^{38}$Ar/$^{40}$Ar = $84.5946/15.3808/0.0246$ and
$^{20}$Ne/$^{21}$Ne/$^{22}$Ne = $92.9431/0.2228/6.8341$, following \cite{wiel02}.
\end{table}

The physical parameters are summarized in Table \ref{table:model}, and
the gas-phase elemental abundances are listed in Table \ref{table:abun}.
Tables \ref{table:model} and \ref{table:abun} contain input parameters for the two models, Model A and Model B.
In Model A, we consider the
physical parameters from \cite{prie17} and initial elemental abundances from the clumpy model VI of \cite{owen15}.
In Model B, we consider the initial elemental abundances and physical input parameters for the ionizing particle model that \cite{rich13} considered. Some significant differences
between the physical parameters of Model A and Model B are that Model A is a constant-density model.
In contrast, we consider a dense core ($\rm{n_{H(core)}} \sim 10^{5.25}$ cm$^{-3}$) by introducing a varying density profile in Model B, and the grain type in both models is different.
For the initial isotopic ratio of argon and neon, we use $^{36}$Ar/$^{38}$Ar/$^{40}$Ar = $84.5946/15.3808/0.0246$ and
$^{20}$Ne/$^{21}$Ne/$^{22}$Ne = $92.9431/0.2228/6.8341$, following \cite{wiel02}.

\subsection{Radiative Transfer Model}

\begin{figure}
\begin{center}
\includegraphics[width=0.65\textwidth,angle=270]{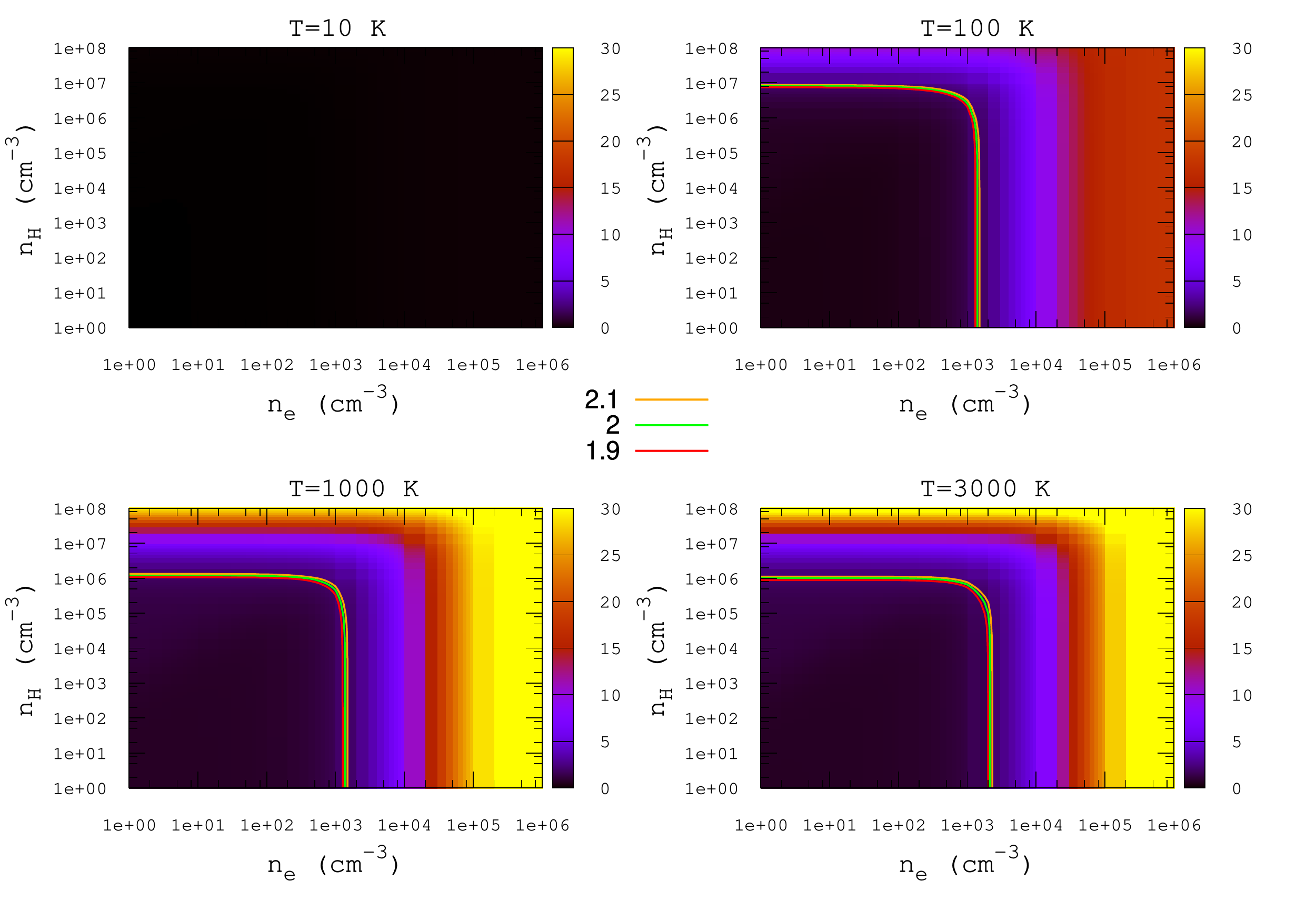}
\includegraphics[width=0.65\textwidth,angle=270]{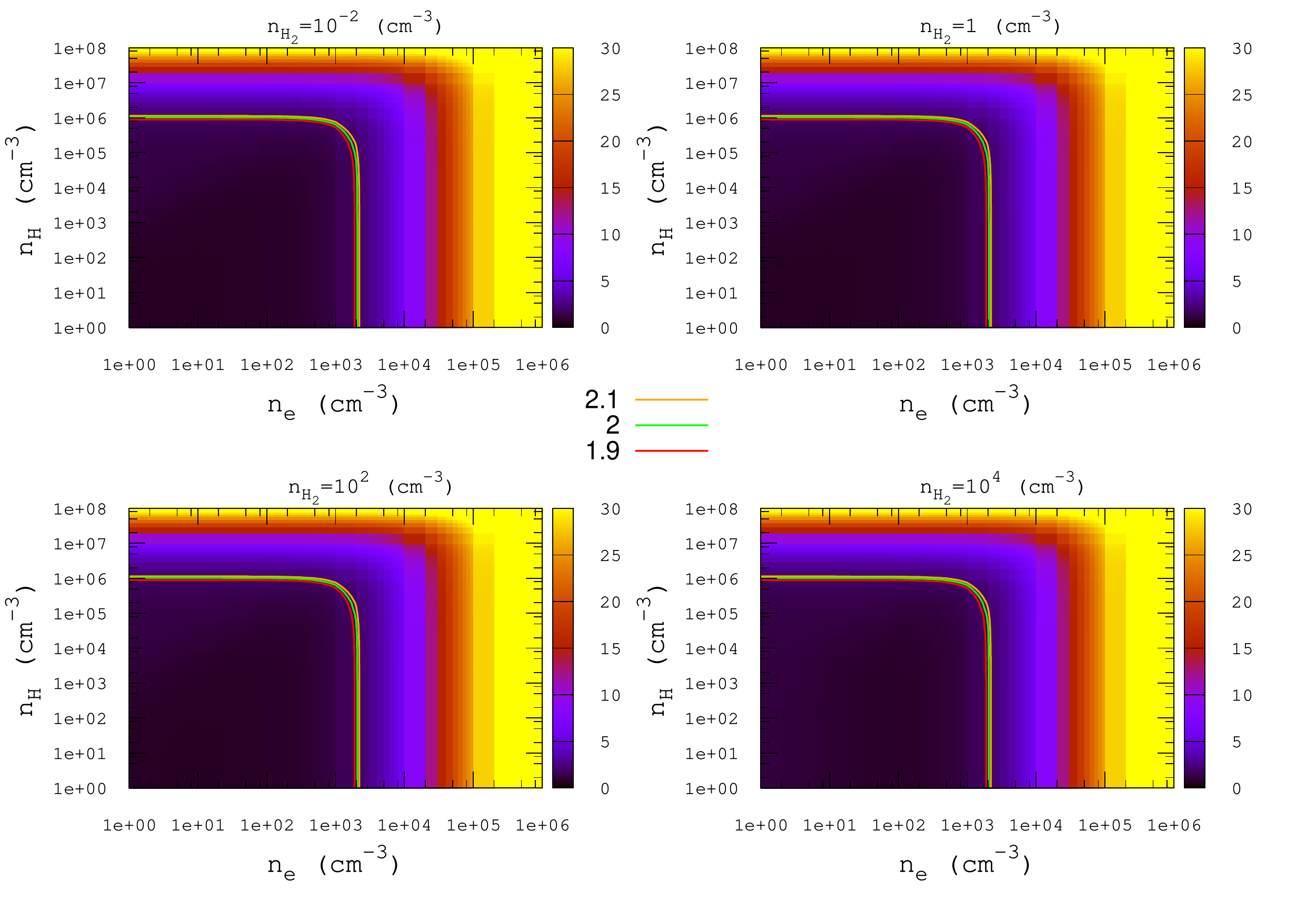}
\caption{Surface brightness (SB) ratio between the $2-1$ and $1-0$ transitions of $\rm{^{36}ArH^+}$ by considering a column density of
$1.7\times10^{12}$ cm$^{-2}$ \citep{das20}. The upper four panels
show the cases with fixed temperatures ($T=10,\ 100,\ 1000$, and $3000$ K respectively) whereas
the lower four panels show the cases with fixed H$_2$ density ($\rm{n_{H_2}=10^{-2}}$, $1$, $10^2$, and $10^4$ cm$^{-3}$ respectively).
The contours are highlighted near the observed SB ratio (of $\sim 2$).}
\label{fig:ratio}
\end{center}
\end{figure}

The $J = 1$ and $J = 2$ levels of $^{36}$ArH$^+$ are at $29.6$ K and $88.9$ K, respectively.
The measured electron temperature \citep[$7500-15000$ K;][]{davi85} for the ionized gas and measured excitation temperature of H$_2$
\cite[$2000-3000$ K;][]{loh11} in the Crab region is much higher than that at these energy levels. If the region where
ArH$^+$ transitions were observed has the density of the colliding partner exceeding the critical density
and temperature $>100$ K, the level populations would be in Boltzmann equilibrium and yield a
$2-1/1-0$ ratio of $\sim 30$. Because the observed ratio is $\sim2$, it is expected that the density of the colliding partner is
much lower than their critical densities.
\cite{barl13} also attributed this difference to the density of the collisional partners being below the
critical density of the $\rm{ArH^+}$ rotational levels. They used a radiative transfer model to determine the densities
of $\rm{H_2}$ and $e^-$ from the observational ratio.
They obtained a critical density of
electrons of $\sim 10^4$ cm$^{-3}$ and H$_2$ $\sim 10^8$ cm$^{-3}$.

ArH$^+$ favors regions where H$_2$/H is small. If there are any significant H$_2$ densities, then the reactive collision
with ArH$^+$ may be high enough to affect the excitation. By including the reactive collision rate with H$_2$, it might be
possible to compare models and observed fluxes to limit the H$_2$/H ratio in the emitting region.
However, with the public version of RADEX, it is not possible to study this feature. Moreover,
around the Crab filamentary region, where ArH$^+$ was identified, the abundance of H atoms and electrons is $>$ $10^4-10^5$ times higher than
that of H$_2$ \citep[see Figure 16 of][and Figures \ref{fig:abun_best} and \ref{fig:abun1-rich} in the latter part of this Chapter]{prie17}. It suggests that a nonreactive collision might be the primary
source of excitation of ArH$^+$.

\cite{barl13} used the MADEX code \citep{cern12}, where they used H$_2$ and electron as the collision partner.
Due to the unavailability of the collisional rate
parameters, they used the collisional deexcitation rate of $\rm{SiH^+ + He}$ and $\rm{CH^+} + e^-$ in place of the interaction
of $\rm{H_2}$ and $\rm{e^-}$ with ArH$^+$, respectively.
Because the electron-impact
rate coefficient for the dipolar transitions is roughly $10^4-10^5$ times larger than the neutrals (H and H$_2$), \cite{hami16}
used electrons as the only colliding partner. Because reactive collisions are not implemented in the public version of RADEX,
we only take the nonreactive collisions into account. We assume that due to the low abundance of H$_2$ in the ArH$^+$
formation region and high electron-impact rate, reactive collision with H$_2$ would have minimal effect in this condition.
Here, we consider three colliders, H, H$_2$, and electrons, in RADEX. Collisional rates with H and H$_2$ are scaled \citep{scho05} from the available collisional rates of $\rm{ ArH^+-He}$ obtained from \cite{garc19}, and collisional rates with electrons are taken from \cite{hami16}.

Here, we use the RADEX code \citep{vand07} for non-LTE computation to explain the observational results.
We prepare this collisional data file using the spectroscopic
parameters available in the JPL\footnote{\url{https://spec.jpl.nasa.gov}} \citep{pick98} or CDMS\footnote{\url{https://cdms.astro.uni-koeln.de}} \citep{mull01,mull05,endr16}
database and include the electron-impact excitation rates from
\cite{hami16}. Collisional data files for the other hydride/hydroxyl cations are unavailable in the \textsc{Cloudy} code as well.
We use our approximated data files to calculate the SB and emissivity discussed in the latter part of this Chapter.
We consider Figure \ref{fig:sed}c as the input of the background radiation field in the radiative transfer calculations reported here.
We prepare the self-made background radiation field in the format prescribed in \url{https://personal.sron.nl/~vdtak/radex/index.shtml}. This file contains three columns. The first column is the wavenumber (in cm$^{-1}$), the second is the intensity (in Jy nsr$^{-1}$), and the third is the dilution factor which varies between 0 and 1. Here, for the estimation, we use an average dilution factor of $0.5$. We do not find a significant difference while considering a different dilution factor in our calculations.

We draw a parameter space with a wide range of H densities ($1-10^{8}$ cm$^{-3}$),
H$_2$ densities ($10^{-2}-10^4$ cm$^{-3}$), electron number densities ($1-10^6$ cm$^{-3}$), and excitation temperatures ($10-3000$ K).
Figure \ref{fig:ratio} shows the SB ratio between $2\rightarrow1$ and $1\rightarrow0$ transitions of $^{36}$ArH$^+$.
For this computation, we consider the column density of $^{36}$ArH$^+$ $\sim1.7\times10^{12}$ cm$^{-2}$ as obtained from
\cite{hami16}, and a line width (FWHM) of $5$ km s$^{-1}$. For the upper four panels,
we consider the H$_2$ density of 1 cm$^{-3}$ and temperature fixed at
$10$ K, $100$ K, $1000$ K, and $3000$ K, respectively.
Some contours near the observed SB ratio ($\sim 2$) are highlighted in all the panels.
The top-left panel of Figure \ref{fig:ratio} shows that at $10$ K, the SB ratio between these two transitions is $\sim 0$.
This is because the
excitation temperature is below the upstate energy of these two transitions.
Thus, for the higher temperature, energy levels are gradually
populated, and the ratio increases.
The upper four panels of Figure \ref{fig:ratio} depict how the observed ratio
is obtained with an
electron density of $1000-3000$ cm$^{-3}$ when the number density of H atoms is $< 10^6-10^7$ cm$^{-3}$, and the temperature is beyond
the upstate energy of $2 \rightarrow 1$ and $1 \rightarrow 0$.
For the case with a temperature of $100$ K,
when H density is below $\sim 10^7$ cm$^{-3}$, the observed ratio is obtained with an electron density
of $\sim 1000$ cm$^{-3}$.
For $\rm{n_H \sim 10^7}$ cm$^{-3}$, the observed ratio is obtained with $\rm{n_e=1-1000}$ cm$^{-3}$.
As we gradually increase the temperature,
the observed ratio is obtained at a lower H density (for example, at $1000$ K, it is $\sim$ a few times $ \times 10^6$ cm$^{-3}$)
and a little higher electron density range ($1-2000$ cm$^{-3}$).
If the temperature is further increased from here (i.e., at $3000$ K),
a minimal decrease of $\rm{n_H}$, and a slight increase in $\rm{n_e}$ range are required to reproduce the observed ratio.
For the higher temperature ($\sim 3000$ K) and higher
electron density ($> 10^5$), the highest value of the ratio $\sim 30$ is achieved.
This value is also obtained when the H density is around $10^8$ cm$^{-3}$.
Thus, the critical density of electrons and hydrogen atoms are $10^5$ cm$^{-3}$ and 10$^8$ cm$^{-3}$, respectively.
In the lower four panels of Figure \ref{fig:ratio}, we keep the temperature fixed at $2700$ K and the H$_2$ density set at $10^{-2}$ cm$^{-3}$, 1 cm$^{-3}$, $10^2$ cm$^{-3}$, and
$10^4$ cm$^{-3}$, respectively.
All four panels give a similar result, implying that the excitation is independent of
the H$_2$ collision.
The upper four panels of Figure \ref{fig:ratio} remain unchanged when the H$_2$ is omitted as a collider.
The right four panels show that it is independent of the collision of H$_2$ when the H$_2$ density
is $< 10^4$ cm$^{-3}$. However, the reactive collisions with H$_2$ may show differences that
are not considered here due to the limitations of the public version of the RADEX code.
In brief, we find that
it is only the nonreactive collision with electrons that can successfully
explain the excitation of
ArH$^+$ when the temperature is beyond the upstate energy of these two levels discussed here.
\cite{loh12} estimated the electron number density and total hydrogen number density ($n(H^+)+n(H)+2n(H_2)$) in the
filaments and knots to be around $1400-2500$ cm$^{-3}$ and $14000-25000$ cm$^{-3}$ respectively.
\cite{barl13} estimated the electron number density of $\sim$ a few times $100$ cm$^{-3}$.
Our results shown in the top four panels of Figure \ref{fig:ratio} require a $n_e$ of $\sim 2000 - 3000$ cm$^{-3}$ to reproduce the observed ratio around the measured excitation temperature of H$_2$.
Only the nonreactive collision with electrons can explain the
ArH$^+$ excitation in the Crab.

\section{Chemical pathways} \label{sec:chem_path}

Following the reaction network of ArH$^+$ presented in \cite{prie17}, here,
we prepare similar pathways for NeH$^+$ and HeH$^+$. Additionally,
we prepare the pathways for the hydroxyl cations
of the noble gas species (ArOH$^+$, NeOH$^+$, and HeOH$^+$) under similar environments. In Table \ref{table:reaction}, the reaction network adopted here to study the
chemical evolution of the related hydride and hydroxyl cations is listed along with the corresponding rate coefficients. The enlisted rate coefficients are
either estimated or taken from the literature as mentioned in the footnote. In the following subsections, we present an extensive discussion to prepare or adapt the rate coefficients of various kinds of reactions considered. We use the reaction rates of UMIST as the default for the other reactions. For H$_2$ formation on grains, we use the modified ``Jura rate'' \citep{ster99} for Model A. The default ``Jura rate'' of H$_2$ formation
is $3 \times 10^{-17}$ cm$^3$ s$^{-1}$ \citep{jura75}. In Model B, the chemical pathways are the same as discussed above, except the
H$_2$ formation rate is through grain catalysis. This rate is taken from \cite{caza02} as it was considered by \cite{rich13}.

\begin{table}
\scriptsize
\centering
\caption{Reaction pathways for the formation and destruction of some noble gas ions \citep{das20}. \label{table:reaction}}
\vskip 0.2cm
\begin{tabular}{cccc}
\hline
{\bf Reaction} & {\bf Reactions} & {\bf Rate coefficient} & {\bf References} \\
{\bf Number (Type)} &  &  & {\bf and comments} \\ 
\hline
\multicolumn{4}{c}{\bf Ar chemistry} \\
\hline
1 (CR)&$\rm{Ar + CR \rightarrow Ar^{+} + e^{-}}$ & $\rm{10\zeta_{H,cr}} \ s^{-1}$&a, d\\
2 (CRPHOT)&$\rm{Ar + CRPHOT \rightarrow Ar^{+} + e^{-}}$ &$\rm{0.8\frac{\zeta_{H_2,cr}}{1-\omega}} \ s^{-1}$&a, d\\
3 (IN)&$\rm{Ar + H_2^{+} \rightarrow ArH^{+} + H}$ &$\rm{10^{-9}}\ cm^3\ s^{-1} $&a\\
4 (IN)&$\rm{Ar + H_3^{+} \rightarrow ArH^{+} + H_2}$&$\rm{8\times10^{-10}exp(\frac{-6019\ K}{T})} \ cm^3\ s^{-1}$ & This work\\
5 (IN)&$\rm{Ar^{+} + H_2 \rightarrow ArH^{+} + H}$&$\rm{8.4\times10^{-10}(\frac{T}{300\ K})^{0.16}} \ cm^3\ s^{-1}$&a\\
6 (IN)&$\rm{ArH^{+} + H_2 \rightarrow Ar + H_3^{+}}$&$\rm{8\times10^{-10}}\ cm^3\ s^{-1}$&a\\
7 (IN)&$\rm{ArH^{+} + CO \rightarrow Ar + HCO^{+}}$&$\rm{1.25\times10^{-9}}\ cm^3\ s^{-1}$&a\\
8 (IN)&$\rm{ArH^{+} + O \rightarrow Ar + OH^{+}}$&$\rm{8\times10^{-10}}\ cm^3\ s^{-1}$&a\\
9 (IN)&$\rm{ArH^{+} + C \rightarrow Ar + CH^{+}}$&$\rm{8\times10^{-10}}\ cm^3\ s^{-1}$&a\\
10 (IN)&$\rm{Ar^{++} + H \rightarrow Ar^{+} + H^{+}}$&$\rm{10^{-15}}\ cm^3\ s^{-1}$&b\\
11 (RA)&$\rm{Ar + OH^+ \rightarrow ArOH^{+} +\ h\nu}$& $\rm{1.9\times10^{-17}}\ cm^3\ s^{-1}$& c, m \\
12 (RA) &$\rm{Ar^{+} + OH \rightarrow ArOH^{+} +\ h\nu}$ &  $\rm{1.5\times10^{-17}}\ cm^3\ s^{-1}$ & c, m \\
13 (RA) &$\rm{ArH^+ + O \rightarrow ArOH^{+} +\ h\nu}$& $\rm{3.0\times10^{-17}}\ cm^3\ s^{-1}$& c, m \\
14 (IN)&$\rm{Ar + N_2^+ \rightarrow Ar^+ + N_2}$&$\rm{3.65\times10^{-10}}\ cm^3\ s^{-1}$&d\\
15 (IN)&$\rm{Ar^+ + H_2 \rightarrow Ar + H_2^+}$&$\rm{2.00\times10^{-12}}\ cm^3\ s^{-1}$&d\\
16 (IN)&$\rm{Ar^+ + O_2 \rightarrow Ar + O_2^+}$&$\rm{3.50\times10^{-11}}\ cm^3\ s^{-1}$&d\\
17 (IN)&$\rm{Ar^+ + CH_4 \rightarrow CH_2^+ + Ar + H_2}$&$\rm{1.40\times10^{-10}}\ cm^3\ s^{-1}$&d\\
18 (IN)&$\rm{Ar^+ + CH_4 \rightarrow CH_3^+ + Ar + H}$&$\rm{7.90\times10^{-10}}\ cm^3\ s^{-1}$&d\\
19 (IN)&$\rm{Ar^+ + HCl \rightarrow Ar + HCl^+}$&$\rm{2.90\times10^{-10}}\ cm^3\ s^{-1}$&d\\
20 (IN)&$\rm{Ar^+ + HCl \rightarrow ArH^+ + Cl}$&$\rm{6.00\times10^{-11}}\ cm^3\ s^{-1}$&d\\
21 (IN)&$\rm{Ar^+ + CO \rightarrow Ar + CO^+}$&$\rm{2.80\times10^{-11}}\ cm^3\ s^{-1}$&d\\
22 (IN)&$\rm{Ar^+ + NH_3 \rightarrow Ar + NH_3^+}$&$\rm{1.60\times10^{-9}}\ cm^3\ s^{-1}$&d\\
23 (IN)&$\rm{Ar^+ + N_2 \rightarrow Ar + N_2^+}$&$\rm{1.20\times10^{-11}}\ cm^3\ s^{-1}$&d\\
24 (IN)&$\rm{Ar^+ + H_2O \rightarrow Ar + H_2O^+}$&$\rm{1.30\times10^{-9}}\ cm^3\ s^{-1}$&d\\
25 (XR)&$\rm{Ar + XR \rightarrow Ar^{++} + e^{-} + e^{-}}$&$\rm{\zeta_{XR}\ s^{-1}}$&d, e\\
26 (XR)&$\rm{Ar^{+} + XR \rightarrow Ar^{++} + e^{-}}$&$\rm{\zeta_{XR}\ s^{-1}}$&d, e\\
27 (XRSEC)&$\rm{Ar + XRSEC \rightarrow Ar^{+} + e^{-}}$&$\rm{5.53}\zeta_{H,XRPHOT}\ s^{-1}$&d, l\\
28 (XRPHOT)&$\rm{Ar + XRPHOT \rightarrow Ar^{+} + e^{-}}$&$\rm{0.8\frac{\zeta_{H2,XRPHOT}}{1-\omega}}\ s^{-1}$&d, l\\
29 (ER)&$\rm{Ar^+ + e^{-} \rightarrow Ar + h\nu}$&   & d \\
30 (ER)&$\rm{Ar^{++} + e^{-} \rightarrow Ar^{+} + h\nu}$&   & d \\
31 (DR)&$\rm{ArH^{+} + e^{-} \rightarrow Ar + H}$& $\rm{10^{-11}}\ cm^3\ s^{-1}$&a, k\\
32 (DR)&$\rm{ArOH^+ + e^-\rightarrow Ar +OH}$&$\rm{10^{-11}}\ cm^3\ s^{-1}$&This work\\
33 (PH)&$\rm{ArH^{+} + h\nu \rightarrow Ar^+ + H}$&$\rm{4.20\times10^{-12}exp(-3.27A_V)\ s^{-1}}$&h \\
34 (PH)&$\rm{ArOH^{+} + h\nu \rightarrow Ar + OH^{+}}$&$\rm{4.20\times10^{-12}exp(-3.27A_V)\ s^{-1}}$& This work\\
\hline
\end{tabular}
\end{table}

\begin{table}
\scriptsize
\centering
\begin{tabular}{cccc}
\hline
{\bf  Reaction} & {\bf  Reactions} & {\bf  Rate coefficient} & {\bf  References} \\
{\bf  Number (Type)} &  &  & {\bf  and comments} \\
\hline
\multicolumn{4}{c}{\bf  Ne chemistry}\\
\hline
1 (CR)&$\rm{Ne + CR \rightarrow Ne^{+} + e^{-}}$ & $\rm{10\zeta_{H,cr}}\ s^{-1}$&This work, d\\
2 (CRPHOT)&$\rm{Ne + CRPHOT \rightarrow Ne^{+} + e^{-}}$ &$\rm{0.8\frac{\zeta_{H_2,cr}}{1-\omega}}\ s^{-1}$&This work, d\\
3 (IN)&$\rm{Ne + H_2^{+} \rightarrow NeH^{+} + H}$ & $\rm{2.58\times10^{-10}exp(\frac{-6717\ K}{T})}\ cm^3\ s^{-1}$ & This work \\
4 (IN)&$\rm{Ne + H_3^{+} \rightarrow NeH^{+} + H_2}$&$\rm{8\times10^{-10}exp(\frac{-27456\ K}{T})}\ cm^3\ s^{-1}$&This work\\
5a (IN) &$\rm{Ne^{+} + H_2 \rightarrow NeH^{+} + H}$& $\rm{3.2\times10^{-9}(\frac{T}{300\ K})^{0.16}} \ cm^3\ s^{-1}$& This work \\
5b (IN) & $\rm{Ne^{+} + H_2 \rightarrow Ne + H + H^+}$ & $\rm{1.98\times10^{-14}exp(-35\ K/T)}$ cm$^3$\ s$^{-1}$ & This work \\
5c (IN) & $\rm{Ne^{+} + H_2 \rightarrow Ne + H_2^+}$ & $\rm{4.84\times10^{-15}}$ cm$^3$ s$^{-1}$ & This work \\
6 (IN)&$\rm{NeH^{+} + H_2 \rightarrow Ne + H_3^{+}}$&$\rm{3.65\times10^{-9}}\ cm^3\ s^{-1}$& This work\\
7 (IN)&$\rm{NeH^{+} + CO \rightarrow Ne + HCO^{+}}$&$\rm{2.26\times10^{-9}}\ cm^3\ s^{-1}$& This work\\
8 (IN)&$\rm{NeH^{+} + O \rightarrow Ne + OH^{+}}$&$\rm{2.54\times10^{-9}}\ cm^3\ s^{-1}$& This work\\
9 (IN)&$\rm{NeH^{+} + C \rightarrow Ne + CH^{+}}$&$\rm{1.15\times10^{-9}}\ cm^3\ s^{-1}$& This work\\
10 (IN)&$\rm{Ne^{++} + H \rightarrow Ne^{+} + H^{+}}$&$\rm{1.94\times10^{-15}}\ cm^3\ s^{-1}$& This work\\
11 (RA) & $\rm{Ne + OH^+ \rightarrow NeOH^{+} +\ h\nu}$ & $\rm{1.4\times10^{-18}}\ cm^3\ s^{-1}$ & c, m \\
12 (RA) & $\rm{Ne^{+} + OH \rightarrow NeOH^{+} +\ h\nu}$ & $\rm{7.5\times10^{-17}}\ cm^3\ s^{-1}$ & c, m \\
13 (RA) &$\rm{NeH^+ + O \rightarrow NeOH^{+} +\ h\nu}$ & $\rm{2.3\times10^{-17}}\ cm^3\ s^{-1}$ & c, m \\
14 (IN)&$\rm{HeH^+ + Ne \rightarrow NeH^+ +He }$&$\rm{1.25 \times 10^{-9}}\ cm^3\ s^{-1}$ & d \\
15 (IN)&$\rm{NeH^+ + He\rightarrow HeH^{+} + Ne }$&$\rm{3.8 \times 10^{-14}}\ cm^3\ s^{-1}$ & d \\
16 (IN)&$\rm{Ne^{+} + CH_4 \rightarrow CH^{+} + Ne + H_2+ H}$&$\rm{8.4\times10^{-13}}\ cm^3\ s^{-1}$&d\\
17 (IN)&$\rm{Ne^{+} + CH_4 \rightarrow {CH_2}{^+} + Ne + H_2}$&$\rm{4.2\times10^{-12}}\ cm^3\ s^{-1}$&d\\
18 (IN)&$\rm{Ne^{+} + CH_4 \rightarrow {CH_3}{^+} + Ne + H}$&$\rm{4.7\times10^{-12}}\ cm^3\ s^{-1}$&d\\
19 (IN)&$\rm{Ne^{+} + CH_4 \rightarrow {CH_4}{^+} + Ne}$&$\rm{1.1\times10^{-11}}\ cm^3\ s^{-1}$&d\\
20 (IN)&$\rm{Ne^{+} + NH_3 \rightarrow {NH}{^+} + Ne+H_2}$&$\rm{4.5\times10^{-12}}\ cm^3\ s^{-1}$&d\\
21 (IN)&$\rm{Ne^{+} + NH_3 \rightarrow {NH_2}{^+} + Ne+H}$&$\rm{1.9\times10^{-10}}\ cm^3\ s^{-1}$&d\\
22 (IN)&$\rm{Ne^{+} + NH_3 \rightarrow {NH_3}{^+} + Ne}$&$\rm{2.7\times10^{-11}}\ cm^3\ s^{-1}$&d\\
23 (IN)&$\rm{Ne^{+} + N_2 \rightarrow {N_2}{^+} + Ne}$&$\rm{1.1\times10^{-13}}\ cm^3\ s^{-1}$&d\\
24 (IN)&$\rm{Ne^{+} + H_2O \rightarrow {H_2O}{^+} + Ne}$&$\rm{8.0\times10^{-10}}\ cm^3\ s^{-1}$&d\\
25 (IN)&$\rm{Ne^{+} + O_2 \rightarrow {O}{^+} + Ne + O}$&$\rm{6.0\times10^{-11}}\ cm^3\ s^{-1}$&d\\
26 (XR)&$\rm{Ne + XR \rightarrow Ne^{++} + e^{-} + e^{-}}$&$\rm{\zeta_{XR}\ s^{-1}}$&d, e\\
27 (XR)&$\rm{Ne^{+} + XR \rightarrow Ne^{++} + e^{-}}$&$\rm{\zeta_{XR}\ s^{-1}}$&d, e\\
28 (XRSEC)&$\rm{Ne + XRSEC \rightarrow Ne^{+} + e^{-}}$&$\rm{1.84}\zeta_{H,XRPHOT}\ s^{-1}$&d, l\\
29 (XRPHOT)&$\rm{Ne + XRPHOT \rightarrow Ne^{+} + e^{-}}$&$\rm{0.8\frac{\zeta_{H2,XRPHOT}}{1-\omega}}\ s^{-1}$&d, l\\
30 (ER)&$\rm{Ne^+ + e^{-} \rightarrow Ne + h\nu}$&   & d\\
31 (ER)&$\rm{Ne^{++} + e^{-} \rightarrow Ne^{+} + h\nu}$&   & d\\
32 (DR)&$\rm{NeH^{+} + e^{-} \rightarrow Ne + H}$& $\rm{10^{-11}\ cm^3\ s^{-1}}$& This work\\
33 (DR)&$\rm{NeOH^+ + e^-\rightarrow Ne +OH }$& $\rm{10^{-11}\ cm^3\ s^{-1}}$& This work\\
34 (PH)&$\rm{NeH^{+} + h\nu \rightarrow Ne^{+} + H}$&$\rm{4.20\times10^{-12}exp(-3.27A_V)\ s^{-1}}$& This work \\
35 (PH)&$\rm{NeOH^{+} + h\nu \rightarrow Ne + OH^{+}}$&$\rm{4.20\times10^{-12}exp(-3.27A_V)\ s^{-1}}$& This work \\
\hline
\end{tabular}
\end{table}

\begin{table}
\scriptsize
{\centering
\begin{tabular}{cccc}
\hline
{\bf  Reaction} & {\bf  Reactions} & {\bf  Rate coefficient} & {\bf  References} \\
{\bf  Number (Type)} &  &  & {\bf  and comments} \\
\hline
\multicolumn{4}{c}{\bf  He chemistry}\\
\hline
1 (CR)&$\rm{He + CR \rightarrow He^{+} + e^{-}}$ & $\rm{10\zeta_{H,cr}}\ s^{-1}$&This work, d\\
2 (CRPHOT)&$\rm{He + CRPHOT \rightarrow He^{+} + e^{-}}$ &$\rm{0.8\frac{\zeta_{H_2,cr}}{1-\omega}} \ s^{-1}$&This work, d\\
3 (IN)&$\rm{He + H_2^{+} \rightarrow HeH^{+} + H}$ & $\rm{3\times10^{-10}exp(\frac{-6717\ K}{T})}\ cm^3\ s^{-1}$ & n \\
4 (IN)&$\rm{He + H_3^{+} \rightarrow HeH^{+} + H_2}$&$\rm{8\times10^{-10}exp(\frac{-29110\ K}{T})}\ cm^3\ s^{-1}$&This work\\
5a (IN) & $\rm{He^{+} + H_2 \rightarrow HeH^{+} + H}$&   & Not considered\\
5b (IN) & $\rm{He^{+} + H_2 \rightarrow He + H + H^+}$ & $\rm{3.70\times10^{-14}exp(-35\ K/T)}$ cm$^3$ s$^{-1}$ & This work, UMIST \\
5c (IN) & $\rm{He^{+} + H_2 \rightarrow He + H_2^+}$ & $\rm{7.20\times10^{-15}}$ cm$^3$ s$^{-1}$ & This work, UMIST \\
6 (IN)&$\rm{HeH^{+} + H_2 \rightarrow He + H_3^{+}}$&$\rm{1.26\times10^{-9}}\ cm^3\ s^{-1}$&j\\
7 (IN)&$\rm{HeH^{+} + CO \rightarrow He + HCO^{+}}$&$\rm{2.33\times10^{-9}}\ cm^3\ s^{-1}$&This work\\
8 (IN)&$\rm{HeH^{+} + O \rightarrow He + OH^{+}}$&$\rm{2.68\times10^{-9}}\ cm^3\ s^{-1}$&This work\\
9 (IN)&$\rm{HeH^{+} + C \rightarrow He + CH^{+}}$&$\rm{1.18\times10^{-9}}\ cm^3\ s^{-1}$&This work\\
10 (IN)&$\rm{He^{++} + H \rightarrow He^{+} + H^{+}}$&$\rm{2.45\times10^{-15}}\ cm^3\ s^{-1}$&This work\\
11 (RA) &$\rm{He + OH^+ \rightarrow HeOH^{+} +\ h\nu}$ & $\rm{2.2\times10^{-18}}\ cm^3\ s^{-1}$ & c, m \\
12 (RA) & $\rm{He^{+} + OH \rightarrow HeOH^{+} +\ h\nu}$ & $\rm{1.7\times10^{-16}}\ cm^3\ s^{-1}$ & c, m \\
13 (RA) &$\rm{HeH^+ + O \rightarrow HeOH^{+} +\ h\nu}$ & $\rm{2.8\times10^{-17}}\ cm^3\ s^{-1}$ & c, m \\
14 (IN) &$\rm{HeH^{+} + H \rightarrow He + H_2^{+}}$&$\rm{1.7\times10^{-9}}\ cm^3\ s^{-1}$ & n \\
15 (RA) &$\rm{He^+ + H \rightarrow HeH^{+} + h\nu}$ &$\rm{1.44\times10^{-16}}\ cm^3\ s^{-1}$& i, n \\
16 (RA) &$\rm{He + H^+ \rightarrow HeH^{+} + h\nu}$ & $\rm{5.6\times10^{-21}(\frac{T}{10^4K})^{-1.25}}\ cm^3\ s^{-1}$ & d,  n \\
17 (XR)&$\rm{He + XR \rightarrow He^{++} + e^{-} + e^{-}}$&$\rm{\zeta_{XR}\ s^{-1}}$&d, e\\
18 (XR)&$\rm{He^{+} + XR \rightarrow He^{++} + e^{-}}$&$\rm{\zeta_{XR}\ s^{-1}}$&d, e\\
19 (XRSEC)&$\rm{He + XRSEC \rightarrow He^{+} + e^{-}}$&$\rm{0.84\zeta_{H,XRPHOT}\ s^{-1}}$&d, l\\
20 (XRPHOT)&$\rm{He + XRPHOT \rightarrow He^{+} + e^{-}}$&$\rm{0.8\frac{\zeta_{H2,XRPHOT}}{1-\omega}\ s^{-1}}$&d, l\\
21 (ER)&$\rm{He^+ + e^{-} \rightarrow He + h\nu}$&   & d\\
22 (ER)&$\rm{He^{++} + e^{-} \rightarrow He^{+} + h\nu}$&   &d \\
23 (DR)&$\rm{HeH^{+} + e^{-} \rightarrow He + H}$& $\rm{4.3\times10^{-10}(\frac{T}{10^4\ K})^{-0.5}}\ cm^3\ s^{-1}$ & n \\
24 (DR)&$\rm{HeOH^+ + e^-\rightarrow He +OH}$& $\rm{4.3\times10^{-10}(\frac{T}{10^4\ K})^{-0.5}}\ cm^3\ s^{-1}$ &This work\\
25 (PH)&$\rm{HeH^{+} + h\nu \rightarrow He^{+} + H}$&   & d, n \\
26 (PH)&$\rm{HeOH^{+} + h\nu \rightarrow He + OH^{+}}$&$\rm{4.20\times10^{-12}exp(-3.27A_V)\ s^{-1}}$& This work \\
27& $\rm{He^+ + H^- \rightarrow HeH^+ + e^-}$ & $\rm{3.2\times10^{-11}(\frac{T}{10^4\ K})^{-0.34}}\ cm^3\ s^{-1}$ & n \\
\hline
\multicolumn{4}{c}{\bf  Additional modified chemistry}\\
\hline
1 (RA) & $\rm{H^+ + H \rightarrow H_2^+ + h\nu}$ & $\rm{2.3\times10^{-16}(\frac{T}{10^4\ K})^{1.5}}\ cm^3\ s^{-1}$ & d, n \\
2 (DR) & $\rm{H_2^+ + e^- \rightarrow H + H}$ & $\rm{3\times10^{-9}(\frac{T}{10^4\ K})^{-0.4}}\ cm^3\ s^{-1}$ & d, n \\
3 (IN) & $\rm{H_2^+ + H \rightarrow H_2 + H^+}$ & $\rm{6.4\times10^{-10}}\ cm^3\ s^{-1}$ & d, n \\
\hline
\end{tabular}} \\
\vskip 0.2 cm
{\bf Note:} \\
CR refers to cosmic-rays, CRPHOT to secondary photons produced by cosmic-rays, XR to direct X-rays, XRSEC to secondary electrons which are
produced by X-rays, XRPHOT to secondary photons from X-rays, IN to ion-neutral reactions, RA to radiative association reactions, ER to electronic recombination reactions for atomic ions, DR to dissociative recombination reactions for molecular ions, PH to photodissociation reactions, h$\nu$ to a photon, $\zeta$ to cosmic-ray or X-ray ionization rates, and $\omega$ is the dust albedo. \\
$^a$ \cite{schi14}.\\
$^b$ \cite{king96}.\\
$^c$ This lower limit of the rate is calculated following \cite{bate83} described in Section \ref{rad_ass}.\\
$^d$ Reaction pathways are already included or automatically calculated in \textsc{Cloudy} by default.\\
$^e$ \cite{meij05}.\\
$^h$ \cite{roue14}.\\
$^i$ \cite{gust19}.\\
$^j$ \cite{orie77}.\\
$^k$ \cite{prie17}.\\
$^l$ See Appendix \ref{chap:xray_ionization} for the calculation details. Here, we are not considering this rate because we are using default values in \textsc{Cloudy}. In the \textsc{Cloudy} code, these values are automatically calculated without any special actions being required.\\
$^m$ This upper limit of the rate is of $\sim 10^{-10}$ cm$^3$ s$^{-1}$. See Section \ref{rad_ass} for a more detailed discussion regarding this upper limit.\\
$^n$ \cite{neuf20} and references therein.
\end{table}

\subsection{Cosmic-ray ionization rate}
\label{cosmic_ray}
The cosmic-ray ionization rate affects the chemical
and ionization state of the gas. The \textsc{Cloudy} code deals with the cosmic-ray density. It automatically converts the
given cosmic-ray ionization rates into the cosmic-ray density.
It considers the cosmic-ray ionization rate to be $2 \times 10^{-16}$ s$^{-1}$ per H ($\zeta_H'$)
and $4.6 \times 10^{-16}$ s$^{-1}$ per H$_2$ ($\zeta_{H_2}'$) by default.
Thus, the default rate per H$_2$ ($\zeta_{H_2}'$) is $2.3$ times higher than that of H ($\zeta_{H}'$).
The factor $2.3$ instead of $2$ in the relation arises because of the secondary ionization. This ionization is produced when the suprathermal electrons are knocked off in the primary ionization.
Here, we use the cosmic-ray ionization rate per H$_2$ as $\zeta_{H_2}=\zeta_0=1.3 \times 10^{-17}$ s$^{-1}$
(\textsc{Cloudy} scales it with respect to $\zeta_H'$ to consider the cosmic-ray density).
Our standard rate is varied (in between $\zeta_0$ and $10^8\zeta_0$) relative to it.
This means our standard $\zeta_{H}$ is $5.65 \times 10^{-18}$ s$^{-1}$.
In Table \ref{table:reaction}, reaction number 1 (CR) of the Ar network represents the cosmic-ray ionization rate
by $\zeta_H$ and reaction number 2 (CRPHOT) by $\zeta_{H_2}$.
For the similar cosmic-ray ionization reactions with He and Ne,
we consider the same leading coefficient as Ar \citep{schi14,prie17}.
In \textsc{Cloudy}, the direct ionization by cosmic-rays is automatically considered for all the ionization stages and all elements.

\subsection{Ion-neutral reaction rate}
\label{ion-neutral}
The rate coefficients of the ion-neutral (IN) reaction of the Ar-related species were already discussed
in \cite{prie17}. In constructing the reaction network with He and Ne, we either assume the same rate
constants as used for the IN reactions of Ar or use
some educated guess.
We also include the reaction pathways and rate constants from \cite{gust19}, \cite{neuf20}, and \cite{orie77}.
In Table \ref{table:reaction}, the IN rates are given in reaction numbers $3-10$, $14-24$ for Ar, $3-10$, $14-25$ for Ne, and $3-10$, $14$ for He netwok.
Reaction numbers $14-24$ of Ar, $14-25$ of Ne network were not considered in \cite{prie17}.
However, these pathways are included in the \textsc{Cloudy} default network, and thus, we use it.

For reaction 3 ($\rm{Ar+{H_2}^+ \rightarrow ArH^+ + H}$) of Ar, we consider a rate coefficient of $10^{-9}$ cm$^3$ s$^{-1}$
following \cite{prie17}. We also use quantum-chemical calculations (DFT B3LYP/6-311++G(d,p) level of theory) with the \textsc{Gaussian} 09 suite of programs \citep{fris13} and find that this
reaction is highly exothermic. Similar calculations for NeH$^+$ formation ($\rm{Ne + {H_2}^+ \rightarrow NeH^+ + H}$) and HeH$^+$ formation ($\rm{He+ {H_2}^+ \rightarrow HeH^+ + H}$) show a highly endothermic nature.
\cite{neuf20} considered a rate coefficient of $\sim \rm{3\times10^{-10}exp(\frac{-6717\ K}{T})}\ cm^3\ s^{-1}$ for the HeH$^+$ formation by this reaction.
We notice that the endothermicity of NeH$^+$ formation by this reaction is smaller than that of the
endothermicity of HeH$^+$. Because no reference is available for $\rm{Ne + {H_2}^+ \rightarrow NeH^+ + H}$, we scale the
HeH$^+$ formation rate here and use $\rm{\sim 2.58\times10^{-10}exp(\frac{-6717\ K}{T})}\ cm^3\ s^{-1}$ in our network.

In the case of reaction 4 ($\rm{X+{H_3}^+ \rightarrow XH^+ + H_2}$) of Ar, an endothermicity of about $6400$ K was noted by \cite{vill82}. We use the same empirical relation for the reaction between
$\rm{H_3}^+$ and He/Ne.
From our quantum-chemical calculations, we obtain endothermicities of about
$6019$ K, $27456$ K, and $29110$ K for reaction 4 of the Ar, Ne, and He-related pathways, respectively, and use these values for the computation of the rate constant of reaction 4 shown in Table \ref{table:reaction}.

We calculate the enthalpies of reactions for reaction numbers $5-10$ of Table \ref{table:reaction} and find all reactions are exothermic.
The rate constants of some of these reactions containing Ar were already given in \cite{prie17}. Here, we use the same rate for our calculation.
To estimate the rate constant for Ne, we derive a scaling factor depending on our computed exothermicity values.
Because the earlier studies did not consider reaction 5a of the He network \citep{gust19,neuf20}, we are not considering
this reaction here. We consider two other routes of Ne and He, having the possible product channels 5(b) $\rm{X^{+} + H_2 \rightarrow X + H + H^+}$ and 5(c) $\rm{X^{+} + H_2 \rightarrow X + H_2^+}$. In the
case of X=Ne, channel 5(b) is considered because the ionization potential of Ne ($21.56$ eV) is greater
than the sum of the ionization potential of H and the dissociation energy of H$_2$, i.e. ($13.60 + 4.48$) eV = $18.08$ eV.
In the UMIST network, similar reaction channels (5b and 5c) are available for the X=He network.
By calculating the enthalpies of reactions and comparing them between reactions 5b and 5c of the Ne and He networks, we again obtain scaling factors to estimate the rate coefficients of reactions 5b and 5c of the Ne network.

For the rate coefficient for the destruction of ArH$^+$ with H$_2$, we consider the same one used in \cite{prie17}. To destroy HeH$^+$ by H$_2$ (i.e., reaction number 6 of the He network), we use the rate coefficient measured by \cite{orie77}. For the NeH$^+$
destruction by H$_2$, we use a scaling technique similar to that mentioned earlier.
We prepare the IN reaction network of He according to the very recent work by \cite{neuf20}. For the sake of completeness, they updated the reaction network developed by \cite{gust19} and added several formation and destruction reactions related to He.
We include the HeH$^+$ destruction by H (reaction 14 of the He network)
with a constant rate coefficient $\rm{1.7\times10^{-9}}\ cm^3 s^{-1}$.

\subsection{Radiative association}
\label{rad_ass}
Recently, \cite{thei16} studied the formation of ArOH$^+$ and NeOH$^+$ quantum-chemically. They considered three
channels for the formation of NeOH$^+$ (by $\rm{Ne^+ + OH}$, $\rm{NeO+H^+}$, and
$\rm{NeH^+ + O}$) and three channels for the formation of ArOH$^+$ (by $\rm{Ar^+ + OH}$, $\rm{ArO+H^+}$, and
$\rm{ArH^+ + O}$). According to their relative energy calculations, ArOH$^+$ remains in an energy state lower
than the total relative energy of their reactants and products \citep[see Figure 2 of][]{thei16},
whereas NeOH$^+$ leads to a likely spontaneous dissociation into Ne and OH$^+$ \citep[see Figure 1 of][]{thei16}.
Because the reactants have higher energy, some energy is released during its formation. These reactions could be
treated as radiative association reactions (reaction numbers 11-13 of Table \ref{table:reaction}). We calculate the rate constant of these reactions using the method described below \citep{bate83}:
\begin{equation} \label{eq:bates}
K=1\times10^{-21}A_r\frac{(6E_0+N-2)^{3N-7}}{(3N-7)!}\ cm^3s^{-1}.
\end{equation}
This temperature-independent semiempirical relation provided by \cite{bate83}
requires the association
energy ($E_0$) in eV, numbers of nuclei ($N$) in the complex,
and transition probability ($A_r$) in s$^{-1}$, which is taken to be $100$, as suggested by \cite{bate83}.
The calculated rates for reactions $11-13$ are noted in Table \ref{table:reaction}.
But this semiempirical relation is temperature-independent and
estimated at a temperature of $\sim 30$ K. Here, we are dealing with Crab knots, where the temperature is much higher.
Keeping this in mind, additionally, we consider an upper limit ($10^{-10}$ cm$^3$ s$^{-1}$)
to these reactions.
Although \cite{thei16} did not consider
the reaction between X ($=$ Ar, Ne, and He) and OH$^+$ for the formation of
XOH$^+$, we consider reaction number 11 of each network because we find it to be exothermic.

We adopt the value of $1.44\times10^{-16}$ cm$^3$ s$^{-1}$
as the rate coefficient of the HeH$^+$ formation reaction (He-related reaction number 15, i.e., $\rm{He^+ + H \rightarrow HeH^{+} + h\nu}$).
\cite{gust19} ignored $\rm{He + H^+ \rightarrow HeH^{+}}$ $\rm{+ h\nu}$ (reaction 16 of He-related reactions) in the planetary nebula environment, which dominates HeH$^+$ formation in the early universe.
But \cite{neuf20} considered the same formation of HeH$^+$ through the radiative association reaction using a
temperature-dependent rate of $\rm{5.6\times10^{-21}(\frac{T}{10^4K})^{-1.25}}\ cm^3\ s^{-1}$. Here also, we use the same
rate coefficient for reaction 16 of the He network.

\subsection{X-ray ionization rate}
\label{x-ray}
X-ray photoionization, including inner-shell ionization and Auger cascades, collisional ionization by secondary electrons
coming from the inner-shell photoionization, are fully treated in \textsc{Cloudy} for all basic elements without any unique action being required.
However, the physical conditions adopted here demand a chemical network that takes the effect of X-ray ionization into
account. We consider the three types of X-ray-induced reactions, namely (a) ionization by direct X-rays ($\zeta_{XR}$),
(b) secondary ionization by X-rays ($\zeta_{XRPHOT}$), and (c) electron-impact X-ray ionization ($\zeta_{XRSEC}$).
The X-ray can mainly ionize the heavy elements by removing the K-shell electron. Auger transitions then fill the vacancy created by the
removal of the K-shell electron. In addition, other electrons
and X-ray photons are emitted by the ion during this process, resulting in multiply ionized species. Thus, x-ray ionization is
an essential means to dictate the chemistry around the Crab environment.
Here, we compute various X-ray ionization rates by adopting the method used in \cite{meij05}. Though these calculated rates are not directly used in the \textsc{Cloudy} model, building the noble-gas-related pathways from scratch will be very useful. Please
see Appendix Chapter \ref{chap:xray_ionization} for the detailed process of the estimation of the X-ray ionization rate.

\subsection{Electronic and dissociative recombination}
\label{ER_DR}
We consider the electronic recombination (ER) reactions of the noble gas atomic ions (X$^+$, X$^{++}$ for X =Ar, Ne, He) and dissociative recombination (DR) reactions of the noble gas molecular ions (XH$^+$, XOH$^+$ for X =Ar, Ne, He). The ER reactions with numbers 29-30 for Ar,
30-31 for Ne, and 21-22 for He are treated automatically in \textsc{Cloudy} to make sure that they correctly balance
the inverse photoionization processes. So, we do not include them again.
We list them in Table \ref{table:reaction} for the sake of completeness.
\cite{prie17} consider a temperature-dependent rate coefficient for ER of Ar$^+$ \citep{schi14} and Ar$^{++}$ \citep{shul82}.

For the DR of ArH$^+$, \cite{prie17} considered a typical rate of
about $10^{-9}$ cm$^3$ s$^{-1}$ for their initial model following \cite{schi14} and
a reduced rate $10^{-11}$ cm$^3$ s$^{-1}$ for their
final models. \cite{abdo18} presented the cross sections for DR and electron-impact vibrational
excitation of ArH$^+$ at electron energies appropriate for the interstellar environment.
They found very low values of the DR rate
coefficients at temperatures below $1000$ K.
They concluded that DR plays a minor role. In contrast, the collisions with $\rm{H_2}$ molecules
and the photodissociation are the only significant ArH$^+$ destruction mechanisms in the ISM.
Here, we consider a temperature-independent
rate constant of $10^{-11}$ cm$^3$ s$^{-1}$, similar to the final models of \cite{prie17} for the DR of ArH$^+$. In addition,
we assume that the same rate constant of
$10^{-11}$ is valid for the DR of ArOH$^+$, NeH$^+$, and NeOH$^+$. For HeH$^+$, we use the very recently updated
temperature-dependent rate of $\rm{4.3\times10^{-10}(T/10^4\ K)^{-0.5}\ cm^3\ s^{-1}}$ following \cite{neuf20}. For HeOH$^+$,
we consider the same DR rate as it was considered for HeH$^+$.

\subsection{Photodissociation}
\label{photo}
We consider the rate coefficients of the photodissociation (PH) reactions of the hydride and hydroxyl cations following the PH reaction of ArH$^+$ \citep{roue14,prie17}.
\cite{prie17} did not consider the PH reaction of HeH$^+$ (i.e., He network reaction number 25) because their input SED has negligible flux
beyond the Lyman limit relevant for the cross-section given by \cite{robe82}. \cite{gust19} also ignored it as the reaction progresses very slowly. According to \cite{neuf20}, we consider this reaction, which is automatically controlled in \textsc{Cloudy} default network.

\section{Results and discussions on chemical modeling}
\label{results_discussions}
Based on the reactions network described in the early Section \ref{sec:chem_path}, we study the hydride and hydroxyl cations
of Ar, Ne, and He. \cite{schi14} assigned absorption lines of ArH$^+$ to the previously unidentified absorption lines.
Though we mainly focus here on the Crab environment, it will be very useful to first check our model with the model described
in \cite{schi14} and \cite{prie17} for the diffuse ISM. It will also be
useful to look at the predicted abundances of other hydride and hydroxyl cations in diffuse cloud conditions.

\subsection{Diffuse interstellar medium}
\label{diff_ISM}
Here, we assume a cloud with the initial number density of total hydrogen nuclei ($\rm{n_H}$) of $50$ cm$^{-3}$ and a
primary cosmic-ray ionization rate for atomic hydrogen of $\rm{\zeta_H = 2 \times 10^{-16}}$ s$^{-1}$ \citep{schi14}.
We consider the default ISM elemental abundances of \textsc{Cloudy} shown in Table \ref{table:init-diffuse}.
The unextinguished local ISRF is generated with the command {\it Table ISM} in \textsc{Cloudy}.
We use the mean ISRF \citep{drai78} of $1$ Draine unit, and the resultant shape of the incident SED is further modified by including the extinction due to photoelectric
absorption by a cold neutral slab with a column density of $N(H) = 10^{20}$ cm$^{-2}$ (Figure \ref{fig:sed_richardson}).
Using the default ISM grain and the H$_2$ grain formation rate of $\rm{3 \times 10^{-17}\ cm^3 s^{-1}}$ \citep{jura75} and
by considering the default PAH treatment in \textsc{Cloudy}, we obtain an extinction-to-gas ratio of $A_V/N(H) = 5.412 \times 10^{-22}$ mag cm$^2$ for this region.

\begin{table}
\scriptsize
{\centering
\caption{Gas-phase elemental abundances of species with respect to total hydrogen nuclei in all forms for the modeling of diffuse ISM using \textsc{Cloudy} \citep{das20}. \label{table:init-diffuse}}
\vskip 0.2 cm
\begin{tabular}{cccc}
\hline
{\bf  Element} & {\bf  Abundance} & {\bf  Element} & {\bf  Abundance} \\
\hline
H & 1.00 & $^{36}$Ar & $2.82 \times 10^{-6}$ \\
He & 0.098 & $^{38}$Ar & $5.13 \times 10^{-7}$ \\
C & $2.51 \times 10^{-4}$ & $^{40}$Ar & $8.20 \times 10^{-10}$ \\
N & $7.94 \times 10^{-5}$ & $^{20}$Ne & $1.23 \times 10^{-4}$ \\
O & $3.19 \times 10^{-4}$ & $^{22}$Ne & $9.04 \times 10^{-6}$ \\
Cl & $1.00 \times 10^{-7}$ & S & $3.24 \times 10^{-5}$ \\
Mg & $1.26 \times 10^{-5}$ & Fe & $6.31 \times 10^{-7}$ \\
Si & $3.16 \times 10^{-6}$ & & \\
\hline
\end{tabular} \\
}
\vskip 0.2 cm
{\bf Note:} For the initial isotopic ratio of argon and neon, we have used $^{36}$Ar/$^{38}$Ar/$^{40}$Ar = $84.5946/15.3808/0.0246$ and
$^{20}$Ne/$^{21}$Ne/$^{22}$Ne = $92.9431/0.2228/6.8341$, following \cite{wiel02}.
\end{table}

\begin{figure}
\begin{center}
\includegraphics[width=0.9\textwidth]{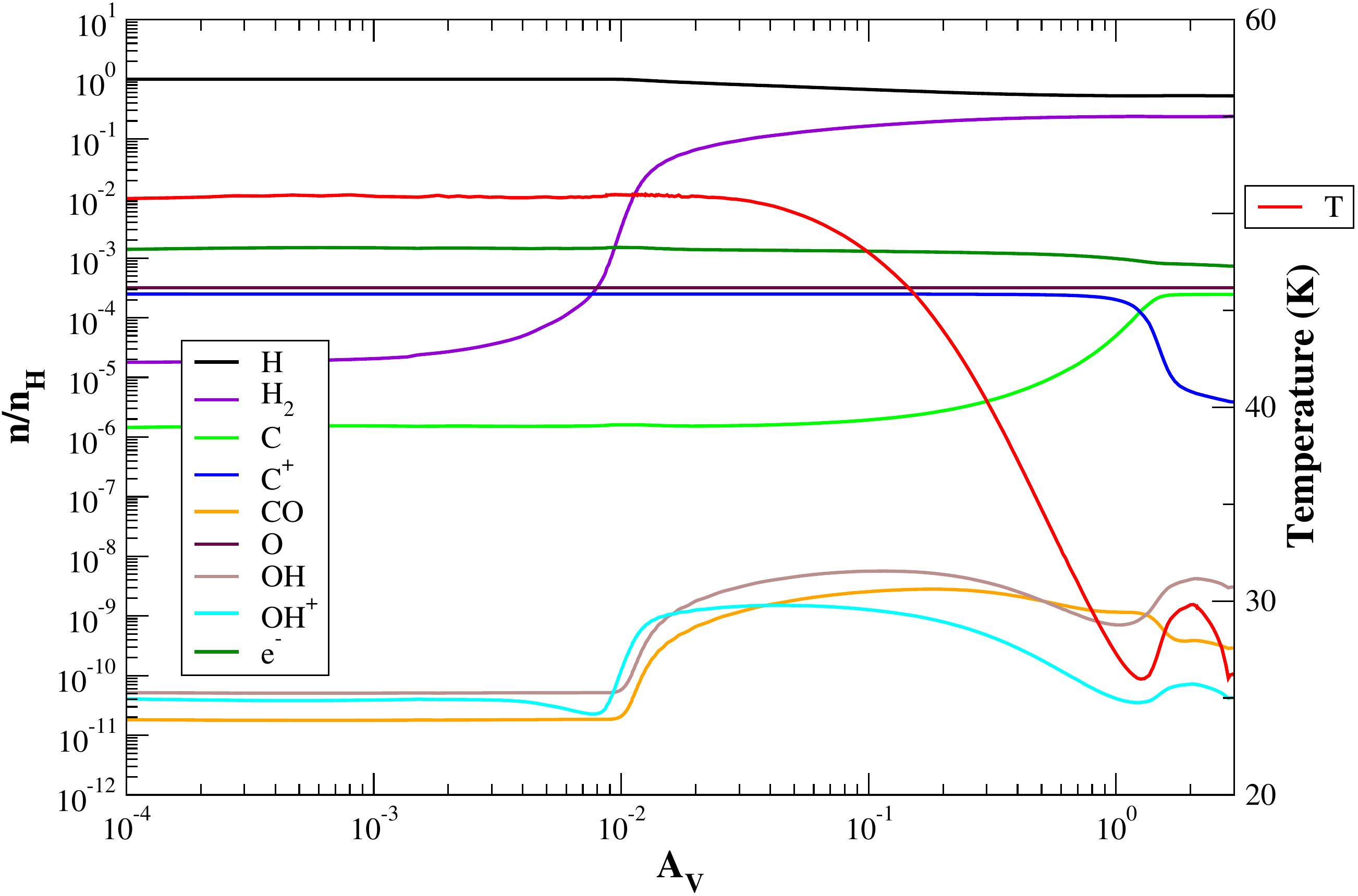}
\includegraphics[height=0.6\textwidth]{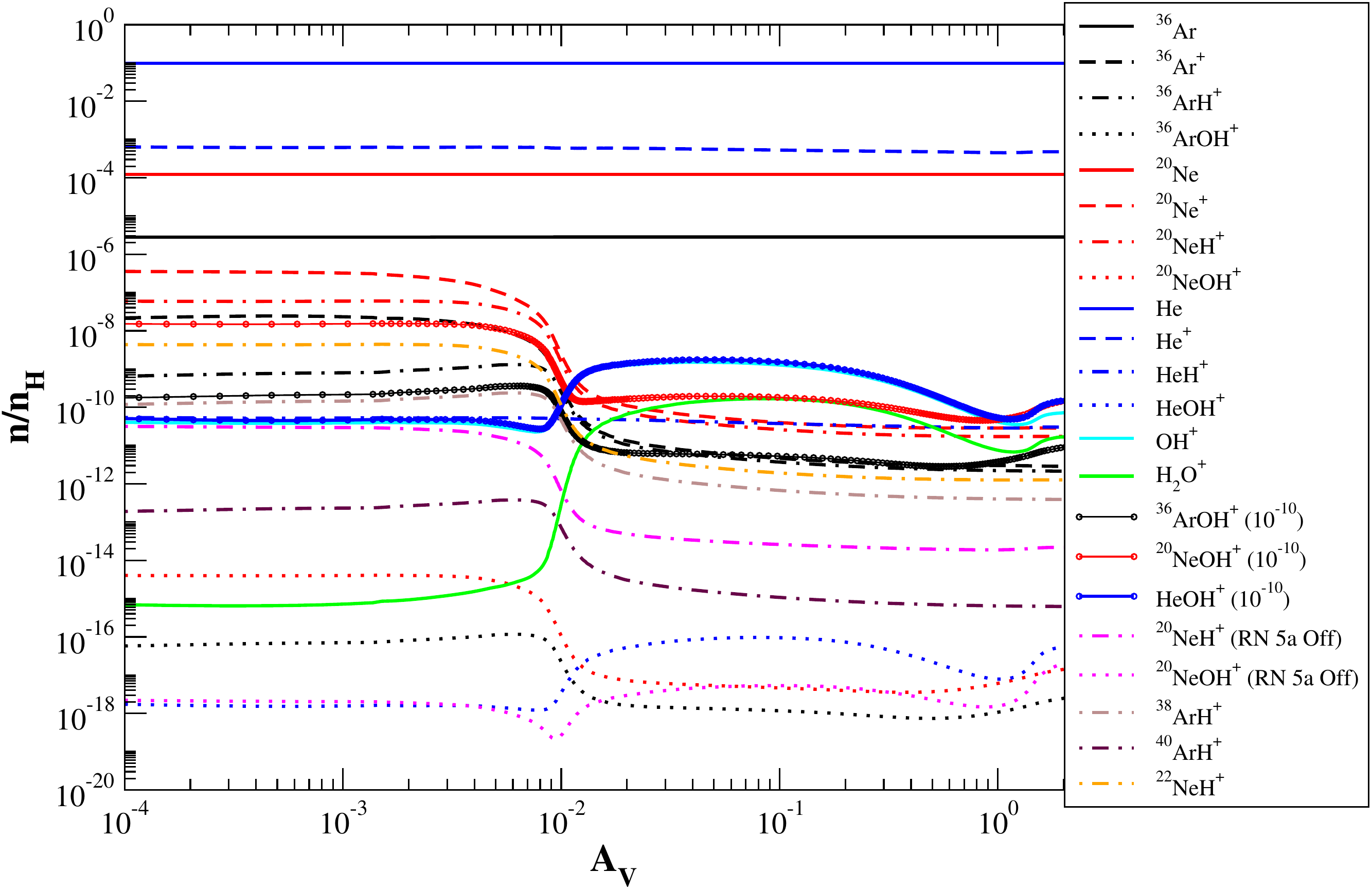}
\caption{Variation of abundances for simple species with a diffuse ISM model shown in the upper panel.
The right side of the upper panel, the electron temperature variation is shown. In the lower panel, the variation of isotopic abundances for noble gas species is shown. The abundances of $^{36}$ArOH$^+$, $^{20}$NeOH$^+$, and HeOH$^+$ considering the upper limit of their formation rate by radiative association reactions ($\sim 10^{-10}$ cm$^3$ s$^{-1}$), are noted [XOH$^+$ ($10^{-10}$)]. The abundance profiles of $^{20}$NeH$^+$ and $^{20}$NeOH$^+$ are also shown when reaction 5a of the Ne network is off \citep{das20}.}
\label{fig:ISMabun}
\end{center}
\end{figure}

Figure \ref{fig:ISMabun} shows the abundances of some of the essential species considered in our
network as a function of the visual extinction, $A_V$. The cloud remains in atomic form throughout the region, and the H$_2$ fractional abundance varies between $2 \times 10^{-5}$ and 10$^{-1}$. The electron temperature ranges between $25$ and $50$ K, and electron fractional abundance remains
roughly constant at $\sim 10^{-3}$. The peak abundance of ArH$^+$ is around $1.3 \times 10^{-9}$, decreasing with increasing $A_V$ deep inside the cloud.
ArH$^+$ is a unique tracer of the atomic gas, having an H$_2$ fractional abundance of $10^{-4}-10^{-3}$ \citep{schi14}. We
find a very similar result here. Deep inside the filament, where the H$_2$ density is sufficiently increased, a strong anticorrelation is present between ArH$^+$ and H$_2$.
The abundance profile of ArH$^+$ shows a strong anticorrelation with OH$^+$ and  H$_2$O$^+$. It implies that
while ArH$^+$ traces the region with lower H$_2$/H ratio, OH$^+$ and H$_2$O$^+$ favor the higher H$_2$/H ratio region.
The obtained abundances of Ar$^+$ and ArH$^+$ match those measured by \cite{schi14} and present a similar variation to $A_V$.
For similar conditions, \cite{prie17} found a slightly lower abundance of these species. NeH$^+$ also follows the similar behavior of
ArH$^+$ and a strong anticorrelation with H$_2$ is observed. Moreover, we obtain a peak fractional abundance of NeH$^+$ $\sim 5 \times 10^{-8}$.
Table \ref{table:init-diffuse}
shows that Ne has a higher initial elemental abundance than Ar ($\rm{\frac{Ne}{Ar}=43.6}$). This is also reflected in the obtained peak
abundance ratio between NeH$^+$ and ArH$^+$ ($\sim 38$). However, the much higher initial
elemental abundance of
He than that of the Ar and Ne is not reflected in the obtained abundance of HeH$^+$. The obtained HeH$^+$ fractional abundance is
smaller (peak abundance $5 \times 10^{-11}$) than that of ArH$^+$ and NeH$^+$. This is because ArH$^+$ and NeH$^+$ formation by
$\rm{X^+ + H_2 \rightarrow XH^+ + H}$ (reaction numbers 5 of Ar and 5a of the Ne network) is considered, which is avoided in HeH$^+$ formation here.

\cite{thei15} questioned the formation of NeH$^+$ by reaction 5a. They also found that the possible product of this reaction
would be Ne and $\rm{H_2}^+$ (i.e., reaction 5(c)).
Here, for the diffuse cloud model, we find that the reaction between Ne$^+$ and H$_2$ (i.e., reaction 5a) forms the
majority of NeH$^+$, and the abundance of NeH$^+$ is higher than that of ArH$^+$. However, NeH$^+$ is yet to be identified in the
diffuse region. This also suggests an overestimation of the NeH$^+$ abundance in our model.
To check the effect of reaction 5a,
we consider the case where this reaction is switched off.
In this case, we find that the abundance of NeH$^+$ significantly dropped
and is consistent with its absence in the observed spectra (having a peak fractional abundance of $\sim 3 \times 10^{-11}$).
The formation of the majority of NeH$^+$ in this case happens via reaction 14 ($\rm{HeH^+ + Ne \rightarrow NeH^+ + He}$).
However, in this case, we also see the anticorrelation between NeH$^+$ and H$_2$.
Unless otherwise stated, the reaction 5a is on by default in all the cases
reported here.

According to the recent work by \cite{thei16}, the hydroxyl cations of noble gas are the most stable small noble gas
molecules analyzed, besides their respective hydride diatomic cation cousins. So, we include them in our network
and plot them here to show the comparison between them. When reaction 5a of the Ne network is on, the abundance profile of ArOH$^+$ and NeOH$^+$ follows the
ArH$^+$ and NeH$^+$ abundance profiles because most of them form by $\rm{ArH^+ + O}$ and $\rm{NeH^+ + O}$ (reaction 13 of the Ar and Ne network), respectively.
The abundance profile of HeOH$^+$ follows that of OH due to the formation of the majority of HeOH$^+$ by He$^+$ and OH.
When reaction 5a of the Ne network is off, we find a similar abundance profile of NeOH$^+$ with HeOH$^+$.
Figure \ref{fig:ISMabun} also shows the abundances of ArOH$^+$, NeOH$^+$, and
HeOH$^+$ by considering the upper limit of their formation rate by radiative association reactions ($\sim 10^{-10}$ cm$^3$ s$^{-1}$;
see Section \ref{rad_ass} for the justification).
A significant production of hydroxyl ions is observed only when the upper limit of the rate coefficients is used.
A comparison between the obtained column densities of some atomic
and molecular ions with the observation
of a diffuse cloud toward W51 is shown in Table \ref{table:column-diffuse}. We find that our results are very close to the observed results.

We also include the $^{38}$Ar, $^{40}$Ar, $^{20}$Ne, and $^{22}$Ne isotopes in our network. $^{21}$NeH$^+$ is not considered here
because the corresponding spectral information is absent in the CDMS/JPL database.
For the initial isotopic ratio of argon and neon, we use $^{36}$Ar/$^{38}$Ar/$^{40}$Ar $= 84.5946/15.3808/0.0246$
and $^{20}$Ne/$^{22}$Ne $= 13.6$ \citep{wiel02}.
We find that the peak fractional abundance of $^{38}$ArH$^+$, $^{40}$ArH$^+$, and $^{22}$NeH$^+$ is $2.2 \times 10^{-10}$, $3.8 \times 10^{-13}$, and
$4.5 \times 10^{9}$, respectively. This yields a ratio of the peak abundance of $^{36}$ArH$^+$/$^{38}$ArH$^+$/$^{40}$ArH$^+$
$=  84.5946/14.32/0.0247$ and $^{20}$NeH$^+$/$^{22}$NeH$^+$
$= 11.11/1.0$ (reaction 5a of the Ne network is considered here). Because no fractionation reactions are considered in this work, initial elemental abundances are roughly reflected in the abundances of their respective hydride ions.

\begin{table}
\scriptsize
\centering
\caption{Comparison between the obtained column densities of some atomic and molecular ions with the observation of diffuse cloud toward W51 \citep{indr12,das20}. \label{table:column-diffuse}}
\vskip 0.2 cm
\begin{tabular}{cccc}
\hline
{\bf  Species} & \multicolumn{2}{c}{\bf  Column density (cm$^{-2}$)} \\
 & {\bf Model} & {\bf Observation} \\
\hline
H & $3.02\times10^{21}$ & $(1.39\pm0.3)\times10^{21}$ \\
$\rm{H_2}$ & $1.26\times10^{21}$ & $(1.06\pm0.52)\times10^{21}$ \\
$\rm{H_3^+}$ & $3.52\times10^{13}$ & $(2.89\pm0.37)\times10^{14}$ \\
OH$^+$ & $9.04\times10^{11}$ & $(2.97\pm0.13)\times10^{13}$  \\
$\rm{H_2O^+}$ & $1.43\times10^{11}$ & $(6.09\pm0.96)\times10^{12}$ \\
C$^+$ & $5.61\times10^{17}$ & $(4.0\pm0.4)\times10^{17}$ \\ 
\hline
\end{tabular}
\end{table}

\subsection{The Crab nebula filament}
\label{crab_nebula}

\begin{figure}
\begin{center}
\includegraphics[width=\textwidth]{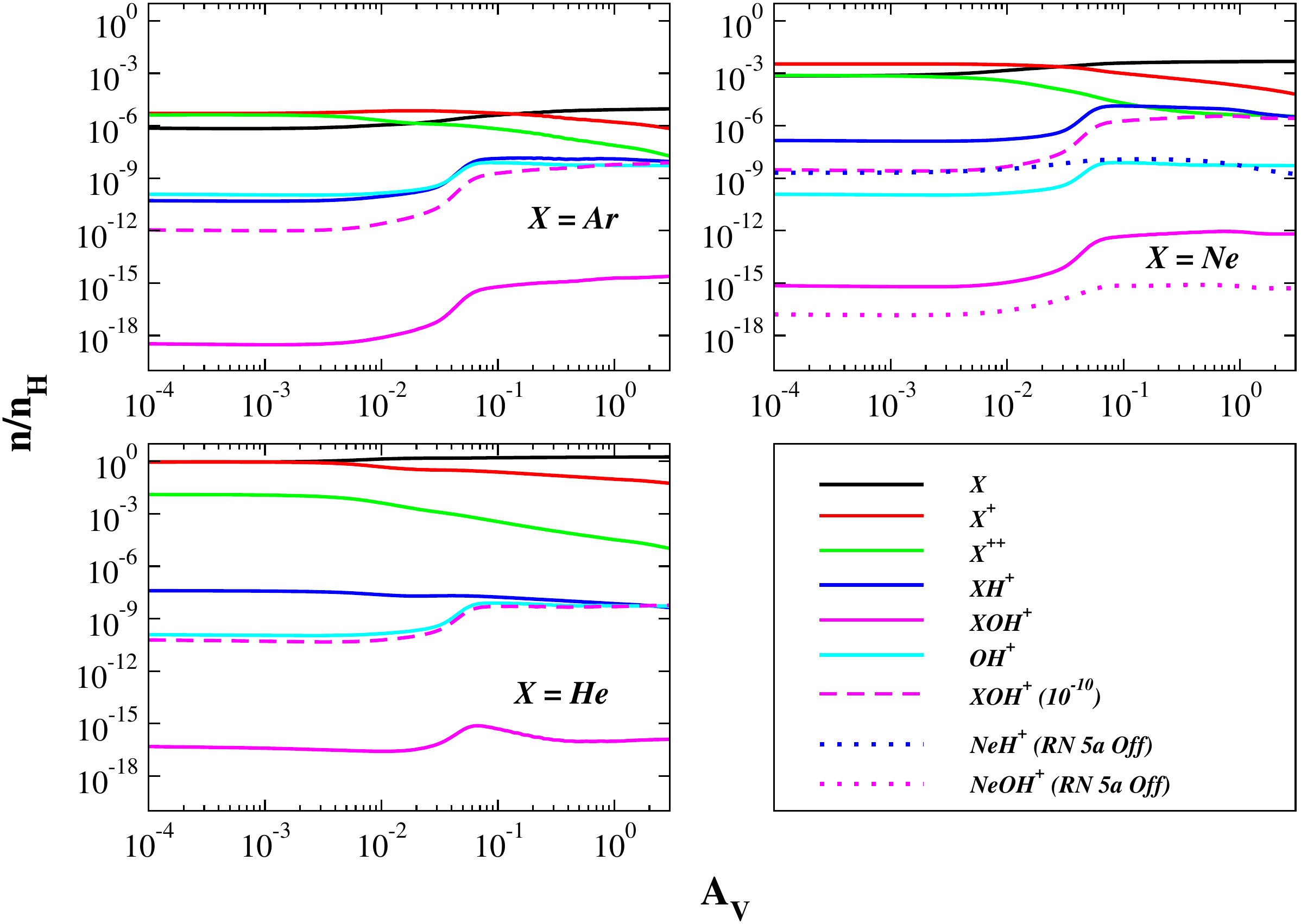}
\caption{Abundances of various ionized states of noble gas (X = $^{36}$Ar, $^{20}$Ne, and He)
along with their respective hydride and hydroxyl cations as a function of $A_V$ considering Crab Model A with $\rm{n_H}=1900$ cm$^{-3}$ and $\zeta_{H_2}=\zeta_0=1.3 \times 10^{-17}$ s$^{-1}$ \citep{das20}. The dashed pink lines denote the abundance of XOH$^+$ considering the upper limit of forming XOH$^+$ ($\sim 10^{-10}$ cm$^3$ s$^{-1}$; see Section \ref{rad_ass} for the justification). Abundances of NeH$^+$ and NeOH$^+$ are shown in dotted blue and dotted magenta lines respectively when Ne network reaction 5a is switched off.}
\label{fig:abun1}
\end{center}
\end{figure}

Physical conditions suitable for the
Crab environment are already presented in Section \ref{sec:physical_cond}.
Figure \ref{fig:abun1} shows the variation of the abundances of different ionization
states of the primary isotope of the noble gas ions (X = $^{36}$Ar, $^{20}$Ne, and He) as a function of the visual extinction ($A_V$) for Model A.
For this case, we consider the initial model of the Crab with a total hydrogen nuclei density $\rm{n_H}=1900$ cm$^{-3}$ and cosmic-ray ionization rate per H$_2$ $\zeta=\zeta_0=1.3\times10^{-17}$ s$^{-1}$.
This $\zeta$ value is too low for an SNR; more realistic values are explored in the following sections.
Here, we use this value because it is the standard value used in chemical models of molecular clouds and used in the initial model of \cite{prie17}.
In the three blocks of Figure \ref{fig:abun1}, we show three noble-gas-related (Ar, Ne, and He) species. We find that
reaction numbers $1-2$ of all reaction sets in Table \ref{table:reaction} and reaction numbers $27-28$ of Ar, $28-29$ of Ne, and $19-20$ of He are responsible for producing X$^+$ from X.
X$^+$ is further converted into X$^{++}$ by direct X-ray ionization.
X$^{++}$ can additionally be produced directly from X-ray ionization.
In all blocks of Figure \ref{fig:abun1}, we obtain a higher abundance of X$^+$ compared to X$^{++}$.
Here, we use the initial elemental abundance of $^{36}$Ar, $^{20}$Ne, and
He of $1.0 \times 10^{-5}$, $4.9 \times 10^{-3}$, and $1.85$, respectively, relative to total hydrogen nuclei in all forms
(see Table \ref{table:abun}). This initial elemental abundance ratio between the noble gases is not maintained after forming their respective hydride ions.
If they followed their initial abundances, then the abundance of the ArH$^+$ would have been $\sim 10^5$ times lower than
that of the HeH$^+$ ion. Instead, from Figure \ref{fig:abun1}, we obtain the peak abundance of ArH$^+$, NeH$^+$ (when Ne reaction 5a is off), and HeH$^+$
in a similar range. The reason behind this is due to (i) the lower ionization potential of
$^{36}$Ar ($15.76$ eV) compared to $^{20}$Ne ($21.5645$ eV) and He ($24.5874$ eV), (ii) the high proton affinity of Ar ($3.85$ eV)
compared to Ne ($2.08$ eV) and He \citep[$1.85$ eV;][]{joll91}, and (iii) the reaction pathways adopted.

In the early universe, HeH$^+$ formation was dominated by the reaction between
He and H$^+$. Due to their high ionization potential, helium ions (He$^+$ and He$^{+2}$)
recombined with electrons to produce the neutral helium first. Neutral helium was indeed the
first neutral atom of the universe. In such a metal-free situation, He then reacted with H$^+$ to
form the first chemical bond of the universe ($\rm{He + H^+ \rightarrow HeH^+ + h\nu}$) and thus the first molecule, HeH$^+$.
Recently, \cite{gust19} identified the pure rotational ($J=1-0$) transition of HeH$^+$ in the planetary nebula NGC 7027.
But the formation of HeH$^+$
in the planetary environment progresses in a very different manner.
Looking at the environment of NGC 7027 and its age, they ignored the HeH$^+$ formation by the reactions $\rm{He + H_2^+ \rightarrow HeH^+ + H}$
and $\rm{He + H^+ \rightarrow HeH^+ + h\nu}$
(reaction numbers 3 and 16, respectively, of the He network in
Table \ref{table:reaction}).
\cite{neuf20} considered these two reactions in their network. Here, we use their adopted rate in our simulation.
Additionally, we also consider $\rm{He^++H \rightarrow HeH^+ + h\nu}$ (reaction number 15)
following \cite{gust19}.
The reaction between Ar and $\rm{H_3}^+$ (reaction number 4) is considered by \cite{prie17} in their model.
We examine XH$^+$ formation by this reaction quantum-chemically (discussed in Section \ref{ion-neutral}) and
find an endothermicity of $\approx 6019$ K for the formation of ArH$^+$.
For the formation of NeH$^+$ and HeH$^+$, it is $\sim 5$ times higher than that of
the ArH$^+$. It shows that the appearance of HeH$^+$ and NeH$^+$ by reaction number 4 is only possible at high temperatures ($>1000$ K).
The consideration of very different chemical pathways for the formation of ArH$^+$ compared to HeH$^+$ and NeH$^+$
thus plays a significant role in the mismatch between the initial elemental ratio considered and the ratio obtained after forming their hydride ions.

\begin{figure}
\begin{center}
\includegraphics[width=0.7\textwidth]{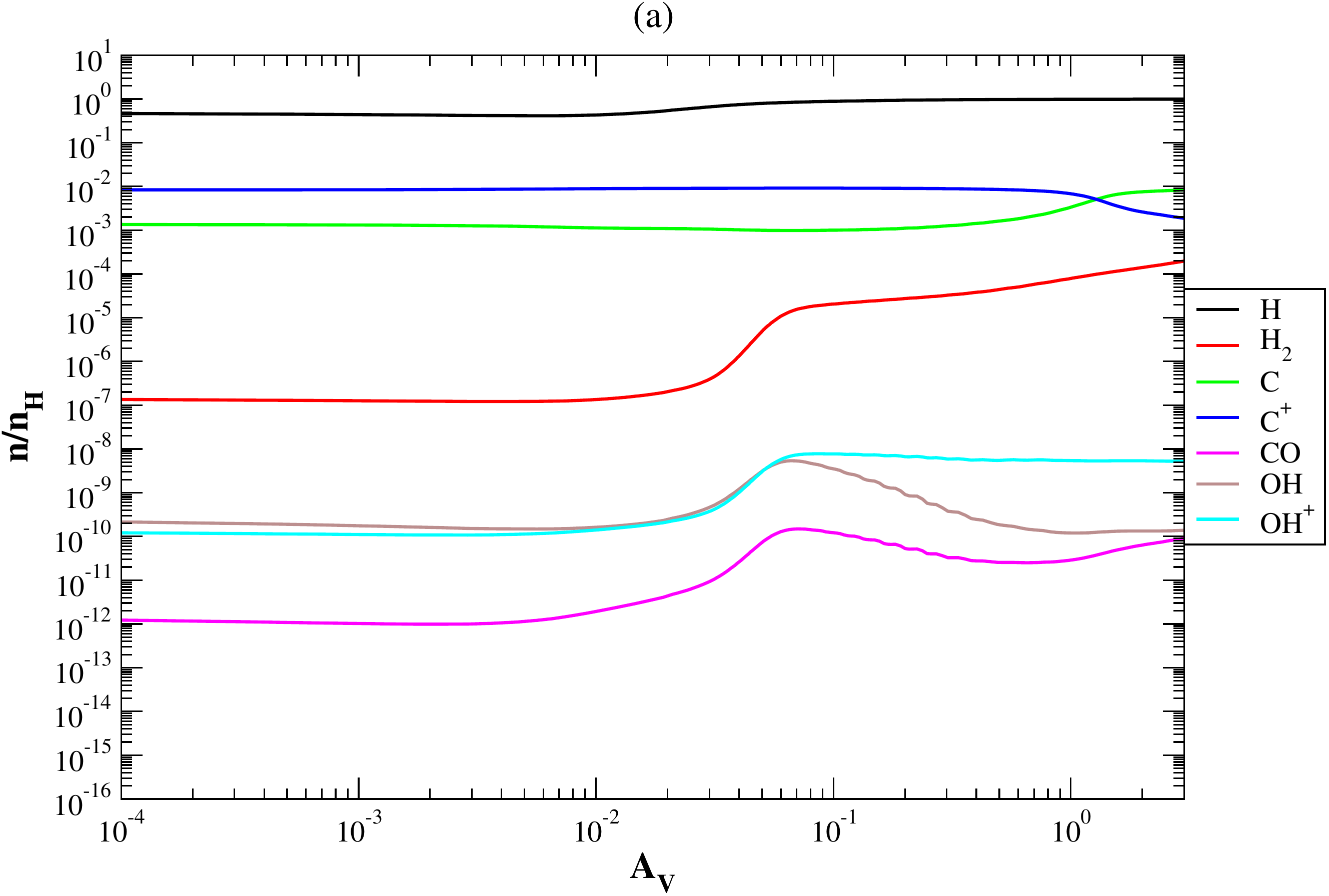}
\includegraphics[width=0.7\textwidth]{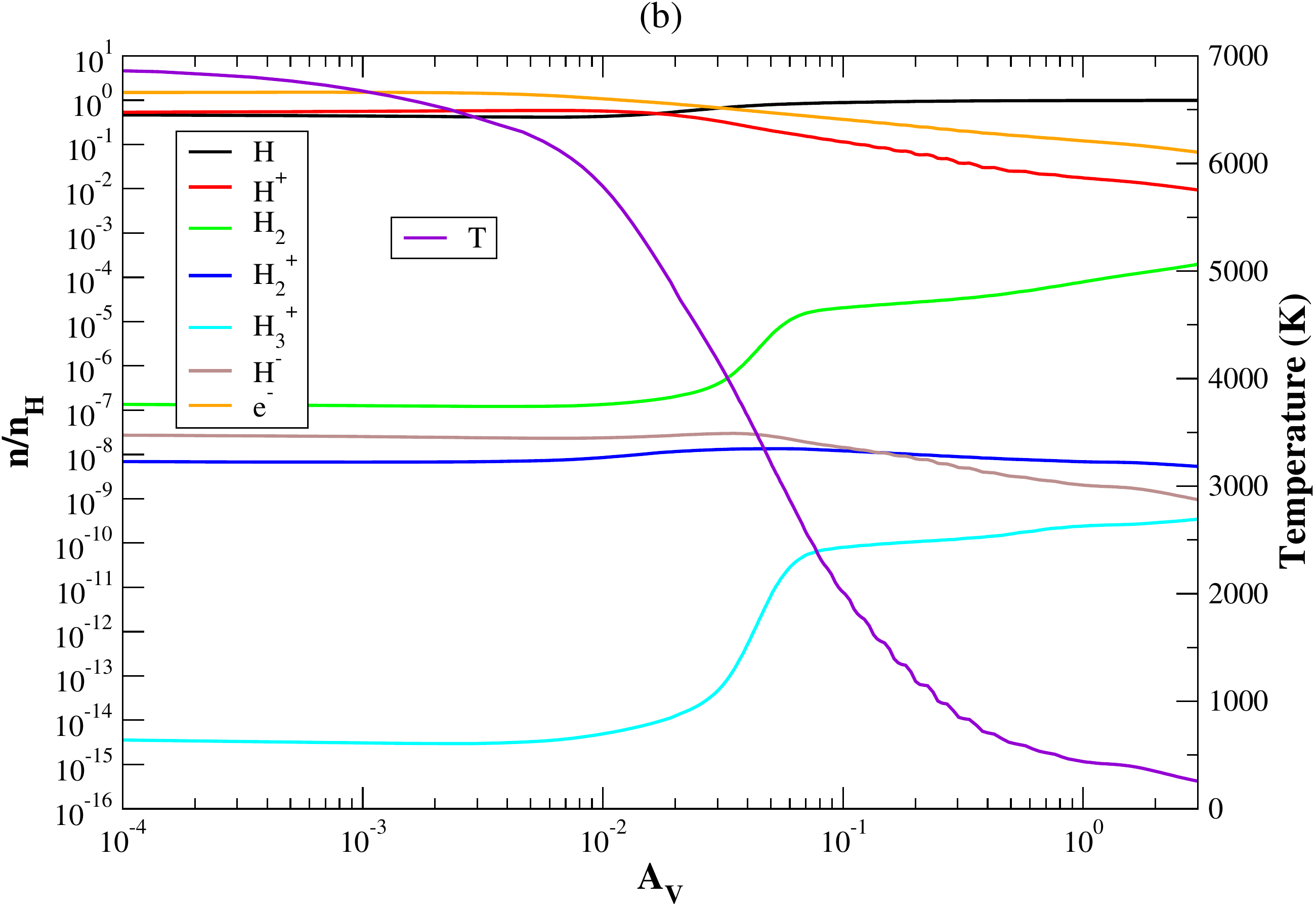}
\caption{Fractional abundance variation of the simple species with $A_V$ by considering $\rm{n_H}=1900$ cm$^{-3}$ and $\zeta_{H_2}=\zeta_0=1.3 \times 10^{-17}$ s$^{-1}$ \citep[Model A;][]{das20}. The right side of the lower panel shows the electron temperature variation.}
\label{fig:abunsimp}
\end{center}
\end{figure}

The lower limit of the detected OH$^+$ transition in the Crab can be used to set the lower
observational limit for the noble gas ions modeled here. To show the comparison between the
OH$^+$ abundance and the abundances of other noble-gas-related species, we plot
the abundance of OH$^+$ in all panels of Figure \ref{fig:abun1}.
We obtain a lower peak abundance of OH$^+$ than \cite{prie17}. This is indeed required
because \cite{barl13}
observed the ArH$^+$ transition to be significantly stronger than that of the OH$^+$. Figure \ref{fig:abun1} shows that ArH$^+$ is initially less abundant than OH$^+$, and finally, deep inside the filament, it shows the opposite trend.

The abundance variation for some important species is shown in Figure \ref{fig:abunsimp},
considering the same physical condition as in Figure \ref{fig:abun1}.
The left panel shows the
abundance variation of
H, H$_2$, C, C$^+$, CO, OH, and OH$^+$ and the right panel shows the simple ions of H (H$^+$, {H$_2$}$^+$, and {H$_3$}$^+$), electrons, and the variation of the electron temperature. The left panel shows that
most of the hydrogen is in atomic form, and thus, the cloud remains entirely atomic. In the outer part ($A_V<1$ mag) of the cloud, carbon remains in the ionized form (C$^+$), but it is converted into the neutral form inside ($A_V>1$ mag) the cloud.
Because the cloud is mostly in diffuse atomic form, the CO fractional abundance is $\sim 10^{-10}$.
Figure \ref{fig:abunsimp} shows that the abundance of H$_2$ is increasing deep inside the cloud.
Figure \ref{fig:abun1} shows that the abundance of ArH$^+$ is also growing deep inside the cloud. Thus, the
anticorrelation that has been seen between the abundance profile of ArH$^+$ and H$_2$ in Figure \ref{fig:ISMabun}
is not reflected here. This might be due to the consideration of completely different physical-chemical conditions
between these two cases.
The right panel shows that H$^+$ is very abundant, and the electron abundance varies within a few times $10^{-1}$
(i.e., electron number density $\sim$ few times $10^2$ cm$^{-3}$ for $\rm{n_H} = 1900$ cm$^{-3}$), which matches with that of the predicted electron number density
in the knot of the Crab \citep{barl13}. In this effort, it is thus essential to find out the physical conditions that can explain most of the observational results of \cite{barl13}.

\subsubsection{Comparison with observations: Model A} \label{favourable_zone}

\begin{figure}
\begin{center}
\includegraphics[width=\textwidth]{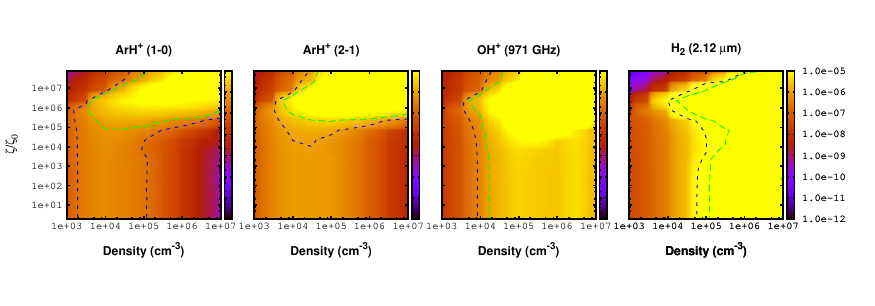}
\caption{Parameter space for the intrinsic line surface brightness (SB) of the $1-0$ and $2-1$ transitions of ArH$^+$, the $971$ GHz/$308$ $\mu$m 
transition of OH$^+$, and the $2.12$ $\mu$m transition of H$_2$ considering Model A \citep{das20}. The right panel is marked with color-coded 
values of the intrinsic line SB (in units erg cm$^{-2}$ s$^{-1}$ sr$^{-1}$). 
The contours are highlighted in the range of observational limits noted in Table \ref{table:summary_1} (Column 2).}
\label{fig:sb}
\end{center}
\end{figure}

To find a suitable favorable zone to explain the observed features, we vary the physical parameters ($\rm{n_H}$ and $\zeta$). Our parameter space consists of $\rm{n_H}$ variation of
about $10^3-10^7$ cm$^{-3}$ and $\zeta/\zeta_0$ ($\zeta_0=1.3 \times 10^{-17}$ s$^{-1}$) variation of about $1-10^8$.
In Table \ref{table:summary_1}, we summarize the results obtained from Model A and Model B in explaining the observed absolute SB of the two transitions of ArH$^+$ ($2 \rightarrow 1$ and $1 \rightarrow 0$), the 308 $\mu$m (971 GHz, $J=2\rightarrow 1,\ F=5/2\rightarrow3/2$) transition of OH$^+$, and the 2.12 $\mu$m transition of H$_2$ \citep{barl13,loh11}.
Figure \ref{fig:sb} shows the absolute SB variation of various transitions with a wide range of parameter space for Model A.
We obtain a reasonable match of the absolute SB of these transitions with the observation
when high values of $\frac{\zeta}{\zeta_0} \sim 10^6-10^8$ and $\rm{n_H}\sim 10^4-10^{5.3}$ cm$^{-3}$ are considered.
In Figure \ref{fig:sb_rich}, we show the variation of the absolute SB of these transitions
with respect to the variation of a wide range of parameter space (varying $\frac{\zeta}{\zeta_0}$ and the core density $\rm{n_{H(core)}}$) by considering Model B.

\begin{figure}
\begin{center}
\includegraphics[width=\textwidth]{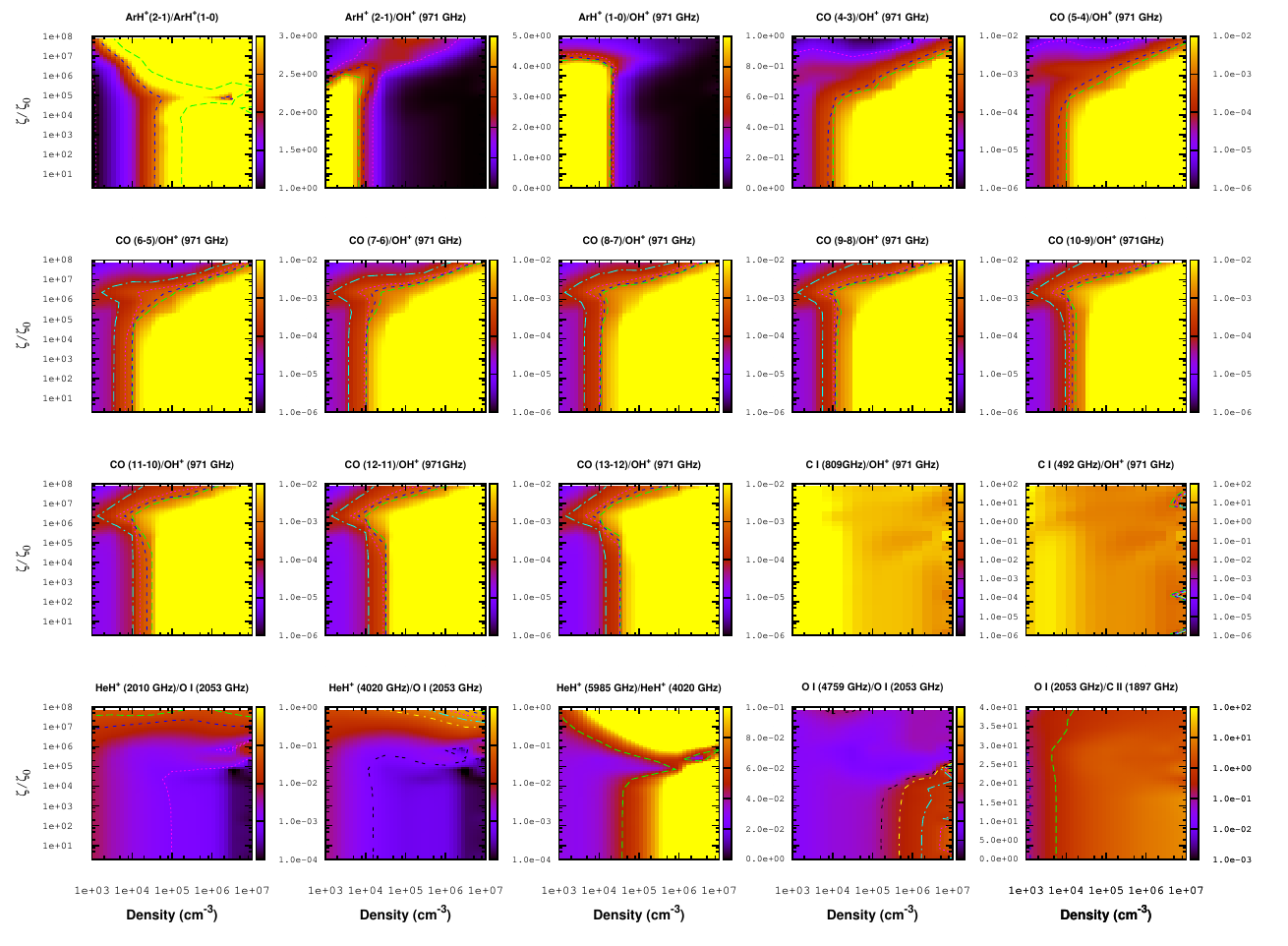}
\caption{Intrinsic line surface brightness (SB) ratio of various molecular and atomic transition fluxes considering Model A \citep{das20}. The right side of each panel is marked with color-coded values of the intrinsic line SB ratio. The contours are highlighted around the previously observed or estimated SB ratios noted in Table \ref{table:summary_2} (Column 2).}
\label{fig:sb-rat}
\end{center}
\end{figure}

The results obtained from Model A and Model B in
explaining the observed SB ratio of several transitions are summarized in Table \ref{table:summary_2}. Figure \ref{fig:sb-rat} shows these ratios for a wide range of parameter space for Model A.
The observed ratio of $ \sim 1-17$ (obtained by taking the minimum and maximum values from the observed two transitions
of ArH$^+$) between the two transitions of ArH$^+$
and the ratio between these two ArH$^+$ transitions with respect to the OH$^+$ 971 GHz transition is
best reproduced when we consider $\frac{\zeta}{\zeta_0}\sim 10^{7}$ with $\rm{n_H}=10^{4-6}$ cm$^{-3}$.
Since no transitions of CO are detected, it is expected that the
SB ratio of the various transitions of CO with respect to the OH$^+$ $971$ GHz transition would be $<1$.
We also obtain a lower SB ratio between all the transitions of CO and the $971$ GHz transition of OH$^+$.
One of the major drawbacks of our Model A is that we cannot reproduce the lack of
[C\,{\sc i}] emissions found by \cite{barl13}. This mismatch is due to the high abundance of neutral carbon [C\,{\sc i}] as compared to OH$^+$ in our Model A.
However, our model can successfully explain the lack of CO emission, the $158$ $\mu$m transition of C$^+$ [C\,{\sc ii}], and the relative line strengths between [O\,{\sc i}] and [C\,{\sc ii}].
Similarly, the results obtained with Model B are shown in Figure \ref{fig:sb-rat_rich}, and the most suitable zone
is highlighted in Table \ref{table:summary_2}. With Model B, we can successfully explain most of the observed features. Even the lack of [C\,{\sc i}] emission is well illustrated by this model.

From Tables \ref{table:summary_1}-\ref{table:summary_2} and Figures \ref{fig:sb}-\ref{fig:sb-rat}, it is challenging to arrive at the best suitable parameter for $\rm{n_H}$ and $\frac{\zeta}{\zeta_0}$ that can reproduce all the observational results
simultaneously.
However, from Model A, we have two favorable matching zones at
$\rm{n_H} \sim 10^{4-5}$ cm$^{-3}$ and $\frac{\zeta}{\zeta_0} \sim 10^{6-7}$, and for Model B, we find that the values used by \cite{rich13} for their ionizing particle model with $\rm{n_{H(core)}} \sim10^{5-6}$ cm$^{-3}$ and $\frac{\zeta}{\zeta_0} \sim 10^{6-7}$ are favorable. So, in general, in terms of the absolute intrinsic SB and SB ratio, we find our favorable parameter
space with $\rm{n_H}\sim 10^{4-6}$ and higher $\frac{\zeta}{\zeta_0}=10^{6-7}$.

\begin{table}
\scriptsize{
\centering
\caption{Summary of the previously observed surface brightness (SB) values in erg cm$^{-2}$ s$^{-1}$ sr$^{-1}$ \citep{das20}.
\label{table:summary_1}
}
\vskip 0.2 cm
\hskip -1.0 cm
\begin{tabular}{cccc}
\hline
{\bf  Molecular} & {\bf  Observational SB} & \multicolumn{2}{c}{\bf  Matching zone with $\frac{\zeta}{\zeta_0}$ and $\rm{n_H}$ (cm$^{-3}$)} \\
\cline{3-4}
{\bf transitions} & {\bf limits$^a$} & {\bf  Model A} & {\bf  Model B ($\rm{n_H = n_{H(core)}})^b$}  \\
\hline
ArH$^+ (1-0)$ &$(2.2-9.9)\times10^{-7}$ &
$\frac{\zeta}{\zeta_0}\sim 10^{0-5}$ for $\rm{n_H} \sim 10^{3-5}$ &
$\frac{\zeta}{\zeta_0}\sim 10^{0-6}$ for $\rm{n_H} \sim (3.16\times10^3)-10^5$\\
(617 GHz/485 $\mu$m) && $\frac{\zeta}{\zeta_0}\sim 10^{6-7}$ for $\rm{n_H} \sim 3.16\times10^4$ &$\frac{\zeta}{\zeta_0}\sim 10^{0-7}$ for $\rm{n_H} \sim (3.16\times10^5)-10^6$\\
&& $\frac{\zeta}{\zeta_0}\sim 10^7$ for $\rm{n_H} \sim 10^5$ & \\
&& $\frac{\zeta}{\zeta_0}\sim 10^5$ for $\rm{n_H} \sim 10^{6-7}$ & \\
&&&\\
ArH$^+ (2-1)$ &$(1-3.8)\times 10^{-6}$ &
$\frac{\zeta}{\zeta_0}\sim 10^{4-7}$ for $\rm{n_H} \sim 10^{4-5}$ &
$\frac{\zeta}{\zeta_0}\sim 10^{0-6}$ for $\rm{n_H} \sim (3.16\times10^3)-10^5$\\
(1234 GHz/242 $\mu$m)&& $\frac{\zeta}{\zeta_0}\sim 10^{5-6}$ for $\rm{n_H} \sim 10^{6-7}$ & $\frac{\zeta}{\zeta_0}\sim 10^{0-7}$ for $\rm{n_H} \sim (3.16\times10^5)-10^6$ \\
&&&\\
OH$^+$ &$(3.4-10.3)\times 10^{-7}$ & $\frac{\zeta}{\zeta_0} \sim 10^{0-4}$ for $\rm{n_H} \sim 10^{4-7}$ &
$\frac{\zeta}{\zeta_0}\sim 10^{0-6}$ for $\rm{n_H} \sim 3.16\times10^{3-5}$ \\
(971 GHz/308 $\mu$m)&& $\frac{\zeta}{\zeta_0} \sim 10^{5-7}$ for $\rm{n_H} \sim 10^4$ & $\frac{\zeta}{\zeta_0}\sim 10^{0-7}$ for $\rm{n_H} \sim 10^6$ \\
&& $\frac{\zeta}{\zeta_0} \sim 10^7$ for $\rm{n_H} \sim 10^5$ & \\
&&&\\
H$_2$ (2.12 $\mu$m)&$(1-4.8)\times 10^{-5}$ & $\frac{\zeta}{\zeta_0}\sim 10^6$ for $\rm{n_H} \sim 10^4$ & 
$\frac{\zeta}{\zeta_0}\sim 3.54\times10^6$ for $\rm{n_H} \sim (3.16\times10^3)-10^6$ \\
&& $\frac{\zeta}{\zeta_0}\sim 10^{0-5}$ for $\rm{n_H} \sim 10^5$ & \\
\hline
\end{tabular}} \\
\vskip 0.2 cm
{\bf Note:} The most suitable values of 
$\rm{n_H}$ and $\dfrac{\zeta}{\zeta_0}$ to explain the observed values are also mentioned. \\
$^a$ \cite{prie17} and references therein. \\
$^b$ $\rm{n_H = n_{H(core)}}$ indicates the core density for Model B (see Section \ref{sec:physical_cond} for details).
\end{table}

\begin{table}
\tiny
\caption{Summary of the previously observed or estimated line surface brightness (SB) ratios \citep{das20}.
\label{table:summary_2}
}
\vskip 0.2 cm
\hskip -0.5 cm
\begin{tabular}{cccc}
\hline
{\bf  Transition ratios}&{\bf  Observed or estimated}&\multicolumn{2}{c}{\bf  Matching zone with $\frac{\zeta}{\zeta_0}$ and $\rm{n_H}$ (cm$^{-3}$)} \\
\cline{3-4}
& {\bf  SB ratios}& {\bf  Model A} & {\bf  Model B ( $\rm{n_H = n_{H(core)}})^a$} \\
\hline
${\dfrac{\rm ArH^+ (2-1)}{\rm ArH^+ (1-0)}}$&2$^b$ $(1-17)^c$ & $\frac{\zeta}{\zeta_0}\sim10^{6-8}$ for $\rm{n_H}\sim10^3$ & $\frac{\zeta}{\zeta_0}\sim10^{0-7}$ for $\rm{n_H} \sim (3.16 \times 10^3) - 10^5$ \\
&&$\frac{\zeta}{\zeta_0}\sim10^{0-7}$ for $\rm{n_H}\sim 10^4$ &
$\frac{\zeta}{\zeta_0}\sim10^{0-6}$ for $\rm{n_H} \sim (3.16 \times 10^5) - 10^6$\\
&&$\frac{\zeta}{\zeta_0}\sim10^{0-5}$ for $\rm{n_H}\sim10^5$ & \\
&& $\frac{\zeta}{\zeta_0}\sim 10^{4-5}$ for $\rm{n_H} \sim 10^{6-7}$ & \\
&&&\\
${\dfrac{\rm ArH^+ (2-1)}{\rm OH^+ (971 \ GHz/308 \ \mu m)}}$& $1.66-3.9^b$ $(1-11)^c$&
$\frac{\zeta}{\zeta_0}\sim10^5$ for $\rm{n_H}\sim10^3$ &
$\frac{\zeta}{\zeta_0}\sim10^{0-4}$ for $\rm{n_H} \sim (3.16\times10^3)-10^4$\\
&& $\frac{\zeta}{\zeta_0}\sim 10^{0-7}$ for $\rm{n_H} \sim 10^4$ &
$\frac{\zeta}{\zeta_0}\sim10^{0-6}$ for $\rm{n_H} \sim 10^{5-6}$\\
&&$\frac{\zeta}{\zeta_0}\sim 10^{6-8}$ for $\rm{n_H} \sim 10^{5-6}$& \\
&&&\\
${\dfrac{\rm ArH^+ (1-0)}{\rm OH^+ (971 \ GHz/308 \ \mu m)}}$&$0.56-0.8^b$ $(0.21-2.91)^c$&
$\frac{\zeta}{\zeta_0}\sim10^6$ for $\rm{n_H}\sim10^{3-4}$ &
$\frac{\zeta}{\zeta_0}\sim10^7$ for $\rm{n_H}\sim(3.16\times10^3)-10^6$\\
&& &
$\frac{\zeta}{\zeta_0} \sim 10^{4-5}$ for $\rm{n_H} \sim 3.16 \times 10^5$\\
&&& $\frac{\zeta}{\zeta_0}\sim10^{0-7}$ for $\rm{n_H}\sim10^6$ \\
&&&\\
${\dfrac{\rm CO \ (4-3,5-4,...,13-12)}{\rm OH^+ (971 \ GHz/308 \ \mu m)}}$&$<<1^d$
&$\frac{\zeta}{\zeta_0}\sim 10^{0-6}$ for $n_H\sim10^{3-4}$ &
$\frac{\zeta}{\zeta_0}\sim10^{0-7}$ for $\rm{n_H}\sim(3.16\times10^3)-10^5$ \\
&&$\frac{\zeta}{\zeta_0}\sim 10^{5-8}$ for $n_H\sim10^{5-7}$ & $\frac{\zeta}{\zeta_0}\sim10^{5-7}$ for $\rm{n_H}\sim10^6$ \\
&&&\\
${\dfrac{\rm C \ I \ (809 \ GHz/ 370 \ \mu m)}{\rm OH^+ (971 \ GHz/308 \ \mu m)}}$&$<1^d$& $\frac{\zeta}{\zeta_0}\sim 3.13 \times 10^2$ for $\rm{n_H}\sim10^7$ &
$\frac{\zeta}{\zeta_0}\sim 10^{0-6}$ for $\rm{n_H}\sim(3.16\times10^3)-10^6$\\
&&& \\
${\dfrac{\rm C \ I \ (492 \ GHz/609 \ \mu m)}{\rm OH^+ (971 \ GHz)/308 \ \mu m}}$&$<1^d$&$\frac{\zeta}{\zeta_0}\sim 10^{3,5,7}$ for $\rm{n_H}\sim10^7$ &
$\frac{\zeta}{\zeta_0}\sim 10^{0-6}$ for $\rm{n_H}\sim(3.16\times10^3)-10^6$\\
&&& \\
$\dfrac{\rm HeH^+ \ (1-0, \ 2010 \ GHz/149 \ \mu m)}{\rm O \ I \ (2053 \ GHz/146 \ \mu m)}$&$<1^e$&
$\zeta/\zeta_0 \sim 10^{0-8}$ for $\rm{n_H}\sim10^{3-7}$ &
$\frac{\zeta}{\zeta_0}\sim 10^{0-8}$ for $\rm{n_H}\sim(3.16\times10^3)-10^6$ \\
&&&\\
${\dfrac{\rm HeH^+ \ (2-1, \ 4020 GHz/74 \ \mu m)}{\rm O \ I \ (2053 \ GHz/146 \ \mu m)}}$&$<1^e$&
$\frac{\zeta}{\zeta_0}\sim 10^{0-8}$ for $\rm{n_H}\sim10^{3-7}$ &
$\frac{\zeta}{\zeta_0}\sim 10^{0-8}$ for $\rm{n_H}\sim(3.16\times10^3)-10^6$ \\
&&&\\
${\dfrac{\rm HeH^+(3-2,5985 \ GHz/50 \ \mu m)}{\rm HeH^+(2-1,4020 \ GHz/74 \ \mu m)}}$&$\sim 0.05^e$& $\frac{\zeta}{\zeta_0}\sim 10^{4-6}$ for $\rm{n_H}\sim10^{6-7}$ & $\frac{\zeta}{\zeta_0}\sim 10^{5}$ for $\rm{n_H}\sim 3.16 \times 10^{3-4}$ \\
&&&$\frac{\zeta}{\zeta_0}\sim 10^{5-6}$ for $\rm{n_H}\sim 10^{5-6}$ \\
&&&\\
${\dfrac{\rm O \ I \ (4758 \ GHz/63 \ \mu m)}{\rm O \ I \ (2053 \ GHz/ 146 \ \mu m)}}$&$16.4-38.7^f$&
$\frac{\zeta}{\zeta_0}\sim 10^8$ for $\rm{n_H}\sim10^{4-5}$ &
$\frac{\zeta}{\zeta_0}\sim 10^{0-8}$ for $\rm{n_H}\sim (3.16\times10^3)-10^6$ \\
&& $\frac{\zeta}{\zeta_0}\sim 10^{0-4}$ for $\rm{n_H} \sim 10^{6-7}$ & \\
&&&\\
${\dfrac{\rm O \ I \ (2053 \ GHz/ 146 \ \mu m)}{\rm C \ II \ (1897 \ GHz/158 \ \mu m)}}$&$0.125-0.323^f$&
$\frac{\zeta}{\zeta_0}\sim10^{5-8}$ for $\rm{n_H}\sim10^{3-4}$ &
$\frac{\zeta}{\zeta_0}\sim 10^{0-4}$ for $\rm{n_H}\sim(3.16 \times 10^3)-10^6$ \\
\hline
\end{tabular} \\
\vskip 0.2 cm
{\bf Note:} The most suitable values of
$\rm{n_H}$ and $\dfrac{\zeta}{\zeta_0}$ to explain the listed SB values are also pointed out. \\
$^a$ $\rm{n_H = n_{H(core)}}$ indicates the core density for Model B (see Section \ref{sec:physical_cond} for details). \\
$^b$ \cite{prie17} and references therein.\\
$^c$ Taking the ratio with the observed maximum and minimum SB between the two transitions noted in Table \ref{table:summary_1}. \\
$^d$ \cite{prie17}; weak enough to be consistent with the observation. \\
$^e$ Prediction from the model of \cite{prie17}. \\
$^f$ \cite{gome12}.
\end{table}

\begin{figure}
\begin{center}
\includegraphics[width=0.7\textwidth]{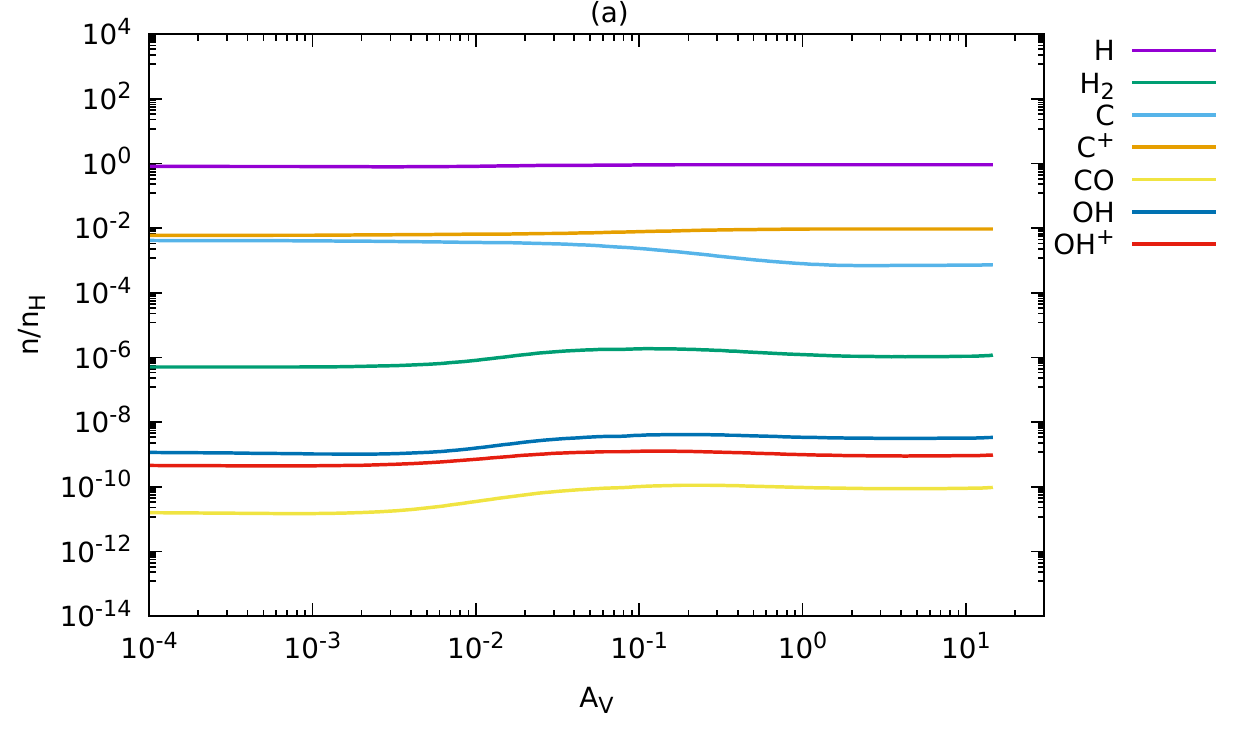}
\includegraphics[width=0.7\textwidth]{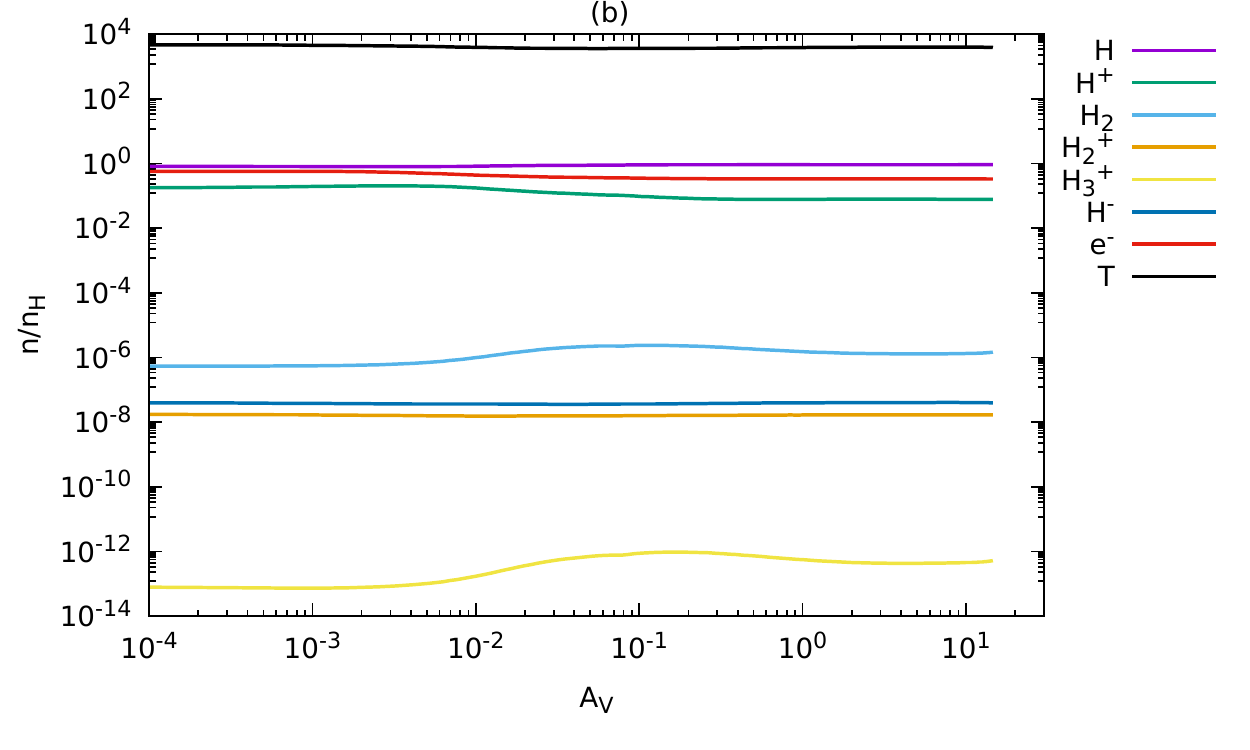}
\caption{Abundance variation of simple species with $A_V$ considering $\rm{n_H} = 2.00 \times 10^4$ cm$^{-3}$ and $\zeta/\zeta_0 = 9.07 \times 10^{6}$ \citep[Model A1;][]{das20}.}
\label{fig:abun_best}
\end{center}
\end{figure}

\begin{figure}
\begin{center}
\includegraphics[width=0.9\textwidth]{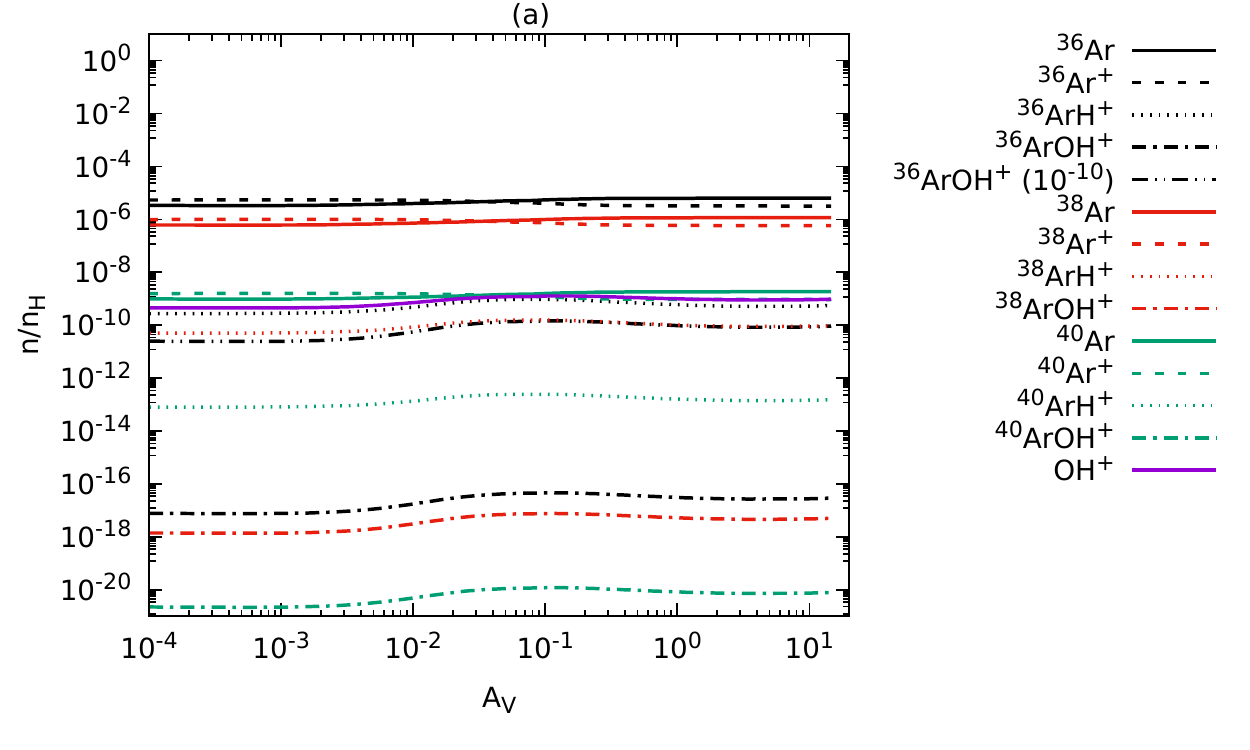}
\includegraphics[width=0.9\textwidth]{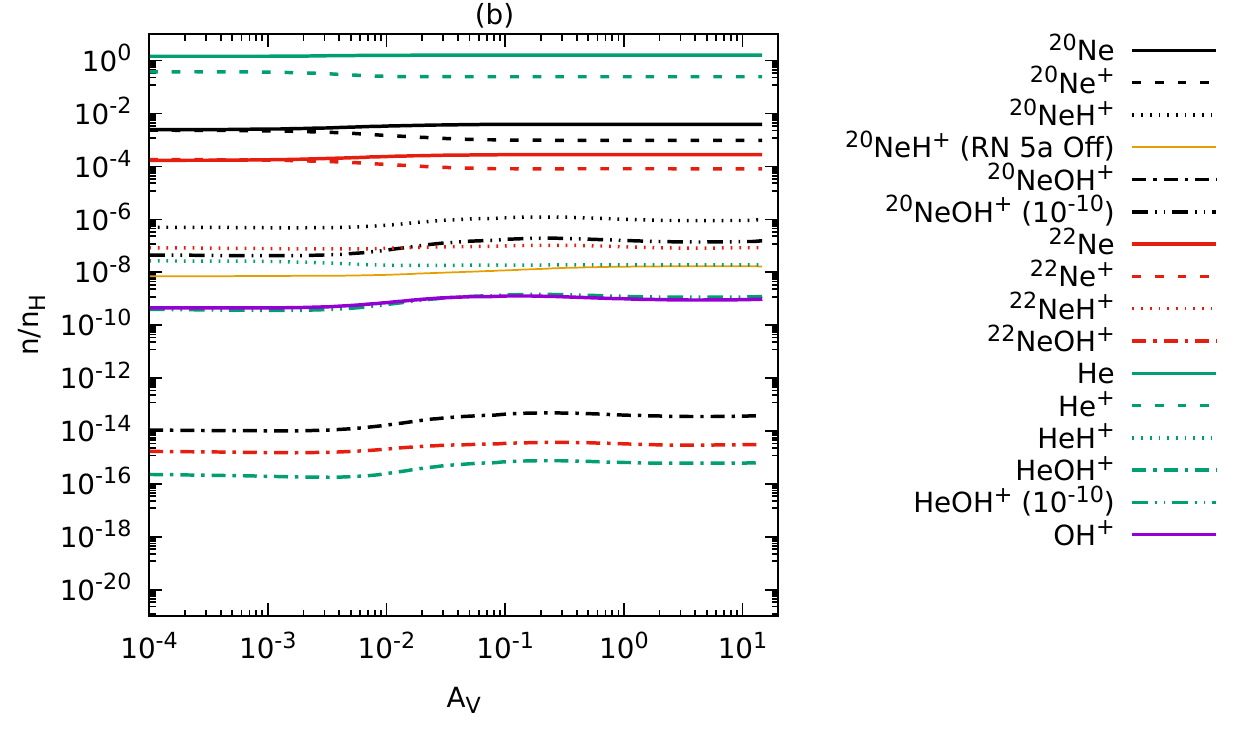}
\caption{Abundance variation of all the hydride and hydroxyl cations considered in this work by considering $\rm{n_H} = 2.00 \times 10^4$ cm$^{-3}$ and $\zeta/\zeta_0 = 9.07 \times 10^{6}$ \citep[Model A1;][]{das20}. In the upper panel (a) Ar-related species are shown, and in the lower panel (b) the cases of Ne and He are shown. The abundance variation of OH$^+$ is shown in both the panels for comparison. The abundances of $^{36}$ArOH$^+$, $^{20}$NeOH$^+$, and HeOH$^+$ by considering the upper limit of their formation rate ($\sim 10^{-10}$ cm$^3$ s$^{-1}$) are noted [XOH$^+$ ($10^{-10}$)]. The abundance profile of $^{20}$NeH$^+$ is also shown when reaction 5a of the Ne network is off.}
\label{fig:abun2}
\end{center}
\end{figure}

In the favorable zone of Model A, we further consider $\rm{n_H} = 2.00 \times 10^4$ cm$^{-3}$ and $\zeta = 9.07 \times 10^6 \zeta_0$ as Model A1 to suitably match the absolute SB of the two transitions of $^{36}$ArH$^+$ (242 and 485 $\mu$m) and the 308 $\mu$m transition of OH$^+$ simultaneously, and $\rm{n_H} = 3.16 \times 10^4$ cm$^{-3}$ and $\zeta = 4.55 \times 10^6 \zeta_0$ as Model A2 to suitably match the absolute SB of $\rm{H_2}$ of 2.12 $\mu$m separately. Unless otherwise stated, Model A1 is always used in all the cases reported throughout this Chapter.
Figure \ref{fig:abun_best} shows the abundance variation of the simple species, electron, and the electron temperature of the Crab. The Figure depicts that the
temperature of the Crab region is $4000$ K and the electron abundance is $>0.1$, which is in line
with the observation \citep{barl13}. A suitably high fractional abundance of H$_2$ ($\sim 10^{-6}$)
is observed, which can explain the H$_2$ SB in the knots of the Crab. Additionally, we show the abundances of H$_2^+$ and H$_3^+$. Figure \ref{fig:abun2}a shows the abundances of Ar-related species along with their isotopologs. Figure \ref{fig:abun2}b shows the abundances of the He- and Ne- related (and its isotopologs) species. We do not consider any fractionation reaction between the isotopologs of Ar and Ne. Due to this reason, the elemental abundance ratio is reflected in the molecular abundances of the various isotopologs. OH$^+$ had been identified in the emitting knots of the Crab. So, the observability of the species may be compared to the OH$^+$ abundance. Both the
panels of Figure \ref{fig:abun2} show the OH$^+$ abundance to understand the fate of other chemical species for future identification in the Crab knots. Figure \ref{fig:abun2}(a) and (b) depict that the abundances of $^{36}$ArH$^+$, $^{20}$NeH$^+$, and HeH$^+$ are higher than that of OH$^+$.
Even in the absence of reaction 5a, we obtain a comparable abundance of $^{20}$NeH$^+$ with OH$^+$ (see Figure \ref{fig:abun2}b). Thus $^{20}$NeH$^+$ and HeH$^+$ could have been observed in
the Crab knots. However, even with the upper limit of the rate coefficient, we always obtain
a lower abundance of hydroxyl cations ($^{36}$ArOH$^+$, $^{20}$NeOH$^+$, and HeOH$^+$) than OH$^+$.

Similarly, the abundance profiles obtained with Model B are shown in Figures \ref{fig:abun1-rich} and \ref{fig:abun2-rich}.
For this case, we obtain a much higher electron temperature ($>10000$ K) that can yield a better estimation for the various atomic transitions listed in Table \ref{table:comp-value}.

\begin{figure}
\begin{center}
\includegraphics[width=0.8\textwidth]{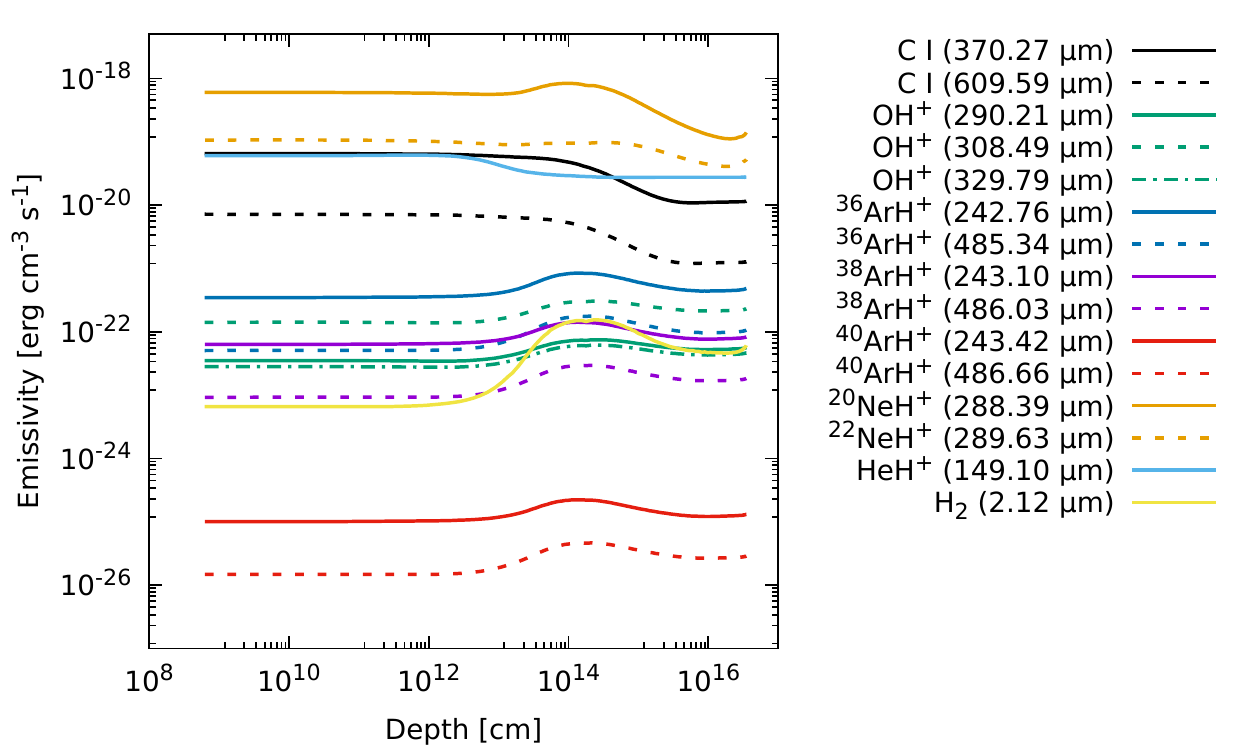}
\caption{Emissivity of some of the strongest transitions that fall in the range of the frequency limit of Herschel's SPIRE and PACS spectrometer, and SOFIA with respect to the depth into the filament by considering $\rm{n_H} = 2.00 \times 10^4$ cm$^{-3}$ and $\zeta/\zeta_0 = 9.07 \times 10^{6}$ \citep[Model A1;][]{das20}.}
\label{fig:emis1}
\end{center}
\end{figure}

\begin{figure}
\begin{center}
\includegraphics[width=0.8\textwidth]{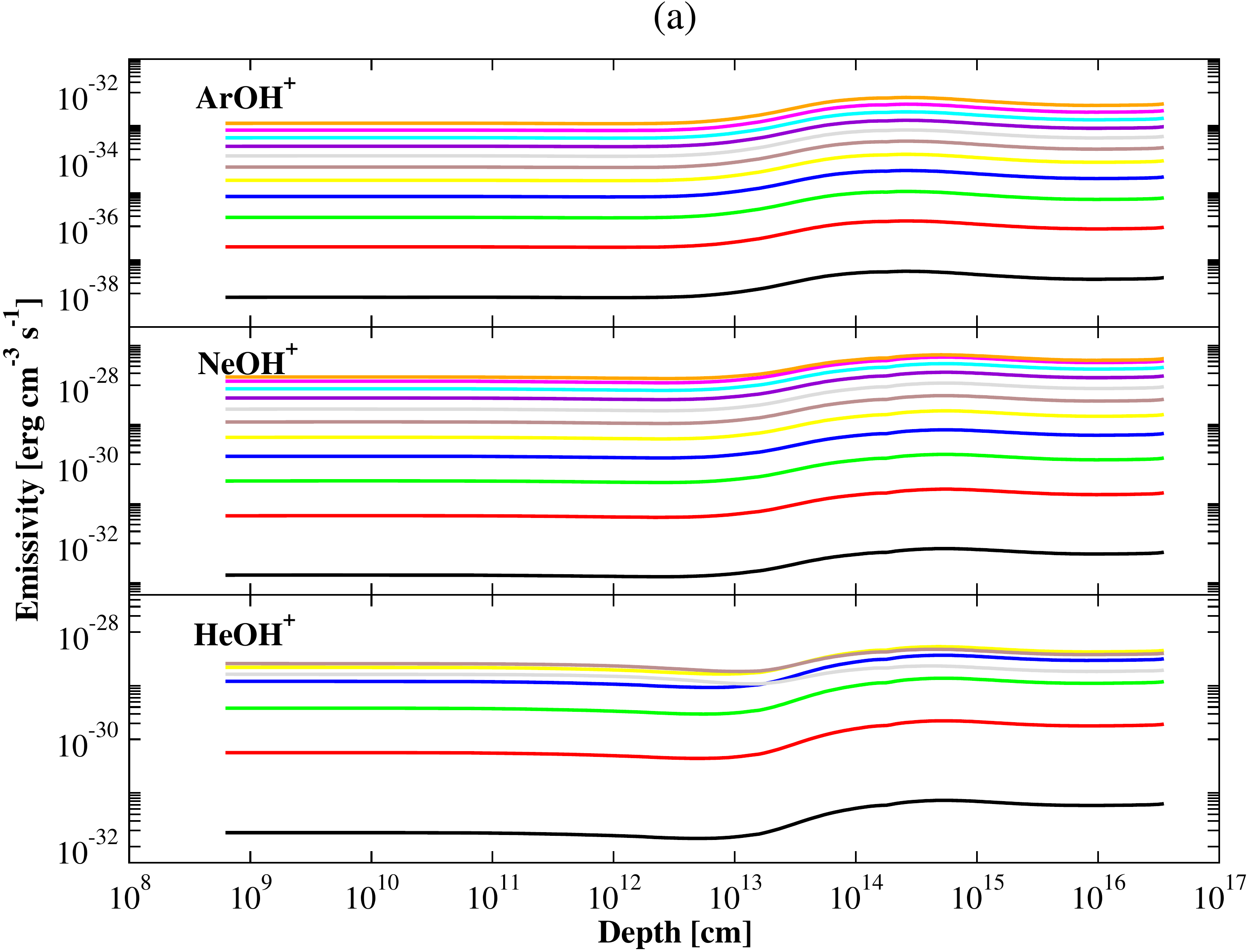}
\vskip 0.5cm
\includegraphics[width=0.8\textwidth]{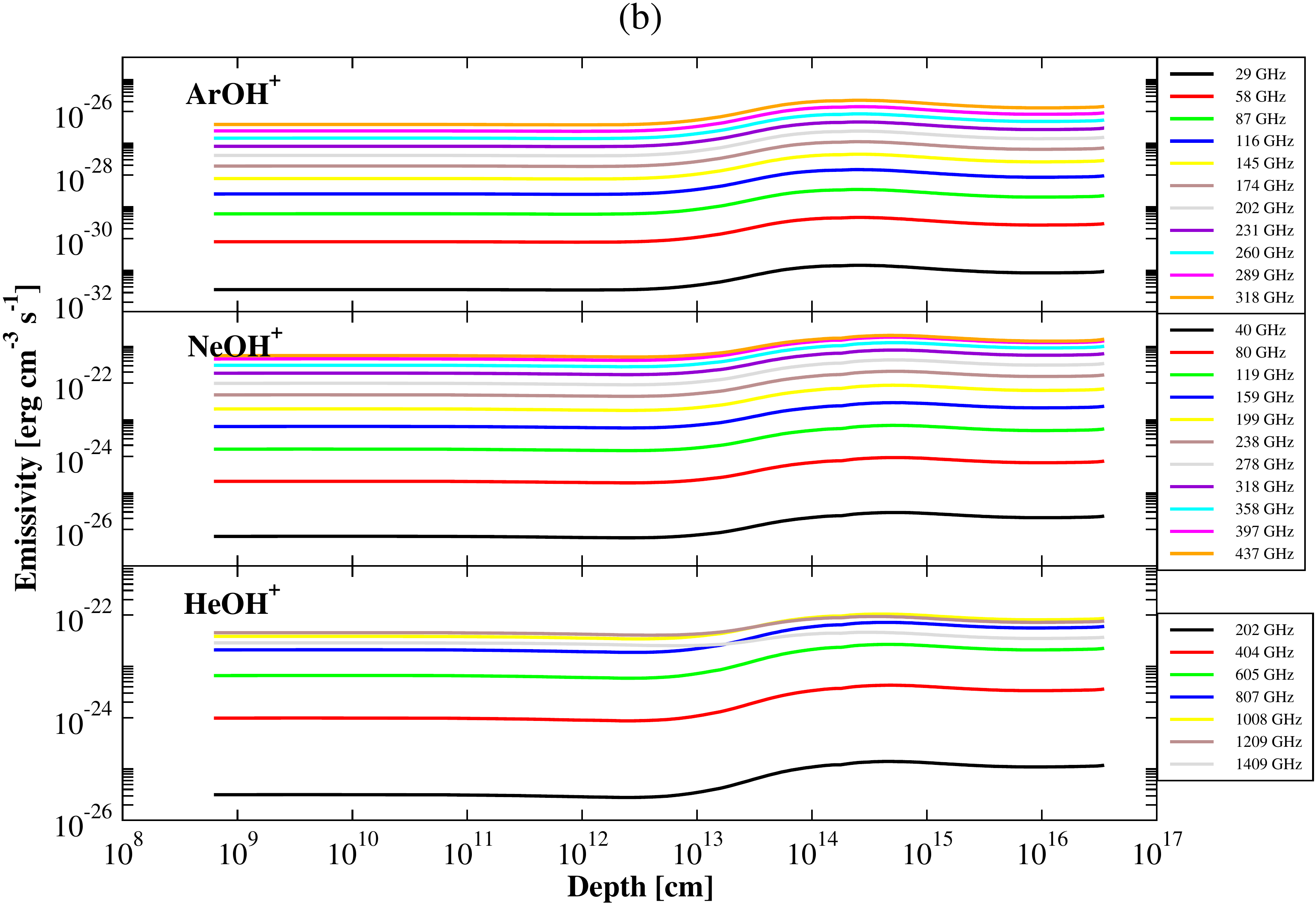}
\caption{Calculated emissivity of various XOH$^+$ transitions (X=$^{36}$Ar, $^{20}$Ne, He) noted in Table \ref{table:optical} lying in the frequency limit of Herschel's SPIRE and PACS spectrometer, SOFIA, ALMA, VLA, IRAM 30m, and NOEMA by considering $\rm{n_H} = 2.00 \times 10^4$ cm$^{-3}$ and $\zeta/\zeta_0 = 9.07 \times 10^{6}$ \citep[Model A1;][]{das20}. The upper panel (a) shows the emissivity considering the formation rates following \cite{bate83} mentioned in Section \ref{rad_ass}, whereas the lower panel (b) considers the upper limit of $\sim 10^{-10}$ cm$^3$ s$^{-1}$.}
\label{fig:emis2}
\end{center}
\end{figure}

\begin{table}
\scriptsize
\caption{Comparison between the observed and our modeling results \citep{das20}.
\label{table:comp-value}
}
\vskip 0.2 cm
\hskip -0.7 cm
\begin{tabular}{ccccccc}
\hline
{\bf Atomic}& \multicolumn{2}{c}{\bf  Flux (erg cm$^{-2}$ s$^{-1}$)} & {\bf  Predicted/observed} & \multicolumn{3}{c}{\bf  Predicted/observed ratio} \\
{\bf lines}&{\bf Observed} & {\bf  Dereddened} & {\bf ratio$^a$} & {\bf Model A1} & {\bf Model A2} & {\bf Model B} \\
\hline
$\rm{H_2}$ $\lambda 2.12$ $\mu$m &$6.5\times 10^{-15a}$ &$7.6\times 10^{-15a}$&1.1$^a$& $5.3\times10^{-4}$ & 0.080 & 0.022 \\
...&$(4.05 \times 10^{-15b})$&...&...& $(8.5\times10^{-4})^c$ & (0.127)$^c$ & (0.036)$^c$ \\
O\,{\sc ii} $\lambda 3727$ &$7.7\times 10^{-14a}$&$6.7\times 10^{-13a}$&1.0$^a$&{  0.17}&{  0.005}&{  1.053}\\
Ne\,{\sc iii} $\lambda 3869$ &$1.7\times 10^{-14a}$&$1.4\times 10^{-13a}$&1.1$^a$& {  0.004} &{  1.7$\times10^{-4}$}&{  1.144}\\
H\,{\sc i} $\lambda 4340$ &$4.4\times 10^{-15a}$&$2.9\times 10^{-14a}$&2.0$^a$&{  20.728}&{  16.056}&{  4.330}\\
He\,{\sc i} $\lambda 4471$ &$1.7\times 10^{-15a}$&$1.0\times 10^{-14a}$&1.2$^a$&{  187.452}&{  189.495}& {  13.029}\\
He\,{\sc ii} $\lambda 4686$ &$2.9\times 10^{-15a}$&$1.7\times 10^{-14a}$&1.2$^a$&{  1.697}&{  0.965}&{  1.013}\\
H\,{\sc i} $\lambda 4861$ &$1.04\times 10^{-14a}$&$5.4\times 10^{-14a}$&2.3$^a$&{  18.826}&{  14.675}&{  3.931}\\
O\,{\sc iii} $\lambda 5007$ &$7.6\times 10^{-14a}$&$3.7\times 10^{-13a}$&1.2$^a$&{  $1.8\times10^{-5}$}&{  $1.13\times10^{-6}$}&{  0.958}\\
N\,{\sc i} $\lambda 5198$ &$1.8\times 10^{-15a}$&$8.1\times 10^{-15a}$&1.6$^a$&{  7.261}&{  1.096}&{  1.301}\\
He\,{\sc i} $\lambda 5876$ &$6.8\times 10^{-15a}$&$2.5\times 10^{-14a}$&1.6$^a$&{  125.885}&{  128.481}&{  8.751}\\
O\,{\sc i} $\lambda 6300$ &$5.3\times 10^{-14a}$&$1.8\times 10^{-13a}$&0.7$^a$&{  7.120}&{  0.981}&{  0.357}\\
H\,{\sc i} $\lambda 6563$ &$5.0\times 10^{-14a}$&$1.6\times 10^{-13a}$&2.5$^a$&{  11.400}&{  9.020}&{  2.384}\\
N\,{\sc ii} $\lambda 6584$ &$9.7\times 10^{-14a}$&$3.1\times 10^{-13a}$&0.5$^a$&{  0.357}&{  0.029}&{  0.296}\\
S\,{\sc ii} $\lambda 6716$ &$9.0\times 10^{-14a}$&$2.8\times 10^{-13a}$&0.8$^a$&  ... & ... &{  0.468}\\
S\,{\sc ii} $\lambda 6731$ &$1.2\times 10^{-13a}$&$3.6\times 10^{-13a}$&0.9$^a$& ...  & ... &{  0.582}\\
He\,{\sc i} $\lambda 7065$ &$2.6\times 10^{-15a}$&$7.6\times 10^{-15a}$&1.3$^a$&{  204.777}&{  200.769}&{  10.875}\\
Ar\,{\sc iii} $\lambda 7136$ &$1.3\times 10^{-14a}$&$3.7\times 10^{-14a}$&1.0$^a$&{  0.285}&{  0.026}&{  0.542}\\
Fe\,{\sc ii} $\lambda 7155$ &$2.7\times 10^{-15a}$&$7.6\times 10^{-15a}$&2.0$^a$&  ... &  ... &{  1.608}\\
O\,{\sc ii} $\lambda 7320$ &$3.6\times 10^{-15a}$&$9.7\times 10^{-15a}$&2.3$^a$& {  0.017} &{  $1.3\times10^{-4}$}&{  0.849}\\
\hline
O\,{\sc iii} (52 $\mu$m) &$4.2\times 10^{-15d}$& ...  & ...  & {  0.001} &{  $7.3\times10^{-4}$}& {  1.629}\\
N\,{\sc iii} (57 $\mu$m) &$4.0\times 10^{-16d}$& ...  & ...  & {  $3.0\times10^{-4}$} &{  $1.1\times10^{-4}$}& {  1.610} \\
O\,{\sc i} (63 $\mu$m) &$1.7\times 10^{-15d}$& ...  & ...  & {  1089.851} &{  1651.994}& {  109.574} \\
O\,{\sc iii} (88 $\mu$m) &$3.6\times 10^{-15d}$& ...  & ...  & {  $2.1\times10^{-4}$} &{  $1.2\times10^{-4}$}& {  0.613} \\
N\,{\sc ii} (122 $\mu$m) &$1.2\times 10^{-16d}$& ...  &...   & {  9.104} &{  4.311}& {  1.451} \\
O\,{\sc i} (145 $\mu$m) &$8.0\times 10^{-17d}$&  ... & ...  & {  1742.984} &{  2981.480}& {  83.172} \\
C\,{\sc ii} (158 $\mu$m) &$2.9\times 10^{-16d}$& ...  & ...  & {  742.966}&{  877.732}& {  16.426} \\
\hline
\end{tabular} \\
\vskip 0.2cm
{\bf Note:}\\
$^a$ \cite{rich13}.\\
$^b$ \cite{loh11}.\\
$^c$ Taking the ratio with the observed values of \cite{loh11}. \\
$^d$ \cite{gome12}.
\end{table}

\begin{table}
\tiny
\caption{Strongest transitions falling in the range of Herschel's SPIRE and PACS spectrometer, SOFIA, ALMA, VLA, IRAM 30m, and NOEMA considering $\rm{n_H} = 2.00 \times 10^4$ cm$^{-3}$ and $\zeta/\zeta_0 = 9.07 \times 10^{6}$ \citep[Model A1;][]{das20}.
\label{table:optical}}
\vskip 0.2 cm
\hskip -1.5 cm
\begin{tabular}{ccccccc}
\hline
{\bf  Species} & {\bf Transitions} & {\bf  $\rm{E_U}$ (K)}& {\bf  Frequency (GHz) ($\mu$m)} & {\bf  Total column} & {\bf  Optical depth ($\tau$)}& {\bf  Surface brightness} \\
&&&& {\bf density (cm$^{-2}$)} & & {\bf (erg cm$^{-2}$ s$^{-1}$ sr$^{-1}$)} \\
\hline
$^{36}$ArH$^+$&J $=1\rightarrow0$ &29.64&617.52 (485.34)&{  $3.80\times10^{11}$}& { $2.557\times10^{-2}$}& $2.84\times10^{-7}$  \\
&&&&&& ($(2.2-9.9) \times10^{-7})^a$ \\
$^{36}$ArH$^+$&J $=2\rightarrow1$ &88.89&1234.60 (242.76)&{  $3.80\times10^{11}$}& { $7.547\times10^{-3}$}& { $1.29\times10^{-6}$ } \\
&&&&&& ($(1.0-3.8) \times10^{-6})^a$ \\
$^{36}$ArH$^+$& {  J $=3\rightarrow2$} & {  177.71} & {  1850.78 (161.94)} & {  $3.80\times10^{11}$} &{ $4.258\times10^{-4}$} & {  $1.15\times10^{-6}$} \\
$^{36}$ArH$^+$& {  J $=4\rightarrow3$} & {  296.04} & {  2465.62 (121.56)} & {  $3.80\times10^{11}$} &  { $5.405\times10^{-5}$}& {  $7.76\times10^{-7}$} \\
$^{36}$ArH$^+$& {  J $=5\rightarrow4$} & {  443.80} & {  3078.68 (97.35)} & {  $3.80\times10^{11}$} &  { $1.287\times10^{-5}$}& {  $3.86\times10^{-7}$} \\
$^{36}$ArH$^+$& {  J $=6\rightarrow5$} & {  620.86} & {  3689.50 (81.23)} & {  $3.80\times10^{11}$} &  { $1.203\times10^{-6}$}& {  $8.63\times10^{-8}$} \\
$^{36}$ArH$^+$& {  J $=7\rightarrow6$} & {  827.12} & {  4297.65 (69.74)} & {  $3.80\times10^{11}$} &  { $4.792\times10^{-8}$}& {  $1.32\times10^{-8}$}\\
\hline
$^{38}$ArH$^+$&J $=1\rightarrow0$ &29.39&616.65 (486.03)&{ $6.57\times10^{10}$}&{ $4.431\times10^{-3}$} & {  $4.92\times10^{-8}$}\\
$^{38}$ArH$^+$&J $=2\rightarrow1$ &88.14&1232.85 (243.10)&{ $6.57\times10^{10}$}& { $1.297\times10^{-3}$}& {  $2.24\times10^{-7}$}\\
$^{38}$ArH$^+$& {  J $=3\rightarrow2$} & {  176.23} & {  1848.16 (162.17)} & { $6.57\times10^{10}$} &{ $7.320\times10^{-5}$}  & {  $2.00\times10^{-7}$} \\
$^{38}$ArH$^+$& {  J $=4\rightarrow3$} & {  293.57} & {  2462.13 (121.73)} & { $6.57\times10^{10}$} &{ $9.492\times10^{-6}$}  & {  $1.36\times10^{-7}$} \\
$^{38}$ArH$^+$& {  J $=5\rightarrow4$} & {  440.09} & {  3074.32 (97.49)} & { $6.57\times10^{10}$} &{ $2.255\times10^{-6}$} & {  $6.76\times10^{-8}$} \\
$^{38}$ArH$^+$& {  J $=6\rightarrow5$} & {  615.68} & {  3684.29 (81.35)} & { $6.57\times10^{10}$}&{ $2.080\times10^{-7}$}& {  $1.50\times10^{-8}$} \\
$^{38}$ArH$^+$& {  J $=7\rightarrow6$} & {  820.22} & {  4291.58 (69.84)} & { $6.57\times10^{10}$}&{ $8.343\times10^{-9}$}  & {  $2.33\times10^{-9}$}\\
\hline
$^{40}$ArH$^+$&J $=1\rightarrow0$ &{  29.35}&615.86 (486.66)&{  $1.04\times10^{8}$}& { $7.012\times10^{-6}$}& {  $7.76\times10^{-11}$}\\
$^{40}$ArH$^+$&J $=2\rightarrow1$ &{  88.03}&1231.27 (243.42)&{  $ 1.04\times10^{8}$}& { $2.061\times10^{-6}$}& {  $3.35\times10^{-10}$} \\
$^{40}$ArH$^+$& {  J $=3\rightarrow2$} & {  176.00} & {  1845.79 (162.38)} & {  $ 1.04\times10^{8}$} &  { $1.160\times10^{-7}$}& {  $3.17\times10^{-10}$} \\
$^{40}$ArH$^+$& {  J $=4\rightarrow3$} & {  293.20} & {  2458.98 (121.88)} & {  $ 1.04\times10^{8}$} &  { $1.516\times10^{-8}$}& {  $2.15\times10^{-10}$} \\
$^{40}$ArH$^+$& {  J $=5\rightarrow4$} & {  439.53} & {  3070.39 (97.61)} & {  $ 1.04\times10^{8}$} &  { $3.578\times10^{-9}$}& {  $1.07\times10^{-10}$} \\
$^{40}$ArH$^+$& {  J $=6\rightarrow5$} & {  614.890} & {  3679.58 (81.45)} & {  $ 1.04\times10^{8}$} &  { $3.328\times10^{-10}$}& {  $2.38\times10^{-11}$} \\
$^{40}$ArH$^+$& {  J $=7\rightarrow6$} & {  819.17} & {  4286.11 (69.93)} & {  $ 1.04\times10^{8}$} &  { $1.323\times10^{-11}$}& {  $3.71\times10^{-12}$} \\
\hline
$^{20}$NeH$^+$&J $=1\rightarrow0$ &49.53&1039.26 (288.39)&{  $6.51\times10^{14}$ }& $4.246\times10^{1}$  & $4.20\times10^{-4}$ \\
&&&& $(1.16\times10^{13})^b$ & $(2.175)^b$ & $(3.97\times10^{-5})^b$ \\
$^{20}$NeH$^+$& {  J $=2\rightarrow1$} & {  148.50} & {  2076.57 (144.33)} & {  $6.51\times10^{14}$ } &{ $4.022\times10^{1}$ }  & {  $2.41\times10^{-3}$ } \\
&&&& $(1.16\times10^{13})^b$ & $(1.352\times10^{-1})^b$ & $(6.74\times10^{-5})^b$ \\
$^{20}$NeH$^+$& {  J $=3\rightarrow2$} & {  296.72} & {  3110.02 (96.37)} & {  $6.51\times10^{14}$ } &  { $4.794$ }& {  $3.67\times10^{-3}$ } \\
&&&& $(1.16\times10^{13})^b$ & $(2.352\times10^{-3})^b$ & $(4.95\times10^{-5})^b$ \\
$^{20}$NeH$^+$& {  J $=4\rightarrow3$} & {  493.92} & {  4137.67 (72.43)} & {  $6.51\times10^{14}$ } & { $8.114\times10^{-2}$ }& {  $2.40\times10^{-3}$ } \\
&&&&$(1.16\times10^{13})^b$ & $(3.061\times10^{-4})^b$ & $(3.44\times10^{-5})^b$ \\
$^{20}$NeH$^+$& {  J $=5\rightarrow4$} & {  739.73} & {  5157.61 (58.11)} & {  $6.51\times10^{14}$ } &  { $4.225\times10^{-3}$ }& {  $1.26\times10^{-3}$ } \\
&&&& $(1.16\times10^{13})^b$ &$(8.033\times10^{-5})^b$& $(9.92\times10^{-6})^b$ \\
$^{20}$NeH$^+$& {  J $=6\rightarrow5$} & {  1033.68} & {  6167.92 (48.59)} & {  $6.51\times10^{14}$ } &  { $6.559\times10^{-4}$ }& {  $4.36\times10^{-4}$ }\\
&&&& $(1.16\times10^{13})^b$ & $(2.499\times10^{-6})^b$ & $(7.74\times10^{-7})^b$ \\
$^{20}$NeH$^+$& {  J $=7\rightarrow6$} & {  1375.24} & {  7166.70 (41.82)} & {  $6.51\times10^{14}$ } &  { $3.649\times10^{-5}$ }& {  $5.49\times10^{-5}$ }\\
&&&& $(1.16\times10^{13})^b$ & $(4.035\times10^{-8})^b$ &  $(6.29\times10^{-8})^b$ \\
\hline
$^{22}$NeH$^+$&J $=1\rightarrow0$ &49.32&1034.79(289.63)&{  $5.94\times10^{13}$}& { $8.939$}& {  $1.34\times10^{-4}$} \\
$^{22}$NeH$^+$& {  J $=2\rightarrow1$} & {  147.86} & {  2067.67 (144.95)} & {  $5.94\times10^{13}$}&{ $1.771$}  & {  $4.00\times10^{-4}$} \\
$^{22}$NeH$^+$& {  J $=3\rightarrow2$} & {  295.45} & {  3096.70 (96.78)} & {  $5.94\times10^{13}$} & { $2.453\times10^{-2}$}& {  $3.11\times10^{-4}$} \\
$^{22}$NeH$^+$& {  J $=4\rightarrow3$} & {  491.80} & {  4119.99 (72.74)} & {  $5.94\times10^{13}$} & { $1.659\times10^{-3}$}& {  $2.05\times10^{-4}$} \\
$^{22}$NeH$^+$& {  J $=5\rightarrow4$} & {  736.56} & {  5135.64 (58.36)} & {  $5.94\times10^{13}$} & { $4.031\times10^{-4}$}& {  $8.45\times10^{-5}$} \\
$^{22}$NeH$^+$& {  J $=6\rightarrow5$} & {  1029.28} & {  6141.73 (48.80)} & {  $5.94\times10^{13}$} &  { $2.677\times10^{-5}$}& {  $1.08\times10^{-5}$}\\
$^{22}$NeH$^+$& {  J $=7\rightarrow6$} & {  1369.39} & {  7136.41 (41.99)} & {  $5.94\times10^{13}$} &  { $4.307\times10^{-7}$}& {  $7.22\times10^{-7}$} \\
\hline
HeH$^+$&J$=1\rightarrow0$ &95.80&2010.18 (149.10)&{  $1.33\times10^{13}$}& { $8.473\times10^{-1}$}& {  $7.68\times10^{-5}$}\\
HeH$^+$& {  J $=2\rightarrow1$} & {  286.86} & {  4008.73 (74.76)} & {  $1.33\times10^{13}$} &{ $7.901\times10^{-3}$} & {  $6.51\times10^{-5}$} \\
HeH$^+$& {  J $=3\rightarrow2$} & {  572.06} & {  5984.14 (50.08)} & {  $1.33\times10^{13}$} &{ $2.080\times10^{-4}$}  & {  $6.69\times10^{-6}$}\\
HeH$^+$& {  J $=4\rightarrow3$} & {  949.76} & {  7925.15 (37.82)} & {  $1.33\times10^{13}$} &{ $1.454\times10^{-6}$} & {  $3.18\times10^{-7}$}\\
HeH$^+$& {  J $=5\rightarrow4$} & {  1417.82} & {  9820.88 (30.52)} & {  $1.33\times10^{13}$} & { $1.289\times10^{-8}$} & {  $1.22\times10^{-8}$}\\
HeH$^+$& {  J $=6\rightarrow5$} & {  1973.57} & {  11660.90 (25.70)} & {  $1.33\times10^{13}$} &{ $7.291\times10^{-11}$} & {  $2.77\times10^{-9}$} \\
HeH$^+$& {  J $=7\rightarrow6$} & {  2613.89} & {  13435.35 (22.31)} & {  $1.33\times10^{13}$} &{ $1.356\times10^{-12}$}  & {  $1.47\times10^{-9}$} \\
\hline
OH$^+$&J $=2\rightarrow1$ (F$=5/2\rightarrow3/2$)&46.64&971.80 (308.41)& {  $6.53\times10^{11}$}& { $2.370\times10^{-2}$}& {  $6.17\times10^{-7}$ } \\
&&&&&& $[(3.4-10.3) \times10^{-7}]^a$ \\
\hline
\end{tabular}
\end{table}

\begin{table}
\tiny
\hskip -2.0 cm
\begin{tabular}{ccccccc}
\hline
{\bf  Species} & {\bf Transitions} & {\bf  $\rm{E_U}$ (K)}& {\bf  Frequency (GHz) ($\mu$m)} & {\bf  Total column} & {\bf  Optical depth ($\tau$)}& {\bf  Surface brightness} \\
&&&& {\bf density (cm$^{-2}$)} & & {\bf (erg cm$^{-2}$ s$^{-1}$ sr$^{-1}$0} \\
\hline
{  $^{36}$ArOH$^+$} & { J $=1\rightarrow0$ (K$_- = 1\rightarrow0$)}& {  1.38} & {  28.94 (10358)} & {  $6.19\times10^{10}$} & { $6.617\times10^{-10}$} & {  $7.76\times10^{-23}$} \\
{  $^{36}$ArOH$^+$} & { J $=2\rightarrow1$ (K$_- = 2\rightarrow1$)}& {  4.14} & {  57.88 (5179)} & {  $6.19\times10^{10}$} & { $2.740\times10^{-9}$} & {  $2.48\times10^{-21}$} \\
{  $^{36}$ArOH$^+$} &{  J $=3\rightarrow2$ (K$_- = 3\rightarrow2$)}& {  8.28} & {  86.82 (3453)} & {  $6.19\times10^{10}$} &{ $6.561\times10^{-9}$} & {  $1.88\times10^{-20}$}\\
{  $^{36}$ArOH$^+$} & { J $=4\rightarrow3$ (K$_- = 4\rightarrow3$)}& {  13.79} & {  115.76 (2590)} & {  $6.19\times10^{10}$} & { $1.225\times10^{-8}$} & {  $7.91\times10^{-20}$} \\
{  $^{36}$ArOH$^+$} & { J $=5\rightarrow4$ (K$_- = 5\rightarrow4$)}& {  20.69} & {  144.70 (2072)} & {  $6.19\times10^{10}$} & { $2.183\times10^{-8}$} & {  $2.41\times10^{-19}$} \\
{  $^{36}$ArOH$^+$} & { J $=6\rightarrow5$ (K$_- = 6\rightarrow5$)}& {  28.96} & {  173.63 (1727)} & {  $6.19\times10^{10}$} & { $3.602\times10^{-8}$} & {  $5.97\times10^{-19}$}\\
{  $^{36}$ArOH$^+$} & { J $=7\rightarrow6$ (K$_- = 7\rightarrow6$)}& {  38.62} & {  202.56 (1480)} & {  $6.19\times10^{10}$} & { $5.809\times10^{-8}$}& {  $1.29\times10^{-18}$}\\
{  $^{36}$ArOH$^+$} & { J $=8\rightarrow7$ (K$_- = 8\rightarrow7$)}& {  49.65} & {  231.48 (1295)} & {  $6.19\times10^{10}$} & { $5.900\times10^{-8}$} & {  $2.50\times10^{-18}$}\\
{  $^{36}$ArOH$^+$} & { J $=9\rightarrow8$ (K$_- = 9\rightarrow8$)}& {  62.06} & {  260.40 (1151)} & {  $6.19\times10^{10}$} & { $1.088\times10^{-7}$} & {  $4.49\times10^{-18}$}\\
{  $^{36}$ArOH$^+$} & { J $=10\rightarrow9$ (K$_- = 10\rightarrow9$)}& {  75.85} & {  289.32 (1036)} & {  $6.19\times10^{10}$} & { $1.845\times10^{-7}$} & {  $7.57\times10^{-18}$}\\
$^{36}$ArOH$^+$&J $=11\rightarrow10$ (K$_- = 11\rightarrow10$)&{  91.02} & {  318.22 (942)}&{  $6.19\times10^{10}$}&{ $4.923\times10^{-7}$} & {  $1.21\times10^{-17}$} \\
\hline
{  $^{20}$NeOH$^+$} &{  J $=1\rightarrow0$ (K$_- = 1\rightarrow0$)}& {  1.89} & {  39.76 (7540)} & {  $1.02\times10^{14}$} &   & {  $1.56\times10^{-17}$}\\
{  $^{20}$NeOH$^+$} &{  J $=2\rightarrow1$ (K$_- = 2\rightarrow1$)}& {  5.68} & {  79.52 (3770)} & {  $1.02\times10^{14}$} &   & {  $4.97\times10^{-16}$} \\
{  $^{20}$NeOH$^+$} &{  J $=3\rightarrow2$ (K$_- = 3\rightarrow2$)}& {  11.37} & {  119.27 (2514)} & {  $1.02\times10^{14}$} & { $3.914\times10^{-3}$} & {  $3.78\times10^{-15}$} \\
{  $^{20}$NeOH$^+$} &{  J $=4\rightarrow3$ (K$_- = 4\rightarrow3$)}& {  18.95} & {  159.01 (1885)} & {  $1.02\times10^{14}$} &{ $1.895\times10^{-2}$} & {  $1.58\times10^{-14}$}\\
{  $^{20}$NeOH$^+$} &{  J $=5\rightarrow4$ (K$_- = 5\rightarrow4$)}& {  28.42} & {  198.75 (1508)} & {  $1.02\times10^{14}$} & { $5.306\times10^{-2}$} & {  $4.77\times10^{-14}$} \\
{  $^{20}$NeOH$^+$} &{  J $=6\rightarrow5$ (K$_- = 6\rightarrow5$)}& {  39.78} & {  238.47 (1257)} & {  $1.02\times10^{14}$} &{ $1.047\times10^{-1}$} & {  $1.16\times10^{-13}$}\\
{  $^{20}$NeOH$^+$} &{  J $=7\rightarrow6$ (K$_- = 7\rightarrow6$)}& {  53.04} & {  278.18 (1078)} & {  $1.02\times10^{14}$} & { $1.603\times10^{-1}$} & {  $2.41\times10^{-13}$}\\
{  $^{20}$NeOH$^+$} &{  J $=8\rightarrow7$ (K$_- = 8\rightarrow7$)}& {  68.19} & {  317.88 (943)} & {  $1.02\times10^{14}$} & { $1.998\times10^{-1}$} & {  $4.52\times10^{-13}$}\\
{  $^{20}$NeOH$^+$} &{  J $=9\rightarrow8$ (K$_- = 9\rightarrow8$)}& {  85.23} & {  357.56 (838)} & {  $1.02\times10^{14}$} & { $2.322\times10^{-1}$} & {  $7.46\times10^{-13}$}\\
{  $^{20}$NeOH$^+$} &{ J $=10\rightarrow9$ (K$_- = 10\rightarrow9$)} & {  104.16} & {  397.21 (755)} & {  $1.02\times10^{14}$} & { $2.176\times10^{-1}$} & {  $1.09\times10^{-12}$}\\
$^{20}$NeOH$^+$&J $=11\rightarrow10$ (K$_- = 11\rightarrow10$)& {  124.98}& {  436.84 (686)} &{  $1.02\times10^{14}$}& { $1.674\times10^{-1}$}& {  $1.24\times10^{-12}$} \\
\hline
{  HeOH$^+$} & { J $=1\rightarrow0$ (K$_- = 1\rightarrow0$)}& {  9.62} & {  201.89 (1485)} & {  $8.19\times10^{11}$} &   & {  $1.67\times10^{-16}$} \\
{  HeOH$^+$} &{ J $=2\rightarrow1$ (K$_- = 2\rightarrow1$)} & {  28.86} & {  403.71 (742)} & {  $8.19\times10^{11}$}& { $1.013\times10^{-3}$}& {  $5.12\times10^{-15}$}\\
HeOH$^+$&J $=3\rightarrow2$ (K$_- = 3\rightarrow2$)&57.71&605.39 (495)&{  $8.19\times10^{11}$}&{ $2.158\times10^{-3}$} & {  $3.19\times10^{-14}$} \\
{  HeOH$^+$} & { J $=4\rightarrow3$ (K$_- = 4\rightarrow3$)}& {  96.17} & {  806.85 (372)} & {  $8.19\times10^{11}$} & { $1.330\times10^{-3}$}& {  $8.53\times10^{-14}$} \\
{  HeOH$^+$} &{  J $=5\rightarrow4$ (K$_- = 5\rightarrow4$)}& {  144.21} & {  1008.02 (297)} &{  $8.19\times10^{11}$} &{ $4.246\times10^{-4}$} & {  $1.23\times10^{-13}$} \\
{  HeOH$^+$} & { J $=6\rightarrow5$ (K$_- = 6\rightarrow5$)}& {  201.82} & {  1208.84 (248)} & {  $8.19\times10^{11}$} & { $8.820\times10^{-5}$}& {  $1.09\times10^{-13}$} \\
HeOH$^+$&J $=7\rightarrow6$ (K$_- = 7\rightarrow6$)& 268.98 & 1409.22 (213)&{  $8.19\times10^{11}$}&{ $1.331\times10^{-5}$} & {  $5.30\times10^{-14}$} \\
\hline
\end{tabular} \\
\vskip 0.2 cm
{\bf Note:}\\
Hydride cations of noble gases and OH$^+$ are calculated using the lower limit of the formation rate, whereas hydroxyl cations of noble gases are calculated using the upper limit of the formation rate mentioned in Section \ref{rad_ass}. Following Bates's \citeyear{bate83} formation rate, the total column density of the hydroxyl cations of noble gases are ArOH$^+$ = $1.97\times 10^4$ cm$^{-2}$, NeOH$^+$ = $2.59\times10^7$ cm$^{-2}$, and HeOH$^+$ = $4.34\times10^5$ cm$^{-2}$. \\
$^a$ \cite{barl13}. \\
$^b$ The total column density, optical depth, and SB of $^{20}$NeH$^+$ transitions are also provided in the parentheses when reaction 5a of the Ne network is off.
\end{table}

The emissivity of some of the prominent transitions that fall in between the frequency regime of Herschel's SPIRE and Photodetecting Array Camera and Spectrometer (PACS) and SOFIA are shown in Figure \ref{fig:emis1} for Model A1. \cite{barl13} found that the $2-1$ and $1-0$ transitions of $\rm{^{36}ArH^+}$
were significantly stronger than those of OH$^+$.
Figure \ref{fig:emis1} depicts that in most regions, 971 GHz (308 $\mu$m) transition of OH$^+$ (the strongest transition of OH$^+$ in such a condition) is comparatively stronger than that of the $1-0$ transition. But it is weaker than the $2-1$ transition of $^{36}$ArH$^+$. This is partly consistent with the observation of \cite{barl13}.
\cite{barl13} also found the $J=2-1$ transition of ArH$^+$ stronger than the
$J=1-0$ transition. We find the same trend in Figure \ref{fig:emis1}.
\cite{barl13} detected only the 971 GHz (308 $\mu$m) transition, which was comparable to the
$J=1-0$ transition of $^{36}$ArH$^+$. Our model shows that the $1-0$
transition of $\rm{^{36}ArH^+}$ is comparable to the $971$ GHz transition of OH$^+$
deep inside the filament. The emissivity of the XOH$^+$ (X = Ar, Ne, and He) transitions, which fall in
between the $29-1409$ GHz region, is shown in Figure \ref{fig:emis2}.
These transitions could be beneficial for the future
astronomical detection of these species around similar environments, where strong OH$^+$ emissions had already been identified.

In Table \ref{table:optical}, we list the strongest transitions that fall in the observed range of Herschel's SPIRE and PACS spectrometer and also within the range of SOFIA, ALMA, Very Large Array (VLA), Institute for Radio Astronomy in the Millimeter Range (IRAM) 30m, and Northern Extended Millimeter Array (NOEMA). The optical depth of all these transitions is also noted.
For this calculation, we use the RADEX program by considering only electrons as colliding partners.
We consider $n_e=10^3$ cm$^{-3}$ and temperature $2700$ K. The radiation field shown in Figure \ref{fig:sed}c is considered as the background radiation field. The total column density of the species is also noted from the calculation with $\rm{n_H} = 2.00 \times 10^4$ cm$^{-3}$ and $\zeta/\zeta_0 = 9.07 \times 10^{6}$ (Model A1). Similarly, the emissivity obtained with Model B is shown in Figures \ref{fig:emis1-rich} and \ref{fig:emis2-rich}.

\cite{barl13} obtained an SB of $\sim(2.2 - 9.9) \times 10^{-7}$ erg cm$^{-2}$ s$^{-1}$ sr$^{-1}$ for  the $1 \rightarrow 0$ transition of
$^{36}$ArH$^+$ whereas our best-fitted Model A (i.e., Model A1) finds $\sim 2.84 \times 10^{-7}$ erg cm$^{-2}$ s$^{-1}$ sr$^{-1}$.
For the $2 \rightarrow 1$ transition of $^{36}$ArH$^+$, \cite{barl13} obtained an SB of $\sim(1.0 - 3.8) \times 10^{-6}$ erg cm$^{-2}$ s$^{-1}$ sr$^{-1}$,
whereas our best-fitted model finds $\sim1.29\times10^{-6}$ erg cm$^{-2}$ s$^{-1}$ sr$^{-1}$. \cite{prie17} checked the detectability of these transitions
based on the observed SB of the 971 GHz (308 $\mu$m) transition of OH$^+$. \cite{barl13} obtained the SB of the
971 GHz transition  of $\sim (3.4 - 10.3) \times 10^{-7}$ erg cm$^{-2}$ s$^{-1}$ sr$^{-1}$ whereas our best-fitted model finds it to be $\sim 6.17 \times 10^{-7}$ erg cm$^{-2}$ s$^{-1}$ sr$^{-1}$. Thus, our
best-fitted model (Model A1) always predicts a comparable or stronger SB of $^{36}$ArH$^+$ transitions (242 and 485 $\mu$m) in comparison with the 308 $\mu$m transition of OH$^+$, which is consistent with the results. Now, to examine the detectability of the other transitions of $^{36}$ArH$^+$ and for other hydride ions along with their isotopic forms considered in this study, we check three criteria for each transition:
(i) whether the SB of that transition is comparable to or stronger than the observed SB of the 308 $\mu$m transition of OH$^+$, (ii) the presence of atmospheric transmission \citep[calculated by the ATRAN program of][]{lord92} at the height of $\sim 41000$ ft (i.e., at the height of SOFIA), and
(iii) the optical depth of that transition. With the ground-based telescope, transitions falling between $30$ and $650$ $\mu$m are heavily affected by
atmospheric transmission. For example, at the ALMA site, the amount of precipitable water vapor is typically 1.0 mm, falling below 0.25 mm up to 5\%
of the time. All transitions of $^{36}$ArH$^+$ reported here are falling in this range (69-486 $\mu$m), and thus, it is difficult
to observe these transitions with any ground-based telescope. However, with a space-based telescope, it is possible to detect some more
transitions of this species.

\begin{figure}
\begin{center}
\includegraphics[width=0.6\textwidth]{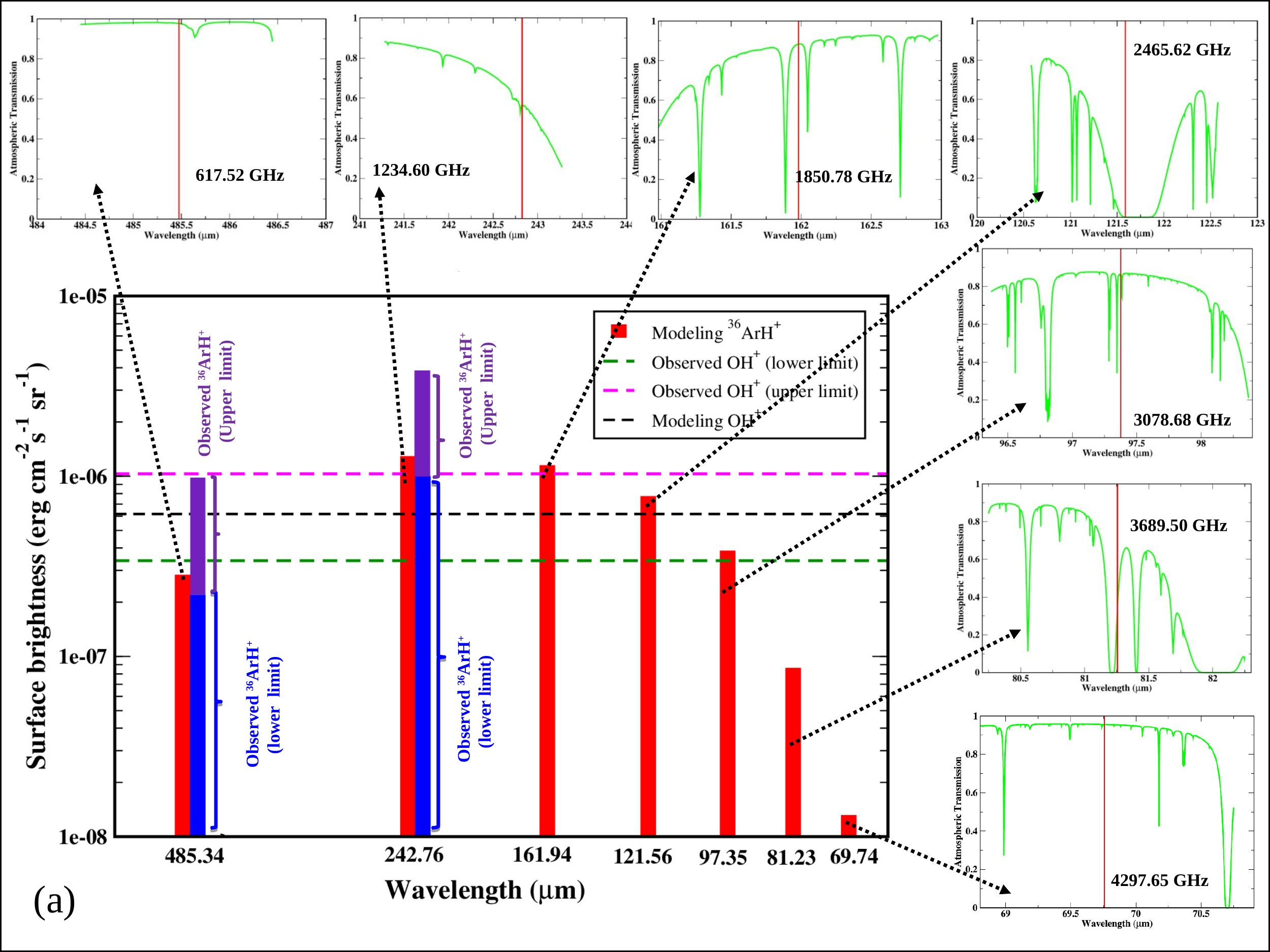}
\includegraphics[width=0.6\textwidth]{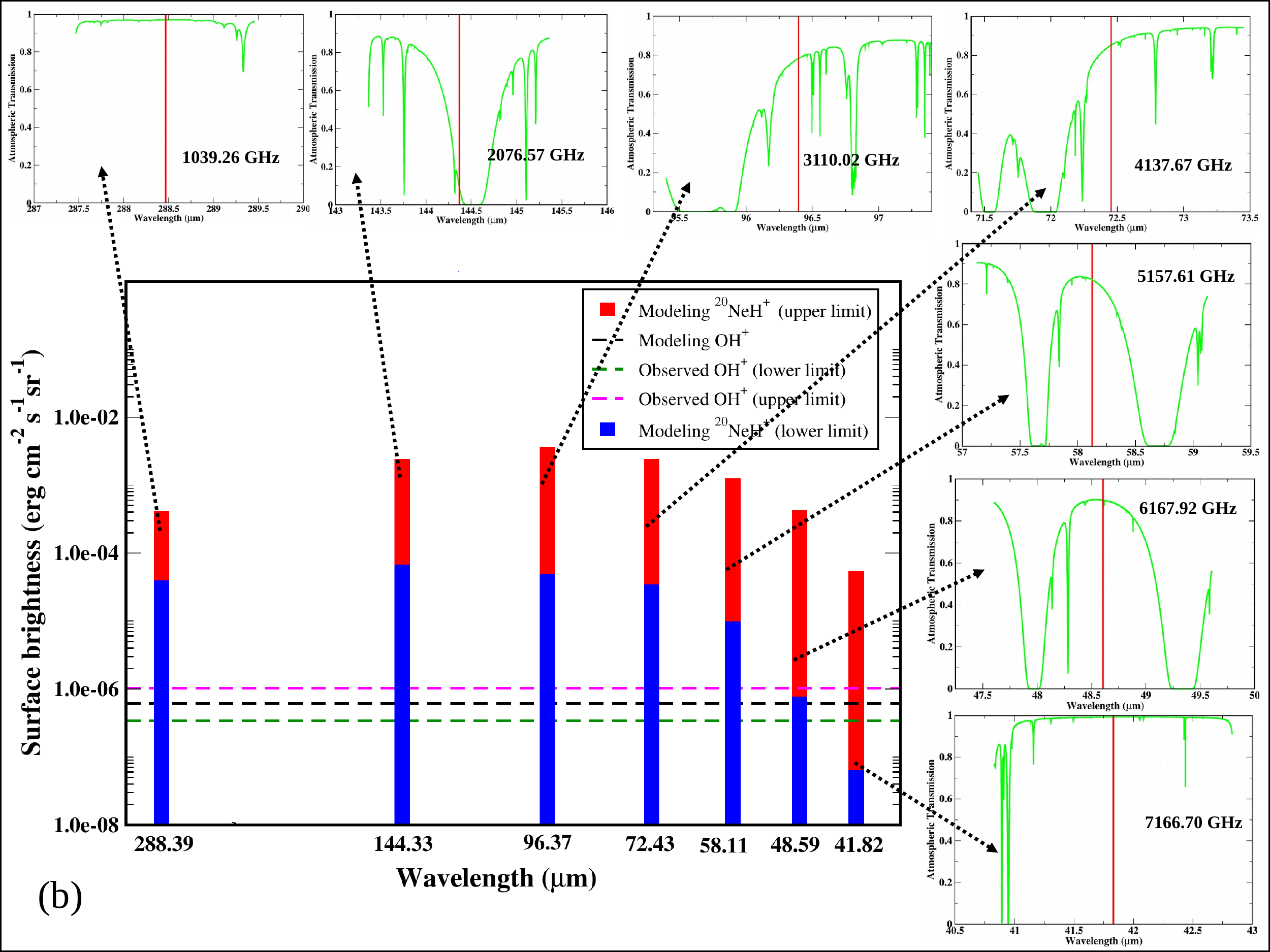}
\includegraphics[width=0.6\textwidth]{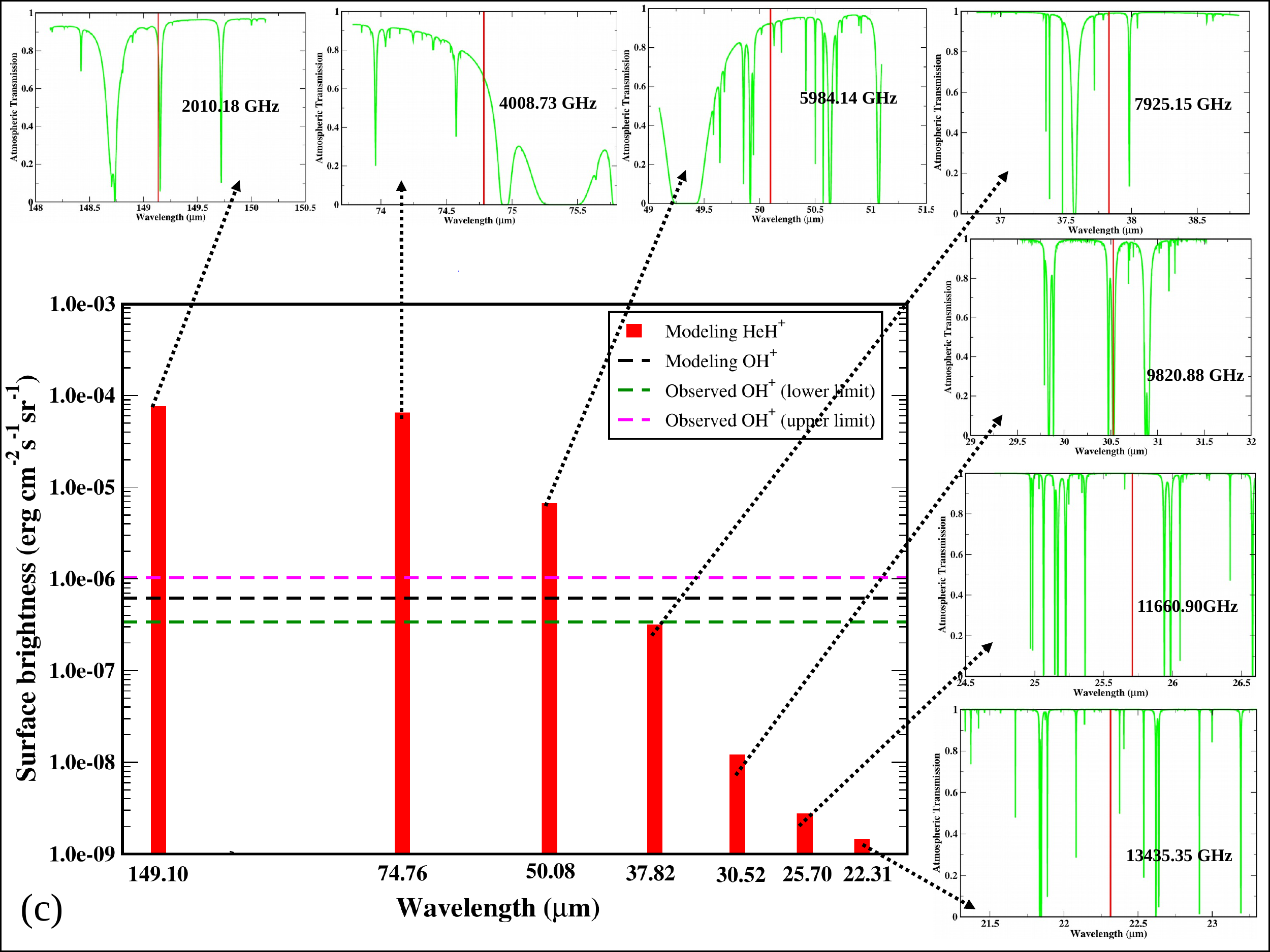}
\caption{A comparison between the observed surface brightness of the $308$ $\mu$m transition of OH$^+$ and the transitions of (a) $^{36}$ArH$^+$, (b) $^{20}$NeH$^+$, and
(c) HeH$^+$ is shown \citep{das20}. The atmospheric transmission for each transition is shown to check the fate of their identification.}
\label{fig:arhp-atm}
\end{center}
\end{figure}

In Figure \ref{fig:arhp-atm}a, we show the SB of these
transitions obtained from our best-fitted Model A1 along with the observed 308 $\mu$m transition of OH$^+$ to show the detectability of these transitions clearly. Table \ref{table:optical}
clearly shows that all these transitions have optical depth $<1$. Figure \ref{fig:arhp-atm}a shows that the first five transitions are
stronger relative to the observed 917 GHz (308 $\mu$m) transition of OH$^+$. The 617 GHz (485 $\mu$m) and 1234 GHz (242 $\mu$m)
transitions were already observed by Herschel, which is no longer operational. Among the other three transitions of $^{36}$ArH$^+$, we can see that
2465 GHz (121 $\mu$m) and 3078 GHz (97 $\mu$m) are heavily affected by the atmospheric transmission and thus challenging to observe.
But the $3 \rightarrow 2$ transition at 1850 GHz (162 $\mu$m) is far from atmospheric absorption features and falls in the range of the
LFA receiver of the modular heterodyne instrument GREAT of SOFIA. However, we find a long integration
time required for this transition with the SOFIA instrument time estimator. We expect that with Herschel, the chance of detection would have been higher.

A similar analysis is carried out for $^{20}$NeH$^+$ and HeH$^+$. When we consider $\rm{Ne^+ + H_2 \rightarrow NeH^+ + H}$ (reaction 5a) for the
formation of NeH$^+$, we obtain a higher abundance of $^{20}$NeH$^+$ and called
it an upper limit. In the absence of this reaction, we obtain a lower limit of the
NeH$^+$ formation. With the upper limit of its formation, Table \ref{table:optical} shows that the 1039 GHz (288 $\mu$m), 2076 GHz (144 $\mu$m),
and 3110 GHz (96 $\mu$m) transitions have an optical depth $>1$. For the other four transitions, it is $<1$. Figure \ref{fig:arhp-atm}b
shows that the other four transitions at 4137 GHz (72 $\mu$m), 5157 GHz (58 $\mu$m), 6167 (48 $\mu$m), and 7166 GHz (42 $\mu$m) are
showing a comparatively stronger SB than that of the observed 308 $\mu$m transition of OH$^+$.
With the lower limit of its formation, Table \ref{table:optical} shows that the 7166 GHz (42 $\mu$m) transition is below, and the 6167 GHz (48 $\mu$m) transition
is comparable to the observed 308 $\mu$m transition of OH$^+$. However, the optical depths of the 2076 and 3110 GHz transitions are found to be $<1$ with the lower limit. But the 2076 GHz transition is very much affected by the atmospheric transmission, as
shown in Figure \ref{fig:arhp-atm}b, which calls into question its detectability.

In the case of HeH$^+$, we find that the optical depths of all the transitions are $<1$. But, Figure \ref{fig:arhp-atm}c shows that only three
transitions show a stronger SB than the 308 $\mu$m transition of OH$^+$. Among them, the 2010 GHz (149 $\mu$m)
transition is heavily affected by atmospheric transmission. The other two transitions at 4008 GHz (75 $\mu$m) and 5984 GHz (50 $\mu$m)
are free from atmospheric features and produce a strong SB.

Table \ref{table:optical} shows that even with the upper limit of the formation, the SB of all the transitions of XOH$^+$ (X = $^{36}$Ar, $^{20}$Ne, and He) is less than the SB of the 308 $\mu$m transition of OH$^+$. So the chance of their detection in the Crab environment is challenging. Thus, we do not carry out any similar analysis for them.

\subsubsection{Comparison with observations: Model B} \label{sec:comp_B}
\begin{figure}
\begin{center}
\includegraphics[width=\textwidth]{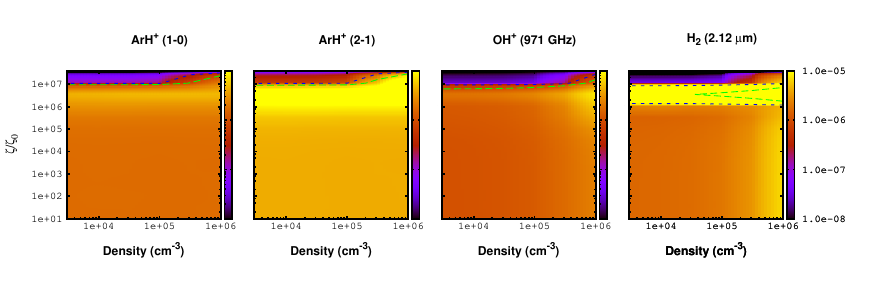}
\caption{Parameter space for the intrinsic line surface brightness (SB) of $1-0$ and $2-1$ transitions of ArH$^+$, the $971$ GHz/$308$ $\mu$m 
transition of OH$^+$, and $2.12$ $\mu$m transition of H$_2$ considering Model B \citep{das20}. Extreme right panel is marked with color coded 
values of the intrinsic line SB (in units erg cm$^{-2}$ s$^{-1}$ sr$^{-1}$). 
The contours are highlighted in the range of observational limits noted in Table \ref{table:summary_1} (Column 2).}
\label{fig:sb_rich}
\end{center}
\end{figure}

\begin{figure}
\begin{center}
\includegraphics[width=\textwidth]{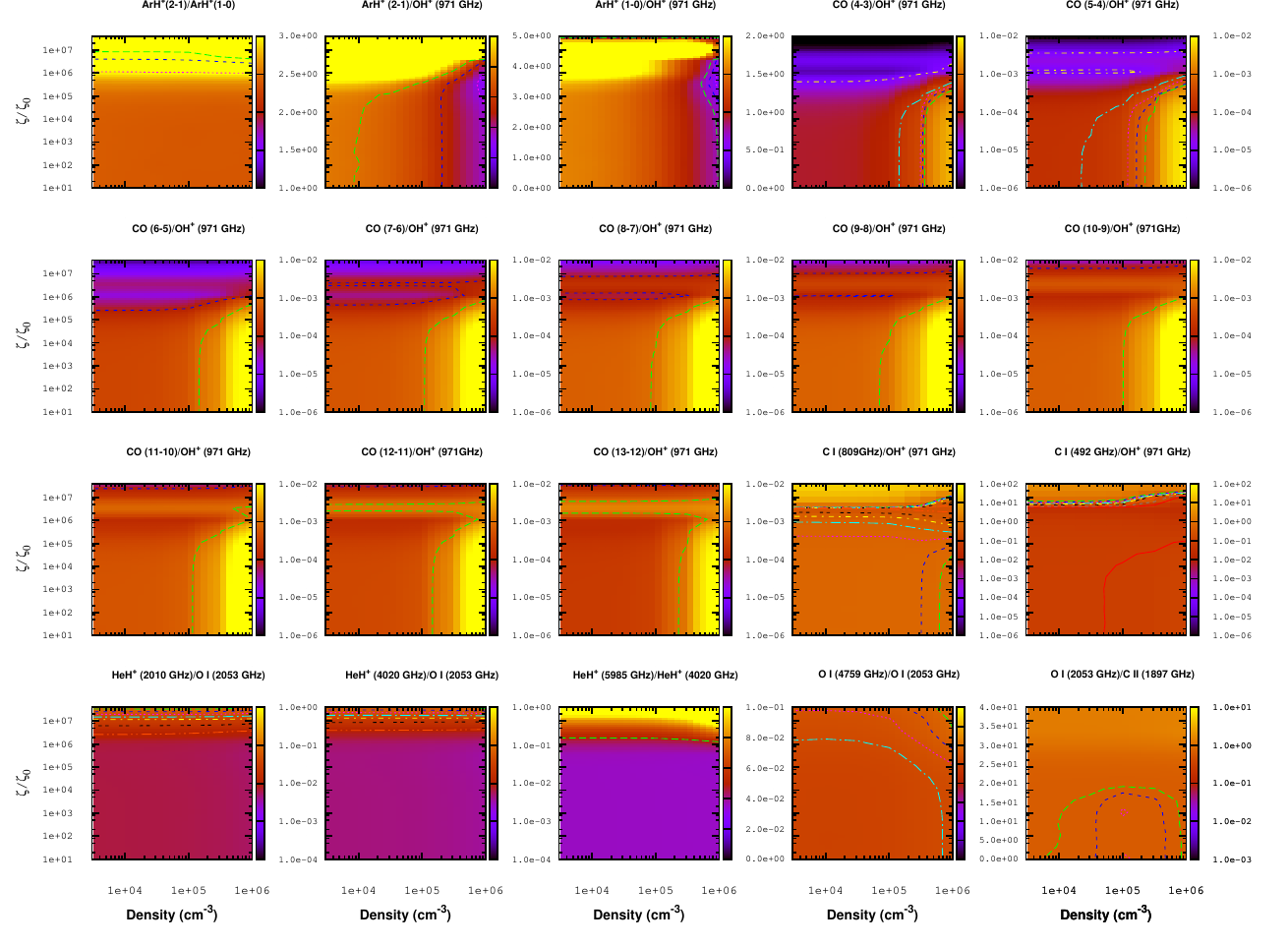}
\caption{Intrinsic line surface brightness (SB) ratio of various molecular and atomic transition fluxes considering Model B \citep{das20}. Contours are highlighted around the observed or previously estimated values shown in Table \ref{table:summary_2}.}
\label{fig:sb-rat_rich}
\end{center}
\end{figure}

\begin{figure}
\begin{center}
\includegraphics[width=\textwidth]{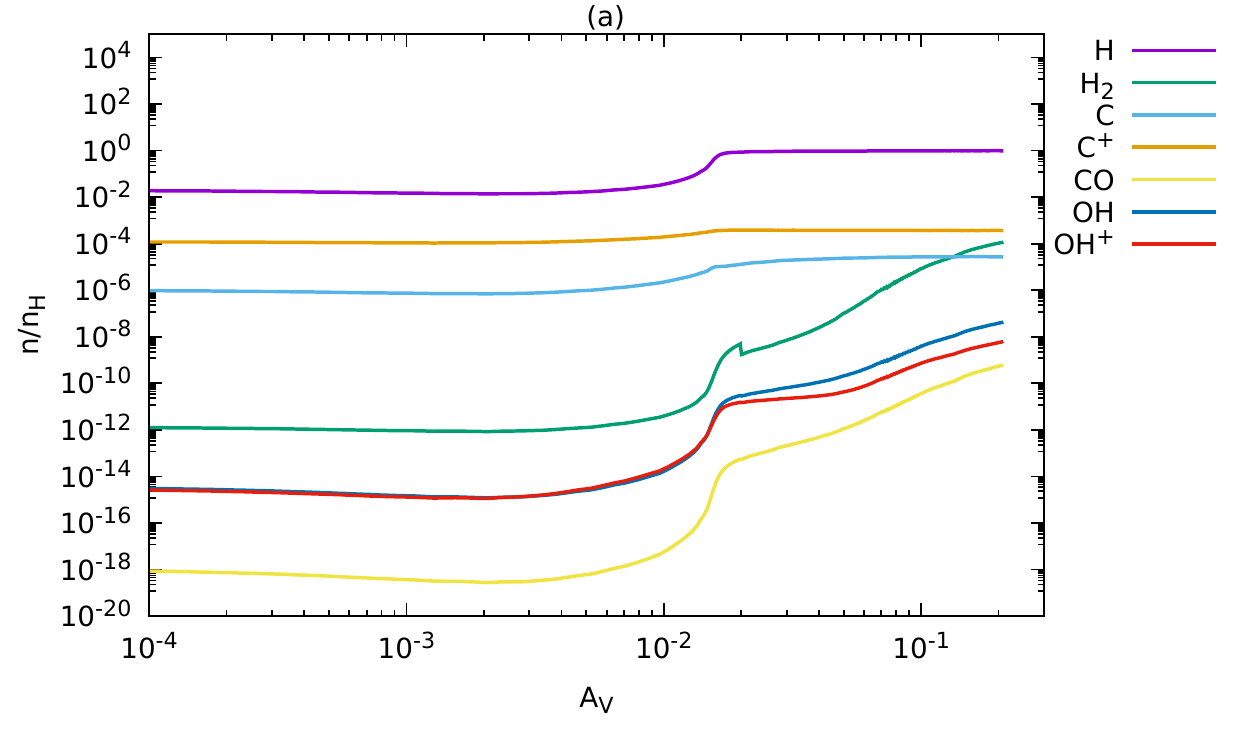}
\includegraphics[width=\textwidth]{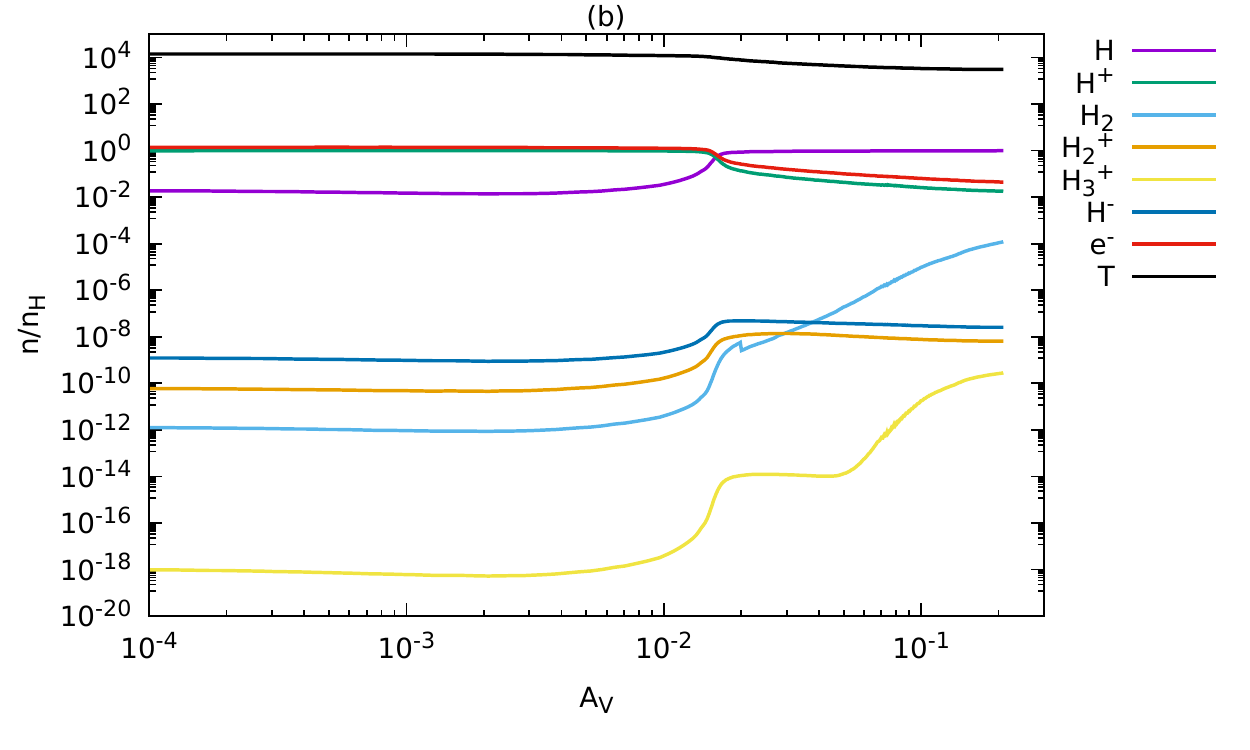}
\caption{Abundance variation of simple species with A$_V$ considering Model B \citep{das20}.}
\label{fig:abun1-rich}
\end{center}
\end{figure}

\begin{figure}
\begin{center}
\includegraphics[width=\textwidth]{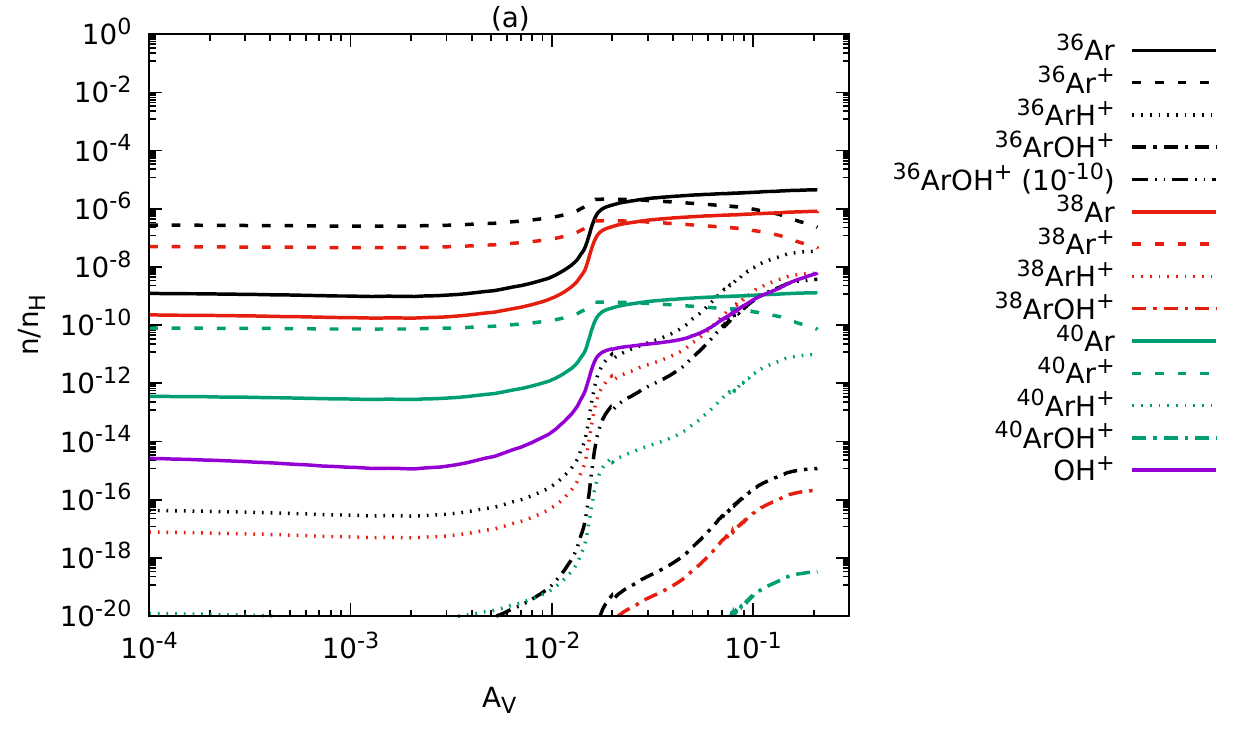}
\includegraphics[width=\textwidth]{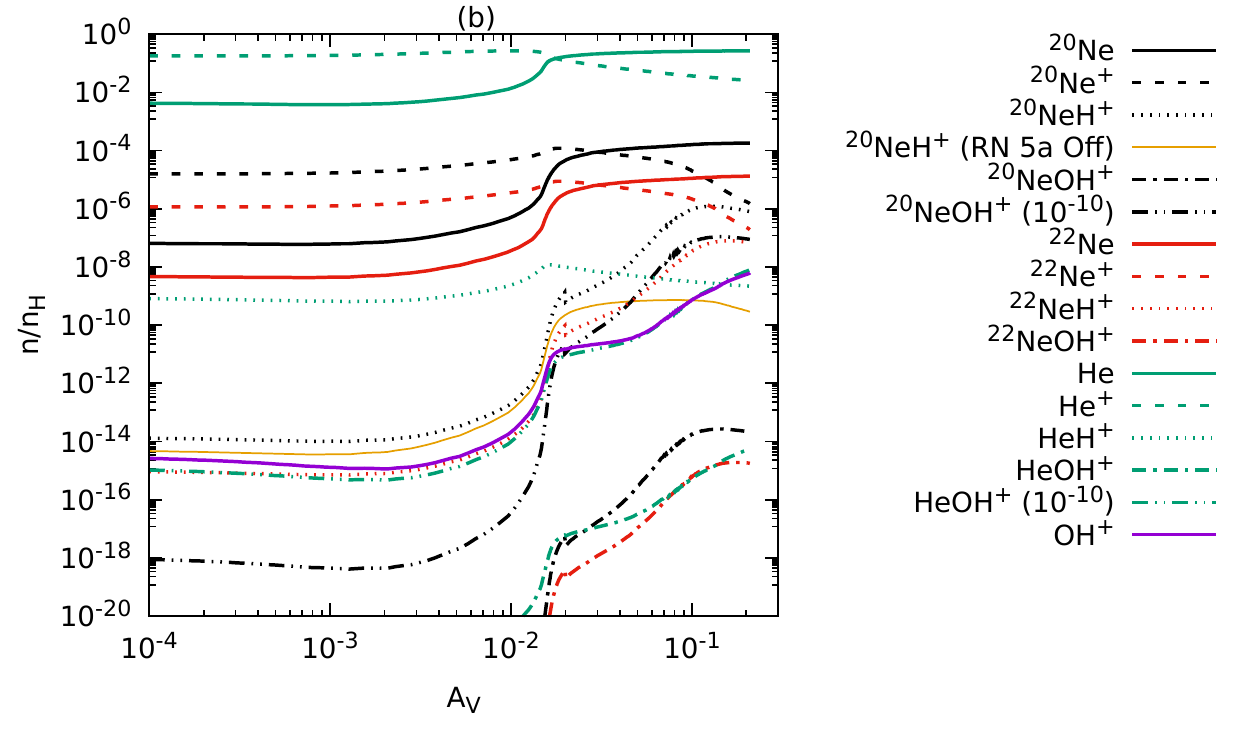}
\caption{Abundance variation of all the hydride and hydroxyl cations considered in this work by considering Model B \citep{das20}. In the upper panel (a) Ar related ions are shown whereas in the lower panel (b) the cases of Ne and He are shown both along with OH$^+$ for comparison.}
\label{fig:abun2-rich}
\end{center}
\end{figure}

\begin{figure}
\begin{center}
\includegraphics[width=\textwidth]{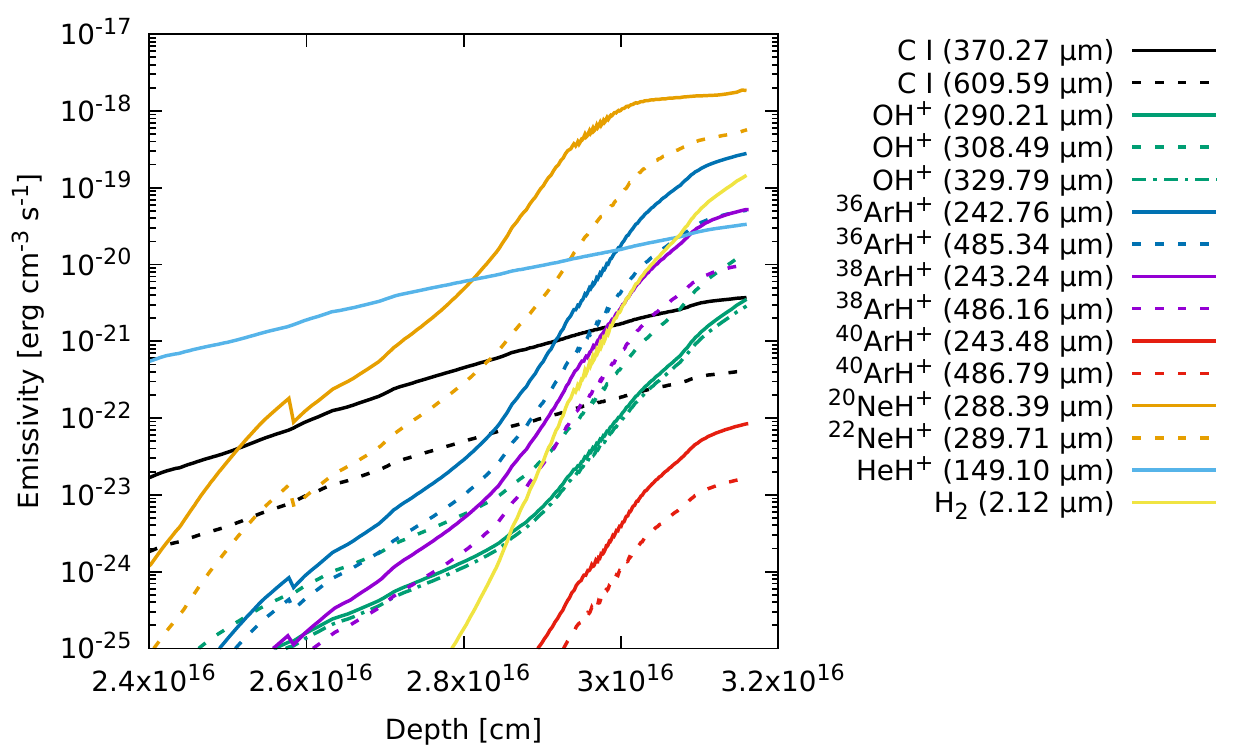}
\caption{Emissivity of some of the strongest transitions which are falling in the frequency limit of Herschel's SPIRE and PACS spectrometer, and SOFIA with respect to the depth into the filament by considering Model B \citep{das20}.}
\label{fig:emis1-rich}
\end{center}
\end{figure}

\begin{figure}
\begin{center}
\includegraphics[width=0.8\textwidth]{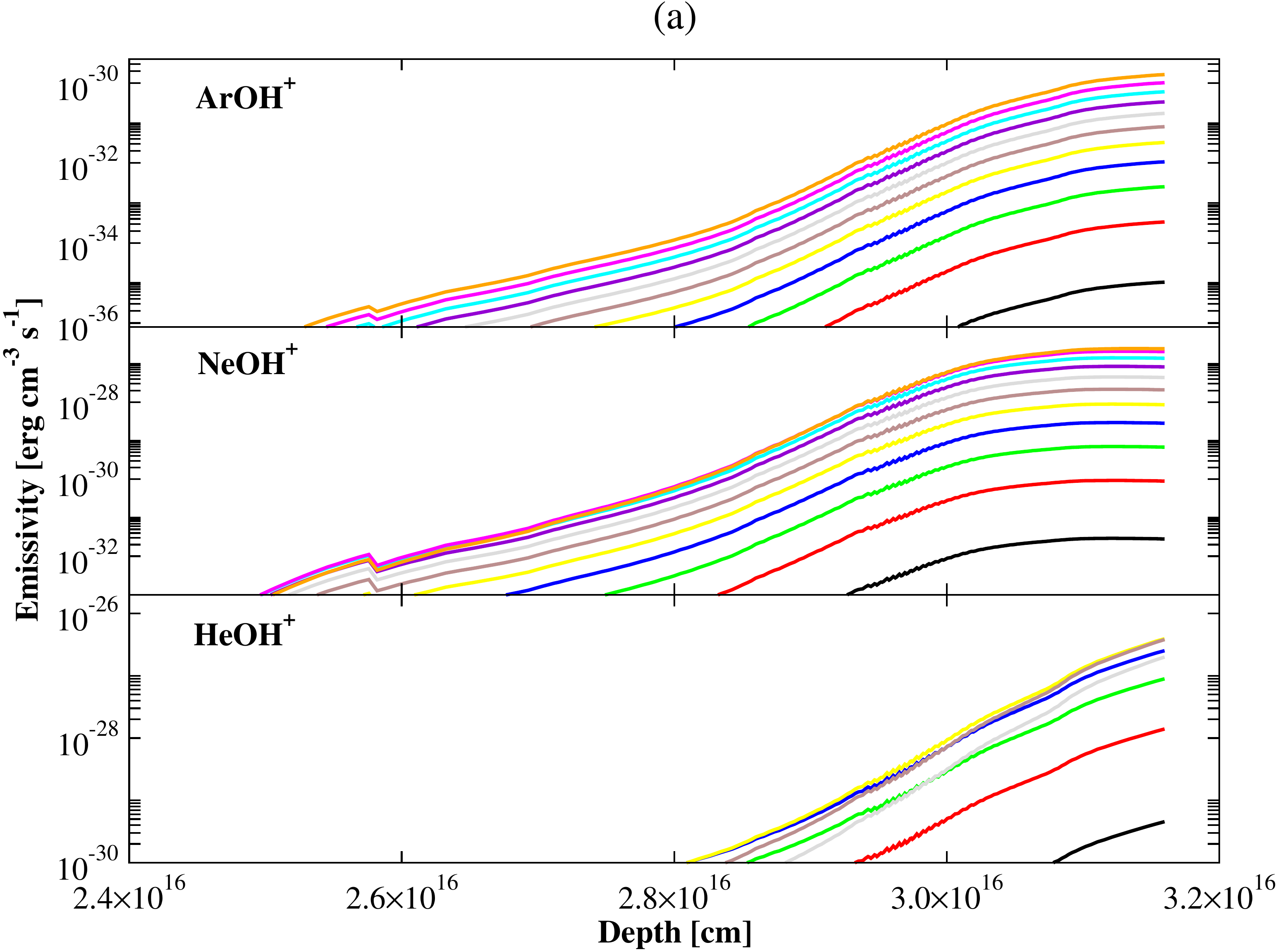}
\vskip 0.5cm
\includegraphics[width=0.8\textwidth]{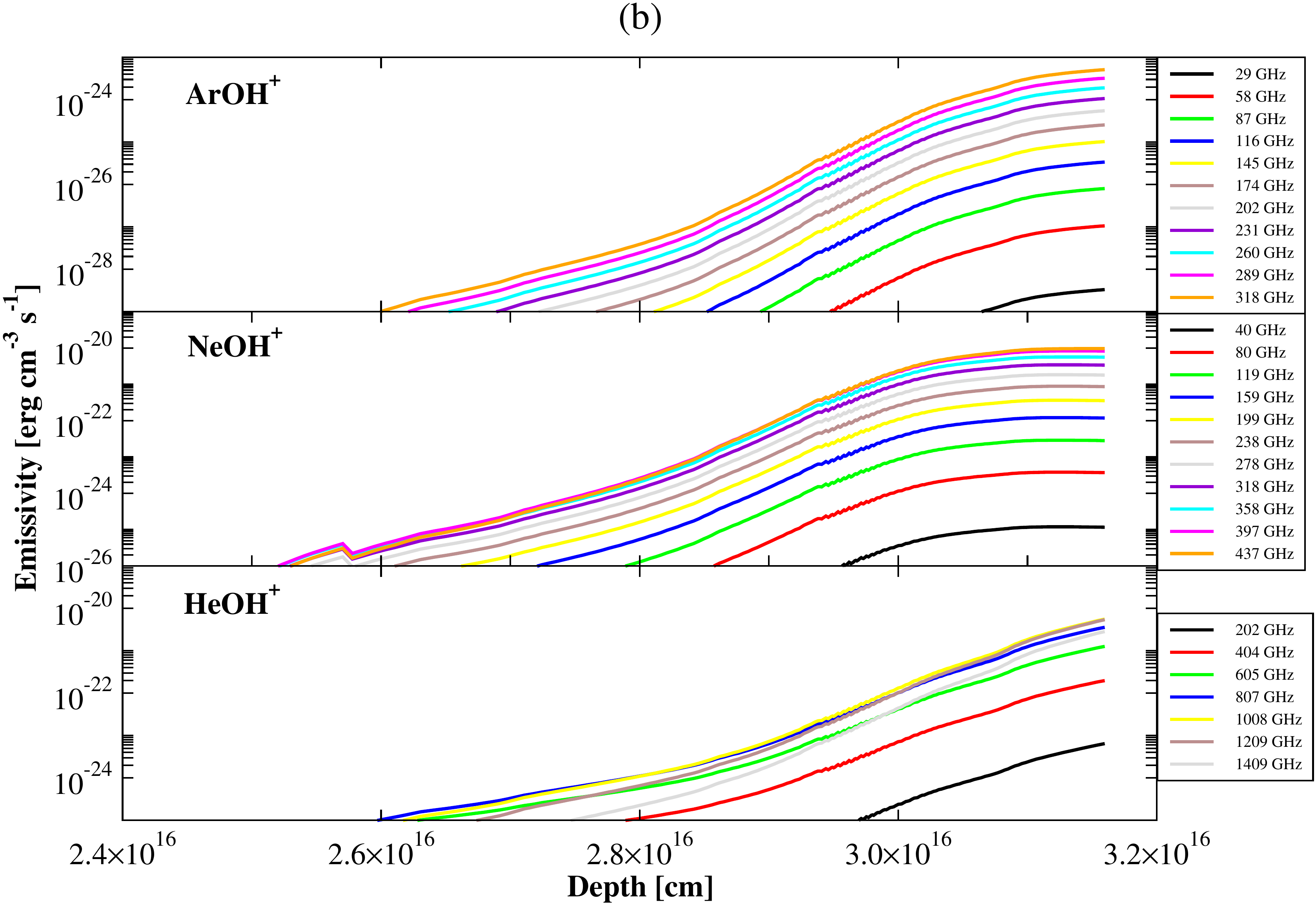}
\caption{Calculated emissivity of various XOH$^+$ transitions (X=$^{36}$Ar, $^{20}$Ne, and He) lying in the frequency limit of Herschel's SPIRE and PACS spectrometer, SOFIA, ALMA, VLA, IRAM 30m, and NOEMA by considering Model B \citep{das20}. (a) Upper panel shows the emissivity considering the formation rates following \cite{bate83} mentioned in Section \ref{rad_ass}, whereas (b) lower panel considering upper limit of $\sim 10^{-10}$ cm$^3$ s$^{-1}$.}
\label{fig:emis2-rich}
\end{center}
\end{figure}

Table \ref{table:comp-value} compares our obtained values with the observational \citep{loh11,loh12,gome12,rich13,prie17} as well as with the previous
modeling results \citep{rich13}. The adopted physical parameters and the gas-phase elemental abundances relative to total hydrogen nuclei in all forms are summarized in Tables \ref{table:model} and \ref{table:abun} for Model B. Though in Model B, we use similar parameters as used in \cite{rich13}, we find very little difference. This small difference is due to the changes in the associative detachment reactions
between the \textsc{Cloudy} version 10.00 \citep{ferl98} used in \cite{rich13} and version 17.02 (used in this work). In the case of Model A, we do not obtain any transition of
sulfur (S) and iron (Fe) because, for this case, we do not consider any initial elemental abundance for these two elements (see Table \ref{table:abun}). 
For Model A, we consider $\rm{n_H} = 2.00 \times 10^4$ cm$^{-3}$ and $\zeta/\zeta_0 = 9.07 \times 10^{6}$ (Model A1) and $\rm{n_H = 3.16 \times 10^4}$ cm$^{-3}$ and $\zeta/\zeta_0 = 4.55 \times 10^6$ (Model A2), whereas, for Model B, we consider the ionizing particle model of \cite{rich13}, which yields a core density $ \rm{n_{H(core)}} = 10^{5.25}$ cm$^{-3}$ and $\frac{\zeta}{\zeta_0}=7.06 \times 10^6$. The striking differences between Model A and Model B are the consideration of a very high abundance of He and a dust to gas ratio of $0.027$ in Model A, whereas in Model B, by considering the initial elemental abundance pointed out in 
Table \ref{table:abun}, we obtain (from the \textsc{Cloudy} output) a dust-to-gas mass ratio $\sim 8$ times lower than that of Model A.
Table \ref{table:h2_lines} provides $\rm{H_2}$ vibrational line SB relative to the 1-0 S(1) line for knot 51 for both Model A and Model B and compared with the observed values \citep{loh12}. We find that our Model A1 can reproduce the observed line strength ratio except for the 2-1 S(X) (X = 1, 2, 3) lines, whereas our Model A2 and Model B are efficient enough to reproduce the 2-1 S(X) lines. All the results obtained with Model B are shown in Figures \ref{fig:sb_rich}-\ref{fig:emis2-rich}.

\begin{table}
\tiny
\caption{$\rm{H_2}$ vibrational line surface brightnesses (SB) relative to the 1-0 S(1) line for Knot 51 from \cite{loh12} and for our final models \citep{das20}. \label{table:h2_lines}}
\vskip 0.2 cm
\hskip -1.5 cm
\begin{tabular}{ccccccccc}
\hline
{\bf  $\rm{H_2}$ Lines} & {\bf  Wavelength ($\mu$m)}&\multicolumn{3}{c}{\bf SB (erg cm$^{-2}$ s$^{-1}$ sr$^{-1}$)}&\multicolumn{3}{c}{\bf  SB relative to the 1-0 S(1) line} & {\bf  Observed SB relative} \\
\cline{3-5}
\cline{6-8}
&&{\bf  Model A1}&{\bf  Model A2}&{\bf  Model B}& {\bf  Model A1} & {\bf  Model A2} & {\bf  Model B} & {\bf to the 1-0 S(1) line} \\
&&&&&&&& {\bf for Knot 51} \\
\hline
{  1-0 S(0)} & 2.22269 & {  $3.13\times10^{-8}$} & {  $4.38\times10^{-6}$}  & {  $1.24\times10^{-6}$} & {  0.214}  & {  0.200} & {  0.200} & $0.23\pm0.04^a$ \\
{  1-0 S(1)} & 2.12125 & {  $1.46\times10^{-7}$} & {  $2.18\times10^{-5}$} & {  $6.18\times10^{-6}$} &  {  1.000} & 1.000  & 1.000 &$1\pm0.04^a$ \\
{  1-0 S(2)} & 2.03320 & {  $7.50\times10^{-8}$} & {  $9.35\times10^{-6}$}  & {  $2.69\times10^{-6}$} & {  0.513} & {  0.428}  & {  0.436} &$0.52\pm0.09^a$  \\
{  2-1 S(1)} & 2.24711 &  {  $1.17\times10^{-7}$} & {  $5.47\times10^{-6}$} & {  $1.49\times10^{-6}$} & {  0.798} & {  0.251}  & {  0.242} & $0.19\pm0.03^a$ \\
{  2-1 S(2)} & 2.15364 &  {  $6.25\times10^{-8}$} & {  $2.40\times10^{-6}$} & {  $6.64\times10^{-7}$} & {  0.428} & {  0.110} & {  0.107} & $<0.13^a$ \\
{  2-1 S(3)} & 2.07294 & {  $1.90\times10^{-7}$} & {  $7.31\times10^{-6}$} & {  $1.97\times10^{-6}$} & {  1.300} & {  0.335} & {  0.319} &  $<0.28^a$ \\
\hline
\end{tabular} \\
\vskip 0.2cm
{\bf Note:}
$^a$ \cite{loh12}.
\end{table}

\clearpage
\subsection{Time scales of molecule formation}
\cite{rich13} and \cite{prie17} mentioned that steady-state chemistry might not be applicable because of
the H$_2$ formation time scale and mass-loss rate of the Crab knot. \cite{rich13} used \textsc{Cloudy} version 10.00 for their study, and \cite{prie17} used the UCL PDR code \citep{bell05,bell06,baye11} for their study. Here, we use \textsc{Cloudy} version 17.02.
To check whether the computation is time steady or not,
we run our model with the `$age$' command available in the \textsc{Cloudy} code.
This command checks whether the microphysics is time steady or not.
We find that both of our best-fitted models show that the longest time scale is below the
age of the Cloud (for the best-fitted case of Model A, it is $\sim 9$ years, and for Model B, it is $\sim 134$ years). Thus, we are not overestimating the abundance of H$_2$ by considering the
radiative attachment of H and then the associative detachment reaction. Because a time-dependent simulation is out of scope
for this work, we discuss the time scale of their formation relevant to the Crab environment.

\subsubsection{\rm ArH$^+$}
ArH$^+$ is mainly formed by the reaction between Ar$^+$ and H$_2$ \citep[][also reported a similar observation]{prie17} with a rate coefficient of $\sim 10^{-9}$ cm$^3$ s$^{-1}$.
This yields a time of $\sim 10^9$ sec $\sim 30$ years (sufficiently smaller than the age of the Crab) by considering an H$_2$ density of $\sim 1$ cm$^{-3}$.
Our best-fitted zone is also within the limit of the observed SB of H$_2$. In the observed region, we have an H$_2$ number density $< 1$ cm$^{-3}$. This rules out the overestimation of the formation of ArH$^+$ considered here.
Our obtained intrinsic absolute line SB and line SB ratio match the observations.

\subsubsection{\rm NeH$^+$}
In the case of NeH$^+$ formation, if we include the reaction between Ne$^+$ and H$_2$ (reaction 5a; see Table \ref{table:reaction}) in our network, that controls the formation.
By considering an H$_2$ number density of $\sim 1$ cm$^{-3}$, the formation time scale is well within the Crab age, as discussed in the context of ArH$^+$. However, we find that its formation depends on the $\rm{HeH^+ + Ne}$ route (reaction 14).
The rate coefficient for the reaction is $\sim 10^{-9}$ cm$^3$ s$^{-1}$.
Because the number density of Ne is $\sim 1$ cm$^{-3}$, it is very fast.
However, its formation depends on the HeH$^+$ produced by a comparatively slower process than ArH$^+$.

\subsubsection{\rm HeH$^+$}
In the best-fitted model, the dominant pathway for the formation of HeH$^+$ is the reaction between He$^+$ and H. \cite{prie17}
also found this pathway to be the dominant one in their network. The rate coefficient used for this reaction is $\sim 1.44 \times 10^{-16}$ cm$^3$ s$^{-1}$ \citep[][found the best fit with a rate constant of $\sim 6 \times 10^{-16}$ cm$^3$ s$^{-1}$]{gust19}.
By considering the H density of $\sim 10^3 - 10^5$ cm$^{-3}$ used here, the time scale for the formation of HeH$^+$ seems to be much shorter ($\sim 10^3$ years by considering
the lowest He$^+$ abundance) than that of the ArH$^+$.
Thus, it is possible to form HeH$^+$ within the lifetime of the Crab.
The recent observation of HeH$^+$ in NGC 7027 (age of $\sim 600$ years) by \cite{gust19} might be a good reason to look for HeH$^+$ in the Crab as well.

Based on the formation time scales of the hydride ions, it is expected that all these molecules would likely be observed in the filamentary region of the Crab.

\subsubsection{\rm ArOH$^+$, NeOH$^+$, and HeOH$^+$}
These three noble gas hydroxyl cations are mainly formed in our network by radiative association reactions (see Section \ref{rad_ass}).
The rate coefficients of these reactions are calculated using a temperature-independent semiempirical formula proposed by \cite{bate83}. Unfortunately, this yields a prolonged formation rate and is thus very unlikely to be formed in the Crab environment.
However, the formula provided by \cite{bate83} to calculate the rate coefficients
is temperature-independent and approximated for $\sim 30$ K.
This semiempirical relation might underestimate the rate in the condition relevant
to the Crab (temperature $\sim 2000-3000$ K). Therefore, to have an
educated estimation of their formation,
we consider an upper limit of these rates ($\sim 10^{-10}$ cm$^3$ s$^{-1}$).
In the case of ArOH$^+$ and NeOH$^+$ formation, the dominant pathway in our network is the reaction between ArH$^+$ and O and NeH$^+$ and O, respectively (reaction 13; see Table \ref{table:reaction}).
For HeOH$^+$ formation, the reaction between He$^+$ and OH dominates (reaction 12). Due to this reason, the ArOH$^+$ and NeOH$^+$ abundance profiles
follow the ArH$^+$ and NeH$^+$ abundance profiles, respectively, whereas HeOH$^+$ roughly follows the abundance profile of OH.
We notice that only with the upper limit of the formation, the abundances of these species are significant.
Otherwise, the formation time scale is much slower and very unlikely to form in the Crab environment.
But the pathways proposed here are beneficial to study their formation in the other sources where they have a much longer time for their appearance.

\begin{table}
\scriptsize{
\caption{Ground vibrational and equilibrium rotational constants and asymmetrically reduced quartic centrifugal distortion constants of ArOH$^+$, NeOH$^+$, and HeOH$^+$ with the DFT-B3LYP/6-311++G(d,p) level of theory \citep{das20}. \label{table:rot}}
\vskip 0.2 cm
\begin{tabular}{cccccc}
\hline
{\bf Sl. No.} & {\bf Species}& {\bf Rotational}& {\bf Calculated values}& {\bf Distortion}& {\bf Calculated values} \\
& & {\bf constants} & {\bf (in MHz)}  & {\bf constants} & {\bf (in MHz)} \\
\hline
&& {  $\rm{A_0}$} & {  606170.618 (574419.7$^a$)} & $D_N$& {  0.026258855} \\
&& {  $\rm{B_0}$} & {  13423.202 (14538.2$^a$)} &$D_{K}$& {  2846.358531040} \\
1. & $\rm{^{36}ArOH^+}$ (Singlet)& {  $\rm{C_0}$} & {  12929.814 (14157.4$^a$)} &$D_{NK}$& {  30.956851344}   \\
& & {  $\rm{A_e}$} & {  568404.429} (577984.9$^a$) & $d_N$& {  -0.001548795} \\
&&{  $\rm{B_e}$} & { 13362.883} (14652.2$^a$)& $d_K$& {  7.374941060} \\
&&{  $\rm{C_e}$} & {  13055.944} (14290.0$^a$)&& \\
\hline
&& {  $\rm{A_0}$} & {  607114.959 (574400.2$^a$)} & {  $D_N$} & {  0.025404061} \\
&& {  $\rm{B_0}$} & {  13198.879 (14293.6$^a$)} & {  $D_{K}$}& {  2929.193961459} \\
{  2.} & {  $\rm{^{38}ArOH^+}$ (Singlet)} & {  $\rm{C_0}$} & {  12717.473 (13925.4$^a$)} &{ $D_{NK}$}& {  30.950234568}   \\
& & {  $\rm{A_e}$} & {  568391.892 (577970.7$^a$)} & {  $d_N$} & {  -0.001507393} \\
&&{  $\rm{B_e}$} & {  13137.742 (14405.4$^a$)} & {  $d_K$} & {  7.371572618} \\
&&{  $\rm{C_e}$} & {  12840.938 (14055.1$^a$)} && \\
\hline
&& {  $\rm{A_0}$} & {  608006.144 (574382.6$^a$)} & {  $D_N$} & {  0.024644498} \\
&& {  $\rm{B_0}$} & {  12996.499 (14073.0$^a$)} & {  $D_{K}$} & {  3007.592807161} \\
{  3.} & {  $\rm{^{40}ArOH^+}$ (Singlet)} & {  $\rm{C_0}$} & {  12525.768 (13715.9$^a$)} & {  $D_{NK}$} & {  30.944237144}   \\
& & {  $\rm{A_e}$} & {  568380.591 (577958.0$^a$)} & {  $d_N$} & {  -0.001470202} \\
&&{  $\rm{B_e}$} & {  12934.645 (14182.7$^a$)} & {  $d_K$} & {  7.368596645} \\
&&{  $\rm{C_e}$} & {  12646.841 (13843.0$^a$)} && \\
\hline
&& {  $\rm{A_0}$}& {  523937.941 (525452.4$^a$)} & $D_N$& {  0.095861623} \\
&& {  $\rm{B_0}$}& {  18963.535 (19702.7$^a$)} &$D_{K}$& {  1279.215533495} \\
4. & $\rm{^{20}NeOH^+}$ (Singlet) & {  $\rm{C_0}$} & {  18045.404 (18942.7$^a$)} &$D_{NK}$& {  38.200509306} \\
& & {  $\rm{A_e}$} & 525035.970 (530275.0$^a$) & $d_N$& {  -0.002683004} \\
&& {  $\rm{B_e}$} & 19104.672 (20252.3$^a$)& $d_K$& {  9.480927416} \\
&&{  $\rm{C_e}$} & 18433.910 (19507.3$^a$)&& \\
\hline
&& {  $\rm{A_0}$}& {  524108.356 (525436.6$^a$)} & {  $D_N$} & {  0.088272895} \\
&& {  $\rm{B_0}$}& {  18178.763 (18884.4$^a$)} & {  $D_{K}$} & {  1366.928818198} \\
{  5.} & {  $\rm{^{22}NeOH^+}$ (Singlet)} & {  $\rm{C_0}$} & {  17320.737 (18185.1$^a$)} & {  $D_{NK}$} & {  38.205763489} \\
& & {  $\rm{A_e}$} & {  525022.539 (530266.0$^a$)} & {  $d_N$} & {  -0.002605291} \\
&& {  $\rm{B_e}$} & {  18307.032 (19406.6$^a$)} & {  $d_K$} & {  9.452753621} \\
&&{  $\rm{C_e}$} & {  17690.192 (18721.4$^a$)} && \\
\hline
&& {  $\rm{A_0}$} & {  526770.350} & $D_N$& {  2.987029963} \\
&& {  $\rm{B_0}$} & {  108480.244} & $D_{K}$ & {  294.469427824} \\
6. & HeOH$^+$ (Singlet) & {  $\rm{C_0}$} & {  88444.204} & $D_{NK}$& {  78.618941712} \\
& & {  $\rm{A_e}$} & 530435.668 & $d_N$& {  0.215953242} \\
&& {  $\rm{B_e}$} & 110472.442 & $d_K$& {  24.945899641} \\
&&{  $\rm{C_e}$} & 91430.461 && \\
\hline
\end{tabular}} \\
\vskip 0.2cm
{\bf Note:} $^a$ \cite{thei16}
\end{table}

\section{Spectroscopic information} \label{chap:spectroscopic_info}
Spectroscopic information of ArH$^+$, NeH$^+$, and HeH$^+$ is already available in the CDMS catalog. However, NeH$^+$ and HeH$^+$ are yet to be identified in the Crab environment.
The $1 \rightarrow 0$ ($2010.18$ GHz) and $2 \rightarrow 1$ ($4008.73$ GHz) transitions of HeH$^+$ fall in the range of SOFIA and the PACS instrument of Herschel.
The $1 \rightarrow 0$ transition of NeH$^+$ ($1039.25$ GHz) is well within the range of the SPIRE instrument of Herschel and SOFIA, whereas
the $2\rightarrow 1$ transition of NeH$^+$ ($2076.57$ GHz) falls in the PACS and SOFIA limit.
We prepare the collisional data files for NeH$^+$ and HeH$^+$ to study the observability of their transitions. To prepare the collisional data file, we consider that electrons are the only colliding
partners. Therefore, we use the electron-impact excitation of HeH$^+$ from \cite{hami16} for this collisional data file.
No collisional rates are available for NeH$^+$.
Thus we approximate the same by considering the collisional rates of ArH$^+-e^-$.

One of the aims of this work is to study the emission lines of hydroxyl ions of noble gases.
Recently, \cite{thei16} calculated rotational constants for the various isotopologs of
ArOH$^+$ and NeOH$^+$. However, the spectroscopic information of HeOH$^+$ is not yet available.
We carry out quantum-chemical calculations using the \textsc{Gaussian} 09 program to find out these rotational parameters. We compute the
rotational constants and asymmetrically reduced quartic centrifugal distortion constants with the DFT-B3LYP/6-311++G(d,p) level
of theory, useful for providing spectral information in the THz domain.
Obtained ground vibrational and equilibrium values of the
rotational constants and asymmetrically reduced quartic centrifugal distortion constants along with the ground vibrational and equilibrium values calculated by \cite{thei16} for comparison are given in Table \ref{table:rot}.
Moreover, we use the SPCAT \citep{pick91} program to determine the rotational transitions of these species, which fall between the THz domain.
The obtained spectral information files are supplied on Zenodo under a Creative Commons Attribution license: \url{https://doi.org/10.5281/zenodo.3998450}. As per the JPL catalog style, we rename the cat files of  $^{36}$ArOH$^+$ as c053009.cat, $^{38}$ArOH$^+$ as c055003.cat, $^{40}$ArOH$^+$ as c057004.cat, $^{20}$NeOH$^+$ as c037006.cat, $^{22}$NeOH$^+$ as c039007.cat, and HeOH$^+$ as c021003.cat.
To prepare the spectral information for ArOH$^+$ and NeOH$^+$, we use both the ground vibrational and equilibrium values of the rotational constants calculated by \cite{thei16}, whereas, in the case of HeOH$^+$, we use our computed parameters.
To prepare the collisional data file,
we consider the interaction between their first $11$ levels. This upper limit of the level is because of the absence of collisional rates of ArH$^+$ for the upper levels \citep{hami16}.
We do not have any first-hand approximate values for the collisional rates of all the species. \cite{hami16} provided the collision rate only for ArH$^+$, and we consider the same collisional rate for all these hydroxyl ions. Therefore, we assess their transitions further for the modeling. However, for the transitions of the first $12$ levels, we obtain
the highest frequency at $318$ GHz for ArOH$^+$ and $437$ GHz for NeOH$^+$.
These frequencies are not in the range of SPIRE or PACS.
However, these transitions fall within the observed range
of the ALMA, IRAM 30m, and NOEMA.
In the case of HeOH$^+$, most of the frequencies that arise fall within the range of Herschel SPIRE, SOFIA, ALMA, IRAM 30m, and NOEMA.

\section{Summary}
We model a Crab filament using the spectral synthesis code, \textsc{Cloudy}
to study the hydride and hydroxyl cations of some noble gas species.
A wide range of parameter space is used to explain the observational aspects suitably. We check that under the Crab filamentary conditions, steady-state chemistry is justified for our best-fitted models.
Our findings are highlighted below:

\begin{itemize}

\item
A realistic chemical network is prepared to study the fate of
hydride and hydroxyl cations of the various isotopes of Ar, Ne, and He.
No fractionation reactions between the isotopologs are considered. We find that
the abundances of $^{36}$ArH$^+$, $^{20}$NeH$^+$, and HeH$^+$ are comparable to the abundance of
OH$^+$ in the Crab filament.
Considering the upper limit of the formation rate, we obtain a reasonably high abundance of
$^{36}$ArOH$^+$, $^{20}$NeOH$^+$, and HeOH$^+$. However, using the realistic rates of these reactions,
we obtain very low abundances of these hydroxyl cations.
It is thus important to accurately measure/estimate these rates.

\item
In the diffuse ISM, we find that the XH$^+$ (X=Ar, Ne, and He) fractional abundances are reasonably high and could have been identified.
For example, we find a
peak fractional abundance of $\sim 1.3 \times 10^{-9}$ for $^{36}$ArH$^+$. $^{20}$NeH$^+$ seems to be also highly abundant
(peak abundance $\sim 5\times10^{-8}$) when reaction 5a ($\rm{Ne^++H_2 \rightarrow NeH^+ + H}$) is
considered. However, its peak fractional abundance significantly drops ($\sim 3 \times 10^{-11}$) in the absence of this pathway.
The peak fractional abundance of HeH$^+$ $\sim 3 \times 10^{-11}$ is obtained.

\item We find that a high value of the cosmic-ray ionization rate ($\frac{\zeta}{\zeta_0}\sim 10^6-10^7$)
with a total hydrogen density a few times $10^4-10^6$ cm$^{-3}$ can successfully reproduce
the absolute SB of the two transitions of $^{36}$ArH$^+$ ($242$ and $485$ $\mu$m), the 308 $\mu$m transition of OH$^+$, and the 2.12 $\mu$m transition of H$_2$.

\item
With the favorable values of $\rm{n_H}$ and $\zeta/\zeta_0$, we can successfully explain the observed
SB ratio between (a) the $2-1$ and $1-0$ transitions of $^{36}$ArH$^+$, (b) two transitions ($2-1$ and $1-0$)
of $^{36}$ArH$^+$ and the 308 $\mu$m transition of OH$^+$, and (c) various transitions of
CO concerning the $308$ $\mu$m transition of OH$^+$. Our most suitable case can explain the
SB ratio obtained by \cite{prie17}
between the transitions (a) HeH$^+$ and 146 $\mu$m of [O\,{\sc i}], and (b) $3-2$ and $2-1$ of HeH$^+$. It can also explain the SB ratio between the transitions (a) 63 $\mu$m and 146 $\mu$m of [O\,{\sc i}], and (b) 146 $\mu$m of [O\,{\sc i}] and 158 $\mu$m of [C\,{\sc ii}] observed by \cite{gome12} using Herschel PACS and ISO Long Wavelength Spectrometer (LWS) fluxes for IR fine-structure emission lines. However, our Model A always overproduces the SB of [C\,{\sc i}], and even around the low $A_V$ region, we have the fractional abundance of CO and OH $\sim 10^{-11}-10^{-9}$. A major reason for this is the obtained electron temperature ($\sim 4000$ K) with Model A which is low.
We find that our Model B
requires a much higher electron temperature ($>10000$ K) to explain most of the observed features in the Crab filamentary region.

\item
The optical depth of the most probable transitions of the XH$^+$ and XOH$^+$ (where X=Ar, Ne, and He) are calculated under the condition of the Crab nebula filamentary region. Analyzing the results, we notice that the 485 $\mu$m, 242 $\mu$m, and 162 $\mu$m transitions of $^{36}$ArH$^+$; 96 $\mu$m, 72$\mu$m, 58$\mu$m, and 48 $\mu$m transitions of $^{20}$NeH$^+$; and 75 $\mu$m and 50 $\mu$m transitions of HeH$^+$ are most likely would be identified with space-based observation. However, the fate of detecting XOH$^+$ in a similar environment with a similar facility could be challenging.

\item
The ground vibrational and equilibrium values of rotational constants are calculated.
Also, asymmetrically reduced quartic centrifugal distortion constants for various isotopologs of ArOH$^+$ and NeOH$^+$ are computed. Moreover, we compare them with the theoretically calculated values
of \cite{thei16}. We also provide these constants for HeOH$^+$, which are not available until now. Additionally, we provide
the catalog files as per JPL style for various isotopologs of ArOH$^+$, and NeOH$^+$ \cite[with both the ground vibrational and equilibrium rotational constants of][]{thei16}, and HeOH$^+$ (with our calculated ground vibrational and equilibrium values), which might enable their future astronomical detection in other sources.

\end{itemize}

%% file: chap4.tex
\chapter{Interstellar Prebiotic Molecules in Radiation Shielded Region} \label{chap:prebiotic}

\section*{Overview}
In the last two decades, astronomers have made a great attempt to explain the evolutionary history of biomolecules in the ISM.
Historically, \cite{chak00a,chak00b} first attempted to obtain the
abundances of 421 species (including several biologically important amino acids)
through detailed time dependent hydrochemical evolution of a collapsing
molecular cloud using the largest network till date with several thousand
reactions. Some unavailable reaction cross-sections were included with certain
assumptions. Subsequently, \cite{ehre00} discussed the interstellar origin of
amino acids and noted that the \cite{chak00a} computation should be relooked as the chemical
environment in ISM could be far from equilibrium.
Glycine ($\rm{NH_2CH_2COOH}$) is the simplest amino
acid and an essential building block for life formation.
It has been extensively searched in space \citep{kuan03}. Though it has been found
(together with many other amino acids and nucleic bases) in
some meteorites on the Earth \citep{kven70} and in the coma of comets \citep{glav08,elsi09,altw16,hadr19},
efforts to detect it in the ISM have so far unsuccessful.
The rotational transitions arising out of glycine are below the detection limit of astronomical surveys \citep{puzz20}.
The lack of its detection raises many speculations about its formation mechanism.
\cite{sahu20} tentatively identified one of its isomer methylcarbamate in the ISM.
Following the procedure clearly laid down by \cite{chak00a},
\cite{das08b,das13b,garr13,maju12,maju13,chak15} studied the formation of glycine in the star-forming region,
with improved rate coefficients and pathways.
Various interdisciplinary studies are involved in the search for the origin of life on the Earth. Whether life evolved
$ab\ initio$ here on the Earth or came from another part of space is debatable.
Still, it is believed that our single-celled ancestors
formed from the raw materials present at that time somewhere
in the universe. When, where, and how the first life came to be is not straightforward
to answer. However, it is necessary to explain how the building blocks of life
(simple molecule $\rightarrow$ complex molecule $\rightarrow$ prebiotic $\rightarrow$ biomolecule)
could be indigenously produced in the universe. 
In the event prebiotic molecules are below detection limits, attempts are made to observe their precursors to estimate their abundances.
This Chapter discusses the chemical studies of some COMs in various parts of the star-forming region.
This study includes the formation of a) aldimines and amines \citep{sil18}, b) three nitrogen (N)-bearing species containing peptide-like bonds \citep{gora20b}, and b) phosphorous (P) bearing \citep{sil21} species.

\clearpage
\section{Aldimines and amines: the building-block of amino acids}

Aldimines and amines are essential ingredients of amino acids \citep{godf73,holt05}.
Thus their discovery under astrophysical
circumstances could be treated as important clues leading to the origin of life.
Aldimines are seen within the reactions of Strecker-type synthesis,
which prepares $\alpha$-aminonitriles, which are versatile
intermediates for synthesizing amino acids via hydrolysis of nitriles.
However, the contribution of Strecker synthesis towards the formation
of these species is less significant  \citep{elsi07}.
A total of $34$ molecules from six isomeric groups
($\rm{CH_3N}$, ~ $\rm{CH_5N}$, ~$\rm{C_2H_5N}$, ~$\rm{C_2H_7N}$, $\rm{C_3H_7N}$, and $\rm{C_3H_9N}$), each contains at least one aldimine or amine, are studied
to determine their possibility of detection in the ISM.
From the $\mathrm{CH_3N}$ isomeric group, methanimine ($\mathrm{CH_2NH}$)
was observed using Parkes $64$ m
telescope toward Sgr B2 \citep{godf73}.
From the $\mathrm{CH_5N}$ isomeric group, methylamine ($\mathrm{CH_3NH_2}$) was detected \citep{kaif74, four74} in both Sgr B2 and Orion A. Glycine could have been formed by
the reaction between methanimine and formic acid \citep{godf73}.
Thus, methanimine could be treated as the precursor of glycine \citep{suzu16}.
Furthermore, \cite{holt05} showed that glycine
could have been formed by reacting with another precursor molecule (reaction between
methylamine and $\mathrm{CO_2}$ under UV irradiation on an icy grain mantle).
Finally, \cite{woon02} recommended that glycine could be formed by the reaction
between $\mathrm{CH_2NH_2}$ and the COOH radical. On the other hand, methylamine
could be produced by two successive H addition reactions with methanimine.
It can also be formed by four subsequent H additions to HCN on the
surface of grains \citep{godf73, woon02, theu11}. 
Both these precursor molecules (methylamine and ethylamine) of glycine were observed
in comet 81P/Wild, also known as Wild 2 \citep{glav08}, and the coma of $\rm{67P/C-G}$ \citep{altw16}.
Microwave and millimeter-wave spectra of the two conformers of ethanimine (E- and Z-ethanimine)
were characterized to guide the astronomical searches \citep{brow80,lova80}.
Finally, from the $\mathrm{C_2H_5N}$ isomeric group, ethanimine has been detected
with both forms in the same sources where methanimine has
already been observed \citep{loom13}.

However, any species from the following $\rm{C_2H_7N}$, $\rm{C_3H_7N}$, and $\rm{C_3H_9N}$ isomeric groups in this sequence have yet to be detected.
This prompts us to study their possibility of detection from these three isomeric groups.
Ethylamine, propanimine, and trimethylamine are of particular interest from these isomeric groups because
they could play a role in forming amino acids and other prebiotic molecules.
\cite{marg15} performed the first spectroscopic study of the propanimine molecule and
found its two conformers, E-propanimine and Z-propanimine.
It is essential to know the spectroscopic details and the chemical abundances of these
species to detect these species under astrophysical conditions.
High-level quantum chemical calculations are employed
to estimate the accurate energies of all the species.
A large gas-grain chemical model is used to study the presence of these species in the ISM.

\subsection{Computational details and methodology}
\subsubsection{Quantum chemical calculations}
For the quantum chemical calculations, we use the \textsc{Gaussian} 09 suite of programs \citep{fris13}.
We consider some ice-phase reactions leading to the formation of
various interstellar amines and aldimines listed in Table \ref{tab:amine_1}.
The reactions include $\rm{radical-radical}$ (RR), which can happen at each encounter,
and $\rm{neutral-radical}$ (NR) that often possess activation barriers.
The Quadratic Synchronous Transit (QST2 method) approach is employed to search for transition state (TS) structures and
to determine reaction pathways using Synchronous Transit-Guided Quasi-Newton (STQN) method \citep{peng93,peng96}.
The QST2 method with DFT-B3LYP/6-311++G(d,p) level of theory is employed to calculate the activation barrier and Gibbs free energy of activation.

\begin{table}
\scriptsize
{\centering
\caption{Ice-phase formation pathways \citep{sil18}.}
\label{tab:amine_1}
\vskip 0.2 cm
\begin{tabular}{ccc}
\hline
{\bf Reaction number (type)}&{\bf Reaction}&{\bf Activation barrier (K)}\\
\hline
R1(RR)$^a$&$\mathrm{N + CH_3 \rightarrow CH_2NH}$&0.0\\
R2(RR)$^a$&$\mathrm{NH  + CH_2 \rightarrow CH_2NH}$&0.0\\
R3(RR)$^a$&$\mathrm{NH_2  + CH \rightarrow CH_2NH}$&0.0\\
R4(NR)$^a$&$\mathrm{HCN+H \rightarrow H_2CN}$&3647$^e$ \\
R5(NR)$^a$&$\mathrm{HCN+H \rightarrow HCNH}$&6440$^e$ \\
R6(RR)$^a$&$\mathrm{H_2CN+H \rightarrow CH_2NH}$&0.0\\
R7(RR)$^a$&$\mathrm{HCNH+H \rightarrow CH_2NH}$&0.0\\
R8(NR)$^a$&$\mathrm{CH_2NH + H \rightarrow CH_3NH} $&2134$^e$ \\
R9(NR)$^a$&$\mathrm{CH_2NH + H \rightarrow CH_2NH_2}$&3170$^e$ \\
R10(RR)$^a$&$\mathrm{CH_3NH+H \rightarrow CH_3NH_2}$&0.0\\
R11(RR)$^a$&$\mathrm{CH_2NH_2+H \rightarrow CH_3NH_2}$&0.0\\
R12(RR)$^b$&$\mathrm{CH_2CN+H \rightarrow CH_3CN}$&0.0\\
R13(RR)$^c$&$\mathrm{CH_3+CN \rightarrow CH_3CN}$&0.0\\
R14(NR)$^d$&$\mathrm{CH_3CN+H \rightarrow CH_3CNH}$& 1400$^e$ \\
R15(RR)$^d$&$\mathrm{CH_3CNH+H \rightarrow CH_3CHNH}$&0.0\\
R16(RR)$^d$&$\mathrm{CH_3+H_2CN \rightarrow CH_3CHNH}$&0.0\\
R17(NR)&$\mathrm{CH_3CHNH+H \rightarrow CH_3CH_2NH}$& 1846\\
R18(RR)&$\mathrm{CH_3CH_2NH+H \rightarrow CH_3CH_2NH_2}$&0.0\\
R19(RR)&$\mathrm{C_2H_5+H_2CN \rightarrow CH_3CH_2CHNH}$&0.0\\
R20(RR)&$\mathrm{C_2H_5+CN \rightarrow CH_3CH_2CN}$&0.0\\
R21(NR)&$\mathrm{CH_3CH_2CN+H \rightarrow CH_3CH_2CNH}$& 2712 \\
R22(RR)&$\mathrm{CH_3CH_2CNH+H \rightarrow CH_3CH_2CHNH}$&0.0\\
\hline
\end{tabular} \\
}
\vskip 0.2cm
{\bf Note:} \\
NR refers to $\rm{neutral-radical}$ reactions, RR to $\rm{radical-radical}$ reactions.\\
$^a$ \cite{suzu16}. \\
$^b$ \cite{hase92}. \\
$^c$ \cite{quan10}. \\
$^d$ \cite{quan16}. \\
$^e$ \cite{woon02}.
\end{table}

The Gaussian-4 (G4) theory is employed to estimate the accurate enthalpies
of formation of all the species. In arriving at precise total
energy for a given species, the G4 composite method performs a sequence of well-defined
$ab\ initio$ molecular calculations \citep{curt07}.
A fully optimized ground-state structure is verified as a stationary point (having non-imaginary frequency) by harmonic vibrational frequency analysis.
To compute the enthalpy of formation,
we calculate the atomization energy of molecules. Experimental
enthalpies of formation of atoms are taken from \cite{curt97}.
Table \ref{tab:amine_2} summarizes the present astronomical status and enthalpy of
formation ($\Delta_fH^O$) of all the considered species.
Subsequently, we arrange the species according to the ascending order of the enthalpy of formation in all the tables.
Some experimental enthalpy of formation values (if available) are also shown
for comparison. Relative energies of each isomeric group member
are also shown with the G4 level of theory.
\cite{osmo07} found that this approach is also suitable for the computation
of the enthalpies of formation. Moreover, in Table \ref{tab:amine_2}, we also include our calculated enthalpies of formation with the DFT-B3LYP/6-31G(d,p) level of theory.
In our case, we find that the computed enthalpies of
formation with the DFT-B3LYP/6-31G(d,p) level of theory are closer to
the experimental values than those with the G4 composite method.

Rotational spectroscopy is the most convenient and the most reliable method
for identifying molecules in the ISM. Species having permanent electric dipole moments are
generally detected from their rotational transitions.
About $80\%$ of all the known interstellar and circumstellar molecules are discovered by these transitions. The intensity of any rotational transition is mainly dependent on the temperature
and the components (a-type, b-type, and c-type) of the dipole moment \citep{fort14,mcmi14}.
The relative signs of the dipole moment components may induce the
change of intensities of some transitions \citep{mull16}. These intensities are directly
proportional to the square of the dipole moment and inversely proportional to the rotational partition function.
Thus, in general, for a fixed temperature, the higher the dipole moments, the higher the intensities. All the molecules considered here have a nonzero permanent electric
dipole moment. Dipole moment components along the
inertial axis ($\mu_a,\mu_b$, and $\mu_c$) are summarized in Table \ref{tab:amine_3}.
For the computation of the dipole moment components, we use various levels of theory.
Among them, our calculations at the HF level yielded excellent
agreement with the existing experimental results.
\cite{laka03} analyzed permanent electric dipole moments of
some primary aliphatic amines. They used various models for comparing
their calculated results with the experimentally obtained results.
They found that the HF/6-31G(3df) level of theory is more reliable for the aliphatic amines.
According to their calculations, on average, this level of theory can predict values of permanent
electric dipole moments with a deviation of only $2.1\%$ of its experimental values.
Concerning their results, here we use the same level of theory to compute the dipole moment components.
Table \ref{tab:amine_3} shows our calculated dipole moment components along with the experimental values, whenever available.
Table \ref{tab:amine_3} shows that our estimated total dipole moments are in good agreement with the experimentally available data for most of the cases. For example, in the case of E-ethanimine, we find a maximum deviation of $11.5\%$ between our calculated and experimental values of total dipole moments. On average, we find a $5.35\%$ variation between our calculated and experimented values.

\begin{table}
\tiny
\caption{Enthalpy of formation and electronic energy (E$_0$) with zero-point vibrational energy (ZPE) and relative energy (in bracket) with G4 composite method \citep{sil18}.
\label{tab:amine_2}}
\vskip 0.2 cm
\hskip -1.0 cm
\begin{tabular}{cccccc}
\hline
\hline
{\bf Number} & {\bf Species} & {\bf Astronomical} & {\bf E$_0$+ZPE} &
{\bf Calculated $\mathrm{\bf \Delta_fH^0}$}
& {\bf Experimental} \\
&&{\bf Status}&{\bf in Hartree/particle}&{\bf using G4 composite method} &{\bf $\mathrm{\bf \Delta_fH^0}$} \\
&&&{\bf (Relative Energy}&{\bf (using B3LYP/6-31G(d,p) method)} & {\bf (in kcal/mol)}\\
&&&{\bf  in kcal/mol)}&{\bf (in kcal/mol)}& \\
\hline
\multicolumn{6}{c}{$\mathrm{\bf CH_3N}$ {\bf Isomeric Group}} \\
\hline
 1 & Methanimine & observed$^b$ & -94.596377 (0.00)  & 18.2604366 (20.0748878) & --- \\
 2 & $\lambda^1$-Azanylmethane & not observed & $-94.519754$ (48.08) & 66.3715977 (66.9874996) & --- \\
\hline
\multicolumn{6}{c}{$\mathrm{\bf CH_5N}$ {\bf Isomeric Group}} \\
\hline
 1 & Methylamine & observed$^{c,d}$ & $-95.802182$ (0.00) & $-9.00194363$ ($-7.3082602$) &  $-5.37763^a$ \\
\hline
\multicolumn{6}{c}{$\mathrm{\bf C_2H_5N}$ {\bf Isomeric Group}} \\
\hline
1 & E-ethanimine & observed$^e$ & $-133.896198$ (0.00) & 5.90189865 (7.9892830) & 5.74$^f$  \\
2 & Z-ethanimine & observed$^e$ & $-133.895732$ (0.29) & 6.20797719 (8.3293932) & ---  \\
3 & Ethenamine & not observed & $-133.889919$ (3.94) & 9.8284533 (12.7953785) & --- \\
4 & N-methylmethanimine & not observed & $-133.884403$ (7.40) & 13.275162 (15.5507728) & 10.51625$^f$  \\
5 & Aziridine & not observed & $-133.862508$ (21.14) & 26.62315 (29.3566098) &  30.11472$^a$  \\
\hline
\multicolumn{6}{c}{$\mathrm{\bf C_2H_7N}$ {\bf Isomeric Group}} \\
\hline
1 & Ethylamine (trans)  & not observed & $-135.094044$ (0.00) & $-16.366079$ ($-14.5306661$) &  --- \\
2 & Ethylamine (gauche) & not observed & $-135.09341$ (0.40) & $-15.955933$ ($-14.1008221$) & $-11.3528^a$ \\
3 & Dimethylamine & not observed & $-135.084612$ (5.92) & $-10.437412$ ($-8.4344111$) &  $-4.445507^a$ \\
\hline
\multicolumn{6}{c}{$\mathrm{\bf C_3H_7N}$ {\bf Isomeric Group}} \\
\hline
1 & 2-Propanimine & not observed & $-173.193699$ (0.00) & $-4.7991787$ ($-2.2671314$) & ---  \\
2 & 2-Propenamine & not observed & $-173.18563$ (5.06) & 0.18444024 (3.5818848) & --- \\
3 & (1E)-1-Propanimine & not observed & $-173.183877$ (6.163) & 1.287921 (3.7186819) & --- \\
4 & (1Z)-1-Propen-1-amine & not observed & $-173.183875$ (6.164) & 1.2981179 (3.7205644) & --- \\
5 & (1E)-N-Methylethanimine & not observed & $-173.183821$ (6.20) & 1.4335242 (4.0067087) & --- \\
6 & (1Z)-1-Propanimine & not observed & $-173.183423$ (6.45) & 1.59211071 (4.0393392) & --- \\
7 & (1E)-1-Propen-1-amine & not observed & $-173.181835$ (7.44) & 2.6644726 (5.9243778) & --- \\
8 & N-Ethylmethanimine & not observed & $-173.17536$ (11.51) & 6.5932994 (12.4893824) & ---  \\
9 & N-Methylethenamine & not observed & $-173.171735$ (13.78) & 9.0000089 (12.4893824) & --- \\
10 & Allylamine & not observed & $-173.168277$ (15.95) & 11.096984 (14.2338589) & ---  \\
11 & Cyclopropanamine & not observed & $-173.159802$ (21.27) & 15.985053 (18.7632226) & 18.475$^a$  \\
12 & S-2-Methylaziridine & not observed & $-173.157389$ (22.7846) & 17.531164 (20.4380455) & --- \\
13 & (2S)-2-Methylaziridine & not observed & $-173.157388$ (22.7853) & 17.535559 (20.4405556) & --- \\
14 & 2-Methylaziridine (trans) & not observed & $-173.157386$ (22.7865) & 17.543877 (20.4393006) & --- \\
15 & 2-Methylaziridine (cis) & not observed & $-173.156991$ (23.03) & 17.7562713 (20.7574479) & --- \\
16 & Azetidine & not observed & $-173.1536$ (25.16) & 19.607872 (22.3205741) & --- \\
17 & Methylaziridine & not observed & $-173.147784$ (28.81) & 23.520139 (26.5719511) & --- \\
18 & N-Methylethanamine & not observed & $-173.126259$ (42.32) & 37.837958 (41.5192279) & --- \\
19 & (Dimethyliminio)methanide & not observed & $-173.112784$ (50.77) & 45.881231 (50.7837785) & --- \\
\hline
\multicolumn{6}{c}{$\mathrm{\bf C_3H_9N}$ {\bf Isomeric Group}} \\
\hline
1 & 2-Aminopropane & not observed & $-174.385779$ (0.00) & $-23.5149351$ ($-21.4656727$) & $-20.0048^a$ \\
2 & Propylamine & not observed & $-174.381773$ (2.51) & $-20.9566988$ ($-18.7880896$) & $-16.7543^a$  \\
3 & Ethylmethylamine & not observed & $-174.375953$ (6.16) & $-17.3159833$ ($-15.0857834$) & ---  \\
4 & Trimethylamine & not observed & $-174.369667$ (10.11) & $-13.4808824$ ($-10.9573983$) & $-5.64054^a$  \\
\hline
\hline
\end{tabular} \\
\\
{\bf Note:} \\
Additional computation of enthalpy of formation by the B3LYP/6-31G(d,p) level of theory is
pointed out in parentheses. \\
$^a$ \cite{fren94}.\\
$^b$ \cite{godf73}. \\
$^c$ \cite{kaif74}. \\
$^d$ \cite{four74}. \\
$^e$ \cite{loom13}. \\
$^f$ NIST Chemistry Webbook (\url{http://webbook.nist.gov/chemistry}).
\end{table}

\begin{table}
\tiny
\caption{Calculated dipole moment components with HF/6-31G(3df) level of theory \citep{sil18}.
\label{tab:amine_3}}
\vskip 0.2 cm
\begin{tabular}{cccccc}
\hline
\hline
{\bf Number} & {\bf Species} & 
{\bf $\mu_a$ (D)} &
{\bf $\mu_b$ (D)} &
{\bf $\mu_c$ (D)} &
{\bf $\mu_{tot}$ (D)}\\
\hline
\multicolumn{6}{c}{$\mathrm{\bf CH_3N}$ {\bf Isomeric Group}} \\
\hline
 1 & Methanimine & $-1.5115$ ($-1.300^g$)  & $-1.4556$ ($-1.500^g$) & 0.0000 (0.000$^g$) & 2.0985 (2.000$^g$)  \\
 2 & $\lambda^1$-azanylmethane & $-1.9499$ & $-0.0176$ & 0.1483 & 1.9556 \\
\hline
\multicolumn{6}{c}{$\mathrm{\bf CH_5N}$ {\bf Isomeric Group}} \\
 \hline
 1 & Methylamine & 0.4410 & 0.2771 & 1.1774 & 1.2874 (1.310$^h$) \\
 \hline
\multicolumn{6}{c}{$\mathrm{\bf C_2H_5N}$ {\bf Isomeric Group}} \\
\hline
1 & (E)-Ethanimine & 0.2063 & $-2.0884$ & 0.2912 & 2.1187 (1.900$^h$)\\
2 & (Z)-Ethanimine & 0.6062 & $-2.3957$ & 0.5926 & 2.5412 \\
3 & Ethenamine & 0.5514 & 1.0539 & $-0.7109$ & 1.3857 \\
4 & N-Methylmethanimine & $-0.2812$ & 1.0514 & 1.2499 & 1.6573 (1.530$^i$) \\
5 & Aziridine & 1.6649 & $-0.2522$ & 0.1768 & 1.6931 (1.90$\pm$0.01$^h$) \\
 \hline
\multicolumn{6}{c}{$\mathrm{\bf C_2H_7N}$ {\bf Isomeric Group}} \\
 \hline
1 & Ethylamine (trans) & 0.8802 & $-0.1949$ & 0.8894 & 1.2664 (1.304$\pm$0.011$^h$) \\
2 & Ethylamine (gauche) & $-0.5839$ & $-1.0489$ & $-0.2731$ & 1.2312 (1.220$^h$)\\
3 & Dimethylamine & 0.1499 & $-0.9771$ & $-0.1444$ & 0.9991 (1.030$^h$) \\
 \hline
\multicolumn{6}{c}{$\mathrm{\bf C_3H_7N}$ {\bf Isomeric Group}} \\
\hline
1 & 2-Propanimine & $-0.8107$ & 1.8357 & 1.4047 & 2.4495 \\
2 & 2-Propenamine & 0.4017 & 0.4670 & $-1.1975$ & 1.3467 \\
3 & (1E)-1-Propanimine & $-1.0918$ & $-1.5598$ & 0.8831 & 2.0987 \\
4 & (1Z)-1-Propen-1-amine & $-0.1209$ & 1.0920 & 1.7882 & 2.0987 \\
5 & (1E)-N-Methylethanimine & $-1.3457$ & $-0.2385$ & 0.8994 & 1.6360 \\
6 & (1Z)-1-Propanimine & $-2.1065$ & 1.5775 & $-0.0300$ & 2.6318 \\
7 & (1E)-1-Propen-1-amine & 1.0898 & $-0.4583$ & $-0.2748$ & 1.2138 \\
8 & N-Ethylmethanimine & $-1.4418$ & $-0.6595$ & $-0.0489$ & 1.5862 \\
9 & N-Methylethenamine & $-1.2679$ & $-0.0583$ & $-0.0886$ & 1.2723 \\
10 & Allylamine & $-0.1645$ & $-1.1175$ & $-0.2388$ & 1.1545 ($\approx$1.2$^h$) \\
11 & Cyclopropanamine & 0.0564 & 0.9405 & $-0.8402$ & 1.2624 (1.190$^g$) \\
12 & S-2-Methylaziridine & 1.3228 & 0.6207 & 0.7330 & 1.6347 \\
13 & (2S)-2-Methylaziridine & 0.5335 & 0.4331 & 1.4840 & 1.6354 \\
14 & 2-Methylaziridine (trans) & 0.6744 & 0.0897 & $-1.4863$ & 1.6346 (1.57$\pm$0.03$^h$) \\
15 & 2-Methylaziridine (cis) & 0.6744 & 0.0897 & $-1.4863$ & 1.6346 (1.77$\pm$0.09$^h$) \\
16 & Azetidine & 0.2802 & 0.2805 & 1.1993 & 1.2632 \\
17 & Methylaziridine & 0.8549 & 0.1523 & 0.9647 & 1.2980 \\
18 & N-methylethanamine & $-0.5108$ & $-1.0613$ & 0.4630 & 1.2655 \\
19 & (Dimethyliminio)methanide & $-1.4791$ & $-2.9791$ & 0.3350 & 3.3429 \\
 \hline
 \multicolumn{6}{c}{$\mathrm{\bf C_3H_9N}$ {\bf Isomeric Group}} \\
 \hline
1 & 2-Aminopropane & $-0.0991$ & $-0.3068$ & 1.1727 & 1.2162 (1.190$^j$) \\
2 & Propylamine & $-0.9663$ & 0.5630 & 0.3766 & 1.1800 (1.170$^h$) \\
3 & Ethylmethylamine & 0.1756 & 0.2227 & $-0.8964$ & 0.9402 \\
4 & Trimethylamine & 0.1174 & 0.1270 & $-0.6598$ & 0.6821 (0.612$^h$) \\
\hline
\hline
\end{tabular} \\
\vskip 0.2cm
{\bf Note:} \\
Experimental values are also shown in parentheses. \\
$^g$ \cite{dema74}. \\
$^h$ \cite{nels67}. \\
$^i$ \cite{sast64}. \\
$^j$ \cite{mehr77}.
\end{table}

\begin{table}
\tiny
\caption{Calculated rotational constants and rotational partition functions at $200$ K with the MP2/6-311++G(d,p) level of theory \citep{sil18}.
\label{tab:amine_4}}
\vskip 0.2 cm
\hskip -1.0 cm
\begin{tabular}{cccccc}
\hline
\hline
{\bf Number} & {\bf Species} & {\bf  A (in GHz)} & 
{\bf B (in GHz)} & 
{\bf C (in GHz)} & {\bf Rotational Partition } \\
&&&&&{\bf function  at $200$ K}\\
\hline
\multicolumn{6}{c}{$\mathrm{\bf CH_3N}$ {\bf Isomeric Group}} \\
\hline
 1 & Methanimine &  195.72173 (196.21116$^k$) & 34.45869 (34.64252$^k$) &  29.30013 (29.35238$^k$) & 0.107265(+04) \\
 2 & $\lambda^1$-azanylmethane & 157.58498 & 27.56460 & 27.56460 & 0.459337(+03) \\
\hline
\multicolumn{6}{c}{$\mathrm{\bf CH_5N}$ {\bf Isomeric Group}} \\
 \hline
 1 & Methylamine & 103.42705 (103.12861$^o$) & 22.75135 (22.62234$^o$) & 21.85872 (21.69598$^o$) & 0.210246(+04) \\
 \hline
\multicolumn{6}{c}{$\mathrm{\bf C_2H_5N}$ {\bf Isomeric Group}} \\
\hline
1 & (E)-Ethanimine & 52.91394 (52.83537$^k$) &  9.76090 (10.07601$^k$) & 8.68503 (8.70427$^k$) & 0.711944(+04) \\
2 & (Z)-Ethanimine & 50.17305 (49.5815$^k$) &  9.76932 (10.15214$^k$)  & 8.61433 (8.644814$^k$)  & 0.733810(+04) \\
3 & Ethenamine & 55.91347 &  9.98960 &  8.55386 & 0.689840(+04) \\
4 & N-Methylmethanimine & 51.70697 (52.52375$^k$)  & 10.71490 (10.66613$^k$) & 9.39306 (9.37719$^k$)  &  0.660982(+04) \\
5 & Aziridine &  22.81302 (22.73612$^p$ ) & 21.19888 (21.19238$^p$) & 13.43101 (13.38307$^p$) & 0.591642(+04) \\
 \hline
\multicolumn{6}{c}{$\mathrm{\bf C_2H_7N}$ {\bf Isomeric Group}} \\
 \hline
1 & Ethylamine (trans) & 31.90275 (31.75833$^m$) & 8.75819 (8.749157$^m$) & 7.82305 (7.798905$^m$) & 0.101989(+05) \\
2 & Ethylamine (gauche) & 32.46287 (32.423470$^n$) & 8.99003 (8.942086$^n$) & 7.86715 (7.825520$^n$) & 0.995128(+04) \\
3 & Dimethylamine & 34.22904 (34.24222$^q$) & 9.38988 (9.33403$^q$)  & 8.26707 (8.21598$^q$)  & 0.925036(+04) \\
 \hline
\multicolumn{6}{c}{$\mathrm{\bf C_3H_7N}$ {\bf Isomeric Group}} \\
\hline
1 & 2-Propanimine & 9.64694 & 8.48897 & 4.78348 & 0.240917(+05) \\
2 & 2-Propenamine & 9.54753 & 8.97855 & 4.79187 & 0.235267(+05) \\
3 & (1E)-1-Propanimine & 23.30832 & 4.33503 & 4.20885 & 0.231221(+05) \\
4 & (1Z)-1-Propen-1-amine & 23.30833 & 4.33547 & 4.20908 & 0.231203(+05) \\
5 & (1E)-N-Methylethanimine & 38.04146 & 4.07938 & 3.86089 & 0.194801(+05) \\
6 & (1Z)-1-Propanimine & 23.19882 (24.1852684$^l$) & 4.28693 (4.2923639$^l$) & 4.17097 (4.1567893$^l$) & 0.234119(+05) \\
7 & (1E)-1-Propen-1-amine & 38.31516 & 3.85080 & 3.59371 &  0.207076(+05) \\
8 & N-Ethylmethanimine & 24.05782 & 4.60272 & 4.48201 & 0.214037(+05) \\
9 & N-Methylethenamine & 32.00012 & 4.30235 & 4.04446 & 0.202070(+05) \\
10 & Allylamine & 23.65103 & 4.23494 & 4.17205 & 0.233259(+05) \\
11 & Cyclopropanamine &  16.28786 (16.26995$^r$) & 6.72692 (6.72300$^r$) &  5.80201 (5.79533$^r$) & 0.189118(+05) \\
12 & S-2-Methylaziridine & 16.91977 & 6.53608 & 5.76664 & 0.188818(+05) \\
13 & (2S)-2-Methylaziridine & 16.91889 & 6.53525 & 5.76592 & 0.188847(+05) \\
14 & 2-Methylaziridine (trans) & 16.92200 & 6.53613 & 5.76677 &  0.188803(+05) \\
15 & 2-Methylaziridine (cis) & 16.68599 & 6.56078 & 5.81794 & 0.188940(+05) \\
16 & Azetidine & 11.54225 & 11.36812 & 6.70581 & 0.160748(+05) \\
17 & Methylaziridine & 16.41594 & 7.25710 & 6.19112 & 0.175575(+05) \\
18 & N-methylethanamine & 36.98588 & 4.14324 & 3.84501 & 0.196438(+05) \\
19 & (Dimethyliminio)methanide & 10.15847 & 9.14934 & 5.12560 & 0.218464(+05) \\
 \hline
\multicolumn{6}{c}{$\mathrm{\bf C_3H_9N}$ {\bf Isomeric Group}} \\
 \hline
1 & 2-Aminopropane & 8.37627 (8.33183$^j$) & 7.99371 (7.97718$^j$) & 4.67889 (4.63719$^j$) & 0.269396(+05) \\
2 & Propylamine & 25.18613 & 3.74012 & 3.49778 & 0.262689(+05) \\
3 & Ethylmethylamine & 26.06033 & 3.92898 & 3.67799 & 0.245712(+05) \\
4 & Trimethylamine & 8.75934 & 8.75934 & 4.99056 & 0.812259(+04) \\
\hline
\hline
\end{tabular} \\
\vskip 0.2cm
{\bf Note:}\\
Experimentally obtained rotational constants are shown in parentheses. \\
$^j$ \cite{mehr77}. \\
$^k$ \cite{pear77}. \\
$^l$ \cite{marg15}. \\
$^m$ \cite{fisc82}. \\
$^n$ \cite{fisc84}. \\
$^o$ \cite{herz66}. \\
$^p$ \cite{bak71}. \\
$^q$ \cite{woll68}. \\
$^r$ \cite{hend69}. 
\end{table}

Accurate quantum chemical studies provide reliable spectroscopic constants
to aid laboratory microwave studies and interstellar
detections with confidence. We use the MP2/6-311++G(d,p) level of theory
to produce spectroscopic constants close to the experimental values.
Corrections for the interaction between rotational and vibrational motions, along with
corrections for vibrational averaging
and anharmonic corrections to the vibrational motion, are considered in our calculations.
Table \ref{tab:amine_4} summarizes our calculated theoretical values of rotational constants for all the species considered here. A comparison with the existing experimental results, whenever available, is also made. These spectroscopic constants can be used to generate catalog files of spectroscopic frequencies by using the SPCAT program \citep{pick91} in the JPL/CDMS format. Table \ref{tab:amine_4} also contains the rotational partition function of a temperature relevant to the hot-core condition ($\sim 200$ K).
These partition functions are calculated quantum chemically using the ``freqchk'' utility, which is used to retrieve frequency and thermochemistry data from a checkpoint file, with the optional specification of an alternate temperature, pressure, scale factor, and/or isotope substitutions.
Among all the species considered here, $\lambda^1$-azanylmethane is a prolate symmetric top and trimethylamine is an oblate symmetric top, and both have three rotational symmetries. The rest of the species in this study
are asymmetric top having rotational symmetry $1$. One can appropriate the ground vibrational state rotational partition function for the asymmetric top molecules \citep{cern16} with
\begin{equation}
 Q_{rot}=5.3311 \times 10^6 \sqrt(T^3/ABC)/\sigma,
\end{equation}
where A, B, and C are the rotational constants for the molecule
(in MHz), T is the temperature (in K), and $\sigma$ is the rotational symmetry number. The rotational partition function for the prolate symmetric top molecule can be approximated with
\begin{equation}
 Q_{rot}=5.3311 \times 10^6 \sqrt(T^3/B^2A)/\sigma.
\end{equation}
For the oblate symmetric top molecule, the rotational partition function can be approximated with
\begin{equation}
 Q_{rot}=5.3311 \times 10^6 \sqrt(T^3/A^2C)/\sigma.
\end{equation}

\begin{table}
\tiny
\caption{Gas-phase formation and destruction pathways \citep{sil18}.
\label{tab:amine_5}}
\vskip 0.2 cm
\hskip -1.2 cm
\begin{tabular}{cccccc}
\hline
{\bf Reaction number} & {\bf Reactions} & $\alpha$ & $\beta$ & $\gamma$  & {\bf Rate coefficient} \\
\hline
{\bf  (type)}&{\bf Formation pathways}& & & & {\bf at $10$ K}\\
\hline
&& & & &\\
G1(RR) & $\mathrm{N+CH_3\rightarrow CH_2NH+PHOTON}$ & $1.00\times10^{-15}$ & $-3.0$ & 0.0 & $2.70\times10^{-11}$ \\
G2(RR) & $\mathrm{NH+CH_2\rightarrow CH_2NH+PHOTON}$ & $1.00\times10^{-15}$ & $-3.0$ & 0.0 & $2.70\times10^{-11}$ \\
G3(RR) & $\mathrm{NH_2+CH\rightarrow CH_2NH+PHOTON}$ & $1.00\times10^{-15}$ & $-3.0$ & 0.0 & $2.70\times10^{-11}$ \\
G4(NR) & $\mathrm{H+HCN\rightarrow H_2CN+PHOTON}$ ($\Delta {G}\ddag=8.37^a$ kcal/mol) & --- & --- & - & --- \\
G5(NR) & $\mathrm{H+HCN\rightarrow HCNH+PHOTON}$ ($\Delta {G}\ddag=10.06^a$ kcal/mol)& --- & --- & --- & ---  \\
G6(RR) & $\mathrm{H+H_2CN\rightarrow CH_2NH+PHOTON}$ & $1.00\times10^{-15}$ & $-3.0$ & 0.0 & $2.70\times10^{-11}$ \\
G7(RR) & $\mathrm{H+HCNH\rightarrow CH_2NH+PHOTON}$ & $1.00\times10^{-15}$ & $-3.0$ & 0.0 & $2.70\times10^{-11}$ \\
G8(NR) & $\mathrm{H+CH_2NH\rightarrow CH_3NH+PHOTON}$ ($\Delta {G}\ddag=7.84^a$ kcal/mol) & --- & --- & --- & --- \\
G9(NR) & $\mathrm{H+CH_2NH\rightarrow CH_2NH_2+PHOTON}$ ($\Delta {G}\ddag=11.64^a$ kcal/mol) & --- & --- & --- & --- \\
G10(RR) & $\mathrm{H+CH_3NH\rightarrow CH_3NH_2+PHOTON}$ & $1.00\times10^{-15}$ & $-3.0$ & 0.0 & $2.70\times10^{-11}$ \\
G11(RR) & $\mathrm{H+CH_2NH_2\rightarrow CH_3NH_2+PHOTON}$ & $1.00\times10^{-15}$ & $-3.0$ & 0.0 & $2.70\times10^{-11}$ \\
G12(RR) & $\mathrm{H+CH_2CN\rightarrow CH_3CN+PHOTON}$ & $1.00\times10^{-15}$ & $-3.0$ & 0.0 & $2.70\times10^{-11}$ \\
G13(RR) & $\mathrm{CH_3+CN\rightarrow CH_3CN+PHOTON}$ & $1.00\times10^{-15}$ & $-3.0$ & 0.0 & $2.70\times10^{-11}$ \\
G14(NR) & $\mathrm{H+CH_3CN\rightarrow CH_3CNH+PHOTON}$ ($\Delta {G}\ddag=10.32^a$ kcal/mol)& --- & --- & --- & ---  \\
G15(RR) & $\mathrm{H+CH_3CNH\rightarrow CH_3CHNH+PHOTON}$ & $1.00\times10^{-15}$ & $-3.0$ & 0.0 & $2.70\times10^{-11}$ \\
G16(RR) & $\mathrm{CH_3+H_2CN\rightarrow CH_3CHNH+PHOTON}$ & $1.00\times10^{-15}$ & $-3.0$ & 0.0 & $2.70\times10^{-11}$ \\
G17(NR) & $\mathrm{H+CH_3CHNH\rightarrow CH_3CH_2NH+PHOTON}$ ($\Delta {G}\ddag=9.98^a$ kcal/mol)& --- & $-3.0$ & 0.0 & ---  \\
G18(RR) & $\mathrm{H+CH_3CH_2NH\rightarrow CH_3CH_2NH_2+PHOTON}$ & $1.00\times10^{-15}$ & $-3.0$ & 0.0 & $2.70\times10^{-11}$ \\
G19(RR) & $\mathrm{C_2H_5+H_2CN\rightarrow CH_3CH_2CHNH+PHOTON}$ & $1.00\times10^{-15}$ & $-3.0$ & 0.0 & $2.70\times10^{-11}$ \\
G20(RR) & $\mathrm{C_2H_5+CN\rightarrow CH_3CH_2CN+PHOTON}$ & $1.00\times10^{-15}$ & $-3.0$ & 0.0 & $2.70\times10^{-11}$ \\
G21(NR) & $\mathrm{H+CH_3CH_2CN\rightarrow CH_3CH_2CNH}$ ($\Delta {G}\ddag=11.03^a$ kcal/mol)& --- & $-3.0$ & 0.0 & ---  \\
G22(RR) & $\mathrm{H+CH_3CH_2CNH\rightarrow CH_3CH_2CHNH}$  & $1.00\times10^{-15}$ & $-3.0$ & 0.0 & $2.70\times10^{-11}$ \\
G23(RR) & $\mathrm{C_2H_5+NH\rightarrow CH_3CHNH+H}$  & $2.75\times10^{-12}$ & 0.0 & 0.0 & $2.75\times10^{-12}$ \\
&& & & &\\
\hline
&{\bf Destruction pathways} & & & &\\
\hline
&& & & &\\
G24(IN) & $\mathrm{C^++CH_3CNH\rightarrow C_2H_3^++HNC}$ & $8.10\times10^{-10}$ & $-0.5$ & 0.0 & $4.44\times10^{-9}$ \\
G25(IN) & $\mathrm{H^++CH_3CNH\rightarrow C2H_4N^++H}$ & $2.50\times10^{-9}$ & $-0.5$ & 0.0 & $1.37\times10^{-8}$ \\
G26(IN) & $\mathrm{H^++CH_3CNH\rightarrow CH_3CN^++H_2}$  & $2.50\times10^{-9}$ & $-0.5$ & 0.0 & $1.37\times10^{-8}$ \\
G27(IN) & $\mathrm{He^++CH_3CNH\rightarrow He+HNC^++CH_3}$  & $1.80\times10^{-9}$ & $-0.5$ & 0.0 & $9.86\times10^{-9}$ \\
G28(IN) & $\mathrm{He^++CH_3CNH\rightarrow He+HNC+CH_3^+}$ & $1.80\times10^{-9}$ & $-0.5$ & 0.0 & $9.86\times10^{-9}$ \\
G29(IN) & $\mathrm{H_3^++CH_3CNH\rightarrow C_2H_5N^++H_2}$  & $1.50\times10^{-9}$ & $-0.5$ & 0.0 & $8.22\times10^{-9}$ \\
G30(IN) & $\mathrm{H_3O^++CH_3CNH\rightarrow C_2H_5N^++H_2O}$ & $6.80\times10^{-10}$ & $-0.5$ & 0.0 & $3.72\times10^{-9}$ \\
G31(IN) & $\mathrm{HCO^++CH_3CNH\rightarrow C_2H_5N^++CO}$ & $6.00\times10^{-10}$ & $-0.5$ & 0.0 & $3.29\times10^{-9}$ \\
G32(IN) & $\mathrm{HCO_2^++CH_3CNH\rightarrow C_2H_5N^++CO_2}$  & $5.30\times10^{-10}$ & $-0.5$ & 0.0 & $2.90\times10^{-9}$ \\
G33(IN) & $\mathrm{C^++CH_3CHNH\rightarrow C_2H_4^++HNC}$ & $1.70\times10^{-9}$ & $-0.5$ & 0.0 & $9.31\times10^{-9}$ \\
G34(IN) & $\mathrm{H^++CH_3CHNH\rightarrow C_2H_4N^++H_2}$  & $5.10\times10^{-9}$ & $-0.5$ & 0.0 & $2.79\times10^{-8}$ \\
G35(IN) & $\mathrm{H^++CH_3CHNH\rightarrow C_2H_5N^++H}$  & $5.10\times10^{-9}$ & $-0.5$ & 0.0 & $2.79\times10^{-8}$  \\
G36(IN) & $\mathrm{H_3^++CH_3CHNH\rightarrow C_2H_6N^++H_2}$  & $3.00\times10^{-9}$ & $-0.5$ & 0.0 &  $1.64\times10^{-8}$  \\
G37(IN) & $\mathrm{H_3O^++CH_3CHNH\rightarrow C_2H_6N^++H_2O}$  & $1.40\times10^{-9}$ & $-0.5$ & 0.0 &  $7.67\times10^{-9}$ \\
G38(IN) & $\mathrm{HCO^++CH_3CHNH\rightarrow C_2H_6N^++CO}$ & $1.20\times10^{-9}$ & $-0.5$ & 0.0 &  $6.57\times10^{-9}$  \\
G39(IN) & $\mathrm{HCO_2^++CH_3CHNH\rightarrow C_2H_6N^++CO_2}$ & $1.10\times10^{-9}$ & $-0.5$ & 0.0 &  $6.02\times10^{-9}$  \\
G40(IN) & $\mathrm{H^++CH_2NH_2\rightarrow NH_2+CH_3^+}$  & $1.00\times10^{-9}$ & 0.0 & 0.0 &  $1.00\times10^{-9}$ \\
G41(IN) & $\mathrm{H^++CH_2NH_2\rightarrow NH_2^++CH_3}$ & $1.00\times10^{-9}$ & 0.0 & 0.0 & $1.00\times10^{-9}$ \\
G42(IN) & $\mathrm{H_3O^++CH_2NH_2\rightarrow CH_2NH_3^++H_2O}$ & $1.00\times10^{-9}$ & 0.0 & 0.0 & $1.00\times10^{-9}$ \\
G43(IN) & $\mathrm{HCO^++CH_2NH_2\rightarrow CH_2NH_3^++CO}$ & $1.00\times10^{-9}$ & 0.0 & 0.0 & $1.00\times10^{-9}$ \\
G44(IN) & $\mathrm{He^++CH_2NH_2\rightarrow NH+CH_3^++He}$  & $1.00\times10^{-9}$ & 0.0 & 0.0 & $1.00\times10^{-9}$ \\
G45(IN) & $\mathrm{H^++CH_3NH\rightarrow NH_2^++CH_3}$ & $1.00\times10^{-9}$ & 0.0 & 0.0 & $1.00\times10^{-9}$ \\
G46(IN) & $\mathrm{H^++CH_3NH\rightarrow NH_2+CH_3^+}$ & $1.00\times10^{-9}$ & 0.0 & 0.0 & $1.00\times10^{-9}$ \\
G47(IN) & $\mathrm{H_3O^++CH_3NH\rightarrow CH_2NH_3^++H_2O}$  & $1.00\times10^{-9}$ & 0.0 & 0.0 & $1.00\times10^{-9}$ \\
G48(IN) & $\mathrm{HCO^++CH_3NH\rightarrow CH_2NH_3^++CO}$  & $1.00\times10^{-9}$ & 0.0 & 0.0 & $1.00\times10^{-9}$ \\
G49(IN) & $\mathrm{He^++CH_3NH\rightarrow NH+CH_3^++He}$ & $1.00\times10^{-9}$ & 0.0 & 0.0 & $1.00\times10^{-9}$ \\
G50(IN) & $\mathrm{H^++CH_3NH_2\rightarrow NH_2+CH_4^+}$ & $1.00\times10^{-9}$ & 0.0 & 0.0 & $1.00\times10^{-9}$ \\
G51(IN) & $\mathrm{H^++CH_3NH_2\rightarrow NH_2^++CH_4}$ & $1.00\times10^{-9}$ & 0.0 & 0.0 & $1.00\times10^{-9}$ \\
G52(IN) & $\mathrm{H_3O^++CH_3NH_2\rightarrow CH_3NH_3^++H_2O}$ & $1.00\times10^{-9}$ & 0.0 & 0.0 & $1.00\times10^{-9}$ \\
G53(IN) & $\mathrm{HCO^++CH_3NH_2\rightarrow CH_3NH_3^++CO}$ & $1.00\times10^{-9}$ & 0.0 & 0.0 & $1.00\times10^{-9}$ \\
G54(IN) & $\mathrm{He^++CH_3NH_2\rightarrow NH_2+CH_3^++He}$ & $1.00\times10^{-9}$ & 0.0 & 0.0 & $1.00\times10^{-9}$ \\
G55(IN) & $\mathrm{C^++CH_3CH_2NH_2\rightarrow C_2H_5^++H_2CN}$ & $1.70\times10^{-9}$ & $-0.5$ & 0.0 & $9.31\times10^{-9}$ \\
G56(IN) & $\mathrm{H^++CH_3CH_2NH_2\rightarrow CH_3CNH^++H_2+H_2}$ & $5.10\times10^{-9}$ & $-0.5$ & 0.0 & $2.79\times10^{-8}$ \\
G57(IN) & $\mathrm{H^++CH_3CH_2NH_2\rightarrow C_2H_4N^++H_2+H_2}$ & $5.10\times10^{-9}$ & $-0.5$ & 0.0 & $2.79\times10^{-8}$ \\
G58(IN) & $\mathrm{H_3^++CH_3CH_2NH_2\rightarrow C_2H_6N^++H_2+H_2}$ & $3.00\times10^{-9}$ & $-0.5$ & 0.0 & $1.64\times10^{-8}$ \\
G59(IN) & $\mathrm{H_3O^++CH_3CH_2NH_2\rightarrow C_2H_6N^++H_2O+H_2}$ & $1.40\times10^{-9}$ & $-0.5$ & 0.0 & $7.67\times10^{-9}$ \\
G60(IN) & $\mathrm{HCO^++CH_3CH_2NH_2\rightarrow C_2H_6N^++CO+H_2}$ & $1.20\times10^{-9}$ & $-0.5$ & 0.0 & $6.57\times10^{-9}$ \\
\hline
\end{tabular}
\end{table}

\begin{table}
\tiny
\hskip -1.2 cm
\begin{tabular}{cccccc}
\hline
{\bf Reaction number (type)} & {\bf Reaction}  & {\bf $\alpha$} & {\bf $\beta$} & {\bf $\gamma$} & {\bf Rate coefficient@$10$K} \\
\hline
G61(IN) & $\mathrm{HCO_2^++CH_3CH_2NH_2\rightarrow C_2H_6N^++CO_2+H_2}$  & $1.10\times10^{-9}$ & $-0.5$ & 0.0 & $6.02\times10^{-9}$ \\
G62(IN) & $\mathrm{C^++CH_3CH_2CN\rightarrow C_2H_5^++C_2N}$  & $1.70\times10^{-9}$ & $-0.5$ & 0.0 & $9.31\times10^{-9}$ \\
G63(IN) & $\mathrm{H^++CH_3CH_2CN\rightarrow CH_3CNH^++CH_2}$  & $5.10\times10^{-9}$ & $-0.5$ & 0.0 & $2.79\times10^{-8}$ \\
G64(IN) & $\mathrm{H^++CH_3CH_2CN\rightarrow C_2H_4N^++CH_2}$  & $5.10\times10^{-9}$ & $-0.5$ & 0.0 & $2.79\times10^{-8}$ \\
G65(IN) & $\mathrm{H_3^++CH_3CH_2CN\rightarrow C_2H_6N^++CH_2}$  & $3.00\times10^{-9}$ & $-0.5$ & 0.0 & $1.64\times10^{-8}$\\
G66(IN) & $\mathrm{H_3O^++CH_3CH_2CN\rightarrow C_2H_6N^++H_2CO}$ & $1.40\times10^{-9}$ & $-0.5$ & 0.0 & $7.67\times10^{-9}$ \\
G67(IN) & $\mathrm{HCO^++CH_3CH_2CN\rightarrow C_2H_6N^++C_2O}$  & $1.20\times10^{-9}$ & $-0.5$ & 0.0 & $6.57\times10^{-9}$\\
G68(IN) & $\mathrm{HCO_2^++CH_3CH_2CN\rightarrow C_2H_5^++CO_2+HCN}$ & $1.10\times10^{-9}$ & $-0.5$ & 0.0 & $6.02\times10^{-9}$ \\
G69(IN) & $\mathrm{C^++CH_3CH_2CNH\rightarrow C_2H_5^++C_2N+H}$  & $1.70\times10^{-9}$ & $-0.5$ & 0.0 & $9.31\times10^{-9}$ \\
G70(IN) & $\mathrm{H^++CH_3CH_2CNH\rightarrow CH_3CNH^++CH_2+H}$  & $5.10\times10^{-9}$ & $-0.5$ & 0.0 & $2.79\times10^{-8}$ \\
G71(IN) & $\mathrm{H^++CH_3CH_2CNH\rightarrow C_2H_4N^++CH_3}$  & $5.10\times10^{-9}$ & $-0.5$ & 0.0 & $2.79\times10^{-8}$ \\
G72(IN) & $\mathrm{H_3^++CH_3CH_2CNH\rightarrow C_2H_6N^++CH_3}$  & $3.00\times10^{-9}$ & $-0.5$ & 0.0 & $1.64\times10^{-8}$ \\
G73(IN) & $\mathrm{H_3O^++CH_3CH_2CNH\rightarrow C_2H_6N^++H_2CO+H}$ & $1.40\times10^{-9}$ & $-0.5$ & 0.0 & $7.67\times10^{-9}$ \\
G74(IN) & $\mathrm{HCO^++CH_3CH_2CNH\rightarrow C_2H_6N^++C_2O+H}$  & $1.20\times10^{-9}$ & $-0.5$ & 0.0 & $6.57\times10^{-9}$ \\
G75(IN) & $\mathrm{HCO_2^++CH_3CH_2CNH\rightarrow C_2H_6N^++C_2O+OH}$  & $1.10\times10^{-9}$ & $-0.5$ & 0.0 & $6.02\times10^{-9}$ \\
G76(IN) & $\mathrm{C^++CH_3CH_2CHNH\rightarrow C_2H_5^++C_2N+H_2}$ & $1.70\times10^{-9}$ & $-0.5$ & 0.0 & $9.31\times10^{-9}$ \\
G77(IN) & $\mathrm{H^++CH_3CH_2CHNH\rightarrow CH_3CNH^++CH_3+H}$  & $5.10\times10^{-9}$ & $-0.5$ & 0.0 & $2.79\times10^{-8}$ \\
G78(IN) & $\mathrm{H^++CH_3CH_2CHNH\rightarrow C_2H_4N^++CH_3+H}$ & $5.10\times10^{-9}$ & $-0.5$ & 0.0 & $2.79\times10^{-8}$ \\
G79(IN) & $\mathrm{H_3^++CH_3CH_2CHNH\rightarrow C_2H_6N^++CH_3+H}$ & $3.00\times10^{-9}$ & $-0.5$ & 0.0 & $1.64\times10^{-8}$ \\
G80(IN) & $\mathrm{H_3O^++CH_3CH_2CHNH\rightarrow C_2H_6N^++H_2CO+H_2}$  & $1.40\times10^{-9}$ & $-0.5$ & 0.0 & $7.67\times10^{-9}$ \\
G81(IN) & $\mathrm{HCO^++CH_3CH_2CHNH\rightarrow C_2H_6N^++C_2O+H_2}$ & $1.20\times10^{-9}$ & $-0.5$ & 0.0 & $6.57\times10^{-9}$ \\
G82(IN) & $\mathrm{HCO_2^++CH_3CH_2CHNH\rightarrow C_2H_6N^++C_2O+H_2O}$ & $1.10\times10^{-9}$ & $-0.5$ & 0.0 & $6.02\times10^{-9}$ \\
G83(IN) & $\mathrm{C^++CH_3CH_2NH\rightarrow C_2H_5^++HCN}$ & $1.70\times10^{-9}$ & $-0.5$ & 0.0 & $9.31\times10^{-9}$ \\
G84(IN) & $\mathrm{H^++CH_3CH_2NH\rightarrow CH_3CNH^++H_2+H}$ & $5.10\times10^{-9}$ & $-0.5$ & 0.0 & $2.79\times10^{-8}$ \\
G85(IN) & $\mathrm{H^++CH_3CH_2NH\rightarrow C_2H_4N^++H_2+H}$ & $5.10\times10^{-9}$ & $-0.5$ & 0.0 & $2.79\times10^{-8}$ \\
G86(IN) & $\mathrm{H_3^++CH_3CH_2NH\rightarrow C_2H_6N^++H_2+H}$  & $3.00\times10^{-9}$ & $-0.5$ & 0.0 &  $1.64\times10^{-8}$ \\
G87(IN) & $\mathrm{H_3O^++CH_3CH_2NH\rightarrow C_2H_6N^++H_2O+H}$ & $1.40\times10^{-9}$ & $-0.5$ & 0.0 & $7.67\times10^{-9}$ \\
G88(IN) & $\mathrm{HCO^++CH_3CH_2NH\rightarrow C_2H_6N^++CO+H}$ & $1.20\times10^{-9}$ & $-0.5$ & 0.0 & $6.57\times10^{-9}$ \\
G89(IN) & $\mathrm{HCO_2^++CH_3CH_2NH\rightarrow C_2H_6N^++CO_2+H}$ & $1.10\times10^{-9}$ & $-0.5$ & 0.0 &  $6.02\times10^{-9}$ \\
G90(NN) & $\mathrm{CH_3CNH+H\rightarrow CH_3CN+H_2}$ & $1.28\times10^{-11}$ & 0.5 & 0.0 & $2.37\times10^{-12}$ \\
G91(NN) & $\mathrm{CH_3CHNH+H\rightarrow CH_3CNH+H_2}$ & $1.28\times10^{-11}$ & 0.5 & 1050.0 & $2.37\times10^{-12}$ \\
G92(NN) & $\mathrm{CH_3CNH+C\rightarrow CH_3CN+CH}$ & $4.18\times10^{-12}$ & 0.5 & 0.0 & $7.67\times10^{-13}$ \\
G93(NN) & $\mathrm{CH_3CHNH+C\rightarrow CH_3CNH+CH}$  & $4.18\times10^{-12}$ & 0.5 & 0.0 & $7.67\times10^{-13}$ \\
G94(NN) & $\mathrm{C_2H_5+N\rightarrow CH_3CNH+H}$  & $8.30\times10^{-12}$ & 0.0 & 0.0 & $8.30\times10^{-12}$ \\
G95(DR) & $\mathrm{C_2H_4N^++e^-\rightarrow CH_3CN+H}$ & $1.50\times10^{-7}$ & $-0.5$ & 0.0 & $8.22\times10^{-7}$ \\
G96(DR) & $\mathrm{C_2H_4N^++e^-\rightarrow CH_2CN+H+H}$  & $1.50\times10^{-7}$ & $-0.5$ & 0.0 & $8.22\times10^{-7}$ \\
G97(DR) & $\mathrm{C_2H_5N^++e^-\rightarrow CH_3CNH+H}$ & $1.50\times10^{-7}$ & $-0.5$ & 0.0 & $8.22\times10^{-7}$ \\
G98(DR) & $\mathrm{C_2H_5N^++e^-\rightarrow CH_3+H_2CN}$  & $1.50\times10^{-7}$ & $-0.5$ & 0.0 & $8.22\times10^{-7}$ \\
G99(DR) & $\mathrm{C_2H_6N^++e^-\rightarrow CH_3CHNH+H}$  & $1.50\times10^{-7}$ & $-0.5$ & 0.0 & $8.22\times10^{-7}$ \\
G100(DR) & $\mathrm{C_2H_6N^++e^-\rightarrow CH_4+H_2CN}$ & $1.50\times10^{-7}$ & $-0.5$ & 0.0 & $8.22\times10^{-7}$ \\
G101(DR) & $\mathrm{CH_2NH_3^++e^-\rightarrow CH_4+NH}$  & $1.50\times10^{-7}$ & $-0.5$ & 0.0 & $8.22\times10^{-7}$ \\
G102(DR) & $\mathrm{CH_3NH_3^++e^-\rightarrow CH_4+NH_2}$ & $1.50\times10^{-7}$ & $-0.5$ & 0.0 & $8.22\times10^{-7}$ \\
G103(PH) & $\mathrm{CH_3CNH+PHOTON\rightarrow CH_3+HNC}$ & $1.00\times10^{-9}$ & 0.0 & 1.9 & $5.60\times10^{-18}$ \\
G104(PH) & $\mathrm{CH_3CNH+PHOTON\rightarrow CH_3CN+H}$ & $1.00\times10^{-9}$ & 0.0 & 1.9 & $5.60\times10^{-18}$ \\
G105(PH) & $\mathrm{CH_3CHNH+PHOTON\rightarrow CH_3+H_2CN}$  & $1.00\times10^{-9}$ & 0.0 & 1.9 & $5.60\times10^{-18}$ \\
G106(PH) & $\mathrm{CH_3CHNH+PHOTON\rightarrow CH_3CNH+H}$ & $1.00\times10^{-9}$ & 0.0 & 1.9 & $5.60\times10^{-18}$ \\
G107(PH) & $\mathrm{CH_3CH_2CN+PHOTON\rightarrow CH_3CNH+CH}$ & $1.00\times10^{-9}$ & 0.0 & 1.9 & $5.60\times10^{-18}$ \\
G108(PH) & $\mathrm{CH_3CH_2NH_2+PHOTON\rightarrow CH_3NH_2+CH_2}$ & $1.00\times10^{-9}$ & 0.0 & 1.9 & $5.60\times10^{-18}$ \\
G109(PH) & $\mathrm{CH_3CH_2CNH+PHOTON\rightarrow CH_3CNH+CH_2}$  & $1.00\times10^{-9}$ & 0.0 & 1.9 & $5.60\times10^{-18}$ \\
G110(PH) & $\mathrm{CH_3CH_2CHNH+PHOTON\rightarrow CH_3CNH+CH_3}$ & $1.00\times10^{-9}$ & 0.0 & 1.9 & $5.60\times10^{-18}$ \\
G111(PH) & $\mathrm{CH_2NH_2+PHOTON\rightarrow H_2CN+H_2}$  & $1.00\times10^{-9}$ & 0.0 & 1.6 & $3.94\times10^{-16}$ \\
G112(PH) & $\mathrm{CH_3NH+PHOTON\rightarrow H_2CN+H_2}$ & $1.00\times10^{-9}$ & 0.0 & 1.6 & $3.94\times10^{-16}$ \\
G113(PH) & $\mathrm{CH_3NH_2+PHOTON\rightarrow H_2CN+H_2+H}$  & $3.50\times10^{-9}$ & 0.0 & 1.6 & $3.94\times10^{-16}$ \\
G114(PH) & $\mathrm{CH_3CH_2NH+PHOTON\rightarrow CH_3CHNH+H}$ & $3.50\times10^{-9}$ & 0.0 & 1.6 & $3.94\times10^{-16}$ \\
G115(PH) & $\mathrm{CH_3CH_2NH+PHOTON\rightarrow C_2H_5+NH}$  & $3.50\times10^{-9}$ & 0.0 & 1.6 & $3.94\times10^{-16}$\\
G116(CR) & $\mathrm{CH_3CNH+CRPHOT\rightarrow CH_3+HNC}$ & $1.30\times10^{-17}$ & 0.0 & 1.9 & $1.95\times10^{-14}$ \\
G117(CR) & $\mathrm{CH_3CNH+CRPHOT\rightarrow CH_3CN+H}$ & $1.30\times10^{-17}$ & 0.0 & 1.9 & $1.95\times10^{-14}$ \\
G118(CR) & $\mathrm{CH_3CHNH+CRPHOT\rightarrow CH_3+H_2CN}$  & $1.30\times10^{-17}$ & 0.0 & 1.9 & $1.95\times10^{-14}$ \\
G119(CR) & $\mathrm{CH_3CHNH+CRPHOT\rightarrow CH_3CNH+H}$ & $1.30\times10^{-17}$ & 0.0 & 1.9 & $1.95\times10^{-14}$ \\
G120(CR) & $\mathrm{CH_3CH_2NH_2+CRPHOT\rightarrow CH_3NH_2+CH_2}$ & $1.30\times10^{-17}$ & 0.0 & 1.9 & $1.95\times10^{-14}$ \\
G121(CR) & $\mathrm{CH_3CH_2CN+CRPHOT\rightarrow CH_3CNH+CH}$  & $1.30\times10^{-17}$ & 0.0 & 1.9 & $1.95\times10^{-14}$ \\
G122(CR) & $\mathrm{CH_3CH_2CNH+CRPHOT\rightarrow CH_3CNH+CH_2}$ & $1.30\times10^{-17}$ & 0.0 & 1.9 & $1.95\times10^{-14}$ \\
G123(CR) & $\mathrm{CH_3CH_2CHNH+CRPHOT\rightarrow CH_3CNH+CH_3}$  & $1.30\times10^{-17}$ & 0.0 & 1.9 & $1.95\times10^{-14}$ \\
G124(CR) & $\mathrm{CH_2NH_2+CRPHOT\rightarrow NH+CH_3}$  & $1.30\times10^{-17}$ & 0.0 & 500.0 & $1.95\times10^{-14}$ \\
G125(CR) & $\mathrm{CH_3NH+CRPHOT\rightarrow NH+CH_3}$  & $1.30\times10^{-17}$ & 0.0 & 500.0 & $1.95\times10^{-14}$ \\
G126(CR) & $\mathrm{CH_3NH_2+CRPHOT\rightarrow NH_2+CH_3}$ & $1.30\times10^{-17}$ & 0.0 & 500.0 & $1.95\times10^{-14}$ \\
G127(CR) & $\mathrm{CH_3CH2NH+CRPHOT\rightarrow CH_3CHNH+H}$ & $1.30\times10^{-17}$ & 0.0 & 1.9 & $1.95\times10^{-14}$ \\
G128(CR) & $\mathrm{CH_3CH2NH+CRPHOT\rightarrow C_2H_5+NH}$ & $1.30\times10^{-17}$ & 0.0 & 1.9 & $1.95\times10^{-14}$ \\
\hline
\end{tabular} \\
\vskip 0.2cm
{\bf Note:} \\
CR refers to cosmic-rays, IN to $\rm{ion-neutral}$ reactions, NR to $\rm{neutral-radical}$ reactions, NN to $\rm{neutral-neutral}$ reactions, RR to $\rm{radical-radical}$ reactions, DR to dissociative recombination reactions for molecular ions, PH to photodissociation reactions. \\
$^a$This work
\end{table}

\subsubsection{Chemical modeling}
Our large gas-grain chemical model \citep{das15a,das15b,gora17a,gora17b,sil18}
is employed for chemical modeling.
Gas and grains are considered as coupled through accretion and thermal/non-thermal
desorption. Unless otherwise stated, a moderate value of the non-thermal desorption factor of $\sim 0.03$ is assumed, as mentioned in \cite{garr07}.
A visual extinction of $150$ and a cosmic-ray ionization rate of
$1.3\times 10^{-17}$ $\rm s^{-1}$ are used.
The initial condition is adopted from \cite{leun84}. To mimic
actual physical conditions of the star-forming region, we consider the warm-up method that
was established by \cite{garr06b}.
Initially, we assume that the cloud remains in the isothermal ($T=10$ K) stage for
$10^6$ years, followed by a subsequent warm-up stage where the temperature can
gradually increase up to $200$ K in $10^5$ years.
Hence, our simulation time is restricted to $1.1 \times 10^6$ years.
Furthermore, we assume that each stage has the same constant density ($n_H=10^7$ cm$^{-3}$).

Our gas-phase chemical network is mainly adapted from the UMIST 2012 database \citep{mcel13}.
For the grain surface reaction network, we primarily follow \cite{ruau16}.
In addition to the above network, our network includes some needed reactions for the formation/destruction of interstellar amines and aldimines (see Tables \ref{tab:amine_1} and \ref{tab:amine_5}).

For the computation of the gas-phase rate coefficients of some
additional gas-phase $\rm{neutral-radical}$ (NR) reactions with a barrier,
we use the transition state theory (TST), which leads to the Eyring equation \citep{eyri35}:
\begin{equation} \label{eqn:eyring}
        k= (k_BT/hc) \exp(-\Delta {G} \ddag/RT) \ s^{-1},
\end{equation}
where $\Delta {G}\ddag$ is the Gibbs free energy of activation and $c$ is the concentration, set to $1$. $\Delta {G}\ddag$ is calculated by the quantum chemical calculation (QST2 method with B3LYP/6-311++G(d,p) level of theory).
Equation \ref{eqn:eyring} suggests that the rate
coefficient is exponentially increasing with the temperature. Thus, to avoid any
unattainable rate coefficient around the high-temperature domain, we use an upper limit
($10^{-10}$ cm$^3$s$^{-1}$) for Equation \ref{eqn:eyring}.

Usually, a $\rm{radical-radical}$ addition reaction with a single product can occur
through the radiative association.
\cite{vasy13} outlined the rate coefficient for the formation of larger molecules
by gas-phase radiative association reactions. According to them, a larger molecule such as
$\mathrm{CH_3OCH_3}$ can be formed by
$$
\mathrm{CH_3+ CH_3O\rightarrow CH_3OCH_3 + Photon}.
$$
They considered the following temperature-dependent rate coefficient for the above reaction:
\begin{equation}
 k = 10^{-15} (T/300)^{-3}.
\end{equation}

We also consider similar rate coefficients for the $\rm{radical-radical}$
gas-phase reactions leading to a single product.
In our model, we consider the formation and destruction of these
species in both phases.

To compute the rate coefficients of ice-phase reaction pathways, we use
diffusive reactions with a barrier against diffusion ($\kappa \times R_{diff}$), which is
based on thermal diffusion \citep{hase92}. $\kappa$ is the quantum mechanical probability
of tunneling through a rectangular barrier of thickness $d$.
$\kappa$ is unity in the absence of a barrier. For reactions with activation energy
barriers ($E_a$), $\kappa$ is defined as the quantum mechanical probability for
tunneling through the rectangular barrier of thickness $d(= 1 \AA$) and is calculated by
\begin{equation}
\kappa= \exp [-2(d/ \hbar)(2\mu E_a)^{1/2}].
\end{equation}

Chemical enrichment of interstellar grain mantles depends on the desorption energies ($E_d$)
and barriers against diffusion ($E_b$) of the adsorbed species.
In the low-temperature regime,
the mobility of the lighter species such as $\rm{H, \ D, \ N}$, and $\rm{O}$ mainly
controls the chemical composition of the interstellar grain mantle.
Here we use $E_b = 0.50 E_d$ \citep{garr13}.
BEs are mostly taken from the KIDA. However, BEs of some of the newly
added ice-phase species are not available in the KIDA. For these species,
we add the BEs of the reactants required to
form these species. A similar technique was also employed in \cite{garr13}.
For example, for calculating the BE of $\rm{CH_3CNH}$, we add the
BEs of $\rm{CH_3CN}$ and H.

We assume various $\rm{ion-neutral}$ (IN) and
photodissociative pathways for the destruction of gaseous amines and aldimines.
Different IN and photodissociative destruction pathways are
already available in \cite{quan16} (for ethanimine) and \cite{mcel13} (for methanimine).
We follow similar pathways and the same rate coefficients for the destructions
of other amines, aldimines, and their associated species.
In Table \ref{tab:amine_5}, we point out all the gas-phase formation and
destruction reactions that are considered here. In analogy, we assume similar
photodissociative reactions to destroy ice-phase amines, aldimines, and their associated neutrals. Rate coefficients for the photodissociative reactions are assumed to be the same in both phases.
Abundances of the gas-phase species can also decrease via
adsorption onto the ice. However, the reverse process (desorption) also occurs.

\subsection{Results and discussion}
Here, the results of high-level quantum chemical calculations,
together with our chemical model, are presented and discussed for each isomeric group.

\begin{figure}
\centering
\includegraphics[width=0.5\textwidth]{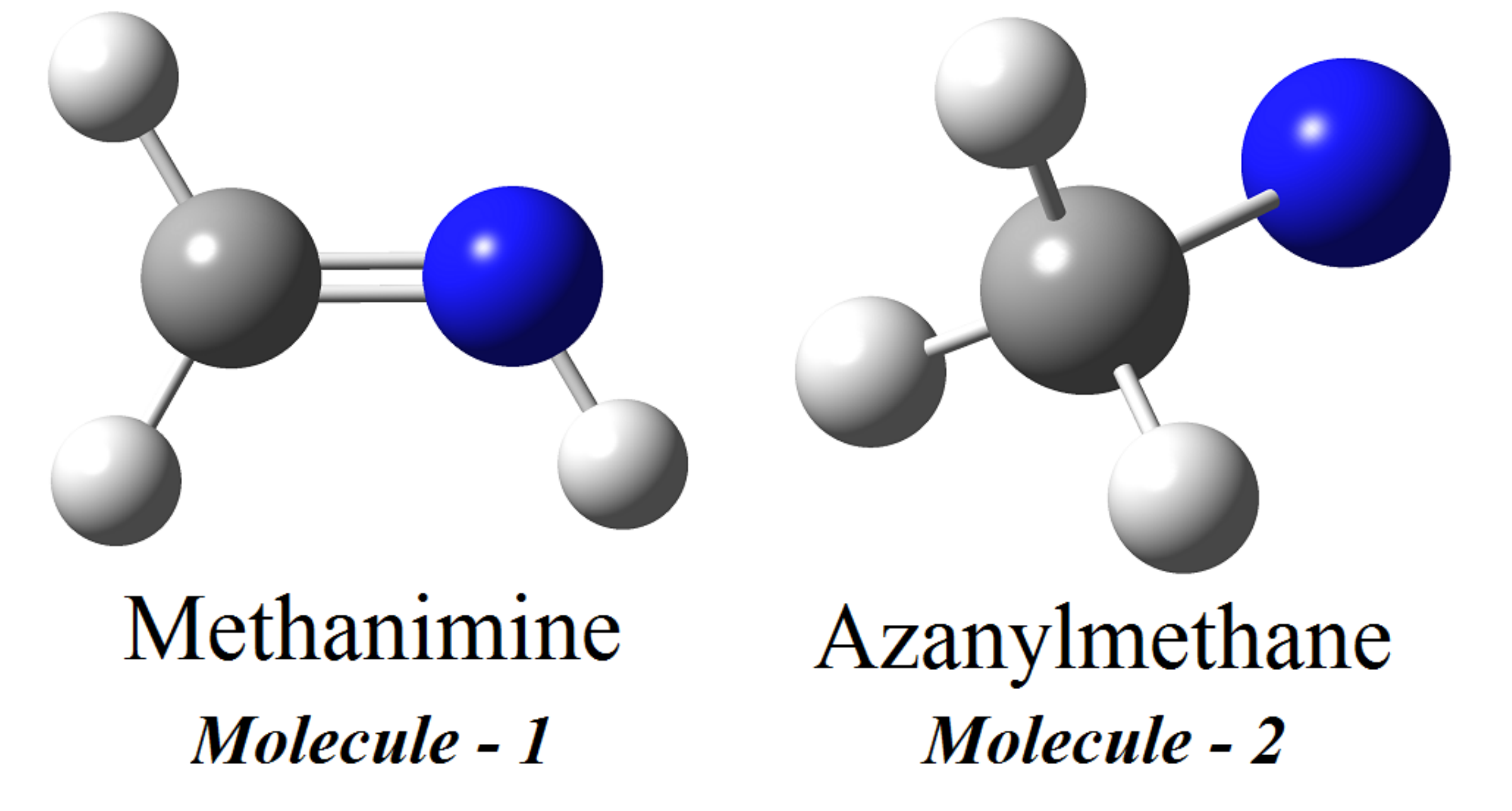}
\caption{$\mathrm{CH_3N}$ isomers \citep{sil18}.
\label{fig:amine_1}}
\end{figure}

\subsubsection{$\rm{CH_3N}$ Isomeric Group}
This group contains two molecular species (see Figure \ref{fig:amine_1}), methanimine and $\lambda^1$-azanylmethane.
Though the presence of methanimine ($\mathrm{CH_2NH}$) in ISM has already been confirmed \citep{godf73}, $\lambda^1$-azanylmethane is
yet to be ascertained.
Based on the enthalpy of formation and relative energy values shown in Table \ref{tab:amine_2}, methanimine appears to be the most stable candidate of the $\rm{CH_3N}$ isomeric group.
But enthalpy of formation is not only a factor
to dictate the abundance of these species, specifically, when it is far away from the equilibrium.
It is only the reaction pathways that can dictate the final abundance of any species in the ISM.
Our calculated dipole moment components (shown in Table \ref{tab:amine_3}) of methanimine are very close to the available experimental values. From our calculated dipole moment components, it is found that for methanimine, ``a'' and ``b'' type rotational
transitions are the strongest, whereas ``c'' type transition is absent.
In the case of $\lambda^1$-azanylmethane, the strongest component of dipole moment is
found to be the ``a'' component, whereas the ``b'' component is found to be the weakest.
The average dipole moment component of methanimine is slightly higher than that of the $\lambda^1$-azanylmethane.
Our calculated rotational constants for methanimine are shown in Table \ref{tab:amine_4},
which are very close to the overall experimental values.

It is believed that methanimine is primarily created within the ice phase.
The dominated pathways are shown in the reaction range $\rm{R4-R7}$ of Table \ref{tab:amine_1}.
$\rm{CH_2NH}$ may form through the successive H addition reaction in the ice phase, starting with the cyanide radical. Subsequent H addition may occur in two ways: H addition with HCN
could result in $\rm{H_2CN}$ (R4) or HCNH (R5). \cite{woon02} pointed out that
reactions R4 and R5 possess activation energy barriers of about
$3647$ K and $6440$ K, respectively. $\rm{H_2CN}$ and HCNH can further produce
$\rm{CH_2NH}$ by the H addition reaction (R6 and R7, respectively).
The surface network of KIDA already considers the reactions enlisted in \cite{gran14},
and thus HCN/HNC-related chemistry is consistent.
The gas-phase pathways of \cite{gran14} are also considered in our gas-phase network.
Near the higher temperatures, methanimine may be produced by the decomposition of
methylamine \citep[$\rm{CH_3NH_2}$;][]{john72}.
\cite{suzu16} pointed out that this species could be produced on the interstellar ice
by other reactions $\rm{(R1-R3)}$ shown in Table \ref{tab:amine_1}.

\begin{figure}
\centering
\includegraphics[width=0.8\textwidth]{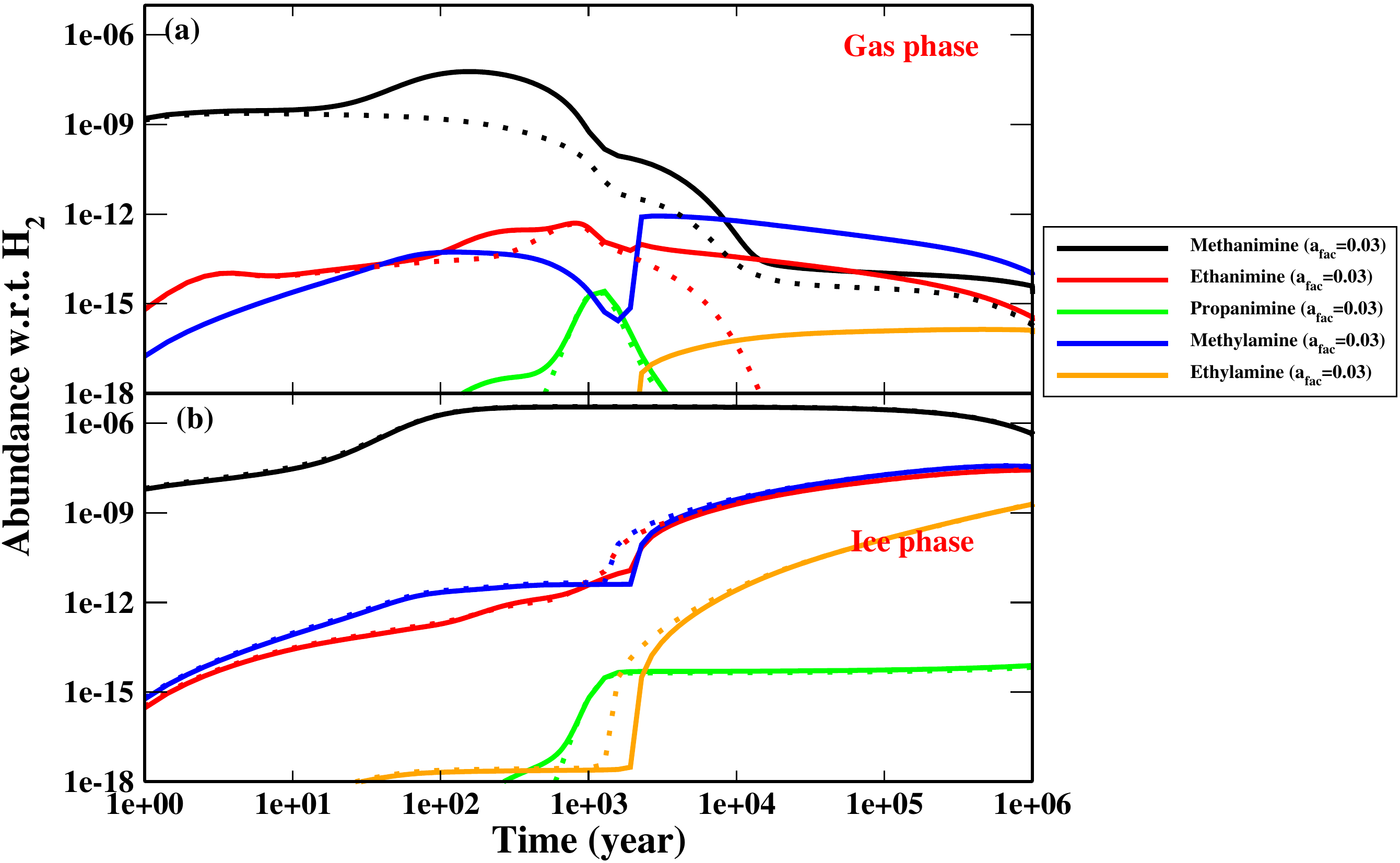}
\caption{Chemical evolution of the aldimines and amines in the isothermal stage for $a_{fac}=0.03$ (solid) and $0$ (dashed) \citep{sil18}.
\label{fig:amine_2}}
\end{figure}

\begin{figure}
\centering
\includegraphics[width=0.8\textwidth]{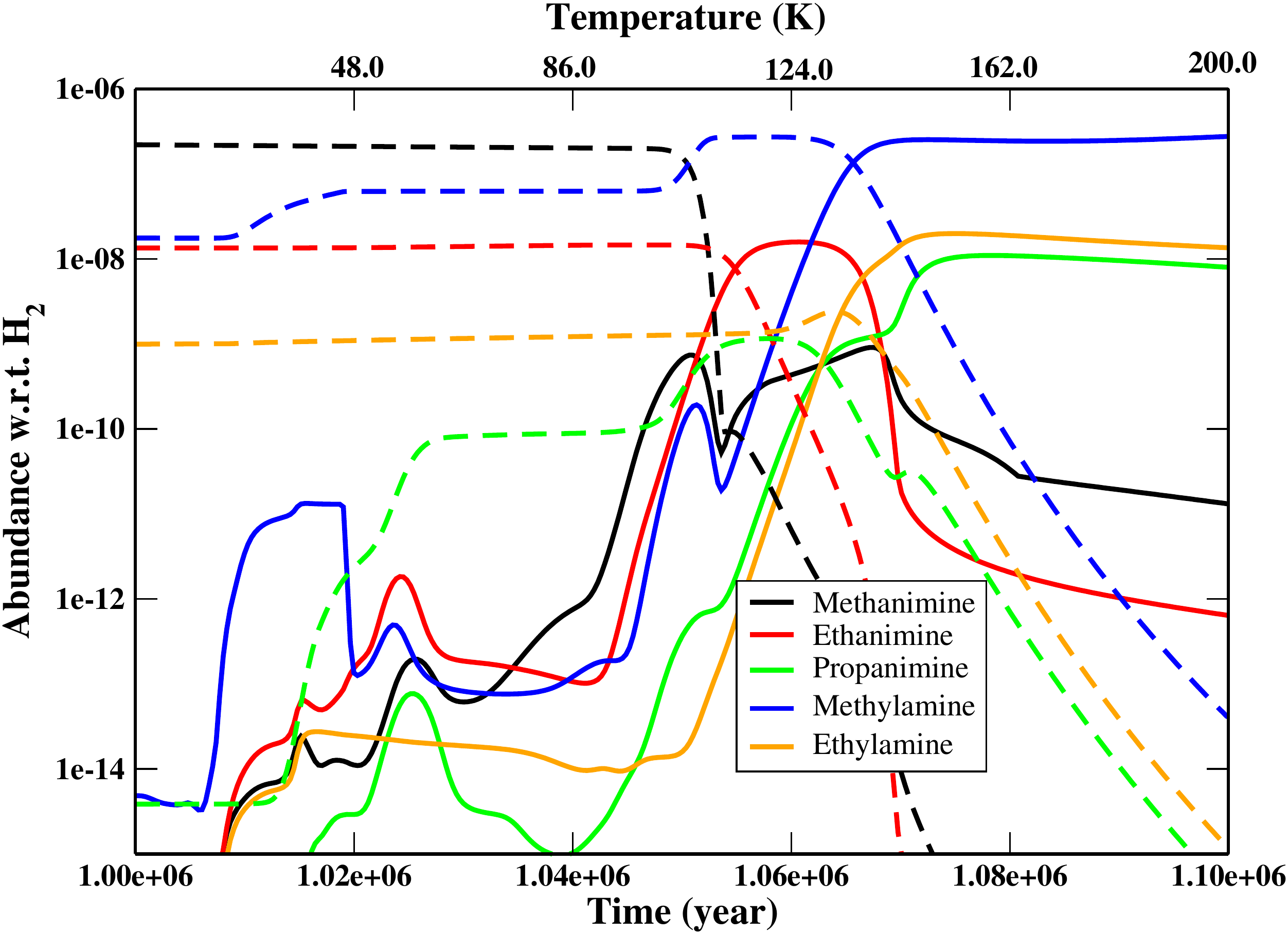}
\caption{Chemical evolution of the aldimines and amines in the warm-up stage. 
Solid lines represent the gas-phase species, whereas corresponding dashed lines represent the ice-phase species \citep{sil18}.
\label{fig:amine_3}}
\end{figure}

For the gas-phase reactions G4 and G5 of Table \ref{tab:amine_5},
we obtain $\Delta G\ddag$ of $8.37$ kcal/mol and $10.06$ kcal/mol, respectively.
The chemical evolution of methanimine within the cold isothermal stage and the subsequent
warm-up stage is shown in Figures \ref{fig:amine_2} and \ref{fig:amine_3}, respectively. Abundances are shown relative to $\rm{H_2}$ molecules.
During the isothermal stage, methanimine is significantly abundant
in both phases and has a peak abundance of $5.93 \times 10^{-08}$ in the gas phase and
$3.49 \times 10^{-06}$ in the ice phase. A strong decreasing slope of gas-phase methanimine
is observed from Figure \ref{fig:amine_2}a. Figure \ref{fig:amine_2}b also depicts a
decreasing slope of ice-phase methanimine (during the end of the isothermal regime) due to the production of methylamine by successive H addition reactions $\rm{(R8-R9)}$.
Dashed curves in Figure \ref{fig:amine_2}a are shown for the gas-phase abundances
considering a non-thermal desorption factor, $a_{fac} = 0$.
The gas-phase abundances of methanimine with
$a_{fac}=0$ (dashed line in Figure \ref{fig:amine_2}a) and $a_{fac}=0.03$ (solid line in Figure \ref{fig:amine_2}a) differ significantly.
This is because, in the isothermal stage, the gas-phase contribution of methanimine is mainly coming from the ice phase via the non-thermal desorption mechanism. Following the KIDA, the BE of methanimine is assumed to be $5534$ K.
It is evident from Figure \ref{fig:amine_3} that sublimation of methanimine occurs around
$110$ K in the warm-up stage and the abundance of gas-phase methanimine is significantly increased owing to the efficient gas-phase formation by reactions $\rm{G1-G7}$.
The peak abundance of gas-phase methanimine is found to be around
$1.81 \times 10^{-09}$. Our obtained abundances can be compared with the hot-core observation of methanimine of $\sim 7.0 \times 10^{-08}$ in G10.47+0.03 (hereafter, G10), and $3.0 \times 10^{-9}$ in NGC 6334F by \cite{suzu16}.

\begin{figure}
\centering
\includegraphics[width=0.3\textwidth]{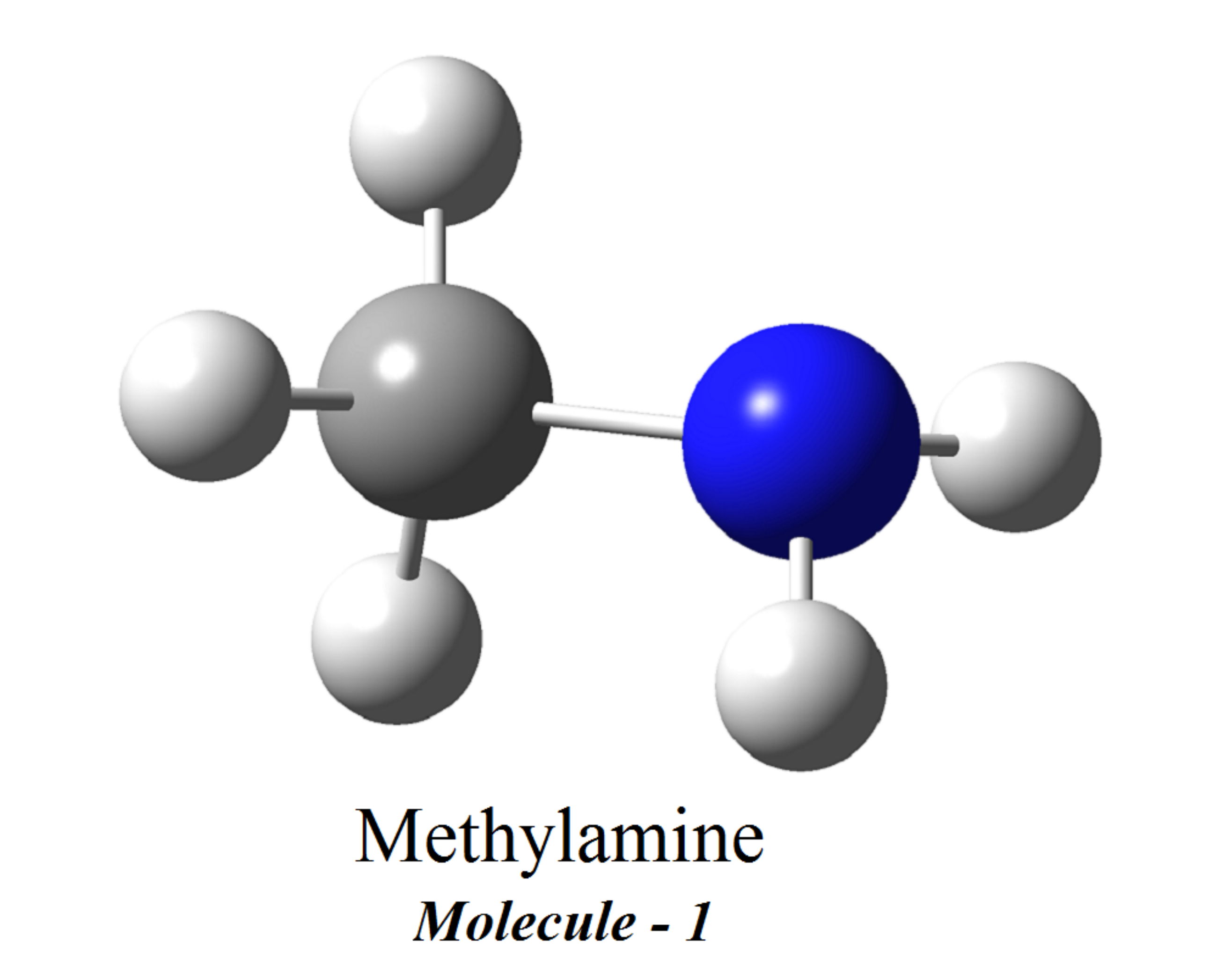}
\caption{$\mathrm{CH_5N}$ isomer \citep{sil18}.
\label{fig:amine_4}}
\end{figure}

\subsubsection{${CH_5N}$ Isomeric Group}
Only methylamine ($\mathrm{CH_3NH_2}$) belongs to this isomeric group (see Figure \ref{fig:amine_4}), and this was
already observed long ago in Sgr B2 and Ori A \citep{four74,kaif74}.
In Table \ref{tab:amine_2}, we compare our calculated enthalpies of formation with that of the current experimental
value. We find that our computed value with the DFT-B3LYP/6-31G(d,p) level of theory is closer to the experimentally obtained value than computed from the G4 composite method.
\cite{taka73,kaif74} found that the c-type transitions of methylamine are four times stronger than the a-type transitions.
We also find a very strong c-component of dipole moment
noted in Table \ref{tab:amine_3}. The calculated total dipole moment component for methylamine is $1.2874$ D,
whereas the experimentally obtained value is $1.31 \pm 0.03$ D.
Also, an excellent correlation between our calculated rotational constants and experimentally obtained values can be seen in Table \ref{tab:amine_4}.

In the ISM, methylamine may be formed via two successive
H addition reactions of methanimine in the ice phase \citep{godf73,suzu16}.
\cite{woon02} determined that the primary step of this H addition reaction may proceed in
two ways. First, hydrogenation of methanimine yields $\rm{CH_3NH}$ (R8) having an activation
barrier of $2134$ K; second, it may produce $\rm{CH_2NH_2}$ (R9) having an activation barrier of $3170$ K.
Reactions R8 and R9 may also occur in the gas phase. From our TST calculation, we have $\Delta G\ddag= 7.84$ kcal/mol for reaction G8 of Table \ref{tab:amine_5}. However, we do not find a suitable TS for gas-phase reaction G9 of Table \ref{tab:amine_5}.
In the ice phase, R9 possesses a higher ($1.485$ times) activation barrier than R8.
We assume that a similar trend is followed for the gas-phase reaction G9, so we consider $\Delta G\ddag=11.64$ kcal/mol for the gas-phase reaction number G9.
Methylamine may further be produced by the hydrogenation reaction of these two products by
reactions R10 and R11, respectively.

The reaction number G8 and G9 possess high $\Delta G \ddag$.
Thus, during the isothermal stage, the production of
gas-phase methylamine is inadequate. However, despite a high activation barrier ($E_a$),
reactions R8 and R9 would be efficiently processed on interstellar ice by quantum
mechanical tunneling and populate the gas phase by the non-thermal desorption.
Mainly due to the non-thermal chemical desorption phenomenon,
ice-phase methylamine populates the gas phase.
It is visible from Figure \ref{fig:amine_2}a that with $a_{fac}=0$, the gas-phase
contribution of methylamine is negligible (the dashed line, which corresponds
to the methylamine is absent). Hence, in the isothermal stage, the contribution for the
gas-phase methylamine mainly comes from the ice phase.
In the isothermal stage, methylamine attains a peak abundance
of $8.46 \times 10^{-13}$ in the gas phase
and  $3.70 \times 10^{-08}$ in the ice phase.
From Figure \ref{fig:amine_3}, it is observed that the ice-phase abundance
initially increases owing to the increase in the mobility of the reactants.
Peak ice-phase abundance of methylamine is obtained to be $5.44 \times 10^{-07}$.
Near the high temperature, the production of gas-phase methylamine significantly
contributed to (a) the enhancement of the temperature-dependent rate coefficient
of reactions G8 and G9 and (b) the increase in the gas-phase methanimine formation.
The peak gas-phase abundance of methylamine is found to be  $5.54 \times 10^{-07}$.
Our obtained values may be compared with the recent observation of methylamine \citep{ohis17}.
They predicted methylamine abundance of $\sim 1.2 \times 10^{-08}$ in G10.

\subsubsection{${C_2H_5N}$ Isomeric Group}
\begin{figure}
\centering
\includegraphics[width=0.7\textwidth]{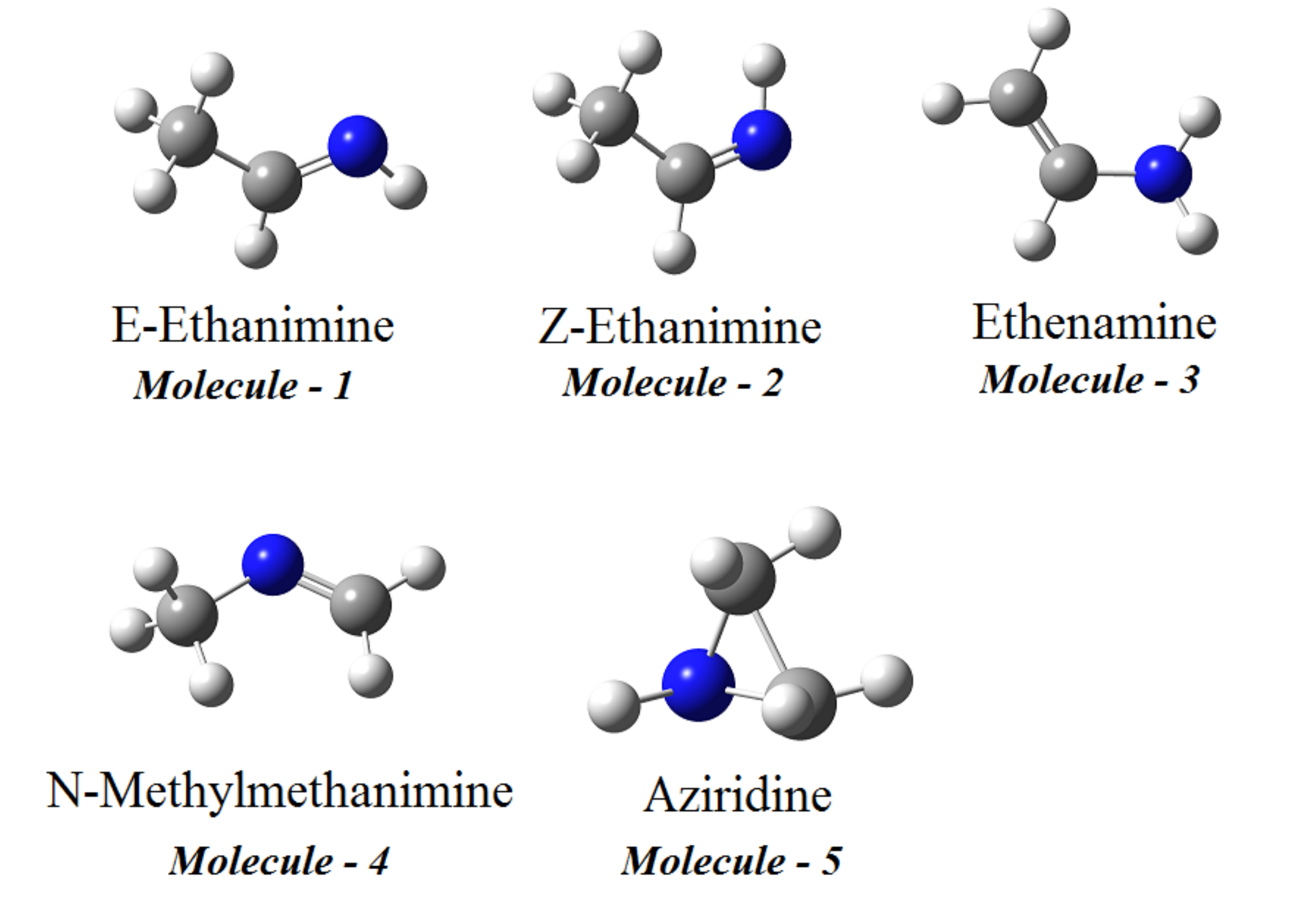}
\caption{$\mathrm{C_2H_5N}$ isomers \citep{sil18}.
\label{fig:amine_5}}
\end{figure}

\begin{figure}
\centering
\includegraphics[width=0.6\textwidth]{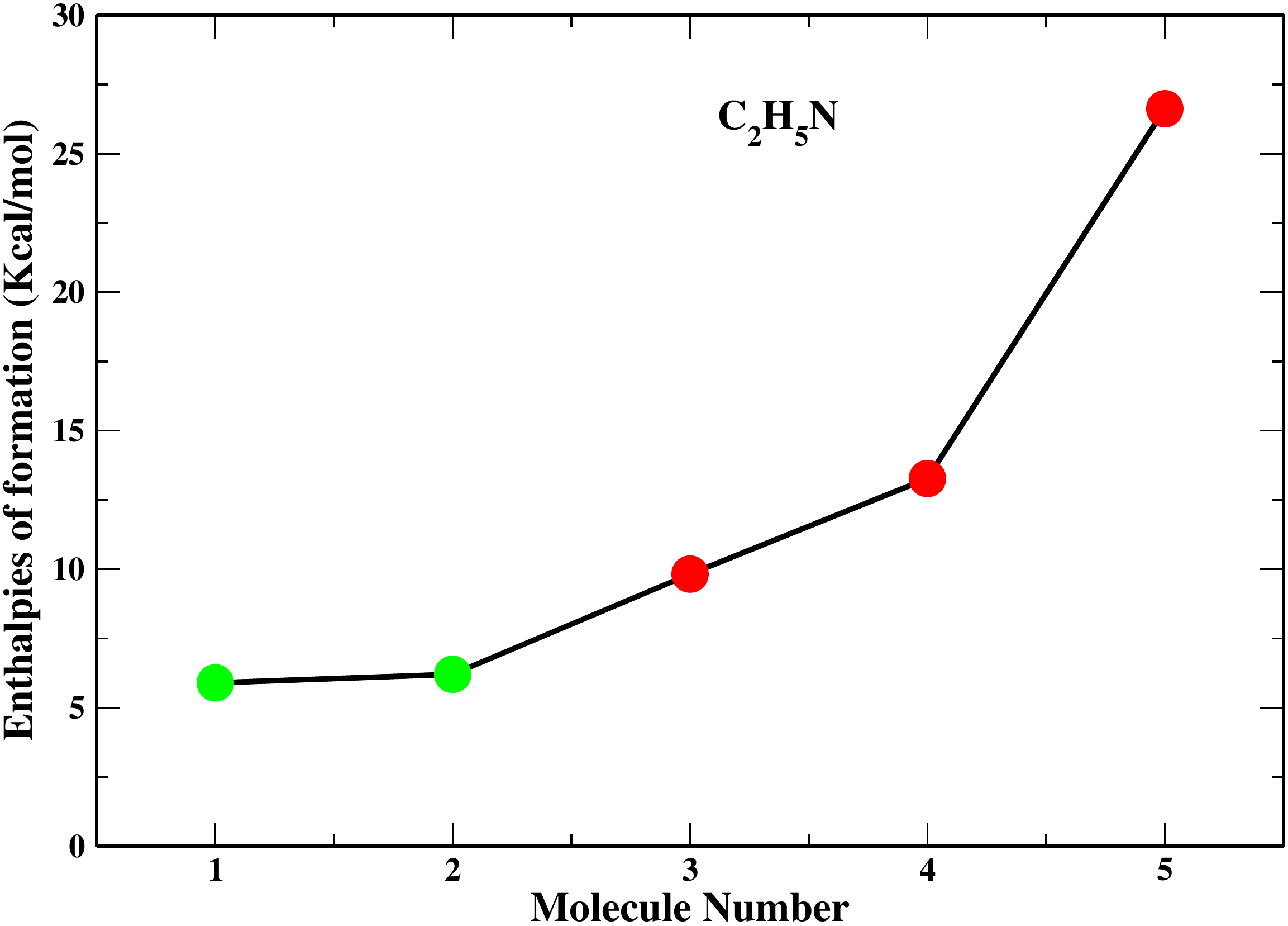}
\caption{Enthalpy of formation of the $\rm{C_2H_5N}$ isomeric group. Molecules already observed are marked as green circles, and those yet to be observed are marked as red circles \citep{sil18}.
\label{fig:amine_6}}
\end{figure}

\begin{figure}
\centering
\includegraphics[width=0.6\textwidth]{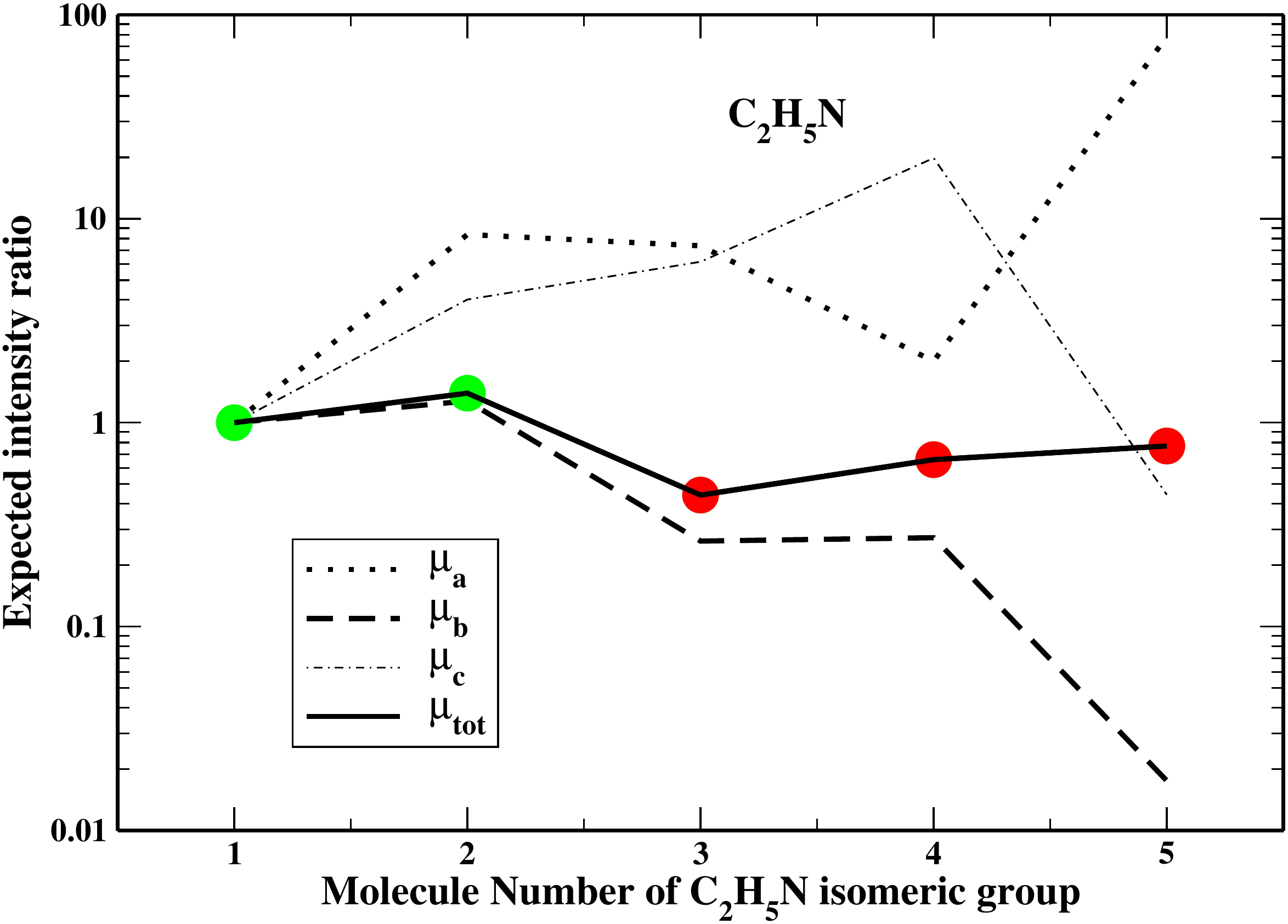}
\caption{Expected intensity ratio of the $\rm{C_2H_5N}$ isomeric group by considering three 
components of the dipole moment and effective dipole moment of all the species \citep{sil18}.
\label{fig:amine_7}}
\end{figure}

Five isomers belong to this isomeric group (see Figure \ref{fig:amine_5}): E-ethanimine, Z-ethanimine, ethenamine, N-methylmethanimine, and aziridine. Out of these five isomers, both conformers (E and Z) of ethanimine ($\rm{CH_3CHNH}$)
had been detected in Sgr B2 \citep{loom13}. From our quantum chemical calculation,
we find that E-ethanimine is
energetically more stable ($4.35$ kJ/mol by using MP2/6-311G++(d,p) level of theory
and $1.2$ kJ/mol by using the G4 composite method)
than Z-ethanimine. \cite{quan16} and \cite{loom13} obtained
an energy difference of $4.60$ kJ/mol and $4.24$ kJ/mol, respectively, between these two conformers.
We show the enthalpy of formation values for all the species in Table \ref{tab:amine_2},
along with the experimentally obtained values, where available.
Our calculated enthalpies of formation using the DFT-B3LYP/6-31G(d,p) level of theory
are in good agreement with the experimentally obtained values of E-ethanimine, N-methylmethanimine, and aziridine.
For a better assessment, in Figure \ref{fig:amine_6}, we show the enthalpy of formation
with the molecule number noted in Table \ref{tab:amine_2} and Figure \ref{fig:amine_5}
for the $\rm{C_2H_5N}$ isomeric group. E-ethanimine has
the minimum enthalpy of formation, followed by Z-ethanimine.
The energy difference between these two is smaller than the other isomers of this isomeric group.
Observed isomers are marked as green circles in Figures \ref{fig:amine_6} and \ref{fig:amine_7}, and the unobserved isomers are marked as red circles.

All the dipole moment components are presented in Table \ref{tab:amine_3}. It is found that in both the cases (E and Z), the b-type transition dominates. Our calculated dipole moments are compared with the existing values \citep{lova80}.
The effective dipole moment of Z-ethanimine is found to be higher than that of E-ethanimine. \cite{char95} pointed out that an estimate of the antenna temperature could be made by calculating the intensity of a given rotational transition for an optically thin emission.
This intensity is proportional to $\mu^2/Q(T_{rot})$, where $\mu$ is the electric
dipole moment, and $Q (T_{rot})$ is the partition function at rotational temperature ($T_{rot}$). In Table \ref{tab:amine_4}, we point out the rotational constants
and the rotational partition function for $T_{rot}=200$ K.
In Figure \ref{fig:amine_7}, we show the plot of the expected intensity ratio
(assuming that all the species of this isomeric group have the same abundances)
concerning the most stable isomer (E-ethanimine) of this isomeric
group. Relative energy values with molecule number of all isomers
of this isomeric group are noted in Table \ref{tab:amine_2}).
In Figure \ref{fig:amine_7}, the expected intensity ratios for all the species of this isomeric group are shown by considering the three
components of the dipole moment along with the effective dipole moments.
In this isomeric group, Z-ethanimine has the largest effective dipole moment.
So considering the same abundances, the probability of detecting the
Z isomer of ethanimine will be more favorable than the other isomers of this isomeric group.
From Figure \ref{fig:amine_5}, we can see that after E and Z isomers of ethanimine,
aziridine has the most robust transition. Still, due to its higher relative energy
compared to E-ethanimine, it is the less probable candidate (if reaction pathways do not
influence it another way) for astronomical detection.

\cite{loom13} mentioned that ethanimine might be produced via
two consecutive H addition reactions with $\rm{CH_3CN}$ in the ice phase.
\cite{quan16} recently proposed that the first step (R14) of the H addition
reaction with $\rm{CH_3CN}$ has a barrier of $1400$ K, and the second step (R15) is
a $\rm{radical-radical}$ reaction and assumed to be barrierless. For the gas-phase
reaction of G14 of Table \ref{tab:amine_5}, our calculated value
of $\Delta G \ddag$ is $10.32$ kcal/mol.
\cite{quan16} additionally suggested that ethanimine can even be produced by the reaction
between $\rm{CH_3}$ and $\rm{H_2CN}$ in the ice phase. Among the gas-phase pathways, the reaction between $\rm{C_2H_5}$ and NH (reaction G23 of Table \ref{tab:amine_5}) may lead to $\rm{CH_3CHNH}$. In our network, we assume only one form of ethanimine (E-ethanimine) for our modeling.

In the isothermal stage (see Figure \ref{fig:amine_2}), ethanimine has a peak value of
$4.99 \times 10^{-13}$ in the gas phase and $2.69 \times 10^{-08}$ in the ice phase.
During the warm-up stage (see Figure \ref{fig:amine_3}), gas-phase ethanimine has a peak
value of $3.17 \times 10^{-08}$.
In the warm-up stage, gas-phase production of ethanimine is also contributing owing to the
enhancement of the temperature-dependent rate coefficient of reaction G14.
The abundance of ethanimine attains a peak around 125 K, and then starts
to decrease due to the formation of ethylamine by the successive hydrogenation
reactions $\rm{(R17-R18)}$.
Since reaction R17 has a barrier, we use Equation \ref{eqn:eyring} for the computation
of its rate coefficient. Equation \ref{eqn:eyring} clearly says that as we are increasing
the temperature, the rate coefficient rises exponentially.
Thus, in the warm-up stage, the rate coefficient of reaction G17 increases
exponentially and attains a reasonable rate ($\sim 10^{-10}$ cm$^3$ s$^{-1}$),
which means that the destruction of ethanimine by the hydrogenation reaction also gradually increases and reaches a quasi-steady state.
At the end of our simulation (after $1.1 \times 10^6$ years), we note
an abundance of $1.27 \times 10^{-12}$ for ethanimine, whereas the
predicted abundance is $\sim 6.0 \times 10^{-11}$ from \cite{quan16}.

\subsubsection{${C_2H_7N}$ Isomeric Group}
\begin{figure}
\centering
\includegraphics[width=0.6\textwidth]{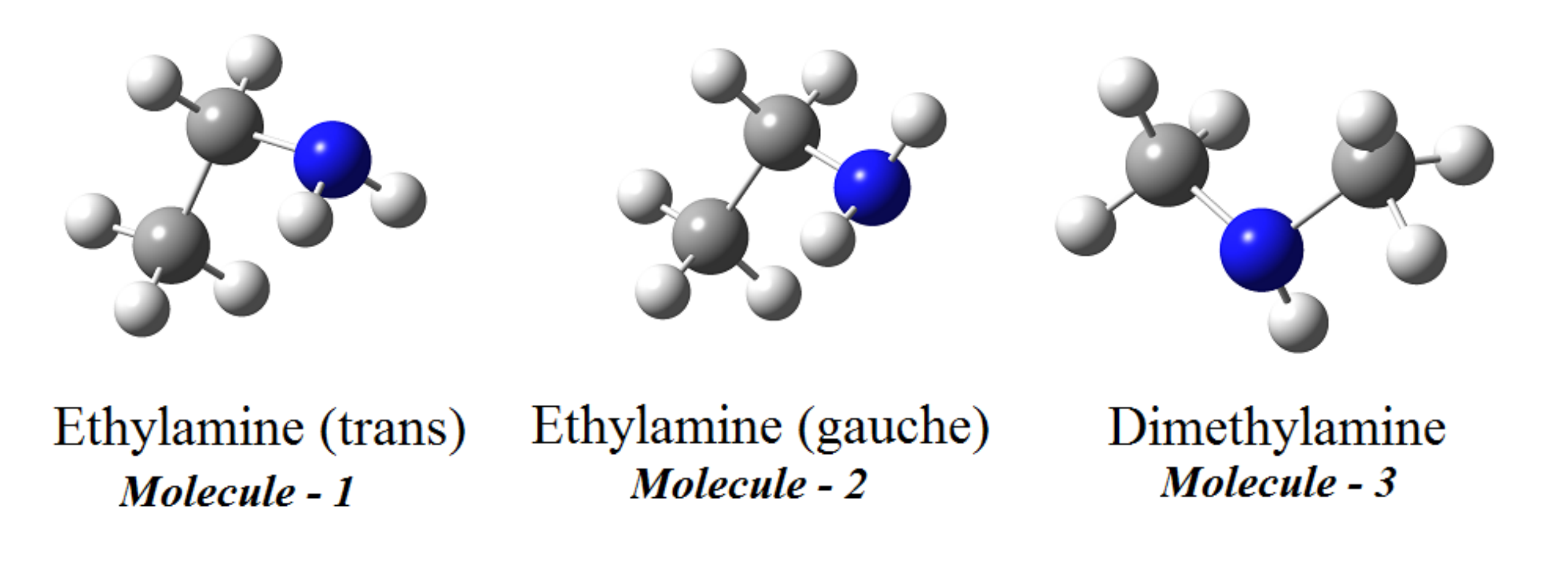}
\caption{$\mathrm{C_2H_7N}$ isomers \citep{sil18}.
\label{fig:amine_8}}
\end{figure}

\begin{figure}
\centering
\includegraphics[width=0.6\textwidth]{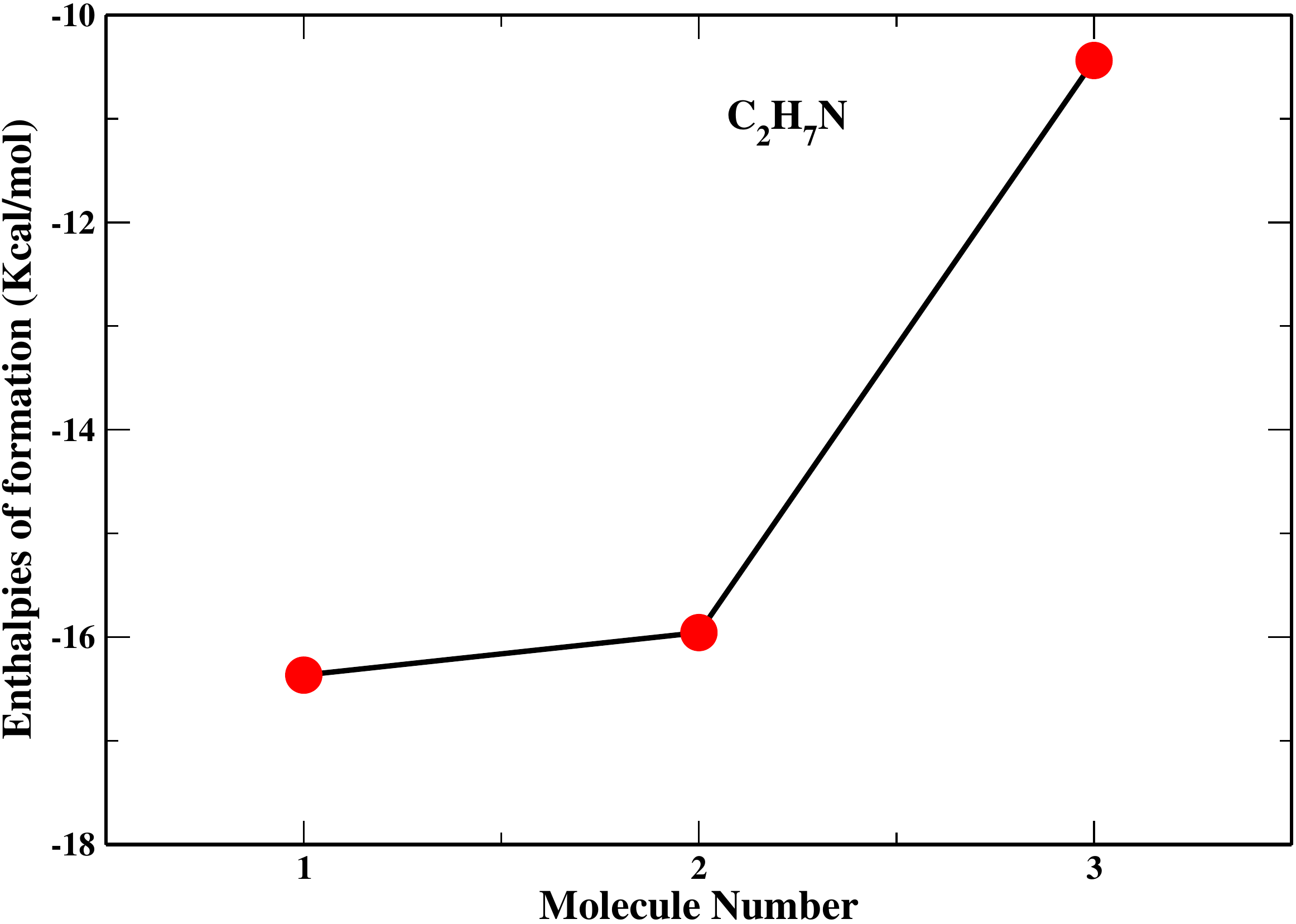}
\caption{Enthalpy of formation of the $\rm{C_2H_7N}$ isomeric group \citep{sil18}.
\label{fig:amine_9}}
\end{figure}

Trans-ethylamine, gauche-ethylamine, and dimethylamine belong to the $\rm{C_2H_7N}$ isomeric
group (see Figure \ref{fig:amine_8}).
Interestingly, no species of this isomeric group is detected in the ISM.
However, the presence of ethylamine, which is the precursor of glycine,
was traced in comet Wild 2 \citep{glav08}.
Ethylamine can exist in the form of two stable conformers: gauche and trans.
An experiment by \cite{hama86} shows that the trans conformer is slightly more stable
than the gauche conformer. Our calculated values are
also in line with this result. We obtain that the gauche conformer has $1.67$ kJ/mol
higher energy than the trans conformer.
Hence, according to the enthalpy of formation and relative energies
as shown in Table \ref{tab:amine_2}, trans-ethylamine has the least enthalpy of formation
and is most stable among this isomeric group. In Figure \ref{fig:amine_9}, the enthalpy of
formation of this isomeric group is depicted with the molecule
number, and the enthalpy of formation is noted in Table \ref{tab:amine_2}.
Compared to the experimentally measured enthalpies of formation, the G4 composite method overestimates it for ethylamine and dimethylamine.

\begin{figure}
\centering
\includegraphics[width=0.6\textwidth]{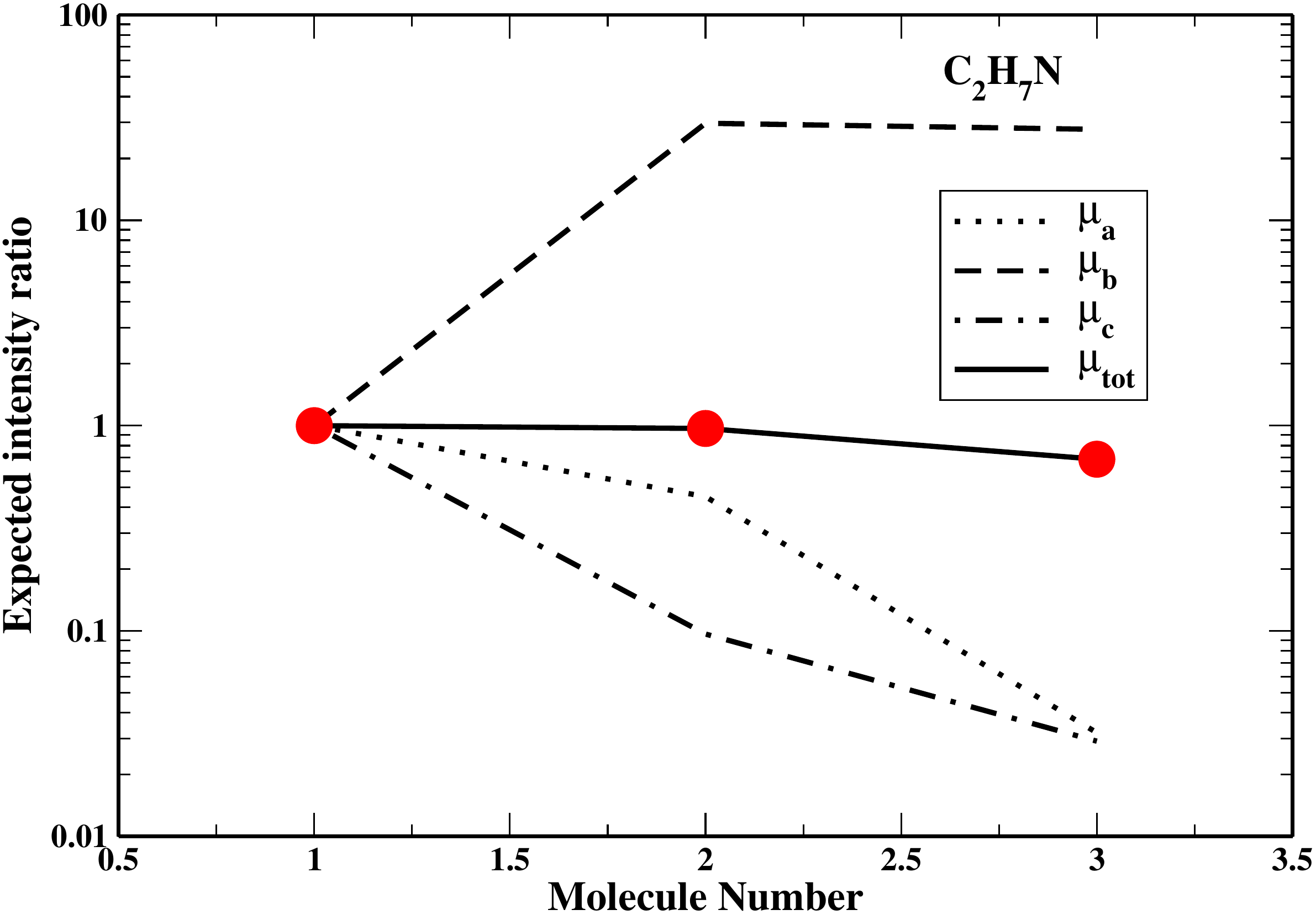}
\caption{Expected intensity ratio of the $\rm{C_2H_7N}$ isomeric group by considering three components of 
the dipole moment and effective dipole moment of all the species \citep{sil18}.
\label{fig:amine_10}}
\end{figure}

In the case of the trans conformer of ethylamine, components ``a'' and ``c'' of
dipole moments are more substantial, whereas, for the gauche form, the b-component
is found to be the strongest. For dimethylamine, also the b-component of the dipole moment dominates, and the ``a'' and ``b'' components have minor contributions.
Based on the data available from our quantum chemical calculations, in Figure \ref{fig:amine_10}, we show the expected intensity ratio concerning the species having the least enthalpy of formation. Figure \ref{fig:amine_10} depicts that the trans-ethylamine has the highest
expected intensity ratio ($\sim 1$) compared to the other two members
(gauche-ethylamine has $0.97$ and dimethylamine has $0.69$) of this isomeric group,
and thus trans-ethylamine has the highest probability of its astronomical detection from this isomeric group.

Ethylamine could be formed on the grain surface via two successive H additions of
ethanimine. Our calculation reveals that the first step of this hydrogenation reaction
has an activation barrier of $1846$ K (R17). For the gas-phase hydrogenation reaction (G17),
our calculated $\Delta G \ddag$ parameter is found to be $9.98$ kcal/mol.
Since the second step of this reaction is $\rm{radical-radical}$,
we assume that reaction R18 may be treated as a barrierless process.
In the isothermal stage (see Figure \ref{fig:amine_2}), we have a peak abundance
of $1.19 \times 10^{-16}$ and $1.99 \times 10^{-09}$ in the gas and ice phase, respectively.
In the warm-up stage (see Figure \ref{fig:amine_3}), ice-phase abundance roughly remains unchanged up to $125$ K,
and then it starts to decrease sharply. Its gas-phase abundance is
higher and has a peak abundance of $3.98 \times 10^{-08}$.

Ethanimine and ethylamine are more complex than methanimine and methylamine, respectively.
Thus, the abundance of ethanimine/ethylamine would be less than methanimine/methylamine.
But Figures \ref{fig:amine_2} and \ref{fig:amine_3} depict that this trend is not
universal for all circumstances. It is because their formation is processed
through totally different channels. Their destruction rates are also different.
For example, the ice-phase formation of methanimine mainly occurs by successive hydrogenation reactions with HCN (reactions R4 to R7). In contrast, ethanimine appearance is primarily controlled by consecutive hydrogenation reactions with CH$_3$CN (reactions R14 and R15).
Now, reactions R4 and R5 contain a much higher barrier than that of reaction R14.
Since HCN is more abundant than CH$_3$CN, despite high barriers involved
in forming methanimine, in the isothermal stage, most of the time,
the ice-phase abundance of methanimine remains higher than in ethanimine.
However, in the hot-core region, due to the lower activation barrier of reaction R14,
ethanimine formation becomes more favorable. Methylamine is mainly formed from methanimine
(by reactions $\rm{R8-R10}$), whereas ethylamine is formed from ethanimine
(by reactions $\rm{R17-R18}$). Once again, the activation barrier engaged in
methylamine formation (reaction R9) is higher than the barrier involved in
forming ethylamine (reaction R17). Since very complex chemistry is going on,
the abundances of these species should be compared very carefully.

\cite{altw16} showed that in $\rm{67P/C-G}$, relative abundance between
methylamine and glycine is $1.0\pm0.5$, and the ethylamine-to-glycine ratio is $0.3 \pm0.2$.
Taking the maximum and minimum values from this observation, we can see that
the methylamine-to-ethylamine ratio may vary in the range of $\sim 1-15$.
We focus on the ice-phase evolution (at $T=10$) to check the correlation (if any) between the cometary and interstellar ice.
From our modeling results (see Figure \ref{fig:amine_2}b), we find that
in the isothermal stage (at time $10^6$ years), the methylamine-to-ethylamine ratio
is $\sim 17.7$ in the ice phase is close to the observed value \citep{altw16}.
It suggests that a more in-depth study is required to confirm this linkage between the interstellar and cometary origins of these molecules.

\subsubsection{${C_3H_7N}$ Isomeric Group}

\begin{figure}
\centering
\includegraphics[width=\textwidth]{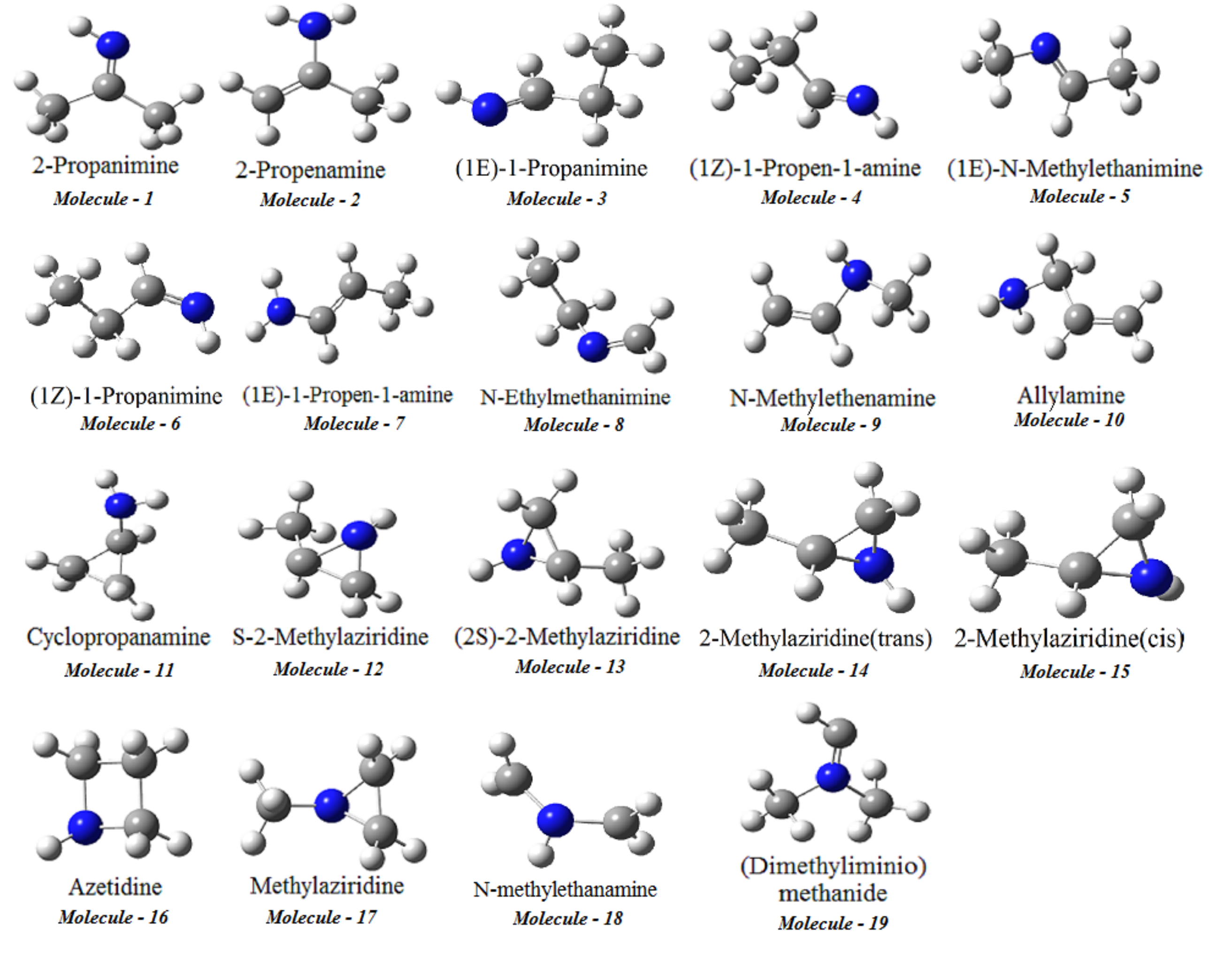}
\caption{$\rm{C_3H_7N}$ isomers \citep{sil18}.
\label{fig:amine_11}}
\end{figure}

\begin{figure}
\centering
\includegraphics[width=0.6\textwidth]{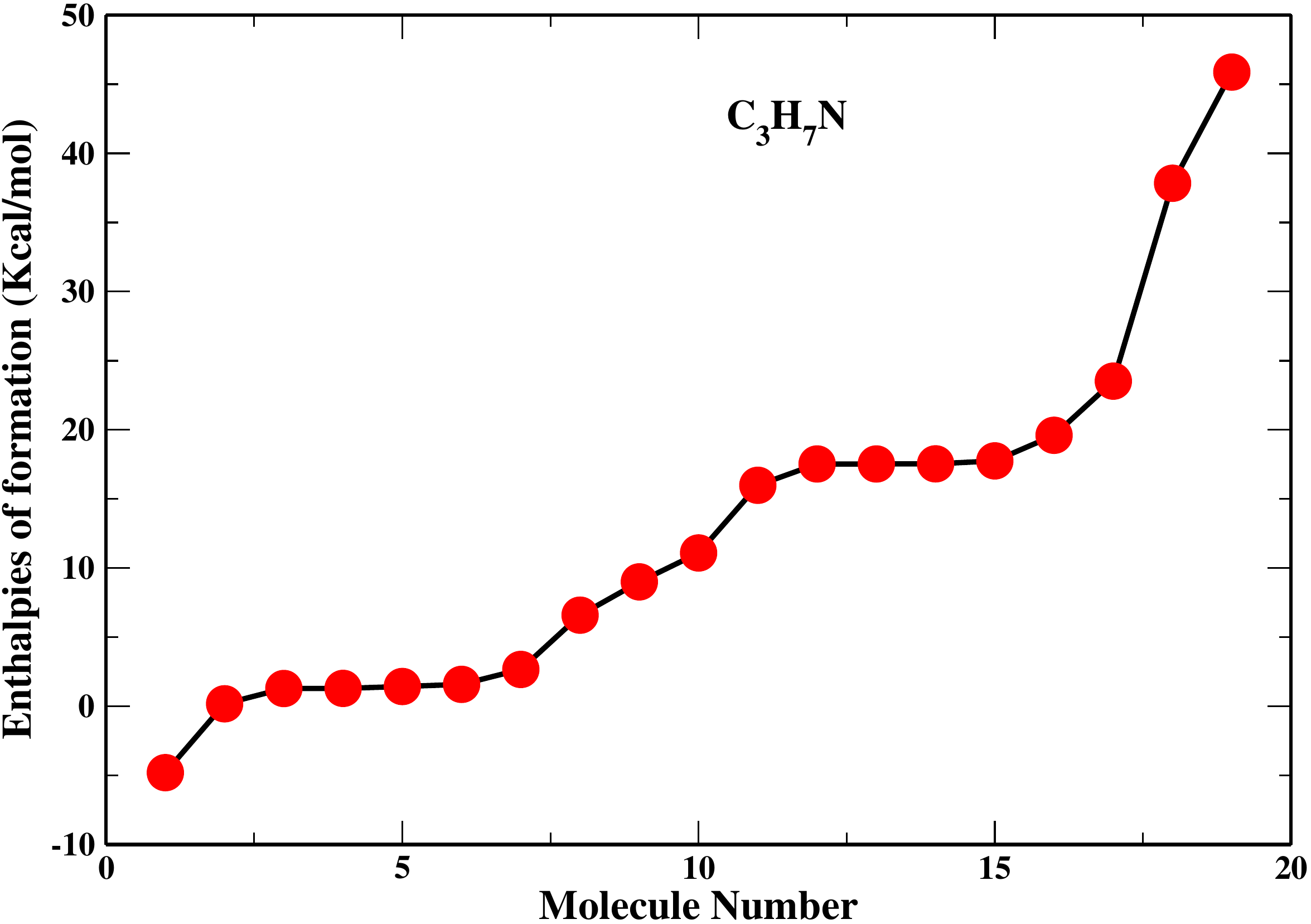}
\caption{Enthalpy of formation of the $\rm{C_3H_7N}$ isomeric group \citep{sil18}.
\label{fig:amine_12}}
\end{figure}

Nineteen isomers (see Figure \ref{fig:amine_11}) belong to this isomeric group.
In Table \ref{tab:amine_2}, we show the relative energies and enthalpy of formation
of various isomers. Figure \ref{fig:amine_12} depicts the enthalpy of formation
($\rm{\Delta_fH^O}$) of this isomeric group.
Only for cyclopropanamine, experimentally obtained enthalpy of formation is available, and
this is in close agreement with our calculated enthalpies of formation with the DFT-B3LYP/6-31G(d,p) level of theory.
2-propanimine is the most stable isomer of this group, followed by 2-propenamine.
The following species in this sequence is (1E)-1-propanimine.
(1E)-1-propanimine has lower energy than that of the (1Z)-1-propanimine.

\begin{figure}
\centering
\includegraphics[width=0.6\textwidth]{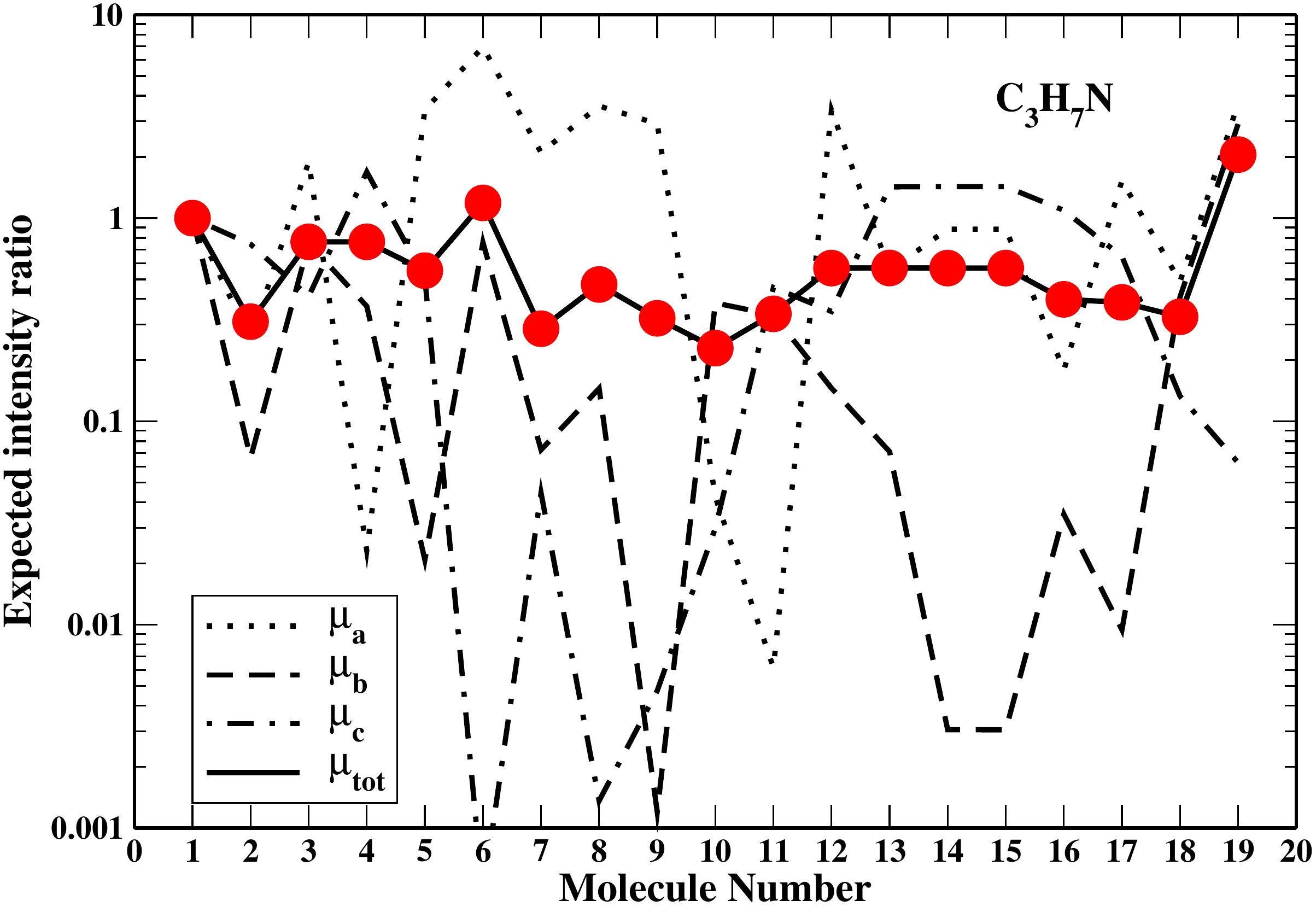}
\caption{Expected intensity ratio of the $\rm{C_3H_7N}$ isomeric group by considering three components of the dipole moment and the
effective dipole moment of all the species \citep{sil18}.
\label{fig:amine_13}}
\end{figure}

Though (dimethyliminio)methanide (molecule no. 19) 
is the least stable species in this isomeric group
(containing the highest enthalpy of formation), our calculation listed
in Table \ref{tab:amine_4} shows that it possesses the highest effective electric dipole moment.
(1Z)-1-propanimine (molecule no. 6) is found to have the second-highest effective dipole moment
in this group. However, it is found that the a-type transitions of (1Z)-1-propanimine are the strongest among all the species of this group. Figure \ref{fig:amine_13} shows the expected intensity ratio (by considering the three components of the dipole moment along with the
effective dipole moment) concerning the most stable (and the species is having the least enthalpy of formation) isomer. Figure \ref{fig:amine_13} shows that if the abundances of all
these isomeric species are same, (dimethyliminio)methanide and (1Z)-1-propanimine
would be the most probable candidates for the astronomical detection from this group.
Since (dimethyliminio)methanide is not a very stable species, it does not have a high
probability of detection. Thus, based on the stability, enthalpy of formation,
and expected intensity ratio, (1Z)-1-propanimine is the most suitable species for future astronomical detection from this isomeric group. However, it is the reaction pathways
that can ultimately decide the fate of this species.

Methanimine and ethanimine have already been observed in the ISM, and (1Z)-1-propanimine
maybe the next probable candidate for astronomical detection. To the best of our knowledge,
the astronomical searches of (1Z)-1-propanimine are yet to be reported in the literature.
Hence, (1Z)-1-propanimine remains the best candidate for astronomical observation among all the
isomers of the $\mathrm{C_3H_7N}$ isomeric group.

(1Z)-1-propanimine may be formed via two sequential H addition reactions on ice
with propionitrile ($\rm{CH_3CH_2CN}$),
where propionitrile may be produced by the $\rm{radical-radical}$ barrierless interaction
between $\mathrm{C_2H_5}$ and CN. Instead of radical reactant CN, $\mathrm{H_2CN}$ may
react with $\mathrm{C_2H_5}$ to form propanimine directly by reaction R19.
This reaction is assumed to be barrierless.
We find that the first step (R21) of H addition with propionitrile has a barrier of
$2712$ K and the second step (R22) is a $\rm{radical-radical}$ barrierless interaction.
For the gas-phase hydrogenation reaction (G21),
our calculated $\Delta G \ddag$ parameter is  $11.03$ kcal/mol.
In Figures \ref{fig:amine_2} and \ref{fig:amine_3},
we show the time evolution of (1Z)-1-propanimine.
These suggest that the production of (1Z)-1-propanimine is only
favorable in the hot-core region. Here we obtain a peak gas-phase abundance
of (1Z)-1-propanimine $\sim 2.20 \times 10^{-08}$ from our model.

\subsubsection{${C_3H_9N}$ Isomeric Group}

\begin{figure}
\centering
\includegraphics[width=0.5\textwidth]{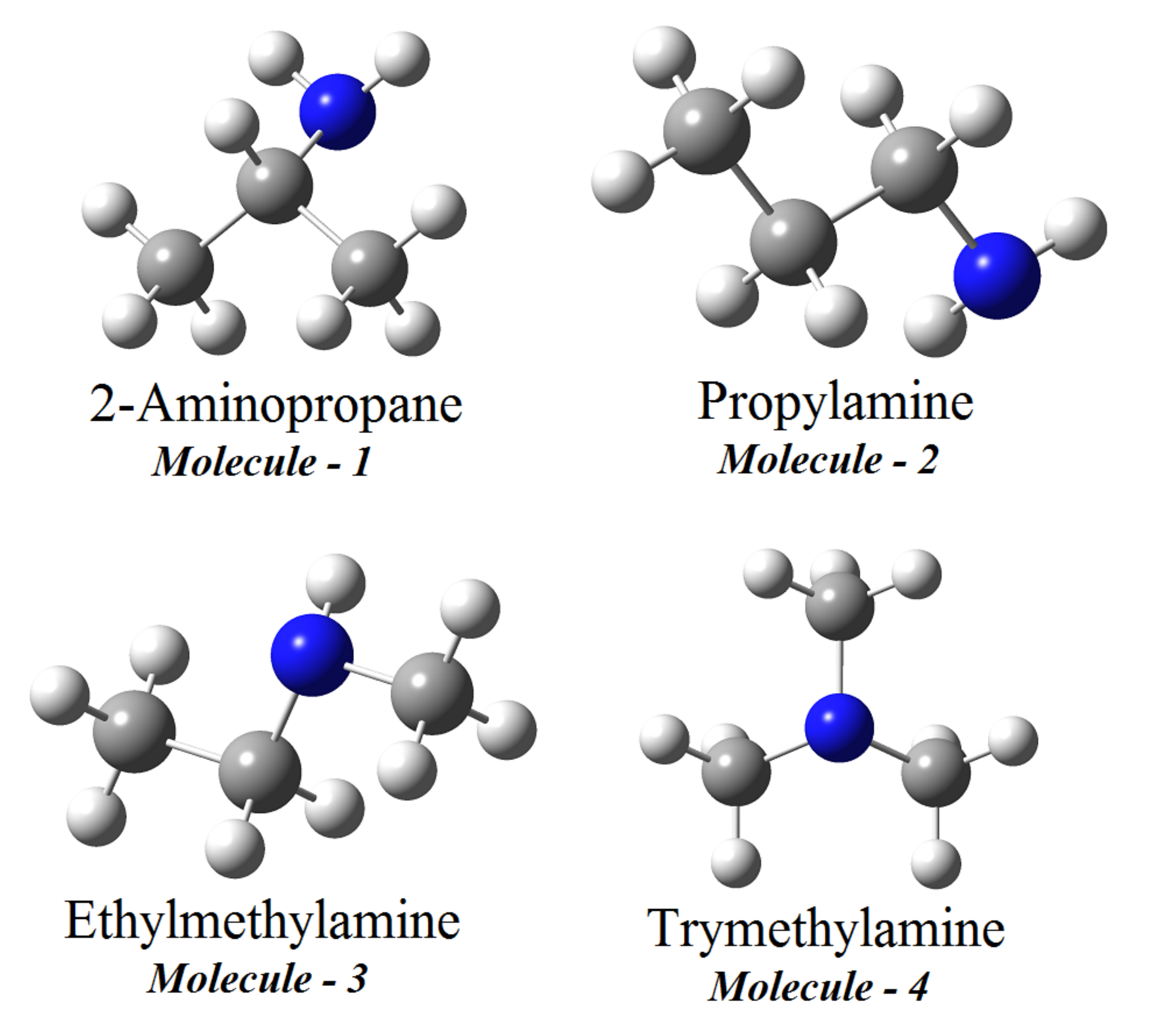}
\caption{$\rm{C_3H_9N}$ isomers \citep{sil18}.
\label{fig:amine_14}}
\end{figure}

\begin{figure}
\centering
\includegraphics[width=0.6\textwidth]{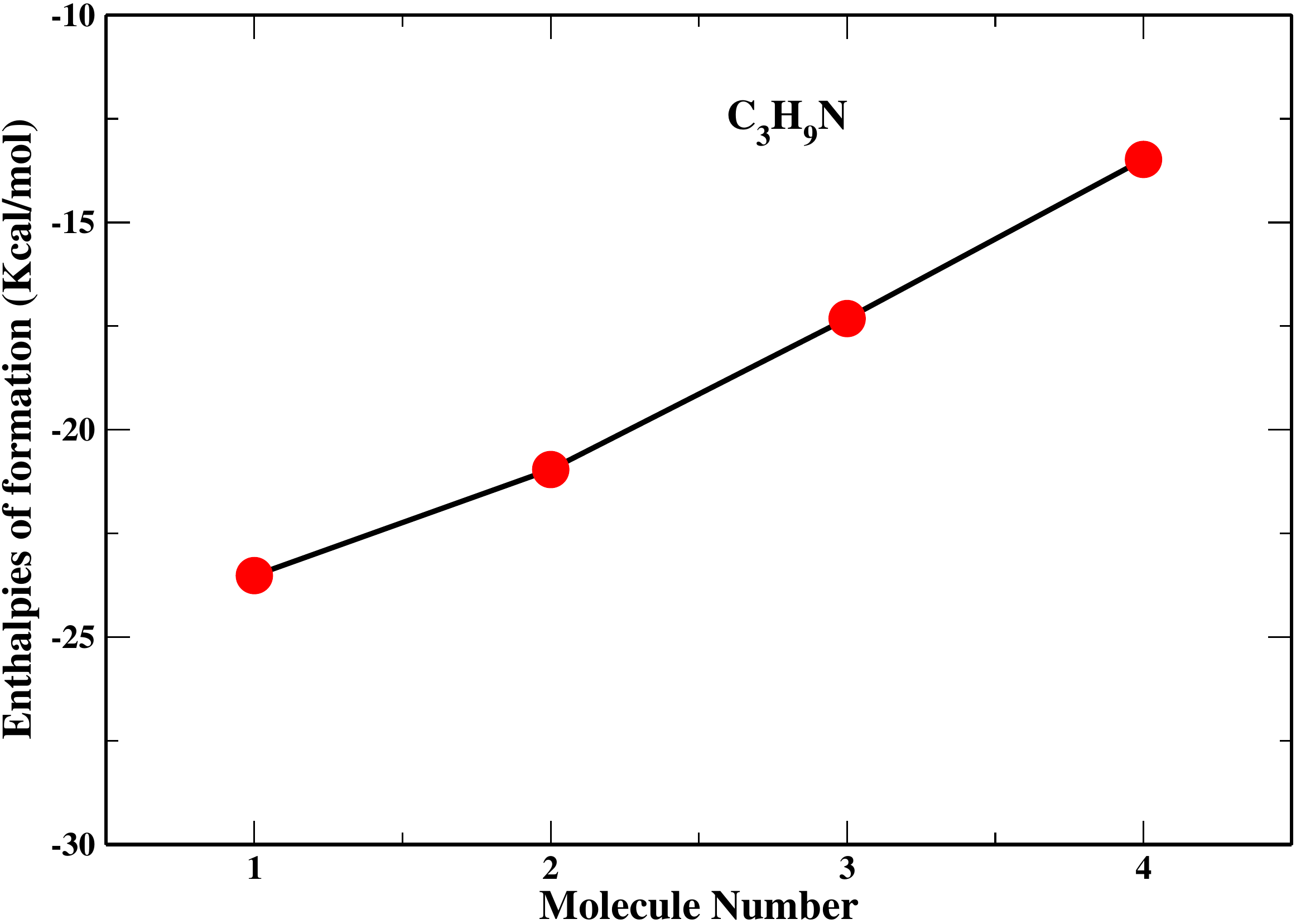}
\caption{Enthalpy of formation of the $\rm{C_3H_9N}$ isomeric group \citep{sil18}.
\label{fig:amine_15}}
\end{figure}

\begin{figure}
\centering
\includegraphics[width=0.6\textwidth]{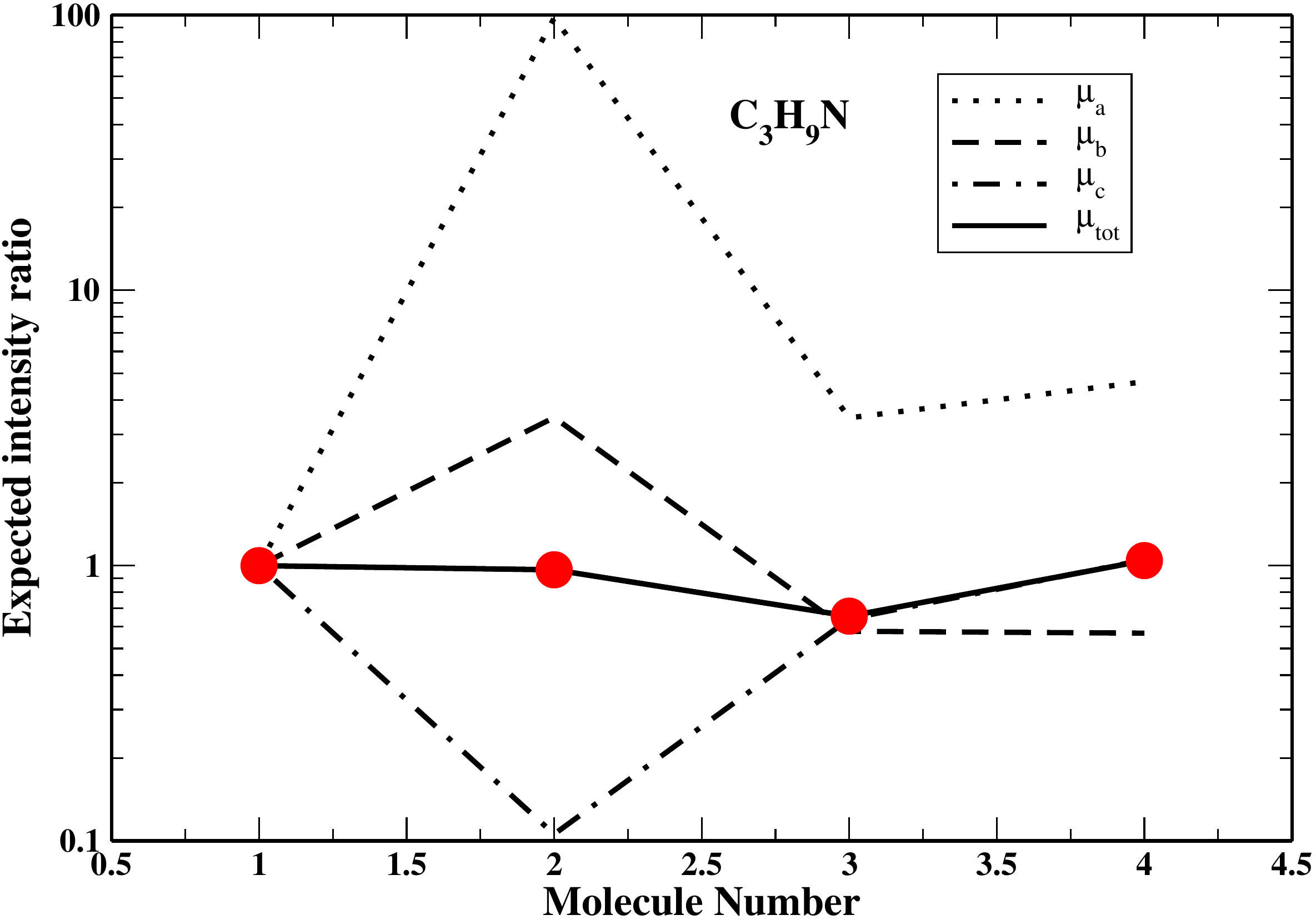}
\caption{Expected intensity ratio of the $\rm{C_3H_9N}$ isomeric group
by considering three components of the dipole moment and effective dipole moment of all the species \citep{sil18}.
\label{fig:amine_16}}
\end{figure}

Four species (2-aminopropane, propylamine, ethylmethylamine, and trimethylamine) are considered (Figure \ref{fig:amine_14}) from this isomeric group.
Methylamine is the most stable isomer of the $\rm{CH_5N}$ group, and ethylamine is
the most stable isomer of the $\rm{C_2H_7N}$ group.
In general, it is expected that the branched-chain molecules
would be comparatively more stable than the other species of an isomeric group.
\cite{etim17} showed that isopropyl cyanide, a branched-chain molecule,
is the most stable within the $\rm{C_4H_7N}$ isomeric group.
Tert-butyl cyanide, another branched-chain molecule, is the most stable
species within the $\rm{C_5H_9N}$ isomeric group. Following a similar trend, we find
that 2-aminopropane, a branched-chain molecule of the $\rm{C_3H_9N}$ isomeric group, is the most stable isomer. 2-aminopropane is found to be $2.51$ kcal/mol more stable than propylamine.
In Figure \ref{fig:amine_15}, we show the enthalpy of formation of these four species.
Relative energy and enthalpy of formation of these four isomers are shown in Table \ref{tab:amine_2} and arranged based on their enthalpy of formation.
Theoretically calculated and experimentally obtained enthalpy of formation values
have a similar trend. From Table \ref{tab:amine_2}, it is evident that the computed
enthalpies of formation with the DFT-B3LYP/6-31G(d,p) level of theory appear
to be comparatively closer to the experimental values than those of the G4 composite method.
The expected intensity ratio concerning the species having the minimum enthalpy of
formation is shown in Figure \ref{fig:amine_16}. Interestingly, though trimethylamine
has the lowest total dipole moment value among this group, the rotational intensity
is maximum because of its lower partition function. More interestingly, due to its unique
structure, rotational constants A and B have the same value ($8.75934$ GHz).
We also reconfirm this unique nature of trimethylamine (an oblate symmetric top species)
using the G4 composite method and the HF/6-31G(3df) level of theory. Since the production
of propanimine is not significantly higher, we do not prepare any reaction pathways
to study the formation of species from this isomeric group. \\

From the above discussed results, we find that
the imines (methanimine, ethanimine, and propanimine)
and amines (methylamine and ethylamine) are efficiently produced in the ice phase.
Depending on the barrier energy considered, Figure \ref{fig:amine_3} depicts a nice trend
of sublimation. For example, we adopt a BE of ethylamine and methylamine of $6480$ K and
$6584$ K, respectively. Thus, ethylamine starts to sublime faster than methylamine.
Similarly, methanimine, ethanimine, and propanimine having BEs $5534$, $5580$, and $6337$ K, respectively, sublime sequentially.

\begin{table}
\scriptsize{
\centering
\caption{Line parameters of trans-ethylamine in the millimeter and submillimeter regime using ALMA (LTE) \citep{sil18}.
\label{tab:amine_6}}
\vskip 0.2 cm
\begin{tabular}{cccc}
\hline
\hline
Frequency range (GHz) & Frequency (GHz)$^a$ & Transition (J$^\prime$ k$_a^\prime$ k$_c^\prime$ v$^\prime$ - J$^{\prime\prime}$ k$_a^{\prime\prime}$ k$_c^{\prime\prime}$ v$^{\prime\prime}$) &Intensity (K)\\
\hline
\hline
    & 31.4005744 & 5 1 4 0 $-$ 5 0 5 0 & 0.0058\\
    & 33.0673762 & 2 0 2 1 $-$ 1 0 1 1 &0.0063\\
$31-45$ (ALMA Band 1)  &34.9850027 & 6 1 5 0 $-$ 6 0 6 0& 0.008\\
    & 39.4293083 & 7 1 6 0 $-$ 7 0 7 0&0.011 \\
    & 44.8063428 & 8 1 7 0 $-$ 8 0 8 0& 0.015\\
   \hline
    & 80.241995 & 5 1 5 1 $-$ 4 1 4 1& 0.098\\
    &82.168784 & 5 0 5 1 $-$ 4 0 4 1 & 0.106\\
$67-90$ (ALMA Band 2)  & 82.674301 & 5 2 4 1 $-$ 4 2 3 1&0.088\\
and    & 83.24287 & 5 2 3 1 $-$ 4 2 2 1 &0.089\\
$84-116$ (ALMA Band 3)       & 84.980785 &5 1 4 1 $-$ 4 1 3 1& 0.107 \\
     & 101.886754 & 6 1 5 1 $-$ 5 1 4 1& 0.169\\
     &112.158369 &7 1 7 1 $-$ 6 1 6 1& 0.222\\
     &114.294235 & 7 0 7 1 $-$ 6 0 6 1 & 0.234\\
     & 115.604884 & 7 2 6 1 $-$ 6 2 5 1 & 0.211\\
     & 115.9771874 & 7 4 3 1 $-$ 6 4 2 1 & 0.256\\
   \hline
    & 145.871741 & 9 0 9 1 $-$ 8 0 8 1 & 0.38 \\
    &149.11269 & 9 5 5 1 $-$ 8 5 4 1 (9 5 4 1 $-$ 8 5 3 1) & 0.436\\
$125-163$ (ALMA Band 4)      &152.224777 &9 1 8 1 $-$ 8 1 7 1   & 0.391\\
    & 159.753196 & 10 1 10 1 $-$ 9 1 9 1 & 0.445\\
    & 161.496431 & 10 0 10 1 $-$ 9 0 9 1 & 0.45 \\
   \hline
    &182.3234429 &11 5 7 0 $-$ 10 5 6 0 & 0.684\\
    & 198.94447 &12 5 8 1 $-$ 11 5 7 1   &0.767\\
$163-211$ (ALMA Band 5) &198.945801 &12 5 7 1 $-$ 11 5 6 1&0.768\\
    &207.02715 & 13 1 13 1 $-$ 12 1 12 1  & 0.64 \\
    & 208.053547 & 13 0 13 1 $-$ 12 0 12 1 & 0.65\\
   \hline
   \hline
 \end{tabular} \\
 }
\vskip 0.2 cm
 {\bf Note:} \\
$^a$ For the transitions with same J$^\prime$ k$_a^\prime$ k$_c^\prime - J ^{\prime\prime}$ k$_a^{\prime\prime}$ k$_c^{\prime\prime}$ but having different
vibrational state please see the cat file available at \url{https://www.astro.uni-koeln.de/cdms/catalog}.
\end{table}

\begin{table}
\tiny
{\centering
\caption{Line parameters of (1Z)-1-propanimine in the millimeter and submillimeter regime using ALMA (LTE) \citep{sil18}.
\label{tab:amine_7}}
\vskip 0.2 cm
\begin{tabular}{cccc}
\hline
\hline
Frequency range (GHz) & Frequency (GHz)$^a$ & Transition (J$^\prime$ k$_a^\prime$ k$_c^\prime$v$^\prime$ - J$^{\prime\prime}$
k$_a^{\prime\prime}$ k$_c^{\prime\prime}$ v$^{\prime\prime}$) &Intensity (K)\\
\hline
\hline
&33.5198821&4 1 4 3 $-$ 3 1 3 2&0.117\\
&33.7888724&4 0 4 5 $-$ 3 0 3 4&0.136\\
&33.7897526&4 2 3 4 $-$ 3 2 2 3&0.105\\
&34.0663468&4 1 3 4 $-$ 3 1 2 3&0.123\\
  $31-45$ (ALMA Band 1) &41.897946&5 1 5 6 $-$ 4 1 4 5&0.261\\
&42.2302861&5 0 5 6 $-$ 4 0 4 5&0.290\\
&42.2318919& 5 3 3 5 $-$ 4 3 1 5&0.326\\
&42.2361126&5 2 4 6 $-$ 4 2 3 5&0.227\\
&42.5812455& 5 1 4 6 $-$ 4 1 3 5&0.270\\
\hline
&67.0236193&8 1 8 8 $-$ 7 1 7 7&1.110\\
&67.5282969&8 0 8 8 $-$ 7 0 7 7 &2.252\\
&67.5717125& 8 3 6 7 $-$ 7 3 5 8&1.725\\
&68.117366&8 1 7 8 $-$ 7 1 6 7 &1.145\\
&75.3952267&9 1 9 9 $-$ 8 1 8 8&1.510\\
&75.9707109&9 5 4 8 $-$ 8 5 3 7&1.720\\
&75.9497207&9 0 9 10 $-$ 8 2 6 9&1.440\\
   $67-90$ (ALMA Band 2)&76.0013965&9 4 6 10 $-$ 8 4 49&2.252\\
and &76.0187663&9 3 6 9 $-$ 8 3 5 9&2.244\\
$84-116$ (ALMA Band 3)   &76.6257696&9 1 8 9 $-$ 8 1 7 8&1.555\\
&83.7646502&10 11 0 10 $-$ 9 1 9 9&1.955\\
&84.364287&10 0 10 9 $-$ 9 1 8 9&2.000\\
&84.4134387& 10 5 6 11 $-$ 9 5 4 10&2.40\\
&84.4483753&10 4 6 9 $-$ 9 4 5 10&4.650\\
&84.4661717&10 3 7 10 $-$ 9 3 6 10&2.450\\
&84.5650434&10 2 8 10 $-$ 9 2 7 9&1.870\\
&85.1318785&10 0 10 9 $-$ 9 1 8 9&2.080\\
&92.1316784&11 1 11 11 $-$ 10 1 10 10&2.420\\
&92.7713594& 11 0 11 12 $-$ 10 0 10 11&2.480\\
&92.8567065&11 5 6 10 $-$ 10 5 6 9&3.150\\
&92.8959487&11 4 8 10 $-$ 10 4 6 9&3.720\\
&93.0411029&11 2 9 11 $-$ 10 2 8 10&3.720\\
&93.6354011&11 1 10 12 $-$ 10 1 9 11&2.480\\
&100.4962033&12 1 12 11 $-$ 11 1 11 12&2.900\\
&101.1702803&12 0 12 12 $-$ 11 0 11 11&2.977\\
&101.3005235&12 5 8 13 $-$ 11 5 6 12&3.925\\
&101.3444369&12 4 8 12 $-$ 12 4 7 11&4.550\\
&101.5225577&12 2 10 11 $-$ 12 2 9 12&2.800\\
&102.1359978&12 1 11 12 $-$ 11 1 10 11&2.960\\
&109.7449013&13 5 9 13 $-$ 12 5 7 12&4.703\\
&109.7938556&13 4 9 13 $-$ 12 4 8 12&5.395\\
&110.0094549&13 2 11 13 $-$ 12 2 10 12&3.300\\
&110.6333677&13 1 12 14 $-$ 12 1 11 13&3.455\\
\hline
&126.6358855&15 5 10 15 $-$ 14 5 10 14&6.200\\
&126.6960852&15 4 11 14 $-$ 14 4 11 13&10.080\\
&133.9242154&16 1 16 15 $-$ 15 0 15 16&4.730\\
&134.6750035&16 0 16 17 $-$ 15 0 15 16&4.800\\
&135.148249&16 4 13 15 $-$ 15 4 12 16&7.750\\
  $125-163$ (ALMA Band 4)  &143.6023263&17 4 13 17 $-$ 16 4 12 16&8.350\\
&151.9783682& 18 5 14 18 $-$ 17 5 12 17&8.100\\
&152.0568371&18 4 15 17 $-$ 17 4 14 18&8.750\\
&160.4277999&19 5 14 18 $-$ 18 5 13 19&8.575\\
&160.5135369&19 4 15 19 $-$ 18 4 14 19&9.200\\
\hline
&168.9718077&20 4 16 20 $-$ 19 4 15 20&9.250\\
&168.878091& 20 5 16 20 $-$ 19 5 15 19&9.00\\
&177.3296583& 21 5 16 20 $-$ 20 5 16 19&9.350\\
  $163-211$ (ALMA Band 5) &185.7821453&22 5 18 23 $-$ 21 5 17 22&9.550\\
&185.8934615&22 4 18 22 $-$ 21 4 17 22&8.350\\
&194.23659&23 5 19 22 $-$ 22 5 17 22 &10.520\\
&202.6910348& 24 5 20 23 $-$ 23 5 19 24&9.770\\
\hline
\hline
\end{tabular} \\
}
\vskip 0.2 cm
{\bf Note:} \\
$^a$ For the transitions with the same J$^\prime$ k$_a^\prime$ k$_c^\prime - J ^{\prime\prime}$ k$_a^{\prime\prime}$ k$_c^{\prime\prime}$ 
but having different vibrational state, please see the catalog in the \href{https://cfn-live-content-bucket-iop-org.s3.amazonaws.com/journals/0004-637X/853/2/139/1/apjaa984d.tar.gz?AWSAccessKeyId=AKIAYDKQL6LTV7YY2HIK&Expires=1625643085&Signature=glu8vNxKvb8lm7XmnfdYA34zx64\%3D}{propanimine.tar.gz} package provided in \cite{sil18}.
\end{table}

 \begin{table}
 \scriptsize
 {\centering
 \caption{Non-LTE modeling line parameters of trans-ethylamine \citep{sil18}.
 \label{tab:amine_8}}
 \vskip 0.2 cm
\begin{tabular}{cccc}
  \hline
  \hline
  Frequency range (GHz)& Frequency (GHz)$^a$ & Transition (J$^\prime$ k$_a^\prime$ k$_c^\prime$v$^\prime$ - J$^{\prime\prime}$
k$_a^{\prime\prime}$ k$_c^{\prime\prime}$ v$^{\prime\prime}$) & Intensity (K) \\
  \hline
  \hline
  
 & 31.4005744 & 5 1 4 0 $-$ 5 0 5 0 & 0.0108\\
    & 33.06739330 & 2 0 2 0 $-$ 1 0 1 0&0.0096\\
  $31-45$ (ALMA Band 1)    & 34.98500270 & 6 1 5 0 $-$ 6 0 6 0 & 0.0129 \\
      & 39.42887460 & 7 1 6 1 $-$ 7 0 7 1  & 0.0152 \\
   & 44.80634280 & 8 1 7 0 $-$ 8 0 8 0 &0.0173 \\
   \hline   
& 80.24199500 & 5 1 5 0 $-$ 4 1 4 0& 0.0534\\
   &82.16878400 & 5 0 5 0 $-$ 4 0 4 0& 0.0569\\
   $67-90$ (ALMA Band 2)    & 82.67430100 & 5 2 4 0 $-$ 4 2 3 0 & 0.0483\\
and   & 83.2428700 & 5 2 3 1 $-$ 4 2 2 1 &0.0486 \\
$84-116$ (ALMA Band 3)      & 84.98078500 & 5 1 4 1 $-$ 4 1 3 1 &0.0566\\
   & 98.30233700& 6 0 6 0 $-$ 5 0 5 0& 0.0796\\
  &101.886754 &6 1 5 0 $-$ 5 1 4 0&0.0806 \\
 & 112.1583690 &7 1 7 0 $-$ 6 1 6 0&0.1016 \\
  &114.294235 & 7 0 7 0 $-$ 6 0 6 0  & 0.1052\\
  & 115.604884 & 7 2 6 0 $-$ 6 2 5 0 & 0.0996\\
   \hline
   & 143.9269253 & 9 1 9 0 $-$ 8 1 8 0& 0.1586\\
   & 145.871741 & 9 0 9 0 $-$ 8 0 8 0 & 0.1620\\
 $125-163$ (ALMA Band 4)   &148.397949 &9 2 8 1 $-$ 8 2 7 1&0.1598\\
    &159.753196 & 10 1 10 0 $-$ 9 1 9 0& 0.1878\\
  & 161.49643100 & 10 0 10 0 $-$ 9 0 9 0 & 0.1914 \\
   \hline
  & 192.561480 & 12 0 12 0 $-$ 11 0 11 0& 0.2432\\
  & 197.25631220 & 12 2 11 1 $-$ 11 2 10 1 & 0.2436\\
$163-211$ (ALMA Band 5)  & 201.671390 &12 1 11 0 $-$ 11 1 10 0&0.2587\\
  &203.166657 &12 2 10 1 $-$ 11 2 9 1 & 0.2534 \\
  & 208.05354700 &13 0 13 0 $-$ 12 0 12 0& 0.2565\\
   \hline
   \hline
\end{tabular} \\
}
\vskip 0.2 cm
{\bf Note:} \\
$^a$ For the transitions with same J$^\prime$ k$_a^\prime$ k$_c^\prime - J ^{\prime\prime}$ k$_a^{\prime\prime}$ k$_c^{\prime\prime}$  but having different
vibrational state please see the cat file available at \url{https://www.astro.uni-koeln.de/cdms/catalog}.
 \end{table} 
 
 \begin{table}
 \scriptsize
 {\centering
 \caption{Non-LTE modeling line parameters of (1Z)-1-propanimine \citep{sil18}.
 \label{tab:amine_9}}
 \vskip 0.2 cm
 \begin{tabular}{cccc}
  \hline
  \hline
  Frequency range (GHz) & Frequency (GHz)$^a$ & Transition (J$^\prime$ k$_a^\prime$ k$_c^\prime$v$^\prime$ - J$^{\prime\prime}$ k$_a^{\prime\prime}$ k$_c^{\prime\prime}$ v$^{\prime\prime}$) 
  & Intensity (K)\\
  \hline
  \hline
 & 33.51992460 & 4 1 4 5 $-$ 3 1 3 4 & 0.1192\\
  & 33.78887240 & 4 0 4 5 $-$ 3 0 3 4&0.1269\\
$31-45$ (ALMA Band 1)    & 41.89797460 & 5 1 5 6 $-$ 4 1 4 5& 0.1925\\
 & 42.25035000 & 5 2 3 6 $-$ 4 2 2 5 &0.1752\\
 & 42.58124550 & 5 1 4 6 $-$ 4 1 3 5   &0.1978 \\
   \hline
  & 76.62578210 & 9 1 8 10 $-$ 8 1 7 9& 0.6728\\
   &83.76466100 & 10 1 10 11 $-$ 9 1 9 10 & 0.7793 \\
  $67-90$ (ALMA Band 2) & 84.36429660 & 10 0 10 11 $-$ 9 0 9 10  & 0.7906\\
and   & 84.56506570 & 10 2 8 11 $-$ 9 2 7 10&0.7710 \\
$84-116$ (ALMA Band 3)    & 85.13188840 & 10 1 9 1 $-$ 9 1 8 10   &0.8215\\
   & 93.63540110 & 11 1 10 12 $-$ 10 1 9 11& 0.9333\\
    &100.49611750 &12 1 12 13 $-$ 11 1 11 12&0.9232\\
      & 101.17028780 &12 0 12 13 $-$ 11 0 11 12&0.9826\\
     &102.13600470 & 12 1 11 13 $-$ 11 1 10 12& 0.9689\\
    & 109.56052950 & 13 0 13 14 $-$ 12 0 12 13& 0.9616\\
   \hline
   & 125.51588020 & 8 2 7 9 $-$ 7 1 6 8& 0.0591\\
   & 126.31319130 & 15 0 15 16 $-$ 14 0 14 15 & 0.6378\\
   $125-163$ (ALMA Band 4)  &127.00092650 & 15 2 13 16 $-$ 14 2 12 15&0.3568\\ 
  &127.61694790 & 15 1 14 15 $-$ 14 1 13 14 & 0.3922 \\
  & 134.67500350 & 16 0 16 17 $-$ 15 0 15 16   & 0.3738 \\
   \hline
    & 175.64941420 & 9 3 7 10 $-$ 8 2 6 9& 0.09671 \\
   & 175.80109580 & 9 3 6 10 $-$ 8 2 7 9 & 0.09373\\
   $163-211$ (ALMA Band 5) & 184.02040330 &10 3 8 11 $-$ 9 2 7 10   &0.09697\\
   & 184.25941520 &10 3 7 10 $-$ 9 2 8 9& 0.09647\\
    & 207.19739490 &8 4 5 9 $-$ 7 3 4 8   & 0.09612\\
   \hline
   \hline
\end{tabular} \\
}
\vskip 0.2 cm
{\bf Note:} \\
$^a$ For the transitions with same J$^\prime$ k$_a^\prime$ k$_c^\prime - J ^{\prime\prime}$ k$_a^{\prime\prime}$ k$_c^{\prime\prime}$ but having different
vibrational state, please see the catalog in the \href{https://cfn-live-content-bucket-iop-org.s3.amazonaws.com/journals/0004-637X/853/2/139/1/apjaa984d.tar.gz?AWSAccessKeyId=AKIAYDKQL6LTV7YY2HIK&Expires=1625643085&Signature=glu8vNxKvb8lm7XmnfdYA34zx64\%3D}{propanimine.tar.gz} package provided in \cite{sil18}.
\end{table}

\begin{figure}
\centering
\includegraphics[width=0.8\textwidth]{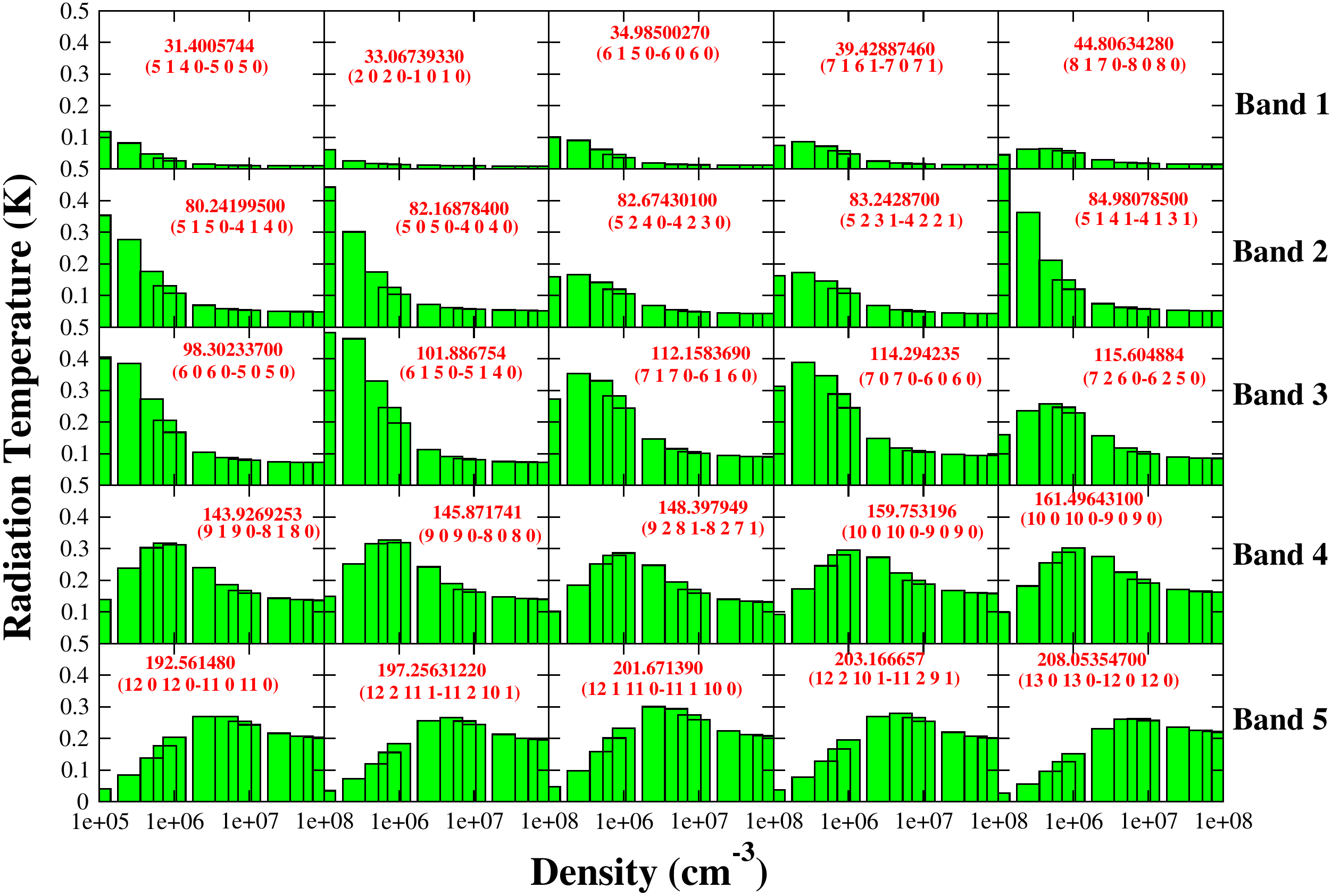}
\caption{Density variation of the intensity of various transitions of ethylamine by considering a non-LTE 
approximation \citep{sil18}.
\label{fig:amine_17}}
\end{figure}

\begin{figure}
\centering
\includegraphics[width=0.8\textwidth]{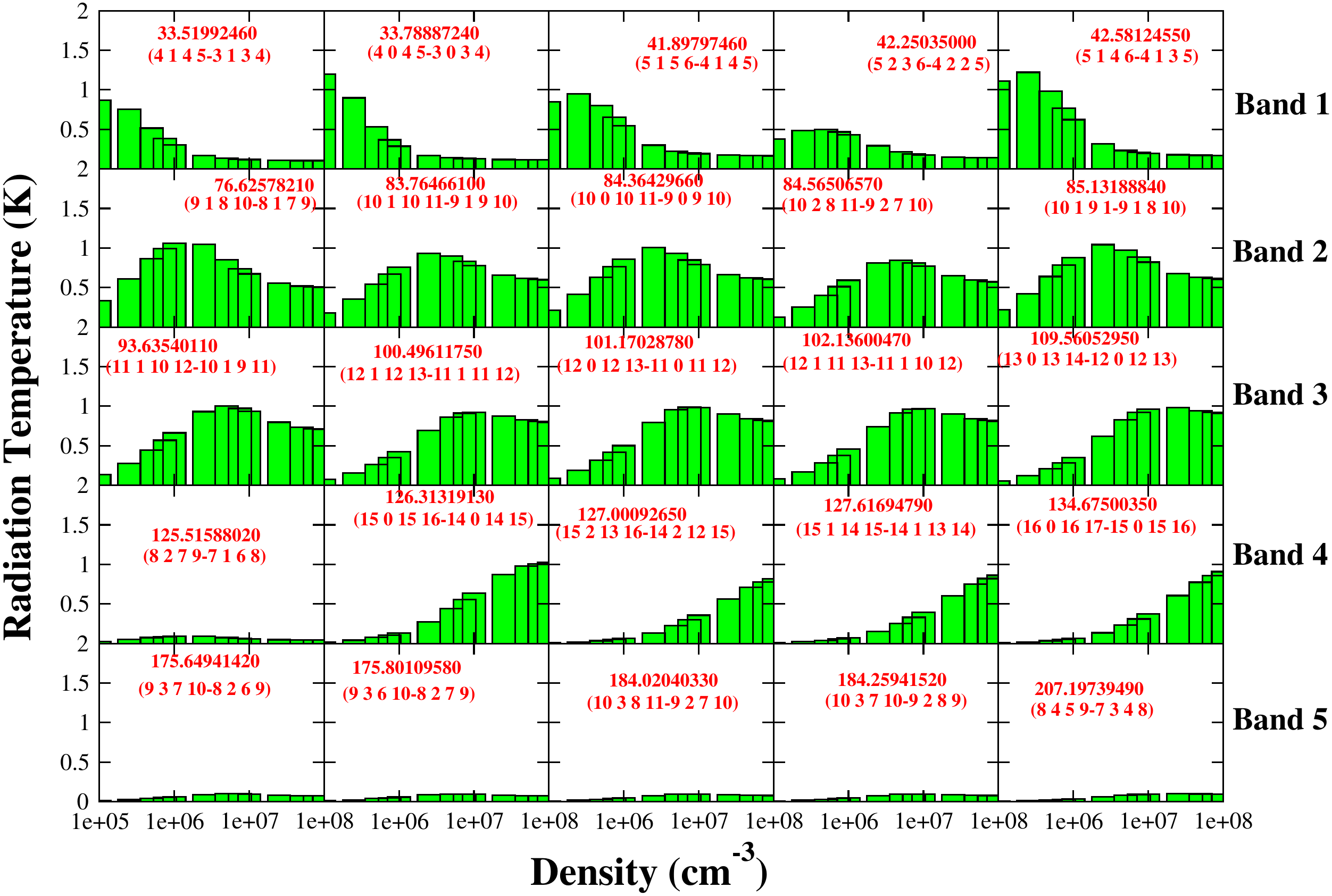}
\caption{Density variation of the intensity of various transitions of propanimine by considering a non-LTE 
approximation \citep{sil18}.
\label{fig:amine_18}}
\end{figure}

\subsection{Astrophysical implications}
Methanimine is a crucial prebiotic molecule that is believed to be the precursor
molecule for the formation of glycine. This species is already identified in the ISM.
Also, in the upper atmosphere of Titan \citep[the massive moon of Saturn;][]{vuit06}, this is detected.
The present atmosphere of Titan resembles the primeval atmosphere of the Earth
and thus is thought to be important for abiotic synthesis.
Our present study finds that methanimine may be further processed
to form methylamine, which is yet to be observed in Titan's atmosphere.
Modeling Titan's atmosphere is beyond the scope of this work. However, the inclusion of
proposed pathways in the modeling of Titan atmosphere may come up with the higher mixing
ratios of higher-order imines and amines in Titan's atmosphere.

We perform radiative transfer modeling (with both LTE and non-LTE
consideration), which may be helpful for the future astronomical observation
of ethylamine and propanimine in the ISM.
For the calculation of the line parameters using an LTE approximation, we use the CASSIS interactive spectrum analyzer\footnote{\url{http://cassis.irap.omp.eu}}.
In Table \ref{tab:amine_6}, we point out some of the most intense ethylamine transitions that falls in ALMA bands $1-5$.
Required spectroscopic details for ethylamine are available in CDMS database\footnote{\url{https://www.astro.uni-koeln.de/cdms/catalog}}.
Similarly, in Table \ref{tab:amine_7}, intense transitions of propanimine are shown.
Required spectroscopic details for (1Z)-1-propanimine are obtained
by including the experimentally obtained rotational and distortional
constants in the SPCAT program \citep{pick91}.
For preparing these tables, we use a column density of
H$_2= 10^{23}$ cm$^{-2}$, $n_H=10^7$ cm$^{-3}$,
excitation temperature $= 130$ K, FWHM $= 10$ km s$^{-1}$, $V_{LSR}=64$ km s$^{-1}$,
source size $=3''$, abundance of ethylamine $=4.0 \times 10^{-08}$, and abundance of
propanimine $=2 \times 10^{-8}$.

We also perform a non-LTE calculation using the RADEX program \citep{vand07}.
The collisional data files for ethylamine and propanimine are unavailable.
Thus, we prepare the collisional data file in the appropriate
format from the spectral information available in JPL (for trans-ethylamine) and 
our calculation [for (1Z)-1-propanimine] \href{https://cfn-live-content-bucket-iop-org.s3.amazonaws.com/journals/0004-637X/853/2/139/1/apjaa984d.tar.gz?AWSAccessKeyId=AKIAYDKQL6LTV7YY2HIK&Expires=1625643085&Signature=glu8vNxKvb8lm7XmnfdYA34zx64\%3D}{propanimine.tar.gz}.
Altogether we consider transitions between $251$ energy levels.
Here, we assume H$_2$ as the colliding partner.
We approximate the collisional rate of ethylamine and propanimine
following the relation mentioned in \cite{shar01} to estimate the line profile with non-LTE.
\cite{shar01} estimated the collisional rate coefficient for a downward transition
of an asymmetric top molecule, cyclopropene, at temperature ``T'', by
\begin{equation}
C(J''K_a''K_c'' \rightarrow J' K_a' K_c')= [1 \times 10^{-11}/(2J''+1)]\sqrt{T/30}.
\end{equation}
In Tables \ref{tab:amine_8} and \ref{tab:amine_9}, we point out the most intense
transitions of trans-ethylamine and (1Z)-1-propanimine, respectively,
which fall within ALMA bands $1-5$.
For the non-LTE calculations, we use a column density of ethylamine of $10^{15}$ cm$^{-2}$
and a column density of propanimine of $5.0 \times 10^{14}$ cm$^{-2}$, $n_H=10^7$ cm$^{-3}$,
excitation temperature $= 130$ K, $FWHM = 10$ km s$^{-1}$.

For the transitions pointed out in Tables \ref{tab:amine_8} and \ref{tab:amine_9},
the density variation of ethylamine (Figure \ref{fig:amine_17}) and propanimine (Figure \ref{fig:amine_18}) is studied with the non-LTE consideration.
Figures \ref{fig:amine_17} and \ref{fig:amine_18} would serve as a good starting point
for observing ethylamine and propanimine in the ISM.
It is to be noted that we use our estimated collisional rate in the absence of the measured or calculated collisional data file.
Still, it is known that
the non-LTE transitions are heavily dependent on collisional rates,
and consideration of random rates may end up with some misleading results.

\subsection{Summary}
In this work, the possibility of detecting various molecules which
belong to six specific isomeric groups is examined.
For this purpose, the chemical abundance, enthalpy of formation, optimized
energy, and expected intensity ratio are used to shortlist some species that
might be viable candidates for future astronomical detection in the ISM.
Our modeling results suggest that ethylamine and propanimine are being produced efficiently in the hot-core condition. So we propose to observe these molecules in a hot molecular core.
Furthermore, LTE and non-LTE models are employed to determine the most probable transitions of ethylamine and propanimine in the millimeter and submillimeter domains.
Some spectroscopic data are provided to aid their identification.
We also find a clue about the linkage between the interstellar and cometary chemical compositions.

\clearpage
\section{Prebiotic molecules containing peptide-like bonds}
Protein synthesis occurs through the peptide-bond ($\rm{-NH-C(=O)-}$) formation \citep{gold10}.
Nitrogen is one of the most chemically active species in the ISM
after hydrogen, oxygen, and carbon. N-bearing
molecules are vital as they are actively involved in the formation of biomolecules.
Therefore, it is essential to look for N-bearing species in various astrophysical sources.
Looking around the high-mass star-forming regions is particularly important because the evolutionary history is comparatively poorly understood.
CN is the first observed N-bearing species in space \citep{mcke40}.
Since then, various N-bearing species are identified in numerous astronomical objects. Hot-core regions are unique laboratories of COMs. Forests of molecular lines are identified in several hot molecular
cores \citep[HMCs; e.g.,][]{bell16,garr17}. Here, we focus
on the observation of an HMC, G10 \citep{gora20b}, which is located at a distance of 8.6 kpc \citep{sann14} with a luminosity $\rm{5\times 10^{5}\ L_{\odot}}$ \citep{cesa10}.

Isocyanic acid (HNCO) is a simple molecule containing four biogenic elements (C, N, O, and H),
making a peptide bond. HNCO was observed long ago toward the high-mass star-forming region
Sgr B2 \citep{snyd72}. In addition, it has been observed in various astronomical objects
such as a translucent molecular cloud \citep{turn99}, a dense core \citep{marc18}, and
the low-mass protostar IRAS $16293-2422$ \citep{biss08}.
It was also previously detected in G10 \citep{wyro99}.

Formamide (NH$_2$CHO) is the simplest possible amide.
It is a potential prebiotic molecule containing a
peptide bond that can link with amino acids and form proteins.
NH$_2$CHO is also a precursor of genetic and metabolic molecules \citep{sala12}.
This molecule is one of the key species for the formation of nucleobases and nucleobase analogs. NH$_2$CHO was observed for the first time toward the high-mass star-forming region, Sgr B2 \citep{rubi71}. Subsequently, it was identified in other hot-cores, such as Orion KL,
$\rm{G327.3-0.6,\ G34-3+0.15,\ NGC\ 6334}$ \citep{turn91,boge19}, the solar-type low-mass protostar IRAS $16293-2422$ \citep{kaha13}, and in the shock of the prestellar
core L1157-B1 \citep{code17}. NH$_2$CHO was previously detected in G10 using millimeter
and submillimeter wavelengths with Submillimeter Array (SMA) observation \citep{rolf11}.

Methyl isocyanate (CH$_3$NCO) is another potential prebiotic molecule, which also
has a peptide-like bond. It has been
observed in the high-mass star-forming region Sgr B2 \citep{cern16}
and the low-mass star-forming region IRAS $16293-2422$ \citep{ligt17,mart17}.
\cite{gora20b} have reported the identification of CH$_3$NCO in G10 for the first time.
HNCO has firmly been identified in G10, but NH$_2$CHO was tentatively detected \citep{rolf11}.
HNCO and NH$_2$CHO have recently been identified,
but $\rm{CH_3NCO}$ has been tentatively identified
in the $\rm{67P/C-G}$ comet by the double-focusing mass spectrometer (DFMS)
of the ROSINA experiment on ESA's Rosetta mission \citep{altw17}.

Here, detailed chemical modeling of three peptide-like bond containing molecules
(HNCO, $\rm{NH_2CHO}$, and $\rm{CH_3NCO}$) is presented to aid their identification in G10 \citep{gora20b}.

\subsection{Chemical Modeling \label{sec:model-results}}
To study the chemical evolution of three peptide-bond-related species,
we use our gas-grain chemical network \citep{das15a,das15b,sil18,gora20b}.
Gas-phase pathways are mainly adopted from the UMIST database \citep{mcel13},
whereas ice-phase paths and BEs of the surface species are taken from the KIDA \citep{ruau16}
unless otherwise stated. We consider that the diffusion energy of a species
is $0.5$ times its desorption energy and non-thermal desorption rate with a fiducial parameter
of $0.01$. A cosmic-ray rate of $1.3 \times 10^{-17}$ s$^{-1}$ is considered.
For the formation and destruction reactions of these species, we mainly follow \cite{quen18}.
In addition, following the recent study of \cite{haup19}, we exclusively
include dual-cyclic H addition and abstraction reactions. These connect NH$_2$CHO,
NH$_2$CO, and HNCO. We show the chemical linkages among HNCO, NH$_2$CHO, and CH$_3$NCO.
Initial abundances of the model are provided in Table \ref{table:initial}.

\begin{table}
\scriptsize
\centering
\caption{Initial abundances with respect to total hydrogen nuclei \citep{gora20b}.
\label{table:initial}}
\vskip 0.2 cm
\begin{tabular}{cc}
\hline
{\bf Species} & {\bf Abundance} \\
\hline
$\mathrm{H_2}$ &    $5.00 \times 10^{-1}$\\
$\mathrm{He}$  &    $9.00 \times 10^{-2}$\\
$\mathrm{C}$ &    $7.30 \times 10^{-5}$\\
$\mathrm{O}$   &    $1.76 \times 10^{-4}$\\
$\mathrm{N}$   &    $2.14 \times 10^{-5}$\\
$\mathrm{Cl}$   &    $1.00 \times 10^{-9}$\\
$\mathrm{Fe}$&    $3.00 \times 10^{-9}$\\
$\mathrm{Mg}$&    $7.00 \times 10^{-9}$\\
$\mathrm{Na}$&    $2.00 \times 10^{-9}$\\
$\mathrm{S}$ &    $8.00 \times 10^{-8}$\\
$\mathrm{Si^+}$&    $8.00 \times 10^{-9}$\\
$\mathrm{e^-}$ &    $7.31 \times 10^{-5}$\\
\hline
\end{tabular}
\end{table}

\subsubsection{Physical condition of the adopted model}
We consider a three-stage model to study the chemical evolution of these
species \citep{garr13}. This model is best suited because G10 is a high-mass star-forming region.
The detailed considerations of each stage are discussed below.

{\it First stage:} In the first stage, we consider that the cloud collapses
from a low total hydrogen density ($\rho_{min}=10^3$ cm$^{-3}$) to a high value ($\rho_{max}$).
The initial gas temperature ($T_{gas}$) is assumed to be $40$ K, whereas the dust 
temperature is supposed to remain fixed at an initial ice temperature ($T_{ice}$).
We consider a time interval of $t_{coll}$ years to 
reach from $\rho_{min}$ to $\rho_{max}$. Since the gas and dust are well coupled
at the highest density, we consider $T_{gas}=T_{ice}$ at the highest
density ($\rho_{max}$), i.e., at $t=t_{coll}$. 
From this stage onward, we assume that the temperature of the dust and the gas are the same.
Thus, we consider a negative slope for $T_{gas}$ for the collapsing stage.
Throughout the first stage, the visual extinction parameter is considered to be constantly
increasing from $A_V={A_V}_{min}=10$ to finally at ${A_V}_{max}=200$ in $t=t_{coll}$.

{\it Second stage:} The second stage of the simulation corresponds to a warm-up stage.
Since G10 is a high-mass star-forming region, we consider a moderate warm-up
time-scale ($t_w$) $5 \times 10^4$ years \citep{garr13}. Therefore, during this short period,
the cloud temperature from $T_{ice}$ can reach the highest hot-core temperature $T_{max}$. 
The density, temperature, and visual extinction parameter remain constant at $\rho_{max}$, $T_{max}$, and ${A_V}_{max}$, respectively.

{\it Third stage:} This stage belongs to the post-warm-up time. Here, we consider
a post-warm-up time scale ($t_{pw}$) of $10^5$ years, so the total simulation
time is $t_{tot}=t_{coll}+t_w+t_{pw}$.
The parameters such as density and visual extinction are assumed to be the same as in
the warm-up stage. The temperature of the cloud is kept at $T_{max}$ throughout the last stage.

\subsubsection{Binding energies of peptide-bond containing species}

\begin{table}
\tiny
{\centering
\caption{Computed BEs \citep{gora20b}.
\label{table:BE}}
\vskip 0.2 cm
\hskip -1.5 cm
\begin{tabular}{|c|c|c|c|c|c|c|}
\hline
{\bf Species}& {\bf Optimized} & {\bf Calculated BE (K)} & {\bf Average BE (K)} & {\bf Scaled BE (K)} & {\bf Calculated BE (K)} & {\bf Available BE (K)} \\
&{\bf Structure}& {\bf using H$_2$O monomer} && {\bf ($\times 1.416$)} & {\bf using H$_2$O hexamer} & {\bf in literatures} \\
\hline
\multicolumn{7}{|c|}{\bf CHNO}\\
\hline
&&&&&& \\
HNCO& \includegraphics[height=0.8cm, width=1.5cm]{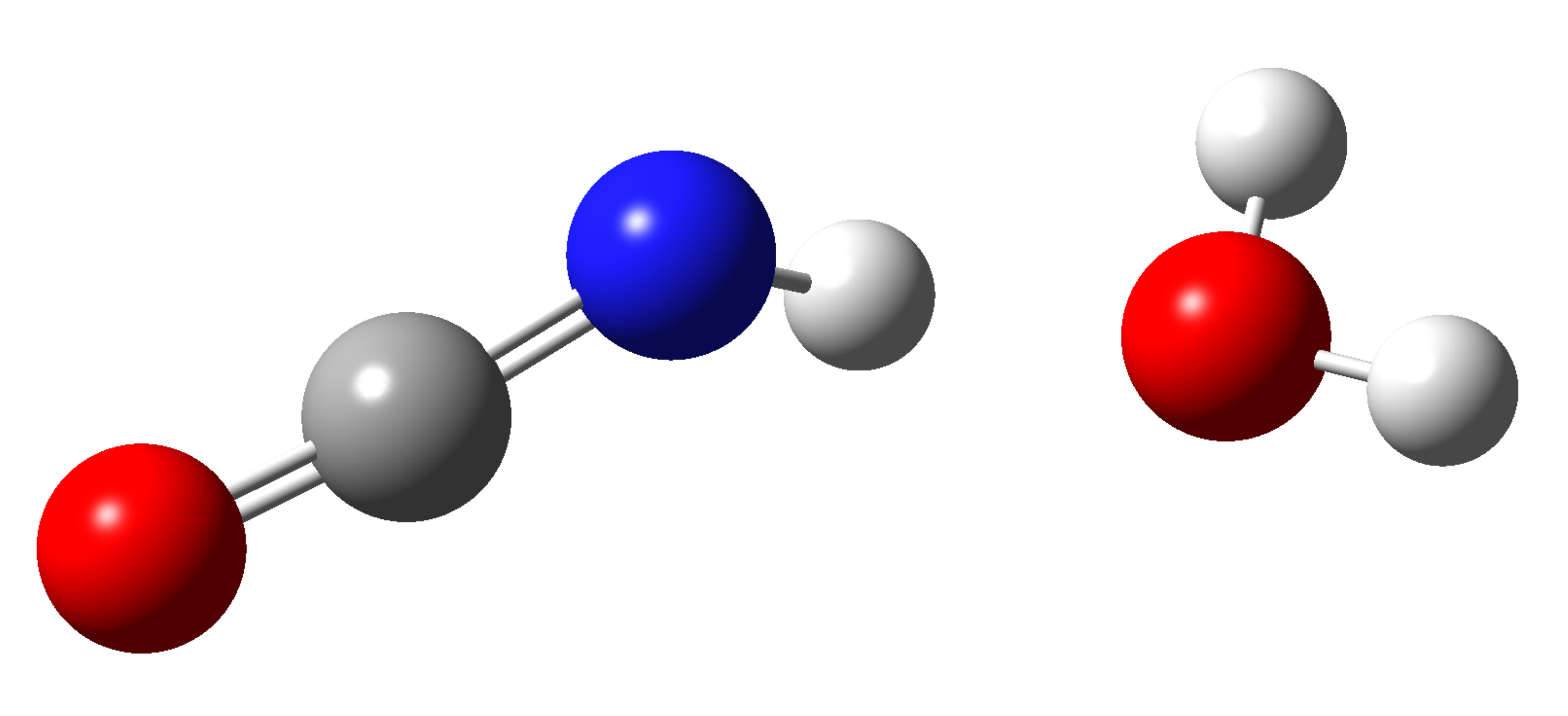}& 3308 & 3308 & 4684 & 6310, 5554 & $4400 \pm 1320$$^{a}$ \\
\hline
HCNO&\includegraphics[height=0.8cm, width=1.5cm]{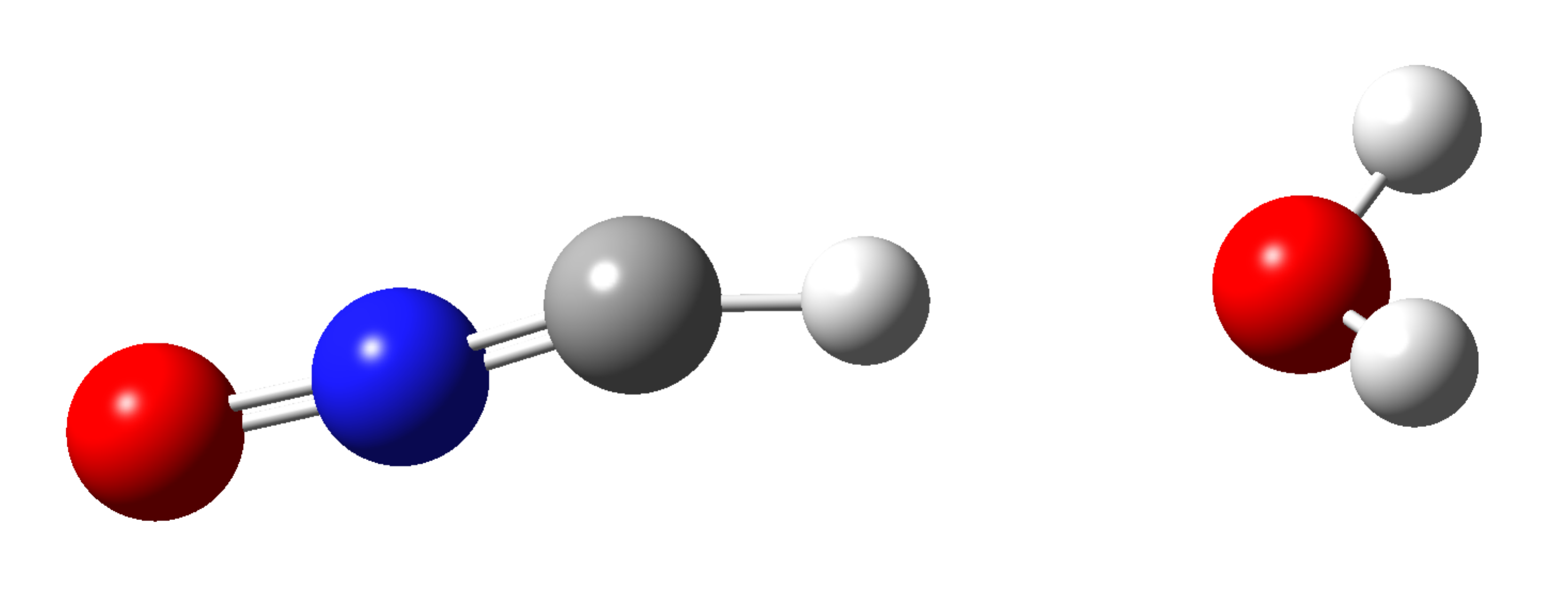}& 2640 & 2345 & 3320 & 6046 & 2800$^{b}$ \\
&\includegraphics[height=1cm, width=1.5cm]{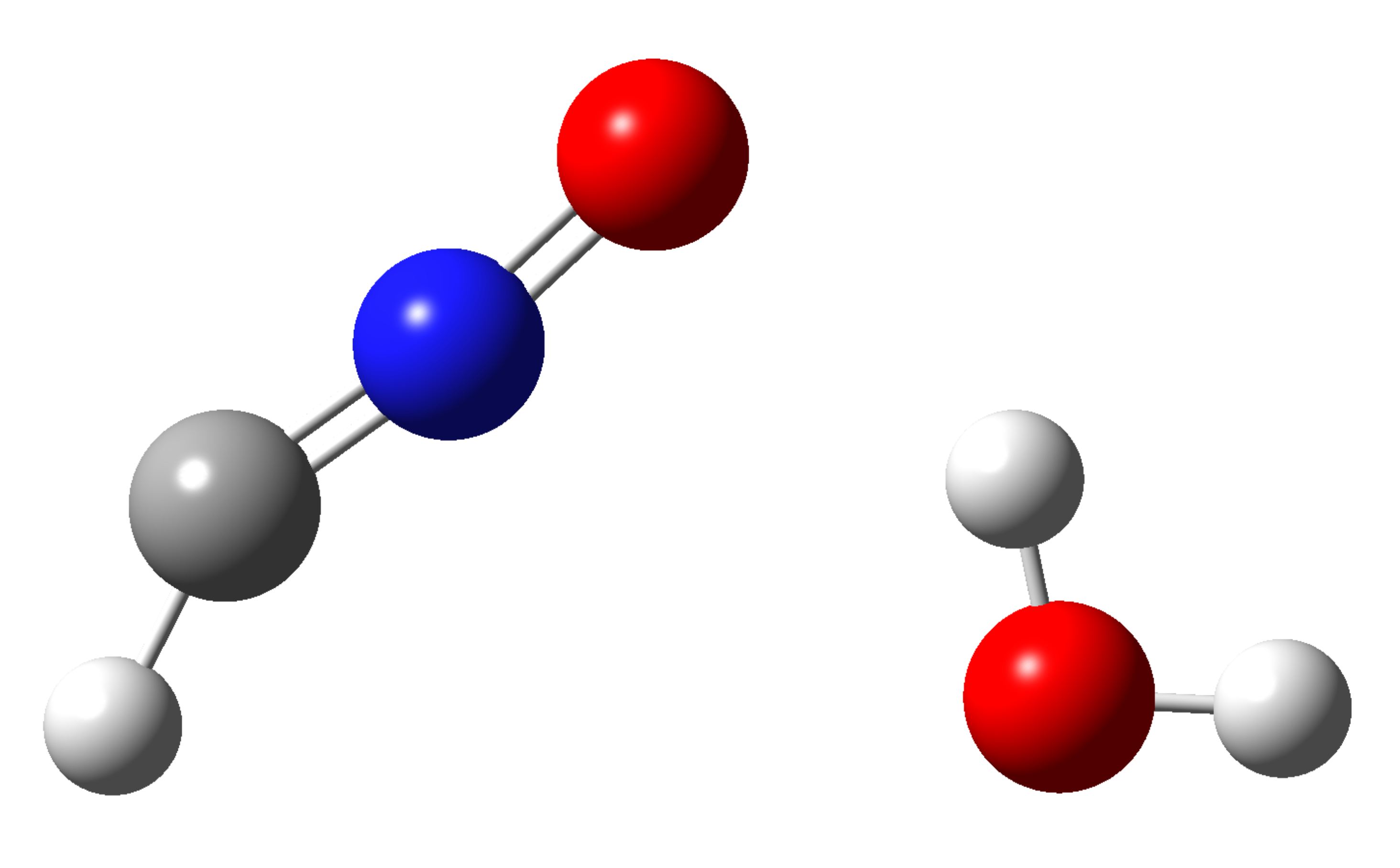}& 2050 &&&& \\
\hline
HOCN&\includegraphics[height=1cm, width=1.5cm]{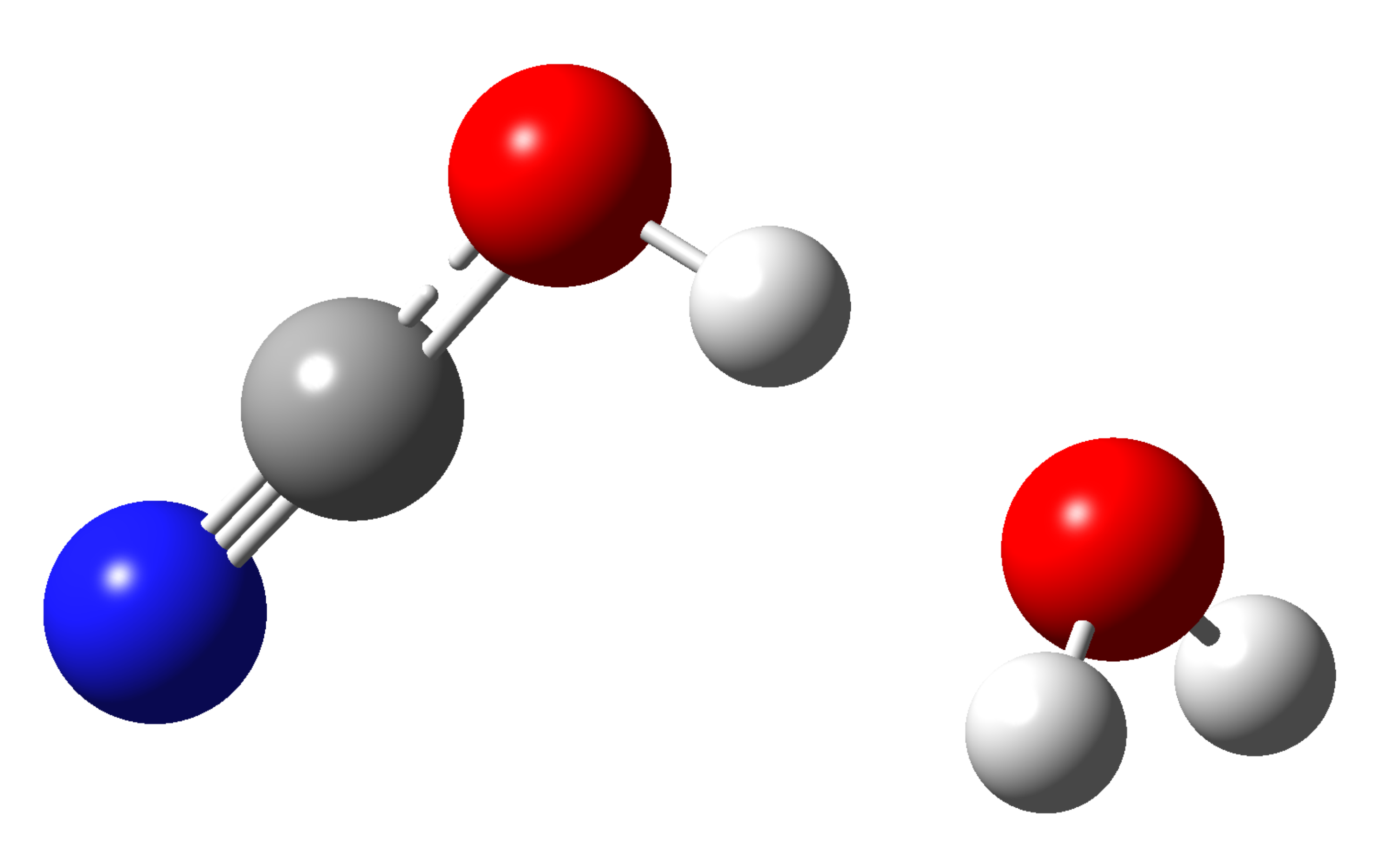}& 5936 & 4250 & 6018 & 2153, 8404 & 2800$^{b}$ \\
&\includegraphics[height=0.7cm, width=1.5cm]{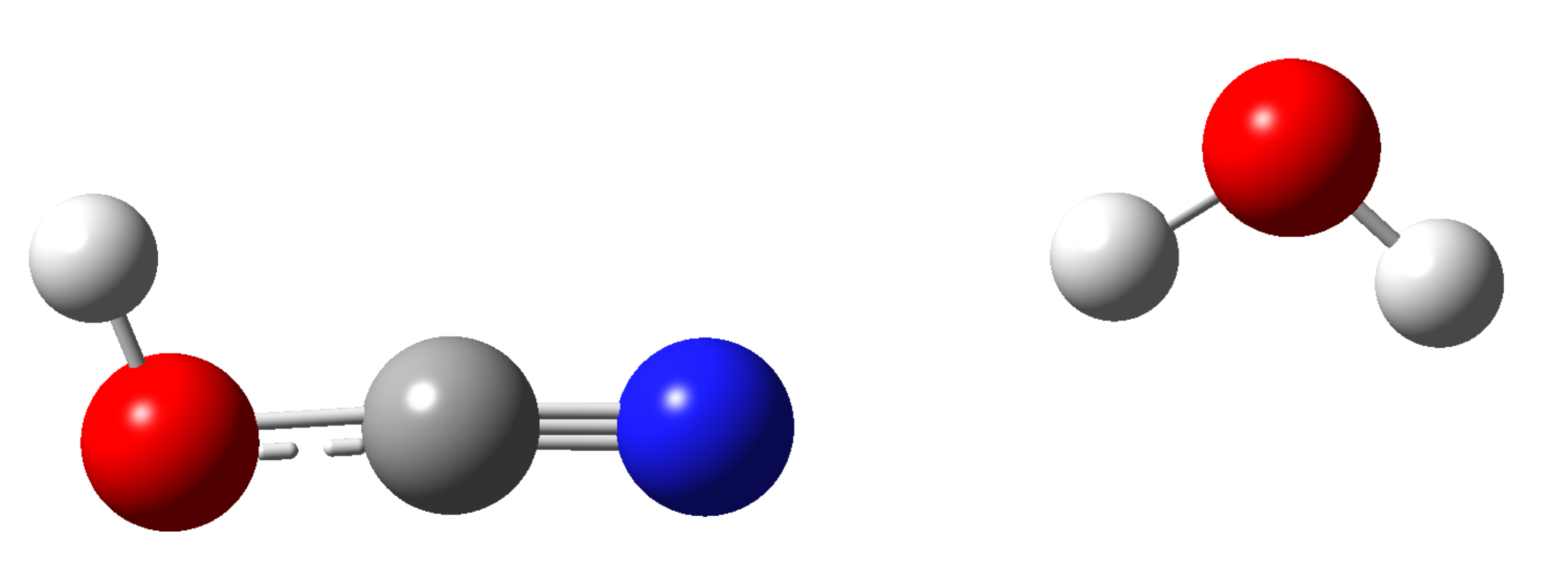}& 2563 &&&& \\
\hline
HONC&\includegraphics[height=1cm, width=1.5cm]{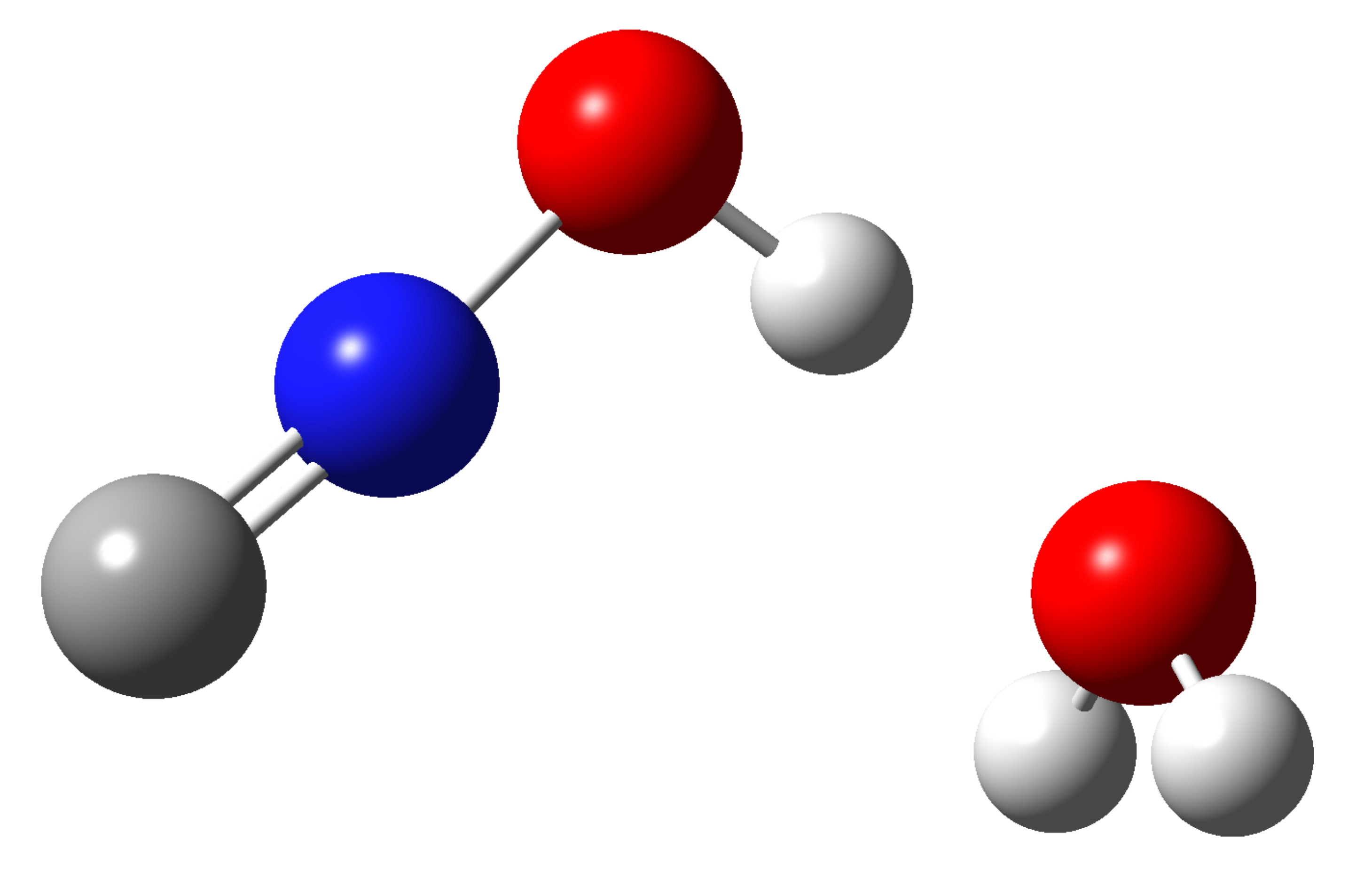}&5874& 4122 & 5837 & 3387, 8727 & 2800$^{b}$ \\
&\includegraphics[height=0.7cm, width=1.5cm]{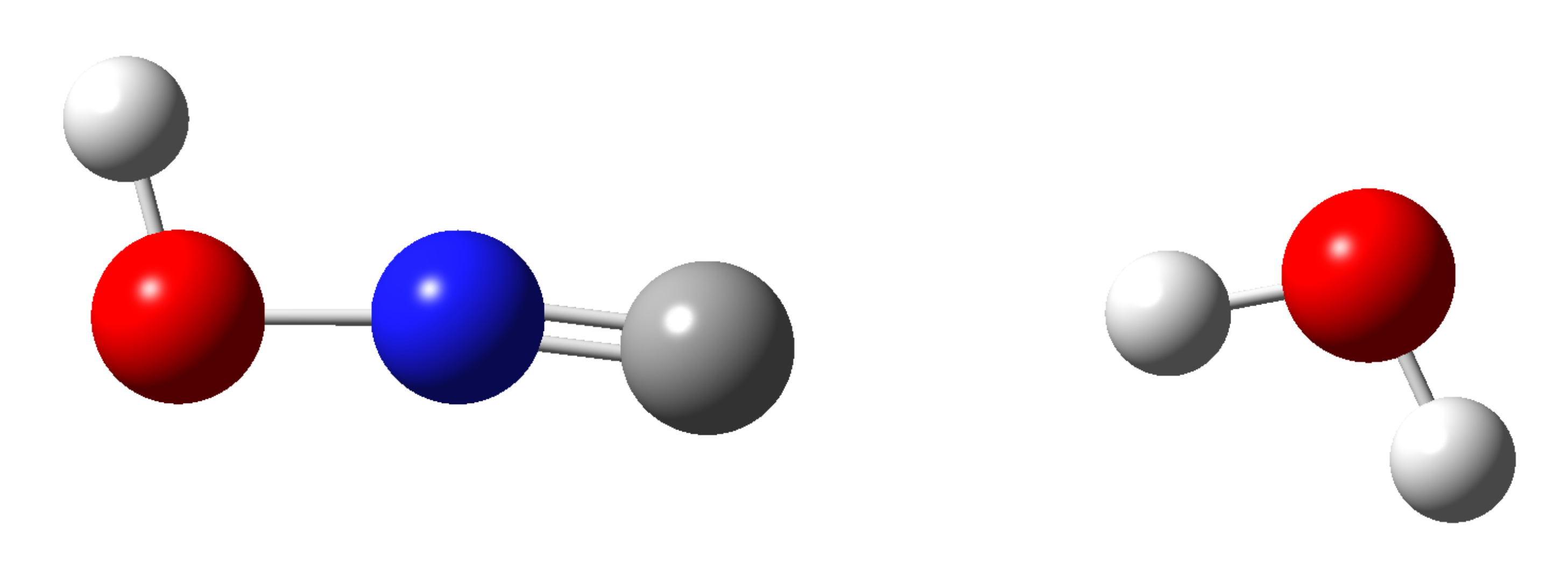}&2370&&&&\\
\hline
\multicolumn{7}{|c|}{$\rm{\bf C_2NH_3O}$}\\
\hline
&&&&&& \\
CH$_3$NCO&\includegraphics[height=1.2cm, width=1.5cm]{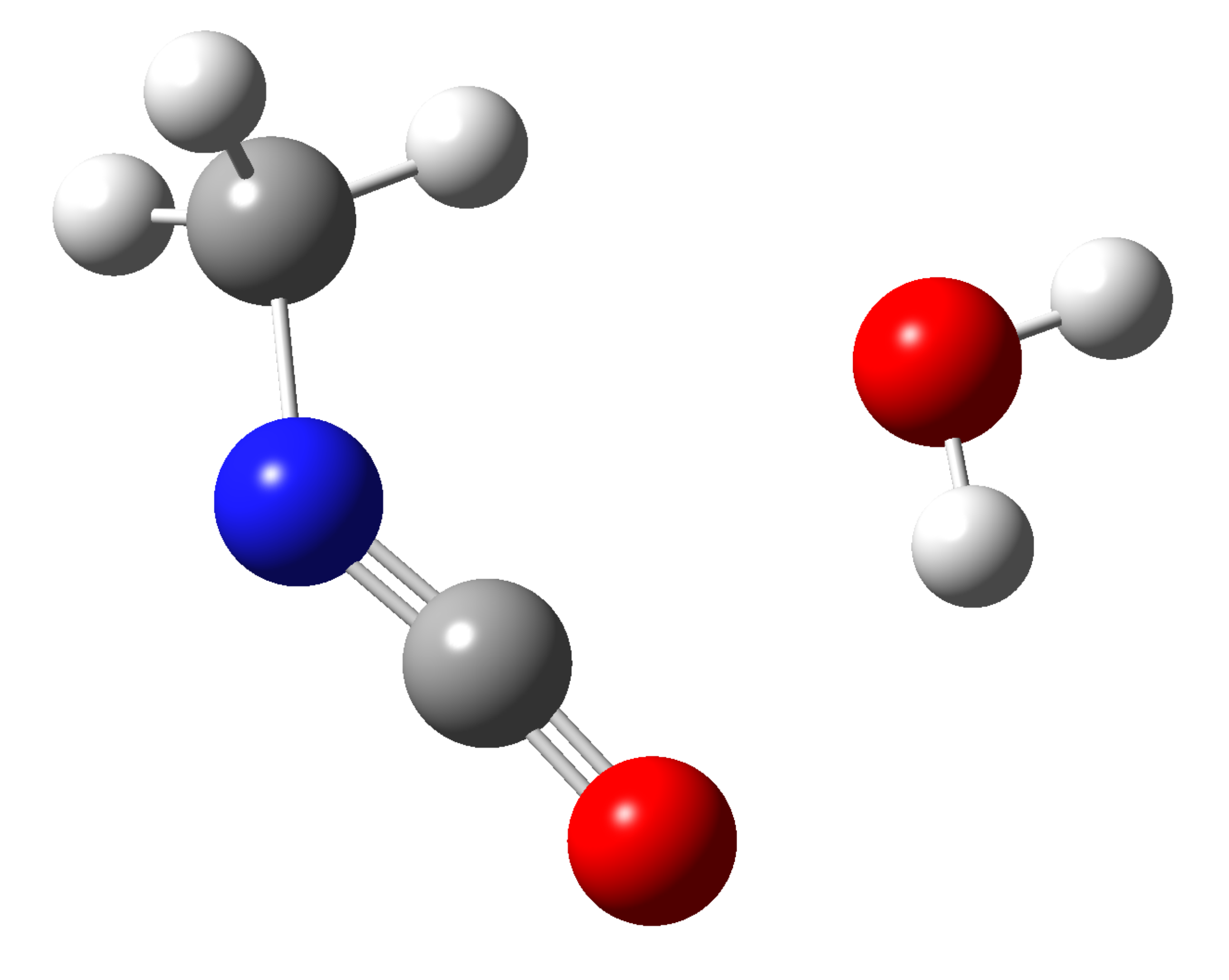}&3627& 3091 & 4377 & 4309 & $4700 \pm 1410$$^{a}$ \\
&\includegraphics[height=1.2cm, width=1.2cm]{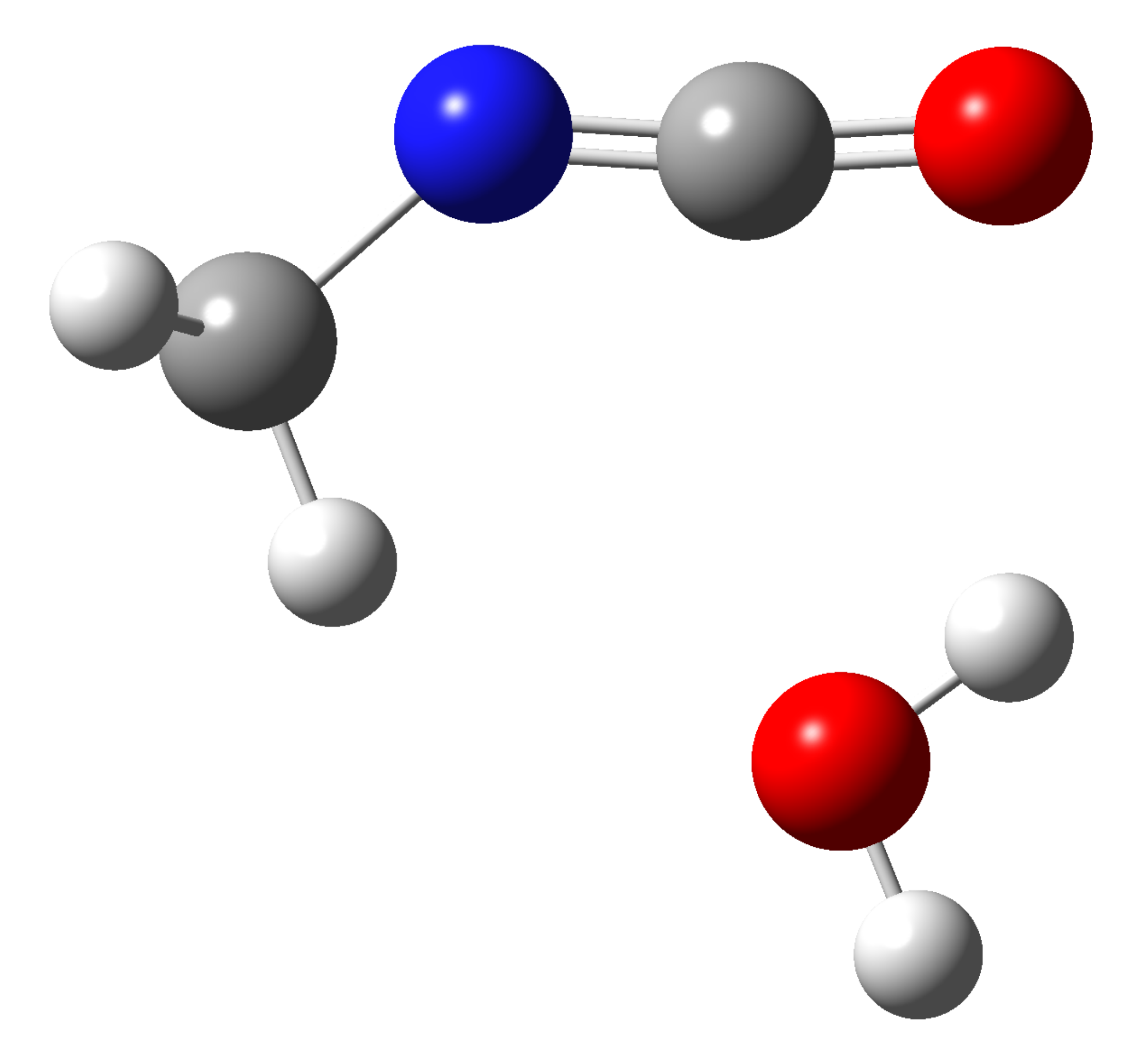}&2555&&&& \\
\hline
CH$_3$CNO&\includegraphics[height=1.2cm, width=1.5cm]{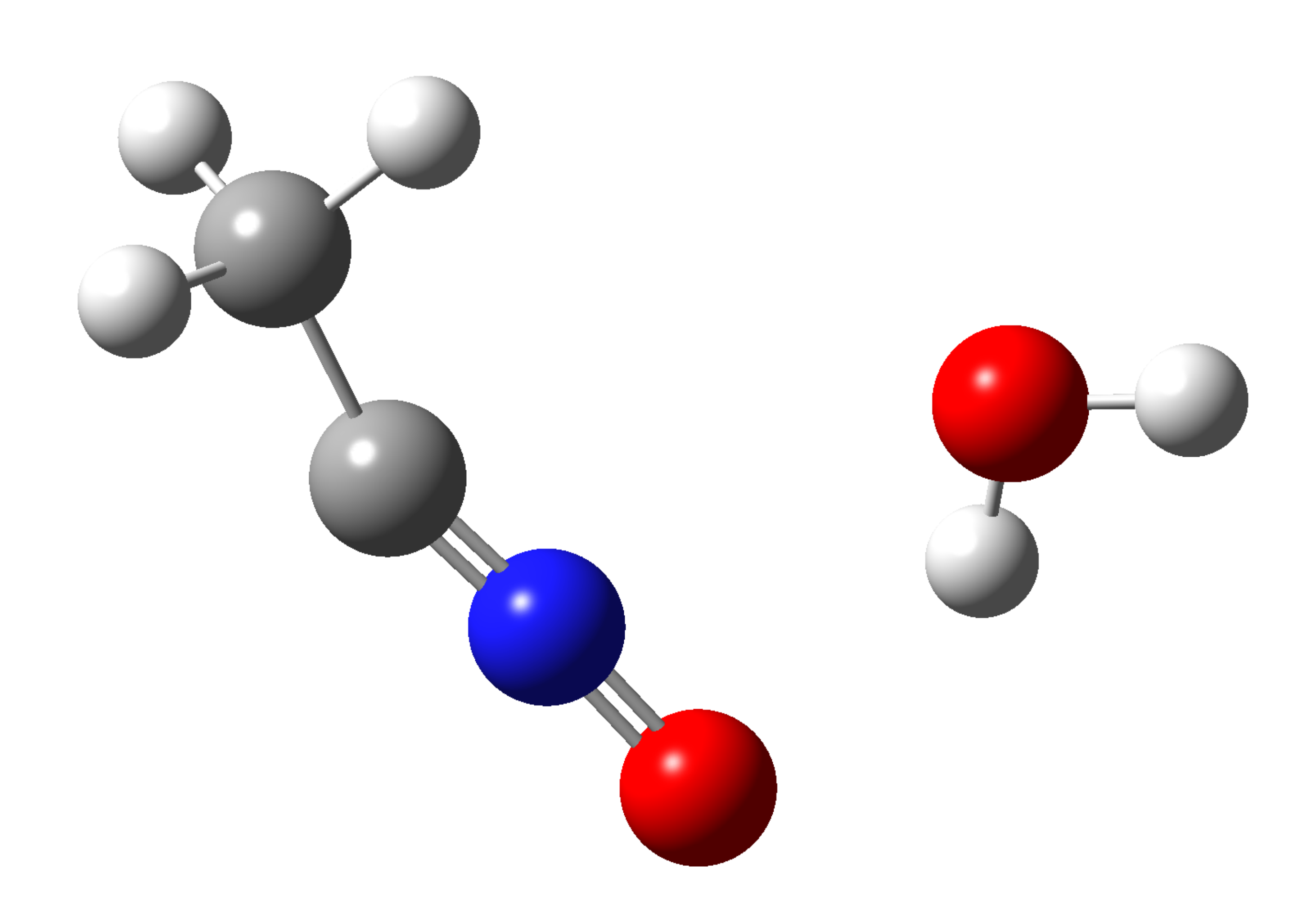}&2786& 2786 & 3945 & --- & --- \\
\hline
CH$_3$OCN&\includegraphics[height=1.2cm, width=1.5cm]{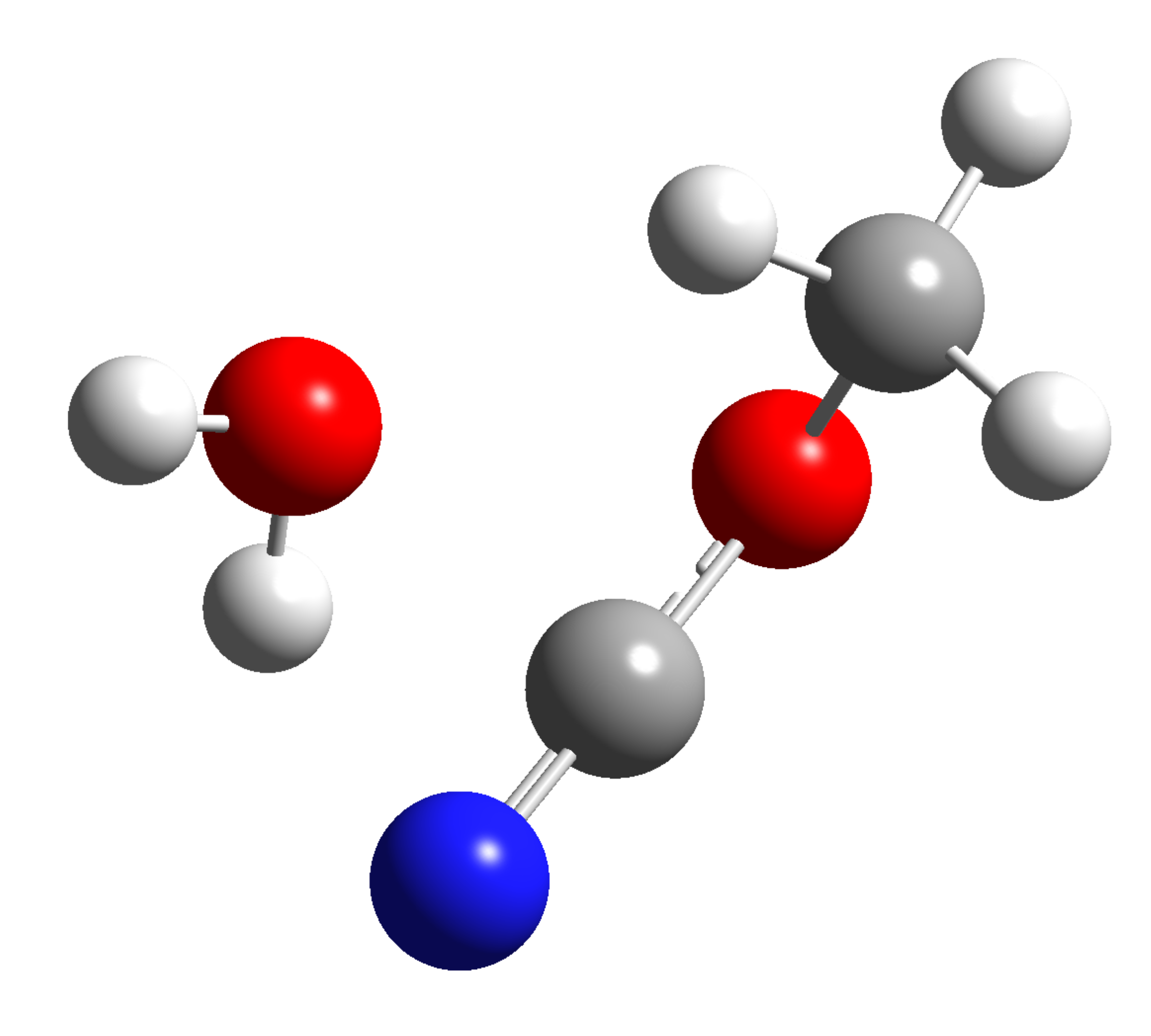}& 3534 & 3535 & 5006 & 6530 & --- \\
&\includegraphics[height=1.2cm, width=1.5cm]{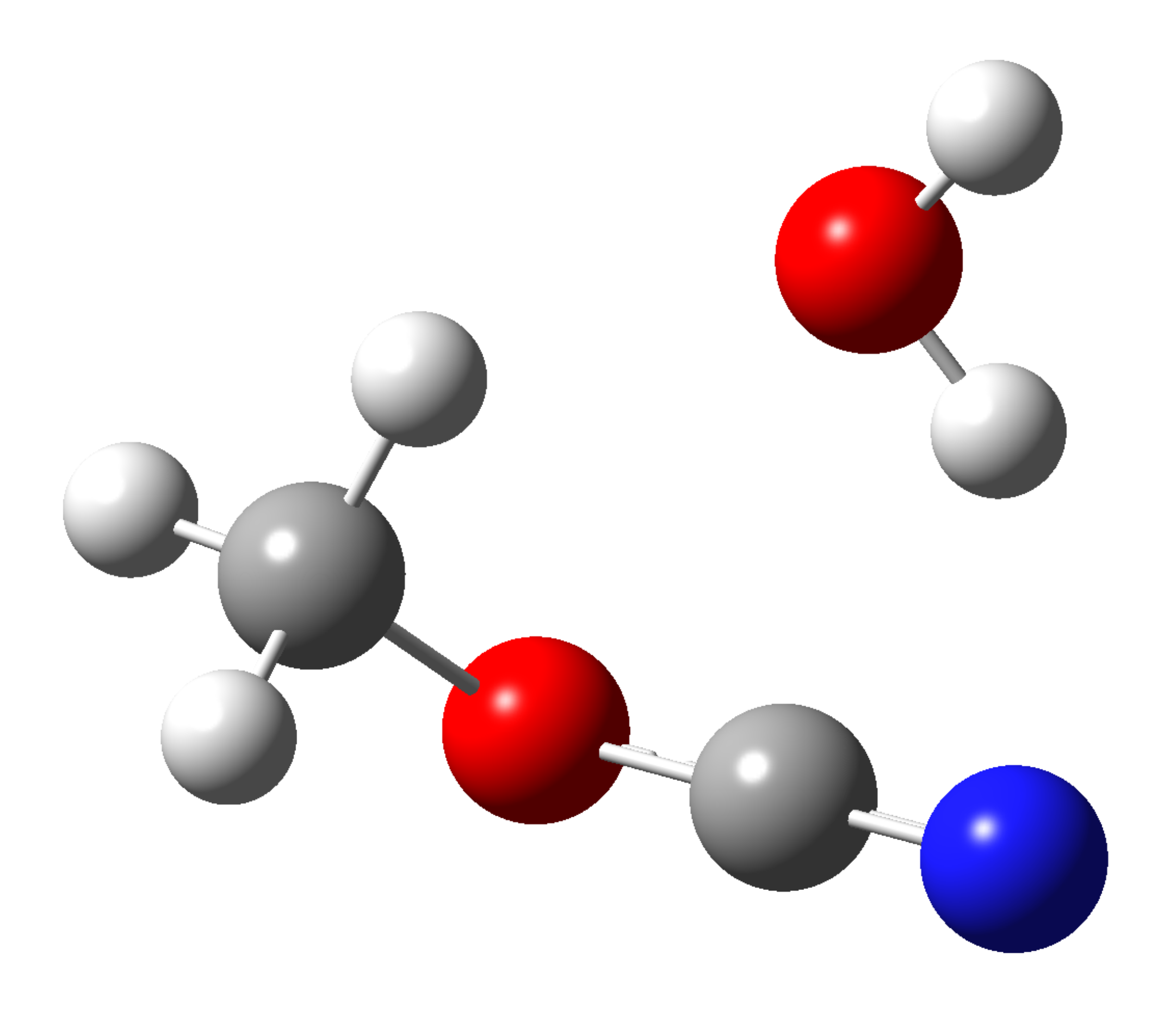}& 3536 &&&& \\
\hline
CH$_3$ONC&\includegraphics[height=1.2cm, width=1.5cm]{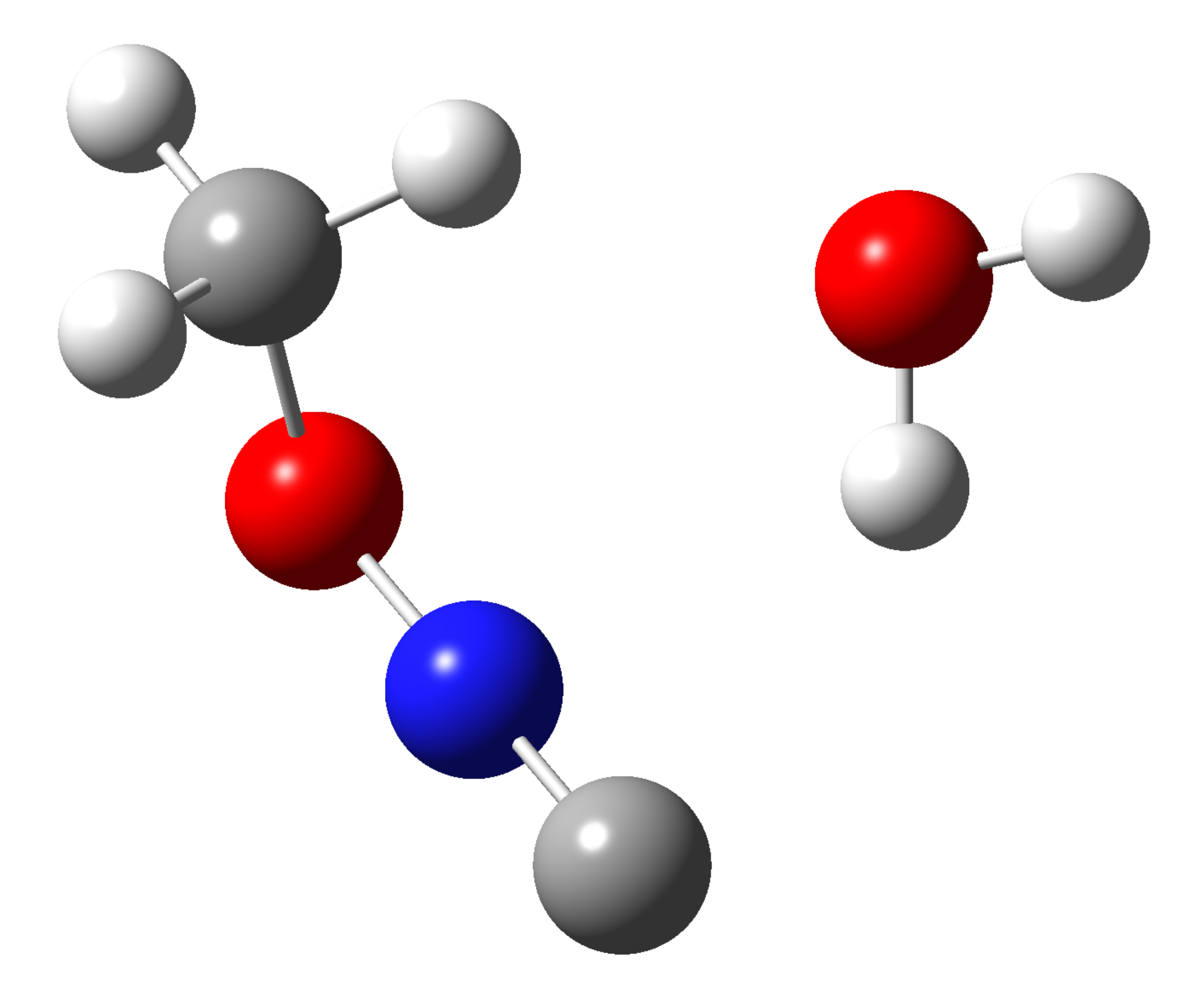}& 2939 & 2752 & 3897 & 4652 & --- \\
&\includegraphics[height=1cm, width=2.2cm]{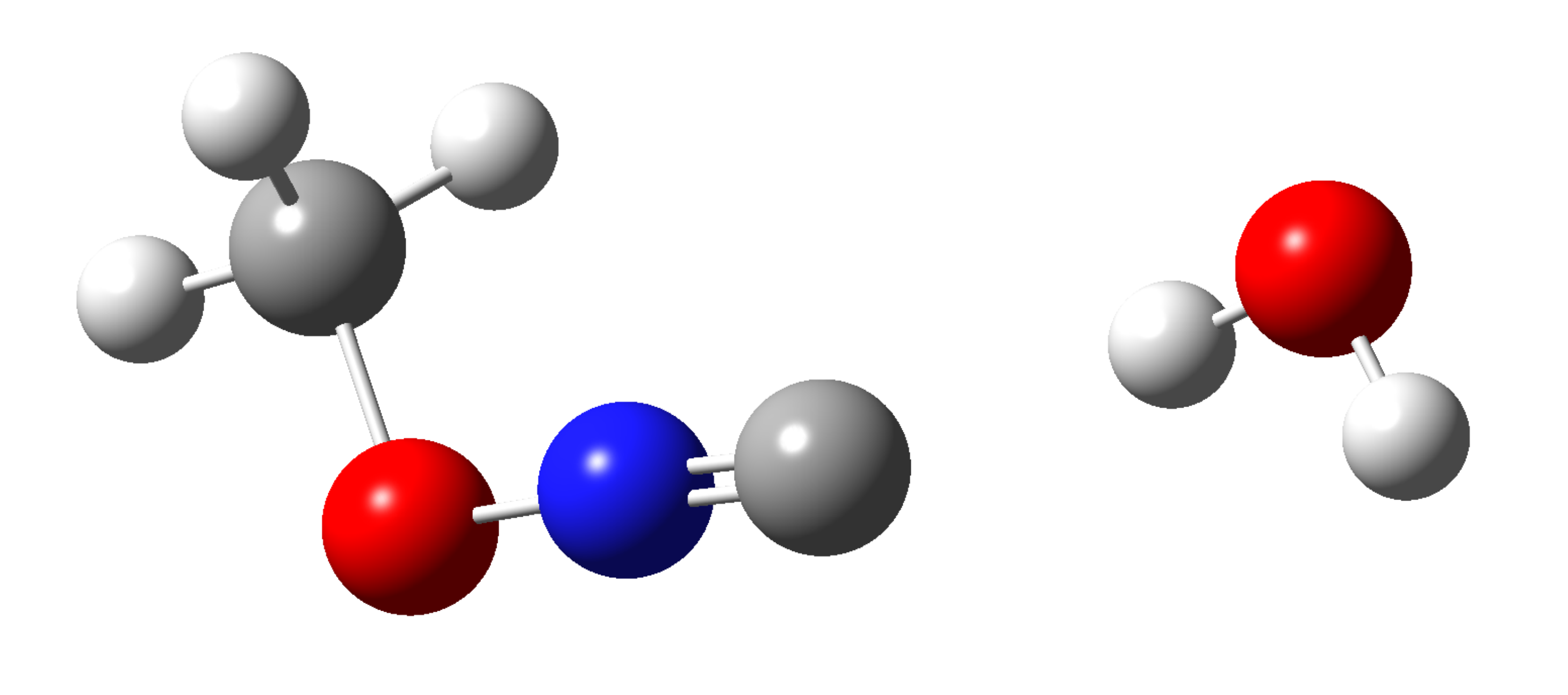}& 2565 &&&& \\
\hline
\multicolumn{7}{|c|}{$\rm{\bf CNH_3O}$}\\
\hline
&&&&&& \\
$\rm{NH_2CHO}$&\includegraphics[height=0.8cm, width=1.5cm]{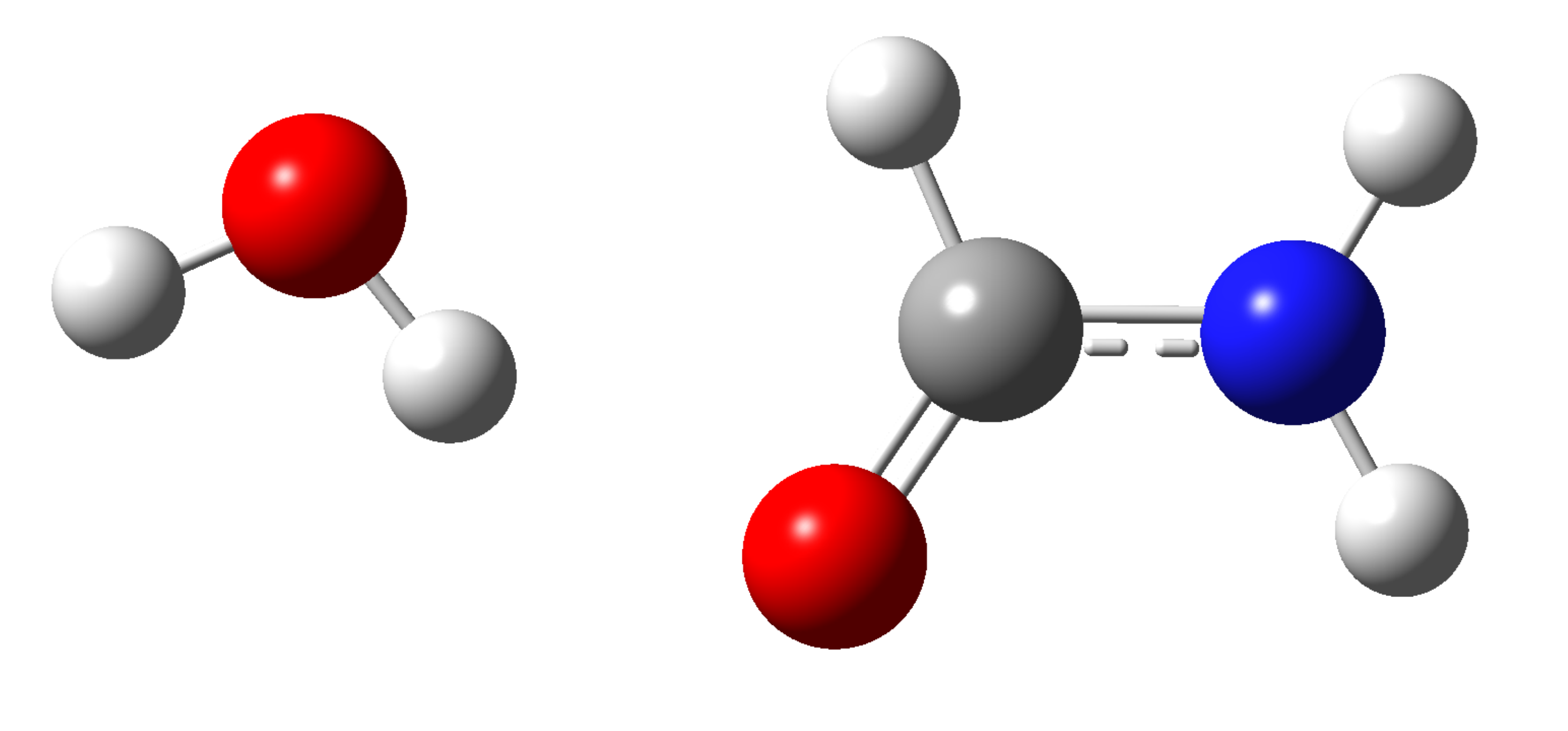}&3627&&&& \\
&\includegraphics[height=1cm, width=1.5cm]{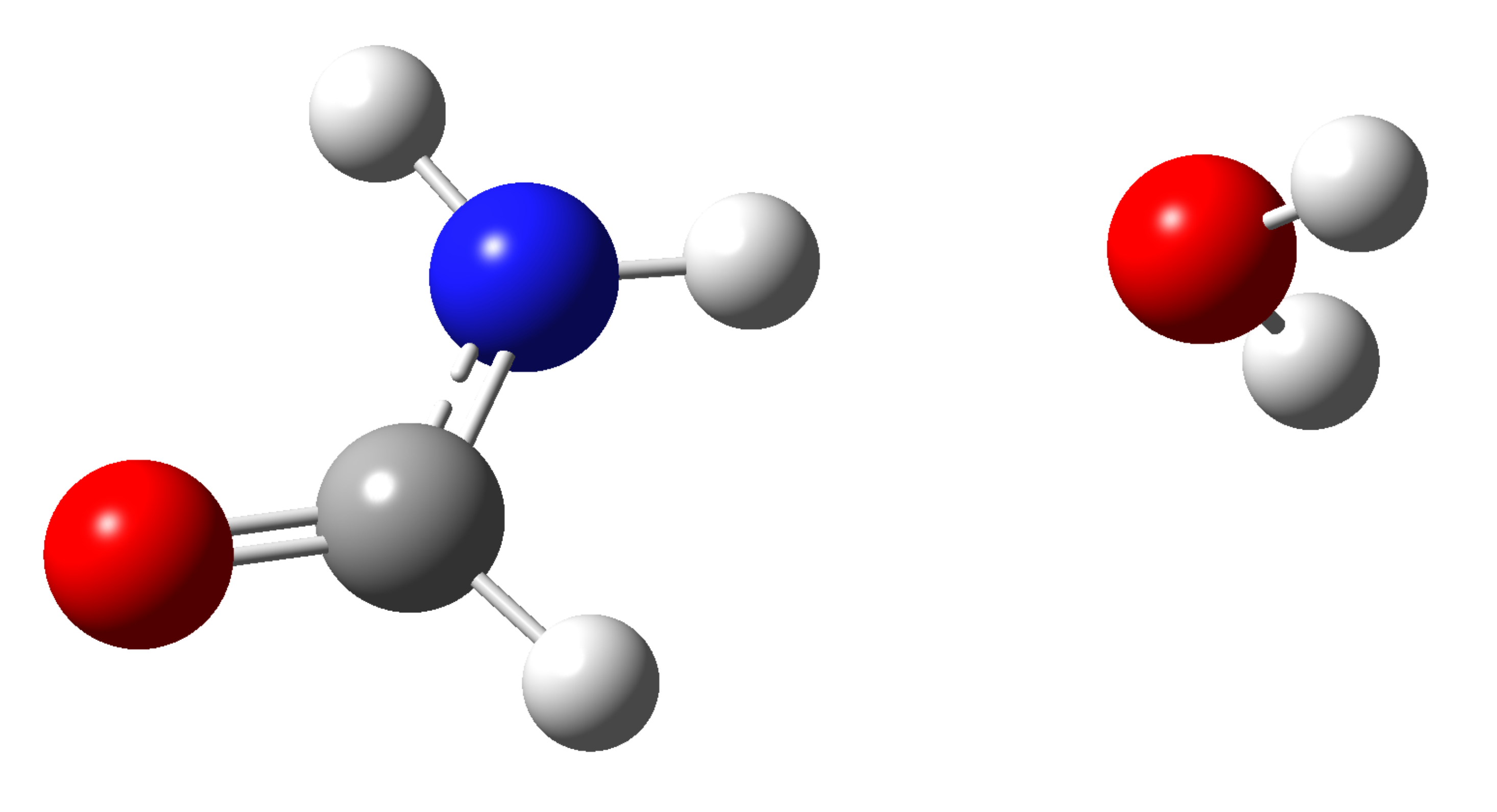}&2880& 3862 & 5468 & 6602 & $6300 \pm 1890$$^{a}$ \\
&\includegraphics[height=1cm, width=1.5cm]{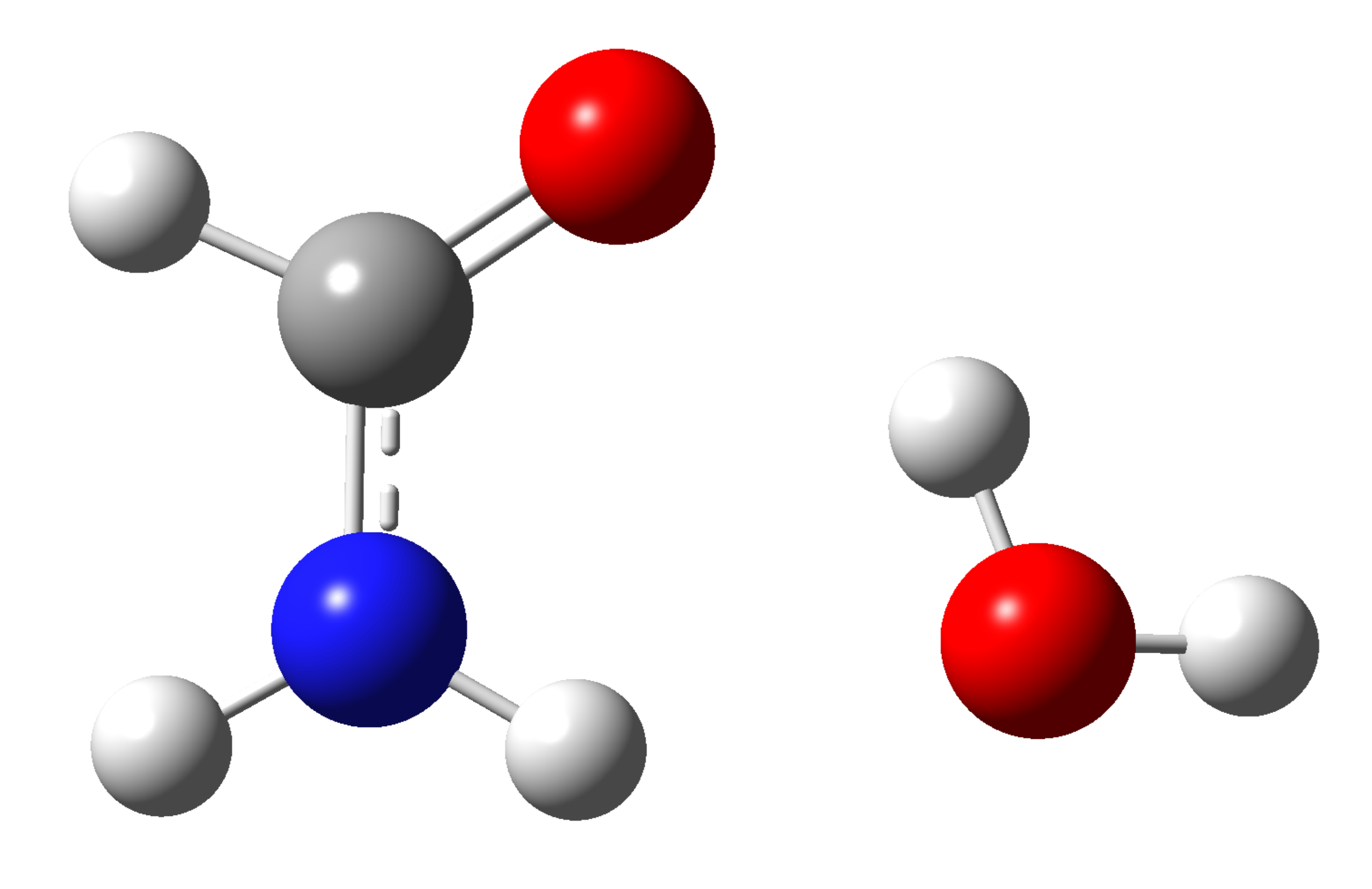}&5079&&&& \\
\hline
\end{tabular}} \\
\vskip 0.2 cm
{\bf Note:} \\
We use our calculated scaled BE values of monomers for astrochemical modeling. \\
$^{a}$ \cite{wake17}. \\
$^{b}$ \cite{quan10}.
\end{table}

To study the BE and reaction pathways of three
peptide-bond-like species HNCO, NH$_2$CHO, CH$_3$NCO,
and their isomers/precursors, we use the \textsc{Gaussian} 09 suite of programs \citep{fris13}.
To estimate the BEs of these species, we use the MP2/aug-cc-pVDZ level of theory.
Our calculated BE values are given in Table \ref{table:BE}. For some cases, we find multiple
probable sites for the adsorption and thus obtained multiple BE values.
In that case, we take the average of the numerous BEs.
Calculated BEs with a single water molecule are
then scaled up by a factor of $1.416$ \citep{das18} to have a realistic estimation.
Additionally, in Table \ref{table:BE}, we present the BE values for some of these species
with a hexamer configuration of water clusters.
Since the BEs with the pentamer/hexamer configurations show minimum deviation \citep{das18},
one can use these BE values in the model without scaling.
However, we cannot provide the BE values of all the
species with a hexamer configuration and all the probable adsorption sites.
Thus, we use BE values obtained with a single water molecule with appropriate
scaling for the modeling.

\subsubsection{Reaction dynamics of peptide-bond-like species}

To form ice-phase HNCO, \cite{quen18} considered the reaction between NH and CO.
They considered an activation barrier of $4200$ K for this reaction \citep{himm02}.
However, for the formation of its other isomers, no such
reactions are available. Due to this reason, for the sake of completeness,
here we run a quantum chemical calculation to check the reaction enthalpy
of the following reactions:
\begin{equation} \label{eqn:peptide_8}
\rm{NH+CO\rightarrow HNCO},
\end{equation}
\begin{equation} \label{eqn:peptide_9}
\rm{CH+NO\rightarrow HCNO},
\end{equation}
\begin{equation} \label{eqn:peptide_10}
\rm{CN+OH\rightarrow HOCN},
\end{equation}
\begin{equation} \label{eqn:peptide_11}
\rm{CN+OH\rightarrow HONC}.
\end{equation}

We find that the above four reactions are exothermic.
Exothermicity values are given in Table \ref{table:Enthalpy}.
The activation barrier for the reaction between NH and CO was known to be $4200$ K \citep{himm02},
but for the others, it was unknown.
The above reactions are mostly between radicals, and finding an actual TS is a difficult task.
Instead, we calculate the reaction enthalpy of these four reactions.
Based on the reaction enthalpies,
we prepare the most probable reaction sequence in between these four reactions.
Since the activation barrier of the first reaction was known to be $4200$ K,
we scale the activation barriers of the rest of the reactions.
Though the reaction enthalpy (exothermicity values) is not directly related
to the activation barrier of the reaction, it is eventually a better-educated approximation
compared to using any other crude approximation.
Scaled activation barriers are provided in Table \ref{table:Enthalpy}.

\begin{table}
\scriptsize{
\centering
\caption{Calculated reaction enthalpies, type of reactions, and activation barriers (with ZPVE correction) of various reactions \citep{gora20b}.
\label{table:Enthalpy}}
\vskip 0.2 cm
\begin{tabular}{lcccc}
\hline
{\bf Reactions} & {\bf Reaction enthalpy} & {\bf Types of} & {\bf Activation barrier (K)} \\
& {\bf (kcal/mol)} & {\bf reactions} & \\
\hline
\multicolumn{4}{c}{\bf Ice-phase reactions} \\
\hline
$\rm{NH + CO \rightarrow HNCO}$ & $-139.19$ & Exothermic & 4200$^{a}$ \\
$\rm{CH + NO \rightarrow HCNO}$ & $-124.62$ & Exothermic & 4691 \\
$\rm{CN + OH \rightarrow HOCN}$ & $-120.35$ & Exothermic & 4857 \\
$\rm{CN + OH \rightarrow HONC}$ & $-59.32$ & Exothermic & 9855 \\
$\rm{HNCO + H \rightarrow H_2NCO}$ & $-31.35$ & Exothermic & 1962 \\
$\rm{H_2NCO + H \rightarrow NH_2CHO}$ & $-92.76$ & Exothermic & 0 \\
$\rm{H_2NCO + H \rightarrow HNCO + H_2}$ & $-73.15$ & Exothermic & 0 \\
$\rm{HCNO + H \rightarrow H_2CNO}$ & $-57.29$ & Exothermic & 1073 \\
$\rm{H_2CNO + H \rightarrow HCNO + H_2}$ & $-47.20$ & Exothermic & 0 \\
\hline
\multicolumn{4}{c}{\bf Gas-phase reactions} \\
\hline
$\rm{HNCO + O \rightarrow CO + HNO}$ & $-23.70$ & Exothermic & ... \\
$\rm{HNCO + O \rightarrow OH + OCN}$ & $5.70$ & Endothermic & ... \\
$\rm{HCNO + O \rightarrow CO + HNO}$ & $-92.52$ & Exothermic & ... \\
$\rm{HCNO + O \rightarrow OH + CNO}$ & 0.064 & Endothermic & ... \\
$\rm{HOCN + O \rightarrow OH + OCN}$ & $-22.90$ & Exothermic & ... \\
$\rm{HONC + O \rightarrow OH + CNO}$ & $-18.80$ & Exothermic & ... \\
\hline
\end{tabular} \\
}
\vskip 0.2 cm
{\bf Note:} \\
$^{a}$ \cite{himm02}
\end{table}

\cite{quen18} studied the peptide-bond-related molecules in the protostar
IRAS $16293-2422$ and the prestellar core L1554 by using a chemical model.
Earlier, it was claimed that HNCO and $\rm{NH_2CHO}$ are
chemically linked because $\rm{NH_2CHO}$ could be formed by the successive hydrogenation
reactions of HNCO \citep[HNCO $\rightarrow$ H$_2$NCO $\rightarrow$ NH$_2$CHO;][]{mend14,lope15,song16,lope19}.
The first step of this hydrogenation sequence has an activation barrier of $1962$ K,
and the second step is a $\rm{radical-radical}$
reaction and thus could be barrierless. However, recent experimental studies by \cite{nobl15} and \cite{fedo15} questioned this fact. They opposed
the formation of $\rm{NH_2CHO}$ by the reaction between $\rm{H_2NCO}$ and H;
instead, they proposed that eventually,
it would return to HNCO again ($\rm{H_2NCO+H\rightarrow HNCO+H_2}$).
Here, we consider only the formation of $\rm{NH_2CHO}$ in our ice-phase network.
In order to continue a comparative study between the various isomers of HNCO,
we are interested in checking the hydrogenation
reactions with the various isomeric forms of HNCO.
Thus, the reaction enthalpies of the following reactions are studied:
\begin{equation} \label{eqn:peptide_12}
\rm{HNCO+H\rightarrow H_2NCO},
\end{equation}
\begin{equation} \label{eqn:peptide_13}
\rm{HCNO+H\rightarrow H_2CNO},
\end{equation}
\begin{equation} \label{eqn:peptide_14}
\rm{HOCN+H\rightarrow H_2OCN},
\end{equation}
\begin{equation} \label{eqn:peptide_15}
\rm{HONC+H\rightarrow H_2ONC}.
\end{equation}

However, we do not obtain a valid neutral structure for $\rm{H_2OCN}$ and $\rm{H_2ONC}$, and
thus we do not consider the last two hydrogenation reactions of this sequence.
In Table \ref{table:Enthalpy}, we summarize the obtained reaction
enthalpies of reactions \ref{eqn:peptide_12} and \ref{eqn:peptide_13}.
Based on the obtained reaction enthalpy for the second reaction relative to the
first reaction, we scale the activation barrier of the second reaction to $1073$ K.

Recently, \cite{haup19} proposed the successive H abstraction reactions
to $\rm{NH_2CHO}$ for the formation of HNCO:
\begin{equation} \label{eqn:peptide_16}
\rm{NH_2CHO+H\rightarrow H_2+H_2NCO},
\end{equation}
\begin{equation} \label{eqn:peptide_17}
\rm{H_2NCO+H\rightarrow H_2+HNCO}.
\end{equation}
They pointed out that reaction \ref{eqn:peptide_16} has an activation barrier of $240-3130$ K
depending on the level of theory used for the quantum chemical calculation.
They found that reaction \ref{eqn:peptide_17} is barrierless.
This reaction is exciting as it might support the earlier
claim of chemical linkage between HNCO and ${\rm NH_2CHO}$.
They also performed quantum chemical calculations for the H addition
reactions to H$_2$NCO and HNCO:
\begin{equation} \label{eqn:peptide_18}
\rm{H + H_2NCO\rightarrow NH_2CHO},
\end{equation}
\begin{equation} \label{eqn:peptide_19}
\rm{H+HNCO\rightarrow H_2NCO}.
\end{equation}
They found that reaction \ref{eqn:peptide_18} is barrierless, whereas reaction \ref{eqn:peptide_19} has an activation barrier of $2530-5050$ K,
depending on the level of theory used for the computation.

For the calculation of the gas-phase reaction rate of these
four reactions, we use
\begin{equation}
rate=\alpha \Big(\frac{T}{300}\Big)^{\beta}exp{({-\gamma/T})},
\end{equation}
where $\alpha$, $\beta$, and $\gamma$ are the three constants of the reaction.
We consider $\alpha=10^{-10}$, $\beta=0$ and
$\gamma=240-3130$ for reaction \ref{eqn:peptide_16}.
For the reaction \ref{eqn:peptide_17} and \ref{eqn:peptide_18}, we consider
$\alpha=10^{-10}$, $\beta=0$, and $\gamma=0$, and for reaction \ref{eqn:peptide_19},
we consider $\alpha=10^{-10}$, $\beta=0$, and $\gamma=2530-5050$.
Since a proper structure for H$_2$CNO is obtained, we consider
the reaction $\rm{H +H_2CNO \rightarrow HCNO + H_2}$ in both gas and ice phases.

\cite{quen18} used the gas-phase destruction of HOCN, HCNO, and HONC by an oxygen atom.
For all three destruction reactions, they considered an activation barrier of $195$ K.
However, \cite{quan10} considered the activation barriers
of $2470$ K, $195$ K, and $3570$ K, respectively, for these three destruction
reactions by oxygen atoms, and these are the default
in the UMIST 2012 network.
Here, we consider the default destruction reactions as in UMIST 2012.
The destruction of HNCO by the oxygen atom is not considered. To this effect,
we calculate the reaction enthalpies for the reactions $\rm{HNCO+O\rightarrow CO+ HNO}$
and $\rm{HNCO+O\rightarrow OH+ OCN}$. We find that the second reaction in this sequence is
endothermic, whereas the first one is exothermic, and thus we do not consider
the second one. The reaction $\rm{HNCO+O\rightarrow CO+ HNO}$ is very similar to the reaction $\rm{HCNO+O \rightarrow CO +HNO}$, for which a $195$ K activation barrier is considered in UMIST 2012.
Based on the exothermicity values between $\rm{HCNO + O}$ and $\rm{HNCO + O}$,
we use a scaling factor and obtain an activation barrier of $765$ K for $\rm{HNCO + O}$.
Calculated reaction enthalpies and the activation barriers with ZPVE correction are noted in Table \ref{table:Enthalpy}.

\subsection{Modeling results}
The observed abundances of HNCO, $\rm{NH_2CHO}$, and $\rm{CH_3NCO}$ toward G10 \citep{gora20b} are shown in Table \ref{table:abunobs}.
The abundances are very sensitive to the physical parameters ($T_{ice}$, $\rho_{max}$, $T_{max}$, $t_{coll}$, and $t_{pw}$) and adopted rate constants.
Here, we make an extensive effort to find out the simultaneous appearance of these three
N-bearing species by varying the sensitive physical parameters and rate constants
of some of the key reactions.
More precisely, we prepare two models: Model A and Model B.
The difference between the two models is highlighted in Table \ref{tab:Model}.

\subsubsection{Results obtained with Model A}
\begin{figure}
\centering
\includegraphics[width=0.5\textwidth, angle=-90]{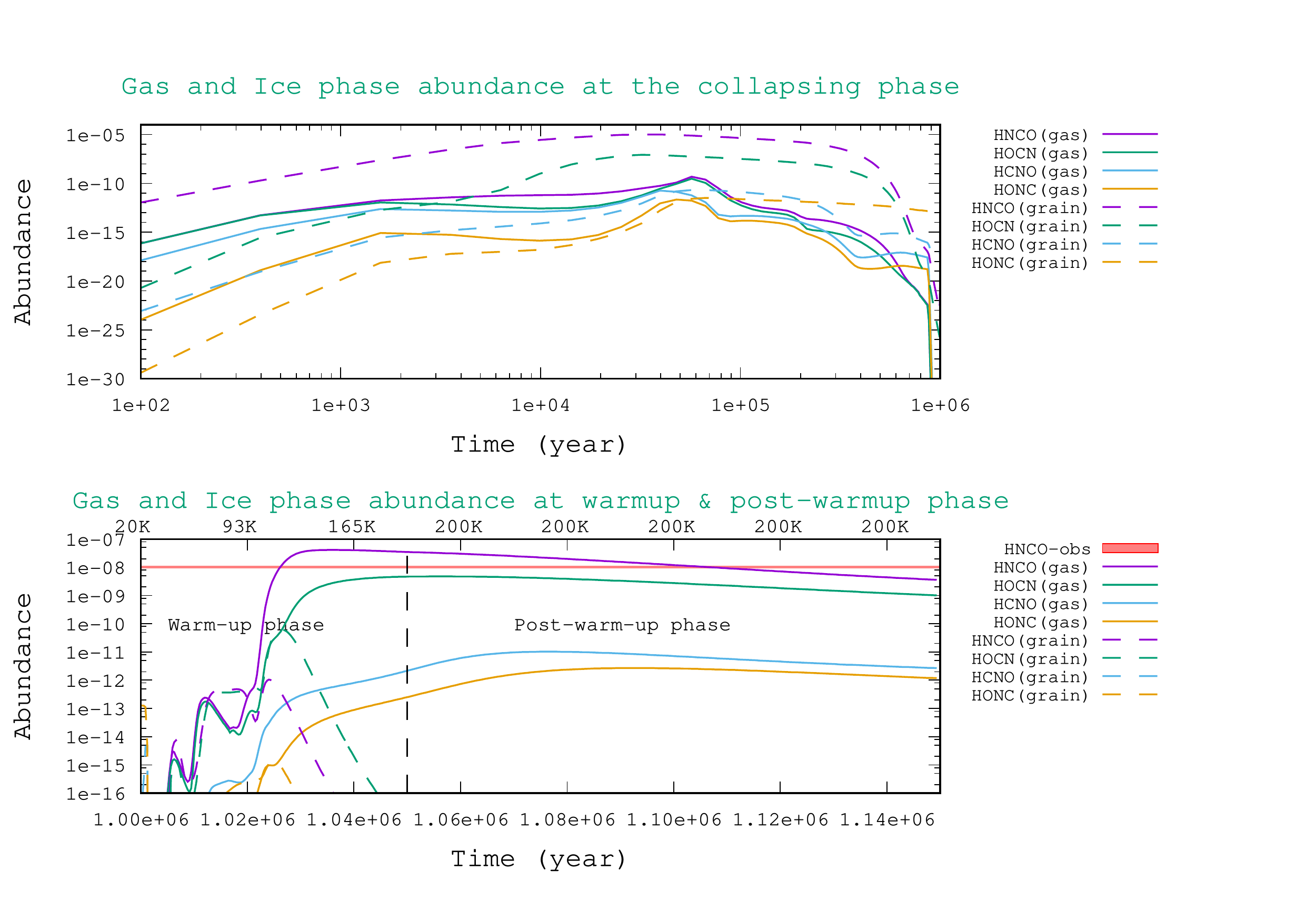}
\caption{Chemical evolution of the peptide-bond-related molecule HNCO and 
corresponding isomers. This is shown for $\rho_{max}=1.0 \times 10^7$ cm$^{-3}$ and $T_{ice}=20$ K by considering the best-fitted parameters with Model A. Red shaded lines represent the observed abundances obtained in G10 \citep{gora20b}.}
\label{fig:ModelA-CHNO}
\end{figure}

\begin{figure}
\centering
\includegraphics[width=0.5\textwidth, angle=-90]{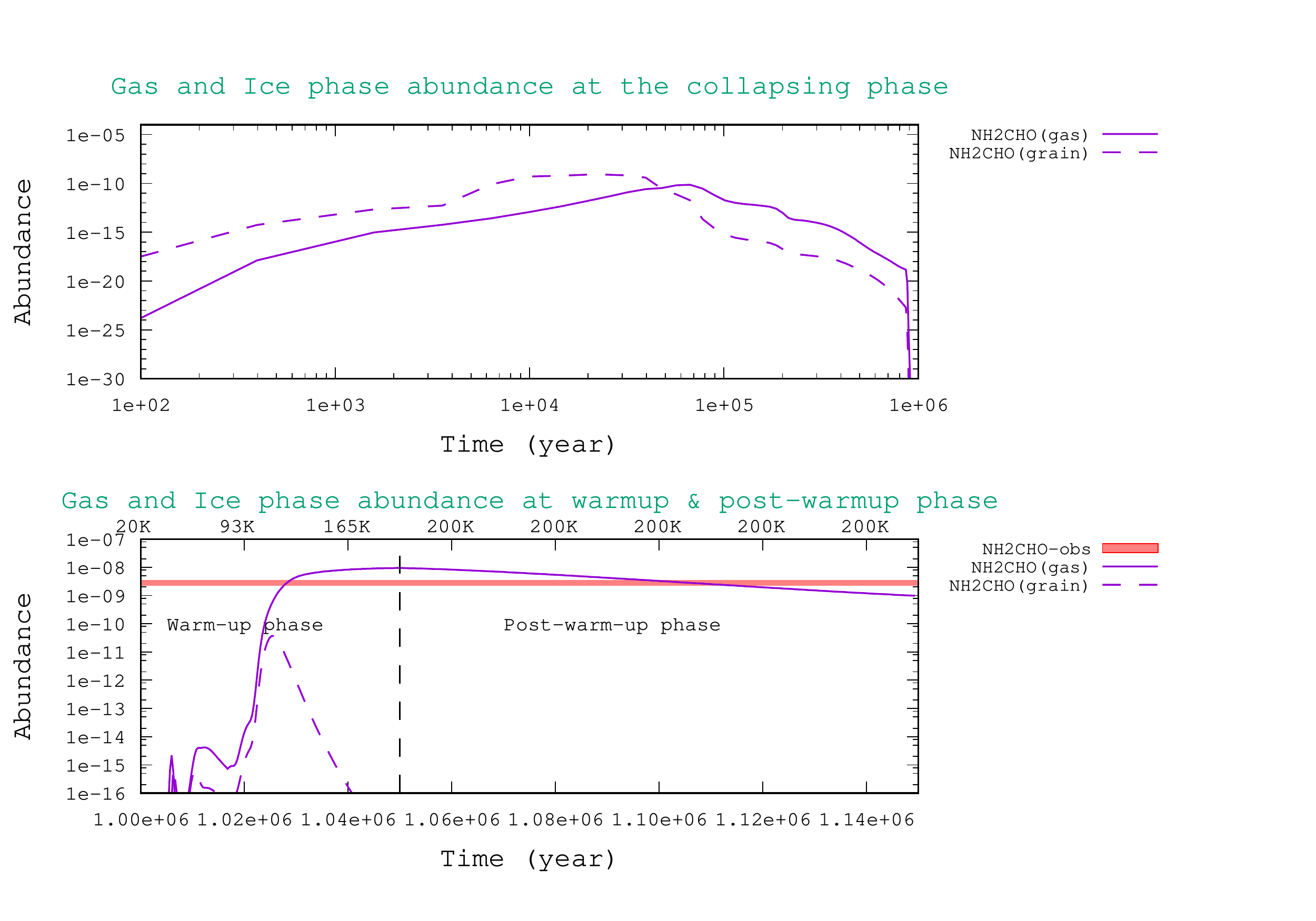}
\caption{Chemical evolution of the peptide-bond-related molecule NH$_2$CHO. This is shown for $\rho_{max}=1.0 \times 10^7$ cm$^{-3}$ and $T_{ice}=20$ K by considering the best-fitted parameters with Model A. Red shaded lines represent the observed abundances obtained in G10 \citep{gora20b}.}
\label{fig:ModelA-NH2CHO}
\end{figure}

\begin{figure}
\centering
\includegraphics[width=0.5\textwidth, angle=-90]{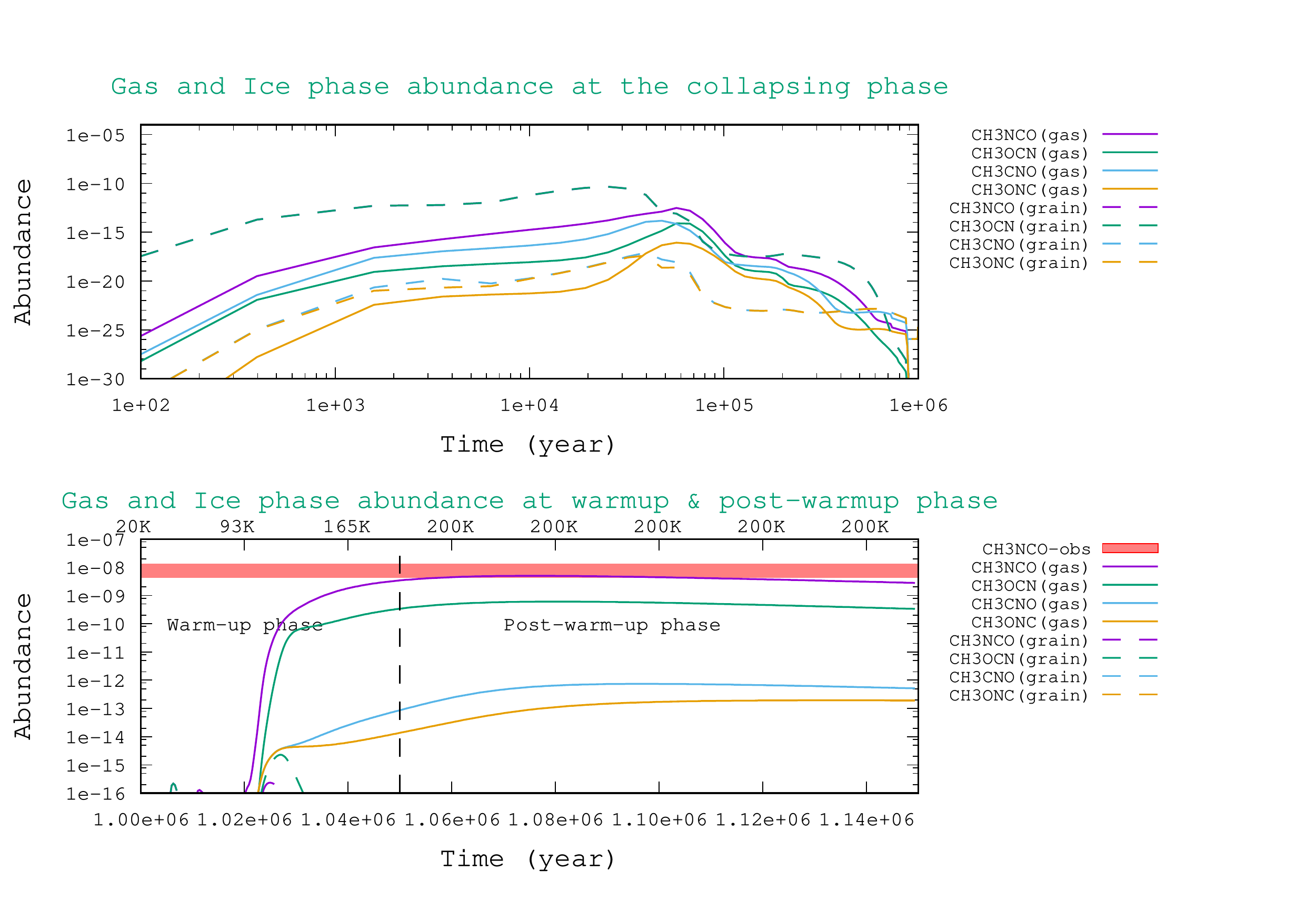}
\caption{Chemical evolution of the peptide-bond-related molecule CH$_3$NCO and 
corresponding isomers. This is shown for $\rho_{max}=1.0 \times 10^7$ cm$^{-3}$ and $T_{ice}=20$ K by considering the best-fitted parameters with Model A. Red shaded lines represent the observed abundances obtained in G10 \citep{gora20b}.}
\label{fig:ModelA-CH3NCO}
\end{figure}

\begin{figure}
\centering
\includegraphics[width=0.5\textwidth, angle=-90]{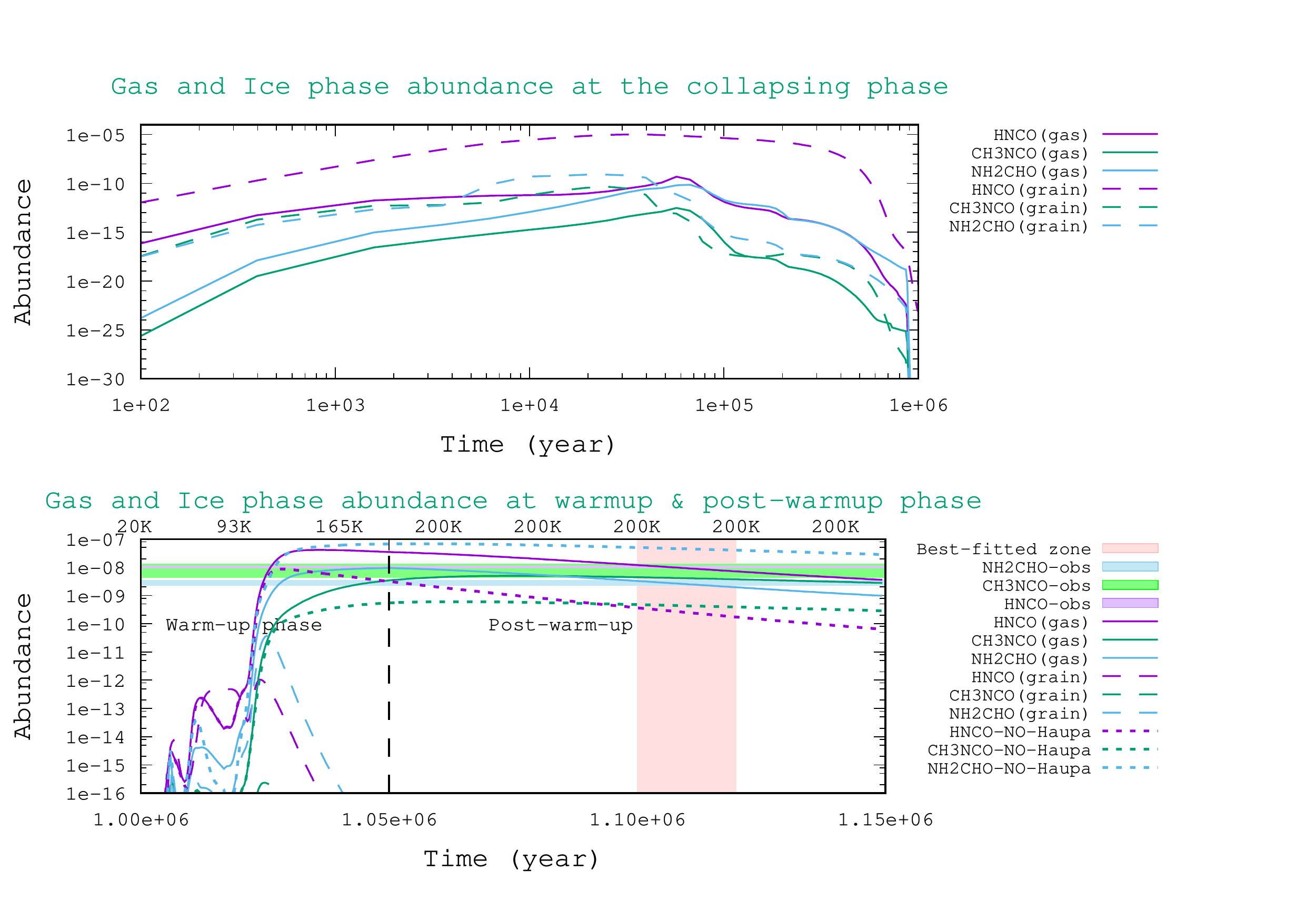}
\caption{Chemical evolution of HNCO, CH$_3$NCO, and NH$_2$CHO during the three stages by considering the best-fit parameters of Model A. The best-fitted time zone is also highlighted.
Abundance variation by avoiding \cite{haup19} pathways in the gas phase is also shown
\citep{gora20b}.
\label{fig:best}}
\end{figure}

\begin{figure}
\hskip -1.2 cm
\includegraphics[width=0.33\textwidth,angle=-90]{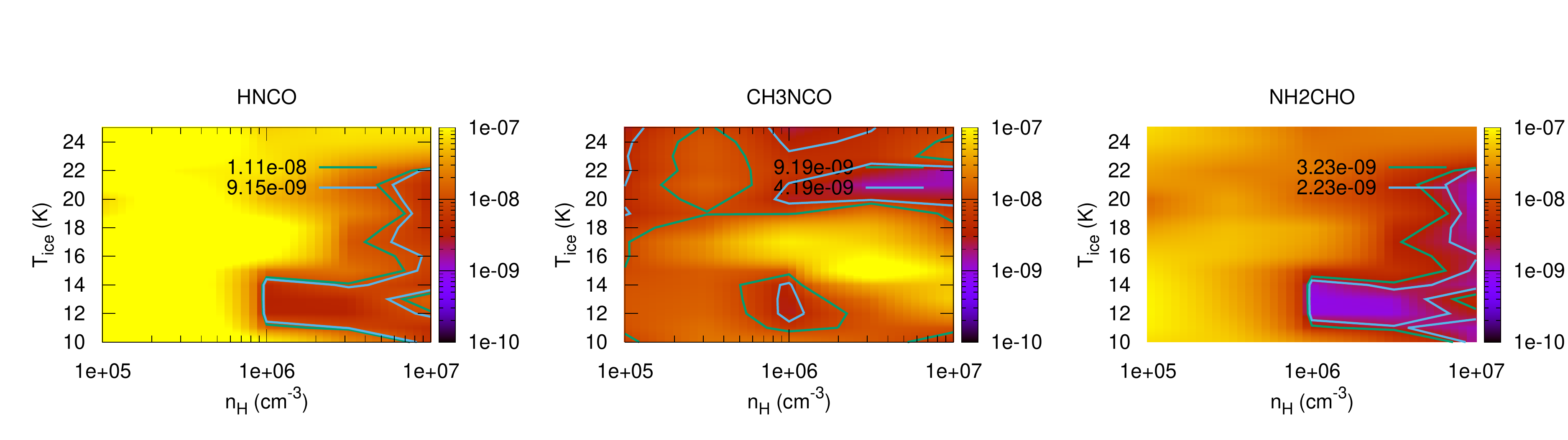}
\caption{Parameter space obtained with Model A by considering the best-fitted parameters noted in Table \ref{tab:Model} at an age
position of $1.12 \times 10^6$ years. Color coding in the right side of the each panel represents the abundance with respect to 
H$_2$ \citep{gora20b}. \label{fig:param}}
\end{figure}

To constrain the best possible model, we explore the parameter space around which the modeling
results agree with the observational results. To this effect, we run several cases by varying the
initial dust temperature ($T_{ice}$) between $10$ K and $25$ K and $\rho_{max}$ between $10^5$ cm$^{-3}$ and $10^7$ cm$^{-3}$ for Model A.
Since G10 is a hot-core, a higher density ($10^6-10^7$ cm$^{-3}$) is preferable.
G10 is extended by roughly $0.1$ pc and has $\sim10^3$
solar masses of matter \citep{cesa94}. From that estimation, the average density of the source
is around $10^7$ cm$^{-3}$. Interestingly, continuum temperatures vary between $19$ K and $27$ K
from observational analysis \citep{gora20b}. Based on the observational results as a preliminary
guess, we use $\rho_{max}=10^7$ cm$^{-3}$ and $T_{ice}=20$ K for Model A.
Initially, we start with Model A with the rate constants of the
gas-phase reactions available in the literature \citep{skou17,quen18,haup19}.
Based on some preliminary iterations of our simulation, we vary the rate constants of some essential gas-phase reactions, which control the abundances of the three targeted species.
We obtain an excellent correlation between these three species when the rate constants
listed in Table \ref{tab:Model} are used.
\cite{quen18} considered the reaction between HNCO and CH$_3$ in the ice phase to form
the isomers CH$_3$NCO and CH$_3$OCN at the same rate. However, for the gas-phase formation
of other isomers of CH$_3$NCO (CH$_3$CNO, CH$_3$OCN, CH$_3$ONC), \cite{quen18} considered some
rate coefficients of $\sim 10^{-20}$ and $5 \times 10^{-11}$ cm$^3$ s$^{-1}$.
Here, instead of the rate constant of $5 \times 10^{-11}$ cm$^3$ s$^{-1}$,
we consider $10^{-12}$ cm$^3$ s$^{-1}$ for some gas-phase reactions and
use a rate of $10^{-20}$ cm$^3$ s$^{-1}$ for reactions as used in \cite{quen18}.
We keep it as it was considered by \cite{quen18} for the ice-phase formation reactions.
To study the abundances of various isomers considered in the network,
we choose our best-fitted parameters listed in Table \ref{tab:Model} for Model A.
Time evolution of the abundances of HNCO isomers, $\rm{CH_3NCO}$ isomers, and
$\rm{NH_2CHO}$ is shown in Figures \ref{fig:ModelA-CHNO}, \ref{fig:ModelA-NH2CHO}, and \ref{fig:ModelA-CH3NCO}, respectively.
Results obtained with the best-fitted rate constants are shown separately
in Figure \ref{fig:best}, which clearly show
a reasonable correlation among HNCO, CH$_3$NCO, and NH$_2$CHO
around the age of $\sim 1.12 \times 10^6$ years.
Parameter space obtained with the best-fitted rate constants after a suitable
age position ($1.12 \times 10^6$ years) is shown in Figure \ref{fig:param}.
For ease of understanding, abundances closer to the observed
values are indicated with contours.

\begin{table}
\scriptsize
{\centering
\caption{Estimated rotational temperatures, column densities, and fractional abundances
of the observed species \citep{gora20b}.
\label{table:abunobs}}
\vskip 0.2 cm
\begin{tabular}{cccc}
\hline
\hline
Species& Rotational& Column & Fractional \\
&temperature& density &abundance\\
&(K)&(cm$^{-2}$)&\\
\hline
HNCO&317 $\pm$ 25&$\rm{1.37\times10^{17}}$&$\rm{1.02\times10^{-8}}$\\
NH$_2$CHO&439 $\pm$ 100&$\rm{3.88\times10^{16}}$&$\rm{2.87\times10^{-9}}$\\
CH$_3$NCO&248 $\pm$ 19&$\rm{1.20\times10^{17}}$&$\rm{8.88\times10^{-9}}$\\
\hline
\end{tabular} \\
}
\vskip 0.2cm
{\bf Note:}
Assuming the mean value of $\rm{N_{H_2}=1.35\times10^{25}}$ cm$^{-2}$ as estimated
in Table 4 of \cite{gora20b}.
\end{table}

\begin{table}
\scriptsize
\caption{Key differences between the Model A and Model B \citep{gora20b}. \label{tab:Model}}
\hskip -2.0 cm
\begin{tabular}{ccc}
\hline
{\bf Physical parameters}& {\bf Model A}&{\bf Model B} \\
\hline
${\rm{\rho_{max}}}$ ($cm^{-3}$)&$10^7$&$10^{5-7}$\\
${\rm{T_{max}}}$ (K)&200&$100-400$\\
$t_{coll}$ (years)&$10^6$&$10^{5-6}$\\
$t_{w}$ (years)&$5 \times 10^4$&$5 \times 10^4$\\
$t_{pw}$ (years)&$(6.2-10) \times 10^4$&$10^5$\\
$T_{ice}$ (K)&$10-25$&20\\
\hline
&Gas-phase reactions parameterized&\\
&Gas-phase rate constants used in Model A and Model B& Rate constant used in literature\\
\hline
$\rm{NH_2+H_2CO \rightarrow NH_2CHO +H}$&$\alpha=5.00 \times 10^{-12}, \ \beta=-2.56, \ \gamma=4.88$&$\alpha=7.79 \times 10^{-15}, \ \beta=-2.56, \ \gamma=4.88$ (a) \\
$\rm{CH_3+HNCO \rightarrow CH_3NCO +H}$&$\alpha= 1.0 \times 10^{-12}, \ \beta=0, \ \gamma=0$&$\alpha= 5 \times 10^{-11}, \ \beta=0, \ \gamma=0$ (b) \\
$\rm{H + NH_2CHO \rightarrow H_2NCO+ H}$&$\alpha= 1 \times 10^{-10}, \ \beta=0, \ \gamma=240$&$\alpha=-, \ \beta=-, \ \gamma=240-3130$ (c) \\
$\rm{H + H_2NCO \rightarrow HNCO+ H_2}$&$\alpha= 1 \times 10^{-10}, \ \beta=0, \ \gamma=0$&$\alpha=-, \ \beta=-, \ \gamma=0$ (c) \\
$\rm{H + H_2NCO \rightarrow NH_2CHO}$&$\alpha= 1 \times 10^{-10}, \ \beta=0, \ \gamma=0$&$\alpha=-, \ \beta=-, \ \gamma=0$ (c)\\
$\rm{H + HNCO \rightarrow H_2NCO}$&$\alpha= 1 \times 10^{-10}, \ \beta=0, \ \gamma=5050$&$\alpha=-, \ \beta=-, \ \gamma=2530-5050$ (c) \\
\hline
\end{tabular} \\
\vskip 0.2 cm
{\bf References.} \\
(a) \cite{skou17}. \\
(b) \cite{quen18}. \\
(c) \cite{haup19}.
\end{table}

Among the other isomers of HNCO, HOCN is found to be significantly abundant.
During the warm-up and post-warm-up stages, it attains a peak value of
$4.8 \times 10^{-9}$ for the best-fitted parameters of Model A.
It is $\sim 10$ times lower than the lowest energy isomer,
HNCO (peak abundance $4.13 \times 10^{-8}$). Similarly, among all the other isomers of
$\rm{CH_3NCO}$, the abundance of $\rm{CH_3OCN}$ is higher.
It is because of the gas-phase formation of CH$_3$OCN
by the reaction between CH$_3$ and HOCN.
With the best-fitted parameters, we obtain a peak abundance of $\rm{CH_3OCN}$ as
$6.1 \times 10^{-10}$, which is $\sim 8$ times lower than that of $\rm{CH_3NCO}$
(peak abundance $5.0 \times 10^{-9}$). We \citep{gora20b} report the
identification of HNCO and CH$_3$NCO in G10. However, HOCN and CH$_3$OCN should also be
considered as the potential candidates for future astronomical
detection in G10, looking at their abundances.
\cite{cern16} predicted an upper limit of the column density of $6 \times 10^{13}$ cm$^{-2}$
for another isomer, CH$_3$CNO, in Orion.
Here, we find its peak abundance to be $7.4 \times 10^{-13}$.
Converting this peak abundance in terms of the column density,
we have $\sim 10^{13}$ cm$^{-2}$ (using a hydrogen column density of $1.35 \times 10^{25}$ cm$^{-2}$), which is in line with the observed upper limit.

\begin{figure}
\hskip -1.5 cm
\includegraphics[width=0.33\textwidth,angle=-90]{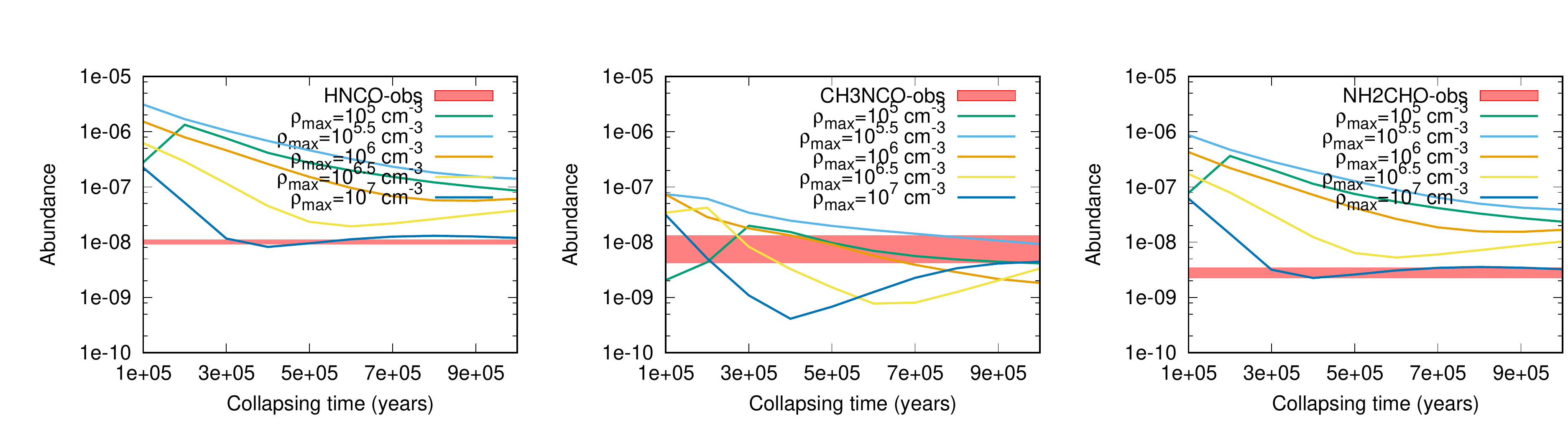}
\caption{Abundances of HNCO, CH$_3$NCO, and NH$_2$CHO by considering different $T_{coll}$ and $\rho_{max}$ with Model B \citep{gora20b}. \label{fig:ModelB-den-time}}
\end{figure}

\begin{figure}
\hskip -1.5 cm
\includegraphics[width=0.33\textwidth,angle=-90]{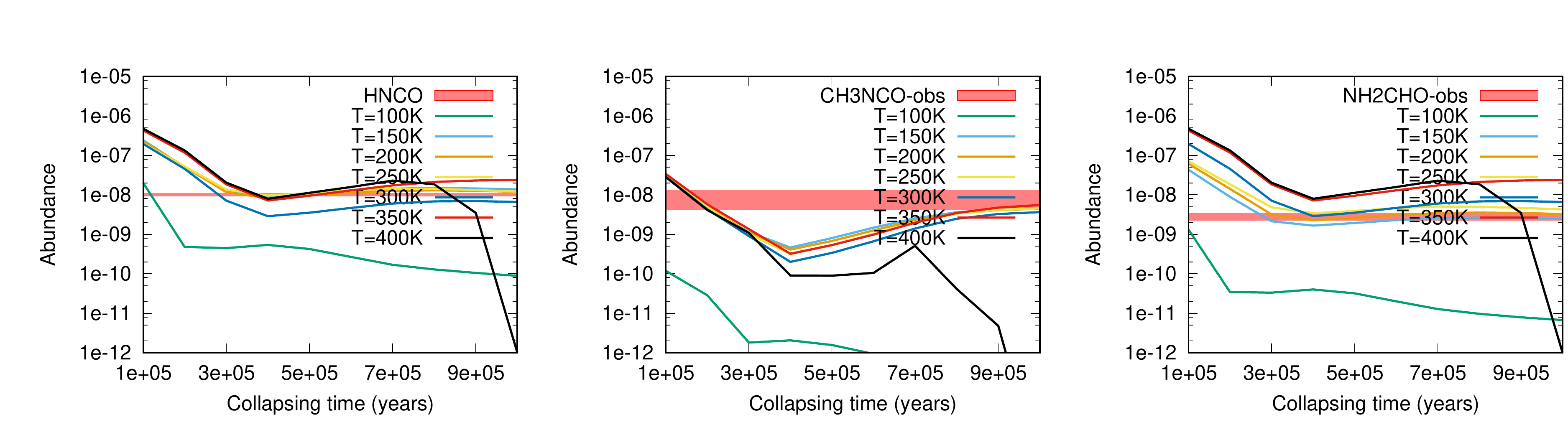}
\caption{Abundances of HNCO, CH$_3$NCO, and NH$_2$CHO by considering $\rho_{max}=10^7$ cm$^{-3}$, and different $T_{max}$ and $T_{coll}$ with Model B \citep{gora20b}.\label{fig:ModelB-temp}}
\end{figure}

\subsubsection{Results obtained with Model B}
\begin{figure}[b!]
\centering
\includegraphics[width=0.6\textwidth]{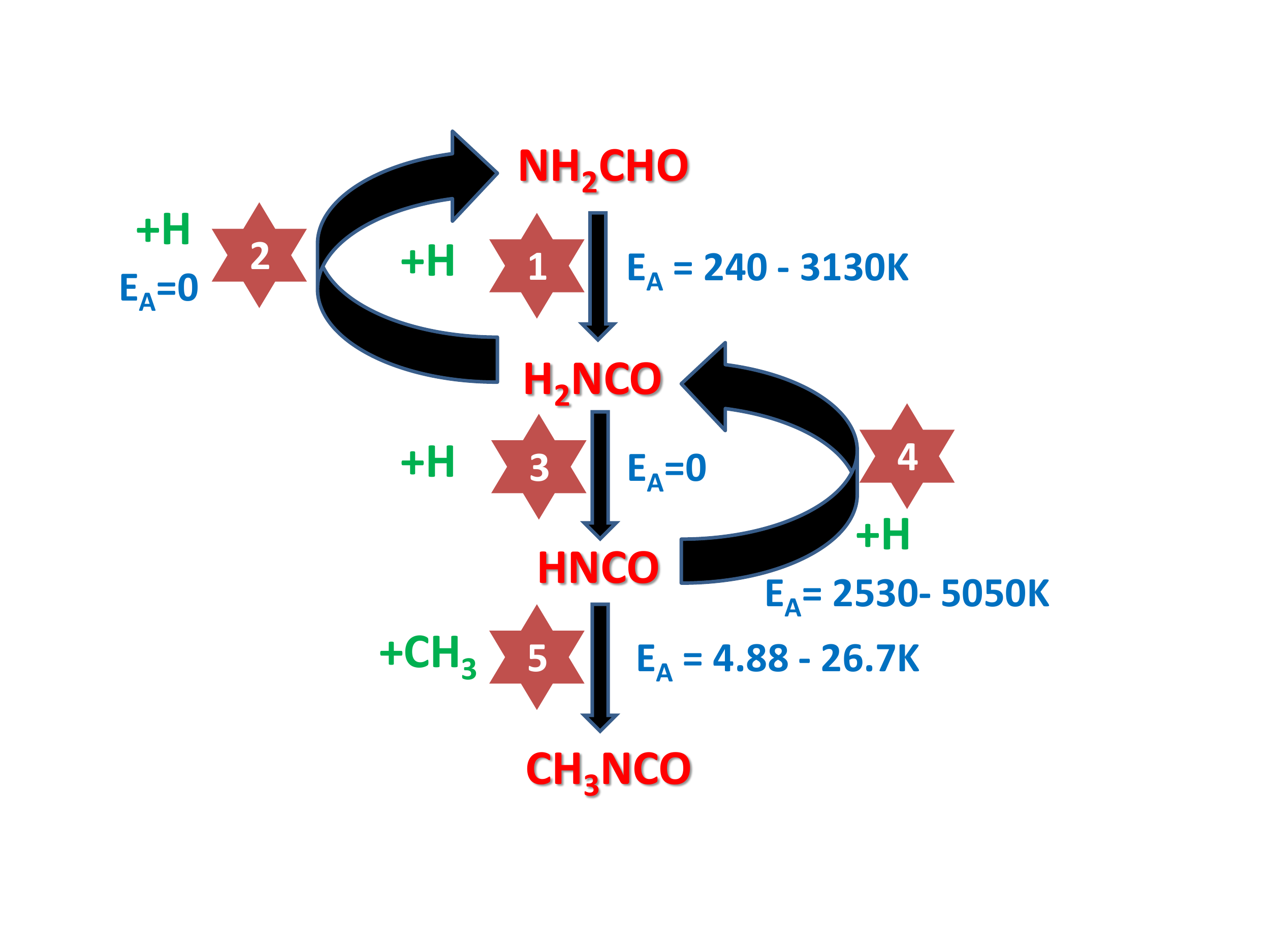}
\caption{Chemical linkage between the three N-bearing molecules \citep{gora20b}.\label{fig:path}}
\end{figure}

For Model B, we do not vary any rate constants. Instead, we keep it as it is obtained with
the best-fitted Model A, noted in Table \ref{tab:Model}. To find out the best-fit physical parameters for Model B, we start with $T_{ice}=20$ K and vary $\rho_{max}$.
Figure \ref{fig:ModelB-den-time} shows the variation of HNCO, CH$_3$NCO, and NH$_2$CHO
abundance by considering a post-warm-up time ($t_{pw}$) of $10^5$ years.
Observed abundances are also marked in each panel.
We find that the abundance of these three species is highly sensitive to the chosen
collapsing time scale ($t_{coll}$) and the
maximum density ($\rho_{max}$) achieved during the collapsing stage.
As we increase $\rho_{max}$, abundance significantly decreases.
Similarly, as we increase $t_{coll}$, abundance gradually reduces.
Based on Figure \ref{fig:ModelB-den-time}, we find that $\rho_{max}=10^7$ cm$^{-3}$ and
$T_{coll} \sim 2-3 \times 10^5$ years are most suitable for explaining the abundance
of these three species simultaneously. We further vary $T_{max}$ between $100$ K and $400$ K
by considering $\rho_{max} =10^7$ cm$^{-3}$. Figure \ref{fig:ModelB-temp} shows
a strong increasing trend in the abundance profile with an increase in $T_{max}$
from $100$ K to $150$ K. A suitable match is found when we use $T_{max}=200$ K.
In between $T_{max}=150$ K and $350$ K, the abundance profile
shows moderate changes. Beyond $350$ K, the abundances drastically decrease,
while we consider a comparatively longer collapsing time scale.
Thus, by considering all types of variation with Model B, we obtain
a good fit among the three targeted N-bearing species using the parameters listed in Table \ref{tab:Model}.
We find that our Model B with $T_{coll}=(2-3) \times 10^5$ years, $T_w=5 \times 10^4$ years,
and $T_{pw}=10^5$ years can explain the observation of these three species simultaneously
considering $\rho_{max}=10^7$ cm$^{-3}$, $T_{max}=200$ K, and $T_{ice}=20$ K.
The obtained lower timescale with Model B is very interesting because
G10 is a high-mass star-forming region.
The required gas-phase pathways to establish the linkage among these three species
are summarized in Figure \ref{fig:path}.

\subsubsection{Chemical linkage between $\rm{HNCO}$, $\rm{NH_2CHO}$, and $\rm{CH_3NCO}$}
Earlier, it was proposed that HNCO and NH$_2$CHO are chemically linked.
The successive hydrogenation reactions of HNCO were proposed for the formation of $\rm{NH_2CHO}$.
However, the validity of the second hydrogenation reaction
was ruled out by the experimental study of \cite{nobl15}.
Recently, a theoretical work by \cite{haup19} proposed dual-cyclic H addition
and abstraction reactions to support the chemical linkage between HNCO and NH$_2$CHO.
The chemical evolution of HNCO, $\rm{NH_2CHO}$, and CH$_3$NCO with Model A is shown
in Figure \ref{fig:best}. Gradual enhancement in the abundance of ice-phase HNCO
and its isomers arises because radicals become
mobile enough to increase temperature. Beyond $80$ K $-$ $90$ K, the diffusion timescale of
the radicals becomes comparable to their desorption timescale, and thus they desorb
back to the gas phase very quickly. Also,
HNCO starts to sublimate beyond $90$ K, resulting in a sharp decrease in the ice phase.
The gas-phase production of CH$_3$NCO mainly occurs by the reaction between
CH$_3$ and HNCO. The formation rate of CH$_3$NCO enhances during the later stages of the simulation.
In the case of $\rm{NH_2CHO}$, ice-phase production is sufficient in the collapsing stage,
but gas-phase production is not adequate. A smooth transfer of $\rm{NH_2CHO}$ from the
ice phase to the gas phase can occur in the warm-up period. The location of this transfer
depends on the adopted BE of $\rm{NH_2CHO}$. A major portion of $\rm{NH_2CHO}$ is formed
by the gas-phase reaction between $\rm{NH_2}$ and $\rm{H_2CO}$ in the warm-up and
post-warm-up stage. Due to the increased temperature, the activation barrier for the hydrogen abstraction reaction of $\rm{NH_2CHO}$ (by reaction \ref{eqn:peptide_16}) becomes probable and thus produces HNCO by the barrierless reaction \ref{eqn:peptide_17}.
To see the effects of adding the \cite{haup19} pathways, we check
with $\alpha=0$ for the gas-phase reactions \ref{eqn:peptide_16}$-$\ref{eqn:peptide_19}.
Figure \ref{fig:best} shows the abundances of these three species by considering $\alpha=0$ (marked as ``NO-Haupa'').
We notice that the abundance of gas-phase HNCO is significantly affected by the inclusion
of the gas-phase pathways of \cite{haup19}.
Consideration of reactions \ref{eqn:peptide_16}$-$\ref{eqn:peptide_19} shows more $\rm{HNCO}$ at the end,
and an absence of these pathways (i.e., with $\alpha=0$) reflects comparatively
lower $\rm{HNCO}$.
In brief, we find that the pathways proposed by \cite{haup19} are relevant
for the gas-phase production of HNCO around the post-warm-up period.
In the first stage, $\rm{CH_3NCO}$ is mainly formed on the grain surface by the reaction
between CH$_3$ and OCN. $\rm{CH_3NCO}$ may also form by the reaction between $\rm{CH_3}$
and HNCO \citep{quen18} in the ice phase. However, it is clear from the
warm-up and post-warm-up stage that the major contribution of the gas-phase $\rm{CH_3NCO}$
is not coming from the ice phase. Instead, it is produced inside the
gas phase itself. The gas-phase formation is efficient using the HNCO channel at the warm-up and post-warm-up stages.

\subsection{Summary}

From our chemical modeling results, we notice that the three N-bearing species
are chemically linked. HNCO and NH$_2$CHO are
chemically linked by dual-cyclic H addition and abstraction reactions proposed
by \cite{haup19} during the warm-up and post-warm-up stage.
HNCO and CH$_3$NCO are also chemically related because HNCO reacts with CH$_3$ to
form $\rm{CH_3NCO}$ (Figure \ref{fig:path}).
Our modeling results suggest that the abundances of HOCN and $\rm{CH_3OCN}$ are
significantly higher and could be observed in G10.

\clearpage
\section{Phosphorous-bearing species: precursor of biomolecules}
Phosphorus (P) and its compounds play an essential role in the chemical evolution in galaxies.
Phosphorus is a crucial element of life and is one of the main biogenic elements.
However, its origin in the terrestrial system is yet to be fully understood.
The ALMA and ESA probe Rosetta suggest that P-related species might have traveled
from star-forming regions to the early Earth \citep[e.g.,][]{altw16,rivi20} through comets.
The interstellar chemistry of P-bearing molecules has significant astrophysical relevance,
and very little has been explored so far. P-bearing compounds are the major ingredients of any
living system, where they carry out numerous biochemical functions.
P-bearing molecules play a significant role in forming large biomolecules or living organisms.
They store and transmit the genetic information in nucleic acids and work in nucleotides as precursors synthesize RNA and DNA \citep{maci97}.
Moreover, these molecules are the essential components of
phospholipids \citep[main characteristic features of cellular membranes;][]{maci05}.

Phosphorus is relatively rare in the ISM. However, it is ubiquitous in various
meteorites \citep{jaro90,lodd03,pase19}. On an average, P is the 13th most abundant
element in a typical meteoritic material and the 11th most abundant element
in the Earth's crust \citep{maci05}. 

P$^+$ was identified with a cosmic abundance
of $\sim 2\times10^{-7}$ in hot regions \citep[$\sim 1200$ K;][]{jura78}. 
PN, PO, HCP, CP, CCP, and PH$_3$ are observed in CSEs around
evolved stars \citep{guel90,agun07,agun08,agun14,tene07,half08,mila08,tene08,debe13,ziur18}.
PN was detected in several star-forming regions \citep{turn87,ziur87,turn90,caux11,yama11,font16,font19,mini18}.
PN remained the only P-bearing species observed in the dense ISM for many years \citep{ziur87,turn90,font16}. \cite{rivi16} reported the first identification of PO toward
W51 1e/2e and W3(OH), along with PN using the IRAM 30 m telescope.
Thereafter, PO and PN were simultaneously observed in several low- and high-mass
star-forming regions and the Galactic center \citep{lefl16,rivi18,rivi20,berg19,bern21}.

Several studies \citep{mill91,char94,aota12,lefl16,rivi16,jime18} reported their theoretical investigation on P-chemistry modeling. However, the study of the
dense cloud region is not well constrained for P-chemistry.
The significant uncertainty lies in the depletion factor
of the initial abundance of P.
The complexity of gas-phase abundance within the gas and the exact depletion of different
elements onto the grains are still uncertain \citep{jenk09}.
The chemical desorption of phosphine (PH$_3$) was recently studied by \cite{nguy20}.
This study is indeed important in constraining the modeling parameters.

PH$_3$ is a relatively stable
molecule that could hold a significant fraction of P in various astronomical environments.
Scientists consider PH$_3$ to be a biosignature \citep{sous20}.
PH$_3$ was detected in the planets of our solar system containing a reducing atmosphere.
PH$_3$ could be produced deep inside the reducing atmospheres of
giant planets \citep{breg75,tarr92} at high temperatures and pressures
and dredged upward by the convection \citep{noll97,viss06}.
Very surprisingly, there is no such reducing atmosphere in Venus,
but $\sim 20$ ppb of PH$_3$ was inferred \citep{grea20}.
Such a high amount of PH$_3$ could not be explained with the steady-state chemical models,
including the photochemical pathways. \cite{grea20} also investigated the other abiotic routes
for explaining this high abundance. But no suitable justification was found.
They speculated some unknown photo or geo-chemical root of it.
The possible reason behind its high abundance by some of the biological means could not be ruled out.
This discovery initiated a series of debates.
The presence of PH$_3$ in Venus's clouds could also not be explained by any abiotic mechanism \citep{bain20}. \cite{vill20} recently questioned the analysis and interpretation of the spectroscopic data used in \cite{grea20}.
\cite{snell20} also did not find any statistical evidence for PH$_3$ in the atmosphere of Venus. 

PH$_3$ was tentatively observed (J = $1 \rightarrow 0$, $266.9$ GHz) in the envelope
of the carbon-rich stars IRC +10216 and CRL 2688 \citep{agun08,tene08}.
Later, \cite{agun14} confirmed the presence of PH$_3$.
They detected the J = $2 \rightarrow 1$ rotational transition of PH$_3$
(at $534$ GHz) with a very high abundance toward IRC +10216 using the HIFI instrument onboard Herschel.

\cite{chan20} recently attempted to observe HCP (2–1), CP (2–1), PN (2–1), and PO (2–1) along the line of sight to the compact extragalactic quasar B0355+508 with the IRAM 30 m telescope.
Unfortunately, they could not detect these transitions along this line of sight predicting $3\sigma$ upper limits. However, they successfully observed HNC (1–0), CN (1–0), and C$^{34}$S (2–1) in absorption and $^{13}$CO (1–0) in emission along the same line of sight.

Here, we implement various state-of-the-art chemical models to understand the PH$_3$ formation under different interstellar conditions such as diffuse cloud, interstellar PDRs, and hot-core regions.

\subsection{The chemical network of phosphorus} \label{sec:chem_network}
N and P have five electrons in the valance shell. Due to this reason, P-chemistry is sometimes
considered analogous to N-chemistry. For example, successive hydrogenation of N yields NH$_3$,
whereas for P it yields PH$_3$. However, the presence of NH$_3$ is ubiquitous in the ISM
\citep{cheu68,wils79,keen83,maue88}, but the presence of PH$_3$ is not so universal \citep{thor83}.
On the other hand, P-related chemistry can significantly differ from the N-related chemistry under
the physical conditions prevailing in the different star-forming regions.
In this work, we prepare an extensive network of P following the chemical pathways
explained in \cite{thor84,adam90,mill91,font16,jime18,sous20,rivi16,chan20}.
\cite{char94} and \cite{anic93} differentiated the formation/destruction mechanism
of PH$_n$ (n=1, 2, 3) and their cationic species. \cite{thor84} presented the reaction
pathways for P, PO, P$^+$, PO$^+$, PH$^+$, HPO$^+$, and H$_2$PO$^+$. Furthermore,
in a recent study, \cite{chan20} extended the gas-phase chemical network of PN and PO in accordance with \cite{mill87,jime18}.
We use the kinetic data of chemical reactions from the KIDA \citep{wake15} and UMIST \citep{mcel13} database.
In Table \ref{tab:reaction_path}, we show all the reactions considered in our P-network.

\begin{table}
\tiny
\caption{Reaction pathways for P-chemistry \citep{sil21}.
\label{tab:reaction_path}}
\vskip 0.2cm
\begin{tabular}{ l c c c c c}
\hline
{\bf Reaction}   & {\bf Reactions} & \multicolumn{3}{c}{\bf Rate coefficient} & {\bf References} \\
{\bf Number (Type)} &  & {$\alpha$}    & {$\beta$} & {$\gamma$} & \\
\hline
\multicolumn{6}{c}{Gas-phase pathways} \\
\hline
 R1 (IN) & $\rm{C^+ + P \rightarrow P^+ + C}$ & $1.0\times10^{-09}$ & 0.0 & 0.0 & \cite{mcel13} \\
 R2 (IN) & $\rm{H^+ + P \rightarrow P^+ + H}$ & $1.0\times10^{-09}$ & 0.0 & 0.0 & \cite{mcel13} \\ 
 R3 (IN) & $\rm{He^+ + P \rightarrow P^+ + He}$ & $1.0\times10^{-09}$ & 0.0 & 0.0 & \cite{mcel13} \\
 R4 (IN) & $\rm{NH_3 + P^+ \rightarrow P + NH_3^+}$ & $3.08\times10^{-10}$ & $-0.5$ & 0.0 & \cite{mcel13} \\
 R5 (IN) & $\rm{Si + P^+ \rightarrow P + Si^+}$ & $1.0\times10^{-09}$ & 0.0 & 0.0 & \cite{mcel13} \\
 R6 (CR) & $\rm{P + CRPHOT \rightarrow P^+ + e^-}$ & $1.30\times10^{-17}$ & 0.0 & 750 & \cite{mcel13} \\
 R7 (PH) & $\rm{P + h\nu \rightarrow P^+ + e^-}$ & $1.0\times10^{-09}$ & 0.0 & 2.7 & \cite{mcel13} \\
 R8 (ER) & $\rm{P^+ + e^- \rightarrow P + h\nu}$ & $3.41\times10^{-12}$ & -0.65 & 0.0 & \cite{mcel13} \\
 R9 (NR) & $\rm{P + CN \rightarrow PN + C}$ & $3.0\times10^{-10}$ & - & - & \cite{jime18} \\
 R10 (NR) & $\rm{N + PH \rightarrow PN + H}$ & $5.0\times10^{-11}$ & 0.0 & 0.0 & \cite{smit04} \\
 R11 (NR) & $\rm{N + PO \rightarrow PN + O}$ & $3.0\times10^{-11}$ & $-0.6$ & 0.0 & \cite{smit04} \\
 R12 (NR) & $\rm{N + PO \rightarrow P + NO}$ & $2.55\times10^{-12}$ & 0.0 & 0.0 & \cite{mill87} \\
 R13 (NR) & $\rm{N + CP \rightarrow PN + C}$ & $3.0\times10^{-10}$ & - & - & \cite{jime18} \\
 R14 (NR) & $\rm{C + PN^+ \rightarrow PN + C^+}$ & $1.0\times10^{-09}$ & 0.0 & 0.0 & \cite{mcel13} \\
 R15 (DR) & $\rm{PN^+ + e^- \rightarrow P + N}$ & $1.8\times10^{-7}$ & $-0.5$ & 0.0 & \cite{mcel13} \\
 R16 (DR) & $\rm{HPN^+ + e^- \rightarrow PN + H}$ & $1.0\times10^{-7}$ & $-0.5$ & 0.0 & \cite{mill91} \\
 R17 (DR) & $\rm{HPN^+ + e^- \rightarrow P + NH}$ & $1.0\times10^{-7}$ & $-0.5$ & 0.0 & \cite{mcel13} \\
 R18 (DR) & $\rm{PNH_2^+ + e^- \rightarrow PN + H_2}$ & $1.5\times10^{-7}$ & $-0.5$ & 0.0 & \cite{mcel13} \\
 R19 (DR) & $\rm{PNH_3^+ + e^- \rightarrow PN + H_2}$ & $1.5\times10^{-7}$ & $-0.5$ & 0.0 & \cite{mcel13} \\
 R20 (CR) & $\rm{PN + CRPHOT \rightarrow P + N}$ & $1.30\times10^{-17}$ & 0.0 & 250 & \cite{mcel13} \\
 R21 (IN) & $\rm{H^+ + PN \rightarrow PN^+ + H}$ & $1.0\times10^{-9}$ & $-0.5$ & 0.0 & \cite{mill91} \\
 R22 (IN) & $\rm{He^+ + PN \rightarrow P^+ + N + He}$ & $1.0\times10^{-9}$ & $-0.5$ & 0.0 & \cite{mill91} \\
 R23 (IN) & $\rm{H_3^+ + PN \rightarrow HPN^+ + H_2}$ & $1.0\times10^{-9}$ & $-0.5$ & 0.0 & \cite{mill91} \\
 R24 (IN) & $\rm{H_3O^+ + PN \rightarrow HPN^+ + H_2O}$ & $1.0\times10^{-9}$ & $-0.5$ & 0.0 & \cite{mill91} \\
 R25 (IN) & $\rm{HCO^+ + PN \rightarrow HPN^+ + CO}$ & $1.0\times10^{-9}$ & $-0.5$ & 0.0 & \cite{mill91} \\
 R26 (NR) & $\rm{N + PN \rightarrow P + N_2}$ & $1.0\times10^{-18}$ & 0.0 & 0.0 & \cite{mill87} \\
 R27 (PH) & $\rm{PN + h\nu \rightarrow P + N}$ & $5.0\times10^{-12}$ & 0.0 & 3.0 & \cite{mcel13} \\
 R28 (CR) & $\rm{HPO + CRPHOT \rightarrow PO + H}$ & $1.30\times10^{-17}$ & 0.0 & 750 & \cite{mcel13} \\
 R29  (DR) & $\rm{HPO^+ + e^- \rightarrow PO + H}$ & $1.50\times10^{-7}$ & $-0.5$ & 0.0 & \cite{thor84} \\
 R30 (DR) & $\rm{HPO^+ + e^- \rightarrow O + PH}$ & $1.50\times10^{-7}$ & $-0.5$ & 0.0 & \cite{mill91} \\
 R31 (DR) & $\rm{HPO^+ + e^- \rightarrow P + O + H}$ & $1.00\times10^{-7}$ & $-0.5$ & 0.0 & \cite{mcel13} \\
 R32 (DR) & $\rm{PO^+ + e^- \rightarrow P + O}$ & $1.80\times10^{-7}$ & $-0.5$ & 0.0 & \cite{mcel13} \\
 R33 (IN) & $\rm{H_2O + HPO^+ \rightarrow PO + H_3O^+}$ & $3.40\times10^{-10}$ & $-0.5$ & 0.0 & \cite{mcel13} \\
 R34 (IN) & $\rm{H_2O + P^+ \rightarrow PO^+ + H_2}$ & $5.50\times10^{-11}$ & $-0.5$ & 0.0 & \cite{mcel13} \\
 R35 (NR) & $\rm{O + HPO \rightarrow PO + OH}$ & $3.80\times10^{-11}$ & 0.0 & 0.0 & \cite{smit04} \\
 R36 (NR) & $\rm{O + PH_2 \rightarrow PO + H_2}$ & $4.0\times10^{-11}$ & 0.0 & 0.0 & \cite{mill87} \\
 R37 (NR) & $\rm{O + PH_2 \rightarrow H + HPO}$ & $8.0\times10^{-11}$ & 0.0 & 0.0 & \cite{smit04} \\
 R38 (NR) & $\rm{O + PH \rightarrow PO + H}$ & $1.0\times10^{-10}$ & 0.0 & 0.0 & \cite{smit04} \\
 R39 (NR) & $\rm{P + OH \rightarrow PO + H}$ & $6.1\times10^{-11}$ & $-0.23$ & 14.9 & \cite{jime18} \\
 R40 (NN) & $\rm{O_2 + P \rightarrow O + PO}$ & $1.0\times10^{-13}$ & 0.0 & 0.0 & \cite{mcel13} \\
 R41 (NN) & $\rm{O_2 + PH_2^+ \rightarrow PO^+ + H_2O}$ & $7.8\times10^{-11}$ & 0.0 & 0.0 & \cite{mcel13} \\
 R42 (IN) & $\rm{P^+ + CO_2 \rightarrow PO^+ + CO}$ & $4.60\times10^{-10}$ & 0.0 & 0.0 & \cite{mcel13} \\
 R43 (IN) & $\rm{P^+ + O_2 \rightarrow PO^+ + O}$ & $5.60\times10^{-10}$ & 0.0 & 0.0 & \cite{mcel13} \\
 R44 (IN) & $\rm{P^+ + OCS \rightarrow PO^+ + CS}$ & $4.18\times10^{-10}$ & 0.0 & 0.0 & \cite{mcel13} \\
 R45 (PH) & $\rm{HPO + h\nu \rightarrow PO + H}$ & $1.70\times10^{-10}$ & 0.0 & $5.3$ & \cite{mcel13} \\
 R46 (CR) & $\rm{PO + CRPHOT \rightarrow P + O}$ & $1.30\times10^{-17}$ & 0.0 & 250 & \cite{mcel13} \\
 R47 (IN) & $\rm{C^+ + PO \rightarrow PO^+ + C}$ & $1.0\times10^{-9}$ & $-0.5$ & 0.0 & \cite{thor84} \\
 R48 (IN) & $\rm{H^+ + PO \rightarrow PO^+ + H}$ & $1.0\times10^{-9}$ & $-0.5$ & 0.0 & \cite{thor84} \\ 
 R49 (IN) & $\rm{H_3^+ + PO \rightarrow HPO^+ + H_2}$ & $1.0\times10^{-9}$ & $-0.5$ & 0.0 & \cite{thor84} \\
 R50 (IN) & $\rm{HCO^+ + PO \rightarrow HPO^+ + CO}$ & $1.0\times10^{-9}$ & $-0.5$ & 0.0 & \cite{thor84} \\ 
 R51 (IN) & $\rm{He^+ + PO \rightarrow P^+ + O + He}$ & $1.0\times10^{-9}$ & $-0.5$ & 0.0 & \cite{thor84} \\
 R52 (PH) & $\rm{PO + h\nu \rightarrow P + O}$ & $3.0\times10^{-10}$ & 0.0 & 2.0 & \cite{mcel13} \\
 R53 (DR) & $\rm{PCH_2^+ + e^- \rightarrow HCP + H}$ & $1.50\times10^{-7}$ & $-0.5$ & 0.0 & \cite{mill91} \\
 R54 (CR) & $\rm{HCP + CRPHOT \rightarrow CP + H}$ & $1.30\times10^{-17}$ & 0.0 & 750 & \cite{mcel13} \\
 R55 (IN) & $\rm{C^+ + HCP \rightarrow CCP^+ + H}$ & $5.0\times10^{-10}$ & 0.0 & 0.0 & \cite{mill91} \\
 R56 (IN) & $\rm{C^+ + HCP \rightarrow HCP^+ + C}$ & $5.0\times10^{-10}$ & 0.0 & 0.0 & \cite{mill91} \\
 R57 (IN) & $\rm{C + HCP^+ \rightarrow CCP^+ + H}$ & $2.0\times10^{-10}$ & 0.0 & 0.0 & \cite{mcel13} \\
 R58 (IN) & $\rm{H^+ + HCP \rightarrow HCP^+ + H}$ & $1.0\times10^{-9}$ & 0.0 & 0.0 & \cite{mill91} \\
 R59 (IN) & $\rm{H_3^+ + HCP \rightarrow PCH_2^+ + H_2}$ & $1.0\times10^{-9}$ & 0.0 & 0.0 & \cite{mill91} \\
 R60 (IN) & $\rm{H_3O^+ + HCP \rightarrow PCH_2^+ + H_2O}$ & $1.0\times10^{-9}$ & 0.0 & 0.0 & \cite{mill91} \\
 R61 (IN) & $\rm{HCO^+ + HCP \rightarrow PCH_2^+ + CO}$ & $1.0\times10^{-9}$ & 0.0 & 0.0 & \cite{adam90} \\
 R62 (IN) & $\rm{He^+ + HCP \rightarrow P^+ + CH + He}$ & $5.0\times10^{-10}$ & 0.0 & 0.0 & \cite{mill91} \\
 R63 (IN) & $\rm{He^+ + HCP \rightarrow PH + C^+ + He}$ & $5.0\times10^{-10}$ & 0.0 & 0.0 & \cite{mill91} \\
 R64 (NR) & $\rm{O + HCP \rightarrow PH + CO}$ & $3.61\times10^{-13}$ & 2.1 & 3080.0 & \cite{smit04} \\
 R65 (PH) & $\rm{HCP + h\nu \rightarrow CP + H}$ & $5.48\times10^{-10}$ & 0.0 & 2.0 & \cite{mcel13} \\
 R66 (NR) & $\rm{CCP + O \rightarrow CO + CP}$ & $6.0\times10^{-12}$ & 0.0 & 0.0 & \cite{smit04} \\
 R67 (IN) & $\rm{CCP + He^+ \rightarrow He + CP + C^+}$ & $5.0\times10^{-10}$ & $-0.5$ & 0.0 & \cite{mill91} \\
 R68 (IN) & $\rm{CCP + He^+ \rightarrow He + C_2 + P^+}$ & $5.0\times10^{-10}$ & $-0.5$ & 0.0 & \cite{mill91} \\
 R69 (DR) & $\rm{PCH_4^+ + e^- \rightarrow CP + C_3H}$ & $7.5\times10^{-8}$ & $-0.5$ & 0.0 & \cite{mcel13} \\
 R70 (DR) & $\rm{CCP^+ + e^- \rightarrow C + CP}$ & $1.5\times10^{-7}$ & $-0.5$ & 0.0 & \cite{mill91} \\
 \hline
 \end{tabular}
 \end{table}
 
 \begin{table}
\tiny
\begin{tabular}{ l c c c c c}
\hline
\hline
{\bf Reaction}   & {\bf Reactions} & \multicolumn{3}{c}{\bf Rate coefficient} & {\bf References} \\
{\bf Number (Type)} &  & {$\alpha$}    & {$\beta$} & {$\gamma$} & \\
\hline
 R71 (DR) & $\rm{CCP^+ + e^- \rightarrow P + C_2}$ & $1.5\times10^{-7}$ & $-0.5$ & 0.0 & \cite{mill91} \\
 R72 (DR) & $\rm{PCH_2^+ + e^- \rightarrow CP + H_2}$ & $1.50\times10^{-7}$ & $-0.5$ & 0.0 & \cite{mill91} \\
 R73 (DR) & $\rm{HCP^+ + e^- \rightarrow CP + H}$ & $1.50\times10^{-7}$ & $-0.5$ & 0.0 & \cite{mill91} \\
 R74 (DR) & $\rm{HCP^+ + e^- \rightarrow P + CH}$ & $1.50\times10^{-7}$ & $-0.5$ & 0.0 & \cite{mill91} \\
 R75 (DR) & $\rm{CP^+ + e^- \rightarrow P + C}$ & $1.00\times10^{-7}$ & $-0.5$ & 0.0 & \cite{mcel13} \\ 
 R76 (NR) & $\rm{C + PH \rightarrow CP + H}$ & $7.50\times10^{-11}$ & 0.0 & 0.0 & \cite{smit04} \\
 R77 (CR) & $\rm{CP + CRPHOT \rightarrow C + P}$ & $1.30\times10^{-17}$ & 0.0 & 250 & \cite{mcel13} \\
 R78 (IN) & $\rm{C^+ + CP \rightarrow CP^+ + C}$ & $1.0\times10^{-9}$ & 0.0 & 0.0 & \cite{mill91} \\
 R79 (IN) & $\rm{H^+ + CP \rightarrow CP^+ + H}$ &$1.0\times10^{-9}$ & 0.0 & 0.0 & \cite{mill91} \\
 R80 (IN) & $\rm{H_3^+ + CP \rightarrow HCP^+ + H_2}$ & $1.0\times10^{-9}$ & 0.0 & 0.0 & \cite{mill91} \\
 R81 (IN) & $\rm{H_2 + CP^+ \rightarrow HCP^+ + H}$ & $1.0\times10^{-9}$ & 0.0 & 0.0 & \cite{mcel13} \\
 R82 (IN) & $\rm{H_3O^+ + CP \rightarrow HCP^+ + H_2O}$ & $1.0\times10^{-9}$ & 0.0 & 0.0 & \cite{mill91} \\
 R83 (IN) & $\rm{HCO^+ + CP \rightarrow HCP^+ + CO}$ & $1.0\times10^{-9}$ & 0.0 & 0.0 & \cite{adam90} \\
 R84 (IN) & $\rm{He^+ + CP \rightarrow P^+ + C + He}$ & $5.0\times10^{-10}$ & 0.0 & 0.0 & \cite{mill91} \\
 R85 (IN) & $\rm{He^+ + CP \rightarrow P + C^+ + He}$ & $5.0\times10^{-10}$ & 0.0 & 0.0 & \cite{mill91} \\
 R86 (NR) & $\rm{O + CP \rightarrow P + CO}$ & $4.0\times10^{-11}$ & 0.0 & 0.0 & \cite{mill91} \\
 R87 (IN) & $\rm{O + CP^+ \rightarrow P^+ + CO}$ & $2.0\times10^{-10}$ & 0.0 & 0.0 & \cite{mcel13} \\
 R88 (PH) & $\rm{CP + h\nu \rightarrow C + P}$ & $1.0\times10^{-9}$ & 0.0 & 2.8 & \cite{mcel13} \\
 R89 (PH) & $\rm{C + P \rightarrow CP + h\nu}$ & $1.41\times10^{-18}$ & 0.03 & 55.0 & \cite{mcel13} \\
 R90 (DR) & $\rm{PC_2H_3^+ + e^- \rightarrow PH + C_2H_2}$ & $1.00\times10^{-7}$ & $-0.50$ & 0.0 & \cite{mcel13} \\ 
 R91 (DR) & $\rm{PNH_2^+ + e^- \rightarrow PH + NH}$ & $1.50\times10^{-7}$ & $-0.50$ & 0.0 & \cite{mcel13} \\ 
 R92 (DR) & $\rm{PNH_3^+ + e^- \rightarrow PH + NH_2}$ & $1.50\times10^{-7}$ & $-0.50$ & 0.0 & \cite{mcel13} \\
 R93 (DR) & $\rm{PNH_2^+ + e^- \rightarrow P + NH_2}$ & $1.50\times10^{-7}$ & $-0.50$ & 0.0 & \cite{mcel13} \\
 R94 (DR) & $\rm{PNH_3^+ + e^- \rightarrow P + NH_3}$ & $3.0\times10^{-7}$ & $-0.50$ & 0.0 & \cite{mcel13} \\
 R95 (DR) & $\rm{PH_2^+ + e^- \rightarrow PH + H}$ & $9.38\times10^{-8}$ & $-0.64$ & 0.0 & \cite{mcel13} \\
 R96 (DR) & $\rm{PH_3^+ + e^- \rightarrow PH + H_2}$ & $1.5\times10^{-7}$ & $-0.5$ & 0.0 & \cite{mill91} \\
 R97 (DR) & $\rm{HPN^+ + e^- \rightarrow PH + N}$ & $1.0\times10^{-7}$ & $-0.5$ & 0.0 & \cite{mill91} \\
 R98 (DR) & $\rm{H_2PO^+ + e^- \rightarrow PH + OH}$ & $1.5\times10^{-7}$ & $-0.5$ & 0.0 & \cite{mcel13} \\
 R99 (DR) & $\rm{H_2PO^+ + e^- \rightarrow HPO + H}$ & $1.5\times10^{-7}$ & $-0.5$ & 0.0 & \cite{mcel13} \\
 R100 (IN) & $\rm{H_2O + PH_2^+ \rightarrow PH + H_3O^+}$ & $1.62\times10^{-10}$ & $-0.5$ & 0.0 & \cite{adam90} \\
 R101 (IN) & $\rm{NH_3 + PH^+ \rightarrow P + NH_4^+}$ & $3.99\times10^{-10}$ & $-0.5$ & 0.0 & \cite{mcel13} \\
 R102 (IN) & $\rm{NH_3 + PH^+ \rightarrow PNH_2^+ + H_2}$ & $5.88\times10^{-10}$ & $-0.5$ & 0.0 & \cite{mcel13} \\
 R103 (IN) & $\rm{NH_3 + PH^+ \rightarrow PNH_3^+ + H}$ & $1.11\times10^{-9}$ & $-0.5$ & 0.0 & \cite{mcel13} \\
 R104 (IN) & $\rm{NH_3 + P^+ \rightarrow PNH_2^+ + H}$ & $2.67\times10^{-10}$ & $-0.5$ & 0.0 & \cite{mcel13} \\
 R105 (IN) & $\rm{NH_3 + PH_2^+ \rightarrow PH + NH_4^+}$ & $3.8\times10^{-10}$ & $-0.5$ & 0.0 & \cite{adam90} \\
 R106 (IN) & $\rm{NH_3 + PH_2^+ \rightarrow PNH_3^+ + H_2}$ & $1.62\times10^{-9}$ & $-0.5$ & 0.0 & \cite{mcel13} \\
 R107 (NR) & $\rm{O + PH_2 \rightarrow PH + OH}$ & $2.0\times10^{-11}$ & 0.0 & 0.0 & \cite{smit04} \\
 R108 (NR) & $\rm{C + HPO \rightarrow PH + CO}$ & $4.0\times10^{-11}$ & 0.0 & 0.0 & \cite{mill91} \\
 R109 (CR) & $\rm{PH + CRPHOT \rightarrow P + H}$ & $1.3\times10^{-17}$ & 0.0 & 250 & \cite{mcel13} \\
 R110 (IN) & $\rm{C^+ + PH \rightarrow PH^+ + C}$ & $1.0\times10^{-9}$ & 0.0 & 0.0 & \cite{mill91} \\
 R111 (IN) & $\rm{H^+ + PH \rightarrow PH^+ + H}$ & $1.0\times10^{-9}$ & 0.0 & 0.0 & \cite{mill91} \\
 R112 (IN) & $\rm{H_3^+ + P \rightarrow PH^+ + H_2}$ & $1.0\times10^{-9}$ & 0.0 & 0.0 & \cite{adam90} \\
 R113 (IN) & $\rm{H_3^+ + PH \rightarrow PH_2^+ + H_2}$ & $2.0\times10^{-9}$ & 0.0 & 0.0 & \cite{mill91} \\
 R114 (IN) & $\rm{HCO^+ + P \rightarrow PH^+ + CO}$ & $1.0\times10^{-9}$ & 0.0 & 0.0 & \cite{adam90} \\
 R115 (IN) & $\rm{O + HCP^+ \rightarrow PH^+ + CO}$ & $2.0\times10^{-10}$ & 0.0 & 0.0 & \cite{mill91} \\
 R116 (IN) & $\rm{O + HPO^+ \rightarrow PH^+ + O_2}$ & $2.0\times10^{-10}$ & 0.0 & 0.0 & \cite{thor84} \\
 R117 (IN) & $\rm{HCO^+ + PH \rightarrow PH_2^+ + CO}$ & $1.0\times10^{-9}$ & 0.0 & 0.0 & \cite{mill91} \\
 R118 (IN) & $\rm{He^+ + PH \rightarrow P^+ + He + H}$ & $1.0\times10^{-9}$ & 0.0 & 0.0 & \cite{mill91} \\
 R119 (IN) & $\rm{He^+ + HPO \rightarrow PH^+ + O + He}$ & $5.0\times10^{-10}$ & $-0.5$ & 0.0 & \cite{mill91} \\
 R120 (IN) & $\rm{He^+ + HPO \rightarrow PO^+ + H + He}$ & $5.0\times10^{-10}$ & $-0.5$ & 0.0 & \cite{mcel13} \\
 R121 (NR) & $\rm{H + PH \rightarrow P + H_2}$ & $1.50\times10^{-10}$ & 0.0 & 416 & \cite{kaye83} \\
 R122 (IN) & $\rm{O + PH^+ \rightarrow PO^+ + H}$ & $1.0\times10^{-9}$ & 0.0 & 0.0 & \cite{thor84} \\
 R123 (IN) & $\rm{HCN + PH^+ \rightarrow HCNH^+ + P}$ & $3.06\times10^{-10}$ & $-0.5$ & 0.0 & \cite{mcel13} \\
 R124 (IN) & $\rm{O + PN^+ \rightarrow PO^+ + N}$ & $2.0\times10^{-10}$ & 0.0 & 0.0 & \cite{mcel13} \\
 R125 (IN) & $\rm{OH + P^+ \rightarrow PO^+ + H}$ & $5.0\times10^{-10}$ & $-0.5$ & 0.0 & \cite{mcel13} \\
 R126 (IN) & $\rm{PH^+ + O_2 \rightarrow PO^+ + OH}$ & $5.4\times10^{-10}$ & 0.0 & 0.0 & \cite{adam90} \\
 R127 (DR) & $\rm{PH^+ + e^- \rightarrow P + H}$ & $1.0\times10^{-7}$ & $-0.5$ & 0.0 & \cite{thor84} \\
 R128 (PH) & $\rm{PH + h\nu \rightarrow P + H}$ & $4.0\times10^{-10}$ & 0.0 & 1.5 & \cite{mcel13} \\
 R129 (DR) & $\rm{PH_3^+ + e^- \rightarrow H + PH_2}$ & $1.5\times10^{-7}$ & $-0.5$ & 0.0 & \cite{mcel13} \\
 R130 (DR) & $\rm{PH_2^+ + e^- \rightarrow P + H_2}$ & $5.36\times10^{-8}$ & $-0.64$ & 0.0 & \cite{mcel13} \\
 R131 (DR) & $\rm{PH_2^+ + e^- \rightarrow P + H + H}$ & $5.23\times10^{-7}$ & $-0.64$ & 0.0 & \cite{mcel13} \\
 R132 (IN) & $\rm{NH_3 + PH_3^+ \rightarrow PH_2 + NH_4^+}$ & $2.3\times10^{-9}$ & $-0.5$ & 0.0 & \cite{adam90} \\
 R133 (DR) & $\rm{PH_4^+ + e^- \rightarrow PH_2 + H_2}$ & $1.50\times10^{-7}$ & $-0.5$ & 0.0 & \cite{char94} \\
 R134 (CR) & $\rm{PH_2 + CRPHOT \rightarrow PH + H}$ & $1.3\times10^{-17}$ & 0.0 & 750.0 & \cite{mcel13} \\
 R135 (IN) & $\rm{H^+ + PH_2 \rightarrow PH_2^+ + H}$ & $1.0\times10^{-9}$ & 0.0 & 0.0 & \cite{mill91} \\
 R136 (NR) & $\rm{H + PH_2 \rightarrow PH + H_2}$ & $6.20\times10^{-11}$ & 0.0 & 318 & \cite{kaye83} \\
 R137 (IN) & $\rm{H_3^+ + PH_2 \rightarrow PH_3^+ + H_2}$ & $2.0\times10^{-9}$ & 0.0 & 0.0 & \cite{mill91} \\
 R138 (IN) & $\rm{HCO^+ + PH_2 \rightarrow PH_3^+ + CO}$ & $1.0\times10^{-9}$ & 0.0 & 0.0 & \cite{mill91} \\
 R139 (IN) & $\rm{He^+ + PH_2 \rightarrow P^+ + He + H_2}$ & $1.0\times10^{-9}$ & 0.0 & 0.0 & \cite{mill91} \\
 R140 (IN) & $\rm{PH_2^+ + O_2 \rightarrow PO^+ + H_2O}$ & $7.8\times10^{-11}$ & 0.0 & 0.0 & \cite{adam90} \\
 R141 (IN) & $\rm{H_2 + P^+ \rightarrow PH_2^+ + h\nu}$ & $7.5\times10^{-18}$ & $-1.3$ & 0.0 & \cite{adam90} \\
 R142 (PH) & $\rm{PH_2 + h\nu \rightarrow PH_2^+ + e^-}$ & $1.73\times10^{-10}$ & 0.0 & 2.6 & \cite{mcel13} \\
 R143 (PH) & $\rm{PH_2 + h\nu \rightarrow PH + H}$ & $2.11\times10^{-10}$ & 0.0 & 1.5 & \cite{mcel13} \\
 R144 (DR) & $\rm{PH_4^+ + e^- \rightarrow PH_3 + H}$ & $1.50\times10^{-7}$ & $-0.5$ & 0.0 & \cite{char94} \\
 R145 (IN) & $\rm{PH_4^+ + NH_3 \rightarrow PH_3 + NH_4^+}$ & $2.10\times10^{-9}$ & 0.0 & 0.0 & \cite{thor83} \\
  \hline
 \end{tabular}
 \end{table}
 
 \begin{table}
\tiny
\hskip -1.0 cm
\begin{tabular}{ l c c c c c}
\hline
\hline
{\bf Reaction}   & {\bf Reactions} & \multicolumn{3}{c}{\bf Rate coefficient} & {\bf References} \\
{\bf Number (Type)} &  & {$\alpha$}    & {$\beta$} & {$\gamma$} & \\
\hline
 R146 (NR) & $\rm{PH_2 + H \rightarrow PH_3}$ & $3.7\times10^{-10}$ & 0.0 & 340 & \cite{kaye83} \\
 R147 (RA) & $\rm{H_2 + PH^+ \rightarrow PH_3^+ + h\nu}$ & $2.40\times10^{-17}$ & $-1.4$ & 0.0 & \cite{adam90} \\
 R148 (NR) & $\rm{H^+ + PH_3 \rightarrow PH_3^+ + H}$ & $7.22\times10^{-11}$ & 0.0 & 886 & \cite{sous20} \\
 R149 (NR) & $\rm{H + PH_3 \rightarrow PH_2 + H_2}$ & $7.22\times10^{-11}$ & 0.0 & 886 & \cite{sous20} \\
 R150 (RR) & $\rm{NH_2 + PH_3 \rightarrow PH_2 + NH_3}$ & $1.50\times10^{-12}$ & 0.0 & 928 & \cite{kaye83} \\
 R151 (IN) & $\rm{H^+ + PH_3 \rightarrow PH_3^+  + H}$ & $2.00\times10^{-9}$ & 0.0 & 0.0 & \cite{char94} \\
 R152 (IN) & $\rm{He^+ + PH_3 \rightarrow PH_2^+ + H + He}$ & $2.00\times10^{-9}$ & 0.0 & 0.0 & \cite{char94} \\
 R153 (IN) & $\rm{C^+ + PH_3 \rightarrow PH_3^+ + C}$ & $2.00\times10^{-9}$ & 0.0 & 0.0 & \cite{char94} \\
 R154 (IN) & $\rm{H_3^+ + PH_3 \rightarrow PH_4^+ + H_2}$ & $2.00\times10^{-9}$ & 0.0 & 0.0 & \cite{char94} \\
 R155 (IN) & $\rm{HCO^+ + PH_3 \rightarrow PH_4^+ + CO}$ & $2.00\times10^{-9}$ & 0.0 & 0.0 & \cite{char94} \\
 R156 (IN) & $\rm{H_3O^+ + PH_3 \rightarrow PH_4^+ + H_2O}$ & $2.00\times10^{-9}$ & 0.0 & 0.0 & \cite{char94} \\
 R157 (IN) & $\rm{PH_3^+ + PH_3 \rightarrow PH_4^+ + PH_2}$ & $1.10\times10^{-9}$ & 0.0 & 0.0 & \cite{smit89} \\
 R158 (PH) & $\rm{PH_3 + h\nu \rightarrow PH_2 + H}$ & $9.23\times10^{-10}$ & 0.0 & 2.1 & \cite{mcel13} [UMIST, following NH$_3$] \\
 R159 (PH) & $\rm{PH_3 + h\nu \rightarrow PH + H_2}$ & $2.76\times10^{-10}$ & 0.0 & 2.1 & \cite{mcel13} [UMIST, following NH$_3$] \\
 R160 (PH) & $\rm{PH_3 + h\nu \rightarrow PH_3^+ + e^-}$ & $2.80\times10^{-10}$ & 0.0 & 3.1 & \cite{mcel13} [UMIST, following NH$_3$] \\
 R161 (NR) & $\rm{PH_3 + OH \rightarrow H_2O + PH_2}$ & $2.71\times10^{-11}$ & 0.0 & 155 & \cite{sous20} \\
 R162 (DR) & $\rm{PC_2H_2^+ + e^- \rightarrow CCP + H_2}$ & $1.0\times10^{-7}$ & $-0.5$ & 0.0 & \cite{mcel13} \\
 R163 (DR) & $\rm{PC_2H_3^+ + e^- \rightarrow CCP + H_2 + H}$ & $1.0\times10^{-7}$ & $-0.5$ & 0.0 & \cite{mcel13} \\
 R164 (DR) & $\rm{PC_2H_4^+ + e^- \rightarrow CCP + H_2 + H_2}$ & $1.0\times10^{-7}$ & $-0.5$ & 0.0 & \cite{mcel13} \\
 R165 (DR) & $\rm{PC_3H^+ + e^- \rightarrow CCP + CH}$ & $1.0\times10^{-7}$ & $-0.5$ & 0.0 & \cite{mcel13} \\
 R166 (DR) & $\rm{PC_4H^+ + e^- \rightarrow CCP + C_2H}$ & $7.5\times10^{-8}$ & $-0.5$ & 0.0 & \cite{mcel13} \\
 R167 (CR) & $\rm{CCP + CRPHOT \rightarrow C_2 + P}$ & $1.3\times10^{-17}$ & 0.0 & 375.0 & \cite{mcel13} \\
 R168 (CR) & $\rm{CCP + CRPHOT \rightarrow CP + C}$ & $1.3\times10^{-17}$ & 0.0 & 375.0 & \cite{mcel13} \\
 R169 (IN) & $\rm{C^+ + CCP \rightarrow CCP^+ + C}$ & $5.0\times10^{-10}$ & $-0.5$ & 0.0 & \cite{mcel13} \\
 R170 (IN) & $\rm{C^+ + CCP \rightarrow CP^+ + C_2}$ & $5.0\times10^{-10}$ & $-0.5$ & 0.0 & \cite{mcel13} \\
 R171 (IN) & $\rm{H^+ + CCP \rightarrow CCP^+ + H}$ & $1.0\times10^{-9}$ & $-0.5$ & 0.0 & \cite{mcel13} \\
 R172 (PH) & $\rm{CCP + h\nu \rightarrow C_2 + P}$ & $1.0\times10^{-10}$ & 0.0 & 1.7 & \cite{mcel13} \\
 R173 (PH) & $\rm{CCP + h\nu \rightarrow CP + C}$ & $1.0\times10^{-9}$ & 0.0 & 1.7 & \cite{mcel13} \\
 R174 (IN) & $\rm{C^+ + HPO \rightarrow HPO^+ + C}$ & $1.0\times10^{-9}$ & $-0.5$ & 0.0 & \cite{mcel13} \\
 R175 (IN) & $\rm{H^+ + HPO \rightarrow HPO^+ + H}$ & $1.0\times10^{-9}$ & $-0.5$ & 0.0 & \cite{mcel13} \\
 R176 (IN) & $\rm{H_2O + P^+ \rightarrow HPO^+ + H}$ & $4.95\times10^{-10}$ & $-0.5$ & 0.0 & \cite{mcel13} \\
 R177 (IN) & $\rm{H_2O + PH^+ \rightarrow HPO^+ + H_2}$ & $7.44\times10^{-10}$ & $-0.5$ & 0.0 & \cite{mcel13} \\
 R178 (IN) & $\rm{H_3O^+ + P \rightarrow HPO^+ + H_2}$ & $1.00\times10^{-9}$ & 0.0 & 0.0 & \cite{mcel13} \\
 R179 (IN) & $\rm{H_2O + PH^+ \rightarrow H_2PO^+ + H}$ & $2.04\times10^{-10}$ & $-0.5$ & 0.0 & \cite{mcel13} \\
 R180 (IN) & $\rm{H_2O + PH^+ \rightarrow P + H_3O^+}$ & $2.52\times10^{-10}$ & $-0.5$ & 0.0 & \cite{mcel13} \\
 R181 (IN) & $\rm{H_2O + PH_2^+ \rightarrow H_2PO^+ + H_2}$ & $3.28\times10^{-10}$ & $-0.5$ & 0.0 & \cite{mcel13} \\
 R182 (IN) & $\rm{H_3^+ + HPO \rightarrow H_2PO^+ + H_2}$ & $1.00\times10^{-9}$ & $-0.5$ & 0.0 & \cite{mcel13} \\
 R183 (IN) & $\rm{H_3O^+ + HPO \rightarrow H_2PO^+ + H_2O}$ & $1.00\times10^{-9}$ & $-0.5$ & 0.0 & \cite{mcel13} \\
 R184 (IN) & $\rm{HCO^+ + HPO \rightarrow H_2PO^+ + CO}$ & $1.00\times10^{-9}$ & $-0.5$ & 0.0 & \cite{mcel13} \\
 R185 (IN) & $\rm{CH_3^+ + P \rightarrow PCH_2^+ + H}$ & $1.0\times10^{-9}$ & 0.0 & 0.0 & \cite{mcel13} \\
 R186 (IN) & $\rm{CH_4 + P^+ \rightarrow PCH_2^+ + H_2}$ & $9.6\times10^{-10}$ & 0.0 & 0.0 & \cite{mcel13} \\
 R187 (IN) & $\rm{H_2 + HCP^+ \rightarrow PCH_2^+ + H}$ & $1.00\times10^{-9}$ & 0.0 & 0.0 & \cite{mcel13} \\
 R188 (DR) & $\rm{PCH_3^+ + e^- \rightarrow CP + H_2 + H}$ & $1.5\times10^{-7}$ & $-0.5$ & 0.0 & \cite{mcel13} \\
 R189 (DR) & $\rm{PCH_3^+ + e^- \rightarrow HCP + H_2}$ & $1.5\times10^{-7}$ & $-0.5$ & 0.0 & \cite{mcel13} \\
 R190 (DR) & $\rm{PCH_3^+ + e^- \rightarrow P + CH_3}$ & $3.0\times10^{-7}$ & $-0.5$ & 0.0 & \cite{mcel13} \\
 R191 (IN) & $\rm{CH_4 + PH^+ \rightarrow PCH_3^+ + H_2}$ & $1.05\times10^{-9}$ & 0.0 & 0.0 & \cite{mcel13} \\
 R192 (IN) & $\rm{H^+ + CH_2PH \rightarrow PCH_3^+ + H}$ & $1.00\times10^{-9}$ & 0.0 & 0.0 & \cite{mcel13} \\
 R193 (DR) & $\rm{PCH_4^+ + e^- \rightarrow CH_2PH + H_2 + H}$ & $1.5\times10^{-7}$ & $-0.5$ & 0.0 & \cite{mcel13} \\
 R194 (DR) & $\rm{PCH_4^+ + e^- \rightarrow HCP + H_2 + H}$ & $1.5\times10^{-7}$ & $-0.5$ & 0.0 & \cite{mcel13} \\
 R195 (DR) & $\rm{PCH_4^+ + e^- \rightarrow P + CH_4}$ & $3.0\times10^{-7}$ & $-0.5$ & 0.0 & \cite{mcel13} \\
 R196 (IN) & $\rm{CH_4 + PH^+ \rightarrow PCH_4^+ + H}$ & $5.5\times10^{-11}$ & 0.0 & 0.0 & \cite{mcel13} \\
 R197 (IN) & $\rm{CH_4 + PH_2^+ \rightarrow PCH_4^+ + H_2}$ & $1.1\times10^{-9}$ & 0.0 & 0.0 & \cite{mcel13} \\
 R198 (IN) & $\rm{H_3^+ + CH_2PH \rightarrow PCH_4^+ + H_2}$ & $1.00\times10^{-9}$ & 0.0 & 0.0 & \cite{mcel13} \\
 R199 (IN) & $\rm{H_3O^+ + CH_2PH \rightarrow PCH_4^+ + H_2O}$ & $1.00\times10^{-9}$ & 0.0 & 0.0 & \cite{mcel13} \\
 R200 (IN) & $\rm{HCO^+ + CH_2PH \rightarrow PCH_4^+ + CO}$ & $1.00\times10^{-9}$ & 0.0 & 0.0 & \cite{mcel13} \\
 R201 (IN) & $\rm{H^+ + HC_2P \rightarrow HC_2P^+ + H}$ & $1.00\times10^{-9}$ & 0.0 & 0.0 & \cite{mcel13} \\
 R202 (CR) & $\rm{HC_2P + CRPHOT \rightarrow CCP + H}$ & $1.3\times10^{-17}$ & 0.0 & 750.0 & \cite{mcel13} \\
 R203 (DR) & $\rm{PC_2H_2^+ + e^- \rightarrow HC_2P + H}$ & $1.0\times10^{-7}$ & $-0.5$ & 0.0 & \cite{mcel13} \\
 R204 (DR) & $\rm{PC_2H_3^+ + e^- \rightarrow HC_2P + H_2}$ & $1.0\times10^{-7}$ & $-0.5$ & 0.0 & \cite{mcel13} \\
 R205 (DR) & $\rm{PC_2H_4^+ + e^- \rightarrow HC_2P + H_2 + H}$ & $1.0\times10^{-7}$ & $-0.5$ & 0.0 & \cite{mcel13} \\
 R206 (DR) & $\rm{PC_3H^+ + e^- \rightarrow HC_2P + C}$ & $1.0\times10^{-7}$ & $-0.5$ & 0.0 & \cite{mcel13} \\
 R207 (IN) & $\rm{C^+ + HC_2P \rightarrow CCP^+ + CH}$ & $5.00\times10^{-10}$ & 0.0 & 0.0 & \cite{mcel13} \\
 R208 (IN) & $\rm{C^+ + HC_2P \rightarrow CP^+ + C_2H}$ & $5.00\times10^{-10}$ & 0.0 & 0.0 & \cite{mcel13} \\
 R209 (IN) & $\rm{H_3^+ + HC_2P \rightarrow PC_2H_2^+ + H_2}$ & $1.00\times10^{-9}$ & 0.0 & 0.0 & \cite{mcel13} \\
 R210 (IN) & $\rm{H_3O^+ + HC_2P \rightarrow PC_2H_2^+ + H_2O}$ & $1.00\times10^{-9}$ & 0.0 & 0.0 & \cite{mcel13} \\
 R211 (IN) & $\rm{HCO^+ + HC_2P \rightarrow PC_2H_2^+ + CO}$ & $1.00\times10^{-9}$ & 0.0 & 0.0 & \cite{mcel13} \\
 R212 (IN) & $\rm{He^+ + HC_2P \rightarrow CP^+ + CH + He}$ & $5.00\times10^{-10}$ & 0.0 & 0.0 & \cite{mcel13} \\
 R213 (IN) & $\rm{He^+ + HC_2P \rightarrow CP + CH^+ + He}$ & $5.00\times10^{-10}$ & 0.0 & 0.0 & \cite{mcel13} \\
 R214 (NN) & $\rm{O + HC_2P \rightarrow HCP + CO}$ & $4.00\times10^{-11}$ & 0.0 & 0.0 & \cite{mcel13} \\
 R215 (PH) & $\rm{HC_2P + h\nu \rightarrow CCP + H}$ & $5.48\times10^{-10}$ & 0.0 & 2.0 & \cite{mcel13} \\
 R216 (IN) & $\rm{H^+ + C_4P \rightarrow C_4P^+ + H}$ & $1.00\times10^{-9}$ & $-0.5$ & 0.0 & \cite{mcel13} \\
 R217 (CR) & $\rm{C_4P + CRPHOT \rightarrow C_3P + C}$ & $1.3\times10^{-17}$ & 0.0 & 750.0 & \cite{mcel13} \\
 R218 (DR) & $\rm{PC_4H^+ + e^- \rightarrow C_4P + H}$ & $7.50\times10^{-8}$ & $-0.5$ & 0.0 & \cite{mcel13} \\
 R219 (IN) & $\rm{H_3^+ + C_4P \rightarrow PC_4H^+ + H_2}$ & $1.00\times10^{-9}$ & $-0.5$ & 0.0 & \cite{mcel13} \\
  \hline
 \end{tabular}
 \end{table}
 
 \begin{table}
\tiny
\hskip -1.0 cm
\begin{tabular}{ l c c c c c}
\hline
\hline
{\bf Reaction}   & {\bf Reactions} & \multicolumn{3}{c}{\bf Rate coefficient} & {\bf References} \\
{\bf Number (Type)} &  & {$\alpha$}    & {$\beta$} & {$\gamma$} & \\
\hline
 R220 (IN) & $\rm{H_3O^+ + C_4P \rightarrow PC_4H^+ + H_2O}$ & $1.00\times10^{-9}$ & $-0.5$ & 0.0 & \cite{mcel13} \\
 R221 (IN) & $\rm{HCO^+ + C_4P \rightarrow PC_4H^+ + CO}$ & $1.00\times10^{-9}$ & $-0.5$ & 0.0 & \cite{mcel13} \\
 R222 (IN) & $\rm{He^+ + C_4P \rightarrow CCP^+ + C_2 + He}$ & $5.00\times10^{-10}$ & $-0.5$ & 0.0 & \cite{mcel13} \\
 R223 (IN) & $\rm{He^+ + C_4P \rightarrow CCP + C_2^+ + He}$ & $5.00\times10^{-10}$ & $-0.5$ & 0.0 & \cite{mcel13} \\
 R224 (NN) & $\rm{O + C_4P \rightarrow C_3P + CO}$ & $1.00\times10^{-11}$ & 0.0 & 0.0 & \cite{mcel13} \\
 R225 (PH) & $\rm{C_4P + h\nu \rightarrow C_3 + CP}$ & $5.48\times10^{-10}$ & 0.0 & 1.7 & \cite{mcel13} \\
 R226 (CR) & $\rm{C_3P + CRPHOT \rightarrow CCP + C}$ & $1.3\times10^{-17}$ & 0.0 & 750.0 & \cite{mcel13} \\
 R227 (DR) & $\rm{C_4P^+ + e^- \rightarrow C_3P + C}$ & $1.50\times10^{-7}$ & $-0.5$ & 0.0 & \cite{mcel13} \\
 R228 (DR) & $\rm{PC_3H^+ + e^- \rightarrow C_3P + H}$ & $1.00\times10^{-7}$ & $-0.5$ & 0.0 & \cite{mcel13} \\
 R229 (DR) & $\rm{PC_4H^+ + e^- \rightarrow C_3P + CH}$ & $7.50\times10^{-8}$ & $-0.5$ & 0.0 & \cite{mcel13} \\
 R230 (IN) & $\rm{H_3^+ + C_3P \rightarrow PC_3H^+ + H_2}$ & $1.00\times10^{-9}$ & $-0.5$ & 0.0 & \cite{mcel13} \\
 R231 (IN) & $\rm{H_3O^+ + C_3P \rightarrow PC_3H^+ + H_2O}$ & $1.00\times10^{-9}$ & $-0.5$ & 0.0 & \cite{mcel13} \\
 R232 (IN) & $\rm{HCO^+ + C_3P \rightarrow PC_3H^+ + CO}$ & $1.00\times10^{-9}$ & $-0.5$ & 0.0 & \cite{mcel13} \\
 R233 (IN) & $\rm{He^+ + C_3P \rightarrow C_3^+ + P + He}$ & $5.00\times10^{-10}$ & $-0.5$ & 0.0 & \cite{mcel13} \\
 R234 (IN) & $\rm{He^+ + C_3P \rightarrow C_3 + P^+ + He}$ & $5.00\times10^{-10}$ & $-0.5$ & 0.0 & \cite{mcel13} \\
 R235 (NN) & $\rm{O + C_3P \rightarrow CCP + CO}$ & $4.00\times10^{-11}$ & 0.0 & 0.0 & \cite{mcel13} \\
 R236 (PH) & $\rm{C_3P + h\nu \rightarrow C_2 + CP}$ & $5.00\times10^{-10}$ & 0.0 & 1.8 & \cite{mcel13} \\
 R237 (CR) & $\rm{CH_2PH + CRPHOT \rightarrow HCP + H_2}$ & $1.3\times10^{-17}$ & 0.0 & 750.0 & \cite{mcel13} \\
 R238 (IN) & $\rm{C^+ + CH_2PH \rightarrow HC_2P^+ + H_2}$ & $1.00\times10^{-9}$ & 0.0 & 0.0 & \cite{mcel13} \\
 R239 (IN) & $\rm{He^+ + CH_2PH \rightarrow PH^+ + CH_2 + He}$ & $5.00\times10^{-10}$ & 0.0 & 0.0 & \cite{mcel13} \\
 R240 (IN) & $\rm{He^+ + CH_2PH \rightarrow PH + CH_2^+ + He}$ & $5.00\times10^{-10}$ & 0.0 & 0.0 & \cite{mcel13} \\
 R241 (NN) & $\rm{O + CH_2PH \rightarrow PH_2 + CO + H}$ & $4.00\times10^{-11}$ & 0.0 & 0.0 & \cite{mcel13} \\
 R242 (PH) & $\rm{CH_2PH + h\nu \rightarrow CH_2 + PH}$ & $9.54\times10^{-10}$ & 0.0 & 1.8 & \cite{mcel13} \\
 R243 (DR) & $\rm{PC_2H_2^+ + e^- \rightarrow P + C_2H_2}$ & $1.00\times10^{-7}$ & $-0.5$ & 0.0 & \cite{mcel13} \\
 R244 (IN) & $\rm{C_2H_2 + PH^+ \rightarrow PC_2H_2^+ + H}$ & $1.30\times10^{-9}$ & 0.0 & 0.0 & \cite{mcel13} \\
 R245 (IN) & $\rm{C_2H_2 + PH_2^+ \rightarrow PC_2H_2^+ + H_2}$ & $1.40\times10^{-9}$ & 0.0 & 0.0 & \cite{mcel13} \\
 R246 (IN) & $\rm{C_2H_2 + PH_3^+ \rightarrow PC_2H_3^+ + H_2}$ & $5.80\times10^{-10}$ & 0.0 & 0.0 & \cite{mcel13} \\
 \hline
\multicolumn{6}{c}{Ice-phase/grain-surface pathways} \\
 \hline
 R1 & $\rm{gH + gP \rightarrow gPH}$ & - & - & - & \cite{chan20} \\
 R2 & $\rm{gH + gPH \rightarrow gPH_2}$ & - & - & - & \cite{chan20} \\
 R3 & $\rm{gH + gPH_2 \rightarrow gPH_3}$ & - & - & - & \cite{chan20} \\
 R4 & $\rm{gH + gPH_3 \rightarrow gPH_2 + gH_2}$ & - & - & - & This work, \cite{sous20} \\
 R5 & $\rm{gOH + gPH_3 \rightarrow gPH_2 + gH_2O}$ & - & - & - & This work, \cite{sous20} \\
 R6 & $\rm{gPH_3 \rightarrow PH_3}$ & - & - & - & \cite{chan20} \\
\hline
\end{tabular} \\
\vskip 0.2cm
{\bf Note:} \\
CR refers to cosmic-rays, IN to $\rm{ion-neutral}$ reactions, NR to $\rm{neutral-radical}$ reactions, NN to $\rm{neutral-neutral}$ reactions, RR to $\rm{radical-radical}$ reactions, RA to radiative association reactions, ER to electronic recombination reactions for atomic ions, DR to dissociative recombination reactions for molecular ions, PH to photodissociation reactions, h$\nu$ to a photon. \\
Grain surface species are denoted by the letter ``g''.
\end{table}

\begin{landscape}
\begin{table}
\scriptsize
\caption{Calculated BE (with MP2/aug-cc-pVDZ level of theory) and enthalpy of formation [with DFT-B3LYP/6-31G(d,p) level of theory] of P-bearing species \citep{sil21}. \label{tab:binding}}
\vskip 0.2cm
\hskip -1.2cm
\begin{tabular}{clcccccccccc}
\hline
{\bf Serial} & {\bf Species} & {\bf Astronomical} & {\bf Ground} & \multicolumn{6}{c}{\bf Binding Energy [Kelvin]} & \multicolumn{2}{c}{\bf Enthalpy of formation [kJ/mol]} \\
\cline{5-10}
{\bf No.} & & {\bf status} & {\bf state} & {\bf CO} & {\bf CH$_3$OH} & {\bf H$_2$O} & {\bf H$_2$O} & {\bf H$_2$} & {\bf Available$^d$} & {\bf Our calculated} & {\bf Available$^d$} \\
& & & & {\bf monomer} & {\bf monomer} & {\bf monomer} & {\bf tetramer} & {\bf monomer} & & & \\
\hline
1. & P & - & quartet & 170 & 442 & 270 & 616$^c$ & 107 & 1100 & 315.557$^a$, 310.202$^b$ & 315.663$^a$, 316.5$^b$ \\
2. & P$_2$ & not observed & singlet & 378 & 904 & 494 & 1671 & 223 & - &182.260$^a$, 180.434$^b$ & 145.8$^a$ \\
3. & PN & observed & singlet & 324 &2560  & 2326 & 2838 & 399 & 1900 & 218.985$^a$, 217.974$^b$ & 172.48$^a$, 171.487$^b$ \\
4. & PO & observed & doublet & 703 & 4334 & 2818 & 4600 & 508 & 1900 & 13.483$^a$, 12.490$^b$ & -27.548$^a$, -27.344$^b$ \\
5. & HCP & observed & singlet & 572 & 2122 & 1723 & 2468 & 132 & 2350 & 251.466$^a$, 250.054$^b$ & 217.79$^a$, 216.363$^b$ \\
6. & CP & observed & doublet & 300 & 1335 & 1126 & 1699$^c$ & 165 & 1900 & 538.388$^a$, 540.696$^b$ & 517.86$^a$, 520.162$^b$ \\
7. & CCP & observed & doublet & 2181 & 3900 & 2701 & 2868 & 396& 4300 & 660.424$^a$, 664.413$^b$ & - \\
8. & HPO & not observed & singlet & 602 & 4097 & 2838 & 5434 & 521 & 2350 & -38.035$^a$, -41.915$^b$ & -89.9$^a$, -93.7$^b$ \\
9. & PH & not observed & triplet & 270  & 780 & 491 & 944$^c$ & 134& 1550 & 228.785$^a$, 227.878$^b$ & 231.698$^a$, 230.752$^b$ \\
10. & PH$_2$ & not observed & doublet & 285 & 851 & 965 & 1226$^c$ &  138 & 2000 & 128.872$^a$, 125.036$^b$ & 139.333$^a$, 135.474$^b$ \\
11. & PH$_3$ & observed & singlet & 716 & 952 & 606 & 1672 & 545 & - & 12.934$^a$, 5.006$^b$ & 13.4$^a$ \\
\hline
\end{tabular} \\
\vskip 0.2cm
{\bf Note:} \\
$^a$ Enthalpy of formation at T = 0 K and 1 atmosphere pressure. \\
$^b$ Enthalpy of formation at T = 298 K and 1 atmosphere pressure. \\
$^c$ \cite{das18}. \\
$^d$ KIDA (\url{http://kida.astrophy.u-bordeaux.fr}).
\end{table}
\end{landscape}

\subsection{The binding energy of P-bearing species} \label{sec:bind_energy}
A continuous exchange of chemical ingredients within the gas and grain can determine the chemical complexity of the ISM.
BEs of a species with a grain substrate are essential for constructing a chemical model. However, most COMs are primarily produced on the dust surface and further desorbed back to the gas phase \citep{requ06}.
The major drawback in constraining this chemical process by astrochemical modeling is the lack of information about the interaction energy of the chemical species with the grain surface.

A sizable portion ($\sim 60-70\%$ of the surface coverage) of the icy interstellar layers may contain water molecules. That is why the BE of the interstellar species is usually expressed with the H$_2$O surface. However, the rest ($\sim30-40\%$) of the grain mantle would comprise of other impurities such as CO$_2$, CO, and CH$_3$OH \citep{furu18}.
\cite{kean01} summarized the relative abundance of CO, CO$_2$, and CH$_3$OH concerning water ice could vary in the range of $0.4-15$, $0.17-21$, and $1.5-30$, respectively, in different lines of sight. So, the BE with these surface species is also required in the various evolutionary phases. Therefore, the surface coverage-dependent BEs are indeed needed to consider, which are not considered here.

The BEs of the interstellar species with the different substrates are unknown.
Looking at these aspects, in Table \ref{tab:binding}, we report the BEs of some relevant P-bearing species by considering H$_2$O, CO, CH$_3$OH, and H$_2$ as a substrate.
The computed BEs should have a range of values that depend on the position of the molecule on the ice \citep{das18,ferr20}. Whenever we obtain different BE values at other binding sites, we exert the average of some of our calculations (see Table \ref{tab:binding}). In this work, we only consider the BE with water substrate for our simulation. BEs with the other substrates are provided for future usage.
We perform all the BE calculations by the \textsc{Gaussian} 09 suite of programs \citep{fris13} using the formulae \ref{eqn:BE_1} and \ref{eqn:BE_2} mentioned earlier (see section \ref{sec:BE_H_H2_method}). To find the optimized energy of all structures, we use the MP2/aug-cc-pVDZ level of theory \citep{dunn89}. We do not include the ZPVE and BSSE corrections. A fully optimized ground-state structure is verified as a stationary point (having nonimaginary frequency) by harmonic
vibrational frequency analysis. The ground-state spin multiplicity of the species is also noted in Table \ref{tab:binding}.

Due to the similarities between the structures of NH$_3$ and PH$_3$, \cite{chan20} considered the BE of PH$_3$ same as NH$_3$.
The BE of NH$_3$ is $5800$ K, according to the KIDA. Since they also used this as BE of PH$_3$, they did not obtain much PH$_3$ by thermal desorption process in the cold regions. They pointed out that most of the PH$_3$ came to the gas phase by photo-desorption rather than thermal desorption from the colder region. Our calculation finds a noticeable decrease in the BE of PH$_3$, which could enable PH$_3$ to populate the gas phase by thermal desorption even at low temperatures.
\cite{das18} noted BE of NH$_3$ with a c-hexamer configuration of
water $\sim 5163$ K in Table \ref{tab:BE_A1},
whereas we find it $\sim 2395$ K for PH$_3$.
With significant ice constituents (monomer of CO, CH$_3$OH, H$_2$O, and H$_2$),
our computed BE of PH$_3$ seems to be $<1000$ K.
\cite{sous20} noted that PH$_3$ has very low water solubility (at 17 $^\circ$C only 22.8 ml of gaseous PH$_3$ dissolves in 100 ml of water), and
it does not easily stick to aerosols.
The recent claim of PH$_3$ in the Venusian atmosphere \citep{grea20} prompted us to determine the PH$_3$ BE considering some other species such as sulfuric acid (H$_2$SO$_4$) and Benzene ($\rm{C_6H_6}$), which are the principal constituents of the Venus atmosphere. Our computed BE with $\rm{H_2SO_4}$ and benzene is $\sim 2271$ K and $\sim 3094$ K, respectively.
However, \cite{snell20} and \cite{vill20} questioned the identification of \cite{grea20}.
Table \ref{tab:binding} also provides our calculated enthalpy of formation
for several P-bearing species.

\subsection{Chemical model} \label{sec:chem_model}
We use the reaction pathways shown in Table \ref{tab:reaction_path} to check the fate
of the P-bearing species in various parts of the ISM
(diffuse cloud, PDR, hot-core, and hot-corino). Here, we employ mainly two models
using two different codes to study the chemical evolution of these species:
a) the spectral synthesis code \textsc{Cloudy} \citep[version 17.02, last described by][]{ferl17} and b) the CMMC code \citep{sil18,sil21,das19,das21,gora20b,shim20}.

\subsubsection{Spectral synthesis code, \textsc{Cloudy}}
We use a photoionization code, \textsc{Cloudy}, which simulates matter under a broad range of interstellar conditions. Using the \textsc{Cloudy} code, we check the fate
of P-bearing species in the diffuse cloud and PDR environment.

\begin{table}
\scriptsize
\centering
\caption{Initial elemental abundance for the diffuse cloud and PDR model considered in the \textsc{Cloudy} code \citep{sil21}. \label{tab:diff_cloud}}
\vskip 0.2cm
\begin{tabular}{cccc}
\hline
{\bf Element} & {\bf Abundance} & {\bf Element} & {\bf Abundance} \\
\hline
H         & 1.0 & Si         & $3.2 \times 10^{-5}$\\
He         & $8.5 \times 10^{-2}$ & Fe         & $3.2 \times 10^{-5}$\\
N         & $6.8 \times 10^{-5}$ & Na         & $1.7 \times 10^{-6}$\\
O         & $4.9 \times 10^{-4}$ & Mg        & $3.9 \times 10^{-5}$\\
C         & $2.7 \times 10^{-4}$ & Cl        & $3.2 \times 10^{-7}$\\ 
S         & $1.3 \times 10^{-5}$ & P         & $2.6 \times 10^{-7}$ \\
& & F         & $3.6 \times 10^{-8}$\\
\hline
\end{tabular}
\end{table}

\subsubsection{Diffuse cloud model}
\cite{chan20} prepared a diffuse cloud model to explain the observed abundance
of HNC, CN, CS, and CO. We also employ a similar process to explain the observed
abundances of the four molecules and then look at the fate of the P-bearing
molecules under similar circumstances.
We consider the initial elemental abundance for this modeling as given in \cite{chan20}. But in \textsc{Cloudy}, only the atomic elemental abundance is allowed (no ionic or molecular form), so we use these abundances as the initial elemental abundance (see Table \ref{tab:diff_cloud}). For each calculation, we check whether the microphysics considered is time steady or not. We notice that the largest reaction timescale is much shorter than the typical lifetime ($\sim 10^7$ years) of the diffuse cloud. We use the grain size distribution, which is appropriate for the $R_V=3.1$ extinction curve of \cite{math77}. This grain size distribution is called ``ISM'' in \textsc{Cloudy}. We use the anisotropic radiation field, which is appropriate for the ISM to specify the incident local ISRF. Additionally, we include the cosmic-ray microwave background with a redshift value of $1.52$ \citep{weng00}.

\begin{figure}
\centering
\includegraphics[width=\textwidth]{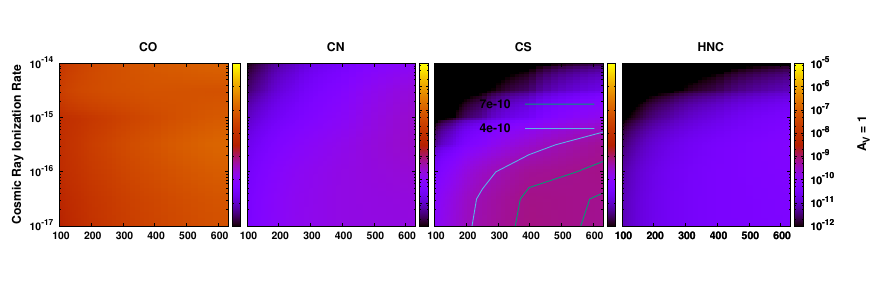}
\vskip -1.8cm
\includegraphics[width=\textwidth]{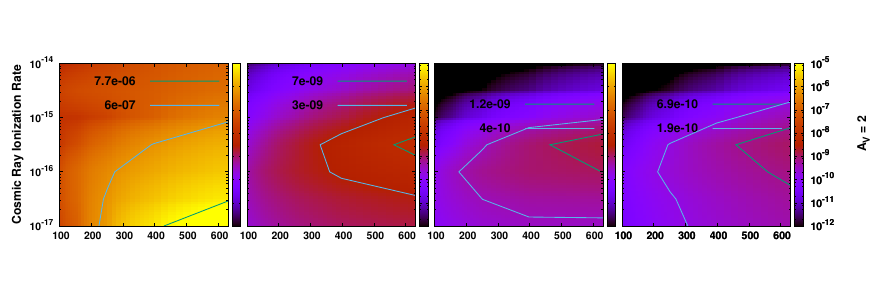}
\vskip -1.8cm
\includegraphics[width=\textwidth]{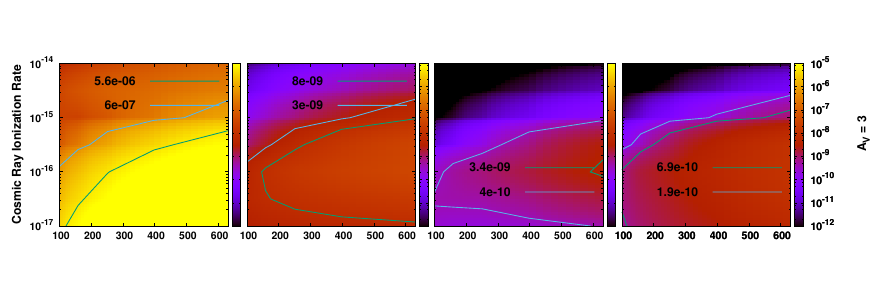}
\vskip -1.8cm
\includegraphics[width=\textwidth]{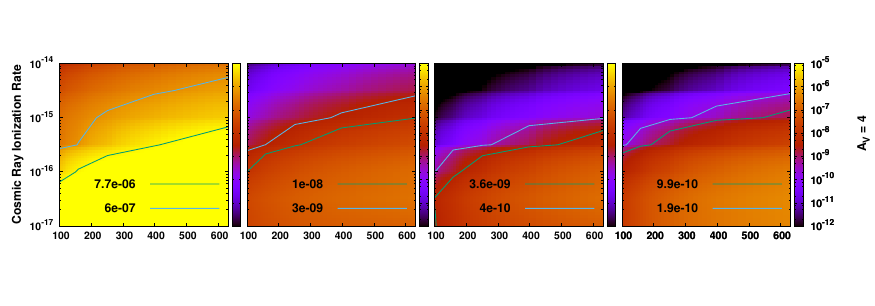}
\vskip -1.8cm
\includegraphics[width=\textwidth]{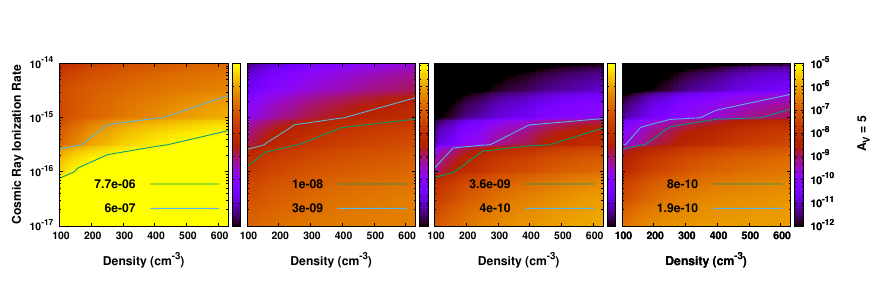}
\caption{Parameter space of the abundances of CO, CN, CS, and HNC for $\rm{A_V}=1,\ 2,\ 3,\ 4,\ 5$ mag for the diffuse cloud model obtained with the \textsc{Cloudy} code \citep{sil21}. The right side of each panel is marked with color-coded values of abundance concerning total hydrogen nuclei. The contours are highlighted around the previously observed abundance limit \citep{chan20} toward the cloud with $v_{LSR}=-17$ km s$^{-1}$, including the inferred uncertainties.}
    \label{fig:param_space}
\end{figure}

\begin{figure}
\centering
\includegraphics[width=0.7\textwidth]{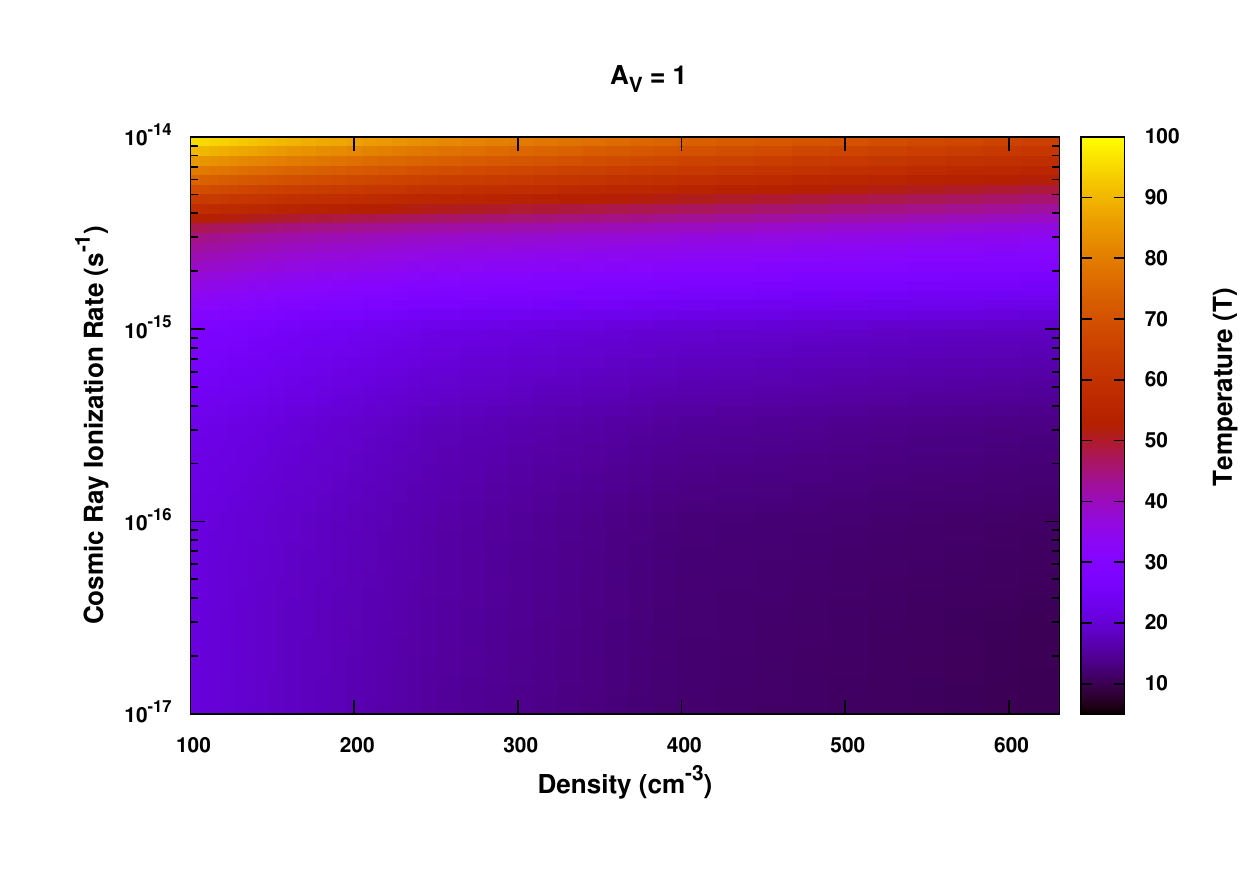}
\vskip -0.5cm
\includegraphics[width=0.7\textwidth]{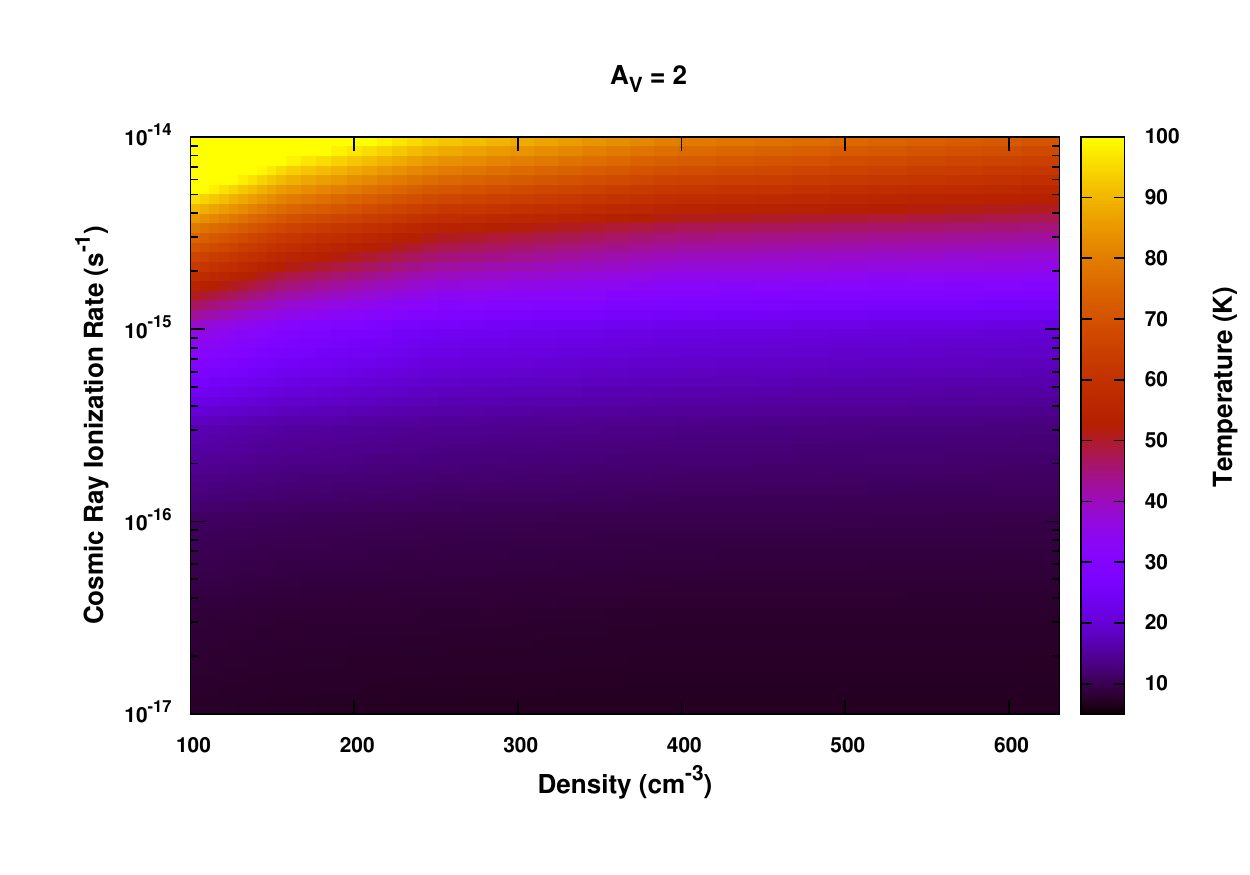}
\vskip -0.5cm
\includegraphics[width=0.65\textwidth]{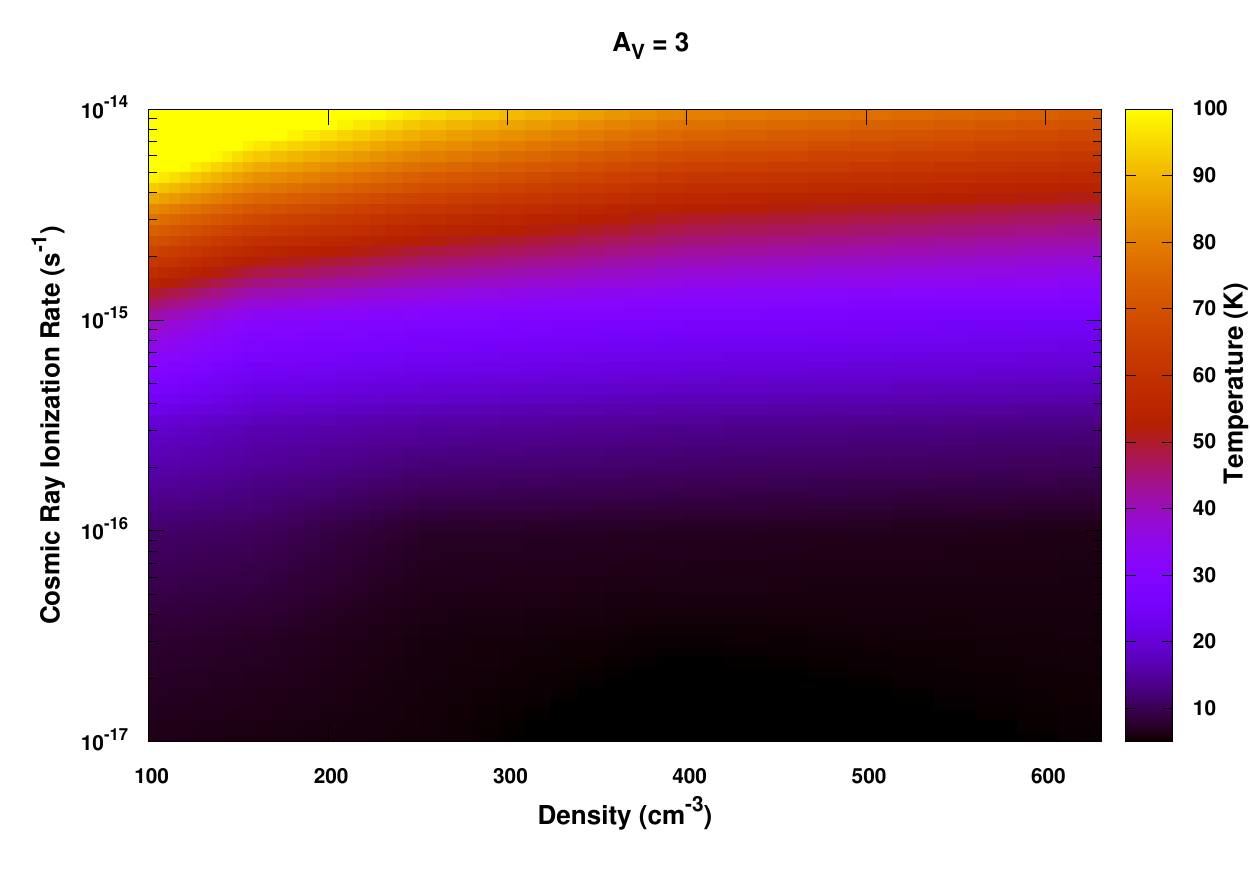}
\caption{Parameter space of the temperature for $\rm{A_V}=1,\ 2,\ 3$ mag for the diffuse cloud model obtained with the \textsc{Cloudy} code \citep{sil21}. The right side of each panel is marked with color-coded values of temperature.}
\label{fig:param_temp}
\end{figure}

To check our model against the observed abundances, we first explain the observed abundance of CO, CN, CS, and HNC as obtained in \cite{chan20}.
For this, we use a parameter space for the formation of these species.
Our parameter space consists of a density variation of $100-600$ cm$^{-3}$ and
a cosmic-ray ionization rate of H$_2$ ($\zeta_{H_2}$)  variation
of $10^{-17}-10^{-14}$ s$^{-1}$. We use a stopping criterion at different $A_V$ values.
The cases obtained by using stopping criterion $A_V= 1-5$ mag are given in Figure \ref{fig:param_space}.
The contours highlight the observed abundances of CO, CN, CS, and HNC of \cite{chan20} toward the cloud having a $V_{LSR}=-17$ km s$^{-1}$.
Figure \ref{fig:param_temp} shows the temperature variation at the end of the calculation for $A_V=1-3$ mag. The parameter space also consists of the variation
of n$_H$ and $\zeta_{H_2}$. Temperature variation is shown with the color bar.

\begin{figure}
\centering
\includegraphics[width=0.8\textwidth]{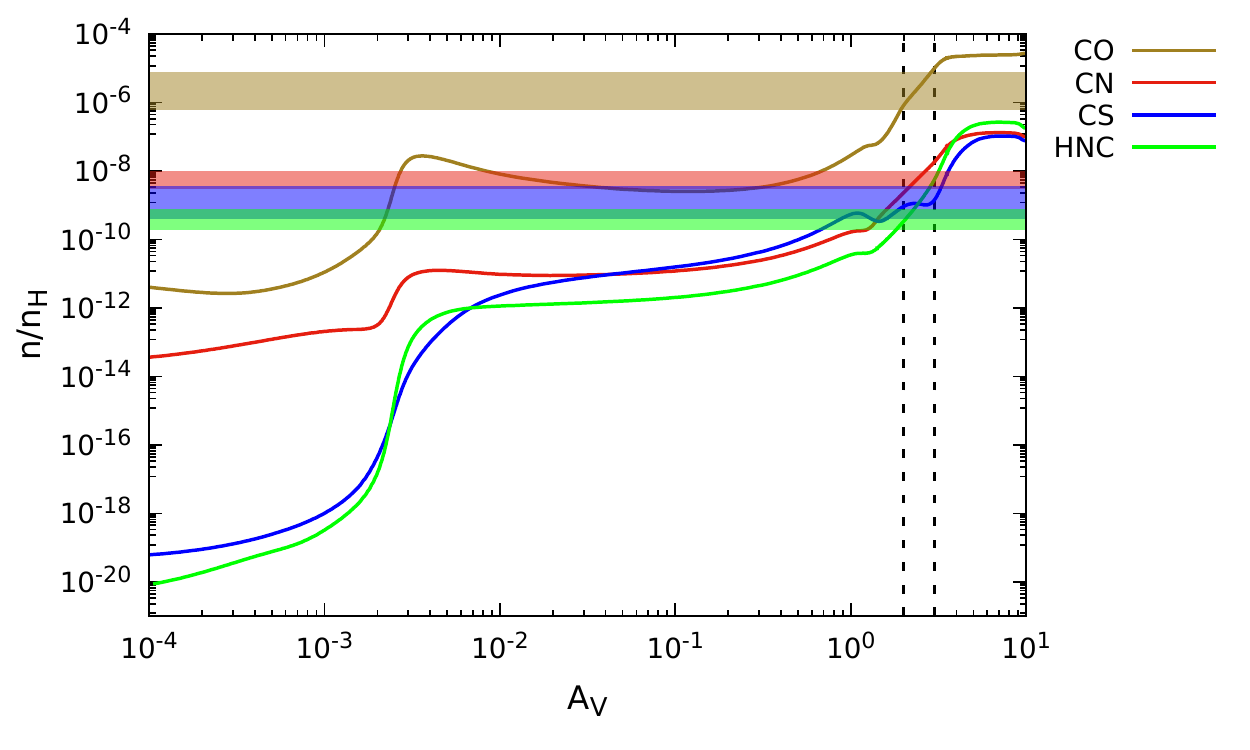}
\caption{Chemical evolution of the abundances of CO, CN, CS, and HNC for the diffuse cloud model ($\rm{n_H}=300$ cm$^{-3}$ and $\rm{\zeta=1.7\times10^{-16}\ s^{-1}}$) with the \textsc{Cloudy} code. The colored horizontal
bands correspond to the observed abundances \citep{chan20} toward the cloud with $v_{LSR}=-17$ km s$^{-1}$, including the inferred uncertainties \citep{sil21}. Here, the vertical dashed line indicates the visual extinction parameter of best agreement between observation and model results.}
\label{fig:diffuse_obs_CLOUDY}
\end{figure}

\begin{figure*}
\centering
\includegraphics[width=0.49\textwidth]{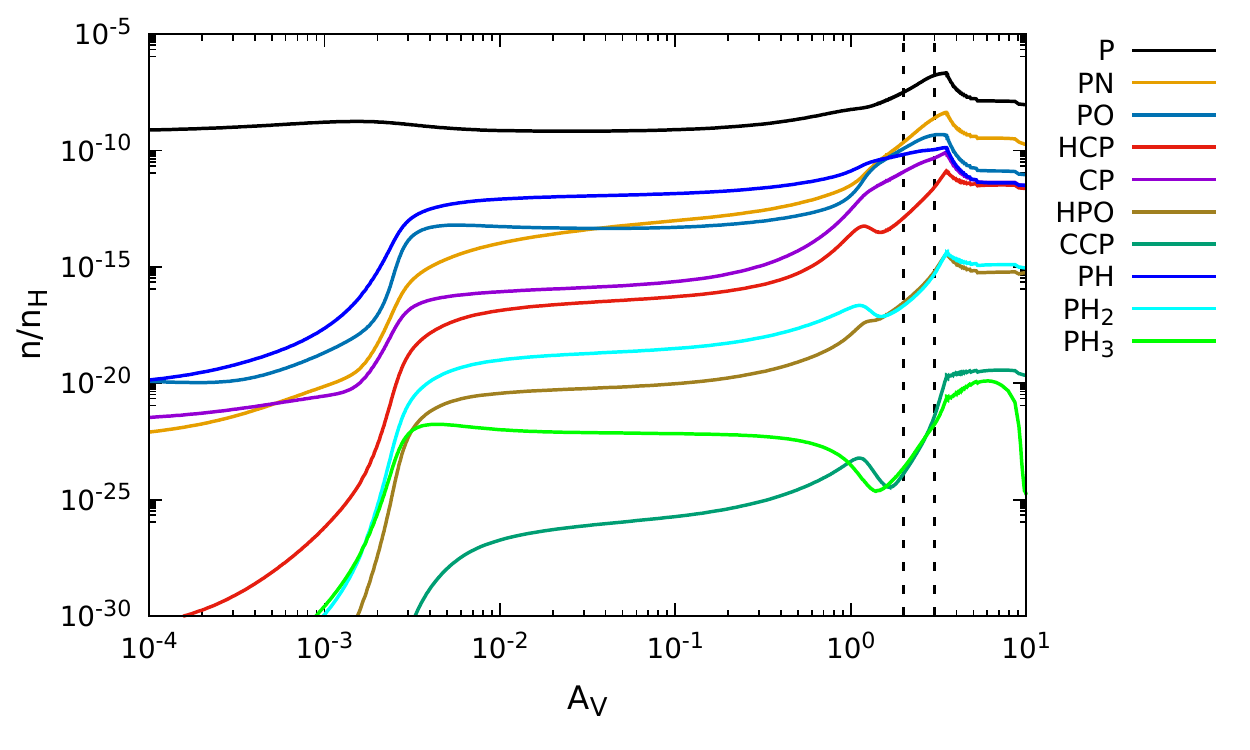}
\includegraphics[width=0.49\textwidth]{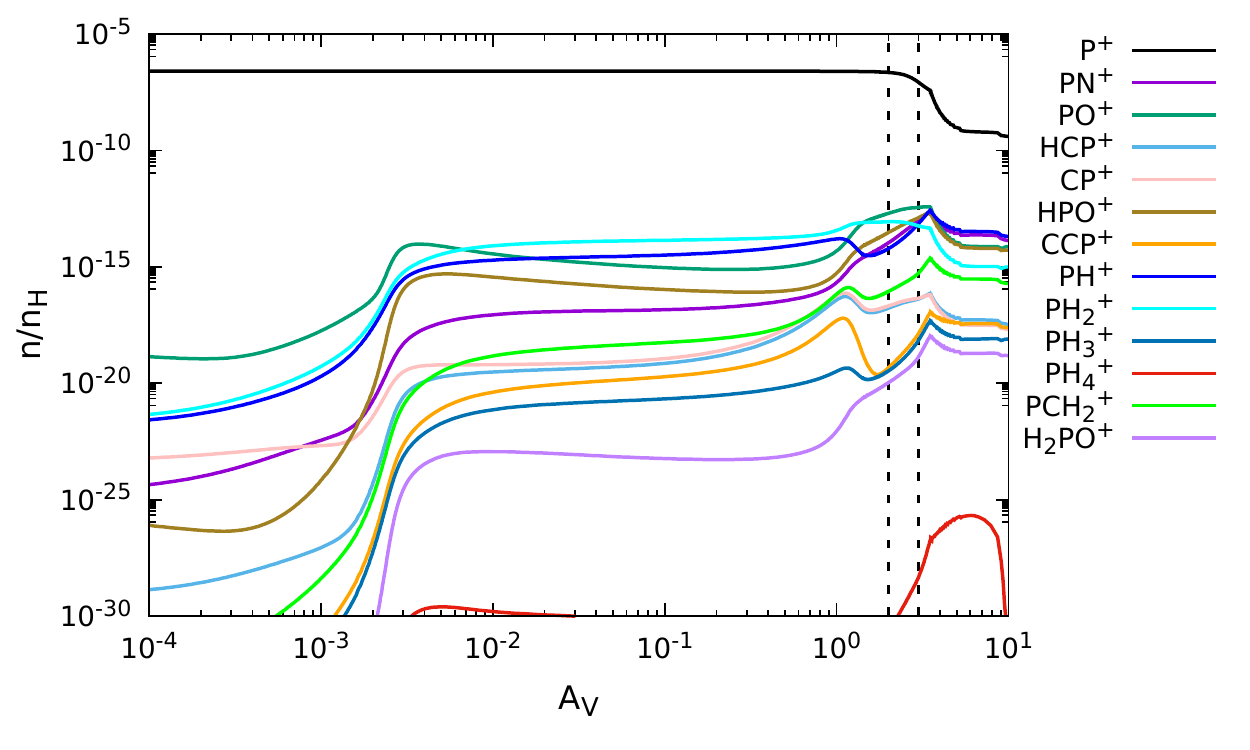}
\caption{Chemical evolution of the abundances of neutral P-bearing species (left panel) and their corresponding cations (right panel) for the diffuse cloud model ($\rm{n_H}=300$ cm$^{-3}$ and $\rm{\zeta=1.7\times10^{-16}\ s^{-1}}$) with the \textsc{Cloudy} code \citep{sil21}. Here, the vertical dashed line indicates the visual extinction parameter of the best agreement between observation and model results.}
\label{fig:diffuse_phosphorus_CLOUDY}
\end{figure*}

From Figures \ref{fig:param_space} and \ref{fig:param_temp}, we see that the parameters that were adopted by \cite{chan20} for the diffuse cloud model (i.e., $\zeta_{H_2}=1.7 \times 10^{-16} \ s^{-1}$, $n_H=300$ cm$^{-3}$, and $T_{gas}=40$ K) can also reproduce the observed abundances of CO, CN, CS, and
HNC with our model. Figure \ref{fig:diffuse_obs_CLOUDY} shows the obtained abundances of the four molecules when we consider $n_H=300$ cm$^{-3}$ and $\zeta_{H_2}$ = $1.7 \times 10^{-16}\ s^{-1}$. The X-axis shows the visual extinction of the cloud, and the Y-axis shows the abundance of n$_H$. Our results are in agreement with the observation between $A_V=2$ and $3$ mag. We highlight the best-suited zone by the black dashed curve.
Thus, based on Figures \ref{fig:param_space}, \ref{fig:param_temp}, and \ref{fig:diffuse_obs_CLOUDY}, we use $A_V=2$ mag, $\zeta_{H_2}= 1.7\times 10^{-16}$, and $n_H=300$ cm$^{-3}$ as the best-fitted parameters to explain the observed abundances of these species. This yields a $A_V/N(H) = 5.398 \times 10^{-22}$ mag cm$^2$ and dust to gas ratio of $6.594 \times 10^{-3}$.
Table \ref{tab:comparison} compares our obtained optical depth and column densities with the observations. Obtained optical depths of CN and HNC are close to the observed values, but the column densities diverge by a few factors. This slight mismatch is because the \textsc{Cloudy} model deals with steady-state values, but the column densities are time dependent in reality.

\begin{table}
\tiny
\caption{Estimated column density and optical depth of the observed molecules for the diffuse cloud model obtained with the \textsc{Cloudy} code \citep{sil21}. \label{tab:comp_obs}}
\vskip 0.2cm
\hskip -2.2cm
\begin{tabular}{cccccccc}
\hline
{\bf Species} & {\bf Transitions} & {\bf E$_{up}$ (K)} & {\bf Frequency (GHz)} & \multicolumn{2}{c}{\bf Optical Depth ($\tau$)} & \multicolumn{2}{c}{\bf Total Column Density (cm$^{-2}$)} \\
 & & & & {\bf Model} & {\bf Observation$^a$} & {\bf Model} & {\bf Observation$^a$} \\
\hline
CN & $N=1-0,\ J=1/2-1/2,\ F=3/2-1/2$ & 5.5 & 113.16867 & 0.256522 & $0.23\pm0.07$ & $1.47\times10^{12}$ & $(0.87\pm0.28)\times10^{13}$ \\
HNC & $J=1-0$ & 4.4 & 90.66357 & 0.33962 & $0.50\pm0.10$ & $2.11\times10^{11}$ & $(0.69\pm0.16)\times10^{12}$ \\
C$^{34}$S & $J=2-1$ & 6.9 & 96.41295 & - & $0.04\pm0.02$ & - & $(1.64\pm0.82)\times10^{11}$ \\
$^{13}$CO & $J=1-0$ & 5.3 & 110.20135 & - & $0.154\pm0.004$ & - & $(3.98\pm0.16)\times10^{14}$ \\
\hline
HCP & $J=2-1$ & 5.8 & 79.90329 & $2.56\times10^{-6}$ & $<0.02$ & $1.10\times10^{8}$ & $<2.27\times10^{12}$ \\
PN & $J=2-1$ & 6.8 & 93.97977 & 0.238015 & $<0.02$ & $1.62\times10^{11}$ & $<4.20\times10^{10}$ \\
CP & $N=2-1,\ J=3/2-1/2,\ F=2-1$ & 6.8 & 95.16416 & $4.92\times10^{-4}$ & $<0.02$ & $1.16\times10^{10}$ & $<1.26\times10^{12}$ \\
PO & $J=5/2-3/2,\ \Omega=1/2,\ F=3-2,\ e$ & 8.4 & 108.99845 & 0.0138432 & $<0.02$ & $9.08\times10^{10}$ & $<4.29\times10^{11}$\\
PO & $J=5/2-3/2,\ \Omega=1/2,\ F=2-1,\ e$ & 8.4 & 109.04540 & 0.0125113 & $<0.02$ & $9.08\times10^{10}$ & $<6.70\times10^{11}$ \\
PO & $J=5/2-3/2,\ \Omega=1/2,\ F=3-2,\ f$ & 8.4 & 109.20620 & 0.0055508 & $<0.02$ & $9.08\times10^{10}$ & $<4.34\times10^{11}$ \\
PO & $J=5/2-3/2,\ \Omega=1/2,\ F=2-1,\ f$ & 8.4 & 109.28119 & 0.0182699 & $<0.02$ & $9.08\times10^{10}$ & $<6.69\times10^{11}$ \\
\hline
\end{tabular} \\
\vskip 0.2cm
{\bf Note:}
$^a$\cite{chan20}
\label{tab:comparison}
\end{table}

Figure \ref{fig:diffuse_phosphorus_CLOUDY} depicts the abundances of most of
the P-bearing species considered in this study. The left panel shows the neutral
P-bearing species, whereas the right panel shows the P-bearing ions.
Interestingly, although we consider a neutral atomic abundance in our
model, the abundance of P$^+$ is high owing to the strong cosmic-ray ionization and
presence of a nonextinguished ISRF. The abundances of the comparatively larger P-bearing
species are very low in the diffuse environment, and it would be pretty challenging to observe them. We notice that some simple neutral P-bearing species, such as PN, PO, HCP, CP,
and PH are significantly more abundant than PH$_3$. The abundances of PO and PN appear
to be comparable to each other (with PO/PN $<1$ when A$_V \geq $ 2).
The obtained column densities for this case are
listed in Table \ref{tab:comparison}. There is an excellent agreement between
the observed and the modeled optical depths of CN and HNC with the \textsc{Cloudy} code.

\begin{figure}
\centering
\includegraphics[width=0.8\textwidth]{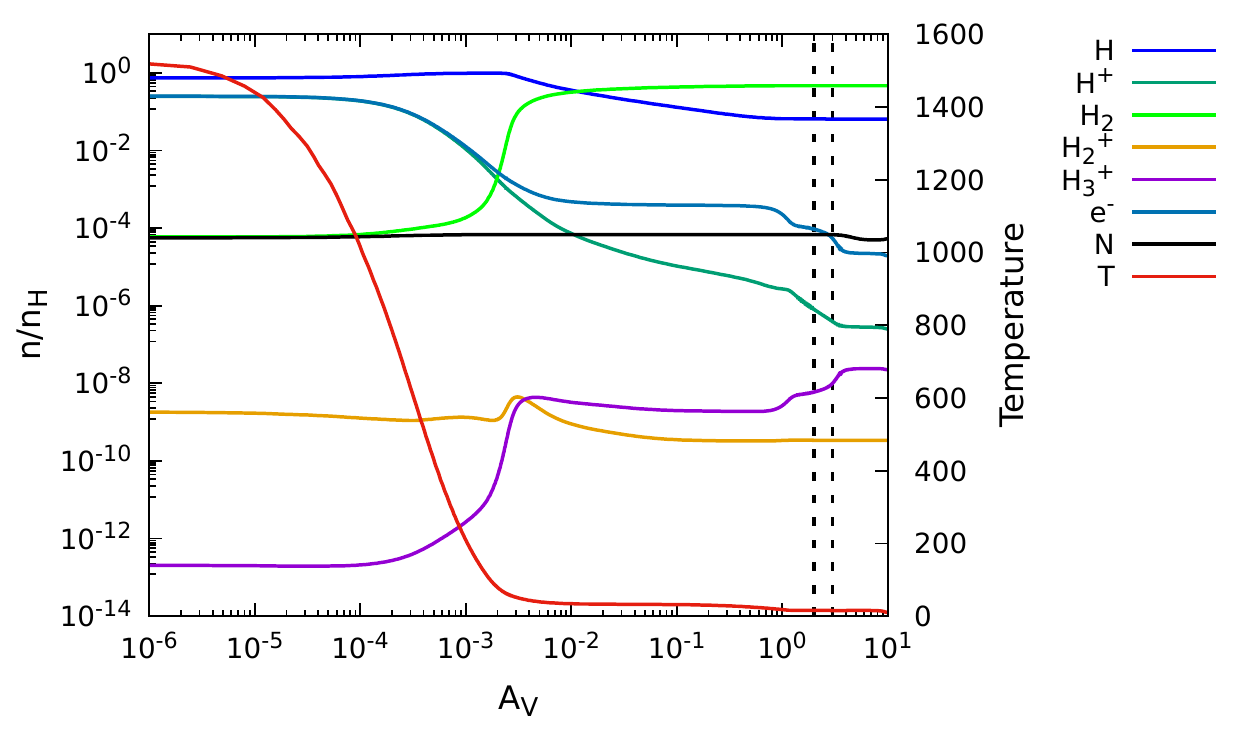}
\caption{Abundance profiles of H, H$^+$, H$_2$, H$_2^+$, H$_3^+$, e$^-$, and N and temperature profile for the diffuse cloud model ($\rm{n_H}=300$ cm$^{-3}$ and $\rm{\zeta=1.7\times10^{-16}\ s^{-1}}$) with the \textsc{Cloudy} code \citep{sil21}. The vertical dashed line indicates the visual extinction parameter of best agreement between observation and model results.}
\label{fig:DIFF_HH2}
\end{figure}

In Figure \ref{fig:DIFF_HH2}, H, H$_2$, and N abundances are shown along with the major hydrogen-related ions, H$^+$, ${\rm H_2}^+$, ${\rm H_3}^+$. In \textsc{Cloudy}, the cloud temperature is calculated self-consistently, depending on the known excitations.
The primary source of such excitation is cosmic-ray ionization. Figure \ref{fig:DIFF_HH2} depicts the gas temperature variation with $A_V$ with the solid red curve.
Deep inside the cloud, the temperature drops sufficiently and enhances the formation of molecular hydrogen.
For $A_V=2$ mag, we obtain a temperature of $\sim 10$ K when we use a cosmic-ray ionization rate of $1.7 \times 10^{-16}$ s$^{-1}$. It is a little lower than that used in \cite{chan20}.

\begin{figure}
\centering
\includegraphics[width=0.49\textwidth]{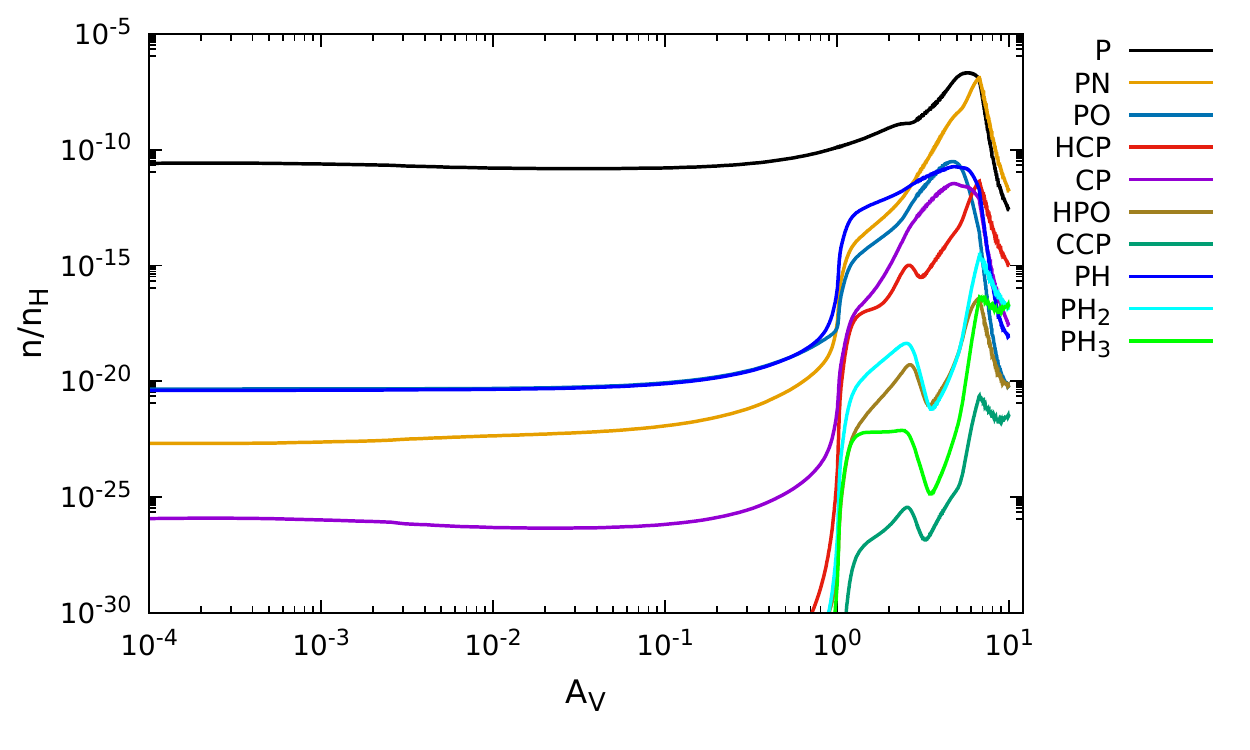}
\includegraphics[width=0.49\textwidth]{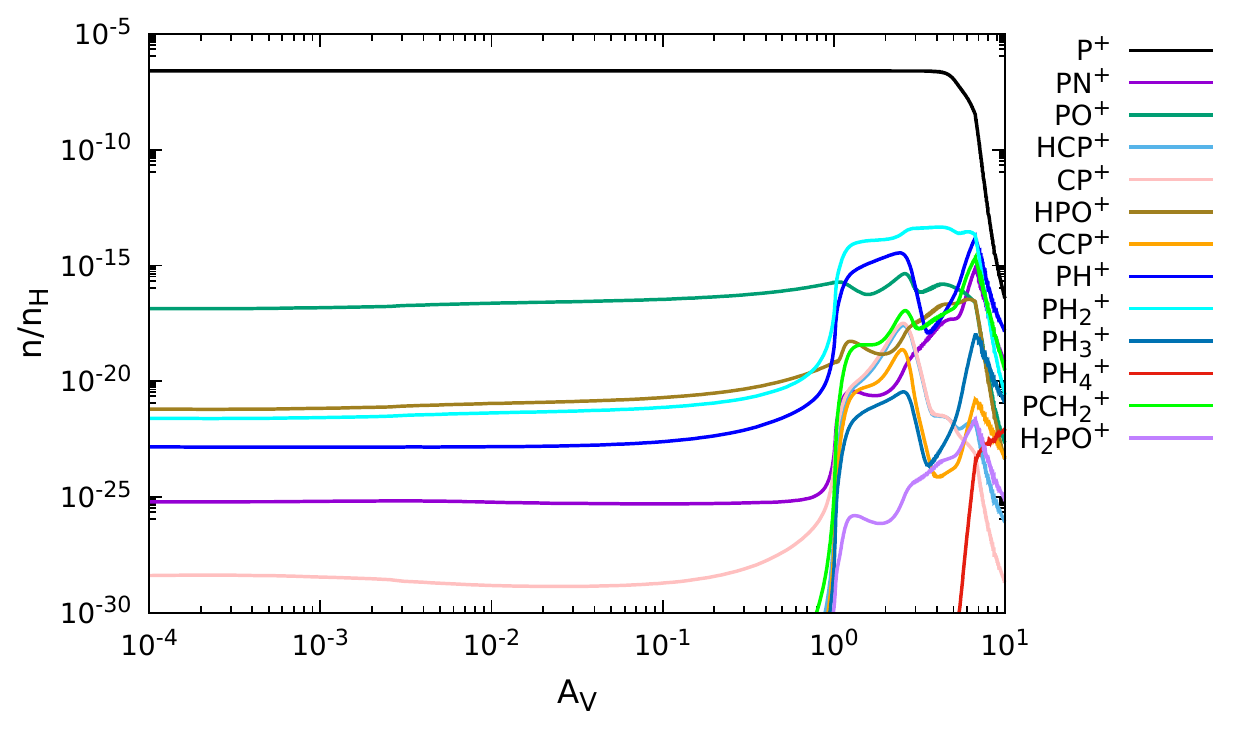}
\caption{Chemical evolution of the abundances of important P-bearing species with the \textsc{Cloudy} code considering \cite{roll07} F4 PDR model \citep{sil21}.}
\label{fig:PDR_phosphorus_CLOUDY}
\end{figure}

\begin{figure}
\centering
\includegraphics[width=0.7\textwidth]{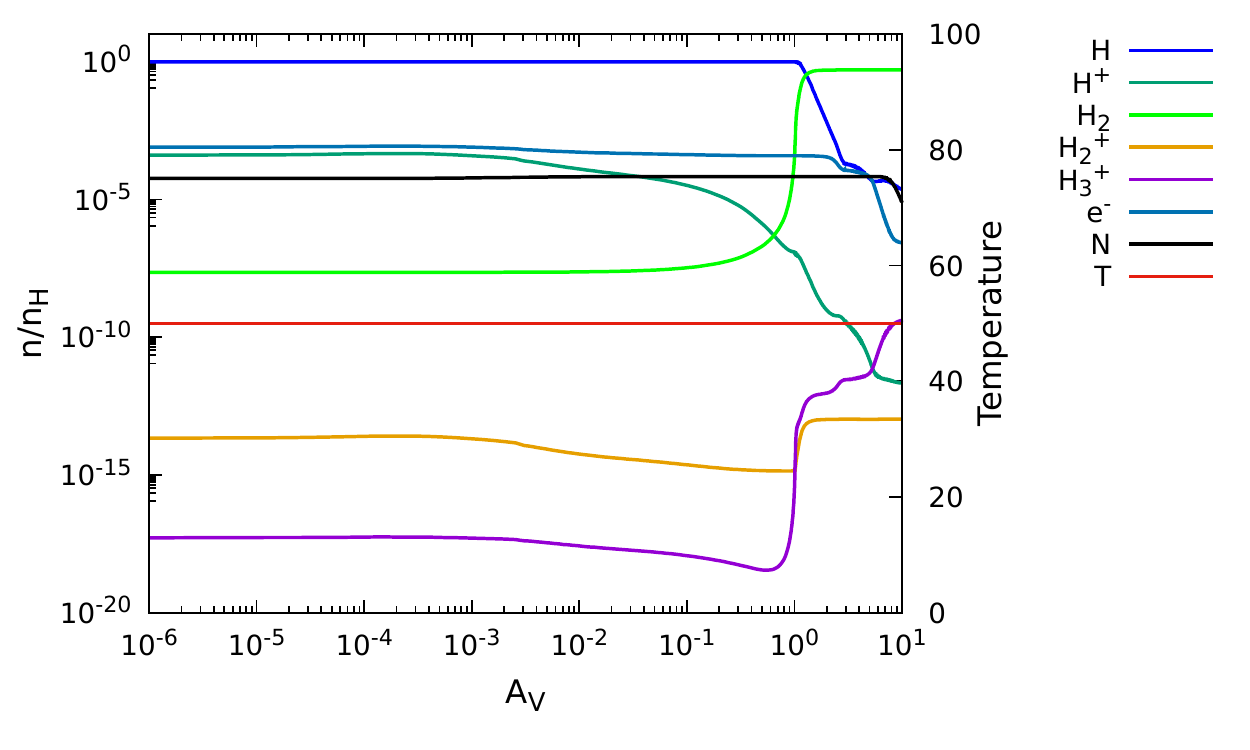}
\caption{Abundance profiles of H, H$^+$, H$_2$, H$_2^+$, H$_3^+$, e$^-$, N, and temperature profile with the \textsc{Cloudy} code considering the \cite{roll07} F4 PDR model \citep{sil21}.}
\label{fig:PDR_HH2}
\end{figure}

\subsubsection{PDR model}
We use the \textsc{Cloudy} code to explain the abundances of the P-bearing species in the PDRs. The \textsc{Cloudy} code is capable of considering realistic physical conditions around this region.
\cite{roll07} included various PDR benchmark models, giving a range of physical conditions associated with PDR environments.
We chose the F4 model. This model considers a constant gas temperature of $50$ K and
dust temperature of $20$ K. A  plane-parallel, semi-infinite cloud with
total constant density of $10^{5.5}$ cm$^{-3}$, and standard
UV field as $\chi = 10^5$ times the \cite{drai78} field ($\chi=1.71G_0$) is considered.
The cosmic-ray H ionization rate $\zeta = 5\times10^{-17}$ s$^{-1}$,
the visual extinction $\rm{A_V = 6.289 \times 10^{-22}N_{H,tot}}$, and a dust-to-gas
ratio of $7.646 \times 10^{-3}$ are assumed. We also consider these physical parameters
and the initial elemental abundances as considered in our diffuse cloud model (Table \ref{tab:diff_cloud}). Figure \ref{fig:PDR_phosphorus_CLOUDY} shows the abundances of important P-bearing species. In the diffuse cloud region (see Figure \ref{fig:diffuse_phosphorus_CLOUDY}), we obtain PO and PN comparable abundances. However, for the PDR model, we find a very high abundance of PN compared to PO.
Its peak abundance appears to be higher by several orders of magnitude than PO (i.e., PO/PN $<<1$). We use the photodestruction rate of PO and PN from \cite{jime18}.
The photodissociation rate of PO is estimated from the photodissociation rate of NO. Similarly, the photodissociation rate of PN is estimated from the photodissociation rate of N$_2$. These values reflect a higher photodissociation rate of PO
(with $\alpha$ $= 3 \times 10^{-10}$ s$^{-1}$ and $\gamma = 2.0$) than PN (with $\alpha$ $= 5 \times 10^{-12}$ s$^{-1}$ and $\gamma = 3.0$).
As in \cite{jime18}, here, we also find that even if PN is photodissociated, it revert again via $\rm{P+CN \rightarrow PN+C}$ owing to the presence of a large amount of atomic P in the gas phase.
The rapid conversion of PO to PN by the reaction $\rm{PO + N \rightarrow PN + O}$ is also crucial for maintaining a higher abundance of PN than PO.

It is not straightforward to relate the diffuse cloud region cases with the PDR because of different physical circumstances. In the diffuse cloud model, we obtain the peak abundances of PO and PN at $\rm{A_V=2-3}$ mag, whereas in the PDR model,
we receive these peaks around $\rm{A_V=6-7}$ mag. In both cases, the PO/PN ratio is $<1$,
but this ratio seems to be $<<1$ for the PDR. \cite{jime18} also modeled the effect of intense UV photon illumination by varying the ISRF within ranges typical of PDRs. They showed for higher extinctions ($\rm{A_V=7.5}$ mag), and under high-UV radiation fields ($\chi=10^4$ Habing) that the abundance of PN always remains above that of PO for both long-lived and short-lived collapse stages. Figure \ref{fig:PDR_HH2} shows H, H$^+$, H$_2$, ${\rm H_2}^+$, ${\rm H_3}^+$, N, and electrons abundances.

\subsubsection{CMMC code}

In the previous section, we implement \textsc{Cloudy} code to study the chemical evolution of the P-bearing species in the diffuse cloud region. Comparatively larger P-bearing species are not very profuse in space, which creates a burden constraining the understanding of the P-bearing species of the ISM. The major drawback in the P-chemistry modeling is the uncertainty of the P depletion factor. The grain-surface chemistry plays a significant role in shaping the chemical complexity in these regions.

Since \textsc{Cloudy} only considers the surface reactions of some key species,
it would be impractical to apply the \textsc{Cloudy} code in the dense cloud region.
Thus, we use our gas-grain CMMC code \citep{sil18,sil21,das19,das21,gora20b} to explore the fate of P-bearing species in the denser region. In the next section, we first test our model for the
diffuse region to validate our results and extend it for the more evolved stage.

\subsubsection{Diffuse cloud model}

\begin{table}
\scriptsize
\centering
\caption{Initial elemental abundance for the diffuse cloud model considered in the CMMC code \citep{chan20,sil21}. \label{tab:diff_cloud_CMMC}}
\vskip 0.2cm
\begin{tabular}{cccc}
\hline
{\bf Element} & {\bf Abundance} & {\bf Element} & {\bf Abundance} \\
\hline
H         & 1.0 & Si$^+$         & $3.2 \times 10^{-5}$\\
He         & $8.5 \times 10^{-2}$ & Fe$^+$         & $3.2 \times 10^{-5}$\\
N         & $6.8 \times 10^{-5}$ & Na$^+$         & $1.7 \times 10^{-6}$\\
O         & $4.9 \times 10^{-4}$ & Mg$^+$        & $3.9 \times 10^{-5}$\\
C$^+$         & $2.7 \times 10^{-4}$ & Cl$^+$        & $3.2 \times 10^{-7}$\\ 
S$^+$         & $1.3 \times 10^{-5}$ & P$^+$         & $2.6 \times 10^{-7}$ \\
& & F$^+$         & $3.6 \times 10^{-8}$\\
\hline
\end{tabular}
\end{table}

\begin{figure}
\centering
\includegraphics[width=8cm, height=12cm, angle=270]{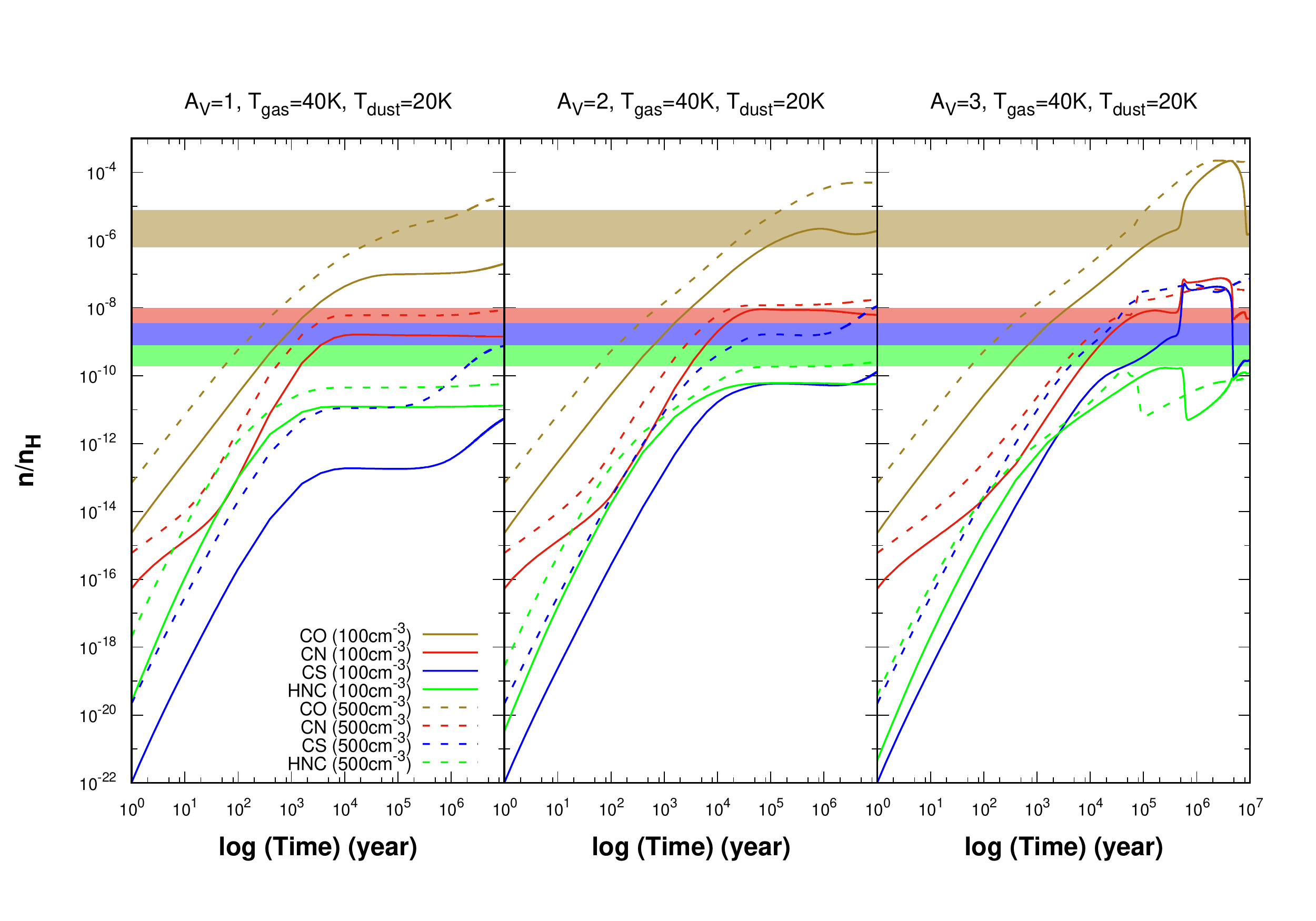}
\caption{Time evolution of the abundances of CO, CN, CS, and HNC with the CMMC code for diffuse cloud. Observed abundances are also highlighted for better understanding. The solid curves represent the case with $n_H=100$ cm$^{-3}$ and dashed curves represent the case with $n_H=500$ cm$^{-3}$ \citep{sil21}.}
\label{fig:diff_CMMC}
\end{figure}

\begin{figure}
\centering
\includegraphics[width=0.49\textwidth]{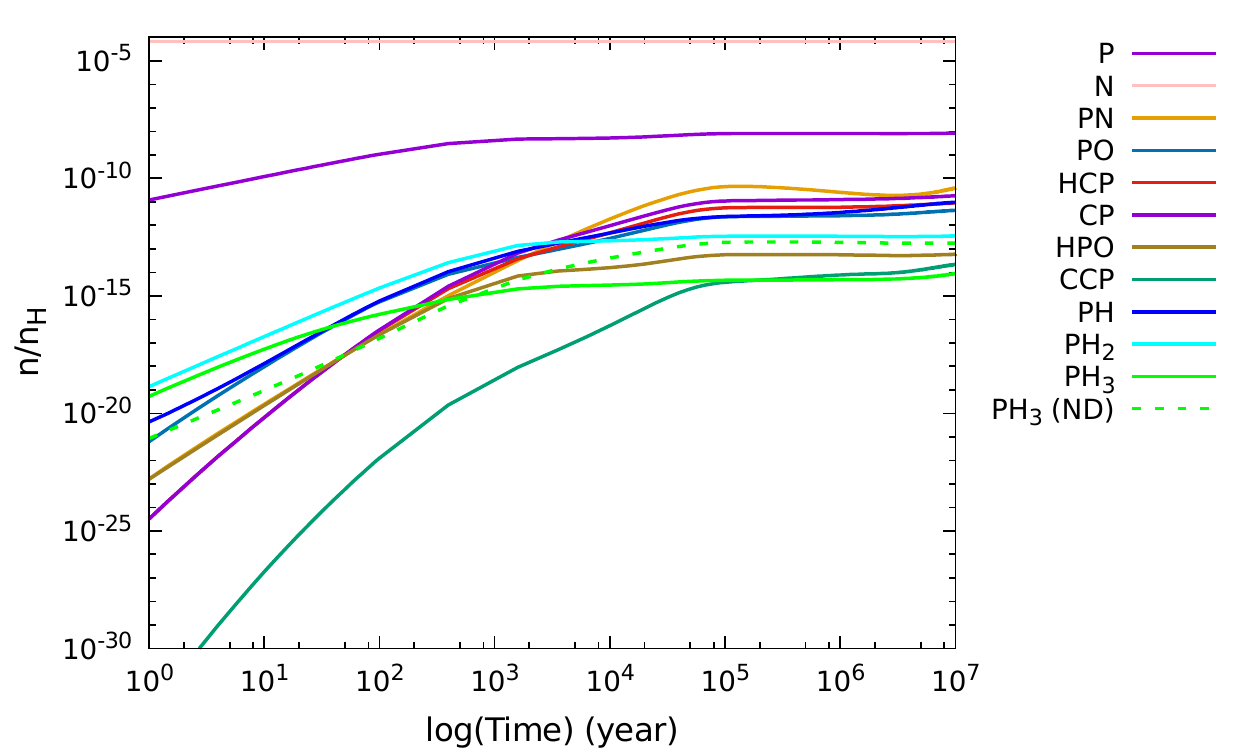}
\includegraphics[width=0.49\textwidth]{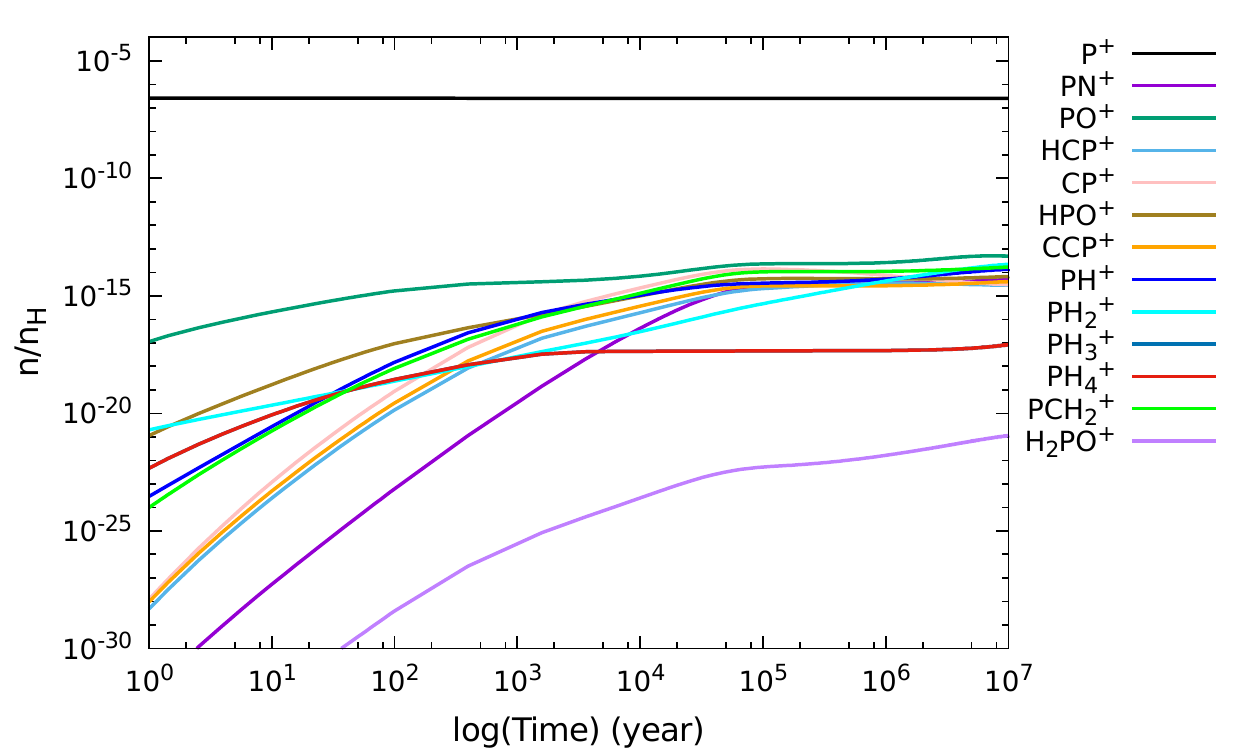}
\caption {Chemical evolution of the P-bearing molecules for $A_V=2$ mag, $n_H=300$ cm$^{-3}$, $T_{gas}=40$ K, and $T_{dust}=20$ K with the CMMC code for diffuse cloud is shown. The evolution of the neutrals and radicals is shown in the left panel, whereas ions are shown in the right panel. Derived abundances of all the species are found to be $<10^{-10}$ for such conditions \citep{sil21}.}
\label{fig:diff_CMMC_ION_NEUTRAL}
\end{figure}

\begin{figure}
\centering
\includegraphics[width=6cm, height=8cm, angle=270]{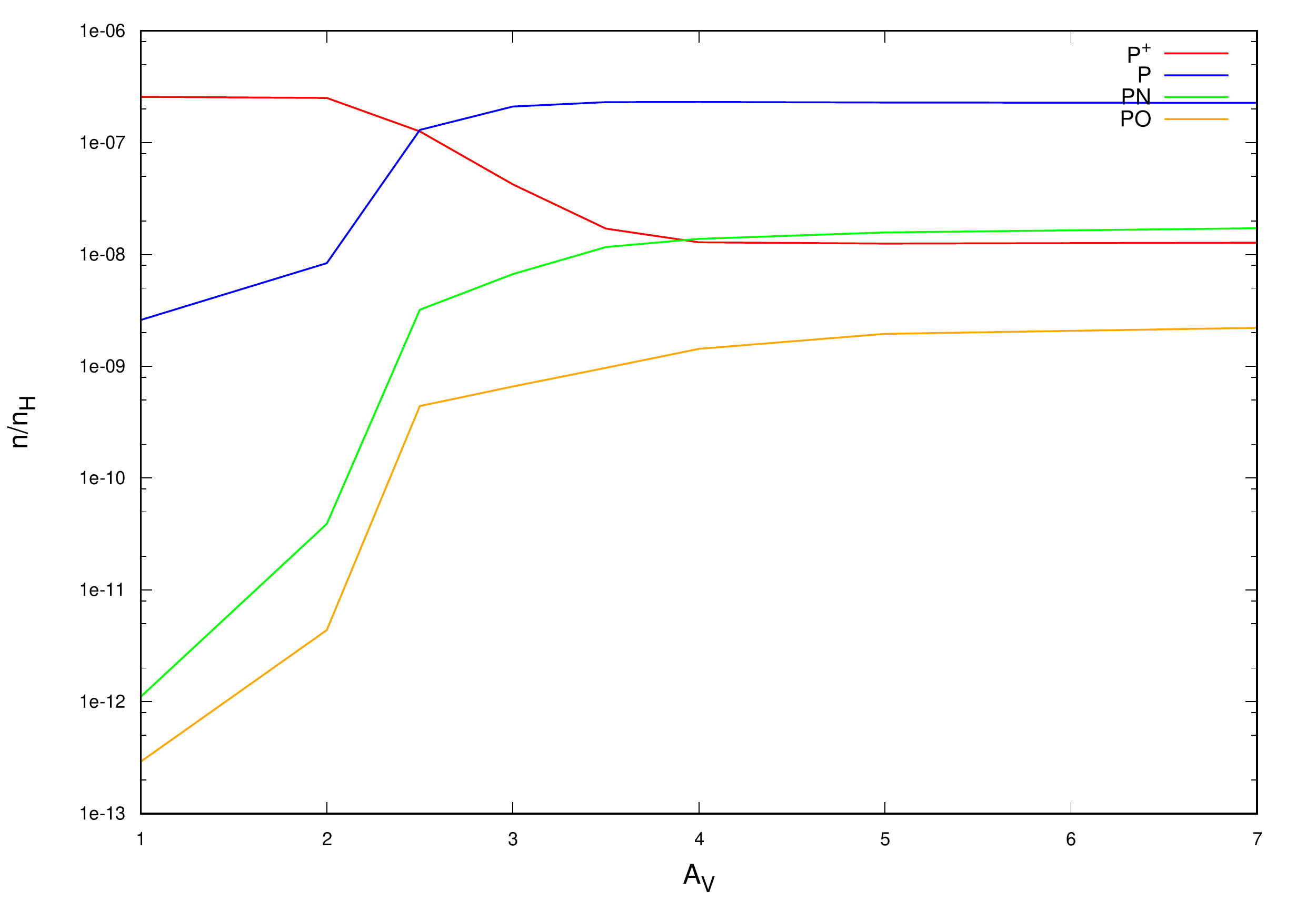}
\caption{The final abundances of P$^+$, P, PO, and PN are shown with A$_V$ with the CMMC code for diffuse cloud. The conversion of P$^+$ to P takes place in the range $A_V=2-3$ mag. Beyond A$_V=4$ mag no changes in the abundances of P$^+$ and P are obtained. The abundance ratio of PO and PN is found to be always $<1$ \citep{sil21}.}
\label{fig:pconv}
\end{figure}

We consider the initial elemental abundances for the diffuse cloud model as in \cite{chan20} (see Table \ref{tab:diff_cloud_CMMC}). The dust-to-gas ratio of $0.01$ is considered in our model. We consider a photodesorption rate of $3 \times 10^{-3}$ molecules per incident UV photon for all the molecules. \cite{ober07} experimentally derived this rate from the laboratory measurement of CO ice. Non-thermal desorption constant $a=0.01$ and a cosmic-ray ionization rate of $1.7 \times 10^{-16}$ s$^{-1}$ are considered. We keep the gas and dust temperatures constant at $40$ K and $20$ K, respectively.

Figure \ref{fig:diff_CMMC} represents the results obtained with the CMMC code for the diffuse cloud region. Here, the solid curve in the figure represents the case when we consider $n_H=100$ cm$^{-3}$, and the dotted curve represents the case when we consider $n_H=500$ cm$^{-3}$. Thus, Figure \ref{fig:diff_CMMC} depicts that in the ranges of $A_V=2-3$ mag and n$_H=100-500$ cm$^{-3}$, we have a better agreement with the observed abundance.

Figure \ref{fig:diff_CMMC_ION_NEUTRAL} shows the chemical evolution of the P-bearing species based on the obtained results.
Here, we do not have a high abundance of PH$_3$ under this situation. This is because of the inclusion of the destruction pathways of PH$_3$ by H and OH in our network. The dashed green curve shows the PH$_3$ when we do not consider its destruction by H and OH in both the gas and ice phases.
Interestingly, we have a couple of orders higher abundance of PH$_3$ in the absence of these destruction pathways. However, these destruction pathways are essential and need to consider before constraining the PH$_3$ abundance.
The right panel of Figure \ref{fig:diff_CMMC_ION_NEUTRAL} depicts that the abundance of P$^+$ remains constant for this case. Figure \ref{fig:pconv} shows the final abundance of P and P$^+$ with a variation of $A_V$. It shows that P$^+$-to-P conversion is possible between $A_V=2$ and $3$ mag and remains unchanged beyond $A_V>4$ mag. For this case, we always get a higher peak abundance of PN compared to PO. At the end of the simulation, \cite{chan20} also obtain a comparatively higher abundance of PN than PO. With the \textsc{Cloudy} code, in Figure \ref{fig:diffuse_phosphorus_CLOUDY},
we also obtain PO/PN $<1$.  The major difference between the models using CMMC code or by \cite{chan20} and the \textsc{Cloudy} code is the consideration of the physical conditions. The \textsc{Cloudy} code considers physical conditions more realistically. Figure \ref{fig:DIFF_HH2} (diffuse cloud results with \textsc{Cloudy}) shows a substantial temperature variation with $A_V$, whereas in the case of the CMMC code, we assume a constant  temperature ($T_{gas}=40$ K and $T_{ice}=20$ K) for all $A_V$. In the CMMC code, dust-to-gas ratio of $\sim 0.01$ is considered, whereas, in the diffuse cloud model using \textsc{Cloudy}, it gives $6.594 \times 10^{-3}$.

\begin{table}
\scriptsize
\centering
\caption{Initial elemental abundance for the hot-core/corino model considered in the CMMC code \citep{wake08,sil21}. \label{tab:init_dense}}
\vskip 0.2cm
\begin{tabular}{cccc}
 \hline
{\bf Element} & {\bf Abundance} & {\bf Element} & {\bf Abundance} \\
\hline
H$_2$     & $0.5$ & Fe$^+$       & $3.00 \times 10^{-9}$\\
He        & $1.40 \times 10^{-1}$ & Na$^+$       & $2.00 \times 10^{-9}$\\
N         & $2.14 \times 10^{-5}$ & Mg$^+$       & $7.00 \times 10^{-9}$\\
O         & $1.76 \times 10^{-4}$ & Cl$^+$       & $1.00 \times 10^{-9}$\\
C$^+$        & $7.30 \times 10^{-5}$ & P$^+$        & $2.00 \times 10^{-10}$\\
S$^+$        & $8.00 \times 10^{-8}$ & F$^+$        & $6.68 \times 10^{-9}$\\
Si$^+$       & $8.00 \times 10^{-9}$ & e$^-$        & $7.31 \times 10^{-5}$\\
\hline
\end{tabular}
\end{table}

\subsubsection{Hot-core / Hot-corino model \label{HC}}
Here, we consider a three-stage starless collapsing cloud model described in \cite{gora20b}. The first stage corresponds to an isothermal ($T_{dust}=T_{gas}=10$ K) collapsing stage where density can increase from $3 \times 10^3$ cm$^{-3}$ to $10^7$ cm$^{-3}$ and the visual extinction parameter ($A_V$) can increase from $2$ to $200$. The second stage corresponds to a warm-up stage where temperature rises from 10 to 200 K, keeping $A_V$ constant at its maximum value in $5 \times 10^4$ years. The last stage is a post-warm-up stage where density, temperature, and visual extinction are constant at their respective highest values and continue for $10^5$ years. Based on the timescale of the initial isothermal collapsing stage, we define the hot-core and hot-corino. In the hot-core case, we use an isothermal collapsing timescale $\sim 10^5$ years, whereas, for the hot-corino case a relatively longer timescale ($\sim 10^6$ years) is used. Thus, the total simulation timescale of $2.5 \times 10^5$ years and $1.15 \times 10^6$ years is considered for the hot-core and hot-corino cases, respectively.
Table \ref{tab:init_dense} shows the low metallic initial elemental abundance \citep{wake08}, which is considered here.

\begin{figure}
\centering
\includegraphics[width=6cm, height=6cm, angle=270]{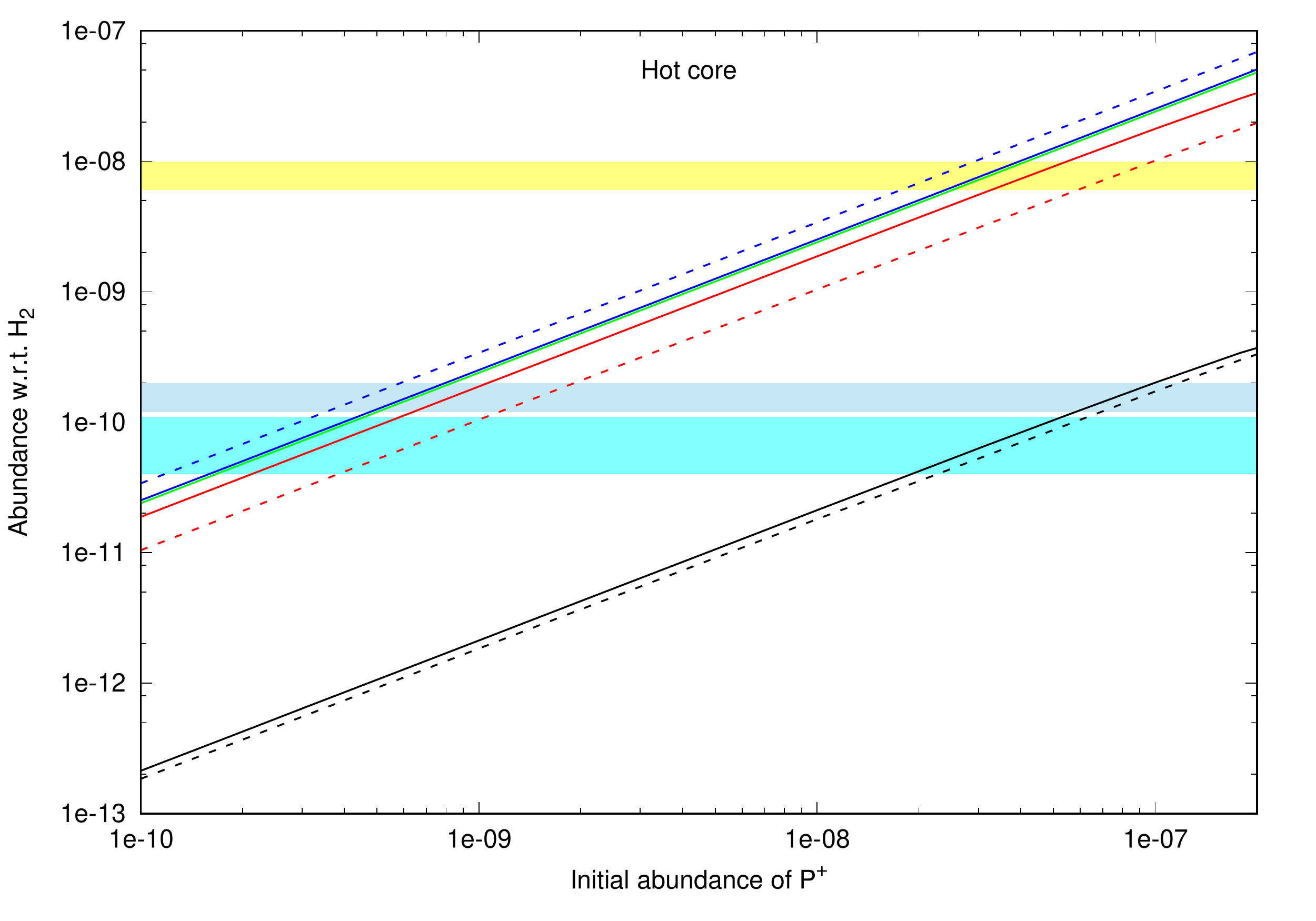}
\includegraphics[width=6cm, height=8.5cm, angle=270]{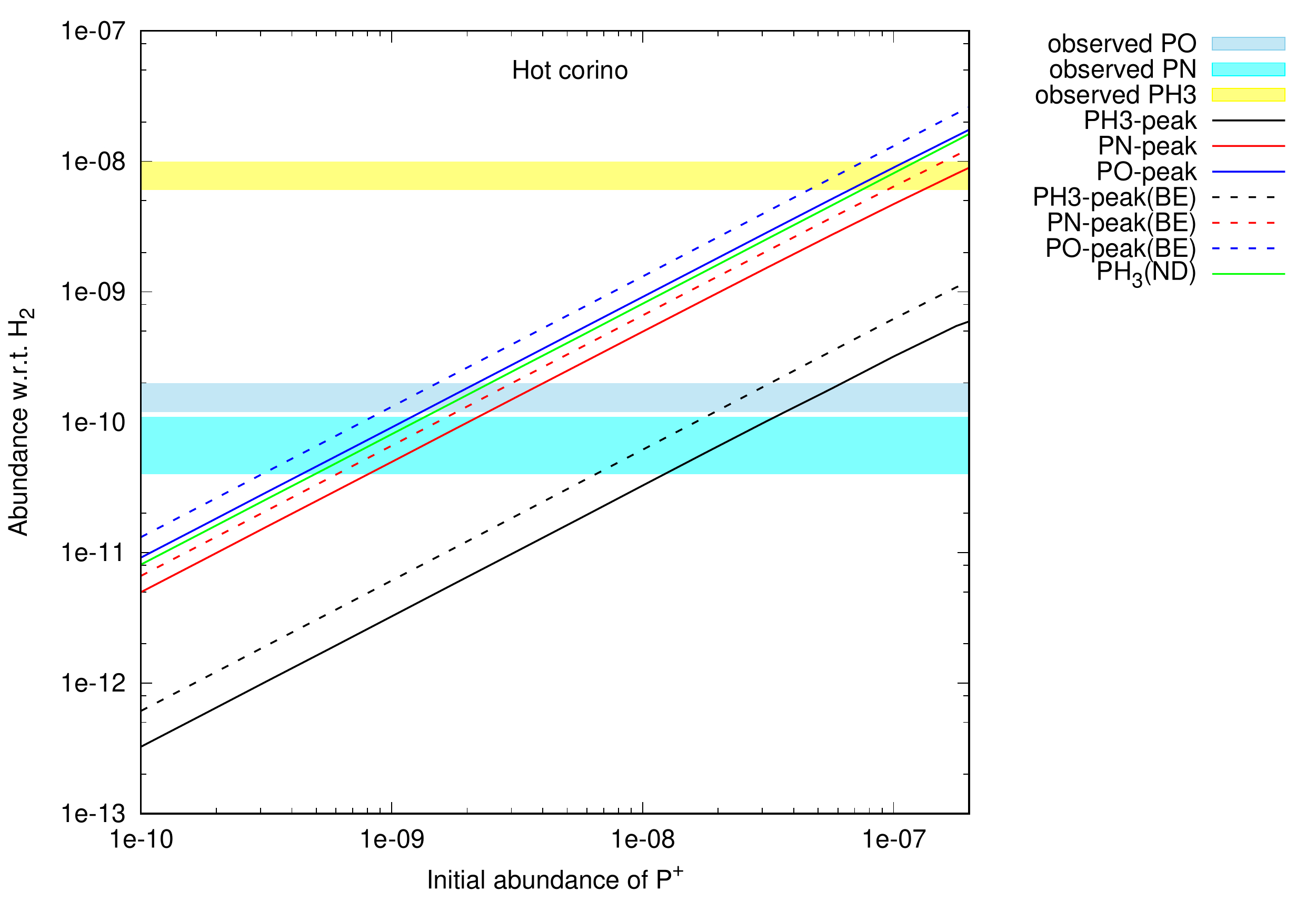}
 \caption{The observed abundances of PO and PN in the high-mass star-forming region W51 and W3(OH) \citep{rivi16} are shown along with our modeled peak abundances (taken beyond the isothermal stage) with hot-core (left panel) and hot-corino (right panel) cases using CMMC code \citep{sil21}. The peak abundance variation of the PH$_3$ is also shown with the initial abundance of P$^+$. PH$_3$ is yet to be observed in hot-core/corino. The obtained peak abundance of PH$_3$ is far below the observed limit of PH$_3$ in C-star envelope IRC +10216 \citep{agun08,agun14}. Peak abundances of PO, PN, and PH$_3$ are also shown with the dashed lines when the BE of the P-bearing species is considered from the tetramer configuration noted in Table \ref{tab:binding}.
The solid green curve shows the peak abundance of PH$_3$ in the absence of its destruction by H and OH. We obtained a significantly higher peak abundance of PH$_3$ with the lack of these destruction pathways.}
\label{fig:depletion}
\end{figure}

\cite{rivi16} reported the observed molecular abundance with an uncertainty of $\sim (0.5-5)\times10^{-10}$ for PO and PN. The ratio between PO and PN is in the range of $1-5$. Figure \ref{fig:depletion} shows the variation of the peak abundance of PO and PN in the case of hot-core and hot-corino by considering various initial elemental P abundances.
The peak abundances of PO and PN from our hot-core (left panel) model show a good match with the observation toward the massive star-forming region when an initial elemental P abundance (P$^+$ abundance) of
$2 \times 10^{-10}- 2 \times 10^{-9}$ is considered.

\begin{figure}
\centering
\includegraphics[width=8cm, height=12cm, angle=270]{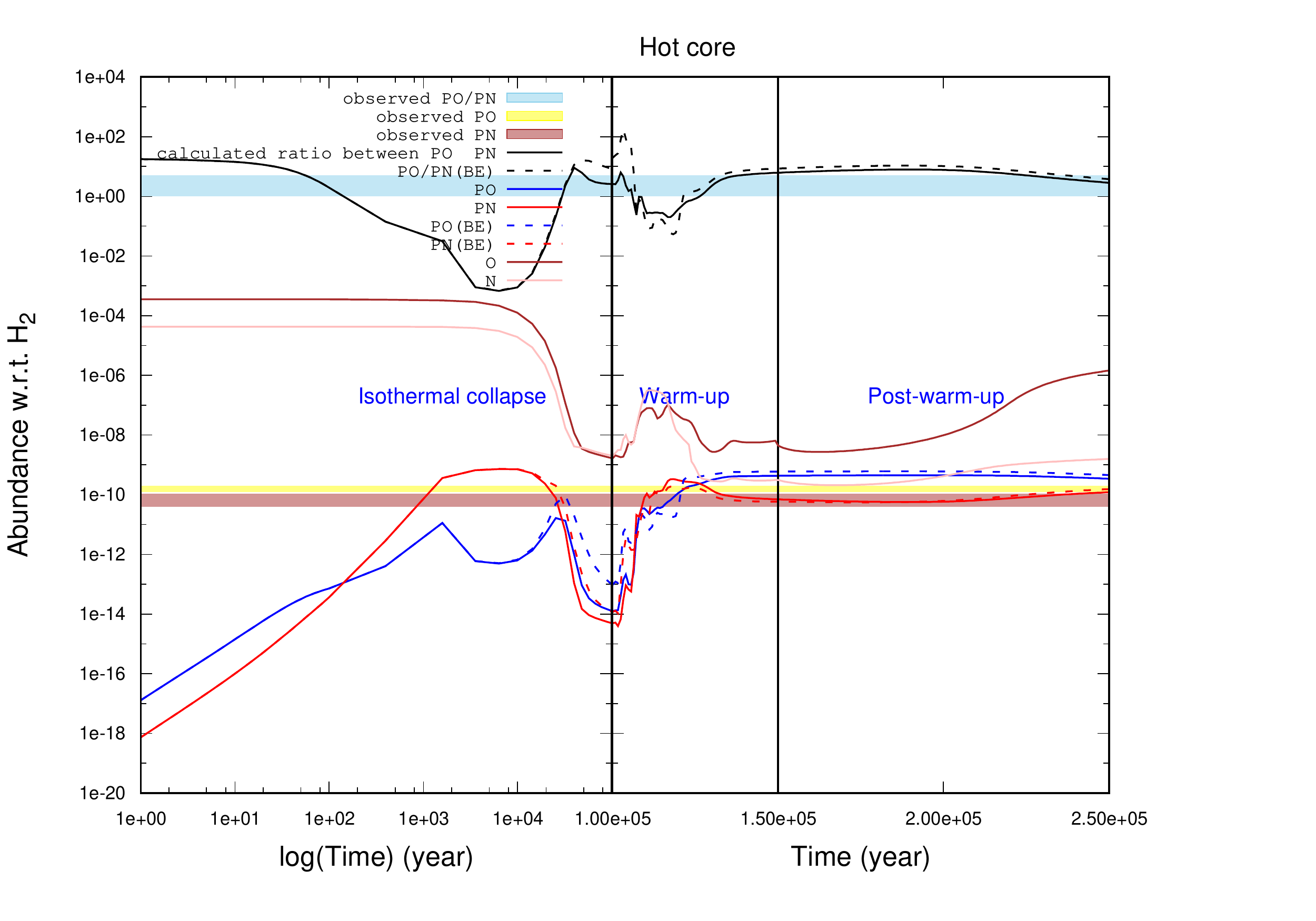}
\includegraphics[width=8cm, height=12cm, angle=270]{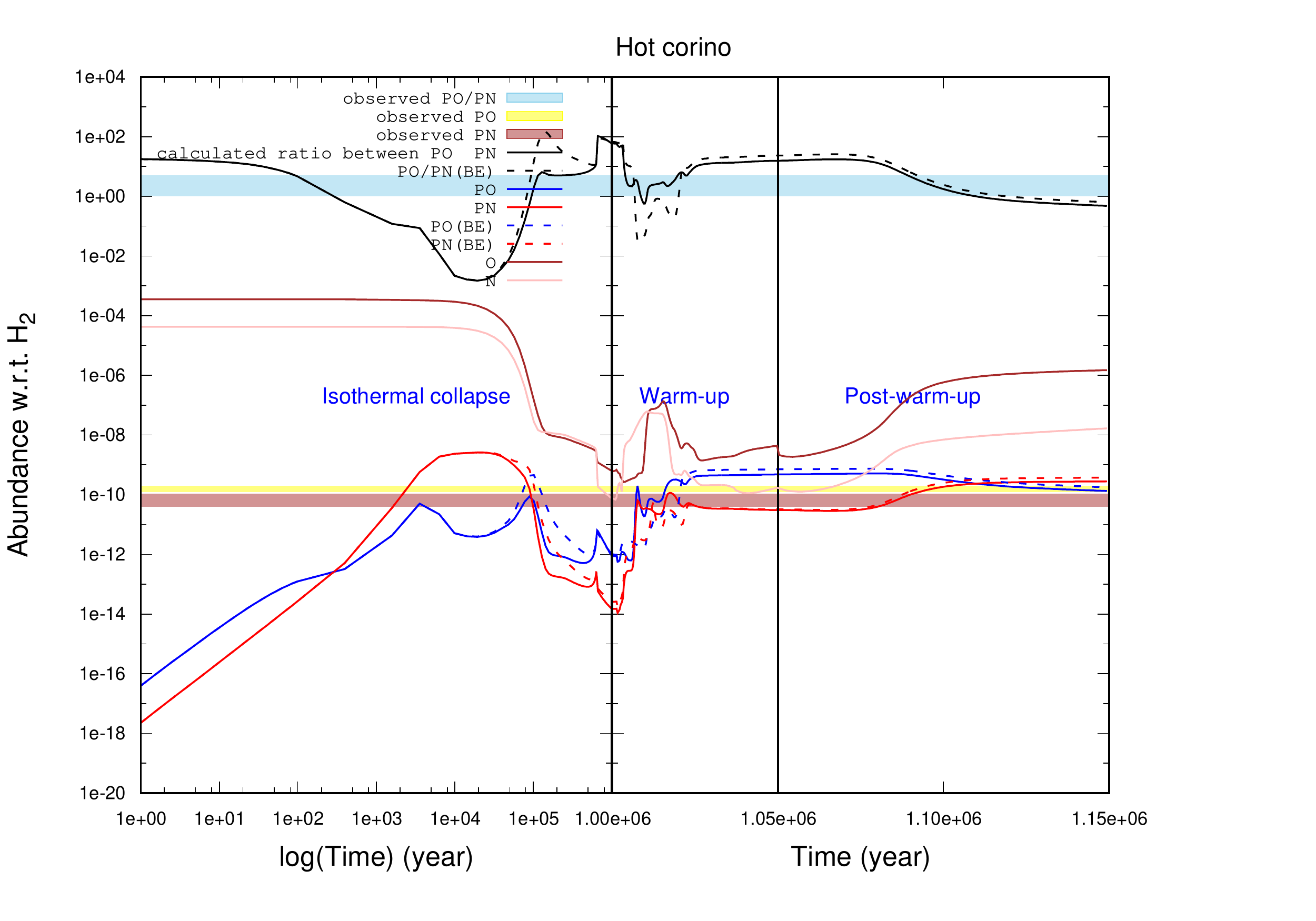}
\caption{Abundance variation of PO, PN, and PO/PN ratio obtained using CMMC code by considering the initial elemental abundance of P$^+$ as $1.8 \times 10^{-9}$ for hot-core and $5.6 \times 10^{-9}$ for the hot-corino case \citep{sil21}. Solid lines represent the cases with the BEs from KIDA, and the dashed lines represent the values with the tetramer configuration of water. During the warm-up and post-warm-up stages, we have a good correlation with our results.}
\label{fig:PObyPN}
\end{figure}

\begin{figure}
\centering
\includegraphics[width=8cm, height=12cm, angle=270]{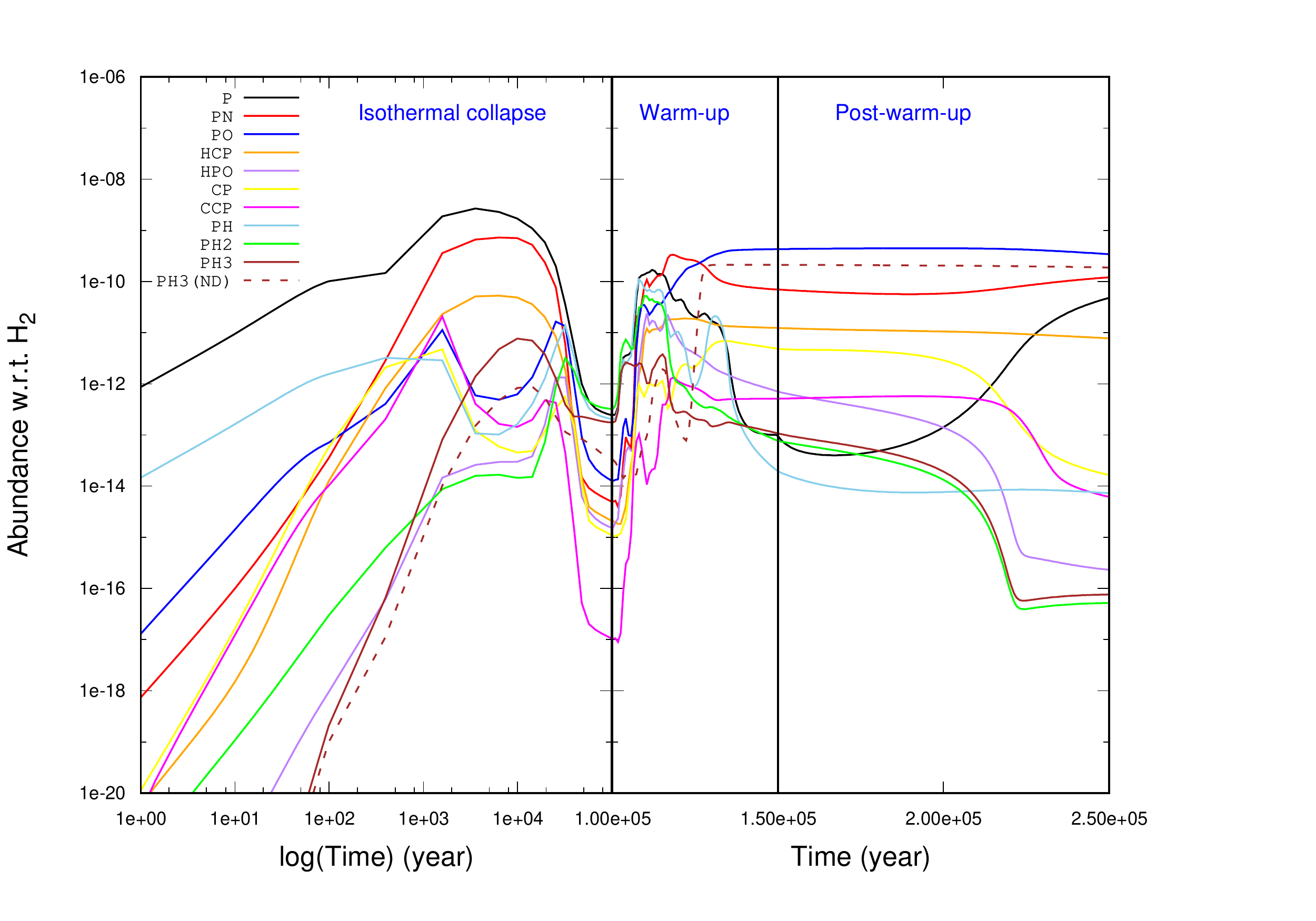}
\includegraphics[width=8cm, height=12cm, angle=270]{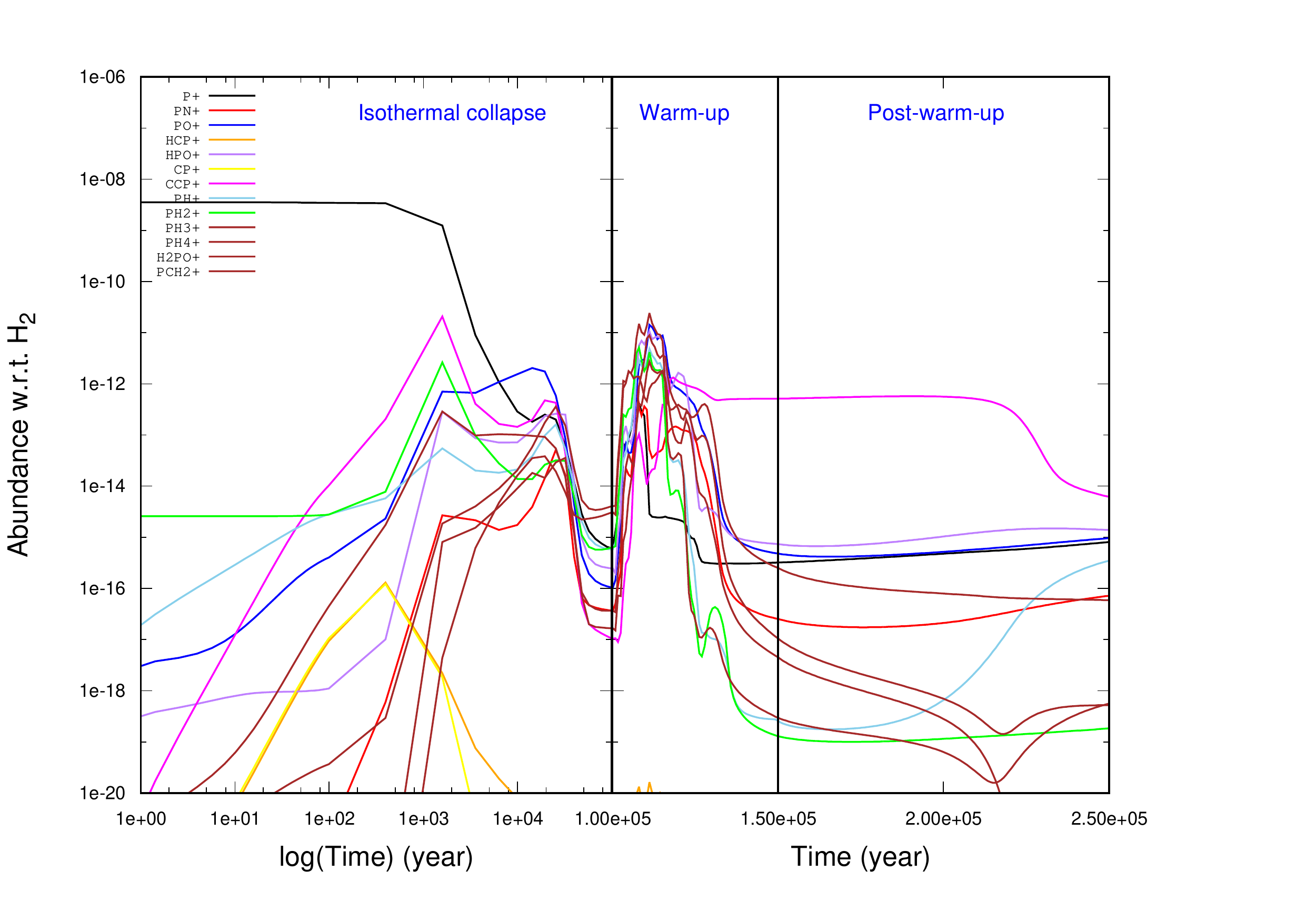}
\caption{Abundance variations of most of the P-bearing neutral and ionic species are shown using CMMC code considering initial elemental abundance of P$^+$ as $1.8 \times 10^{-9}$ and for the hot-core region \citep{sil21}. In the absence of the destruction of PH$_3$ by H and OH, PH$_3$ is highly abundant.}
\label{fig:PObyPN-evo}
\end{figure}

The solid curve of Figure \ref{fig:depletion} represents the BE of the P-bearing species as it is in KIDA.
The dashed curve represents our computed BE values (BE with water c-tetramer) of phosphorous (see the BE values in Table \ref{tab:binding}) and keeps all other BE values the same as in KIDA. We notice an increase in the peak abundances of PO and PN for the inclusion of our calculated BEs for the hot-corino case. We obtain that the peak abundance of PN increases, whereas it decreases for PO for the hot-core case.
The peak abundance of PO is always more prominent than the peak abundance (beyond the isothermal stage) of PN in both cases.
Variation of the PH$_3$ abundance is also shown, which is far below the derived upper limit for the C-rich envelop \citep{agun08,agun14}.
The changes in BE reflect a significant increase in the gas-phase abundance of PH$_3$ for the hot-corino case, but
a marginal decrease in the peak abundance of PH$_3$ is obtained in the hot-core case.
Additionally, in Figure \ref{fig:depletion}, PH$_3$ abundance is shown with a solid green curve (with the BE of KIDA) when its destruction by H and OH is not included. It offers a significantly higher abundance of PH$_3$.

Unless otherwise stated, we always use a nonthermal desorption factor of $a=0.01$. Additionally, we test with a reactive desorption factor of $0.023$ \citep{nguy20} for PH$_3$, which yields a roughly $2.3$ times higher abundance of PH$_3$ in our case.

The solid lines of Figure \ref{fig:PObyPN} show the time evolution of the abundances of PO and PN and PO/PN for the hot-core (top panel) and hot-corino (bottom panel) cases. The dashed lines represent the time evolution of the abundances of PO and PN and the ratio PO/PN by considering the BEs of P-bearing species with the tetramer configuration of water. Finally, we highlight the observed abundances and observed abundance ratio in the high-mass star-forming region \citep{rivi16} to better understand.
For the hot-core case, we consider the initial P$^+$ abundance $\sim 1.8 \times 10^{-9}$, and for the hot-corino case, a comparatively higher initial abundance of P$^+$ ($\sim 5.6 \times 10^{-9}$) is considered to match the observational result.
We obtain an exciting difference between the PO and PN abundances during the later stages of the simulation of the post-warm-up stage.

We obtain PO/PN $> 1$ in the late post-warm-up stage of the hot-core, whereas it shows an opposite trend in the hot-corino case. Thus, the higher abundance of PN in the hot-corino case happens as a result of the presence of a comparatively more elevated amount of atomic nitrogen (shown in Figure \ref{fig:PObyPN} with the pink curve) in this case.
This high abundance of atomic nitrogen undergoes $\rm{PO+ N \rightarrow PN + O}$ and yields PO/PN $<1$ in the hot-corino case. Also, the abundance of atomic P increases owing to the destruction of P-compounds.
Our modeling results for hot-core (with PO/PN $>1$) and hot-corino (with PO/PN $<1$) agree well with the modeling results presented by \cite{jime18}.
The peak abundance of Figure \ref{fig:PObyPN-evo} shows the time evolution of most of the P-bearing species with an initial P$^+$ abundance of $1.8 \times 10^{-9}$. Interestingly, P is mainly locked in PO and PN at the end of the simulation timescale.

\begin{figure}
\centering
\includegraphics[width=0.55\textwidth,angle=270]{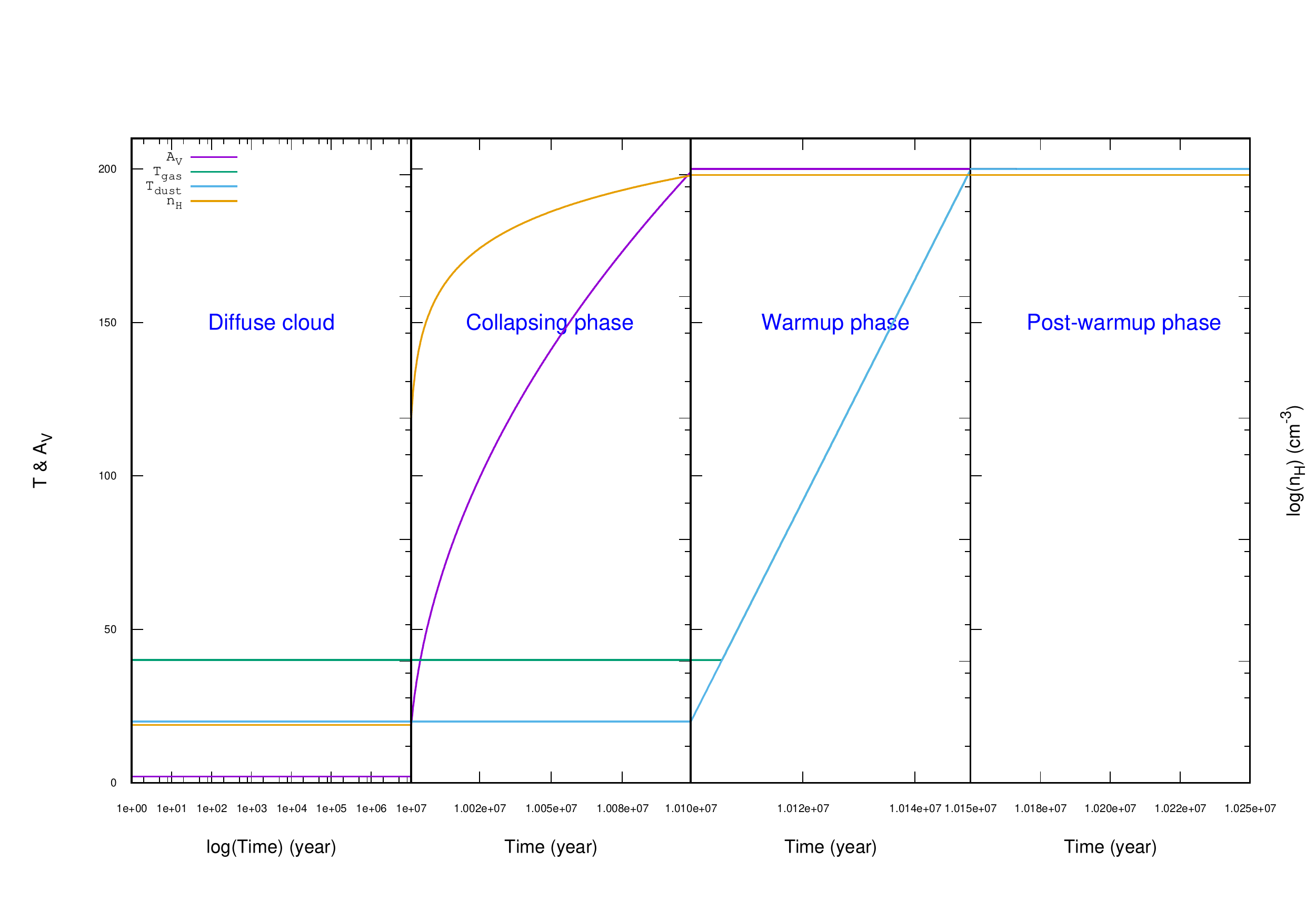}
\caption{Four distinct stages considered for the simulation using CMMC code \citep{sil21}. In the first stage, the cloud remains in the diffuse stage for $10^7$ years. It starts to collapse in the next step, which continues for $10^5$ years. The collapsing stage is followed by a warm-up and post-warm-up stage, which continues for another $1.5 \times 10^5$ years.}
\label{fig:diff_den}
\end{figure}

\begin{figure}
\centering
\includegraphics[width=0.55\textwidth,angle=270]{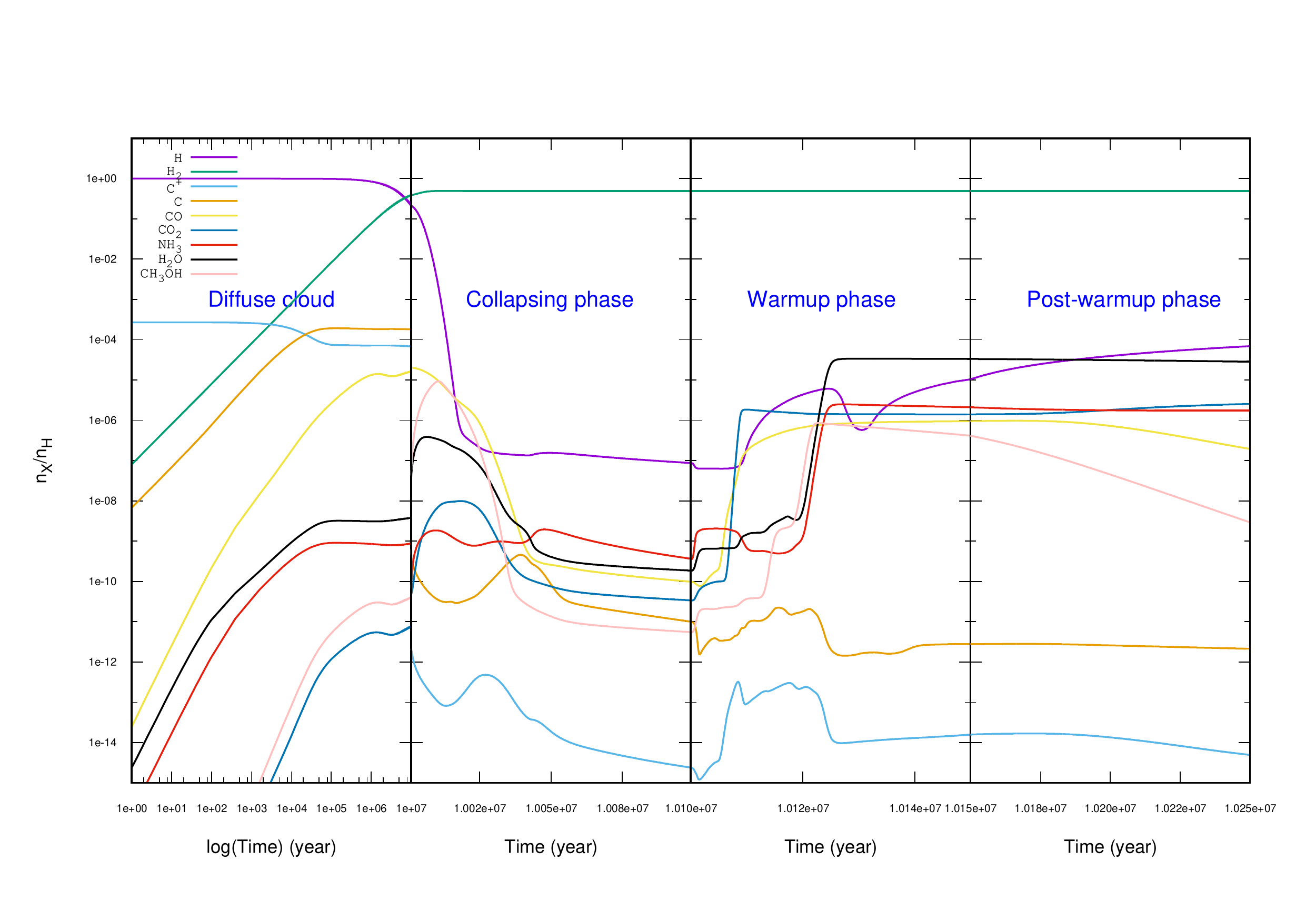}
\caption{Time evolution of H, H$_2$, C$^+$, C, CO, CO$_2$, NH$_3$, H$_2$O, and CH$_3$OH obtained using CMMC code is shown \citep{sil21}. During the lifetime of the diffuse cloud H converts into H$_2$ and C$^+$ converts into C. During the collapsing stage, C atom is heavily depleted and converts into CO. }
\label{fig:mol_diff_den}
\end{figure}

\begin{figure}
\centering
\includegraphics[width=0.55\textwidth,angle=270]{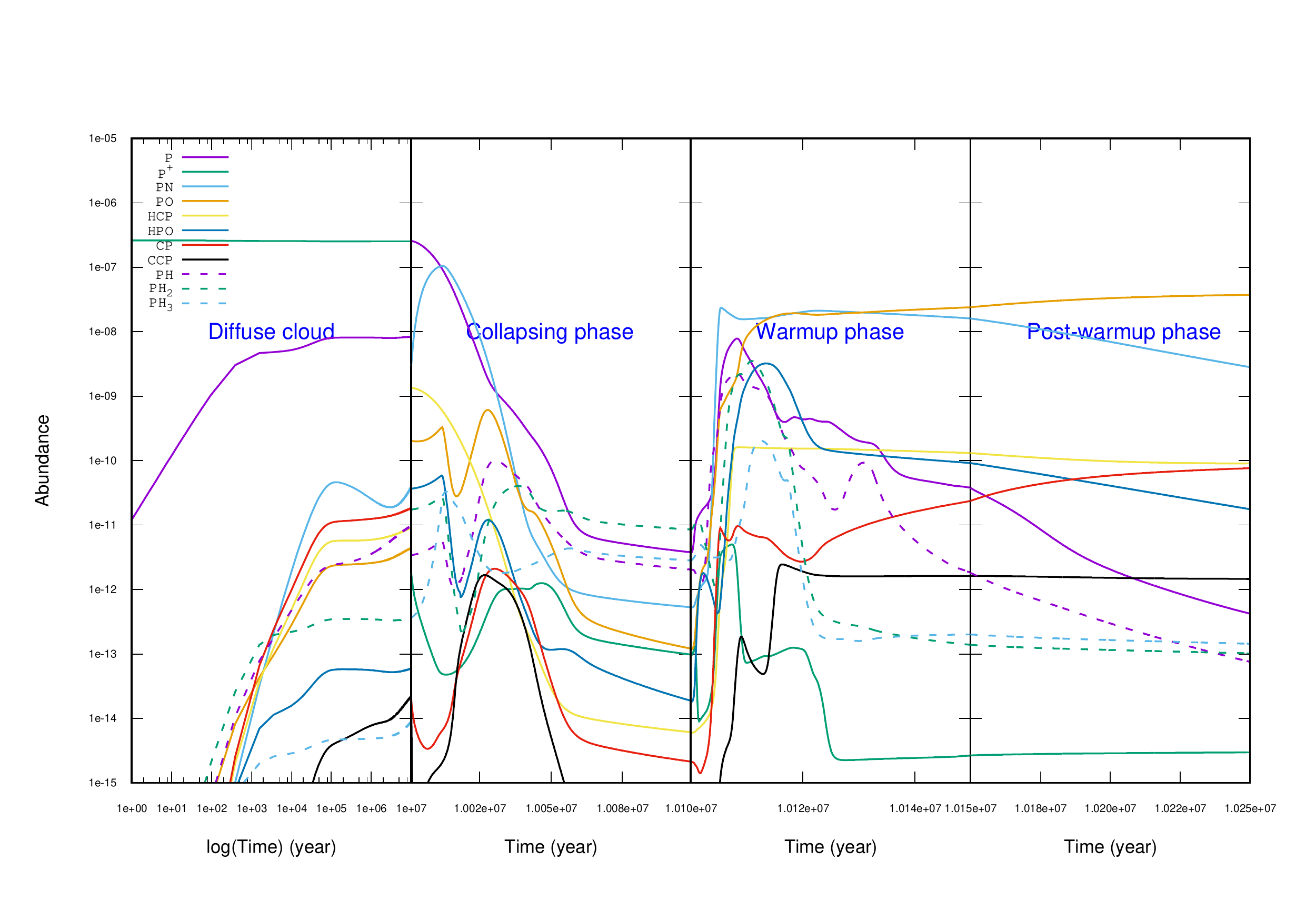}
\caption{Time evolution of the abundances (concerning total hydrogen nuclei in all forms) of major P-bearing species using CMMC code is shown \citep{sil21}.}
\label{fig:P_diff_den}
\end{figure}

\subsubsection{Diffuse to Dense cloud model}
To avoid any ambiguity for considering the depletion factor, we consider that the diffuse cloud converts into a molecular cloud after a sufficient time. The adopted physical condition for this model is shown in Figure \ref{fig:diff_den}.
It depicts that the total simulation time is divided into four steps. The first step is the diffuse stage, and it continues for the initial $10^7$ years. The second step is the collapsing stage, whose span is for $10^5$ years. Then, the warm-up stage starts after the collapsing stage, and it lasts for $5 \times 10^4$ years. Finally, the post-warm-up stage continues for another $10^5$ years. In total, the simulation timescale is $1.025 \times 10^7$ years.
Initial elemental abundance for this case is taken from \cite{chan20} (see Table \ref{tab:diff_cloud_CMMC}). All the BEs are used from the KIDA.
Figure \ref{fig:mol_diff_den} shows the time evolution of the H, H$_2$, C$^+$, C, and CO. Interestingly, most of the H atoms convert into H$_2$ during the lifetime of the diffuse cloud ($\sim 10^7$ years). Ionized carbon is converted into neutral carbon. During the collapsing stage, the conversion of C to CO takes place. Once CO has formed, it starts to deplete on the grain and forms COMs. In the warm-up stage, these CO and its related molecules are desorbed back to the gas phase. Figure \ref{fig:P_diff_den} shows the chemical evolution of the notable P-bearing species. Figure \ref{fig:P_diff_den} shows that in the warm-up stage we have very high abundances of PN and PO. The PH$_3$ abundance is also significantly higher ($\sim 10^{-10}$ concerning total H nuclei), comparable with \cite{jime18}. At the end of the simulation, most of the P is locked as PO and PN.

\subsection{An outline of our modeling results and a comparison with the earlier results}
Table \ref{tab:abundances} shows the peak abundances of the significant P-bearing species obtained with the two model codes (\textsc{Cloudy} and CMMC). However, due to more realistic physical conditions, the \textsc{Cloudy} code is preferred for the diffuse cloud region.
Both the diffuse cloud models with \textsc{Cloudy} and CMMC codes show higher abundance of PN than PO. \cite{chan20} also predicted
a higher abundance of PN ($4.8 \times 10^{-11}$) compared to PO  ($1.4 \times 10^{-11}$) in the diffuse cloud region.
The abundances of PH$ _3$ dramatically vary between the two models. This is because \textsc{Cloudy} does not consider the grain surface reactions for the formation of PH$_3$, whereas the grain chemistry is adequately considered in CMMC. Moreover, the adopted physical condition differs between the two models. We obtain a higher abundance of P-bearing species for our diffuse-dense model because of the consideration of the initially high abundance of P$^+$ ($\sim 2.6 \times 10^{-7}$).
In general, we obtain a nice trend with the peak abundance ratio between PO and PN. We find PO/PN ratio $<1$ for the diffuse cloud region, $<<1$ for the PDR, and $>1$ for the hot-core/cornio (late warm-up stage) region. We have PO/PN $>1$ for the hot-core and $<1$ for the hot-corino in the late post-warm-up stage. We notice that the reaction
$\rm{PO+N \rightarrow PN + O}$ is mainly responsible for controlling this ratio.

Our diffuse cloud modeling results with the \textsc{Cloudy} and CMMC code are in line with those obtained by \cite{chan20}. The significant difference between our diffuse cloud model obtained with the \textsc{Cloudy} code and the model presented in \cite{chan20} is in consideration of realistic physical conditions instead of just considering the constant temperature. In contrast, the diffuse cloud model obtained with the CMMC code adopts similar physical parameters to those considered in \cite{chan20}. Since \textsc{Cloudy} is not equipped with adequate surface chemistry treatment, we do not have a notable abundance of PH$_3$. Our CMMC results show lower PH$_3$ because of its destruction by H and OH.

\cite{char94} studied the chemistry of P in hot molecular cores. They considered that the gas-phase P-chemistry in the hot-core starts from PH$_3$ release from the interstellar grains. In their case, PH$_3$ was gradually destroyed and transformed into P, PO, and PN. They noticed that the formation timescale of other P-bearing species is longer than those of the hot-core. We consider the in situ formation of PH$_3$ via chemical reaction on interstellar grains. Like \cite{char94}, in Figure \ref{fig:PObyPN-evo}, we also notice that at the end of the warm-up stage the abundances of most of the P-bearing species remain in the form of P, PO, and PN.

\cite{rivi16} found that PO and PN are chemically associated and formed during the cold collapse stage by the gas-phase reactions. At the end of the collapsing stage, these two species are deposited to the interstellar grain. These again desorb back to the gas phase during the warm-up stage when the temperature is around 35 K. Our model result using CMMC code shown in Figure \ref{fig:PObyPN} reflects the similar behavior of PO and PN. In the isothermal collapsing stage, the gas-phase abundance of these two species offers a higher abundance, while at the end of this stage, these two are depleted to the grain. Once the temperature has become $>35$ K in the warm-up stage, it desorbed back to the gas phase.

\begin{landscape}
\begin{table}
\scriptsize
\caption{Peak abundances of the P-bearing species under various physical conditions \citep{sil21}. \label{tab:abundances}}
\vskip 0.2cm
\hskip -3.5cm
\begin{tabular}{ccccccc}
\hline
{\bf Species} & \multicolumn{2}{c}{\bf \textsc{Cloudy} output}&\multicolumn{4}{c}{\bf CMMC output$^a$} \\
& {\bf Diffuse cloud ($n/n_H$)} & {\bf PDR ($n/n_H$)} & {\bf Diffuse cloud ($n/n_H$)}  & {\bf Hot-core $(n/n_{H_2}$)} & {\bf Hot-corino $(n/n_{H_2}$)} & {\bf Diffuse-Dense ($n/n_{H_2}$)} \\
 &{\bf ($A_V=2$, $n_H=300$ cm$^{-3}$)} & {\bf ($A_V=5$, $T_{gas}=50$ K,} & {\bf ($A_V=2$, $T_{gas}=40$ K,} & {\bf (initial P$^+ \sim 1.8 \times 10^{-9}$)} & {\bf (initial P$^+ \sim 5.6 \times 10^{-9}$)} & {\bf (initial P$^+ \sim 2.6 \times 10^{-7}$)} \\
 &  & {\bf $T_{ice}=20$ K, $n_H=10^{5.5}$ cm$^{-3}$)} & {\bf $T_{ice}=20$ K, $n_H=300$ cm$^{-3}$)} & & \\
\hline
 PN&$3.33\times10^{-10}$& $4.02\times10^{-09}$ & $4.6 \times 10^{-11}$ & $1.7 \times 10^{-10} (2.9 \times 10^{-11})$ & $1.4 \times 10^{-10} (8.0 \times 10^{-11})$ & $1.04 \times 10^{-7}$ \\
 PO & $1.47\times10^{-10}$ & $2.60\times10^{-11}$ & $4.4 \times 10^{-12}$ & $2.3 \times 10^{-10} (4.9 \times 10^{-11})$ & $2.6 \times 10^{-10} (6.6 \times 10^{-11})$ & $3.7 \times 10^{-8}$ \\
PH & $1.13\times10^{-10}$ & $1.82\times10^{-11}$ & $9.6 \times 10^{-12}$ & $5.5 \times 10^{-11} (1.5 \times 10^{-11}$)& $6.5 \times 10^{-11} (8.9 \times 10^{-12})$ & $2.3 \times 10^{-9}$ \\
PH$_3$ & $6.30\times10^{-31}$& $1.16\times10^{-22}$ & $8.9 \times 10^{-15}$ & $1.9 \times 10^{-12} (2.1 \times 10^{-10})$ & $9.2 \times 10^{-12} (2.3 \times 10^{-10})$ & $2.0 \times 10^{-10}$ \\
CP & $2.03\times10^{-11}$ & $3.24\times10^{-12}$ & $1.8 \times 10^{-11}$ & $3.5 \times 10^{-12} (6.7 \times 10^{-13})$ & $3.1 \times 10^{-11} (3.2 \times 10^{-12})$ & $1.2 \times 10^{-10}$ \\
HCP & $1.38\times10^{-13}$ & $3.85\times10^{-14}$ & $9.1 \times 10^{-12}$ & $9.5 \times 10^{-12} (3.7 \times 10^{-12})$ & $3.5 \times 10^{-11} (4.3 \times 10^{-12})$ & $1.3 \times 10^{-9}$ \\
HPO & $4.43\times10^{-17}$ & $1.28\times10^{-19}$ & $5.9 \times 10^{-14}$ & $1.2 \times 10^{-11} (1.6 \times 10^{-12})$ & $1.5 \times 10^{-11} (9.8 \times 10^{-13})$ & $3.2 \times 10^{-9}$ \\
\hline
\end{tabular} \\
\vskip 0.2 cm
{\bf Note:} \\
$^a$ The bracketed values from the CMMC output denote the case when the destruction of PH$_3$ by H and OH are ignored.
\end{table}
\end{landscape}

\cite{jime18} constructed their model to explain the abundances of P-bearing species in a wide range of astrophysical conditions.
They constructed a short-lived and long-lived chemical model depending on the time of the collapse. In the short-lived collapse, they stopped their calculations when the gas density reaches its maximum value. In the long-lived collapse, they considered some additional time after getting the final density. They noticed that the P and PN are the most abundant P-bearing species in the collapsing stage. Their maximum abundance is in the range $(5-10) \times 10^{-10}$. From our hot-core model, in the collapsing stage (see Figure \ref{fig:PObyPN-evo}), we also obtain that P and PN remain the most abundant P-bearing species, and their peak abundance varies in the range $(7-27) \times 10^{-10}$.

\cite{jime16} carried out a high-sensitivity observation toward the L1544 pre-stellar core. Unfortunately, they were not able to identify the PN transitions. However, they predicted an upper limit of $\sim 4.6 \times 10^{-13}$ for the abundance of PN. The first stage of Figure \ref{fig:PObyPN-evo} represents the isothermal collapsing stage of a hot-core. At the end of the collapsing stage, we obtain a PN abundance of $\sim 5.4 \times 10^{-15}$, consistent with this upper limit.

\cite{font16} identified PN in various dense cores. These dense cores belong to various evolutionary stages (starless, with a protostellar object,
and with ultracompact H\,{\sc ii} region) of intermediate- and high-mass stars. They obtained all the transitions of PN where the temperature is $<100$ K, and line widths are $<5$ km s$^{-1}$, suggesting that the origin of PN is the quiescent and cold region. The abundance of PN was not derived because of the lack of thermal dust continuum emission (at the millimeter/submillimeter regime for all these sources). \cite{mini18} identified a few transitions of PN toward some of these sources (a sample of nine massive dense cores in different evolutionary stages). They calculated the H$_2$ total column densities of the sources and derived the abundances of PN.
For the slightly warmer region ($25-30$ K), \cite{font16} and \cite{mini18} obtained the abundances of PN of $10^{-11}$ and $5 \times 10^{-12}$, respectively. Figure \ref{fig:PObyPN-evo} depicts that when the temperature is $\sim 35$ K in the warm-up stage, we have a little higher PN abundance $\sim 2.08 \times 10^{-11}$ with respect to H$_2$.
This is because \cite{mini18} found the excitation temperature of PN $\sim 5-30$ K. Since the average density ($\sim10^{4-5}$ cm$^{-3}$) of their targeted regions are below the critical density ($10^{5-6}$ cm$^{-3}$) of the PN, they suggested a subthermal excitation of PN. This is because the total hydrogen density can reach as high as $10^7$ cm$^{-3}$ in our isothermal stage.
In the warm-up stage, we consider the same density.
So, in our case, we have an average density, which is greater than the critical
density of PN. So, a direct comparison between our model and the observation of \cite{mini18} and \cite{font16} would not be appropriate.
Here, we refer to these observations to infer the enhancement of the PN abundance with an increase in temperature from $10$ K to $35$ K only.

After releasing to the gas phase, PH$_3$ can be destroyed rapidly \citep{jime18}.
The gas-phase formation of PH$_3$ can continue by the reactions R144 and R145 of Table \ref{tab:reaction_path}. These two reactions were considered in \cite{jime18} and were
found to contribute marginally. Here, we have some additional destruction reactions of PH$_3$ by H (grain-phase reaction R4 and gas-phase reaction R149 of Table \ref{tab:reaction_path}) and
OH (grain-phase reaction R5 and gas-phase reaction R161 of Table \ref{tab:reaction_path}), which yields a much lower PH$_3$ in the gas phase.

In Table \ref{tab:PO/PN}, a summary of the PO/PN abundance ratio is listed in the different astrophysical environments, along with a comparison with the earlier literature \citep{tene07,aota12,debe13,font16,lefl16,rivi16,rivi18,rivi20,jime18,berg19,chan20,bern21}.
Our modeling results agree well with the observed \citep{rivi16} and modeled \citep{jime18,chan20} PO/PN ratios.

\begin{table}
\tiny
\caption{Summary of the obtained abundance ratio between PO and PN in different astrophysical environments \citep{sil21}. \label{tab:PO/PN}}
\vskip 0.2cm
\hskip -1.5cm
\begin{tabular}{ccccccccc}
\hline
{\bf References} & {\bf Type of study} & {\bf PDR} & {\bf Diffuse cloud} & {\bf Hot-core} & {\bf Hot-corino} & {\bf Shock} & {\bf Comet} & {\bf Other} \\
\hline
This work & Modeling & $<<1$ & $<1$         & $\ge 1^l$& $\ge 1^m$, $\le 1^n$ & - & - & - \\
\cite{bern21} & Observation & - & - & - & - & - & - & 2.7$^a$ \\
\cite{rivi20}    & Observation & -  & -  & - & - & - & $>10^b$ & 1.7$^c$ \\
\cite{chan20} & Modeling & & $<1$ & & & & & \\
\cite{berg19} & Observation & - & - & - & - & - & - & $(1-3)^d$ \\
\cite{rivi18} & Observation & - & - & - & - & - & - & $\sim1.5^e$ \\
\cite{jime18} & Modeling & $<<1$ & - & $\geq 1$ & $<1$ & $\leq 1$ & - & - \\
\cite{lefl16} & Observation & - & - & - & - & $\approx3^f$ & - & - \\
\cite{rivi16} & Observation & - & - & $1.8^g$, $3^h$ & - & - & - & - \\
\cite{font16} & Observation & - & - & - & - & - & - & $<(1.3-4.5)^i$ \\
\cite{debe13} & Observation & - & - & - & - & - & - & $0.17-2^j$ \\
\cite{aota12} & Modeling  & - & - & - & - & $\sim(0.5-1.3)^{f,i}$ & - & - \\
\cite{tene07} & Observation & - & - & - & - & - & - & $2.2^k$ \\
\hline
\end{tabular} \\
\vskip 0.2cm
{\bf Note:} \\
$^a$ Orion-KL. \\
$^b$ 67P/C-G \citep{altw16}. \\
$^c$ Average over multiple positions and velocity components toward AFGL 5142. \\
$^d$ Class I low-mass protostar B1-a. \\
$^e$ G+0.693-0.03. \\
$^f$ L1157-B1. \\
$^g$ W51 e1/e2. \\
$^h$ W3(OH). \\
$^i$ Taking minimum and maximum values of upper limit of beam-averagd column densities of multiple sources. \\
$^j$ IK Tauri. \\
$^k$ VY Canis Majoris.\\
$^l$ For the late warm-up to post-warm-up stage of the hot-core model.\\
$^m$ During the late warm-up to the initial post-warm-up of hot-corino stage.\\
$^n$ During the late post warm-up stage.
\end{table}

\subsection{IR spectroscopy of PH$_3$} \label{sec:IR_spectra}

\begin{figure}
\centering
\includegraphics[width=0.7\textwidth]{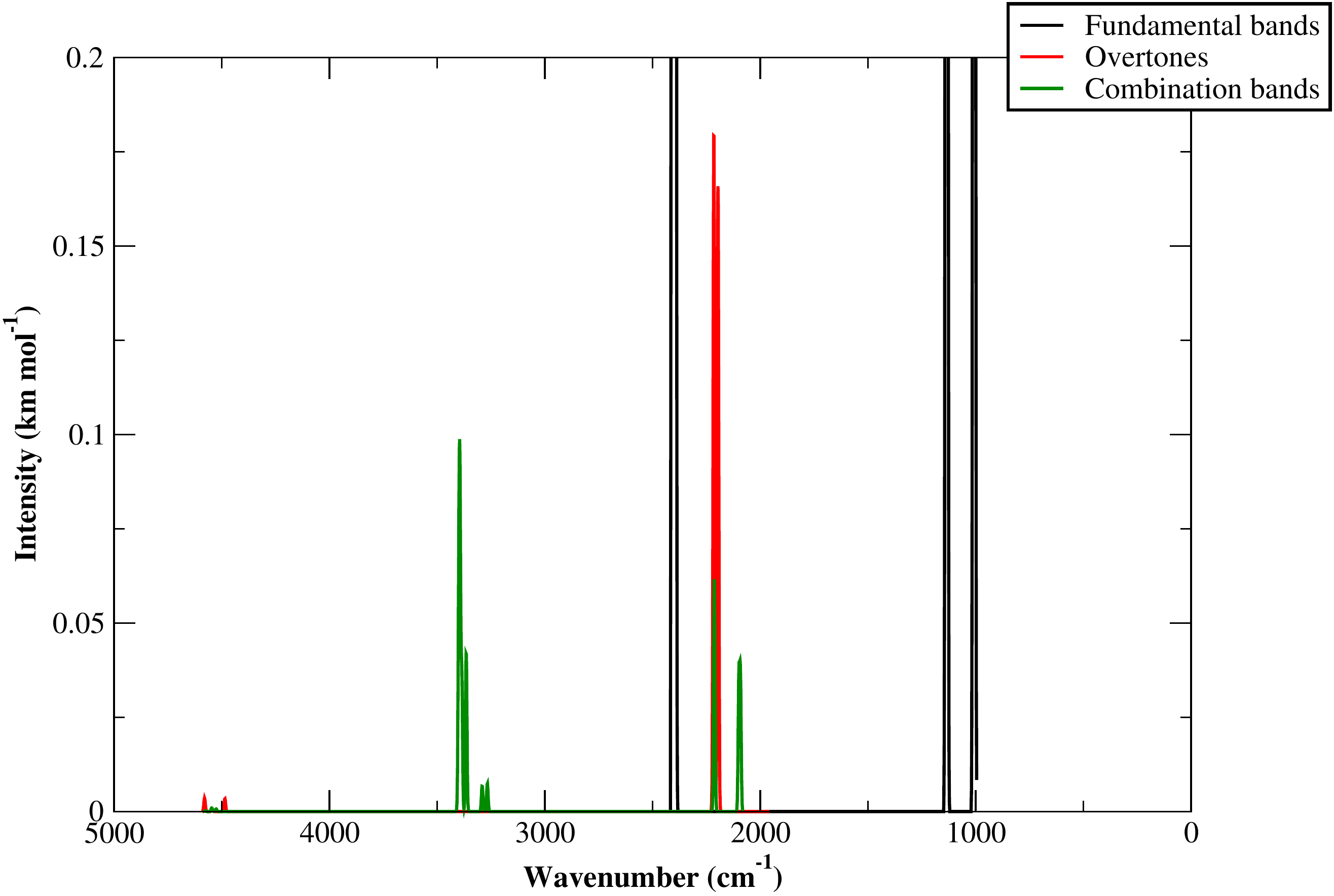}
\caption{Ice-phase IR absorption spectra of PH$_3$ including the harmonic fundamental bands and anharmonic overtones and combination bands \citep{sil21}.}
\label{fig:IR-PH3}
\end{figure}

Phosphine is an important reservoir of interstellar phosphorous \citep{char94}.
Since it is of particular interest to astronomers, the IR vibrational spectra analysis of PH$_3$ could be helpful to the community.
The vibrational spectra of PH$_3$ would be beneficial for future astronomical observations with the JWST. To precisely estimate the frequency and interpret the intensities, it is necessary to go beyond the harmonic approximation. The anharmonic calculations show a significant deviation from the harmonic calculations. Another advantage of the anharmonic analysis is that overtones and combination bands can also be analyzed with this. When anharmonicity is considered, its intensity vanishes at the harmonic level.
Within the DFT approach, the standard B3LYP functional \citep{beck93} is used in conjunction with the 6-311G(d,p) basis set to investigate the IR feature of PH$_3$. In addition, all DFT computations are performed employing the \textsc{Gaussian} 09 suite of programs for quantum chemistry \citep{fris13}.
Figure \ref{fig:IR-PH3} shows the calculated IR feature of the ice-phase PH$_3$.
The PH$_3$ molecule is embedded in a continuum solvation field to represent local effects to mimic the ice features. The IEF variant of the PCM as a default SCRF method is employed with water as a solvent \citep{canc97,toma05}. The implicit solvation
model places the molecule of interest inside a cavity in a continuous homogeneous dielectric medium representing the solvent.
The fundamental modes of vibration, along with the overtones and combination bands, are shown in Figure \ref{fig:IR-PH3}. Table \ref{tab:IR} also notes down the wavenumbers (in cm$^{-1}$) and corresponding absorption coefficients (IR intensities in cm molecule$^{-1}$) of the fundamental bands, overtones, and combination bands.

A comparison between our computed spectra and those obtained experimentally by \cite{turn15} is shown in Table \ref{tab:IR}.
The computed vibrational frequencies are often scaled to resemble the experimental results. This scaling factor varies with the choice of the basis sets and implemented method. The NIST database\footnote{\url{https://cccbdb.nist.gov/vibscalejust.asp}} noted some of the scaling factors, which are helpful to compare with the experimental values.
Based on our method and the chosen basis set, we use a vibrational scaling factor of $\sim 0.967$. \cite{puzz14b} demonstrated that the best state-of-the-art theoretical estimation accuracy is about $\rm{5 - 10 \ cm^{-1} \ and \ 10 - 20 \ cm^{-1}}$ for fundamental
and nonfundamental transitions, respectively. This accuracy allows for reliable simulations of IR
spectra supporting astronomical observations.
The harmonic frequencies presented in Table \ref{tab:IR} agree with the experimental values noted in \cite{turn15} within a limit of about $\rm{1 - 15 \ cm^{-1}}$ after scaling. Our calculated values agree well with the experimented values for the overtones and combination bands, even without scaling.

\begin{figure}
\centering
\includegraphics[width=0.7\textwidth]{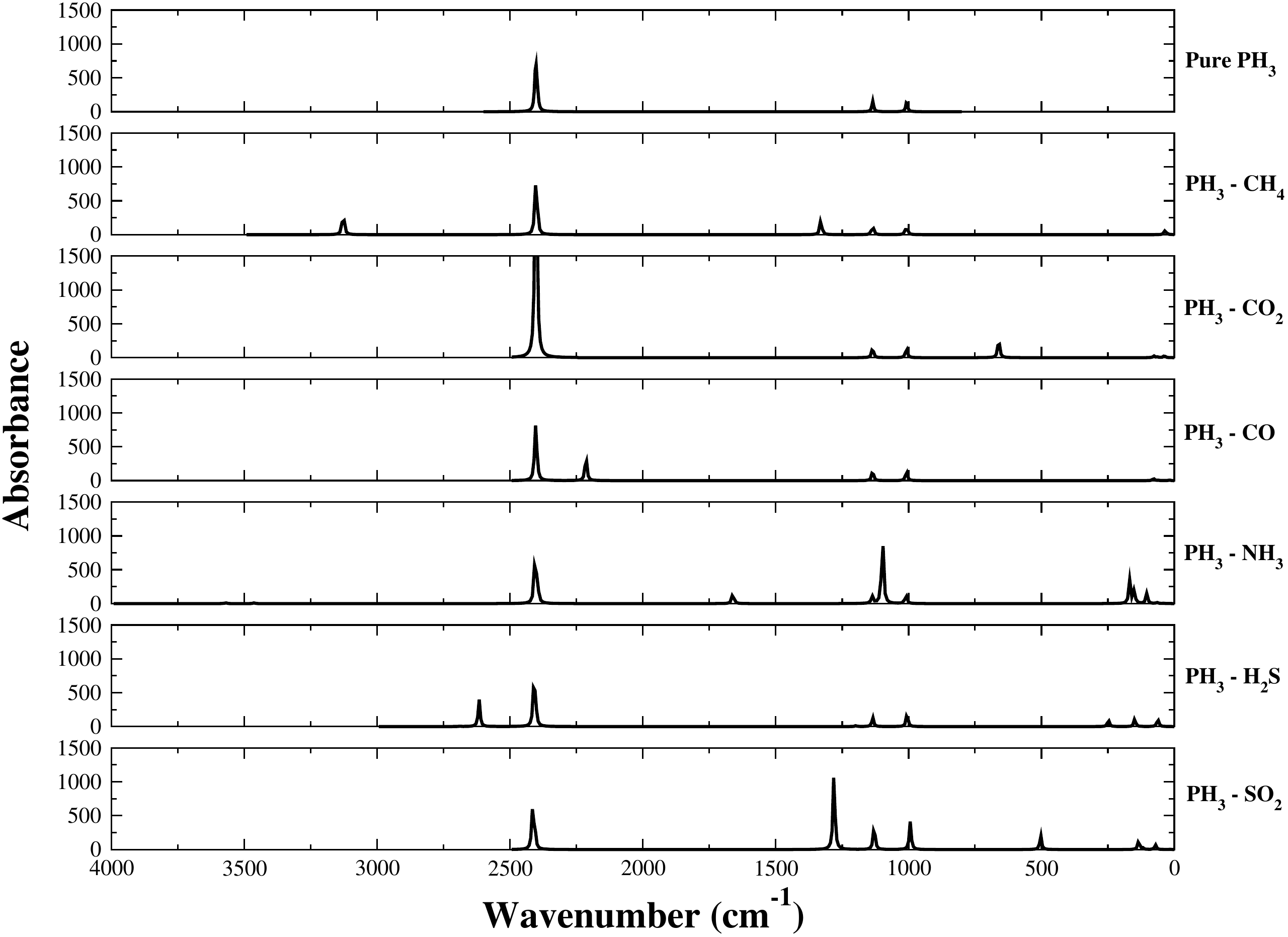}
\caption{Comparison of ice-phase IR absorption spectra of pure PH$_3$ and mixture of PH$_3$ with other volatiles (CH$_4$, CO$_2$, CO, NH$_3$, H$_2$S, and SO$_2$) considering the harmonic fundamental bands \citep{sil21}.}
\label{fig:IR-imp}
\end{figure}

\begin{figure}
\centering
\includegraphics[width=0.7\textwidth]{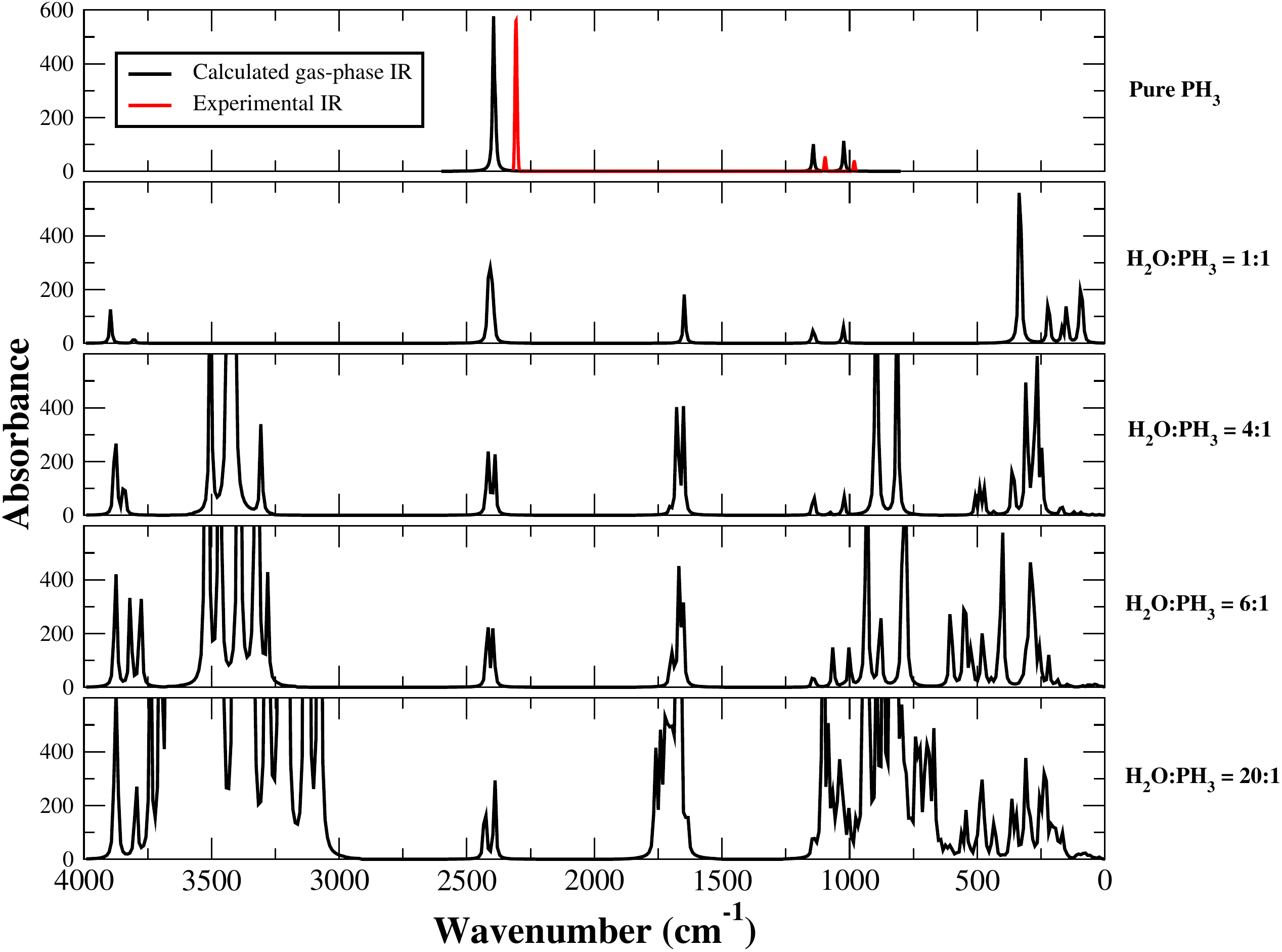}
\caption{Gas-phase IR absorption spectra of PH$_3$ with increasing concentration of H$_2$O \citep{sil21}.}
\label{fig:IR-H2O}
\end{figure}

Figure \ref{fig:IR-imp} shows the IR feature of PH$_3$ in the presence of various impurities. H$_2$O molecules would cover a significant portion of the interstellar ice mantles in the dense interstellar region. Some other species, such as CO, CO$_2$, CH$_4$, NH$_3$, etc. \citep{boog15}, can also constitute a sizable portion of the grain mantle. These molecules can influence the band profile of PH$_3$. We notice that the stretching of PH$_3$ around 2400 cm$^{-1}$ is getting more robust in the presence of  CO$_2$. H$_2$S and SO$_2$ are among the main components of Venusian atmospheres \citep{grea20,sous20}. It is important that a fundamental mode of SO$_2$ coincides with the bending-scissoring modes of PH$_3$ around $\sim 1000-1100$ cm$^{-1}$, which can blend the PH$_3$ transitions. Since H$_2$O is a major component of the interstellar dense cloud region, we check the effect of increasing concentration of H$_2$O on PH$_3$ separately in Figure \ref{fig:IR-H2O}. We notice that with increasing H$_2$O concentration, PH$_3$ fundamental bands are highly affected.

\begin{landscape}
\begin{table}
\scriptsize
\caption{Experimental and calculated IR data for PH$_3$ \citep{sil21}. \label{tab:IR}}
\vskip 0.2cm
\begin{tabular}{ccccccccc}
\hline
{\bf Assignment} & \multicolumn{5}{c}{\bf Frequency (cm$^{-1}$)} & \multicolumn{3}{c}{\bf Absorption Coefficients (cm molecule$^{-1}$)} \\
 & {\bf Experimental$^a$} &\multicolumn{2}{c}{\bf Calculated$^b$} & \multicolumn{2}{c}{\bf Calculated$^b$ $\times$ 0.967} & {\bf Experimental$^a$} & \multicolumn{2}{c}{\bf Calculated$^b$} \\
 & & {\bf Harmonic} & {\bf Anharmonic} & {\bf Harmonic} & {\bf Anharmonic} & & {\bf Harmonic} & {\bf Anharmonic} \\
 \hline
\multicolumn{9}{c}{Fundamental Bands} \\
\hline
$\nu_2$ (bending) & 982 & 1007.267 & 990.267 & 974.027 & 957.588 & $0.51\times10^{-18}$ & $5.40\times10^{-18}$ & $5.11\times10^{-18}$ \\
$\nu_4$ (scissoring) & 1096 & 1134.859 & 1110.02 & 1097.409 & 1073.389 & $0.71\times10^{-18}$ & $2.78\times10^{-18}$ & $2.55\times10^{-18}$\\
scissoring & & 1134.967 & 1103.681 & 1097.513 & 1067.260 & & $2.78\times10^{-18}$ & $2.58\times10^{-18}$\\
$\nu_1$ (symmetric stretching) & 2303 & 2396.887 & 2296.619 & 2317.790 & 2220.831 & $2.4\times10^{-18}$ & $7.82\times10^{-18}$ & $7.86\times10^{-18}$ \\
$\nu_3$ (asymmetric stretching) & 2316 & 2401.893 & 2303.377 & 2322.630 & 2227.366 & $8.4\times10^{-18}$ & $1.38\times10^{-17}$ & $1.47\times10^{-17}$ \\
asymmetric stretching & & 2405.153 & 2283.634 & 2325.783 & 2208.274 & & $1.34\times10^{-17}$ & $1.45\times10^{-17}$ \\
\hline
\multicolumn{9}{c}{Overtones} \\
\hline
2$\nu_4$ & 2195 & & 2213.024 & & 2139.994 & & & $2.98\times10^{-19}$ \\
3$\nu_2$ & 2905 & & & & & & & \\
2$\nu_1$ & 4536 & & 4540.981 & & 4391.129 & & & $1.35\times10^{-22}$ \\
\hline
\multicolumn{9}{c}{Combination Bands} \\
\hline
$\nu_2+\nu_4$ & 2067, 2083 & & 2093.256 & & 2024.178 & & & $2.86\times10^{-20}$ \\
$\nu_1/\nu_3+\nu_L$ & 2376, 2426, 2461 & & & & & & & \\
$\nu_1+\nu_2$ & 3288 & & 3280.701 & & 3172.438 & & & $6.50\times10^{-22}$ \\
$\nu_1+\nu_4$ & 3392 & & 3391.715 & & 3279.788 & & & $7.12\times10^{-20}$ \\
$\nu_3+\nu_4$ & 3405 & & 3392.344 & & 3280.397 & & & $5.67\times10^{-20}$ \\
$\nu_1+\nu_3$ & 4621 & & 4523.242 & & 4373.975 & & & $1.35\times10^{-21}$ \\
\hline
\end{tabular} \\
\vskip 0.2cm
{\bf Note:} \\
$^a$ \cite{turn15}. \\
$^b$ This work.
\end{table}
\end{landscape}

\clearpage
\subsection{Summary}
In this work, the abundance of the P-bearing species is examined under various interstellar circumstances.
PH$_3$ abundance is low in the diffuse cloud region and PDR, whereas a detectable amount of PH$_3$ could be produced in the hot-core region.
The abundance of PH$_3$ could be significantly affected by the destruction of H and OH. Without the inclusion of these pathways, any inference on the abundance of PH$_3$ in astrophysical environments would be inaccurate.
The BE of PH$_3$ with water is found to be lower than that of NH$_3$.
A good agreement between the calculated IR wavenumbers of PH$_3$ and the experimental feature of \cite{turn15} is seen. The stretching and bending-scissoring modes of PH$_3$ could be affected by the CO$_2$ and SO$_2$, respectively, in the ice. PH$_3$ fundamental bands are highly influenced by increasing H$_2$O concentration in the gas phase.
We map the PO/PN abundance ratio, which is mainly governed by the reaction $\rm{PO + N \rightarrow PN + O}$. A high abundance of atomic N in the diffuse clouds and PDRs compared to hot-cores yields PO/PN $<1$ for the diffuse cloud region and $<<1$ for the PDR. For the late warm-up to the post-warm-up part of the hot-core, we obtain PO/PN $\ge 1$.
We notice PO/PN $\ge 1$ for the late warm-up to the initial post-warm-up stage for the hot-corino case, whereas it is $\le 1$ for the late post-warm-up stage.

%% file: chap5.tex
\chapter{Summary, Conclusions and Future Plans} \label{chap:conclusions}
In this thesis, the synthesis of various simple and complex molecules is studied under different astrophysical conditions.
The spectral properties of several species and the band profiles of interstellar ices in the presence of impurities are examined.
Quantum chemical calculations using different theories are carried out to investigate various reactions, BEs, and various spectroscopic parameters of chemical species.
Various physical conditions are prepared by varying density, temperature, and timescale of molecular clouds to study the time-dependent and depth-dependent chemical evolution of different exotic, simple, and complex species.
The possibility of detecting several species in star-forming regions is discussed based on simple radiative transfer modeling. \\

The principal findings of this thesis are summarized below:

\section{Summary and Conclusions}

\begin{itemize}

\item Quantum-chemical calculations are performed for estimating BEs of different interstellar relevant species considering various substrates like water, methanol, carbon monoxide, hydrogen molecule.
 
\item An exciting trend is noticed that BE of $\rm{H_2}$ molecule always remains higher than H atom considering benzene, silica, and water cluster as substrates.

\item The BEs of $100$ interstellar species with the water tetramer as substrate are provided. The calculated BE values tend to converge to the experimental values with increasing cluster sizes of water molecules.

\item The BEs of $95$ interstellar species with $\rm{H_2}$ monomer substrate are provided.
Almost $\sim 10$ times lower BE values are found than those considering $\rm{H_2O}$ tetramer as a substrate.

\item These new BE values have significant consequences, such as locating the so-called snow lines of protoplanetary disks or obtaining the observed abundances toward hot cores and corinos or prestellar cores. 

 \item A series of laboratory experiments and an extensive computational investigation are carried out to explain the effect of different amounts of representative impurities on the band strengths and absorption band features of interstellar ice. Based on the observational, experimental, and theoretical results, the most abundant contaminants are chosen to study their effects on four fundamental vibrational bands of pure water ice. 

\item The fate of noble gas hydride and hydroxyl cations are studied in the diffuse ISM and
the Crab filamentary region. A wide range of parameter space and realistic chemical networks are involved in explaining the observational aspects suitably. Spectroscopic constants are calculated quantum chemically for the noble gas hydroxyl cations.

\item Detectability of trans-ethylamine and (1Z)-1-propanimine
as the most probable candidates from $\rm{C_2H_7N}$ and $\rm{C_3H_7N}$ aldimines and amines
isomeric groups respectively are discussed based on
the enthalpy of formation, chemical abundances, rotational transition along with the
radiative transfer modeling. It is to be noted that interstellar chemistry is not controlled by thermodynamics alone. Instead, it depends on various kinetic parameters.
We provide the spectroscopic parameters and the dipole moments that could provide guidance for future
searches of these molecular species.

\item Observed methylamine/ethylamine ratio is compared with that obtained from the
chemical model in dark cloud condition.
A good agreement is found between interstellar and cometary chemical compositions.
However, a more in-depth study is required to understand whether the cometary chemical compositions are inherited from dense cloud conditions via protoplanetary disk or not. 

\item Three pre-biotic molecules, namely, HNCO, NH$_2$CHO, and CH$_3$NCO containing peptide-like
bonds are investigated in a hot molecular core G10.47+0.03 and their chemical linkage is proposed.  

\item The abundances of important P-bearing species are modeled under various interstellar circumstances. For example, the PH$_3$ abundance is low in the diffuse cloud region and PDR, whereas a detectable amount of PH$_3$ can be produced in the hot core region.
The abundance of PH$_3$ can be significantly affected by the destruction of H and OH. Without the inclusion of these, any inference on the abundance of PH$_3$ in astrophysical environments would be inaccurate.

\item The BE of PH$_3$ with water is found to be lower as compared to that of NH$_3$.
A good agreement between the calculated IR wavenumbers of PH$_3$ and the experimental feature of \cite{turn15} is noticed. Thus, the stretching and bending-scissoring modes of PH$_3$ can be affected by the CO$_2$ and SO$_2$, respectively, in the ice. Furthermore, PH$_3$ fundamental bands are highly influenced by increasing H$_2$O concentration in the gas-phase.

\item The PO/PN abundance ratio is mapped in different astrophysical environments and compared with the earlier studies. The ratio is found to be mainly governed by the reaction $\rm{PO + N \rightarrow PN + O}$. A high abundance of atomic N in the diffuse clouds and PDRs compared to hot cores yields PO/PN $<1$ for the diffuse cloud region and $<<1$ for the PDR. For the late warm-up to the post-warm-up region of the hot core, we obtain PO/PN $\ge 1$.  For the hot corino case, we notice PO/PN $\ge 1$ for the late warm-up to the initial post-warm-up stage, whereas it is $\le 1$ for the late post-warm-up stage. So, PO/PN ratio can be used as a tracer of different evolutionary stages of star formation.

\end{itemize}

\section{Future Research Plans}

Presently, I have started work related to the properties of complex interstellar species along
with some noble-gas-related molecular ions and metallic compounds exposed to intense interstellar radiation environments (e.g., PDR, XDR, H\,{\sc ii} regions, plus the transition to dense molecular regions from diffuse regions). For this purpose, I am using the spectral synthesis code \textsc{Cloudy} \citep{ferl17}. I want to use this code to understand the more realistic physical and chemical properties for the particular environment.

Interstellar gas at low densities and low temperatures, of cosmic composition, left on its own will not generate extensive chemistry to produce useful tracer molecules. Instead, the chemistry
must be driven by starlight, cosmic rays, grain catalysis, and gas dynamics. For example, diffuse clouds are readily penetrated by starlight, which is largely excluded from molecular
clouds. Still, cosmic rays are almost unattenuated as they pass through such regions, and they drive much of the chemistry. Grain surfaces become essential drivers of complex chemistry in very dense regions of massive star formation. The gas dynamics in
regions such as stellar jets can inject energy and drive characteristic chemistry at the interface between the jet and the ambient gas. An interstellar shock arises when an external
event – such as a collision of two interstellar clouds or the impact of a rapidly expanding
H\,{\sc ii} region or SNR on a nearby cloud of cold neutral gas – drives a
perturbation faster than the local sound speed.

Consequently, the kinetic energy of bulk motion is converted to internal thermal energy, and the shocked gas is heated and
compressed. Such dynamic events are common in interstellar and circumstellar media and maybe detected spectroscopically. I am very much interested in carrying forward this type of investigation.

I want to model the properties of and the emission from COMs,
star formation processes, shocked region, and the molecular emission from cometary knots in planetary nebulae (PNe). PNe is the end-stage of the life cycle of low- and intermediate-mass stars,
consisting of the gas and dust ejected during previous stages of stellar evolution, 
surrounding a hot central star. My work would also focus on the modeling of the thermal emission from dust in supernova ejecta.
The large amounts of dust observed in some high-redshift galaxies may have been produced by supernovae (SNe) from massive stars.
In addition, core-collapse supernovae (CCSNe) play a vital role in the evolution of the ISM of galaxies.

%% file: appA-chap3.tex
\chapter{X-ray ionization} \label{chap:xray_ionization}

\section{Direct X-ray ionization}

In Table \ref{table:reaction}, we have pointed out the direct X-ray ionization rates in reaction numbers 25-26 for Ar, 26-27 for Ne, and
17-18 for He. Rate constants are computed using the method discussed in the following. \\

We used the direct (or primary) ionization rate of species $i$ at a certain depth $z$ into the filament as:
\begin{equation} \label{eqn:A1}
\rm
 \zeta_{XR}=\zeta_{i,prim}={\int_{E_{min}}^{E_{max}}} \sigma_i(E) \frac{F(E,z)}{E}dE \ s^{-1}\ ,
\end{equation}
where the integration bound is the spectral range of the emitted energy ($[E_{min},\ E_{max}]=[1,10]$ keV \citep{meij05} for the entire X-ray rate calculations). 
The ionization cross section $\sigma_i(E)$ at energy $E$ is calculated by using
the eqn. \ref{eqn:A2} and \ref{eqn:A3} and the parameters provided in Table \ref{table:param}. 
\cite{vern95} used a fitting procedure proposed by \cite{kamr83} for partial photoionization cross section
$\sigma_{nl}(E)$ for different atoms and ions:
\begin{equation} \label{eqn:A2}
 \sigma_i(E)=\sigma_{nl}(E) = \sigma_0F(y), \ y=E/E_0,
\end{equation}
\begin{equation} \label{eqn:A3}
 F(y)=[(y-1)^2+y_w^2]y^{-Q}\Big(1+\sqrt{\frac{y}{y_a}}\Big)^{-P}, \ Q=5.5+l-0.5P,
\end{equation}
where $n$ is the principal quantum number of the shell, $l=0, 1, 2$ (or s, p, d) is the subshell orbital
quantum number, $E$ is the photon energy in eV, $\sigma_0$=$\sigma_0 (nl, Z, N)$, $E_0=E_0(nl, Z, N)$,
$y_w,\ y_a$, and $P$ are the fitting parameters given in Table \ref{table:param} ($Z$ and $N$ are the atomic number and 
number of electrons, respectively). \cite{vern95} noticed that $F(y)$ is a ``nearly universal'' function 
for all species $(Z,\ N)$ at a fixed shell $nl$.

The flux $F(E,z)$ in Equation (\ref{eqn:A1}) at depth $z$ into the filament is given by
\begin{equation} \label{eqn:A4}
 F(E,z)=F(E,z=0)exp\big(-\sigma_{pa}(E)N_H\big),
\end{equation}
where $N_H\sim4.77\times10^{21}$ cm$^{-2}$ is considered as the total column density of hydrogen nuclei 
and $F(E,z=0)=0.35$ erg cm$^{-2}$ s$^{-1}$ is considered as the flux at the surface of the cloud.
The photoelectric absorption cross section per hydrogen nucleus,
$\sigma_{pa}$ used in Equation (\ref{eqn:A4}) is given by
\begin{equation} \label{eqn:A5}
 \sigma_{pa}(E)=\sum_iA_i(total)\sigma_i(E),
\end{equation}
where $A_i(total)$ is the total (gas and dust) elemental abundance used.

\section{Secondary X-ray ionization}
Part of the kinetic energy of fast photoelectrons is lost by ionizations. 
These secondary ionizations are far more important for H, $\rm{H_2}$, and He than direct ionization.
The energy carried away by the fast photoelectrons and Auger electrons is very efficient in ionizing the other species.
For example, these electrons can readily ionize H, He, and H$_2$ and decay back to ground state
by the removal of UV photons. These photons can trigger the induced chemistry and are very important for the
chemical network. The secondary ionization rate per hydrogen molecule at depth z into the filament can be calculated using
\begin{equation} \label{eqn:A6}
 \zeta_{H_2,XRPHOT}=\zeta_{i,sec}=\int_{E_{min}}^{E_{max}} \sigma_{pa}(E)F(E,z)\frac{E}{Wx(H_2)}dE \ s^{-1},
\end{equation}
where $x(H_2)$ is the fractional abundance of $\rm{H_2}$ with respect to total hydrogen nuclei
and $W$ is the mean energy per ion pair. 
For our calculations, we considered $\rm{x(H_2)\sim 2 \times 10^{-4}}$, which means that most of the hydrogen is in atomic form.
\cite{dalg99} calculated $W$ for pure ionized $\rm{H-He}$ and $\rm{H_2-He}$ mixtures
for $E$ between 30 eV and 1 keV and parameterized $W$ as
\begin{equation} \label{eqn:A7}
 W=W_0(1+Cx^\alpha),
\end{equation}
where $x = 0.1$ is considered as the ionization fraction and $W_0$ is the value for neutral gas. $W_0,\ C$, and $\alpha$ are given in Table 4 
of \cite{dalg99}. We took those values ($W_0=48.6$ eV, $C=9.13$, and $\alpha$=0.807) only for pure He
 gas for 1 keV.  
Following \cite{meij05}, we integrated over the range $1-10$ keV and $W$ goes to a limiting value ($42.69$ eV). We considered the parameters for the 1 keV electron to determine the electron energy
deposition, because these parameters do not change for higher energies. 
The X-ray photoionization rate then simplifies to,
\begin{equation} \label{eqn:A8}
\zeta_{H_2,XRPHOT}=\zeta_{i,sec}=\frac{1 keV}{W(1 keV)x(H_2)} {\int_{E_{min}}^{E_{max}}} \sigma_{pa}(E)F(E,z)dE \ s^{-1} \ .
\end{equation}
The photon energy absorbed per hydrogen nucleus $H_X$ is given by
\begin{equation} \label{eqn:A9}
 H_X={\int_{E_{min}}^{E_{max}}}\sigma_{pa}(E)F(E,z)dE. 
\end{equation}
Hence, the X-ray photoionization rate is given by
\begin{equation} \label{eqn:A10}
  \zeta_{H_2,XRPHOT}=\zeta_{i,sec}=\frac{1 keV}{W(1 keV)x(H_2)}H_X \ s^{-1}.
\end{equation}
Following \cite{prie17}, we multiplied $\rm{\zeta_{H_2,XRPHOT}}$ by 
$\frac{0.8}{1-\omega}$, where $\omega$ is the grain albedo ($\sim 0.5$). 

\section{Electron-impact X-ray ionization}
The electron-impact ionization rate ($\zeta_{XRSEC}$) of other atoms or molecules can be calculated as a first approximation by
\begin{equation} \label{eqn:A11}
 \zeta_{XRSEC}=\zeta_{H_2,XRPHOT} \times R_\sigma,
\end{equation}
where $R_{\sigma}$ is the  
ratio of electron-impact cross sections of that species to H$_2$ at a specific energy \citep{stau05}.
For simplicity, here we assumed $\rm{\zeta_{H2,XRPHOT}=\zeta_{H,XRPHOT}}$.
Following \cite{lenn88}, we determined the rate coefficients $\rm{\langle\sigma v\rangle}$ (cross sections at a given
energy multiplied by electron velocity $v$ at the same energy,
evaluated over a Maxwellian velocity distribution) given by
\begin{equation} \label{eqn:A12}
 \langle\sigma v\rangle=\Big(\frac{8kT}{\pi m}\Big)^{1/2}{\int_{I/kT}^{\infty}}\sigma(E)\Big(\frac{E}{kT}\Big)exp\Big(\frac{-E}{kT}\Big)d\Big(\frac{E}{kT}\Big),
\end{equation}
where $m$ is the electron mass.
For the temperature range $I/10 \leq kT \leq 10I$, they fitted the rate coefficient with the following functional form,
\begin{equation} \label{eqn:A13}
 \langle\sigma v\rangle=exp\Big(\frac{-I}{kT}\Big)\Big(\frac{kT}{I}\Big)^{1/2}\sum_{n=0}^5a_n\Big[log_{10}\Big(\frac{kT}{I}\Big)\Big]^n,
\end{equation}
and for $kT > 10I$, they used the formula
\begin{equation} \label{eqn:A14}
 \langle\sigma v\rangle=\Big(\frac{kT}{I}\Big)^{-1/2}\Big[\alpha ln\Big(\frac{kT}{I}\Big)+\sum_{n=0}^2\beta_n\Big(\frac{I}{kT}\Big)^n\Big].
\end{equation}
Following \cite{lenn88}, the coefficients $a_0$, ..., $a_5$ and $\alpha$, $\beta_0$, $\beta_1$, and $\beta_2$ are given in Table \ref{table:rate1}. For $T$ in K, $I$ in eV, and $k = 0.8617 \times 10^{-4}$ eV K$^{-1}$, these coefficients 
provide the
rate $\langle\sigma v\rangle$ in cm$^3$ s$^{-1}$. 
Using Equations (\ref{eqn:A12}) and (\ref{eqn:A13}), we have determined $\langle\sigma v\rangle$ for Ar, Ne, and He. The obtained values are shown in the
last row of Table \ref{table:rate1} and the calculated values of $R_{\sigma}$ are 5.53, 1.84, and 0.84 
for Ar, Ne, and He, respectively. All of the calculated values of different
X-ray ionization rates of argon, neon, and helium are provided in Table \ref{table:rate2}.

\begin{table}
\scriptsize
\centering
\caption{The parameters taken from \cite{vern95} for calculating ionization cross sections $\sigma_i$(E) \citep{das20}.
\label{table:param}}
\vskip 0.2 cm 
\begin{tabular}{cccccc}
\hline
{\bf  Species} & {\bf  $E_0$ (eV)} & {\bf $\sigma_0$ (cm$^2$)} & {\bf $y_a$} & {\bf $P$} \\
\hline
He\,{\sc i} & $0.2024 \times 10^1$ & $0.2578 \times 10^{-14}$ & $0.9648 \times 10^1$ & $0.6218 \times 10^1$ \\
Ne\,{\sc i} & $0.3144 \times 10^3$ & $0.1664\times 10^{-16}$ & $0.2042 \times 10^6$ & $0.8450 \times 10^0$ \\
Ar\,{\sc i} & $0.1135 \times 10^{4}$ & $0.4280 \times 10^{-17}$ & $0.3285 \times 10^{8}$ & $0.7631 \times 10^0$ \\
\hline
\end{tabular}
\end{table}

\begin{table}
\scriptsize
\centering
\caption{The parameters taken from \cite{lenn88} to calculate the rate coefficients $\langle\sigma v\rangle$ \citep{das20}. \label{table:rate1}}
\vskip 0.2 cm 
\begin{tabular}{ccccc}
\hline
{\bf Parameters} & \multicolumn{4}{c}{\bf Species} \\
\cline{2-5}
{\bf (cm$^3$s$^{-1}$) } & {\bf H\,{\sc i}} & {\bf He\,{\sc i}} & {\bf  Ne\,{\sc i}} & {\bf  Ar\,{\sc i}} \\
\hline
a$_0$ & $2.3743 \times 10^{-08}$ &  $1.4999 \times 10^{-08}$ & $2.5262 \times 10^{-08}$ & $9.4727 \times 10^{-08}$ \\
a$_1$ & $-3.6867 \times 10^{-09}$ & $5.6657 \times 10^{-10}$ &  $1.6088 \times 10^{-09}$ & $1.4910 \times 10^{-09}$ \\
a$_2$&  $-1.0366 \times 10^{-08}$ & $-6.0822 \times 10^{-09}$ & $1.5446 \times 10^{-08}$ & $-5.9294 \times 10^{-08}$ \\
a$_3$& $-3.8010 \times 10^{-09}$ & $-3.5594 \times 10^{-09}$ &  $-3.5149 \times 10^{-08}$ & $1.7977 \times 10^{08}$ \\
a$_4$& $3.4159 \times 10^{-09}$ & $1.5529 \times 10^{-09}$ & $-1.0676 \times 10^{-09}$ & $1.2962 \times 10^{-08}$ \\
a$_5$& $1.6834 \times 10^{-09}$ & $1.3207 \times 10^{-09}$ & $1.2656 \times 10^{-08}$ & $-9.7203 \times 10^{-09}$ \\
$\alpha$ &  $2.4617 \times 10^{-08}$ &  $3.1373 \times 10^{-08}$ & $1.4653 \times 10^{-07}$ &  $4.2289 \times 10^{-07}$ \\
$\beta_0$ & $9.5987 \times 10^{-08}$ & $4.7094 \times 10^{-08}$ & $-1.8777 \times 10^{-07}$ &  $-5.8297 \times 10^{-07}$ \\
$\beta_1$ & $-9.2464 \times 10^{-07}$ & $-7.7361 \times 10^{-07}$ &  $1.5661 \times 10^{-08}$ &  $1.2344 \times 10^{-06}$ \\
$\beta_2$ & $3.9974 \times 10^{-06}$ & $3.7367 \times 10^{-06}$ & $1.9135 \times 10^{-06}$ & $-7.2826 \times 10^{-07}$ \\
\hline
$\langle\sigma v\rangle$ & $3.00 \times 10^{-08}$ & $2.53 \times 10^{-08}$ & $5.51 \times 10^{-08}$ & $1.66 \times 10^{-07}$ \\
\hline
\end{tabular}
\end{table}

\begin{table}
\scriptsize
\centering
\caption{Calculated values of X-ray ionization rates \citep{das20}. \label{table:rate2}}
\vskip 0.2 cm 
\begin{tabular}{ccccc}
\hline
{\bf  Species} & {\bf $\rm{\zeta_{XR}}$ (s$^{-1}$)} & {\bf $\rm{\zeta_{XRPHOT}}$ (s$^{-1}$)} & {\bf $\rm{\zeta_{XRSEC}}$ (s$^{-1}$)} \\
\hline
$^{36}$Ar & $3.85 \times 10^{-13}$ & $1.67 \times 10^{-10}$ & $5.79 \times 10^{-10}$ \\
$^{38}$Ar & $1.53 \times 10^{-12}$ & $3.31 \times 10^{-10}$ & $1.14 \times 10^{-9}$ \\
$^{40}$Ar & $1.35 \times 10^{-11}$ & $4.57 \times 10^{-11}$ & $1.58 \times 10^{-10}$ \\
$^{20}$Ne & $2.47 \times 10^{-17}$ & $8.28 \times 10^{-15}$ & $9.52 \times 10^{-15}$ \\
$^{22}$Ne & $9.41 \times 10^{-15}$ & $7.27 \times 10^{-13}$ & $8.36 \times 10^{-13}$\\
He & $1.31 \times 10^{-19}$ & $1.67 \times 10^{-14}$ & $8.76 \times 10^{-15}$ \\
\hline
\end{tabular}
\end{table}

%% file: glossary.tex
\chapter{Glossary}

A convenient list of covering units, constants, terms
and acronyms used frequently in this thesis or in cited literature are provided here.

\section{Units and constants}

\begin{quote}
\hskip -2.5 cm
\begin{tabular}{llll}
\hline
\hline
{\bf Symbol} & {\bf Description} & {\bf SI Units} & {\bf CGS Units}\\
\hline
\AA & Angstrom & $\rm{10^{-10}}$ m & $\rm{10^{-8}}$ cm  \\
$\mu$m & Micron & $10^{-6}$ m & $10^{-4}$ cm \\
$G$ &Gravitational constant & $\rm{6.673\times10^{-11} \ N \ m^{2} \ kg^{-2}}$&  $\rm{6.673\times10^{-8} \ dyn \ cm^{2} \ gm^{-2}}$\\
$k_B$ & Boltzmann constant &$\rm{1.3807\times10^{-23}}$ J/K&$\rm{1.3807\times10^{-16}}$ erg/K\\
$h$ & Planck's constant & $\rm{6.6262\times10^{-34}}$ J s&$\rm{6.6262\times10^{-27}}$ erg s\\
$c$ & Speed of light &$\rm{2.997925\times10^{8}\ m \ s^{-1}}$ &$\rm{2.997925\times10^{10} \ cm \ s^{-1}}$\\
eV & Electronvolt & $ 1.602 \times 10^{-19}$ J & $ 1.602 \times 10^{-12}$ erg \\
D & Debye & $3.336\times10^{-30}$ C m & $10^{-18}$ esu cm \\
$m_e$ & Electron mass &$\rm{9.10956\times10^{-31}}$ kg&$\rm{9.10956\times10^{-28}}$ gm\\
$m_p$ & Proton mass &$\rm{1.6726231\times10^{-27}}$ kg&$\rm{1.6726231\times10^{-24}}$ gm\\
$m_H$ & Hydrogen mass &$\rm{1.673534\times10^{-27}}$ kg&$\rm{1.673534\times10^{-24}}$ gm\\
$u$ & Atomic mass unit &$\rm{1.6605402\times10^{-27}}$ kg&$\rm{1.6605402\times10^{-24}}$ gm\\
$E_{\bigoplus}$ & Earth radius & $\rm{6.378\times10^{6}}$ m&$\rm{6.378\times10^{8}}$ cm\\
AU & Astronomical unit & $\rm{1.4959\times10^{11}}$ m&$\rm{1.4959\times10^{13}}$ cm\\
$R_{\odot}$ & Solar radius &$\rm{6.9599\times10^{8}}$ m&$\rm{6.9599\times10^{10}}$ cm\\
ly & Light year &$\rm{9.463\times10^{15}}$ m&$\rm{9.463\times10^{17}}$cm\\
pc & Parsec &$\rm{3.085678\times10^{16}}$ m&$\rm{3.085678\times10^{18}}$ cm\\
$L_{\odot}$ & Solar luminosity & $\rm{3.826\times10^{26} \ J \ s^{-1}}$&$\rm{3.826\times10^{33} \ erg \ s^{-1}}$\\
$M_{\bigoplus}$ & Mass of the Earth & $\rm{5.977\times10^{24}}$ kg&$\rm{5.977\times10^{27}}$ gm\\
$M_{\odot}$ & Solar Mass &$\rm{1.989\times10^{30}}\ kg$&$\rm{1.989\times10^{33}}\ gm$\\
Jy & Jansky & $\rm{1.00\times10^{-26} \ W \ m^{-2} \ Hz^{-1}}$& $\rm{1.00\times10^{-23} \ erg \ sec^{-1}\ cm^{-2} \ Hz^{-1}}$\\
\hline
\hline
\end{tabular} 
\end{quote}

\section{Acronyms}
\begin{quote}
\begin {tabular}{l l}
{\bf 67P/C-G}&{        67P/Churyumov-Gerasimenko} \\
{\bf $A_V$}&{       Visual extinction, measured in magnitudes (mag)} \\
{\bf $ab\ initio$}&{       Latin word for ``from the beginning''} \\
{\bf ALMA  }&{         Atacama Large Millimeter/submillimeter Array}\\
{\bf APEX  }&{         Atacama Pathfinder Experiment}\\ 
{\bf Ariel }&{         Atmospheric Remote-sensing Infrared Exoplanet Large-survey} \\
{\bf ASTE    }&{         Atacama Submillimeter Telescope Experiment}\\
{\bf ASW    }&{         Amorphous Solid Water}\\
{\bf BE    }&{         Binding Enegry}\\
{\bf BSSE    }&{         Basis Set Superposition Error}\\
{\bf CCSD(T) }&{         Coupled Cluster Single-Double and perturbative Triple excitation}\\
{\bf CDMS  }&{         The Cologne Database for Molecular Spectroscopy}\\
{\bf CMB   }&{         Cosmic Microwave Background}\\
{\bf CMMC   }&{         Chemical Model for Molecular Cloud}\\
{\bf CP   }&{         Counterpoise} \\
{\bf COMs  }&{         Complex Organic Molecules}\\
{\bf CR    }&{         Cosmic Rays}\\
{\bf CRESU }&{         $\rm{Cin\acute{e}tique\ de\ R\acute{e}action\ en\ Ecoulement\ Supersonique\ Uniforme}$ (in French)} \\
{\bf CSE   }&{         Circumstellar Envelop}\\
{\bf CT    }&{         Classical Trajectory}\\
{\bf DFT   }&{         Density Functional Theory} \\
{\bf DFMS   }&{         Double Focusing Mass Spectrometer} \\
{\bf ESA   }&{         European Space Agency}\\
{\bf FTIR  }&{         Fourier Transform Infrared}\\
{\bf ELT  }&{          Extremely Large Telescope}\\
{\bf ESD   }&{         Electron-Stimulated Desorption}\\
{\bf FTIR   }&{        Fourier Transform InfraRed}\\
{\bf FWHM  }&{         Full Width at Half Maximum}\\
{\bf GREAT   }&{        German REceiver for Astronomy at Terahertz Frequencies}\\
{\bf HF   }&{         Hartree-Fock}\\
{\bf HIFI  }&{         Heterodyne Instrument for the Far Infrared}\\
{\bf HITRAN}&{         High-resolution Transmission Molecular Absorption Database}\\
{\bf HMC   }&{         Hot Molecular Core}\\
{\bf HSO   }&{         Herschel Space Observatory}\\
{\bf HST  }&{          Hubble Space Telescope}\\
{\bf IR   }&{         Infrared}\\
{\bf IRAM  }&{         Institute for Radio Astronomy in the Millimeter Range}\\
{\bf IRTF  }&{         Infrared Telescope Facility}\\ 
{\bf ISM   }&{         Interstellar Medium}\\
{\bf ISO   }&{         Infrared Space Observatory}\\
{\bf ISRF  }&{         Interstellar Radiation Field}\\ 
{\bf ICSP  }&{         Indian Centre for Space Physics}\\
\end {tabular}
\end{quote}

\begin{table}
\centering
\vskip -1.0cm
\begin{tabular}{l l}
{\bf IRC   }&{         Intrinsic Reaction Coordinate}\\
{\bf JPL   }&{         Jet Propulsion Laboratory}\\
{\bf JWST  }&{         James Webb Space Telescope}\\
{\bf KIDA  }&{         KInetic Database for Astrochemistry}\\
{\bf KMC    }&{        Kinetic Monte Carlo}\\
{\bf LAMDA }&{         Leiden Atomic and Molecular Database}\\
{\bf LWS  }&{         Long Wavelength Spectrometer}\\
{\bf LTE  }&{         Local Thermodynamic Equillibrium}\\
{\bf MC    }&{         Monte Carlo}\\
{\bf MCT    }&{         Mercury Cadmium Telluride}\\
{\bf MP    }&{         M\o ller-Plesset perturbation}\\
{\bf MD    }&{         Molecular Dynamics}\\
{\bf NASA  }&{         National Aeronautics and Space Administration}\\ 
{\bf NOEMA }&{         Northern Extended Millimeter Array}\\
{\bf NGC   }&{         New General Catalog}\\
{\bf NIST  }&{         National Institute of Standards and Technology}\\
{\bf PAC  }&{         Portable Astrochemistry Chamber}\\
{\bf PACS  }&{         Photodetector Array Camera and Spectrometer}\\
{\bf PAH   }&{         Polycyclic Aromatic Hydrocarbons}\\
{\bf PCM   }&{         Polarizable Continuum Model}\\
{\bf PDR   }&{         Photon-dominated/Photodissociation Region}\\
{\bf QM/MM   }&{        Quantum Mechanics/Molecular Mechanics}\\
{\bf QST   }&{         Quadratic Synchronous Transit}\\
{\bf RAIRS }&{         Reflection Absorption Infra-Red Spectroscopy} \\
{\bf ROSINA   }&{      Rosetta Orbiter Spectrometer for Ion and Neutral Analysis}\\ 
{\bf SB   }&{         Surface Brightness}\\
{\bf SCRF  }&{         Self-consistent Reaction Field}\\
{\bf SED   }&{         Spectral Energy Distribution} \\
{\bf SIFT  }&{         Selected Ion Flow Tube} \\
{\bf SMA   }&{         Submillimeter Array}\\
{\bf SNR   }&{         Supernova remnant} \\
{\bf SOFIA }&{         The Stratospheric Observatory for Infrared Astronomy}\\
{\bf SPIRE }&{         Spectral and Photometric Imaging Receiver}\\
{\bf STQN }&{          Synchronous Transit-Guided Quasi-Newton}\\
{\bf TPD   }&{         Temperature-Programmed Desorption}\\
{\bf TS    }&{         Transition State}\\
{\bf TST   }&{         Transition-State Theory}\\
{\bf UHV   }&{         Ultra-High Vacuum}\\ 
{\bf UDfA  }&{         UMIST Database for Astrochemistry}\\
{\bf UV    }&{         Ultraviolet}\\
{\bf VLA }&{           Very Large Array} \\
{\bf VUV    }&{         Vacuum Ultaviolet}\\
{\bf XMM    }&{         X-ray Multi-Mirror Mission} \\
{\bf XRISM  }&{         X-Ray Imaging and Spectroscopy Mission}\\
{\bf YSO   }&{         Young Stellar Object}\\
{\bf ZPVE  }&{         Zero Point Vibrational Energy}\\
\end {tabular}
\end{table}

